\documentclass[preprint,prd,aps,showpacs,showkeys,nofootinbib]{revtex4}
\usepackage{graphicx}
\begin{document}

\title{GKZ-system of the 2-loop self energy with 4 propagators}

\author{
Tai-Fu Feng\footnote{email:fengtf@hbu.edu.cn(corresponding author)}$^{a,b,c,d,e}$,
Hai-Bin Zhang\footnote{email:hbzhang@hbu.edu.cn(corresponding author)}$^{a,b,c,d}$,
Yan-Qing Dong$^{a,b}$,
Yang Zhou$^{a,b}$}

\affiliation{$^a$Department of Physics, Hebei University, Baoding, 071002, China}
\affiliation{$^b$Hebei Key Laboratory of High-precision Computation and Application of
Quantum Field Theory, Baoding, 071002, China}
\affiliation{$^c$Research Center for Computational Physics of Hebei Province, Baoding, 071002, China}
\affiliation{$^d$Department of Physics, Guangxi University, Nanning, 530004, China}
\affiliation{$^e$Department of Physics, Chongqing University, Chongqing, 401331, China}

\begin{abstract}
Applying the system of linear partial differential equations derived from the Mellin-Barnes representation and the Miller transformation, we present the GKZ-system of the Feynman integral of the 2-loop self energy diagram with 4 propagators. The codimension of the derived GKZ-system equals the number of independent dimensionless ratios among the external momentum squared and virtual mass squared. In total 536 hypergeometric functions are obtained in the neighborhoods of the origin and infinity, in which 30 linearly independent hypergeometric functions whose convergent regions have nonempty intersection constitute a fundamental solution system in a proper subset of the whole parameter space.
\end{abstract}


\maketitle

\newpage
\tableofcontents

\newpage

\section{Introduction\label{sec1}}
\indent\indent
One of the main goals of high energy physics is to test the standard model (SM)
and to search for new physics (NP) beyond the SM \cite{CEPC-SPPC,ILC,HI-LHC} after the
discovery of the Higgs boson~\cite{CMS2012,ATLAS2012}.
Before the collision energy of the center of mass of the running collider
reaches the threshold energy of new physics, the SM theoretical
predictions of various physical observations should be accurately calculated~\cite{Heinrich2021}.
In order to predict the observables precisely, one should evaluate Feynman integrals
exactly in the spacetime dimension $D=4-2\varepsilon$~\cite{tHooft1979} at first.

Taking Feynman integrals as the generalized hypergeometric functions~\cite{Regge1967},
one finds that the $D-$module of a Feynman diagram~\cite{Nasrollahpoursamami2016} is isomorphic to
Gel'fand-Kapranov-Zelevinsky (GKZ) $D-$module~\cite{Gelfand1987,Gelfand1988,Gelfand1988a,Gelfand1989,Gelfand1990}.
Correspondingly the Feynman integral satisfies a system of
holonomic linear partial differential equations (PDEs)~\cite{Kashiwara1976} whose singularities
are determined by the Landau singularities.

Some progress has been made on the subject in the past decades.
Under the assumption of zero virtual masses, the Feynman integral of the 1-loop triangle
diagram is formulated as a linear combination of the fourth kind of Appell function~\cite{Davydychev1} whose
arguments are the dimensionless ratios among the external momentum squared, and is simplified
further as the linear combination of the Gauss function $_2F_1$ through the quadratic
transformation in the literature~\cite{Davydychev2000}. An algorithm to obtain the power
series in the external momentum of 2-loop self-energy diagrams with arbitrary masses of
the internal particles is examined in Ref.~\cite{Davydychev1993NPB}.
Taking some special assumptions on the parameter space, the authors of Ref.~\cite{Davydychev3}
obtain the multiple hypergeometric expressions of the scalar integral $C_{_0}$ through the
Mellin-Barnes representation. An algorithm to evaluate the scalar integral of
the 1-loop vertex-type Feynman diagram is given in Ref.~\cite{Davydychev1992JPA},
and the geometrical interpretation of the analytic expression of the scalar integral
of the 1-loop $N-$point diagram is also presented in Ref.~\cite{Davydychev2006}.
Certainly the expressions of the function $C_{_0}$ can also be derived from the expression of
the 1-loop massive $N-$point diagram~\cite{Davydychev1991JMP,Davydychev1992JMP}.
Furthermore the Feynman integrals of ladder diagrams with 3 or 4 external lines
can be evaluated through the Feynman parametric representation and the Mellin-Barnes contour
integrals~\cite{Davydychev1993}.
In fact the 1-loop 2-point function $B_{_0}$ can be formulated as a linear combination
of the Gauss function $_2F_1$ through the recurrence relations respecting the spacetime
dimension, the 1-loop 3-point function $C_{_0}$ similarly is presented as a linear
combination of the Appell function $F_{_1}$, and the 1-loop 4-point function $D_{_0}$
is given as a linear combination of the Lauricella function
$F_{_s}$~\cite{Tarasov2000,Tarasov2003}, respectively.
The expression of $C_{_0}$ is convenient
for analytic continuation and numerical evaluation because the continuation of $F_{_1}$
has been analyzed thoroughly. However, how to perform continuation
of the Lauricella function $F_{_s}$ outside its convergent domain is still a challenge.
Taking Feynman integrals as the generalized hypergeometric functions in dimension regularization,
the authors of Ref.~\cite{Kalmykov2009} analyze the Laurent expansion of these hypergeometric
functions around $D=4$. The differential-reduction algorithm to evaluate those hypergeometric
functions can be found in Refs.~\cite{Bytev2010,Kalmykov2011,Bytev2013,Bytev2015,Bytev2016}.
A system of linear partial differential equations (PDEs) is given through the Mellin-Barnes
representation~\cite{Kalmykov2012},
and some irreducible master integrals of the sunset and bubble Feynman diagrams
are explicitly evaluated through the Mellin-Barnes contour~\cite{Kalmykov2017}.
In addition, some problems related to the constructions of the $\varepsilon$
expansion of dimensionally regulated Feynman integrals are discussed in
Refs.~\cite{Davydychev2000a,Davydychev2001a}. Adopting the negative dimensional
integration method, Refs.~\cite{Suzuki2002,Suzuki2003a} present the complete massless
and massive 1-loop triangle diagrams results. Proof of the equivalence
between the Mellin-Barnes representation and the Feynman parametric representation
is given in Ref.~\cite{Suzuki2003b}, and the inverse binomial sums relating
to the $\varepsilon$ expansion of massive Feynman integrals are presented in
Ref.~\cite{Davydychev2004}. Those results are applied to calculate the
${\cal O}(\alpha\alpha_{_s})$ corrections to the relationship between the $\overline{MS}$
mass and the pole of top propagator in the SM~\cite{Jegerlehner2004}.

Some GKZ-systems of the Feynman integrals with codimension$=0,\;1$
are presented in Refs.~\cite{Cruz2019,Klausen2019}
through the Lee-Pomeransky parametric representation~\cite{Lee2013}.
To obtain hypergeometric series solutions with suitable independent variables,
one should compute the restricted $D$-module of the GKZ-system originating from
the Lee-Pomeransky representation on the corresponding hyperplane in the parameter
space~\cite{Oaku1997,Walther1999,Oaku2001}.
In addition, the GKZ-systems of some Feynman diagrams
can also be derived from their Mellin-Barnes representation~\cite{Feng2018,Feng2019,Feng2020}
through the Miller transformation~\cite{Miller68,Miller72}.
Ref.~\cite{Loebbert2020} explores the idea of bootstrap Feynman integrals using integrability,
the authors of Ref.~\cite{Klemm2020} apply the GKZ description of periods to solve the
$l$-loop banana amplitudes with their general mass dependence,  and completely clarify
the analytic structure of all banana amplitudes with arbitrary masses~\cite{Bonisch2021}.
The updated summary on some classical and modern aspects of hypergeometric
differential equations is given in literature~\cite{Reichelt2020}.
A Mathematica package for integrating families of Feynman integrals
order by order in the dimensional regulator from their GKZ-systems
is also given in literature~\cite{Hidding2021}.
A specialized integration algorithm for parametric Feynman integrals
with tame kinematics is also presented in literature~\cite{Borinsky2020}.
Actually the Cohen-Macaulay property of the Feynman integrals indicates
that the process of finding a series representation of these
integrals is fully algorithmic~\cite{Tellander2021}, and regular
singularities of the Feynman integrals are determined by its Landau discriminant~\cite{Mizera2021}.
Ref.~\cite{Arkani-Hamed2022} introduces a class of polytopes to analyze
the structure of UV and IR divergences of general Feynman integrals in the Schwinger parameter space.
Furthermore, Ref.~\cite{Chestnov2022} elaborates on the connection among the GKZ-systems, de
Rham theory for twisted cohomology groups, and the Pfaffian equations for Feynman integrals,
and Ref.~\cite{Ananthanarayan2022} presents a Mathematica package to find the linear transformations
for some classes of multivariable hypergeometric functions. In literature~\cite{Berends1994}
a class of $N$-loop massive scalar self-energy diagrams with $N+1$ propagators
is studied, and the new convergent series representation for the 2-loop sunset diagram
with three different propagator masses and external momentum is obtained in
Ref.~\cite{Ananthanarayan2019}.  The relationship between Feynman diagrams
and generalized hypergeometric functions
is reviewed in Ref.~\cite{Kalmykov2021}. Using the GKZ system and its relation
to D-module theory, Ref.~\cite{Munch2022} proposes a novel method to obtain differential
equations for master integrals. Adopting some assumptions on the virtual masses, Ref.~\cite{Gu2020}
investigates the analytical expression of a 3-loop vacuum integral.
A new methodology to perform the expansion of multi-variable hypergeometric
functions around the spacetime dimension $D=4$ is given in Ref.~\cite{Bera2022}.
Using an algorithm that extends the Griffiths-Dwork reduction for the case
of projective hypersurfaces with singularities, the authors of Ref.\cite{Lairez2022}
derive a system of Fuchsian linear differential equations with respect to kinematic parameters
for a large class of massive multi-loop Feynman integrals.

Basing on the Mellin-Barnes representation of the 2-loop self energy with 4 propagators,
we derive the GKZ-system of the diagram through the Miller transformation,
where the codimension of the GKZ-system equals the number of independent dimensionless
ratios among the external momentum squared and virtual mass squared. As
the Feynman integral is regarded as an analytic function of the square of external momentum,
the thresholds are the branch-points of the analytic function
in the complex $p^2$-plane~\cite{Eden1966}.
Applying the residue theorem, one can only construct the analytic
solutions of the kinematic regions where the external energy is lower than the minimum
threshold or higher than the maximum threshold. This is the fundamental reason
why the combinatorial information contained in the GKZ-system derived here can only
be used to obtain the hypergeometric solutions in the neighborhoods of $m_{_i}^2/p^2=0$ and
$p^2/m_{_i}^2=0$, respectively. Applying the alpha parametric representation in the Ref.~\cite{Feng2022},
we can embed the integral in some subvarieties of the Grassmannians $G_{_{5,11}}$,
and derive the GKZ-systems on those subvarieties.
Abundant combinatorial information carried by the GKZ-systems
on those subvarieties enables us to obtain the fundamental solution systems
in the neighborhoods of all regular singularities, especially to obtain
the fundamental solution systems when the absolute value of external momentum
is located in the range surrounded by two adjacent thresholds (pseudo-thresholds).
As the threshold coincides with the regular singularity of the Feynman integral,
we can apply this approach to derive the analytical
expression of the Feynman integral in the corresponding parameter space.
Generally we should investigate further whether or not the obtained
hypergeometric functions can be analytically continued to the threshold
hypersurface. Nevertheless, we can always apply the heavy mass and large
momentum expansion~\cite{V.A.Smirnov2012}
to approximate the expression of the Feynman integral on the threshold hypersurface.

The most general case of the 2-loop 2-point functions can be approximated
through the expansion of the external momentum squared $p^2$, where the coefficients
of the expansion can be expressed in terms of the 2-loop vacuum integrals~\cite{Davydychev1993NPB}.
For the 2-loop self energy which is obtained through inserting the self energy
into one virtual particle of the 1-loop self energy, the corresponding Feynman
integral can be reduced to the considered topology here by partial fractioning.
Using the Mellin-Barnes representation of the general 5-propagator topology with
a rung, we similarly derive the GKZ-system satisfied by the Feynman integral.
Unfortunately the codimension of the GKZ-system is larger than the number
of independent dimensionless ratios among the external momentum squared
and virtual mass squared. This is the reason why we cannot construct the
fundamental solution system composed of canonical series through the GKZ-system
of the general 5-propagator topology which originates from the Mellin-Barnes representation.
Applying the alpha parametric representation in Ref.~\cite{Feng2022},
we can embed the integral in some subvarieties of the Grassmannian $G_{_{6,12}}$,
and derive the GKZ-systems on those subvarieties.
In the neighborhoods of all regular singularities, the fundamental solution systems
are expressed in canonical series.

The general strategy for analysing the Feynman integral includes three steps here.
Firstly we obtain the holonomic system of the linear PDEs satisfied by the Feynman integral through
its Mellin-Barnes representation, then find the GKZ-system via the Miller transformation,
finally construct the hypergeometric series solutions. The integration constants, i.e.
the combination coefficients, are determined from the Feynman integral at an ordinary
point or some regular singularities.

Our presentation is organized as following. Through the Miller
transformation~\cite{Kalmykov2012,Feng2018,Feng2019},
we derive the GKZ-system of the Feynman integral of the 2-loop
self energy with 4 propagates in section \ref{sec2}.
Then we present in detail how to obtain the hypergeometric series
solutions of the GKZ-system in section \ref{sec3}.
Assuming only one virtual nonzero mass, we elucidate how to obtain
the combination coefficients in section \ref{sec4}.
The conclusions are summarized in section \ref{sec5}, and some tedious formulas
are presented in the appendices.

\section{Mellin-Barnes representation of the 2-loop self energy\label{sec2}}
\indent\indent

\begin{figure}[ht]
\setlength{\unitlength}{1cm}
\centering
\vspace{0.0cm}\hspace{-.0cm}
\includegraphics[height=6cm,width=10.0cm]{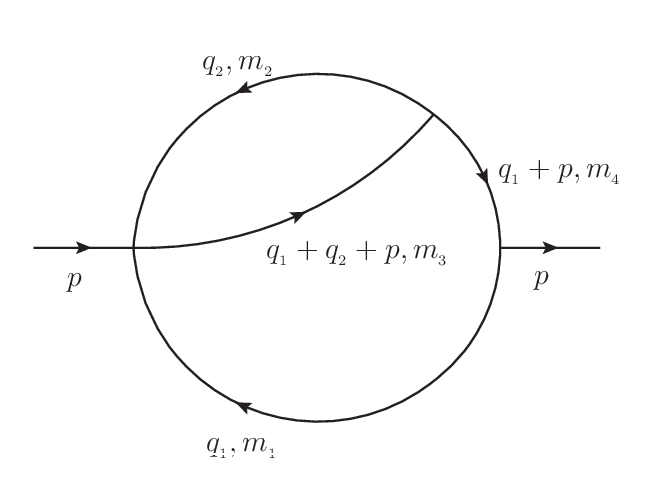}
\vspace{0cm}
\caption[]{The 2-loop self energy with 4 propagators.}
\label{fig1}
\end{figure}

The Feynman diagram for the 2-loop self energy with 4 propagators is drawn in Fig.\ref{fig1}.
The general analytic expression for the Feynman integral of the 2-loop self energy can be written as
\begin{eqnarray}
&&\Sigma_{_{1234}}(p^2)=\Big(\Lambda_{_{\rm RE}}^2\Big)^{4-D}\int{d^Dq_{_1}\over(2\pi)^D}
{d^Dq_{_2}\over(2\pi)^D}
\nonumber\\
&&\hspace{2.2cm}\times
{1\over(q_{_1}^2-m_{_1}^2)(q_{_2}^2-m_{_2}^2)((q_{_1}+q_{_2}+p)^2-m_{_3}^2)
((q_{_1}+p)^2-m_{_4}^2)},
\label{GKZ0-1}
\end{eqnarray}
where $D=4-2\varepsilon$ is the dimension in dimensional
regularization and $\Lambda_{_{\rm RE}}$ denotes the renormalization energy scale.
Adopting the notation of Refs.~\cite{Feng2018,Feng2019}, one can write
\begin{eqnarray}
&&\Sigma_{_{1234}}(p^2)=
{(\Lambda_{_{\rm RE}}^2)^{4-D}\over(2\pi i)^4}
\int_{-i\infty}^{+i\infty}ds_{_1}ds_{_2}ds_{_3}ds_{_4}
\Big[\prod\limits_{i=1}^4(-m_{_i}^2)^{s_{_i}}\Gamma(-s_{_i})\Gamma(1+s_{_i})\Big]
\nonumber\\
&&\hspace{2.2cm}\times
\int{d^Dq_{_1}\over(2\pi)^D}{d^Dq_{_2}\over(2\pi)^D}{1\over(q_{_1}^2)^{1+s_{_1}}(q_{_2}^2)^{1+s_{_2}}
((q_{_1}+q_{_2}+p)^2)^{1+s_{_3}}((q_{_1}+p)^2)^{1+{s_{_4}}}}\;.
\label{GKZ0-2}
\end{eqnarray}
Here, the Feynman integral for the two-loop massless self energy can be calculated as
\begin{eqnarray}
&&\int{d^Dq_{_1}\over(2\pi)^D}{d^Dq_{_2}\over(2\pi)^D}
{1\over(q_{_1}^2)^{1+s_{_1}}(q_{_2}^2)^{1+s_{_2}}((q_{_1}+q_{_2}+p)^2)^{1+s_{_3}}
((q_{_1}+p)^2)^{1+{s_{_4}}}}
\nonumber\\
&&\hspace{-0.5cm}=
-{(-)^{D}\Gamma(2-{D\over2}+s_{_2}+s_{_3})\Gamma({D\over2}-1-s_{_2})\Gamma({D\over2}-1-s_{_3})
\over(4\pi)^{D}\Gamma(1+s_{_2})\Gamma(1+s_{_3})\Gamma(D-2-s_{_2}-s_{_3})}
\nonumber\\
&&\hspace{-0.0cm}\times
{\Gamma(4-D+\sum\limits_{i=1}^4s_{_i})\Gamma({D\over2}-1-s_{_1})\Gamma(D-3-\sum\limits_{i=2}^4s_{_i})
\over\Gamma(1+s_{_1})\Gamma(3-{D\over2}+\sum\limits_{i=2}^4s_{_i})\Gamma({3D\over2}-4-\sum\limits_{i=1}^4s_{_i})}
(p^2)^{D-4-\sum\limits_{i=1}^4s_{_i}}\;.
\label{GKZ0-3}
\end{eqnarray}
The Mellin-Barnes representation of the 2-loop self energy is
\begin{eqnarray}
&&\Sigma_{_{1234}}(p^2)=
-{1\over(2\pi i)^4(4\pi)^4}\Big({4\pi\Lambda_{_{\rm RE}}^2\over-p^2}\Big)^{4-D}
\int_{-i\infty}^{+i\infty}ds_{_1}ds_{_2}ds_{_3}ds_{_4}
\nonumber\\
&&\hspace{2.2cm}\times
\Big[\prod\limits_{i=1}^4\Big({m_{_i}^2\over-p^2}\Big)^{s_{_i}}\Gamma(-s_{_i})\Big]\Gamma(1+s_{_4})
\prod\limits_{i=1}^3\Gamma({D\over2}-1-s_{_i})
\nonumber\\
&&\hspace{2.2cm}\times
{\Gamma(2-{D\over2}+s_{_2}+s_{_3})\Gamma(D-3-\sum\limits_{i=2}^4s_{_i})
\Gamma(4-D+\sum\limits_{i=1}^4s_{_i})
\over\Gamma(D-2-s_{_2}-s_{_3})\Gamma(3-{D\over2}+\sum\limits_{i=2}^4s_{_i})
\Gamma({3D\over2}-4-\sum\limits_{i=1}^4s_{_i})}\;.
\label{GKZ1}
\end{eqnarray}
It should be emphasized that this expression directly follows from the method
presented in the Refs~\cite{Davydychev3},
just by substituting the contour variables $s_{_i}\rightarrow 1-s_{_i}$.

It is well known that negative integers and zero are simple poles of the function
$\Gamma(z)$. As all $s_{_i}$ contours are closed to the right in corresponding
complex planes, one finds that the analytic expression of the two-loop self-energy
can be written as a linear combination of generalized hypergeometric functions,
in which an independent linear term is
\begin{eqnarray}
&&\Sigma_{_{1234}}(p^2)\ni
-{1\over(4\pi)^4}\Big({4\pi\Lambda_{_{\rm RE}}^2\over-p^2}\Big)^{4-D}
{\pi^2\sin{3\pi D\over2} \over \sin^3{\pi D\over2}}
\nonumber\\
&&\hspace{2.2cm}\times
\sum\limits_{n_{_1}=0}^\infty\sum_{n_{_2}=0}^\infty\sum_{n_{_3}=0}^\infty\sum_{n_{_4}=0}^\infty
x_{_1}^{n_{_1}}x_{_2}^{n_{_2}}x_{_3}^{n_{_3}}x_{_4}^{n_{_4}}
{\Gamma(3-D+n_{_2}+n_{_3})\over\prod\limits_{i=1}^3\Big[n_{_i}!\Gamma(2-{D\over2}+n_{_i})\Big]}
\nonumber\\
&&\hspace{2.2cm}\times
{\Gamma(2-{D\over2}+n_{_2}+n_{_3})\Gamma(4-D+\sum\limits_{i=1}^4n_{_i})\Gamma(5-{3D\over2}+\sum\limits_{i=1}^4n_{_i})
\over\Gamma(4-D+\sum\limits_{i=2}^4n_{_i})\Gamma(3-{D\over2}+\sum\limits_{i=2}^4n_{_i})},
\label{GKZ2}
\end{eqnarray}
with
\begin{eqnarray}
x_{_i}={m_{_i}^2\over p^2},\;i=1,\;2,\;3,\;4\;.
\label{GKZ2a}
\end{eqnarray}
For convenience in the following discussion,
we reformulate the term as
\begin{eqnarray}
&&\Sigma_{_{1234}}(p^2)\ni
-{1\over(4\pi)^4}\Big({4\pi\Lambda_{_{\rm RE}}^2\over-p^2}\Big)^{4-D}
{\pi^2\sin{3\pi D\over2} \over \sin^3{\pi D\over2}}
F({\bf a},\;{\bf b}\;\Big|\;{\bf x})\;,
\label{GKZ4}
\end{eqnarray}
where
\begin{eqnarray}
F({\bf a},\;{\bf b}\;\Big|\;{\bf x})=
\sum\limits_{n_{_1}=0}^\infty\sum_{n_{_2}=0}^\infty\sum_{n_{_3}=0}^\infty\sum_{n_{_4}=0}^\infty
A_{_{n_{_1}n_{_2}n_{_3}n_{_4}}}x_{_1}^{n_{_1}}x_{_2}^{n_{_2}}x_{_3}^{n_{_3}}x_{_4}^{n_{_4}}.
\label{GKZ4-1}
\end{eqnarray}
Here, the coefficient is
\begin{eqnarray}
&&A_{_{n_{_1}n_{_2}n_{_3}n_{_4}}}=
{\Gamma(a_{_1}+\sum\limits_{i=1}^4n_{_i}) \Gamma(a_{_2}+\sum\limits_{i=1}^4n_{_i})
\Gamma(a_{_3}+n_{_2}+n_{_3})\Gamma(a_{_4}+n_{_2}+n_{_3})\Gamma(a_{_5}+n_{_4})
\over  n_{_1}!n_{_2}!n_{_3}!n_{_4}!\prod\limits_{i=1}^3\Gamma(b_{_i}+n_{_i})
\Gamma(b_{_4}+\sum\limits_{i=2}^4n_{_i})\Gamma(b_{_5}+\sum\limits_{i=2}^4n_{_i})},
\label{GKZ4-2}
\end{eqnarray}
with ${\bf a}=(a_{_1},\;\cdots,a_{_5})$, ${\bf b}=(b_{_1},\;\cdots,b_{_5})$,
${\bf x}=(x_{_1},\;\cdots,\;x_{_4})$, and
\begin{eqnarray}
&&a_{_1}=4-D,\;a_{_2}=5-{3D\over2},\;a_{_3}=2-{D\over2},\;
a_{_4}=3-D,\;a_{_5}=1,
\nonumber\\
&&b_{_1}=b_{_2}=b_{_3}=2-{D\over2},\;b_{_4}=3-{D\over2},\;b_{_5}=4-D.
\label{GKZ5}
\end{eqnarray}

Through adjacent relations of the coefficient $A_{_{n_{_1}n_{_2}n_{_3}n_{_4}}}$,
the difference-differential operators can be written as
\begin{eqnarray}
&&(\sum\limits_{i=1}^4\vartheta_{_{x_{_i}}}+a_{_1})F({\bf a},\;{\bf b}\;\Big|\;{\bf x})
=a_{_1}F({\bf a}+{\bf e}_{_1},\;{\bf b}\;\Big|\;{\bf x})
\;,\nonumber\\
&&(\sum\limits_{i=1}^4\vartheta_{_{x_{_i}}}+a_{_2})F({\bf a},\;{\bf b}\;\Big|\;{\bf x})
=a_{_2}F({\bf a}+{\bf e}_{_2},\;{\bf b}\;\Big|\;{\bf x})
\;,\nonumber\\
&&(\vartheta_{_{x_{_2}}}+\vartheta_{_{x_{_3}}}+a_{_3})F({\bf a},\;{\bf b}\;\Big|\;{\bf x})
=a_{_3}F({\bf a}+{\bf e}_{_3},\;{\bf b}\;\Big|\;{\bf x})
\;,\nonumber\\
&&(\vartheta_{_{x_{_2}}}+\vartheta_{_{x_{_3}}}+a_{_4})F({\bf a},\;{\bf b}\;\Big|\;{\bf x})
=a_{_4}F({\bf a}+{\bf e}_{_4},\;{\bf b}\;\Big|\;{\bf x})
\;,\nonumber\\
&&(\vartheta_{_{x_{_4}}}+a_{_5})F({\bf a},\;{\bf b}\;\Big|\;{\bf x})
=a_{_5}F({\bf a}+{\bf e}_{_5},\;{\bf b}\;\Big|\;{\bf x})
\;,\nonumber\\
&&(\vartheta_{_{x_{_k}}}+b_{_k}-1)F({\bf a},\;{\bf b}\;\Big|\;{\bf x})
=(b_{_k}-1)F({\bf a},\;{\bf b}-{\bf e}_{_k}\;\Big|\;{\bf x}),\;k=1,2,3
\;,\nonumber\\
&&(\sum\limits_{i=2}^4\vartheta_{_{x_{_i}}}+b_{_k}-1)F({\bf a},\;{\bf b}\;\Big|\;{\bf x})
=(b_{_k}-1)F({\bf a},\;{\bf b}-{\bf e}_{_k}\;\Big|\;{\bf x}),\;k=4,5\;,
\nonumber\\
&&\partial_{_{x_{_1}}}F({\bf a},\;{\bf b}\;\Big|\;{\bf x})={a_{_1}a_{_2}\over b_{_1}}
F({\bf a}+{\bf e}_{_{1}}+{\bf e}_{_{2}},\;{\bf b}+{\bf e}_{_{1}}\;\Big|\;{\bf x})
\;,\nonumber\\
&&\partial_{_{x_{_2}}}F({\bf a},\;{\bf b}\;\Big|\;{\bf x})={a_{_1}a_{_2}a_{_3}a_{_4}\over b_{_2}b_{_4}b_{_5}}
F({\bf a}+{\bf e}_{_{1}}+{\bf e}_{_{2}}+{\bf e}_{_{3}}+{\bf e}_{_{4}},\;{\bf b}+{\bf e}_{_{2}}
+{\bf e}_{_{4}}+{\bf e}_{_{5}}\;\Big|\;{\bf x})
\;,\nonumber\\
&&\partial_{_{x_{_3}}}F({\bf a},\;{\bf b}\;\Big|\;{\bf x})={a_{_1}a_{_2}a_{_3}a_{_4}\over b_{_3}b_{_4}b_{_5}}
F({\bf a}+{\bf e}_{_{1}}+{\bf e}_{_{2}}+{\bf e}_{_{3}}+{\bf e}_{_{4}},\;{\bf b}+{\bf e}_{_{3}}
+{\bf e}_{_{4}}+{\bf e}_{_{5}}\;\Big|\;{\bf x})
\;,\nonumber\\
&&\partial_{_{x_{_4}}}F({\bf a},\;{\bf b}\;\Big|\;{\bf x})={a_{_1}a_{_2}a_{_5}\over b_{_4}b_{_5}}
F({\bf a}+{\bf e}_{_{1}}+{\bf e}_{_{2}}+{\bf e}_{_5},\;{\bf b}+{\bf e}_{_{4}}
+{\bf e}_{_{5}}\;\Big|\;{\bf x})\;,
\label{GKZ9}
\end{eqnarray}
where ${\bf e}_{_{j}}\in{\bf R}^5$ denotes the row vector whose entry is zero
except that the $j-$th entry is $1$, $\vartheta_{_{x_{_j}}}=x_{_j}\partial_{_{x_{_j}}}$
denotes the Euler operator, and $\partial_{_{x_{_j}}}=\partial/\partial x_{_j}$, respectively.
In order to proceed with our analysis, we define an auxiliary function,
\begin{eqnarray}
&&\Phi({\bf a},\;{\bf b}\;\Big|\;{\bf x},\;{\bf u},\;{\bf v})={\bf u}^{\bf a}{\bf v}^{{\bf b}-{\bf e}}
F({\bf a},\;{\bf b}\;\Big|\;{\bf x})\;,
\label{GKZ10}
\end{eqnarray}
with the intermediate variables ${\bf u}={\bf v}={\bf e}=(1,\;1,\;1,\;1,\;1)$.
To avoid cumbersome symbols, here we adopt the multiple index notations
${\bf u}^{\bf a}=\prod\limits_{i=1}^5u_{_i}^{a_{_i}}$, etc.
Then the relations can be obtained easily
\begin{eqnarray}
&&\vartheta_{_{u_j}}\Phi({\bf a},\;{\bf b}\;\Big|\;{\bf x},\;{\bf u},\;{\bf v})=a_{_j}
\Phi({\bf a},\;{\bf b}\;\Big|\;{\bf x},\;{\bf u},\;{\bf v})
\;,\nonumber\\
&&\vartheta_{_{v_j}}\Phi({\bf a},\;{\bf b}\;\Big|\;{\bf x},\;{\bf u},\;{\bf v})=(b_{_j}-1)
\Phi({\bf a},\;{\bf b}\;\Big|\;{\bf x},\;{\bf u},\;{\bf v})\;.
\label{GKZ11}
\end{eqnarray}
In addition, the contiguous relations of Eq.~(\ref{GKZ9}) can be rewritten as
\begin{eqnarray}
&&\hat{{\cal O}}_{_1}\Phi({\bf a},\;{\bf b}\;\Big|\;{\bf x},\;{\bf u},\;{\bf v})
=a_{_1}\Phi({\bf a}+{\bf e}_{_1},\;{\bf b}\;\Big|\;{\bf x},\;{\bf u},\;{\bf v})
\;,\nonumber\\
&&\hat{{\cal O}}_{_2}\Phi({\bf a},\;{\bf b}\;\Big|\;{\bf x},\;{\bf u},\;{\bf v})
=a_{_2}\Phi({\bf a}+{\bf e}_{_2},\;{\bf b}\;\Big|\;{\bf x},\;{\bf u},\;{\bf v})
\;,\nonumber\\
&&\hat{{\cal O}}_{_3}\Phi({\bf a},\;{\bf b}\;\Big|\;{\bf x},\;{\bf u},\;{\bf v})
=a_{_3}\Phi({\bf a}+{\bf e}_{_3},\;{\bf b}\;\Big|\;{\bf x},\;{\bf u},\;{\bf v})
\;,\nonumber\\
&&\hat{{\cal O}}_{_4}\Phi({\bf a},\;{\bf b}\;\Big|\;{\bf x},\;{\bf u},\;{\bf v})
=a_{_4}\Phi({\bf a}+{\bf e}_{_4},\;{\bf b}\;\Big|\;{\bf x},\;{\bf u},\;{\bf v})
\;,\nonumber\\
&&\hat{{\cal O}}_{_5}\Phi({\bf a},\;{\bf b}\;\Big|\;{\bf x},\;{\bf u},\;{\bf v})
=a_{_5}\Phi({\bf a}+{\bf e}_{_5},\;{\bf b}\;\Big|\;{\bf x},\;{\bf u},\;{\bf v})
\;,\nonumber\\
&&\hat{{\cal O}}_{_{5+k}}\Phi({\bf a},\;{\bf b}\;\Big|\;{\bf x},\;{\bf u},\;{\bf v})
=(b_{_k}-1)\Phi({\bf a},\;{\bf b}-{\bf e}_{_k}\;\Big|\;{\bf x},\;{\bf u},\;{\bf v}),\;k=1,2,3
\;,\nonumber\\
&&\hat{{\cal O}}_{_{5+k}}\Phi({\bf a},\;{\bf b}\;\Big|\;{\bf x},\;{\bf u},\;{\bf v})
=(b_{_k}-1)\Phi({\bf a},\;{\bf b}-{\bf e}_{_k}\;\Big|\;{\bf x},\;{\bf u},\;{\bf v}),\;k=4,5
\;,\nonumber\\
&&\hat{{\cal O}}_{_{11}}\Phi({\bf a},\;{\bf b}\;\Big|\;{\bf x},\;{\bf u},\;{\bf v})={a_{_1}a_{_2}\over b_{_1}}
\Phi({\bf a}+{\bf e}_{_{1}}+{\bf e}_{_{2}},\;{\bf b}+{\bf e}_{_{1}}\;\Big|\;{\bf x},\;{\bf u},\;{\bf v})
\;,\nonumber\\
&&\hat{{\cal O}}_{_{12}}\Phi({\bf a},\;{\bf b}\;\Big|\;{\bf x},\;{\bf u},\;{\bf v})={a_{_1}a_{_2}a_{_3}a_{_4}\over b_{_2}b_{_4}b_{_5}}
\Phi({\bf a}+{\bf e}-{\bf e}_{_{5}},\;{\bf b}+{\bf e}_{_{2}}
+{\bf e}_{_{4}}+{\bf e}_{_{5}}\;\Big|\;{\bf x},\;{\bf u},\;{\bf v})
\;,\nonumber\\
&&\hat{{\cal O}}_{_{13}}\Phi({\bf a},\;{\bf b}\;\Big|\;{\bf x},\;{\bf u},\;{\bf v})={a_{_1}a_{_2}a_{_3}a_{_4}\over b_{_3}b_{_4}b_{_5}}
\Phi({\bf a}+{\bf e}-{\bf e}_{_{5}},\;{\bf b}+{\bf e}_{_{3}}
+{\bf e}_{_{4}}+{\bf e}_{_{5}}\;\Big|\;{\bf x},\;{\bf u},\;{\bf v})
\;,\nonumber\\
&&\hat{{\cal O}}_{_{14}}\Phi({\bf a},\;{\bf b}\;\Big|\;{\bf x},\;{\bf u},\;{\bf v})={a_{_1}a_{_2}a_{_5}\over b_{_4}b_{_5}}
\Phi({\bf a}+{\bf e}_{_{1}}+{\bf e}_{_{2}}+{\bf e}_{_5},\;{\bf b}+{\bf e}_{_{4}}
+{\bf e}_{_{5}}\;\Big|\;{\bf x},\;{\bf u},\;{\bf v})\;,
\label{GKZ12}
\end{eqnarray}
where the operators are
\begin{eqnarray}
&&\hat{{\cal O}}_{_1}=u_{_1}(\sum\limits_{i=1}^4\vartheta_{_{x_{_i}}}+\vartheta_{_{u_1}})
\;,\nonumber\\
&&\hat{{\cal O}}_{_2}=u_{_2}(\sum\limits_{i=1}^4\vartheta_{_{x_{_i}}}+\vartheta_{_{u_2}})
\;,\nonumber\\
&&\hat{{\cal O}}_{_3}=u_{_3}(\vartheta_{_{x_{_2}}}+\vartheta_{_{x_{_3}}}+\vartheta_{_{u_3}})
\;,\nonumber\\
&&\hat{{\cal O}}_{_4}=u_{_4}(\vartheta_{_{x_{_2}}}+\vartheta_{_{x_{_3}}}+\vartheta_{_{u_4}})
\;,\nonumber\\
&&\hat{{\cal O}}_{_5}=u_{_5}(\vartheta_{_{x_{_4}}}+\vartheta_{_{u_5}})
\;,\nonumber\\
&&\hat{{\cal O}}_{_{5+k}}={1\over v_{_k}}(\vartheta_{_{x_{_k}}}+\vartheta_{_{v_k}}),\;k=1,2,3
\;,\nonumber\\
&&\hat{{\cal O}}_{_{5+k}}={1\over v_{_k}}(\sum\limits_{i=2}^4\vartheta_{_{x_{_i}}}+\vartheta_{_{v_k}}),\;k=4,5
\;,\nonumber\\
&&\hat{{\cal O}}_{_{11}}=u_{_1}u_{_2}v_{_1}\partial_{_{x_{_1}}}
\;,\nonumber\\
&&\hat{{\cal O}}_{_{12}}=u_{_1}u_{_2}u_{_3}u_{_4}v_{_2}v_{_4}v_{_5}\partial_{_{x_{_2}}}
\;,\nonumber\\
&&\hat{{\cal O}}_{_{13}}=u_{_1}u_{_2}u_{_3}u_{_4}v_{_3}v_{_4}v_{_5}\partial_{_{x_{_3}}}
\;,\nonumber\\
&&\hat{{\cal O}}_{_{14}}=u_{_1}u_{_2}u_{_5}v_{_4}v_{_5}\partial_{_{x_{_4}}}\;.
\label{GKZ13}
\end{eqnarray}
The operators above together with $\vartheta_{_{u_j}}$, $\vartheta_{_{v_j}}$
define the Lie algebra of the GKZ-system~\cite{Miller68,Miller72}.
Through the transformations of indeterminacy
\begin{eqnarray}
&&z_{_j}={1\over u_{_j}},\;\;z_{_{5+j}}=v_{_j},\;\;(j=1,\cdots,5),
\nonumber\\
&&z_{_{11}}={x_{_1}\over u_{_1}u_{_2}v_{_1}},\;\;z_{_{12}}={x_{_2}\over u_{_1}u_{_2}u_{_3}u_{_4}v_{_2}v_{_4}v_{_5}},
\nonumber\\
&&z_{_{13}}={x_{_3}\over u_{_1}u_{_2}u_{_3}u_{_4}v_{_3}v_{_4}v_{_5}},
\;\;z_{_{14}}={x_{_4}\over u_{_1}u_{_2}u_{_5}v_{_4}v_{_5}}\;,
\label{GKZ14}
\end{eqnarray}
one derives
\begin{eqnarray}
&&\vartheta_{_{x_{_1}}}=\vartheta_{_{z_{_{11}}}},\;\;\;\vartheta_{_{x_{_2}}}=\vartheta_{_{z_{_{12}}}},
\;\;\;\vartheta_{_{x_{_3}}}=\vartheta_{_{z_{_{13}}}},\;\;\;\vartheta_{_{x_{_4}}}=\vartheta_{_{z_{_{14}}}}
\;,\nonumber\\
&&\vartheta_{_{u_{_1}}}=-\vartheta_{_{z_{_1}}}-\vartheta_{_{z_{_{11}}}}-\vartheta_{_{z_{_{12}}}}
-\vartheta_{_{z_{_{13}}}}-\vartheta_{_{z_{_{14}}}}
\;,\nonumber\\
&&\vartheta_{_{u_{_2}}}=-\vartheta_{_{z_{_2}}}-\vartheta_{_{z_{_{11}}}}-\vartheta_{_{z_{_{12}}}}
-\vartheta_{_{z_{_{13}}}}-\vartheta_{_{z_{_{14}}}}
\;,\nonumber\\
&&\vartheta_{_{u_{_3}}}=-\vartheta_{_{z_{_3}}}-\vartheta_{_{z_{_{12}}}}-\vartheta_{_{z_{_{13}}}},\;\;\;
\vartheta_{_{u_{_4}}}=-\vartheta_{_{z_{_4}}}-\vartheta_{_{z_{_{12}}}}-\vartheta_{_{z_{_{13}}}}
\;,\nonumber\\
&&\vartheta_{_{u_{_5}}}=-\vartheta_{_{z_{_5}}}-\vartheta_{_{z_{_{14}}}},\;\;\;
\vartheta_{_{v_{_1}}}=\vartheta_{_{z_{_6}}}-\vartheta_{_{z_{_{11}}}},\;\;\;
\vartheta_{_{v_{_2}}}=\vartheta_{_{z_{_7}}}-\vartheta_{_{z_{_{12}}}}
\;,\nonumber\\
&&\vartheta_{_{v_{_3}}}=\vartheta_{_{z_{_8}}}-\vartheta_{_{z_{_{13}}}},\;\;\;
\vartheta_{_{v_{_4}}}=\vartheta_{_{z_{_9}}}-\vartheta_{_{z_{_{12}}}}-\vartheta_{_{z_{_{13}}}}-\vartheta_{_{z_{_{14}}}}
\;,\nonumber\\
&&\vartheta_{_{v_{_5}}}=\vartheta_{_{z_{_{10}}}}-\vartheta_{_{z_{_{12}}}}-\vartheta_{_{z_{_{13}}}}-\vartheta_{_{z_{_{14}}}}
,\;\;\;\partial_{_{x_{_1}}}={1\over u_{_1}u_{_2}v_{_1}}{\partial\over\partial z_{_{11}}}
\;,\nonumber\\
&&\partial_{_{x_{_2}}}={1\over u_{_1}u_{_2}u_{_3}u_{_4}v_{_2}v_{_4}v_{_5}}{\partial\over\partial z_{_{12}}}
,\;\;\;\partial_{_{x_{_3}}}={1\over u_{_1}u_{_2}u_{_3}u_{_4}v_{_3}v_{_4}v_{_5}}{\partial\over\partial z_{_{13}}}
\;,\nonumber\\
&&\partial_{_{x_{_4}}}={1\over u_{_1}u_{_2}u_{_5}v_{_4}v_{_5}}{\partial\over\partial z_{_{14}}}\;.
\label{GKZ15}
\end{eqnarray}
Furthermore, the operators in Eq.~(\ref{GKZ13}) are transformed into
\begin{eqnarray}
&&\hat{{\cal O}}_{_j}=-{\partial\over\partial_{_{z_{_j}}}}=-\partial_{_{z_{_j}}},\;\;(j=1,\cdots,5)
\;,\nonumber\\
&&\hat{{\cal O}}_{_j}={\partial\over\partial_{_{z_{_j}}}}=\partial_{_{z_{_j}}},\;\;(j=6,\cdots,14),
\label{GKZ16}
\end{eqnarray}
and the relations in Eq.~(\ref{GKZ11}) are changed as
\begin{eqnarray}
&&\mathbf{A}\cdot\vec{\vartheta}\Phi=\mathbf{B}\Phi\;,
\label{GKZ17}
\end{eqnarray}
where
\begin{eqnarray}
&&\mathbf{A}=\left(\begin{array}{cccccccccccccc}
1\;\;&0\;\;&0\;\;&0\;\;&0\;\;&0\;\;&0\;\;&0\;\;&0\;\;&0\;\;&1\;\;&1\;\;&1\;\;&1\;\;\\
0\;\;&1\;\;&0\;\;&0\;\;&0\;\;&0\;\;&0\;\;&0\;\;&0\;\;&0\;\;&1\;\;&1\;\;&1\;\;&1\;\;\\
0\;\;&0\;\;&1\;\;&0\;\;&0\;\;&0\;\;&0\;\;&0\;\;&0\;\;&0\;\;&0\;\;&1\;\;&1\;\;&0\;\;\\
0\;\;&0\;\;&0\;\;&1\;\;&0\;\;&0\;\;&0\;\;&0\;\;&0\;\;&0\;\;&0\;\;&1\;\;&1\;\;&0\;\;\\
0\;\;&0\;\;&0\;\;&0\;\;&1\;\;&0\;\;&0\;\;&0\;\;&0\;\;&0\;\;&0\;\;&0\;\;&0\;\;&1\;\;\\
0\;\;&0\;\;&0\;\;&0\;\;&0\;\;&1\;\;&0\;\;&0\;\;&0\;\;&0\;\;&-1\;\;&0\;\;&0\;\;&0\;\;\\
0\;\;&0\;\;&0\;\;&0\;\;&0\;\;&0\;\;&1\;\;&0\;\;&0\;\;&0\;\;&0\;\;&-1\;\;&0\;\;&0\;\;\\
0\;\;&0\;\;&0\;\;&0\;\;&0\;\;&0\;\;&0\;\;&1\;\;&0\;\;&0\;\;&0\;\;&0\;\;&-1\;\;&0\;\;\\
0\;\;&0\;\;&0\;\;&0\;\;&0\;\;&0\;\;&0\;\;&0\;\;&1\;\;&0\;\;&0\;\;&-1\;\;&-1\;\;&-1\;\;\\
0\;\;&0\;\;&0\;\;&0\;\;&0\;\;&0\;\;&0\;\;&0\;\;&0\;\;&1\;\;&0\;\;&-1\;\;&-1\;\;&-1\;\;\\
\end{array}\right)
\;,\nonumber\\
&&\vec{\vartheta}^{\;T}=(\vartheta_{_{z_{_1}}},\;\vartheta_{_{z_{_{2}}}},\;\vartheta_{_{z_{_{3}}}},\;
\vartheta_{_{z_{_{4}}}},\;\vartheta_{_{z_{_{5}}}},\;\vartheta_{_{z_{_{6}}}},\;\vartheta_{_{z_{_{7}}}}
,\;\vartheta_{_{z_{_{8}}}},\;\vartheta_{_{z_{_{9}}}},\;\vartheta_{_{z_{_{10}}}},\;\vartheta_{_{z_{_{11}}}}
,\;\vartheta_{_{z_{_{12}}}},\;\vartheta_{_{z_{_{13}}}},\;\vartheta_{_{z_{_{14}}}})
\;,\nonumber\\
&&\mathbf{B}^{\;T}=(-a_{_1},\;-a_{_2},\;-a_{_3},\;-a_{_4},\;-a_{_5},\;b_{_1}-1,\;b_{_2}-1
,\;b_{_3}-1,\;b_{_4}-1,\;b_{_5}-1)\;.
\label{GKZ18}
\end{eqnarray}
Correspondingly the dual matrix $\mathbf{\tilde A}$ of $\mathbf{A}$ is
\begin{eqnarray}
&&\mathbf{\tilde A}=\left(\begin{array}{cccccccccccccc}
-1\;\;&-1\;\;&0\;\;&0\;\;&0\;\;&1\;\;&0\;\;&0\;\;&0\;\;&0\;\;&1\;\;&0\;\;&0\;\;&0\;\;\\
-1\;\;&-1\;\;&-1\;\;&-1\;\;&0\;\;&0\;\;&1\;\;&0\;\;&1\;\;&1\;\;&0\;\;&1\;\;&0\;\;&0\;\;\\
-1\;\;&-1\;\;&-1\;\;&-1\;\;&0\;\;&0\;\;&0\;\;&1\;\;&1\;\;&1\;\;&0\;\;&0\;\;&1\;\;&0\;\;\\
-1\;\;&-1\;\;&0\;\;&0\;\;&-1\;\;&0\;\;&0\;\;&0\;\;&1\;\;&1\;\;&0\;\;&0\;\;&0\;\;&1\;\;
\end{array}\right)\;.
\label{GKZ19}
\end{eqnarray}
The row vectors of the matrix $\mathbf{\tilde A}$ induce the integer sublattice
which can be used to obtain the formal solutions in the hypergeometric series. Actually the integer sublattice
indicates that the solutions of the system should satisfy the Euler equations in Eq.~(\ref{GKZ17}) and the following
hyperbolic equations simultaneously
\begin{eqnarray}
&&{\partial^2\Phi\over\partial z_{_1}\partial z_{_2}}={\partial^2\Phi\over\partial z_{_6}\partial z_{_{11}}}
\;,\nonumber\\
&&{\partial^2\Phi\over\partial z_{_7}\partial z_{_{12}}}={\partial^2\Phi\over\partial z_{_8}\partial z_{_{13}}}
\;,\nonumber\\
&&{\partial^3\Phi\over\partial z_{_3}\partial z_{_4}\partial z_{_{14}}}=
{\partial^3\Phi\over\partial z_{_5}\partial z_{_{8}}\partial z_{_{13}}}
\;,\nonumber\\
&&{\partial^3\Phi\over\partial z_{_5}\partial z_{_{6}}\partial z_{_{11}}}=
{\partial^3\Phi\over\partial z_{_9}\partial z_{_{10}}\partial z_{_{14}}}
\;,\nonumber\\
&&{\partial^4\Phi\over\partial z_{_3}\partial z_{_4}\partial z_{_6}\partial z_{_{11}}}=
{\partial^4\Phi\over\partial z_{_8}\partial z_{_9}\partial z_{_{10}}\partial z_{_{13}}}\;.
\label{GKZ19-1}
\end{eqnarray}
Actually the PDEs constitute a Gr\"obner basis of the toric ideal of the matrix
$\mathbf{\tilde A}$ presented in Eq.(\ref{GKZ19}). Defining the combined variables
\begin{eqnarray}
&&y_{_1}={z_{_6}z_{_{11}}\over z_{_1}z_{_2}}
\;,\nonumber\\
&&y_{_2}={z_{_7}z_{_9}z_{_{10}}z_{_{12}}\over z_{_1}z_{_2}z_{_3}z_{_4}}
\;,\nonumber\\
&&y_{_3}={z_{_8}z_{_9}z_{_{10}}z_{_{13}}\over z_{_1}z_{_2}z_{_3}z_{_4}}
\;,\nonumber\\
&&y_{_4}={z_{_9}z_{_{10}}z_{_{14}}\over z_{_1}z_{_2}z_{_5}}\;,
\label{GKZ19-2}
\end{eqnarray}
we write the solutions satisfying Eq.(\ref{GKZ17}) and Eq.(\ref{GKZ19-1}) as
\begin{eqnarray}
&&\Phi({\mathbf z})=\Big(\prod\limits_{i=1}^{14}z_{_i}^{\alpha_{_i}}\Big)
\varphi(y_{_1},\;y_{_2},\;y_{_3},\;y_{_4})\;,
\label{GKZ19-3}
\end{eqnarray}
where $\alpha^T=(\alpha_{_1},\;\alpha_{_2},\;\cdots,\;\alpha_{_{14}})$
denotes a sequence of complex number such that
\begin{eqnarray}
&&\mathbf{A}\cdot\alpha=\mathbf{B}\;.
\label{GKZ19-4}
\end{eqnarray}
Substituting Eq.~(\ref{GKZ19-3}) into Eq.~(\ref{GKZ19-1}), we obtain the
independent PDEs $\hat{L}_{_i}\varphi=0,\;\;(i=1,\cdots,5)$,
where the linear partial differential operators $\hat{L}_{_i}$ are presented in Eq.~(\ref{GKZ19-5}).
It should be emphasized that the system of PDEs can also be derived through
the approach presented in the literature~\cite{Kalmykov2012}.
Although the above system of linear PDEs is too complicated to construct the solutions
in the neighborhoods of the ordinary points, it gives the number of linear independent solutions of the system.
Differentiating those equations in Eq.(\ref{GKZ19-5}), one finds easily that any derivative of $\varphi$ can be
formulated as a linear combination of $\varphi$ and the following $29$ partial derivatives
\begin{eqnarray}
&&{\partial\varphi\over\partial y_{_i}},\;{\partial^2\varphi\over\partial y_{_3}^2},\;
{\partial^2\varphi\over\partial y_{_4}^2}
,\;{\partial^2\varphi\over\partial y_{_1}\partial y_{_i}}\;(i\neq1)
,\;{\partial^2\varphi\over\partial y_{_2}\partial y_{_3}},\;
{\partial^2\varphi\over\partial y_{_2}\partial y_{_4}}
,\;{\partial^2\varphi\over\partial y_{_3}\partial y_{_4}}
,\nonumber\\
&&{\partial^3\varphi\over\partial y_{_3}^3},\;
{\partial^3\varphi\over\partial y_{_1}\partial y_{_3}^2},\;
{\partial^3\varphi\over\partial y_{_2}\partial y_{_3}^2},\;
{\partial^3\varphi\over\partial y_{_1}\partial y_{_4}^2},\;
{\partial^3\varphi\over\partial y_{_2}\partial y_{_4}^2},\;
{\partial^3\varphi\over\partial y_{_3}\partial y_{_4}^2},\;
{\partial^3\varphi\over\partial y_{_1}\partial y_{_2}\partial y_{_3}}
,\nonumber\\
&&{\partial^3\varphi\over\partial y_{_1}\partial y_{_2}\partial y_{_4}},
\;{\partial^3\varphi\over\partial y_{_1}\partial y_{_3}\partial y_{_4}},
\;{\partial^3\varphi\over\partial y_{_2}\partial y_{_3}\partial y_{_4}},\;
{\partial^4\varphi\over\partial y_{_1}\partial y_{_3}^3},\;
{\partial^4\varphi\over\partial y_{_2}\partial y_{_3}^3},
\nonumber\\
&&{\partial^4\varphi\over\partial y_{_1}\partial y_{_2}\partial y_{_3}^2},
\;{\partial^4\varphi\over\partial y_{_1}\partial y_{_2}\partial y_{_4}^2},\;
{\partial^4\varphi\over\partial y_{_1}\partial y_{_3}\partial y_{_4}^2},
\;{\partial^4\varphi\over\partial y_{_2}\partial y_{_3}\partial y_{_4}^2},\;
{\partial^4\varphi\over\partial y_{_1}\partial y_{_2}\partial y_{_3}\partial y_{_4}},
\label{GKZ20}
\end{eqnarray}
with $i=1,\;\cdots,\;4$. In other words, the dimension of the solution space of the
system in Eq.(\ref{GKZ19-5}) is 30. Furthermore,
the dimension of the solution space of the PDEs decreases in the singular locus.
At the regular singularity, the dimension of the solution space is one. The fact implies
that we can determine the linear combination constants through the Feynman
integral at those regular singularities.

\section{The hypergeometric solutions of GKZ-system of the 2-loop self energy\label{sec3}}
\indent\indent
To construct the hypergeometric series solutions of the GKZ-system of Eq.(\ref{GKZ17})
and the hyperbolic equations Eq.(\ref{GKZ19-1}) through triangulation
is equivalent to choosing a set of linearly independent column vectors of the matrix in Eq.(\ref{GKZ19})
which spans the dual space. We denote the submatrix
composed of the first, third, fifth, and seventh column vectors of the dual
matrix of Eq.(\ref{GKZ19}) as $\mathbf{\tilde A}_{_{1357}}$, i.e.
\begin{eqnarray}
&&\mathbf{\tilde A}_{_{1357}}=\left(\begin{array}{cccc}
-1\;\;&0\;\;&0\;\;&0\;\;\\
-1\;\;&-1\;\;&0\;\;&1\;\;\\
-1\;\;&-1\;\;&0\;\;&0\;\;\\
-1\;\;&0\;\;&-1\;\;&0\;\;
\end{array}\right)\;.
\label{GKZ21-a}
\end{eqnarray}
Obviously $\det\mathbf{\tilde A}_{_{1357}}=-1\neq0$, and
\begin{eqnarray}
&&\mathbf{B}_{_{1357}}=\mathbf{\tilde A}_{_{1357}}^{-1}\cdot\mathbf{\tilde A}
\nonumber\\
&&\hspace{1.0cm}=\left(\begin{array}{cccccccccccccc}
1\;\;& 1\;\;& 0\;\;& 0\;\;& 0\;\;& -1\;\;& 0\;\;& 0\;\;& 0\;\;& 0\;\;& -1\;\;& 0\;\;& 0\;\;& 0\;\;\\
0\;\;& 0\;\;& 1\;\;& 1\;\;& 0\;\;& 1\;\;& 0\;\;& -1\;\;& -1\;\;& -1\;\;& 1\;\;& 0\;\;& -1\;\;& 0\;\;\\
0\;\;&0\;\;&0\;\;& 0\;\;& 1\;\;& 1\;\;& 0\;\;& 0\;\;& -1\;\;& -1\;\;& 1\;\;& 0\;\;& 0\;\;& -1\;\;\\
0\;\;&0\;\;& 0\;\;& 0\;\;& 0\;\;& 0\;\;& 1\;\;& -1\;\;& 0\;\;& 0\;\;& 0\;\;& 1\;\;& -1\;\;& 0\;\;
\end{array}\right)\;.
\label{GKZ21}
\end{eqnarray}
Taking 4 row vectors of the matrix $\mathbf{B}_{_{1357}}$ as the basis of integer lattice,
one constructs the hypergeometric series solution in the neighborhoods of the regular singularities $0$
and $\infty$ in the parameter space. For example, we take the set of column indices
$I_{_1}=[2,4,6,8,\cdots,14]$, i.e. the implement $J_{_1}=[1,14]\setminus I_{_1}=[1,3,5,7]$.
The choice of the set of indices implies the exponent numbers
$\alpha_{_1}=\alpha_{_3}=\alpha_{_5}=\alpha_{_7}=0$, and
\begin{eqnarray}
&&\alpha_{_2}=a_{_1}-a_{_2},\;\alpha_{_4}=a_{_3}-a_{_4},\;\alpha_{_6}=a_{_3}+a_{_5}+b_{_1}-a_{_1}-1,\;
\nonumber\\
&&\alpha_{_8}=b_{_2}+b_{_3}-a_{_3}-2,\;\alpha_{_9}=b_{_4}-a_{_3}-a_{_5}-1,\;
\nonumber\\
&&\alpha_{_{10}}=b_{_5}-a_{_3}-a_{_5}-1,\;\alpha_{_{11}}=a_{_3}+a_{_5}-a_{_1},\;\alpha_{_{12}}=1-b_{_2},\;
\nonumber\\
&&\alpha_{_{13}}=b_{_2}-a_{_3}-1,\;\alpha_{_{14}}=-a_{_5}\;.
\label{GKZ21-1-1}
\end{eqnarray}
The corresponding hypergeometric series solutions with quadruple independent variables can be written as
\begin{eqnarray}
&&\Phi_{_{[1357]}}^{(1),a}(\alpha,z)=
{y_{_1}^{{D\over2}-1}y_{_2}^{{D\over2}-1}y_{_3}^{-1}y_{_4}^{-1}}\sum\limits_{n_{_1}=0}^\infty
\sum\limits_{n_{_2}=0}^\infty\sum\limits_{n_{_3}=0}^\infty\sum\limits_{n_{_4}=0}^\infty
c_{_{[1357]}}^{(1),a}(\alpha,{\bf n})
\nonumber\\
&&\hspace{2.5cm}\times
\Big({1\over y_{_4}}\Big)^{n_{_1}}\Big({y_{_4}\over y_{_3}}\Big)^{n_{_2}}
\Big({y_{_1}\over y_{_4}}\Big)^{n_{_3}}\Big({y_{_2}\over y_{_3}}\Big)^{n_{_4}}
\;,\nonumber\\
&&\Phi_{_{[1357]}}^{(1),b}(\alpha,z)=
{y_{_1}^{{D\over2}-2}y_{_2}^{{D\over2}-1}y_{_3}^{-1}y_{_4}^{-1}}
\sum\limits_{n_{_1}=0}^\infty
\sum\limits_{n_{_2}=0}^\infty\sum\limits_{n_{_3}=0}^\infty\sum\limits_{n_{_4}=0}^\infty
c_{_{[1357]}}^{(1),b}(\alpha,{\bf n})
\nonumber\\
&&\hspace{2.5cm}\times
\Big({1\over y_{_1}}\Big)^{n_{_1}}\Big({1\over y_{_3}}\Big)^{n_{_2}}
\Big({1\over y_{_4}}\Big)^{n_{_3}}\Big({y_{_2}\over y_{_3}}\Big)^{n_{_4}}\;.
\label{GKZ21-1-2a}
\end{eqnarray}
Where the coefficients are
\begin{eqnarray}
&&c_{_{[1357]}}^{(1),a}(\alpha,{\bf n})=
(-)^{n_{_4}}\Gamma(1+n_{_1}+n_{_3})\Gamma(1+n_{_2}+n_{_4})
\Big\{n_{_1}!n_{_2}!n_{_3}!n_{_4}!
\nonumber\\
&&\hspace{2.5cm}\times
\Gamma({D\over2}+n_{_1})\Gamma({D\over2}+n_{_2})\Gamma(1-{D\over2}-n_{_2}-n_{_4})
\nonumber\\
&&\hspace{2.5cm}\times
\Gamma(1-{D\over2}-n_{_1}-n_{_3})\Gamma({D\over2}+n_{_3})\Gamma({D\over2}+n_{_4})\Big\}^{-1}
\;,\nonumber\\
&&c_{_{[1357]}}^{(1),b}(\alpha,{\bf n})=
(-)^{n_{_1}+n_{_4}}\Gamma(1+n_{_1})\Gamma(1+n_{_2}+n_{_3})\Gamma(1+n_{_2}+n_{_4})
\Big\{n_{_2}!n_{_4}!
\nonumber\\
&&\hspace{2.5cm}\times
\Gamma(2+n_{_1}+n_{_2}+n_{_3})\Gamma({D\over2}+1+n_{_1}+n_{_2}+n_{_3})
\nonumber\\
&&\hspace{2.5cm}\times
\Gamma({D\over2}+n_{_2})\Gamma(1-{D\over2}-n_{_2}-n_{_3})
\Gamma(1-{D\over2}-n_{_2}-n_{_4})
\nonumber\\
&&\hspace{2.5cm}\times
\Gamma({D\over2}-1-n_{_1})\Gamma({D\over2}+n_{_4})\Big\}^{-1}\;.
\label{GKZ21-1-3}
\end{eqnarray}
Obviously the intersection of the convergent regions of two hypergeometric series
is nonempty. Furthermore $\Phi_{_{[1357]}}^{(1),b}$ is equal to
$\Phi_{_{[1357]}}^{(1),a}$ up to a constant scalar multiple in the nonempty intersection
because they originate from the same exponent vector presented in Eq.(\ref{GKZ21-1-1}).
In a similar way, we can give other 46 hypergeometric solutions which
are consistent with the basis of the integer lattice $\mathbf{B}_{_{1357}}$.
In order to shorten the length of text, we collect all the hypergeometric solutions
of the integer lattice $\mathbf{B}_{_{1357}}$ together in the supplementary material.

Multiplying one or more of the row vectors of the matrix $\mathbf{B}_{_{1357}}$ by -1,
the induced integer matrix can also be chosen as a basis of the integer lattice
space of certain hypergeometric series. Taking 4 row vectors of the following matrix as the
basis of the integer lattice, for example,
\begin{eqnarray}
&&\mathbf{B}_{_{\tilde{1}357}}={\rm diag}(-1,1,1,1)\cdot\mathbf{B}_{_{1357}}
\nonumber\\
&&\hspace{1.0cm}=\left(\begin{array}{cccccccccccccc}
-1\;\;& -1\;\;& 0\;\;& 0\;\;& 0\;\;& 1\;\;& 0\;\;& 0\;\;& 0\;\;& 0\;\;& 1\;\;& 0\;\;& 0\;\;& 0\;\;\\
0\;\;& 0\;\;& 1\;\;& 1\;\;& 0\;\;& 1\;\;& 0\;\;& -1\;\;& -1\;\;& -1\;\;& 1\;\;& 0\;\;& -1\;\;& 0\;\;\\
0\;\;&0\;\;&0\;\;& 0\;\;& 1\;\;& 1\;\;& 0\;\;& 0\;\;& -1\;\;& -1\;\;& 1\;\;& 0\;\;& 0\;\;& -1\;\;\\
0\;\;&0\;\;& 0\;\;& 0\;\;& 0\;\;& 0\;\;& 1\;\;& -1\;\;& 0\;\;& 0\;\;& 0\;\;& 1\;\;& -1\;\;& 0\;\;
\end{array}\right)\;,
\label{GKZ21a}
\end{eqnarray}
one obtains 14 hypergeometric series solutions in the neighborhoods of
regular singularities $0$ and $\infty$ similarly.
In order to shorten the length of text, we collect the expressions in
the  supplementary material.

Choosing 4 row vectors of the following matrix as the
basis of the integer lattice,
\begin{eqnarray}
&&\mathbf{B}_{_{1\tilde{3}57}}={\rm diag}(1,-1,1,1)\cdot\mathbf{B}_{_{1357}}
\nonumber\\
&&\hspace{1.0cm}=\left(\begin{array}{cccccccccccccc}
1\;\;& 1\;\;& 0\;\;& 0\;\;& 0\;\;& -1\;\;& 0\;\;& 0\;\;& 0\;\;& 0\;\;& -1\;\;& 0\;\;& 0\;\;& 0\;\;\\
0\;\;& 0\;\;& -1\;\;& -1\;\;& 0\;\;& -1\;\;& 0\;\;& 1\;\;& 1\;\;& 1\;\;& -1\;\;& 0\;\;& 1\;\;& 0\;\;\\
0\;\;&0\;\;&0\;\;& 0\;\;& 1\;\;& 1\;\;& 0\;\;& 0\;\;& -1\;\;& -1\;\;& 1\;\;& 0\;\;& 0\;\;& -1\;\;\\
0\;\;&0\;\;& 0\;\;& 0\;\;& 0\;\;& 0\;\;& 1\;\;& -1\;\;& 0\;\;& 0\;\;& 0\;\;& 1\;\;& -1\;\;& 0\;\;
\end{array}\right)\;,
\label{GKZ21b}
\end{eqnarray}
we obtain 48 hypergeometric series solutions in the neighborhoods of regular singularities
$0$ and $\infty$. In order to shorten the length of text,
we collect the tedious expressions in the  supplementary material.

Taking 4 row vectors of the following matrix as the
basis of the integer lattice,
\begin{eqnarray}
&&\mathbf{B}_{_{13\tilde{5}7}}={\rm diag}(1,1,-1,1)\cdot\mathbf{B}_{_{1357}}
\nonumber\\
&&\hspace{1.0cm}=\left(\begin{array}{cccccccccccccc}
1\;\;& 1\;\;& 0\;\;& 0\;\;& 0\;\;& -1\;\;& 0\;\;& 0\;\;& 0\;\;& 0\;\;& -1\;\;& 0\;\;& 0\;\;& 0\;\;\\
0\;\;& 0\;\;& 1\;\;& 1\;\;& 0\;\;& 1\;\;& 0\;\;& -1\;\;& -1\;\;& -1\;\;& 1\;\;& 0\;\;& -1\;\;& 0\;\;\\
0\;\;&0\;\;&0\;\;& 0\;\;& -1\;\;& -1\;\;& 0\;\;& 0\;\;& 1\;\;& 1\;\;& -1\;\;& 0\;\;& 0\;\;& 1\;\;\\
0\;\;&0\;\;& 0\;\;& 0\;\;& 0\;\;& 0\;\;& 1\;\;& -1\;\;& 0\;\;& 0\;\;& 0\;\;& 1\;\;& -1\;\;& 0\;\;
\end{array}\right)\;,
\label{GKZ21c}
\end{eqnarray}
one derives 46 hypergeometric series solutions in the neighborhoods of regular singularities
$0$ and $\infty$. The concrete expressions are collected in the  supplementary material.

Taking 4 row vectors of the following matrix as the
basis of the integer lattice,
\begin{eqnarray}
&&\mathbf{B}_{_{135\tilde{7}}}={\rm diag}(1,1,1,-1)\cdot\mathbf{B}_{_{1357}}
\nonumber\\
&&\hspace{1.0cm}=\left(\begin{array}{cccccccccccccc}
1\;\;& 1\;\;& 0\;\;& 0\;\;& 0\;\;& -1\;\;& 0\;\;& 0\;\;& 0\;\;& 0\;\;& -1\;\;& 0\;\;& 0\;\;& 0\;\;\\
0\;\;& 0\;\;& 1\;\;& 1\;\;& 0\;\;& 1\;\;& 0\;\;& -1\;\;& -1\;\;& -1\;\;& 1\;\;& 0\;\;& -1\;\;& 0\;\;\\
0\;\;&0\;\;&0\;\;& 0\;\;& 1\;\;& 1\;\;& 0\;\;& 0\;\;& -1\;\;& -1\;\;& 1\;\;& 0\;\;& 0\;\;& -1\;\;\\
0\;\;&0\;\;& 0\;\;& 0\;\;& 0\;\;& 0\;\;& -1\;\;& 1\;\;& 0\;\;& 0\;\;& 0\;\;& -1\;\;& 1\;\;& 0\;\;
\end{array}\right)\;,
\label{GKZ21d}
\end{eqnarray}
one obtains 48 hypergeometric series solutions in the neighborhoods of regular singularities
$0$ and $\infty$. Similarly we collect those concrete expressions in the supplementary material.

Taking 4 row vectors of the following matrix as the basis of the integer lattice,
\begin{eqnarray}
&&\mathbf{B}_{_{\tilde{1}\tilde{3}57}}={\rm diag}(-1,-1,1,1)\cdot\mathbf{B}_{_{1357}}
\nonumber\\
&&\hspace{1.0cm}=\left(\begin{array}{cccccccccccccc}
-1\;\;& -1\;\;& 0\;\;& 0\;\;& 0\;\;& 1\;\;& 0\;\;& 0\;\;& 0\;\;& 0\;\;& 1\;\;& 0\;\;& 0\;\;& 0\;\;\\
0\;\;& 0\;\;& -1\;\;& -1\;\;& 0\;\;& -1\;\;& 0\;\;& 1\;\;& 1\;\;& 1\;\;& -1\;\;& 0\;\;& 1\;\;& 0\;\;\\
0\;\;&0\;\;&0\;\;& 0\;\;& 1\;\;& 1\;\;& 0\;\;& 0\;\;& -1\;\;& -1\;\;& 1\;\;& 0\;\;& 0\;\;& -1\;\;\\
0\;\;&0\;\;& 0\;\;& 0\;\;& 0\;\;& 0\;\;& 1\;\;& -1\;\;& 0\;\;& 0\;\;& 0\;\;& 1\;\;& -1\;\;& 0\;\;
\end{array}\right)\;,
\label{GKZ21e}
\end{eqnarray}
one derives 32 hypergeometric series solutions in the neighborhoods of regular singularities
$0$ and $\infty$, which are collected in the supplementary material.

Choosing 4 row vectors of the following matrix as the basis of the integer lattice,
\begin{eqnarray}
&&\mathbf{B}_{_{\tilde{1}3\tilde{5}7}}={\rm diag}(-1,1,-1,1)\cdot\mathbf{B}_{_{1357}}
\nonumber\\
&&\hspace{1.0cm}=\left(\begin{array}{cccccccccccccc}
-1\;\;& -1\;\;& 0\;\;& 0\;\;& 0\;\;& 1\;\;& 0\;\;& 0\;\;& 0\;\;& 0\;\;& 1\;\;& 0\;\;& 0\;\;& 0\;\;\\
0\;\;& 0\;\;& 1\;\;& 1\;\;& 0\;\;& 1\;\;& 0\;\;& -1\;\;& -1\;\;& -1\;\;& 1\;\;& 0\;\;& -1\;\;& 0\;\;\\
0\;\;&0\;\;&0\;\;& 0\;\;& -1\;\;& -1\;\;& 0\;\;& 0\;\;& 1\;\;& 1\;\;& -1\;\;& 0\;\;& 0\;\;& 1\;\;\\
0\;\;&0\;\;& 0\;\;& 0\;\;& 0\;\;& 0\;\;& 1\;\;& -1\;\;& 0\;\;& 0\;\;& 0\;\;& 1\;\;& -1\;\;& 0\;\;
\end{array}\right)\;,
\label{GKZ21f}
\end{eqnarray}
we obtain 38 hypergeometric series solutions in the neighborhoods of regular singularities
$0$ and $\infty$, whose concrete expressions are collected in the  supplementary material.

Choosing 4 row vectors of the following matrix as the basis of the integer lattice,
\begin{eqnarray}
&&\mathbf{B}_{_{\tilde{1}35\tilde{7}}}={\rm diag}(-1,1,1,-1)\cdot\mathbf{B}_{_{1357}}
\nonumber\\
&&\hspace{1.0cm}=\left(\begin{array}{cccccccccccccc}
-1\;\;& -1\;\;& 0\;\;& 0\;\;& 0\;\;& 1\;\;& 0\;\;& 0\;\;& 0\;\;& 0\;\;& 1\;\;& 0\;\;& 0\;\;& 0\;\;\\
0\;\;& 0\;\;& 1\;\;& 1\;\;& 0\;\;& 1\;\;& 0\;\;& -1\;\;& -1\;\;& -1\;\;& 1\;\;& 0\;\;& -1\;\;& 0\;\;\\
0\;\;&0\;\;&0\;\;& 0\;\;& 1\;\;& 1\;\;& 0\;\;& 0\;\;& -1\;\;& -1\;\;& 1\;\;& 0\;\;& 0\;\;& -1\;\;\\
0\;\;&0\;\;& 0\;\;& 0\;\;& 0\;\;& 0\;\;& -1\;\;& 1\;\;& 0\;\;& 0\;\;& 0\;\;& -1\;\;& 1\;\;& 0\;\;
\end{array}\right)\;,
\label{GKZ21g}
\end{eqnarray}
we construct 14 hypergeometric series solutions in the neighborhoods of regular singularities
$0$ and $\infty$, whose concrete expressions are collected in the  supplementary material.

Taking 4 row vectors of the following matrix as the basis of the integer lattice,
\begin{eqnarray}
&&\mathbf{B}_{_{1\tilde{3}\tilde{5}7}}={\rm diag}(1,-1,-1,1)\cdot\mathbf{B}_{_{1357}}
\nonumber\\
&&\hspace{1.0cm}=\left(\begin{array}{cccccccccccccc}
1\;\;& 1\;\;& 0\;\;& 0\;\;& 0\;\;& -1\;\;& 0\;\;& 0\;\;& 0\;\;& 0\;\;& -1\;\;& 0\;\;& 0\;\;& 0\;\;\\
0\;\;& 0\;\;& -1\;\;& -1\;\;& 0\;\;& -1\;\;& 0\;\;& 1\;\;& 1\;\;& 1\;\;& -1\;\;& 0\;\;& 1\;\;& 0\;\;\\
0\;\;&0\;\;&0\;\;& 0\;\;& -1\;\;& -1\;\;& 0\;\;& 0\;\;& 1\;\;& 1\;\;& -1\;\;& 0\;\;& 0\;\;& 1\;\;\\
0\;\;&0\;\;& 0\;\;& 0\;\;& 0\;\;& 0\;\;& 1\;\;& -1\;\;& 0\;\;& 0\;\;& 0\;\;& 1\;\;& -1\;\;& 0\;\;
\end{array}\right)\;,
\label{GKZ21h}
\end{eqnarray}
we obtain 46 hypergeometric series solutions in  the neighborhoods of regular singularities
$0$ and $\infty$. In order to shorten the length of text,
we put those tedious expressions in the supplementary material.

Choosing 4 row vectors of the following matrix as the basis of the integer lattice,
\begin{eqnarray}
&&\mathbf{B}_{_{1\tilde{3}5\tilde{7}}}={\rm diag}(1,-1,1,-1)\cdot\mathbf{B}_{_{1357}}
\nonumber\\
&&\hspace{1.0cm}=\left(\begin{array}{cccccccccccccc}
1\;\;& 1\;\;& 0\;\;& 0\;\;& 0\;\;& -1\;\;& 0\;\;& 0\;\;& 0\;\;& 0\;\;& -1\;\;& 0\;\;& 0\;\;& 0\;\;\\
0\;\;& 0\;\;& -1\;\;& -1\;\;& 0\;\;& -1\;\;& 0\;\;& 1\;\;& 1\;\;& 1\;\;& -1\;\;& 0\;\;& 1\;\;& 0\;\;\\
0\;\;&0\;\;&0\;\;& 0\;\;& 1\;\;& 1\;\;& 0\;\;& 0\;\;& -1\;\;& -1\;\;& 1\;\;& 0\;\;& 0\;\;& -1\;\;\\
0\;\;&0\;\;& 0\;\;& 0\;\;& 0\;\;& 0\;\;& -1\;\;& 1\;\;& 0\;\;& 0\;\;& 0\;\;& -1\;\;& 1\;\;& 0\;\;
\end{array}\right)\;,
\label{GKZ21i}
\end{eqnarray}
we obtain 14 hypergeometric series solutions in the neighborhoods of regular singularities
$0$ and $\infty$, whose concrete expressions are presented in the  supplementary material.

Choosing 4 row vectors of the following matrix as the basis of the integer lattice,
\begin{eqnarray}
&&\mathbf{B}_{_{13\tilde{5}\tilde{7}}}={\rm diag}(1,1,-1,-1)\cdot\mathbf{B}_{_{1357}}
\nonumber\\
&&\hspace{1.0cm}=\left(\begin{array}{cccccccccccccc}
1\;\;& 1\;\;& 0\;\;& 0\;\;& 0\;\;& -1\;\;& 0\;\;& 0\;\;& 0\;\;& 0\;\;& -1\;\;& 0\;\;& 0\;\;& 0\;\;\\
0\;\;& 0\;\;& 1\;\;& 1\;\;& 0\;\;& 1\;\;& 0\;\;& -1\;\;& -1\;\;& -1\;\;& 1\;\;& 0\;\;& -1\;\;& 0\;\;\\
0\;\;&0\;\;&0\;\;& 0\;\;& -1\;\;& -1\;\;& 0\;\;& 0\;\;& 1\;\;& 1\;\;& -1\;\;& 0\;\;& 0\;\;& 1\;\;\\
0\;\;&0\;\;& 0\;\;& 0\;\;& 0\;\;& 0\;\;& -1\;\;& 1\;\;& 0\;\;& 0\;\;& 0\;\;& -1\;\;& 1\;\;& 0\;\;
\end{array}\right)\;,
\label{GKZ21j}
\end{eqnarray}
one derives 46 hypergeometric series solutions in the neighborhoods of regular singularities
$0$ and $\infty$.
In order to shorten the length of text, we presented the concrete expressions in the
supplementary material.

Taking 4 row vectors of the following matrix as the basis of the integer lattice,
\begin{eqnarray}
&&\mathbf{B}_{_{\tilde{1}\tilde{3}\tilde{5}7}}={\rm diag}(-1,-1,-1,1)\cdot\mathbf{B}_{_{1357}}
\nonumber\\
&&\hspace{1.0cm}=\left(\begin{array}{cccccccccccccc}
-1\;\;& -1\;\;& 0\;\;& 0\;\;& 0\;\;& 1\;\;& 0\;\;& 0\;\;& 0\;\;& 0\;\;& 1\;\;& 0\;\;& 0\;\;& 0\;\;\\
0\;\;& 0\;\;& -1\;\;& -1\;\;& 0\;\;& -1\;\;& 0\;\;& 1\;\;& 1\;\;& 1\;\;& -1\;\;& 0\;\;& 1\;\;& 0\;\;\\
0\;\;&0\;\;&0\;\;& 0\;\;& -1\;\;& -1\;\;& 0\;\;& 0\;\;& 1\;\;& 1\;\;& -1\;\;& 0\;\;& 0\;\;& 1\;\;\\
0\;\;&0\;\;& 0\;\;& 0\;\;& 0\;\;& 0\;\;& 1\;\;& -1\;\;& 0\;\;& 0\;\;& 0\;\;& 1\;\;& -1\;\;& 0\;\;
\end{array}\right)\;,
\label{GKZ21k}
\end{eqnarray}
one obtains totally 46 hypergeometric series solutions in the neighborhoods of regular singularities
$0$ and $\infty$, whose concrete expressions are presented in the supplementary material.

Choosing 4 row vectors of the following matrix as the basis of the integer lattice,
\begin{eqnarray}
&&\mathbf{B}_{_{\tilde{1}\tilde{3}5\tilde{7}}}={\rm diag}(-1,-1,1,-1)\cdot\mathbf{B}_{_{1357}}
\nonumber\\
&&\hspace{1.0cm}=\left(\begin{array}{cccccccccccccc}
-1\;\;& -1\;\;& 0\;\;& 0\;\;& 0\;\;& 1\;\;& 0\;\;& 0\;\;& 0\;\;& 0\;\;& 1\;\;& 0\;\;& 0\;\;& 0\;\;\\
0\;\;& 0\;\;& -1\;\;& -1\;\;& 0\;\;& -1\;\;& 0\;\;& 1\;\;& 1\;\;& 1\;\;& -1\;\;& 0\;\;& 1\;\;& 0\;\;\\
0\;\;&0\;\;&0\;\;& 0\;\;& 1\;\;& 1\;\;& 0\;\;& 0\;\;& -1\;\;& -1\;\;& 1\;\;& 0\;\;& 0\;\;& -1\;\;\\
0\;\;&0\;\;& 0\;\;& 0\;\;& 0\;\;& 0\;\;& -1\;\;& 1\;\;& 0\;\;& 0\;\;& 0\;\;& -1\;\;& 1\;\;& 0\;\;
\end{array}\right)\;,
\label{GKZ21l}
\end{eqnarray}
one obtains 14 hypergeometric series solutions in the neighborhoods of regular singularities
$0$ and $\infty$, whose concrete expressions are presented in the supplementary material.

Choosing 4 row vectors of the following matrix as the basis of the integer lattice,
\begin{eqnarray}
&&\mathbf{B}_{_{\tilde{1}3\tilde{5}\tilde{7}}}={\rm diag}(-1,1,-1,-1)\cdot\mathbf{B}_{_{1357}}
\nonumber\\
&&\hspace{1.0cm}=\left(\begin{array}{cccccccccccccc}
-1\;\;& -1\;\;& 0\;\;& 0\;\;& 0\;\;& 1\;\;& 0\;\;& 0\;\;& 0\;\;& 0\;\;& 1\;\;& 0\;\;& 0\;\;& 0\;\;\\
0\;\;& 0\;\;& 1\;\;& 1\;\;& 0\;\;& 1\;\;& 0\;\;& -1\;\;& -1\;\;& -1\;\;& 1\;\;& 0\;\;& -1\;\;& 0\;\;\\
0\;\;&0\;\;&0\;\;& 0\;\;& -1\;\;& -1\;\;& 0\;\;& 0\;\;& 1\;\;& 1\;\;& -1\;\;& 0\;\;& 0\;\;& 1\;\;\\
0\;\;&0\;\;& 0\;\;& 0\;\;& 0\;\;& 0\;\;& -1\;\;& 1\;\;& 0\;\;& 0\;\;& 0\;\;& -1\;\;& 1\;\;& 0\;\;
\end{array}\right)\;,
\label{GKZ21m}
\end{eqnarray}
one derives totally 38 hypergeometric series solutions in the neighborhoods of regular singularities
$0$ and $\infty$. In order to shorten the length of text, we put those concrete expressions
in the supplementary material.

Choosing 4 row vectors of the following matrix as the basis of the integer lattice,
\begin{eqnarray}
&&\mathbf{B}_{_{1\tilde{3}\tilde{5}\tilde{7}}}={\rm diag}(1,-1,-1,-1)\cdot\mathbf{B}_{_{1357}}
\nonumber\\
&&\hspace{1.0cm}=\left(\begin{array}{cccccccccccccc}
1\;\;& 1\;\;& 0\;\;& 0\;\;& 0\;\;& -1\;\;& 0\;\;& 0\;\;& 0\;\;& 0\;\;& -1\;\;& 0\;\;& 0\;\;& 0\;\;\\
0\;\;& 0\;\;& -1\;\;& -1\;\;& 0\;\;& -1\;\;& 0\;\;& 1\;\;& 1\;\;& 1\;\;& -1\;\;& 0\;\;& 1\;\;& 0\;\;\\
0\;\;&0\;\;&0\;\;& 0\;\;& -1\;\;& -1\;\;& 0\;\;& 0\;\;& 1\;\;& 1\;\;& -1\;\;& 0\;\;& 0\;\;& 1\;\;\\
0\;\;&0\;\;& 0\;\;& 0\;\;& 0\;\;& 0\;\;& -1\;\;& 1\;\;& 0\;\;& 0\;\;& 0\;\;& -1\;\;& 1\;\;& 0\;\;
\end{array}\right)\;,
\label{GKZ21n}
\end{eqnarray}
one obtains 22 hypergeometric series solutions in the neighborhoods of regular singularities
$0$ and $\infty$.
In order to shorten the length of text, we present those concrete expressions
in the supplementary material.

Choosing 4 row vectors of the following matrix as the basis of the integer lattice,
\begin{eqnarray}
&&\mathbf{B}_{_{\tilde{1}\tilde{3}\tilde{5}\tilde{7}}}={\rm diag}(-1,-1,-1,-1)\cdot\mathbf{B}_{_{1357}}
\nonumber\\
&&\hspace{1.0cm}=\left(\begin{array}{cccccccccccccc}
-1\;\;& -1\;\;& 0\;\;& 0\;\;& 0\;\;& 1\;\;& 0\;\;& 0\;\;& 0\;\;& 0\;\;& 1\;\;& 0\;\;& 0\;\;& 0\;\;\\
0\;\;& 0\;\;& -1\;\;& -1\;\;& 0\;\;& -1\;\;& 0\;\;& 1\;\;& 1\;\;& 1\;\;& -1\;\;& 0\;\;& 1\;\;& 0\;\;\\
0\;\;&0\;\;&0\;\;& 0\;\;& -1\;\;& -1\;\;& 0\;\;& 0\;\;& 1\;\;& 1\;\;& -1\;\;& 0\;\;& 0\;\;& 1\;\;\\
0\;\;&0\;\;& 0\;\;& 0\;\;& 0\;\;& 0\;\;& -1\;\;& 1\;\;& 0\;\;& 0\;\;& 0\;\;& -1\;\;& 1\;\;& 0\;\;
\end{array}\right)\;,
\label{GKZ21o}
\end{eqnarray}
we obtain 22 hypergeometric series solutions in the neighborhoods of regular singularities
$0$ and $\infty$.
Similarly we present those tedious expressions in the supplementary material.

In summary, we construct totally 536 hypergeometric functions in some proper
subsets of the whole parameter space. In a proper subset of the whole parameter space,
we choose 30 linearly independent hypergeometric functions constituting the fundamental solution system.
The Feynman integral can be expressed as a linear combination of hypergeometric functions of
the fundamental solution system in the parameter space.
The combination coefficients can be uniquely determined by the values of the Feynman integral
together with its some partial derivatives at an ordinary point, or some values
at the regular singularities. In order to obtain the fundamental solution system of
the GKZ-system, we investigate the convergent region
of the hypergeometric function constructed above. For example, the convergent region
of the hypergeometric functions $\Phi_{_{[1357]}}^{(1),a}(\alpha,z)$ in Eq. (\ref{GKZ21-1-2a}) is
\begin{eqnarray}
&&\Xi_{_{[1357]}}^{1}=\{(y_{_1},\;y_{_2},\;y_{_3},\;y_{_4})
\Big|1<|y_{_4}|,\;|y_{_4}|<|y_{_3}|,\;|y_{_1}|<|y_{_4}|,\;|y_{_2}|<|y_{_3}|\}\;.
\label{GKZ22-1-1a}
\end{eqnarray}
In order to shorten the length of text, we present all  convergent regions
of the hypergeometric functions together in the supplementary material.

The above analysis implies that the fundamental solution system is composed of
30 linear independent hypergeometric solutions in a nonempty proper subset of
the whole parameter space, and the scalar integral
can be expressed as a linear combination of the 30 hypergeometric solutions.
Those nonempty proper subsets
together with the fundamental solution systems are presented as follows.
\begin{itemize}
\item In the nonempty proper subset of the whole parameter space,
\begin{eqnarray}
&&\hspace{-0.5cm}\Xi^{1}=\Xi_{_{[1357]}}^{1}\bigcap\Xi_{_{[1357]}}^{3}
\nonumber\\
&&\hspace{0.06cm}=\{(y_{_1},\;y_{_2},\;y_{_3},\;y_{_4})
\Big||y_{_2}|<|y_{_3}|,\;1<|y_{_1}|<|y_{_4}|<|y_{_3}|\}\;,
\label{GKZ22a-1}
\end{eqnarray}
the fundamental solution system is composed of the 30 hypergeometric functions
$\Phi_{_{[1357]}}^{(i),a}$ and $\Phi_{_{[1357]}}^{(j)}$, with
$i=1,\cdots,4,7,\cdots,13,15$, $j=5,6,17,\cdots,32$.

\item In the nonempty proper subset of the whole parameter space,
\begin{eqnarray}
&&\hspace{-0.5cm}\Xi^{2}=\Xi_{_{[1357]}}^{2}\bigcap\Xi_{_{[1357]}}^{3}
\bigcap\Xi_{_{[1357]}}^{5}\bigcap\Xi_{_{[\tilde{1}357]}}^{1}
\bigcap\Xi_{_{[1\tilde{3}57]}}^{6}\bigcap\Xi_{_{[13\tilde{5}7]}}^{3}
\nonumber\\
&&\hspace{0.06cm}=\{(y_{_1},\;y_{_2},\;y_{_3},\;y_{_4})
\Big||y_{_2}|<|y_{_1}|,\;1<|y_{_1}|<|y_{_4}|<|y_{_3}|\}\;,
\label{GKZ22a-2}
\end{eqnarray}
the fundamental solution system is composed of the 30 hypergeometric functions
$\Phi_{_{[1357]}}^{(i),b}$, $\Phi_{_{[1357]}}^{(j)}$,
$\Phi_{_{[1357]}}^{(k),c}$, $\Phi_{_{[\tilde{1}357]}}^{(l),a}$,
$\Phi_{_{[1\tilde{3}57]}}^{(m),b}$ and $\Phi_{_{[13\tilde{5}7]}}^{(n),c}$,
with $i=1,\cdots,4,7,8$, $j=5,6$, $k=9,10,11,13$, $l=1,\cdots,4,7,8$, $m=25,26,27,29,30,31$,
and $n=17,\cdots,20,23,24$, respectively.

\item In the nonempty proper subset of the whole parameter space,
\begin{eqnarray}
&&\hspace{-0.5cm}\Xi^{3}=\Xi_{_{[1357]}}^{2}
\bigcap\Xi_{_{[1357]}}^{5}\bigcap\Xi_{_{[1357]}}^{6}\bigcap\Xi_{_{[1\tilde{3}57]}}^{3}
\bigcap\Xi_{_{[1\tilde{3}57]}}^{6}\bigcap\Xi_{_{[135\tilde{7}]}}^{6}
\bigcap\Xi_{_{[\tilde{1}\tilde{3}57]}}^{2}
\nonumber\\
&&\hspace{0.06cm}=\{(y_{_1},\;y_{_2},\;y_{_3},\;y_{_4})
\Big||y_{_2}|<|y_{_1}|,\;1<|y_{_1}|<|y_{_3}|<|y_{_4}|\}\;,
\label{GKZ22a-3}
\end{eqnarray}
the fundamental solution system is composed of the 30 hypergeometric functions
$\Phi_{_{[1357]}}^{(i),b}$, $\Phi_{_{[1357]}}^{(j)}$, $\Phi_{_{[1357]}}^{(k),c}$,
$\Phi_{_{[1\tilde{3}57]}}^{(l),c}$, $\Phi_{_{[1\tilde{3}57]}}^{(m),b}$,
$\Phi_{_{[135\tilde{7}]}}^{(n),b}$ and $\Phi_{_{[\tilde{1}\tilde{3}57]}}^{(l),b}$,
with $i=1,\cdots,4,7,8$, $j=14,16$, $k=9,10,11,13$, $l=1,2,3,5$,
$m=25,26,27,29,30,31$, and $n=25,26,27,29$, respectively.

\item In the nonempty proper subset of the whole parameter space,
\begin{eqnarray}
&&\hspace{-0.5cm}\Xi^{4}=\Xi_{_{[1357]}}^{2}\bigcap\Xi_{_{[1357]}}^{5}
\bigcap\Xi_{_{[1357]}}^{7}\bigcap\Xi_{_{[1\tilde{3}57]}}^{4}
\bigcap\Xi_{_{[1\tilde{3}57]}}^{5}\bigcap\Xi_{_{[1\tilde{3}57]}}^{6}
\bigcap\Xi_{_{[135\tilde{7}]}}^{6}
\bigcap\Xi_{_{[\tilde{1}\tilde{3}57]}}^{2}
\nonumber\\
&&\hspace{0.06cm}=\{(y_{_1},\;y_{_2},\;y_{_3},\;y_{_4})
\Big||y_{_2}|<|y_{_3}|,\;1<|y_{_3}|<|y_{_1}|<|y_{_4}|\}\;,
\label{GKZ22a-4}
\end{eqnarray}
the fundamental solution system is composed of the 30 hypergeometric functions
$\Phi_{_{[1357]}}^{(i),b}$, $\Phi_{_{[1357]}}^{(j),c}$, $\Phi_{_{[1\tilde{3}57]}}^{(k)}$,
$\Phi_{_{[1\tilde{3}57]}}^{(l),b}$, $\Phi_{_{[135\tilde{7}]}}^{(m),b}$,
and $\Phi_{_{[\tilde{1}\tilde{3}57]}}^{(n),b}$,
where $i=1,\cdots,4,7,8,15$, $j=9,10,11,13$, $k=4,8$, $l=5,6,7,25,26,27,29,30,31$, $m=25,26,27,29$,
and $n=1,2,3,5$, respectively.

\item In the nonempty proper subset of the whole parameter space,
\begin{eqnarray}
&&\hspace{-0.5cm}\Xi^{5}=\Xi_{_{[1357]}}^{2}\bigcap\Xi_{_{[1357]}}^{3}
\bigcap\Xi_{_{[1357]}}^{5}\bigcap\Xi_{_{[1\tilde{3}57]}}^{6}
\bigcap\Xi_{_{[13\tilde{5}7]}}^{3}\bigcap\Xi_{_{[\tilde{1}3\tilde{5}7]}}^{2}
\nonumber\\
&&\hspace{0.06cm}=\{(y_{_1},\;y_{_2},\;y_{_3},\;y_{_4})
\Big||y_{_2}|<|y_{_1}|,\;1<|y_{_1}|<|y_{_4}|<|y_{_3}|\}\;,
\label{GKZ22a-5}
\end{eqnarray}
the fundamental solution system is composed of the 30 hypergeometric functions
$\Phi_{_{[1357]}}^{(i),b}$, $\Phi_{_{[1357]}}^{(j)}$, $\Phi_{_{[1357]}}^{(k),c}$,
$\Phi_{_{[1\tilde{3}57]}}^{(l),b}$, $\Phi_{_{[13\tilde{5}7]}}^{(m),c}$
and $\Phi_{_{[\tilde{1}3\tilde{5}7]}}^{(m),b}$,
where $i=1,\cdots,4,7,8$, $j=5,6$, $k=9,10,11,13$, $l=25,26,27,29,30,31$,
and $m=17,\cdots,20,23,24$, respectively.

\item In the nonempty proper subset of the whole parameter space,
\begin{eqnarray}
&&\hspace{-0.5cm}\Xi^{6}=\Xi_{_{[1357]}}^{2}
\bigcap\Xi_{_{[1\tilde{3}57]}}^{7}\bigcap\Xi_{_{[1\tilde{3}\tilde{5}7]}}^{2}
\bigcap\Xi_{_{[1\tilde{3}\tilde{5}7]}}^{5}\bigcap\Xi_{_{[1\tilde{3}\tilde{5}\tilde{7}]}}^{2}
\nonumber\\
&&\hspace{0.06cm}=\{(y_{_1},\;y_{_2},\;y_{_3},\;y_{_4})
\Big||y_{_2}|<|y_{_3}|,\;1<|y_{_3}|<|y_{_4}|<|y_{_1}|\}\;,
\label{GKZ22a-6}
\end{eqnarray}
the fundamental solution system is composed of the 30 hypergeometric functions
$\Phi_{_{[1357]}}^{(i),b}$, $\Phi_{_{[1\tilde{3}57]}}^{(j)}$,
$\Phi_{_{[1\tilde{3}\tilde{5}7]}}^{(k)}$, $\Phi_{_{[1\tilde{3}\tilde{5}7]}}^{(l),b}$,
and $\Phi_{_{[1\tilde{3}\tilde{5}\tilde{7}]}}^{(m),b}$,
where $i=1,\cdots,4,7,8$, $j=28,32$, $k=17,\cdots,24$, $l=25,\cdots,32$,
and $m=1,2,3,5,6,7$, respectively.

\item In the nonempty proper subset of the whole parameter space,
\begin{eqnarray}
&&\hspace{-0.5cm}\Xi^{7}=\Xi_{_{[1357]}}^{4}\bigcap\Xi_{_{[1357]}}^{7}
\bigcap\Xi_{_{[\tilde{1}357]}}^{2}\bigcap\Xi_{_{[135\tilde{7}]}}^{6}
\bigcap\Xi_{_{[\tilde{1}\tilde{3}57]}}^{3}\bigcap\Xi_{_{[\tilde{1}\tilde{3}57]}}^{4}
\bigcap\Xi_{_{[\tilde{1}\tilde{3}57]}}^{5}\bigcap\Xi_{_{[\tilde{1}\tilde{3}57]}}^{6}
\nonumber\\
&&\hspace{0.06cm}=\{(y_{_1},\;y_{_2},\;y_{_3},\;y_{_4})
\Big||y_{_1}|<|y_{_3}|,\;|y_{_2}|<|y_{_3}|<1<|y_{_4}|\}\;,
\label{GKZ22a-7}
\end{eqnarray}
the fundamental solution system is composed of the 30 hypergeometric functions
$\Phi_{_{[1357]}}^{(i),b}$, $\Phi_{_{[\tilde{1}357]}}^{(j),b}$, $\Phi_{_{[135\tilde{7}]}}^{(k),b}$ and $\Phi_{_{[\tilde{1}\tilde{3}57]}}^{(l),c}$, $\Phi_{_{[\tilde{1}\tilde{3}57]}}^{(m)}$,
$\Phi_{_{[\tilde{1}\tilde{3}57]}}^{(n),b}$,
where $i=9,\cdots,13,15$, $j=1,\cdots,4,7,8$, $k=25,26,27,29$, $l=1,2,3,5$,
$m=4,8$, and $n=6,7,9,10,11,13,14,15$, respectively.

\item In the nonempty proper subset of the whole parameter space,
\begin{eqnarray}
&&\hspace{-0.5cm}\Xi^{8}=\Xi_{_{[1357]}}^{4}\bigcap\Xi_{_{[1\tilde{3}57]}}^{2}
\bigcap\Xi_{_{[135\tilde{7}]}}^{4}\bigcap\Xi_{_{[\tilde{1}\tilde{3}57]}}^{2}
\nonumber\\
&&\hspace{0.06cm}=\{(y_{_1},\;y_{_2},\;y_{_3},\;y_{_4})
\Big|1<|y_{_4}|,\;|y_{_1}|<|y_{_4}|,\;|y_{_3}|<|y_{_2}|<|y_{_4}||\}\;,
\label{GKZ22a-8}
\end{eqnarray}
the fundamental solution system is composed of the 30 hypergeometric functions
$\Phi_{_{[1357]}}^{(i),b}$, $\Phi_{_{[1\tilde{3}57]}}^{(j),b}$,
$\Phi_{_{[1\tilde{3}57]}}^{(k)}$, $\Phi_{_{[135\tilde{7}]}}^{(l),b}$
and $\Phi_{_{[\tilde{1}\tilde{3}57]}}^{(m),b}$, where $i=9,\cdots,13$,
$j=1,2,3$, $k=9,\cdots,24$, $l=28,30$, and $m=1,2,3,5$, respectively.

\item In the nonempty proper subset of the whole parameter space,
\begin{eqnarray}
&&\hspace{-0.5cm}\Xi^{9}=\Xi_{_{[1357]}}^{4}\bigcap\Xi_{_{[1\tilde{3}57]}}^{5}
\bigcap\Xi_{_{[135\tilde{7}]}}^{2}\bigcap\Xi_{_{[135\tilde{7}]}}^{6}
\bigcap\Xi_{_{[135\tilde{7}]}}^{7}\bigcap\Xi_{_{[\tilde{1}\tilde{3}57]}}^{2}
\bigcap\Xi_{_{[1\tilde{3}5\tilde{7}]}}^{2}
\nonumber\\
&&\hspace{0.06cm}=\{(y_{_1},\;y_{_2},\;y_{_3},\;y_{_4})
\Big||y_{_3}|<|y_{_1}|,\;1<|y_{_1}|<|y_{_2}|<|y_{_4}|\}\;,
\label{GKZ22a-9}
\end{eqnarray}
the fundamental solution system is composed of the 30 hypergeometric functions
$\Phi_{_{[1357]}}^{(i),b}$, $\Phi_{_{[135\tilde{7}]}}^{(j),b}$,
$\Phi_{_{[135\tilde{7}]}}^{(k)}$, $\Phi_{_{[1\tilde{3}57]}}^{(l),b}$,
$\Phi_{_{[\tilde{1}\tilde{3}57]}}^{(m),b}$ and
$\Phi_{_{[1\tilde{3}5\tilde{7}]}}^{(n),b}$,
where $i=9,\cdots,13$, $j=1,2,3,4,7,8,25,26,27,29$, $k=31,32$,
$l=5,6,7$, $m=1,2,3,5$, and $n=1,2,3,5,6,7$, respectively.

\item In the nonempty proper subset of the whole parameter space,
\begin{eqnarray}
&&\hspace{-0.5cm}\Xi^{10}=\Xi_{_{[1357]}}^{4}\bigcap\Xi_{_{[1\tilde{3}57]}}^{2}
\bigcap\Xi_{_{[\tilde{1}35\tilde{7}]}}^{2}
\nonumber\\
&&\hspace{0.21cm}=\{(y_{_1},\;y_{_2},\;y_{_3},\;y_{_4})
\Big||y_{_1}|<1<|y_{_4}|,\;|y_{_3}|<|y_{_2}|<|y_{_4}|,\;|y_{_1}|<|y_{_2}|\}\;,
\label{GKZ22a-10}
\end{eqnarray}
the fundamental solution system is composed of the 30 hypergeometric functions
$\Phi_{_{[1357]}}^{(i),b}$, $\Phi_{_{[1\tilde{3}57]}}^{(j),b}$,
$\Phi_{_{[1\tilde{3}57]}}^{(k)}$, and $\Phi_{_{[\tilde{1}35\tilde{7}]}}^{(l),b}$,
where $i=9,\cdots,13$, $j=1,2,3$, $k=9,\cdots,24$, $l=1,\cdots,4,7,8$, respectively.

\item In the nonempty proper subset of the whole parameter space,
\begin{eqnarray}
&&\hspace{-0.5cm}\Xi^{11}=\Xi_{_{[1357]}}^{4}\bigcap\Xi_{_{[1\tilde{3}57]}}^{2}
\bigcap\Xi_{_{[\tilde{1}35\tilde{7}]}}^{3}\bigcap\Xi_{_{[\tilde{1}\tilde{3}57]}}^{3}
\nonumber\\
&&\hspace{0.21cm}=\{(y_{_1},\;y_{_2},\;y_{_3},\;y_{_4})
\Big||y_{_1}|<1<|y_{_2}|,\;|y_{_3}|<|y_{_2}|<|y_{_4}|\}\;,
\label{GKZ22a-11}
\end{eqnarray}
the fundamental solution system is composed of the 30 hypergeometric functions
$\Phi_{_{[1357]}}^{(i),b}$, $\Phi_{_{[1\tilde{3}57]}}^{(j),b}$,
$\Phi_{_{[1\tilde{3}57]}}^{(k)}$, $\Phi_{_{[\tilde{1}35\tilde{7}]}}^{(l),b}$
and $\Phi_{_{[\tilde{1}\tilde{3}57]}}^{(m),c}$, where $i=9,\cdots,13$,
$j=1,2,3$, $k=9,\cdots,24$, $l=5,6$, and $m=1,2,3,5$, respectively.

\item In the nonempty proper subset of the whole parameter space,
\begin{eqnarray}
&&\hspace{-0.5cm}\Xi^{12}=\Xi_{_{[\tilde{1}357]}}^{2}
\bigcap\Xi_{_{[\tilde{1}357]}}^{3}\bigcap\Xi_{_{[13\tilde{5}7]}}^{2}
\bigcap\Xi_{_{[\tilde{1}\tilde{3}57]}}^{6}
\nonumber\\
&&\hspace{0.21cm}=\{(y_{_1},\;y_{_2},\;y_{_3},\;y_{_4})
\Big||y_{_1}|<1,\;|y_{_2}|<1<|y_{_4}|<|y_{_3}|\}\;,
\label{GKZ22a-12}
\end{eqnarray}
the fundamental solution system is composed of the 30 hypergeometric functions
$\Phi_{_{[\tilde{1}357]}}^{(i),b}$, $\Phi_{_{[\tilde{1}357]}}^{(j)}$,
$\Phi_{_{[13\tilde{5}7]}}^{(k),b}$, $\Phi_{_{[13\tilde{5}7]}}^{(l)}$,
and $\Phi_{_{[\tilde{1}\tilde{3}57]}}^{(m),b}$, where
$i=1,\cdots,4,7,8$, $j=5,6$, $k=17,\cdots,24$, $l=25,\cdots,32$,
and $m=9,10,11,13,14,15$, respectively.

\item In the nonempty proper subset of the whole parameter space,
\begin{eqnarray}
&&\hspace{-0.5cm}\Xi^{13}=\Xi_{_{[\tilde{1}357]}}^{2}
\bigcap\Xi_{_{[\tilde{1}\tilde{3}57]}}^{3}
\bigcap\Xi_{_{[\tilde{1}\tilde{3}\tilde{5}7]}}^{2}
\bigcap\Xi_{_{[\tilde{1}\tilde{3}\tilde{5}\tilde{7}]}}^{2}
\nonumber\\
&&\hspace{0.21cm}=\{(y_{_1},\;y_{_2},\;y_{_3},\;y_{_4})
\Big||y_{_1}|<|y_{_3}|,\;|y_{_2}|<|y_{_3}|<|y_{_4}|<1\}\;,
\label{GKZ22a-13}
\end{eqnarray}
the fundamental solution system is composed of the 30 hypergeometric functions
$\Phi_{_{[\tilde{1}357]}}^{(i),b}$,  $\Phi_{_{[\tilde{1}\tilde{3}57]}}^{(j),c}$, $\Phi_{_{[\tilde{1}\tilde{3}\tilde{5}7]}}^{(k)}$,
$\Phi_{_{[\tilde{1}\tilde{3}\tilde{5}7]}}^{(l),b}$, and
$\Phi_{_{[\tilde{1}\tilde{3}\tilde{5}\tilde{7}]}}^{(m),b}$,
where $i=1,\cdots,4,7,8$, $j=1,2,3,5$, $k=17,\cdots,24$, $l=25,26,27,29,30,31$,
and $m=1,2,3,5,6,7$, respectively.

\item In the nonempty proper subset of the whole parameter space,
\begin{eqnarray}
&&\hspace{-0.5cm}\Xi^{14}=\Xi_{_{[\tilde{1}357]}}^{2}
\bigcap\Xi_{_{[\tilde{1}\tilde{3}57]}}^{3}\bigcap\Xi_{_{[\tilde{1}\tilde{3}57]}}^{6}
\bigcap\Xi_{_{[\tilde{1}\tilde{3}\tilde{5}7]}}^{4}
\bigcap\Xi_{_{[\tilde{1}\tilde{3}\tilde{5}\tilde{7}]}}^{2}
\nonumber\\
&&\hspace{0.21cm}=\{(y_{_1},\;y_{_2},\;y_{_3},\;y_{_4})
\Big||y_{_2}|<|y_{_3}|,\;|y_{_1}|<|y_{_3}|<|y_{_4}|<1\}\;,
\label{GKZ22a-14}
\end{eqnarray}
the fundamental solution system is composed of the 30 hypergeometric functions
$\Phi_{_{[\tilde{1}357]}}^{(i),b}$,  $\Phi_{_{[\tilde{1}\tilde{3}57]}}^{(j),c}$,
$\Phi_{_{[\tilde{1}\tilde{3}57]}}^{(k),b}$,
$\Phi_{_{[\tilde{1}\tilde{3}\tilde{5}7]}}^{(l),c}$,
$\Phi_{_{[\tilde{1}\tilde{3}\tilde{5}7]}}^{(m),b}$, and
$\Phi_{_{[\tilde{1}\tilde{3}\tilde{5}\tilde{7}]}}^{(n),b}$,
where $i=1,\cdots,4,7,8$, $j=1,2,3,5$, $k=9,10,11,13,14,15$,
$l=25,26,26,29,30,31$, $m=28,32$, and $n=1,2,3,5,6,7$, respectively.

\item In the nonempty proper subset of the whole parameter space,
\begin{eqnarray}
&&\hspace{-0.5cm}\Xi^{15}=\Xi_{_{[1\tilde{3}57]}}^{1}
\bigcap\Xi_{_{[1\tilde{3}57]}}^{7}
\bigcap\Xi_{_{[1\tilde{3}\tilde{5}7]}}^{1}
\nonumber\\
&&\hspace{0.21cm}=\{(y_{_1},\;y_{_2},\;y_{_3},\;y_{_4})
\Big|1<|y_{_1}|,\;|y_{_2}|<|y_{_3}|<|y_{_4}|<|y_{_1}|\}\;,
\label{GKZ22a-15}
\end{eqnarray}
the fundamental solution system is composed of the 30 hypergeometric functions
$\Phi_{_{[1\tilde{3}57]}}^{(i),a}$, $\Phi_{_{[1\tilde{3}57]}}^{(j)}$,
and $\Phi_{_{[1\tilde{3}\tilde{5}7]}}^{(1-16)}$,
where $i=1,2,3,5,6,7,25,26,27,29,30,31$, $j=28,32$, and $k=1,\cdots,16$, respectively.

\item In the nonempty proper subset of the whole parameter space,
\begin{eqnarray}
&&\hspace{-0.5cm}\Xi^{16}=\Xi_{_{[1\tilde{3}57]}}^{3}
\bigcap\Xi_{_{[1\tilde{3}57]}}^{4}\bigcap\Xi_{_{[135\tilde{7}]}}^{2}
\bigcap\Xi_{_{[135\tilde{7}]}}^{4}\bigcap\Xi_{_{[135\tilde{7}]}}^{6}
\bigcap\Xi_{_{[\tilde{1}\tilde{3}57]}}^{2}
\bigcap\Xi_{_{[1\tilde{3}5\tilde{7}]}}^{2}
\bigcap\Xi_{_{[1\tilde{3}5\tilde{7}]}}^{3}
\nonumber\\
&&\hspace{0.21cm}=\{(y_{_1},\;y_{_2},\;y_{_3},\;y_{_4})
\Big||y_{_3}|<|y_{_2}|,\;1<|y_{_2}|<|y_{_1}|<|y_{_4}|\}\;,
\label{GKZ22a-16}
\end{eqnarray}
the fundamental solution system is composed of the 30 hypergeometric functions
$\Phi_{_{[1\tilde{3}57]}}^{(i),c}$, $\Phi_{_{[1\tilde{3}57]}}^{(j)}$,
$\Phi_{_{[135\tilde{7}]}}^{(k),b}$, $\Phi_{_{[\tilde{1}\tilde{3}57]}}^{(i),b}$,
$\Phi_{_{[1\tilde{3}5\tilde{7}]}}^{(l),b}$,
and $\Phi_{_{[1\tilde{3}5\tilde{7}]}}^{(m)}$,
where $i=1,2,3,5$, $j=4,8$, $k=1,\cdots,4,7,8,25,\cdots,30$, $l=1,2,3,5,6,7$,
and $m=4,8$, respectively.

\item In the nonempty proper subset of the whole parameter space,
\begin{eqnarray}
&&\hspace{-0.5cm}\Xi^{17}=\Xi_{_{[1\tilde{3}57]}}^{6}
\bigcap\Xi_{_{[1\tilde{3}57]}}^{7}
\bigcap\Xi_{_{[1\tilde{3}\tilde{5}7]}}^{1}
\bigcap\Xi_{_{[1\tilde{3}\tilde{5}7]}}^{4}
\bigcap\Xi_{_{[1\tilde{3}\tilde{5}7]}}^{6}
\nonumber\\
&&\hspace{0.21cm}=\{(y_{_1},\;y_{_2},\;y_{_3},\;y_{_4})
\Big||y_{_2}|<|y_{_3}|<|y_{_4}|<1<|y_{_1}|\}\;,
\label{GKZ22a-17}
\end{eqnarray}
the fundamental solution system is composed of the 30 hypergeometric functions
$\Phi_{_{[1\tilde{3}57]}}^{(i),b}$, $\Phi_{_{[1\tilde{3}57]}}^{(j)}$,
$\Phi_{_{[1\tilde{3}\tilde{5}7]}}^{(k)}$, and $\Phi_{_{[1\tilde{3}\tilde{5}7]}}^{(i),c}$,
where $i=25,26,27,29,30,31$, $j=28,32$, and $k=1,\cdots,16$, respectively.

\item In the nonempty proper subset of the whole parameter space,
\begin{eqnarray}
&&\hspace{-0.5cm}\Xi^{18}=\Xi_{_{[1\tilde{3}57]}}^{6}
\bigcap\Xi_{_{[13\tilde{5}7]}}^{1}
\nonumber\\
&&\hspace{0.21cm}=\{(y_{_1},\;y_{_2},\;y_{_3},\;y_{_4})
\Big||y_{_2}|<|y_{_4}|<|y_{_3}|,\;|y_{_4}|<|y_{_1}|,\;1<|y_{_1}|\}\;,
\label{GKZ22a-18}
\end{eqnarray}
the fundamental solution system is composed of the 30 hypergeometric functions
$\Phi_{_{[1\tilde{3}57]}}^{(i),b}$, $\Phi_{_{[13\tilde{5}7]}}^{(j)}$,
and $\Phi_{_{[13\tilde{5}7]}}^{(k),a}$,
where $i=25,26,27,29,30,31$, $j=1,\cdots,16$, and $k=17,\cdots,24$, respectively.

\item In the nonempty proper subset of the whole parameter space,
\begin{eqnarray}
&&\hspace{-0.5cm}\Xi^{19}=\Xi_{_{[1\tilde{3}57]}}^{6}
\bigcap\Xi_{_{[13\tilde{5}7]}}^{2}\bigcap\Xi_{_{[1\tilde{3}\tilde{5}7]}}^{3}
\nonumber\\
&&\hspace{0.21cm}=\{(y_{_1},\;y_{_2},\;y_{_3},\;y_{_4})
\Big|1<|y_{_1}|,\;|y_{_2}|<|y_{_4}|<|y_{_1}|<|y_{_3}|\}\;,
\label{GKZ22a-19}
\end{eqnarray}
the fundamental solution system is composed of the 30 hypergeometric functions
$\Phi_{_{[1\tilde{3}57]}}^{(i),b}$, $\Phi_{_{[13\tilde{5}7]}}^{(j),b}$,
$\Phi_{_{[13\tilde{5}7]}}^{(k)}$, and $\Phi_{_{[1\tilde{3}\tilde{5}7]}}^{(k),a}$,
where $i=25,26,27,29,30,31$, $j=17,\cdots,24$, $k=25,\cdots,32$, respectively.

\item In the nonempty proper subset of the whole parameter space,
\begin{eqnarray}
&&\hspace{-0.5cm}\Xi^{20}\equiv\Xi_{_{[13\tilde{5}7]}}^{2}
\bigcap\Xi_{_{[\tilde{1}\tilde{3}57]}}^{6}
\bigcap\Xi_{_{[\tilde{1}3\tilde{5}7]}}^{3}
\bigcap\Xi_{_{[\tilde{1}3\tilde{5}7]}}^{4}
\nonumber\\
&&\hspace{0.21cm}=\{(y_{_1},\;y_{_2},\;y_{_3},\;y_{_4})
\Big||y_{_1}|<1,\;|y_{_2}|<1,\;|y_{_2}|<|y_{_4}|<|y_{_3}|,\;1<|y_{_3}|\}\;,
\label{GKZ22a-20}
\end{eqnarray}
the fundamental solution system is composed of the 30 hypergeometric functions
$\Phi_{_{[13\tilde{5}7]}}^{(i),b}$, $\Phi_{_{[13\tilde{5}7]}}^{(j)}$,
$\Phi_{_{[\tilde{1}\tilde{3}57]}}^{(k),b}$,
$\Phi_{_{[\tilde{1}3\tilde{5}7]}}^{(l),c}$,
and $\Phi_{_{[\tilde{1}3\tilde{5}7]}}^{(m),b}$,
where $i=17,\cdots,24$, $j=25,\cdots,32$, $k=9,10,11,13,14,15$,
$l=17,\cdots,20,23,24$, and $m=21,22$, respectively.

\item In the nonempty proper subset of the whole parameter space,
\begin{eqnarray}
&&\hspace{-0.5cm}\Xi^{21}\equiv\Xi_{_{[13\tilde{5}7]}}^{2}
\bigcap\Xi_{_{[\tilde{1}\tilde{3}57]}}^{6}
\bigcap\Xi_{_{[\tilde{1}\tilde{3}\tilde{5}7]}}^{3}
\nonumber\\
&&\hspace{0.21cm}=\{(y_{_1},\;y_{_2},\;y_{_3},\;y_{_4})
\Big||y_{_1}|<1,\;|y_{_2}|<|y_{_4}|<1<|y_{_3}|\}\;,
\label{GKZ22a-21}
\end{eqnarray}
the fundamental solution system is composed of the 30 hypergeometric functions
$\Phi_{_{[13\tilde{5}7]}}^{(i),b}$, $\Phi_{_{[13\tilde{5}7]}}^{(j)}$,
$\Phi_{_{[\tilde{1}\tilde{3}57]}}^{(k),b}$ and $\Phi_{_{[\tilde{1}\tilde{3}\tilde{5}7]}}^{(j),b}$,
where $i=17,\cdots,24$, $j=25,\cdots,32$, $k=9,10,11,13,14,15$, respectively.

\item In the nonempty proper subset of the whole parameter space,
\begin{eqnarray}
&&\hspace{-0.5cm}\Xi^{22}=\Xi_{_{[135\tilde{7}]}}^{1}
\bigcap\Xi_{_{[135\tilde{7}]}}^{3}
\nonumber\\
&&\hspace{0.21cm}=\{(y_{_1},\;y_{_2},\;y_{_3},\;y_{_4})
\Big||y_{_3}|<|y_{_2}|,\;1<|y_{_1}|<|y_{_4}|<|y_{_2}|\}\;,
\label{GKZ22a-22}
\end{eqnarray}
the fundamental solution system is composed of the 30 hypergeometric functions
$\Phi_{_{[135\tilde{7}]}}^{(i),a}$, $\Phi_{_{[135\tilde{7}]}}^{(j)}$,
where $i=1,\cdots,4,7,8,25,\cdots,30$, and $j=5,6,9,\cdots,24$, respectively.

\item In the nonempty proper subset of the whole parameter space,
\begin{eqnarray}
&&\hspace{-0.5cm}\Xi^{23}=\Xi_{_{[135\tilde{7}]}}^{2}
\bigcap\Xi_{_{[1\tilde{3}\tilde{5}7]}}^{1}\bigcap\Xi_{_{[1\tilde{3}5\tilde{7}]}}^{1}
\bigcap\Xi_{_{[1\tilde{3}5\tilde{7}]}}^{3}
\nonumber\\
&&\hspace{0.21cm}=\{(y_{_1},\;y_{_2},\;y_{_3},\;y_{_4})
\Big||y_{_3}|<|y_{_2}|,\;1<|y_{_2}|<|y_{_4}|<|y_{_1}|\}\;,
\label{GKZ22a-23}
\end{eqnarray}
the fundamental solution system is composed of the 30 hypergeometric functions
$\Phi_{_{[135\tilde{7}]}}^{(i),b}$, $\Phi_{_{[1\tilde{3}\tilde{5}7]}}^{(j)}$,
$\Phi_{_{[1\tilde{3}5\tilde{7}]}}^{(k),a}$, and $\Phi_{_{[1\tilde{3}5\tilde{7}]}}^{(l)}$,
where $i=1,\cdots,4,7,8$, $j=1,\cdots,16$, $k=1,2,3,5,6,7$,
and $l=4,8$, respectively.

\item In the nonempty proper subset of the whole parameter space,
\begin{eqnarray}
&&\hspace{-0.5cm}\Xi^{24}=\Xi_{_{[135\tilde{7}]}}^{2}
\Xi_{_{[135\tilde{7}]}}^{3}\bigcap\Xi_{_{[135\tilde{7}]}}^{5}
\bigcap\Xi_{_{[1\tilde{3}5\tilde{7}]}}^{2}
\bigcap\Xi_{_{[13\tilde{5}\tilde{7}]}}^{3}
\bigcap\Xi_{_{[\tilde{1}3\tilde{5}\tilde{7}]}}^{2}
\nonumber\\
&&\hspace{0.21cm}=\{(y_{_1},\;y_{_2},\;y_{_3},\;y_{_4})
\Big||y_{_3}|<|y_{_1}|,\;1<|y_{_1}|<|y_{_4}|<|y_{_2}|\}\;,
\label{GKZ22a-24}
\end{eqnarray}
the fundamental solution system is composed of the 30 hypergeometric functions
$\Phi_{_{[135\tilde{7}]}}^{(i),b}$, $\Phi_{_{[135\tilde{7}]}}^{(j)}$,
$\Phi_{_{[135\tilde{7}]}}^{(k),c}$, $\Phi_{_{[1\tilde{3}5\tilde{7}]}}^{(l),b}$,
$\Phi_{_{[13\tilde{5}\tilde{7}]}}^{(m),c}$, and $\Phi_{_{[\tilde{1}3\tilde{5}\tilde{7}]}}^{(m),b}$,
where $i=1,\cdots,4,7,8$, $j=5,6$, $k=25,\cdots,27,29$, $l=1,2,3,5,6,7$,
and $m=17,\cdots,20,23,24$, respectively.

\item In the nonempty proper subset of the whole parameter space,
\begin{eqnarray}
&&\hspace{-0.5cm}\Xi^{25}=\Xi_{_{[135\tilde{7}]}}^{2}
\bigcap\Xi_{_{[13\tilde{5}\tilde{7}]}}^{1}
\nonumber\\
&&\hspace{0.21cm}=\{(y_{_1},\;y_{_2},\;y_{_3},\;y_{_4})
\Big|1<|y_{_4}|,\;|y_{_4}|<|y_{_1}|,\;|y_{_4}|<|y_{_2}|,\;|y_{_3}|<|y_{_2}|\}\;,
\label{GKZ22a-25}
\end{eqnarray}
the fundamental solution system is composed of the 30 hypergeometric functions
$\Phi_{_{[135\tilde{7}]}}^{(i),b}$, $\Phi_{_{[13\tilde{5}\tilde{7}]}}^{(j)}$,
and $\Phi_{_{[13\tilde{5}\tilde{7}]}}^{(k),a}$,
where $i=1,\cdots,4,7,8$, $j=1,\cdots,16$, and $k=17,\cdots,24$, respectively.

\item In the nonempty proper subset of the whole parameter space,
\begin{eqnarray}
&&\hspace{-0.5cm}\Xi^{26}=\Xi_{_{[135\tilde{7}]}}^{2}
\Xi_{_{[135\tilde{7}]}}^{3}
\bigcap\Xi_{_{[13\tilde{5}\tilde{7}]}}^{2}
\bigcap\Xi_{_{[13\tilde{5}\tilde{7}]}}^{4}
\bigcap\Xi_{_{[\tilde{1}3\tilde{5}\tilde{7}]}}^{2}
\nonumber\\
&&\hspace{0.21cm}=\{(y_{_1},\;y_{_2},\;y_{_3},\;y_{_4})
\Big||y_{_3}|<|y_{_2}|,\;1<|y_{_1}|<|y_{_2}|,\;1<|y_{_4}|<|y_{_2}|\}\;,
\label{GKZ22a-26}
\end{eqnarray}
the fundamental solution system is composed of the 30 hypergeometric functions
$\Phi_{_{[135\tilde{7}]}}^{(i),b}$, $\Phi_{_{[135\tilde{7}]}}^{(j)}$,
$\Phi_{_{[13\tilde{5}\tilde{7}]}}^{(k),b}$, $\Phi_{_{[13\tilde{5}\tilde{7}]}}^{(l)}$,
and $\Phi_{_{[\tilde{1}3\tilde{5}\tilde{7}]}}^{(m),b}$,
where $i=1,\cdots,4,7,8$, $j=5,6$, $k=17,\cdots,24$, $l=25,\cdots,32$,
and $m=17,\cdots,20,23,24$, respectively.

\item In the nonempty proper subset of the whole parameter space,
\begin{eqnarray}
&&\hspace{-0.5cm}\Xi^{27}=\Xi_{_{[135\tilde{7}]}}^{2}
\Xi_{_{[135\tilde{7}]}}^{3}\bigcap\Xi_{_{[\tilde{1}35\tilde{7}]}}^{1}
\bigcap\Xi_{_{[13\tilde{5}\tilde{7}]}}^{2}
\bigcap\Xi_{_{[13\tilde{5}\tilde{7}]}}^{4}
\nonumber\\
&&\hspace{0.21cm}=\{(y_{_1},\;y_{_2},\;y_{_3},\;y_{_4})
\Big||y_{_3}|<|y_{_2}|,\;1<|y_{_1}|<|y_{_4}|<|y_{_2}|\}\;,
\label{GKZ22a-27}
\end{eqnarray}
the fundamental solution system is composed of the 30 hypergeometric functions
$\Phi_{_{[135\tilde{7}]}}^{(i),b}$, $\Phi_{_{[135\tilde{7}]}}^{(j)}$,
$\Phi_{_{[\tilde{1}35\tilde{7}]}}^{(i),b}$, $\Phi_{_{[13\tilde{5}\tilde{7}]}}^{(k),b}$,
and $\Phi_{_{[13\tilde{5}\tilde{7}]}}^{(l)}$,
where $i=1,\cdots,4,7,8$, $j=5,6$, $k=17,\cdots,24$, and $l=25,\cdots,32$, respectively.

\item In the nonempty proper subset of the whole parameter space,
\begin{eqnarray}
&&\hspace{-0.5cm}\Xi^{28}=\Xi_{_{[135\tilde{7}]}}^{2}
\bigcap\Xi_{_{[1\tilde{3}\tilde{5}\tilde{7}]}}^{1}
\bigcap\Xi_{_{[13\tilde{5}\tilde{7}]}}^{2}
\bigcap\Xi_{_{[13\tilde{5}\tilde{7}]}}^{4}
\nonumber\\
&&\hspace{0.21cm}=\{(y_{_1},\;y_{_2},\;y_{_3},\;y_{_4})
\Big||y_{_3}|<|y_{_2}|,\;1<|y_{_4}|<|y_{_1}|<|y_{_2}|\}\;,
\label{GKZ22a-28}
\end{eqnarray}
the fundamental solution system is composed of the 30 hypergeometric functions
$\Phi_{_{[135\tilde{7}]}}^{(i),b}$, $\Phi_{_{[1\tilde{3}\tilde{5}\tilde{7}]}}^{(j),a}$ and $\Phi_{_{[13\tilde{5}\tilde{7}]}}^{(k),b}$, and $\Phi_{_{[13\tilde{5}\tilde{7}]}}^{(l)}$,
where $i=1,\cdots,4,7,8$, $j=1,\cdots,8$, $k=17,\cdots,24$, and $l=25,\cdots,32$, respectively.

\item In the nonempty proper subset of the whole parameter space,
\begin{eqnarray}
&&\hspace{-0.5cm}\Xi^{29}=\Xi_{_{[\tilde{1}\tilde{3}57]}}^{1}
\bigcap\Xi_{_{[\tilde{1}\tilde{3}57]}}^{7}
\bigcap\Xi_{_{[\tilde{1}\tilde{3}\tilde{5}7]}}^{1}
\nonumber\\
&&\hspace{0.21cm}=\{(y_{_1},\;y_{_2},\;y_{_3},\;y_{_4})
\Big||y_{_1}|<1,\;|y_{_2}|<|y_{_3}|<|y_{_4}|<1\}\;,
\label{GKZ22a-29}
\end{eqnarray}
the fundamental solution system is composed of the 30 hypergeometric functions
$\Phi_{_{[\tilde{1}\tilde{3}57]}}^{(i),a}$, $\Phi_{_{[\tilde{1}\tilde{3}57]}}^{(j)}$,
and $\Phi_{_{[\tilde{1}\tilde{3}\tilde{5}7]}}^{(k)}$,
where $i=1,2,3,5,6,7,9,10,11,13,14,15$, $j=12,16$,
and $k=1,\cdots,16$, respectively.

\item In the nonempty proper subset of the whole parameter space,
\begin{eqnarray}
&&\hspace{-0.5cm}\Xi^{30}=\Xi_{_{[\tilde{1}\tilde{3}57]}}^{6}
\bigcap\Xi_{_{[\tilde{1}3\tilde{5}7]}}^{1}
\nonumber\\
&&\hspace{0.21cm}=\{(y_{_1},\;y_{_2},\;y_{_3},\;y_{_4})
\Big||y_{_1}|<1,\;|y_{_2}|<|y_{_4}|<1,\;|y_{_4}|<|y_{_3}|\}\;,
\label{GKZ22a-30}
\end{eqnarray}
the fundamental solution system is composed of the 30 hypergeometric functions
$\Phi_{_{[\tilde{1}\tilde{3}57]}}^{(i),b}$, $\Phi_{_{[\tilde{1}3\tilde{5}7]}}^{(j)}$,
and $\Phi_{_{[\tilde{1}3\tilde{5}7]}}^{(k),a}$,
where $i=9,10,11,13,14,15$, $j=1,\cdots,16$, and $k=17,\cdots,14$, respectively.

\item In the nonempty proper subset of the whole parameter space,
\begin{eqnarray}
&&\hspace{-0.5cm}\Xi^{31}=\Xi_{_{[\tilde{1}\tilde{3}57]}}^{6}
\bigcap\Xi_{_{[\tilde{1}\tilde{3}57]}}^{7}
\bigcap\Xi_{_{[\tilde{1}\tilde{3}\tilde{5}7]}}^{1}
\bigcap\Xi_{_{[\tilde{1}\tilde{3}5\tilde{7}]}}^{1}
\nonumber\\
&&\hspace{0.21cm}=\{(y_{_1},\;y_{_2},\;y_{_3},\;y_{_4})
\Big||y_{_1}|<1,\;|y_{_2}|<|y_{_3}|<|y_{_4}|<1\}\;,
\label{GKZ22a-31}
\end{eqnarray}
the fundamental solution system is composed of the 30 hypergeometric functions
$\Phi_{_{[\tilde{1}\tilde{3}57]}}^{(i),b}$, $\Phi_{_{[\tilde{1}\tilde{3}57]}}^{(j)}$,
$\Phi_{_{[\tilde{1}\tilde{3}\tilde{5}7]}}^{(k)}$,
and $\Phi_{_{[\tilde{1}\tilde{3}5\tilde{7}]}}^{(l),a}$,
where $i=9,10,11,13,14,15$, $j=12,16$, $k=1,\cdots,16$, and
$l=1,2,3,5,6,7$, respectively.

\item In the nonempty proper subset of the whole parameter space,
\begin{eqnarray}
&&\hspace{-0.5cm}\Xi^{32}=\Xi_{_{[\tilde{1}35\tilde{7}]}}^{2}
\bigcap\Xi_{_{[\tilde{1}3\tilde{5}\tilde{7}]}}^{1}
\nonumber\\
&&\hspace{0.21cm}=\{(y_{_1},\;y_{_2},\;y_{_3},\;y_{_4})
\Big||y_{_1}|<|y_{_4}|<1,\;|y_{_4}|<|y_{_2}|,\;|y_{_3}|<|y_{_2}|\}\;,
\label{GKZ22a-32}
\end{eqnarray}
the fundamental solution system is composed of the 30 hypergeometric functions
$\Phi_{_{[\tilde{1}35\tilde{7}]}}^{(i),b}$,
$\Phi_{_{[\tilde{1}3\tilde{5}\tilde{7}]}}^{(j)}$,
and $\Phi_{_{[\tilde{1}3\tilde{5}\tilde{7}]}}^{(k),a}$,
where $i=1,\cdots,4,7,8$, $j=1,\cdots,16$, and $k=17,\cdots,24$, respectively.

\item In the nonempty proper subset of the whole parameter space,
\begin{eqnarray}
&&\hspace{-0.5cm}\Xi^{33}=\Xi_{_{[1\tilde{3}\tilde{5}7]}}^{1}
\bigcap\Xi_{_{[1\tilde{3}5\tilde{7}]}}^{2}
\bigcap\Xi_{_{[1\tilde{3}\tilde{5}\tilde{7}]}}^{3}
\bigcap\Xi_{_{[1\tilde{3}\tilde{5}\tilde{7}]}}^{4}
\nonumber\\
&&\hspace{0.21cm}=\{(y_{_1},\;y_{_2},\;y_{_3},\;y_{_4})
\Big|1<|y_{_1}|,\;|y_{_3}|<|y_{_2}|<|y_{_4}|<|y_{_1}|\}\;,
\label{GKZ22a-33}
\end{eqnarray}
the fundamental solution system is composed of the 30 hypergeometric functions
$\Phi_{_{[1\tilde{3}\tilde{5}7]}}^{(i)}$, $\Phi_{_{[1\tilde{3}5\tilde{7}]}}^{(j),b}$,
$\Phi_{_{[1\tilde{3}\tilde{5}\tilde{7}]}}^{(j),c}$,
and $\Phi_{_{[1\tilde{3}\tilde{5}\tilde{7}]}}^{(k),b}$,
where $i=1,\cdots,16$, $j=1,2,3,5,6,7$, and $k=4,8$, respectively.

\item In the nonempty proper subset of the whole parameter space,
\begin{eqnarray}
&&\hspace{-0.5cm}\Xi^{34}=\Xi_{_{[13\tilde{5}\tilde{7}]}}^{4}
\bigcap\Xi_{_{[\tilde{1}\tilde{3}5\tilde{7}]}}^{2}
\bigcap\Xi_{_{[\tilde{1}3\tilde{5}\tilde{7}]}}^{3}
\bigcap\Xi_{_{[\tilde{1}3\tilde{5}\tilde{7}]}}^{4}
\bigcap\Xi_{_{[\tilde{1}\tilde{3}\tilde{5}\tilde{7}]}}^{1}
\nonumber\\
&&\hspace{0.21cm}=\{(y_{_1},\;y_{_2},\;y_{_3},\;y_{_4})
\Big||y_{_1}|<1,\;|y_{_3}|<|y_{_4}|<1<|y_{_2}|\}\;,
\label{GKZ22a-34}
\end{eqnarray}
the fundamental solution system is composed of the 30 hypergeometric functions
$\Phi_{_{[13\tilde{5}\tilde{7}]}}^{(i)}$,
$\Phi_{_{[\tilde{1}\tilde{3}5\tilde{7}]}}^{(j)b}$,
$\Phi_{_{[\tilde{1}3\tilde{5}\tilde{7}]}}^{(k),c}$,
$\Phi_{_{[\tilde{1}3\tilde{5}\tilde{7}]}}^{(l),b}$,
and $\Phi_{_{[\tilde{1}\tilde{3}\tilde{5}\tilde{7}]}}^{(m),a}$,
where $i=25,\cdots,32$, $j=1,2,3,5,6,7$, $k=17,\cdots,20,23,24$,
$l=21,22$, and $m=1,\cdots,8$, respectively.

\item In the nonempty proper subset of the whole parameter space,
\begin{eqnarray}
&&\hspace{-0.5cm}\Xi^{35}=\Xi_{_{[\tilde{1}\tilde{3}\tilde{5}7]}}^{2}
\bigcap\Xi_{_{[\tilde{1}\tilde{3}5\tilde{7}]}}^{2}
\bigcap\Xi_{_{[\tilde{1}\tilde{3}5\tilde{7}]}}^{3}
\bigcap\Xi_{_{[\tilde{1}\tilde{3}\tilde{5}\tilde{7}]}}^{3}
\bigcap\Xi_{_{[\tilde{1}\tilde{3}\tilde{5}\tilde{7}]}}^{4}
\nonumber\\
&&\hspace{0.21cm}=\{(y_{_1},\;y_{_2},\;y_{_3},\;y_{_4})
\Big||y_{_1}|<1,\;|y_{_3}|<|y_{_2}|<|y_{_4}|<1\}\;,
\label{GKZ22a-35}
\end{eqnarray}
the fundamental solution system is composed of the 30 hypergeometric functions
$\Phi_{_{[\tilde{1}\tilde{3}\tilde{5}7]}}^{(i)}$,
$\Phi_{_{[\tilde{1}\tilde{3}\tilde{5}7]}}^{(j),b}$,
$\Phi_{_{[\tilde{1}\tilde{3}5\tilde{7}]}}^{(k),b}$
$\Phi_{_{[\tilde{1}\tilde{3}5\tilde{7}]}}^{(l)}$
$\Phi_{_{[\tilde{1}\tilde{3}\tilde{5}\tilde{7}]}}^{(k),c}$,
and $\Phi_{_{[\tilde{1}\tilde{3}\tilde{5}\tilde{7}]}}^{(l),b}$,
where $i=17,\cdots,24$, $j=25,26,27,29,30,31$, $k=1,2,3,5,6,7$,
and $l=4,8$, respectively.

\end{itemize}

Some analytic expressions of the 2-loop self energy are given in Ref.~\cite{Bauberger1995},
where the multiple power series expression of the small $p^2$-region is derived from the dispersion
relation, and that of the large $p^2$-region is derived from the Mellin-Barnes representation, respectively.
Certainly the expressions satisfy the system of PDEs
$\hat{L}_{_i}\varphi=0,\;\;(i=1,\cdots,5)$. Obviously the convergent region
of the multiple power series expression of the small $p^2$-region in Ref.~\cite{Bauberger1995}
has a nonempty intersection with the proper subset
$\Xi^{3}$ defined in Eq.(\ref{GKZ22a-3}). The analytic expression of the small $p^2$-region
in Ref.~\cite{Bauberger1995} can be written as a linear combination of the
generalized hypergeometric functions of the fundamental solution system presented
below Eq.(\ref{GKZ22a-3}) in the nonempty intersection. Requiring that the expression
of the small $p^2$-region in Ref.~\cite{Bauberger1995} and its
partial derivatives presented in Eq.(\ref{GKZ20}) equal that of the linear combination
of the generalized hypergeometric functions, one obtains the combination coefficients uniquely.
Similarly the convergent region of the multiple power series of the large $p^2$-region
in Ref.~\cite{Bauberger1995} has a nonempty intersection with the proper subset
$\Xi^{14}$ defined in Eq.(\ref{GKZ22a-14}).
The analytic expression of the large $p^2$-region
in Ref.~\cite{Bauberger1995} can be written as a linear combination of the
generalized hypergeometric functions of the fundamental solution system presented
below Eq.(\ref{GKZ22a-14}) in the nonempty intersection.
Assuming only one virtual particle with non-zero mass, we briefly elucidate how to
determine the coefficients in the following section \ref{sec4}.

\section{The analytical expressions of the integral with one nonzero virtual mass\label{sec4}}
\indent\indent
When the masses of virtual particles are all zero, we can
use the properties of the Bessel function to obtain the analytical
expression of any self energy integral~\cite{Feng2018,Feng2019}. On the other hand,
one can also use the Bessel function to obtain the analytical expression of the
Feynman integral when the mass squared of a virtual particle
is much larger than the absolute value of the 4-momentum squared of the
outgoing particle. In the parameter space, any solution of the GKZ
hypergeometric system in Eq.(\ref{GKZ17}) can be expanded as a linear combination
of 30 functionally independent solutions.
Two analytical expressions mentioned above can be taken as the boundary conditions to
determine those linear combination coefficients concretely.

In order to elucidate how to obtain the analytical expression clearly, we assume
that there is only one nonzero virtual mass in the 2-loop self energy.
The corresponding scalar integral can be expressed as a linear combination of
Gauss functions or a linear combination of Pochammer functions.

\subsection{The analytical expressions with $m_{_1}\neq0$, $m_{_2}=m_{_3}=m_{_4}=0$\label{sec4a}}
\indent\indent
In this case, the GKZ-hypergeometric system is simplified as
\begin{eqnarray}
&&\mathbf{A}_{_1}\cdot\vec{\vartheta}_{_1}\Phi=\mathbf{B}\Phi\;,
\label{GKZ34a}
\end{eqnarray}
where the vector of Euler operators is defined as
\begin{eqnarray}
&&\vec{\vartheta}_{_1}^{\;T}=(\vartheta_{_{z_{_1}}},\;\vartheta_{_{z_{_{2}}}},\;\vartheta_{_{z_{_{3}}}},\;
\vartheta_{_{z_{_{4}}}},\;\vartheta_{_{z_{_{5}}}},\;\vartheta_{_{z_{_{6}}}},\;\vartheta_{_{z_{_{7}}}}
,\;\vartheta_{_{z_{_{8}}}},\;\vartheta_{_{z_{_{9}}}},\;\vartheta_{_{z_{_{10}}}},\;\vartheta_{_{z_{_{11}}}})\;,
\label{GKZ35a}
\end{eqnarray}
and the matrix $\mathbf{A}_{_1}$ is
\begin{eqnarray}
&&\mathbf{A}_{_1}=\left(\begin{array}{ccccccccccc}
1\;\;&0\;\;&0\;\;&0\;\;&0\;\;&0\;\;&0\;\;&0\;\;&0\;\;&0\;\;&1\;\\
0\;\;&1\;\;&0\;\;&0\;\;&0\;\;&0\;\;&0\;\;&0\;\;&0\;\;&0\;\;&1\;\\
0\;\;&0\;\;&1\;\;&0\;\;&0\;\;&0\;\;&0\;\;&0\;\;&0\;\;&0\;\;&0\;\\
0\;\;&0\;\;&0\;\;&1\;\;&0\;\;&0\;\;&0\;\;&0\;\;&0\;\;&0\;\;&0\;\\
0\;\;&0\;\;&0\;\;&0\;\;&1\;\;&0\;\;&0\;\;&0\;\;&0\;\;&0\;\;&0\;\\
0\;\;&0\;\;&0\;\;&0\;\;&0\;\;&1\;\;&0\;\;&0\;\;&0\;\;&0\;\;&-1\;\\
0\;\;&0\;\;&0\;\;&0\;\;&0\;\;&0\;\;&1\;\;&0\;\;&0\;\;&0\;\;&0\;\\
0\;\;&0\;\;&0\;\;&0\;\;&0\;\;&0\;\;&0\;\;&1\;\;&0\;\;&0\;\;&0\;\\
0\;\;&0\;\;&0\;\;&0\;\;&0\;\;&0\;\;&0\;\;&0\;\;&1\;\;&0\;\;&0\;\\
0\;\;&0\;\;&0\;\;&0\;\;&0\;\;&0\;\;&0\;\;&0\;\;&0\;\;&1\;\;&0\;\\
\end{array}\right)\;.
\label{GKZ36a}
\end{eqnarray}
The integer sublattice is determined by
the dual matrix $\mathbf{\tilde A}_{_1}$ of $\mathbf{A}_{_1}$,
\begin{eqnarray}
&&\mathbf{\tilde A}_{_1}=\left(\begin{array}{ccccccccccc}
-1\;\;&-1\;\;&0\;\;&0\;\;&0\;\;&1\;\;&0\;\;&0\;\;&0\;\;&0\;\;&1\;\;
\end{array}\right)\;.
\label{GKZ37a}
\end{eqnarray}
This integer sublattice induces the first hyperbolic equation
in Eq.(\ref{GKZ19-1}), which implies that the system of fundamental solutions
is composed of two linear independent hypergeometric functions.
In this case, the only threshold is $|p|=m_{_1}$.
In the region $|y_{_1}|<1$, the
fundamental solution system is composed of two Gauss functions:
\begin{eqnarray}
&&\Phi_{_{A_1}}^1(y_{_1})=\;_{_2}F_{_1}
\left(\left.\begin{array}{cc}a_{_1},\;&a_{_2}\\
\;\;&b_{_1}\end{array}\right|y_{_1}\right)=\;_{_2}F_{_1}
\left(\left.\begin{array}{cc}4-D,\;&5-{3D\over2}\\
\;\;&2-{D\over2}\end{array}\right|y_{_1}\right)\;,
\nonumber\\
&&\Phi_{_{A_1}}^2(y_{_1})=(-y_{_1})^{1-b_{_1}}\;_{_2}F_{_1}
\left(\left.\begin{array}{cc}1+a_{_1}-b_{_1},\;&1+a_{_2}-b_{_1}\\
\;\;&2-b_{_1}\end{array}\right|y_{_1}\right)
\nonumber\\
&&\hspace{1.5cm}=
(-y_{_1})^{D/2-1}\;_{_2}F_{_1}
\left(\left.\begin{array}{cc}3-{D\over2},\;&4-D\\
\;\;&{D\over2}\end{array}\right|y_{_1}\right)\;,
\label{GKZ38a}
\end{eqnarray}
with $y_{_1}=x_{_1}=m_{_1}^2/p^2$.
Correspondingly the scalar integral is a linear combination of two fundamental solutions:
\begin{eqnarray}
&&\Phi_{_{A_1}}(y_{_1})=C_{_{A_1}}^1\Phi_{_{A_1}}^1(y_{_1})
+C_{_{A_1}}^2\Phi_{_{A_1}}^2(y_{_1})\;.
\label{GKZ38a1}
\end{eqnarray}

In the region $|y_{_1}|>1$, the fundamental solution system is similarly
composed of two Gauss functions:
\begin{eqnarray}
&&\Phi_{_{A_1}}^3(y_{_1})=(-y_{_1})^{-a_{_1}}\;_{_2}F_{_1}
\left(\left.\begin{array}{cc}a_{_1},\;&1+a_{_1}-b_{_1}\\
\;\;&1+a_{_1}-a_{_2}\end{array}\right|{1\over y_{_1}}\right)
\nonumber\\
&&\hspace{1.5cm}=
(-y_{_1})^{D-4}\;_{_2}F_{_1}
\left(\left.\begin{array}{cc}4-D,\;&3-{D\over2}\\
\;\;&{D\over2}\end{array}\right|{1\over y_{_1}}\right)\;,
\nonumber\\
&&\Phi_{_{A_1}}^4(y_{_1})=(-y_{_1})^{-a_{_2}}\;_{_2}F_{_1}
\left(\left.\begin{array}{cc}a_{_2},\;&1+a_{_2}-b_{_1}\\
\;\;&1-a_{_1}+a_{_2}\end{array}\right|{1\over y_{_1}}\right)
\nonumber\\
&&\hspace{1.5cm}=
(-y_{_1})^{3D/2-5}\;_{_2}F_{_1}
\left(\left.\begin{array}{cc}5-{3D\over2},\;&4-D\\
\;\;&2-{D\over2}\end{array}\right|{1\over y_{_1}}\right)\;.
\label{GKZ39a}
\end{eqnarray}
Correspondingly the scalar integral is a linear combination of two fundamental solutions:
\begin{eqnarray}
&&\Phi_{_{A_1}}(y_{_1})=C_{_{A_1}}^3\Phi_{_{A_1}}^3(y_{_1})
+C_{_{A_1}}^4\Phi_{_{A_1}}^4(y_{_1})\;.
\label{GKZ39a1}
\end{eqnarray}

As $m_{_1}^2\ll|p^2|,\;m_{_2}=m_{_3}=m_{_4}=0$,
\begin{eqnarray}
&&I_{_1}=\mu^{4\varepsilon}\int{d^Dq_{_1}\over(2\pi)^D}{d^Dq_{_2}\over(2\pi)^D}
{1\over(q_{_1}^2-m_{_1}^2)(q_{_1}+p)^2q_{_2}^2(q_{_1}+q_{_2}+p)^2}
\nonumber\\
&&\hspace{0.5cm}=I_{_{1,0}}+\cdots\;,
\label{GKZ40a}
\end{eqnarray}
where
\begin{eqnarray}
&&I_{_{1,0}}=\mu^{4\varepsilon}\int{d^Dq_{_1}\over(2\pi)^D}{d^Dq_{_2}\over(2\pi)^D}
{1\over q_{_1}^2(q_{_1}+p)^2q_{_2}^2(q_{_1}+q_{_2}+p)^2}
\nonumber\\
&&\hspace{0.7cm}=
-{1\over(4\pi)^4}\Big({4\pi\mu^2\over-p^2}\Big)^{2\varepsilon}
{\pi\Gamma^2({D\over2}-1)\Gamma(3-D)\over\sin\pi(2-{D\over2})\Gamma(3-{D\over2})
\Gamma({3\over2}D-4)}\;.
\label{GKZ41a}
\end{eqnarray}

As $m_{_1}^2\gg|p^2|,\;m_{_2}=m_{_3}=m_{_4}=0$,
\begin{eqnarray}
&&I_{_1}^\prime=\mu^{4\varepsilon}\int{d^Dq_{_1}\over(2\pi)^D}{d^Dq_{_2}\over(2\pi)^D}
{1\over(q_{_1}^2-m_{_1}^2)(q_{_1}+p)^2q_{_2}^2(q_{_1}+q_{_2}+p)^2}
\nonumber\\
&&\hspace{0.5cm}=I_{_{1,0}}^\prime+\cdots\;,
\label{GKZ42a}
\end{eqnarray}
where
\begin{eqnarray}
&&I_{_{1,0}}^\prime=\mu^{4\varepsilon}\int{d^Dq_{_1}\over(2\pi)^D}{d^Dq_{_2}\over(2\pi)^D}
{1\over q_{_1}^2(q_{_1}^2-m_{_1}^2)q_{_2}^2(q_{_1}+q_{_2})^2}
\nonumber\\
&&\hspace{0.7cm}=
{1\over(4\pi)^4}\Big({4\pi\mu^2\over m_{_1}^2}\Big)^{2\varepsilon}{2\pi\Gamma(2-D)
\over\sin\pi(2-{D\over2})}\;.
\label{GKZ44a}
\end{eqnarray}
Since the unique threshold is $p^2=m_{_1}^2$ under our assumption on the parameter space,
the Feynman integral is an analytic function of $p^2$ in the neighborhood of $p^2=0$.
The fact indicates
\begin{eqnarray}
&&C_{_{A_1}}^3={1\over(4\pi)^4}\Big({4\pi\mu^2\over -p^2}\Big)^{2\varepsilon}
{2\pi\Gamma(2-D)\over\sin\pi(2-{D\over2})}\;,
\nonumber\\
&&C_{_{A_1}}^4=0\;.
\label{GKZ44a-1}
\end{eqnarray}
Using the well-known relation
\begin{eqnarray}
&&{\Gamma(a)\Gamma(b)\over\Gamma(c)}\;_{_2}F_{_1}
\left(\left.\begin{array}{cc}a,\;&b\\
\;\;&c\end{array}\right|z\right)
={\Gamma(a)\Gamma(b-a)\over\Gamma(c-a)}(-z)^{-a}\;_{_2}F_{_1}
\left(\left.\begin{array}{cc}a,\;&1+a-c\\
\;\;&1+a-b\end{array}\right|{1\over z}\right)
\nonumber\\
&&\hspace{5.0cm}
+{\Gamma(b)\Gamma(a-b)\over\Gamma(c-b)}(-z)^{-b}\;_{_2}F_{_1}
\left(\left.\begin{array}{cc}b,\;&1+b-c\\
\;\;&1-a+b\end{array}\right|{1\over z}\right)\;,
\label{GKZ44a-2}
\end{eqnarray}
we obtain
\begin{eqnarray}
&&C_{_{A_1}}^1=-{1\over(4\pi)^4}\Big({4\pi\mu^2\over-p^2}\Big)^{2\varepsilon}
{\pi\Gamma^2({D\over2}-1)\Gamma(3-D)\over\sin\pi(2-{D\over2})\Gamma(3-{D\over2})
\Gamma({3\over2}D-4)}\;,
\nonumber\\
&&C_{_{A_1}}^2={1\over(4\pi)^4}\Big({4\pi\mu^2\over-p^2}\Big)^{2\varepsilon}
{2\pi^2\over\sin{\pi D\over2}\sin\pi(2-{D\over2})\Gamma(D-1)}\;,
\label{GKZ44a-3}
\end{eqnarray}
where $C_{_{A_1}}^1=I_{_{1,0}}$ obviously. Actually the Mellin-Barnes representation
of the Feynman integral can be obtained from Eq.(\ref{GKZ1}) as
\begin{eqnarray}
&&\Sigma_{_{1234}}(p^2)=
-{1\over(4\pi)^4}\Big({4\pi\Lambda_{_{\rm RE}}^2\over-p^2}\Big)^{4-D}
{\pi\Gamma({D\over2}-1)\over(D-3)\sin\pi(2-{D\over2})\Gamma(3-{D\over2})}
\nonumber\\
&&\hspace{2.2cm}\times
{1\over2\pi i}\int_{-i\infty}^{+i\infty}ds_{_1}
\Big({m_{_1}^2\over-p^2}\Big)^{s_{_1}}{\Gamma(-s_{_1})
\Gamma({D\over2}-1-s_{_1})\Gamma(4-D+s_{_1})
\over\Gamma({3\over2}D-4-s_{_1})}\;,
\label{GKZ44a-4}
\end{eqnarray}
under this special assumption on the parameter space.
The residue of simple pole of $\Gamma(-s_{_1})$ provides $C_{_{A_1}}^1\Phi_{_{A_1}}^1(y_{_1})$,
that of simple pole of $\Gamma(D/2-1-s_{_1})$ provides $C_{_{A_1}}^2\Phi_{_{A_1}}^2(y_{_1})$,
and that of simple pole of $\Gamma(4-D+s_{_1})$ provides $C_{_{A_1}}^3\Phi_{_{A_1}}^3(y_{_1})$,
respectively.

Combining the transformation
\begin{eqnarray}
&&\;_{_2}F_{_1}\left(\left.\begin{array}{cc}a,\;&b\\
\;\;&c\end{array}\right|z\right)
=(1-z)^{-b}\;_{_2}F_{_1}
\left(\left.\begin{array}{cc}c-a,\;&b\\
\;\;&c\end{array}\right|{z\over z-1}\right)
\label{GKZ44a-5}
\end{eqnarray}
with the relation Eq.(\ref{GKZ44a-2}), one derives the analytic expression
in the neighborhood of $y_{_1}=1$, i.e. the analytic expression in
the neighborhood of the threshold. Note that the threshold exactly coincides
with the regular singularity $z=1$ of the PDEs satisfied by the Feynman integral
in this special case.
Taking the Feynman integral as a function of some subvarieties of Grassmannians~\cite{Feng2022},
we can get the similar relations of Eq.(\ref{GKZ44a-5}) among the generalized hypergeometric functions
through the algorithm in combinatorial geometry.

\subsection{The analytical expressions with $m_{_4}\neq0$, $m_{_1}=m_{_2}=m_{_3}=0$\label{sec4b}}
\indent\indent
In the parameter space, the GKZ hypergeometric system is simplified as
\begin{eqnarray}
&&\mathbf{A}_{_4}\cdot\vec{\vartheta}_{_4}\Phi=\mathbf{B}\Phi\;,
\label{GKZ34b}
\end{eqnarray}
where the vector of Euler operators is defined as
\begin{eqnarray}
&&\vec{\vartheta}_{_4}^{\;T}=(\vartheta_{_{z_{_1}}},\;\vartheta_{_{z_{_{2}}}},\;\vartheta_{_{z_{_{3}}}},\;
\vartheta_{_{z_{_{4}}}},\;\vartheta_{_{z_{_{5}}}},\;\vartheta_{_{z_{_{6}}}},\;\vartheta_{_{z_{_{7}}}}
,\;\vartheta_{_{z_{_{8}}}},\;\vartheta_{_{z_{_{9}}}},\;\vartheta_{_{z_{_{10}}}},\;\vartheta_{_{z_{_{14}}}})\;,
\label{GKZ35b}
\end{eqnarray}
and the matrix $\mathbf{A}_{_4}$ is
\begin{eqnarray}
&&\mathbf{A}_{_4}=\left(\begin{array}{ccccccccccc}
1\;\;&0\;\;&0\;\;&0\;\;&0\;\;&0\;\;&0\;\;&0\;\;&0\;\;&0\;\;&1\;\;\\
0\;\;&1\;\;&0\;\;&0\;\;&0\;\;&0\;\;&0\;\;&0\;\;&0\;\;&0\;\;&1\;\;\\
0\;\;&0\;\;&1\;\;&0\;\;&0\;\;&0\;\;&0\;\;&0\;\;&0\;\;&0\;\;&0\;\;\\
0\;\;&0\;\;&0\;\;&1\;\;&0\;\;&0\;\;&0\;\;&0\;\;&0\;\;&0\;\;&0\;\;\\
0\;\;&0\;\;&0\;\;&0\;\;&1\;\;&0\;\;&0\;\;&0\;\;&0\;\;&0\;\;&1\;\;\\
0\;\;&0\;\;&0\;\;&0\;\;&0\;\;&1\;\;&0\;\;&0\;\;&0\;\;&0\;\;&0\;\;\\
0\;\;&0\;\;&0\;\;&0\;\;&0\;\;&0\;\;&1\;\;&0\;\;&0\;\;&0\;\;&0\;\;\\
0\;\;&0\;\;&0\;\;&0\;\;&0\;\;&0\;\;&0\;\;&1\;\;&0\;\;&0\;\;&0\;\;\\
0\;\;&0\;\;&0\;\;&0\;\;&0\;\;&0\;\;&0\;\;&0\;\;&1\;\;&0\;\;&-1\;\;\\
0\;\;&0\;\;&0\;\;&0\;\;&0\;\;&0\;\;&0\;\;&0\;\;&0\;\;&1\;\;&-1\;\;\\
\end{array}\right)\;.
\label{GKZ36b}
\end{eqnarray}

The integer sublattice is determined by
the dual matrix $\mathbf{\tilde A}_{_4}$ of $\mathbf{A}_{_4}$,
\begin{eqnarray}
&&\mathbf{\tilde A}_{_4}=\left(\begin{array}{ccccccccccc}
-1\;\;&-1\;\;&0\;\;&0\;\;&-1\;\;&0\;\;&0\;\;&0\;\;&1\;\;&1\;\;&1\;\;
\end{array}\right)
\label{GKZ37b}
\end{eqnarray}
This integer sublattice $\mathbf{B}$ induces the fourth hyperbolic equation
in Eq.(\ref{GKZ19-1}). With the assumption on the parameter space, the Feynman
integral contains a zero threshold $|p|_{_{min}}=0$ and a nonzero threshold
$|p|_{_{max}}=m_{_4}$. In the region $|y_{_4}|<1$, the
fundamental solution system is composed of three Pochammer functions $_{_3}F_{_2}$
which are simplified as the Gauss functions under the special exponents
of Eq.(\ref{GKZ5}):
\begin{eqnarray}
&&\Phi_{_{A_4}}^1(y_{_4})=\;_{_3}F_{_2}
\left(\left.\begin{array}{ccc}a_{_1},\;&a_{_2},\;&a_{_5}\\
\;\;&b_{_4},\;&b_{_5}\end{array}\right|y_{_4}\right)
\nonumber\\
&&\hspace{1.5cm}=
\;_{_2}F_{_1}\left(\left.\begin{array}{cc}1,\;&5-{3D\over2}\\
\;\;&3-{D\over2}\end{array}\right|y_{_4}\right)\;,
\nonumber\\
&&\Phi_{_{A_4}}^2(y_{_4})=(-y_{_4})^{1-b_{_5}}\;_{_3}F_{_2}
\left(\left.\begin{array}{ccc}1+a_{_1}-b_{_5},\;&1+a_{_2}-b_{_5},\;&1+a_{_5}-b_{_5}\\
\;\;&2-b_{_5},\;&1+b_{_4}-b_{_5}\end{array}\right|y_{_4}\right)
\nonumber\\
&&\hspace{1.5cm}=
(-y_{_4})^{D-3}\;_{_2}F_{_1}\left(\left.\begin{array}{cc}1,\;&2-{D\over2}\\
\;\;&{D\over2}\end{array}\right|y_{_4}\right)\;,
\nonumber\\
&&\Phi_{_{A_4}}^3(y_{_4})=(-y_{_4})^{1-b_{_4}}\;_{_3}F_{_2}
\left(\left.\begin{array}{ccc}1+a_{_1}-b_{_4},\;&1+a_{_2}-b_{_4},\;&1+a_{_5}-b_{_4}\\
\;\;&2-b_{_4},\;&1-b_{_4}+b_{_5}\end{array}\right|y_{_4}\right)
\nonumber\\
&&\hspace{1.5cm}=
(-y_{_4})^{D/2-2}(1-y_{_4})^{D-3}\;,
\label{GKZ38b}
\end{eqnarray}
with $y_{_4}=x_{_4}=m_{_4}^2/p^2$.
Correspondingly the Feynman integral is formulated as a linear combination
\begin{eqnarray}
&&\Phi_{_{A_4}}(y_{_4})=C_{_{A_4}}^1\Phi_{_{A_4}}^1(y_{_4})
+C_{_{A_4}}^2\Phi_{_{A_4}}^2(y_{_4})+C_{_{A_4}}^3\Phi_{_{A_4}}^3(y_{_4})
\label{GKZ38b-1}
\end{eqnarray}
in the region $|y_{_4}|<1$.

In the region $|y_{_4}|>1$, the fundamental solution system is similarly
composed of three Pochammer functions $_{_3}F_{_2}$
which are simplified as the Gauss functions under the special exponents
of Eq.(\ref{GKZ5}):
\begin{eqnarray}
&&\Phi_{_{A_4}}^4(y_{_4})=(-y_{_4})^{-a_{_1}}\;_{_3}F_{_2}
\left(\left.\begin{array}{ccc}a_{_1},\;&1+a_{_1}-b_{_4},\;&1+a_{_1}-b_{_5}\\
\;\;&1+a_{_1}-a_{_2},\;&1+a_{_1}-a_{_5}\end{array}\right|{1\over y_{_4}}\right)
\nonumber\\
&&\hspace{1.5cm}=
(-y_{_4})^{D-4}\;_{_2}F_{_1}
\left(\left.\begin{array}{cc}1,\;&2-{D\over2}\\
\;\;&{D\over2}\end{array}\right|{1\over y_{_4}}\right)\;,
\nonumber\\
&&\Phi_{_{A_4}}^5(y_{_4})=(-y_{_4})^{-a_{_5}}\;_{_3}F_{_2}
\left(\left.\begin{array}{ccc}a_{_5},\;&1+a_{_5}-b_{_4},\;&1+a_{_5}-b_{_5}\\
\;\;&1-a_{_2}+a_{_5},\;&1-a_{_1}+a_{_5}\end{array}\right|{1\over y_{_4}}\right)
\nonumber\\
&&\hspace{1.5cm}=
{-1\over y_{_4}}\;_{_2}F_{_1}
\left(\left.\begin{array}{cc}1,\;&{D\over2}-1\\
\;\;&{3D\over2}-3\end{array}\right|{1\over y_{_4}}\right)\;,
\nonumber\\
&&\Phi_{_{A_4}}^6(y_{_4})=(-y_{_4})^{-a_{_2}}\;_{_3}F_{_2}
\left(\left.\begin{array}{ccc}a_{_2},\;&1+a_{_2}-b_{_4},\;&1+a_{_2}-b_{_5}\\
\;\;&1-a_{_1}+a_{_2},\;&1+a_{_2}-a_{_5}\end{array}\right|{1\over y_{_4}}\right)
\nonumber\\
&&\hspace{1.5cm}=
(-y_{_4})^{D/2-2}(1-y_{_4})^{D-3}\;.
\label{GKZ39b}
\end{eqnarray}
The Feynman integral is formulated as a linear combination
\begin{eqnarray}
&&\Phi_{_{A_4}}(y_{_4})=C_{_{A_4}}^4\Phi_{_{A_4}}^4(y_{_4})
+C_{_{A_4}}^5\Phi_{_{A_4}}^5(y_{_4})+C_{_{A_4}}^3\Phi_{_{A_4}}^6(y_{_4})
\label{GKZ39b-1}
\end{eqnarray}
in the region $|y_{_4}|>1$.

As $m_{_4}^2\gg|p^2|,\;m_{_1}=m_{_2}=m_{_3}=0$,
\begin{eqnarray}
&&I_{_4}^\prime=\mu^{4\varepsilon}\int{d^Dq_{_1}\over(2\pi)^D}{d^Dq_{_2}\over(2\pi)^D}
{1\over q_{_1}^2((q_{_1}+p)^2-m_{_4}^2)q_{_2}^2(q_{_1}+q_{_2}+p)^2}
\nonumber\\
&&\hspace{0.5cm}=I_{_{4,0}}^\prime+\cdots\;,
\label{GKZ42b}
\end{eqnarray}
where
\begin{eqnarray}
&&I_{_{4,0}}^\prime={1\over(4\pi)^4}
\Big({4\pi\mu^2\over m_{_4}^2}\Big)^{2\varepsilon}{2\pi\Gamma(2-D)\over\sin\pi(2-{D\over2})}\;.
\label{GKZ44b}
\end{eqnarray}

The results of Eq.(\ref{GKZ41a}) and Eq.(\ref{GKZ44b}) induce
\begin{eqnarray}
&&C_{_{A_4}}^1=-{1\over(4\pi)^4}\Big({4\pi\mu^2\over-p^2}\Big)^{2\varepsilon}
{\pi\Gamma^2({D\over2}-1)\Gamma(3-D)\over\sin\pi(2-{D\over2})\Gamma(3-{D\over2})
\Gamma({3\over2}D-4)}\;,
\nonumber\\
&&C_{_{A_4}}^4={1\over(4\pi)^4}
\Big({4\pi\mu^2\over -p^2}\Big)^{2\varepsilon}{2\pi\Gamma(2-D)\over\sin\pi(2-{D\over2})}\;.
\label{GKZ44b-1}
\end{eqnarray}
Furthermore, the relation Eq.(\ref{GKZ44a-2}) indicates
\begin{eqnarray}
&&C_{_{A_4}}^2=-{1\over(4\pi)^4}\Big({4\pi\mu^2\over-p^2}\Big)^{2\varepsilon}
{2\pi\Gamma(2-D)\over\sin\pi(2-{D\over2})}\;,
\nonumber\\
&&C_{_{A_4}}^5={1\over(4\pi)^4}\Big({4\pi\mu^2\over-p^2}\Big)^{2\varepsilon}
{\pi\Gamma^2({D\over2}-1)\Gamma(3-D)\over\sin\pi(2-{D\over2})\Gamma(2-{D\over2})
\Gamma({3\over2}D-3)}\;.
\label{GKZ44b-2}
\end{eqnarray}
In this case, the values of the Feynman integral at $p^2\neq0,\;m_{_4}=0$
and $p^2=0,\;m_{_4}\neq0$ can not uniquely determine the combination coefficient
$C_{_{A_4}}^3$. Nevertheless, the Mellin-Barnes representation
of the Feynman integral gives $C_{_{A_4}}^3=0$ under the assumption on the whole
parameter space. In fact, the Mellin-Barnes representation of the Feynman integral
is written as
\begin{eqnarray}
&&\Sigma_{_{1234}}(p^2)=
-{1\over(4\pi)^4}\Big({4\pi\Lambda_{_{\rm RE}}^2\over-p^2}\Big)^{4-D}
{\pi\Gamma^2({D\over2}-1)\over\sin\pi(2-{D\over2})\Gamma(D-2)}
{1\over2\pi i}\int_{-i\infty}^{+i\infty}ds_{_4}
\nonumber\\
&&\hspace{2.2cm}\times
\Big({m_{_4}^2\over-p^2}\Big)^{s_{_4}}{\Gamma(-s_{_4})\Gamma(1+s_{_4})
\Gamma(D-3-s_{_4})\Gamma(4-D+s_{_4})
\over\Gamma(3-{D\over2}+s_{_4})\Gamma({3\over2}D-4-s_{_4})}\;.
\label{GKZ44b-3}
\end{eqnarray}
The residue of simple pole of $\Gamma(-s_{_4})$ provides $C_{_{A_4}}^1\Phi_{_{A_4}}^1(y_{_4})$,
that of simple pole of $\Gamma(D-3-s_{_4})$ provides $C_{_{A_4}}^2\Phi_{_{A_4}}^2(y_{_4})$,
that of simple pole of $\Gamma(4-D+s_{_4})$ provides $C_{_{A_4}}^4\Phi_{_{A_4}}^4(y_{_4})$,
and that of simple pole of $\Gamma(1+s_{_4})$ provides $C_{_{A_4}}^5\Phi_{_{A_4}}^5(y_{_4})$,
respectively. Applying the relations Eq.(\ref{GKZ44a-2}) and Eq.(\ref{GKZ44a-5}), we
obtain the analytic expression in the neighborhood of $y_{_4}=1$, i.e. the analytic expression of
the neighborhood of the maximum threshold.
It should be emphasized that the analytic expression Eq. (\ref{GKZ39b-1}) in the
neighborhood of the zero-threshold is consistent with that obtained by the heavy
mass expansion approach~\cite{Berends1995}.

\subsection{The analytical expressions with $m_{_2}\neq0$, $m_{_1}=m_{_3}=m_{_4}=0$\label{sec4c}}
\indent\indent
In the parameter space, the GKZ hypergeometric system is simplified as
\begin{eqnarray}
&&\mathbf{A}_{_2}\cdot\vec{\vartheta}_{_2}\Phi=\mathbf{B}\Phi\;,
\label{GKZ34c}
\end{eqnarray}
where the vector of Euler operators is defined as
\begin{eqnarray}
&&\vec{\vartheta}_{_2}^{\;T}=(\vartheta_{_{z_{_1}}},\;\vartheta_{_{z_{_{2}}}},\;\vartheta_{_{z_{_{3}}}},\;
\vartheta_{_{z_{_{4}}}},\;\vartheta_{_{z_{_{5}}}},\;\vartheta_{_{z_{_{6}}}},\;\vartheta_{_{z_{_{7}}}}
,\;\vartheta_{_{z_{_{8}}}},\;\vartheta_{_{z_{_{9}}}},\;\vartheta_{_{z_{_{10}}}},\;\vartheta_{_{z_{_{12}}}})\;,
\label{GKZ35c}
\end{eqnarray}
and the matrix $\mathbf{A}_{_2}$ is
\begin{eqnarray}
&&\mathbf{A}_{_2}=\left(\begin{array}{ccccccccccc}
1\;\;&0\;\;&0\;\;&0\;\;&0\;\;&0\;\;&0\;\;&0\;\;&0\;\;&0\;\;&1\;\;\\
0\;\;&1\;\;&0\;\;&0\;\;&0\;\;&0\;\;&0\;\;&0\;\;&0\;\;&0\;\;&1\;\;\\
0\;\;&0\;\;&1\;\;&0\;\;&0\;\;&0\;\;&0\;\;&0\;\;&0\;\;&0\;\;&1\;\;\\
0\;\;&0\;\;&0\;\;&1\;\;&0\;\;&0\;\;&0\;\;&0\;\;&0\;\;&0\;\;&1\;\;\\
0\;\;&0\;\;&0\;\;&0\;\;&1\;\;&0\;\;&0\;\;&0\;\;&0\;\;&0\;\;&0\;\;\\
0\;\;&0\;\;&0\;\;&0\;\;&0\;\;&1\;\;&0\;\;&0\;\;&0\;\;&0\;\;&0\;\;\\
0\;\;&0\;\;&0\;\;&0\;\;&0\;\;&0\;\;&1\;\;&0\;\;&0\;\;&0\;\;&-1\;\;\\
0\;\;&0\;\;&0\;\;&0\;\;&0\;\;&0\;\;&0\;\;&1\;\;&0\;\;&0\;\;&0\;\;\\
0\;\;&0\;\;&0\;\;&0\;\;&0\;\;&0\;\;&0\;\;&0\;\;&1\;\;&0\;\;&-1\;\;\\
0\;\;&0\;\;&0\;\;&0\;\;&0\;\;&0\;\;&0\;\;&0\;\;&0\;\;&1\;\;&-1\;\;\\
\end{array}\right)\;.
\label{GKZ36c}
\end{eqnarray}
Correspondingly the dual matrix $\mathbf{\tilde A}_{_2}$ of $\mathbf{A}_{_2}$ is
\begin{eqnarray}
&&\mathbf{\tilde A}_{_2}=\left(\begin{array}{ccccccccccc}
-1\;\;&-1\;\;&-1\;\;&-1\;\;&0\;\;&0\;\;&1\;\;&0\;\;&1\;\;&1\;\;&1\;\;
\end{array}\right)
\label{GKZ37c}
\end{eqnarray}
This integer sublattice $\mathbf{B}$ induces the second hyperbolic equation
in Eq.(\ref{GKZ19-1}). Under the assumption, the Feynman
integral contains a zero threshold $|p|_{_{min}}=0$ and a nonzero threshold
$|p|_{_{max}}=m_{_2}$. In the region $|y_{_2}|<1$, the
fundamental solution system is composed of four Pochammer functions $_{_4}F_{_3}$:
\begin{eqnarray}
&&\Phi_{_{A_2}}^1(y_{_2})=\;_{_4}F_{_3}
\left(\left.\begin{array}{cccc}a_{_1},\;&a_{_2},\;&a_{_3},\;&a_{_4}\\
\;\;&b_{_2},\;&b_{_4},\;&b_{_5}\end{array}\right|y_{_2}\right)
=\;_{_2}F_{_1}\left(\left.\begin{array}{cc}3-D,\;&5-{3D\over2}\\
\;\;&3-{D\over2}\end{array}\right|y_{_2}\right)\;,
\nonumber\\
&&\Phi_{_{A_2}}^2(y_{_2})=(-y_{_2})^{1-b_{_2}}\;_{_4}F_{_3}
\left(\left.\begin{array}{cccc}1+a_{_1}-b_{_2},\;&1+a_{_2}-b_{_2},\;&1+a_{_3}-b_{_2},\;&1+a_{_4}-b_{_2}\\
\;\;&2-b_{_2},\;&1-b_{_2}+b_{_4},\;&1-b_{_2}+b_{_5}\end{array}\right|y_{_2}\right)
\nonumber\\
&&\hspace{1.5cm}=
(-y_{_2})^{D/2-1}\;_{_3}F_{_2}
\left(\left.\begin{array}{ccc}1,\;&2-{D\over2},\;&4-D\\
\;\;&2,\;&{D\over2}\end{array}\right|y_{_2}\right)
\;,\nonumber\\
&&\Phi_{_{A_2}}^{3}(y_{_2})=(-y_{_2})^{1-b_{_4}}\;_{_4}F_{_3}
\left(\left.\begin{array}{cccc}1+a_{_1}-b_{_4},\;&1+a_{_2}-b_{_4},\;&1+a_{_3}-b_{_4},\;&1+a_{_4}-b_{_4}\\
\;\;&2-b_{_4},\;&1+b_{_2}-b_{_4},\;&1-b_{_4}+b_{_5}\end{array}\right|y_{_2}\right)
\nonumber\\
&&\hspace{1.5cm}=
(-y_{_2})^{D/2-2}\;_{_2}F_{_1}\left(\left.\begin{array}{cc}1-{D\over2},\;&3-D\\
\;\;&{D\over2}-1\end{array}\right|y_{_2}\right)
\;,\nonumber\\
&&\Phi_{_{A_2}}^{4}(y_{_2})=(-y_{_2})^{1-b_{_5}}\;_{_4}F_{_3}
\left(\left.\begin{array}{cccc}1+a_{_1}-b_{_5},\;&1+a_{_2}-b_{_5},\;&1+a_{_3}-b_{_5},\;&1+a_{_4}-b_{_5}\\
\;\;&2-b_{_5},\;&1+b_{_2}-b_{_5},\;&1+b_{_4}-b_{_5}\end{array}\right|y_{_2}\right)
\nonumber\\
&&\hspace{1.5cm}=
(-y_{_2})^{D-3}\;,
\label{GKZ38c}
\end{eqnarray}
where $y_{_2}=x_{_2}=m_{_2}^2/p^2$.
Correspondingly the Feynman integral is formulated as a linear combination
\begin{eqnarray}
&&\Phi_{_{A_2}}(y_{_2})=C_{_{A_2}}^1\Phi_{_{A_2}}^1(y_{_2})
+C_{_{A_2}}^2\Phi_{_{A_2}}^2(y_{_2})
+C_{_{A_2}}^3\Phi_{_{A_2}}^3(y_{_2})
+C_{_{A_2}}^4\Phi_{_{A_2}}^4(y_{_2})
\label{GKZ38c-1}
\end{eqnarray}
in the region $|y_{_2}|<1$.

In the region $|y_{_2}|>1$, the
fundamental solution system is similarly composed of four Pochammer functions $_{_3}F_{_2}$:
\begin{eqnarray}
&&\Phi_{_{A_2}}^5(y_{_2})=(-y_{_2})^{-a_{_1}}\;_{_4}F_{_3}
\left(\left.\begin{array}{cccc}a_{_1},\;&1+a_{_1}-b_{_2},\;&1+a_{_1}-b_{_4},\;&1+a_{_1}-b_{_5}\\
\;\;&1+a_{_1}-a_{_2},\;&1+a_{_1}-a_{_3},\;&1+a_{_1}-a_{_4}\end{array}\right|{1\over y_{_2}}\right)
\nonumber\\
&&\hspace{1.5cm}=
(-y_{_2})^{D-4}\;_{_3}F_{_2}
\left(\left.\begin{array}{ccc}1,\;&4-D,\;&2-{D\over2}\\
\;\;&2,\;&{D\over2}\end{array}\right|{1\over y_{_2}}\right)
\;,\nonumber\\
&&\Phi_{_{A_2}}^6(y_{_2})=(-y_{_2})^{-a_{_3}}\;_{_4}F_{_3}
\left(\left.\begin{array}{cccc}a_{_3},\;&1+a_{_3}-b_{_2},\;&1+a_{_3}-b_{_4},\;&1+a_{_3}-b_{_5}\\
\;\;&1-a_{_2}+a_{_3},\;&1-a_{_1}+a_{_3},\;&1+a_{_3}-a_{_4}\end{array}\right|{1\over y_{_2}}\right)
\nonumber\\
&&\hspace{1.5cm}=
(-y_{_2})^{D/2-2}
\;,\nonumber\\
&&\Phi_{_{A_2}}^{7}(y_{_2})=(-y_{_2})^{-a_{_2}}\;_{_4}F_{_3}
\left(\left.\begin{array}{cccc}a_{_2},\;&1+a_{_2}-b_{_2},\;&1+a_{_2}-b_{_4},\;&1+a_{_2}-b_{_5}\\
\;\;&1-a_{_1}+a_{_2},\;&1+a_{_2}-a_{_3},\;&1+a_{_2}-a_{_4}\end{array}\right|{1\over y_{_2}}\right)
\nonumber\\
&&\hspace{1.5cm}=
(-y_{_2})^{3D/2-5}\;_{_2}F_{_1}
\left(\left.\begin{array}{cc}3-D,\;&5-{3D\over2}\\
\;\;&3-{D\over2}\end{array}\right|{1\over y_{_2}}\right)
\;,\nonumber\\
&&\Phi_{_{A_2}}^{8}(y_{_2})=(-y_{_2})^{-a_{_4}}\;_{_4}F_{_3}
\left(\left.\begin{array}{cccc}a_{_4},\;&1+a_{_4}-b_{_2},\;&1+a_{_4}-b_{_4},\;&1+a_{_4}-b_{_5}\\
\;\;&1-a_{_2}+a_{_4},\;&1-a_{_1}+a_{_4},\;&1-a_{_3}+a_{_4}\end{array}\right|{1\over y_{_2}}\right)
\nonumber\\
&&\hspace{1.5cm}=
(-y_{_2})^{D-3}\;_{_2}F_{_1}
\left(\left.\begin{array}{cc}3-D,\;&1-{D\over2}\\
\;\;&{D\over2}-1\end{array}\right|{1\over y_{_2}}\right)\;.
\label{GKZ39c}
\end{eqnarray}
The Feynman integral is formulated as a linear combination
\begin{eqnarray}
&&\Phi_{_{A_2}}(y_{_2})=C_{_{A_2}}^5\Phi_{_{A_2}}^5(y_{_2})
+C_{_{A_2}}^6\Phi_{_{A_2}}^6(y_{_2})+C_{_{A_2}}^7\Phi_{_{A_2}}^7(y_{_2})
+C_{_{A_2}}^8\Phi_{_{A_2}}^8(y_{_2})
\label{GKZ39c-1}
\end{eqnarray}
in the region $|y_{_2}|>1$. The value of the Feynman integral
at $p^2\neq0$, $m_{_i}^2=0,\;(i=1,\cdots,4)$ induces
\begin{eqnarray}
&&C_{_{A_2}}^1=-{1\over(4\pi)^4}\Big({4\pi\mu^2\over-p^2}\Big)^{2\varepsilon}
{\pi\Gamma^2({D\over2}-1)\Gamma(3-D)\over\sin\pi(2-{D\over2})\Gamma(3-{D\over2})
\Gamma({3\over2}D-4)}\;,
\label{GKZ39c-2}
\end{eqnarray}
and the value of  the Feynman integral
at $m_{_2}^2\neq0$, $p^2=m_{_i}^2=0,\;(i=1,3,4)$ induces
\begin{eqnarray}
&&C_{_{A_2}}^5={1\over(4\pi)^4}\Big({4\pi\mu^2\over -p^2}\Big)^{2\varepsilon}
{2\pi\Gamma(2-D)\over\sin\pi(2-{D\over2})}\;,
\label{GKZ39c-3}
\end{eqnarray}
respectively.
Because there is no relation among the Pochammer functions $\;_{_3}F_{_2}$
which is similar to the relation among the Gauss functions $\;_{_2}F_{_1}$
presented in Eq.(\ref{GKZ44a-2}), one cannot derive the constraints on the
combination coefficients.
However, the combination coefficients can be obtained through the Mellin-Barnes
representation:
\begin{eqnarray}
&&\Sigma_{_{1234}}(p^2)=
-{\Gamma^2({D\over2}-1)\over(4\pi)^4}\Big({4\pi\Lambda_{_{\rm RE}}^2\over-p^2}\Big)^{4-D}
{1\over2\pi i}\int_{-i\infty}^{+i\infty}ds_{_2}
\Big({m_{_2}^2\over-p^2}\Big)^{s_{_2}}\Gamma(-s_{_2})
\nonumber\\
&&\hspace{2.2cm}\times
{\Gamma({D\over2}-1-s_{_2})
\Gamma(2-{D\over2}+s_{_2})\Gamma(D-3-s_{_2})\Gamma(4-D+s_{_2})
\over\Gamma(D-2-s_{_2})\Gamma(3-{D\over2}+s_{_2})\Gamma({3\over2}D-4-s_{_2})}\;,
\label{GKZ39c-4}
\end{eqnarray}
where the coefficient $C_{_{A_2}}^1$ is determined from the residue of the simple
pole of $\Gamma(-s_{_2})$, and the coefficient $C_{_{A_2}}^5$ is determined from
the residue of the simple pole of $\Gamma(4-D+s_{_2})$, respectively.
In addition, the residue of the simple pole of $\Gamma(D/2-1-s_{_2})$ induces
\begin{eqnarray}
&&C_{_{A_2}}^2={1\over(4\pi)^4}\Big({4\pi\mu^2\over-p^2}\Big)^{2\varepsilon}
{2\pi\Gamma(4-D)\sin\pi(4-D)\over(D-2)\sin^2\pi(2-{D\over2})}\;,
\label{GKZ39c-5}
\end{eqnarray}
the residue of the simple pole of $\Gamma(D-3-s_{_2})$ induces
\begin{eqnarray}
&&C_{_{A_2}}^4={1\over(4\pi)^4}\Big({4\pi\mu^2\over-p^2}\Big)^{2\varepsilon}
{2\pi\Gamma(2-D)\over\sin\pi(2-{D\over2})}\;,
\label{GKZ39c-6}
\end{eqnarray}
and the residue of the simple pole of $\Gamma(2-D/2+s_{_2})$ induces
\begin{eqnarray}
&&C_{_{A_2}}^6={1\over(4\pi)^4}\Big({4\pi\mu^2\over-p^2}\Big)^{2\varepsilon}
{2\pi^2\over(D-2)\Gamma(D-2)\sin^2\pi(2-{D\over2})}\;,
\label{GKZ39c-7}
\end{eqnarray}
and $C_{_{A_2}}^3=C_{_{A_2}}^7=C_{_{A_2}}^8=0$ simultaneously.
Certainly the analytic expression Eq. (\ref{GKZ39c-1}) in the
neighborhood of the zero-threshold is consistent with that obtained by the heavy
mass expansion~\cite{Berends1995} also.
A conclusion of the case $m_{_3}\neq0$, $m_{_1}=m_{_2}=m_{_4}=0$ is analogous
to that of $m_{_2}\neq0$, $m_{_1}=m_{_3}=m_{_4}=0$.

\section{Conclusions\label{sec5}}
\indent\indent
Using the Miller transformation, we derive the GKZ-hypergeometric system of
the 2-loop self energy with 4 propagators from its Mellin-Barnes representation.
The dimension of dual space of the GKZ-system equals the number of independent
dimensionless ratios among the external momentum squared and virtual mass squared.
In the neighborhoods of the origin and infinity, we obtain 536 hypergeometric series
solutions totally. In the nonempty intersections of the convergent
regions of those hypergeometric series, the fundamental solution system is composed of 30
linear independent hypergeometric functions. In other words, the Feynman integral
of the 2-loop self energy can be formulated as a linear combination of
those hypergeometric functions from the fundamental solution system
in the convergent region. The combination coefficients are determined by
the Feynman integral at some ordinary points or regular singularities.
We elucidate how to derive the combination coefficients in some special
cases with one nonzero virtual mass.

Using the GKZ-systems on general manifold, we only obtain the hypergeometric solutions in the
neighborhoods of the origin and infinity. In order to derive the fundamental solutions
in the neighborhoods of all possible regular singularities, we should embed the
Feynman integral in some subvarieties of the Grassmannians $G_{_{5,11}}$
through its alpha parametrization. We will present the results elsewhere in detail.

\begin{acknowledgments}
\indent\indent
The work has been supported partly by the National Natural Science Foundation
of China (NNSFC) with Grant No. 12075074, No. 11535002, No. 12235008, Natural Science
Foundation of Guangxi Autonomous Region with Grant No. 2022GXNSFDA035068,
Natural Science Foundation of Hebei Province with Grant No. A2022201017,
and the youth top-notch talent support program of Hebei province.
\end{acknowledgments}

\appendix
\section{Some partial derivatives of $\Phi$\label{app1}}
\indent\indent
\begin{eqnarray}
&&{\partial\Phi\over\partial z_{_1}}={1\over z_{_1}}\prod\limits_{i=1}^{14}z_{_i}^{\alpha_{_i}}
\Big\{\alpha_{_1}\varphi-\sum\limits_{i=1}^4y_{_i}{\partial\varphi\over\partial y_{_i}}\Big\}
\;,\nonumber\\
&&{\partial^2\Phi\over\partial z_{_1}\partial z_{_2}}={1\over z_{_1}z_{_2}}\prod\limits_{i=1}^{14}z_{_i}^{\alpha_{_i}}
\Big\{\alpha_{_1}\alpha_{_2}\varphi+(1-\alpha_{_1}-\alpha_{_2})\sum\limits_{i=1}^4y_{_i}{\partial\varphi\over\partial y_{_i}}
+\sum\limits_{i=1}^4\sum\limits_{j=1}^4y_{_i}y_{_j}{\partial^2\varphi\over\partial y_{_i}\partial y_{_j}}\Big\}
\;,\nonumber\\
&&{\partial\Phi\over\partial z_{_6}}={1\over z_{_6}}\prod\limits_{i=1}^{14}z_{_i}^{\alpha_{_i}}
\Big\{\alpha_{_6}\varphi+y_{_1}{\partial\varphi\over\partial y_{_1}}\Big\}
\;,\nonumber\\
&&{\partial^2\Phi\over\partial z_{_6}\partial z_{_{11}}}={1\over z_{_6}z_{_{11}}}\prod\limits_{i=1}^{14}z_{_i}^{\alpha_{_i}}
\Big\{\alpha_{_6}\alpha_{_{11}}\varphi+(1+\alpha_{_6}+\alpha_{_{11}})y_{_1}{\partial\varphi\over\partial y_{_1}}
+y_{_1}^2{\partial^2\varphi\over\partial y_{_1}^2}\Big\}
\;,\nonumber\\
&&{\partial^3\Phi\over\partial z_{_5}\partial z_{_6}\partial z_{_{11}}}={1\over z_{_5}z_{_6}z_{_{11}}}\prod\limits_{i=1}^{14}z_{_i}^{\alpha_{_i}}
\Big\{\alpha_{_{5}}\alpha_{_6}\alpha_{_{11}}\varphi+(1+\alpha_{_6}+\alpha_{_{11}})\alpha_{_{5}}y_{_1}{\partial\varphi\over\partial y_{_1}}
-\alpha_{_{6}}\alpha_{_{11}}y_{_4}{\partial\varphi\over\partial y_{_4}}
\nonumber\\
&&\hspace{2.6cm}
+\alpha_{_{5}}y_{_1}^2{\partial^2\varphi\over\partial y_{_1}^2}
-(1+\alpha_{_{6}}+\alpha_{_{11}})y_{_1}y_{_4}{\partial^2\varphi\over\partial y_{_1}\partial y_{_4}}
-y_{_1}^2y_{_4}{\partial^3\varphi\over\partial y_{_1}^2\partial y_{_4}}\Big\}
\;,\nonumber\\
&&{\partial\Phi\over\partial z_{_9}}={1\over z_{_9}}\prod\limits_{i=1}^{14}z_{_i}^{\alpha_{_i}}
\Big\{\alpha_{_9}\varphi+\sum\limits_{i=2}^4y_{_i}{\partial\varphi\over\partial y_{_i}}\Big\}
\;,\nonumber\\
&&{\partial^2\Phi\over\partial z_{_9}\partial z_{_{10}}}={1\over z_{_9}z_{_{10}}}\prod\limits_{i=1}^{14}z_{_i}^{\alpha_{_i}}
\Big\{\alpha_{_9}\alpha_{_{10}}\varphi+(1+\alpha_{_9}+\alpha_{_{10}})\sum\limits_{i=2}^4y_{_i}{\partial\varphi\over\partial y_{_i}}
+\sum\limits_{i=2}^4\sum\limits_{j=2}^4y_{_i}y_{_j}{\partial^2\varphi\over\partial y_{_i}\partial y_{_j}}\Big\}
\;,\nonumber\\
&&{\partial\Phi\over\partial z_{_{12}}}={1\over z_{_{12}}}\prod\limits_{i=1}^{14}z_{_i}^{\alpha_{_i}}
\Big\{\alpha_{_{12}}\varphi+y_{_2}{\partial\varphi\over\partial y_{_2}}\Big\}
\;,\nonumber\\
&&{\partial^2\Phi\over\partial z_{_7}\partial z_{_{12}}}={1\over z_{_7}z_{_{12}}}\prod\limits_{i=1}^{14}z_{_i}^{\alpha_{_i}}
\Big\{\alpha_{_{7}}\alpha_{_{12}}\varphi+(1+\alpha_{_{7}}+\alpha_{_{12}})y_{_2}{\partial\varphi\over\partial y_{_2}}
+y_{_2}^2{\partial^2\varphi\over\partial y_{_2}^2}\Big\}
\;,\nonumber\\
&&{\partial^3\Phi\over\partial z_{_5}\partial z_{_7}\partial z_{_{12}}}={1\over z_{_5}z_{_7}z_{_{12}}}\prod\limits_{i=1}^{14}z_{_i}^{\alpha_{_i}}
\Big\{\alpha_{_{5}}\alpha_{_{7}}\alpha_{_{12}}\varphi+(1+\alpha_{_{7}}+\alpha_{_{12}})\alpha_{_{5}}y_{_2}{\partial\varphi\over\partial y_{_2}}
-\alpha_{_{7}}\alpha_{_{12}}y_{_4}{\partial\varphi\over\partial y_{_4}}
\nonumber\\
&&\hspace{2.6cm}
+\alpha_{_{5}}y_{_2}^2{\partial^2\varphi\over\partial y_{_2}^2}
-(1+\alpha_{_{7}}+\alpha_{_{12}})y_{_2}y_{_4}{\partial^2\varphi\over\partial y_{_2}\partial y_{_4}}
-y_{_2}^2y_{_4}{\partial^3\varphi\over\partial y_{_2}^2\partial y_{_4}}\Big\}
\;,\nonumber\\
&&{\partial\Phi\over\partial z_{_{13}}}={1\over z_{_{13}}}\prod\limits_{i=1}^{14}z_{_i}^{\alpha_{_i}}
\Big\{\alpha_{_{13}}\varphi+y_{_3}{\partial\varphi\over\partial y_{_3}}\Big\}
\;,\nonumber\\
&&{\partial^2\Phi\over\partial z_{_8}\partial z_{_{13}}}={1\over z_{_8}z_{_{13}}}\prod\limits_{i=1}^{14}z_{_i}^{\alpha_{_i}}
\Big\{\alpha_{_{8}}\alpha_{_{13}}\varphi+(1+\alpha_{_{8}}+\alpha_{_{13}})y_{_3}{\partial\varphi\over\partial y_{_3}}
+y_{_3}^2{\partial^2\varphi\over\partial y_{_3}^2}\Big\}
\;,\nonumber\\
&&{\partial^3\Phi\over\partial z_{_5}\partial z_{_8}\partial z_{_{13}}}={1\over z_{_5}z_{_8}z_{_{13}}}\prod\limits_{i=1}^{14}z_{_i}^{\alpha_{_i}}
\Big\{\alpha_{_{5}}\alpha_{_{8}}\alpha_{_{13}}\varphi+(1+\alpha_{_{8}}+\alpha_{_{13}})\alpha_{_{5}}y_{_3}{\partial\varphi\over\partial y_{_3}}
-\alpha_{_{8}}\alpha_{_{13}}y_{_4}{\partial\varphi\over\partial y_{_4}}
\nonumber\\
&&\hspace{2.6cm}
+\alpha_{_{5}}y_{_3}^2{\partial^2\varphi\over\partial y_{_3}^2}
-(1+\alpha_{_{8}}+\alpha_{_{13}})y_{_3}y_{_4}{\partial^2\varphi\over\partial y_{_3}\partial y_{_4}}
-y_{_3}^2y_{_4}{\partial^3\varphi\over\partial y_{_3}^2\partial y_{_4}}\Big\}
\;,\nonumber\\
&&{\partial\Phi\over\partial z_{_{14}}}={1\over z_{_{14}}}\prod\limits_{i=1}^{14}z_{_i}^{\alpha_{_i}}
\Big\{\alpha_{_{14}}\varphi+y_{_4}{\partial\varphi\over\partial y_{_4}}\Big\}
\;,\nonumber\\
&&{\partial^2\Phi\over\partial z_{_4}\partial z_{_{14}}}={1\over z_{_4}z_{_{14}}}\prod\limits_{i=1}^{14}z_{_i}^{\alpha_{_i}}
\Big\{\alpha_{_{4}}\alpha_{_{14}}\varphi+\alpha_{_{4}}y_{_4}{\partial\varphi\over\partial y_{_4}}
-\alpha_{_{14}}y_{_2}{\partial\varphi\over\partial y_{_2}}-\alpha_{_{14}}y_{_3}{\partial\varphi\over\partial y_{_3}}
\nonumber\\
&&\hspace{2.1cm}
-y_{_2}y_{_4}{\partial^2\varphi\over\partial y_{_2}\partial y_{_4}}-y_{_3}y_{_4}{\partial^2\varphi\over\partial y_{_3}\partial y_{_4}}\Big\}
\;,\nonumber\\
&&{\partial^3\Phi\over\partial z_{_3}\partial z_{_4}\partial z_{_{14}}}={1\over z_{_3}z_{_4}z_{_{14}}}\prod\limits_{i=1}^{14}z_{_i}^{\alpha_{_i}}
\Big\{\alpha_{_{3}}\alpha_{_{4}}\alpha_{_{14}}\varphi+\alpha_{_{3}}\alpha_{_{4}}y_{_4}{\partial\varphi\over\partial y_{_4}}
+(1-\alpha_{_{3}}-\alpha_{_{4}})\alpha_{_{14}}\Big[y_{_2}{\partial\varphi\over\partial y_{_2}}
\nonumber\\
&&\hspace{2.6cm}
+y_{_3}{\partial\varphi\over\partial y_{_3}}\Big]+(1-\alpha_{_{3}}-\alpha_{_{4}})\Big[
y_{_2}y_{_4}{\partial^2\varphi\over\partial y_{_2}\partial y_{_4}}+y_{_3}y_{_4}{\partial^2\varphi\over\partial y_{_3}\partial y_{_4}}\Big]
\nonumber\\
&&\hspace{2.6cm}
+\alpha_{_{14}}\Big[y_{_2}^2{\partial^2\varphi\over\partial y_{_2}^2}
+2y_{_2}y_{_3}{\partial^2\varphi\over\partial y_{_2}\partial y_{_3}}+y_{_3}^2{\partial^2\varphi\over\partial y_{_3}^2}\Big]
+y_{_2}^2y_{_4}{\partial^3\varphi\over\partial y_{_2}^2\partial y_{_4}}
\nonumber\\
&&\hspace{2.6cm}
+2y_{_2}y_{_3}y_{_4}{\partial^3\varphi\over\partial y_{_2}\partial y_{_3}\partial y_{_4}}
+y_{_3}^2y_{_4}{\partial^3\varphi\over\partial y_{_3}^2\partial y_{_4}}\Big\}
\;,\nonumber\\
&&{\partial^3\Phi\over\partial z_{_1}\partial z_{_2}\partial z_{_3}}={1\over z_{_1}z_{_2}z_{_3}}\prod\limits_{i=1}^{14}z_{_i}^{\alpha_{_i}}
\Big\{\alpha_{_1}\alpha_{_2}\alpha_{_3}\varphi+(1-\alpha_{_1}-\alpha_{_2})\alpha_{_3}\sum\limits_{i=1}^4y_{_i}{\partial\varphi\over\partial y_{_i}}
\nonumber\\
&&\hspace{2.5cm}
-(1-\alpha_{_1})(1-\alpha_{_2})\Big[y_{_2}{\partial\varphi\over\partial y_{_2}}+y_{_3}{\partial\varphi\over\partial y_{_3}}\Big]
+\alpha_{_3}\sum\limits_{i=1}^4\sum\limits_{j=1}^4y_{_i}y_{_j}{\partial^2\varphi\over\partial y_{_i}\partial y_{_j}}
\nonumber\\
&&\hspace{2.5cm}
-(3-\alpha_{_1}-\alpha_{_2})\sum\limits_{i=1}^4\Big[y_{_2}y_{_i}{\partial^2\varphi\over\partial y_{_2}\partial y_{_i}}
+y_{_3}y_{_i}{\partial^2\varphi\over\partial y_{_3}\partial y_{_i}}\Big]
\nonumber\\
&&\hspace{2.5cm}
-\sum\limits_{i=1}^4\sum\limits_{j=1}^4\Big[y_{_2}y_{_i}y_{_j}{\partial^3\varphi\over\partial y_{_2}\partial y_{_i}\partial y_{_j}}
+y_{_3}y_{_i}y_{_j}{\partial^3\varphi\over\partial y_{_3}\partial y_{_i}\partial y_{_j}}\Big]\Big\}
\;,\nonumber\\
&&{\partial^4\Phi\over\partial z_{_1}\partial z_{_2}\partial z_{_3}\partial z_{_4}}={1\over z_{_1}z_{_2}z_{_3}z_{_4}}\prod\limits_{i=1}^{14}z_{_i}^{\alpha_{_i}}
\Big\{\alpha_{_1}\alpha_{_2}\alpha_{_3}\alpha_{_4}\varphi+(1-\alpha_{_1}-\alpha_{_2})\alpha_{_3}\alpha_{_4}\sum\limits_{i=1}^4y_{_i}{\partial\varphi\over\partial y_{_i}}
\nonumber\\
&&\hspace{3.0cm}
+(1-\alpha_{_1})(1-\alpha_{_2})(1-\alpha_{_3}-\alpha_{_4})\Big[y_{_2}{\partial\varphi\over\partial y_{_2}}+y_{_3}{\partial\varphi\over\partial y_{_3}}\Big]
\nonumber\\
&&\hspace{3.0cm}
+(4-2\alpha_{_1}-2\alpha_{_2}+\alpha_{_1}\alpha_{_2})\Big[y_{_2}^2{\partial^2\varphi\over\partial y_{_2}^2}
+2y_{_2}y_{_3}{\partial^2\varphi\over\partial y_{_2}\partial y_{_3}}+y_{_3}^2{\partial^2\varphi\over\partial y_{_3}^2}\Big]
\nonumber\\
&&\hspace{3.0cm}
+(3-\alpha_{_1}-\alpha_{_2})(1-\alpha_{_3}-\alpha_{_4})\sum\limits_{i=1}^4\Big[y_{_2}y_{_i}{\partial^2\varphi\over\partial y_{_2}\partial y_{_i}}
+y_{_3}y_{_i}{\partial^2\varphi\over\partial y_{_3}\partial y_{_i}}\Big]
\nonumber\\
&&\hspace{3.0cm}
+(1-\alpha_{_3}-\alpha_{_4})\sum\limits_{i=1}^4\sum\limits_{j=1}^4\Big[y_{_2}y_{_i}y_{_j}{\partial^3\varphi\over\partial y_{_2}\partial y_{_i}\partial y_{_j}}
+y_{_3}y_{_i}y_{_j}{\partial^3\varphi\over\partial y_{_3}\partial y_{_i}\partial y_{_j}}\Big]
\nonumber\\
&&\hspace{3.0cm}
+(5-\alpha_{_1}-\alpha_{_2})\sum\limits_{i=1}^4\Big[y_{_2}^2y_{_i}{\partial^3\varphi\over\partial y_{_2}^2\partial y_{_i}}
+2y_{_2}y_{_3}y_{_i}{\partial^3\varphi\over\partial y_{_2}\partial y_{_3}\partial y_{_i}}+y_{_3}^2y_{_i}{\partial^3\varphi\over\partial y_{_3}^2\partial y_{_i}}\Big]
\nonumber\\
&&\hspace{3.0cm}
+\alpha_{_3}\alpha_{_4}\sum\limits_{i=1}^4\sum\limits_{j=1}^4y_{_i}y_{_j}{\partial^2\varphi\over\partial y_{_i}\partial y_{_j}}
+\sum\limits_{i=1}^4\sum\limits_{j=1}^4\Big[y_{_2}^2y_{_i}y_{_j}{\partial^4\varphi\over\partial y_{_2}^2\partial y_{_i}\partial y_{_j}}
\nonumber\\
&&\hspace{3.0cm}
+2y_{_2}y_{_3}y_{_i}y_{_j}{\partial^4\varphi\over\partial y_{_2}\partial y_{_3}\partial y_{_i}\partial y_{_j}}
+y_{_3}^2y_{_i}y_{_j}{\partial^4\varphi\over\partial y_{_3}^2\partial y_{_i}\partial y_{_j}}\Big]\Big\}
\;,\nonumber\\
&&{\partial^3\Phi\over\partial z_{_7}\partial z_{_9}\partial z_{_{10}}}={1\over z_{_7}z_{_9}z_{_{10}}}\prod\limits_{i=1}^{14}z_{_i}^{\alpha_{_i}}
\Big\{\alpha_{_7}\alpha_{_9}\alpha_{_{10}}\varphi+\alpha_{_7}(1+\alpha_{_9}+\alpha_{_{10}})\sum\limits_{i=2}^4y_{_i}{\partial\varphi\over\partial y_{_i}}
\nonumber\\
&&\hspace{2.7cm}
+(1+\alpha_{_9})(1+\alpha_{_{10}})y_{_2}{\partial\varphi\over\partial y_{_2}}
+(3+\alpha_{_9}+\alpha_{_{10}})\sum\limits_{i=2}^4y_{_2}y_{_i}{\partial^2\varphi\over\partial y_{_2}\partial y_{_i}}
\nonumber\\
&&\hspace{2.7cm}
+\alpha_{_7}\sum\limits_{i=2}^4\sum\limits_{j=2}^4y_{_i}y_{_j}{\partial^2\varphi\over\partial y_{_i}\partial y_{_j}}
+\sum\limits_{i=2}^4\sum\limits_{j=2}^4y_{_2}y_{_i}y_{_j}{\partial^3\varphi\over\partial y_{_2}\partial y_{_i}\partial y_{_j}}\Big\}
\;,\nonumber\\
&&{\partial^4\Phi\over\partial z_{_7}\partial z_{_9}\partial z_{_{10}}\partial z_{_{12}}}={1\over z_{_7}z_{_9}z_{_{10}}z_{_{12}}}\prod\limits_{i=1}^{14}z_{_i}^{\alpha_{_i}}
\Big\{\alpha_{_7}\alpha_{_9}\alpha_{_{10}}\alpha_{_{12}}\varphi+\alpha_{_7}\alpha_{_{12}}(1+\alpha_{_9}+\alpha_{_{10}})\sum\limits_{i=2}^4y_{_i}{\partial\varphi\over\partial y_{_i}}
\nonumber\\
&&\hspace{3.2cm}
+(1+\alpha_{_7}+\alpha_{_{12}})(1+\alpha_{_9})(1+\alpha_{_{10}})y_{_2}{\partial\varphi\over\partial y_{_2}}
+\alpha_{_7}\alpha_{_{12}}\sum\limits_{i=2}^4\sum\limits_{j=2}^4y_{_i}y_{_j}{\partial^2\varphi\over\partial y_{_i}\partial y_{_j}}
\nonumber\\
&&\hspace{3.2cm}
+(1+\alpha_{_7}+\alpha_{_{12}})(3+\alpha_{_9}+\alpha_{_{10}})\sum\limits_{i=2}^4y_{_2}y_{_i}{\partial^2\varphi\over\partial y_{_2}\partial y_{_i}}
\nonumber\\
&&\hspace{3.2cm}
+(4+2\alpha_{_9}+2\alpha_{_{10}}+\alpha_{_{9}}\alpha_{_{10}})y_{_2}^2{\partial^2\varphi\over\partial y_{_2}^2}
\nonumber\\
&&\hspace{3.2cm}
+(1+\alpha_{_7}+\alpha_{_{12}})\sum\limits_{i=2}^4\sum\limits_{j=2}^4y_{_2}y_{_i}y_{_j}{\partial^3\varphi\over\partial y_{_2}\partial y_{_i}\partial y_{_j}}
\nonumber\\
&&\hspace{3.2cm}
+(5+\alpha_{_9}+\alpha_{_{10}})\sum\limits_{i=2}^4y_{_2}^2y_{_i}{\partial^3\varphi\over\partial y_{_2}^2\partial y_{_i}}
+\sum\limits_{i=2}^4\sum\limits_{j=2}^4y_{_2}^2y_{_i}y_{_j}{\partial^4\varphi\over\partial y_{_2}^2\partial y_{_i}\partial y_{_j}}\Big\}
\;,\nonumber\\
&&{\partial^4\Phi\over\partial z_{_8}\partial z_{_9}\partial z_{_{10}}\partial z_{_{13}}}={1\over z_{_8}z_{_9}z_{_{10}}z_{_{13}}}\prod\limits_{i=1}^{14}z_{_i}^{\alpha_{_i}}
\Big\{\alpha_{_8}\alpha_{_9}\alpha_{_{10}}\alpha_{_{13}}\varphi+\alpha_{_8}\alpha_{_{13}}(1+\alpha_{_9}+\alpha_{_{10}})\sum\limits_{i=2}^4y_{_i}{\partial\varphi\over\partial y_{_i}}
\nonumber\\
&&\hspace{3.2cm}
+(1+\alpha_{_8}+\alpha_{_{13}})(1+\alpha_{_9})(1+\alpha_{_{10}})y_{_3}{\partial\varphi\over\partial y_{_3}}
+\alpha_{_8}\alpha_{_{13}}\sum\limits_{i=2}^4\sum\limits_{j=2}^4y_{_i}y_{_j}{\partial^2\varphi\over\partial y_{_i}\partial y_{_j}}
\nonumber\\
&&\hspace{3.2cm}
+(1+\alpha_{_8}+\alpha_{_{13}})(3+\alpha_{_9}+\alpha_{_{10}})\sum\limits_{i=2}^4y_{_3}y_{_i}{\partial^2\varphi\over\partial y_{_3}\partial y_{_i}}
\nonumber\\
&&\hspace{3.2cm}
+(4+2\alpha_{_9}+2\alpha_{_{10}}+\alpha_{_{9}}\alpha_{_{10}})y_{_3}^2{\partial^2\varphi\over\partial y_{_3}^2}
\nonumber\\
&&\hspace{3.2cm}
+(1+\alpha_{_8}+\alpha_{_{13}})\sum\limits_{i=2}^4\sum\limits_{j=2}^4y_{_3}y_{_i}y_{_j}{\partial^3\varphi\over\partial y_{_3}\partial y_{_i}\partial y_{_j}}
\nonumber\\
&&\hspace{3.2cm}
+(5+\alpha_{_9}+\alpha_{_{10}})\sum\limits_{i=2}^4y_{_3}^2y_{_i}{\partial^3\varphi\over\partial y_{_3}^2\partial y_{_i}}
+\sum\limits_{i=2}^4\sum\limits_{j=2}^4y_{_3}^2y_{_i}y_{_j}{\partial^4\varphi\over\partial y_{_3}^2\partial y_{_i}\partial y_{_j}}\Big\}
\;,\nonumber\\
&&{\partial^3\Phi\over\partial z_{_1}\partial z_{_2}\partial z_{_5}}={1\over z_{_1}z_{_2}z_{_5}}\prod\limits_{i=1}^{14}z_{_i}^{\alpha_{_i}}
\Big\{\alpha_{_1}\alpha_{_2}\alpha_{_5}\varphi+(1-\alpha_{_1}-\alpha_{_2})\alpha_{_5}\sum\limits_{i=1}^4y_{_i}{\partial\varphi\over\partial y_{_i}}
\nonumber\\
&&\hspace{2.6cm}
-(1-\alpha_{_1})(1-\alpha_{_2})y_{_4}{\partial\varphi\over\partial y_{_4}}
-(3-\alpha_{_1}-\alpha_{_2})\sum\limits_{i=1}^4y_{_4}y_{_i}{\partial^2\varphi\over\partial y_{_4}\partial y_{_i}}
\nonumber\\
&&\hspace{2.6cm}
+\alpha_{_5}\sum\limits_{i=1}^4\sum\limits_{j=1}^4y_{_i}y_{_j}{\partial^2\varphi\over\partial y_{_i}\partial y_{_j}}
-\sum\limits_{i=1}^4\sum\limits_{j=1}^4y_{_4}y_{_i}y_{_j}{\partial^3\varphi\over\partial y_{_4}\partial y_{_i}\partial y_{_j}}\Big\}
\;,\nonumber\\
&&{\partial^3\Phi\over\partial z_{_9}\partial z_{_{10}}\partial z_{_{14}}}={1\over z_{_9}z_{_{10}}z_{_{14}}}\prod\limits_{i=1}^{14}z_{_i}^{\alpha_{_i}}
\Big\{\alpha_{_9}\alpha_{_{10}}\alpha_{_{14}}\varphi+(1+\alpha_{_9}+\alpha_{_{10}})\alpha_{_{14}}\sum\limits_{i=2}^4y_{_i}{\partial\varphi\over\partial y_{_i}}
\nonumber\\
&&\hspace{2.7cm}
+(1+\alpha_{_9})(1+\alpha_{_{10}})y_{_4}{\partial\varphi\over\partial y_{_4}}
+(3+\alpha_{_9}+\alpha_{_{10}})\sum\limits_{i=2}^4y_{_4}y_{_i}{\partial^2\varphi\over\partial y_{_4}\partial y_{_i}}
\nonumber\\
&&\hspace{2.7cm}
+\alpha_{_{14}}\sum\limits_{i=2}^4\sum\limits_{j=2}^4y_{_i}y_{_j}{\partial^2\varphi\over\partial y_{_i}\partial y_{_j}}
+\sum\limits_{i=2}^4\sum\limits_{j=2}^4y_{_4}y_{_i}y_{_j}{\partial^3\varphi\over\partial y_{_4}\partial y_{_i}\partial y_{_j}}\Big\}
\;,\nonumber\\
&&{\partial^3\Phi\over\partial z_{_4}\partial z_{_6}\partial z_{_{11}}}={1\over z_{_4}z_{_6}z_{_{11}}}\prod\limits_{i=1}^{14}z_{_i}^{\alpha_{_i}}
\Big\{\alpha_{_{4}}\alpha_{_6}\alpha_{_{11}}\varphi+(1+\alpha_{_6}+\alpha_{_{11}})\alpha_{_{4}}y_{_1}{\partial\varphi\over\partial y_{_1}}
\nonumber\\
&&\hspace{2.6cm}
-\alpha_{_{6}}\alpha_{_{11}}\Big[y_{_2}{\partial\varphi\over\partial y_{_2}}+y_{_3}{\partial\varphi\over\partial y_{_3}}\Big]
+\alpha_{_{4}}y_{_1}^2{\partial^2\varphi\over\partial y_{_1}^2}
-(1+\alpha_{_{6}}+\alpha_{_{11}})\Big[y_{_1}y_{_2}{\partial^2\varphi\over\partial y_{_1}\partial y_{_2}}
\nonumber\\
&&\hspace{2.6cm}
+y_{_1}y_{_3}{\partial^2\varphi\over\partial y_{_1}\partial y_{_3}}\Big]
-y_{_1}^2y_{_2}{\partial^3\varphi\over\partial y_{_1}^2\partial y_{_2}}-y_{_1}^2y_{_3}{\partial^3\varphi\over\partial y_{_1}^2\partial y_{_3}}\Big\}
\;,\nonumber\\
&&{\partial^4\Phi\over\partial z_{_3}\partial z_{_4}\partial z_{_6}\partial z_{_{11}}}={1\over z_{_3}z_{_4}z_{_6}z_{_{11}}}\prod\limits_{i=1}^{14}z_{_i}^{\alpha_{_i}}
\Big\{\alpha_{_{3}}\alpha_{_{4}}\alpha_{_6}\alpha_{_{11}}\varphi+(1+\alpha_{_6}+\alpha_{_{11}})\alpha_{_{3}}\alpha_{_{4}}y_{_1}{\partial\varphi\over\partial y_{_1}}
\nonumber\\
&&\hspace{3.2cm}
+(1-\alpha_{_{3}}-\alpha_{_{4}})\alpha_{_{6}}\alpha_{_{11}}\Big[y_{_2}{\partial\varphi\over\partial y_{_2}}+y_{_3}{\partial\varphi\over\partial y_{_3}}\Big]
+\alpha_{_{3}}\alpha_{_{4}}y_{_1}^2{\partial^2\varphi\over\partial y_{_1}^2}
\nonumber\\
&&\hspace{3.2cm}
+(1-\alpha_{_{3}}-\alpha_{_{4}})(1+\alpha_{_{6}}+\alpha_{_{11}})\Big[y_{_1}y_{_2}{\partial^2\varphi\over\partial y_{_1}\partial y_{_2}}
+y_{_1}y_{_3}{\partial^2\varphi\over\partial y_{_1}\partial y_{_3}}\Big]
\nonumber\\
&&\hspace{3.2cm}
+\alpha_{_{6}}\alpha_{_{11}}\Big[y_{_2}^2{\partial^2\varphi\over\partial y_{_2}^2}+y_{_3}^2{\partial^2\varphi\over\partial y_{_3}^2}
+2y_{_2}y_{_3}{\partial^2\varphi\over\partial y_{_2}\partial y_{_3}}\Big]
\nonumber\\
&&\hspace{3.2cm}
+(1-\alpha_{_{3}}-\alpha_{_{4}})\Big[y_{_1}^2y_{_2}{\partial^3\varphi\over\partial y_{_1}^2\partial y_{_2}}
+y_{_1}^2y_{_3}{\partial^3\varphi\over\partial y_{_1}^2\partial y_{_3}}\Big]
\nonumber\\
&&\hspace{3.2cm}
+(1+\alpha_{_{6}}+\alpha_{_{11}})\Big[y_{_1}y_{_2}^2{\partial^3\varphi\over\partial y_{_1}\partial y_{_2}^2}
+y_{_1}y_{_3}^2{\partial^3\varphi\over\partial y_{_1}\partial y_{_3}^2}
+2y_{_1}y_{_2}y_{_3}{\partial^3\varphi\over\partial y_{_1}\partial y_{_2}\partial y_{_3}}\Big]
\nonumber\\
&&\hspace{3.2cm}
+y_{_1}^2y_{_2}^2{\partial^4\varphi\over\partial y_{_1}^2\partial y_{_2}^2}
+y_{_1}^2y_{_3}^2{\partial^4\varphi\over\partial y_{_1}^2\partial y_{_3}^2}
+2y_{_1}^2y_{_2}y_{_3}{\partial^4\varphi\over\partial y_{_1}^2\partial y_{_2}\partial y_{_3}}\Big\}
\;.
\label{GKZ-app-1}
\end{eqnarray}

Correspondingly the linear partial differential operators $\hat{L}_{_i}$
$(i=1,\cdots,5)$ are
\begin{eqnarray}
&&\hat{L}_{_1}=(1-y_{_1})y_{_1}^2{\partial^2\over\partial y_{_1}^2}-y_{_1}\sum\limits_{i=2}^4y_{_i}^2{\partial^2\over\partial y_{_i}^2}
-y_{_1}\sum\limits_{i=1}^4\sum\limits_{j\neq i}^4y_{_i}y_{_j}{\partial^2\over\partial y_{_i}\partial y_{_j}}
\nonumber\\
&&\hspace{1.1cm}
+\Big[(1+\alpha_{_6}+\alpha_{_{11}})-(1-\alpha_{_1}-\alpha_{_2})y_{_1}\Big]y_{_1}{\partial\over\partial y_{_1}}
\nonumber\\
&&\hspace{1.1cm}
-(1-\alpha_{_1}-\alpha_{_2})y_{_1}\sum\limits_{i=2}^4y_{_i}{\partial\over\partial y_{_i}}
+\Big[\alpha_{_6}\alpha_{_{11}}-\alpha_{_1}\alpha_{_2}y_{_1}\Big]
\;,\nonumber\\
&&\hat{L}_{_2}=y_{_2}^2y_{_3}{\partial^2\over\partial y_{_2}^2}-y_{_2}y_{_3}^2{\partial^2\over\partial y_{_3}^2}
+(1+\alpha_{_{7}}+\alpha_{_{12}})y_{_2}y_{_3}{\partial\over\partial y_{_2}}
\nonumber\\
&&\hspace{1.1cm}
-(1+\alpha_{_{8}}+\alpha_{_{13}})y_{_2}y_{_3}{\partial\over\partial y_{_3}}
+(\alpha_{_{7}}\alpha_{_{12}}y_{_3}-\alpha_{_{8}}\alpha_{_{13}}y_{_2})
\;,\nonumber\\
&&\hat{L}_{_3}=y_{_2}^2y_{_3}y_{_4}{\partial^3\over\partial y_{_2}^2\partial y_{_4}}
+2y_{_2}y_{_3}^2y_{_4}{\partial^3\over\partial y_{_2}\partial y_{_3}\partial y_{_4}}
+\Big(y_{_3}^3y_{_4}+y_{_3}^2y_{_4}^2\Big){\partial^3\over\partial y_{_3}^2\partial y_{_4}}
\nonumber\\
&&\hspace{1.1cm}
+\alpha_{_{14}}\Big[y_{_2}^2y_{_3}{\partial^2\over\partial y_{_2}^2}
+2y_{_2}y_{_3}^2{\partial^2\over\partial y_{_2}\partial y_{_3}}\Big]
+(1-\alpha_{_{3}}-\alpha_{_{4}})y_{_2}y_{_3}y_{_4}{\partial^2\over\partial y_{_2}\partial y_{_4}}
\nonumber\\
&&\hspace{1.1cm}
+\Big[(1-\alpha_{_{3}}-\alpha_{_{4}})y_{_3}^2y_{_4}
+(1+\alpha_{_{8}}+\alpha_{_{13}})y_{_3}y_{_4}^2\Big]{\partial^2\over\partial y_{_3}\partial y_{_4}}
\nonumber\\
&&\hspace{1.1cm}
+\Big(\alpha_{_{14}}y_{_3}^3-\alpha_{_{5}}y_{_3}^2y_{_4}\Big){\partial^2\over\partial y_{_3}^2}
+(1-\alpha_{_{3}}-\alpha_{_{4}})\alpha_{_{14}}y_{_2}y_{_3}{\partial\over\partial y_{_2}}
\nonumber\\
&&\hspace{1.1cm}
+\Big[(1-\alpha_{_{3}}-\alpha_{_{4}})\alpha_{_{14}}y_{_3}^2
-(1+\alpha_{_{8}}+\alpha_{_{13}})\alpha_{_{5}}y_{_3}y_{_4}\Big]{\partial\over\partial y_{_3}}
\nonumber\\
&&\hspace{1.1cm}
+\Big[\alpha_{_{3}}\alpha_{_{4}}y_{_3}y_{_4}+\alpha_{_{8}}\alpha_{_{13}}y_{_4}^2\Big]{\partial\over\partial y_{_4}}
+\Big(\alpha_{_{3}}\alpha_{_{4}}\alpha_{_{14}}y_{_3}-\alpha_{_{5}}\alpha_{_{8}}\alpha_{_{13}}y_{_4}\Big)
\;,\nonumber\\
&&\hat{L}_{_4}=y_{_1}^2y_{_4}^2{\partial^3\over\partial y_{_1}^2\partial y_{_4}}
+\sum\limits_{i=2}^4\sum\limits_{j=2}^4y_{_1}y_{_4}y_{_i}y_{_j}{\partial^3\over\partial y_{_4}\partial y_{_i}\partial y_{_j}}
-\alpha_{_{5}}y_{_1}^2y_{_4}{\partial^2\over\partial y_{_1}^2}
\nonumber\\
&&\hspace{1.1cm}
+(1+\alpha_{_{6}}+\alpha_{_{11}})y_{_1}y_{_4}^2{\partial^2\over\partial y_{_1}\partial y_{_4}}
+(3+\alpha_{_9}+\alpha_{_{10}})\sum\limits_{i=2}^4y_{_1}y_{_4}y_{_i}{\partial^2\over\partial y_{_4}\partial y_{_i}}
\nonumber\\
&&\hspace{1.1cm}
+\alpha_{_{14}}\sum\limits_{i=2}^4\sum\limits_{j=2}^4y_{_1}y_{_i}y_{_j}{\partial^2\over\partial y_{_i}\partial y_{_j}}
-(1+\alpha_{_6}+\alpha_{_{11}})\alpha_{_{5}}y_{_1}y_{_4}{\partial\over\partial y_{_1}}
\nonumber\\
&&\hspace{1.1cm}
+\Big[\alpha_{_{6}}\alpha_{_{11}}y_{_4}^2+(1+\alpha_{_9})(1+\alpha_{_{10}})y_{_1}y_{_4}\Big]{\partial\over\partial y_{_4}}
+(1+\alpha_{_9}+\alpha_{_{10}})\alpha_{_{14}}\sum\limits_{i=2}^4y_{_1}y_{_i}{\partial\over\partial y_{_i}}
\nonumber\\
&&\hspace{1.1cm}
+\Big(\alpha_{_9}\alpha_{_{10}}\alpha_{_{14}}y_{_1}-\alpha_{_{5}}\alpha_{_6}\alpha_{_{11}}y_{_4}\Big)
\;,\nonumber\\
&&\hat{L}_{_5}=y_{_1}^2y_{_2}^2y_{_3}{\partial^4\over\partial y_{_1}^2\partial y_{_2}^2}
+y_{_1}^2y_{_3}^3{\partial^4\over\partial y_{_1}^2\partial y_{_3}^2}
+2y_{_1}^2y_{_2}y_{_3}^2{\partial^4\over\partial y_{_1}^2\partial y_{_2}\partial y_{_3}}
\nonumber\\
&&\hspace{1.1cm}
-\sum\limits_{i=2}^4\sum\limits_{j=2}^4y_{_1}y_{_3}^2y_{_i}y_{_j}{\partial^4\over\partial y_{_3}^2\partial y_{_i}\partial y_{_j}}
+(1-\alpha_{_{3}}-\alpha_{_{4}})\Big[y_{_1}^2y_{_2}y_{_3}{\partial^3\over\partial y_{_1}^2\partial y_{_2}}
\nonumber\\
&&\hspace{1.1cm}
+y_{_1}^2y_{_3}^2{\partial^3\over\partial y_{_1}^2\partial y_{_3}}\Big]
+(1+\alpha_{_{6}}+\alpha_{_{11}})\Big[y_{_1}y_{_2}^2y_{_3}{\partial^3\over\partial y_{_1}\partial y_{_2}^2}
+y_{_1}y_{_3}^3{\partial^3\over\partial y_{_1}\partial y_{_3}^2}
\nonumber\\
&&\hspace{1.1cm}
+2y_{_1}y_{_2}y_{_3}^2{\partial^3\over\partial y_{_1}\partial y_{_2}\partial y_{_3}}\Big]
-(1+\alpha_{_8}+\alpha_{_{13}})\sum\limits_{i=2}^4\sum\limits_{j=2}^4y_{_1}y_{_3}y_{_i}y_{_j}{\partial^3\over\partial y_{_3}\partial y_{_i}\partial y_{_j}}
\nonumber\\
&&\hspace{1.1cm}
-(5+\alpha_{_9}+\alpha_{_{10}})\sum\limits_{i=2}^4y_{_1}y_{_3}^2y_{_i}{\partial^3\over\partial y_{_3}^2\partial y_{_i}}
+\alpha_{_{3}}\alpha_{_{4}}y_{_1}^2y_{_3}{\partial^2\over\partial y_{_1}^2}
\nonumber\\
&&\hspace{1.1cm}
+(1-\alpha_{_{3}}-\alpha_{_{4}})(1+\alpha_{_{6}}+\alpha_{_{11}})\Big[y_{_1}y_{_2}y_{_3}{\partial^2\over\partial y_{_1}\partial y_{_2}}
+y_{_1}y_{_3}^2{\partial^2\over\partial y_{_1}\partial y_{_3}}\Big]
\nonumber\\
&&\hspace{1.1cm}
+\alpha_{_{6}}\alpha_{_{11}}\Big[y_{_2}^2y_{_3}{\partial^2\over\partial y_{_2}^2}
+2y_{_2}y_{_3}^2{\partial^2\over\partial y_{_2}\partial y_{_3}}\Big]
-\alpha_{_8}\alpha_{_{13}}\sum\limits_{i=2}^4\sum\limits_{j=2}^4y_{_1}y_{_i}y_{_j}{\partial^2\over\partial y_{_i}\partial y_{_j}}
\nonumber\\
&&\hspace{1.1cm}
+\Big[\alpha_{_{6}}\alpha_{_{11}}y_{_3}^3-(2+\alpha_{_9})(2+\alpha_{_{10}})
y_{_1}y_{_3}^2\Big]{\partial^2\over\partial y_{_3}^2}
\nonumber\\
&&\hspace{1.1cm}
-(1+\alpha_{_8}+\alpha_{_{13}})(3+\alpha_{_9}+\alpha_{_{10}})\sum\limits_{i=2}^4y_{_1}y_{_3}y_{_i}{\partial^2\over\partial y_{_3}\partial y_{_i}}
\nonumber\\
&&\hspace{1.1cm}
+(1+\alpha_{_6}+\alpha_{_{11}})\alpha_{_{3}}\alpha_{_{4}}y_{_1}y_{_3}{\partial\over\partial y_{_1}}
+(1-\alpha_{_{3}}-\alpha_{_{4}})\alpha_{_{6}}\alpha_{_{11}}y_{_2}y_{_3}{\partial\over\partial y_{_2}}
\nonumber\\
&&\hspace{1.1cm}
+\Big[(1-\alpha_{_{3}}-\alpha_{_{4}})\alpha_{_{6}}\alpha_{_{11}}y_{_3}^2
-(1+\alpha_{_8}+\alpha_{_{13}})(1+\alpha_{_9})(1+\alpha_{_{10}})y_{_1}y_{_3}\Big]{\partial\over\partial y_{_3}}
\nonumber\\
&&\hspace{1.1cm}
-(1+\alpha_{_9}+\alpha_{_{10}})\alpha_{_8}\alpha_{_{13}}\sum\limits_{i=2}^4y_{_1}y_{_i}{\partial\over\partial y_{_i}}
+\Big(\alpha_{_{3}}\alpha_{_{4}}\alpha_{_6}\alpha_{_{11}}y_{_3}-\alpha_{_8}\alpha_{_9}\alpha_{_{10}}\alpha_{_{13}}y_{_1}\Big)\;.
\label{GKZ19-5}
\end{eqnarray}

\section*{
\bf{Supplementary material:\\
GKZ-system of the 2-loop self energy with 4 propagators}}

\section{The hypergeometric solutions of the integer lattice ${\bf B}_{_{1357}}$\label{app1a}}
\indent\indent

\begin{itemize}
\item The set of column indices
$I_{_1}=[2,4,6,8,\cdots,14]$, i.e. the implement $J_{_1}=[1,14]\setminus I_{_1}=[1,3,5,7]$.
The choice on the set of indices implies the exponent numbers
$\alpha_{_1}=\alpha_{_3}=\alpha_{_5}=\alpha_{_7}=0$, and
\begin{eqnarray}
&&\alpha_{_2}=a_{_1}-a_{_2},\;\alpha_{_4}=a_{_3}-a_{_4},\;\alpha_{_6}=a_{_3}+a_{_5}+b_{_1}-a_{_1}-1,\;
\nonumber\\
&&\alpha_{_8}=b_{_2}+b_{_3}-a_{_3}-2,\;\alpha_{_9}=b_{_4}-a_{_3}-a_{_5}-1,\;
\nonumber\\
&&\alpha_{_{10}}=b_{_5}-a_{_3}-a_{_5}-1,\;\alpha_{_{11}}=a_{_3}+a_{_5}-a_{_1},\;\alpha_{_{12}}=1-b_{_2},\;
\nonumber\\
&&\alpha_{_{13}}=b_{_2}-a_{_3}-1,\;\alpha_{_{14}}=-a_{_5}\;.
\label{GKZ21-1-1}
\end{eqnarray}
The corresponding hypergeometric series solutions with quadruple independent variables are written as
\begin{eqnarray}
&&\Phi_{_{[1357]}}^{(1),a}(\alpha,z)=
{y_{_1}^{{D\over2}-1}y_{_2}^{{D\over2}-1}y_{_3}^{-1}y_{_4}^{-1}}\sum\limits_{n_{_1}=0}^\infty
\sum\limits_{n_{_2}=0}^\infty\sum\limits_{n_{_3}=0}^\infty\sum\limits_{n_{_4}=0}^\infty
c_{_{[1357]}}^{(1),a}(\alpha,{\bf n})
\nonumber\\
&&\hspace{2.5cm}\times
\Big({1\over y_{_4}}\Big)^{n_{_1}}\Big({y_{_4}\over y_{_3}}\Big)^{n_{_2}}
\Big({y_{_1}\over y_{_4}}\Big)^{n_{_3}}\Big({y_{_2}\over y_{_3}}\Big)^{n_{_4}}
\;,\nonumber\\
&&\Phi_{_{[1357]}}^{(1),b}(\alpha,z)=
{y_{_1}^{{D\over2}-2}y_{_2}^{{D\over2}-1}y_{_3}^{-1}y_{_4}^{-1}}
\sum\limits_{n_{_1}=0}^\infty
\sum\limits_{n_{_2}=0}^\infty\sum\limits_{n_{_3}=0}^\infty\sum\limits_{n_{_4}=0}^\infty
c_{_{[1357]}}^{(1),b}(\alpha,{\bf n})
\nonumber\\
&&\hspace{2.5cm}\times
\Big({1\over y_{_1}}\Big)^{n_{_1}}\Big({1\over y_{_3}}\Big)^{n_{_2}}
\Big({1\over y_{_4}}\Big)^{n_{_3}}\Big({y_{_2}\over y_{_3}}\Big)^{n_{_4}}\;.
\label{GKZ21-1-2a}
\end{eqnarray}
Where the coefficients are
\begin{eqnarray}
&&c_{_{[1357]}}^{(1),a}(\alpha,{\bf n})=
(-)^{n_{_4}}\Gamma(1+n_{_1}+n_{_3})\Gamma(1+n_{_2}+n_{_4})
\Big\{n_{_1}!n_{_2}!n_{_3}!n_{_4}!
\nonumber\\
&&\hspace{2.5cm}\times
\Gamma({D\over2}+n_{_1})\Gamma({D\over2}+n_{_2})\Gamma(1-{D\over2}-n_{_2}-n_{_4})
\nonumber\\
&&\hspace{2.5cm}\times
\Gamma(1-{D\over2}-n_{_1}-n_{_3})\Gamma({D\over2}+n_{_3})\Gamma({D\over2}+n_{_4})\Big\}^{-1}
\;,\nonumber\\
&&c_{_{[1357]}}^{(1),b}(\alpha,{\bf n})=
(-)^{n_{_1}+n_{_4}}\Gamma(1+n_{_1})\Gamma(1+n_{_2}+n_{_3})\Gamma(1+n_{_2}+n_{_4})
\Big\{n_{_2}!n_{_4}!
\nonumber\\
&&\hspace{2.5cm}\times
\Gamma(2+n_{_1}+n_{_2}+n_{_3})\Gamma({D\over2}+1+n_{_1}+n_{_2}+n_{_3})
\nonumber\\
&&\hspace{2.5cm}\times
\Gamma({D\over2}+n_{_2})\Gamma(1-{D\over2}-n_{_2}-n_{_3})
\Gamma(1-{D\over2}-n_{_2}-n_{_4})
\nonumber\\
&&\hspace{2.5cm}\times
\Gamma({D\over2}-1-n_{_1})\Gamma({D\over2}+n_{_4})\Big\}^{-1}\;.
\label{GKZ21-1-3}
\end{eqnarray}

\item The set of column indices $I_{_2}=[2,4,6,\cdots,11,13,14]$, i.e.
the implement $J_{_2}=[1,14]\setminus I_{_2}=[1,3,5,12]$.
The choice implies the exponent numbers $\alpha_{_1}=\alpha_{_3}=\alpha_{_5}=\alpha_{_{12}}=0$, and
\begin{eqnarray}
&&\alpha_{_2}=a_{_1}-a_{_2},\;\alpha_{_4}=a_{_3}-a_{_4},\;\alpha_{_6}=a_{_3}+a_{_5}+b_{_1}-a_{_1}-1,\;
\nonumber\\
&&\alpha_{_7}=b_{_2}-1,\;\alpha_{_8}=b_{_3}-a_{_3}-1,\;\alpha_{_9}=b_{_4}-a_{_3}-a_{_5}-1,\;
\nonumber\\
&&\alpha_{_{10}}=b_{_5}-a_{_3}-a_{_5}-1,\;\alpha_{_{11}}=a_{_3}+a_{_5}-a_{_1},\;\alpha_{_{13}}=-a_{_3},\;
\nonumber\\
&&\alpha_{_{14}}=-a_{_5}\;.
\label{GKZ21-2-1}
\end{eqnarray}
The corresponding hypergeometric solutions are written as
\begin{eqnarray}
&&\Phi_{_{[1357]}}^{(2),a}(\alpha,z)=
{y_{_1}^{{D\over2}-1}y_{_3}^{{D\over2}-2}y_{_4}^{-1}}\sum\limits_{n_{_1}=0}^\infty
\sum\limits_{n_{_2}=0}^\infty\sum\limits_{n_{_3}=0}^\infty\sum\limits_{n_{_4}=0}^\infty
c_{_{[1357]}}^{(2),a}(\alpha,{\bf n})
\nonumber\\
&&\hspace{2.5cm}\times
\Big({1\over y_{_4}}\Big)^{n_{_1}}\Big({y_{_4}\over y_{_3}}\Big)^{n_{_2}}
\Big({y_{_1}\over y_{_4}}\Big)^{n_{_3}}\Big({y_{_2}\over y_{_3}}\Big)^{n_{_4}}
\;,\nonumber\\
&&\Phi_{_{[1357]}}^{(2),b}(\alpha,z)=
{y_{_1}^{{D\over2}-2}y_{_3}^{{D\over2}-2}y_{_4}^{-1}}\sum\limits_{n_{_1}=0}^\infty
\sum\limits_{n_{_2}=0}^\infty\sum\limits_{n_{_3}=0}^\infty\sum\limits_{n_{_4}=0}^\infty
c_{_{[1357]}}^{(2),b}(\alpha,{\bf n})
\nonumber\\
&&\hspace{2.5cm}\times
\Big({1\over y_{_1}}\Big)^{n_{_1}}\Big({1\over y_{_3}}\Big)^{n_{_2}}
\Big({1\over y_{_4}}\Big)^{n_{_3}}\Big({y_{_2}\over y_{_3}}\Big)^{n_{_4}}\;.
\label{GKZ21-2-2a}
\end{eqnarray}
Where the coefficients are
\begin{eqnarray}
&&c_{_{[1357]}}^{(2),a}(\alpha,{\bf n})=
(-)^{n_{_4}}\Gamma(1+n_{_1}+n_{_3})\Gamma(1+n_{_2}+n_{_4})
\Big\{n_{_1}!n_{_2}!n_{_3}!n_{_4}!
\nonumber\\
&&\hspace{2.5cm}\times
\Gamma({D\over2}+n_{_1})\Gamma({D\over2}+n_{_2})\Gamma(2-{D\over2}+n_{_4})
\Gamma({D\over2}+n_{_3})
\nonumber\\
&&\hspace{2.5cm}\times
\Gamma({D\over2}-1-n_{_2}-n_{_4})
\Gamma(1-{D\over2}-n_{_1}-n_{_3})\Big\}^{-1}
\;,\nonumber\\
&&c_{_{[1357]}}^{(2),b}(\alpha,{\bf n})=
(-)^{n_{_1}+n_{_4}}\Gamma(1+n_{_1})\Gamma(1+n_{_2}+n_{_3})\Gamma(1+n_{_2}+n_{_4})
\Big\{n_{_2}!n_{_4}!
\nonumber\\
&&\hspace{2.5cm}\times
\Gamma(2+n_{_1}+n_{_2}+n_{_3})\Gamma({D\over2}+1+n_{_1}+n_{_2}+n_{_3})
\nonumber\\
&&\hspace{2.5cm}\times
\Gamma({D\over2}+n_{_2})\Gamma({D\over2}-1-n_{_1})
\Gamma(2-{D\over2}+n_{_4})
\nonumber\\
&&\hspace{2.5cm}\times
\Gamma(1-{D\over2}-n_{_2}-n_{_3})\Gamma({D\over2}-1-n_{_2}-n_{_4})\Big\}^{-1}\;.
\label{GKZ21-2-3}
\end{eqnarray}
Certainly the intersection of the convergent regions of two hypergeometric series
is nonempty. Additionally $\Phi_{_{[1357]}}^{(2),b}$ is equal to
$\Phi_{_{[1357]}}^{(2),a}$ up to a constant scalar multiple in the nonempty intersection
because they originate from the same exponent vector presented in Eq.(\ref{GKZ21-2-1}).

\item The set of column indices $I_{_3}=[2,3,6,8,\cdots,14]$, i.e.
the implement $J_{_3}=[1,14]\setminus I_{_3}=[1,4,5,7]$.
The set of column indices gives the exponent numbers $\alpha_{_1}=\alpha_{_4}=\alpha_{_5}=\alpha_{_{7}}=0$, and
\begin{eqnarray}
&&\alpha_{_2}=a_{_1}-a_{_2},\;\alpha_{_3}=a_{_4}-a_{_3},\;\alpha_{_6}=a_{_4}+a_{_5}+b_{_1}-a_{_1}-1,\;
\nonumber\\
&&\alpha_{_8}=b_{_2}+b_{_3}-a_{_4}-2,\;\alpha_{_9}=b_{_4}-a_{_4}-a_{_5}-1,
\nonumber\\
&&\alpha_{_{10}}=b_{_5}-a_{_4}-a_{_5}-1,\;\alpha_{_{11}}=a_{_4}+a_{_5}-a_{_1},\;\alpha_{_{12}}=1-b_{_2},
\nonumber\\
&&\alpha_{_{13}}=b_{_2}-a_{_4}-1,\;\alpha_{_{14}}=-a_{_5}\;.
\label{GKZ21-3-1}
\end{eqnarray}
The constructed hypergeometric functions are
\begin{eqnarray}
&&\Phi_{_{[1357]}}^{(3),a}(\alpha,z)=
{y_{_2}^{{D\over2}-1}y_{_3}^{{D\over2}-2}y_{_4}^{-1}}\sum\limits_{n_{_1}=0}^\infty
\sum\limits_{n_{_2}=0}^\infty\sum\limits_{n_{_3}=0}^\infty\sum\limits_{n_{_4}=0}^\infty
c_{_{[1357]}}^{(3),a}(\alpha,{\bf n})
\nonumber\\
&&\hspace{2.5cm}\times
\Big({1\over y_{_4}}\Big)^{n_{_1}}\Big({y_{_4}\over y_{_3}}\Big)^{n_{_2}}
\Big({y_{_1}\over y_{_4}}\Big)^{n_{_3}}\Big({y_{_2}\over y_{_3}}\Big)^{n_{_4}}
\;,\nonumber\\
&&\Phi_{_{[1357]}}^{(3),b}(\alpha,z)=
{y_{_1}^{-1}y_{_2}^{{D\over2}-1}y_{_3}^{{D\over2}-2}y_{_4}^{-1}}\sum\limits_{n_{_1}=0}^\infty
\sum\limits_{n_{_2}=0}^\infty\sum\limits_{n_{_3}=0}^\infty\sum\limits_{n_{_4}=0}^\infty
c_{_{[1357]}}^{(3),b}(\alpha,{\bf n})
\nonumber\\
&&\hspace{2.5cm}\times
\Big({1\over y_{_1}}\Big)^{n_{_1}}\Big({1\over y_{_3}}\Big)^{n_{_2}}
\Big({1\over y_{_4}}\Big)^{n_{_3}}\Big({y_{_2}\over y_{_3}}\Big)^{n_{_4}}\;.
\label{GKZ21-3-2a}
\end{eqnarray}
Where the coefficients are
\begin{eqnarray}
&&c_{_{[1357]}}^{(3),a}(\alpha,{\bf n})=
(-)^{n_{_4}}\Gamma(1+n_{_1}+n_{_3})\Gamma(1+n_{_2}+n_{_4})
\Big\{n_{_1}!n_{_2}!n_{_3}!n_{_4}!\Gamma({D\over2}+n_{_1})
\nonumber\\
&&\hspace{2.5cm}\times
\Gamma(2-{D\over2}+n_{_2})
\Gamma(2-{D\over2}+n_{_3})\Gamma({D\over2}-1-n_{_1}-n_{_3})\Gamma({D\over2}+n_{_4})
\nonumber\\
&&\hspace{2.5cm}\times
\Gamma({D\over2}-1-n_{_2}-n_{_4})\Big\}^{-1}
\;,\nonumber\\
&&c_{_{[1357]}}^{(3),b}(\alpha,{\bf n})=
(-)^{n_{_1}+n_{_4}}\Gamma(1+n_{_1})\Gamma(1+n_{_2}+n_{_3})\Gamma(1+n_{_2}+n_{_4})
\Big\{n_{_2}!n_{_4}!
\nonumber\\
&&\hspace{2.5cm}\times
\Gamma(2+n_{_1}+n_{_2}+n_{_3})\Gamma({D\over2}+1+n_{_1}+n_{_2}+n_{_3})
\Gamma(2-{D\over2}+n_{_2})
\nonumber\\
&&\hspace{2.5cm}\times
\Gamma(1-{D\over2}-n_{_1})\Gamma({D\over2}-1-n_{_2}-n_{_3})\Gamma({D\over2}+n_{_4})
\nonumber\\
&&\hspace{2.5cm}\times
\Gamma({D\over2}-1-n_{_2}-n_{_4})\Big\}^{-1}\;.
\label{GKZ21-3-3}
\end{eqnarray}
The intersection of the convergent regions of two hypergeometric series
is nonempty. Obviously $\Phi_{_{[1357]}}^{(3),b}$ is equal to
$\Phi_{_{[1357]}}^{(3),a}$ up to a constant scalar multiple in the nonempty intersection
because they originate from the same exponent vector presented in Eq.(\ref{GKZ21-3-1}).

\item The set of column indices $I_{_4}=[2,3,6,\cdots,11,13,14]$, i.e.
the implement $J_{_4}=[1,14]\setminus I_{_4}=[1,4,5,12]$.
The choice on the set of column indices implies
$\alpha_{_1}=\alpha_{_4}=\alpha_{_5}=\alpha_{_{12}}=0$, and
\begin{eqnarray}
&&\alpha_{_2}=a_{_1}-a_{_2},\;\alpha_{_3}=a_{_4}-a_{_3},\;\alpha_{_6}=a_{_4}+a_{_5}+b_{_1}-a_{_1}-1,\;
\nonumber\\
&&\alpha_{_7}=b_{_2}-1,\;\alpha_{_8}=b_{_3}-a_{_4}-1,\;\alpha_{_9}=b_{_4}-a_{_4}-a_{_5}-1,
\nonumber\\
&&\alpha_{_{10}}=b_{_5}-a_{_4}-a_{_5}-1,\;\alpha_{_{11}}=a_{_4}+a_{_5}-a_{_1},
\nonumber\\
&&\alpha_{_{13}}=-a_{_4},\;\alpha_{_{14}}=-a_{_5}\;.
\label{GKZ21-4-1}
\end{eqnarray}
Using the integer lattice $\mathbf{B}_{_{1357}}$ and the corresponding power numbers above,
we formulate the hypergeometric series solutions as
\begin{eqnarray}
&&\Phi_{_{[1357]}}^{(4),a}(\alpha,z)=
{y_{_3}^{D-3}y_{_4}^{-1}}\sum\limits_{n_{_1}=0}^\infty
\sum\limits_{n_{_2}=0}^\infty\sum\limits_{n_{_3}=0}^\infty\sum\limits_{n_{_4}=0}^\infty
c_{_{[1357]}}^{(4),a}(\alpha,{\bf n})
\nonumber\\
&&\hspace{2.5cm}\times
\Big({1\over y_{_4}}\Big)^{n_{_1}}\Big({y_{_4}\over y_{_3}}\Big)^{n_{_2}}
\Big({y_{_1}\over y_{_4}}\Big)^{n_{_3}}\Big({y_{_2}\over y_{_3}}\Big)^{n_{_4}}
\;,\nonumber\\
&&\Phi_{_{[1357]}}^{(4),b}(\alpha,z)=
{y_{_1}^{-1}y_{_3}^{D-3}y_{_4}^{-1}}\sum\limits_{n_{_1}=0}^\infty
\sum\limits_{n_{_2}=0}^\infty\sum\limits_{n_{_3}=0}^\infty\sum\limits_{n_{_4}=0}^\infty
c_{_{[1357]}}^{(4),b}(\alpha,{\bf n})
\nonumber\\
&&\hspace{2.5cm}\times
\Big({1\over y_{_1}}\Big)^{n_{_1}}\Big({1\over y_{_3}}\Big)^{n_{_2}}
\Big({1\over y_{_4}}\Big)^{n_{_3}}\Big({y_{_2}\over y_{_3}}\Big)^{n_{_4}}\;.
\label{GKZ21-4-2a}
\end{eqnarray}
Where the coefficients are
\begin{eqnarray}
&&c_{_{[1357]}}^{(4),a}(\alpha,{\bf n})=
(-)^{n_{_2}}\Gamma(1+n_{_1}+n_{_3})\Big\{n_{_1}!n_{_2}!n_{_3}!n_{_4}!
\Gamma({D\over2}+n_{_1})\Gamma(2-{D\over2}+n_{_2})
\nonumber\\
&&\hspace{2.5cm}\times
\Gamma(2-{D\over2}+n_{_3})\Gamma(2-{D\over2}+n_{_4})\Gamma({D\over2}-1-n_{_2}-n_{_4})
\nonumber\\
&&\hspace{2.5cm}\times
\Gamma({D\over2}-1-n_{_1}-n_{_3})\Gamma(D-2-n_{_2}-n_{_4})\Big\}^{-1}
\;,\nonumber\\
&&c_{_{[1357]}}^{(4),b}(\alpha,{\bf n})=
(-)^{n_{_1}+n_{_2}}\Gamma(1+n_{_1})\Gamma(1+n_{_2}+n_{_3})\Big\{n_{_2}!n_{_4}!
\Gamma(2+n_{_1}+n_{_2}+n_{_3})
\nonumber\\
&&\hspace{2.5cm}\times
\Gamma({D\over2}+1+n_{_1}+n_{_2}+n_{_3})\Gamma(2-{D\over2}+n_{_2})
\Gamma(1-{D\over2}-n_{_1})
\nonumber\\
&&\hspace{2.5cm}\times
\Gamma(2-{D\over2}+n_{_4})
\Gamma({D\over2}-1-n_{_2}-n_{_4})\Gamma({D\over2}-1-n_{_2}-n_{_3})
\nonumber\\
&&\hspace{2.5cm}\times
\Gamma(D-2-n_{_2}-n_{_4})\Big\}^{-1}\;.
\label{GKZ21-4-3}
\end{eqnarray}
Because two solutions originate from the same exponent vector presented
in Eq.(\ref{GKZ21-4-1}), $\Phi_{_{[1357]}}^{(4),b}$ is equal to
$\Phi_{_{[1357]}}^{(4),a}$ up to a constant scalar multiple in the nonempty intersection
of their convergent regions. In other words, two solutions are analytic continuation
of each other on the union of their convergent regions.

\item The set of column indices $I_{_5}=[1,4,6,8,\cdots,14]$, i.e.
the implement $J_{_5}=[1,14]\setminus I_{_5}=[2,3,5,7]$.
The set of column indices implies the exponent numbers
$\alpha_{_2}=\alpha_{_3}=\alpha_{_5}=\alpha_{_7}=0$, and
\begin{eqnarray}
&&\alpha_{_1}=a_{_2}-a_{_1},\;\alpha_{_4}=a_{_3}-a_{_4},\;\alpha_{_6}=a_{_3}+a_{_5}+b_{_1}-a_{_2}-1,\;
\nonumber\\
&&\alpha_{_8}=b_{_2}+b_{_3}-a_{_3}-2,\;\alpha_{_9}=b_{_4}-a_{_3}-a_{_5}-1,\;
\nonumber\\
&&\alpha_{_{10}}=b_{_5}-a_{_3}-a_{_5}-1,\;\alpha_{_{11}}=a_{_3}+a_{_5}-a_{_2},\;\alpha_{_{12}}=1-b_{_2},\;
\nonumber\\
&&\alpha_{_{13}}=b_{_2}-a_{_3}-1,\;\alpha_{_{14}}=-a_{_5}\;.
\label{GKZ21-5-1}
\end{eqnarray}
The exponent numbers above and the integer lattice $\mathbf{B}_{_{1357}}$
give the hypergeometric solution as
\begin{eqnarray}
&&\Phi_{_{[1357]}}^{(5)}(\alpha,z)=
{y_{_1}^{{D}-2}y_{_2}^{{D\over2}-1}y_{_3}^{-1}y_{_4}^{-1}}\sum\limits_{n_{_1}=0}^\infty
\sum\limits_{n_{_2}=0}^\infty\sum\limits_{n_{_3}=0}^\infty\sum\limits_{n_{_4}=0}^\infty
c_{_{[1357]}}^{(5)}(\alpha,{\bf n})
\nonumber\\
&&\hspace{2.5cm}\times
\Big({1\over y_{_1}}\Big)^{n_{_1}}\Big({y_{_1}\over y_{_3}}\Big)^{n_{_2}}
\Big({y_{_1}\over y_{_4}}\Big)^{n_{_3}}\Big({y_{_2}\over y_{_3}}\Big)^{n_{_4}}\;,
\label{GKZ21-5-2}
\end{eqnarray}
with
\begin{eqnarray}
&&c_{_{[1357]}}^{(5)}(\alpha,{\bf n})=
(-)^{n_{_4}}\Gamma(1+n_{_2}+n_{_3})\Gamma(1+n_{_2}+n_{_4})
\nonumber\\
&&\hspace{2.5cm}\times
\Big\{n_{_1}!n_{_2}!n_{_4}!\Gamma(2-{D\over2}+n_{_1})\Gamma({D\over2}+n_{_2})
\nonumber\\
&&\hspace{2.5cm}\times
\Gamma({D\over2}-n_{_1}+n_{_2}+n_{_3})\Gamma(1-{D\over2}-n_{_2}-n_{_4})
\nonumber\\
&&\hspace{2.5cm}\times
\Gamma(1-{D\over2}-n_{_2}-n_{_3})\Gamma(D-1-n_{_1}+n_{_2}+n_{_3})
\Gamma({D\over2}+n_{_4})\Big\}^{-1}\;.
\label{GKZ21-5-3}
\end{eqnarray}

\item The set of column indices $I_{_6}=[1,4,6,\cdots,11,13,14]$, i.e.
the implement $J_{_6}=[1,14]\setminus I_{_6}=[2,3,5,12]$.
The choice gives the exponent numbers $\alpha_{_2}=\alpha_{_3}=\alpha_{_5}=\alpha_{_{12}}=0$, and
\begin{eqnarray}
&&\alpha_{_1}=a_{_2}-a_{_1},\;\alpha_{_4}=a_{_3}-a_{_4},\;\alpha_{_6}=a_{_3}+a_{_5}+b_{_1}-a_{_2}-1,\;
\nonumber\\
&&\alpha_{_7}=b_{_2}-1,\;\alpha_{_8}=b_{_3}-a_{_3}-1,\;\alpha_{_9}=b_{_4}-a_{_3}-a_{_5}-1,\;
\nonumber\\
&&\alpha_{_{10}}=b_{_5}-a_{_3}-a_{_5}-1,\;\alpha_{_{11}}=a_{_3}+a_{_5}-a_{_2},\;\alpha_{_{13}}=-a_{_3},\;
\nonumber\\
&&\alpha_{_{14}}=-a_{_5}\;.
\label{GKZ21-6-1}
\end{eqnarray}
Using the integer lattice and those power numbers above, we write the
corresponding hypergeometric series solution as
\begin{eqnarray}
&&\Phi_{_{[1357]}}^{(6)}(\alpha,z)=
{y_{_1}^{{D}-2}y_{_3}^{{D\over2}-2}y_{_4}^{-1}}\sum\limits_{n_{_1}=0}^\infty
\sum\limits_{n_{_2}=0}^\infty\sum\limits_{n_{_3}=0}^\infty\sum\limits_{n_{_4}=0}^\infty
c_{_{[1357]}}^{(6)}(\alpha,{\bf n})
\nonumber\\
&&\hspace{2.5cm}\times
\Big({1\over y_{_1}}\Big)^{n_{_1}}\Big({y_{_1}\over y_{_3}}\Big)^{n_{_2}}
\Big({y_{_1}\over y_{_4}}\Big)^{n_{_3}}\Big({y_{_2}\over y_{_3}}\Big)^{n_{_4}}\;,
\label{GKZ21-6-2}
\end{eqnarray}
with
\begin{eqnarray}
&&c_{_{[1357]}}^{(6)}(\alpha,{\bf n})=
(-)^{n_{_4}}\Gamma(1+n_{_2}+n_{_4})\Gamma(1+n_{_2}+n_{_3})
\nonumber\\
&&\hspace{2.5cm}\times
\Big\{n_{_1}!n_{_2}!n_{_4}!
\Gamma(2-{D\over2}+n_{_1})\Gamma({D\over2}+n_{_2})
\nonumber\\
&&\hspace{2.5cm}\times
\Gamma({D\over2}-n_{_1}+n_{_2}+n_{_3})\Gamma(2-{D\over2}+n_{_4})
\nonumber\\
&&\hspace{2.5cm}\times
\Gamma(1-{D\over2}-n_{_2}-n_{_3})\Gamma(D-1-n_{_1}+n_{_2}+n_{_3})
\nonumber\\
&&\hspace{2.5cm}\times
\Gamma({D\over2}-1-n_{_2}-n_{_4})\Big\}^{-1}\;.
\label{GKZ21-6-3}
\end{eqnarray}

\item The set of column indices $I_{_7}=[1,3,6,8,\cdots,14]$, i.e.
the implement $J_{_7}=[1,14]\setminus I_{_7}=[2,4,5,7]$.
The choice implies the exponents $\alpha_{_2}=\alpha_{_4}=\alpha_{_5}=\alpha_{_{7}}=0$, and
\begin{eqnarray}
&&\alpha_{_1}=a_{_2}-a_{_1},\;\alpha_{_3}=a_{_4}-a_{_3},\;\alpha_{_6}=a_{_4}+a_{_5}+b_{_1}-a_{_2}-1,\;
\nonumber\\
&&\alpha_{_8}=b_{_2}+b_{_3}-a_{_4}-2,\;\alpha_{_9}=b_{_4}-a_{_4}-a_{_5}-1,
\nonumber\\
&&\alpha_{_{10}}=b_{_5}-a_{_4}-a_{_5}-1,\;\alpha_{_{11}}=a_{_4}+a_{_5}-a_{_2},\;\alpha_{_{12}}=1-b_{_2},
\nonumber\\
&&\alpha_{_{13}}=b_{_2}-a_{_4}-1,\;\alpha_{_{14}}=-a_{_5}\;.
\label{GKZ21-7-1}
\end{eqnarray}
Similarly the corresponding hypergeometric series solutions are written as
\begin{eqnarray}
&&\Phi_{_{[1357]}}^{(7),a}(\alpha,z)=
{y_{_1}^{{D\over2}-1}y_{_2}^{{D\over2}-1}y_{_3}^{{D\over2}-2}y_{_4}^{-1}}\sum\limits_{n_{_1}=0}^\infty
\sum\limits_{n_{_2}=0}^\infty\sum\limits_{n_{_3}=0}^\infty\sum\limits_{n_{_4}=0}^\infty
c_{_{[1357]}}^{(7),a}(\alpha,{\bf n})
\nonumber\\
&&\hspace{2.5cm}\times
\Big({1\over y_{_4}}\Big)^{n_{_1}}\Big({y_{_4}\over y_{_3}}\Big)^{n_{_2}}
\Big({y_{_1}\over y_{_4}}\Big)^{n_{_3}}\Big({y_{_2}\over y_{_3}}\Big)^{n_{_4}}
\;,\nonumber\\
&&\Phi_{_{[1357]}}^{(7),b}(\alpha,z)=
{y_{_1}^{{D\over2}-2}y_{_2}^{{D\over2}-1}y_{_3}^{{D\over2}-2}y_{_4}^{-1}}\sum\limits_{n_{_1}=0}^\infty
\sum\limits_{n_{_2}=0}^\infty\sum\limits_{n_{_3}=0}^\infty\sum\limits_{n_{_4}=0}^\infty
c_{_{[1357]}}^{(7),b}(\alpha,{\bf n})
\nonumber\\
&&\hspace{2.5cm}\times
\Big({1\over y_{_1}}\Big)^{n_{_1}}\Big({1\over y_{_3}}\Big)^{n_{_2}}
\Big({1\over y_{_4}}\Big)^{n_{_3}}\Big({y_{_2}\over y_{_3}}\Big)^{n_{_4}}\;.
\label{GKZ21-7-2a}
\end{eqnarray}
Where the coefficients are
\begin{eqnarray}
&&c_{_{[1357]}}^{(7),(a)}(\alpha,{\bf n})=
(-)^{n_{_4}}\Gamma(1+n_{_1}+n_{_3})\Gamma(1+n_{_2}+n_{_4})
\Big\{n_{_1}!n_{_2}!n_{_3}!n_{_4}!
\nonumber\\
&&\hspace{2.5cm}\times
\Gamma(2-{D\over2}+n_{_1})\Gamma(2-{D\over2}+n_{_2})
\Gamma({D\over2}+n_{_4})
\nonumber\\
&&\hspace{2.5cm}\times
\Gamma({D\over2}-1-n_{_1}-n_{_3})\Gamma({D\over2}+n_{_3})
\Gamma({D\over2}-1-n_{_2}-n_{_4})\Big\}^{-1}
\;,\nonumber\\
&&c_{_{[1357]}}^{(7),(b)}(\alpha,{\bf n})=
(-)^{n_{_1}+n_{_4}}\Gamma(1+n_{_1})\Gamma(1+n_{_2}+n_{_3})\Gamma(1+n_{_2}+n_{_4})
\nonumber\\
&&\hspace{2.5cm}\times
\Big\{n_{_2}!n_{_4}!\Gamma(2+n_{_1}+n_{_2}+n_{_3})
\Gamma(3-{D\over2}+n_{_1}+n_{_2}+n_{_3})
\nonumber\\
&&\hspace{2.5cm}\times
\Gamma(2-{D\over2}+n_{_2})\Gamma({D\over2}+n_{_4})
\Gamma({D\over2}-1-n_{_2}-n_{_3})
\nonumber\\
&&\hspace{2.5cm}\times
\Gamma({D\over2}-1-n_{_1})
\Gamma({D\over2}-1-n_{_2}-n_{_4})\Big\}^{-1}\;.
\label{GKZ21-7-3}
\end{eqnarray}
Because two solutions originate from the same exponent vector presented
in Eq.(\ref{GKZ21-7-1}), $\Phi_{_{[1357]}}^{(7),b}$ is equal to
$\Phi_{_{[1357]}}^{(7),a}$ up to a constant scalar multiple in the nonempty intersection
of their convergent regions. In other words, two solutions are analytic continuation
of each other on the union of their convergent regions.

\item The set of column indices $I_{_8}=[1,3,6,\cdots,11,13,14]$, i.e.
the implement $J_{_8}=[1,14]\setminus I_{_8}=[2,4,5,12]$.
The choice implies the powers $\alpha_{_2}=\alpha_{_4}=\alpha_{_5}=\alpha_{_{12}}=0$, and
\begin{eqnarray}
&&\alpha_{_1}=a_{_2}-a_{_1},\;\alpha_{_3}=a_{_4}-a_{_3},\;\alpha_{_6}=a_{_4}+a_{_5}+b_{_1}-a_{_2}-1,\;
\nonumber\\
&&\alpha_{_7}=b_{_2}-1,\;\alpha_{_8}=b_{_3}-a_{_4}-1,\;\alpha_{_9}=b_{_4}-a_{_4}-a_{_5}-1,
\nonumber\\
&&\alpha_{_{10}}=b_{_5}-a_{_4}-a_{_5}-1,\;\alpha_{_{11}}=a_{_4}+a_{_5}-a_{_2},
\nonumber\\
&&\alpha_{_{13}}=-a_{_4},\;\alpha_{_{14}}=-a_{_5}\;.
\label{GKZ21-8-1}
\end{eqnarray}
Then the corresponding hypergeometric solutions are
\begin{eqnarray}
&&\Phi_{_{[1357]}}^{(8),a}(\alpha,z)=
{y_{_1}^{{D\over2}-1}y_{_3}^{{D}-3}y_{_4}^{-1}}\sum\limits_{n_{_1}=0}^\infty
\sum\limits_{n_{_2}=0}^\infty\sum\limits_{n_{_3}=0}^\infty\sum\limits_{n_{_4}=0}^\infty
c_{_{[1357]}}^{(8),a}(\alpha,{\bf n})
\nonumber\\
&&\hspace{2.5cm}\times
\Big({1\over y_{_4}}\Big)^{n_{_1}}\Big({y_{_4}\over y_{_3}}\Big)^{n_{_2}}
\Big({y_{_1}\over y_{_4}}\Big)^{n_{_3}}\Big({y_{_2}\over y_{_3}}\Big)^{n_{_4}}
\;,\nonumber\\
&&\Phi_{_{[1357]}}^{(8),b}(\alpha,z)=
{y_{_1}^{{D\over2}-2}y_{_3}^{{D}-3}y_{_4}^{-1}}\sum\limits_{n_{_1}=0}^\infty
\sum\limits_{n_{_2}=0}^\infty\sum\limits_{n_{_3}=0}^\infty\sum\limits_{n_{_4}=0}^\infty
c_{_{[1357]}}^{(8),b}(\alpha,{\bf n})
\nonumber\\
&&\hspace{2.5cm}\times
\Big({1\over y_{_1}}\Big)^{n_{_1}}\Big({1\over y_{_3}}\Big)^{n_{_2}}
\Big({1\over y_{_4}}\Big)^{n_{_3}}\Big({y_{_2}\over y_{_3}}\Big)^{n_{_4}}\;.
\label{GKZ21-8-2a}
\end{eqnarray}
Where the coefficients are
\begin{eqnarray}
&&c_{_{[1357]}}^{(8),a}(\alpha,{\bf n})=(-)^{n_{_2}}\Gamma(1+n_{_1}+n_{_3})
\Big\{n_{_1}!n_{_2}!n_{_3}!n_{_4}!
\Gamma(2-{D\over2}+n_{_1})
\nonumber\\
&&\hspace{2.5cm}\times
\Gamma(2-{D\over2}+n_{_2})\Gamma(2-{D\over2}+n_{_4})
\nonumber\\
&&\hspace{2.5cm}\times
\Gamma({D\over2}-1-n_{_2}-n_{_4})\Gamma({D\over2}-1-n_{_1}-n_{_3})
\nonumber\\
&&\hspace{2.5cm}\times
\Gamma({D\over2}+n_{_3})\Gamma(D-2-n_{_2}-n_{_4})\Big\}^{-1}
\;,\nonumber\\
&&c_{_{[1357]}}^{(8),b}(\alpha,{\bf n})=(-)^{n_{_1}+n_{_2}}\Gamma(1+n_{_1})
\Gamma(1+n_{_2}+n_{_3})\Big\{n_{_2}!n_{_4}!
\nonumber\\
&&\hspace{2.5cm}\times
\Gamma(2+n_{_1}+n_{_2}+n_{_3})\Gamma(3-{D\over2}+n_{_1}+n_{_2}+n_{_3})
\nonumber\\
&&\hspace{2.5cm}\times
\Gamma(2-{D\over2}+n_{_2})\Gamma(2-{D\over2}+n_{_4})
\nonumber\\
&&\hspace{2.5cm}\times
\Gamma({D\over2}-1-n_{_2}-n_{_4})\Gamma({D\over2}-1-n_{_2}-n_{_3})
\nonumber\\
&&\hspace{2.5cm}\times
\Gamma({D\over2}-n_{_1})\Gamma(D-2-n_{_2}-n_{_4})\Big\}^{-1}\;.
\label{GKZ21-8-3}
\end{eqnarray}
The intersection of the convergent regions of two hypergeometric series
is a proper subset of the whole parameter space. Obviously $\Phi_{_{[1357]}}^{(8),b}$ is equal to
$\Phi_{_{[1357]}}^{(8),a}$ up to a constant scalar multiple in the nonempty intersection
because they originate from the same exponent vector presented in Eq.(\ref{GKZ21-8-1}).

\item The set of column indices $I_{_9}=[2,3,4,8,\cdots,14]$, i.e.
the implement $J_{_9}=[1,14]\setminus I_{_9}=[1,5,6,7]$.
The choice implies the exponents $\alpha_{_1}=\alpha_{_5}=\alpha_{_6}=\alpha_{_{7}}=0$, and
\begin{eqnarray}
&&\alpha_{_2}=a_{_1}-a_{_2},\;\alpha_{_3}=a_{_1}-a_{_3}-a_{_5}-b_{_1}+1,
\nonumber\\
&&\alpha_{_4}=a_{_1}-a_{_4}-a_{_5}-b_{_1}+1,
\nonumber\\
&&\alpha_{_8}=b_{_1}+b_{_2}+b_{_3}+a_{_5}-a_{_1}-3,\;\alpha_{_9}=b_{_1}+b_{_4}-a_{_1}-2,
\nonumber\\
&&\alpha_{_{10}}=b_{_1}+b_{_5}-a_{_1}-2,\;\alpha_{_{11}}=1-b_{_1},\;\alpha_{_{12}}=1-b_{_2},
\nonumber\\
&&\alpha_{_{13}}=b_{_1}+b_{_2}+a_{_5}-a_{_1}-2,\;\alpha_{_{14}}=-a_{_5}\;.
\label{GKZ21-9-1}
\end{eqnarray}
Thus the corresponding hypergeometric functions are similarly written as
\begin{eqnarray}
&&\Phi_{_{[1357]}}^{(9),a}(\alpha,z)=
{y_{_1}^{{D\over2}-1}y_{_2}^{{D\over2}-1}y_{_3}^{-1}y_{_4}^{-1}}\sum\limits_{n_{_1}=0}^\infty
\sum\limits_{n_{_2}=0}^\infty\sum\limits_{n_{_3}=0}^\infty\sum\limits_{n_{_4}=0}^\infty
c_{_{[1357]}}^{(9),a}(\alpha,{\bf n})
\nonumber\\
&&\hspace{2.5cm}\times
\Big({1\over y_{_4}}\Big)^{n_{_1}}\Big({y_{_1}\over y_{_4}}\Big)^{n_{_2}}
\Big({y_{_4}\over y_{_3}}\Big)^{n_{_3}}\Big({y_{_2}\over y_{_3}}\Big)^{n_{_4}}
\;,\nonumber\\
&&\Phi_{_{[1357]}}^{(9),b}(\alpha,z)=
{y_{_1}^{{D\over2}-1}y_{_2}^{{D\over2}-1}y_{_4}^{-2}}\sum\limits_{n_{_1}=0}^\infty
\sum\limits_{n_{_2}=0}^\infty\sum\limits_{n_{_3}=0}^\infty\sum\limits_{n_{_4}=0}^\infty
c_{_{[1357]}}^{(9),b}(\alpha,{\bf n})
\nonumber\\
&&\hspace{2.5cm}\times
\Big({1\over y_{_4}}\Big)^{n_{_1}}\Big({y_{_1}\over y_{_4}}\Big)^{n_{_2}}
\Big({y_{_3}\over y_{_4}}\Big)^{n_{_3}}\Big({y_{_2}\over y_{_4}}\Big)^{n_{_4}}
\;,\nonumber\\
&&\Phi_{_{[1357]}}^{(9),c}(\alpha,z)=
{y_{_1}^{{D\over2}-1}y_{_2}^{{D\over2}}y_{_3}^{-1}y_{_4}^{-2}}\sum\limits_{n_{_1}=0}^\infty
\sum\limits_{n_{_2}=0}^\infty\sum\limits_{n_{_3}=0}^\infty\sum\limits_{n_{_4}=0}^\infty
c_{_{[1357]}}^{(9),c}(\alpha,{\bf n})
\nonumber\\
&&\hspace{2.5cm}\times
\Big({1\over y_{_4}}\Big)^{n_{_1}}\Big({y_{_1}\over y_{_4}}\Big)^{n_{_2}}
\Big({y_{_2}\over y_{_4}}\Big)^{n_{_3}}\Big({y_{_2}\over y_{_3}}\Big)^{n_{_4}}\;.
\label{GKZ21-9-2a}
\end{eqnarray}
Where the coefficients are
\begin{eqnarray}
&&c_{_{[1357]}}^{(9),a}(\alpha,{\bf n})=
(-)^{n_{_4}}\Gamma(1+n_{_1}+n_{_2})\Gamma(1+n_{_3}+n_{_4})
\Big\{n_{_1}!n_{_2}!n_{_3}!n_{_4}!
\nonumber\\
&&\hspace{2.5cm}\times
\Gamma({D\over2}+n_{_1})\Gamma({D\over2}+n_{_3})\Gamma(1-{D\over2}-n_{_3}-n_{_4})
\nonumber\\
&&\hspace{2.5cm}\times
\Gamma(1-{D\over2}-n_{_1}-n_{_2})
\Gamma({D\over2}+n_{_2})\Gamma({D\over2}+n_{_4})\Big\}^{-1}
\;,\nonumber\\
&&c_{_{[1357]}}^{(9),b}(\alpha,{\bf n})=
-\Gamma(1+n_{_1}+n_{_2})\Gamma(1+n_{_3}+n_{_4})
\Big\{n_{_1}!n_{_2}!n_{_3}!n_{_4}!
\nonumber\\
&&\hspace{2.5cm}\times
\Gamma({D\over2}+n_{_1})\Gamma({D\over2}-1-n_{_3}-n_{_4})
\Gamma(2-{D\over2}+n_{_3})
\nonumber\\
&&\hspace{2.5cm}\times
\Gamma(1-{D\over2}-n_{_1}-n_{_2})
\Gamma({D\over2}+n_{_2})\Gamma({D\over2}+n_{_4})\Big\}^{-1}
\;,\nonumber\\
&&c_{_{[1357]}}^{(9),c}(\alpha,{\bf n})=
(-)^{1+n_{_4}}\Gamma(1+n_{_1}+n_{_2})\Gamma(1+n_{_3})\Gamma(1+n_{_4})
\Big\{n_{_1}!n_{_2}!
\nonumber\\
&&\hspace{2.5cm}\times
\Gamma(2+n_{_3}+n_{_4})\Gamma({D\over2}+n_{_1})\Gamma({D\over2}-1-n_{_3})\Gamma(1-{D\over2}-n_{_4})
\nonumber\\
&&\hspace{2.5cm}\times
\Gamma(1-{D\over2}-n_{_1}-n_{_2})
\Gamma({D\over2}+n_{_2})\Gamma({D\over2}+1+n_{_3}+n_{_4})\Big\}^{-1}\;.
\label{GKZ21-9-3}
\end{eqnarray}
Obviously the intersection of the convergent regions of $\Phi_{_{[1357]}}^{(9),a}$
and $\Phi_{_{[1357]}}^{(9),b}$ is empty. However, the intersection of the convergent
regions of $\Phi_{_{[1357]}}^{(9),a}$ and $\Phi_{_{[1357]}}^{(9),c}$ is nonempty.
In other words, $\Phi_{_{[1357]}}^{(9),a}$ is equal to $\Phi_{_{[1357]}}^{(9),c}$
up to a constant scalar factor in the nonempty intersection because they originate
from the exponent vector presented in Eq.(\ref{GKZ21-9-1}) simultaneously.
Similarly the intersection of the convergent regions of $\Phi_{_{[1357]}}^{(9),b}$
and $\Phi_{_{[1357]}}^{(9),c}$ is nonempty, and $\Phi_{_{[1357]}}^{(9),b}$ is equal
to $\Phi_{_{[1357]}}^{(9),c}$ up to a constant scalar factor in the nonempty intersection.
Because of the reason mentioned above, three solutions are analytic continuation
of each other on the union of their convergent regions.

\item The set of column indices $I_{_{10}}=[2,3,4,6,8,9,10,12,13,14]$, i.e.
the implement $J_{_{10}}=[1,14]\setminus I_{_{10}}=[1,5,7,11]$.
The choice of column indices implies the exponent numbers
$\alpha_{_1}=\alpha_{_5}=\alpha_{_7}=\alpha_{_{11}}=0$, and
\begin{eqnarray}
&&\alpha_{_2}=a_{_1}-a_{_2},\;\alpha_{_3}=a_{_1}-a_{_3}-a_{_5},\;\alpha_{_4}=a_{_1}-a_{_4}-a_{_5},
\nonumber\\
&&\alpha_{_6}=b_{_1}-1,\;
\alpha_{_8}=a_{_5}+b_{_2}+b_{_3}-a_{_1}-2,\;\alpha_{_9}=b_{_4}-a_{_1}-1,
\nonumber\\
&&\alpha_{_{10}}=b_{_5}-a_{_1}-1,\;\alpha_{_{12}}=1-b_{_2},
\nonumber\\
&&\alpha_{_{13}}=a_{_5}+b_{_2}-a_{_1}-1,\;\alpha_{_{14}}=-a_{_5}\;.
\label{GKZ21-10-1}
\end{eqnarray}
Applying the integer lattice and exponents above, we construct
the corresponding hypergeometric series solutions as
\begin{eqnarray}
&&\Phi_{_{[1357]}}^{(10),a}(\alpha,z)=
{y_{_2}^{{D\over2}-1}y_{_3}^{{D\over2}-2}y_{_4}^{-1}}\sum\limits_{n_{_1}=0}^\infty
\sum\limits_{n_{_2}=0}^\infty\sum\limits_{n_{_3}=0}^\infty\sum\limits_{n_{_4}=0}^\infty
c_{_{[1357]}}^{(10),a}(\alpha,{\bf n})
\nonumber\\
&&\hspace{2.5cm}\times
\Big({1\over y_{_4}}\Big)^{n_{_1}}\Big({y_{_1}\over y_{_4}}\Big)^{n_{_2}}
\Big({y_{_4}\over y_{_3}}\Big)^{n_{_3}}\Big({y_{_2}\over y_{_3}}\Big)^{n_{_4}}
\;,\nonumber\\
&&\Phi_{_{[1357]}}^{(10),b}(\alpha,z)=
{y_{_2}^{{D\over2}-1}y_{_3}^{{D\over2}-1}y_{_4}^{-2}}\sum\limits_{n_{_1}=0}^\infty
\sum\limits_{n_{_2}=0}^\infty\sum\limits_{n_{_3}=0}^\infty\sum\limits_{n_{_4}=0}^\infty
c_{_{[1357]}}^{(10),b}(\alpha,{\bf n})
\nonumber\\
&&\hspace{2.5cm}\times
\Big({1\over y_{_4}}\Big)^{n_{_1}}\Big({y_{_1}\over y_{_4}}\Big)^{n_{_2}}
\Big({y_{_3}\over y_{_4}}\Big)^{n_{_3}}\Big({y_{_2}\over y_{_4}}\Big)^{n_{_4}}
\;,\nonumber\\
&&\Phi_{_{[1357]}}^{(10),c}(\alpha,z)=
{y_{_2}^{{D\over2}}y_{_3}^{{D\over2}-2}y_{_4}^{-2}}\sum\limits_{n_{_1}=0}^\infty
\sum\limits_{n_{_2}=0}^\infty\sum\limits_{n_{_3}=0}^\infty\sum\limits_{n_{_4}=0}^\infty
c_{_{[1357]}}^{(10),c}(\alpha,{\bf n})
\nonumber\\
&&\hspace{2.5cm}\times
\Big({1\over y_{_4}}\Big)^{n_{_1}}\Big({y_{_1}\over y_{_4}}\Big)^{n_{_2}}
\Big({y_{_2}\over y_{_4}}\Big)^{n_{_3}}\Big({y_{_2}\over y_{_3}}\Big)^{n_{_4}}\;.
\label{GKZ21-10-2a}
\end{eqnarray}
Where the coefficients are
\begin{eqnarray}
&&c_{_{[1357]}}^{(10),a}(\alpha,{\bf n})=
(-)^{n_{_4}}\Gamma(1+n_{_1}+n_{_2})\Gamma(1+n_{_3}+n_{_4})
\Big\{n_{_1}!n_{_2}!n_{_3}!n_{_4}!
\nonumber\\
&&\hspace{2.5cm}\times
\Gamma({D\over2}+n_{_1})\Gamma(2-{D\over2}+n_{_3})\Gamma({D\over2}-1-n_{_3}-n_{_4})
\nonumber\\
&&\hspace{2.5cm}\times
\Gamma({D\over2}-1-n_{_1}-n_{_2})
\Gamma(2-{D\over2}+n_{_2})\Gamma({D\over2}+n_{_4})\Big\}^{-1}
\;,\nonumber\\
&&c_{_{[1357]}}^{(10),b}(\alpha,{\bf n})=
-\Gamma(1+n_{_1}+n_{_2})\Gamma(1+n_{_3}+n_{_4})
\Big\{n_{_1}!n_{_2}!n_{_3}!n_{_4}!
\nonumber\\
&&\hspace{2.5cm}\times
\Gamma({D\over2}+n_{_1})\Gamma(1-{D\over2}-n_{_3}-n_{_4})
\Gamma({D\over2}+n_{_3})
\nonumber\\
&&\hspace{2.5cm}\times
\Gamma({D\over2}-1-n_{_1}-n_{_2})
\Gamma(2-{D\over2}+n_{_2})\Gamma({D\over2}+n_{_4})\Big\}^{-1}
\;,\nonumber\\
&&c_{_{[1357]}}^{(10),c}(\alpha,{\bf n})=
(-)^{1+n_{_4}}\Gamma(1+n_{_1}+n_{_2})\Gamma(1+n_{_3})\Gamma(1+n_{_4})
\Big\{n_{_1}!n_{_2}!
\nonumber\\
&&\hspace{2.5cm}\times
\Gamma(2+n_{_3}+n_{_4})\Gamma({D\over2}+n_{_1})\Gamma(1-{D\over2}-n_{_3})\Gamma({D\over2}-1-n_{_4})
\nonumber\\
&&\hspace{2.5cm}\times
\Gamma({D\over2}-1-n_{_1}-n_{_2})
\Gamma(2-{D\over2}+n_{_2})\Gamma({D\over2}+1+n_{_3}+n_{_4})\Big\}^{-1}\;.
\label{GKZ21-10-3}
\end{eqnarray}
The intersection of the convergent regions of $\Phi_{_{[1357]}}^{(10),a}$
and $\Phi_{_{[1357]}}^{(10),b}$ is empty. Nevertheless, the intersection of the convergent
regions of $\Phi_{_{[1357]}}^{(10),a}$ and $\Phi_{_{[1357]}}^{(10),c}$ is nonempty.
In other words, $\Phi_{_{[1357]}}^{(10),a}$ is equal to $\Phi_{_{[1357]}}^{(10),c}$
up to a constant scalar factor in the nonempty intersection of their convergent regions because they originate
from the exponent vector presented in Eq.(\ref{GKZ21-10-1}) simultaneously.
Similarly the intersection of the convergent regions of $\Phi_{_{[1357]}}^{(10),b}$
and $\Phi_{_{[1357]}}^{(10),c}$ is nonempty, and $\Phi_{_{[1357]}}^{(10),b}$ is equal
to $\Phi_{_{[1357]}}^{(10),c}$ up to a constant scalar factor in the nonempty intersection
of their convergent regions. Because of the reason mentioned above, three solutions
are analytic continuation of each other on the union of their convergent regions.

\item The set of column indices $I_{_{11}}=[2,3,4,7,\cdots,11,13,14]$, i.e.
the implement $J_{_{11}}=[1,14]\setminus I_{_{11}}=[1,5,6,12]$.
The choice of the set of indices induces the exponents
$\alpha_{_1}=\alpha_{_5}=\alpha_{_6}=\alpha_{_{12}}=0$, and
\begin{eqnarray}
&&\alpha_{_2}=a_{_1}-a_{_2},\;\alpha_{_3}=a_{_1}-a_{_3}-a_{_5}-b_{_1}+1,\;
\alpha_{_4}=a_{_1}-a_{_4}-a_{_5}-b_{_1}+1,
\nonumber\\
&&\alpha_{_7}=b_{_2}-1,\;
\alpha_{_8}=a_{_5}+b_{_1}+b_{_3}-a_{_1}-2,\;\alpha_{_9}=b_{_1}+b_{_4}-a_{_1}-2,
\nonumber\\
&&\alpha_{_{10}}=b_{_1}+b_{_5}-a_{_1}-2,\;\alpha_{_{11}}=1-b_{_1},
\nonumber\\
&&\alpha_{_{13}}=a_{_5}+b_{_1}-a_{_1}-1,\;\alpha_{_{14}}=-a_{_5}\;.
\label{GKZ21-11-1}
\end{eqnarray}
The corresponding hypergeometric series functions are
\begin{eqnarray}
&&\Phi_{_{[1357]}}^{(11),a}(\alpha,z)=
{y_{_1}^{{D\over2}-1}y_{_3}^{{D\over2}-2}y_{_4}^{-1}}\sum\limits_{n_{_1}=0}^\infty
\sum\limits_{n_{_2}=0}^\infty\sum\limits_{n_{_3}=0}^\infty\sum\limits_{n_{_4}=0}^\infty
c_{_{[1357]}}^{(11),a}(\alpha,{\bf n})
\nonumber\\
&&\hspace{2.5cm}\times
\Big({1\over y_{_4}}\Big)^{n_{_1}}\Big({y_{_1}\over y_{_4}}\Big)^{n_{_2}}
\Big({y_{_4}\over y_{_3}}\Big)^{n_{_3}}\Big({y_{_2}\over y_{_3}}\Big)^{n_{_4}}
\;,\nonumber\\
&&\Phi_{_{[1357]}}^{(11),b}(\alpha,z)=
{y_{_1}^{{D\over2}-1}y_{_3}^{{D\over2}-1}y_{_4}^{-2}}\sum\limits_{n_{_1}=0}^\infty
\sum\limits_{n_{_2}=0}^\infty\sum\limits_{n_{_3}=0}^\infty\sum\limits_{n_{_4}=0}^\infty
c_{_{[1357]}}^{(11),b}(\alpha,{\bf n})
\nonumber\\
&&\hspace{2.5cm}\times
\Big({1\over y_{_4}}\Big)^{n_{_1}}\Big({y_{_1}\over y_{_4}}\Big)^{n_{_2}}
\Big({y_{_3}\over y_{_4}}\Big)^{n_{_3}}\Big({y_{_2}\over y_{_4}}\Big)^{n_{_4}}
\;,\nonumber\\
&&\Phi_{_{[1357]}}^{(11),c}(\alpha,z)=
{y_{_1}^{{D\over2}-1}y_{_2}y_{_3}^{{D\over2}-2}y_{_4}^{-2}}\sum\limits_{n_{_1}=0}^\infty
\sum\limits_{n_{_2}=0}^\infty\sum\limits_{n_{_3}=0}^\infty\sum\limits_{n_{_4}=0}^\infty
c_{_{[1357]}}^{(11),c}(\alpha,{\bf n})
\nonumber\\
&&\hspace{2.5cm}\times
\Big({1\over y_{_4}}\Big)^{n_{_1}}\Big({y_{_1}\over y_{_4}}\Big)^{n_{_2}}
\Big({y_{_2}\over y_{_4}}\Big)^{n_{_3}}\Big({y_{_2}\over y_{_3}}\Big)^{n_{_4}}\;.
\label{GKZ21-11-2a}
\end{eqnarray}
Where the coefficients are
\begin{eqnarray}
&&c_{_{[1357]}}^{(11),a}(\alpha,{\bf n})=
(-)^{n_{_4}}\Gamma(1+n_{_1}+n_{_2})\Gamma(1+n_{_3}+n_{_4})\Big\{n_{_1}!n_{_2}!n_{_3}!n_{_4}!
\Gamma({D\over2}+n_{_1})
\nonumber\\
&&\hspace{2.5cm}\times
\Gamma({D\over2}+n_{_3})\Gamma(2-{D\over2}+n_{_4})
\Gamma(1-{D\over2}-n_{_1}-n_{_2})\Gamma({D\over2}+n_{_2})
\nonumber\\
&&\hspace{2.5cm}\times
\Gamma({D\over2}-1-n_{_3}-n_{_4})\Big\}^{-1}
\;,\nonumber\\
&&c_{_{[1357]}}^{(11),b}(\alpha,{\bf n})=
-\Gamma(1+n_{_1}+n_{_2})\Gamma(1+n_{_3}+n_{_4})\Big\{n_{_1}!n_{_2}!n_{_3}!n_{_4}!
\Gamma({D\over2}+n_{_1})
\nonumber\\
&&\hspace{2.5cm}\times
\Gamma({D\over2}-1-n_{_3}-n_{_4})\Gamma(2-{D\over2}+n_{_4})
\Gamma(1-{D\over2}-n_{_1}-n_{_2})
\nonumber\\
&&\hspace{2.5cm}\times
\Gamma({D\over2}+n_{_2})\Gamma({D\over2}+n_{_3})\Big\}^{-1}
\;,\nonumber\\
&&c_{_{[1357]}}^{(11),c}(\alpha,{\bf n})=
(-)^{1+n_{_4}}\Gamma(1+n_{_1}+n_{_2})\Gamma(1+n_{_3})\Gamma(1+n_{_4})
\Big\{n_{_1}!n_{_2}!
\nonumber\\
&&\hspace{2.5cm}\times
\Gamma(2+n_{_3}+n_{_4})\Gamma({D\over2}+n_{_1})\Gamma({D\over2}-1-n_{_3})
\Gamma(3-{D\over2}+n_{_3}+n_{_4})
\nonumber\\
&&\hspace{2.5cm}\times
\Gamma(1-{D\over2}-n_{_1}-n_{_2})\Gamma({D\over2}+n_{_2})\Gamma({D\over2}-1-n_{_4})\Big\}^{-1}\;.
\label{GKZ21-11-3}
\end{eqnarray}
Obviously the intersection of the convergent regions of $\Phi_{_{[1357]}}^{(11),a}$
and $\Phi_{_{[1357]}}^{(11),b}$ is empty. However, the intersection of the convergent
regions of $\Phi_{_{[1357]}}^{(11),a}$ and $\Phi_{_{[1357]}}^{(11),c}$ is nonempty.
In other words, $\Phi_{_{[1357]}}^{(11),a}$ is equal to $\Phi_{_{[1357]}}^{(11),c}$
up to a constant scalar factor in the nonempty intersection because they originate
from the exponent vector presented in Eq.(\ref{GKZ21-11-1}) simultaneously.
Similarly the intersection of the convergent regions of $\Phi_{_{[1357]}}^{(11),b}$
and $\Phi_{_{[1357]}}^{(11),c}$ is nonempty, and $\Phi_{_{[1357]}}^{(11),b}$ is equal
to $\Phi_{_{[1357]}}^{(11),c}$ up to a constant scalar factor in the nonempty intersection.
Because of the reason mentioned above, three solutions are analytic continuation
of each other on the union of their convergent regions.

\item The set of column indices $I_{_{12}}=[2,3,4,6,\cdots,10,13,14]$, i.e.
the implement $J_{_{12}}=[1,14]\setminus I_{_{12}}=[1,5,11,12]$.
Basing on the choice of the set of indices, we derive the exponents
as $\alpha_{_1}=\alpha_{_5}=\alpha_{_{11}}=\alpha_{_{12}}=0$, and
\begin{eqnarray}
&&\alpha_{_2}=a_{_1}-a_{_2},\;\alpha_{_3}=a_{_1}-a_{_3}-a_{_5},\;\alpha_{_4}=a_{_1}-a_{_4}-a_{_5},
\nonumber\\
&&\alpha_{_6}=b_{_1}-1,\;\alpha_{_7}=b_{_2}-1,\;
\alpha_{_8}=a_{_5}+b_{_3}-a_{_1}-1,
\nonumber\\
&&\alpha_{_9}=b_{_4}-a_{_1}-1,\;\alpha_{_{10}}=b_{_5}-a_{_1}-1,\;\alpha_{_{13}}=a_{_5}-a_{_1},
\nonumber\\
&&\alpha_{_{14}}=-a_{_5}\;.
\label{GKZ21-12-1}
\end{eqnarray}
The hypergeometric series solutions are correspondingly written as
\begin{eqnarray}
&&\Phi_{_{[1357]}}^{(12),a}(\alpha,z)=
y_{_3}^{{D}-3}y_{_4}^{-1}\sum\limits_{n_{_1}=0}^\infty
\sum\limits_{n_{_2}=0}^\infty\sum\limits_{n_{_3}=0}^\infty\sum\limits_{n_{_4}=0}^\infty
c_{_{[1357]}}^{(12),a}(\alpha,{\bf n})
\nonumber\\
&&\hspace{2.5cm}\times
\Big({1\over y_{_4}}\Big)^{n_{_1}}\Big({y_{_1}\over y_{_4}}\Big)^{n_{_2}}
\Big({y_{_4}\over y_{_3}}\Big)^{n_{_3}}\Big({y_{_2}\over y_{_3}}\Big)^{n_{_4}}
\;,\nonumber\\
&&\Phi_{_{[1357]}}^{(12),b}(\alpha,z)=
y_{_3}^{{D}-2}y_{_4}^{-2}\sum\limits_{n_{_1}=0}^\infty
\sum\limits_{n_{_2}=0}^\infty\sum\limits_{n_{_3}=0}^\infty\sum\limits_{n_{_4}=0}^\infty
c_{_{[1357]}}^{(12),b}(\alpha,{\bf n})
\nonumber\\
&&\hspace{2.5cm}\times
\Big({1\over y_{_4}}\Big)^{n_{_1}}\Big({y_{_1}\over y_{_4}}\Big)^{n_{_2}}
\Big({y_{_3}\over y_{_4}}\Big)^{n_{_3}}\Big({y_{_2}\over y_{_3}}\Big)^{n_{_4}}\;.
\label{GKZ21-12-2a}
\end{eqnarray}
Where the coefficients are
\begin{eqnarray}
&&c_{_{[1357]}}^{(12),a}(\alpha,{\bf n})=
(-)^{n_{_3}}\Gamma(1+n_{_1}+n_{_2})\Big\{n_{_1}!n_{_2}!n_{_3}!n_{_4}!
\Gamma({D\over2}+n_{_1})\Gamma(2-{D\over2}+n_{_3})
\nonumber\\
&&\hspace{2.5cm}\times
\Gamma(2-{D\over2}+n_{_2})
\Gamma(2-{D\over2}+n_{_4})\Gamma({D\over2}-1-n_{_3}-n_{_4})
\nonumber\\
&&\hspace{2.5cm}\times
\Gamma({D\over2}-1-n_{_1}-n_{_2})\Gamma(D-2-n_{_3}-n_{_4})\Big\}^{-1}
\;,\nonumber\\
&&c_{_{[1357]}}^{(12),b}(\alpha,{\bf n})=
-\Gamma(1+n_{_1}+n_{_2})\Gamma(1+n_{_3})\Big\{n_{_1}!n_{_2}!n_{_4}!\Gamma({D\over2}+n_{_1})
\nonumber\\
&&\hspace{2.5cm}\times
\Gamma(1-{D\over2}-n_{_3})\Gamma(2-{D\over2}+n_{_2})
\Gamma({D\over2}+n_{_3}-n_{_4})
\nonumber\\
&&\hspace{2.5cm}\times
\Gamma(2-{D\over2}+n_{_4})\Gamma({D\over2}-1-n_{_1}-n_{_2})\Gamma(D-1+n_{_3}-n_{_4})\Big\}^{-1}\;.
\label{GKZ21-12-3}
\end{eqnarray}
The intersection of the convergent regions of $\Phi_{_{[1357]}}^{(12),a}$
and $\Phi_{_{[1357]}}^{(12),b}$ is empty. Because they originate
from the exponent vector presented in Eq.(\ref{GKZ21-12-1}) simultaneously,
the solutions are analytic continuation of each other
on the union of their convergent regions.

\item The set of column indices $I_{_{13}}=[1,3,4,8,\cdots,14]$, i.e.
the implement $J_{_{13}}=[1,14]\setminus I_{_{13}}=[2,5,6,7]$.
The choice of column indices implies the power numbers
$\alpha_{_2}=\alpha_{_5}=\alpha_{_6}=\alpha_{_{7}}=0$, and
\begin{eqnarray}
&&\alpha_{_1}=a_{_2}-a_{_1},\;\alpha_{_3}=a_{_2}-a_{_3}-a_{_5}-b_{_1}+1,
\nonumber\\
&&\alpha_{_4}=a_{_2}-a_{_4}-a_{_5}-b_{_1}+1,
\nonumber\\
&&\alpha_{_8}=b_{_1}+b_{_2}+b_{_3}+a_{_5}-a_{_2}-3,\;\alpha_{_9}=b_{_1}+b_{_4}-a_{_2}-2,
\nonumber\\
&&\alpha_{_{10}}=b_{_1}+b_{_5}-a_{_2}-2,\;\alpha_{_{11}}=1-b_{_1},\;\alpha_{_{12}}=1-b_{_2},
\nonumber\\
&&\alpha_{_{13}}=b_{_1}+b_{_2}+a_{_5}-a_{_2}-2,\;\alpha_{_{14}}=-a_{_5}\;.
\label{GKZ21-13-1}
\end{eqnarray}
The corresponding hypergeometric series solutions are written as
\begin{eqnarray}
&&\Phi_{_{[1357]}}^{(13),a}(\alpha,z)=
y_{_1}^{{D\over2}-1}y_{_2}^{{D\over2}-1}y_{_3}^{{D\over2}-2}y_{_4}^{-1}\sum\limits_{n_{_1}=0}^\infty
\sum\limits_{n_{_2}=0}^\infty\sum\limits_{n_{_3}=0}^\infty\sum\limits_{n_{_4}=0}^\infty
c_{_{[1357]}}^{(13),a}(\alpha,{\bf n})
\nonumber\\
&&\hspace{2.5cm}\times
\Big({1\over y_{_4}}\Big)^{n_{_1}}\Big({y_{_1}\over y_{_4}}\Big)^{n_{_2}}
\Big({y_{_4}\over y_{_3}}\Big)^{n_{_3}}\Big({y_{_2}\over y_{_3}}\Big)^{n_{_4}}
\;,\nonumber\\
&&\Phi_{_{[1357]}}^{(13),b}(\alpha,z)=
y_{_1}^{{D\over2}-1}y_{_2}^{{D\over2}-1}y_{_3}^{{D\over2}-1}y_{_4}^{-2}\sum\limits_{n_{_1}=0}^\infty
\sum\limits_{n_{_2}=0}^\infty\sum\limits_{n_{_3}=0}^\infty\sum\limits_{n_{_4}=0}^\infty
c_{_{[1357]}}^{(13),b}(\alpha,{\bf n})
\nonumber\\
&&\hspace{2.5cm}\times
\Big({1\over y_{_4}}\Big)^{n_{_1}}\Big({y_{_1}\over y_{_4}}\Big)^{n_{_2}}
\Big({y_{_3}\over y_{_4}}\Big)^{n_{_3}}\Big({y_{_2}\over y_{_4}}\Big)^{n_{_4}}
\;,\nonumber\\
&&\Phi_{_{[1357]}}^{(13),c}(\alpha,z)=
y_{_1}^{{D\over2}-1}y_{_2}^{{D\over2}}y_{_3}^{{D\over2}-2}y_{_4}^{-2}\sum\limits_{n_{_1}=0}^\infty
\sum\limits_{n_{_2}=0}^\infty\sum\limits_{n_{_3}=0}^\infty\sum\limits_{n_{_4}=0}^\infty
c_{_{[1357]}}^{(13),c}(\alpha,{\bf n})
\nonumber\\
&&\hspace{2.5cm}\times
\Big({1\over y_{_4}}\Big)^{n_{_1}}\Big({y_{_1}\over y_{_4}}\Big)^{n_{_2}}
\Big({y_{_2}\over y_{_4}}\Big)^{n_{_3}}\Big({y_{_2}\over y_{_3}}\Big)^{n_{_4}}\;.
\label{GKZ21-13-2a}
\end{eqnarray}
Where the coefficients are
\begin{eqnarray}
&&c_{_{[1357]}}^{(13),a}(\alpha,{\bf n})=
(-)^{n_{_4}}\Gamma(1+n_{_1}+n_{_2})\Gamma(1+n_{_3}+n_{_4})
\Big\{n_{_1}!n_{_2}!n_{_3}!n_{_4}!
\nonumber\\
&&\hspace{2.5cm}\times
\Gamma(2-{D\over2}+n_{_1})\Gamma({D\over2}+n_{_2})\Gamma({D\over2}+n_{_4})
\Gamma(2-{D\over2}+n_{_3})
\nonumber\\
&&\hspace{2.5cm}\times
\Gamma({D\over2}-1-n_{_1}-n_{_2})\Gamma({D\over2}-1-n_{_3}-n_{_4})\Big\}^{-1}
\;,\nonumber\\
&&c_{_{[1357]}}^{(13),b}(\alpha,{\bf n})=
-\Gamma(1+n_{_1}+n_{_2})\Gamma(1+n_{_3}+n_{_4})\Big\{n_{_1}!n_{_2}!n_{_3}!n_{_4}!
\nonumber\\
&&\hspace{2.5cm}\times
\Gamma(2-{D\over2}+n_{_1})\Gamma({D\over2}+n_{_2})
\Gamma(1-{D\over2}-n_{_3}-n_{_4})
\nonumber\\
&&\hspace{2.5cm}\times
\Gamma({D\over2}+n_{_3})\Gamma({D\over2}-1-n_{_1}-n_{_2})
\Gamma({D\over2}+n_{_4})\Big\}^{-1}
\;,\nonumber\\
&&c_{_{[1357]}}^{(13),c}(\alpha,{\bf n})=
(-)^{1+n_{_4}}\Gamma(1+n_{_1}+n_{_2})\Gamma(1+n_{_3})\Gamma(1+n_{_4})
\nonumber\\
&&\hspace{2.5cm}\times
\Big\{n_{_1}!n_{_2}!\Gamma(2+n_{_3}+n_{_4})\Gamma(2-{D\over2}+n_{_1})
\Gamma({D\over2}+n_{_2})
\nonumber\\
&&\hspace{2.5cm}\times
\Gamma({D\over2}+1+n_{_3}+n_{_4})\Gamma(1-{D\over2}-n_{_3})
\nonumber\\
&&\hspace{2.5cm}\times
\Gamma({D\over2}-1-n_{_1}-n_{_2})\Gamma({D\over2}-1-n_{_4})\Big\}^{-1}\;.
\label{GKZ21-13-3}
\end{eqnarray}
Certainly the intersection of the convergent regions of $\Phi_{_{[1357]}}^{(13),a}$
and $\Phi_{_{[1357]}}^{(13),b}$ is empty. However, the intersection of the convergent
regions of $\Phi_{_{[1357]}}^{(13),a}$ and $\Phi_{_{[1357]}}^{(13),c}$ is nonempty.
In other words, $\Phi_{_{[1357]}}^{(13),a}$ is proportional to $\Phi_{_{[1357]}}^{(13),c}$
in the nonempty intersection because they originate
from the exponent vector presented in Eq.(\ref{GKZ21-13-1}) simultaneously.
Similarly the intersection of the convergent regions of $\Phi_{_{[1357]}}^{(13),b}$
and $\Phi_{_{[1357]}}^{(13),c}$ is nonempty, and $\Phi_{_{[1357]}}^{(13),b}$ is proportional
to $\Phi_{_{[1357]}}^{(13),c}$ in the nonempty intersection.
Because of the reason mentioned above, three solutions are analytic continuation
of each other on the union of their convergent regions.

\item The set of column indices $I_{_{14}}=[1,3,4,6,8,9,10,12,13,14]$, i.e.
the implement $J_{_{14}}=[1,14]\setminus I_{_{14}}=[2,5,7,11]$.
The set of column indices induces the power numbers
$\alpha_{_2}=\alpha_{_5}=\alpha_{_7}=\alpha_{_{11}}=0$, and
\begin{eqnarray}
&&\alpha_{_1}=a_{_2}-a_{_1},\;\alpha_{_3}=a_{_2}-a_{_3}-a_{_5},\;\alpha_{_4}=a_{_2}-a_{_4}-a_{_5},
\nonumber\\
&&\alpha_{_6}=b_{_1}-1,\;
\alpha_{_8}=a_{_5}+b_{_2}+b_{_3}-a_{_2}-2,
\nonumber\\
&&\alpha_{_9}=b_{_4}-a_{_2}-1,\;\alpha_{_{10}}=b_{_5}-a_{_2}-1,\;\alpha_{_{12}}=1-b_{_2},
\nonumber\\
&&\alpha_{_{13}}=a_{_5}+b_{_2}-a_{_2}-1,\;\alpha_{_{14}}=-a_{_5}\;.
\label{GKZ21-14-1}
\end{eqnarray}
Using the integer lattice and exponents above, we write the
corresponding hypergeometric series solution as
\begin{eqnarray}
&&\Phi_{_{[1357]}}^{(14)}(\alpha,z)=
y_{_2}^{{D\over2}-1}y_{_3}^{{D}-3}y_{_4}^{-1}\sum\limits_{n_{_1}=0}^\infty
\sum\limits_{n_{_2}=0}^\infty\sum\limits_{n_{_3}=0}^\infty\sum\limits_{n_{_4}=0}^\infty
c_{_{[1357]}}^{(14)}(\alpha,{\bf n})
\nonumber\\
&&\hspace{2.5cm}\times
\Big({1\over y_{_3}}\Big)^{n_{_1}}\Big({y_{_1}\over y_{_3}}\Big)^{n_{_2}}
\Big({y_{_3}\over y_{_4}}\Big)^{n_{_3}}\Big({y_{_2}\over y_{_3}}\Big)^{n_{_4}}\;,
\label{GKZ21-14-2}
\end{eqnarray}
with
\begin{eqnarray}
&&c_{_{[1357]}}^{(14)}(\alpha,{\bf n})=
(-)^{n_{_3}}\Big\{n_{_1}!n_{_2}!n_{_4}!\Gamma(2-{D\over2}+n_{_1})
\Gamma(2-{D\over2}+n_{_2})
\nonumber\\
&&\hspace{2.5cm}\times
\Gamma(3-D+n_{_1}+n_{_2}-n_{_3})\Gamma(2-{D\over2}+n_{_1}+n_{_2}-n_{_3})
\nonumber\\
&&\hspace{2.5cm}\times
\Gamma({D\over2}-1-n_{_1}-n_{_2}+n_{_3}-n_{_4})
\Gamma(D-2-n_{_1}-n_{_2})
\nonumber\\
&&\hspace{2.5cm}\times
\Gamma({D\over2}-1-n_{_1}-n_{_2})
\Gamma({D\over2}+n_{_4})
\nonumber\\
&&\hspace{2.5cm}\times
\Gamma(D-2-n_{_1}-n_{_2}+n_{_3}-n_{_4})\Big\}^{-1}\;.
\label{GKZ21-14-3}
\end{eqnarray}

\item The set of column indices $I_{_{15}}=[1,3,4,7,\cdots,11,13,14]$, i.e.
the implement $J_{_{15}}=[1,14]\setminus I_{_{15}}=[2,5,6,12]$.
Basing on the choice of the set of column indices, one derives the power numbers
$\alpha_{_2}=\alpha_{_5}=\alpha_{_6}=\alpha_{_{12}}=0$, and
\begin{eqnarray}
&&\alpha_{_1}=a_{_2}-a_{_1},\;\alpha_{_3}=a_{_2}-a_{_3}-a_{_5}-b_{_1}+1,\;\alpha_{_4}=a_{_2}-a_{_4}-a_{_5}-b_{_1}+1,
\nonumber\\
&&\alpha_{_7}=b_{_2}-1,\;
\alpha_{_8}=a_{_5}+b_{_1}+b_{_3}-a_{_2}-2,\;\alpha_{_9}=b_{_1}+b_{_4}-a_{_2}-2,
\nonumber\\
&&\alpha_{_{10}}=b_{_1}+b_{_5}-a_{_2}-2,\;\alpha_{_{11}}=1-b_{_1},
\nonumber\\
&&\alpha_{_{13}}=a_{_5}+b_{_1}-a_{_2}-1,\;\alpha_{_{14}}=-a_{_5}\;.
\label{GKZ21-15-1}
\end{eqnarray}
In a similar way the hypergeometric series solutions are formulated as
\begin{eqnarray}
&&\Phi_{_{[1357]}}^{(15),a}(\alpha,z)=
y_{_1}^{{D\over2}-1}y_{_3}^{{D}-3}y_{_4}^{-1}\sum\limits_{n_{_1}=0}^\infty
\sum\limits_{n_{_2}=0}^\infty\sum\limits_{n_{_3}=0}^\infty\sum\limits_{n_{_4}=0}^\infty
c_{_{[1357]}}^{(15),a}(\alpha,{\bf n})
\nonumber\\
&&\hspace{2.5cm}\times
\Big({1\over y_{_4}}\Big)^{n_{_1}}\Big({y_{_1}\over y_{_4}}\Big)^{n_{_2}}
\Big({y_{_4}\over y_{_3}}\Big)^{n_{_3}}\Big({y_{_2}\over y_{_3}}\Big)^{n_{_4}}
\;,\nonumber\\
&&\Phi_{_{[1357]}}^{(15),b}(\alpha,z)=
y_{_1}^{{D\over2}-1}y_{_3}^{{D}-2}y_{_4}^{-2}\sum\limits_{n_{_1}=0}^\infty
\sum\limits_{n_{_2}=0}^\infty\sum\limits_{n_{_3}=0}^\infty\sum\limits_{n_{_4}=0}^\infty
c_{_{[1357]}}^{(15),b}(\alpha,{\bf n})
\nonumber\\
&&\hspace{2.5cm}\times
\Big({1\over y_{_4}}\Big)^{n_{_1}}\Big({y_{_1}\over y_{_4}}\Big)^{n_{_2}}
\Big({y_{_3}\over y_{_4}}\Big)^{n_{_3}}\Big({y_{_2}\over y_{_3}}\Big)^{n_{_4}}\;.
\label{GKZ21-15-2a}
\end{eqnarray}
Where the coefficients are
\begin{eqnarray}
&&c_{_{[1357]}}^{(15),a}(\alpha,{\bf n})=(-)^{n_{_3}}
\Gamma(1+n_{_1}+n_{_2})\Big\{n_{_1}!n_{_2}!n_{_3}!n_{_4}!
\Gamma(2-{D\over2}+n_{_1})
\nonumber\\
&&\hspace{2.5cm}\times
\Gamma(2-{D\over2}+n_{_3})\Gamma(2-{D\over2}+n_{_4})
\Gamma({D\over2}-1-n_{_3}-n_{_4})
\nonumber\\
&&\hspace{2.5cm}\times
\Gamma({D\over2}-1-n_{_1}-n_{_2})\Gamma({D\over2}+n_{_2})
\Gamma(D-2-n_{_3}-n_{_4})\Big\}^{-1}
\;,\nonumber\\
&&c_{_{[1357]}}^{(15),b}(\alpha,{\bf n})=
-\Gamma(1+n_{_1}+n_{_2})\Gamma(1+n_{_3})\Big\{n_{_1}!n_{_2}!n_{_4}!
\Gamma(2-{D\over2}+n_{_1})
\nonumber\\
&&\hspace{2.5cm}\times
\Gamma(1-{D\over2}-n_{_3})\Gamma(2-{D\over2}+n_{_4})
\Gamma({D\over2}+n_{_3}-n_{_4})
\nonumber\\
&&\hspace{2.5cm}\times
\Gamma({D\over2}-1-n_{_1}-n_{_2})\Gamma({D\over2}+n_{_2})
\Gamma(D-1+n_{_3}-n_{_4})\Big\}^{-1}\;.
\label{GKZ21-15-3}
\end{eqnarray}
The intersection of the convergent regions of two hypergeometric series
is empty. Two solutions are analytic continuation of each other
on the union of their convergent regions because they are constructed
from the same exponent vector presented in Eq.(\ref{GKZ21-15-1}).

\item The set of column indices $I_{_{16}}=[1,3,4,6,\cdots,10,13,14]$, i.e.
the implement $J_{_{16}}=[1,14]\setminus I_{_{16}}=[2,5,11,12]$.
The choice of the set of column indices induces the exponents
$\alpha_{_2}=\alpha_{_5}=\alpha_{_{11}}=\alpha_{_{12}}=0$, and
\begin{eqnarray}
&&\alpha_{_1}=a_{_2}-a_{_1},\;\alpha_{_3}=a_{_2}-a_{_3}-a_{_5},\;\alpha_{_4}=a_{_2}-a_{_4}-a_{_5},
\nonumber\\
&&\alpha_{_6}=b_{_1}-1,\;\alpha_{_7}=b_{_2}-1,\;
\alpha_{_8}=a_{_5}+b_{_3}-a_{_2}-1,
\nonumber\\
&&\alpha_{_9}=b_{_4}-a_{_2}-1,\;\alpha_{_{10}}=b_{_5}-a_{_2}-1,\;\alpha_{_{13}}=a_{_5}-a_{_2},
\nonumber\\
&&\alpha_{_{14}}=-a_{_5}\;.
\label{GKZ21-16-1}
\end{eqnarray}
Using the integer lattice and the exponents above, we write
the hypergeometric series solution as
\begin{eqnarray}
&&\Phi_{_{[1357]}}^{(16)}(\alpha,z)=
y_{_3}^{{3D\over2}-4}y_{_4}^{-1}\sum\limits_{n_{_1}=0}^\infty
\sum\limits_{n_{_2}=0}^\infty\sum\limits_{n_{_3}=0}^\infty\sum\limits_{n_{_4}=0}^\infty
c_{_{[1357]}}^{(16)}(\alpha,{\bf n})
\nonumber\\
&&\hspace{2.5cm}\times
\Big({1\over y_{_3}}\Big)^{n_{_1}}\Big({y_{_1}\over y_{_3}}\Big)^{n_{_2}}
\Big({y_{_3}\over y_{_4}}\Big)^{n_{_3}}\Big({y_{_2}\over y_{_3}}\Big)^{n_{_4}}\;,
\label{GKZ21-16-2}
\end{eqnarray}
with
\begin{eqnarray}
&&c_{_{[1357]}}^{(16)}(\alpha,{\bf n})=
(-)^{n_{_3}}\Big\{n_{_1}!n_{_2}!n_{_4}!\Gamma(2-{D\over2}+n_{_1})
\Gamma(2-{D\over2}+n_{_2})
\nonumber\\
&&\hspace{2.5cm}\times
\Gamma(3-D+n_{_1}+n_{_2}-n_{_3})\Gamma(2-{D\over2}+n_{_1}+n_{_2}-n_{_3})
\nonumber\\
&&\hspace{2.5cm}\times
\Gamma(2-{D\over2}+n_{_4})
\Gamma(D-2-n_{_1}-n_{_2}+n_{_3}-n_{_4})
\nonumber\\
&&\hspace{2.5cm}\times
\Gamma(D-2-n_{_1}-n_{_2})
\Gamma({D\over2}-1-n_{_1}-n_{_2})
\nonumber\\
&&\hspace{2.5cm}\times
\Gamma({3D\over2}-3-n_{_1}-n_{_2}+n_{_3}-n_{_4})\Big\}^{-1}\;.
\label{GKZ21-16-3}
\end{eqnarray}

\item The set of column indices $I_{_{17}}=[2,4,5,8,\cdots,14]$, i.e.
the implement $J_{_{17}}=[1,14]\setminus I_{_{17}}=[1,3,6,7]$.
The choice of the set of column indices gives the exponents
$\alpha_{_1}=\alpha_{_3}=\alpha_{_{6}}=\alpha_{_{7}}=0$, and
\begin{eqnarray}
&&\alpha_{_2}=a_{_1}-a_{_2},\;\alpha_{_4}=a_{_3}-a_{_4},\;\alpha_{_5}=a_{_1}-a_{_3}-a_{_5}-b_{_1}+1,
\nonumber\\
&&\alpha_{_8}=b_{_2}+b_{_3}-a_{_3}-2,\;\alpha_{_9}=b_{_1}+b_{_4}-a_{_1}-2,
\nonumber\\
&&\alpha_{_{10}}=b_{_1}+b_{_5}-a_{_1}-2,\;
\alpha_{_{11}}=1-b_{_1},\;\alpha_{_{12}}=1-b_{_2},
\nonumber\\
&&\alpha_{_{13}}=b_{_2}-a_{_3}-1,\;\alpha_{_{14}}=a_{_3}+b_{_1}-a_{_1}-1\;.
\label{GKZ21-17-1}
\end{eqnarray}
Thus the hypergeometric series solution is
\begin{eqnarray}
&&\Phi_{_{[1357]}}^{(17)}(\alpha,z)=
y_{_1}^{{D\over2}-1}y_{_2}^{{D\over2}-1}y_{_3}^{-1}y_{_4}^{-1}\sum\limits_{n_{_1}=0}^\infty
\sum\limits_{n_{_2}=0}^\infty\sum\limits_{n_{_3}=0}^\infty\sum\limits_{n_{_4}=0}^\infty
c_{_{[1357]}}^{(17)}(\alpha,{\bf n})
\nonumber\\
&&\hspace{2.5cm}\times
\Big({1\over y_{_4}}\Big)^{n_{_1}}\Big({y_{_4}\over y_{_3}}\Big)^{n_{_2}}
\Big({y_{_1}\over y_{_4}}\Big)^{n_{_3}}\Big({y_{_2}\over y_{_3}}\Big)^{n_{_4}}\;,
\label{GKZ21-17-2}
\end{eqnarray}
with
\begin{eqnarray}
&&c_{_{[1357]}}^{(17)}(\alpha,{\bf n})=
(-)^{n_{_4}}\Gamma(1+n_{_1}+n_{_3})\Gamma(1+n_{_2}+n_{_4})
\Big\{n_{_1}!n_{_2}!n_{_3}!n_{_4}!
\nonumber\\
&&\hspace{2.5cm}\times
\Gamma({D\over2}+n_{_1})\Gamma({D\over2}+n_{_2})
\Gamma(1-{D\over2}-n_{_2}-n_{_4})
\nonumber\\
&&\hspace{2.5cm}\times
\Gamma(1-{D\over2}-n_{_1}-n_{_3})\Gamma({D\over2}+n_{_3})
\Gamma({D\over2}+n_{_4})\Big\}^{-1}\;.
\label{GKZ21-17-3}
\end{eqnarray}

\item The set of column indices $I_{_{18}}=[2,4,5,7,\cdots,11,13,14]$, i.e.
the implement $J_{_{18}}=[1,14]\setminus I_{_{18}}=[1,3,6,12]$.
The set of column indices implies the power numbers
$\alpha_{_1}=\alpha_{_3}=\alpha_{_{6}}=\alpha_{_{12}}=0$, and
\begin{eqnarray}
&&\alpha_{_2}=a_{_1}-a_{_2},\;\alpha_{_4}=a_{_3}-a_{_4},\;\alpha_{_5}=a_{_1}-a_{_3}-a_{_5}-b_{_1}+1,
\nonumber\\
&&\alpha_{_7}=b_{_2}-1,\;\alpha_{_8}=b_{_3}-a_{_3}-1,\;\alpha_{_9}=b_{_1}+b_{_4}-a_{_1}-2,
\nonumber\\
&&\alpha_{_{10}}=b_{_1}+b_{_5}-a_{_1}-2,\;
\alpha_{_{11}}=1-b_{_1},
\nonumber\\
&&\alpha_{_{13}}=-a_{_3},\;\alpha_{_{14}}=a_{_3}+b_{_1}-a_{_1}-1\;.
\label{GKZ21-18-1}
\end{eqnarray}
Applying the integer lattice and the exponents above, we obtain
the hypergeometric series solution as
\begin{eqnarray}
&&\Phi_{_{[1357]}}^{(18)}(\alpha,z)=
y_{_1}^{{D\over2}-1}y_{_3}^{{D\over2}-2}y_{_4}^{-1}\sum\limits_{n_{_1}=0}^\infty
\sum\limits_{n_{_2}=0}^\infty\sum\limits_{n_{_3}=0}^\infty\sum\limits_{n_{_4}=0}^\infty
c_{_{[1357]}}^{(18)}(\alpha,{\bf n})
\nonumber\\
&&\hspace{2.5cm}\times
\Big({1\over y_{_4}}\Big)^{n_{_1}}\Big({y_{_4}\over y_{_3}}\Big)^{n_{_2}}
\Big({y_{_1}\over y_{_4}}\Big)^{n_{_3}}\Big({y_{_2}\over y_{_3}}\Big)^{n_{_4}}\;,
\label{GKZ21-18-2}
\end{eqnarray}
with
\begin{eqnarray}
&&c_{_{[1357]}}^{(18)}(\alpha,{\bf n})=
(-)^{n_{_4}}\Gamma(1+n_{_1}+n_{_3})\Gamma(1+n_{_2}+n_{_4})
\Big\{n_{_1}!n_{_2}!n_{_3}!n_{_4}!
\nonumber\\
&&\hspace{2.5cm}\times
\Gamma({D\over2}+n_{_1})\Gamma({D\over2}+n_{_2})\Gamma(2-{D\over2}+n_{_4})
\Gamma(1-{D\over2}-n_{_1}-n_{_3})
\nonumber\\
&&\hspace{2.5cm}\times
\Gamma({D\over2}+n_{_3})
\Gamma({D\over2}-1-n_{_2}-n_{_4})\Big\}^{-1}\;.
\label{GKZ21-18-3}
\end{eqnarray}

\item The set of column indices $I_{_{19}}=[2,4,5,6,8,9,10,12,13,14]$,
i.e. the implement $J_{_{19}}=[1,14]\setminus I_{_{19}}=[1,3,7,11]$.
The choice of column indices implies the exponents
$\alpha_{_1}=\alpha_{_3}=\alpha_{_{7}}=\alpha_{_{11}}=0$, and
\begin{eqnarray}
&&\alpha_{_2}=a_{_1}-a_{_2},\;\alpha_{_4}=a_{_3}-a_{_4},\;\alpha_{_5}=a_{_1}-a_{_3}-a_{_5},
\nonumber\\
&&\alpha_{_{6}}=b_{_1}-1,\;\alpha_{_8}=b_{_2}+b_{_3}-a_{_3}-2,\;\alpha_{_9}=b_{_4}-a_{_1}-1,
\nonumber\\
&&\alpha_{_{10}}=b_{_5}-a_{_1}-1,\;
\alpha_{_{12}}=1-b_{_2},
\nonumber\\
&&\alpha_{_{13}}=b_{_2}-a_{_3}-1,\;\alpha_{_{14}}=a_{_3}-a_{_1}\;.
\label{GKZ21-19-1}
\end{eqnarray}
Basing on the integer lattice ${\bf B}_{_{1357}}$ and those powers derived above,
one derives the hypergeometric series solution as
\begin{eqnarray}
&&\Phi_{_{[1357]}}^{(19)}(\alpha,z)=
y_{_2}^{{D\over2}-1}y_{_3}^{-1}y_{_4}^{{D\over2}-2}\sum\limits_{n_{_1}=0}^\infty
\sum\limits_{n_{_2}=0}^\infty\sum\limits_{n_{_3}=0}^\infty\sum\limits_{n_{_4}=0}^\infty
c_{_{[1357]}}^{(19)}(\alpha,{\bf n})
\nonumber\\
&&\hspace{2.5cm}\times
\Big({1\over y_{_4}}\Big)^{n_{_1}}\Big({y_{_4}\over y_{_3}}\Big)^{n_{_2}}
\Big({y_{_1}\over y_{_4}}\Big)^{n_{_3}}\Big({y_{_2}\over y_{_3}}\Big)^{n_{_4}}\;,
\label{GKZ21-19-2}
\end{eqnarray}
with
\begin{eqnarray}
&&c_{_{[1357]}}^{(19)}(\alpha,{\bf n})=
(-)^{1+n_{_4}}\Gamma(1+n_{_1}+n_{_3})\Gamma(1+n_{_2}+n_{_4})
\nonumber\\
&&\hspace{2.5cm}\times
\Big\{n_{_1}!n_{_2}!n_{_3}!n_{_4}!\Gamma({D\over2}+n_{_1})\Gamma({D\over2}+n_{_2})
\Gamma(2-{D\over2}+n_{_3})
\nonumber\\
&&\hspace{2.5cm}\times
\Gamma(1-{D\over2}-n_{_2}-n_{_4})\Gamma({D\over2}-1-n_{_1}-n_{_3})
\Gamma({D\over2}+n_{_4})\Big\}^{-1}\;.
\label{GKZ21-19-3}
\end{eqnarray}

\item The set of column indices $I_{_{20}}=[2,4,\cdots,10,13,14]$,
i.e. the implement $J_{_{20}}=[1,14]\setminus I_{_{20}}=[1,3,11,12]$.
The set of column indices implies the power numbers
$\alpha_{_1}=\alpha_{_3}=\alpha_{_{11}}=\alpha_{_{12}}=0$, and
\begin{eqnarray}
&&\alpha_{_2}=a_{_1}-a_{_2},\;\alpha_{_4}=a_{_3}-a_{_4},\;\alpha_{_5}=a_{_1}-a_{_3}-a_{_5},
\nonumber\\
&&\alpha_{_6}=b_{_1}-1,\;\alpha_{_7}=b_{_2}-1,\;\alpha_{_8}=b_{_3}-a_{_3}-1,
\nonumber\\
&&\alpha_{_9}=b_{_4}-a_{_1}-1,\;\alpha_{_{10}}=b_{_5}-a_{_1}-1,\;
\nonumber\\
&&\alpha_{_{13}}=-a_{_3},\;\alpha_{_{14}}=a_{_3}-a_{_1}\;.
\label{GKZ21-20-1}
\end{eqnarray}
Thus we obtain the hypergeometric series as
\begin{eqnarray}
&&\Phi_{_{[1357]}}^{(20)}(\alpha,z)=
y_{_3}^{{D\over2}-2}y_{_4}^{{D\over2}-2}\sum\limits_{n_{_1}=0}^\infty
\sum\limits_{n_{_2}=0}^\infty\sum\limits_{n_{_3}=0}^\infty\sum\limits_{n_{_4}=0}^\infty
c_{_{[1357]}}^{(20)}(\alpha,{\bf n})
\nonumber\\
&&\hspace{2.5cm}\times
\Big({1\over y_{_4}}\Big)^{n_{_1}}\Big({y_{_4}\over y_{_3}}\Big)^{n_{_2}}
\Big({y_{_1}\over y_{_4}}\Big)^{n_{_3}}\Big({y_{_2}\over y_{_3}}\Big)^{n_{_4}}\;,
\label{GKZ21-20-2}
\end{eqnarray}
with
\begin{eqnarray}
&&c_{_{[1357]}}^{(20)}(\alpha,{\bf n})=
(-)^{1+n_{_4}}\Gamma(1+n_{_1}+n_{_3})\Gamma(1+n_{_2}+n_{_4})
\nonumber\\
&&\hspace{2.5cm}\times
\Big\{n_{_1}!n_{_2}!n_{_3}!n_{_4}!\Gamma({D\over2}+n_{_1})\Gamma({D\over2}+n_{_2})
\Gamma(2-{D\over2}+n_{_3})
\nonumber\\
&&\hspace{2.5cm}\times
\Gamma(2-{D\over2}+n_{_4})\Gamma({D\over2}-1-n_{_1}-n_{_3})
\Gamma({D\over2}-1-n_{_2}-n_{_4})\Big\}^{-1}\;.
\label{GKZ21-20-3}
\end{eqnarray}

\item The set of column indices $I_{_{21}}=[2,3,5,8,\cdots,14]$,
i.e. the implement $J_{_{21}}=[1,14]\setminus I_{_{21}}=[1,4,6,7]$.
The choice of the set of column indices implies the exponents
$\alpha_{_1}=\alpha_{_4}=\alpha_{_{6}}=\alpha_{_{7}}=0$, and
\begin{eqnarray}
&&\alpha_{_2}=a_{_1}-a_{_2},\;\alpha_{_3}=a_{_4}-a_{_3},\;\alpha_{_5}=a_{_1}-a_{_4}-a_{_5}-b_{_1}+1,
\nonumber\\
&&\alpha_{_8}=b_{_2}+b_{_3}-a_{_4}-2,\;\alpha_{_9}=b_{_1}+b_{_4}-a_{_1}-2,
\nonumber\\
&&\alpha_{_{10}}=b_{_1}+b_{_5}-a_{_1}-2,\;
\alpha_{_{11}}=1-b_{_1},\;\alpha_{_{12}}=1-b_{_2},
\nonumber\\
&&\alpha_{_{13}}=b_{_2}-a_{_4}-1,\;\alpha_{_{14}}=a_{_4}+b_{_1}-a_{_1}-1\;.
\label{GKZ21-21-1}
\end{eqnarray}
Using the integer lattice ${\bf B}_{_{1357}}$ and the exponents obtained above,
we write the hypergeometric series solution as
\begin{eqnarray}
&&\Phi_{_{[1357]}}^{(21)}(\alpha,z)=
y_{_1}^{{D\over2}-1}y_{_2}^{{D\over2}-1}y_{_3}^{{D\over2}-2}y_{_4}^{-{D\over2}}\sum\limits_{n_{_1}=0}^\infty
\sum\limits_{n_{_2}=0}^\infty\sum\limits_{n_{_3}=0}^\infty\sum\limits_{n_{_4}=0}^\infty
c_{_{[1357]}}^{(21)}(\alpha,{\bf n})
\nonumber\\
&&\hspace{2.5cm}\times
\Big({1\over y_{_4}}\Big)^{n_{_1}}\Big({y_{_4}\over y_{_3}}\Big)^{n_{_2}}
\Big({y_{_1}\over y_{_4}}\Big)^{n_{_3}}\Big({y_{_2}\over y_{_3}}\Big)^{n_{_4}}\;,
\label{GKZ21-21-2}
\end{eqnarray}
with
\begin{eqnarray}
&&c_{_{[1357]}}^{(21)}(\alpha,{\bf n})=
(-)^{n_{_4}}\Gamma(1+n_{_1}+n_{_3})\Gamma(1+n_{_2}+n_{_4})
\nonumber\\
&&\hspace{2.5cm}\times
\Big\{n_{_1}!n_{_2}!n_{_3}!n_{_4}!\Gamma({D\over2}+n_{_1})
\Gamma(2-{D\over2}+n_{_2})\Gamma(1-{D\over2}-n_{_1}-n_{_3})
\nonumber\\
&&\hspace{2.5cm}\times
\Gamma({D\over2}+n_{_3})\Gamma({D\over2}+n_{_4})
\Gamma({D\over2}-1-n_{_2}-n_{_4})\Big\}^{-1}\;.
\label{GKZ21-21-3}
\end{eqnarray}

\item The set of column indices $I_{_{22}}=[2,3,5,7,\cdots,11,13,14]$,
i.e. the implement $J_{_{22}}=[1,14]\setminus I_{_{22}}=[1,4,6,12]$.
The set of column indices gives the power numbers
$\alpha_{_1}=\alpha_{_4}=\alpha_{_{6}}=\alpha_{_{12}}=0$, and
\begin{eqnarray}
&&\alpha_{_2}=a_{_1}-a_{_2},\;\alpha_{_3}=a_{_4}-a_{_3},\;\alpha_{_5}=a_{_1}-a_{_4}-a_{_5}-b_{_1}+1,
\nonumber\\
&&\alpha_{_7}=b_{_2}-1,\;\alpha_{_8}=b_{_3}-a_{_4}-1,\;\alpha_{_9}=b_{_1}+b_{_4}-a_{_1}-2,
\nonumber\\
&&\alpha_{_{10}}=b_{_1}+b_{_5}-a_{_1}-2,\;
\alpha_{_{11}}=1-b_{_1},
\nonumber\\
&&\alpha_{_{13}}=-a_{_4},\;\alpha_{_{14}}=a_{_4}+b_{_1}-a_{_1}-1\;.
\label{GKZ21-22-1}
\end{eqnarray}
Using the integer lattice ${\bf B}_{_{1357}}$ and those exponents above,
one obtains the hypergeometric series as
\begin{eqnarray}
&&\Phi_{_{[1357]}}^{(22)}(\alpha,z)=
y_{_1}^{{D\over2}-1}y_{_3}^{{D}-3}y_{_4}^{-{D\over2}}\sum\limits_{n_{_1}=0}^\infty
\sum\limits_{n_{_2}=0}^\infty\sum\limits_{n_{_3}=0}^\infty\sum\limits_{n_{_4}=0}^\infty
c_{_{[1357]}}^{(22)}(\alpha,{\bf n})
\nonumber\\
&&\hspace{2.5cm}\times
\Big({1\over y_{_4}}\Big)^{n_{_1}}\Big({y_{_4}\over y_{_3}}\Big)^{n_{_2}}
\Big({y_{_1}\over y_{_4}}\Big)^{n_{_3}}\Big({y_{_2}\over y_{_3}}\Big)^{n_{_4}}\;,
\label{GKZ21-22-2}
\end{eqnarray}
with
\begin{eqnarray}
&&c_{_{[1357]}}^{(22)}(\alpha,{\bf n})=
(-)^{n_{_2}}\Gamma(1+n_{_1}+n_{_3})\Big\{n_{_1}!n_{_2}!n_{_3}!n_{_4}!
\nonumber\\
&&\hspace{2.5cm}\times
\Gamma({D\over2}+n_{_1})\Gamma(2-{D\over2}+n_{_2})
\Gamma(2-{D\over2}+n_{_4})\Gamma({D\over2}+n_{_3})
\nonumber\\
&&\hspace{2.5cm}\times
\Gamma({D\over2}-1-n_{_2}-n_{_4})\Gamma(1-{D\over2}-n_{_1}-n_{_3})
\Gamma(D-2-n_{_2}-n_{_4})\Big\}^{-1}\;.
\label{GKZ21-22-3}
\end{eqnarray}

\item The set of column indices $I_{_{23}}=[2,3,5,6,8,9,10,12,13,14]$,
i.e. the implement $J_{_{23}}=[1,14]\setminus I_{_{23}}=[1,4,7,11]$.
The set of column indices implies the power numbers
$\alpha_{_1}=\alpha_{_4}=\alpha_{_{7}}=\alpha_{_{11}}=0$, and
\begin{eqnarray}
&&\alpha_{_2}=a_{_1}-a_{_2},\;\alpha_{_3}=a_{_4}-a_{_3},\;\alpha_{_5}=a_{_1}-a_{_4}-a_{_5},
\nonumber\\
&&\alpha_{_{6}}=b_{_1}-1,\;\alpha_{_8}=b_{_2}+b_{_3}-a_{_4}-2,\;\alpha_{_9}=b_{_4}-a_{_1}-1,
\nonumber\\
&&\alpha_{_{10}}=b_{_5}-a_{_1}-1,\;
\alpha_{_{12}}=1-b_{_2},
\nonumber\\
&&\alpha_{_{13}}=b_{_2}-a_{_4}-1,\;\alpha_{_{14}}=a_{_4}-a_{_1}\;.
\label{GKZ21-23-1}
\end{eqnarray}
Basing on the integer lattice and the exponents above, we derive
the hypergeometric series solution as
\begin{eqnarray}
&&\Phi_{_{[1357]}}^{(23)}(\alpha,z)=
y_{_2}^{{D\over2}-1}y_{_3}^{{D\over2}-2}y_{_4}^{-1}\sum\limits_{n_{_1}=0}^\infty
\sum\limits_{n_{_2}=0}^\infty\sum\limits_{n_{_3}=0}^\infty\sum\limits_{n_{_4}=0}^\infty
c_{_{[1357]}}^{(23)}(\alpha,{\bf n})
\nonumber\\
&&\hspace{2.5cm}\times
\Big({1\over y_{_4}}\Big)^{n_{_1}}\Big({y_{_4}\over y_{_3}}\Big)^{n_{_2}}
\Big({y_{_1}\over y_{_4}}\Big)^{n_{_3}}\Big({y_{_2}\over y_{_3}}\Big)^{n_{_4}}\;,
\label{GKZ21-23-2}
\end{eqnarray}
with
\begin{eqnarray}
&&c_{_{[1357]}}^{(23)}(\alpha,{\bf n})=
(-)^{n_{_4}}\Gamma(1+n_{_1}+n_{_3})\Gamma(1+n_{_2}+n_{_4})
\nonumber\\
&&\hspace{2.5cm}\times
\Big\{n_{_1}!n_{_2}!n_{_3}!n_{_4}!
\Gamma({D\over2}+n_{_1})\Gamma(2-{D\over2}+n_{_2})\Gamma(2-{D\over2}+n_{_3})
\nonumber\\
&&\hspace{2.5cm}\times
\Gamma({D\over2}-1-n_{_1}-n_{_3})\Gamma({D\over2}+n_{_4})
\Gamma({D\over2}-1-n_{_2}-n_{_4})\Big\}^{-1}\;.
\label{GKZ21-23-3}
\end{eqnarray}

\item The set of column indices $I_{_{24}}=[2,3,5,\cdots,10,13,14]$,
i.e. the implement $J_{_{24}}=[1,14]\setminus I_{_{24}}=[1,4,11,12]$.
The choice of the set of column indices implies the power numbers
$\alpha_{_1}=\alpha_{_4}=\alpha_{_{11}}=\alpha_{_{12}}=0$, and
\begin{eqnarray}
&&\alpha_{_2}=a_{_1}-a_{_2},\;\alpha_{_3}=a_{_4}-a_{_3},\;\alpha_{_5}=a_{_1}-a_{_4}-a_{_5},
\nonumber\\
&&\alpha_{_6}=b_{_1}-1,\;\alpha_{_7}=b_{_2}-1,\;\alpha_{_8}=b_{_3}-a_{_4}-1,
\nonumber\\
&&\alpha_{_9}=b_{_4}-a_{_1}-1,\;\alpha_{_{10}}=b_{_5}-a_{_1}-1,\;
\nonumber\\
&&\alpha_{_{13}}=-a_{_4},\;\alpha_{_{14}}=a_{_4}-a_{_1}\;.
\label{GKZ21-24-1}
\end{eqnarray}
Thus the hypergeometric series solution is
\begin{eqnarray}
&&\Phi_{_{[1357]}}^{(24)}(\alpha,z)=
y_{_3}^{{D}-3}y_{_4}^{-1}\sum\limits_{n_{_1}=0}^\infty
\sum\limits_{n_{_2}=0}^\infty\sum\limits_{n_{_3}=0}^\infty\sum\limits_{n_{_4}=0}^\infty
c_{_{[1357]}}^{(24)}(\alpha,{\bf n})
\nonumber\\
&&\hspace{2.5cm}\times
\Big({1\over y_{_4}}\Big)^{n_{_1}}\Big({y_{_4}\over y_{_3}}\Big)^{n_{_2}}
\Big({y_{_1}\over y_{_4}}\Big)^{n_{_3}}\Big({y_{_2}\over y_{_3}}\Big)^{n_{_4}}\;,
\label{GKZ21-24-2}
\end{eqnarray}
with
\begin{eqnarray}
&&c_{_{[1357]}}^{(24)}(\alpha,{\bf n})=
(-)^{n_{_2}}\Gamma(1+n_{_1}+n_{_3})\Big\{n_{_1}!n_{_2}!n_{_3}!n_{_4}!\Gamma({D\over2}+n_{_1})
\nonumber\\
&&\hspace{2.5cm}\times
\Gamma(2-{D\over2}+n_{_2})\Gamma(2-{D\over2}+n_{_3})
\Gamma({D\over2}-1-n_{_2}-n_{_4})
\nonumber\\
&&\hspace{2.5cm}\times
\Gamma(2-{D\over2}+n_{_4})
\Gamma({D\over2}-1-n_{_1}-n_{_3})\Gamma(D-2-n_{_2}-n_{_4})\Big\}^{-1}\;.
\label{GKZ21-24-3}
\end{eqnarray}

\item The set of column indices $I_{_{25}}=[1,4,5,8,\cdots,14]$,
i.e. the implement $J_{_{25}}=[1,14]\setminus I_{_{25}}=[2,3,6,7]$.
The choice of column indices implies the exponents
$\alpha_{_2}=\alpha_{_3}=\alpha_{_{6}}=\alpha_{_{7}}=0$, and
\begin{eqnarray}
&&\alpha_{_1}=a_{_2}-a_{_1},\;\alpha_{_4}=a_{_3}-a_{_4},\;\alpha_{_5}=a_{_2}-a_{_3}-a_{_5}-b_{_1}+1,
\nonumber\\
&&\alpha_{_8}=b_{_2}+b_{_3}-a_{_3}-2,\;\alpha_{_9}=b_{_1}+b_{_4}-a_{_2}-2,
\nonumber\\
&&\alpha_{_{10}}=b_{_1}+b_{_5}-a_{_2}-2,\;
\alpha_{_{11}}=1-b_{_1},\;\alpha_{_{12}}=1-b_{_2},
\nonumber\\
&&\alpha_{_{13}}=b_{_2}-a_{_3}-1,\;\alpha_{_{14}}=a_{_3}+b_{_1}-a_{_2}-1\;.
\label{GKZ21-25-1}
\end{eqnarray}
Using the integer lattice ${\bf B}_{_{1357}}$ and those exponents,
we write the hypergeometric series solution as
\begin{eqnarray}
&&\Phi_{_{[1357]}}^{(25)}(\alpha,z)=
y_{_1}^{{D\over2}-1}y_{_2}^{{D\over2}-1}y_{_3}^{-1}y_{_4}^{{D\over2}-2}\sum\limits_{n_{_1}=0}^\infty
\sum\limits_{n_{_2}=0}^\infty\sum\limits_{n_{_3}=0}^\infty\sum\limits_{n_{_4}=0}^\infty
c_{_{[1357]}}^{(25)}(\alpha,{\bf n})
\nonumber\\
&&\hspace{2.5cm}\times
\Big({1\over y_{_4}}\Big)^{n_{_1}}\Big({y_{_4}\over y_{_3}}\Big)^{n_{_2}}
\Big({y_{_1}\over y_{_4}}\Big)^{n_{_3}}\Big({y_{_2}\over y_{_3}}\Big)^{n_{_4}}\;,
\label{GKZ21-25-2}
\end{eqnarray}
with
\begin{eqnarray}
&&c_{_{[1357]}}^{(25)}(\alpha,{\bf n})=
(-)^{1+n_{_4}}\Gamma(1+n_{_1}+n_{_3})\Gamma(1+n_{_2}+n_{_4})
\Big\{n_{_1}!n_{_2}!n_{_3}!n_{_4}!
\nonumber\\
&&\hspace{2.5cm}\times
\Gamma(2-{D\over2}+n_{_1})\Gamma({D\over2}+n_{_2})\Gamma(1-{D\over2}-n_{_2}-n_{_4})
\nonumber\\
&&\hspace{2.5cm}\times
\Gamma({D\over2}-1-n_{_1}-n_{_3})\Gamma({D\over2}+n_{_3})
\Gamma({D\over2}+n_{_4})\Big\}^{-1}\;.
\label{GKZ21-25-3}
\end{eqnarray}

\item The set of column indices $I_{_{26}}=[1,4,5,7,\cdots,11,13,14]$, i.e.
the implement $J_{_{26}}=[1,14]\setminus I_{_{26}}=[2,3,6,12]$.
The set of column indices induces the power numbers as
$\alpha_{_2}=\alpha_{_3}=\alpha_{_{6}}=\alpha_{_{12}}=0$, and
\begin{eqnarray}
&&\alpha_{_1}=a_{_2}-a_{_1},\;\alpha_{_4}=a_{_3}-a_{_4},\;\alpha_{_5}=a_{_2}-a_{_3}-a_{_5}-b_{_1}+1,
\nonumber\\
&&\alpha_{_7}=b_{_2}-1,\;\alpha_{_8}=b_{_3}-a_{_3}-1,\;\alpha_{_9}=b_{_1}+b_{_4}-a_{_2}-2,
\nonumber\\
&&\alpha_{_{10}}=b_{_1}+b_{_5}-a_{_2}-2,\;
\alpha_{_{11}}=1-b_{_1},
\nonumber\\
&&\alpha_{_{13}}=-a_{_3},\;\alpha_{_{14}}=a_{_3}+b_{_1}-a_{_2}-1\;.
\label{GKZ21-26-1}
\end{eqnarray}
Then the hypergeometric series solution is
\begin{eqnarray}
&&\Phi_{_{[1357]}}^{(26)}(\alpha,z)=
y_{_1}^{{D\over2}-1}y_{_3}^{{D\over2}-2}y_{_4}^{{D\over2}-2}\sum\limits_{n_{_1}=0}^\infty
\sum\limits_{n_{_2}=0}^\infty\sum\limits_{n_{_3}=0}^\infty\sum\limits_{n_{_4}=0}^\infty
c_{_{[1357]}}^{(26)}(\alpha,{\bf n})
\nonumber\\
&&\hspace{2.5cm}\times
\Big({1\over y_{_4}}\Big)^{n_{_1}}\Big({y_{_4}\over y_{_3}}\Big)^{n_{_2}}
\Big({y_{_1}\over y_{_4}}\Big)^{n_{_3}}\Big({y_{_2}\over y_{_3}}\Big)^{n_{_4}}\;,
\label{GKZ21-26-2}
\end{eqnarray}
with
\begin{eqnarray}
&&c_{_{[1357]}}^{(26)}(\alpha,{\bf n})=
(-)^{1+n_{_4}}\Gamma(1+n_{_1}+n_{_3})\Gamma(1+n_{_2}+n_{_4})
\nonumber\\
&&\hspace{2.5cm}\times
\Big\{n_{_1}!n_{_2}!n_{_3}!n_{_4}!\Gamma(2-{D\over2}+n_{_1})
\Gamma({D\over2}+n_{_2})\Gamma(2-{D\over2}+n_{_4})
\nonumber\\
&&\hspace{2.5cm}\times
\Gamma({D\over2}-1-n_{_1}-n_{_3})
\Gamma({D\over2}+n_{_3})\Gamma({D\over2}-1-n_{_2}-n_{_4})\Big\}^{-1}\;.
\label{GKZ21-26-3}
\end{eqnarray}

\item The set of column indices $I_{_{27}}=[1,4,5,6,8,9,10,12,13,14]$,
i.e. the implement $J_{_{27}}=[1,14]\setminus I_{_{27}}=[2,3,7,11]$.
The set of column indices implies the exponents as
$\alpha_{_2}=\alpha_{_3}=\alpha_{_{7}}=\alpha_{_{11}}=0$, and
\begin{eqnarray}
&&\alpha_{_1}=a_{_2}-a_{_1},\;\alpha_{_4}=a_{_3}-a_{_4},\;\alpha_{_5}=a_{_2}-a_{_3}-a_{_5},
\nonumber\\
&&\alpha_{_{6}}=b_{_1}-1,\;\alpha_{_8}=b_{_2}+b_{_3}-a_{_3}-2,\;\alpha_{_9}=b_{_4}-a_{_2}-1,
\nonumber\\
&&\alpha_{_{10}}=b_{_5}-a_{_2}-1,\;
\alpha_{_{12}}=1-b_{_2},
\nonumber\\
&&\alpha_{_{13}}=b_{_2}-a_{_3}-1,\;\alpha_{_{14}}=a_{_3}-a_{_2}\;.
\label{GKZ21-27-1}
\end{eqnarray}
Using the integer lattice ${\bf B}_{_{1357}}$ and the exponents, we
write the hypergeometric series solution as
\begin{eqnarray}
&&\Phi_{_{[1357]}}^{(27)}(\alpha,z)=
y_{_2}^{{D\over2}-1}y_{_3}^{-1}y_{_4}^{D-3}\sum\limits_{n_{_1}=0}^\infty
\sum\limits_{n_{_2}=0}^\infty\sum\limits_{n_{_3}=0}^\infty\sum\limits_{n_{_4}=0}^\infty
c_{_{[1357]}}^{(27)}(\alpha,{\bf n})
\nonumber\\
&&\hspace{2.5cm}\times
\Big({1\over y_{_4}}\Big)^{n_{_1}}\Big({y_{_4}\over y_{_3}}\Big)^{n_{_2}}
\Big({y_{_1}\over y_{_4}}\Big)^{n_{_3}}\Big({y_{_2}\over y_{_3}}\Big)^{n_{_4}}\;,
\label{GKZ21-27-2}
\end{eqnarray}
with
\begin{eqnarray}
&&c_{_{[1357]}}^{(27)}(\alpha,{\bf n})=
(-)^{n_{_1}+n_{_3}+n_{_4}}\Gamma(1+n_{_2}+n_{_4})\Big\{n_{_1}!n_{_2}!n_{_3}!n_{_4}!
\nonumber\\
&&\hspace{2.5cm}\times
\Gamma(2-{D\over2}+n_{_1})\Gamma({D\over2}+n_{_2})\Gamma(2-{D\over2}+n_{_3})
\nonumber\\
&&\hspace{2.5cm}\times
\Gamma(1-{D\over2}-n_{_2}-n_{_4})\Gamma(D-2-n_{_1}-n_{_3})
\nonumber\\
&&\hspace{2.5cm}\times
\Gamma({D\over2}-1-n_{_1}-n_{_3})\Gamma({D\over2}+n_{_4})\Big\}^{-1}\;.
\label{GKZ21-27-3}
\end{eqnarray}

\item The set of column indices $I_{_{28}}=[1,4,\cdots,10,13,14]$,
i.e. the implement $J_{_{28}}=[1,14]\setminus I_{_{28}}=[2,3,11,12]$.
The choice of column indices gives the power numbers as
$\alpha_{_2}=\alpha_{_3}=\alpha_{_{11}}=\alpha_{_{12}}=0$, and
\begin{eqnarray}
&&\alpha_{_1}=a_{_2}-a_{_1},\;\alpha_{_4}=a_{_3}-a_{_4},\;\alpha_{_5}=a_{_2}-a_{_3}-a_{_5},
\nonumber\\
&&\alpha_{_6}=b_{_1}-1,\;\alpha_{_7}=b_{_2}-1,\;\alpha_{_8}=b_{_3}-a_{_3}-1,
\nonumber\\
&&\alpha_{_9}=b_{_4}-a_{_2}-1,\;\alpha_{_{10}}=b_{_5}-a_{_2}-1,\;
\nonumber\\
&&\alpha_{_{13}}=-a_{_3},\;\alpha_{_{14}}=a_{_3}-a_{_2}\;.
\label{GKZ21-28-1}
\end{eqnarray}
Basing on the integer lattice and those exponent numbers, we
write the hypergeometric series solution as
\begin{eqnarray}
&&\Phi_{_{[1357]}}^{(28)}(\alpha,z)=
y_{_3}^{{D\over2}-2}y_{_4}^{{D}-3}\sum\limits_{n_{_1}=0}^\infty
\sum\limits_{n_{_2}=0}^\infty\sum\limits_{n_{_3}=0}^\infty\sum\limits_{n_{_4}=0}^\infty
c_{_{[1357]}}^{(28)}(\alpha,{\bf n})
\nonumber\\
&&\hspace{2.5cm}\times
\Big({1\over y_{_4}}\Big)^{n_{_1}}\Big({y_{_4}\over y_{_3}}\Big)^{n_{_2}}
\Big({y_{_1}\over y_{_4}}\Big)^{n_{_3}}\Big({y_{_2}\over y_{_3}}\Big)^{n_{_4}}\;,
\label{GKZ21-28-2}
\end{eqnarray}
with
\begin{eqnarray}
&&c_{_{[1357]}}^{(28)}(\alpha,{\bf n})=
(-)^{n_{_1}+n_{_3}+n_{_4}}\Gamma(1+n_{_2}+n_{_4})\Big\{n_{_1}!n_{_2}!n_{_3}!n_{_4}!
\nonumber\\
&&\hspace{2.5cm}\times
\Gamma(2-{D\over2}+n_{_1})\Gamma({D\over2}+n_{_2})
\Gamma(2-{D\over2}+n_{_3})\Gamma(2-{D\over2}+n_{_4})
\nonumber\\
&&\hspace{2.5cm}\times
\Gamma(D-2-n_{_1}-n_{_3})\Gamma({D\over2}-1-n_{_1}-n_{_3})
\Gamma({D\over2}-1-n_{_2}-n_{_4})\Big\}^{-1}\;.
\label{GKZ21-28-3}
\end{eqnarray}

\item The set of column indices $I_{_{29}}=[1,3,5,8,\cdots,14]$,
i.e. the implement $J_{_{29}}=[1,14]\setminus I_{_{29}}=[2,4,6,7]$.
The set of column indices implies the power numbers as
$\alpha_{_2}=\alpha_{_4}=\alpha_{_{6}}=\alpha_{_{7}}=0$, and
\begin{eqnarray}
&&\alpha_{_1}=a_{_2}-a_{_1},\;\alpha_{_3}=a_{_4}-a_{_3},\;\alpha_{_5}=a_{_2}-a_{_4}-a_{_5}-b_{_1}+1,
\nonumber\\
&&\alpha_{_8}=b_{_2}+b_{_3}-a_{_4}-2,\;\alpha_{_9}=b_{_1}+b_{_4}-a_{_2}-2,
\nonumber\\
&&\alpha_{_{10}}=b_{_1}+b_{_5}-a_{_2}-2,\;
\alpha_{_{11}}=1-b_{_1},\;\alpha_{_{12}}=1-b_{_2},
\nonumber\\
&&\alpha_{_{13}}=b_{_2}-a_{_4}-1,\;\alpha_{_{14}}=a_{_4}+b_{_1}-a_{_2}-1\;.
\label{GKZ21-29-1}
\end{eqnarray}
With the integer lattice ${\bf B}_{_{1357}}$ and those power number above, one derives
the hypergeometric series solution as
\begin{eqnarray}
&&\Phi_{_{[1357]}}^{(29)}(\alpha,z)=
y_{_1}^{{D\over2}-1}y_{_2}^{{D\over2}-1}y_{_3}^{{D\over2}-2}y_{_4}^{-1}\sum\limits_{n_{_1}=0}^\infty
\sum\limits_{n_{_2}=0}^\infty\sum\limits_{n_{_3}=0}^\infty\sum\limits_{n_{_4}=0}^\infty
c_{_{[1357]}}^{(29)}(\alpha,{\bf n})
\nonumber\\
&&\hspace{2.5cm}\times
\Big({1\over y_{_4}}\Big)^{n_{_1}}\Big({y_{_4}\over y_{_3}}\Big)^{n_{_2}}
\Big({y_{_1}\over y_{_4}}\Big)^{n_{_3}}\Big({y_{_2}\over y_{_3}}\Big)^{n_{_4}}\;,
\label{GKZ21-29-2}
\end{eqnarray}
with
\begin{eqnarray}
&&c_{_{[1357]}}^{(29)}(\alpha,{\bf n})=
(-)^{n_{_4}}\Gamma(1+n_{_1}+n_{_3})\Gamma(1+n_{_2}+n_{_4})
\Big\{n_{_1}!n_{_2}!n_{_3}!n_{_4}!
\nonumber\\
&&\hspace{2.5cm}\times
\Gamma(2-{D\over2}+n_{_1})\Gamma(2-{D\over2}+n_{_2})
\Gamma({D\over2}-1-n_{_1}-n_{_3})
\nonumber\\
&&\hspace{2.5cm}\times
\Gamma({D\over2}+n_{_3})
\Gamma({D\over2}+n_{_4})\Gamma({D\over2}-1-n_{_2}-n_{_4})\Big\}^{-1}\;.
\label{GKZ21-29-3}
\end{eqnarray}

\item The set of column indices $I_{_{30}}=[1,3,5,7,\cdots,11,13,14]$,
i.e. the implement $J_{_{30}}=[1,14]\setminus I_{_{30}}=[2,4,6,12]$.
The choice of the set of column indices implies the power numbers as
$\alpha_{_2}=\alpha_{_4}=\alpha_{_{6}}=\alpha_{_{12}}=0$, and
\begin{eqnarray}
&&\alpha_{_1}=a_{_2}-a_{_1},\;\alpha_{_3}=a_{_4}-a_{_3},\;\alpha_{_5}=a_{_2}-a_{_4}-a_{_5}-b_{_1}+1,
\nonumber\\
&&\alpha_{_7}=b_{_2}-1,\;\alpha_{_8}=b_{_3}-a_{_4}-1,\;\alpha_{_9}=b_{_1}+b_{_4}-a_{_2}-2,
\nonumber\\
&&\alpha_{_{10}}=b_{_1}+b_{_5}-a_{_2}-2,\;
\alpha_{_{11}}=1-b_{_1},
\nonumber\\
&&\alpha_{_{13}}=-a_{_4},\;\alpha_{_{14}}=a_{_4}+b_{_1}-a_{_2}-1\;.
\label{GKZ21-30-1}
\end{eqnarray}
Similarly the hypergeometric series solution is
\begin{eqnarray}
&&\Phi_{_{[1357]}}^{(30)}(\alpha,z)=
y_{_1}^{{D\over2}-1}y_{_3}^{{D}-3}y_{_4}^{-1}\sum\limits_{n_{_1}=0}^\infty
\sum\limits_{n_{_2}=0}^\infty\sum\limits_{n_{_3}=0}^\infty\sum\limits_{n_{_4}=0}^\infty
c_{_{[1357]}}^{(30)}(\alpha,{\bf n})
\nonumber\\
&&\hspace{2.5cm}\times
\Big({1\over y_{_4}}\Big)^{n_{_1}}\Big({y_{_4}\over y_{_3}}\Big)^{n_{_2}}
\Big({y_{_1}\over y_{_4}}\Big)^{n_{_3}}\Big({y_{_2}\over y_{_3}}\Big)^{n_{_4}}\;,
\label{GKZ21-30-2}
\end{eqnarray}
with
\begin{eqnarray}
&&c_{_{[1357]}}^{(30)}(\alpha,{\bf n})=
(-)^{n_{_2}}\Gamma(1+n_{_1}+n_{_3})\Big\{n_{_1}!n_{_2}!n_{_3}!n_{_4}!
\Gamma(2-{D\over2}+n_{_1})
\nonumber\\
&&\hspace{2.5cm}\times
\Gamma(2-{D\over2}+n_{_2})\Gamma(2-{D\over2}+n_{_4})
\Gamma({D\over2}-1-n_{_2}-n_{_4})
\nonumber\\
&&\hspace{2.5cm}\times
\Gamma({D\over2}-1-n_{_1}-n_{_3})\Gamma({D\over2}+n_{_3})
\Gamma(D-2-n_{_2}-n_{_4})\Big\}^{-1}\;.
\label{GKZ21-30-3}
\end{eqnarray}

\item The set of column indices $I_{_{31}}=[1,3,5,6,8,9,10,12,13,14]$,
i.e. the implement $J_{_{31}}=[1,14]\setminus I_{_{31}}=[2,4,7,11]$.
The choice of column indices gives the power numbers as
$\alpha_{_2}=\alpha_{_4}=\alpha_{_{7}}=\alpha_{_{11}}=0$, and
\begin{eqnarray}
&&\alpha_{_1}=a_{_2}-a_{_1},\;\alpha_{_3}=a_{_4}-a_{_3},\;\alpha_{_5}=a_{_2}-a_{_4}-a_{_5},
\nonumber\\
&&\alpha_{_{6}}=b_{_1}-1,\;\alpha_{_8}=b_{_2}+b_{_3}-a_{_4}-2,\;\alpha_{_9}=b_{_4}-a_{_2}-1,
\nonumber\\
&&\alpha_{_{10}}=b_{_5}-a_{_2}-1,\;
\alpha_{_{12}}=1-b_{_2},
\nonumber\\
&&\alpha_{_{13}}=b_{_2}-a_{_4}-1,\;\alpha_{_{14}}=a_{_4}-a_{_2}\;.
\label{GKZ21-31-1}
\end{eqnarray}
Using the integer lattice ${\bf B}_{_{1357}}$ and those exponents,
we formulate the hypergeometric series solution as
\begin{eqnarray}
&&\Phi_{_{[1357]}}^{(31)}(\alpha,z)=
y_{_2}^{{D\over2}-1}y_{_3}^{{D\over2}-2}y_{_4}^{{D\over2}-2}\sum\limits_{n_{_1}=0}^\infty
\sum\limits_{n_{_2}=0}^\infty\sum\limits_{n_{_3}=0}^\infty\sum\limits_{n_{_4}=0}^\infty
c_{_{[1357]}}^{(31)}(\alpha,{\bf n})
\nonumber\\
&&\hspace{2.5cm}\times
\Big({1\over y_{_4}}\Big)^{n_{_1}}\Big({y_{_4}\over y_{_3}}\Big)^{n_{_2}}
\Big({y_{_1}\over y_{_4}}\Big)^{n_{_3}}\Big({y_{_2}\over y_{_3}}\Big)^{n_{_4}}\;,
\label{GKZ21-31-2}
\end{eqnarray}
with
\begin{eqnarray}
&&c_{_{[1357]}}^{(31)}(\alpha,{\bf n})=
(-)^{1+n_{_1}+n_{_3}+n_{_4}}\Gamma(1+n_{_2}+n_{_4})\Big\{n_{_1}!n_{_2}!n_{_3}!n_{_4}!
\nonumber\\
&&\hspace{2.5cm}\times
\Gamma(2-{D\over2}+n_{_1})\Gamma(2-{D\over2}+n_{_2})\Gamma(2-{D\over2}+n_{_3})
\Gamma(D-2-n_{_1}-n_{_3})
\nonumber\\
&&\hspace{2.5cm}\times
\Gamma({D\over2}-1-n_{_1}-n_{_3})\Gamma({D\over2}+n_{_4})
\Gamma({D\over2}-1-n_{_2}-n_{_4})\Big\}^{-1}\;.
\label{GKZ21-31-3}
\end{eqnarray}

\item The set of column indices $I_{_{32}}=[1,3,5,\cdots,10,13,14]$,
i.e. the implement $J_{_{32}}=[1,14]\setminus I_{_{32}}=[2,4,11,12]$.
The set of column indices implies the power numbers as
$\alpha_{_2}=\alpha_{_4}=\alpha_{_{11}}=\alpha_{_{12}}=0$, and
\begin{eqnarray}
&&\alpha_{_1}=a_{_2}-a_{_1},\;\alpha_{_3}=a_{_4}-a_{_3},\;\alpha_{_5}=a_{_2}-a_{_4}-a_{_5},
\nonumber\\
&&\alpha_{_6}=b_{_1}-1,\;\alpha_{_7}=b_{_2}-1,\;\alpha_{_8}=b_{_3}-a_{_4}-1,
\nonumber\\
&&\alpha_{_9}=b_{_4}-a_{_2}-1,\;\alpha_{_{10}}=b_{_5}-a_{_2}-1,\;
\nonumber\\
&&\alpha_{_{13}}=-a_{_4},\;\alpha_{_{14}}=a_{_4}-a_{_2}\;.
\label{GKZ21-32-1}
\end{eqnarray}
In a similar way, we formulate the hypergeometric series solution as
\begin{eqnarray}
&&\Phi_{_{[1357]}}^{(32)}(\alpha,z)=
y_{_3}^{{D}-3}y_{_4}^{{D\over2}-2}\sum\limits_{n_{_1}=0}^\infty
\sum\limits_{n_{_2}=0}^\infty\sum\limits_{n_{_3}=0}^\infty\sum\limits_{n_{_4}=0}^\infty
c_{_{[1357]}}^{(32)}(\alpha,{\bf n})
\nonumber\\
&&\hspace{2.5cm}\times
\Big({1\over y_{_4}}\Big)^{n_{_1}}\Big({y_{_4}\over y_{_3}}\Big)^{n_{_2}}
\Big({y_{_1}\over y_{_4}}\Big)^{n_{_3}}\Big({y_{_2}\over y_{_3}}\Big)^{n_{_4}}\;,
\label{GKZ21-32-2}
\end{eqnarray}
with
\begin{eqnarray}
&&c_{_{[1357]}}^{(32)}(\alpha,{\bf n})=
(-)^{1+n_{_1}+n_{_2}+n_{_3}}
\Big\{n_{_1}!n_{_2}!n_{_3}!n_{_4}!\Gamma(2-{D\over2}+n_{_1})
\nonumber\\
&&\hspace{2.5cm}\times
\Gamma(2-{D\over2}+n_{_2})\Gamma(2-{D\over2}+n_{_3})
\Gamma(2-{D\over2}+n_{_4})
\nonumber\\
&&\hspace{2.5cm}\times
\Gamma({D\over2}-1-n_{_2}-n_{_4})\Gamma(D-2-n_{_1}-n_{_3})
\nonumber\\
&&\hspace{2.5cm}\times
\Gamma({D\over2}-1-n_{_1}-n_{_3})\Gamma(D-2-n_{_2}-n_{_4})\Big\}^{-1}\;.
\label{GKZ21-32-3}
\end{eqnarray}
\end{itemize}

\section{The hypergeometric solutions of the integer lattice ${\bf B}_{_{\tilde{1}357}}$\label{app2}}
\indent\indent

\begin{itemize}
\item $I_{_{1}}=[1,2,4,8,\cdots,14]$, i.e. the implement $J_{_{1}}=[1,14]\setminus I_{_{1}}=[3,5,6,7]$.
The choice implies the power numbers $\alpha_{_3}=\alpha_{_5}=\alpha_{_{6}}=\alpha_{_{7}}=0$, and
\begin{eqnarray}
&&\alpha_{_1}=a_{_3}+a_{_5}+b_{_1}-a_{_1}-1,\;\alpha_{_2}=a_{_3}+a_{_5}+b_{_1}-a_{_2}-1,
\nonumber\\
&&\alpha_{_4}=a_{_3}-a_{_4},\;\alpha_{_8}=b_{_2}+b_{_3}-a_{_3}-2,\;
\nonumber\\
&&\alpha_{_9}=b_{_4}-a_{_3}-a_{_5}-1,\;\alpha_{_{10}}=b_{_5}-a_{_3}-a_{_5}-1,\;
\nonumber\\
&&\alpha_{_{11}}=1-b_{_1},\;\alpha_{_{12}}=1-b_{_2},
\nonumber\\
&&\alpha_{_{13}}=b_{_2}-a_{_3}-1,\;\alpha_{_{14}}=-a_{_5}\;.
\label{GKZ21a-1-1}
\end{eqnarray}
The corresponding hypergeometric series solutions are written as
\begin{eqnarray}
&&\Phi_{_{[\tilde{1}357]}}^{(1),a}(\alpha,z)=
y_{_1}^{{D\over2}-1}y_{_2}^{{D\over2}-1}y_{_3}^{-1}y_{_4}^{-1}\sum\limits_{n_{_1}=0}^\infty
\sum\limits_{n_{_2}=0}^\infty\sum\limits_{n_{_3}=0}^\infty\sum\limits_{n_{_4}=0}^\infty
c_{_{[\tilde{1}357]}}^{(1),a}(\alpha,{\bf n})
\nonumber\\
&&\hspace{2.5cm}\times
\Big({y_{_1}\over y_{_4}}\Big)^{n_{_1}}\Big({y_{_4}\over y_{_3}}\Big)^{n_{_2}}
\Big({1\over y_{_4}}\Big)^{n_{_3}}\Big({y_{_2}\over y_{_3}}\Big)^{n_{_4}}
\;,\nonumber\\
&&\Phi_{_{[\tilde{1}357]}}^{(1),b}(\alpha,z)=
y_{_1}^{{D\over2}}y_{_2}^{{D\over2}-1}y_{_3}^{-1}y_{_4}^{-1}\sum\limits_{n_{_1}=0}^\infty
\sum\limits_{n_{_2}=0}^\infty\sum\limits_{n_{_3}=0}^\infty\sum\limits_{n_{_4}=0}^\infty
c_{_{[\tilde{1}357]}}^{(1),b}(\alpha,{\bf n})
\nonumber\\
&&\hspace{2.5cm}\times
y_{_1}^{n_{_1}}\Big({y_{_1}\over y_{_3}}\Big)^{n_{_2}}
\Big({y_{_1}\over y_{_4}}\Big)^{n_{_3}}\Big({y_{_2}\over y_{_3}}\Big)^{n_{_4}}\;.
\label{GKZ21a-1-2a}
\end{eqnarray}
Where the coefficients are
\begin{eqnarray}
&&c_{_{[\tilde{1}357]}}^{(1),a}(\alpha,{\bf n})=
(-)^{n_{_4}}\Gamma(1+n_{_1}+n_{_3})\Gamma(1+n_{_2}+n_{_4})
\Big\{n_{_1}!n_{_2}!n_{_3}!n_{_4}!
\nonumber\\
&&\hspace{2.5cm}\times
\Gamma({D\over2}+n_{_3})\Gamma({D\over2}+n_{_2})
\Gamma(1-{D\over2}-n_{_2}-n_{_4})
\nonumber\\
&&\hspace{2.5cm}\times
\Gamma(1-{D\over2}-n_{_1}-n_{_3})\Gamma({D\over2}+n_{_1})
\Gamma({D\over2}+n_{_4})\Big\}^{-1}
\;,\nonumber\\
&&c_{_{[\tilde{1}357]}}^{(1),b}(\alpha,{\bf n})=
(-)^{n_{_1}+n_{_4}}\Gamma(1+n_{_1})\Gamma(1+n_{_2}+n_{_3})\Gamma(1+n_{_2}+n_{_4})
\nonumber\\
&&\hspace{2.5cm}\times
\Big\{n_{_2}!n_{_4}!\Gamma(2+n_{_1}+n_{_2}+n_{_3})
\Gamma({D\over2}-1-n_{_1})\Gamma({D\over2}+n_{_2})
\nonumber\\
&&\hspace{2.5cm}\times
\Gamma(1-{D\over2}-n_{_2}-n_{_4})\Gamma(1-{D\over2}-n_{_2}-n_{_3})
\nonumber\\
&&\hspace{2.5cm}\times
\Gamma({D\over2}+1+n_{_1}+n_{_2}+n_{_3})\Gamma({D\over2}+n_{_4})\Big\}^{-1}\;.
\label{GKZ21a-1-3}
\end{eqnarray}

\item $I_{_{2}}=[1,2,4,7,\cdots,11,13,14]$, i.e. the implement $J_{_{2}}=[1,14]\setminus I_{_{2}}=[3,5,6,12]$.
The choice implies the power numbers $\alpha_{_3}=\alpha_{_5}=\alpha_{_{6}}=\alpha_{_{12}}=0$, and
\begin{eqnarray}
&&\alpha_{_1}=a_{_3}+a_{_5}+b_{_1}-a_{_1}-1,\;\alpha_{_2}=a_{_3}+a_{_5}+b_{_1}-a_{_2}-1,
\nonumber\\
&&\alpha_{_4}=a_{_3}-a_{_4},\;\alpha_{_7}=b_{_2}-1,\;\alpha_{_8}=b_{_3}-a_{_3}-1,\;
\nonumber\\
&&\alpha_{_9}=b_{_4}-a_{_3}-a_{_5}-1,\;\alpha_{_{10}}=b_{_5}-a_{_3}-a_{_5}-1,\;
\nonumber\\
&&\alpha_{_{11}}=1-b_{_1},\;\alpha_{_{13}}=-a_{_3},\;\alpha_{_{14}}=-a_{_5}\;.
\label{GKZ21a-2-1}
\end{eqnarray}
The corresponding hypergeometric series solutions are
\begin{eqnarray}
&&\Phi_{_{[\tilde{1}357]}}^{(2),a}(\alpha,z)=
y_{_1}^{{D\over2}-1}y_{_3}^{{D\over2}-2}y_{_4}^{-1}\sum\limits_{n_{_1}=0}^\infty
\sum\limits_{n_{_2}=0}^\infty\sum\limits_{n_{_3}=0}^\infty\sum\limits_{n_{_4}=0}^\infty
c_{_{[\tilde{1}357]}}^{(2),a}(\alpha,{\bf n})
\nonumber\\
&&\hspace{2.5cm}\times
\Big({y_{_1}\over y_{_4}}\Big)^{n_{_1}}\Big({y_{_4}\over y_{_3}}\Big)^{n_{_2}}
\Big({1\over y_{_4}}\Big)^{n_{_3}}\Big({y_{_2}\over y_{_3}}\Big)^{n_{_4}}
\;,\nonumber\\
&&\Phi_{_{[\tilde{1}357]}}^{(2),b}(\alpha,z)=
y_{_1}^{{D\over2}}y_{_3}^{{D\over2}-2}y_{_4}^{-1}\sum\limits_{n_{_1}=0}^\infty
\sum\limits_{n_{_2}=0}^\infty\sum\limits_{n_{_3}=0}^\infty\sum\limits_{n_{_4}=0}^\infty
c_{_{[\tilde{1}357]}}^{(2),a}(\alpha,{\bf n})
\nonumber\\
&&\hspace{2.5cm}\times
y_{_1}^{n_{_1}}\Big({y_{_1}\over y_{_3}}\Big)^{n_{_2}}
\Big({y_{_1}\over y_{_4}}\Big)^{n_{_3}}\Big({y_{_2}\over y_{_3}}\Big)^{n_{_4}}\;.
\label{GKZ21a-2-2a}
\end{eqnarray}
Where the coefficients are
\begin{eqnarray}
&&c_{_{[\tilde{1}357]}}^{(2),a}(\alpha,{\bf n})=
(-)^{n_{_4}}\Gamma(1+n_{_1}+n_{_3})\Gamma(1+n_{_2}+n_{_4})
\Big\{n_{_1}!n_{_2}!n_{_3}!n_{_4}!
\nonumber\\
&&\hspace{2.5cm}\times
\Gamma({D\over2}+n_{_3})\Gamma({D\over2}+n_{_2})
\Gamma({D\over2}-1-n_{_2}-n_{_4})
\nonumber\\
&&\hspace{2.5cm}\times
\Gamma(1-{D\over2}-n_{_1}-n_{_3})\Gamma({D\over2}+n_{_1})
\Gamma(2-{D\over2}+n_{_4})\Big\}^{-1}
\;,\nonumber\\
&&c_{_{[\tilde{1}357]}}^{(2),b}(\alpha,{\bf n})=
(-)^{n_{_1}+n_{_4}}\Gamma(1+n_{_1})\Gamma(1+n_{_2}+n_{_3})\Gamma(1+n_{_2}+n_{_4})
\nonumber\\
&&\hspace{2.5cm}\times
\Big\{n_{_2}!n_{_4}!\Gamma(2+n_{_1}+n_{_2}+n_{_3})
\Gamma({D\over2}-1-n_{_1})\Gamma({D\over2}+n_{_2})
\nonumber\\
&&\hspace{2.5cm}\times
\Gamma({D\over2}-1-n_{_2}-n_{_4})\Gamma(1-{D\over2}-n_{_2}-n_{_3})
\nonumber\\
&&\hspace{2.5cm}\times
\Gamma({D\over2}+1+n_{_1}+n_{_2}+n_{_3})\Gamma(2-{D\over2}+n_{_4})\Big\}^{-1}\;.
\label{GKZ21a-2-3}
\end{eqnarray}

\item $I_{_{3}}=[1,2,3,8,\cdots,14]$, i.e. the implement $J_{_{3}}=[1,14]\setminus I_{_{3}}=[4,5,6,7]$.
The choice implies the power numbers $\alpha_{_4}=\alpha_{_5}=\alpha_{_{6}}=\alpha_{_{7}}=0$, and
\begin{eqnarray}
&&\alpha_{_1}=a_{_4}+a_{_5}+b_{_1}-a_{_1}-1,\;\alpha_{_2}=a_{_4}+a_{_5}+b_{_1}-a_{_2}-1,
\nonumber\\
&&\alpha_{_3}=a_{_4}-a_{_3},\;\alpha_{_8}=b_{_2}+b_{_3}-a_{_4}-2,\;
\nonumber\\
&&\alpha_{_9}=b_{_4}-a_{_4}-a_{_5}-1,\;\alpha_{_{10}}=b_{_5}-a_{_4}-a_{_5}-1,\;
\nonumber\\
&&\alpha_{_{11}}=1-b_{_1},\;\alpha_{_{12}}=1-b_{_2},\;\alpha_{_{13}}=b_{_2}-a_{_4}-1,\;\alpha_{_{14}}=-a_{_5}\;.
\label{GKZ21a-3-1}
\end{eqnarray}
The corresponding hypergeometric series solutions are written as
\begin{eqnarray}
&&\Phi_{_{[\tilde{1}357]}}^{(3),a}(\alpha,z)=
y_{_1}^{{D\over2}-1}y_{_2}^{{D\over2}-1}y_{_3}^{{D\over2}-2}y_{_4}^{-1}\sum\limits_{n_{_1}=0}^\infty
\sum\limits_{n_{_2}=0}^\infty\sum\limits_{n_{_3}=0}^\infty\sum\limits_{n_{_4}=0}^\infty
c_{_{[\tilde{1}357]}}^{(3),a}(\alpha,{\bf n})
\nonumber\\
&&\hspace{2.5cm}\times
\Big({y_{_1}\over y_{_4}}\Big)^{n_{_1}}\Big({y_{_4}\over y_{_3}}\Big)^{n_{_2}}
\Big({1\over y_{_4}}\Big)^{n_{_3}}\Big({y_{_2}\over y_{_3}}\Big)^{n_{_4}}
\;,\nonumber\\
&&\Phi_{_{[\tilde{1}357]}}^{(3),b}(\alpha,z)=
y_{_1}^{{D\over2}}y_{_2}^{{D\over2}-1}y_{_3}^{{D\over2}-2}y_{_4}^{-1}\sum\limits_{n_{_1}=0}^\infty
\sum\limits_{n_{_2}=0}^\infty\sum\limits_{n_{_3}=0}^\infty\sum\limits_{n_{_4}=0}^\infty
c_{_{[\tilde{1}357]}}^{(3),b}(\alpha,{\bf n})
\nonumber\\
&&\hspace{2.5cm}\times
y_{_1}^{n_{_1}}\Big({y_{_1}\over y_{_3}}\Big)^{n_{_2}}
\Big({y_{_1}\over y_{_4}}\Big)^{n_{_3}}\Big({y_{_2}\over y_{_3}}\Big)^{n_{_4}}\;.
\label{GKZ21a-3-2a}
\end{eqnarray}
Where the coefficients are
\begin{eqnarray}
&&c_{_{[\tilde{1}357]}}^{(3),a}(\alpha,{\bf n})=
(-)^{n_{_4}}\Gamma(1+n_{_1}+n_{_3})\Gamma(1+n_{_2}+n_{_4})\Big\{n_{_1}!n_{_2}!n_{_3}!n_{_4}!
\nonumber\\
&&\hspace{2.5cm}\times
\Gamma(2-{D\over2}+n_{_3})
\Gamma(2-{D\over2}+n_{_2})\Gamma({D\over2}+n_{_4})
\nonumber\\
&&\hspace{2.5cm}\times
\Gamma({D\over2}-1-n_{_2}-n_{_4})\Gamma({D\over2}-1-n_{_1}-n_{_3})
\Gamma({D\over2}+n_{_1})\Big\}^{-1}
\;,\nonumber\\
&&c_{_{[\tilde{1}357]}}^{(3),b}(\alpha,{\bf n})=
(-)^{n_{_1}+n_{_4}}\Gamma(1+n_{_1})\Gamma(1+n_{_2}+n_{_3})\Gamma(1+n_{_2}+n_{_3})
\Big\{n_{_2}!n_{_4}!
\nonumber\\
&&\hspace{2.5cm}\times
\Gamma(2+n_{_1}+n_{_2}+n_{_3})\Gamma(1-{D\over2}-n_{_1})\Gamma(2-{D\over2}+n_{_2})
\nonumber\\
&&\hspace{2.5cm}\times
\Gamma({D\over2}+n_{_4})
\Gamma({D\over2}-1-n_{_2}-n_{_4})\Gamma({D\over2}-1-n_{_2}-n_{_3})
\nonumber\\
&&\hspace{2.5cm}\times
\Gamma({D\over2}+1+n_{_1}+n_{_2}+n_{_3})\Big\}^{-1}\;.
\label{GKZ21a-3-3}
\end{eqnarray}

\item $I_{_{4}}=[1,2,3,7,\cdots,11,13,14]$, i.e. the implement $J_{_{4}}=[1,14]\setminus I_{_{4}}=[4,5,6,12]$.
The choice implies the power numbers $\alpha_{_4}=\alpha_{_5}=\alpha_{_{6}}=\alpha_{_{12}}=0$, and
\begin{eqnarray}
&&\alpha_{_1}=a_{_4}+a_{_5}+b_{_1}-a_{_1}-1,\;\alpha_{_2}=a_{_4}+a_{_5}+b_{_1}-a_{_2}-1,
\nonumber\\
&&\alpha_{_3}=a_{_4}-a_{_3},\;\alpha_{_{7}}=b_{_2}-1,\;\alpha_{_8}=b_{_3}-a_{_4}-1,\;
\nonumber\\
&&\alpha_{_9}=b_{_4}-a_{_4}-a_{_5}-1,\;\alpha_{_{10}}=b_{_5}-a_{_4}-a_{_5}-1,\;
\nonumber\\
&&\alpha_{_{11}}=1-b_{_1},\;\alpha_{_{13}}=-a_{_4},\;\alpha_{_{14}}=-a_{_5}\;.
\label{GKZ21a-4-1}
\end{eqnarray}
The corresponding hypergeometric series solutions are
\begin{eqnarray}
&&\Phi_{_{[\tilde{1}357]}}^{(4),a}(\alpha,z)=
y_{_1}^{{D\over2}-1}y_{_3}^{{D}-3}y_{_4}^{-1}\sum\limits_{n_{_1}=0}^\infty
\sum\limits_{n_{_2}=0}^\infty\sum\limits_{n_{_3}=0}^\infty\sum\limits_{n_{_4}=0}^\infty
c_{_{[\tilde{1}357]}}^{(4),a}(\alpha,{\bf n})
\nonumber\\
&&\hspace{2.5cm}\times
\Big({y_{_1}\over y_{_4}}\Big)^{n_{_1}}\Big({y_{_4}\over y_{_3}}\Big)^{n_{_2}}
\Big({1\over y_{_4}}\Big)^{n_{_3}}\Big({y_{_2}\over y_{_3}}\Big)^{n_{_4}}
\;,\nonumber\\
&&\Phi_{_{[\tilde{1}357]}}^{(4),b}(\alpha,z)=
y_{_1}^{{D\over2}}y_{_3}^{{D}-3}y_{_4}^{-1}\sum\limits_{n_{_1}=0}^\infty
\sum\limits_{n_{_2}=0}^\infty\sum\limits_{n_{_3}=0}^\infty\sum\limits_{n_{_4}=0}^\infty
c_{_{[\tilde{1}357]}}^{(4),b}(\alpha,{\bf n})
\nonumber\\
&&\hspace{2.5cm}\times
y_{_1}^{n_{_1}}\Big({y_{_1}\over y_{_3}}\Big)^{n_{_2}}
\Big({y_{_1}\over y_{_4}}\Big)^{n_{_3}}\Big({y_{_2}\over y_{_3}}\Big)^{n_{_4}}\;.
\label{GKZ21a-4-2a}
\end{eqnarray}
Where the coefficients are
\begin{eqnarray}
&&c_{_{[\tilde{1}357]}}^{(4),a}(\alpha,{\bf n})=
(-)^{n_{_2}}\Gamma(1+n_{_1}+n_{_3})\Big\{n_{_1}!n_{_2}!n_{_3}!n_{_4}!
\Gamma(2-{D\over2}+n_{_3})
\nonumber\\
&&\hspace{2.5cm}\times
\Gamma(2-{D\over2}+n_{_2})\Gamma(2-{D\over2}+n_{_4})
\nonumber\\
&&\hspace{2.5cm}\times
\Gamma({D\over2}-1-n_{_2}-n_{_4})\Gamma({D\over2}-1-n_{_1}-n_{_3})
\nonumber\\
&&\hspace{2.5cm}\times
\Gamma({D\over2}+n_{_1})\Gamma(D-2-n_{_2}-n_{_4})\Big\}^{-1}
\;,\nonumber\\
&&c_{_{[\tilde{1}357]}}^{(4),b}(\alpha,{\bf n})=
(-)^{n_{_1}+n_{_2}}\Gamma(1+n_{_1})\Gamma(1+n_{_2}+n_{_3})\Big\{n_{_2}!n_{_4}!
\nonumber\\
&&\hspace{2.5cm}\times
\Gamma(2+n_{_1}+n_{_2}+n_{_3})\Gamma(1-{D\over2}-n_{_1})
\nonumber\\
&&\hspace{2.5cm}\times
\Gamma(2-{D\over2}+n_{_2})\Gamma(2-{D\over2}+n_{_4})
\nonumber\\
&&\hspace{2.5cm}\times
\Gamma({D\over2}-1-n_{_2}-n_{_4})\Gamma({D\over2}-1-n_{_2}-n_{_3})
\nonumber\\
&&\hspace{2.5cm}\times
\Gamma({D\over2}+1+n_{_1}+n_{_2}+n_{_3})\Gamma(D-2-n_{_2}-n_{_4})\Big\}^{-1}\;.
\label{GKZ21a-4-3}
\end{eqnarray}

\item $I_{_{5}}=[1,2,4,6,8,9,10,12,13,14]$, i.e. the implement $J_{_{5}}=[1,14]\setminus I_{_{5}}=[3,5,7,11]$.
The choice implies the power numbers $\alpha_{_3}=\alpha_{_5}=\alpha_{_{7}}=\alpha_{_{11}}=0$, and
\begin{eqnarray}
&&\alpha_{_1}=a_{_3}+a_{_5}-a_{_1},\;\alpha_{_2}=a_{_3}+a_{_5}-a_{_2},
\nonumber\\
&&\alpha_{_4}=a_{_3}-a_{_4},\;\alpha_{_{6}}=b_{_1}-1,\;\alpha_{_8}=b_{_2}+b_{_3}-a_{_3}-2,\;
\nonumber\\
&&\alpha_{_9}=b_{_4}-a_{_3}-a_{_5}-1,\;\alpha_{_{10}}=b_{_5}-a_{_3}-a_{_5}-1,\;
\nonumber\\
&&\alpha_{_{12}}=1-b_{_2},\;\alpha_{_{13}}=b_{_2}-a_{_3}-1,\;\alpha_{_{14}}=-a_{_5}\;.
\label{GKZ21a-5-1}
\end{eqnarray}
The corresponding hypergeometric series solution is written as
\begin{eqnarray}
&&\Phi_{_{[\tilde{1}357]}}^{(5)}(\alpha,z)=
y_{_2}^{{D\over2}-1}y_{_3}^{-1}y_{_4}^{-1}\sum\limits_{n_{_1}=0}^\infty
\sum\limits_{n_{_2}=0}^\infty\sum\limits_{n_{_3}=0}^\infty\sum\limits_{n_{_4}=0}^\infty
c_{_{[\tilde{1}357]}}^{(5)}(\alpha,{\bf n})
\nonumber\\
&&\hspace{2.5cm}\times
y_{_1}^{n_{_1}}\Big({1\over y_{_3}}\Big)^{n_{_2}}
\Big({1\over y_{_4}}\Big)^{n_{_3}}\Big({y_{_2}\over y_{_3}}\Big)^{n_{_4}}\;,
\label{GKZ21a-5-2}
\end{eqnarray}
with
\begin{eqnarray}
&&c_{_{[\tilde{1}357]}}^{(5)}(\alpha,{\bf n})=
(-)^{n_{_4}}\Gamma(1+n_{_2}+n_{_3})\Gamma(1+n_{_2}+n_{_4})
\Big\{n_{_1}!n_{_2}!n_{_4}!
\nonumber\\
&&\hspace{2.5cm}\times
\Gamma({D\over2}-n_{_1}+n_{_2}+n_{_3})\Gamma(D-1-n_{_1}+n_{_2}+n_{_3})
\nonumber\\
&&\hspace{2.5cm}\times
\Gamma({D\over2}+n_{_2})\Gamma(2-{D\over2}+n_{_1})\Gamma(1-{D\over2}-n_{_2}-n_{_4})
\nonumber\\
&&\hspace{2.5cm}\times
\Gamma(1-{D\over2}-n_{_2}-n_{_3})\Gamma({D\over2}+n_{_4})\Big\}^{-1}\;.
\label{GKZ21a-5-3}
\end{eqnarray}

\item $I_{_{6}}=[1,2,4,6,\cdots,10,13,14]$, i.e. the implement $J_{_{6}}=[1,14]\setminus I_{_{6}}=[3,5,11,12]$.
The choice implies the power numbers $\alpha_{_3}=\alpha_{_5}=\alpha_{_{11}}=\alpha_{_{12}}=0$, and
\begin{eqnarray}
&&\alpha_{_1}=a_{_3}+a_{_5}-a_{_1},\;\alpha_{_2}=a_{_3}+a_{_5}-a_{_2},
\nonumber\\
&&\alpha_{_4}=a_{_3}-a_{_4},\;\alpha_{_6}=b_{_1}-1,\;\alpha_{_7}=b_{_2}-1,\;\alpha_{_8}=b_{_3}-a_{_3}-1,\;
\nonumber\\
&&\alpha_{_9}=b_{_4}-a_{_3}-a_{_5}-1,\;\alpha_{_{10}}=b_{_5}-a_{_3}-a_{_5}-1,\;
\nonumber\\
&&\alpha_{_{13}}=-a_{_3},\;\alpha_{_{14}}=-a_{_5}\;.
\label{GKZ21a-6-1}
\end{eqnarray}
The corresponding hypergeometric series solution is
\begin{eqnarray}
&&\Phi_{_{[\tilde{1}357]}}^{(6)}(\alpha,z)=
y_{_3}^{{D\over2}-2}y_{_4}^{-1}\sum\limits_{n_{_1}=0}^\infty
\sum\limits_{n_{_2}=0}^\infty\sum\limits_{n_{_3}=0}^\infty\sum\limits_{n_{_4}=0}^\infty
c_{_{[\tilde{1}357]}}^{(6)}(\alpha,{\bf n})
\nonumber\\
&&\hspace{2.5cm}\times
y_{_1}^{n_{_1}}\Big({1\over y_{_3}}\Big)^{n_{_2}}
\Big({1\over y_{_4}}\Big)^{n_{_3}}\Big({y_{_2}\over y_{_3}}\Big)^{n_{_4}}\;,
\label{GKZ21a-6-2}
\end{eqnarray}
with
\begin{eqnarray}
&&c_{_{[\tilde{1}357]}}^{(6)}(\alpha,{\bf n})=
(-)^{n_{_4}}\Gamma(1+n_{_2}+n_{_3})\Gamma(1+n_{_2}+n_{_4})
\Big\{n_{_1}!n_{_2}!n_{_4}!
\nonumber\\
&&\hspace{2.5cm}\times
\Gamma({D\over2}-n_{_1}+n_{_2}+n_{_3})\Gamma(D-1-n_{_1}+n_{_2}+n_{_3})
\nonumber\\
&&\hspace{2.5cm}\times
\Gamma({D\over2}+n_{_2})\Gamma(2-{D\over2}+n_{_1})\Gamma(2-{D\over2}+n_{_4})
\nonumber\\
&&\hspace{2.5cm}\times
\Gamma(1-{D\over2}-n_{_2}-n_{_3})\Gamma({D\over2}-1-n_{_2}-n_{_4})\Big\}^{-1}\;.
\label{GKZ21a-6-3}
\end{eqnarray}

\item $I_{_{7}}=[1,2,3,6,8,9,10,12,13,14]$, i.e. the implement $J_{_{7}}=[1,14]\setminus I_{_{7}}=[4,5,7,11]$.
The choice implies the power numbers $\alpha_{_4}=\alpha_{_5}=\alpha_{_{7}}=\alpha_{_{11}}=0$, and
\begin{eqnarray}
&&\alpha_{_1}=a_{_4}+a_{_5}-a_{_1},\;\alpha_{_2}=a_{_4}+a_{_5}-a_{_2},
\nonumber\\
&&\alpha_{_3}=a_{_4}-a_{_3},\;\alpha_{_6}=b_{_1}-1,\;\alpha_{_8}=b_{_2}+b_{_3}-a_{_4}-2,\;
\nonumber\\
&&\alpha_{_9}=b_{_4}-a_{_4}-a_{_5}-1,\;\alpha_{_{10}}=b_{_5}-a_{_4}-a_{_5}-1,\;
\nonumber\\
&&\alpha_{_{12}}=1-b_{_2},\;\alpha_{_{13}}=b_{_2}-a_{_4}-1,\;\alpha_{_{14}}=-a_{_5}\;.
\label{GKZ21a-7-1}
\end{eqnarray}
The corresponding hypergeometric series solutions are written as
\begin{eqnarray}
&&\Phi_{_{[\tilde{1}357]}}^{(7),a}(\alpha,z)=
y_{_2}^{{D\over2}-1}y_{_3}^{{D\over2}-2}y_{_4}^{-1}\sum\limits_{n_{_1}=0}^\infty
\sum\limits_{n_{_2}=0}^\infty\sum\limits_{n_{_3}=0}^\infty\sum\limits_{n_{_4}=0}^\infty
c_{_{[\tilde{1}357]}}^{(7),a}(\alpha,{\bf n})
\nonumber\\
&&\hspace{2.5cm}\times
\Big({y_{_1}\over y_{_4}}\Big)^{n_{_1}}\Big({y_{_4}\over y_{_3}}\Big)^{n_{_2}}
\Big({1\over y_{_4}}\Big)^{n_{_3}}\Big({y_{_2}\over y_{_3}}\Big)^{n_{_4}}
\;,\nonumber\\
&&\Phi_{_{[\tilde{1}357]}}^{(7),b}(\alpha,z)=y_{_1}
y_{_2}^{{D\over2}-1}y_{_3}^{{D\over2}-2}y_{_4}^{-1}\sum\limits_{n_{_1}=0}^\infty
\sum\limits_{n_{_2}=0}^\infty\sum\limits_{n_{_3}=0}^\infty\sum\limits_{n_{_4}=0}^\infty
c_{_{[\tilde{1}357]}}^{(7),b}(\alpha,{\bf n})
\nonumber\\
&&\hspace{2.5cm}\times
y_{_1}^{n_{_1}}\Big({y_{_1}\over y_{_3}}\Big)^{n_{_2}}
\Big({y_{_1}\over y_{_4}}\Big)^{n_{_3}}\Big({y_{_2}\over y_{_3}}\Big)^{n_{_4}}\;.
\label{GKZ21a-7-2a}
\end{eqnarray}
Where the coefficients are
\begin{eqnarray}
&&c_{_{[\tilde{1}357]}}^{(7),a}(\alpha,{\bf n})=
(-)^{n_{_4}}\Gamma(1+n_{_1}+n_{_3})\Gamma(1+n_{_2}+n_{_4})
\Big\{n_{_1}!n_{_2}!n_{_3}!n_{_4}!\Gamma({D\over2}+n_{_3})
\nonumber\\
&&\hspace{2.5cm}\times
\Gamma(2-{D\over2}+n_{_2})
\Gamma(2-{D\over2}+n_{_1})\Gamma({D\over2}-1-n_{_1}-n_{_3})
\nonumber\\
&&\hspace{2.5cm}\times
\Gamma({D\over2}+n_{_4})\Gamma({D\over2}-1-n_{_2}-n_{_4})\Big\}^{-1}
\;,\nonumber\\
&&c_{_{[\tilde{1}357]}}^{(7),b}(\alpha,{\bf n})=
(-)^{n_{_1}+n_{_4}}\Gamma(1+n_{_1})\Gamma(1+n_{_2}+n_{_3})\Gamma(1+n_{_2}+n_{_4})
\Big\{n_{_2}!n_{_4}!
\nonumber\\
&&\hspace{2.5cm}\times
\Gamma(2+n_{_1}+n_{_2}+n_{_3})\Gamma({D\over2}-1-n_{_1})
\Gamma(2-{D\over2}+n_{_2})
\nonumber\\
&&\hspace{2.5cm}\times
\Gamma(3-{D\over2}+n_{_1}+n_{_2}+n_{_3})\Gamma({D\over2}-1-n_{_2}-n_{_3})
\nonumber\\
&&\hspace{2.5cm}\times
\Gamma({D\over2}+n_{_4})\Gamma({D\over2}-1-n_{_2}-n_{_4})\Big\}^{-1}\;.
\label{GKZ21a-7-3}
\end{eqnarray}

\item $I_{_{8}}=[1,2,3,6,\cdots,10,13,14]$, i.e. the implement $J_{_{8}}=[1,14]\setminus I_{_{8}}=[4,5,11,12]$.
The choice implies the power numbers $\alpha_{_4}=\alpha_{_5}=\alpha_{_{11}}=\alpha_{_{12}}=0$, and
\begin{eqnarray}
&&\alpha_{_1}=a_{_4}+a_{_5}-a_{_1},\;\alpha_{_2}=a_{_4}+a_{_5}-a_{_2},
\;\alpha_{_3}=a_{_4}-a_{_3},
\nonumber\\
&&\alpha_{_{6}}=b_{_1}-1,\;\alpha_{_{7}}=b_{_2}-1,\;\alpha_{_8}=b_{_3}-a_{_4}-1,\;
\nonumber\\
&&\alpha_{_9}=b_{_4}-a_{_4}-a_{_5}-1,\;\alpha_{_{10}}=b_{_5}-a_{_4}-a_{_5}-1,\;
\nonumber\\
&&\alpha_{_{13}}=-a_{_4},\;\alpha_{_{14}}=-a_{_5}\;.
\label{GKZ21a-8-1}
\end{eqnarray}
The corresponding hypergeometric series solutions are
\begin{eqnarray}
&&\Phi_{_{[\tilde{1}357]}}^{(8),a}(\alpha,z)=
y_{_3}^{{D}-3}y_{_4}^{-1}\sum\limits_{n_{_1}=0}^\infty
\sum\limits_{n_{_2}=0}^\infty\sum\limits_{n_{_3}=0}^\infty\sum\limits_{n_{_4}=0}^\infty
c_{_{[\tilde{1}357]}}^{(8),a}(\alpha,{\bf n})
\nonumber\\
&&\hspace{2.5cm}\times
\Big({y_{_1}\over y_{_4}}\Big)^{n_{_1}}\Big({y_{_4}\over y_{_3}}\Big)^{n_{_2}}
\Big({1\over y_{_4}}\Big)^{n_{_3}}\Big({y_{_2}\over y_{_3}}\Big)^{n_{_4}}
\;,\nonumber\\
&&\Phi_{_{[\tilde{1}357]}}^{(8),b}(\alpha,z)=y_{_1}
y_{_3}^{{D}-3}y_{_4}^{-1}\sum\limits_{n_{_1}=0}^\infty
\sum\limits_{n_{_2}=0}^\infty\sum\limits_{n_{_3}=0}^\infty\sum\limits_{n_{_4}=0}^\infty
c_{_{[\tilde{1}357]}}^{(8),b}(\alpha,{\bf n})
\nonumber\\
&&\hspace{2.5cm}\times
y_{_1}^{n_{_1}}\Big({y_{_1}\over y_{_3}}\Big)^{n_{_2}}
\Big({y_{_1}\over y_{_4}}\Big)^{n_{_3}}\Big({y_{_2}\over y_{_3}}\Big)^{n_{_4}}\;.
\label{GKZ21a-8-2a}
\end{eqnarray}
Where the coefficients are
\begin{eqnarray}
&&c_{_{[\tilde{1}357]}}^{(8),a}(\alpha,{\bf n})=
(-)^{n_{_2}}\Gamma(1+n_{_1}+n_{_3})\Big\{n_{_1}!n_{_2}!n_{_3}!n_{_4}!
\Gamma({D\over2}+n_{_3})\Gamma(2-{D\over2}+n_{_2})
\nonumber\\
&&\hspace{2.5cm}\times
\Gamma(2-{D\over2}+n_{_1})\Gamma(2-{D\over2}+n_{_4})
\Gamma({D\over2}-1-n_{_2}-n_{_4})
\nonumber\\
&&\hspace{2.5cm}\times
\Gamma({D\over2}-1-n_{_1}-n_{_3})\Gamma(D-2-n_{_2}-n_{_4})\Big\}^{-1}
\;,\nonumber\\
&&c_{_{[\tilde{1}357]}}^{(8),b}(\alpha,{\bf n})=
(-)^{n_{_1}+n_{_2}}\Gamma(1+n_{_1})\Gamma(1+n_{_2}+n_{_3})
\Big\{n_{_2}!n_{_4}!\Gamma(2+n_{_1}+n_{_2}+n_{_3})
\nonumber\\
&&\hspace{2.5cm}\times
\Gamma({D\over2}-1-n_{_1})\Gamma(2-{D\over2}+n_{_2})
\Gamma(3-{D\over2}+n_{_1}+n_{_2}+n_{_3})
\nonumber\\
&&\hspace{2.5cm}\times
\Gamma(2-{D\over2}+n_{_4})\Gamma({D\over2}-1-n_{_2}-n_{_4})
\Gamma({D\over2}-1-n_{_2}-n_{_3})
\nonumber\\
&&\hspace{2.5cm}\times
\Gamma(D-2-n_{_2}-n_{_4})\Big\}^{-1}\;.
\label{GKZ21a-8-3}
\end{eqnarray}
\end{itemize}

\section{The hypergeometric solutions of the integer lattice ${\bf B}_{_{1\tilde{3}57}}$\label{app3}}
\indent\indent
\begin{itemize}
\item  $I_{_{1}}=\{2,3,4,6,9,\cdots,14\}$, i.e. the implement $J_{_{1}}=[1,14]\setminus I_{_{1}}=\{1,5,7,8\}$.
The choice implies the power numbers $\alpha_{_1}=\alpha_{_5}=\alpha_{_{7}}=\alpha_{_{8}}=0$, and
\begin{eqnarray}
&&\alpha_{_2}=a_{_1}-a_{_2},\;\alpha_{_3}=b_{_2}+b_{_3}-a_{_3}-2,
\;\alpha_{_4}=b_{_2}+b_{_3}-a_{_4}-2,
\nonumber\\
&&\alpha_{_{6}}=a_{_5}+b_{_1}+b_{_2}+b_{_3}-a_{_1}-3,\;\alpha_{_9}=b_{_4}-a_{_5}-b_{_2}-b_{_3}+1,
\nonumber\\
&&\alpha_{_{10}}=b_{_5}-a_{_5}-b_{_2}-b_{_3}+1,\;\alpha_{_{11}}=a_{_5}+b_{_2}+b_{_3}-a_{_1}-2,
\nonumber\\
&&\alpha_{_{12}}=1-b_{_2},\;\alpha_{_{13}}=1-b_{_3},\;\alpha_{_{14}}=-a_{_5}\;.
\label{GKZ21b-1-1}
\end{eqnarray}
The corresponding hypergeometric series solutions are written as
\begin{eqnarray}
&&\Phi_{_{[1\tilde{3}57]}}^{(1),a}(\alpha,z)=
y_{_1}^{-1}y_{_2}^{{D\over2}-1}y_{_3}^{{D\over2}-1}y_{_4}^{-1}\sum\limits_{n_{_1}=0}^\infty
\sum\limits_{n_{_2}=0}^\infty\sum\limits_{n_{_3}=0}^\infty\sum\limits_{n_{_4}=0}^\infty
c_{_{[1\tilde{3}57]}}^{(1),a}(\alpha,{\bf n})
\nonumber\\
&&\hspace{2.5cm}\times
\Big({1\over y_{_1}}\Big)^{n_{_1}}\Big({y_{_3}\over y_{_4}}\Big)^{n_{_2}}
\Big({y_{_4}\over y_{_1}}\Big)^{n_{_3}}\Big({y_{_2}\over y_{_4}}\Big)^{n_{_4}}
\;,\nonumber\\
&&\Phi_{_{[1\tilde{3}57]}}^{(1),b}(\alpha,z)=
y_{_2}^{{D\over2}-1}y_{_3}^{{D\over2}-1}y_{_4}^{-2}\sum\limits_{n_{_1}=0}^\infty
\sum\limits_{n_{_2}=0}^\infty\sum\limits_{n_{_3}=0}^\infty\sum\limits_{n_{_4}=0}^\infty
c_{_{[1\tilde{3}57]}}^{(1),b}(\alpha,{\bf n})
\nonumber\\
&&\hspace{2.5cm}\times
\Big({1\over y_{_4}}\Big)^{n_{_1}}\Big({y_{_3}\over y_{_4}}\Big)^{n_{_2}}
\Big({y_{_1}\over y_{_4}}\Big)^{n_{_3}}\Big({y_{_2}\over y_{_4}}\Big)^{n_{_4}}
\;,\nonumber\\
&&\Phi_{_{[1\tilde{3}57]}}^{(1),c}(\alpha,z)=
y_{_1}^{-1}y_{_2}^{{D\over2}-1}y_{_3}^{{D\over2}-1}y_{_4}^{-2}\sum\limits_{n_{_1}=0}^\infty
\sum\limits_{n_{_2}=0}^\infty\sum\limits_{n_{_3}=0}^\infty\sum\limits_{n_{_4}=0}^\infty
c_{_{[1\tilde{3}57]}}^{(1),c}(\alpha,{\bf n})
\nonumber\\
&&\hspace{2.5cm}\times
\Big({1\over y_{_1}}\Big)^{n_{_1}}\Big({y_{_3}\over y_{_4}}\Big)^{n_{_2}}
\Big({1\over y_{_4}}\Big)^{n_{_3}}\Big({y_{_2}\over y_{_4}}\Big)^{n_{_4}}\;.
\label{GKZ21b-1-2a}
\end{eqnarray}
Where the coefficients are
\begin{eqnarray}
&&c_{_{[1\tilde{3}57]}}^{(1),a}(\alpha,{\bf n})=(-)^{n_{_1}}
\Gamma(1+n_{_1}+n_{_3})\Gamma(1+n_{_2}+n_{_4})
\Big\{n_{_1}!n_{_2}!n_{_3}!n_{_4}!
\nonumber\\
&&\hspace{2.5cm}\times
\Gamma({D\over2}+n_{_1})
\Gamma(1-{D\over2}-n_{_2}-n_{_4})\Gamma(1-{D\over2}-n_{_1}-n_{_3})
\nonumber\\
&&\hspace{2.5cm}\times
\Gamma({D\over2}+n_{_3})\Gamma({D\over2}+n_{_4})\Gamma({D\over2}+n_{_2})\Big\}^{-1}
\;,\nonumber\\
&&c_{_{[1\tilde{3}57]}}^{(1),b}(\alpha,{\bf n})=-\Gamma(1+n_{_1}+n_{_3})
\Gamma(1+n_{_2}+n_{_4})\Big\{n_{_1}!n_{_2}!n_{_3}!n_{_4}!\Gamma({D\over2}+n_{_1})
\nonumber\\
&&\hspace{2.5cm}\times
\Gamma(1-{D\over2}-n_{_2}-n_{_4})
\Gamma(2-{D\over2}+n_{_3})\Gamma({D\over2}-1-n_{_1}-n_{_3})
\nonumber\\
&&\hspace{2.5cm}\times
\Gamma({D\over2}+n_{_4})\Gamma({D\over2}+n_{_2})\Big\}^{-1}
\;,\nonumber\\
&&c_{_{[1\tilde{3}57]}}^{(1),c}(\alpha,{\bf n})=
(-)^{1+n_{_1}}\Gamma(1+n_{_1})\Gamma(1+n_{_3})
\Gamma(1+n_{_2}+n_{_4})\Big\{n_{_2}!n_{_4}!
\nonumber\\
&&\hspace{2.5cm}\times
\Gamma(2+n_{_1}+n_{_3})\Gamma({D\over2}+1+n_{_1}+n_{_3})
\Gamma(1-{D\over2}-n_{_2}-n_{_4})
\nonumber\\
&&\hspace{2.5cm}\times
\Gamma(1-{D\over2}-n_{_1})\Gamma({D\over2}-1-n_{_3})
\Gamma({D\over2}+n_{_4})\Gamma({D\over2}+n_{_2})\Big\}^{-1}\;.
\label{GKZ21b-1-3}
\end{eqnarray}

\item  $I_{_{2}}=\{2,3,4,6,7,9,10,11,13,14\}$, i.e. the implement $J_{_{2}}=[1,14]\setminus I_{_{2}}=\{1,5,8,12\}$.
The choice implies the power numbers $\alpha_{_1}=\alpha_{_5}=\alpha_{_{8}}=\alpha_{_{12}}=0$, and
\begin{eqnarray}
&&\alpha_{_2}=a_{_1}-a_{_2},\;\alpha_{_3}=b_{_3}-a_{_3}-1,
\;\alpha_{_4}=b_{_3}-a_{_4}-1,
\nonumber\\
&&\alpha_{_{6}}=a_{_5}+b_{_1}+b_{_3}-a_{_1}-2,\;\alpha_{_7}=b_{_2}-1,
\nonumber\\
&&\alpha_{_9}=b_{_4}-b_{_3}-a_{_5},\;\alpha_{_{10}}=b_{_5}-b_{_3}-a_{_5},
\nonumber\\
&&\alpha_{_{11}}=a_{_5}+b_{_3}-a_{_1}-1,\;
\alpha_{_{13}}=1-b_{_3},\;\alpha_{_{14}}=-a_{_5}\;.
\label{GKZ21b-2-1}
\end{eqnarray}
The corresponding hypergeometric series solutions are
\begin{eqnarray}
&&\Phi_{_{[1\tilde{3}57]}}^{(2),a}(\alpha,z)=
y_{_1}^{{D\over2}-2}y_{_3}^{{D\over2}-1}y_{_4}^{-1}\sum\limits_{n_{_1}=0}^\infty
\sum\limits_{n_{_2}=0}^\infty\sum\limits_{n_{_3}=0}^\infty\sum\limits_{n_{_4}=0}^\infty
c_{_{[1\tilde{3}57]}}^{(2),a}(\alpha,{\bf n})
\nonumber\\
&&\hspace{2.5cm}\times
\Big({1\over y_{_1}}\Big)^{n_{_1}}\Big({y_{_3}\over y_{_4}}\Big)^{n_{_2}}
\Big({y_{_4}\over y_{_1}}\Big)^{n_{_3}}\Big({y_{_2}\over y_{_4}}\Big)^{n_{_4}}
\;,\nonumber\\
&&\Phi_{_{[1\tilde{3}57]}}^{(2),b}(\alpha,z)=
y_{_1}^{{D\over2}-1}y_{_3}^{{D\over2}-1}y_{_4}^{-2}\sum\limits_{n_{_1}=0}^\infty
\sum\limits_{n_{_2}=0}^\infty\sum\limits_{n_{_3}=0}^\infty\sum\limits_{n_{_4}=0}^\infty
c_{_{[1\tilde{3}57]}}^{(2),b}(\alpha,{\bf n})
\nonumber\\
&&\hspace{2.5cm}\times
\Big({1\over y_{_4}}\Big)^{n_{_1}}\Big({y_{_3}\over y_{_4}}\Big)^{n_{_2}}
\Big({y_{_1}\over y_{_4}}\Big)^{n_{_3}}\Big({y_{_2}\over y_{_4}}\Big)^{n_{_4}}
\;,\nonumber\\
&&\Phi_{_{[1\tilde{3}57]}}^{(2),c}(\alpha,z)=
y_{_1}^{{D\over2}-2}y_{_3}^{{D\over2}-1}y_{_4}^{-2}\sum\limits_{n_{_1}=0}^\infty
\sum\limits_{n_{_2}=0}^\infty\sum\limits_{n_{_3}=0}^\infty\sum\limits_{n_{_4}=0}^\infty
c_{_{[1\tilde{3}57]}}^{(2),c}(\alpha,{\bf n})
\nonumber\\
&&\hspace{2.5cm}\times
\Big({1\over y_{_1}}\Big)^{n_{_1}}\Big({y_{_3}\over y_{_4}}\Big)^{n_{_2}}
\Big({1\over y_{_4}}\Big)^{n_{_3}}\Big({y_{_2}\over y_{_4}}\Big)^{n_{_4}}\;.
\label{GKZ21b-2-2a}
\end{eqnarray}
Where the coefficients are
\begin{eqnarray}
&&c_{_{[1\tilde{3}57]}}^{(2),a}(\alpha,{\bf n})=
(-)^{n_{_1}}\Gamma(1+n_{_1}+n_{_3})\Gamma(1+n_{_2}+n_{_4})
\Big\{n_{_1}!n_{_2}!n_{_3}!n_{_4}!\Gamma({D\over2}+n_{_1})
\nonumber\\
&&\hspace{2.5cm}\times
\Gamma({D\over2}-1-n_{_2}-n_{_4})\Gamma({D\over2}-1-n_{_1}-n_{_3})
\Gamma(2-{D\over2}+n_{_4})
\nonumber\\
&&\hspace{2.5cm}\times
\Gamma(2-{D\over2}+n_{_3})\Gamma({D\over2}+n_{_2})\Big\}^{-1}
\;,\nonumber\\
&&c_{_{[1\tilde{3}57]}}^{(2),b}(\alpha,{\bf n})=
-\Gamma(1+n_{_1}+n_{_3})\Gamma(1+n_{_2}+n_{_4})
\Big\{n_{_1}!n_{_2}!n_{_3}!n_{_4}!\Gamma({D\over2}+n_{_1})
\nonumber\\
&&\hspace{2.5cm}\times
\Gamma({D\over2}-1-n_{_2}-n_{_4})\Gamma({D\over2}+n_{_3})\Gamma(2-{D\over2}+n_{_4})
\nonumber\\
&&\hspace{2.5cm}\times
\Gamma(1-{D\over2}-n_{_1}-n_{_3})\Gamma({D\over2}+n_{_2})\Big\}^{-1}
\;,\nonumber\\
&&c_{_{[1\tilde{3}57]}}^{(2),c}(\alpha,{\bf n})=
(-)^{1+n_{_1}}\Gamma(1+n_{_1})\Gamma(1+n_{_3})\Gamma(1+n_{_2}+n_{_4})
\Big\{n_{_2}!n_{_4}!
\nonumber\\
&&\hspace{2.5cm}\times
\Gamma(2+n_{_1}+n_{_3})\Gamma({D\over2}+1+n_{_1}+n_{_3})
\Gamma({D\over2}-1-n_{_2}-n_{_4})
\nonumber\\
&&\hspace{2.5cm}\times
\Gamma({D\over2}-1-n_{_1})\Gamma(2-{D\over2}+n_{_4})
\Gamma(1-{D\over2}-n_{_3})\Gamma({D\over2}+n_{_2})\Big\}^{-1}\;.
\label{GKZ21b-2-3}
\end{eqnarray}

\item  $I_{_{3}}=\{2,3,4,6,8,\cdots,12,14\}$, i.e. the implement $J_{_{3}}=[1,14]\setminus I_{_{3}}=\{1,5,7,13\}$.
The choice implies the power numbers $\alpha_{_1}=\alpha_{_5}=\alpha_{_{7}}=\alpha_{_{13}}=0$, and
\begin{eqnarray}
&&\alpha_{_2}=a_{_1}-a_{_2},\;\alpha_{_3}=b_{_2}-a_{_3}-1,
\;\alpha_{_4}=b_{_2}-a_{_4}-1,
\nonumber\\
&&\alpha_{_{6}}=a_{_5}+b_{_1}+b_{_2}-a_{_1}-2,\;\alpha_{_{8}}=b_{_3}-1,
\nonumber\\
&&\alpha_{_9}=b_{_4}-a_{_5}-b_{_2},\;
\alpha_{_{10}}=b_{_5}-a_{_5}-b_{_2},
\nonumber\\
&&\alpha_{_{11}}=a_{_5}+b_{_2}-a_{_1}-1,\;
\alpha_{_{12}}=1-b_{_2},\;\alpha_{_{14}}=-a_{_5}\;.
\label{GKZ21b-11-1}
\end{eqnarray}
The corresponding hypergeometric series solutions are written as
\begin{eqnarray}
&&\Phi_{_{[1\tilde{3}57]}}^{(3),a}(\alpha,z)=
y_{_1}^{{D\over2}-2}y_{_2}^{{D\over2}-1}y_{_4}^{-1}\sum\limits_{n_{_1}=0}^\infty
\sum\limits_{n_{_2}=0}^\infty\sum\limits_{n_{_3}=0}^\infty\sum\limits_{n_{_4}=0}^\infty
c_{_{[1\tilde{3}57]}}^{(3),a}(\alpha,{\bf n})
\nonumber\\
&&\hspace{2.5cm}\times
\Big({1\over y_{_1}}\Big)^{n_{_1}}\Big({y_{_3}\over y_{_4}}\Big)^{n_{_2}}
\Big({y_{_4}\over y_{_1}}\Big)^{n_{_3}}\Big({y_{_2}\over y_{_4}}\Big)^{n_{_4}}
\;,\nonumber\\
&&\Phi_{_{[1\tilde{3}57]}}^{(3),b}(\alpha,z)=
y_{_1}^{{D\over2}-1}y_{_2}^{{D\over2}-1}y_{_4}^{-2}\sum\limits_{n_{_1}=0}^\infty
\sum\limits_{n_{_2}=0}^\infty\sum\limits_{n_{_3}=0}^\infty\sum\limits_{n_{_4}=0}^\infty
c_{_{[1\tilde{3}57]}}^{(3),b}(\alpha,{\bf n})
\nonumber\\
&&\hspace{2.5cm}\times
\Big({1\over y_{_4}}\Big)^{n_{_1}}\Big({y_{_3}\over y_{_4}}\Big)^{n_{_2}}
\Big({y_{_1}\over y_{_4}}\Big)^{n_{_3}}\Big({y_{_2}\over y_{_4}}\Big)^{n_{_4}}
\;,\nonumber\\
&&\Phi_{_{[1\tilde{3}57]}}^{(3),c}(\alpha,z)=
y_{_1}^{{D\over2}-2}y_{_2}^{{D\over2}-1}y_{_4}^{-2}\sum\limits_{n_{_1}=0}^\infty
\sum\limits_{n_{_2}=0}^\infty\sum\limits_{n_{_3}=0}^\infty\sum\limits_{n_{_4}=0}^\infty
c_{_{[1\tilde{3}57]}}^{(3),c}(\alpha,{\bf n})
\nonumber\\
&&\hspace{2.5cm}\times
\Big({1\over y_{_1}}\Big)^{n_{_1}}\Big({y_{_3}\over y_{_4}}\Big)^{n_{_2}}
\Big({1\over y_{_4}}\Big)^{n_{_3}}\Big({y_{_2}\over y_{_4}}\Big)^{n_{_4}}\;.
\label{GKZ21b-11-2a}
\end{eqnarray}
Where the coefficients are
\begin{eqnarray}
&&c_{_{[1\tilde{3}57]}}^{(3),a}(\alpha,{\bf n})=
(-)^{n_{_1}}\Gamma(1+n_{_1}+n_{_3})
\Gamma(1+n_{_2}+n_{_4})\Big\{n_{_1}!n_{_2}!n_{_3}!n_{_4}!
\nonumber\\
&&\hspace{2.5cm}\times
\Gamma({D\over2}+n_{_1})\Gamma({D\over2}-1-n_{_2}-n_{_4})
\Gamma({D\over2}-1-n_{_1}-n_{_3})
\nonumber\\
&&\hspace{2.5cm}\times
\Gamma(2-{D\over2}+n_{_2})\Gamma(2-{D\over2}+n_{_3})
\Gamma({D\over2}+n_{_4})\Big\}^{-1}
\;,\nonumber\\
&&c_{_{[1\tilde{3}57]}}^{(3),b}(\alpha,{\bf n})=
-\Gamma(1+n_{_1}+n_{_3})\Gamma(1+n_{_2}+n_{_4})\Big\{n_{_1}!n_{_2}!n_{_3}!n_{_4}!
\nonumber\\
&&\hspace{2.5cm}\times
\Gamma({D\over2}+n_{_1})\Gamma({D\over2}-1-n_{_2}-n_{_4})
\Gamma(1-{D\over2}-n_{_1}-n_{_3})
\nonumber\\
&&\hspace{2.5cm}\times
\Gamma(2-{D\over2}+n_{_2})\Gamma({D\over2}+n_{_3})
\Gamma({D\over2}+n_{_4})\Big\}^{-1}
\;,\nonumber\\
&&c_{_{[1\tilde{3}57]}}^{(3),c}(\alpha,{\bf n})=
(-)^{1+n_{_1}}\Gamma(1+n_{_1})\Gamma(1+n_{_2}+n_{_4})
\Gamma(1+n_{_3})\Big\{n_{_2}!n_{_4}!
\nonumber\\
&&\hspace{2.5cm}\times
\Gamma(2+n_{_1}+n_{_3})
\Gamma({D\over2}+1+n_{_1}+n_{_3})\Gamma({D\over2}-1-n_{_2}-n_{_4})
\nonumber\\
&&\hspace{2.5cm}\times
\Gamma({D\over2}-1-n_{_1})\Gamma(2-{D\over2}+n_{_2})\Gamma(1-{D\over2}-n_{_3})
\Gamma({D\over2}+n_{_4})\Big\}^{-1}\;.
\label{GKZ21b-11-3}
\end{eqnarray}

\item  $I_{_{4}}=\{2,3,4,6,\cdots,11,14\}$, i.e. the implement $J_{_{4}}=[1,14]\setminus I_{_{4}}=\{1,5,12,13\}$.
The choice implies the power numbers $\alpha_{_1}=\alpha_{_5}=\alpha_{_{12}}=\alpha_{_{13}}=0$, and
\begin{eqnarray}
&&\alpha_{_2}=a_{_1}-a_{_2},\;\alpha_{_3}=-a_{_3},
\;\alpha_{_4}=-a_{_4},
\nonumber\\
&&\alpha_{_{6}}=a_{_5}+b_{_1}-a_{_1}-1,\;\alpha_{_7}=b_{_2}-1,\;\alpha_{_8}=b_{_3}-1,
\nonumber\\
&&\alpha_{_9}=b_{_4}-a_{_5}-1,\;\alpha_{_{10}}=b_{_5}-a_{_5}-1,
\nonumber\\
&&\alpha_{_{11}}=a_{_5}-a_{_1},\;\alpha_{_{14}}=-a_{_5}\;.
\label{GKZ21b-12-1}
\end{eqnarray}
The corresponding hypergeometric series solution is written as
\begin{eqnarray}
&&\Phi_{_{[1\tilde{3}57]}}^{(4)}(\alpha,z)=
y_{_1}^{{D}-3}y_{_4}^{-1}\sum\limits_{n_{_1}=0}^\infty
\sum\limits_{n_{_2}=0}^\infty\sum\limits_{n_{_3}=0}^\infty\sum\limits_{n_{_4}=0}^\infty
c_{_{[1\tilde{3}57]}}^{(4)}(\alpha,{\bf n})
\nonumber\\
&&\hspace{2.5cm}\times
\Big({1\over y_{_1}}\Big)^{n_{_1}}\Big({y_{_3}\over y_{_1}}\Big)^{n_{_2}}
\Big({y_{_1}\over y_{_4}}\Big)^{n_{_3}}\Big({y_{_2}\over y_{_1}}\Big)^{n_{_4}}\;,
\label{GKZ21b-12-2}
\end{eqnarray}
with
\begin{eqnarray}
&&c_{_{[1\tilde{3}57]}}^{(4)}(\alpha,{\bf n})=
(-)^{n_{_3}}\Big\{n_{_1}!n_{_2}!n_{_4}!\Gamma({D\over2}+n_{_1})
\Gamma({D\over2}-1-n_{_2}-n_{_4})
\nonumber\\
&&\hspace{2.5cm}\times
\Gamma(D-2-n_{_2}-n_{_4})\Gamma({D\over2}-1-n_{_1}-n_{_2}+n_{_3}-n_{_4})
\nonumber\\
&&\hspace{2.5cm}\times
\Gamma(2-{D\over2}+n_{_4})\Gamma(2-{D\over2}+n_{_2})
\nonumber\\
&&\hspace{2.5cm}\times
\Gamma(2-{D\over2}+n_{_2}-n_{_3}+n_{_4})\Gamma(3-D+n_{_2}-n_{_3}+n_{_4})
\nonumber\\
&&\hspace{2.5cm}\times
\Gamma(D-2-n_{_1}-n_{_2}+n_{_3}-n_{_4})\Big\}^{-1}\;.
\label{GKZ21b-12-3}
\end{eqnarray}

\item  $I_{_{5}}=\{1,3,4,6,9,\cdots,14\}$, i.e. the implement $J_{_{5}}=[1,14]\setminus I_{_{5}}=\{2,5,7,8\}$.
The choice implies the power numbers $\alpha_{_2}=\alpha_{_5}=\alpha_{_{7}}=\alpha_{_{8}}=0$, and
\begin{eqnarray}
&&\alpha_{_1}=a_{_2}-a_{_1},\;\alpha_{_3}=b_{_2}+b_{_3}-a_{_3}-2,
\;\alpha_{_4}=b_{_2}+b_{_3}-a_{_4}-2,
\nonumber\\
&&\alpha_{_{6}}=a_{_5}+b_{_1}+b_{_2}+b_{_3}-a_{_2}-3,\;\alpha_{_9}=b_{_4}-a_{_5}-b_{_2}-b_{_3}+1,
\nonumber\\
&&\alpha_{_{10}}=b_{_5}-a_{_5}-b_{_2}-b_{_3}+1,\;\alpha_{_{11}}=a_{_5}+b_{_2}+b_{_3}-a_{_2}-2,
\nonumber\\
&&\alpha_{_{12}}=1-b_{_2},\;\alpha_{_{13}}=1-b_{_3},\;\alpha_{_{14}}=-a_{_5}\;.
\label{GKZ21b-17-1}
\end{eqnarray}
The corresponding hypergeometric series solutions are written as
\begin{eqnarray}
&&\Phi_{_{[1\tilde{3}57]}}^{(5),a}(\alpha,z)=
y_{_1}^{{D\over2}-2}y_{_2}^{{D\over2}-1}y_{_3}^{{D\over2}-1}y_{_4}^{-1}\sum\limits_{n_{_1}=0}^\infty
\sum\limits_{n_{_2}=0}^\infty\sum\limits_{n_{_3}=0}^\infty\sum\limits_{n_{_4}=0}^\infty
c_{_{[1\tilde{3}57]}}^{(5),a}(\alpha,{\bf n})
\nonumber\\
&&\hspace{2.5cm}\times
\Big({1\over y_{_1}}\Big)^{n_{_1}}\Big({y_{_3}\over y_{_4}}\Big)^{n_{_2}}
\Big({y_{_4}\over y_{_1}}\Big)^{n_{_3}}\Big({y_{_2}\over y_{_4}}\Big)^{n_{_4}}
\;,\nonumber\\
&&\Phi_{_{[1\tilde{3}57]}}^{(5),b}(\alpha,z)=
y_{_1}^{{D\over2}-1}y_{_2}^{{D\over2}-1}y_{_3}^{{D\over2}-1}y_{_4}^{-2}\sum\limits_{n_{_1}=0}^\infty
\sum\limits_{n_{_2}=0}^\infty\sum\limits_{n_{_3}=0}^\infty\sum\limits_{n_{_4}=0}^\infty
c_{_{[1\tilde{3}57]}}^{(5),b}(\alpha,{\bf n})
\nonumber\\
&&\hspace{2.5cm}\times
\Big({1\over y_{_1}}\Big)^{n_{_1}}\Big({y_{_3}\over y_{_4}}\Big)^{n_{_2}}
\Big({y_{_1}\over y_{_4}}\Big)^{n_{_3}}\Big({y_{_2}\over y_{_4}}\Big)^{n_{_4}}
\;,\nonumber\\
&&\Phi_{_{[1\tilde{3}57]}}^{(5),c}(\alpha,z)=
y_{_1}^{{D\over2}-2}y_{_2}^{{D\over2}-1}y_{_3}^{{D\over2}-1}y_{_4}^{-2}\sum\limits_{n_{_1}=0}^\infty
\sum\limits_{n_{_2}=0}^\infty\sum\limits_{n_{_3}=0}^\infty\sum\limits_{n_{_4}=0}^\infty
c_{_{[1\tilde{3}57]}}^{(5),c}(\alpha,{\bf n})
\nonumber\\
&&\hspace{2.5cm}\times
\Big({1\over y_{_1}}\Big)^{n_{_1}}\Big({y_{_3}\over y_{_4}}\Big)^{n_{_2}}
\Big({1\over y_{_4}}\Big)^{n_{_3}}\Big({y_{_2}\over y_{_4}}\Big)^{n_{_4}}\;.
\label{GKZ21b-17-2a}
\end{eqnarray}
Where the coefficients are
\begin{eqnarray}
&&c_{_{[1\tilde{3}57]}}^{(5),a}(\alpha,{\bf n})=
(-)^{n_{_1}}\Gamma(1+n_{_1}+n_{_3})\Gamma(1+n_{_2}+n_{_4})
\Big\{n_{_1}!n_{_2}!n_{_3}!n_{_4}!
\nonumber\\
&&\hspace{2.5cm}\times
\Gamma(2-{D\over2}+n_{_1})\Gamma(1-{D\over2}-n_{_2}-n_{_4})
\Gamma({D\over2}+n_{_3})
\nonumber\\
&&\hspace{2.5cm}\times
\Gamma({D\over2}-1-n_{_1}-n_{_3})
\Gamma({D\over2}+n_{_4})\Gamma({D\over2}+n_{_2})\Big\}^{-1}
\;,\nonumber\\
&&c_{_{[1\tilde{3}57]}}^{(5),b}(\alpha,{\bf n})=
-\Gamma(1+n_{_1}+n_{_3})\Gamma(1+n_{_2}+n_{_4})\Big\{n_{_1}!n_{_2}!n_{_3}!n_{_4}!
\nonumber\\
&&\hspace{2.5cm}\times
\Gamma(2-{D\over2}+n_{_1})
\Gamma(1-{D\over2}-n_{_2}-n_{_4})\Gamma({D\over2}-1-n_{_1}-n_{_3})
\nonumber\\
&&\hspace{2.5cm}\times
\Gamma({D\over2}+n_{_3})
\Gamma({D\over2}+n_{_4})\Gamma({D\over2}+n_{_2})\Big\}^{-1}
\;,\nonumber\\
&&c_{_{[1\tilde{3}57]}}^{(5),c}(\alpha,{\bf n})=
(-)^{1+n_{_1}}\Gamma(1+n_{_1})\Gamma(1+n_{_3})\Gamma(1+n_{_2}+n_{_4})
\Big\{n_{_2}!n_{_4}!
\nonumber\\
&&\hspace{2.5cm}\times
\Gamma(2+n_{_1}+n_{_3})\Gamma(3-{D\over2}+n_{_1}+n_{_3})
\Gamma(1-{D\over2}-n_{_2}-n_{_4})
\nonumber\\
&&\hspace{2.5cm}\times
\Gamma({D\over2}-1-n_{_3})\Gamma({D\over2}-1-n_{_1})
\Gamma({D\over2}+n_{_4})\Gamma({D\over2}+n_{_2})\Big\}^{-1}\;.
\label{GKZ21b-17-3}
\end{eqnarray}

\item  $I_{_{6}}=\{1,3,4,6,7,9,10,11,13,14\}$, i.e. the implement $J_{_{6}}=[1,14]\setminus I_{_{6}}=\{2,5,8,12\}$.
The choice implies the power numbers $\alpha_{_2}=\alpha_{_5}=\alpha_{_{8}}=\alpha_{_{12}}=0$, and
\begin{eqnarray}
&&\alpha_{_1}=a_{_2}-a_{_1},\;\alpha_{_3}=b_{_3}-a_{_3}-1,
\;\alpha_{_4}=b_{_3}-a_{_4}-1,
\nonumber\\
&&\alpha_{_{6}}=a_{_5}+b_{_1}+b_{_3}-a_{_2}-2,\;\alpha_{_7}=b_{_2}-1,
\nonumber\\
&&\alpha_{_9}=b_{_4}-b_{_3}-a_{_5},\;\alpha_{_{10}}=b_{_5}-b_{_3}-a_{_5},
\nonumber\\
&&\alpha_{_{11}}=a_{_5}+b_{_3}-a_{_2}-1,\;
\alpha_{_{13}}=1-b_{_3},\;\alpha_{_{14}}=-a_{_5}\;.
\label{GKZ21b-18-1}
\end{eqnarray}
The corresponding hypergeometric series solutions are
\begin{eqnarray}
&&\Phi_{_{[1\tilde{3}57]}}^{(6),a}(\alpha,z)=
y_{_1}^{{D}-3}y_{_3}^{{D\over2}-1}y_{_4}^{-1}\sum\limits_{n_{_1}=0}^\infty
\sum\limits_{n_{_2}=0}^\infty\sum\limits_{n_{_3}=0}^\infty\sum\limits_{n_{_4}=0}^\infty
c_{_{[1\tilde{3}57]}}^{(6),a}(\alpha,{\bf n})
\nonumber\\
&&\hspace{2.5cm}\times
\Big({1\over y_{_1}}\Big)^{n_{_1}}\Big({y_{_3}\over y_{_4}}\Big)^{n_{_2}}
\Big({y_{_4}\over y_{_1}}\Big)^{n_{_3}}\Big({y_{_2}\over y_{_4}}\Big)^{n_{_4}}
\;,\nonumber\\
&&\Phi_{_{[1\tilde{3}57]}}^{(6),b}(\alpha,z)=
y_{_1}^{{D}-2}y_{_3}^{{D\over2}-1}y_{_4}^{-2}\sum\limits_{n_{_1}=0}^\infty
\sum\limits_{n_{_2}=0}^\infty\sum\limits_{n_{_3}=0}^\infty\sum\limits_{n_{_4}=0}^\infty
c_{_{[1\tilde{3}57]}}^{(6),b}(\alpha,{\bf n})
\nonumber\\
&&\hspace{2.5cm}\times
\Big({1\over y_{_1}}\Big)^{n_{_1}}\Big({y_{_3}\over y_{_4}}\Big)^{n_{_2}}
\Big({y_{_1}\over y_{_4}}\Big)^{n_{_3}}\Big({y_{_2}\over y_{_4}}\Big)^{n_{_4}}\;.
\label{GKZ21b-18-2a}
\end{eqnarray}
Where the coefficients are
\begin{eqnarray}
&&c_{_{[1\tilde{3}57]}}^{(6),a}(\alpha,{\bf n})=
(-)^{n_{_3}}\Gamma(1+n_{_2}+n_{_4})\Big\{n_{_1}!n_{_2}!n_{_3}!n_{_4}!
\Gamma(2-{D\over2}+n_{_1})
\nonumber\\
&&\hspace{2.5cm}\times
\Gamma({D\over2}-1-n_{_2}-n_{_4})\Gamma({D\over2}-1-n_{_1}-n_{_3})
\Gamma(2-{D\over2}+n_{_4})
\nonumber\\
&&\hspace{2.5cm}\times
\Gamma(2-{D\over2}+n_{_3})\Gamma(D-2-n_{_1}-n_{_3})
\Gamma({D\over2}+n_{_2})\Big\}^{-1}
\;,\nonumber\\
&&c_{_{[1\tilde{3}57]}}^{(6),b}(\alpha,{\bf n})=
\Gamma(1+n_{_2}+n_{_4})\Gamma(1+n_{_3})\Big\{n_{_1}!n_{_2}!n_{_4}!
\Gamma(2-{D\over2}+n_{_1})
\nonumber\\
&&\hspace{2.5cm}\times
\Gamma({D\over2}-1-n_{_2}-n_{_4})\Gamma({D\over2}-n_{_1}+n_{_3})
\Gamma(2-{D\over2}+n_{_4})
\nonumber\\
&&\hspace{2.5cm}\times
\Gamma(1-{D\over2}-n_{_3})\Gamma(D-1-n_{_1}+n_{_3})
\Gamma({D\over2}+n_{_2})\Big\}^{-1}\;.
\label{GKZ21b-18-3}
\end{eqnarray}

\item  $I_{_{7}}=\{1,3,4,6,8,\cdots,12,14\}$, i.e. the implement $J_{_{7}}=[1,14]\setminus I_{_{7}}=\{2,5,7,13\}$.
The choice implies the power numbers $\alpha_{_2}=\alpha_{_5}=\alpha_{_{7}}=\alpha_{_{13}}=0$, and
\begin{eqnarray}
&&\alpha_{_1}=a_{_2}-a_{_1},\;\alpha_{_3}=b_{_2}-a_{_3}-1,
\;\alpha_{_4}=b_{_2}-a_{_4}-1,
\nonumber\\
&&\alpha_{_{6}}=a_{_5}+b_{_1}+b_{_2}-a_{_2}-2,\;\alpha_{_{8}}=b_{_3}-1,
\nonumber\\
&&\alpha_{_9}=b_{_4}-a_{_5}-b_{_2},\;
\alpha_{_{10}}=b_{_5}-a_{_5}-b_{_2},
\nonumber\\
&&\alpha_{_{11}}=a_{_5}+b_{_2}-a_{_2}-1,\;
\alpha_{_{12}}=1-b_{_2},\;\alpha_{_{14}}=-a_{_5}\;.
\label{GKZ21b-27-1}
\end{eqnarray}
The corresponding hypergeometric series solutions are written as
\begin{eqnarray}
&&\Phi_{_{[1\tilde{3}57]}}^{(7),a}(\alpha,z)=
y_{_1}^{{D}-3}y_{_2}^{{D\over2}-1}y_{_4}^{-1}\sum\limits_{n_{_1}=0}^\infty
\sum\limits_{n_{_2}=0}^\infty\sum\limits_{n_{_3}=0}^\infty\sum\limits_{n_{_4}=0}^\infty
c_{_{[1\tilde{3}57]}}^{(7),a}(\alpha,{\bf n})
\nonumber\\
&&\hspace{2.5cm}\times
\Big({1\over y_{_1}}\Big)^{n_{_1}}\Big({y_{_3}\over y_{_4}}\Big)^{n_{_2}}
\Big({y_{_4}\over y_{_1}}\Big)^{n_{_3}}\Big({y_{_2}\over y_{_4}}\Big)^{n_{_4}}
\;,\nonumber\\
&&\Phi_{_{[1\tilde{3}57]}}^{(7),b}(\alpha,z)=
y_{_1}^{{D}-2}y_{_2}^{{D\over2}-1}y_{_4}^{-2}\sum\limits_{n_{_1}=0}^\infty
\sum\limits_{n_{_2}=0}^\infty\sum\limits_{n_{_3}=0}^\infty\sum\limits_{n_{_4}=0}^\infty
c_{_{[1\tilde{3}57]}}^{(7),b}(\alpha,{\bf n})
\nonumber\\
&&\hspace{2.5cm}\times
\Big({1\over y_{_1}}\Big)^{n_{_1}}\Big({y_{_3}\over y_{_4}}\Big)^{n_{_2}}
\Big({y_{_1}\over y_{_4}}\Big)^{n_{_3}}\Big({y_{_2}\over y_{_4}}\Big)^{n_{_4}}\;.
\label{GKZ21b-27-2a}
\end{eqnarray}
Where the coefficients are
\begin{eqnarray}
&&c_{_{[1\tilde{3}57]}}^{(7),a}(\alpha,{\bf n})=
(-)^{n_{_3}}\Gamma(1+n_{_2}+n_{_4})\Big\{n_{_1}!n_{_2}!n_{_3}!n_{_4}!
\Gamma(2-{D\over2}+n_{_1})
\nonumber\\
&&\hspace{2.5cm}\times
\Gamma({D\over2}-1-n_{_2}-n_{_4})\Gamma({D\over2}-1-n_{_1}-n_{_3})
\Gamma(2-{D\over2}+n_{_2})
\nonumber\\
&&\hspace{2.5cm}\times
\Gamma(2-{D\over2}+n_{_3})\Gamma(D-2-n_{_1}-n_{_3})
\Gamma({D\over2}+n_{_4})\Big\}^{-1}
\;,\nonumber\\
&&c_{_{[1\tilde{3}57]}}^{(7),b}(\alpha,{\bf n})=
\Gamma(1+n_{_2}+n_{_4})\Gamma(1+n_{_3})\Big\{n_{_1}!n_{_2}!n_{_4}!
\Gamma(2-{D\over2}+n_{_1})
\nonumber\\
&&\hspace{2.5cm}\times
\Gamma({D\over2}-1-n_{_2}-n_{_4})\Gamma({D\over2}-n_{_1}+n_{_3})
\Gamma(2-{D\over2}+n_{_2})
\nonumber\\
&&\hspace{2.5cm}\times
\Gamma(1-{D\over2}-n_{_3})\Gamma(D-1-n_{_1}+n_{_3})
\Gamma({D\over2}+n_{_4})\Big\}^{-1}\;.
\label{GKZ21b-27-3}
\end{eqnarray}

\item  $I_{_{8}}=\{1,3,4,6,\cdots,11,14\}$, i.e. the implement $J_{_{8}}=[1,14]\setminus I_{_{8}}=\{2,5,12,13\}$.
The choice implies the power numbers $\alpha_{_2}=\alpha_{_5}=\alpha_{_{12}}=\alpha_{_{13}}=0$, and
\begin{eqnarray}
&&\alpha_{_1}=a_{_2}-a_{_1},\;\alpha_{_3}=-a_{_3},
\;\alpha_{_4}=-a_{_4},
\nonumber\\
&&\alpha_{_{6}}=a_{_5}+b_{_1}-a_{_2}-1,\;\alpha_{_7}=b_{_2}-1,\;\alpha_{_8}=b_{_3}-1,
\nonumber\\
&&\alpha_{_9}=b_{_4}-a_{_5}-1,\;\alpha_{_{10}}=b_{_5}-a_{_5}-1,
\nonumber\\
&&\alpha_{_{11}}=a_{_5}-a_{_2},\;\alpha_{_{14}}=-a_{_5}\;.
\label{GKZ21b-28-1}
\end{eqnarray}
The corresponding hypergeometric series solution is written as
\begin{eqnarray}
&&\Phi_{_{[1\tilde{3}57]}}^{(8)}(\alpha,z)=
y_{_1}^{{3D\over2}-4}y_{_4}^{-1}\sum\limits_{n_{_1}=0}^\infty
\sum\limits_{n_{_2}=0}^\infty\sum\limits_{n_{_3}=0}^\infty\sum\limits_{n_{_4}=0}^\infty
c_{_{[1\tilde{3}57]}}^{(8)}(\alpha,{\bf n})
\nonumber\\
&&\hspace{2.5cm}\times
\Big({1\over y_{_1}}\Big)^{n_{_1}}\Big({y_{_3}\over y_{_1}}\Big)^{n_{_2}}
\Big({y_{_1}\over y_{_4}}\Big)^{n_{_3}}\Big({y_{_2}\over y_{_1}}\Big)^{n_{_4}}\;,
\label{GKZ21b-28-2}
\end{eqnarray}
with
\begin{eqnarray}
&&c_{_{[1\tilde{3}57]}}^{(8)}(\alpha,{\bf n})=
(-)^{n_{_3}}\Big\{n_{_1}!n_{_2}!n_{_4}!\Gamma(2-{D\over2}+n_{_1})
\Gamma({D\over2}-1-n_{_2}-n_{_4})
\nonumber\\
&&\hspace{2.5cm}\times
\Gamma(D-2-n_{_2}-n_{_4})\Gamma(D-2-n_{_1}-n_{_2}+n_{_3}-n_{_4})
\nonumber\\
&&\hspace{2.5cm}\times
\Gamma(2-{D\over2}+n_{_4})\Gamma(2-{D\over2}+n_{_2})\Gamma(2-{D\over2}+n_{_2}-n_{_3}+n_{_4})
\nonumber\\
&&\hspace{2.5cm}\times
\Gamma(3-D+n_{_2}-n_{_3}+n_{_4})\Gamma({3D\over2}-3-n_{_1}-n_{_2}+n_{_3}-n_{_4})\Big\}^{-1}\;.
\label{GKZ21b-28-3}
\end{eqnarray}

\item  $I_{_{9}}=\{2,3,4,5,9,\cdots,14\}$, i.e. the implement $J_{_{9}}=[1,14]\setminus I_{_{9}}=\{1,6,7,8\}$.
The choice implies the power numbers $\alpha_{_1}=\alpha_{_6}=\alpha_{_{7}}=\alpha_{_{8}}=0$, and
\begin{eqnarray}
&&\alpha_{_2}=a_{_1}-a_{_2},\;\alpha_{_3}=b_{_2}+b_{_3}-a_{_3}-2,
\;\alpha_{_4}=b_{_2}+b_{_3}-a_{_4}-2,
\nonumber\\
&&\alpha_{_{5}}=a_{_1}-a_{_5}-b_{_1}-b_{_2}-b_{_3}+3,\;
\alpha_{_9}=b_{_1}+b_{_4}-a_{_1}-2,
\nonumber\\
&&\alpha_{_{10}}=b_{_1}+b_{_5}-a_{_1}-2,\;
\alpha_{_{11}}=1-b_{_1},\;\alpha_{_{12}}=1-b_{_2},
\nonumber\\
&&\alpha_{_{13}}=1-b_{_3},\;\alpha_{_{14}}=b_{_1}+b_{_2}+b_{_3}-a_{_1}-3\;.
\label{GKZ21b-3-1}
\end{eqnarray}
The corresponding hypergeometric series solution is written as
\begin{eqnarray}
&&\Phi_{_{[1\tilde{3}57]}}^{(9)}(\alpha,z)=
y_{_1}^{{D\over2}-1}y_{_2}^{{D\over2}-1}y_{_3}^{{D\over2}-1}y_{_4}^{-{D\over2}-1}
\sum\limits_{n_{_1}=0}^\infty
\sum\limits_{n_{_2}=0}^\infty\sum\limits_{n_{_3}=0}^\infty\sum\limits_{n_{_4}=0}^\infty
c_{_{[1\tilde{3}57]}}^{(9)}(\alpha,{\bf n})
\nonumber\\
&&\hspace{2.5cm}\times
\Big({1\over y_{_4}}\Big)^{n_{_1}}\Big({y_{_3}\over y_{_4}}\Big)^{n_{_2}}
\Big({y_{_1}\over y_{_4}}\Big)^{n_{_3}}\Big({y_{_2}\over y_{_4}}\Big)^{n_{_4}}\;,
\label{GKZ21b-3-2}
\end{eqnarray}
with
\begin{eqnarray}
&&c_{_{[1\tilde{3}57]}}^{(9)}(\alpha,{\bf n})=
\Gamma(1+n_{_1}+n_{_3})\Gamma(1+n_{_2}+n_{_4})\Big\{n_{_1}!n_{_2}!n_{_3}!n_{_4}!
\nonumber\\
&&\hspace{2.5cm}\times
\Gamma({D\over2}+n_{_1})\Gamma(1-{D\over2}-n_{_2}-n_{_4})
\Gamma(1-{D\over2}-n_{_1}-n_{_3})
\nonumber\\
&&\hspace{2.5cm}\times
\Gamma({D\over2}+n_{_3})\Gamma({D\over2}+n_{_4})\Gamma({D\over2}+n_{_2})\Big\}^{-1}\;.
\label{GKZ21b-3-3}
\end{eqnarray}

\item  $I_{_{10}}=\{2,\cdots,5,7,9,10,11,13,14\}$, i.e. the implement $J_{_{10}}=[1,14]\setminus I_{_{10}}=\{1,6,8,12\}$.
The choice implies the power numbers $\alpha_{_1}=\alpha_{_6}=\alpha_{_{8}}=\alpha_{_{12}}=0$, and
\begin{eqnarray}
&&\alpha_{_2}=a_{_1}-a_{_2},\;\alpha_{_3}=b_{_3}-a_{_3}-1,
\;\alpha_{_4}=b_{_3}-a_{_4}-1,
\nonumber\\
&&\alpha_{_{5}}=a_{_1}-a_{_5}-b_{_1}-b_{_3}+2,\;\alpha_{_{7}}=b_{_2}-1,
\nonumber\\
&&\alpha_{_9}=b_{_1}+b_{_4}-a_{_1}-2,\;
\alpha_{_{10}}=b_{_1}+b_{_5}-a_{_1}-2,
\nonumber\\
&&\alpha_{_{11}}=1-b_{_1},\;
\alpha_{_{13}}=1-b_{_3},\;\alpha_{_{14}}=b_{_1}+b_{_3}-a_{_1}-2\;.
\label{GKZ21b-4-1}
\end{eqnarray}
The corresponding hypergeometric series solution is written as
\begin{eqnarray}
&&\Phi_{_{[1\tilde{3}57]}}^{(10)}(\alpha,z)=
y_{_1}^{{D\over2}-1}y_{_3}^{{D\over2}-1}y_{_4}^{-2}\sum\limits_{n_{_1}=0}^\infty
\sum\limits_{n_{_2}=0}^\infty\sum\limits_{n_{_3}=0}^\infty\sum\limits_{n_{_4}=0}^\infty
c_{_{[1\tilde{3}57]}}^{(10)}(\alpha,{\bf n})
\nonumber\\
&&\hspace{2.5cm}\times
\Big({1\over y_{_4}}\Big)^{n_{_1}}\Big({y_{_3}\over y_{_4}}\Big)^{n_{_2}}
\Big({y_{_1}\over y_{_4}}\Big)^{n_{_3}}\Big({y_{_2}\over y_{_4}}\Big)^{n_{_4}}\;,
\label{GKZ21b-4-2}
\end{eqnarray}
with
\begin{eqnarray}
&&c_{_{[1\tilde{3}57]}}^{(10)}(\alpha,{\bf n})=
\Gamma(1+n_{_1}+n_{_3})\Gamma(1+n_{_2}+n_{_4})
\Big\{n_{_1}!n_{_2}!n_{_3}!n_{_4}!
\nonumber\\
&&\hspace{2.5cm}\times
\Gamma({D\over2}+n_{_1})\Gamma({D\over2}-n_{_2}-n_{_4})
\Gamma(2-{D\over2}+n_{_4})
\nonumber\\
&&\hspace{2.5cm}\times
\Gamma(1-{D\over2}-n_{_1}-n_{_3})\Gamma({D\over2}+n_{_3})
\Gamma({D\over2}+n_{_2})\Big\}^{-1}\;.
\label{GKZ21b-4-3}
\end{eqnarray}

\item  $I_{_{11}}=\{2,\cdots,6,9,10,12,13,14\}$, i.e. the implement $J_{_{11}}=[1,14]\setminus I_{_{11}}=\{1,7,8,11\}$.
The choice implies the power numbers $\alpha_{_1}=\alpha_{_7}=\alpha_{_{8}}=\alpha_{_{11}}=0$, and
\begin{eqnarray}
&&\alpha_{_2}=a_{_1}-a_{_2},\;\alpha_{_3}=b_{_2}+b_{_3}-a_{_3}-2,
\nonumber\\
&&\alpha_{_4}=b_{_2}+b_{_3}-a_{_4}-2,\;
\alpha_{_{5}}=a_{_1}-a_{_5}-b_{_2}-b_{_3}+2,
\nonumber\\
&&\alpha_{_{6}}=b_{_1}-1,\;
\alpha_{_9}=b_{_4}-a_{_1}-1,\;\alpha_{_{10}}=b_{_5}-a_{_1}-1,
\nonumber\\
&&\alpha_{_{12}}=1-b_{_2},\;\alpha_{_{13}}=1-b_{_3},\;\alpha_{_{14}}=b_{_2}+b_{_3}-a_{_1}-2\;.
\label{GKZ21b-5-1}
\end{eqnarray}
The corresponding hypergeometric series solution is
\begin{eqnarray}
&&\Phi_{_{[1\tilde{3}57]}}^{(11)}(\alpha,z)=
y_{_2}^{{D\over2}-1}y_{_3}^{{D\over2}-1}y_{_4}^{-2}\sum\limits_{n_{_1}=0}^\infty
\sum\limits_{n_{_2}=0}^\infty\sum\limits_{n_{_3}=0}^\infty\sum\limits_{n_{_4}=0}^\infty
c_{_{[1\tilde{3}57]}}^{(11)}(\alpha,{\bf n})
\nonumber\\
&&\hspace{2.5cm}\times
\Big({1\over y_{_4}}\Big)^{n_{_1}}\Big({y_{_3}\over y_{_4}}\Big)^{n_{_2}}
\Big({y_{_1}\over y_{_4}}\Big)^{n_{_3}}\Big({y_{_2}\over y_{_4}}\Big)^{n_{_4}}\;,
\label{GKZ21b-5-2}
\end{eqnarray}
with
\begin{eqnarray}
&&c_{_{[1\tilde{3}57]}}^{(11)}(\alpha,{\bf n})=
\Gamma(1+n_{_1}+n_{_3})\Gamma(1+n_{_2}+n_{_4})\Big\{n_{_1}!n_{_2}!n_{_3}!n_{_4}!
\nonumber\\
&&\hspace{2.5cm}\times
\Gamma({D\over2}+n_{_1})\Gamma(1-{D\over2}-n_{_2}-n_{_4})
\Gamma(2-{D\over2}+n_{_3})
\nonumber\\
&&\hspace{2.5cm}\times
\Gamma({D\over2}-1-n_{_1}-n_{_3})
\Gamma({D\over2}+n_{_4})\Gamma({D\over2}+n_{_2})\Big\}^{-1}\;.
\label{GKZ21b-5-3}
\end{eqnarray}

\item  $I_{_{12}}=\{2,\cdots,7,9,10,13,14\}$, i.e. the implement $J_{_{12}}=[1,14]\setminus I_{_{12}}=\{1,8,11,12\}$.
The choice implies the power numbers $\alpha_{_1}=\alpha_{_8}=\alpha_{_{11}}=\alpha_{_{12}}=0$, and
\begin{eqnarray}
&&\alpha_{_2}=a_{_1}-a_{_2},\;\alpha_{_3}=b_{_3}-a_{_3}-1,
\;\alpha_{_4}=b_{_3}-a_{_4}-1,
\nonumber\\
&&\alpha_{_{5}}=a_{_1}-a_{_5}-b_{_3}+1,\;\alpha_{_{6}}=b_{_1}-1,\;\alpha_{_{7}}=b_{_2}-1,
\nonumber\\
&&\alpha_{_9}=b_{_4}-a_{_1}-1,\;
\alpha_{_{10}}=b_{_5}-a_{_1}-1,\;
\alpha_{_{13}}=1-b_{_3},
\nonumber\\
&&\alpha_{_{14}}=b_{_3}-a_{_1}-1\;.
\label{GKZ21b-6-1}
\end{eqnarray}
The corresponding hypergeometric series solution is written as
\begin{eqnarray}
&&\Phi_{_{[1\tilde{3}57]}}^{(12)}(\alpha,z)=
y_{_3}^{{D\over2}-1}y_{_4}^{{D\over2}-3}\sum\limits_{n_{_1}=0}^\infty
\sum\limits_{n_{_2}=0}^\infty\sum\limits_{n_{_3}=0}^\infty\sum\limits_{n_{_4}=0}^\infty
c_{_{[1\tilde{3}57]}}^{(12)}(\alpha,{\bf n})
\nonumber\\
&&\hspace{2.5cm}\times
\Big({1\over y_{_4}}\Big)^{n_{_1}}\Big({y_{_3}\over y_{_4}}\Big)^{n_{_2}}
\Big({y_{_1}\over y_{_4}}\Big)^{n_{_3}}\Big({y_{_2}\over y_{_4}}\Big)^{n_{_4}}\;,
\label{GKZ21b-6-2}
\end{eqnarray}
with
\begin{eqnarray}
&&c_{_{[1\tilde{3}57]}}^{(12)}(\alpha,{\bf n})=
\Gamma(1+n_{_1}+n_{_3})\Gamma(1+n_{_2}+n_{_4})\Big\{n_{_1}!n_{_2}!n_{_3}!n_{_4}!
\nonumber\\
&&\hspace{2.5cm}\times
\Gamma({D\over2}+n_{_1})\Gamma({D\over2}-1-n_{_2}-n_{_4})
\Gamma(2-{D\over2}+n_{_3})
\nonumber\\
&&\hspace{2.5cm}\times
\Gamma(2-{D\over2}+n_{_4})\Gamma({D\over2}-1-n_{_1}-n_{_3})
\Gamma({D\over2}+n_{_2})\Big\}^{-1}\;.
\label{GKZ21b-6-3}
\end{eqnarray}

\item  $I_{_{13}}=\{2,\cdots,5,8,\cdots,12,14\}$, i.e. the implement $J_{_{13}}=[1,14]\setminus I_{_{13}}=\{1,6,7,13\}$.
The choice implies the power numbers $\alpha_{_1}=\alpha_{_6}=\alpha_{_{7}}=\alpha_{_{13}}=0$, and
\begin{eqnarray}
&&\alpha_{_2}=a_{_1}-a_{_2},\;\alpha_{_3}=b_{_2}-a_{_3}-1,
\;\alpha_{_4}=b_{_2}-a_{_4}-1,
\nonumber\\
&&\alpha_{_{5}}=a_{_1}-a_{_5}-b_{_1}-b_{_2}+2,\;\alpha_{_{8}}=b_{_3}-1,
\nonumber\\
&&\alpha_{_9}=b_{_1}+b_{_4}-a_{_1}-2,\;
\alpha_{_{10}}=b_{_1}+b_{_5}-a_{_1}-2,
\nonumber\\
&&\alpha_{_{11}}=1-b_{_1},\;\alpha_{_{12}}=1-b_{_2},\;
\alpha_{_{14}}=b_{_1}+b_{_2}-a_{_1}-2\;.
\label{GKZ21b-13-1}
\end{eqnarray}
The corresponding hypergeometric series is
\begin{eqnarray}
&&\Phi_{_{[1\tilde{3}57]}}^{(13)}(\alpha,z)=
y_{_1}^{{D\over2}-1}y_{_2}^{{D\over2}-1}y_{_4}^{-2}\sum\limits_{n_{_1}=0}^\infty
\sum\limits_{n_{_2}=0}^\infty\sum\limits_{n_{_3}=0}^\infty\sum\limits_{n_{_4}=0}^\infty
c_{_{[1\tilde{3}57]}}^{(13)}(\alpha,{\bf n})
\nonumber\\
&&\hspace{2.5cm}\times
\Big({1\over y_{_4}}\Big)^{n_{_1}}\Big({y_{_3}\over y_{_4}}\Big)^{n_{_2}}
\Big({y_{_1}\over y_{_4}}\Big)^{n_{_3}}\Big({y_{_2}\over y_{_4}}\Big)^{n_{_4}}\;,
\label{GKZ21b-13-2}
\end{eqnarray}
with
\begin{eqnarray}
&&c_{_{[1\tilde{3}57]}}^{(13)}(\alpha,{\bf n})=
\Gamma(1+n_{_1}+n_{_3})\Gamma(1+n_{_2}+n_{_4})\Big\{n_{_1}!n_{_2}!n_{_3}!n_{_4}!
\nonumber\\
&&\hspace{2.5cm}\times
\Gamma({D\over2}+n_{_1})\Gamma({D\over2}-1-n_{_2}-n_{_4})
\Gamma(2-{D\over2}+n_{_2})
\nonumber\\
&&\hspace{2.5cm}\times
\Gamma(1-{D\over2}-n_{_1}-n_{_3})
\Gamma({D\over2}+n_{_3})\Gamma({D\over2}+n_{_4})\Big\}^{-1}\;.
\label{GKZ21b-13-3}
\end{eqnarray}

\item  $I_{_{14}}=\{2,\cdots,5,7,\cdots,11,14\}$, i.e. the implement $J_{_{14}}=[1,14]\setminus I_{_{14}}=\{1,6,12,13\}$.
The choice implies the power numbers $\alpha_{_1}=\alpha_{_6}=\alpha_{_{12}}=\alpha_{_{13}}=0$, and
\begin{eqnarray}
&&\alpha_{_2}=a_{_1}-a_{_2},\;\alpha_{_3}=-a_{_3},
\;\alpha_{_4}=-a_{_4},
\nonumber\\
&&\alpha_{_{5}}=a_{_1}-a_{_5}-b_{_1}+1,\;\alpha_{_{7}}=b_{_2}-1,\;\alpha_{_{8}}=b_{_3}-1,
\nonumber\\
&&\alpha_{_9}=b_{_1}+b_{_4}-a_{_1}-2,\;
\alpha_{_{10}}=b_{_1}+b_{_5}-a_{_1}-2,
\nonumber\\
&&\alpha_{_{11}}=1-b_{_1},\;\alpha_{_{14}}=b_{_1}-a_{_1}-1\;.
\label{GKZ21b-14-1}
\end{eqnarray}
The corresponding hypergeometric series solution is given as
\begin{eqnarray}
&&\Phi_{_{[1\tilde{3}57]}}^{(14)}(\alpha,z)=
y_{_1}^{{D\over2}-1}y_{_4}^{{D\over2}-3}\sum\limits_{n_{_1}=0}^\infty
\sum\limits_{n_{_2}=0}^\infty\sum\limits_{n_{_3}=0}^\infty\sum\limits_{n_{_4}=0}^\infty
c_{_{[1\tilde{3}57]}}^{(14)}(\alpha,{\bf n})
\nonumber\\
&&\hspace{2.5cm}\times
\Big({1\over y_{_4}}\Big)^{n_{_1}}\Big({y_{_3}\over y_{_4}}\Big)^{n_{_2}}
\Big({y_{_1}\over y_{_4}}\Big)^{n_{_3}}\Big({y_{_2}\over y_{_4}}\Big)^{n_{_4}}\;,
\label{GKZ21b-14-2}
\end{eqnarray}
with
\begin{eqnarray}
&&c_{_{[1\tilde{3}57]}}^{(14)}(\alpha,{\bf n})=(-)^{n_{_2}+n_{_4}}
\Gamma(1+n_{_1}+n_{_3})\Big\{n_{_1}!n_{_2}!n_{_3}!n_{_4}!\Gamma({D\over2}+n_{_1})
\nonumber\\
&&\hspace{2.5cm}\times
\Gamma({D\over2}-1-n_{_2}-n_{_4})\Gamma(D-2-n_{_2}-n_{_4})
\Gamma(2-{D\over2}+n_{_4})
\nonumber\\
&&\hspace{2.5cm}\times
\Gamma(2-{D\over2}+n_{_2})\Gamma(1-{D\over2}-n_{_1}-n_{_3})
\Gamma({D\over2}+n_{_3})\Big\}^{-1}\;.
\label{GKZ21b-14-3}
\end{eqnarray}

\item  $I_{_{15}}=\{2,\cdots,6,8,9,10,12,14\}$, i.e. the implement $J_{_{15}}=[1,14]\setminus I_{_{15}}=\{1,7,11,13\}$.
The choice implies the power numbers $\alpha_{_1}=\alpha_{_7}=\alpha_{_{11}}=\alpha_{_{13}}=0$, and
\begin{eqnarray}
&&\alpha_{_2}=a_{_1}-a_{_2},\;\alpha_{_3}=b_{_2}-a_{_3}-1,\;
\alpha_{_4}=b_{_2}-a_{_4}-1,
\nonumber\\
&&\alpha_{_{5}}=a_{_1}-a_{_5}-b_{_2}+1,\;
\alpha_{_{6}}=b_{_1}-1,\;\alpha_{_{8}}=b_{_3}-1,
\nonumber\\
&&\alpha_{_9}=b_{_4}-a_{_1}-1,\;\alpha_{_{10}}=b_{_5}-a_{_1}-1,\;
\alpha_{_{12}}=1-b_{_2},
\nonumber\\
&&\alpha_{_{14}}=b_{_2}-a_{_1}-1\;.
\label{GKZ21b-15-1}
\end{eqnarray}
The corresponding hypergeometric series solution is
\begin{eqnarray}
&&\Phi_{_{[1\tilde{3}57]}}^{(15)}(\alpha,z)=
y_{_2}^{{D\over2}-1}y_{_4}^{{D\over2}-3}\sum\limits_{n_{_1}=0}^\infty
\sum\limits_{n_{_2}=0}^\infty\sum\limits_{n_{_3}=0}^\infty\sum\limits_{n_{_4}=0}^\infty
c_{_{[1\tilde{3}57]}}^{(15)}(\alpha,{\bf n})
\nonumber\\
&&\hspace{2.5cm}\times
\Big({1\over y_{_4}}\Big)^{n_{_1}}\Big({y_{_3}\over y_{_4}}\Big)^{n_{_2}}
\Big({y_{_1}\over y_{_4}}\Big)^{n_{_3}}\Big({y_{_2}\over y_{_4}}\Big)^{n_{_4}}\;,
\label{GKZ21b-15-2}
\end{eqnarray}
with
\begin{eqnarray}
&&c_{_{[1\tilde{3}57]}}^{(15)}(\alpha,{\bf n})=
\Gamma(1+n_{_1}+n_{_3})\Gamma(1+n_{_2}+n_{_4})\Big\{n_{_1}!n_{_2}!n_{_3}!n_{_4}!
\nonumber\\
&&\hspace{2.5cm}\times
\Gamma({D\over2}+n_{_1})\Gamma({D\over2}-1-n_{_2}-n_{_4})
\Gamma(2-{D\over2}+n_{_3})
\nonumber\\
&&\hspace{2.5cm}\times
\Gamma(2-{D\over2}+n_{_2})\Gamma({D\over2}-1-n_{_1}-n_{_3})
\Gamma({D\over2}+n_{_4})\Big\}^{-1}\;.
\label{GKZ21b-15-3}
\end{eqnarray}

\item  $I_{_{16}}=\{2,\cdots,10,14\}$, i.e. the implement $J_{_{16}}=[1,14]\setminus I_{_{16}}=\{1,11,12,13\}$.
The choice implies the power numbers $\alpha_{_1}=\alpha_{_{11}}=\alpha_{_{12}}=\alpha_{_{13}}=0$, and
\begin{eqnarray}
&&\alpha_{_2}=a_{_1}-a_{_2},\;\alpha_{_3}=-a_{_3},
\;\alpha_{_4}=-a_{_4},\;\alpha_{_{5}}=a_{_1}-a_{_5},
\nonumber\\
&&\alpha_{_{6}}=b_{_1}-1,\;\alpha_{_{7}}=b_{_2}-1,\;\alpha_{_{8}}=b_{_3}-1,
\nonumber\\
&&\alpha_{_9}=b_{_4}-a_{_1}-1,\;
\alpha_{_{10}}=b_{_5}-a_{_1}-1,\;
\alpha_{_{14}}=-a_{_1}\;.
\label{GKZ21b-16-1}
\end{eqnarray}
The corresponding hypergeometric series solution is written as
\begin{eqnarray}
&&\Phi_{_{[1\tilde{3}57]}}^{(16)}(\alpha,z)=
y_{_4}^{{D}-4}\sum\limits_{n_{_1}=0}^\infty
\sum\limits_{n_{_2}=0}^\infty\sum\limits_{n_{_3}=0}^\infty\sum\limits_{n_{_4}=0}^\infty
c_{_{[1\tilde{3}57]}}^{(16)}(\alpha,{\bf n})
\nonumber\\
&&\hspace{2.5cm}\times
\Big({1\over y_{_4}}\Big)^{n_{_1}}\Big({y_{_3}\over y_{_4}}\Big)^{n_{_2}}
\Big({y_{_1}\over y_{_4}}\Big)^{n_{_3}}\Big({y_{_2}\over y_{_4}}\Big)^{n_{_4}}\;,
\label{GKZ21b-16-2}
\end{eqnarray}
with
\begin{eqnarray}
&&c_{_{[1\tilde{3}57]}}^{(16)}(\alpha,{\bf n})=
(-)^{1+n_{_2}+n_{_4}}\Gamma(1+n_{_1}+n_{_3})
\Big\{n_{_1}!n_{_2}!n_{_3}!n_{_4}!\Gamma({D\over2}+n_{_1})
\nonumber\\
&&\hspace{2.5cm}\times
\Gamma({D\over2}-1-n_{_2}-n_{_4})\Gamma(D-2-n_{_2}-n_{_4})
\Gamma(2-{D\over2}+n_{_3})
\nonumber\\
&&\hspace{2.5cm}\times
\Gamma(2-{D\over2}+n_{_4})\Gamma(2-{D\over2}+n_{_2})
\Gamma({D\over2}-1-n_{_1}-n_{_3})
\Big\}^{-1}\;.
\label{GKZ21b-16-3}
\end{eqnarray}

\item  $I_{_{17}}=\{1,3,4,5,9,\cdots,14\}$, i.e. the implement $J_{_{17}}=[1,14]\setminus I_{_{17}}=\{2,6,7,8\}$.
The choice implies the power numbers $\alpha_{_2}=\alpha_{_6}=\alpha_{_{7}}=\alpha_{_{8}}=0$, and
\begin{eqnarray}
&&\alpha_{_1}=a_{_2}-a_{_1},\;\alpha_{_3}=b_{_2}+b_{_3}-a_{_3}-2,
\;\alpha_{_4}=b_{_2}+b_{_3}-a_{_4}-2,
\nonumber\\
&&\alpha_{_{5}}=a_{_2}-a_{_5}-b_{_1}-b_{_2}-b_{_3}+3,\;
\alpha_{_9}=b_{_1}+b_{_4}-a_{_2}-2,
\nonumber\\
&&\alpha_{_{10}}=b_{_1}+b_{_5}-a_{_2}-2,\;
\alpha_{_{11}}=1-b_{_1},\;\alpha_{_{12}}=1-b_{_2},
\nonumber\\
&&\alpha_{_{13}}=1-b_{_3},\;\alpha_{_{14}}=b_{_1}+b_{_2}+b_{_3}-a_{_2}-3\;.
\label{GKZ21b-19-1}
\end{eqnarray}
The corresponding hypergeometric series is
\begin{eqnarray}
&&\Phi_{_{[1\tilde{3}57]}}^{(17)}(\alpha,z)=
y_{_1}^{{D\over2}-1}y_{_2}^{{D\over2}-1}y_{_3}^{{D\over2}-1}y_{_4}^{-2}
\sum\limits_{n_{_1}=0}^\infty
\sum\limits_{n_{_2}=0}^\infty\sum\limits_{n_{_3}=0}^\infty\sum\limits_{n_{_4}=0}^\infty
c_{_{[1\tilde{3}57]}}^{(17)}(\alpha,{\bf n})
\nonumber\\
&&\hspace{2.5cm}\times
\Big({1\over y_{_4}}\Big)^{n_{_1}}\Big({y_{_3}\over y_{_4}}\Big)^{n_{_2}}
\Big({y_{_1}\over y_{_4}}\Big)^{n_{_3}}\Big({y_{_2}\over y_{_4}}\Big)^{n_{_4}}\;,
\label{GKZ21b-19-2}
\end{eqnarray}
with
\begin{eqnarray}
&&c_{_{[1\tilde{3}57]}}^{(17)}(\alpha,{\bf n})=
\Gamma(1+n_{_1}+n_{_3})\Gamma(1+n_{_2}+n_{_4})\Big\{n_{_1}!n_{_2}!n_{_3}!n_{_4}!
\nonumber\\
&&\hspace{2.5cm}\times
\Gamma(2-{D\over2}+n_{_1})\Gamma(1-{D\over2}-n_{_2}-n_{_4})
\Gamma({D\over2}-1-n_{_1}-n_{_3})
\nonumber\\
&&\hspace{2.5cm}\times
\Gamma({D\over2}+n_{_3})\Gamma({D\over2}+n_{_4})\Gamma({D\over2}+n_{_2})\Big\}^{-1}\;.
\label{GKZ21b-19-3}
\end{eqnarray}

\item  $I_{_{18}}=\{1,3,4,5,7,9,10,11,13,14\}$, i.e. the implement $J_{_{18}}=[1,14]\setminus I_{_{18}}=\{2,6,8,12\}$.
The choice implies the power numbers $\alpha_{_2}=\alpha_{_6}=\alpha_{_{8}}=\alpha_{_{12}}=0$, and
\begin{eqnarray}
&&\alpha_{_1}=a_{_2}-a_{_1},\;\alpha_{_3}=b_{_3}-a_{_3}-1,
\;\alpha_{_4}=b_{_3}-a_{_4}-1,
\nonumber\\
&&\alpha_{_{5}}=a_{_2}-a_{_5}-b_{_1}-b_{_3}+2,\;\alpha_{_{7}}=b_{_2}-1,
\nonumber\\
&&\alpha_{_9}=b_{_1}+b_{_4}-a_{_2}-2,\;
\alpha_{_{10}}=b_{_1}+b_{_5}-a_{_2}-2,
\nonumber\\
&&\alpha_{_{11}}=1-b_{_1},\;
\alpha_{_{13}}=1-b_{_3},\;\alpha_{_{14}}=b_{_1}+b_{_3}-a_{_2}-2\;.
\label{GKZ21b-20-1}
\end{eqnarray}
The corresponding hypergeometric series solution is written as
\begin{eqnarray}
&&\Phi_{_{[1\tilde{3}57]}}^{(18)}(\alpha,z)=
y_{_1}^{{D\over2}-1}y_{_3}^{{D\over2}-1}y_{_4}^{{D\over2}-3}
\sum\limits_{n_{_1}=0}^\infty
\sum\limits_{n_{_2}=0}^\infty\sum\limits_{n_{_3}=0}^\infty\sum\limits_{n_{_4}=0}^\infty
c_{_{[1\tilde{3}57]}}^{(18)}(\alpha,{\bf n})
\nonumber\\
&&\hspace{2.5cm}\times
\Big({1\over y_{_4}}\Big)^{n_{_1}}\Big({y_{_3}\over y_{_4}}\Big)^{n_{_2}}
\Big({y_{_1}\over y_{_4}}\Big)^{n_{_3}}\Big({y_{_2}\over y_{_4}}\Big)^{n_{_4}}\;,
\label{GKZ21b-20-2}
\end{eqnarray}
with
\begin{eqnarray}
&&c_{_{[1\tilde{3}57]}}^{(18)}(\alpha,{\bf n})=
\Gamma(1+n_{_1}+n_{_3})
\Gamma(1+n_{_2}+n_{_4})\Big\{n_{_1}!n_{_2}!n_{_3}!n_{_4}!
\nonumber\\
&&\hspace{2.5cm}\times
\Gamma(2-{D\over2}+n_{_1})\Gamma({D\over2}-1-n_{_2}-n_{_4})
\Gamma(2-{D\over2}+n_{_4})
\nonumber\\
&&\hspace{2.5cm}\times
\Gamma({D\over2}-1-n_{_1}-n_{_3})
\Gamma({D\over2}+n_{_3})\Gamma({D\over2}+n_{_2})\Big\}^{-1}\;.
\label{GKZ21b-20-3}
\end{eqnarray}

\item  $I_{_{19}}=\{1,3,\cdots,6,9,10,12,13,14\}$, i.e. the implement $J_{_{19}}=[1,14]\setminus I_{_{19}}=\{2,7,8,11\}$.
The choice implies the power numbers $\alpha_{_2}=\alpha_{_7}=\alpha_{_{8}}=\alpha_{_{11}}=0$, and
\begin{eqnarray}
&&\alpha_{_1}=a_{_2}-a_{_1},\;\alpha_{_3}=b_{_2}+b_{_3}-a_{_3}-2,
\nonumber\\
&&\alpha_{_4}=b_{_2}+b_{_3}-a_{_4}-2,\;
\alpha_{_{5}}=a_{_2}-a_{_5}-b_{_2}-b_{_3}+2,
\nonumber\\
&&\alpha_{_{6}}=b_{_1}-1,\;
\alpha_{_9}=b_{_4}-a_{_2}-1,\;\alpha_{_{10}}=b_{_5}-a_{_2}-1,
\nonumber\\
&&\alpha_{_{12}}=1-b_{_2},\;\alpha_{_{13}}=1-b_{_3},\;\alpha_{_{14}}=b_{_2}+b_{_3}-a_{_2}-2\;.
\label{GKZ21b-21-1}
\end{eqnarray}
The corresponding hypergeometric series is written as
\begin{eqnarray}
&&\Phi_{_{[1\tilde{3}57]}}^{(19)}(\alpha,z)=
y_{_2}^{{D\over2}-1}y_{_3}^{{D\over2}-1}y_{_4}^{{D\over2}-3}
\sum\limits_{n_{_1}=0}^\infty
\sum\limits_{n_{_2}=0}^\infty\sum\limits_{n_{_3}=0}^\infty\sum\limits_{n_{_4}=0}^\infty
c_{_{[1\tilde{3}57]}}^{(19)}(\alpha,{\bf n})
\nonumber\\
&&\hspace{2.5cm}\times
\Big({1\over y_{_4}}\Big)^{n_{_1}}\Big({y_{_3}\over y_{_4}}\Big)^{n_{_2}}
\Big({y_{_1}\over y_{_4}}\Big)^{n_{_3}}\Big({y_{_2}\over y_{_4}}\Big)^{n_{_4}}\;,
\label{GKZ21b-21-2}
\end{eqnarray}
with
\begin{eqnarray}
&&c_{_{[1\tilde{3}57]}}^{(19)}(\alpha,{\bf n})=
(-)^{n_{_1}+n_{_3}}\Gamma(1+n_{_2}+n_{_4})\Big\{n_{_1}!n_{_2}!n_{_3}!n_{_4}!
\nonumber\\
&&\hspace{2.5cm}\times
\Gamma(2-{D\over2}+n_{_1})\Gamma(1-{D\over2}-n_{_2}-n_{_4})
\Gamma(2-{D\over2}+n_{_3})
\nonumber\\
&&\hspace{2.5cm}\times
\Gamma(D-2-n_{_1}-n_{_3})\Gamma({D\over2}-1-n_{_1}-n_{_3})\Gamma({D\over2}+n_{_4})
\nonumber\\
&&\hspace{2.5cm}\times
\Gamma({D\over2}+n_{_2})\Big\}^{-1}\;.
\label{GKZ21b-21-3}
\end{eqnarray}

\item  $I_{_{20}}=\{1,3,\cdots,7,9,10,13,14\}$, i.e. the implement $J_{_{20}}=[1,14]\setminus I_{_{20}}=\{2,8,11,12\}$.
The choice implies the power numbers $\alpha_{_2}=\alpha_{_8}=\alpha_{_{11}}=\alpha_{_{12}}=0$, and
\begin{eqnarray}
&&\alpha_{_1}=a_{_2}-a_{_1},\;\alpha_{_3}=b_{_3}-a_{_3}-1,
\;\alpha_{_4}=b_{_3}-a_{_4}-1,
\nonumber\\
&&\alpha_{_{5}}=a_{_2}-a_{_5}-b_{_3}+1,\;\alpha_{_{6}}=b_{_1}-1,\;\alpha_{_{7}}=b_{_2}-1,
\nonumber\\
&&\alpha_{_9}=b_{_4}-a_{_2}-1,\;
\alpha_{_{10}}=b_{_5}-a_{_2}-1,\;
\alpha_{_{13}}=1-b_{_3},
\nonumber\\
&&\alpha_{_{14}}=b_{_3}-a_{_2}-1\;.
\label{GKZ21b-22-1}
\end{eqnarray}
The corresponding hypergeometric series is
\begin{eqnarray}
&&\Phi_{_{[1\tilde{3}57]}}^{(20)}(\alpha,z)=
y_{_3}^{{D\over2}-1}y_{_4}^{{D}-4}\sum\limits_{n_{_1}=0}^\infty
\sum\limits_{n_{_2}=0}^\infty\sum\limits_{n_{_3}=0}^\infty\sum\limits_{n_{_4}=0}^\infty
c_{_{[1\tilde{3}57]}}^{(20)}(\alpha,{\bf n})
\nonumber\\
&&\hspace{2.5cm}\times
\Big({1\over y_{_4}}\Big)^{n_{_1}}\Big({y_{_3}\over y_{_4}}\Big)^{n_{_2}}
\Big({y_{_1}\over y_{_4}}\Big)^{n_{_3}}\Big({y_{_2}\over y_{_4}}\Big)^{n_{_4}}\;,
\label{GKZ21b-22-2}
\end{eqnarray}
with
\begin{eqnarray}
&&c_{_{[1\tilde{3}57]}}^{(20)}(\alpha,{\bf n})=
(-)^{1+n_{_1}+n_{_3}}\Gamma(1+n_{_2}+n_{_4})
\Big\{n_{_1}!n_{_2}!n_{_3}!n_{_4}!
\nonumber\\
&&\hspace{2.5cm}\times
\Gamma(2-{D\over2}+n_{_1})\Gamma({D\over2}-1-n_{_2}-n_{_4})
\Gamma(2-{D\over2}+n_{_3})
\nonumber\\
&&\hspace{2.5cm}\times
\Gamma(2-{D\over2}+n_{_4})\Gamma(D-2-n_{_1}-n_{_3})
\nonumber\\
&&\hspace{2.5cm}\times
\Gamma({D\over2}-1-n_{_1}-n_{_3})\Gamma({D\over2}+n_{_2})\Big\}^{-1}\;.
\label{GKZ21b-22-3}
\end{eqnarray}

\item  $I_{_{21}}=\{1,3,4,5,8,\cdots,12,14\}$, i.e. the implement $J_{_{21}}=[1,14]\setminus I_{_{21}}=\{2,6,7,13\}$.
The choice implies the power numbers $\alpha_{_2}=\alpha_{_6}=\alpha_{_{7}}=\alpha_{_{13}}=0$, and
\begin{eqnarray}
&&\alpha_{_1}=a_{_2}-a_{_1},\;\alpha_{_3}=b_{_2}-a_{_3}-1,
\;\alpha_{_4}=b_{_2}-a_{_4}-1,
\nonumber\\
&&\alpha_{_{5}}=a_{_2}-a_{_5}-b_{_1}-b_{_2}+2,\;\alpha_{_{8}}=b_{_3}-1,
\nonumber\\
&&\alpha_{_9}=b_{_1}+b_{_4}-a_{_2}-2,\;
\alpha_{_{10}}=b_{_1}+b_{_5}-a_{_2}-2,
\nonumber\\
&&\alpha_{_{11}}=1-b_{_1},\;\alpha_{_{12}}=1-b_{_2},\;
\alpha_{_{14}}=b_{_1}+b_{_2}-a_{_2}-2\;.
\label{GKZ21b-29-1}
\end{eqnarray}
The corresponding hypergeometric series is written as
\begin{eqnarray}
&&\Phi_{_{[1\tilde{3}57]}}^{(21)}(\alpha,z)=
y_{_1}^{{D\over2}-1}y_{_2}^{{D\over2}-1}y_{_4}^{{D\over2}-3}\sum\limits_{n_{_1}=0}^\infty
\sum\limits_{n_{_2}=0}^\infty\sum\limits_{n_{_3}=0}^\infty\sum\limits_{n_{_4}=0}^\infty
c_{_{[1\tilde{3}57]}}^{(21)}(\alpha,{\bf n})
\nonumber\\
&&\hspace{2.5cm}\times
\Big({1\over y_{_4}}\Big)^{n_{_1}}\Big({y_{_3}\over y_{_4}}\Big)^{n_{_2}}
\Big({y_{_1}\over y_{_4}}\Big)^{n_{_3}}\Big({y_{_2}\over y_{_4}}\Big)^{n_{_4}}\;,
\label{GKZ21b-29-2}
\end{eqnarray}
with
\begin{eqnarray}
&&c_{_{[1\tilde{3}57]}}^{(21)}(\alpha,{\bf n})=
\Gamma(1+n_{_1}+n_{_3})\Gamma(1+n_{_2}+n_{_4})\Big\{n_{_1}!n_{_2}!n_{_3}!n_{_4}!
\nonumber\\
&&\hspace{2.5cm}\times
\Gamma(2-{D\over2}+n_{_1})\Gamma({D\over2}-1-n_{_2}-n_{_4})
\Gamma(2-{D\over2}+n_{_2})
\nonumber\\
&&\hspace{2.5cm}\times
\Gamma({D\over2}-1-n_{_1}-n_{_3})\Gamma({D\over2}+n_{_3})
\Gamma({D\over2}+n_{_4})\Big\}^{-1}\;.
\label{GKZ21b-29-3}
\end{eqnarray}

\item  $I_{_{22}}=\{1,3,4,5,7,\cdots,11,14\}$, i.e. the implement $J_{_{22}}=[1,14]\setminus I_{_{22}}=\{2,6,12,13\}$.
The choice implies the power numbers $\alpha_{_2}=\alpha_{_6}=\alpha_{_{12}}=\alpha_{_{13}}=0$, and
\begin{eqnarray}
&&\alpha_{_1}=a_{_2}-a_{_1},\;\alpha_{_3}=-a_{_3},
\;\alpha_{_4}=-a_{_4},
\nonumber\\
&&\alpha_{_{5}}=a_{_2}-a_{_5}-b_{_1}+1,\;\alpha_{_{7}}=b_{_2}-1,\;\alpha_{_{8}}=b_{_3}-1,
\nonumber\\
&&\alpha_{_9}=b_{_1}+b_{_4}-a_{_2}-2,\;
\alpha_{_{10}}=b_{_1}+b_{_5}-a_{_2}-2,
\nonumber\\
&&\alpha_{_{11}}=1-b_{_1},\;\alpha_{_{14}}=b_{_1}-a_{_2}-1\;.
\label{GKZ21b-30-1}
\end{eqnarray}
The corresponding hypergeometric series solution is written as
\begin{eqnarray}
&&\Phi_{_{[1\tilde{3}57]}}^{(22)}(\alpha,z)=
y_{_1}^{{D\over2}-1}y_{_4}^{{D}-4}\sum\limits_{n_{_1}=0}^\infty
\sum\limits_{n_{_2}=0}^\infty\sum\limits_{n_{_3}=0}^\infty\sum\limits_{n_{_4}=0}^\infty
c_{_{[1\tilde{3}57]}}^{(22)}(\alpha,{\bf n})
\nonumber\\
&&\hspace{2.5cm}\times
\Big({1\over y_{_4}}\Big)^{n_{_1}}\Big({y_{_3}\over y_{_4}}\Big)^{n_{_2}}
\Big({y_{_1}\over y_{_4}}\Big)^{n_{_3}}\Big({y_{_2}\over y_{_4}}\Big)^{n_{_4}}\;,
\label{GKZ21b-30-2}
\end{eqnarray}
with
\begin{eqnarray}
&&c_{_{[1\tilde{3}57]}}^{(22)}(\alpha,{\bf n})=
(-)^{1+n_{_2}+n_{_4}}\Gamma(1+n_{_1}+n_{_3})\Big\{n_{_1}!n_{_2}!n_{_3}!n_{_4}!
\Gamma(2-{D\over2}+n_{_1})
\nonumber\\
&&\hspace{2.5cm}\times
\Gamma({D\over2}-1-n_{_2}-n_{_4})\Gamma(D-2-n_{_2}-n_{_4})
\Gamma(2-{D\over2}+n_{_4})
\nonumber\\
&&\hspace{2.5cm}\times
\Gamma(2-{D\over2}+n_{_2})
\Gamma({D\over2}-1-n_{_1}-n_{_3})\Gamma({D\over2}+n_{_3})\Big\}^{-1}\;.
\label{GKZ21b-30-3}
\end{eqnarray}

\item  $I_{_{23}}=\{1,3,\cdots,6,8,9,10,12,14\}$, i.e. the implement $J_{_{23}}=[1,14]\setminus I_{_{23}}=\{2,7,11,13\}$.
The choice implies the power numbers $\alpha_{_2}=\alpha_{_7}=\alpha_{_{11}}=\alpha_{_{13}}=0$, and
\begin{eqnarray}
&&\alpha_{_1}=a_{_2}-a_{_1},\;\alpha_{_3}=b_{_2}-a_{_3}-1,\;
\alpha_{_4}=b_{_2}-a_{_4}-1,
\nonumber\\
&&\alpha_{_{5}}=a_{_2}-a_{_5}-b_{_2}+1,\;
\alpha_{_{6}}=b_{_1}-1,\;\alpha_{_{8}}=b_{_3}-1,
\nonumber\\
&&\alpha_{_9}=b_{_4}-a_{_2}-1,\;\alpha_{_{10}}=b_{_5}-a_{_2}-1,\;
\alpha_{_{12}}=1-b_{_2},
\nonumber\\
&&\alpha_{_{14}}=b_{_2}-a_{_2}-1\;.
\label{GKZ21b-31-1}
\end{eqnarray}
The corresponding hypergeometric series is written as
\begin{eqnarray}
&&\Phi_{_{[1\tilde{3}57]}}^{(23)}(\alpha,z)=
y_{_2}^{{D\over2}-1}y_{_4}^{{D}-4}\sum\limits_{n_{_1}=0}^\infty
\sum\limits_{n_{_2}=0}^\infty\sum\limits_{n_{_3}=0}^\infty\sum\limits_{n_{_4}=0}^\infty
c_{_{[1\tilde{3}57]}}^{(23)}(\alpha,{\bf n})
\nonumber\\
&&\hspace{2.5cm}\times
\Big({1\over y_{_4}}\Big)^{n_{_1}}\Big({y_{_3}\over y_{_4}}\Big)^{n_{_2}}
\Big({y_{_1}\over y_{_4}}\Big)^{n_{_3}}\Big({y_{_2}\over y_{_4}}\Big)^{n_{_4}}\;,
\label{GKZ21b-31-2}
\end{eqnarray}
with
\begin{eqnarray}
&&c_{_{[1\tilde{3}57]}}^{(23)}(\alpha,{\bf n})=
(-)^{1+n_{_1}+n_{_3}}\Gamma(1+n_{_2}+n_{_4})\Big\{n_{_1}!n_{_2}!n_{_3}!n_{_4}!
\Gamma(2-{D\over2}+n_{_1})
\nonumber\\
&&\hspace{2.5cm}\times
\Gamma({D\over2}-1-n_{_2}-n_{_4})\Gamma(2-{D\over2}+n_{_3})
\Gamma(2-{D\over2}+n_{_2})
\nonumber\\
&&\hspace{2.5cm}\times
\Gamma(D-2-n_{_1}-n_{_3})\Gamma({D\over2}-1-n_{_1}-n_{_3})
\Gamma({D\over2}+n_{_4})\Big\}^{-1}\;.
\label{GKZ21b-31-3}
\end{eqnarray}

\item  $I_{_{24}}=\{1,3,\cdots,10,14\}$, i.e. the implement $J_{_{24}}=[1,14]\setminus I_{_{24}}=\{2,11,12,13\}$.
The choice implies the power numbers $\alpha_{_2}=\alpha_{_{11}}=\alpha_{_{12}}=\alpha_{_{13}}=0$, and
\begin{eqnarray}
&&\alpha_{_1}=a_{_2}-a_{_1},\;\alpha_{_3}=-a_{_3},
\;\alpha_{_4}=-a_{_4},\;\alpha_{_{5}}=a_{_2}-a_{_5},
\nonumber\\
&&\alpha_{_{6}}=b_{_1}-1,\;\alpha_{_{7}}=b_{_2}-1,\;\alpha_{_{8}}=b_{_3}-1,
\nonumber\\
&&\alpha_{_9}=b_{_4}-a_{_2}-1,\;
\alpha_{_{10}}=b_{_5}-a_{_2}-1,\;
\alpha_{_{14}}=-a_{_2}\;.
\label{GKZ21b-32-1}
\end{eqnarray}
The corresponding hypergeometric series is
\begin{eqnarray}
&&\Phi_{_{[1\tilde{3}57]}}^{(24)}(\alpha,z)=
y_{_4}^{{3D\over2}-5}\sum\limits_{n_{_1}=0}^\infty
\sum\limits_{n_{_2}=0}^\infty\sum\limits_{n_{_3}=0}^\infty\sum\limits_{n_{_4}=0}^\infty
c_{_{[1\tilde{3}57]}}^{(24)}(\alpha,{\bf n})
\nonumber\\
&&\hspace{2.5cm}\times
\Big({1\over y_{_4}}\Big)^{n_{_1}}\Big({y_{_3}\over y_{_4}}\Big)^{n_{_2}}
\Big({y_{_1}\over y_{_4}}\Big)^{n_{_3}}\Big({y_{_2}\over y_{_4}}\Big)^{n_{_4}}\;,
\label{GKZ21b-32-2}
\end{eqnarray}
with
\begin{eqnarray}
&&c_{_{[1\tilde{3}57]}}^{(24)}(\alpha,{\bf n})=
(-)^{n_{_1}+n_{_2}+n_{_3}+n_{_4}}
\Big\{n_{_1}!n_{_2}!n_{_3}!n_{_4}!\Gamma(2-{D\over2}+n_{_1})
\Gamma({D\over2}-1-n_{_2}-n_{_4})
\nonumber\\
&&\hspace{2.5cm}\times
\Gamma(D-2-n_{_2}-n_{_4})\Gamma(2-{D\over2}+n_{_3})
\Gamma(2-{D\over2}+n_{_4})
\nonumber\\
&&\hspace{2.5cm}\times
\Gamma(2-{D\over2}+n_{_2})\Gamma(D-2-n_{_1}-n_{_3})
\Gamma({D\over2}-1-n_{_1}-n_{_3})\Big\}^{-1}\;.
\label{GKZ21b-32-3}
\end{eqnarray}

\item  $I_{_{25}}=\{2,3,4,6,8,10,\cdots,14\}$, i.e. the implement $J_{_{25}}=[1,14]\setminus I_{_{25}}=\{1,5,7,9\}$.
The choice implies the power numbers $\alpha_{_1}=\alpha_{_5}=\alpha_{_{7}}=\alpha_{_{9}}=0$, and
\begin{eqnarray}
&&\alpha_{_2}=a_{_1}-a_{_2},\;\alpha_{_3}=b_{_4}-a_{_3}-a_{_5}-1,
\;\alpha_{_4}=b_{_4}-a_{_4}-a_{_5}-1,
\nonumber\\
&&\alpha_{_{6}}=b_{_1}+b_{_4}-a_{_1}-2,\;\alpha_{_8}=a_{_5}+b_{_2}+b_{_3}-b_{_4}-1,
\nonumber\\
&&\alpha_{_{10}}=b_{_5}-b_{_4},\;\alpha_{_{11}}=b_{_4}-a_{_1}-1,\;\alpha_{_{12}}=1-b_{_2},
\nonumber\\
&&\alpha_{_{13}}=a_{_5}+b_{_2}-b_{_4},\;\alpha_{_{14}}=-a_{_5}\;.
\label{GKZ21b-7-1}
\end{eqnarray}
The corresponding hypergeometric series solutions are written as
\begin{eqnarray}
&&\Phi_{_{[1\tilde{3}57]}}^{(25),a}(\alpha,z)=
y_{_1}^{{D\over2}-2}y_{_2}^{{D\over2}-1}y_{_4}^{-1}\sum\limits_{n_{_1}=0}^\infty
\sum\limits_{n_{_2}=0}^\infty\sum\limits_{n_{_3}=0}^\infty\sum\limits_{n_{_4}=0}^\infty
c_{_{[1\tilde{3}57]}}^{(25),a}(\alpha,{\bf n})
\nonumber\\
&&\hspace{2.5cm}\times
\Big({1\over y_{_1}}\Big)^{n_{_1}}\Big({y_{_4}\over y_{_1}}\Big)^{n_{_2}}
\Big({y_{_3}\over y_{_4}}\Big)^{n_{_3}}\Big({y_{_2}\over y_{_4}}\Big)^{n_{_4}}
\;,\nonumber\\
&&\Phi_{_{[1\tilde{3}57]}}^{(25),b}(\alpha,z)=
y_{_1}^{{D\over2}-2}y_{_2}^{{D\over2}}y_{_3}^{-1}y_{_4}^{-1}\sum\limits_{n_{_1}=0}^\infty
\sum\limits_{n_{_2}=0}^\infty\sum\limits_{n_{_3}=0}^\infty\sum\limits_{n_{_4}=0}^\infty
c_{_{[1\tilde{3}57]}}^{(25),b}(\alpha,{\bf n})
\nonumber\\
&&\hspace{2.5cm}\times
\Big({1\over y_{_1}}\Big)^{n_{_1}}\Big({y_{_2}\over y_{_1}}\Big)^{n_{_2}}
\Big({y_{_2}\over y_{_4}}\Big)^{n_{_3}}\Big({y_{_2}\over y_{_3}}\Big)^{n_{_4}}\;.
\label{GKZ21b-7-2a}
\end{eqnarray}
Where the coefficients are
\begin{eqnarray}
&&c_{_{[1\tilde{3}57]}}^{(25),a}(\alpha,{\bf n})=
(-)^{n_{_1}}\Gamma(1+n_{_1}+n_{_2})\Gamma(1+n_{_3}+n_{_4})
\nonumber\\
&&\hspace{2.5cm}\times
\Big\{n_{_1}!n_{_2}!n_{_3}!n_{_4}!\Gamma({D\over2}+n_{_1})
\Gamma({D\over2}-1-n_{_3}-n_{_4})
\nonumber\\
&&\hspace{2.5cm}\times
\Gamma({D\over2}-1-n_{_1}-n_{_2})
\Gamma(2-{D\over2}+n_{_3})
\nonumber\\
&&\hspace{2.5cm}\times
\Gamma(2-{D\over2}+n_{_2})\Gamma({D\over2}+n_{_4})\Big\}^{-1}
\;,\nonumber\\
&&c_{_{[1\tilde{3}57]}}^{(25),b}(\alpha,{\bf n})=
(-)^{n_{_1}+n_{_4}}\Gamma(1+n_{_1}+n_{_2})\Gamma(1+n_{_2}+n_{_3})\Gamma(1+n_{_4})
\nonumber\\
&&\hspace{2.5cm}\times
\Big\{n_{_1}!n_{_2}!\Gamma(2+n_{_2}+n_{_3}+n_{_4})\Gamma({D\over2}+n_{_1})
\nonumber\\
&&\hspace{2.5cm}\times
\Gamma({D\over2}-1-n_{_2}-n_{_3})\Gamma({D\over2}-1-n_{_1}-n_{_2})
\Gamma(1-{D\over2}-n_{_4})
\nonumber\\
&&\hspace{2.5cm}\times
\Gamma(2-{D\over2}+n_{_2})\Gamma({D\over2}+1+n_{_2}+n_{_3}+n_{_4})\Big\}^{-1}\;.
\label{GKZ21b-7-3}
\end{eqnarray}

\item  $I_{_{26}}=\{2,3,4,6,7,8,10,11,13,14\}$, i.e. the implement $J_{_{26}}=[1,14]\setminus I_{_{26}}=\{1,5,9,12\}$.
The choice implies the power numbers $\alpha_{_1}=\alpha_{_5}=\alpha_{_{9}}=\alpha_{_{12}}=0$, and
\begin{eqnarray}
&&\alpha_{_2}=a_{_1}-a_{_2},\;\alpha_{_3}=b_{_4}-a_{_3}-a_{_5}-1,
\nonumber\\
&&\alpha_{_4}=b_{_4}-a_{_4}-a_{_5}-1,\;
\alpha_{_{6}}=b_{_1}+b_{_4}-a_{_1}-2,
\nonumber\\
&&\alpha_{_{7}}=b_{_2}-1,\;\alpha_{_8}=a_{_5}+b_{_3}-b_{_4},\;
\alpha_{_{10}}=b_{_5}-b_{_4},
\nonumber\\
&&\alpha_{_{11}}=b_{_4}-a_{_1}-1,\;\alpha_{_{13}}=a_{_5}-b_{_4}+1,\;\alpha_{_{14}}=-a_{_5}\;.
\label{GKZ21b-8-1}
\end{eqnarray}
The corresponding hypergeometric series functions are written as
\begin{eqnarray}
&&\Phi_{_{[1\tilde{3}57]}}^{(26),a}(\alpha,z)=
y_{_1}^{{D\over2}-2}y_{_3}^{{D\over2}-1}y_{_4}^{-1}\sum\limits_{n_{_1}=0}^\infty
\sum\limits_{n_{_2}=0}^\infty\sum\limits_{n_{_3}=0}^\infty\sum\limits_{n_{_4}=0}^\infty
c_{_{[1\tilde{3}57]}}^{(26),a}(\alpha,{\bf n})
\nonumber\\
&&\hspace{2.5cm}\times
\Big({1\over y_{_1}}\Big)^{n_{_1}}\Big({y_{_4}\over y_{_1}}\Big)^{n_{_2}}
\Big({y_{_3}\over y_{_4}}\Big)^{n_{_3}}\Big({y_{_2}\over y_{_4}}\Big)^{n_{_4}}
\;,\nonumber\\
&&\Phi_{_{[1\tilde{3}57]}}^{(26),b}(\alpha,z)=
y_{_1}^{{D\over2}-2}y_{_2}y_{_3}^{{D\over2}-2}y_{_4}^{-1}\sum\limits_{n_{_1}=0}^\infty
\sum\limits_{n_{_2}=0}^\infty\sum\limits_{n_{_3}=0}^\infty\sum\limits_{n_{_4}=0}^\infty
c_{_{[1\tilde{3}57]}}^{(26),b}(\alpha,{\bf n})
\nonumber\\
&&\hspace{2.5cm}\times
\Big({1\over y_{_1}}\Big)^{n_{_1}}\Big({y_{_2}\over y_{_1}}\Big)^{n_{_2}}
\Big({y_{_2}\over y_{_4}}\Big)^{n_{_3}}\Big({y_{_2}\over y_{_3}}\Big)^{n_{_4}}\;.
\label{GKZ21b-8-2a}
\end{eqnarray}
Where the coefficients are
\begin{eqnarray}
&&c_{_{[1\tilde{3}57]}}^{(26),a}(\alpha,{\bf n})=
(-)^{n_{_1}}\Gamma(1+n_{_1}+n_{_2})\Gamma(1+n_{_3}+n_{_4})
\Big\{n_{_1}!n_{_2}!n_{_3}!n_{_4}!
\nonumber\\
&&\hspace{2.5cm}\times
\Gamma({D\over2}+n_{_1})\Gamma({D\over2}-1-n_{_3}-n_{_4})
\Gamma({D\over2}-1-n_{_1}-n_{_2})
\nonumber\\
&&\hspace{2.5cm}\times
\Gamma(2-{D\over2}+n_{_4})\Gamma(2-{D\over2}+n_{_2})\Gamma({D\over2}+n_{_3})\Big\}^{-1}
\;,\nonumber\\
&&c_{_{[1\tilde{3}57]}}^{(26),b}(\alpha,{\bf n})=
(-)^{n_{_1}+n_{_4}}\Gamma(1+n_{_1}+n_{_2})\Gamma(1+n_{_2}+n_{_3})\Gamma(1+n_{_4})
\nonumber\\
&&\hspace{2.5cm}\times
\Big\{n_{_1}!n_{_2}!\Gamma(2+n_{_2}+n_{_3}+n_{_4})\Gamma({D\over2}+n_{_1})
\Gamma(2-{D\over2}+n_{_2})
\nonumber\\
&&\hspace{2.5cm}\times
\Gamma({D\over2}-1-n_{_2}-n_{_3})
\Gamma({D\over2}-1-n_{_1}-n_{_2})
\nonumber\\
&&\hspace{2.5cm}\times
\Gamma(3-{D\over2}+n_{_2}+n_{_3}+n_{_4})
\Gamma({D\over2}-1-n_{_4})\Big\}^{-1}\;.
\label{GKZ21b-8-3}
\end{eqnarray}

\item  $I_{_{27}}=\{2,3,4,6,8,9,11,\cdots,14\}$, i.e. the implement $J_{_{27}}=[1,14]\setminus I_{_{27}}=\{1,5,7,10\}$.
The choice implies the power numbers $\alpha_{_1}=\alpha_{_5}=\alpha_{_{7}}=\alpha_{_{10}}=0$, and
\begin{eqnarray}
&&\alpha_{_2}=a_{_1}-a_{_2},\;\alpha_{_3}=b_{_5}-a_{_3}-a_{_5}-1,
\;\alpha_{_4}=b_{_5}-a_{_4}-a_{_5}-1,
\nonumber\\
&&\alpha_{_{6}}=b_{_1}+b_{_5}-a_{_1}-2,\;\alpha_{_8}=a_{_5}+b_{_2}+b_{_3}-b_{_5}-1,
\nonumber\\
&&\alpha_{_{9}}=b_{_4}-b_{_5},\;\alpha_{_{11}}=b_{_5}-a_{_1}-1,\;\alpha_{_{12}}=1-b_{_2},
\nonumber\\
&&\alpha_{_{13}}=a_{_5}+b_{_2}-b_{_5},\;\alpha_{_{14}}=-a_{_5}\;.
\label{GKZ21b-9-1}
\end{eqnarray}
The corresponding hypergeometric series solutions are
\begin{eqnarray}
&&\Phi_{_{[1\tilde{3}57]}}^{(27),a}(\alpha,z)=
y_{_1}^{-1}y_{_2}^{{D\over2}-1}y_{_3}^{{D\over2}-1}y_{_4}^{-1}
\sum\limits_{n_{_1}=0}^\infty
\sum\limits_{n_{_2}=0}^\infty\sum\limits_{n_{_3}=0}^\infty\sum\limits_{n_{_4}=0}^\infty
c_{_{[1\tilde{3}57]}}^{(27),a}(\alpha,{\bf n})
\nonumber\\
&&\hspace{2.5cm}\times
\Big({1\over y_{_1}}\Big)^{n_{_1}}\Big({y_{_4}\over y_{_1}}\Big)^{n_{_2}}
\Big({y_{_3}\over y_{_4}}\Big)^{n_{_3}}\Big({y_{_2}\over y_{_4}}\Big)^{n_{_4}}
\;,\nonumber\\
&&\Phi_{_{[1\tilde{3}57]}}^{(27),b}(\alpha,z)=
y_{_1}^{-1}y_{_2}^{{D\over2}}y_{_3}^{{D\over2}-2}y_{_4}^{-1}\sum\limits_{n_{_1}=0}^\infty
\sum\limits_{n_{_2}=0}^\infty\sum\limits_{n_{_3}=0}^\infty\sum\limits_{n_{_4}=0}^\infty
c_{_{[1\tilde{3}57]}}^{(27),a}(\alpha,{\bf n})
\nonumber\\
&&\hspace{2.5cm}\times
\Big({1\over y_{_1}}\Big)^{n_{_1}}\Big({y_{_2}\over y_{_1}}\Big)^{n_{_2}}
\Big({y_{_2}\over y_{_4}}\Big)^{n_{_3}}\Big({y_{_2}\over y_{_4}}\Big)^{n_{_4}}\;.
\label{GKZ21b-9-2a}
\end{eqnarray}
Where the coefficients are
\begin{eqnarray}
&&c_{_{[1\tilde{3}57]}}^{(27),a}(\alpha,{\bf n})=
(-)^{n_{_1}}\Gamma(1+n_{_1}+n_{_2})\Gamma(1+n_{_3}+n_{_4})
\Big\{n_{_1}!n_{_2}!n_{_3}!n_{_4}!\Gamma({D\over2}+n_{_1})
\nonumber\\
&&\hspace{2.5cm}\times
\Gamma(1-{D\over2}-n_{_3}-n_{_4})\Gamma(1-{D\over2}-n_{_1}-n_{_2})
\Gamma({D\over2}+n_{_2})
\nonumber\\
&&\hspace{2.5cm}\times
\Gamma({D\over2}+n_{_3})\Gamma({D\over2}+n_{_4})\Big\}^{-1}
\;,\nonumber\\
&&c_{_{[1\tilde{3}57]}}^{(27),b}(\alpha,{\bf n})=
(-)^{n_{_1}+n_{_4}}\Gamma(1+n_{_1}+n_{_2})\Gamma(1+n_{_2}+n_{_3})\Gamma(1+n_{_4})
\Big\{n_{_1}!n_{_2}!
\nonumber\\
&&\hspace{2.5cm}\times
\Gamma(2+n_{_2}+n_{_3}+n_{_4})\Gamma({D\over2}+n_{_1})
\Gamma(1-{D\over2}-n_{_2}-n_{_3})
\nonumber\\
&&\hspace{2.5cm}\times
\Gamma(1-{D\over2}-n_{_1}-n_{_2})\Gamma({D\over2}+n_{_2})
\nonumber\\
&&\hspace{2.5cm}\times
\Gamma({D\over2}+1+n_{_2}+n_{_3}+n_{_4})
\Gamma({D\over2}-1-n_{_4})\Big\}^{-1}\;.
\label{GKZ21b-9-3}
\end{eqnarray}

\item  $I_{_{28}}=\{2,3,4,6,\cdots,9,11,13,14\}$, i.e. the implement $J_{_{28}}=[1,14]\setminus I_{_{28}}=\{1,5,10,12\}$.
The choice implies the power numbers $\alpha_{_1}=\alpha_{_5}=\alpha_{_{10}}=\alpha_{_{12}}=0$, and
\begin{eqnarray}
&&\alpha_{_2}=a_{_1}-a_{_2},\;\alpha_{_3}=b_{_5}-a_{_3}-a_{_5}-1,
\nonumber\\
&&\alpha_{_4}=b_{_5}-a_{_4}-a_{_5}-1,\;
\alpha_{_{6}}=b_{_1}+b_{_5}-a_{_1}-2,
\nonumber\\
&&\alpha_{_{7}}=b_{_2}-1,\;\alpha_{_8}=a_{_5}+b_{_3}-b_{_5},\;
\alpha_{_{9}}=b_{_4}-b_{_5},
\nonumber\\
&&\alpha_{_{11}}=b_{_5}-a_{_1}-1,\;\alpha_{_{13}}=a_{_5}-b_{_5}+1,\;\alpha_{_{14}}=-a_{_5}\;.
\label{GKZ21b-10-1}
\end{eqnarray}
The corresponding hypergeometric series is
\begin{eqnarray}
&&\Phi_{_{[1\tilde{3}57]}}^{(28)}(\alpha,z)=
y_{_1}^{-1}y_{_3}^{{D}-2}y_{_4}^{-1}\sum\limits_{n_{_1}=0}^\infty
\sum\limits_{n_{_2}=0}^\infty\sum\limits_{n_{_3}=0}^\infty\sum\limits_{n_{_4}=0}^\infty
c_{_{[1\tilde{3}57]}}^{(28)}(\alpha,{\bf n})
\nonumber\\
&&\hspace{2.5cm}\times
\Big({1\over y_{_1}}\Big)^{n_{_1}}\Big({y_{_3}\over y_{_1}}\Big)^{n_{_2}}
\Big({y_{_3}\over y_{_4}}\Big)^{n_{_3}}\Big({y_{_2}\over y_{_3}}\Big)^{n_{_4}}\;,
\label{GKZ21b-10-2}
\end{eqnarray}
with
\begin{eqnarray}
&&c_{_{[1\tilde{3}57]}}^{(28)}(\alpha,{\bf n})=
(-)^{n_{_1}}\Gamma(1+n_{_1}+n_{_2})\Gamma(1+n_{_2}+n_{_3})
\Big\{n_{_1}!n_{_2}!n_{_4}!
\nonumber\\
&&\hspace{2.5cm}\times
\Gamma({D\over2}+n_{_1})\Gamma(1-{D\over2}-n_{_2}-n_{_3})
\Gamma(1-{D\over2}-n_{_1}-n_{_2})
\nonumber\\
&&\hspace{2.5cm}\times
\Gamma(2-{D\over2}+n_{_4})\Gamma({D\over2}+n_{_2}+n_{_3}-n_{_4})
\Gamma({D\over2}+n_{_2})
\nonumber\\
&&\hspace{2.5cm}\times
\Gamma(D-1+n_{_2}+n_{_3}-n_{_4})\Big\}^{-1}\;.
\label{GKZ21b-10-3}
\end{eqnarray}

\item  $I_{_{29}}=\{1,3,4,6,8,10,\cdots,14\}$, i.e. the implement $J_{_{29}}=[1,14]\setminus I_{_{29}}=\{2,5,7,9\}$.
The choice implies the power numbers $\alpha_{_2}=\alpha_{_5}=\alpha_{_{7}}=\alpha_{_{9}}=0$, and
\begin{eqnarray}
&&\alpha_{_1}=a_{_2}-a_{_1},\;\alpha_{_3}=b_{_4}-a_{_3}-a_{_5}-1,
\;\alpha_{_4}=b_{_4}-a_{_4}-a_{_5}-1,
\nonumber\\
&&\alpha_{_{6}}=b_{_1}+b_{_4}-a_{_2}-2,\;\alpha_{_8}=a_{_5}+b_{_2}+b_{_3}-b_{_4}-1,
\nonumber\\
&&\alpha_{_{10}}=b_{_5}-b_{_4},\;\alpha_{_{11}}=b_{_4}-a_{_2}-1,\;\alpha_{_{12}}=1-b_{_2},
\nonumber\\
&&\alpha_{_{13}}=a_{_5}+b_{_2}-b_{_4},\;\alpha_{_{14}}=-a_{_5}\;.
\label{GKZ21b-23-1}
\end{eqnarray}
The corresponding hypergeometric series solutions are presented as
\begin{eqnarray}
&&\Phi_{_{[1\tilde{3}57]}}^{(29),a}(\alpha,z)=
y_{_1}^{{D}-3}y_{_2}^{{D\over2}-1}y_{_4}^{-1}\sum\limits_{n_{_1}=0}^\infty
\sum\limits_{n_{_2}=0}^\infty\sum\limits_{n_{_3}=0}^\infty\sum\limits_{n_{_4}=0}^\infty
c_{_{[1\tilde{3}57]}}^{(29),a}(\alpha,{\bf n})
\nonumber\\
&&\hspace{2.5cm}\times
\Big({1\over y_{_1}}\Big)^{n_{_1}}\Big({y_{_4}\over y_{_1}}\Big)^{n_{_2}}
\Big({y_{_3}\over y_{_4}}\Big)^{n_{_3}}\Big({y_{_2}\over y_{_4}}\Big)^{n_{_4}}
\;,\nonumber\\
&&\Phi_{_{[1\tilde{3}57]}}^{(29),b}(\alpha,z)=
y_{_1}^{{D}-3}y_{_2}^{{D\over2}}y_{_3}^{-1}y_{_4}^{-1}\sum\limits_{n_{_1}=0}^\infty
\sum\limits_{n_{_2}=0}^\infty\sum\limits_{n_{_3}=0}^\infty\sum\limits_{n_{_4}=0}^\infty
c_{_{[1\tilde{3}57]}}^{(29),b}(\alpha,{\bf n})
\nonumber\\
&&\hspace{2.5cm}\times
\Big({1\over y_{_1}}\Big)^{n_{_1}}\Big({y_{_2}\over y_{_1}}\Big)^{n_{_2}}
\Big({y_{_2}\over y_{_4}}\Big)^{n_{_3}}\Big({y_{_2}\over y_{_3}}\Big)^{n_{_4}}\;.
\label{GKZ21b-23-2a}
\end{eqnarray}
Where the coefficients are
\begin{eqnarray}
&&c_{_{[1\tilde{3}57]}}^{(29),a}(\alpha,{\bf n})=
(-)^{n_{_2}}\Gamma(1+n_{_3}+n_{_4})
\Big\{n_{_1}!n_{_2}!n_{_3}!n_{_4}!\Gamma(2-{D\over2}+n_{_1})
\nonumber\\
&&\hspace{2.5cm}\times
\Gamma({D\over2}-1-n_{_3}-n_{_4})\Gamma({D\over2}-1-n_{_1}-n_{_2})
\Gamma(2-{D\over2}+n_{_3})
\nonumber\\
&&\hspace{2.5cm}\times
\Gamma(2-{D\over2}+n_{_2})\Gamma(D-2-n_{_1}-n_{_2})\Gamma({D\over2}+n_{_4})\Big\}^{-1}
\;,\nonumber\\
&&c_{_{[1\tilde{3}57]}}^{(29),b}(\alpha,{\bf n})=
(-)^{n_{_2}+n_{_4}}\Gamma(1+n_{_2}+n_{_3})\Gamma(1+n_{_4})
\Big\{n_{_1}!n_{_2}!
\nonumber\\
&&\hspace{2.5cm}\times
\Gamma(2+n_{_2}+n_{_3}+n_{_4})\Gamma(2-{D\over2}+n_{_1})
\Gamma({D\over2}-1-n_{_2}-n_{_3})
\nonumber\\
&&\hspace{2.5cm}\times
\Gamma({D\over2}-1-n_{_1}-n_{_2})\Gamma(1-{D\over2}-n_{_4})
\Gamma(2-{D\over2}+n_{_2})
\nonumber\\
&&\hspace{2.5cm}\times
\Gamma(D-2-n_{_1}-n_{_2})
\Gamma({D\over2}+1+n_{_2}+n_{_3}+n_{_4})\Big\}^{-1}\;.
\label{GKZ21b-23-3}
\end{eqnarray}

\item  $I_{_{30}}=\{1,3,4,6,7,8,10,11,13,14\}$, i.e. the implement $J_{_{30}}=[1,14]\setminus I_{_{30}}=\{2,5,9,12\}$.
The choice implies the power numbers $\alpha_{_2}=\alpha_{_5}=\alpha_{_{9}}=\alpha_{_{12}}=0$, and
\begin{eqnarray}
&&\alpha_{_1}=a_{_2}-a_{_1},\;\alpha_{_3}=b_{_4}-a_{_3}-a_{_5}-1,
\nonumber\\
&&\alpha_{_4}=b_{_4}-a_{_4}-a_{_5}-1,\;
\alpha_{_{6}}=b_{_1}+b_{_4}-a_{_2}-2,
\nonumber\\
&&\alpha_{_{7}}=b_{_2}-1,\;\alpha_{_8}=a_{_5}+b_{_3}-b_{_4},\;
\alpha_{_{10}}=b_{_5}-b_{_4},
\nonumber\\
&&\alpha_{_{11}}=b_{_4}-a_{_2}-1,\;\alpha_{_{13}}=a_{_5}-b_{_4}+1,\;\alpha_{_{14}}=-a_{_5}\;.
\label{GKZ21b-24-1}
\end{eqnarray}
The corresponding hypergeometric series solutions are
\begin{eqnarray}
&&\Phi_{_{[1\tilde{3}57]}}^{(30),a}(\alpha,z)=
y_{_1}^{{D}-3}y_{_3}^{{D\over2}-1}y_{_4}^{-1}\sum\limits_{n_{_1}=0}^\infty
\sum\limits_{n_{_2}=0}^\infty\sum\limits_{n_{_3}=0}^\infty\sum\limits_{n_{_4}=0}^\infty
c_{_{[1\tilde{3}57]}}^{(30),a}(\alpha,{\bf n})
\nonumber\\
&&\hspace{2.5cm}\times
\Big({1\over y_{_1}}\Big)^{n_{_1}}\Big({y_{_4}\over y_{_1}}\Big)^{n_{_2}}
\Big({y_{_3}\over y_{_4}}\Big)^{n_{_3}}\Big({y_{_2}\over y_{_4}}\Big)^{n_{_4}}
\;,\nonumber\\
&&\Phi_{_{[1\tilde{3}57]}}^{(30),b}(\alpha,z)=
y_{_1}^{{D}-3}y_{_2}y_{_3}^{{D\over2}-2}y_{_4}^{-1}\sum\limits_{n_{_1}=0}^\infty
\sum\limits_{n_{_2}=0}^\infty\sum\limits_{n_{_3}=0}^\infty\sum\limits_{n_{_4}=0}^\infty
c_{_{[1\tilde{3}57]}}^{(30),b}(\alpha,{\bf n})
\nonumber\\
&&\hspace{2.5cm}\times
\Big({1\over y_{_1}}\Big)^{n_{_1}}\Big({y_{_2}\over y_{_1}}\Big)^{n_{_2}}
\Big({y_{_2}\over y_{_4}}\Big)^{n_{_3}}\Big({y_{_2}\over y_{_3}}\Big)^{n_{_4}}\;.
\label{GKZ21b-24-2a}
\end{eqnarray}
Where the coefficients are
\begin{eqnarray}
&&c_{_{[1\tilde{3}57]}}^{(30),a}(\alpha,{\bf n})=
(-)^{n_{_2}}\Gamma(1+n_{_3}+n_{_4})
\Big\{n_{_1}!n_{_2}!n_{_3}!n_{_4}!\Gamma(2-{D\over2}+n_{_1})
\nonumber\\
&&\hspace{2.5cm}\times
\Gamma({D\over2}-1-n_{_3}-n_{_4})\Gamma({D\over2}-1-n_{_1}-n_{_2})
\Gamma(2-{D\over2}+n_{_4})
\nonumber\\
&&\hspace{2.5cm}\times
\Gamma(2-{D\over2}+n_{_2})\Gamma(D-2-n_{_1}-n_{_2})\Gamma({D\over2}+n_{_3})\Big\}^{-1}
\;,\nonumber\\
&&c_{_{[1\tilde{3}57]}}^{(30),b}(\alpha,{\bf n})=
(-)^{n_{_2}+n_{_4}}\Gamma(1+n_{_2}+n_{_3})\Gamma(1+n_{_4})
\Big\{n_{_1}!n_{_2}!
\nonumber\\
&&\hspace{2.5cm}\times
\Gamma(2+n_{_2}+n_{_3}+n_{_4})\Gamma(2-{D\over2}+n_{_1})
\Gamma({D\over2}-1-n_{_2}-n_{_3})
\nonumber\\
&&\hspace{2.5cm}\times
\Gamma({D\over2}-1-n_{_1}-n_{_2})\Gamma(3-{D\over2}+n_{_2}+n_{_3}+n_{_4})
\nonumber\\
&&\hspace{2.5cm}\times
\Gamma(2-{D\over2}+n_{_2})\Gamma(D-2-n_{_1}-n_{_2})
\Gamma({D\over2}-1-n_{_4})\Big\}^{-1}\;.
\label{GKZ21b-24-3}
\end{eqnarray}

\item  $I_{_{31}}=\{1,3,4,6,8,9,11,\cdots,14\}$, i.e. the implement $J_{_{31}}=[1,14]\setminus I_{_{31}}=\{2,5,7,10\}$.
The choice implies the power numbers $\alpha_{_2}=\alpha_{_5}=\alpha_{_{7}}=\alpha_{_{10}}=0$, and
\begin{eqnarray}
&&\alpha_{_1}=a_{_2}-a_{_1},\;\alpha_{_3}=b_{_5}-a_{_3}-a_{_5}-1,
\;\alpha_{_4}=b_{_5}-a_{_4}-a_{_5}-1,
\nonumber\\
&&\alpha_{_{6}}=b_{_1}+b_{_5}-a_{_2}-2,\;\alpha_{_8}=a_{_5}+b_{_2}+b_{_3}-b_{_5}-1,
\nonumber\\
&&\alpha_{_{9}}=b_{_4}-b_{_5},\;\alpha_{_{11}}=b_{_5}-a_{_2}-1,\;\alpha_{_{12}}=1-b_{_2},
\nonumber\\
&&\alpha_{_{13}}=a_{_5}+b_{_2}-b_{_5},\;\alpha_{_{14}}=-a_{_5}\;.
\label{GKZ21b-25-1}
\end{eqnarray}
The corresponding hypergeometric series solutions are presented as
\begin{eqnarray}
&&\Phi_{_{[1\tilde{3}57]}}^{(31),a}(\alpha,z)=
y_{_1}^{{D\over2}-2}y_{_2}^{{D\over2}-1}y_{_3}^{{D\over2}-1}y_{_4}^{-1}
\sum\limits_{n_{_1}=0}^\infty
\sum\limits_{n_{_2}=0}^\infty\sum\limits_{n_{_3}=0}^\infty\sum\limits_{n_{_4}=0}^\infty
c_{_{[1\tilde{3}57]}}^{(31),a}(\alpha,{\bf n})
\nonumber\\
&&\hspace{2.5cm}\times
\Big({1\over y_{_1}}\Big)^{n_{_1}}\Big({y_{_4}\over y_{_1}}\Big)^{n_{_2}}
\Big({y_{_3}\over y_{_4}}\Big)^{n_{_3}}\Big({y_{_2}\over y_{_4}}\Big)^{n_{_4}}
\;,\nonumber\\
&&\Phi_{_{[1\tilde{3}57]}}^{(31),b}(\alpha,z)=
y_{_1}^{{D\over2}-2}y_{_2}^{{D\over2}}y_{_3}^{{D\over2}-2}y_{_4}^{-1}
\sum\limits_{n_{_1}=0}^\infty
\sum\limits_{n_{_2}=0}^\infty\sum\limits_{n_{_3}=0}^\infty\sum\limits_{n_{_4}=0}^\infty
c_{_{[1\tilde{3}57]}}^{(31),b}(\alpha,{\bf n})
\nonumber\\
&&\hspace{2.5cm}\times
\Big({1\over y_{_1}}\Big)^{n_{_1}}\Big({y_{_2}\over y_{_1}}\Big)^{n_{_2}}
\Big({y_{_2}\over y_{_4}}\Big)^{n_{_3}}\Big({y_{_2}\over y_{_3}}\Big)^{n_{_4}}\;.
\label{GKZ21b-25-2a}
\end{eqnarray}
Where the coefficients are
\begin{eqnarray}
&&c_{_{[1\tilde{3}57]}}^{(31),a}(\alpha,{\bf n})=
(-)^{n_{_1}}\Gamma(1+n_{_1}+n_{_2})\Gamma(1+n_{_3}+n_{_4})
\nonumber\\
&&\hspace{2.5cm}\times
\Big\{n_{_1}!n_{_2}!n_{_3}!n_{_4}!\Gamma(2-{D\over2}+n_{_1})
\nonumber\\
&&\hspace{2.5cm}\times
\Gamma(1-{D\over2}-n_{_3}-n_{_4})\Gamma({D\over2}-1-n_{_1}-n_{_2})
\nonumber\\
&&\hspace{2.5cm}\times
\Gamma({D\over2}+n_{_4})
\Gamma({D\over2}+n_{_2})\Gamma({D\over2}+n_{_3})\Big\}^{-1}
\;,\nonumber\\
&&c_{_{[1\tilde{3}57]}}^{(31),b}(\alpha,{\bf n})=
(-)^{n_{_1}+n_{_4}}\Gamma(1+n_{_1}+n_{_2})\Gamma(1+n_{_2}+n_{_3})\Gamma(1+n_{_4})
\nonumber\\
&&\hspace{2.5cm}\times
\Big\{n_{_1}!n_{_2}!
\Gamma(2+n_{_2}+n_{_3}+n_{_4})\Gamma(2-{D\over2}+n_{_1})
\nonumber\\
&&\hspace{2.5cm}\times
\Gamma(1-{D\over2}-n_{_2}-n_{_3})\Gamma({D\over2}+n_{_2})
\Gamma({D\over2}-1-n_{_1}-n_{_2})
\nonumber\\
&&\hspace{2.5cm}\times
\Gamma({D\over2}+1+n_{_2}+n_{_3}+n_{_4})\Gamma({D\over2}-1-n_{_4})\Big\}^{-1}\;.
\label{GKZ21b-25-3}
\end{eqnarray}

\item  $I_{_{32}}=\{1,3,4,6,\cdots,9,11,13,14\}$, i.e. the implement $J_{_{32}}=[1,14]\setminus I_{_{32}}=\{2,5,10,12\}$.
The choice implies the power numbers $\alpha_{_2}=\alpha_{_5}=\alpha_{_{10}}=\alpha_{_{12}}=0$, and
\begin{eqnarray}
&&\alpha_{_1}=a_{_2}-a_{_1},\;\alpha_{_3}=b_{_5}-a_{_3}-a_{_5}-1,
\nonumber\\
&&\alpha_{_4}=b_{_5}-a_{_4}-a_{_5}-1,\;
\alpha_{_{6}}=b_{_1}+b_{_5}-a_{_2}-2,
\nonumber\\
&&\alpha_{_{7}}=b_{_2}-1,\;\alpha_{_8}=a_{_5}+b_{_3}-b_{_5},\;
\alpha_{_{9}}=b_{_4}-b_{_5},
\nonumber\\
&&\alpha_{_{11}}=b_{_5}-a_{_2}-1,\;\alpha_{_{13}}=a_{_5}-b_{_5}+1,\;\alpha_{_{14}}=-a_{_5}\;.
\label{GKZ21b-26-1}
\end{eqnarray}
The corresponding hypergeometric series solutions is
\begin{eqnarray}
&&\Phi_{_{[1\tilde{3}57]}}^{(32)}(\alpha,z)=
y_{_1}^{{D\over2}-2}y_{_3}^{{D}-2}y_{_4}^{-1}\sum\limits_{n_{_1}=0}^\infty
\sum\limits_{n_{_2}=0}^\infty\sum\limits_{n_{_3}=0}^\infty\sum\limits_{n_{_4}=0}^\infty
c_{_{[1\tilde{3}57]}}^{(32)}(\alpha,{\bf n})
\nonumber\\
&&\hspace{2.5cm}\times
\Big({1\over y_{_1}}\Big)^{n_{_1}}\Big({y_{_3}\over y_{_1}}\Big)^{n_{_2}}
\Big({y_{_3}\over y_{_4}}\Big)^{n_{_3}}\Big({y_{_2}\over y_{_3}}\Big)^{n_{_4}}\;,
\label{GKZ21b-26-2}
\end{eqnarray}
with
\begin{eqnarray}
&&c_{_{[1\tilde{3}57]}}^{(32)}(\alpha,{\bf n})=
(-)^{n_{_1}}\Gamma(1+n_{_1}+n_{_2})\Gamma(1+n_{_2}+n_{_3})
\Big\{n_{_1}!n_{_2}!n_{_4}!
\nonumber\\
&&\hspace{2.5cm}\times
\Gamma(2-{D\over2}+n_{_1})\Gamma(1-{D\over2}-n_{_2}-n_{_3})
\Gamma(2-{D\over2}+n_{_4})
\nonumber\\
&&\hspace{2.5cm}\times
\Gamma({D\over2}+n_{_2}+n_{_3}-n_{_4})\Gamma({D\over2}+n_{_2})\Gamma({D\over2}-1-n_{_1}-n_{_2})
\nonumber\\
&&\hspace{2.5cm}\times
\Gamma(D-1+n_{_2}+n_{_3}-n_{_4})\Big\}^{-1}\;.
\label{GKZ21b-26-3}
\end{eqnarray}
\end{itemize}

\section{The hypergeometric solutions of the integer lattice ${\bf B}_{_{13\tilde{5}7}}$\label{app4}}
\indent\indent

\begin{itemize}
\item   $I_{_{1}}=\{2,4,5,6,8,10,\cdots,14\}$, i.e. the implement $J_{_{1}}=[1,14]\setminus I_{_{1}}=\{1,3,7,9\}$.
The choice implies the power numbers $\alpha_{_1}=\alpha_{_{3}}=\alpha_{_{7}}=\alpha_{_{9}}=0$, and
\begin{eqnarray}
&&\alpha_{_2}=a_{_1}-a_{_2},\;\alpha_{_4}=a_{_3}-a_{_4},
\;\alpha_{_5}=b_{_4}-a_{_3}-a_{_5}-1,
\nonumber\\
&&\alpha_{_{6}}=b_{_1}+b_{_4}-a_{_1}-2,\;\alpha_{_{8}}=b_{_2}+b_{_3}-a_{_3}-2,
\nonumber\\
&&\alpha_{_{10}}=b_{_5}-b_{_4},\;\alpha_{_{11}}=b_{_4}-a_{_1}-1,\;\alpha_{_{12}}=1-b_{_2},
\nonumber\\
&&\alpha_{_{13}}=b_{_2}-a_{_3}-1,\;\alpha_{_{14}}=a_{_3}-b_{_4}+1\;.
\label{GKZ21c-1-1}
\end{eqnarray}
The corresponding hypergeometric series is written as
\begin{eqnarray}
&&\Phi_{_{[13\tilde{5}7]}}^{(1)}(\alpha,z)=
y_{_1}^{{D\over2}-2}y_{_2}^{{D\over2}-1}y_{_3}^{-1}\sum\limits_{n_{_1}=0}^\infty
\sum\limits_{n_{_2}=0}^\infty\sum\limits_{n_{_3}=0}^\infty\sum\limits_{n_{_4}=0}^\infty
c_{_{[13\tilde{5}7]}}^{(1)}(\alpha,{\bf n})
\nonumber\\
&&\hspace{2.5cm}\times
\Big({1\over y_{_1}}\Big)^{n_{_1}}\Big({y_{_4}\over y_{_3}}\Big)^{n_{_2}}
\Big({y_{_4}\over y_{_1}}\Big)^{n_{_3}}\Big({y_{_2}\over y_{_3}}\Big)^{n_{_4}}\;,
\label{GKZ21c-1-2}
\end{eqnarray}
with
\begin{eqnarray}
&&c_{_{[13\tilde{5}7]}}^{(1)}(\alpha,{\bf n})=
(-)^{n_{_1}+n_{_4}}\Gamma(1+n_{_1}+n_{_3})\Gamma(1+n_{_2}+n_{_4})
\Big\{n_{_1}!n_{_2}!n_{_3}!n_{_4}!
\nonumber\\
&&\hspace{2.5cm}\times
\Gamma({D\over2}+n_{_1})\Gamma({D\over2}+n_{_2})
\Gamma(1-{D\over2}-n_{_2}-n_{_4})
\nonumber\\
&&\hspace{2.5cm}\times
\Gamma(2-{D\over2}+n_{_3})
\Gamma({D\over2}-1-n_{_1}-n_{_3})\Gamma({D\over2}+n_{_4})\Big\}^{-1}\;.
\label{GKZ21c-1-3}
\end{eqnarray}

\item   $I_{_{2}}=\{2,4,\cdots,8,10,11,13,14\}$, i.e. the implement $J_{_{2}}=[1,14]\setminus I_{_{2}}=\{1,3,9,12\}$.
The choice implies the power numbers $\alpha_{_1}=\alpha_{_{3}}=\alpha_{_{9}}=\alpha_{_{12}}=0$, and
\begin{eqnarray}
&&\alpha_{_2}=a_{_1}-a_{_2},\;\alpha_{_4}=a_{_3}-a_{_4},
\;\alpha_{_5}=b_{_4}-a_{_3}-a_{_5}-1,
\nonumber\\
&&\alpha_{_{6}}=b_{_1}+b_{_4}-a_{_1}-2,\;\alpha_{_{7}}=b_{_2}-1,\;\alpha_{_{8}}=b_{_3}-a_{_3}-1,
\nonumber\\
&&\alpha_{_{10}}=b_{_5}-b_{_4},\;\alpha_{_{11}}=b_{_4}-a_{_1}-1,\;\alpha_{_{13}}=-a_{_3},
\nonumber\\
&&\alpha_{_{14}}=a_{_3}-b_{_4}+1\;.
\label{GKZ21c-2-1}
\end{eqnarray}
The corresponding hypergeometric series solution is written as
\begin{eqnarray}
&&\Phi_{_{[13\tilde{5}7]}}^{(2)}(\alpha,z)=
y_{_1}^{{D\over2}-2}y_{_3}^{{D\over2}-2}\sum\limits_{n_{_1}=0}^\infty
\sum\limits_{n_{_2}=0}^\infty\sum\limits_{n_{_3}=0}^\infty\sum\limits_{n_{_4}=0}^\infty
c_{_{[13\tilde{5}7]}}^{(2)}(\alpha,{\bf n})
\nonumber\\
&&\hspace{2.5cm}\times
\Big({1\over y_{_1}}\Big)^{n_{_1}}\Big({y_{_4}\over y_{_3}}\Big)^{n_{_2}}
\Big({y_{_4}\over y_{_1}}\Big)^{n_{_3}}\Big({y_{_2}\over y_{_3}}\Big)^{n_{_4}}\;,
\label{GKZ21c-2-2}
\end{eqnarray}
with
\begin{eqnarray}
&&c_{_{[13\tilde{5}7]}}^{(2)}(\alpha,{\bf n})=
(-)^{n_{_1}+n_{_4}}\Gamma(1+n_{_1}+n_{_3})\Gamma(1+n_{_2}+n_{_4})
\nonumber\\
&&\hspace{2.5cm}\times
\Big\{n_{_1}!n_{_2}!n_{_3}!n_{_4}!\Gamma({D\over2}+n_{_1})\Gamma({D\over2}+n_{_2})
\Gamma(2-{D\over2}+n_{_4})
\nonumber\\
&&\hspace{2.5cm}\times
\Gamma(2-{D\over2}+n_{_3})
\Gamma({D\over2}-1-n_{_1}-n_{_3})\Gamma({D\over2}-1-n_{_2}-n_{_4})\Big\}^{-1}\;.
\label{GKZ21c-2-3}
\end{eqnarray}

\item   $I_{_{3}}=\{2,4,5,6,8,9,11,\cdots,14\}$, i.e. the implement $J_{_{3}}=[1,14]\setminus I_{_{3}}=\{1,3,7,10\}$.
The choice implies the power numbers $\alpha_{_1}=\alpha_{_{3}}=\alpha_{_{7}}=\alpha_{_{10}}=0$, and
\begin{eqnarray}
&&\alpha_{_2}=a_{_1}-a_{_2},\;\alpha_{_4}=a_{_3}-a_{_4},
\;\alpha_{_5}=b_{_5}-a_{_3}-a_{_5}-1,
\nonumber\\
&&\alpha_{_{6}}=b_{_1}+b_{_5}-a_{_1}-2,\;\alpha_{_{8}}=b_{_2}+b_{_3}-a_{_3}-2,
\nonumber\\
&&\alpha_{_{9}}=b_{_4}-b_{_5},\;\alpha_{_{11}}=b_{_5}-a_{_1}-1,\;\alpha_{_{12}}=1-b_{_2},
\nonumber\\
&&\alpha_{_{13}}=b_{_2}-a_{_3}-1,\;\alpha_{_{14}}=a_{_3}-b_{_5}+1\;.
\label{GKZ21c-3-1}
\end{eqnarray}
The corresponding hypergeometric series is
\begin{eqnarray}
&&\Phi_{_{[13\tilde{5}7]}}^{(3)}(\alpha,z)=
y_{_1}^{-1}y_{_2}^{{D\over2}-1}y_{_3}^{-1}y_{_4}^{{D\over2}-1}\sum\limits_{n_{_1}=0}^\infty
\sum\limits_{n_{_2}=0}^\infty\sum\limits_{n_{_3}=0}^\infty\sum\limits_{n_{_4}=0}^\infty
c_{_{[13\tilde{5}7]}}^{(3)}(\alpha,{\bf n})
\nonumber\\
&&\hspace{2.5cm}\times
\Big({1\over y_{_1}}\Big)^{n_{_1}}\Big({y_{_4}\over y_{_3}}\Big)^{n_{_2}}
\Big({y_{_4}\over y_{_1}}\Big)^{n_{_3}}\Big({y_{_2}\over y_{_3}}\Big)^{n_{_4}}\;,
\label{GKZ21c-3-2}
\end{eqnarray}
with
\begin{eqnarray}
&&c_{_{[13\tilde{5}7]}}^{(3)}(\alpha,{\bf n})=
(-)^{n_{_1}+n_{_4}}\Gamma(1+n_{_1}+n_{_3})\Gamma(1+n_{_2}+n_{_4})
\nonumber\\
&&\hspace{2.5cm}\times
\Big\{n_{_1}!n_{_2}!n_{_3}!n_{_4}!\Gamma({D\over2}+n_{_1})
\Gamma({D\over2}+n_{_2})\Gamma(1-{D\over2}-n_{_1}-n_{_3})
\nonumber\\
&&\hspace{2.5cm}\times
\Gamma(1-{D\over2}-n_{_2}-n_{_4})\Gamma({D\over2}+n_{_3})\Gamma({D\over2}+n_{_4})\Big\}^{-1}\;.
\label{GKZ21c-3-3}
\end{eqnarray}

\item   $I_{_{4}}=\{2,4,\cdots,9,11,13,14\}$, i.e. the implement $J_{_{4}}=[1,14]\setminus I_{_{4}}=\{1,3,10,12\}$.
The choice implies the power numbers $\alpha_{_1}=\alpha_{_{3}}=\alpha_{_{10}}=\alpha_{_{12}}=0$, and
\begin{eqnarray}
&&\alpha_{_2}=a_{_1}-a_{_2},\;\alpha_{_4}=a_{_3}-a_{_4},
\;\alpha_{_5}=b_{_5}-a_{_3}-a_{_5}-1,
\nonumber\\
&&\alpha_{_{6}}=b_{_1}+b_{_5}-a_{_1}-2,\;\alpha_{_{7}}=b_{_2}-1,\;\alpha_{_{8}}=b_{_3}-a_{_3}-1,
\nonumber\\
&&\alpha_{_{9}}=b_{_4}-b_{_5},\;\alpha_{_{11}}=b_{_5}-a_{_1}-1,\;\alpha_{_{13}}=-a_{_3},\;
\alpha_{_{14}}=a_{_3}-b_{_5}+1\;.
\label{GKZ21c-4-1}
\end{eqnarray}
The corresponding hypergeometric series is written as
\begin{eqnarray}
&&\Phi_{_{[13\tilde{5}7]}}^{(4)}(\alpha,z)=
y_{_1}^{-1}y_{_3}^{{D\over2}-2}y_{_4}^{{D\over2}-1}\sum\limits_{n_{_1}=0}^\infty
\sum\limits_{n_{_2}=0}^\infty\sum\limits_{n_{_3}=0}^\infty\sum\limits_{n_{_4}=0}^\infty
c_{_{[13\tilde{5}7]}}^{(4)}(\alpha,{\bf n})
\nonumber\\
&&\hspace{2.5cm}\times
\Big({1\over y_{_1}}\Big)^{n_{_1}}\Big({y_{_4}\over y_{_3}}\Big)^{n_{_2}}
\Big({y_{_4}\over y_{_1}}\Big)^{n_{_3}}\Big({y_{_2}\over y_{_3}}\Big)^{n_{_4}}\;,
\label{GKZ21c-4-2}
\end{eqnarray}
with
\begin{eqnarray}
&&c_{_{[13\tilde{5}7]}}^{(4)}(\alpha,{\bf n})=
(-)^{n_{_1}+n_{_4}}\Gamma(1+n_{_1}+n_{_3})\Gamma(1+n_{_2}+n_{_4})
\nonumber\\
&&\hspace{2.5cm}\times
\Big\{n_{_1}!n_{_2}!n_{_3}!n_{_4}!\Gamma({D\over2}+n_{_1})\Gamma({D\over2}+n_{_2})
\Gamma(1-{D\over2}-n_{_1}-n_{_3})
\nonumber\\
&&\hspace{2.5cm}\times
\Gamma(2-{D\over2}+n_{_4})
\Gamma({D\over2}+n_{_3})\Gamma({D\over2}-1-n_{_2}-n_{_4})\Big\}^{-1}\;.
\label{GKZ21c-4-3}
\end{eqnarray}

\item   $I_{_{5}}=\{2,3,5,6,8,10,\cdots,14\}$, i.e. the implement $J_{_{5}}=[1,14]\setminus I_{_{5}}=\{1,4,7,9\}$.
The choice implies the power numbers $\alpha_{_1}=\alpha_{_{4}}=\alpha_{_{7}}=\alpha_{_{9}}=0$, and
\begin{eqnarray}
&&\alpha_{_2}=a_{_1}-a_{_2},\;\alpha_{_3}=a_{_4}-a_{_3},
\;\alpha_{_5}=b_{_4}-a_{_4}-a_{_5}-1,
\nonumber\\
&&\alpha_{_{6}}=b_{_1}+b_{_4}-a_{_1}-2,\;\alpha_{_{8}}=b_{_2}+b_{_3}-a_{_4}-2,
\nonumber\\
&&\alpha_{_{10}}=b_{_5}-b_{_4},\;\alpha_{_{11}}=b_{_4}-a_{_1}-1,\;\alpha_{_{12}}=1-b_{_2},
\nonumber\\
&&\alpha_{_{13}}=b_{_2}-a_{_4}-1,\;\alpha_{_{14}}=a_{_4}-b_{_4}+1\;.
\label{GKZ21c-7-1}
\end{eqnarray}
The corresponding hypergeometric series is written as
\begin{eqnarray}
&&\Phi_{_{[13\tilde{5}7]}}^{(5)}(\alpha,z)=
y_{_1}^{{D\over2}-2}y_{_2}^{{D\over2}-1}y_{_3}^{{D\over2}-2}y_{_4}^{1-{D\over2}}
\sum\limits_{n_{_1}=0}^\infty
\sum\limits_{n_{_2}=0}^\infty\sum\limits_{n_{_3}=0}^\infty\sum\limits_{n_{_4}=0}^\infty
c_{_{[13\tilde{5}7]}}^{(5)}(\alpha,{\bf n})
\nonumber\\
&&\hspace{2.5cm}\times
\Big({1\over y_{_1}}\Big)^{n_{_1}}\Big({y_{_4}\over y_{_3}}\Big)^{n_{_2}}
\Big({y_{_4}\over y_{_1}}\Big)^{n_{_3}}\Big({y_{_2}\over y_{_3}}\Big)^{n_{_4}}\;,
\label{GKZ21c-7-2}
\end{eqnarray}
with
\begin{eqnarray}
&&c_{_{[13\tilde{5}7]}}^{(5)}(\alpha,{\bf n})=
(-)^{1+n_{_1}+n_{_4}}\Gamma(1+n_{_1}+n_{_3})\Gamma(1+n_{_2}+n_{_4})
\nonumber\\
&&\hspace{2.5cm}\times
\Big\{n_{_1}!n_{_2}!n_{_3}!n_{_4}!\Gamma({D\over2}+n_{_1})
\Gamma(2-{D\over2}+n_{_2})\Gamma(2-{D\over2}+n_{_3})
\nonumber\\
&&\hspace{2.5cm}\times
\Gamma({D\over2}-1-n_{_1}-n_{_3})\Gamma({D\over2}+n_{_4})
\Gamma({D\over2}-1-n_{_2}-n_{_4})\Big\}^{-1}\;.
\label{GKZ21c-7-3}
\end{eqnarray}

\item   $I_{_{6}}=\{2,3,5,\cdots,8,10,11,13,14\}$, i.e. the implement $J_{_{6}}=[1,14]\setminus I_{_{6}}=\{1,4,9,12\}$.
The choice implies the power numbers $\alpha_{_1}=\alpha_{_{4}}=\alpha_{_{9}}=\alpha_{_{12}}=0$, and
\begin{eqnarray}
&&\alpha_{_2}=a_{_1}-a_{_2},\;\alpha_{_3}=a_{_4}-a_{_3},
\;\alpha_{_5}=b_{_4}-a_{_4}-a_{_5}-1,
\nonumber\\
&&\alpha_{_{6}}=b_{_1}+b_{_4}-a_{_1}-2,\;\alpha_{_{7}}=b_{_2}-1,\;\alpha_{_{8}}=b_{_3}-a_{_4}-1,
\nonumber\\
&&\alpha_{_{10}}=b_{_5}-b_{_4},\;\alpha_{_{11}}=b_{_4}-a_{_1}-1,\;\alpha_{_{13}}=-a_{_4},
\nonumber\\
&&\alpha_{_{14}}=a_{_4}-b_{_4}+1\;.
\label{GKZ21c-8-1}
\end{eqnarray}
The corresponding hypergeometric series is
\begin{eqnarray}
&&\Phi_{_{[13\tilde{5}7]}}^{(6)}(\alpha,z)=
y_{_1}^{{D\over2}-2}y_{_3}^{{D}-3}y_{_4}^{1-{D\over2}}\sum\limits_{n_{_1}=0}^\infty
\sum\limits_{n_{_2}=0}^\infty\sum\limits_{n_{_3}=0}^\infty\sum\limits_{n_{_4}=0}^\infty
c_{_{[13\tilde{5}7]}}^{(6)}(\alpha,{\bf n})
\nonumber\\
&&\hspace{2.5cm}\times
\Big({1\over y_{_1}}\Big)^{n_{_1}}\Big({y_{_4}\over y_{_3}}\Big)^{n_{_2}}
\Big({y_{_4}\over y_{_1}}\Big)^{n_{_3}}\Big({y_{_2}\over y_{_3}}\Big)^{n_{_4}}\;,
\label{GKZ21c-8-2}
\end{eqnarray}
with
\begin{eqnarray}
&&c_{_{[13\tilde{5}7]}}^{(6)}(\alpha,{\bf n})=
(-)^{1+n_{_1}+n_{_2}}\Gamma(1+n_{_1}+n_{_3})\Big\{n_{_1}!n_{_2}!n_{_3}!n_{_4}!
\Gamma({D\over2}+n_{_1})
\nonumber\\
&&\hspace{2.5cm}\times
\Gamma(2-{D\over2}+n_{_2})
\Gamma(2-{D\over2}+n_{_4})\Gamma({D\over2}-1-n_{_2}-n_{_4})
\nonumber\\
&&\hspace{2.5cm}\times
\Gamma(2-{D\over2}+n_{_3})\Gamma({D\over2}-1-n_{_1}-n_{_3})
\Gamma(D-2-n_{_2}-n_{_4})\Big\}^{-1}\;.
\label{GKZ21c-8-3}
\end{eqnarray}

\item   $I_{_{7}}=\{2,3,5,6,8,9,11,\cdots,14\}$, i.e. the implement $J_{_{7}}=[1,14]\setminus I_{_{7}}=\{1,4,7,10\}$.
The choice implies the power numbers $\alpha_{_1}=\alpha_{_{4}}=\alpha_{_{7}}=\alpha_{_{10}}=0$, and
\begin{eqnarray}
&&\alpha_{_2}=a_{_1}-a_{_2},\;\alpha_{_3}=a_{_4}-a_{_3},
\;\alpha_{_5}=b_{_5}-a_{_4}-a_{_5}-1,
\nonumber\\
&&\alpha_{_{6}}=b_{_1}+b_{_5}-a_{_1}-2,\;\alpha_{_{8}}=b_{_2}+b_{_3}-a_{_4}-2,
\nonumber\\
&&\alpha_{_{9}}=b_{_4}-b_{_5},\;\alpha_{_{11}}=b_{_5}-a_{_1}-1,\;\alpha_{_{12}}=1-b_{_2},
\nonumber\\
&&\alpha_{_{13}}=b_{_2}-a_{_4}-1,\;\alpha_{_{14}}=a_{_4}-b_{_5}+1\;.
\label{GKZ21c-9-1}
\end{eqnarray}
The corresponding hypergeometric series is
\begin{eqnarray}
&&\Phi_{_{[13\tilde{5}7]}}^{(7)}(\alpha,z)=
y_{_1}^{-1}y_{_2}^{{D\over2}-1}y_{_3}^{{D\over2}-2}\sum\limits_{n_{_1}=0}^\infty
\sum\limits_{n_{_2}=0}^\infty\sum\limits_{n_{_3}=0}^\infty\sum\limits_{n_{_4}=0}^\infty
c_{_{[13\tilde{5}7]}}^{(7)}(\alpha,{\bf n})
\nonumber\\
&&\hspace{2.5cm}\times
\Big({1\over y_{_1}}\Big)^{n_{_1}}\Big({y_{_4}\over y_{_3}}\Big)^{n_{_2}}
\Big({y_{_4}\over y_{_1}}\Big)^{n_{_3}}\Big({y_{_2}\over y_{_3}}\Big)^{n_{_4}}\;,
\label{GKZ21c-9-2}
\end{eqnarray}
with
\begin{eqnarray}
&&c_{_{[13\tilde{5}7]}}^{(7)}(\alpha,{\bf n})=
(-)^{n_{_1}+n_{_4}}\Gamma(1+n_{_1}+n_{_3})\Gamma(1+n_{_2}+n_{_4})
\Big\{n_{_1}!n_{_2}!n_{_3}!n_{_4}!
\nonumber\\
&&\hspace{2.5cm}\times
\Gamma({D\over2}+n_{_1})\Gamma(2-{D\over2}+n_{_2})
\Gamma(1-{D\over2}-n_{_1}-n_{_3})
\nonumber\\
&&\hspace{2.5cm}\times
\Gamma({D\over2}+n_{_3})\Gamma({D\over2}+n_{_4})
\Gamma({D\over2}-1-n_{_2}-n_{_4})\Big\}^{-1}\;.
\label{GKZ21c-9-3}
\end{eqnarray}

\item   $I_{_{8}}=\{2,3,5,\cdots,9,11,13,14\}$, i.e. the implement $J_{_{8}}=[1,14]\setminus I_{_{8}}=\{1,4,10,12\}$.
The choice implies the power numbers $\alpha_{_1}=\alpha_{_{4}}=\alpha_{_{10}}=\alpha_{_{12}}=0$, and
\begin{eqnarray}
&&\alpha_{_2}=a_{_1}-a_{_2},\;\alpha_{_3}=a_{_4}-a_{_3},
\;\alpha_{_5}=b_{_5}-a_{_4}-a_{_5}-1,
\nonumber\\
&&\alpha_{_{6}}=b_{_1}+b_{_5}-a_{_1}-2,\;\alpha_{_{7}}=b_{_2}-1,\;\alpha_{_{8}}=b_{_3}-a_{_4}-1,
\nonumber\\
&&\alpha_{_{9}}=b_{_4}-b_{_5},\;\alpha_{_{11}}=b_{_5}-a_{_1}-1,\;\alpha_{_{13}}=-a_{_4},\;
\alpha_{_{14}}=a_{_4}-b_{_5}+1\;.
\label{GKZ21c-10-1}
\end{eqnarray}
The corresponding hypergeometric series solution is written as
\begin{eqnarray}
&&\Phi_{_{[13\tilde{5}7]}}^{(8)}(\alpha,z)=
y_{_1}^{-1}y_{_3}^{{D}-3}\sum\limits_{n_{_1}=0}^\infty
\sum\limits_{n_{_2}=0}^\infty\sum\limits_{n_{_3}=0}^\infty\sum\limits_{n_{_4}=0}^\infty
c_{_{[13\tilde{5}7]}}^{(8)}(\alpha,{\bf n})
\nonumber\\
&&\hspace{2.5cm}\times
\Big({1\over y_{_1}}\Big)^{n_{_1}}\Big({y_{_4}\over y_{_3}}\Big)^{n_{_2}}
\Big({y_{_4}\over y_{_1}}\Big)^{n_{_3}}\Big({y_{_2}\over y_{_3}}\Big)^{n_{_4}}\;,
\label{GKZ21c-10-2}
\end{eqnarray}
with
\begin{eqnarray}
&&c_{_{[13\tilde{5}7]}}^{(8)}(\alpha,{\bf n})=
(-)^{n_{_1}+n_{_2}}\Gamma(1+n_{_1}+n_{_3})
\Big\{n_{_1}!n_{_2}!n_{_3}!n_{_4}!\Gamma({D\over2}+n_{_1})
\nonumber\\
&&\hspace{2.5cm}\times
\Gamma(2-{D\over2}+n_{_2})\Gamma(1-{D\over2}-n_{_1}-n_{_3})
\Gamma(2-{D\over2}+n_{_4})
\nonumber\\
&&\hspace{2.5cm}\times
\Gamma({D\over2}-1-n_{_2}-n_{_4})\Gamma({D\over2}+n_{_3})
\Gamma(D-2-n_{_2}-n_{_4})\Big\}^{-1}\;.
\label{GKZ21c-10-3}
\end{eqnarray}

\item   $I_{_{9}}=\{1,4,5,6,8,10,\cdots,14\}$, i.e. the implement $J_{_{9}}=[1,14]\setminus I_{_{9}}=\{2,3,7,9\}$.
The choice implies the power numbers $\alpha_{_2}=\alpha_{_{3}}=\alpha_{_{7}}=\alpha_{_{9}}=0$, and
\begin{eqnarray}
&&\alpha_{_1}=a_{_2}-a_{_1},\;\alpha_{_4}=a_{_3}-a_{_4},
\;\alpha_{_5}=b_{_4}-a_{_3}-a_{_5}-1,
\nonumber\\
&&\alpha_{_{6}}=b_{_1}+b_{_4}-a_{_2}-2,\;\alpha_{_{8}}=b_{_2}+b_{_3}-a_{_3}-2,
\nonumber\\
&&\alpha_{_{10}}=b_{_5}-b_{_4},\;\alpha_{_{11}}=b_{_4}-a_{_2}-1,\;\alpha_{_{12}}=1-b_{_2},
\nonumber\\
&&\alpha_{_{13}}=b_{_2}-a_{_3}-1,\;\alpha_{_{14}}=a_{_3}-b_{_4}+1\;.
\label{GKZ21c-17-1}
\end{eqnarray}
The corresponding hypergeometric series is presented as
\begin{eqnarray}
&&\Phi_{_{[13\tilde{5}7]}}^{(9)}(\alpha,z)=
y_{_1}^{{D}-3}y_{_2}^{{D\over2}-1}y_{_3}^{-1}\sum\limits_{n_{_1}=0}^\infty
\sum\limits_{n_{_2}=0}^\infty\sum\limits_{n_{_3}=0}^\infty\sum\limits_{n_{_4}=0}^\infty
c_{_{[13\tilde{5}7]}}^{(9)}(\alpha,{\bf n})
\nonumber\\
&&\hspace{2.5cm}\times
\Big({1\over y_{_1}}\Big)^{n_{_1}}\Big({y_{_4}\over y_{_3}}\Big)^{n_{_2}}
\Big({y_{_4}\over y_{_1}}\Big)^{n_{_3}}\Big({y_{_2}\over y_{_3}}\Big)^{n_{_4}}\;,
\label{GKZ21c-17-2}
\end{eqnarray}
with
\begin{eqnarray}
&&c_{_{[13\tilde{5}7]}}^{(9)}(\alpha,{\bf n})=
(-)^{n_{_3}+n_{_4}}\Gamma(1+n_{_2}+n_{_4})
\Big\{n_{_1}!n_{_2}!n_{_3}!n_{_4}!\Gamma(2-{D\over2}+n_{_1})
\nonumber\\
&&\hspace{2.5cm}\times
\Gamma({D\over2}+n_{_2})\Gamma({D\over2}-1-n_{_1}-n_{_3})
\Gamma(1-{D\over2}-n_{_2}-n_{_4})
\nonumber\\
&&\hspace{2.5cm}\times
\Gamma(2-{D\over2}+n_{_3})
\Gamma(D-2-n_{_1}-n_{_3})\Gamma({D\over2}+n_{_4})\Big\}^{-1}\;.
\label{GKZ21c-17-3}
\end{eqnarray}

\item   $I_{_{10}}=\{1,4,\cdots,8,10,11,13,14\}$, i.e. the implement $J_{_{10}}=[1,14]\setminus I_{_{10}}=\{2,3,9,12\}$.
The choice implies the power numbers $\alpha_{_2}=\alpha_{_{3}}=\alpha_{_{9}}=\alpha_{_{12}}=0$, and
\begin{eqnarray}
&&\alpha_{_1}=a_{_2}-a_{_1},\;\alpha_{_4}=a_{_3}-a_{_4},
\;\alpha_{_5}=b_{_4}-a_{_3}-a_{_5}-1,
\nonumber\\
&&\alpha_{_{6}}=b_{_1}+b_{_4}-a_{_2}-2,\;\alpha_{_{7}}=b_{_2}-1,\;\alpha_{_{8}}=b_{_3}-a_{_3}-1,
\nonumber\\
&&\alpha_{_{10}}=b_{_5}-b_{_4},\;\alpha_{_{11}}=b_{_4}-a_{_2}-1,\;\alpha_{_{13}}=-a_{_3},
\nonumber\\
&&\alpha_{_{14}}=a_{_3}-b_{_4}+1\;.
\label{GKZ21c-18-1}
\end{eqnarray}
The corresponding hypergeometric series solution is given as
\begin{eqnarray}
&&\Phi_{_{[13\tilde{5}7]}}^{(10)}(\alpha,z)=
y_{_1}^{{D}-3}y_{_3}^{{D\over2}-2}\sum\limits_{n_{_1}=0}^\infty
\sum\limits_{n_{_2}=0}^\infty\sum\limits_{n_{_3}=0}^\infty\sum\limits_{n_{_4}=0}^\infty
c_{_{[13\tilde{5}7]}}^{(10)}(\alpha,{\bf n})
\nonumber\\
&&\hspace{2.5cm}\times
\Big({1\over y_{_1}}\Big)^{n_{_1}}\Big({y_{_4}\over y_{_3}}\Big)^{n_{_2}}
\Big({y_{_4}\over y_{_1}}\Big)^{n_{_3}}\Big({y_{_2}\over y_{_3}}\Big)^{n_{_4}}\;,
\label{GKZ21c-18-2}
\end{eqnarray}
with
\begin{eqnarray}
&&c_{_{[13\tilde{5}7]}}^{(10)}(\alpha,{\bf n})=
(-)^{n_{_3}+n_{_4}}\Gamma(1+n_{_2}+n_{_4})
\Big\{n_{_1}!n_{_2}!n_{_3}!n_{_4}!\Gamma(2-{D\over2}+n_{_1})
\nonumber\\
&&\hspace{2.5cm}\times
\Gamma({D\over2}+n_{_2})\Gamma({D\over2}-1-n_{_1}-n_{_3})
\Gamma(2-{D\over2}+n_{_4})
\nonumber\\
&&\hspace{2.5cm}\times
\Gamma(2-{D\over2}+n_{_3})\Gamma(D-2-n_{_1}-n_{_3})
\Gamma({D\over2}-1-n_{_2}-n_{_4})\Big\}^{-1}\;.
\label{GKZ21c-18-3}
\end{eqnarray}

\item   $I_{_{11}}=\{1,4,5,6,8,9,11,\cdots,14\}$, i.e. the implement $J_{_{11}}=[1,14]\setminus I_{_{11}}=\{2,3,7,10\}$.
The choice implies the power numbers $\alpha_{_2}=\alpha_{_{3}}=\alpha_{_{7}}=\alpha_{_{10}}=0$, and
\begin{eqnarray}
&&\alpha_{_1}=a_{_2}-a_{_1},\;\alpha_{_4}=a_{_3}-a_{_4},
\;\alpha_{_5}=b_{_5}-a_{_3}-a_{_5}-1,
\nonumber\\
&&\alpha_{_{6}}=b_{_1}+b_{_5}-a_{_2}-2,\;\alpha_{_{8}}=b_{_2}+b_{_3}-a_{_3}-2,
\nonumber\\
&&\alpha_{_{9}}=b_{_4}-b_{_5},\;\alpha_{_{11}}=b_{_5}-a_{_2}-1,\;\alpha_{_{12}}=1-b_{_2},
\nonumber\\
&&\alpha_{_{13}}=b_{_2}-a_{_3}-1,\;\alpha_{_{14}}=a_{_3}-b_{_5}+1\;.
\label{GKZ21c-19-1}
\end{eqnarray}
The corresponding hypergeometric series is
\begin{eqnarray}
&&\Phi_{_{[13\tilde{5}7]}}^{(11)}(\alpha,z)=
y_{_1}^{{D\over2}-2}y_{_2}^{{D\over2}-1}y_{_3}^{-1}y_{_4}^{{D\over2}-1}
\sum\limits_{n_{_1}=0}^\infty
\sum\limits_{n_{_2}=0}^\infty\sum\limits_{n_{_3}=0}^\infty\sum\limits_{n_{_4}=0}^\infty
c_{_{[13\tilde{5}7]}}^{(11)}(\alpha,{\bf n})
\nonumber\\
&&\hspace{2.5cm}\times
\Big({1\over y_{_1}}\Big)^{n_{_1}}\Big({y_{_4}\over y_{_3}}\Big)^{n_{_2}}
\Big({y_{_4}\over y_{_1}}\Big)^{n_{_3}}\Big({y_{_2}\over y_{_3}}\Big)^{n_{_4}}\;,
\label{GKZ21c-19-2}
\end{eqnarray}
with
\begin{eqnarray}
&&c_{_{[13\tilde{5}7]}}^{(11)}(\alpha,{\bf n})=
(-)^{n_{_1}+n_{_4}}\Gamma(1+n_{_1}+n_{_3})\Gamma(1+n_{_2}+n_{_4})
\Big\{n_{_1}!n_{_2}!n_{_3}!n_{_4}!
\nonumber\\
&&\hspace{2.5cm}\times
\Gamma(2-{D\over2}+n_{_1})
\Gamma({D\over2}+n_{_2})\Gamma(1-{D\over2}-n_{_2}-n_{_4})
\nonumber\\
&&\hspace{2.5cm}\times
\Gamma({D\over2}+n_{_3})
\Gamma({D\over2}-1-n_{_1}-n_{_3})\Gamma({D\over2}+n_{_4})\Big\}^{-1}\;.
\label{GKZ21c-19-3}
\end{eqnarray}

\item   $I_{_{12}}=\{1,4,\cdots,9,11,13,14\}$, i.e. the implement $J_{_{12}}=[1,14]\setminus I_{_{12}}=\{2,3,10,12\}$.
The choice implies the power numbers $\alpha_{_2}=\alpha_{_{3}}=\alpha_{_{10}}=\alpha_{_{12}}=0$, and
\begin{eqnarray}
&&\alpha_{_1}=a_{_2}-a_{_1},\;\alpha_{_4}=a_{_3}-a_{_4},
\;\alpha_{_5}=b_{_5}-a_{_3}-a_{_5}-1,
\nonumber\\
&&\alpha_{_{6}}=b_{_1}+b_{_5}-a_{_2}-2,\;\alpha_{_{7}}=b_{_2}-1,\;\alpha_{_{8}}=b_{_3}-a_{_3}-1,
\nonumber\\
&&\alpha_{_{9}}=b_{_4}-b_{_5},\;\alpha_{_{11}}=b_{_5}-a_{_2}-1,\;\alpha_{_{13}}=-a_{_3},\;
\alpha_{_{14}}=a_{_3}-b_{_5}+1\;.
\label{GKZ21c-20-1}
\end{eqnarray}
The corresponding hypergeometric series solution is written as
\begin{eqnarray}
&&\Phi_{_{[13\tilde{5}7]}}^{(12)}(\alpha,z)=
y_{_1}^{{D\over2}-2}y_{_3}^{{D\over2}-2}y_{_4}^{{D\over2}-1}\sum\limits_{n_{_1}=0}^\infty
\sum\limits_{n_{_2}=0}^\infty\sum\limits_{n_{_3}=0}^\infty\sum\limits_{n_{_4}=0}^\infty
c_{_{[13\tilde{5}7]}}^{(12)}(\alpha,{\bf n})
\nonumber\\
&&\hspace{2.5cm}\times
\Big({1\over y_{_1}}\Big)^{n_{_1}}\Big({y_{_4}\over y_{_3}}\Big)^{n_{_2}}
\Big({y_{_4}\over y_{_1}}\Big)^{n_{_3}}\Big({y_{_2}\over y_{_3}}\Big)^{n_{_4}}\;,
\label{GKZ21c-20-2}
\end{eqnarray}
with
\begin{eqnarray}
&&c_{_{[13\tilde{5}7]}}^{(12)}(\alpha,{\bf n})=
(-)^{n_{_1}+n_{_4}}\Gamma(1+n_{_1}+n_{_3})\Gamma(1+n_{_2}+n_{_4})
\Big\{n_{_1}!n_{_2}!n_{_3}!n_{_4}!
\nonumber\\
&&\hspace{2.5cm}\times
\Gamma(2-{D\over2}+n_{_1})
\Gamma({D\over2}+n_{_2})\Gamma(2-{D\over2}+n_{_4})
\nonumber\\
&&\hspace{2.5cm}\times
\Gamma({D\over2}+n_{_3})\Gamma({D\over2}-1-n_{_1}-n_{_3})
\Gamma({D\over2}-1-n_{_2}-n_{_4})\Big\}^{-1}\;.
\label{GKZ21c-20-3}
\end{eqnarray}

\item   $I_{_{13}}=\{1,3,5,6,8,10,\cdots,14\}$, i.e. the implement $J_{_{13}}=[1,14]\setminus I_{_{13}}=\{2,4,7,9\}$.
The choice implies the power numbers $\alpha_{_2}=\alpha_{_{4}}=\alpha_{_{7}}=\alpha_{_{9}}=0$, and
\begin{eqnarray}
&&\alpha_{_1}=a_{_2}-a_{_1},\;\alpha_{_3}=a_{_4}-a_{_3},
\;\alpha_{_5}=b_{_4}-a_{_4}-a_{_5}-1,
\nonumber\\
&&\alpha_{_{6}}=b_{_1}+b_{_4}-a_{_2}-2,\;\alpha_{_{8}}=b_{_2}+b_{_3}-a_{_4}-2,
\nonumber\\
&&\alpha_{_{10}}=b_{_5}-b_{_4},\;\alpha_{_{11}}=b_{_4}-a_{_2}-1,\;\alpha_{_{12}}=1-b_{_2},
\nonumber\\
&&\alpha_{_{13}}=b_{_2}-a_{_4}-1,\;\alpha_{_{14}}=a_{_4}-b_{_4}+1\;.
\label{GKZ21c-23-1}
\end{eqnarray}
The corresponding hypergeometric series is
\begin{eqnarray}
&&\Phi_{_{[13\tilde{5}7]}}^{(13)}(\alpha,z)=
y_{_1}^{{D}-3}y_{_2}^{{D\over2}-1}y_{_3}^{{D\over2}-2}y_{_4}^{1-{D\over2}}\sum\limits_{n_{_1}=0}^\infty
\sum\limits_{n_{_2}=0}^\infty\sum\limits_{n_{_3}=0}^\infty\sum\limits_{n_{_4}=0}^\infty
c_{_{[13\tilde{5}7]}}^{(13)}(\alpha,{\bf n})
\nonumber\\
&&\hspace{2.5cm}\times
\Big({1\over y_{_1}}\Big)^{n_{_1}}\Big({y_{_4}\over y_{_3}}\Big)^{n_{_2}}
\Big({y_{_4}\over y_{_1}}\Big)^{n_{_3}}\Big({y_{_2}\over y_{_3}}\Big)^{n_{_4}}\;,
\label{GKZ21c-23-2}
\end{eqnarray}
with
\begin{eqnarray}
&&c_{_{[13\tilde{5}7]}}^{(13)}(\alpha,{\bf n})=
(-)^{1+n_{_3}+n_{_4}}\Gamma(1+n_{_2}+n_{_4})\Big\{n_{_1}!n_{_2}!n_{_3}!n_{_4}!
\Gamma(2-{D\over2}+n_{_1})
\nonumber\\
&&\hspace{2.5cm}\times
\Gamma(2-{D\over2}+n_{_2})\Gamma({D\over2}-1-n_{_1}-n_{_3})
\Gamma(2-{D\over2}+n_{_3})
\nonumber\\
&&\hspace{2.5cm}\times
\Gamma(D-2-n_{_1}-n_{_3})\Gamma({D\over2}+n_{_4})
\Gamma({D\over2}-1-n_{_2}-n_{_4})\Big\}^{-1}\;.
\label{GKZ21c-23-3}
\end{eqnarray}

\item   $I_{_{14}}=\{1,3,5,\cdots,8,10,11,13,14\}$, i.e. the implement $J_{_{14}}=[1,14]\setminus I_{_{14}}=\{2,4,9,12\}$.
The choice implies the power numbers $\alpha_{_2}=\alpha_{_{4}}=\alpha_{_{9}}=\alpha_{_{12}}=0$, and
\begin{eqnarray}
&&\alpha_{_1}=a_{_2}-a_{_1},\;\alpha_{_3}=a_{_4}-a_{_3},
\;\alpha_{_5}=b_{_4}-a_{_4}-a_{_5}-1,
\nonumber\\
&&\alpha_{_{6}}=b_{_1}+b_{_4}-a_{_2}-2,\;\alpha_{_{7}}=b_{_2}-1,\;\alpha_{_{8}}=b_{_3}-a_{_4}-1,
\nonumber\\
&&\alpha_{_{10}}=b_{_5}-b_{_4},\;\alpha_{_{11}}=b_{_4}-a_{_2}-1,\;\alpha_{_{13}}=-a_{_4},
\nonumber\\
&&\alpha_{_{14}}=a_{_4}-b_{_4}+1\;.
\label{GKZ21c-24-1}
\end{eqnarray}
The corresponding hypergeometric series solution is written as
\begin{eqnarray}
&&\Phi_{_{[13\tilde{5}7]}}^{(14)}(\alpha,z)=
y_{_1}^{{D}-3}y_{_3}^{{D}-3}y_{_4}^{1-{D\over2}}\sum\limits_{n_{_1}=0}^\infty
\sum\limits_{n_{_2}=0}^\infty\sum\limits_{n_{_3}=0}^\infty\sum\limits_{n_{_4}=0}^\infty
c_{_{[13\tilde{5}7]}}^{(14)}(\alpha,{\bf n})
\nonumber\\
&&\hspace{2.5cm}\times
\Big({1\over y_{_1}}\Big)^{n_{_1}}\Big({y_{_4}\over y_{_3}}\Big)^{n_{_2}}
\Big({y_{_4}\over y_{_1}}\Big)^{n_{_3}}\Big({y_{_2}\over y_{_3}}\Big)^{n_{_4}}\;,
\label{GKZ21c-24-2}
\end{eqnarray}
with
\begin{eqnarray}
&&c_{_{[13\tilde{5}7]}}^{(14)}(\alpha,{\bf n})=
(-)^{1+n_{_2}+n_{_3}}
\Big\{n_{_1}!n_{_2}!n_{_3}!n_{_4}!\Gamma(2-{D\over2}+n_{_1})
\Gamma(2-{D\over2}+n_{_2})
\nonumber\\
&&\hspace{2.5cm}\times
\Gamma({D\over2}-1-n_{_1}-n_{_3})
\Gamma(2-{D\over2}+n_{_4})\Gamma({D\over2}-1-n_{_2}-n_{_4})
\nonumber\\
&&\hspace{2.5cm}\times
\Gamma(2-{D\over2}+n_{_3})\Gamma(D-2-n_{_1}-n_{_3})
\Gamma(D-2-n_{_2}-n_{_4})\Big\}^{-1}\;.
\label{GKZ21c-24-3}
\end{eqnarray}

\item   $I_{_{15}}=\{1,3,5,6,8,9,11,\cdots,14\}$, i.e. the implement $J_{_{15}}=[1,14]\setminus I_{_{15}}=\{2,4,7,10\}$.
The choice implies the power numbers $\alpha_{_2}=\alpha_{_{4}}=\alpha_{_{7}}=\alpha_{_{10}}=0$, and
\begin{eqnarray}
&&\alpha_{_1}=a_{_2}-a_{_1},\;\alpha_{_3}=a_{_4}-a_{_3},
\;\alpha_{_5}=b_{_5}-a_{_4}-a_{_5}-1,
\nonumber\\
&&\alpha_{_{6}}=b_{_1}+b_{_5}-a_{_2}-2,\;\alpha_{_{8}}=b_{_2}+b_{_3}-a_{_4}-2,
\nonumber\\
&&\alpha_{_{9}}=b_{_4}-b_{_5},\;\alpha_{_{11}}=b_{_5}-a_{_2}-1,\;\alpha_{_{12}}=1-b_{_2},
\nonumber\\
&&\alpha_{_{13}}=b_{_2}-a_{_4}-1,\;\alpha_{_{14}}=a_{_4}-b_{_5}+1\;.
\label{GKZ21c-25-1}
\end{eqnarray}
The corresponding hypergeometric series is
\begin{eqnarray}
&&\Phi_{_{[13\tilde{5}7]}}^{(15)}(\alpha,z)=
y_{_1}^{{D\over2}-2}y_{_2}^{{D\over2}-1}y_{_3}^{{D\over2}-2}\sum\limits_{n_{_1}=0}^\infty
\sum\limits_{n_{_2}=0}^\infty\sum\limits_{n_{_3}=0}^\infty\sum\limits_{n_{_4}=0}^\infty
c_{_{[13\tilde{5}7]}}^{(15)}(\alpha,{\bf n})
\nonumber\\
&&\hspace{2.5cm}\times
\Big({1\over y_{_1}}\Big)^{n_{_1}}\Big({y_{_4}\over y_{_3}}\Big)^{n_{_2}}
\Big({y_{_4}\over y_{_1}}\Big)^{n_{_3}}\Big({y_{_2}\over y_{_3}}\Big)^{n_{_4}}\;,
\label{GKZ21c-25-2}
\end{eqnarray}
with
\begin{eqnarray}
&&c_{_{[13\tilde{5}7]}}^{(15)}(\alpha,{\bf n})=
(-)^{n_{_1}+n_{_4}}\Gamma(1+n_{_1}+n_{_3})\Gamma(1+n_{_2}+n_{_4})
\Big\{n_{_1}!n_{_2}!n_{_3}!n_{_4}!
\nonumber\\
&&\hspace{2.5cm}\times
\Gamma(2-{D\over2}+n_{_1})\Gamma(2-{D\over2}+n_{_2})
\Gamma({D\over2}+n_{_3})
\nonumber\\
&&\hspace{2.5cm}\times
\Gamma({D\over2}-1-n_{_1}-n_{_3})\Gamma({D\over2}+n_{_4})
\Gamma({D\over2}-1-n_{_2}-n_{_4})\Big\}^{-1}\;.
\label{GKZ21c-25-3}
\end{eqnarray}

\item   $I_{_{16}}=\{1,3,5,\cdots,9,11,13,14\}$, i.e. the implement $J_{_{16}}=[1,14]\setminus I_{_{16}}=\{2,4,10,12\}$.
The choice implies the power numbers $\alpha_{_2}=\alpha_{_{4}}=\alpha_{_{10}}=\alpha_{_{12}}=0$, and
\begin{eqnarray}
&&\alpha_{_1}=a_{_2}-a_{_1},\;\alpha_{_3}=a_{_4}-a_{_3},
\;\alpha_{_5}=b_{_5}-a_{_4}-a_{_5}-1,
\nonumber\\
&&\alpha_{_{6}}=b_{_1}+b_{_5}-a_{_2}-2,\;\alpha_{_{7}}=b_{_2}-1,\;\alpha_{_{8}}=b_{_3}-a_{_4}-1,
\nonumber\\
&&\alpha_{_{9}}=b_{_4}-b_{_5},\;\alpha_{_{11}}=b_{_5}-a_{_2}-1,\;\alpha_{_{13}}=-a_{_4},\;
\alpha_{_{14}}=a_{_4}-b_{_5}+1\;.
\label{GKZ21c-26-1}
\end{eqnarray}
The corresponding hypergeometric series is written as
\begin{eqnarray}
&&\Phi_{_{[13\tilde{5}7]}}^{(16)}(\alpha,z)=
y_{_1}^{{D\over2}-2}y_{_3}^{{D}-3}\sum\limits_{n_{_1}=0}^\infty
\sum\limits_{n_{_2}=0}^\infty\sum\limits_{n_{_3}=0}^\infty\sum\limits_{n_{_4}=0}^\infty
c_{_{[13\tilde{5}7]}}^{(16)}(\alpha,{\bf n})
\nonumber\\
&&\hspace{2.5cm}\times
\Big({1\over y_{_1}}\Big)^{n_{_1}}\Big({y_{_4}\over y_{_3}}\Big)^{n_{_2}}
\Big({y_{_4}\over y_{_1}}\Big)^{n_{_3}}\Big({y_{_2}\over y_{_3}}\Big)^{n_{_4}}\;,
\label{GKZ21c-26-2}
\end{eqnarray}
with
\begin{eqnarray}
&&c_{_{[13\tilde{5}7]}}^{(16)}(\alpha,{\bf n})=
(-)^{n_{_1}+n_{_2}}\Gamma(1+n_{_1}+n_{_3})
\Big\{n_{_1}!n_{_2}!n_{_3}!n_{_4}!\Gamma(2-{D\over2}+n_{_1})
\nonumber\\
&&\hspace{2.5cm}\times
\Gamma(2-{D\over2}+n_{_2})
\Gamma(2-{D\over2}+n_{_4})\Gamma({D\over2}-1-n_{_2}-n_{_4})
\nonumber\\
&&\hspace{2.5cm}\times
\Gamma({D\over2}+n_{_3})\Gamma({D\over2}-1-n_{_1}-n_{_3})
\Gamma(D-1-n_{_2}-n_{_4})\Big\}^{-1}\;.
\label{GKZ21c-26-3}
\end{eqnarray}

\item   $I_{_{17}}=\{2,4,5,6,8,\cdots,13\}$, i.e. the implement $J_{_{17}}=[1,14]\setminus I_{_{17}}=\{1,3,7,14\}$.
The choice implies the power numbers $\alpha_{_1}=\alpha_{_{3}}=\alpha_{_{7}}=\alpha_{_{14}}=0$, and
\begin{eqnarray}
&&\alpha_{_2}=a_{_1}-a_{_2},\;\alpha_{_4}=a_{_3}-a_{_4},
\;\alpha_{_5}=-a_{_5},
\nonumber\\
&&\alpha_{_{6}}=a_{_3}+b_{_1}-a_{_1}-1,\;\alpha_{_{8}}=b_{_2}+b_{_3}-a_{_3}-2,
\nonumber\\
&&\alpha_{_{9}}=b_{_4}-a_{_3}-1,\;\alpha_{_{10}}=b_{_5}-a_{_3}-1,
\nonumber\\
&&\alpha_{_{11}}=a_{_3}-a_{_1},\;\alpha_{_{12}}=1-b_{_2},\;\alpha_{_{13}}=b_{_2}-a_{_3}-1.
\label{GKZ21c-5-1}
\end{eqnarray}
The corresponding hypergeometric series solutions are written as
\begin{eqnarray}
&&\Phi_{_{[13\tilde{5}7]}}^{(17),a}(\alpha,z)=
y_{_1}^{{D\over2}-2}y_{_2}^{{D\over2}-1}y_{_3}^{-1}\sum\limits_{n_{_1}=0}^\infty
\sum\limits_{n_{_2}=0}^\infty\sum\limits_{n_{_3}=0}^\infty\sum\limits_{n_{_4}=0}^\infty
c_{_{[13\tilde{5}7]}}^{(17),a}(\alpha,{\bf n})
\nonumber\\
&&\hspace{2.5cm}\times
\Big({1\over y_{_1}}\Big)^{n_{_1}}\Big({y_{_4}\over y_{_3}}\Big)^{n_{_2}}
\Big({y_{_4}\over y_{_1}}\Big)^{n_{_3}}\Big({y_{_2}\over y_{_3}}\Big)^{n_{_4}}
\;,\nonumber\\
&&\Phi_{_{[13\tilde{5}7]}}^{(17),b}(\alpha,z)=
y_{_1}^{{D\over2}-1}y_{_2}^{{D\over2}-1}y_{_3}^{-2}\sum\limits_{n_{_1}=0}^\infty
\sum\limits_{n_{_2}=0}^\infty\sum\limits_{n_{_3}=0}^\infty\sum\limits_{n_{_4}=0}^\infty
c_{_{[13\tilde{5}7]}}^{(17),b}(\alpha,{\bf n})
\nonumber\\
&&\hspace{2.5cm}\times
\Big({1\over y_{_3}}\Big)^{n_{_1}}\Big({y_{_1}\over y_{_3}}\Big)^{n_{_2}}
\Big({y_{_4}\over y_{_3}}\Big)^{n_{_3}}\Big({y_{_2}\over y_{_3}}\Big)^{n_{_4}}
\;,\nonumber\\
&&\Phi_{_{[13\tilde{5}7]}}^{(17),c}(\alpha,z)=
y_{_1}^{{D\over2}-2}y_{_2}^{{D\over2}-1}y_{_3}^{-2}\sum\limits_{n_{_1}=0}^\infty
\sum\limits_{n_{_2}=0}^\infty\sum\limits_{n_{_3}=0}^\infty\sum\limits_{n_{_4}=0}^\infty
c_{_{[13\tilde{5}7]}}^{(17),c}(\alpha,{\bf n})
\nonumber\\
&&\hspace{2.5cm}\times
\Big({1\over y_{_1}}\Big)^{n_{_1}}\Big({1\over y_{_3}}\Big)^{n_{_2}}
\Big({y_{_4}\over y_{_3}}\Big)^{n_{_3}}\Big({y_{_2}\over y_{_3}}\Big)^{n_{_4}}\;.
\label{GKZ21c-5-2a}
\end{eqnarray}
Where the coefficients are
\begin{eqnarray}
&&c_{_{[13\tilde{5}7]}}^{(17),a}(\alpha,{\bf n})=
(-)^{n_{_1}+n_{_4}}\Gamma(1+n_{_1}+n_{_3})\Gamma(1+n_{_2}+n_{_4})
\Big\{n_{_1}!n_{_2}!n_{_3}!n_{_4}!
\nonumber\\
&&\hspace{2.5cm}\times
\Gamma({D\over2}+n_{_1})\Gamma({D\over2}+n_{_2})
\Gamma({D\over2}-1-n_{_1}-n_{_3})
\nonumber\\
&&\hspace{2.5cm}\times
\Gamma(1-{D\over2}-n_{_2}-n_{_4})
\Gamma(2-{D\over2}+n_{_3})\Gamma({D\over2}+n_{_4})\Big\}^{-1}
\;,\nonumber\\
&&c_{_{[13\tilde{5}7]}}^{(17),b}(\alpha,{\bf n})=
(-)^{1+n_{_4}}\Gamma(1+n_{_1}+n_{_2})\Gamma(2+n_{_1}+n_{_2}+n_{_3}+n_{_4})
\Big\{n_{_1}!n_{_2}!n_{_4}!
\nonumber\\
&&\hspace{2.5cm}\times
\Gamma(2+n_{_1}+n_{_2}+n_{_3})
\Gamma({D\over2}+n_{_1})\Gamma({D\over2}+1+n_{_1}+n_{_2}+n_{_3})
\nonumber\\
&&\hspace{2.5cm}\times
\Gamma(-{D\over2}-n_{_1}-n_{_2}-n_{_3}-n_{_4})\Gamma(1-{D\over2}-n_{_1}-n_{_2})
\Gamma({D\over2}+n_{_2})
\nonumber\\
&&\hspace{2.5cm}\times
\Gamma({D\over2}+n_{_4})\Big\}^{-1}
\;,\nonumber\\
&&c_{_{[13\tilde{5}7]}}^{(17),c}(\alpha,{\bf n})=
(-)^{1+n_{_1}+n_{_4}}\Gamma(1+n_{_1})\Gamma(1+n_{_2})
\Gamma(2+n_{_2}+n_{_3}+n_{_4})
\nonumber\\
&&\hspace{2.5cm}\times
\Big\{n_{_4}!\Gamma(2+n_{_1}+n_{_2})\Gamma(2+n_{_2}+n_{_3})
\Gamma({D\over2}+1+n_{_1}+n_{_2})
\nonumber\\
&&\hspace{2.5cm}\times
\Gamma({D\over2}+1+n_{_2}+n_{_3})
\Gamma(-{D\over2}-n_{_2}-n_{_3}-n_{_4})\Gamma(1-{D\over2}-n_{_2})
\nonumber\\
&&\hspace{2.5cm}\times
\Gamma({D\over2}-1-n_{_1})\Gamma({D\over2}+n_{_4})\Big\}^{-1}\;.
\label{GKZ21c-5-3}
\end{eqnarray}

\item   $I_{_{18}}=\{2,4,\cdots,11,13\}$, i.e. the implement $J_{_{18}}=[1,14]\setminus I_{_{18}}=\{1,3,12,14\}$.
The choice implies the power numbers $\alpha_{_1}=\alpha_{_{3}}=\alpha_{_{12}}=\alpha_{_{14}}=0$, and
\begin{eqnarray}
&&\alpha_{_2}=a_{_1}-a_{_2},\;\alpha_{_4}=a_{_3}-a_{_4},\;\alpha_{_5}=-a_{_5},
\nonumber\\
&&\alpha_{_{6}}=a_{_3}+b_{_1}-a_{_1}-1,\;\alpha_{_{7}}=b_{_2}-1,
\nonumber\\
&&\alpha_{_{8}}=b_{_3}-a_{_3}-1,\;\alpha_{_{9}}=b_{_4}-a_{_3}-1,
\nonumber\\
&&\alpha_{_{10}}=b_{_5}-a_{_3}-1,\;\alpha_{_{11}}=a_{_3}-a_{_1},\;\alpha_{_{13}}=-a_{_3}.
\label{GKZ21c-6-1}
\end{eqnarray}
The corresponding hypergeometric series solutions are
\begin{eqnarray}
&&\Phi_{_{[13\tilde{5}7]}}^{(18),a}(\alpha,z)=
y_{_1}^{{D\over2}-2}y_{_3}^{{D\over2}-2}\sum\limits_{n_{_1}=0}^\infty
\sum\limits_{n_{_2}=0}^\infty\sum\limits_{n_{_3}=0}^\infty\sum\limits_{n_{_4}=0}^\infty
c_{_{[13\tilde{5}7]}}^{(18),a}(\alpha,{\bf n})
\nonumber\\
&&\hspace{2.5cm}\times
\Big({1\over y_{_1}}\Big)^{n_{_1}}\Big({y_{_4}\over y_{_3}}\Big)^{n_{_2}}
\Big({y_{_4}\over y_{_1}}\Big)^{n_{_3}}\Big({y_{_2}\over y_{_3}}\Big)^{n_{_4}}
\;,\nonumber\\
&&\Phi_{_{[13\tilde{5}7]}}^{(18),b}(\alpha,z)=
y_{_1}^{{D\over2}-1}y_{_3}^{{D\over2}-3}\sum\limits_{n_{_1}=0}^\infty
\sum\limits_{n_{_2}=0}^\infty\sum\limits_{n_{_3}=0}^\infty\sum\limits_{n_{_4}=0}^\infty
c_{_{[13\tilde{5}7]}}^{(18),b}(\alpha,{\bf n})
\nonumber\\
&&\hspace{2.5cm}\times
\Big({1\over y_{_3}}\Big)^{n_{_1}}\Big({y_{_1}\over y_{_3}}\Big)^{n_{_2}}
\Big({y_{_4}\over y_{_3}}\Big)^{n_{_3}}\Big({y_{_2}\over y_{_3}}\Big)^{n_{_4}}
\;,\nonumber\\
&&\Phi_{_{[13\tilde{5}7]}}^{(18),c}(\alpha,z)=
y_{_1}^{{D\over2}-2}y_{_3}^{{D\over2}-3}\sum\limits_{n_{_1}=0}^\infty
\sum\limits_{n_{_2}=0}^\infty\sum\limits_{n_{_3}=0}^\infty\sum\limits_{n_{_4}=0}^\infty
c_{_{[13\tilde{5}7]}}^{(18),c}(\alpha,{\bf n})
\nonumber\\
&&\hspace{2.5cm}\times
\Big({1\over y_{_1}}\Big)^{n_{_1}}\Big({1\over y_{_3}}\Big)^{n_{_2}}
\Big({y_{_4}\over y_{_3}}\Big)^{n_{_3}}\Big({y_{_2}\over y_{_3}}\Big)^{n_{_4}}\;.
\label{GKZ21c-6-2a}
\end{eqnarray}
Where the coefficients are
\begin{eqnarray}
&&c_{_{[13\tilde{5}7]}}^{(18),a}(\alpha,{\bf n})=
(-)^{n_{_1}+n_{_4}}\Gamma(1+n_{_1}+n_{_3})\Gamma(1+n_{_2}+n_{_4})
\Big\{n_{_1}!n_{_2}!n_{_3}!n_{_4}!
\nonumber\\
&&\hspace{2.5cm}\times
\Gamma({D\over2}+n_{_1})\Gamma({D\over2}+n_{_2})
\Gamma({D\over2}-1-n_{_1}-n_{_3})
\nonumber\\
&&\hspace{2.5cm}\times
\Gamma({D\over2}-1-n_{_2}-n_{_4})
\Gamma(2-{D\over2}+n_{_3})\Gamma(2-{D\over2}+n_{_4})\Big\}^{-1}
\;,\nonumber\\
&&c_{_{[13\tilde{5}7]}}^{(18),b}(\alpha,{\bf n})=
(-)^{1+n_{_4}}\Gamma(1+n_{_1}+n_{_2})\Gamma(2+n_{_1}+n_{_2}+n_{_3}+n_{_4})
\Big\{n_{_1}!n_{_2}!n_{_4}!
\nonumber\\
&&\hspace{2.5cm}\times
\Gamma(2+n_{_1}+n_{_2}+n_{_3})
\Gamma({D\over2}+n_{_1})\Gamma({D\over2}+1+n_{_1}+n_{_2}+n_{_3})
\nonumber\\
&&\hspace{2.5cm}\times
\Gamma({D\over2}-2-n_{_1}-n_{_2}-n_{_3}-n_{_4})\Gamma(1-{D\over2}-n_{_1}-n_{_2})
\Gamma({D\over2}+n_{_2})
\nonumber\\
&&\hspace{2.5cm}\times
\Gamma(2-{D\over2}+n_{_4})\Big\}^{-1}
\;,\nonumber\\
&&c_{_{[13\tilde{5}7]}}^{(18),c}(\alpha,{\bf n})=
(-)^{1+n_{_1}+n_{_4}}\Gamma(1+n_{_1})\Gamma(1+n_{_2})
\Gamma(2+n_{_2}+n_{_3}+n_{_4})
\nonumber\\
&&\hspace{2.5cm}\times
\Big\{n_{_4}!\Gamma(2+n_{_1}+n_{_2})\Gamma(2+n_{_2}+n_{_3})
\Gamma({D\over2}+1+n_{_1}+n_{_2})
\nonumber\\
&&\hspace{2.5cm}\times
\Gamma({D\over2}+1+n_{_2}+n_{_3})
\Gamma({D\over2}-2-n_{_2}-n_{_3}-n_{_4})\Gamma(1-{D\over2}-n_{_2})
\nonumber\\
&&\hspace{2.5cm}\times
\Gamma({D\over2}-1-n_{_1})\Gamma(2-{D\over2}+n_{_4})\Big\}^{-1}\;.
\label{GKZ21c-6-3}
\end{eqnarray}

\item   $I_{_{19}}=\{2,3,5,6,8,\cdots,13\}$, i.e. the implement $J_{_{19}}=[1,14]\setminus I_{_{19}}=\{1,4,7,14\}$.
The choice implies the power numbers $\alpha_{_1}=\alpha_{_{4}}=\alpha_{_{7}}=\alpha_{_{14}}=0$, and
\begin{eqnarray}
&&\alpha_{_2}=a_{_1}-a_{_2},\;\alpha_{_3}=a_{_4}-a_{_3},
\;\alpha_{_5}=-a_{_5},
\nonumber\\
&&\alpha_{_{6}}=a_{_4}+b_{_1}-a_{_1}-1,\;\alpha_{_{8}}=b_{_2}+b_{_3}-a_{_4}-2,
\nonumber\\
&&\alpha_{_{9}}=b_{_4}-a_{_4}-1,\;\alpha_{_{10}}=b_{_5}-a_{_4}-1,
\nonumber\\
&&\alpha_{_{11}}=a_{_4}-a_{_1},\;\alpha_{_{12}}=1-b_{_2},\;\alpha_{_{13}}=b_{_2}-a_{_4}-1.
\label{GKZ21c-11-1}
\end{eqnarray}
The corresponding hypergeometric series are
\begin{eqnarray}
&&\Phi_{_{[13\tilde{5}7]}}^{(19),a}(\alpha,z)=
y_{_1}^{-1}y_{_2}^{{D\over2}-1}y_{_3}^{{D\over2}-2}\sum\limits_{n_{_1}=0}^\infty
\sum\limits_{n_{_2}=0}^\infty\sum\limits_{n_{_3}=0}^\infty\sum\limits_{n_{_4}=0}^\infty
c_{_{[13\tilde{5}7]}}^{(19),a}(\alpha,{\bf n})
\nonumber\\
&&\hspace{2.5cm}\times
\Big({1\over y_{_1}}\Big)^{n_{_1}}\Big({y_{_4}\over y_{_3}}\Big)^{n_{_2}}
\Big({y_{_4}\over y_{_1}}\Big)^{n_{_3}}\Big({y_{_2}\over y_{_3}}\Big)^{n_{_4}}
\;,\nonumber\\
&&\Phi_{_{[13\tilde{5}7]}}^{(19),b}(\alpha,z)=
y_{_2}^{{D\over2}-1}y_{_3}^{{D\over2}-3}\sum\limits_{n_{_1}=0}^\infty
\sum\limits_{n_{_2}=0}^\infty\sum\limits_{n_{_3}=0}^\infty\sum\limits_{n_{_4}=0}^\infty
c_{_{[13\tilde{5}7]}}^{(19),b}(\alpha,{\bf n})
\nonumber\\
&&\hspace{2.5cm}\times
\Big({1\over y_{_3}}\Big)^{n_{_1}}\Big({y_{_1}\over y_{_3}}\Big)^{n_{_2}}
\Big({y_{_4}\over y_{_3}}\Big)^{n_{_3}}\Big({y_{_2}\over y_{_3}}\Big)^{n_{_4}}
\;,\nonumber\\
&&\Phi_{_{[13\tilde{5}7]}}^{(19),c}(\alpha,z)=
y_{_1}^{-1}y_{_2}^{{D\over2}-1}y_{_3}^{{D\over2}-3}\sum\limits_{n_{_1}=0}^\infty
\sum\limits_{n_{_2}=0}^\infty\sum\limits_{n_{_3}=0}^\infty\sum\limits_{n_{_4}=0}^\infty
c_{_{[13\tilde{5}7]}}^{(19),c}(\alpha,{\bf n})
\nonumber\\
&&\hspace{2.5cm}\times
\Big({1\over y_{_1}}\Big)^{n_{_1}}\Big({1\over y_{_3}}\Big)^{n_{_2}}
\Big({y_{_4}\over y_{_3}}\Big)^{n_{_3}}\Big({y_{_2}\over y_{_3}}\Big)^{n_{_4}}\;.
\label{GKZ21c-11-2a}
\end{eqnarray}
Where the coefficients are
\begin{eqnarray}
&&c_{_{[13\tilde{5}7]}}^{(19),a}(\alpha,{\bf n})=
(-)^{n_{_1}+n_{_4}}\Gamma(1+n_{_1}+n_{_3})\Gamma(1+n_{_2}+n_{_4})\Big\{n_{_1}!n_{_2}!n_{_3}!n_{_4}!
\nonumber\\
&&\hspace{2.5cm}\times
\Gamma({D\over2}+n_{_1})\Gamma(2-{D\over2}+n_{_2})\Gamma(1-{D\over2}-n_{_1}-n_{_3})
\nonumber\\
&&\hspace{2.5cm}\times
\Gamma({D\over2}+n_{_3})\Gamma({D\over2}+n_{_4})\Gamma({D\over2}-1-n_{_2}-n_{_4})\Big\}^{-1}
\;,\nonumber\\
&&c_{_{[13\tilde{5}7]}}^{(19),b}(\alpha,{\bf n})=
(-)^{1+n_{_4}}\Gamma(1+n_{_1}+n_{_2})\Gamma(2+n_{_1}+n_{_2}+n_{_3}+n_{_4})\Big\{n_{_1}!n_{_2}!n_{_4}!
\nonumber\\
&&\hspace{2.5cm}\times
\Gamma(2+n_{_1}+n_{_2}+n_{_3})\Gamma({D\over2}+n_{_1})
\Gamma(3-{D\over2}+n_{_1}+n_{_2}+n_{_3})
\nonumber\\
&&\hspace{2.5cm}\times
\Gamma(2-{D\over2}+n_{_2})\Gamma({D\over2}-1-n_{_1}-n_{_2})
\Gamma({D\over2}+n_{_4})
\nonumber\\
&&\hspace{2.5cm}\times
\Gamma({D\over2}-2-n_{_1}-n_{_2}-n_{_3}-n_{_4})\Big\}^{-1}
\;,\nonumber\\
&&c_{_{[13\tilde{5}7]}}^{(19),c}(\alpha,{\bf n})=
(-)^{1+n_{_1}+n_{_4}}\Gamma(1+n_{_1})\Gamma(1+n_{_2})\Gamma(2+n_{_2}+n_{_3}+n_{_4})
\Big\{n_{_4}!
\nonumber\\
&&\hspace{2.5cm}\times
\Gamma(2+n_{_1}+n_{_2})\Gamma(2+n_{_2}+n_{_3})\Gamma({D\over2}+1+n_{_1}+n_{_2})
\nonumber\\
&&\hspace{2.5cm}\times
\Gamma(3-{D\over2}+n_{_2}+n_{_3})\Gamma(1-{D\over2}-n_{_1})\Gamma({D\over2}-1-n_{_2})
\nonumber\\
&&\hspace{2.5cm}\times
\Gamma({D\over2}+n_{_4})\Gamma({D\over2}-2-n_{_2}-n_{_3}-n_{_4})\Big\}^{-1}\;.
\label{GKZ21c-11-3}
\end{eqnarray}

\item   $I_{_{20}}=\{2,3,5,\cdots,11,13\}$, i.e. the implement $J_{_{20}}=[1,14]\setminus I_{_{20}}=\{1,4,12,14\}$.
The choice implies the power numbers $\alpha_{_1}=\alpha_{_{4}}=\alpha_{_{12}}=\alpha_{_{14}}=0$, and
\begin{eqnarray}
&&\alpha_{_2}=a_{_1}-a_{_2},\;\alpha_{_3}=a_{_4}-a_{_3},\;\alpha_{_5}=-a_{_5},
\nonumber\\
&&\alpha_{_{6}}=a_{_4}+b_{_1}-a_{_1}-1,\;\alpha_{_{7}}=b_{_2}-1,
\nonumber\\
&&\alpha_{_{8}}=b_{_3}-a_{_4}-1,\;\alpha_{_{9}}=b_{_4}-a_{_4}-1,
\nonumber\\
&&\alpha_{_{10}}=b_{_5}-a_{_4}-1,\;\alpha_{_{11}}=a_{_4}-a_{_1},\;\alpha_{_{13}}=-a_{_4}.
\label{GKZ21c-12-1}
\end{eqnarray}
The corresponding hypergeometric series functions are written as
\begin{eqnarray}
&&\Phi_{_{[13\tilde{5}7]}}^{(20),a}(\alpha,z)=
y_{_1}^{-1}y_{_3}^{{D}-3}\sum\limits_{n_{_1}=0}^\infty
\sum\limits_{n_{_2}=0}^\infty\sum\limits_{n_{_3}=0}^\infty\sum\limits_{n_{_4}=0}^\infty
c_{_{[13\tilde{5}7]}}^{(20),a}(\alpha,{\bf n})
\nonumber\\
&&\hspace{2.5cm}\times
\Big({1\over y_{_1}}\Big)^{n_{_1}}\Big({y_{_4}\over y_{_3}}\Big)^{n_{_2}}
\Big({y_{_4}\over y_{_1}}\Big)^{n_{_3}}\Big({y_{_2}\over y_{_3}}\Big)^{n_{_4}}
\;,\nonumber\\
&&\Phi_{_{[13\tilde{5}7]}}^{(20),b}(\alpha,z)=
y_{_3}^{{D}-4}\sum\limits_{n_{_1}=0}^\infty
\sum\limits_{n_{_2}=0}^\infty\sum\limits_{n_{_3}=0}^\infty\sum\limits_{n_{_4}=0}^\infty
c_{_{[13\tilde{5}7]}}^{(20),b}(\alpha,{\bf n})
\nonumber\\
&&\hspace{2.5cm}\times
\Big({1\over y_{_3}}\Big)^{n_{_1}}\Big({y_{_1}\over y_{_3}}\Big)^{n_{_2}}
\Big({y_{_4}\over y_{_3}}\Big)^{n_{_3}}\Big({y_{_2}\over y_{_3}}\Big)^{n_{_4}}
\;,\nonumber\\
&&\Phi_{_{[13\tilde{5}7]}}^{(20),c}(\alpha,z)=
y_{_1}^{-1}y_{_3}^{{D}-4}\sum\limits_{n_{_1}=0}^\infty
\sum\limits_{n_{_2}=0}^\infty\sum\limits_{n_{_3}=0}^\infty\sum\limits_{n_{_4}=0}^\infty
c_{_{[13\tilde{5}7]}}^{(20),c}(\alpha,{\bf n})
\nonumber\\
&&\hspace{2.5cm}\times
\Big({1\over y_{_1}}\Big)^{n_{_1}}\Big({1\over y_{_3}}\Big)^{n_{_2}}
\Big({y_{_4}\over y_{_3}}\Big)^{n_{_3}}\Big({y_{_2}\over y_{_3}}\Big)^{n_{_4}}\;.
\label{GKZ21c-12-2a}
\end{eqnarray}
Where the coefficients are
\begin{eqnarray}
&&c_{_{[13\tilde{5}7]}}^{(20),a}(\alpha,{\bf n})=
(-)^{n_{_1}+n_{_2}}\Gamma(1+n_{_1}+n_{_3})\Big\{n_{_1}!n_{_2}!n_{_3}!n_{_4}!
\Gamma({D\over2}+n_{_1})\Gamma(2-{D\over2}+n_{_2})
\nonumber\\
&&\hspace{2.5cm}\times
\Gamma(1-{D\over2}-n_{_1}-n_{_3})\Gamma(2-{D\over2}+n_{_4})
\Gamma({D\over2}-1-n_{_2}-n_{_4})
\nonumber\\
&&\hspace{2.5cm}\times
\Gamma({D\over2}+n_{_3})\Gamma(D-2-n_{_2}-n_{_4})\Big\}^{-1}
\;,\nonumber\\
&&c_{_{[13\tilde{5}7]}}^{(20),b}(\alpha,{\bf n})=
(-)^{n_{_1}+n_{_2}+n_{_3}}\Gamma(1+n_{_1}+n_{_2})\Big\{n_{_1}!n_{_2}!n_{_4}!
\Gamma(2+n_{_1}+n_{_2}+n_{_3})
\nonumber\\
&&\hspace{2.5cm}\times
\Gamma({D\over2}+n_{_1})\Gamma(3-{D\over2}+n_{_1}+n_{_2}+n_{_3})
\Gamma(2-{D\over2}+n_{_2})
\nonumber\\
&&\hspace{2.5cm}\times
\Gamma(2-{D\over2}+n_{_4})\Gamma({D\over2}-2-n_{_1}-n_{_2}-n_{_3}-n_{_4})
\nonumber\\
&&\hspace{2.5cm}\times
\Gamma({D\over2}-1-n_{_1}-n_{_2})\Gamma(D-2-n_{_1}-n_{_2}-n_{_3}-n_{_4})\Big\}^{-1}
\;,\nonumber\\
&&c_{_{[13\tilde{5}7]}}^{(20),c}(\alpha,{\bf n})=
(-)^{n_{_1}+n_{_2}+n_{_3}}\Gamma(1+n_{_1})\Gamma(1+n_{_2})\Big\{n_{_4}!
\Gamma(2+n_{_1}+n_{_2})\Gamma(2+n_{_2}+n_{_3})
\nonumber\\
&&\hspace{2.5cm}\times
\Gamma({D\over2}+1+n_{_1}+n_{_2})\Gamma(3-{D\over2}+n_{_2}+n_{_3})
\Gamma(1-{D\over2}-n_{_1})
\nonumber\\
&&\hspace{2.5cm}\times
\Gamma(2-{D\over2}+n_{_4})\Gamma({D\over2}-2-n_{_2}-n_{_3}-n_{_4})
\nonumber\\
&&\hspace{2.5cm}\times
\Gamma({D\over2}-1-n_{_2})\Gamma(D-3-n_{_2}-n_{_3}-n_{_4})\Big\}^{-1}\;.
\label{GKZ21c-12-3}
\end{eqnarray}

\item   $I_{_{21}}=\{1,4,5,6,8,\cdots,13\}$, i.e. the implement $J_{_{21}}=[1,14]\setminus I_{_{21}}=\{2,3,7,14\}$.
The choice implies the power numbers $\alpha_{_2}=\alpha_{_{3}}=\alpha_{_{7}}=\alpha_{_{14}}=0$, and
\begin{eqnarray}
&&\alpha_{_1}=a_{_2}-a_{_1},\;\alpha_{_4}=a_{_3}-a_{_4},
\;\alpha_{_5}=-a_{_5},
\nonumber\\
&&\alpha_{_{6}}=a_{_3}+b_{_1}-a_{_2}-1,\;\alpha_{_{8}}=b_{_2}+b_{_3}-a_{_3}-2,
\nonumber\\
&&\alpha_{_{9}}=b_{_4}-a_{_3}-1,\;\alpha_{_{10}}=b_{_5}-a_{_3}-1,
\nonumber\\
&&\alpha_{_{11}}=a_{_3}-a_{_2},\;\alpha_{_{12}}=1-b_{_2},\;\alpha_{_{13}}=b_{_2}-a_{_3}-1.
\label{GKZ21c-21-1}
\end{eqnarray}
The corresponding hypergeometric series solutions are given as
\begin{eqnarray}
&&\Phi_{_{[13\tilde{5}7]}}^{(21),a}(\alpha,z)=
y_{_1}^{{D}-3}y_{_2}^{{D\over2}-1}y_{_3}^{-1}\sum\limits_{n_{_1}=0}^\infty
\sum\limits_{n_{_2}=0}^\infty\sum\limits_{n_{_3}=0}^\infty\sum\limits_{n_{_4}=0}^\infty
c_{_{[13\tilde{5}7]}}^{(21),a}(\alpha,{\bf n})
\nonumber\\
&&\hspace{2.5cm}\times
\Big({1\over y_{_1}}\Big)^{n_{_1}}\Big({y_{_4}\over y_{_3}}\Big)^{n_{_2}}
\Big({y_{_4}\over y_{_1}}\Big)^{n_{_3}}\Big({y_{_2}\over y_{_3}}\Big)^{n_{_4}}
\;,\nonumber\\
&&\Phi_{_{[13\tilde{5}7]}}^{(21),b}(\alpha,z)=
y_{_1}^{{D}-2}y_{_2}^{{D\over2}-1}y_{_3}^{-2}\sum\limits_{n_{_1}=0}^\infty
\sum\limits_{n_{_2}=0}^\infty\sum\limits_{n_{_3}=0}^\infty\sum\limits_{n_{_4}=0}^\infty
c_{_{[13\tilde{5}7]}}^{(21),b}(\alpha,{\bf n})
\nonumber\\
&&\hspace{2.5cm}\times
\Big({1\over y_{_1}}\Big)^{n_{_1}}\Big({y_{_1}\over y_{_3}}\Big)^{n_{_2}}
\Big({y_{_4}\over y_{_3}}\Big)^{n_{_3}}\Big({y_{_2}\over y_{_3}}\Big)^{n_{_4}}\;.
\label{GKZ21c-21-2a}
\end{eqnarray}
Where the coefficients are
\begin{eqnarray}
&&c_{_{[13\tilde{5}7]}}^{(21),a}(\alpha,{\bf n})=
(-)^{n_{_3}+n_{_4}}\Gamma(1+n_{_2}+n_{_4})\Big\{n_{_1}!n_{_2}!n_{_3}!n_{_4}!
\Gamma(2-{D\over2}+n_{_1})
\nonumber\\
&&\hspace{2.5cm}\times
\Gamma({D\over2}+n_{_2})\Gamma({D\over2}-1-n_{_1}-n_{_3})
\Gamma(1-{D\over2}-n_{_2}-n_{_4})
\nonumber\\
&&\hspace{2.5cm}\times
\Gamma(2-{D\over2}+n_{_3})\Gamma(D-2-n_{_1}-n_{_3})
\Gamma({D\over2}+n_{_4})\Big\}^{-1}
\;,\nonumber\\
&&c_{_{[13\tilde{5}7]}}^{(21),b}(\alpha,{\bf n})=
(-)^{1+n_{_4}}\Gamma(1+n_{_2})\Gamma(2+n_{_2}+n_{_3}+n_{_4})\Big\{n_{_1}!n_{_4}!
\Gamma(2+n_{_2}+n_{_3})
\nonumber\\
&&\hspace{2.5cm}\times
\Gamma(2-{D\over2}+n_{_1})\Gamma({D\over2}+1+n_{_2}+n_{_3})\Gamma({D\over2}-n_{_1}+n_{_2})
\nonumber\\
&&\hspace{2.5cm}\times
\Gamma(-{D\over2}-n_{_2}-n_{_3}-n_{_4})\Gamma(1-{D\over2}-n_{_2})\Gamma(D-1-n_{_1}+n_{_2})
\nonumber\\
&&\hspace{2.5cm}\times
\Gamma({D\over2}+n_{_4})\Big\}^{-1}\;.
\label{GKZ21c-21-3}
\end{eqnarray}

\item   $I_{_{22}}=\{1,4,\cdots,11,13\}$, i.e. the implement $J_{_{22}}=[1,14]\setminus I_{_{22}}=\{2,3,12,14\}$.
The choice implies the power numbers $\alpha_{_2}=\alpha_{_{3}}=\alpha_{_{12}}=\alpha_{_{14}}=0$, and
\begin{eqnarray}
&&\alpha_{_1}=a_{_2}-a_{_1},\;\alpha_{_4}=a_{_3}-a_{_4},\;\alpha_{_5}=-a_{_5},
\nonumber\\
&&\alpha_{_{6}}=a_{_3}+b_{_1}-a_{_2}-1,\;\alpha_{_{7}}=b_{_2}-1,
\nonumber\\
&&\alpha_{_{8}}=b_{_3}-a_{_3}-1,\;\alpha_{_{9}}=b_{_4}-a_{_3}-1,
\nonumber\\
&&\alpha_{_{10}}=b_{_5}-a_{_3}-1,\;\alpha_{_{11}}=a_{_3}-a_{_2},\;\alpha_{_{13}}=-a_{_3}.
\label{GKZ21c-22-1}
\end{eqnarray}
The corresponding hypergeometric series solutions are written as
\begin{eqnarray}
&&\Phi_{_{[13\tilde{5}7]}}^{(22),a}(\alpha,z)=
y_{_1}^{{D}-3}y_{_3}^{{D\over2}-2}\sum\limits_{n_{_1}=0}^\infty
\sum\limits_{n_{_2}=0}^\infty\sum\limits_{n_{_3}=0}^\infty\sum\limits_{n_{_4}=0}^\infty
c_{_{[13\tilde{5}7]}}^{(22),a}(\alpha,{\bf n})
\nonumber\\
&&\hspace{2.5cm}\times
\Big({1\over y_{_1}}\Big)^{n_{_1}}\Big({y_{_4}\over y_{_3}}\Big)^{n_{_2}}
\Big({y_{_4}\over y_{_1}}\Big)^{n_{_3}}\Big({y_{_2}\over y_{_3}}\Big)^{n_{_4}}
\;,\nonumber\\
&&\Phi_{_{[13\tilde{5}7]}}^{(22),b}(\alpha,z)=
y_{_1}^{{D}-2}y_{_3}^{{D\over2}-3}\sum\limits_{n_{_1}=0}^\infty
\sum\limits_{n_{_2}=0}^\infty\sum\limits_{n_{_3}=0}^\infty\sum\limits_{n_{_4}=0}^\infty
c_{_{[13\tilde{5}7]}}^{(22),b}(\alpha,{\bf n})
\nonumber\\
&&\hspace{2.5cm}\times
\Big({1\over y_{_1}}\Big)^{n_{_1}}\Big({y_{_1}\over y_{_3}}\Big)^{n_{_2}}
\Big({y_{_4}\over y_{_3}}\Big)^{n_{_3}}\Big({y_{_2}\over y_{_3}}\Big)^{n_{_4}}\;.
\label{GKZ21c-22-2a}
\end{eqnarray}
Where the coefficients are
\begin{eqnarray}
&&c_{_{[13\tilde{5}7]}}^{(22),a}(\alpha,{\bf n})=
(-)^{n_{_3}+n_{_4}}\Gamma(1+n_{_2}+n_{_4})\Big\{n_{_1}!n_{_2}!n_{_3}!n_{_4}!
\Gamma(2-{D\over2}+n_{_1})
\nonumber\\
&&\hspace{2.5cm}\times
\Gamma({D\over2}+n_{_2})\Gamma({D\over2}-1-n_{_1}-n_{_3})
\Gamma({D\over2}-1-n_{_2}-n_{_4})
\nonumber\\
&&\hspace{2.5cm}\times
\Gamma(2-{D\over2}+n_{_3})\Gamma(D-2-n_{_1}-n_{_3})
\Gamma(2-{D\over2}+n_{_4})\Big\}^{-1}
\;,\nonumber\\
&&c_{_{[13\tilde{5}7]}}^{(22),b}(\alpha,{\bf n})=
(-)^{1+n_{_4}}\Gamma(1+n_{_2})\Gamma(2+n_{_2}+n_{_3}+n_{_4})\Big\{n_{_1}!n_{_4}!
\Gamma(2+n_{_2}+n_{_3})
\nonumber\\
&&\hspace{2.5cm}\times
\Gamma(2-{D\over2}+n_{_1})\Gamma({D\over2}+1+n_{_2}+n_{_3})\Gamma({D\over2}-n_{_1}+n_{_2})
\nonumber\\
&&\hspace{2.5cm}\times
\Gamma({D\over2}-2-n_{_2}-n_{_3}-n_{_4})\Gamma(1-{D\over2}-n_{_2})\Gamma(D-1-n_{_1}+n_{_2})
\nonumber\\
&&\hspace{2.5cm}\times
\Gamma(2-{D\over2}+n_{_4})\Big\}^{-1}\;.
\label{GKZ21c-22-3}
\end{eqnarray}

\item   $I_{_{23}}=\{1,3,5,6,8,\cdots,13\}$, i.e. the implement $J_{_{23}}=[1,14]\setminus I_{_{23}}=\{2,4,7,14\}$.
The choice implies the power numbers $\alpha_{_2}=\alpha_{_{4}}=\alpha_{_{7}}=\alpha_{_{14}}=0$, and
\begin{eqnarray}
&&\alpha_{_1}=a_{_2}-a_{_1},\;\alpha_{_3}=a_{_4}-a_{_3},
\;\alpha_{_5}=-a_{_5},
\nonumber\\
&&\alpha_{_{6}}=a_{_4}+b_{_1}-a_{_2}-1,\;\alpha_{_{8}}=b_{_2}+b_{_3}-a_{_4}-2,
\nonumber\\
&&\alpha_{_{9}}=b_{_4}-a_{_4}-1,\;\alpha_{_{10}}=b_{_5}-a_{_4}-1,
\nonumber\\
&&\alpha_{_{11}}=a_{_4}-a_{_2},\;\alpha_{_{12}}=1-b_{_2},\;\alpha_{_{13}}=b_{_2}-a_{_4}-1.
\label{GKZ21c-27-1}
\end{eqnarray}
The corresponding hypergeometric functions are written as
\begin{eqnarray}
&&\Phi_{_{[13\tilde{5}7]}}^{(23),a}(\alpha,z)=
y_{_1}^{{D\over2}-2}y_{_2}^{{D\over2}-1}y_{_3}^{{D\over2}-2}\sum\limits_{n_{_1}=0}^\infty
\sum\limits_{n_{_2}=0}^\infty\sum\limits_{n_{_3}=0}^\infty\sum\limits_{n_{_4}=0}^\infty
c_{_{[13\tilde{5}7]}}^{(23),a}(\alpha,{\bf n})
\nonumber\\
&&\hspace{2.5cm}\times
\Big({1\over y_{_1}}\Big)^{n_{_1}}\Big({y_{_4}\over y_{_3}}\Big)^{n_{_2}}
\Big({y_{_4}\over y_{_1}}\Big)^{n_{_3}}\Big({y_{_2}\over y_{_3}}\Big)^{n_{_4}}
\;,\nonumber\\
&&\Phi_{_{[13\tilde{5}7]}}^{(23),b}(\alpha,z)=
y_{_1}^{{D\over2}-1}y_{_2}^{{D\over2}-1}y_{_3}^{{D\over2}-3}\sum\limits_{n_{_1}=0}^\infty
\sum\limits_{n_{_2}=0}^\infty\sum\limits_{n_{_3}=0}^\infty\sum\limits_{n_{_4}=0}^\infty
c_{_{[13\tilde{5}7]}}^{(23),b}(\alpha,{\bf n})
\nonumber\\
&&\hspace{2.5cm}\times
\Big({1\over y_{_3}}\Big)^{n_{_1}}\Big({y_{_1}\over y_{_3}}\Big)^{n_{_2}}
\Big({y_{_4}\over y_{_3}}\Big)^{n_{_3}}\Big({y_{_2}\over y_{_3}}\Big)^{n_{_4}}
\;,\nonumber\\
&&\Phi_{_{[13\tilde{5}7]}}^{(23),c}(\alpha,z)=
y_{_1}^{{D\over2}-2}y_{_2}^{{D\over2}-1}y_{_3}^{{D\over2}-3}\sum\limits_{n_{_1}=0}^\infty
\sum\limits_{n_{_2}=0}^\infty\sum\limits_{n_{_3}=0}^\infty\sum\limits_{n_{_4}=0}^\infty
c_{_{[13\tilde{5}7]}}^{(23),c}(\alpha,{\bf n})
\nonumber\\
&&\hspace{2.5cm}\times
\Big({1\over y_{_1}}\Big)^{n_{_1}}\Big({1\over y_{_3}}\Big)^{n_{_2}}
\Big({y_{_4}\over y_{_3}}\Big)^{n_{_3}}\Big({y_{_2}\over y_{_3}}\Big)^{n_{_4}}\;.
\label{GKZ21c-27-2a}
\end{eqnarray}
Where the coefficients are
\begin{eqnarray}
&&c_{_{[13\tilde{5}7]}}^{(23),a}(\alpha,{\bf n})=
(-)^{n_{_1}+n_{_4}}\Gamma(1+n_{_1}+n_{_3})\Gamma(1+n_{_2}+n_{_4})
\Big\{n_{_1}!n_{_2}!n_{_3}!n_{_4}!
\nonumber\\
&&\hspace{2.5cm}\times
\Gamma(2-{D\over2}+n_{_1})\Gamma(2-{D\over2}+n_{_2})
\Gamma({D\over2}+n_{_3})
\nonumber\\
&&\hspace{2.5cm}\times
\Gamma({D\over2}-1-n_{_1}-n_{_3})
\Gamma({D\over2}+n_{_4})\Gamma({D\over2}-1-n_{_2}-n_{_4})\Big\}^{-1}
\;,\nonumber\\
&&c_{_{[13\tilde{5}7]}}^{(23),b}(\alpha,{\bf n})=
(-)^{1+n_{_4}}\Gamma(1+n_{_1}+n_{_2})\Gamma(2+n_{_1}+n_{_2}+n_{_3}+n_{_4})
\Big\{n_{_1}!n_{_2}!n_{_4}!
\nonumber\\
&&\hspace{2.5cm}\times
\Gamma(2+n_{_1}+n_{_2}+n_{_3})\Gamma(2-{D\over2}+n_{_1})\Gamma({D\over2}-1-n_{_1}-n_{_2})
\nonumber\\
&&\hspace{2.5cm}\times
\Gamma(3-{D\over2}+n_{_1}+n_{_2}+n_{_3})
\Gamma({D\over2}+n_{_2})\Gamma({D\over2}+n_{_4})
\nonumber\\
&&\hspace{2.5cm}\times
\Gamma({D\over2}-2-n_{_1}-n_{_2}-n_{_3}-n_{_4})\Big\}^{-1}
\;,\nonumber\\
&&c_{_{[13\tilde{5}7]}}^{(23),c}(\alpha,{\bf n})=
(-)^{1+n_{_1}+n_{_4}}\Gamma(1+n_{_1})\Gamma(1+n_{_2})\Gamma(2+n_{_2}+n_{_3}+n_{_4})
\nonumber\\
&&\hspace{2.5cm}\times
\Big\{n_{_4}!\Gamma(2+n_{_1}+n_{_2})\Gamma(2+n_{_2}+n_{_3})\Gamma(3-{D\over2}+n_{_1}+n_{_2})
\nonumber\\
&&\hspace{2.5cm}\times
\Gamma(3-{D\over2}+n_{_2}+n_{_3})\Gamma({D\over2}-1-n_{_2})\Gamma({D\over2}-1-n_{_1})
\nonumber\\
&&\hspace{2.5cm}\times
\Gamma({D\over2}+n_{_4})\Gamma({D\over2}-2-n_{_2}-n_{_3}-n_{_4})\Big\}^{-1}\;.
\label{GKZ21c-27-3}
\end{eqnarray}

\item   $I_{_{24}}=\{1,3,5,\cdots,11,13\}$, i.e. the implement $J_{_{24}}=[1,14]\setminus I_{_{24}}=\{2,4,12,14\}$.
The choice implies the power numbers $\alpha_{_2}=\alpha_{_{4}}=\alpha_{_{12}}=\alpha_{_{14}}=0$, and
\begin{eqnarray}
&&\alpha_{_1}=a_{_2}-a_{_1},\;\alpha_{_3}=a_{_4}-a_{_3},\;\alpha_{_5}=-a_{_5},
\nonumber\\
&&\alpha_{_{6}}=a_{_4}+b_{_1}-a_{_2}-1,\;\alpha_{_{7}}=b_{_2}-1,
\nonumber\\
&&\alpha_{_{8}}=b_{_3}-a_{_4}-1,\;\alpha_{_{9}}=b_{_4}-a_{_4}-1,
\nonumber\\
&&\alpha_{_{10}}=b_{_5}-a_{_4}-1,\;\alpha_{_{11}}=a_{_4}-a_{_2},\;\alpha_{_{13}}=-a_{_4}.
\label{GKZ21c-28-1}
\end{eqnarray}
The corresponding hypergeometric functions are written as
\begin{eqnarray}
&&\Phi_{_{[13\tilde{5}7]}}^{(24),a}(\alpha,z)=
y_{_1}^{{D\over2}-2}y_{_3}^{{D}-3}\sum\limits_{n_{_1}=0}^\infty
\sum\limits_{n_{_2}=0}^\infty\sum\limits_{n_{_3}=0}^\infty\sum\limits_{n_{_4}=0}^\infty
c_{_{[13\tilde{5}7]}}^{(24),a}(\alpha,{\bf n})
\nonumber\\
&&\hspace{2.5cm}\times
\Big({1\over y_{_1}}\Big)^{n_{_1}}\Big({y_{_4}\over y_{_3}}\Big)^{n_{_2}}
\Big({y_{_4}\over y_{_1}}\Big)^{n_{_3}}\Big({y_{_2}\over y_{_3}}\Big)^{n_{_4}}
\;,\nonumber\\
&&\Phi_{_{[13\tilde{5}7]}}^{(24),b}(\alpha,z)=
y_{_1}^{{D\over2}-1}y_{_3}^{{D}-4}\sum\limits_{n_{_1}=0}^\infty
\sum\limits_{n_{_2}=0}^\infty\sum\limits_{n_{_3}=0}^\infty\sum\limits_{n_{_4}=0}^\infty
c_{_{[13\tilde{5}7]}}^{(24),b}(\alpha,{\bf n})
\nonumber\\
&&\hspace{2.5cm}\times
\Big({1\over y_{_3}}\Big)^{n_{_1}}\Big({y_{_1}\over y_{_3}}\Big)^{n_{_2}}
\Big({y_{_4}\over y_{_3}}\Big)^{n_{_3}}\Big({y_{_2}\over y_{_3}}\Big)^{n_{_4}}
\;,\nonumber\\
&&\Phi_{_{[13\tilde{5}7]}}^{(24),c}(\alpha,z)=
y_{_1}^{{D\over2}-2}y_{_3}^{{D}-4}\sum\limits_{n_{_1}=0}^\infty
\sum\limits_{n_{_2}=0}^\infty\sum\limits_{n_{_3}=0}^\infty\sum\limits_{n_{_4}=0}^\infty
c_{_{[13\tilde{5}7]}}^{(24),c}(\alpha,{\bf n})
\nonumber\\
&&\hspace{2.5cm}\times
\Big({1\over y_{_1}}\Big)^{n_{_1}}\Big({1\over y_{_3}}\Big)^{n_{_2}}
\Big({y_{_4}\over y_{_3}}\Big)^{n_{_3}}\Big({y_{_2}\over y_{_3}}\Big)^{n_{_4}}\;.
\label{GKZ21c-28-2a}
\end{eqnarray}
Where the coefficients are
\begin{eqnarray}
&&c_{_{[13\tilde{5}7]}}^{(24),a}(\alpha,{\bf n})=
(-)^{n_{_1}}\Gamma(1+n_{_1}+n_{_3})\Big\{n_{_1}!n_{_2}!n_{_3}!n_{_4}!
\Gamma(2-{D\over2}+n_{_1})
\nonumber\\
&&\hspace{2.5cm}\times
\Gamma(2-{D\over2}+n_{_2})\Gamma(2-{D\over2}+n_{_4})\Gamma({D\over2}-1-n_{_2}-n_{_4})
\nonumber\\
&&\hspace{2.5cm}\times
\Gamma({D\over2}+n_{_3})\Gamma({D\over2}-1-n_{_1}-n_{_3})\Gamma(D-2-n_{_2}-n_{_4})\Big\}^{-1}
\;,\nonumber\\
&&c_{_{[13\tilde{5}7]}}^{(24),b}(\alpha,{\bf n})=
(-)^{n_{_1}+n_{_2}+n_{_3}}\Gamma(1+n_{_1}+n_{_2})\Big\{n_{_1}!n_{_2}!n_{_4}!
\Gamma(2+n_{_1}+n_{_2}+n_{_3})
\nonumber\\
&&\hspace{2.5cm}\times
\Gamma(2-{D\over2}+n_{_1})\Gamma(3-{D\over2}+n_{_1}+n_{_2}+n_{_3})\Gamma(2-{D\over2}+n_{_4})
\nonumber\\
&&\hspace{2.5cm}\times
\Gamma({D\over2}-2-n_{_1}-n_{_2}-n_{_3}-n_{_4})\Gamma({D\over2}-1-n_{_1}-n_{_2})
\nonumber\\
&&\hspace{2.5cm}\times
\Gamma({D\over2}+n_{_2})\Gamma(D-3-n_{_1}-n_{_2}-n_{_3}-n_{_4})\Big\}^{-1}
\;,\nonumber\\
&&c_{_{[13\tilde{5}7]}}^{(24),c}(\alpha,{\bf n})=
(-)^{n_{_1}+n_{_2}+n_{_3}}\Gamma(1+n_{_1})\Gamma(1+n_{_2})
\Big\{n_{_4}!\Gamma(2+n_{_1}+n_{_2})
\nonumber\\
&&\hspace{2.5cm}\times
\Gamma(2+n_{_2}+n_{_3})\Gamma(3-{D\over2}+n_{_1}+n_{_2})
\Gamma(3-{D\over2}+n_{_2}+n_{_3})
\nonumber\\
&&\hspace{2.5cm}\times
\Gamma(2-{D\over2}+n_{_4})\Gamma({D\over2}-2-n_{_2}-n_{_3}-n_{_4})
\Gamma({D\over2}-1-n_{_2})
\nonumber\\
&&\hspace{2.5cm}\times
\Gamma({D\over2}-1-n_{_1})\Gamma(D-3-n_{_2}-n_{_3}-n_{_4})\Big\}^{-1}\;.
\label{GKZ21c-28-3}
\end{eqnarray}

\item   $I_{_{25}}=\{2,3,4,5,8,\cdots,13\}$, i.e. the implement $J_{_{25}}=[1,14]\setminus I_{_{25}}=\{1,6,7,14\}$.
The choice implies the power numbers $\alpha_{_1}=\alpha_{_{6}}=\alpha_{_{7}}=\alpha_{_{14}}=0$, and
\begin{eqnarray}
&&\alpha_{_2}=a_{_1}-a_{_2},\;\alpha_{_3}=a_{_1}-a_{_3}-b_{_1}+1,
\nonumber\\
&&\alpha_{_4}=a_{_1}-a_{_4}-b_{_1}+1,\;\alpha_{_5}=-a_{_5},
\nonumber\\
&&\alpha_{_{8}}=b_{_1}+b_{_2}+b_{_3}-a_{_1}-3,\;\alpha_{_{9}}=b_{_1}+b_{_4}-a_{_1}-2,
\nonumber\\
&&\alpha_{_{10}}=b_{_1}+b_{_5}-a_{_1}-2,\;\alpha_{_{11}}=1-b_{_1},
\nonumber\\
&&\alpha_{_{12}}=1-b_{_2},\;\alpha_{_{13}}=b_{_1}+b_{_2}-a_{_1}-2.
\label{GKZ21c-13-1}
\end{eqnarray}
The corresponding hypergeometric function is
\begin{eqnarray}
&&\Phi_{_{[13\tilde{5}7]}}^{(25)}(\alpha,z)=
y_{_1}^{{D\over2}-1}y_{_2}^{{D\over2}-1}y_{_3}^{-2}\sum\limits_{n_{_1}=0}^\infty
\sum\limits_{n_{_2}=0}^\infty\sum\limits_{n_{_3}=0}^\infty\sum\limits_{n_{_4}=0}^\infty
c_{_{[13\tilde{5}7]}}^{(25)}(\alpha,{\bf n})
\nonumber\\
&&\hspace{2.5cm}\times
\Big({1\over y_{_3}}\Big)^{n_{_1}}\Big({y_{_1}\over y_{_3}}\Big)^{n_{_2}}
\Big({y_{_4}\over y_{_3}}\Big)^{n_{_3}}\Big({y_{_2}\over y_{_3}}\Big)^{n_{_4}}\;,
\label{GKZ21c-13-2}
\end{eqnarray}
with
\begin{eqnarray}
&&c_{_{[13\tilde{5}7]}}^{(25)}(\alpha,{\bf n})=
(-)^{1+n_{_4}}\Gamma(2+n_{_1}+n_{_2}+n_{_3}+n_{_4})\Gamma(1+n_{_1}+n_{_2})
\nonumber\\
&&\hspace{2.5cm}\times
\Big\{n_{_1}!n_{_2}!n_{_4}!\Gamma({D\over2}+n_{_1})
\Gamma(2+n_{_1}+n_{_2}+n_{_3})
\nonumber\\
&&\hspace{2.5cm}\times
\Gamma(1+{D\over2}+n_{_1}+n_{_2}+n_{_3})\Gamma(-{D\over2}-n_{_1}-n_{_2}-n_{_3}-n_{_4})
\nonumber\\
&&\hspace{2.5cm}\times
\Gamma(1-{D\over2}-n_{_1}-n_{_2})\Gamma({D\over2}+n_{_2})
\Gamma({D\over2}+n_{_4})\Big\}^{-1}\;.
\label{GKZ21c-13-3}
\end{eqnarray}

\item   $I_{_{26}}=\{2,3,4,5,7,\cdots,11,13\}$, i.e. the implement $J_{_{26}}=[1,14]\setminus I_{_{26}}=\{1,6,12,14\}$.
The choice implies the power numbers $\alpha_{_1}=\alpha_{_{6}}=\alpha_{_{12}}=\alpha_{_{14}}=0$, and
\begin{eqnarray}
&&\alpha_{_2}=a_{_1}-a_{_2},\;\alpha_{_3}=a_{_1}-a_{_3}-b_{_1}+1,
\nonumber\\
&&\alpha_{_4}=a_{_1}-a_{_4}-b_{_1}+1,\;\alpha_{_5}=-a_{_5},\;\alpha_{_{7}}=b_{_2}-1,
\nonumber\\
&&\alpha_{_{8}}=b_{_1}+b_{_3}-a_{_1}-2,\;\alpha_{_{9}}=b_{_1}+b_{_4}-a_{_1}-2,
\nonumber\\
&&\alpha_{_{10}}=b_{_1}+b_{_5}-a_{_1}-2,\;\alpha_{_{11}}=1-b_{_1},\;
\alpha_{_{13}}=b_{_1}-a_{_1}-1.
\label{GKZ21c-14-1}
\end{eqnarray}
The corresponding hypergeometric series is written as
\begin{eqnarray}
&&\Phi_{_{[13\tilde{5}7]}}^{(26)}(\alpha,z)=
y_{_1}^{{D\over2}-1}y_{_3}^{{D\over2}-3}\sum\limits_{n_{_1}=0}^\infty
\sum\limits_{n_{_2}=0}^\infty\sum\limits_{n_{_3}=0}^\infty\sum\limits_{n_{_4}=0}^\infty
c_{_{[13\tilde{5}7]}}^{(26)}(\alpha,{\bf n})
\nonumber\\
&&\hspace{2.5cm}\times
\Big({1\over y_{_3}}\Big)^{n_{_1}}\Big({y_{_1}\over y_{_3}}\Big)^{n_{_2}}
\Big({y_{_4}\over y_{_3}}\Big)^{n_{_3}}\Big({y_{_2}\over y_{_3}}\Big)^{n_{_4}}\;,
\label{GKZ21c-14-2}
\end{eqnarray}
with
\begin{eqnarray}
&&c_{_{[13\tilde{5}7]}}^{(26)}(\alpha,{\bf n})=
(-)^{1+n_{_4}}\Gamma(2+n_{_1}+n_{_2}+n_{_3}+n_{_4})\Gamma(1+n_{_1}+n_{_2})
\nonumber\\
&&\hspace{2.5cm}\times
\Big\{n_{_1}!n_{_2}!n_{_4}!\Gamma({D\over2}+n_{_1})\Gamma(2+n_{_1}+n_{_2}+n_{_3})
\Gamma(2-{D\over2}+n_{_4})
\nonumber\\
&&\hspace{2.5cm}\times
\Gamma(1+{D\over2}+n_{_1}+n_{_2}+n_{_3})\Gamma(1-{D\over2}-n_{_1}-n_{_2})
\Gamma({D\over2}+n_{_2})
\nonumber\\
&&\hspace{2.5cm}\times
\Gamma({D\over2}-2-n_{_1}-n_{_2}-n_{_3}-n_{_4})\Big\}^{-1}\;.
\label{GKZ21c-14-3}
\end{eqnarray}

\item   $I_{_{27}}=\{2,\cdots,6,8,9,10,12,13\}$, i.e. the implement $J_{_{27}}=[1,14]\setminus I_{_{27}}=\{1,7,11,14\}$.
The choice implies the power numbers $\alpha_{_1}=\alpha_{_{7}}=\alpha_{_{11}}=\alpha_{_{14}}=0$, and
\begin{eqnarray}
&&\alpha_{_2}=a_{_1}-a_{_2},\;\alpha_{_3}=a_{_1}-a_{_3},\;\alpha_{_4}=a_{_1}-a_{_4},\;
\alpha_{_5}=-a_{_5},
\nonumber\\
&&\alpha_{_{6}}=b_{_1}-1,\;\alpha_{_{8}}=b_{_2}+b_{_3}-a_{_1}-2,\;\alpha_{_{9}}=b_{_4}-a_{_1}-1,
\nonumber\\
&&\alpha_{_{10}}=b_{_5}-a_{_1}-1,\;
\alpha_{_{12}}=1-b_{_2},\;\alpha_{_{13}}=b_{_2}-a_{_1}-1.
\label{GKZ21c-15-1}
\end{eqnarray}
The corresponding hypergeometric series is
\begin{eqnarray}
&&\Phi_{_{[13\tilde{5}7]}}^{(27)}(\alpha,z)=
y_{_2}^{{D\over2}-1}y_{_3}^{{D\over2}-3}\sum\limits_{n_{_1}=0}^\infty
\sum\limits_{n_{_2}=0}^\infty\sum\limits_{n_{_3}=0}^\infty\sum\limits_{n_{_4}=0}^\infty
c_{_{[13\tilde{5}7]}}^{(27)}(\alpha,{\bf n})
\nonumber\\
&&\hspace{2.5cm}\times
\Big({1\over y_{_3}}\Big)^{n_{_1}}\Big({y_{_1}\over y_{_3}}\Big)^{n_{_2}}
\Big({y_{_4}\over y_{_3}}\Big)^{n_{_3}}\Big({y_{_2}\over y_{_3}}\Big)^{n_{_4}}\;,
\label{GKZ21c-15-2}
\end{eqnarray}
with
\begin{eqnarray}
&&c_{_{[13\tilde{5}7]}}^{(27)}(\alpha,{\bf n})=
(-)^{1+n_{_4}}\Gamma(1+n_{_1}+n_{_2})\Gamma(2+n_{_1}+n_{_2}+n_{_3}+n_{_4})
\Big\{n_{_1}!n_{_2}!n_{_4}!
\nonumber\\
&&\hspace{2.5cm}\times
\Gamma({D\over2}+n_{_1})\Gamma(3-{D\over2}+n_{_1}+n_{_2}+n_{_3})
\Gamma(2-{D\over2}+n_{_2})
\nonumber\\
&&\hspace{2.5cm}\times
\Gamma(2+n_{_1}+n_{_2}+n_{_3})\Gamma({D\over2}-1-n_{_1}-n_{_2})
\Gamma({D\over2}+n_{_4})
\nonumber\\
&&\hspace{2.5cm}\times
\Gamma({D\over2}-2-n_{_1}-n_{_2}-n_{_3}-n_{_4})\Big\}^{-1}\;.
\label{GKZ21c-15-3}
\end{eqnarray}

\item   $I_{_{28}}=\{2,\cdots,10,13\}$, i.e. the implement $J_{_{28}}=[1,14]\setminus I_{_{28}}=\{1,11,12,14\}$.
The choice implies the power numbers $\alpha_{_1}=\alpha_{_{11}}=\alpha_{_{12}}=\alpha_{_{14}}=0$, and
\begin{eqnarray}
&&\alpha_{_2}=a_{_1}-a_{_2},\;\alpha_{_3}=a_{_1}-a_{_3},\;
\alpha_{_4}=a_{_1}-a_{_4},\;\alpha_{_5}=-a_{_5},
\nonumber\\
&&\alpha_{_{6}}=b_{_1}-1,\;\alpha_{_{7}}=b_{_2}-1,\;
\alpha_{_{8}}=b_{_3}-a_{_1}-1,
\nonumber\\
&&\alpha_{_{9}}=b_{_4}-a_{_1}-1,\;\alpha_{_{10}}=b_{_5}-a_{_1}-1,\;
\alpha_{_{13}}=-a_{_1}.
\label{GKZ21c-16-1}
\end{eqnarray}
The corresponding hypergeometric series is written as
\begin{eqnarray}
&&\Phi_{_{[13\tilde{5}7]}}^{(28)}(\alpha,z)=
y_{_3}^{{D}-4}\sum\limits_{n_{_1}=0}^\infty
\sum\limits_{n_{_2}=0}^\infty\sum\limits_{n_{_3}=0}^\infty\sum\limits_{n_{_4}=0}^\infty
c_{_{[13\tilde{5}7]}}^{(28)}(\alpha,{\bf n})
\nonumber\\
&&\hspace{2.5cm}\times
\Big({1\over y_{_3}}\Big)^{n_{_1}}\Big({y_{_1}\over y_{_3}}\Big)^{n_{_2}}
\Big({y_{_4}\over y_{_3}}\Big)^{n_{_3}}\Big({y_{_2}\over y_{_3}}\Big)^{n_{_4}}\;,
\label{GKZ21c-16-2}
\end{eqnarray}
with
\begin{eqnarray}
&&c_{_{[13\tilde{5}7]}}^{(28)}(\alpha,{\bf n})=
(-)^{n_{_1}+n_{_2}+n_{_3}}
\Gamma(1+n_{_1}+n_{_2})\Big\{n_{_1}!n_{_2}!n_{_4}!\Gamma({D\over2}+n_{_1})
\nonumber\\
&&\hspace{2.5cm}\times
\Gamma(3-{D\over2}+n_{_1}+n_{_2}+n_{_3})\Gamma(2+n_{_1}+n_{_2}+n_{_3})
\nonumber\\
&&\hspace{2.5cm}\times
\Gamma(2-{D\over2}+n_{_2})
\Gamma(2-{D\over2}+n_{_4})
\Gamma({D\over2}-2-n_{_1}-n_{_2}-n_{_3}-n_{_4})
\nonumber\\
&&\hspace{2.5cm}\times
\Gamma({D\over2}-1-n_{_1}-n_{_2})\Gamma(D-3-n_{_1}-n_{_2}-n_{_3}-n_{_4})\Big\}^{-1}\;.
\label{GKZ21c-16-3}
\end{eqnarray}

\item   $I_{_{29}}=\{1,3,4,5,8,\cdots,13\}$, i.e. the implement $J_{_{29}}=[1,14]\setminus I_{_{29}}=\{2,6,7,14\}$.
The choice implies the power numbers $\alpha_{_2}=\alpha_{_{6}}=\alpha_{_{7}}=\alpha_{_{14}}=0$, and
\begin{eqnarray}
&&\alpha_{_1}=a_{_2}-a_{_1},\;\alpha_{_3}=a_{_2}-a_{_3}-b_{_1}+1,
\nonumber\\
&&\alpha_{_4}=a_{_2}-a_{_4}-b_{_1}+1,\;\alpha_{_5}=-a_{_5},
\nonumber\\
&&\alpha_{_{8}}=b_{_1}+b_{_2}+b_{_3}-a_{_2}-3,\;\alpha_{_{9}}=b_{_1}+b_{_4}-a_{_2}-2,
\nonumber\\
&&\alpha_{_{10}}=b_{_1}+b_{_5}-a_{_2}-2,\;\alpha_{_{11}}=1-b_{_1},
\nonumber\\
&&\alpha_{_{12}}=1-b_{_2},\;\alpha_{_{13}}=b_{_1}+b_{_2}-a_{_2}-2.
\label{GKZ21c-29-1}
\end{eqnarray}
The corresponding hypergeometric series is written as
\begin{eqnarray}
&&\Phi_{_{[13\tilde{5}7]}}^{(29)}(\alpha,z)=
y_{_1}^{{D\over2}-1}y_{_2}^{{D\over2}-1}y_{_3}^{{D\over2}-3}\sum\limits_{n_{_1}=0}^\infty
\sum\limits_{n_{_2}=0}^\infty\sum\limits_{n_{_3}=0}^\infty\sum\limits_{n_{_4}=0}^\infty
c_{_{[13\tilde{5}7]}}^{(29)}(\alpha,{\bf n})
\nonumber\\
&&\hspace{2.5cm}\times
\Big({1\over y_{_3}}\Big)^{n_{_1}}\Big({y_{_1}\over y_{_3}}\Big)^{n_{_2}}
\Big({y_{_4}\over y_{_3}}\Big)^{n_{_3}}\Big({y_{_2}\over y_{_3}}\Big)^{n_{_4}}\;,
\label{GKZ21c-29-2}
\end{eqnarray}
with
\begin{eqnarray}
&&c_{_{[13\tilde{5}7]}}^{(29)}(\alpha,{\bf n})=
(-)^{1+n_{_4}}\Gamma(1+n_{_1}+n_{_2})
\Gamma(2+n_{_1}+n_{_2}+n_{_3}+n_{_4})
\nonumber\\
&&\hspace{2.5cm}\times
\Big\{n_{_1}!n_{_2}!n_{_4}!\Gamma(2-{D\over2}+n_{_1})
\Gamma(3-{D\over2}+n_{_1}+n_{_2}+n_{_3})
\nonumber\\
&&\hspace{2.5cm}\times
\Gamma(2+n_{_1}+n_{_2}+n_{_3})\Gamma({D\over2}-1-n_{_1}-n_{_2})
\Gamma({D\over2}+n_{_2})
\nonumber\\
&&\hspace{2.5cm}\times
\Gamma({D\over2}+n_{_4})
\Gamma({D\over2}-2-n_{_1}-n_{_2}-n_{_3}-n_{_4})\Big\}^{-1}\;.
\label{GKZ21c-29-3}
\end{eqnarray}

\item   $I_{_{30}}=\{1,3,4,5,7,\cdots,11,13\}$, i.e. the implement $J_{_{30}}=[1,14]\setminus I_{_{30}}=\{2,6,12,14\}$.
The choice implies the power numbers $\alpha_{_2}=\alpha_{_{6}}=\alpha_{_{12}}=\alpha_{_{14}}=0$, and
\begin{eqnarray}
&&\alpha_{_1}=a_{_2}-a_{_1},\;\alpha_{_3}=a_{_2}-a_{_3}-b_{_1}+1,
\nonumber\\
&&\alpha_{_4}=a_{_2}-a_{_4}-b_{_1}+1,\;\alpha_{_5}=-a_{_5},\;\alpha_{_{7}}=b_{_2}-1,
\nonumber\\
&&\alpha_{_{8}}=b_{_1}+b_{_3}-a_{_2}-2,\;\alpha_{_{9}}=b_{_1}+b_{_4}-a_{_2}-2,
\nonumber\\
&&\alpha_{_{10}}=b_{_1}+b_{_5}-a_{_2}-2,\;\alpha_{_{11}}=1-b_{_1},\;
\alpha_{_{13}}=b_{_1}-a_{_2}-1.
\label{GKZ21c-30-1}
\end{eqnarray}
The corresponding hypergeometric series is written as
\begin{eqnarray}
&&\Phi_{_{[13\tilde{5}7]}}^{(30)}(\alpha,z)=
y_{_1}^{{D\over2}-1}y_{_3}^{{D}-4}\sum\limits_{n_{_1}=0}^\infty
\sum\limits_{n_{_2}=0}^\infty\sum\limits_{n_{_3}=0}^\infty\sum\limits_{n_{_4}=0}^\infty
c_{_{[13\tilde{5}7]}}^{(30)}(\alpha,{\bf n})
\nonumber\\
&&\hspace{2.5cm}\times
\Big({1\over y_{_3}}\Big)^{n_{_1}}\Big({y_{_1}\over y_{_3}}\Big)^{n_{_2}}
\Big({y_{_4}\over y_{_3}}\Big)^{n_{_3}}\Big({y_{_2}\over y_{_3}}\Big)^{n_{_4}}\;,
\label{GKZ21c-30-2}
\end{eqnarray}
with
\begin{eqnarray}
&&c_{_{[13\tilde{5}7]}}^{(30)}(\alpha,{\bf n})=
(-)^{n_{_1}+n_{_2}+n_{_3}}
\Gamma(1+n_{_1}+n_{_2})\Big\{n_{_1}!n_{_2}!n_{_4}!\Gamma(2-{D\over2}+n_{_1})
\nonumber\\
&&\hspace{2.5cm}\times
\Gamma(3-{D\over2}+n_{_1}+n_{_2}+n_{_3})\Gamma(2+n_{_1}+n_{_2}+n_{_3})
\Gamma(2-{D\over2}+n_{_4})
\nonumber\\
&&\hspace{2.5cm}\times
\Gamma({D\over2}-2-n_{_1}-n_{_2}-n_{_3}-n_{_4})
\Gamma({D\over2}-1-n_{_1}-n_{_2})
\nonumber\\
&&\hspace{2.5cm}\times
\Gamma({D\over2}+n_{_2})
\Gamma(D-3-n_{_1}-n_{_2}-n_{_3}-n_{_4})\Big\}^{-1}\;.
\label{GKZ21c-30-3}
\end{eqnarray}

\item   $I_{_{31}}=\{1,3,\cdots,6,8,9,10,12,13\}$, i.e. the implement $J_{_{31}}=[1,14]\setminus I_{_{31}}=\{2,7,11,14\}$.
The choice implies the power numbers $\alpha_{_2}=\alpha_{_{7}}=\alpha_{_{11}}=\alpha_{_{14}}=0$, and
\begin{eqnarray}
&&\alpha_{_1}=a_{_2}-a_{_1},\;\alpha_{_3}=a_{_2}-a_{_3},\;\alpha_{_4}=a_{_2}-a_{_4},\;
\alpha_{_5}=-a_{_5},
\nonumber\\
&&\alpha_{_{6}}=b_{_1}-1,\;\alpha_{_{8}}=b_{_2}+b_{_3}-a_{_2}-2,\;\alpha_{_{9}}=b_{_4}-a_{_2}-1,
\nonumber\\
&&\alpha_{_{10}}=b_{_5}-a_{_4}-1,\;
\alpha_{_{12}}=1-b_{_2},\;\alpha_{_{13}}=b_{_2}-a_{_2}-1.
\label{GKZ21c-31-1}
\end{eqnarray}
The corresponding hypergeometric series is written as
\begin{eqnarray}
&&\Phi_{_{[13\tilde{5}7]}}^{(31)}(\alpha,z)=
y_{_2}^{{D\over2}-1}y_{_3}^{{D}-4}\sum\limits_{n_{_1}=0}^\infty
\sum\limits_{n_{_2}=0}^\infty\sum\limits_{n_{_3}=0}^\infty\sum\limits_{n_{_4}=0}^\infty
c_{_{[13\tilde{5}7]}}^{(31)}(\alpha,{\bf n})
\nonumber\\
&&\hspace{2.5cm}\times
\Big({1\over y_{_3}}\Big)^{n_{_1}}\Big({y_{_1}\over y_{_3}}\Big)^{n_{_2}}
\Big({y_{_4}\over y_{_3}}\Big)^{n_{_3}}\Big({y_{_2}\over y_{_3}}\Big)^{n_{_4}}\;,
\label{GKZ21c-31-2}
\end{eqnarray}
with
\begin{eqnarray}
&&c_{_{[13\tilde{5}7]}}^{(31)}(\alpha,{\bf n})=
(-)^{n_{_3}}\Big\{n_{_1}!n_{_2}!n_{_4}!\Gamma(2-{D\over2}+n_{_1})
\nonumber\\
&&\hspace{2.5cm}\times
\Gamma(4-D+n_{_1}+n_{_2}+n_{_3})\Gamma(3-{D\over2}+n_{_1}+n_{_2}+n_{_3})
\nonumber\\
&&\hspace{2.5cm}\times
\Gamma(2-{D\over2}+n_{_2})\Gamma({D\over2}-2-n_{_1}-n_{_2}-n_{_3}-n_{_4})
\nonumber\\
&&\hspace{2.5cm}\times
\Gamma(D-2-n_{_1}-n_{_2})\Gamma({D\over2}-1-n_{_1}-n_{_2})
\nonumber\\
&&\hspace{2.5cm}\times
\Gamma({D\over2}+n_{_4})\Gamma(D-3-n_{_1}-n_{_2}-n_{_3}-n_{_4})\Big\}^{-1}\;.
\label{GKZ21c-31-3}
\end{eqnarray}

\item   $I_{_{32}}=\{1,3,\cdots,10,13\}$, i.e. the implement $J_{_{32}}=[1,14]\setminus I_{_{32}}=\{2,11,12,14\}$.
The choice implies the power numbers $\alpha_{_2}=\alpha_{_{11}}=\alpha_{_{12}}=\alpha_{_{14}}=0$, and
\begin{eqnarray}
&&\alpha_{_1}=a_{_2}-a_{_1},\;\alpha_{_3}=a_{_2}-a_{_3},\;
\alpha_{_4}=a_{_2}-a_{_4},\;\alpha_{_5}=-a_{_5},
\nonumber\\
&&\alpha_{_{6}}=b_{_1}-1,\;\alpha_{_{7}}=b_{_2}-1,\;
\alpha_{_{8}}=b_{_3}-a_{_2}-1,
\nonumber\\
&&\alpha_{_{9}}=b_{_4}-a_{_2}-1,\;\alpha_{_{10}}=b_{_5}-a_{_2}-1,\;
\alpha_{_{13}}=-a_{_2}.
\label{GKZ21c-32-1}
\end{eqnarray}
The corresponding hypergeometric series is written as
\begin{eqnarray}
&&\Phi_{_{[13\tilde{5}7]}}^{(32)}(\alpha,z)=
y_{_3}^{{3D\over2}-5}\sum\limits_{n_{_1}=0}^\infty
\sum\limits_{n_{_2}=0}^\infty\sum\limits_{n_{_3}=0}^\infty\sum\limits_{n_{_4}=0}^\infty
c_{_{[13\tilde{5}7]}}^{(32)}(\alpha,{\bf n})
\nonumber\\
&&\hspace{2.5cm}\times
\Big({1\over y_{_3}}\Big)^{n_{_1}}\Big({y_{_1}\over y_{_3}}\Big)^{n_{_2}}
\Big({y_{_4}\over y_{_3}}\Big)^{n_{_3}}\Big({y_{_2}\over y_{_3}}\Big)^{n_{_4}}\;,
\label{GKZ21c-32-2}
\end{eqnarray}
with
\begin{eqnarray}
&&c_{_{[13\tilde{5}7]}}^{(32)}(\alpha,{\bf n})=
(-)^{n_{_3}}\Big\{n_{_1}!n_{_2}!n_{_4}!\Gamma(2-{D\over2}+n_{_1})
\Gamma(4-D+n_{_1}+n_{_2}+n_{_3})
\nonumber\\
&&\hspace{2.5cm}\times
\Gamma(3-{D\over2}+n_{_1}+n_{_2}+n_{_3})
\Gamma(2-{D\over2}+n_{_2})\Gamma(2-{D\over2}+n_{_4})
\nonumber\\
&&\hspace{2.5cm}\times
\Gamma(D-3-n_{_1}-n_{_2}-n_{_3}-n_{_4})\Gamma(D-2-n_{_1}-n_{_2})
\nonumber\\
&&\hspace{2.5cm}\times
\Gamma({D\over2}-1-n_{_1}-n_{_2})
\Gamma({3D\over2}-4-n_{_1}-n_{_2}-n_{_3}-n_{_4})\Big\}^{-1}\;.
\label{GKZ21c-32-3}
\end{eqnarray}
\end{itemize}

\section{The hypergeometric solutions of the integer lattice ${\bf B}_{_{135\tilde{7}}}$\label{app5}}
\indent\indent

\begin{itemize}
\item   $I_{_{1}}=\{2,4,6,7,9,\cdots,14\}$, i.e. the implement $J_{_{1}}=[1,14]\setminus I_{_{1}}=\{1,3,5,8\}$.
The choice implies the power numbers $\alpha_{_1}=\alpha_{_{3}}=\alpha_{_{5}}=\alpha_{_{8}}=0$, and
\begin{eqnarray}
&&\alpha_{_2}=a_{_1}-a_{_2},\;\alpha_{_4}=a_{_3}-a_{_4},
\;\alpha_{_6}=a_{_3}+a_{_5}+b_{_1}-a_{_1}-1,
\nonumber\\
&&\alpha_{_{7}}=b_{_2}+b_{_3}-a_{_3}-2,\;\alpha_{_{9}}=b_{_4}-a_{_3}-a_{_5}-1,
\nonumber\\
&&\alpha_{_{10}}=b_{_5}-a_{_3}-a_{_5}-1,\;\alpha_{_{11}}=a_{_3}+a_{_5}-a_{_1},
\nonumber\\
&&\alpha_{_{12}}=b_{_3}-a_{_3}-1,\;\alpha_{_{13}}=1-b_{_3},\;\alpha_{_{14}}=-a_{_5}\;.
\label{GKZ21d-1-1}
\end{eqnarray}
The corresponding hypergeometric series solutions are written as
\begin{eqnarray}
&&\Phi_{_{[135\tilde{7}]}}^{(1),a}(\alpha,z)=
y_{_1}^{{D\over2}-1}y_{_2}^{-1}y_{_3}^{{D\over2}-1}y_{_4}^{-1}\sum\limits_{n_{_1}=0}^\infty
\sum\limits_{n_{_2}=0}^\infty\sum\limits_{n_{_3}=0}^\infty\sum\limits_{n_{_4}=0}^\infty
c_{_{[135\tilde{7}]}}^{(1),a}(\alpha,{\bf n})
\nonumber\\
&&\hspace{2.5cm}\times
\Big({1\over y_{_4}}\Big)^{n_{_1}}\Big({y_{_4}\over y_{_2}}\Big)^{n_{_2}}
\Big({y_{_1}\over y_{_4}}\Big)^{n_{_3}}\Big({y_{_3}\over y_{_2}}\Big)^{n_{_4}}
\;,\nonumber\\
&&\Phi_{_{[135\tilde{7}]}}^{(1),b}(\alpha,z)=
y_{_1}^{{D\over2}-2}y_{_2}^{-1}y_{_3}^{{D\over2}-1}y_{_4}^{-1}\sum\limits_{n_{_1}=0}^\infty
\sum\limits_{n_{_2}=0}^\infty\sum\limits_{n_{_3}=0}^\infty\sum\limits_{n_{_4}=0}^\infty
c_{_{[135\tilde{7}]}}^{(1),b}(\alpha,{\bf n})
\nonumber\\
&&\hspace{2.5cm}\times
\Big({1\over y_{_1}}\Big)^{n_{_1}}\Big({1\over y_{_2}}\Big)^{n_{_2}}
\Big({1\over y_{_4}}\Big)^{n_{_3}}\Big({y_{_3}\over y_{_2}}\Big)^{n_{_4}}\;.
\label{GKZ21d-1-2a}
\end{eqnarray}
Where the coefficients are
\begin{eqnarray}
&&c_{_{[135\tilde{7}]}}^{(1),a}(\alpha,{\bf n})=
(-)^{n_{_4}}\Gamma(1+n_{_1}+n_{_3})\Gamma(1+n_{_2}+n_{_4})
\Big\{n_{_1}!n_{_2}!n_{_3}!n_{_4}!
\nonumber\\
&&\hspace{2.5cm}\times
\Gamma({D\over2}+n_{_1})\Gamma({D\over2}+n_{_2})
\Gamma({D\over2}+n_{_3})\Gamma(1-{D\over2}-n_{_2}-n_{_4})
\nonumber\\
&&\hspace{2.5cm}\times
\Gamma(1-{D\over2}-n_{_1}-n_{_3})\Gamma({D\over2}+n_{_4})\Big\}^{-1}
\;,\nonumber\\
&&c_{_{[135\tilde{7}]}}^{(1),b}(\alpha,{\bf n})=
(-)^{n_{_1}+n_{_4}}\Gamma(1+n_{_1})\Gamma(1+n_{_2}+n_{_3})\Gamma(1+n_{_2}+n_{_4})
\nonumber\\
&&\hspace{2.5cm}\times
\Big\{n_{_2}!n_{_4}!\Gamma(2+n_{_1}+n_{_2}+n_{_3})\Gamma({D\over2}+1+n_{_1}+n_{_2}+n_{_3})
\nonumber\\
&&\hspace{2.5cm}\times
\Gamma({D\over2}+n_{_2})\Gamma({D\over2}-1-n_{_1})\Gamma(1-{D\over2}-n_{_2}-n_{_4})
\nonumber\\
&&\hspace{2.5cm}\times
\Gamma(1-{D\over2}-n_{_2}-n_{_3})\Gamma({D\over2}+n_{_4})\Big\}^{-1}\;.
\label{GKZ21d-1-3}
\end{eqnarray}

\item   $I_{_{2}}=\{2,4,6,\cdots,12,14\}$, i.e. the implement $J_{_{2}}=[1,14]\setminus I_{_{2}}=\{1,3,5,13\}$.
The choice implies the power numbers $\alpha_{_1}=\alpha_{_{3}}=\alpha_{_{5}}=\alpha_{_{13}}=0$, and
\begin{eqnarray}
&&\alpha_{_2}=a_{_1}-a_{_2},\;\alpha_{_4}=a_{_3}-a_{_4},
\;\alpha_{_6}=a_{_3}+a_{_5}+b_{_1}-a_{_1}-1,
\nonumber\\
&&\alpha_{_{7}}=b_{_2}-a_{_3}-1,\;\alpha_{_{8}}=b_{_3}-1,\;\alpha_{_{9}}=b_{_4}-a_{_3}-a_{_5}-1,
\nonumber\\
&&\alpha_{_{10}}=b_{_5}-a_{_3}-a_{_5}-1,\;\alpha_{_{11}}=a_{_3}+a_{_5}-a_{_1},\;\alpha_{_{12}}=-a_{_3},
\nonumber\\
&&\alpha_{_{14}}=-a_{_5}\;.
\label{GKZ21d-2-1}
\end{eqnarray}
The corresponding hypergeometric series functions are
\begin{eqnarray}
&&\Phi_{_{[135\tilde{7}]}}^{(2),a}(\alpha,z)=
y_{_1}^{{D\over2}-1}y_{_2}^{{D\over2}-2}y_{_4}^{-1}\sum\limits_{n_{_1}=0}^\infty
\sum\limits_{n_{_2}=0}^\infty\sum\limits_{n_{_3}=0}^\infty\sum\limits_{n_{_4}=0}^\infty
c_{_{[135\tilde{7}]}}^{(2),a}(\alpha,{\bf n})
\nonumber\\
&&\hspace{2.5cm}\times
\Big({1\over y_{_4}}\Big)^{n_{_1}}\Big({y_{_4}\over y_{_2}}\Big)^{n_{_2}}
\Big({y_{_1}\over y_{_4}}\Big)^{n_{_3}}\Big({y_{_3}\over y_{_2}}\Big)^{n_{_4}}
\;,\nonumber\\
&&\Phi_{_{[135\tilde{7}]}}^{(2),b}(\alpha,z)=
y_{_1}^{{D\over2}-2}y_{_2}^{{D\over2}-2}y_{_4}^{-1}\sum\limits_{n_{_1}=0}^\infty
\sum\limits_{n_{_2}=0}^\infty\sum\limits_{n_{_3}=0}^\infty\sum\limits_{n_{_4}=0}^\infty
c_{_{[135\tilde{7}]}}^{(2),b}(\alpha,{\bf n})
\nonumber\\
&&\hspace{2.5cm}\times
\Big({1\over y_{_1}}\Big)^{n_{_1}}\Big({1\over y_{_2}}\Big)^{n_{_2}}
\Big({1\over y_{_4}}\Big)^{n_{_3}}\Big({y_{_3}\over y_{_2}}\Big)^{n_{_4}}\;.
\label{GKZ21d-2-2a}
\end{eqnarray}
Where the coefficients are
\begin{eqnarray}
&&c_{_{[135\tilde{7}]}}^{(2),a}(\alpha,{\bf n})=
(-)^{n_{_4}}\Gamma(1+n_{_1}+n_{_3})\Gamma(1+n_{_2}+n_{_4})
\Big\{n_{_1}!n_{_2}!n_{_3}!n_{_4}!
\nonumber\\
&&\hspace{2.5cm}\times
\Gamma({D\over2}+n_{_1})
\Gamma({D\over2}+n_{_2})\Gamma(2-{D\over2}+n_{_4})\Gamma({D\over2}+n_{_3})
\nonumber\\
&&\hspace{2.5cm}\times
\Gamma(1-{D\over2}-n_{_1}-n_{_3})\Gamma({D\over2}-1-n_{_2}-n_{_4})\Big\}^{-1}
\;,\nonumber\\
&&c_{_{[135\tilde{7}]}}^{(2),b}(\alpha,{\bf n})=
(-)^{n_{_1}+n_{_4}}\Gamma(1+n_{_1})\Gamma(1+n_{_2}+n_{_3})\Gamma(1+n_{_2}+n_{_4})
\nonumber\\
&&\hspace{2.5cm}\times
\Big\{n_{_2}!n_{_4}!\Gamma(2+n_{_1}+n_{_2}+n_{_3})\Gamma({D\over2}+1+n_{_1}+n_{_2}+n_{_3})
\nonumber\\
&&\hspace{2.5cm}\times
\Gamma({D\over2}+n_{_2})\Gamma({D\over2}-1-n_{_1})\Gamma(2-{D\over2}+n_{_4})
\nonumber\\
&&\hspace{2.5cm}\times
\Gamma(1-{D\over2}-n_{_2}-n_{_3})\Gamma({D\over2}-1-n_{_2}-n_{_4})\Big\}^{-1}\;.
\label{GKZ21d-2-3}
\end{eqnarray}

\item   $I_{_{3}}=\{2,3,6,7,9,\cdots,14\}$, i.e. the implement $J_{_{3}}=[1,14]\setminus I_{_{3}}=\{1,4,5,8\}$.
The choice implies the power numbers $\alpha_{_1}=\alpha_{_{4}}=\alpha_{_{5}}=\alpha_{_{8}}=0$, and
\begin{eqnarray}
&&\alpha_{_2}=a_{_1}-a_{_2},\;\alpha_{_3}=a_{_4}-a_{_3},
\;\alpha_{_6}=a_{_4}+a_{_5}+b_{_1}-a_{_1}-1,
\nonumber\\
&&\alpha_{_{7}}=b_{_2}+b_{_3}-a_{_4}-2,\;\alpha_{_{9}}=b_{_4}-a_{_4}-a_{_5}-1,
\nonumber\\
&&\alpha_{_{10}}=b_{_5}-a_{_4}-a_{_5}-1,\;\alpha_{_{11}}=a_{_4}+a_{_5}-a_{_1},
\nonumber\\
&&\alpha_{_{12}}=b_{_3}-a_{_4}-1,\;\alpha_{_{13}}=1-b_{_3},\;\alpha_{_{14}}=-a_{_5}\;.
\label{GKZ21d-7-1}
\end{eqnarray}
The corresponding hypergeometric functions are
\begin{eqnarray}
&&\Phi_{_{[135\tilde{7}]}}^{(3),a}(\alpha,z)=
y_{_2}^{{D\over2}-2}y_{_3}^{{D\over2}-1}y_{_4}^{-1}\sum\limits_{n_{_1}=0}^\infty
\sum\limits_{n_{_2}=0}^\infty\sum\limits_{n_{_3}=0}^\infty\sum\limits_{n_{_4}=0}^\infty
c_{_{[135\tilde{7}]}}^{(3),a}(\alpha,{\bf n})
\nonumber\\
&&\hspace{2.5cm}\times
\Big({1\over y_{_4}}\Big)^{n_{_1}}\Big({y_{_4}\over y_{_2}}\Big)^{n_{_2}}
\Big({y_{_1}\over y_{_4}}\Big)^{n_{_3}}\Big({y_{_3}\over y_{_2}}\Big)^{n_{_4}}
\;,\nonumber\\
&&\Phi_{_{[135\tilde{7}]}}^{(3),b}(\alpha,z)=
y_{_1}^{-1}y_{_2}^{{D\over2}-2}y_{_3}^{{D\over2}-1}y_{_4}^{-1}\sum\limits_{n_{_1}=0}^\infty
\sum\limits_{n_{_2}=0}^\infty\sum\limits_{n_{_3}=0}^\infty\sum\limits_{n_{_4}=0}^\infty
c_{_{[135\tilde{7}]}}^{(3),b}(\alpha,{\bf n})
\nonumber\\
&&\hspace{2.5cm}\times
\Big({1\over y_{_1}}\Big)^{n_{_1}}\Big({1\over y_{_2}}\Big)^{n_{_2}}
\Big({1\over y_{_4}}\Big)^{n_{_3}}\Big({y_{_3}\over y_{_2}}\Big)^{n_{_4}}\;.
\label{GKZ21d-7-2a}
\end{eqnarray}
Where the coefficients are
\begin{eqnarray}
&&c_{_{[135\tilde{7}]}}^{(3),a}(\alpha,{\bf n})=
(-)^{n_{_4}}\Gamma(1+n_{_1}+n_{_3})\Gamma(1+n_{_2}+n_{_4})
\Big\{n_{_1}!n_{_2}!n_{_3}!n_{_4}!
\nonumber\\
&&\hspace{2.5cm}\times
\Gamma({D\over2}+n_{_1})
\Gamma(2-{D\over2}+n_{_2})\Gamma(2-{D\over2}+n_{_3})\Gamma({D\over2}+n_{_4})
\nonumber\\
&&\hspace{2.5cm}\times
\Gamma({D\over2}-1-n_{_1}-n_{_3})\Gamma({D\over2}-1-n_{_2}-n_{_4})\Big\}^{-1}
\;,\nonumber\\
&&c_{_{[135\tilde{7}]}}^{(3),b}(\alpha,{\bf n})=
(-)^{n_{_1}+n_{_4}}\Gamma(1+n_{_1})\Gamma(1+n_{_2}+n_{_3})\Gamma(1+n_{_2}+n_{_4})
\nonumber\\
&&\hspace{2.5cm}\times
\Big\{n_{_2}!n_{_4}!\Gamma(2+n_{_1}+n_{_2}+n_{_3})\Gamma({D\over2}+1+n_{_1}+n_{_2}+n_{_3})
\nonumber\\
&&\hspace{2.5cm}\times
\Gamma(2-{D\over2}+n_{_2})\Gamma(1-{D\over2}-n_{_1})\Gamma({D\over2}+n_{_4})
\nonumber\\
&&\hspace{2.5cm}\times
\Gamma({D\over2}-1-n_{_2}-n_{_3})\Gamma({D\over2}-1-n_{_2}-n_{_4})\Big\}^{-1}\;.
\label{GKZ21d-7-3}
\end{eqnarray}

\item   $I_{_{4}}=\{2,3,6,\cdots,12,14\}$, i.e. the implement $J_{_{4}}=[1,14]\setminus I_{_{4}}=\{1,4,5,13\}$.
The choice implies the power numbers $\alpha_{_1}=\alpha_{_{4}}=\alpha_{_{5}}=\alpha_{_{13}}=0$, and
\begin{eqnarray}
&&\alpha_{_2}=a_{_1}-a_{_2},\;\alpha_{_3}=a_{_4}-a_{_3},
\;\alpha_{_6}=a_{_4}+a_{_5}+b_{_1}-a_{_1}-1,
\nonumber\\
&&\alpha_{_{7}}=b_{_2}-a_{_4}-1,\;\alpha_{_{8}}=b_{_3}-1,\;\alpha_{_{9}}=b_{_4}-a_{_4}-a_{_5}-1,
\nonumber\\
&&\alpha_{_{10}}=b_{_5}-a_{_4}-a_{_5}-1,\;\alpha_{_{11}}=a_{_4}+a_{_5}-a_{_1},\;\alpha_{_{12}}=-a_{_4},
\nonumber\\
&&\alpha_{_{14}}=-a_{_5}\;.
\label{GKZ21d-8-1}
\end{eqnarray}
The corresponding hypergeometric functions are written as
\begin{eqnarray}
&&\Phi_{_{[135\tilde{7}]}}^{(4),a}(\alpha,z)=
y_{_2}^{{D}-3}y_{_4}^{-1}\sum\limits_{n_{_1}=0}^\infty
\sum\limits_{n_{_2}=0}^\infty\sum\limits_{n_{_3}=0}^\infty\sum\limits_{n_{_4}=0}^\infty
c_{_{[135\tilde{7}]}}^{(4),a}(\alpha,{\bf n})
\nonumber\\
&&\hspace{2.5cm}\times
\Big({1\over y_{_4}}\Big)^{n_{_1}}\Big({y_{_4}\over y_{_2}}\Big)^{n_{_2}}
\Big({y_{_1}\over y_{_4}}\Big)^{n_{_3}}\Big({y_{_3}\over y_{_2}}\Big)^{n_{_4}}
\;,\nonumber\\
&&\Phi_{_{[135\tilde{7}]}}^{(4),b}(\alpha,z)=
y_{_1}^{-1}y_{_2}^{{D}-3}y_{_4}^{-1}\sum\limits_{n_{_1}=0}^\infty
\sum\limits_{n_{_2}=0}^\infty\sum\limits_{n_{_3}=0}^\infty\sum\limits_{n_{_4}=0}^\infty
c_{_{[135\tilde{7}]}}^{(4),b}(\alpha,{\bf n})
\nonumber\\
&&\hspace{2.5cm}\times
\Big({1\over y_{_1}}\Big)^{n_{_1}}\Big({1\over y_{_2}}\Big)^{n_{_2}}
\Big({1\over y_{_4}}\Big)^{n_{_3}}\Big({y_{_3}\over y_{_2}}\Big)^{n_{_4}}\;.
\label{GKZ21d-8-2a}
\end{eqnarray}
Where the coefficients are
\begin{eqnarray}
&&c_{_{[135\tilde{7}]}}^{(4),a}(\alpha,{\bf n})=
(-)^{n_{_2}}\Gamma(1+n_{_1}+n_{_3})
\Big\{n_{_1}!n_{_2}!n_{_3}!n_{_4}!\Gamma({D\over2}+n_{_1})
\nonumber\\
&&\hspace{2.5cm}\times
\Gamma(2-{D\over2}+n_{_2})\Gamma(2-{D\over2}+n_{_3})
\Gamma({D\over2}-1-n_{_2}-n_{_4})
\nonumber\\
&&\hspace{2.5cm}\times
\Gamma(2-{D\over2}+n_{_4})\Gamma({D\over2}-1-n_{_1}-n_{_3})
\Gamma(D-2-n_{_2}-n_{_4})\Big\}^{-1}
\;,\nonumber\\
&&c_{_{[135\tilde{7}]}}^{(4),b}(\alpha,{\bf n})=
(-)^{n_{_1}+n_{_2}}\Gamma(1+n_{_1})\Gamma(1+n_{_2}+n_{_3})
\Big\{n_{_2}!n_{_4}!\Gamma(2+n_{_1}+n_{_2}+n_{_3})
\nonumber\\
&&\hspace{2.5cm}\times
\Gamma({D\over2}+1+n_{_1}+n_{_2}+n_{_3})
\Gamma(2-{D\over2}+n_{_2})\Gamma(1-{D\over2}-n_{_1})
\nonumber\\
&&\hspace{2.5cm}\times
\Gamma({D\over2}-1-n_{_2}-n_{_4})\Gamma(2-{D\over2}+n_{_4})\Gamma({D\over2}-1-n_{_2}-n_{_3})
\nonumber\\
&&\hspace{2.5cm}\times
\Gamma(D-2-n_{_2}-n_{_4})\Big\}^{-1}\;.
\label{GKZ21d-8-3}
\end{eqnarray}

\item   $I_{_{5}}=\{1,4,6,7,9,\cdots,14\}$, i.e. the implement $J_{_{5}}=[1,14]\setminus I_{_{5}}=\{2,3,5,8\}$.
The choice implies the power numbers $\alpha_{_2}=\alpha_{_{3}}=\alpha_{_{5}}=\alpha_{_{8}}=0$, and
\begin{eqnarray}
&&\alpha_{_1}=a_{_2}-a_{_1},\;\alpha_{_4}=a_{_3}-a_{_4},
\;\alpha_{_6}=a_{_3}+a_{_5}+b_{_1}-a_{_2}-1,
\nonumber\\
&&\alpha_{_{7}}=b_{_2}+b_{_3}-a_{_3}-2,\;\alpha_{_{9}}=b_{_4}-a_{_3}-a_{_5}-1,
\nonumber\\
&&\alpha_{_{10}}=b_{_5}-a_{_3}-a_{_5}-1,\;\alpha_{_{11}}=a_{_3}+a_{_5}-a_{_2},
\nonumber\\
&&\alpha_{_{12}}=b_{_3}-a_{_3}-1,\;\alpha_{_{13}}=1-b_{_3},\;\alpha_{_{14}}=-a_{_5}\;.
\label{GKZ21d-17-1}
\end{eqnarray}
The corresponding hypergeometric function is given as
\begin{eqnarray}
&&\Phi_{_{[135\tilde{7}]}}^{(5)}(\alpha,z)=
y_{_1}^{{D}-2}y_{_2}^{-1}y_{_3}^{{D\over2}-1}y_{_4}^{-1}\sum\limits_{n_{_1}=0}^\infty
\sum\limits_{n_{_2}=0}^\infty\sum\limits_{n_{_3}=0}^\infty\sum\limits_{n_{_4}=0}^\infty
{c_{_{[135\tilde{7}]}}^{(5)}(\alpha,{\bf n})}
\nonumber\\
&&\hspace{2.5cm}\times
\Big({1\over y_{_1}}\Big)^{n_{_1}}\Big({y_{_1}\over y_{_2}}\Big)^{n_{_2}}
\Big({y_{_1}\over y_{_4}}\Big)^{n_{_3}}\Big({y_{_3}\over y_{_2}}\Big)^{n_{_4}}\;,
\label{GKZ21d-17-2}
\end{eqnarray}
with
\begin{eqnarray}
&&c_{_{[135\tilde{7}]}}^{(5)}(\alpha,{\bf n})=
(-)^{n_{_4}}\Gamma(1+n_{_2}+n_{_3})\Gamma(1+n_{_2}+n_{_4})
\Big\{n_{_1}!n_{_2}!n_{_4}!
\nonumber\\
&&\hspace{2.5cm}\times
\Gamma(2-{D\over2}+n_{_1})\Gamma({D\over2}+n_{_2})\Gamma({D\over2}-n_{_1}+n_{_2}+n_{_3})
\nonumber\\
&&\hspace{2.5cm}\times
\Gamma(1-{D\over2}-n_{_2}-n_{_4})\Gamma(1-{D\over2}-n_{_2}-n_{_3})
\nonumber\\
&&\hspace{2.5cm}\times
\Gamma(D-1-n_{_1}+n_{_2}+n_{_3})\Gamma({D\over2}+n_{_4})\Big\}^{-1}\;.
\label{GKZ21d-17-3}
\end{eqnarray}

\item   $I_{_{6}}=\{1,4,6,\cdots,12,14\}$, i.e. the implement $J_{_{6}}=[1,14]\setminus I_{_{6}}=\{2,3,5,13\}$.
The choice implies the power numbers $\alpha_{_2}=\alpha_{_{3}}=\alpha_{_{5}}=\alpha_{_{13}}=0$, and
\begin{eqnarray}
&&\alpha_{_1}=a_{_2}-a_{_1},\;\alpha_{_4}=a_{_3}-a_{_4},
\;\alpha_{_6}=a_{_3}+a_{_5}+b_{_1}-a_{_2}-1,
\nonumber\\
&&\alpha_{_{7}}=b_{_2}-a_{_3}-1,\;\alpha_{_{8}}=b_{_3}-1,\;\alpha_{_{9}}=b_{_4}-a_{_3}-a_{_5}-1,
\nonumber\\
&&\alpha_{_{10}}=b_{_5}-a_{_3}-a_{_5}-1,\;\alpha_{_{11}}=a_{_3}+a_{_5}-a_{_2},\;\alpha_{_{12}}=-a_{_3},
\nonumber\\
&&\alpha_{_{14}}=-a_{_5}\;.
\label{GKZ21d-18-1}
\end{eqnarray}
The corresponding hypergeometric series is
\begin{eqnarray}
&&\Phi_{_{[135\tilde{7}]}}^{(6)}(\alpha,z)=
y_{_1}^{{D}-2}y_{_2}^{{D\over2}-2}y_{_4}^{-1}\sum\limits_{n_{_1}=0}^\infty
\sum\limits_{n_{_2}=0}^\infty\sum\limits_{n_{_3}=0}^\infty\sum\limits_{n_{_4}=0}^\infty
c_{_{[135\tilde{7}]}}^{(6)}(\alpha,{\bf n})
\nonumber\\
&&\hspace{2.5cm}\times
\Big({1\over y_{_1}}\Big)^{n_{_1}}\Big({y_{_1}\over y_{_2}}\Big)^{n_{_2}}
\Big({y_{_1}\over y_{_4}}\Big)^{n_{_3}}\Big({y_{_3}\over y_{_2}}\Big)^{n_{_4}}\;,
\label{GKZ21d-18-2}
\end{eqnarray}
with
\begin{eqnarray}
&&c_{_{[135\tilde{7}]}}^{(6)}(\alpha,{\bf n})=
(-)^{n_{_4}}\Gamma(1+n_{_2}+n_{_3})\Gamma(1+n_{_2}+n_{_4})
\Big\{n_{_1}!n_{_2}!n_{_4}!
\nonumber\\
&&\hspace{2.5cm}\times
\Gamma(2-{D\over2}+n_{_1})\Gamma({D\over2}+n_{_2})
\Gamma(2-{D\over2}+n_{_4})
\nonumber\\
&&\hspace{2.5cm}\times
\Gamma({D\over2}-n_{_1}+n_{_2}+n_{_3})\Gamma(1-{D\over2}-n_{_2}-n_{_3})
\nonumber\\
&&\hspace{2.5cm}\times
\Gamma(D-1-n_{_1}+n_{_2}+n_{_3})\Gamma({D\over2}-1-n_{_2}-n_{_4})\Big\}^{-1}\;.
\label{GKZ21d-18-3}
\end{eqnarray}

\item   $I_{_{7}}=\{1,3,6,7,9,\cdots,14\}$, i.e. the implement $J_{_{7}}=[1,14]\setminus I_{_{7}}=\{2,4,5,8\}$.
The choice implies the power numbers $\alpha_{_2}=\alpha_{_{4}}=\alpha_{_{5}}=\alpha_{_{8}}=0$, and
\begin{eqnarray}
&&\alpha_{_1}=a_{_2}-a_{_1},\;\alpha_{_3}=a_{_4}-a_{_3},
\;\alpha_{_6}=a_{_4}+a_{_5}+b_{_1}-a_{_2}-1,
\nonumber\\
&&\alpha_{_{7}}=b_{_2}+b_{_3}-a_{_4}-2,\;\alpha_{_{9}}=b_{_4}-a_{_4}-a_{_5}-1,
\nonumber\\
&&\alpha_{_{10}}=b_{_5}-a_{_4}-a_{_5}-1,\;\alpha_{_{11}}=a_{_4}+a_{_5}-a_{_2},
\nonumber\\
&&\alpha_{_{12}}=b_{_3}-a_{_4}-1,\;\alpha_{_{13}}=1-b_{_3},\;\alpha_{_{14}}=-a_{_5}\;.
\label{GKZ21d-23-1}
\end{eqnarray}
The corresponding hypergeometric functions are presented as
\begin{eqnarray}
&&\Phi_{_{[135\tilde{7}]}}^{(7),a}(\alpha,z)=
y_{_1}^{{D\over2}-1}y_{_2}^{{D\over2}-2}y_{_3}^{{D\over2}-1}y_{_4}^{-1}
\sum\limits_{n_{_1}=0}^\infty
\sum\limits_{n_{_2}=0}^\infty\sum\limits_{n_{_3}=0}^\infty\sum\limits_{n_{_4}=0}^\infty
c_{_{[135\tilde{7}]}}^{(7),a}(\alpha,{\bf n})
\nonumber\\
&&\hspace{2.5cm}\times
\Big({1\over y_{_4}}\Big)^{n_{_1}}\Big({y_{_4}\over y_{_2}}\Big)^{n_{_2}}
\Big({y_{_1}\over y_{_4}}\Big)^{n_{_3}}\Big({y_{_3}\over y_{_2}}\Big)^{n_{_4}}
\;,\nonumber\\
&&\Phi_{_{[135\tilde{7}]}}^{(7),b}(\alpha,z)=
y_{_1}^{{D\over2}-2}y_{_2}^{{D\over2}-2}y_{_3}^{{D\over2}-1}y_{_4}^{-1}
\sum\limits_{n_{_1}=0}^\infty
\sum\limits_{n_{_2}=0}^\infty\sum\limits_{n_{_3}=0}^\infty\sum\limits_{n_{_4}=0}^\infty
c_{_{[135\tilde{7}]}}^{(7),b}(\alpha,{\bf n})
\nonumber\\
&&\hspace{2.5cm}\times
\Big({1\over y_{_1}}\Big)^{n_{_1}}\Big({1\over y_{_2}}\Big)^{n_{_2}}
\Big({1\over y_{_4}}\Big)^{n_{_3}}\Big({y_{_3}\over y_{_2}}\Big)^{n_{_4}}\;.
\label{GKZ21d-23-2a}
\end{eqnarray}
Where the coefficients are
\begin{eqnarray}
&&c_{_{[135\tilde{7}]}}^{(7),a}(\alpha,{\bf n})=
(-)^{n_{_4}}\Gamma(1+n_{_1}+n_{_3})\Gamma(1+n_{_2}+n_{_4})
\nonumber\\
&&\hspace{2.5cm}\times
\Big\{n_{_1}!n_{_2}!n_{_3}!n_{_4}!\Gamma(2-{D\over2}+n_{_1})\Gamma(2-{D\over2}+n_{_2})
\Gamma({D\over2}+n_{_4})
\nonumber\\
&&\hspace{2.5cm}\times
\Gamma({D\over2}-1-n_{_1}-n_{_3})\Gamma({D\over2}+n_{_3})
\Gamma({D\over2}-1-n_{_2}-n_{_4})\Big\}^{-1}
\;,\nonumber\\
&&c_{_{[135\tilde{7}]}}^{(7),b}(\alpha,{\bf n})=
(-)^{n_{_1}+n_{_4}}\Gamma(1+n_{_1})\Gamma(1+n_{_2}+n_{_3})\Gamma(1+n_{_2}+n_{_4})
\nonumber\\
&&\hspace{2.5cm}\times
\Big\{n_{_2}!n_{_4}!\Gamma(2+n_{_1}+n_{_2}+n_{_3})\Gamma(3-{D\over2}+n_{_1}+n_{_2}+n_{_3})
\nonumber\\
&&\hspace{2.5cm}\times
\Gamma(2-{D\over2}+n_{_2})\Gamma({D\over2}+n_{_4})\Gamma({D\over2}-1-n_{_2}-n_{_3})
\nonumber\\
&&\hspace{2.5cm}\times
\Gamma({D\over2}-1-n_{_1})\Gamma({D\over2}-1-n_{_2}-n_{_4})\Big\}^{-1}\;.
\label{GKZ21d-23-3}
\end{eqnarray}

\item   $I_{_{8}}=\{1,3,6,\cdots,12,14\}$, i.e. the implement $J_{_{8}}=[1,14]\setminus I_{_{8}}=\{2,4,5,13\}$.
The choice implies the power numbers $\alpha_{_2}=\alpha_{_{4}}=\alpha_{_{5}}=\alpha_{_{13}}=0$, and
\begin{eqnarray}
&&\alpha_{_1}=a_{_2}-a_{_1},\;\alpha_{_3}=a_{_4}-a_{_3},
\;\alpha_{_6}=a_{_4}+a_{_5}+b_{_1}-a_{_2}-1,
\nonumber\\
&&\alpha_{_{7}}=b_{_2}-a_{_4}-1,\;\alpha_{_{8}}=b_{_3}-1,\;\alpha_{_{9}}=b_{_4}-a_{_4}-a_{_5}-1,
\nonumber\\
&&\alpha_{_{10}}=b_{_5}-a_{_4}-a_{_5}-1,\;\alpha_{_{11}}=a_{_4}+a_{_5}-a_{_2},\;\alpha_{_{12}}=-a_{_4},
\nonumber\\
&&\alpha_{_{14}}=-a_{_5}\;.
\label{GKZ21d-24-1}
\end{eqnarray}
The corresponding hypergeometric series functions are
\begin{eqnarray}
&&\Phi_{_{[135\tilde{7}]}}^{(8),a}(\alpha,z)=
y_{_1}^{{D\over2}-1}y_{_2}^{{D}-3}y_{_4}^{-1}\sum\limits_{n_{_1}=0}^\infty
\sum\limits_{n_{_2}=0}^\infty\sum\limits_{n_{_3}=0}^\infty\sum\limits_{n_{_4}=0}^\infty
c_{_{[135\tilde{7}]}}^{(8),a}(\alpha,{\bf n})
\nonumber\\
&&\hspace{2.5cm}\times
\Big({1\over y_{_4}}\Big)^{n_{_1}}\Big({y_{_4}\over y_{_2}}\Big)^{n_{_2}}
\Big({y_{_1}\over y_{_4}}\Big)^{n_{_3}}\Big({y_{_3}\over y_{_2}}\Big)^{n_{_4}}
\;,\nonumber\\
&&\Phi_{_{[135\tilde{7}]}}^{(8),b}(\alpha,z)=
y_{_1}^{{D\over2}-2}y_{_2}^{{D}-3}y_{_4}^{-1}\sum\limits_{n_{_1}=0}^\infty
\sum\limits_{n_{_2}=0}^\infty\sum\limits_{n_{_3}=0}^\infty\sum\limits_{n_{_4}=0}^\infty
c_{_{[135\tilde{7}]}}^{(8),b}(\alpha,{\bf n})
\nonumber\\
&&\hspace{2.5cm}\times
\Big({1\over y_{_1}}\Big)^{n_{_1}}\Big({1\over y_{_2}}\Big)^{n_{_2}}
\Big({1\over y_{_4}}\Big)^{n_{_3}}\Big({y_{_3}\over y_{_2}}\Big)^{n_{_4}}\;.
\label{GKZ21d-24-2a}
\end{eqnarray}
Where the coefficients are
\begin{eqnarray}
&&c_{_{[135\tilde{7}]}}^{(8),a}(\alpha,{\bf n})=
(-)^{n_{_4}}\Gamma(1+n_{_1}+n_{_3})
\Big\{n_{_1}!n_{_2}!n_{_3}!n_{_4}!\Gamma(2-{D\over2}+n_{_1})
\nonumber\\
&&\hspace{2.5cm}\times
\Gamma(2-{D\over2}+n_{_2})
\Gamma({D\over2}-1-n_{_2}-n_{_4})\Gamma(2-{D\over2}+n_{_4})
\nonumber\\
&&\hspace{2.5cm}\times
\Gamma({D\over2}-1-n_{_1}-n_{_3})\Gamma({D\over2}+n_{_3})
\Gamma(D-2-n_{_2}-n_{_4})\Big\}^{-1}
\;,\nonumber\\
&&c_{_{[135\tilde{7}]}}^{(8),b}(\alpha,{\bf n})=
(-)^{n_{_1}+n_{_4}}\Gamma(1+n_{_1})\Gamma(1+n_{_2}+n_{_3})
\Big\{n_{_2}!n_{_4}!
\nonumber\\
&&\hspace{2.5cm}\times
\Gamma(2+n_{_1}+n_{_2}+n_{_3})\Gamma(3-{D\over2}+n_{_1}+n_{_2}+n_{_3})
\nonumber\\
&&\hspace{2.5cm}\times
\Gamma(2-{D\over2}+n_{_2})\Gamma({D\over2}-1-n_{_2}-n_{_4})
\Gamma(2-{D\over2}+n_{_4})
\nonumber\\
&&\hspace{2.5cm}\times
\Gamma({D\over2}-1-n_{_2}-n_{_3})
\Gamma({D\over2}-1-n_{_1})\Gamma(D-2-n_{_2}-n_{_4})\Big\}^{-1}\;.
\label{GKZ21d-24-3}
\end{eqnarray}

\item   $I_{_{9}}=\{2,4,5,7,9,\cdots,14\}$, i.e. the implement $J_{_{9}}=[1,14]\setminus I_{_{9}}=\{1,3,6,8\}$.
The choice implies the power numbers $\alpha_{_1}=\alpha_{_{3}}=\alpha_{_{6}}=\alpha_{_{8}}=0$, and
\begin{eqnarray}
&&\alpha_{_2}=a_{_1}-a_{_2},\;\alpha_{_4}=a_{_3}-a_{_4},
\;\alpha_{_5}=a_{_1}-a_{_3}-a_{_5}-b_{_1}+1,
\nonumber\\
&&\alpha_{_{7}}=b_{_2}+b_{_3}-a_{_3}-2,\;\alpha_{_{9}}=b_{_1}+b_{_4}-a_{_1}-2,
\nonumber\\
&&\alpha_{_{10}}=b_{_1}+b_{_5}-a_{_1}-2,\;\alpha_{_{11}}=1-b_{_1},\;
\alpha_{_{12}}=b_{_3}-a_{_3}-1,
\nonumber\\
&&\alpha_{_{13}}=1-b_{_3},\;\alpha_{_{14}}=a_{_3}+b_{_1}-a_{_1}-1\;.
\label{GKZ21d-3-1}
\end{eqnarray}
The corresponding hypergeometric series solution is written as
\begin{eqnarray}
&&\Phi_{_{[135\tilde{7}]}}^{(9)}(\alpha,z)=
y_{_1}^{{D\over2}-1}y_{_2}^{-1}y_{_3}^{{D\over2}-1}y_{_4}^{-1}\sum\limits_{n_{_1}=0}^\infty
\sum\limits_{n_{_2}=0}^\infty\sum\limits_{n_{_3}=0}^\infty\sum\limits_{n_{_4}=0}^\infty
c_{_{[135\tilde{7}]}}^{(9)}(\alpha,{\bf n})
\nonumber\\
&&\hspace{2.5cm}\times
\Big({1\over y_{_4}}\Big)^{n_{_1}}\Big({y_{_4}\over y_{_2}}\Big)^{n_{_2}}
\Big({y_{_1}\over y_{_4}}\Big)^{n_{_3}}\Big({y_{_3}\over y_{_2}}\Big)^{n_{_4}}\;,
\label{GKZ21d-3-2}
\end{eqnarray}
with
\begin{eqnarray}
&&c_{_{[135\tilde{7}]}}^{(9)}(\alpha,{\bf n})=
(-)^{n_{_4}}\Gamma(1+n_{_1}+n_{_3})\Gamma(1+n_{_2}+n_{_4})
\Big\{n_{_1}!n_{_2}!n_{_3}!n_{_4}!
\nonumber\\
&&\hspace{2.5cm}\times
\Gamma({D\over2}+n_{_1})
\Gamma({D\over2}+n_{_2})\Gamma(1-{D\over2}-n_{_2}-n_{_4})
\nonumber\\
&&\hspace{2.5cm}\times
\Gamma(1-{D\over2}-n_{_1}-n_{_3})\Gamma({D\over2}+n_{_3})
\Gamma({D\over2}+n_{_4})\Big\}^{-1}\;.
\label{GKZ21d-3-3}
\end{eqnarray}

\item   $I_{_{10}}=\{2,4,5,7,\cdots,12,14\}$, i.e. the implement $J_{_{10}}=[1,14]\setminus I_{_{10}}=\{1,3,6,13\}$.
The choice implies the power numbers $\alpha_{_1}=\alpha_{_{3}}=\alpha_{_{6}}=\alpha_{_{13}}=0$, and
\begin{eqnarray}
&&\alpha_{_2}=a_{_1}-a_{_2},\;\alpha_{_4}=a_{_3}-a_{_4},
\;\alpha_{_5}=a_{_1}-a_{_3}-a_{_5}-b_{_1}+1,
\nonumber\\
&&\alpha_{_{7}}=b_{_2}-a_{_3}-1,\;\alpha_{_{8}}=b_{_3}-1,\;\alpha_{_{9}}=b_{_1}+b_{_4}-a_{_1}-2,
\nonumber\\
&&\alpha_{_{10}}=b_{_1}+b_{_5}-a_{_1}-2,\;\alpha_{_{11}}=1-b_{_1},\;
\alpha_{_{12}}=-a_{_3},
\nonumber\\
&&\alpha_{_{14}}=a_{_3}+b_{_1}-a_{_1}-1\;.
\label{GKZ21d-4-1}
\end{eqnarray}
The corresponding hypergeometric series is written as
\begin{eqnarray}
&&\Phi_{_{[135\tilde{7}]}}^{(10)}(\alpha,z)=
y_{_1}^{{D\over2}-1}y_{_2}^{{D\over2}-2}y_{_4}^{-1}\sum\limits_{n_{_1}=0}^\infty
\sum\limits_{n_{_2}=0}^\infty\sum\limits_{n_{_3}=0}^\infty\sum\limits_{n_{_4}=0}^\infty
c_{_{[135\tilde{7}]}}^{(10)}(\alpha,{\bf n})
\nonumber\\
&&\hspace{2.5cm}\times
\Big({1\over y_{_4}}\Big)^{n_{_1}}\Big({y_{_4}\over y_{_2}}\Big)^{n_{_2}}
\Big({y_{_1}\over y_{_4}}\Big)^{n_{_3}}\Big({y_{_3}\over y_{_2}}\Big)^{n_{_4}}\;,
\label{GKZ21d-4-2}
\end{eqnarray}
with
\begin{eqnarray}
&&c_{_{[135\tilde{7}]}}^{(10)}(\alpha,{\bf n})=
(-)^{n_{_4}}\Gamma(1+n_{_1}+n_{_3})\Gamma(1+n_{_2}+n_{_4})
\Big\{n_{_1}!n_{_2}!n_{_3}!n_{_4}!
\nonumber\\
&&\hspace{2.5cm}\times
\Gamma({D\over2}+n_{_1})\Gamma({D\over2}+n_{_2})
\Gamma(2-{D\over2}+n_{_4})
\nonumber\\
&&\hspace{2.5cm}\times
\Gamma(1-{D\over2}-n_{_1}-n_{_3})\Gamma({D\over2}+n_{_3})
\Gamma({D\over2}-1-n_{_2}-n_{_4})\Big\}^{-1}\;.
\label{GKZ21d-4-3}
\end{eqnarray}

\item   $I_{_{11}}=\{2,4,\cdots,7,9,10,12,13,14\}$, i.e. the implement $J_{_{11}}=[1,14]\setminus I_{_{11}}=\{1,3,8,11\}$.
The choice implies the power numbers $\alpha_{_1}=\alpha_{_{3}}=\alpha_{_{8}}=\alpha_{_{11}}=0$, and
\begin{eqnarray}
&&\alpha_{_2}=a_{_1}-a_{_2},\;\alpha_{_4}=a_{_3}-a_{_4},
\;\alpha_{_5}=a_{_1}-a_{_3}-a_{_5},
\nonumber\\
&&\alpha_{_{6}}=b_{_1}-1,\;\alpha_{_{7}}=b_{_2}+b_{_3}-a_{_3}-2,\;\alpha_{_{9}}=b_{_4}-a_{_1}-1,
\nonumber\\
&&\alpha_{_{10}}=b_{_5}-a_{_1}-1,\;
\alpha_{_{12}}=b_{_3}-a_{_3}-1,\;\alpha_{_{13}}=1-b_{_3},
\nonumber\\
&&\alpha_{_{14}}=a_{_3}-a_{_1}\;.
\label{GKZ21d-5-1}
\end{eqnarray}
The corresponding hypergeometric series is
\begin{eqnarray}
&&\Phi_{_{[135\tilde{7}]}}^{(11)}(\alpha,z)=
y_{_2}^{-1}y_{_3}^{{D\over2}-1}y_{_4}^{{D\over2}-2}\sum\limits_{n_{_1}=0}^\infty
\sum\limits_{n_{_2}=0}^\infty\sum\limits_{n_{_3}=0}^\infty\sum\limits_{n_{_4}=0}^\infty
c_{_{[135\tilde{7}]}}^{(11)}(\alpha,{\bf n})
\nonumber\\
&&\hspace{2.5cm}\times
\Big({1\over y_{_4}}\Big)^{n_{_1}}\Big({y_{_4}\over y_{_2}}\Big)^{n_{_2}}
\Big({y_{_1}\over y_{_4}}\Big)^{n_{_3}}\Big({y_{_3}\over y_{_2}}\Big)^{n_{_4}}\;,
\label{GKZ21d-5-2}
\end{eqnarray}
with
\begin{eqnarray}
&&c_{_{[135\tilde{7}]}}^{(11)}(\alpha,{\bf n})=
(-)^{1+n_{_4}}\Gamma(1+n_{_1}+n_{_3})\Gamma(1+n_{_2}+n_{_4})
\nonumber\\
&&\hspace{2.5cm}\times
\Big\{n_{_1}!n_{_2}!n_{_3}!n_{_4}!\Gamma({D\over2}+n_{_1})\Gamma({D\over2}+n_{_2})
\Gamma(2-{D\over2}+n_{_3})
\nonumber\\
&&\hspace{2.5cm}\times
\Gamma(1-{D\over2}-n_{_2}-n_{_4})\Gamma({D\over2}-1-n_{_1}-n_{_3})
\Gamma({D\over2}+n_{_4})\Big\}^{-1}\;.
\label{GKZ21d-5-3}
\end{eqnarray}

\item   $I_{_{12}}=\{2,4,\cdots,10,12,14\}$, i.e. the implement $J_{_{12}}=[1,14]\setminus I_{_{12}}=\{1,3,11,13\}$.
The choice implies the power numbers $\alpha_{_1}=\alpha_{_{3}}=\alpha_{_{11}}=\alpha_{_{13}}=0$, and
\begin{eqnarray}
&&\alpha_{_2}=a_{_1}-a_{_2},\;\alpha_{_4}=a_{_3}-a_{_4},
\;\alpha_{_5}=a_{_1}-a_{_3}-a_{_5},
\nonumber\\
&&\alpha_{_{6}}=b_{_1}-1,\;\alpha_{_{7}}=b_{_2}-a_{_3}-1,\;\alpha_{_{8}}=b_{_3}-1,
\nonumber\\
&&\alpha_{_{9}}=b_{_4}-a_{_1}-1,\;\alpha_{_{10}}=b_{_5}-a_{_1}-1,\;\alpha_{_{12}}=-a_{_3},
\nonumber\\
&&\alpha_{_{14}}=a_{_3}-a_{_1}\;.
\label{GKZ21d-6-1}
\end{eqnarray}
The corresponding hypergeometric series is
\begin{eqnarray}
&&\Phi_{_{[135\tilde{7}]}}^{(12)}(\alpha,z)=
y_{_2}^{{D\over2}-2}y_{_4}^{{D\over2}-2}\sum\limits_{n_{_1}=0}^\infty
\sum\limits_{n_{_2}=0}^\infty\sum\limits_{n_{_3}=0}^\infty\sum\limits_{n_{_4}=0}^\infty
c_{_{[135\tilde{7}]}}^{(12)}(\alpha,{\bf n})
\nonumber\\
&&\hspace{2.5cm}\times
\Big({1\over y_{_4}}\Big)^{n_{_1}}\Big({y_{_4}\over y_{_2}}\Big)^{n_{_2}}
\Big({y_{_1}\over y_{_4}}\Big)^{n_{_3}}\Big({y_{_3}\over y_{_2}}\Big)^{n_{_4}}\;,
\label{GKZ21d-6-2}
\end{eqnarray}
with
\begin{eqnarray}
&&c_{_{[135\tilde{7}]}}^{(12)}(\alpha,{\bf n})=
(-)^{1+n_{_4}}\Gamma(1+n_{_1}+n_{_3})\Gamma(1+n_{_2}+n_{_4})
\nonumber\\
&&\hspace{2.5cm}\times
\Big\{n_{_1}!n_{_2}!n_{_3}!n_{_4}!\Gamma({D\over2}+n_{_1})
\Gamma({D\over2}+n_{_2})\Gamma(2-{D\over2}+n_{_3})
\nonumber\\
&&\hspace{2.5cm}\times
\Gamma(2-{D\over2}+n_{_4})\Gamma({D\over2}-1-n_{_1}-n_{_3})
\Gamma({D\over2}-1-n_{_2}-n_{_4})\Big\}^{-1}\;.
\label{GKZ21d-6-3}
\end{eqnarray}

\item   $I_{_{13}}=\{2,3,5,7,9,\cdots,14\}$, i.e. the implement $J_{_{13}}=[1,14]\setminus I_{_{13}}=\{1,4,6,8\}$.
The choice implies the power numbers $\alpha_{_1}=\alpha_{_{4}}=\alpha_{_{6}}=\alpha_{_{8}}=0$, and
\begin{eqnarray}
&&\alpha_{_2}=a_{_1}-a_{_2},\;\alpha_{_3}=a_{_4}-a_{_3},
\;\alpha_{_5}=a_{_1}-a_{_4}-a_{_5}-b_{_1}+1,
\nonumber\\
&&\alpha_{_{7}}=b_{_2}+b_{_3}-a_{_4}-2,\;\alpha_{_{9}}=b_{_1}+b_{_4}-a_{_1}-2,
\nonumber\\
&&\alpha_{_{10}}=b_{_1}+b_{_5}-a_{_1}-2,\;\alpha_{_{11}}=1-b_{_1},\;
\alpha_{_{12}}=b_{_3}-a_{_4}-1,
\nonumber\\
&&\alpha_{_{13}}=1-b_{_3},\;\alpha_{_{14}}=a_{_4}+b_{_1}-a_{_1}-1\;.
\label{GKZ21d-9-1}
\end{eqnarray}
The corresponding hypergeometric series is written as
\begin{eqnarray}
&&\Phi_{_{[135\tilde{7}]}}^{(13)}(\alpha,z)=
y_{_1}^{{D\over2}-1}y_{_2}^{{D\over2}-2}y_{_3}^{{D\over2}-1}y_{_4}^{-{D\over2}}\sum\limits_{n_{_1}=0}^\infty
\sum\limits_{n_{_2}=0}^\infty\sum\limits_{n_{_3}=0}^\infty\sum\limits_{n_{_4}=0}^\infty
c_{_{[135\tilde{7}]}}^{(13)}(\alpha,{\bf n})
\nonumber\\
&&\hspace{2.5cm}\times
\Big({1\over y_{_4}}\Big)^{n_{_1}}\Big({y_{_4}\over y_{_2}}\Big)^{n_{_2}}
\Big({y_{_1}\over y_{_4}}\Big)^{n_{_3}}\Big({y_{_3}\over y_{_2}}\Big)^{n_{_4}}\;,
\label{GKZ21d-9-2}
\end{eqnarray}
with
\begin{eqnarray}
&&c_{_{[135\tilde{7}]}}^{(13)}(\alpha,{\bf n})=
(-)^{n_{_4}}\Gamma(1+n_{_1}+n_{_3})\Gamma(1+n_{_2}+n_{_4})
\Big\{n_{_1}!n_{_2}!n_{_3}!n_{_4}!
\nonumber\\
&&\hspace{2.5cm}\times
\Gamma({D\over2}+n_{_1})
\Gamma(2-{D\over2}+n_{_2})\Gamma(1-{D\over2}-n_{_1}-n_{_3})
\nonumber\\
&&\hspace{2.5cm}\times
\Gamma({D\over2}+n_{_3})\Gamma({D\over2}-1-n_{_2}-n_{_4})
\Gamma({D\over2}+n_{_4})\Big\}^{-1}\;.
\label{GKZ21d-9-3}
\end{eqnarray}

\item   $I_{_{14}}=\{2,3,5,7,\cdots,12,14\}$, i.e. the implement $J_{_{14}}=[1,14]\setminus I_{_{14}}=\{1,4,6,13\}$.
The choice implies the power numbers $\alpha_{_1}=\alpha_{_{4}}=\alpha_{_{6}}=\alpha_{_{13}}=0$, and
\begin{eqnarray}
&&\alpha_{_2}=a_{_1}-a_{_2},\;\alpha_{_3}=a_{_4}-a_{_3},
\;\alpha_{_5}=a_{_1}-a_{_4}-a_{_5}-b_{_1}+1,
\nonumber\\
&&\alpha_{_{7}}=b_{_2}-a_{_4}-1,\;\alpha_{_{8}}=b_{_3}-1,\;\alpha_{_{9}}=b_{_1}+b_{_4}-a_{_1}-2,
\nonumber\\
&&\alpha_{_{10}}=b_{_1}+b_{_5}-a_{_1}-2,\;\alpha_{_{11}}=1-b_{_1},\;
\alpha_{_{12}}=-a_{_4},
\nonumber\\
&&\alpha_{_{14}}=a_{_4}+b_{_1}-a_{_1}-1\;.
\label{GKZ21d-10-1}
\end{eqnarray}
The corresponding hypergeometric series is written as
\begin{eqnarray}
&&\Phi_{_{[135\tilde{7}]}}^{(14)}(\alpha,z)=
y_{_1}^{{D\over2}-1}y_{_2}^{{D}-3}y_{_4}^{-{D\over2}}\sum\limits_{n_{_1}=0}^\infty
\sum\limits_{n_{_2}=0}^\infty\sum\limits_{n_{_3}=0}^\infty\sum\limits_{n_{_4}=0}^\infty
c_{_{[135\tilde{7}]}}^{(14)}(\alpha,{\bf n})
\nonumber\\
&&\hspace{2.5cm}\times
\Big({1\over y_{_4}}\Big)^{n_{_1}}\Big({y_{_4}\over y_{_2}}\Big)^{n_{_2}}
\Big({y_{_1}\over y_{_4}}\Big)^{n_{_3}}\Big({y_{_3}\over y_{_2}}\Big)^{n_{_4}}\;,
\label{GKZ21d-10-2}
\end{eqnarray}
with
\begin{eqnarray}
&&c_{_{[135\tilde{7}]}}^{(14)}(\alpha,{\bf n})=
(-)^{n_{_2}}\Gamma(1+n_{_1}+n_{_3})
\Big\{n_{_1}!n_{_2}!n_{_3}!n_{_4}!\Gamma({D\over2}+n_{_1})
\nonumber\\
&&\hspace{2.5cm}\times
\Gamma(2-{D\over2}+n_{_2})\Gamma({D\over2}-1-n_{_2}-n_{_4})
\Gamma(2-{D\over2}+n_{_4})
\nonumber\\
&&\hspace{2.5cm}\times
\Gamma(1-{D\over2}-n_{_1}-n_{_3})\Gamma({D\over2}+n_{_3})
\Gamma(D-2-n_{_2}-n_{_4})\Big\}^{-1}\;.
\label{GKZ21d-10-3}
\end{eqnarray}

\item   $I_{_{15}}=\{2,3,\cdots,7,9,10,12,13,14\}$, i.e. the implement $J_{_{15}}=[1,14]\setminus I_{_{15}}=\{1,4,8,11\}$.
The choice implies the power numbers $\alpha_{_1}=\alpha_{_{4}}=\alpha_{_{8}}=\alpha_{_{11}}=0$, and
\begin{eqnarray}
&&\alpha_{_2}=a_{_1}-a_{_2},\;\alpha_{_3}=a_{_4}-a_{_3},
\;\alpha_{_5}=a_{_1}-a_{_4}-a_{_5},
\nonumber\\
&&\alpha_{_{6}}=b_{_1}-1,\;\alpha_{_{7}}=b_{_2}+b_{_3}-a_{_4}-2,\;\alpha_{_{9}}=b_{_4}-a_{_1}-1,
\nonumber\\
&&\alpha_{_{10}}=b_{_5}-a_{_1}-1,\;
\alpha_{_{12}}=b_{_3}-a_{_4}-1,\;\alpha_{_{13}}=1-b_{_3},
\nonumber\\
&&\alpha_{_{14}}=a_{_4}-a_{_1}\;.
\label{GKZ21d-11-1}
\end{eqnarray}
The corresponding hypergeometric series is
\begin{eqnarray}
&&\Phi_{_{[135\tilde{7}]}}^{(15)}(\alpha,z)=
y_{_2}^{{D\over2}-2}y_{_3}^{{D\over2}-1}y_{_4}^{-1}\sum\limits_{n_{_1}=0}^\infty
\sum\limits_{n_{_2}=0}^\infty\sum\limits_{n_{_3}=0}^\infty\sum\limits_{n_{_4}=0}^\infty
c_{_{[135\tilde{7}]}}^{(15)}(\alpha,{\bf n})
\nonumber\\
&&\hspace{2.5cm}\times
\Big({1\over y_{_4}}\Big)^{n_{_1}}\Big({y_{_4}\over y_{_2}}\Big)^{n_{_2}}
\Big({y_{_1}\over y_{_4}}\Big)^{n_{_3}}\Big({y_{_3}\over y_{_2}}\Big)^{n_{_4}}\;,
\label{GKZ21d-11-2}
\end{eqnarray}
with
\begin{eqnarray}
&&c_{_{[135\tilde{7}]}}^{(15)}(\alpha,{\bf n})=
(-)^{n_{_4}}\Gamma(1+n_{_1}+n_{_3})\Gamma(1+n_{_2}+n_{_4})
\Big\{n_{_1}!n_{_2}!n_{_3}!n_{_4}!
\nonumber\\
&&\hspace{2.5cm}\times
\Gamma({D\over2}+n_{_1})\Gamma(2-{D\over2}+n_{_2})
\Gamma(2-{D\over2}+n_{_3})
\nonumber\\
&&\hspace{2.5cm}\times
\Gamma({D\over2}-1-n_{_1}-n_{_3})
\Gamma({D\over2}-1-n_{_2}-n_{_4})\Gamma({D\over2}+n_{_4})\Big\}^{-1}\;.
\label{GKZ21d-11-3}
\end{eqnarray}

\item   $I_{_{16}}=\{2,3,5,\cdots,10,12,14\}$, i.e. the implement $J_{_{16}}=[1,14]\setminus I_{_{16}}=\{1,4,11,13\}$.
The choice implies the power numbers $\alpha_{_1}=\alpha_{_{4}}=\alpha_{_{11}}=\alpha_{_{13}}=0$, and
\begin{eqnarray}
&&\alpha_{_2}=a_{_1}-a_{_2},\;\alpha_{_3}=a_{_4}-a_{_3},
\;\alpha_{_5}=a_{_1}-a_{_4}-a_{_5},
\nonumber\\
&&\alpha_{_{6}}=b_{_1}-1,\;\alpha_{_{7}}=b_{_2}-a_{_4}-1,\;\alpha_{_{8}}=b_{_3}-1,
\nonumber\\
&&\alpha_{_{9}}=b_{_4}-a_{_1}-1,\;\alpha_{_{10}}=b_{_5}-a_{_1}-1,\;\alpha_{_{12}}=-a_{_4},
\nonumber\\
&&\alpha_{_{14}}=a_{_4}-a_{_1}\;.
\label{GKZ21d-12-1}
\end{eqnarray}
The corresponding hypergeometric series is written as
\begin{eqnarray}
&&\Phi_{_{[135\tilde{7}]}}^{(16)}(\alpha,z)=
y_{_2}^{{D}-3}y_{_4}^{-1}\sum\limits_{n_{_1}=0}^\infty
\sum\limits_{n_{_2}=0}^\infty\sum\limits_{n_{_3}=0}^\infty\sum\limits_{n_{_4}=0}^\infty
c_{_{[135\tilde{7}]}}^{(16)}(\alpha,{\bf n})
\nonumber\\
&&\hspace{2.5cm}\times
\Big({1\over y_{_4}}\Big)^{n_{_1}}\Big({y_{_4}\over y_{_2}}\Big)^{n_{_2}}
\Big({y_{_1}\over y_{_4}}\Big)^{n_{_3}}\Big({y_{_3}\over y_{_2}}\Big)^{n_{_4}}\;,
\label{GKZ21d-12-2}
\end{eqnarray}
with
\begin{eqnarray}
&&c_{_{[135\tilde{7}]}}^{(16)}(\alpha,{\bf n})=
(-)^{n_{_2}}\Gamma(1+n_{_1}+n_{_3})
\Big\{n_{_1}!n_{_2}!n_{_3}!n_{_4}!\Gamma({D\over2}+n_{_1})
\nonumber\\
&&\hspace{2.5cm}\times
\Gamma(2-{D\over2}+n_{_2})\Gamma(2-{D\over2}+n_{_3})
\Gamma({D\over2}-1-n_{_2}-n_{_4})
\nonumber\\
&&\hspace{2.5cm}\times
\Gamma(2-{D\over2}+n_{_4})\Gamma({D\over2}-1-n_{_1}-n_{_3})
\Gamma(D-2-n_{_2}-n_{_4})\Big\}^{-1}\;.
\label{GKZ21d-12-3}
\end{eqnarray}

\item   $I_{_{17}}=\{1,4,5,7,9,\cdots,14\}$, i.e. the implement $J_{_{17}}=[1,14]\setminus I_{_{17}}=\{2,3,6,8\}$.
The choice implies the power numbers $\alpha_{_2}=\alpha_{_{3}}=\alpha_{_{6}}=\alpha_{_{8}}=0$, and
\begin{eqnarray}
&&\alpha_{_1}=a_{_2}-a_{_1},\;\alpha_{_4}=a_{_3}-a_{_4},
\;\alpha_{_5}=a_{_2}-a_{_3}-a_{_5}-b_{_1}+1,
\nonumber\\
&&\alpha_{_{7}}=b_{_2}+b_{_3}-a_{_3}-2,\;\alpha_{_{9}}=b_{_1}+b_{_4}-a_{_2}-2,
\nonumber\\
&&\alpha_{_{10}}=b_{_1}+b_{_5}-a_{_2}-2,\;\alpha_{_{11}}=1-b_{_1},\;
\alpha_{_{12}}=b_{_3}-a_{_3}-1,
\nonumber\\
&&\alpha_{_{13}}=1-b_{_3},\;\alpha_{_{14}}=a_{_3}+b_{_1}-a_{_2}-1\;.
\label{GKZ21d-19-1}
\end{eqnarray}
The corresponding hypergeometric series is written as
\begin{eqnarray}
&&\Phi_{_{[135\tilde{7}]}}^{(17)}(\alpha,z)=
y_{_1}^{{D\over2}-1}y_{_2}^{-1}y_{_3}^{{D\over2}-1}y_{_4}^{{D\over2}-2}
\sum\limits_{n_{_1}=0}^\infty
\sum\limits_{n_{_2}=0}^\infty\sum\limits_{n_{_3}=0}^\infty\sum\limits_{n_{_4}=0}^\infty
c_{_{[135\tilde{7}]}}^{(17)}(\alpha,{\bf n})
\nonumber\\
&&\hspace{2.5cm}\times
\Big({1\over y_{_4}}\Big)^{n_{_1}}\Big({y_{_4}\over y_{_2}}\Big)^{n_{_2}}
\Big({y_{_1}\over y_{_4}}\Big)^{n_{_3}}\Big({y_{_3}\over y_{_2}}\Big)^{n_{_4}}\;,
\label{GKZ21d-19-2}
\end{eqnarray}
with
\begin{eqnarray}
&&c_{_{[135\tilde{7}]}}^{(17)}(\alpha,{\bf n})=
(-)^{1+n_{_4}}\Gamma(1+n_{_1}+n_{_3})\Gamma(1+n_{_2}+n_{_4})
\Big\{n_{_1}!n_{_2}!n_{_3}!n_{_4}!
\nonumber\\
&&\hspace{2.5cm}\times
\Gamma(2-{D\over2}+n_{_1})\Gamma({D\over2}+n_{_2})
\Gamma(1-{D\over2}-n_{_2}-n_{_4})
\nonumber\\
&&\hspace{2.5cm}\times
\Gamma({D\over2}-1-n_{_1}-n_{_3})\Gamma({D\over2}+n_{_3})
\Gamma({D\over2}+n_{_4})\Big\}^{-1}\;.
\label{GKZ21d-19-3}
\end{eqnarray}

\item   $I_{_{18}}=\{1,4,5,7,\cdots,12,14\}$, i.e. the implement $J_{_{18}}=[1,14]\setminus I_{_{18}}=\{2,3,6,13\}$.
The choice implies the power numbers $\alpha_{_2}=\alpha_{_{3}}=\alpha_{_{6}}=\alpha_{_{13}}=0$, and
\begin{eqnarray}
&&\alpha_{_1}=a_{_2}-a_{_1},\;\alpha_{_4}=a_{_3}-a_{_4},
\;\alpha_{_5}=a_{_2}-a_{_3}-a_{_5}-b_{_1}+1,
\nonumber\\
&&\alpha_{_{7}}=b_{_2}-a_{_3}-1,\;\alpha_{_{8}}=b_{_3}-1,\;\alpha_{_{9}}=b_{_1}+b_{_4}-a_{_2}-2,
\nonumber\\
&&\alpha_{_{10}}=b_{_1}+b_{_5}-a_{_2}-2,\;\alpha_{_{11}}=1-b_{_1},\;
\alpha_{_{12}}=-a_{_3},
\nonumber\\
&&\alpha_{_{14}}=a_{_3}+b_{_1}-a_{_2}-1\;.
\label{GKZ21d-20-1}
\end{eqnarray}
The corresponding hypergeometric series functions is written as
\begin{eqnarray}
&&\Phi_{_{[135\tilde{7}]}}^{(18)}(\alpha,z)=
y_{_1}^{{D\over2}-1}y_{_2}^{{D\over2}-2}y_{_4}^{{D\over2}-2}\sum\limits_{n_{_1}=0}^\infty
\sum\limits_{n_{_2}=0}^\infty\sum\limits_{n_{_3}=0}^\infty\sum\limits_{n_{_4}=0}^\infty
c_{_{[135\tilde{7}]}}^{(18)}(\alpha,{\bf n})
\nonumber\\
&&\hspace{2.5cm}\times
\Big({1\over y_{_4}}\Big)^{n_{_1}}\Big({y_{_4}\over y_{_2}}\Big)^{n_{_2}}
\Big({y_{_1}\over y_{_4}}\Big)^{n_{_3}}\Big({y_{_3}\over y_{_2}}\Big)^{n_{_4}}\;,
\label{GKZ21d-20-2}
\end{eqnarray}
with
\begin{eqnarray}
&&c_{_{[135\tilde{7}]}}^{(18)}(\alpha,{\bf n})=
(-)^{1+n_{_4}}\Gamma(1+n_{_1}+n_{_3})\Gamma(1+n_{_2}+n_{_4})
\Big\{n_{_1}!n_{_2}!n_{_3}!n_{_4}!
\nonumber\\
&&\hspace{2.5cm}\times
\Gamma(2-{D\over2}+n_{_1})\Gamma({D\over2}+n_{_2})\Gamma(2-{D\over2}+n_{_1}-n_{_2}+n_{_3})
\nonumber\\
&&\hspace{2.5cm}\times
\Gamma(2-{D\over2}+n_{_4})\Gamma({D\over2}-1-n_{_1}-n_{_3})
\Gamma({D\over2}+n_{_3})
\nonumber\\
&&\hspace{2.5cm}\times
\Gamma({D\over2}-1-n_{_2}-n_{_4})\Gamma({D\over2}-1-n_{_1}+n_{_2}-n_{_3})\Big\}^{-1}\;.
\label{GKZ21d-20-3}
\end{eqnarray}

\item   $I_{_{19}}=\{1,4,\cdots,7,9,10,12,13,14\}$, i.e. the implement $J_{_{19}}=[1,14]\setminus I_{_{19}}=\{2,3,8,11\}$.
The choice implies the power numbers $\alpha_{_2}=\alpha_{_{3}}=\alpha_{_{8}}=\alpha_{_{11}}=0$, and
\begin{eqnarray}
&&\alpha_{_1}=a_{_2}-a_{_1},\;\alpha_{_4}=a_{_3}-a_{_4},
\;\alpha_{_5}=a_{_2}-a_{_3}-a_{_5},
\nonumber\\
&&\alpha_{_{6}}=b_{_1}-1,\;\alpha_{_{7}}=b_{_2}+b_{_3}-a_{_3}-2,\;\alpha_{_{9}}=b_{_4}-a_{_2}-1,
\nonumber\\
&&\alpha_{_{10}}=b_{_5}-a_{_2}-1,\;
\alpha_{_{12}}=b_{_3}-a_{_3}-1,\;\alpha_{_{13}}=1-b_{_3},
\nonumber\\
&&\alpha_{_{14}}=a_{_3}-a_{_2}\;.
\label{GKZ21d-21-1}
\end{eqnarray}
The corresponding hypergeometric series is
\begin{eqnarray}
&&\Phi_{_{[135\tilde{7}]}}^{(19)}(\alpha,z)=
y_{_2}^{-1}y_{_3}^{{D\over2}-1}y_{_4}^{{D}-3}\sum\limits_{n_{_1}=0}^\infty
\sum\limits_{n_{_2}=0}^\infty\sum\limits_{n_{_3}=0}^\infty\sum\limits_{n_{_4}=0}^\infty
c_{_{[135\tilde{7}]}}^{(19)}(\alpha,{\bf n})
\nonumber\\
&&\hspace{2.5cm}\times
\Big({1\over y_{_4}}\Big)^{n_{_1}}\Big({y_{_4}\over y_{_2}}\Big)^{n_{_2}}
\Big({y_{_1}\over y_{_4}}\Big)^{n_{_3}}\Big({y_{_3}\over y_{_2}}\Big)^{n_{_4}}\;,
\label{GKZ21d-21-2}
\end{eqnarray}
with
\begin{eqnarray}
&&c_{_{[135\tilde{7}]}}^{(19)}(\alpha,{\bf n})=
(-)^{n_{_1}+n_{_3}+n_{_4}}\Gamma(1+n_{_2}+n_{_4})
\Big\{n_{_1}!n_{_2}!n_{_3}!n_{_4}!\Gamma(2-{D\over2}+n_{_1})
\nonumber\\
&&\hspace{2.5cm}\times
\Gamma({D\over2}+n_{_2})\Gamma(2-{D\over2}+n_{_3})
\Gamma(1-{D\over2}-n_{_2}-n_{_4})
\nonumber\\
&&\hspace{2.5cm}\times
\Gamma(D-2-n_{_1}-n_{_3})\Gamma({D\over2}-1-n_{_1}-n_{_3})
\Gamma({D\over2}+n_{_4})\Big\}^{-1}\;.
\label{GKZ21d-21-3}
\end{eqnarray}

\item   $I_{_{20}}=\{1,4,\cdots,10,12,14\}$, i.e. the implement $J_{_{20}}=[1,14]\setminus I_{_{20}}=\{2,3,11,13\}$.
The choice implies the power numbers $\alpha_{_2}=\alpha_{_{3}}=\alpha_{_{11}}=\alpha_{_{13}}=0$, and
\begin{eqnarray}
&&\alpha_{_1}=a_{_2}-a_{_1},\;\alpha_{_4}=a_{_3}-a_{_4},
\;\alpha_{_5}=a_{_2}-a_{_3}-a_{_5},
\nonumber\\
&&\alpha_{_{6}}=b_{_1}-1,\;\alpha_{_{7}}=b_{_2}-a_{_3}-1,\;\alpha_{_{8}}=b_{_3}-1,
\nonumber\\
&&\alpha_{_{9}}=b_{_4}-a_{_2}-1,\;\alpha_{_{10}}=b_{_5}-a_{_2}-1,\;\alpha_{_{12}}=-a_{_3},
\nonumber\\
&&\alpha_{_{14}}=a_{_3}-a_{_2}\;.
\label{GKZ21d-22-1}
\end{eqnarray}
The corresponding hypergeometric series is written as
\begin{eqnarray}
&&\Phi_{_{[135\tilde{7}]}}^{(20)}(\alpha,z)=
y_{_2}^{{D\over2}-2}y_{_4}^{{D}-3}\sum\limits_{n_{_1}=0}^\infty
\sum\limits_{n_{_2}=0}^\infty\sum\limits_{n_{_3}=0}^\infty\sum\limits_{n_{_4}=0}^\infty
c_{_{[135\tilde{7}]}}^{(20)}(\alpha,{\bf n})
\nonumber\\
&&\hspace{2.5cm}\times
\Big({1\over y_{_4}}\Big)^{n_{_1}}\Big({y_{_4}\over y_{_2}}\Big)^{n_{_2}}
\Big({y_{_1}\over y_{_4}}\Big)^{n_{_3}}\Big({y_{_3}\over y_{_2}}\Big)^{n_{_4}}\;,
\label{GKZ21d-22-2}
\end{eqnarray}
with
\begin{eqnarray}
&&c_{_{[135\tilde{7}]}}^{(20)}(\alpha,{\bf n})=
(-)^{n_{_1}+n_{_3}+n_{_4}}\Gamma(1+n_{_2}+n_{_4})\Big\{n_{_1}!n_{_2}!n_{_3}!n_{_4}!
\Gamma(2-{D\over2}+n_{_1})
\nonumber\\
&&\hspace{2.5cm}\times
\Gamma({D\over2}+n_{_2})
\Gamma(2-{D\over2}+n_{_3})\Gamma(D-2-n_{_1}-n_{_3})
\nonumber\\
&&\hspace{2.5cm}\times
\Gamma(2-{D\over2}+n_{_4})\Gamma({D\over2}-1-n_{_1}-n_{_3})
\Gamma({D\over2}-1-n_{_2}-n_{_4})\Big\}^{-1}\;.
\label{GKZ21d-22-3}
\end{eqnarray}

\item   $I_{_{21}}=\{1,3,5,7,9,\cdots,14\}$, i.e. the implement $J_{_{21}}=[1,14]\setminus I_{_{21}}=\{2,4,6,8\}$.
The choice implies the power numbers $\alpha_{_2}=\alpha_{_{4}}=\alpha_{_{6}}=\alpha_{_{8}}=0$, and
\begin{eqnarray}
&&\alpha_{_1}=a_{_2}-a_{_1},\;\alpha_{_3}=a_{_4}-a_{_3},
\;\alpha_{_5}=a_{_2}-a_{_4}-a_{_5}-b_{_1}+1,
\nonumber\\
&&\alpha_{_{7}}=b_{_2}+b_{_3}-a_{_4}-2,\;\alpha_{_{9}}=b_{_1}+b_{_4}-a_{_2}-2,
\nonumber\\
&&\alpha_{_{10}}=b_{_1}+b_{_5}-a_{_2}-2,\;\alpha_{_{11}}=1-b_{_1},\;
\alpha_{_{12}}=b_{_3}-a_{_4}-1,
\nonumber\\
&&\alpha_{_{13}}=1-b_{_3},\;\alpha_{_{14}}=a_{_4}+b_{_1}-a_{_2}-1\;.
\label{GKZ21d-25-1}
\end{eqnarray}
The corresponding hypergeometric series is
\begin{eqnarray}
&&\Phi_{_{[135\tilde{7}]}}^{(21)}(\alpha,z)=
y_{_1}^{{D\over2}-1}y_{_2}^{{D\over2}-2}y_{_3}^{{D\over2}-1}y_{_4}^{-1}
\sum\limits_{n_{_1}=0}^\infty
\sum\limits_{n_{_2}=0}^\infty\sum\limits_{n_{_3}=0}^\infty\sum\limits_{n_{_4}=0}^\infty
c_{_{[135\tilde{7}]}}^{(21)}(\alpha,{\bf n})
\nonumber\\
&&\hspace{2.5cm}\times
\Big({1\over y_{_4}}\Big)^{n_{_1}}\Big({y_{_4}\over y_{_2}}\Big)^{n_{_2}}
\Big({y_{_1}\over y_{_4}}\Big)^{n_{_3}}\Big({y_{_3}\over y_{_2}}\Big)^{n_{_4}}\;,
\label{GKZ21d-25-2}
\end{eqnarray}
with
\begin{eqnarray}
&&c_{_{[135\tilde{7}]}}^{(21)}(\alpha,{\bf n})=
(-)^{n_{_4}}\Gamma(1+n_{_1}+n_{_3})\Gamma(1+n_{_2}+n_{_4})
\Big\{n_{_1}!n_{_2}!n_{_3}!n_{_4}!
\nonumber\\
&&\hspace{2.5cm}\times
\Gamma(2-{D\over2}+n_{_1})\Gamma(2-{D\over2}+n_{_2})
\Gamma({D\over2}-1-n_{_1}-n_{_3})
\nonumber\\
&&\hspace{2.5cm}\times
\Gamma({D\over2}+n_{_3})\Gamma({D\over2}-1-n_{_2}-n_{_4})
\Gamma({D\over2}+n_{_4})\Big\}^{-1}\;.
\label{GKZ21d-25-3}
\end{eqnarray}

\item   $I_{_{22}}=\{1,3,5,7,\cdots,12,14\}$, i.e. the implement $J_{_{22}}=[1,14]\setminus I_{_{22}}=\{2,4,6,13\}$.
The choice implies the power numbers $\alpha_{_2}=\alpha_{_{4}}=\alpha_{_{6}}=\alpha_{_{13}}=0$, and
\begin{eqnarray}
&&\alpha_{_1}=a_{_2}-a_{_1},\;\alpha_{_3}=a_{_4}-a_{_3},
\;\alpha_{_5}=a_{_2}-a_{_4}-a_{_5}-b_{_1}+1,
\nonumber\\
&&\alpha_{_{7}}=b_{_2}-a_{_4}-1,\;\alpha_{_{8}}=b_{_3}-1,\;\alpha_{_{9}}=b_{_1}+b_{_4}-a_{_2}-2,
\nonumber\\
&&\alpha_{_{10}}=b_{_1}+b_{_5}-a_{_2}-2,\;\alpha_{_{11}}=1-b_{_1},\;
\alpha_{_{12}}=-a_{_4},
\nonumber\\
&&\alpha_{_{14}}=a_{_4}+b_{_1}-a_{_2}-1\;.
\label{GKZ21d-26-1}
\end{eqnarray}
The corresponding hypergeometric function is written as
\begin{eqnarray}
&&\Phi_{_{[135\tilde{7}]}}^{(22)}(\alpha,z)=
y_{_1}^{{D\over2}-1}y_{_2}^{{D}-3}y_{_4}^{-1}\sum\limits_{n_{_1}=0}^\infty
\sum\limits_{n_{_2}=0}^\infty\sum\limits_{n_{_3}=0}^\infty\sum\limits_{n_{_4}=0}^\infty
c_{_{[135\tilde{7}]}}^{(22)}(\alpha,{\bf n})
\nonumber\\
&&\hspace{2.5cm}\times
\Big({1\over y_{_4}}\Big)^{n_{_1}}\Big({y_{_4}\over y_{_2}}\Big)^{n_{_2}}
\Big({y_{_1}\over y_{_4}}\Big)^{n_{_3}}\Big({y_{_3}\over y_{_2}}\Big)^{n_{_4}}\;,
\label{GKZ21d-26-2}
\end{eqnarray}
with
\begin{eqnarray}
&&c_{_{[135\tilde{7}]}}^{(22)}(\alpha,{\bf n})=
(-)^{n_{_2}}\Gamma(1+n_{_1}+n_{_3})\Big\{n_{_1}!n_{_2}!n_{_3}!n_{_4}!
\Gamma(2-{D\over2}+n_{_1})
\nonumber\\
&&\hspace{2.5cm}\times
\Gamma(2-{D\over2}+n_{_2})\Gamma({D\over2}-1-n_{_2}-n_{_4})
\Gamma(2-{D\over2}+n_{_4})
\nonumber\\
&&\hspace{2.5cm}\times
\Gamma({D\over2}-1-n_{_1}-n_{_3})\Gamma({D\over2}+n_{_3})
\Gamma(D-2-n_{_2}-n_{_4})\Big\}^{-1}\;.
\label{GKZ21d-26-3}
\end{eqnarray}

\item   $I_{_{23}}=\{1,3,\cdots,7,9,10,12,13,14\}$, i.e. the implement $J_{_{23}}=[1,14]\setminus I_{_{23}}=\{2,4,8,11\}$.
The choice implies the power numbers $\alpha_{_2}=\alpha_{_{4}}=\alpha_{_{8}}=\alpha_{_{11}}=0$, and
\begin{eqnarray}
&&\alpha_{_1}=a_{_2}-a_{_1},\;\alpha_{_3}=a_{_4}-a_{_3},
\;\alpha_{_5}=a_{_2}-a_{_4}-a_{_5},
\nonumber\\
&&\alpha_{_{6}}=b_{_1}-1,\;\alpha_{_{7}}=b_{_2}+b_{_3}-a_{_4}-2,\;\alpha_{_{9}}=b_{_4}-a_{_2}-1,
\nonumber\\
&&\alpha_{_{10}}=b_{_5}-a_{_2}-1,\;
\alpha_{_{12}}=b_{_3}-a_{_4}-1,\;\alpha_{_{13}}=1-b_{_3},
\nonumber\\
&&\alpha_{_{14}}=a_{_4}-a_{_2}\;.
\label{GKZ21d-27-1}
\end{eqnarray}
The corresponding hypergeometric series is written as
\begin{eqnarray}
&&\Phi_{_{[135\tilde{7}]}}^{(23)}(\alpha,z)=
y_{_2}^{{D\over2}-2}y_{_3}^{{D\over2}-1}y_{_4}^{{D\over2}-2}\sum\limits_{n_{_1}=0}^\infty
\sum\limits_{n_{_2}=0}^\infty\sum\limits_{n_{_3}=0}^\infty\sum\limits_{n_{_4}=0}^\infty
{c_{_{[135\tilde{7}]}}^{(23)}(\alpha,{\bf n})}
\nonumber\\
&&\hspace{2.5cm}\times
\Big({1\over y_{_4}}\Big)^{n_{_1}}\Big({y_{_4}\over y_{_2}}\Big)^{n_{_2}}
\Big({y_{_1}\over y_{_4}}\Big)^{n_{_3}}\Big({y_{_3}\over y_{_2}}\Big)^{n_{_4}}\;,
\label{GKZ21d-27-2}
\end{eqnarray}
with
\begin{eqnarray}
&&c_{_{[135\tilde{7}]}}^{(23)}(\alpha,{\bf n})=
(-)^{1+n_{_1}+n_{_3}+n_{_4}}\Gamma(1+n_{_2}+n_{_4})\Big\{n_{_1}!n_{_2}!n_{_3}!n_{_4}!
\Gamma(2-{D\over2}+n_{_1})
\nonumber\\
&&\hspace{2.5cm}\times
\Gamma(2-{D\over2}+n_{_2})\Gamma(2-{D\over2}+n_{_3})
\Gamma(D-2-n_{_1}-n_{_3})
\nonumber\\
&&\hspace{2.5cm}\times
\Gamma({D\over2}-1-n_{_1}-n_{_3})\Gamma({D\over2}-1-n_{_2}-n_{_4})
\Gamma({D\over2}+n_{_4})\Big\}^{-1}\;.
\label{GKZ21d-27-3}
\end{eqnarray}

\item   $I_{_{24}}=\{1,3,5,\cdots,10,12,14\}$, i.e. the implement $J_{_{24}}=[1,14]\setminus I_{_{24}}=\{2,4,11,13\}$.
The choice implies the power numbers $\alpha_{_2}=\alpha_{_{4}}=\alpha_{_{11}}=\alpha_{_{13}}=0$, and
\begin{eqnarray}
&&\alpha_{_1}=a_{_2}-a_{_1},\;\alpha_{_3}=a_{_4}-a_{_3},
\;\alpha_{_5}=a_{_2}-a_{_4}-a_{_5},
\nonumber\\
&&\alpha_{_{6}}=b_{_1}-1,\;\alpha_{_{7}}=b_{_2}-a_{_4}-1,\;\alpha_{_{8}}=b_{_3}-1,
\nonumber\\
&&\alpha_{_{9}}=b_{_4}-a_{_2}-1,\;\alpha_{_{10}}=b_{_5}-a_{_2}-1,\;\alpha_{_{12}}=-a_{_4},
\nonumber\\
&&\alpha_{_{14}}=a_{_4}-a_{_2}\;.
\label{GKZ21d-28-1}
\end{eqnarray}
The corresponding hypergeometric function is written as
\begin{eqnarray}
&&\Phi_{_{[135\tilde{7}]}}^{(24)}(\alpha,z)=
y_{_2}^{{D}-3}y_{_4}^{{D\over2}-2}\sum\limits_{n_{_1}=0}^\infty
\sum\limits_{n_{_2}=0}^\infty\sum\limits_{n_{_3}=0}^\infty\sum\limits_{n_{_4}=0}^\infty
c_{_{[135\tilde{7}]}}^{(24)}(\alpha,{\bf n})
\nonumber\\
&&\hspace{2.5cm}\times
\Big({1\over y_{_4}}\Big)^{n_{_1}}\Big({y_{_4}\over y_{_2}}\Big)^{n_{_2}}
\Big({y_{_1}\over y_{_4}}\Big)^{n_{_3}}\Big({y_{_3}\over y_{_2}}\Big)^{n_{_4}}\;,
\label{GKZ21d-28-2}
\end{eqnarray}
with
\begin{eqnarray}
&&c_{_{[135\tilde{7}]}}^{(24)}(\alpha,{\bf n})=
(-)^{1+n_{_1}+n_{_2}+n_{_3}}
\Big\{n_{_1}!n_{_2}!n_{_3}!n_{_4}!\Gamma(2-{D\over2}+n_{_1})
\Gamma(2-{D\over2}+n_{_2})
\nonumber\\
&&\hspace{2.5cm}\times
\Gamma(2-{D\over2}+n_{_3})
\Gamma({D\over2}-1-n_{_2}-n_{_4})\Gamma(2-{D\over2}+n_{_4})
\nonumber\\
&&\hspace{2.5cm}\times
\Gamma(D-2-n_{_1}-n_{_3})\Gamma({D\over2}-1-n_{_1}-n_{_3})
\Gamma(D-2-n_{_2}-n_{_4})\Big\}^{-1}\;.
\label{GKZ21d-28-3}
\end{eqnarray}

\item   $I_{_{25}}=\{2,3,4,7,9,\cdots,14\}$, i.e. the implement $J_{_{25}}=[1,14]\setminus I_{_{25}}=\{1,5,6,8\}$.
The choice implies the power numbers $\alpha_{_1}=\alpha_{_{5}}=\alpha_{_{6}}=\alpha_{_{8}}=0$, and
\begin{eqnarray}
&&\alpha_{_2}=a_{_1}-a_{_2},\;\alpha_{_3}=a_{_1}-a_{_3}-a_{_5}-b_{_1}+1,
\nonumber\\
&&\alpha_{_4}=a_{_1}-a_{_4}-a_{_5}-b_{_1}+1,\;
\alpha_{_{7}}=a_{_5}+b_{_1}+b_{_2}+b_{_3}-a_{_1}-3,
\nonumber\\
&&\alpha_{_{9}}=b_{_1}+b_{_4}-a_{_1}-2,\;\alpha_{_{10}}=b_{_1}+b_{_5}-a_{_1}-2,\;
\alpha_{_{11}}=1-b_{_1},
\nonumber\\
&&\alpha_{_{12}}=a_{_5}+b_{_1}+b_{_3}-a_{_1}-2,\;\alpha_{_{13}}=1-b_{_3},\;
\alpha_{_{14}}=-a_{_5}\;.
\label{GKZ21d-13-1}
\end{eqnarray}
The corresponding hypergeometric series functions are written as
\begin{eqnarray}
&&\Phi_{_{[135\tilde{7}]}}^{(25),a}(\alpha,z)=
y_{_1}^{{D\over2}-1}y_{_2}^{-1}y_{_3}^{{D\over2}-1}y_{_4}^{-1}\sum\limits_{n_{_1}=0}^\infty
\sum\limits_{n_{_2}=0}^\infty\sum\limits_{n_{_3}=0}^\infty\sum\limits_{n_{_4}=0}^\infty
c_{_{[135\tilde{7}]}}^{(25),a}(\alpha,{\bf n})
\nonumber\\
&&\hspace{2.5cm}\times
\Big({1\over y_{_4}}\Big)^{n_{_1}}\Big({y_{_1}\over y_{_4}}\Big)^{n_{_2}}
\Big({y_{_4}\over y_{_2}}\Big)^{n_{_3}}\Big({y_{_3}\over y_{_2}}\Big)^{n_{_4}}
\;,\nonumber\\
&&\Phi_{_{[135\tilde{7}]}}^{(25),b}(\alpha,z)=
y_{_1}^{{D\over2}-1}y_{_3}^{{D\over2}-1}y_{_4}^{-2}\sum\limits_{n_{_1}=0}^\infty
\sum\limits_{n_{_2}=0}^\infty\sum\limits_{n_{_3}=0}^\infty\sum\limits_{n_{_4}=0}^\infty
c_{_{[135\tilde{7}]}}^{(25),b}(\alpha,{\bf n})
\nonumber\\
&&\hspace{2.5cm}\times
\Big({1\over y_{_4}}\Big)^{n_{_1}}\Big({y_{_1}\over y_{_4}}\Big)^{n_{_2}}
\Big({y_{_2}\over y_{_4}}\Big)^{n_{_3}}\Big({y_{_3}\over y_{_4}}\Big)^{n_{_4}}
\;,\nonumber\\
&&\Phi_{_{[135\tilde{7}]}}^{(25),c}(\alpha,z)=
y_{_1}^{{D\over2}-1}y_{_2}^{-1}y_{_3}^{{D\over2}}y_{_4}^{-2}\sum\limits_{n_{_1}=0}^\infty
\sum\limits_{n_{_2}=0}^\infty\sum\limits_{n_{_3}=0}^\infty\sum\limits_{n_{_4}=0}^\infty
c_{_{[135\tilde{7}]}}^{(25),c}(\alpha,{\bf n})
\nonumber\\
&&\hspace{2.5cm}\times
\Big({1\over y_{_4}}\Big)^{n_{_1}}\Big({y_{_1}\over y_{_4}}\Big)^{n_{_2}}
\Big({y_{_3}\over y_{_4}}\Big)^{n_{_3}}\Big({y_{_3}\over y_{_2}}\Big)^{n_{_4}}\;.
\label{GKZ21d-13-2a}
\end{eqnarray}
Where the coefficients are
\begin{eqnarray}
&&c_{_{[135\tilde{7}]}}^{(25),a}(\alpha,{\bf n})=
(-)^{n_{_4}}\Gamma(1+n_{_1}+n_{_2})\Gamma(1+n_{_3}+n_{_4})
\Big\{n_{_1}!n_{_2}!n_{_3}!n_{_4}!\Gamma({D\over2}+n_{_1})
\nonumber\\
&&\hspace{2.5cm}\times
\Gamma({D\over2}+n_{_3})
\Gamma(1-{D\over2}-n_{_3}-n_{_4})\Gamma(1-{D\over2}-n_{_1}-n_{_2})
\nonumber\\
&&\hspace{2.5cm}\times
\Gamma({D\over2}+n_{_2})\Gamma({D\over2}+n_{_4})\Big\}^{-1}
\;,\nonumber\\
&&c_{_{[135\tilde{7}]}}^{(25),b}(\alpha,{\bf n})=
-\Gamma(1+n_{_1}+n_{_2})\Gamma(1+n_{_3}+n_{_4})
\Big\{n_{_1}!n_{_2}!n_{_3}!n_{_4}!\Gamma({D\over2}+n_{_1})
\nonumber\\
&&\hspace{2.5cm}\times
\Gamma({D\over2}-1-n_{_3}-n_{_4})
\Gamma(2-{D\over2}+n_{_3})\Gamma(1-{D\over2}-n_{_1}-n_{_2})
\nonumber\\
&&\hspace{2.5cm}\times
\Gamma({D\over2}+n_{_2})\Gamma({D\over2}+n_{_4})\Big\}^{-1}
\;,\nonumber\\
&&c_{_{[135\tilde{7}]}}^{(25),c}(\alpha,{\bf n})=
(-)^{1+n_{_4}}\Gamma(1+n_{_1}+n_{_2})\Gamma(1+n_{_3})\Gamma(1+n_{_4})
\Big\{n_{_1}!n_{_2}!
\nonumber\\
&&\hspace{2.5cm}\times
\Gamma(2+n_{_3}+n_{_4})\Gamma({D\over2}+n_{_1})\Gamma({D\over2}-1-n_{_3})
\Gamma(1-{D\over2}-n_{_4})
\nonumber\\
&&\hspace{2.5cm}\times
\Gamma(1-{D\over2}-n_{_1}-n_{_2})\Gamma({D\over2}+n_{_2})
\Gamma({D\over2}+1+n_{_3}+n_{_4})\Big\}^{-1}\;.
\label{GKZ21d-13-3}
\end{eqnarray}

\item   $I_{_{26}}=\{2,3,4,7,\cdots,12,14\}$, i.e. the implement $J_{_{26}}=[1,14]\setminus I_{_{26}}=\{1,5,6,13\}$.
The choice implies the power numbers $\alpha_{_1}=\alpha_{_{5}}=\alpha_{_{6}}=\alpha_{_{13}}=0$, and
\begin{eqnarray}
&&\alpha_{_2}=a_{_1}-a_{_2},\;\alpha_{_3}=a_{_1}-a_{_3}-a_{_5}-b_{_1}+1,
\nonumber\\
&&\alpha_{_4}=a_{_1}-a_{_4}-a_{_5}-b_{_1}+1,\;
\alpha_{_{7}}=a_{_5}+b_{_1}+b_{_2}-a_{_1}-2,
\nonumber\\
&&\alpha_{_{8}}=b_{_3}-1,\;\alpha_{_{9}}=b_{_1}+b_{_4}-a_{_1}-2,\;\alpha_{_{10}}=b_{_1}+b_{_5}-a_{_1}-2,
\nonumber\\
&&\alpha_{_{11}}=1-b_{_1},\;
\alpha_{_{12}}=a_{_5}+b_{_1}-a_{_1}-1,\;\alpha_{_{14}}=-a_{_5}\;.
\label{GKZ21d-14-1}
\end{eqnarray}
The corresponding hypergeometric functions are
\begin{eqnarray}
&&\Phi_{_{[135\tilde{7}]}}^{(26),a}(\alpha,z)=
y_{_1}^{{D\over2}-1}y_{_2}^{{D\over2}-2}y_{_4}^{-1}\sum\limits_{n_{_1}=0}^\infty
\sum\limits_{n_{_2}=0}^\infty\sum\limits_{n_{_3}=0}^\infty\sum\limits_{n_{_4}=0}^\infty
c_{_{[135\tilde{7}]}}^{(26),a}(\alpha,{\bf n})
\nonumber\\
&&\hspace{2.5cm}\times
\Big({1\over y_{_4}}\Big)^{n_{_1}}\Big({y_{_1}\over y_{_4}}\Big)^{n_{_2}}
\Big({y_{_4}\over y_{_2}}\Big)^{n_{_3}}\Big({y_{_3}\over y_{_2}}\Big)^{n_{_4}}
\;,\nonumber\\
&&\Phi_{_{[135\tilde{7}]}}^{(26),b}(\alpha,z)=
y_{_1}^{{D\over2}-1}y_{_2}^{{D\over2}-1}y_{_4}^{-2}\sum\limits_{n_{_1}=0}^\infty
\sum\limits_{n_{_2}=0}^\infty\sum\limits_{n_{_3}=0}^\infty\sum\limits_{n_{_4}=0}^\infty
c_{_{[135\tilde{7}]}}^{(26),b}(\alpha,{\bf n})
\nonumber\\
&&\hspace{2.5cm}\times
\Big({1\over y_{_4}}\Big)^{n_{_1}}\Big({y_{_1}\over y_{_4}}\Big)^{n_{_2}}
\Big({y_{_2}\over y_{_4}}\Big)^{n_{_3}}\Big({y_{_3}\over y_{_4}}\Big)^{n_{_4}}
\;,\nonumber\\
&&\Phi_{_{[135\tilde{7}]}}^{(26),c}(\alpha,z)=
y_{_1}^{{D\over2}-1}y_{_2}^{{D\over2}-2}y_{_3}y_{_4}^{-2}\sum\limits_{n_{_1}=0}^\infty
\sum\limits_{n_{_2}=0}^\infty\sum\limits_{n_{_3}=0}^\infty\sum\limits_{n_{_4}=0}^\infty
c_{_{[135\tilde{7}]}}^{(26),c}(\alpha,{\bf n})
\nonumber\\
&&\hspace{2.5cm}\times
\Big({1\over y_{_4}}\Big)^{n_{_1}}\Big({y_{_1}\over y_{_4}}\Big)^{n_{_2}}
\Big({y_{_3}\over y_{_4}}\Big)^{n_{_3}}\Big({y_{_3}\over y_{_2}}\Big)^{n_{_4}}\;.
\label{GKZ21d-14-2a}
\end{eqnarray}
Where the coefficients are
\begin{eqnarray}
&&c_{_{[135\tilde{7}]}}^{(26),a}(\alpha,{\bf n})=
(-)^{n_{_4}}\Gamma(1+n_{_1}+n_{_2})\Gamma(1+n_{_3}+n_{_4})
\Big\{n_{_1}!n_{_2}!n_{_3}!n_{_4}!\Gamma({D\over2}+n_{_1})
\nonumber\\
&&\hspace{2.5cm}\times
\Gamma({D\over2}+n_{_3})
\Gamma({D\over2}-1-n_{_3}-n_{_4})\Gamma(1-{D\over2}-n_{_1}-n_{_2})
\nonumber\\
&&\hspace{2.5cm}\times
\Gamma({D\over2}+n_{_2})\Gamma(2-{D\over2}+n_{_4})\Big\}^{-1}
\;,\nonumber\\
&&c_{_{[135\tilde{7}]}}^{(26),b}(\alpha,{\bf n})=
-\Gamma(1+n_{_1}+n_{_2})\Gamma(1+n_{_3}+n_{_4})
\Big\{n_{_1}!n_{_2}!n_{_3}!n_{_4}!\Gamma({D\over2}+n_{_1})
\nonumber\\
&&\hspace{2.5cm}\times
\Gamma({D\over2}-1-n_{_3}-n_{_4})
\Gamma(2-{D\over2}+n_{_4})\Gamma(1-{D\over2}-n_{_1}-n_{_2})
\nonumber\\
&&\hspace{2.5cm}\times
\Gamma({D\over2}+n_{_2})\Gamma({D\over2}+n_{_3})\Big\}^{-1}
\;,\nonumber\\
&&c_{_{[135\tilde{7}]}}^{(26),c}(\alpha,{\bf n})=
(-)^{1+n_{_4}}\Gamma(1+n_{_1}+n_{_2})\Gamma(1+n_{_3})\Gamma(1+n_{_4})
\Big\{n_{_1}!n_{_2}!
\nonumber\\
&&\hspace{2.5cm}\times
\Gamma(2+n_{_3}+n_{_4})\Gamma({D\over2}+n_{_1})\Gamma({D\over2}-1-n_{_3})
\Gamma({D\over2}-1-n_{_4})
\nonumber\\
&&\hspace{2.5cm}\times
\Gamma(1-{D\over2}-n_{_1}-n_{_2})\Gamma({D\over2}+n_{_2})
\Gamma(3-{D\over2}+n_{_3}+n_{_4})\Big\}^{-1}\;.
\label{GKZ21d-14-3}
\end{eqnarray}

\item   $I_{_{27}}=\{2,3,4,6,7,9,10,12,13,14\}$, i.e. the implement $J_{_{27}}=[1,14]\setminus I_{_{27}}=\{1,5,8,11\}$.
The choice implies the power numbers $\alpha_{_1}=\alpha_{_{5}}=\alpha_{_{8}}=\alpha_{_{11}}=0$, and
\begin{eqnarray}
&&\alpha_{_2}=a_{_1}-a_{_2},\;\alpha_{_3}=a_{_1}-a_{_3}-a_{_5},\;
\alpha_{_4}=a_{_1}-a_{_4}-a_{_5},
\nonumber\\
&&\alpha_{_{6}}=b_{_1}-1,\;
\alpha_{_{7}}=a_{_5}+b_{_2}+b_{_3}-a_{_1}-2,
\nonumber\\
&&\alpha_{_{9}}=b_{_4}-a_{_1}-1,\;\alpha_{_{10}}=b_{_5}-a_{_1}-1,
\nonumber\\
&&\alpha_{_{12}}=a_{_5}+b_{_3}-a_{_1}-1,\;\alpha_{_{13}}=1-b_{_3},\;\alpha_{_{14}}=-a_{_5}\;.
\label{GKZ21d-15-1}
\end{eqnarray}
The corresponding hypergeometric series solutions are written as
\begin{eqnarray}
&&\Phi_{_{[135\tilde{7}]}}^{(27),a}(\alpha,z)=
y_{_2}^{{D\over2}-2}y_{_3}^{{D\over2}-1}y_{_4}^{-1}\sum\limits_{n_{_1}=0}^\infty
\sum\limits_{n_{_2}=0}^\infty\sum\limits_{n_{_3}=0}^\infty\sum\limits_{n_{_4}=0}^\infty
c_{_{[135\tilde{7}]}}^{(27),a}(\alpha,{\bf n})
\nonumber\\
&&\hspace{2.5cm}\times
\Big({1\over y_{_4}}\Big)^{n_{_1}}\Big({y_{_1}\over y_{_4}}\Big)^{n_{_2}}
\Big({y_{_4}\over y_{_2}}\Big)^{n_{_3}}\Big({y_{_3}\over y_{_2}}\Big)^{n_{_4}}
\;,\nonumber\\
&&\Phi_{_{[135\tilde{7}]}}^{(27),b}(\alpha,z)=
y_{_2}^{{D\over2}-1}y_{_3}^{{D\over2}-1}y_{_4}^{-2}\sum\limits_{n_{_1}=0}^\infty
\sum\limits_{n_{_2}=0}^\infty\sum\limits_{n_{_3}=0}^\infty\sum\limits_{n_{_4}=0}^\infty
c_{_{[135\tilde{7}]}}^{(27),b}(\alpha,{\bf n})
\nonumber\\
&&\hspace{2.5cm}\times
\Big({1\over y_{_4}}\Big)^{n_{_1}}\Big({y_{_1}\over y_{_4}}\Big)^{n_{_2}}
\Big({y_{_2}\over y_{_4}}\Big)^{n_{_3}}\Big({y_{_3}\over y_{_4}}\Big)^{n_{_4}}
\;,\nonumber\\
&&\Phi_{_{[135\tilde{7}]}}^{(27),c}(\alpha,z)=
y_{_2}^{{D\over2}-2}y_{_3}^{{D\over2}}y_{_4}^{-2}\sum\limits_{n_{_1}=0}^\infty
\sum\limits_{n_{_2}=0}^\infty\sum\limits_{n_{_3}=0}^\infty\sum\limits_{n_{_4}=0}^\infty
c_{_{[135\tilde{7}]}}^{(27),c}(\alpha,{\bf n})
\nonumber\\
&&\hspace{2.5cm}\times
\Big({1\over y_{_4}}\Big)^{n_{_1}}\Big({y_{_1}\over y_{_4}}\Big)^{n_{_2}}
\Big({y_{_3}\over y_{_4}}\Big)^{n_{_3}}\Big({y_{_3}\over y_{_2}}\Big)^{n_{_4}}\;.
\label{GKZ21d-15-2a}
\end{eqnarray}
Where the coefficients are
\begin{eqnarray}
&&c_{_{[135\tilde{7}]}}^{(27),a}(\alpha,{\bf n})=
(-)^{n_{_4}}\Gamma(1+n_{_1}+n_{_2})\Gamma(1+n_{_3}+n_{_4})
\Big\{n_{_1}!n_{_2}!n_{_3}!n_{_4}!
\nonumber\\
&&\hspace{2.5cm}\times
\Gamma({D\over2}+n_{_1})\Gamma(2-{D\over2}+n_{_3})\Gamma(2-{D\over2}+n_{_2})
\nonumber\\
&&\hspace{2.5cm}\times
\Gamma({D\over2}-1-n_{_3}-n_{_4})\Gamma({D\over2}-1-n_{_1}-n_{_2})
\Gamma({D\over2}+n_{_4})\Big\}^{-1}
\;,\nonumber\\
&&c_{_{[135\tilde{7}]}}^{(27),b}(\alpha,{\bf n})=
-\Gamma(1+n_{_1}+n_{_2})\Gamma(1+n_{_3}+n_{_4})\Big\{n_{_1}!n_{_2}!n_{_3}!n_{_4}!
\nonumber\\
&&\hspace{2.5cm}\times
\Gamma({D\over2}+n_{_1})\Gamma(1-{D\over2}-n_{_3}-n_{_4})\Gamma(2-{D\over2}+n_{_2})
\nonumber\\
&&\hspace{2.5cm}\times
\Gamma({D\over2}+n_{_3})
\Gamma({D\over2}-1-n_{_1}-n_{_2})\Gamma({D\over2}+n_{_4})\Big\}^{-1}
\;,\nonumber\\
&&c_{_{[135\tilde{7}]}}^{(27),c}(\alpha,{\bf n})=
(-)^{1+n_{_4}}\Gamma(1+n_{_1}+n_{_2})\Gamma(1+n_{_3})\Gamma(1+n_{_4})
\Big\{n_{_1}!n_{_2}!
\nonumber\\
&&\hspace{2.5cm}\times
\Gamma(2+n_{_3}+n_{_4})\Gamma({D\over2}+n_{_1})\Gamma(1-{D\over2}-n_{_3})\Gamma(2-{D\over2}+n_{_2})
\nonumber\\
&&\hspace{2.5cm}\times
\Gamma({D\over2}-1-n_{_4})\Gamma({D\over2}-1-n_{_1}-n_{_2})
\Gamma({D\over2}+1+n_{_3}+n_{_4})\Big\}^{-1}\;.
\label{GKZ21d-15-3}
\end{eqnarray}

\item   $I_{_{28}}=\{2,3,4,6,\cdots,10,12,14\}$, i.e. the implement $J_{_{28}}=[1,14]\setminus I_{_{28}}=\{1,5,11,13\}$.
The choice implies the power numbers $\alpha_{_1}=\alpha_{_{5}}=\alpha_{_{11}}=\alpha_{_{13}}=0$, and
\begin{eqnarray}
&&\alpha_{_2}=a_{_1}-a_{_2},\;\alpha_{_3}=a_{_1}-a_{_3}-a_{_5},
\;\alpha_{_4}=a_{_1}-a_{_4}-a_{_5},
\nonumber\\
&&\alpha_{_6}=b_{_1}-1,\;\alpha_{_{7}}=a_{_5}+b_{_2}-a_{_1}-1,\;
\alpha_{_{8}}=b_{_3}-1,
\nonumber\\
&&\alpha_{_{9}}=b_{_4}-a_{_1}-1,\;\alpha_{_{10}}=b_{_5}-a_{_1}-1,\;
\alpha_{_{12}}=a_{_5}-a_{_1},\;\alpha_{_{14}}=-a_{_5}\;.
\label{GKZ21d-16-1}
\end{eqnarray}
The corresponding hypergeometric series solutions are presented as
\begin{eqnarray}
&&\Phi_{_{[135\tilde{7}]}}^{(28),a}(\alpha,z)=
y_{_2}^{{D}-3}y_{_4}^{-1}\sum\limits_{n_{_1}=0}^\infty
\sum\limits_{n_{_2}=0}^\infty\sum\limits_{n_{_3}=0}^\infty\sum\limits_{n_{_4}=0}^\infty
c_{_{[135\tilde{7}]}}^{(28),a}(\alpha,{\bf n})
\nonumber\\
&&\hspace{2.5cm}\times
\Big({1\over y_{_4}}\Big)^{n_{_1}}\Big({y_{_1}\over y_{_4}}\Big)^{n_{_2}}
\Big({y_{_4}\over y_{_2}}\Big)^{n_{_3}}\Big({y_{_3}\over y_{_2}}\Big)^{n_{_4}}
\;,\nonumber\\
&&\Phi_{_{[135\tilde{7}]}}^{(28),b}(\alpha,z)=
y_{_2}^{{D}-2}y_{_4}^{-2}\sum\limits_{n_{_1}=0}^\infty
\sum\limits_{n_{_2}=0}^\infty\sum\limits_{n_{_3}=0}^\infty\sum\limits_{n_{_4}=0}^\infty
c_{_{[135\tilde{7}]}}^{(28),b}(\alpha,{\bf n})
\nonumber\\
&&\hspace{2.5cm}\times
\Big({1\over y_{_4}}\Big)^{n_{_1}}\Big({y_{_1}\over y_{_4}}\Big)^{n_{_2}}
\Big({y_{_2}\over y_{_4}}\Big)^{n_{_3}}\Big({y_{_3}\over y_{_2}}\Big)^{n_{_4}}\;.
\label{GKZ21d-16-2a}
\end{eqnarray}
Where the coefficients are
\begin{eqnarray}
&&c_{_{[135\tilde{7}]}}^{(28),a}(\alpha,{\bf n})=
(-)^{n_{_3}}\Gamma(1+n_{_1}+n_{_2})\Big\{n_{_1}!n_{_2}!n_{_3}!n_{_4}!
\Gamma({D\over2}+n_{_1})\Gamma(2-{D\over2}+n_{_3})
\nonumber\\
&&\hspace{2.5cm}\times
\Gamma(2-{D\over2}+n_{_2})\Gamma(2-{D\over2}+n_{_4})
\Gamma({D\over2}-1-n_{_1}-n_{_2})
\nonumber\\
&&\hspace{2.5cm}\times
\Gamma({D\over2}-1-n_{_3}-n_{_4})\Gamma(D-2-n_{_3}-n_{_4})\Big\}^{-1}
\;,\nonumber\\
&&c_{_{[135\tilde{7}]}}^{(28),b}(\alpha,{\bf n})=
-\Gamma(1+n_{_1}+n_{_2})\Gamma(1+n_{_3})\Big\{n_{_1}!n_{_2}!n_{_4}!
\Gamma({D\over2}+n_{_1})\Gamma(1-{D\over2}-n_{_3})
\nonumber\\
&&\hspace{2.5cm}\times
\Gamma(2-{D\over2}+n_{_2})\Gamma(2-{D\over2}+n_{_4})
\Gamma({D\over2}-1-n_{_1}-n_{_2})
\nonumber\\
&&\hspace{2.5cm}\times
\Gamma({D\over2}+n_{_3}-n_{_4})\Gamma(D-1+n_{_3}-n_{_4})\Big\}^{-1}\;.
\label{GKZ21d-16-3}
\end{eqnarray}

\item   $I_{_{29}}=\{1,3,4,7,9,\cdots,14\}$, i.e. the implement $J_{_{29}}=[1,14]\setminus I_{_{29}}=\{2,5,6,8\}$.
The choice implies the power numbers $\alpha_{_2}=\alpha_{_{5}}=\alpha_{_{6}}=\alpha_{_{8}}=0$, and
\begin{eqnarray}
&&\alpha_{_1}=a_{_2}-a_{_1},\;\alpha_{_3}=a_{_2}-a_{_3}-a_{_5}-b_{_1}+1,
\nonumber\\
&&\alpha_{_4}=a_{_2}-a_{_4}-a_{_5}-b_{_1}+1,\;
\alpha_{_{7}}=a_{_5}+b_{_1}+b_{_2}+b_{_3}-a_{_2}-3,
\nonumber\\
&&\alpha_{_{9}}=b_{_1}+b_{_4}-a_{_2}-2,\;\alpha_{_{10}}=b_{_1}+b_{_5}-a_{_2}-2,\;
\alpha_{_{11}}=1-b_{_1},
\nonumber\\
&&\alpha_{_{12}}=a_{_5}+b_{_1}+b_{_3}-a_{_2}-2,\;\alpha_{_{13}}=1-b_{_3},\;
\alpha_{_{14}}=-a_{_5}\;.
\label{GKZ21d-29-1}
\end{eqnarray}
The corresponding hypergeometric functions are written as
\begin{eqnarray}
&&\Phi_{_{[135\tilde{7}]}}^{(29),a}(\alpha,z)=
y_{_1}^{{D\over2}-1}y_{_2}^{{D\over2}-2}y_{_3}^{{D\over2}-1}y_{_4}^{-1}
\sum\limits_{n_{_1}=0}^\infty
\sum\limits_{n_{_2}=0}^\infty\sum\limits_{n_{_3}=0}^\infty\sum\limits_{n_{_4}=0}^\infty
c_{_{[135\tilde{7}]}}^{(29),a}(\alpha,{\bf n})
\nonumber\\
&&\hspace{2.5cm}\times
\Big({1\over y_{_4}}\Big)^{n_{_1}}\Big({y_{_1}\over y_{_4}}\Big)^{n_{_2}}
\Big({y_{_4}\over y_{_2}}\Big)^{n_{_3}}\Big({y_{_3}\over y_{_2}}\Big)^{n_{_4}}
\;,\nonumber\\
&&\Phi_{_{[135\tilde{7}]}}^{(29),b}(\alpha,z)=
y_{_1}^{{D\over2}-1}y_{_2}^{{D\over2}-1}y_{_3}^{{D\over2}-1}y_{_4}^{-2}
\sum\limits_{n_{_1}=0}^\infty
\sum\limits_{n_{_2}=0}^\infty\sum\limits_{n_{_3}=0}^\infty\sum\limits_{n_{_4}=0}^\infty
c_{_{[135\tilde{7}]}}^{(29),b}(\alpha,{\bf n})
\nonumber\\
&&\hspace{2.5cm}\times
\Big({1\over y_{_4}}\Big)^{n_{_1}}\Big({y_{_1}\over y_{_4}}\Big)^{n_{_2}}
\Big({y_{_2}\over y_{_4}}\Big)^{n_{_3}}\Big({y_{_3}\over y_{_4}}\Big)^{n_{_4}}
\;,\nonumber\\
&&\Phi_{_{[135\tilde{7}]}}^{(29),c}(\alpha,z)=
y_{_1}^{{D\over2}-1}y_{_2}^{{D\over2}-2}y_{_3}^{{D\over2}}y_{_4}^{-2}
\sum\limits_{n_{_1}=0}^\infty
\sum\limits_{n_{_2}=0}^\infty\sum\limits_{n_{_3}=0}^\infty\sum\limits_{n_{_4}=0}^\infty
c_{_{[135\tilde{7}]}}^{(29),c}(\alpha,{\bf n})
\nonumber\\
&&\hspace{2.5cm}\times
\Big({1\over y_{_4}}\Big)^{n_{_1}}\Big({y_{_1}\over y_{_4}}\Big)^{n_{_2}}
\Big({y_{_3}\over y_{_4}}\Big)^{n_{_3}}\Big({y_{_3}\over y_{_2}}\Big)^{n_{_4}}\;.
\label{GKZ21d-29-2a}
\end{eqnarray}
Where the coefficients are
\begin{eqnarray}
&&c_{_{[135\tilde{7}]}}^{(29),a}(\alpha,{\bf n})=
(-)^{n_{_4}}\Gamma(1+n_{_1}+n_{_2})\Gamma(1+n_{_3}+n_{_4})
\Big\{n_{_1}!n_{_2}!n_{_3}!n_{_4}!
\nonumber\\
&&\hspace{2.5cm}\times
\Gamma(2-{D\over2}+n_{_1})\Gamma(2-{D\over2}+n_{_3})\Gamma({D\over2}-1-n_{_1}-n_{_2})
\nonumber\\
&&\hspace{2.5cm}\times
\Gamma({D\over2}+n_{_2})
\Gamma({D\over2}-1-n_{_3}-n_{_4})\Gamma({D\over2}+n_{_4})\Big\}^{-1}
\;,\nonumber\\
&&c_{_{[135\tilde{7}]}}^{(29),b}(\alpha,{\bf n})=
-\Gamma(1+n_{_1}+n_{_2})\Gamma(1+n_{_3}+n_{_4})
\Big\{n_{_1}!n_{_2}!n_{_3}!n_{_4}!
\nonumber\\
&&\hspace{2.5cm}\times
\Gamma(2-{D\over2}+n_{_1})\Gamma(1-{D\over2}-n_{_3}-n_{_4})\Gamma({D\over2}+n_{_2})
\nonumber\\
&&\hspace{2.5cm}\times
\Gamma({D\over2}-1-n_{_1}-n_{_2})
\Gamma({D\over2}-2+n_{_3})\Gamma({D\over2}+n_{_4})\Big\}^{-1}
\;,\nonumber\\
&&c_{_{[135\tilde{7}]}}^{(29),c}(\alpha,{\bf n})=
(-)^{1+n_{_4}}\Gamma(1+n_{_1}+n_{_2})\Gamma(1+n_{_3})\Gamma(1+n_{_4})
\Big\{n_{_1}!n_{_2}!
\nonumber\\
&&\hspace{2.5cm}\times
\Gamma(2+n_{_3}+n_{_4})\Gamma(2-{D\over2}+n_{_1})\Gamma(1-{D\over2}-n_{_3})
\nonumber\\
&&\hspace{2.5cm}\times
\Gamma({D\over2}-1-n_{_1}-n_{_2})\Gamma({D\over2}+n_{_2})
\Gamma({D\over2}-3-n_{_4})
\nonumber\\
&&\hspace{2.5cm}\times
\Gamma({D\over2}+1+n_{_3}+n_{_4})\Big\}^{-1}\;.
\label{GKZ21d-29-3}
\end{eqnarray}

\item   $I_{_{30}}=\{1,3,4,7,\cdots,12,14\}$, i.e. the implement $J_{_{30}}=[1,14]\setminus I_{_{30}}=\{2,5,6,13\}$.
The choice implies the power numbers $\alpha_{_2}=\alpha_{_{5}}=\alpha_{_{6}}=\alpha_{_{13}}=0$, and
\begin{eqnarray}
&&\alpha_{_1}=a_{_2}-a_{_1},\;\alpha_{_3}=a_{_2}-a_{_3}-a_{_5}-b_{_1}+1,
\nonumber\\
&&\alpha_{_4}=a_{_2}-a_{_4}-a_{_5}-b_{_1}+1,\;
\alpha_{_{7}}=a_{_5}+b_{_1}+b_{_2}-a_{_2}-2,
\nonumber\\
&&\alpha_{_{8}}=b_{_3}-1,\;\alpha_{_{9}}=b_{_1}+b_{_4}-a_{_2}-2,\;\alpha_{_{10}}=b_{_1}+b_{_5}-a_{_2}-2,
\nonumber\\
&&\alpha_{_{11}}=1-b_{_1},\;
\alpha_{_{12}}=a_{_5}+b_{_1}-a_{_2}-1,\;\alpha_{_{14}}=-a_{_5}\;.
\label{GKZ21d-30-1}
\end{eqnarray}
The corresponding hypergeometric series solutions are written as
\begin{eqnarray}
&&\Phi_{_{[135\tilde{7}]}}^{(30),a}(\alpha,z)=
y_{_1}^{{D\over2}-1}y_{_2}^{{D}-3}y_{_4}^{-1}\sum\limits_{n_{_1}=0}^\infty
\sum\limits_{n_{_2}=0}^\infty\sum\limits_{n_{_3}=0}^\infty\sum\limits_{n_{_4}=0}^\infty
c_{_{[135\tilde{7}]}}^{(30),a}(\alpha,{\bf n})
\nonumber\\
&&\hspace{2.5cm}\times
\Big({1\over y_{_4}}\Big)^{n_{_1}}\Big({y_{_1}\over y_{_4}}\Big)^{n_{_2}}
\Big({y_{_4}\over y_{_2}}\Big)^{n_{_3}}\Big({y_{_3}\over y_{_2}}\Big)^{n_{_4}}
\;,\nonumber\\
&&\Phi_{_{[135\tilde{7}]}}^{(30),b}(\alpha,z)=
y_{_1}^{{D\over2}-1}y_{_2}^{{D}-2}y_{_4}^{-2}\sum\limits_{n_{_1}=0}^\infty
\sum\limits_{n_{_2}=0}^\infty\sum\limits_{n_{_3}=0}^\infty\sum\limits_{n_{_4}=0}^\infty
c_{_{[135\tilde{7}]}}^{(30),b}(\alpha,{\bf n})
\nonumber\\
&&\hspace{2.5cm}\times
\Big({1\over y_{_4}}\Big)^{n_{_1}}\Big({y_{_1}\over y_{_4}}\Big)^{n_{_2}}
\Big({y_{_2}\over y_{_4}}\Big)^{n_{_3}}\Big({y_{_3}\over y_{_2}}\Big)^{n_{_4}}\;.
\label{GKZ21d-30-2a}
\end{eqnarray}
Where the coefficients are
\begin{eqnarray}
&&c_{_{[135\tilde{7}]}}^{(30),a}(\alpha,{\bf n})=
(-)^{n_{_3}}\Gamma(1+n_{_1}+n_{_2})\Big\{n_{_1}!n_{_2}!n_{_3}!n_{_4}!
\Gamma(2-{D\over2}+n_{_1})
\nonumber\\
&&\hspace{2.5cm}\times
\Gamma(2-{D\over2}+n_{_3})\Gamma({D\over2}-1-n_{_3}-n_{_4})
\Gamma(2-{D\over2}+n_{_4})
\nonumber\\
&&\hspace{2.5cm}\times
\Gamma({D\over2}-1-n_{_1}-n_{_2})
\Gamma({D\over2}+n_{_2})\Gamma(D-2-n_{_3}-n_{_4})\Big\}^{-1}
\;,\nonumber\\
&&c_{_{[135\tilde{7}]}}^{(30),b}(\alpha,{\bf n})=
-\Gamma(1+n_{_1}+n_{_2})\Gamma(1+n_{_3})\Big\{n_{_1}!n_{_2}!n_{_4}!
\Gamma(2-{D\over2}+n_{_1})
\nonumber\\
&&\hspace{2.5cm}\times
\Gamma(1-{D\over2}-n_{_3})\Gamma({D\over2}+n_{_3}-n_{_4})
\Gamma(2-{D\over2}+n_{_4})
\nonumber\\
&&\hspace{2.5cm}\times
\Gamma({D\over2}-1-n_{_1}-n_{_2})
\Gamma({D\over2}+n_{_2})\Gamma(D-3+n_{_3}-n_{_4})\Big\}^{-1}\;.
\label{GKZ21d-30-3}
\end{eqnarray}

\item   $I_{_{31}}=\{1,3,4,6,7,9,10,12,13,14\}$, i.e. the implement $J_{_{31}}=[1,14]\setminus I_{_{31}}=\{2,5,8,11\}$.
The choice implies the power numbers $\alpha_{_2}=\alpha_{_{5}}=\alpha_{_{8}}=\alpha_{_{11}}=0$, and
\begin{eqnarray}
&&\alpha_{_1}=a_{_2}-a_{_1},\;\alpha_{_3}=a_{_2}-a_{_3}-a_{_5},\;
\alpha_{_4}=a_{_2}-a_{_4}-a_{_5},
\nonumber\\
&&\alpha_{_{6}}=b_{_1}-1,\;
\alpha_{_{7}}=a_{_5}+b_{_2}+b_{_3}-a_{_2}-2,
\nonumber\\
&&\alpha_{_{9}}=b_{_4}-a_{_2}-1,\;\alpha_{_{10}}=b_{_5}-a_{_2}-1,
\nonumber\\
&&\alpha_{_{12}}=a_{_5}+b_{_3}-a_{_2}-1,\;\alpha_{_{13}}=1-b_{_3},\;\alpha_{_{14}}=-a_{_5}\;.
\label{GKZ21d-31-1}
\end{eqnarray}
The corresponding hypergeometric series is written as
\begin{eqnarray}
&&\Phi_{_{[135\tilde{7}]}}^{(31)}(\alpha,z)=
y_{_2}^{{D}-3}y_{_3}^{{D\over2}-1}y_{_4}^{-1}\sum\limits_{n_{_1}=0}^\infty
\sum\limits_{n_{_2}=0}^\infty\sum\limits_{n_{_3}=0}^\infty\sum\limits_{n_{_4}=0}^\infty
c_{_{[135\tilde{7}]}}^{(31)}(\alpha,{\bf n})
\nonumber\\
&&\hspace{2.5cm}\times
\Big({1\over y_{_2}}\Big)^{n_{_1}}\Big({y_{_1}\over y_{_2}}\Big)^{n_{_2}}
\Big({y_{_2}\over y_{_4}}\Big)^{n_{_3}}\Big({y_{_3}\over y_{_2}}\Big)^{n_{_4}}\;,
\label{GKZ21d-31-2}
\end{eqnarray}
with
\begin{eqnarray}
&&c_{_{[135\tilde{7}]}}^{(31)}(\alpha,{\bf n})=
(-)^{n_{_3}}\Big\{n_{_1}!n_{_2}!n_{_4}!\Gamma(2-{D\over2}+n_{_1})
\nonumber\\
&&\hspace{2.5cm}\times
\Gamma(3-D+n_{_1}+n_{_2}-n_{_3})\Gamma(2-{D\over2}+n_{_1}+n_{_2}-n_{_3})
\nonumber\\
&&\hspace{2.5cm}\times
\Gamma(2-{D\over2}+n_{_2})\Gamma({D\over2}-1-n_{_1}-n_{_2}+n_{_3}-n_{_4})
\nonumber\\
&&\hspace{2.5cm}\times
\Gamma(D-2-n_{_1}-n_{_2})\Gamma({D\over2}-1-n_{_1}-n_{_2})
\nonumber\\
&&\hspace{2.5cm}\times
\Gamma(D-2-n_{_1}-n_{_2}+n_{_3}-n_{_4})\Gamma({D\over2}+n_{_4})\Big\}^{-1}\;.
\label{GKZ21d-31-3}
\end{eqnarray}

\item   $I_{_{32}}=\{1,3,4,6,\cdots,10,12,14\}$, i.e. the implement $J_{_{32}}=[1,14]\setminus I_{_{32}}=\{2,5,11,13\}$.
The choice implies the power numbers $\alpha_{_2}=\alpha_{_{5}}=\alpha_{_{11}}=\alpha_{_{13}}=0$, and
\begin{eqnarray}
&&\alpha_{_1}=a_{_2}-a_{_1},\;\alpha_{_3}=a_{_2}-a_{_3}-a_{_5},
\;\alpha_{_4}=a_{_2}-a_{_4}-a_{_5},
\nonumber\\
&&\alpha_{_6}=b_{_1}-1,\;\alpha_{_{7}}=a_{_5}+b_{_2}-a_{_2}-1,\;
\alpha_{_{8}}=b_{_3}-1,
\nonumber\\
&&\alpha_{_{9}}=b_{_4}-a_{_2}-1,\;\alpha_{_{10}}=b_{_5}-a_{_2}-1,\;
\alpha_{_{12}}=a_{_5}-a_{_2},\;\alpha_{_{14}}=-a_{_5}\;.
\label{GKZ21d-32-1}
\end{eqnarray}
The corresponding hypergeometric series is
\begin{eqnarray}
&&\Phi_{_{[135\tilde{7}]}}^{(32)}(\alpha,z)=
y_{_2}^{{3D\over2}-4}y_{_4}^{-1}\sum\limits_{n_{_1}=0}^\infty
\sum\limits_{n_{_2}=0}^\infty\sum\limits_{n_{_3}=0}^\infty\sum\limits_{n_{_4}=0}^\infty
c_{_{[135\tilde{7}]}}^{(32)}(\alpha,{\bf n})
\nonumber\\
&&\hspace{2.5cm}\times
\Big({1\over y_{_2}}\Big)^{n_{_1}}\Big({y_{_1}\over y_{_2}}\Big)^{n_{_2}}
\Big({y_{_2}\over y_{_4}}\Big)^{n_{_3}}\Big({y_{_3}\over y_{_2}}\Big)^{n_{_4}}\;,
\label{GKZ21d-32-2}
\end{eqnarray}
with
\begin{eqnarray}
&&c_{_{[135\tilde{7}]}}^{(32)}(\alpha,{\bf n})=
(-)^{n_{_3}}\Big\{n_{_1}!n_{_2}!n_{_4}!\Gamma(2-{D\over2}+n_{_1})
\nonumber\\
&&\hspace{2.5cm}\times
\Gamma(3-D+n_{_1}+n_{_2}-n_{_3})\Gamma(2-{D\over2}+n_{_1}+n_{_2}-n_{_3})
\nonumber\\
&&\hspace{2.5cm}\times
\Gamma(2-{D\over2}+n_{_2})\Gamma(D-2-n_{_1}-n_{_2}+n_{_3}-n_{_4})
\nonumber\\
&&\hspace{2.5cm}\times
\Gamma(2-{D\over2}+n_{_4})\Gamma(D-2-n_{_1}-n_{_2})\Gamma({D\over2}-1-n_{_1}-n_{_2})
\nonumber\\
&&\hspace{2.5cm}\times
\Gamma({3D\over2}-3-n_{_1}-n_{_2}+n_{_3}-n_{_4})\Big\}^{-1}\;.
\label{GKZ21d-32-3}
\end{eqnarray}
\end{itemize}

\section{The hypergeometric solutions of the integer lattice ${\bf B}_{_{\widetilde{13}57}}$\label{app6}}
\indent\indent

\begin{itemize}
\item   $I_{_{1}}=\{1,\cdots,4,9,\cdots,14\}$, i.e. the implement $J_{_{1}}=[1,14]\setminus I_{_{1}}=\{5,6,7,8\}$.
The choice implies the power numbers $\alpha_{_5}=\alpha_{_{6}}=\alpha_{_{7}}=\alpha_{_{8}}=0$, and
\begin{eqnarray}
&&\alpha_{_1}=a_{_5}+b_{_1}+b_{_2}+b_{_3}-a_{_1}-3,\;\alpha_{_2}=a_{_5}+b_{_1}+b_{_2}+b_{_3}-a_{_2}-3,
\nonumber\\
&&\alpha_{_3}=b_{_2}+b_{_3}-a_{_3}-2,\;\alpha_{_4}=b_{_2}+b_{_3}-a_{_4}-2,
\nonumber\\
&&\alpha_{_9}=b_{_4}-b_{_2}-b_{_3}-a_{_5}+1,\;\alpha_{_{10}}=b_{_5}-b_{_2}-b_{_3}-a_{_5}+1,
\nonumber\\
&&\alpha_{_{11}}=1-b_{_1},\;\alpha_{_{12}}=1-b_{_2},\;\alpha_{_{13}}=1-b_{_3},\;
\alpha_{_{14}}=-a_{_5}\;.
\label{GKZ21e-1-1}
\end{eqnarray}
The corresponding hypergeometric series solutions are written as
\begin{eqnarray}
&&\Phi_{_{[\tilde{1}\tilde{3}57]}}^{(1),a}(\alpha,z)=
y_{_1}^{{D\over2}-1}y_{_2}^{{D\over2}-1}y_{_3}^{{D\over2}-1}y_{_4}^{-1}
\sum\limits_{n_{_1}=0}^\infty
\sum\limits_{n_{_2}=0}^\infty\sum\limits_{n_{_3}=0}^\infty\sum\limits_{n_{_4}=0}^\infty
c_{_{[\tilde{1}\tilde{3}57]}}^{(1),a}(\alpha,{\bf n})
\nonumber\\
&&\hspace{2.5cm}\times
y_{_1}^{n_{_1}}\Big({y_{_3}\over y_{_4}}\Big)^{n_{_2}}y_{_4}^{n_{_3}}
\Big({y_{_2}\over y_{_4}}\Big)^{n_{_4}}
\;,\nonumber\\
&&\Phi_{_{[\tilde{1}\tilde{3}57]}}^{(1),b}(\alpha,z)=
y_{_1}^{{D\over2}-1}y_{_2}^{{D\over2}-1}y_{_3}^{{D\over2}-1}y_{_4}^{-2}
\sum\limits_{n_{_1}=0}^\infty
\sum\limits_{n_{_2}=0}^\infty\sum\limits_{n_{_3}=0}^\infty\sum\limits_{n_{_4}=0}^\infty
c_{_{[\tilde{1}\tilde{3}57]}}^{(1),b}(\alpha,{\bf n})
\nonumber\\
&&\hspace{2.5cm}\times
\Big({y_{_1}\over y_{_4}}\Big)^{n_{_1}}\Big({y_{_3}\over y_{_4}}\Big)^{n_{_2}}
\Big({1\over y_{_4}}\Big)^{n_{_3}}\Big({y_{_2}\over y_{_4}}\Big)^{n_{_4}}
\;,\nonumber\\
&&\Phi_{_{[\tilde{1}\tilde{3}57]}}^{(1),c}(\alpha,z)=
y_{_1}^{{D\over2}}y_{_2}^{{D\over2}-1}y_{_3}^{{D\over2}-1}y_{_4}^{-2}
\sum\limits_{n_{_1}=0}^\infty
\sum\limits_{n_{_2}=0}^\infty\sum\limits_{n_{_3}=0}^\infty\sum\limits_{n_{_4}=0}^\infty
c_{_{[\tilde{1}\tilde{3}57]}}^{(1),c}(\alpha,{\bf n})
\nonumber\\
&&\hspace{2.5cm}\times
y_{_1}^{n_{_1}}\Big({y_{_3}\over y_{_4}}\Big)^{n_{_2}}
\Big({y_{_1}\over y_{_4}}\Big)^{n_{_3}}\Big({y_{_2}\over y_{_4}}\Big)^{n_{_4}}\;.
\label{GKZ21e-1-2a}
\end{eqnarray}
Where the coefficients are
\begin{eqnarray}
&&c_{_{[\tilde{1}\tilde{3}57]}}^{(1),a}(\alpha,{\bf n})=
(-)^{n_{_1}}\Gamma(1+n_{_1}+n_{_3})\Gamma(1+n_{_2}+n_{_4})
\Big\{n_{_1}!n_{_2}!n_{_3}!n_{_4}!
\nonumber\\
&&\hspace{2.5cm}\times
\Gamma(1-{D\over2}-n_{_1}-n_{_3})
\Gamma(1-{D\over2}-n_{_2}-n_{_4})\Gamma({D\over2}+n_{_3})
\nonumber\\
&&\hspace{2.5cm}\times
\Gamma({D\over2}+n_{_1})\Gamma({D\over2}+n_{_4})\Gamma({D\over2}+n_{_2})\Big\}^{-1}
\;,\nonumber\\
&&c_{_{[\tilde{1}\tilde{3}57]}}^{(1),b}(\alpha,{\bf n})=
-\Gamma(1+n_{_1}+n_{_3})\Gamma(1+n_{_2}+n_{_4})
\Big\{n_{_1}!n_{_2}!n_{_3}!n_{_4}!\Gamma(2-{D\over2}+n_{_3})
\nonumber\\
&&\hspace{2.5cm}\times
\Gamma(1-{D\over2}-n_{_2}-n_{_4})\Gamma({D\over2}-1-n_{_1}-n_{_3})
\Gamma({D\over2}+n_{_1})
\nonumber\\
&&\hspace{2.5cm}\times
\Gamma({D\over2}+n_{_4})\Gamma({D\over2}+n_{_2})\Big\}^{-1}
\;,\nonumber\\
&&c_{_{[\tilde{1}\tilde{3}57]}}^{(1),c}(\alpha,{\bf n})=
(-)^{1+n_{_1}}\Gamma(1+n_{_1})\Gamma(1+n_{_3})\Gamma(1+n_{_2}+n_{_4})
\Big\{n_{_2}!n_{_4}!\Gamma(2+n_{_1}+n_{_3})
\nonumber\\
&&\hspace{2.5cm}\times
\Gamma(1-{D\over2}-n_{_1})
\Gamma(1-{D\over2}-n_{_2}-n_{_4})\Gamma({D\over2}-1-n_{_3})
\nonumber\\
&&\hspace{2.5cm}\times
\Gamma({D\over2}+1+n_{_1}+n_{_3})\Gamma({D\over2}+n_{_4})\Gamma({D\over2}+n_{_2})\Big\}^{-1}\;.
\label{GKZ21e-1-3}
\end{eqnarray}

\item   $I_{_{2}}=\{1,\cdots,4,7,9,10,11,13,14\}$, i.e. the implement $J_{_{2}}=[1,14]\setminus I_{_{2}}=\{5,6,8,12\}$.
The choice implies the power numbers $\alpha_{_5}=\alpha_{_{6}}=\alpha_{_{8}}=\alpha_{_{12}}=0$, and
\begin{eqnarray}
&&\alpha_{_1}=a_{_5}+b_{_1}+b_{_3}-a_{_1}-2,\;\alpha_{_2}=a_{_5}+b_{_1}+b_{_3}-a_{_2}-2,
\nonumber\\
&&\alpha_{_3}=b_{_3}-a_{_3}-1,\;\alpha_{_4}=b_{_3}-a_{_4}-1,\;\alpha_{_{7}}=b_{_2}-1,
\nonumber\\
&&\alpha_{_9}=b_{_4}-a_{_5}-b_{_3},\;\alpha_{_{10}}=b_{_5}-a_{_5}-b_{_3},
\nonumber\\
&&\alpha_{_{11}}=1-b_{_1},\;\alpha_{_{13}}=1-b_{_3},\;
\alpha_{_{14}}=-a_{_5}\;.
\label{GKZ21e-2-1}
\end{eqnarray}
The corresponding hypergeometric functions are
\begin{eqnarray}
&&\Phi_{_{[\tilde{1}\tilde{3}57]}}^{(2),a}(\alpha,z)=
y_{_1}^{{D\over2}-1}y_{_3}^{{D\over2}-1}y_{_4}^{-1}\sum\limits_{n_{_1}=0}^\infty
\sum\limits_{n_{_2}=0}^\infty\sum\limits_{n_{_3}=0}^\infty\sum\limits_{n_{_4}=0}^\infty
c_{_{[\tilde{1}\tilde{3}57]}}^{(2),a}(\alpha,{\bf n})
\nonumber\\
&&\hspace{2.5cm}\times
y_{_1}^{n_{_1}}\Big({y_{_3}\over y_{_4}}\Big)^{n_{_2}}y_{_4}^{n_{_3}}
\Big({y_{_2}\over y_{_4}}\Big)^{n_{_4}}
\;,\nonumber\\
&&\Phi_{_{[\tilde{1}\tilde{3}57]}}^{(2),b}(\alpha,z)=
y_{_1}^{{D\over2}-1}y_{_3}^{{D\over2}-1}y_{_4}^{-2}\sum\limits_{n_{_1}=0}^\infty
\sum\limits_{n_{_2}=0}^\infty\sum\limits_{n_{_3}=0}^\infty\sum\limits_{n_{_4}=0}^\infty
c_{_{[\tilde{1}\tilde{3}57]}}^{(2),b}(\alpha,{\bf n})
\nonumber\\
&&\hspace{2.5cm}\times
\Big({y_{_1}\over y_{_4}}\Big)^{n_{_1}}\Big({y_{_3}\over y_{_4}}\Big)^{n_{_2}}
\Big({1\over y_{_4}}\Big)^{n_{_3}}\Big({y_{_2}\over y_{_4}}\Big)^{n_{_4}}
\;,\nonumber\\
&&\Phi_{_{[\tilde{1}\tilde{3}57]}}^{(2),c}(\alpha,z)=
y_{_1}^{{D\over2}}y_{_3}^{{D\over2}-1}y_{_4}^{-2}\sum\limits_{n_{_1}=0}^\infty
\sum\limits_{n_{_2}=0}^\infty\sum\limits_{n_{_3}=0}^\infty\sum\limits_{n_{_4}=0}^\infty
c_{_{[\tilde{1}\tilde{3}57]}}^{(2),c}(\alpha,{\bf n})
\nonumber\\
&&\hspace{2.5cm}\times
y_{_1}^{n_{_1}}\Big({y_{_3}\over y_{_4}}\Big)^{n_{_2}}
\Big({y_{_1}\over y_{_4}}\Big)^{n_{_3}}\Big({y_{_2}\over y_{_4}}\Big)^{n_{_4}}\;.
\label{GKZ21e-2-2a}
\end{eqnarray}
Where the coefficients are
\begin{eqnarray}
&&c_{_{[\tilde{1}\tilde{3}57]}}^{(2),a}(\alpha,{\bf n})=
(-)^{n_{_1}}\Gamma(1+n_{_1}+n_{_3})\Gamma(1+n_{_2}+n_{_4})
\Big\{n_{_1}!n_{_2}!n_{_3}!n_{_4}!
\nonumber\\
&&\hspace{2.5cm}\times
\Gamma({D\over2}-1-n_{_1}-n_{_3})
\Gamma({D\over2}-1-n_{_2}-n_{_4})\Gamma(2-{D\over2}+n_{_4})
\nonumber\\
&&\hspace{2.5cm}\times
\Gamma(2-{D\over2}+n_{_3})
\Gamma({D\over2}+n_{_1})\Gamma({D\over2}+n_{_2})\Big\}^{-1}
\;,\nonumber\\
&&c_{_{[\tilde{1}\tilde{3}57]}}^{(2),b}(\alpha,{\bf n})=
-\Gamma(1+n_{_1}+n_{_3})\Gamma(1+n_{_2}+n_{_4})
\Big\{n_{_1}!n_{_2}!n_{_3}!n_{_4}!\Gamma({D\over2}+n_{_3})
\nonumber\\
&&\hspace{2.5cm}\times
\Gamma({D\over2}-1-n_{_2}-n_{_4})\Gamma(2-{D\over2}+n_{_4})
\nonumber\\
&&\hspace{2.5cm}\times
\Gamma(1-{D\over2}-n_{_1}-n_{_3})
\Gamma({D\over2}+n_{_1})\Gamma({D\over2}+n_{_2})\Big\}^{-1}
\;,\nonumber\\
&&c_{_{[\tilde{1}\tilde{3}57]}}^{(2),c}(\alpha,{\bf n})=
(-)^{1+n_{_1}}\Gamma(1+n_{_1})\Gamma(1+n_{_3})\Gamma(1+n_{_2}+n_{_4})
\Big\{n_{_2}!n_{_4}!
\nonumber\\
&&\hspace{2.5cm}\times
\Gamma(2+n_{_1}+n_{_3})\Gamma({D\over2}-1-n_{_1})
\Gamma({D\over2}-1-n_{_2}-n_{_4})
\nonumber\\
&&\hspace{2.5cm}\times
\Gamma(2-{D\over2}+n_{_4})\Gamma(1-{D\over2}-n_{_3})
\Gamma({D\over2}+1+n_{_1}+n_{_3})\Gamma({D\over2}+n_{_2})\Big\}^{-1}\;.
\label{GKZ21e-2-3}
\end{eqnarray}

\item   $I_{_{3}}=\{1,\cdots,4,8,\cdots,12,14\}$, i.e. the implement $J_{_{3}}=[1,14]\setminus I_{_{3}}=\{5,6,7,13\}$.
The choice implies the power numbers $\alpha_{_5}=\alpha_{_{6}}=\alpha_{_{7}}=\alpha_{_{13}}=0$, and
\begin{eqnarray}
&&\alpha_{_1}=a_{_5}+b_{_1}+b_{_2}-a_{_1}-2,\;\alpha_{_2}=a_{_5}+b_{_1}+b_{_2}-a_{_2}-2,
\nonumber\\
&&\alpha_{_3}=b_{_2}-a_{_3}-1,\;\alpha_{_4}=b_{_2}-a_{_4}-1,\;\alpha_{_{8}}=b_{_3}-1,
\nonumber\\
&&\alpha_{_9}=b_{_4}-a_{_5}-b_{_2},\;\alpha_{_{10}}=b_{_5}-a_{_5}-b_{_2},
\nonumber\\
&&\alpha_{_{11}}=1-b_{_1},\;\alpha_{_{12}}=1-b_{_2},\;
\alpha_{_{14}}=-a_{_5}\;.
\label{GKZ21e-7-1}
\end{eqnarray}
The corresponding hypergeometric solutions are written as
\begin{eqnarray}
&&\Phi_{_{[\tilde{1}\tilde{3}57]}}^{(3),a}(\alpha,z)=
y_{_1}^{{D\over2}-1}y_{_2}^{{D\over2}-1}y_{_4}^{-1}\sum\limits_{n_{_1}=0}^\infty
\sum\limits_{n_{_2}=0}^\infty\sum\limits_{n_{_3}=0}^\infty\sum\limits_{n_{_4}=0}^\infty
c_{_{[\tilde{1}\tilde{3}57]}}^{(3),a}(\alpha,{\bf n})
\nonumber\\
&&\hspace{2.5cm}\times
y_{_1}^{n_{_1}}\Big({y_{_3}\over y_{_4}}\Big)^{n_{_2}}y_{_4}^{n_{_3}}
\Big({y_{_2}\over y_{_4}}\Big)^{n_{_4}}
\;,\nonumber\\
&&\Phi_{_{[\tilde{1}\tilde{3}57]}}^{(3),b}(\alpha,z)=
y_{_1}^{{D\over2}-1}y_{_2}^{{D\over2}-1}y_{_4}^{-2}\sum\limits_{n_{_1}=0}^\infty
\sum\limits_{n_{_2}=0}^\infty\sum\limits_{n_{_3}=0}^\infty\sum\limits_{n_{_4}=0}^\infty
c_{_{[\tilde{1}\tilde{3}57]}}^{(3),b}(\alpha,{\bf n})
\nonumber\\
&&\hspace{2.5cm}\times
\Big({y_{_1}\over y_{_4}}\Big)^{n_{_1}}\Big({y_{_3}\over y_{_4}}\Big)^{n_{_2}}
\Big({1\over y_{_4}}\Big)^{n_{_3}}\Big({y_{_2}\over y_{_4}}\Big)^{n_{_4}}
\;,\nonumber\\
&&\Phi_{_{[\tilde{1}\tilde{3}57]}}^{(3),c}(\alpha,z)=
y_{_1}^{{D\over2}}y_{_2}^{{D\over2}-1}y_{_4}^{-2}\sum\limits_{n_{_1}=0}^\infty
\sum\limits_{n_{_2}=0}^\infty\sum\limits_{n_{_3}=0}^\infty\sum\limits_{n_{_4}=0}^\infty
c_{_{[\tilde{1}\tilde{3}57]}}^{(3),c}(\alpha,{\bf n})
\nonumber\\
&&\hspace{2.5cm}\times
y_{_1}^{n_{_1}}\Big({y_{_3}\over y_{_4}}\Big)^{n_{_2}}
\Big({y_{_1}\over y_{_4}}\Big)^{n_{_3}}\Big({y_{_2}\over y_{_4}}\Big)^{n_{_4}}\;.
\label{GKZ21e-7-2a}
\end{eqnarray}
Where the coefficients are
\begin{eqnarray}
&&c_{_{[\tilde{1}\tilde{3}57]}}^{(3),a}(\alpha,{\bf n})=
(-)^{n_{_1}}\Gamma(1+n_{_1}+n_{_3})\Gamma(1+n_{_2}+n_{_4})
\Big\{n_{_1}!n_{_2}!n_{_3}!n_{_4}!
\nonumber\\
&&\hspace{2.5cm}\times
\Gamma({D\over2}-1-n_{_1}-n_{_3})
\Gamma({D\over2}-1-n_{_2}-n_{_4})\Gamma(2-{D\over2}+n_{_2})
\nonumber\\
&&\hspace{2.5cm}\times
\Gamma(2-{D\over2}+n_{_3})
\Gamma({D\over2}+n_{_1})\Gamma({D\over2}+n_{_4})\Big\}^{-1}
\;,\nonumber\\
&&c_{_{[\tilde{1}\tilde{3}57]}}^{(3),b}(\alpha,{\bf n})=
-\Gamma(1+n_{_1}+n_{_3})\Gamma(1+n_{_2}+n_{_4})
\Big\{n_{_1}!n_{_2}!n_{_3}!n_{_4}!\Gamma({D\over2}+n_{_3})
\nonumber\\
&&\hspace{2.5cm}\times
\Gamma({D\over2}-1-n_{_2}-n_{_4})\Gamma(2-{D\over2}+n_{_2})
\nonumber\\
&&\hspace{2.5cm}\times
\Gamma(1-{D\over2}-n_{_1}-n_{_3})
\Gamma({D\over2}+n_{_1})\Gamma({D\over2}+n_{_4})\Big\}^{-1}
\;,\nonumber\\
&&c_{_{[\tilde{1}\tilde{3}57]}}^{(3),c}(\alpha,{\bf n})=
(-)^{1+n_{_1}}\Gamma(1+n_{_1})\Gamma(1+n_{_3})\Gamma(1+n_{_2}+n_{_4})
\Big\{n_{_2}!n_{_4}!
\nonumber\\
&&\hspace{2.5cm}\times
\Gamma(2+n_{_1}+n_{_3})\Gamma({D\over2}-1-n_{_1})
\Gamma({D\over2}-1-n_{_2}-n_{_4})
\nonumber\\
&&\hspace{2.5cm}\times
\Gamma(2-{D\over2}+n_{_2})\Gamma(1-{D\over2}-n_{_3})
\Gamma({D\over2}+1+n_{_1}+n_{_3})\Gamma({D\over2}+n_{_4})\Big\}^{-1}\;.
\label{GKZ21e-7-3}
\end{eqnarray}

\item   $I_{_{4}}=\{1,\cdots,4,7\cdots,11,14\}$, i.e. the implement $J_{_{4}}=[1,14]\setminus I_{_{4}}=\{5,6,12,13\}$.
The choice implies the power numbers $\alpha_{_5}=\alpha_{_{6}}=\alpha_{_{12}}=\alpha_{_{13}}=0$, and
\begin{eqnarray}
&&\alpha_{_1}=a_{_5}+b_{_1}-a_{_1}-1,\;\alpha_{_2}=a_{_5}+b_{_1}-a_{_2}-1,
\nonumber\\
&&\alpha_{_3}=-a_{_3},\;\alpha_{_4}=-a_{_4},\;\alpha_{_{7}}=b_{_2}-1,\;\alpha_{_{8}}=b_{_3}-1
\nonumber\\
&&\alpha_{_9}=b_{_4}-a_{_5}-1,\;\alpha_{_{10}}=b_{_5}-a_{_5}-1,\;\alpha_{_{11}}=1-b_{_1},
\nonumber\\
&&\alpha_{_{14}}=-a_{_5}\;.
\label{GKZ21e-8-1}
\end{eqnarray}
The corresponding hypergeometric series is
\begin{eqnarray}
&&\Phi_{_{[\tilde{1}\tilde{3}57]}}^{(4)}(\alpha,z)=
y_{_1}^{{D\over2}-1}y_{_4}^{-1}\sum\limits_{n_{_1}=0}^\infty
\sum\limits_{n_{_2}=0}^\infty\sum\limits_{n_{_3}=0}^\infty\sum\limits_{n_{_4}=0}^\infty
c_{_{[\tilde{1}\tilde{3}57]}}^{(4)}(\alpha,{\bf n})
\nonumber\\
&&\hspace{2.5cm}\times
y_{_1}^{n_{_1}}y_{_3}^{n_{_2}}\Big({1\over y_{_4}}\Big)^{n_{_3}}y_{_2}^{n_{_4}}\;,
\label{GKZ21e-8-2}
\end{eqnarray}
with
\begin{eqnarray}
&&c_{_{[\tilde{1}\tilde{3}57]}}^{(4)}(\alpha,{\bf n})=
(-)^{n_{_3}}\Big\{n_{_1}!n_{_2}!n_{_4}!\Gamma({D\over2}-1-n_{_1}-n_{_2}+n_{_3}-n_{_4})
\nonumber\\
&&\hspace{2.5cm}\times
\Gamma(D-2-n_{_1}-n_{_2}+n_{_3}-n_{_4})
\Gamma({D\over2}-1-n_{_2}-n_{_4})
\nonumber\\
&&\hspace{2.5cm}\times
\Gamma(D-2-n_{_2}-n_{_4})
\Gamma(2-{D\over2}+n_{_4})\Gamma(2-{D\over2}+n_{_2})
\nonumber\\
&&\hspace{2.5cm}\times
\Gamma(2-{D\over2}+n_{_2}-n_{_3}+n_{_4})\Gamma(3-D+n_{_2}-n_{_3}+n_{_4})
\nonumber\\
&&\hspace{2.5cm}\times
\Gamma({D\over2}+n_{_1})\Big\}^{-1}\;.
\label{GKZ21e-8-3}
\end{eqnarray}

\item   $I_{_{5}}=\{1,\cdots,4,9,\cdots,14\}$, i.e. the implement $J_{_{5}}=[1,14]\setminus I_{_{5}}=\{5,7,8,11\}$.
The choice implies the power numbers $\alpha_{_5}=\alpha_{_{7}}=\alpha_{_{8}}=\alpha_{_{11}}=0$, and
\begin{eqnarray}
&&\alpha_{_1}=a_{_5}+b_{_2}+b_{_3}-a_{_1}-2,\;\alpha_{_2}=a_{_5}+b_{_2}+b_{_3}-a_{_2}-2,
\nonumber\\
&&\alpha_{_3}=b_{_2}+b_{_3}-a_{_3}-2,\;\alpha_{_4}=b_{_2}+b_{_3}-a_{_4}-2,\;\alpha_{_{6}}=b_{_1}-1,
\nonumber\\
&&\alpha_{_9}=b_{_4}-a_{_5}-b_{_2}-b_{_3}+1,\;\alpha_{_{10}}=b_{_5}-a_{_5}-b_{_2}-b_{_3}+1,
\nonumber\\
&&\alpha_{_{12}}=1-b_{_2},\;\alpha_{_{13}}=1-b_{_3},\;
\alpha_{_{14}}=-a_{_5}\;.
\label{GKZ21e-9-1}
\end{eqnarray}
The corresponding hypergeometric functions are
\begin{eqnarray}
&&\Phi_{_{[\tilde{1}\tilde{3}57]}}^{(5),a}(\alpha,z)=
y_{_2}^{{D\over2}-1}y_{_3}^{{D\over2}-1}y_{_4}^{-1}\sum\limits_{n_{_1}=0}^\infty
\sum\limits_{n_{_2}=0}^\infty\sum\limits_{n_{_3}=0}^\infty\sum\limits_{n_{_4}=0}^\infty
c_{_{[\tilde{1}\tilde{3}57]}}^{(5),a}(\alpha,{\bf n})
\nonumber\\
&&\hspace{2.5cm}\times
y_{_1}^{n_{_1}}\Big({y_{_3}\over y_{_4}}\Big)^{n_{_2}}y_{_4}^{n_{_3}}
\Big({y_{_2}\over y_{_4}}\Big)^{n_{_4}}
\;,\nonumber\\
&&\Phi_{_{[\tilde{1}\tilde{3}57]}}^{(5),b}(\alpha,z)=
y_{_2}^{{D\over2}-1}y_{_3}^{{D\over2}-1}y_{_4}^{-2}\sum\limits_{n_{_1}=0}^\infty
\sum\limits_{n_{_2}=0}^\infty\sum\limits_{n_{_3}=0}^\infty\sum\limits_{n_{_4}=0}^\infty
c_{_{[\tilde{1}\tilde{3}57]}}^{(5),b}(\alpha,{\bf n})
\nonumber\\
&&\hspace{2.5cm}\times
\Big({y_{_1}\over y_{_4}}\Big)^{n_{_1}}\Big({y_{_3}\over y_{_4}}\Big)^{n_{_2}}
\Big({1\over y_{_4}}\Big)^{n_{_3}}\Big({y_{_2}\over y_{_4}}\Big)^{n_{_4}}
\;,\nonumber\\
&&\Phi_{_{[\tilde{1}\tilde{3}57]}}^{(5),c}(\alpha,z)=y_{_1}
y_{_2}^{{D\over2}-1}y_{_3}^{{D\over2}-1}y_{_4}^{-2}\sum\limits_{n_{_1}=0}^\infty
\sum\limits_{n_{_2}=0}^\infty\sum\limits_{n_{_3}=0}^\infty\sum\limits_{n_{_4}=0}^\infty
c_{_{[\tilde{1}\tilde{3}57]}}^{(5),c}(\alpha,{\bf n})
\nonumber\\
&&\hspace{2.5cm}\times
y_{_1}^{n_{_1}}\Big({y_{_3}\over y_{_4}}\Big)^{n_{_2}}
\Big({y_{_1}\over y_{_4}}\Big)^{n_{_3}}\Big({y_{_2}\over y_{_4}}\Big)^{n_{_4}}\;.
\label{GKZ21e-9-2a}
\end{eqnarray}
Where the coefficients are
\begin{eqnarray}
&&c_{_{[\tilde{1}\tilde{3}57]}}^{(5),a}(\alpha,{\bf n})=
(-)^{n_{_1}}\Gamma(1+n_{_1}+n_{_3})\Gamma(1+n_{_2}+n_{_4})
\Big\{n_{_1}!n_{_2}!n_{_3}!n_{_4}!
\nonumber\\
&&\hspace{2.5cm}\times
\Gamma({D\over2}-1-n_{_1}-n_{_3})\Gamma(1-{D\over2}-n_{_2}-n_{_4})
\Gamma(2-{D\over2}+n_{_1})
\nonumber\\
&&\hspace{2.5cm}\times
\Gamma({D\over2}+n_{_3})\Gamma({D\over2}+n_{_4})\Gamma({D\over2}+n_{_2})\Big\}^{-1}
\;,\nonumber\\
&&c_{_{[\tilde{1}\tilde{3}57]}}^{(5),b}(\alpha,{\bf n})=
-\Gamma(1+n_{_1}+n_{_3})\Gamma(1+n_{_2}+n_{_4})
\Big\{n_{_1}!n_{_2}!n_{_3}!n_{_4}!\Gamma({D\over2}+n_{_3})
\nonumber\\
&&\hspace{2.5cm}\times
\Gamma({D\over2}-1-n_{_1}-n_{_3})\Gamma(1-{D\over2}-n_{_2}-n_{_4})
\Gamma(2-{D\over2}+n_{_1})
\nonumber\\
&&\hspace{2.5cm}\times
\Gamma({D\over2}+n_{_4})\Gamma({D\over2}+n_{_2})\Big\}^{-1}
\;,\nonumber\\
&&c_{_{[\tilde{1}\tilde{3}57]}}^{(5),c}(\alpha,{\bf n})=
(-)^{1+n_{_1}}\Gamma(1+n_{_1})\Gamma(1+n_{_3})\Gamma(1+n_{_2}+n_{_4})
\Big\{n_{_2}!n_{_4}!
\nonumber\\
&&\hspace{2.5cm}\times
\Gamma(2+n_{_1}+n_{_3})\Gamma({D\over2}-1-n_{_1})
\Gamma(3-{D\over2}+n_{_1}+n_{_3})
\nonumber\\
&&\hspace{2.5cm}\times
\Gamma(1-{D\over2}-n_{_2}-n_{_4})\Gamma({D\over2}-1-n_{_3})
\Gamma({D\over2}+n_{_4})\Gamma({D\over2}+n_{_2})\Big\}^{-1}\;.
\label{GKZ21e-9-3}
\end{eqnarray}

\item   $I_{_{6}}=\{1,\cdots,4,6,7,9,10,13,14\}$, i.e. the implement $J_{_{6}}=[1,14]\setminus I_{_{6}}=\{5,8,11,12\}$.
The choice implies the power numbers $\alpha_{_5}=\alpha_{_{8}}=\alpha_{_{11}}=\alpha_{_{12}}=0$, and
\begin{eqnarray}
&&\alpha_{_1}=a_{_5}+b_{_3}-a_{_1}-1,\;\alpha_{_2}=a_{_5}+b_{_3}-a_{_2}-1,
\nonumber\\
&&\alpha_{_3}=b_{_3}-a_{_3}-1,\;\alpha_{_4}=b_{_3}-a_{_4}-1,\;\alpha_{_{6}}=b_{_1}-1,
\nonumber\\
&&\alpha_{_{7}}=b_{_2}-1,\;\alpha_{_9}=b_{_4}-a_{_5}-b_{_3},\;\alpha_{_{10}}=b_{_5}-a_{_5}-b_{_3},
\nonumber\\
&&\alpha_{_{13}}=1-b_{_3},\;\alpha_{_{14}}=-a_{_5}\;.
\label{GKZ21e-10-1}
\end{eqnarray}
The corresponding hypergeometric solutions are written as
\begin{eqnarray}
&&\Phi_{_{[\tilde{1}\tilde{3}57]}}^{(6),a}(\alpha,z)=
y_{_3}^{{D\over2}-1}y_{_4}^{-1}\sum\limits_{n_{_1}=0}^\infty
\sum\limits_{n_{_2}=0}^\infty\sum\limits_{n_{_3}=0}^\infty\sum\limits_{n_{_4}=0}^\infty
c_{_{[\tilde{1}\tilde{3}57]}}^{(6),a}(\alpha,{\bf n})
\nonumber\\
&&\hspace{2.5cm}\times
y_{_1}^{n_{_1}}\Big({y_{_3}\over y_{_4}}\Big)^{n_{_2}}
y_{_4}^{n_{_3}}\Big({y_{_2}\over y_{_4}}\Big)^{n_{_4}}
\;,\nonumber\\
&&\Phi_{_{[\tilde{1}\tilde{3}57]}}^{(6),b}(\alpha,z)=
y_{_3}^{{D\over2}-1}y_{_4}^{-2}\sum\limits_{n_{_1}=0}^\infty
\sum\limits_{n_{_2}=0}^\infty\sum\limits_{n_{_3}=0}^\infty\sum\limits_{n_{_4}=0}^\infty
c_{_{[\tilde{1}\tilde{3}57]}}^{(6),b}(\alpha,{\bf n})
\nonumber\\
&&\hspace{2.5cm}\times
y_{_1}^{n_{_1}}\Big({y_{_3}\over y_{_4}}\Big)^{n_{_2}}
\Big({1\over y_{_4}}\Big)^{n_{_3}}\Big({y_{_2}\over y_{_4}}\Big)^{n_{_4}}\;.
\label{GKZ21e-10-2a}
\end{eqnarray}
Where the coefficients are
\begin{eqnarray}
&&c_{_{[\tilde{1}\tilde{3}57]}}^{(6),a}(\alpha,{\bf n})=
(-)^{n_{_3}}\Gamma(1+n_{_2}+n_{_4})\Big\{n_{_1}!n_{_2}!n_{_3}!n_{_4}!
\Gamma({D\over2}-1-n_{_1}-n_{_3})
\nonumber\\
&&\hspace{2.5cm}\times
\Gamma(D-2-n_{_1}-n_{_3})\Gamma({D\over2}-1-n_{_2}-n_{_4})\Gamma(2-{D\over2}+n_{_1})
\nonumber\\
&&\hspace{2.5cm}\times
\Gamma(2-{D\over2}+n_{_4})
\Gamma(2-{D\over2}+n_{_3})\Gamma({D\over2}+n_{_2})\Big\}^{-1}
\;,\nonumber\\
&&c_{_{[\tilde{1}\tilde{3}57]}}^{(6),b}(\alpha,{\bf n})=
-\Gamma(1+n_{_3})\Gamma(1+n_{_2}+n_{_4})\Big\{n_{_1}!n_{_2}!n_{_4}!
\Gamma({D\over2}-n_{_1}+n_{_3})
\nonumber\\
&&\hspace{2.5cm}\times
\Gamma(D-1-n_{_1}+n_{_3})\Gamma({D\over2}-1-n_{_2}-n_{_4})\Gamma(2-{D\over2}+n_{_1})
\nonumber\\
&&\hspace{2.5cm}\times
\Gamma(2-{D\over2}+n_{_4})
\Gamma(1-{D\over2}-n_{_3})\Gamma({D\over2}+n_{_2})\Big\}^{-1}\;.
\label{GKZ21e-10-3}
\end{eqnarray}

\item   $I_{_{7}}=\{1,\cdots,4,6,8,9,10,12,14\}$, i.e. the implement $J_{_{7}}=[1,14]\setminus I_{_{7}}=\{5,7,11,13\}$.
The choice implies the power numbers $\alpha_{_5}=\alpha_{_{7}}=\alpha_{_{11}}=\alpha_{_{13}}=0$, and
\begin{eqnarray}
&&\alpha_{_1}=a_{_5}+b_{_2}-a_{_1}-1,\;\alpha_{_2}=a_{_5}+b_{_2}-a_{_2}-1,
\nonumber\\
&&\alpha_{_3}=b_{_2}-a_{_3}-1,\;\alpha_{_4}=b_{_2}-a_{_4}-1,\;\alpha_{_{6}}=b_{_1}-1,
\nonumber\\
&&\alpha_{_{8}}=b_{_3}-1,\;\alpha_{_9}=b_{_4}-a_{_5}-b_{_2},\;\alpha_{_{10}}=b_{_5}-a_{_5}-b_{_2},
\nonumber\\
&&\alpha_{_{12}}=1-b_{_2},\;\alpha_{_{14}}=-a_{_5}\;.
\label{GKZ21e-15-1}
\end{eqnarray}
The corresponding hypergeometric series solutions are
\begin{eqnarray}
&&\Phi_{_{[\tilde{1}\tilde{3}57]}}^{(7),a}(\alpha,z)=
y_{_2}^{{D\over2}-1}y_{_4}^{-1}\sum\limits_{n_{_1}=0}^\infty
\sum\limits_{n_{_2}=0}^\infty\sum\limits_{n_{_3}=0}^\infty\sum\limits_{n_{_4}=0}^\infty
c_{_{[\tilde{1}\tilde{3}57]}}^{(7),a}(\alpha,{\bf n})
\nonumber\\
&&\hspace{2.5cm}\times
y_{_1}^{n_{_1}}\Big({y_{_3}\over y_{_4}}\Big)^{n_{_2}}
y_{_4}^{n_{_3}}\Big({y_{_2}\over y_{_4}}\Big)^{n_{_4}}
\;,\nonumber\\
&&\Phi_{_{[\tilde{1}\tilde{3}57]}}^{(7),b}(\alpha,z)=
y_{_2}^{{D\over2}-1}y_{_4}^{-2}\sum\limits_{n_{_1}=0}^\infty
\sum\limits_{n_{_2}=0}^\infty\sum\limits_{n_{_3}=0}^\infty\sum\limits_{n_{_4}=0}^\infty
c_{_{[\tilde{1}\tilde{3}57]}}^{(7),b}(\alpha,{\bf n})
\nonumber\\
&&\hspace{2.5cm}\times
y_{_1}^{n_{_1}}\Big({y_{_3}\over y_{_4}}\Big)^{n_{_2}}
\Big({1\over y_{_4}}\Big)^{n_{_3}}\Big({y_{_2}\over y_{_4}}\Big)^{n_{_4}}\;.
\label{GKZ21e-15-2a}
\end{eqnarray}
Where the coefficients are
\begin{eqnarray}
&&c_{_{[\tilde{1}\tilde{3}57]}}^{(7),a}(\alpha,{\bf n})=(-)^{n_{_3}}
\Gamma(1+n_{_2}+n_{_4})\Big\{n_{_1}!n_{_2}!n_{_3}!n_{_4}!\Gamma({D\over2}-1-n_{_1}-n_{_3})
\nonumber\\
&&\hspace{2.5cm}\times
\Gamma(D-2-n_{_1}-n_{_3})\Gamma({D\over2}-1-n_{_2}-n_{_4})
\Gamma(2-{D\over2}+n_{_1})
\nonumber\\
&&\hspace{2.5cm}\times
\Gamma(2-{D\over2}+n_{_2})
\Gamma(2-{D\over2}+n_{_3})\Gamma({D\over2}+n_{_4})\Big\}^{-1}
\;,\nonumber\\
&&c_{_{[\tilde{1}\tilde{3}57]}}^{(7),b}(\alpha,{\bf n})=
-\Gamma(1+n_{_3})\Gamma(1+n_{_2}+n_{_4})
\Big\{n_{_1}!n_{_2}!n_{_4}!\Gamma({D\over2}-n_{_1}+n_{_3})
\nonumber\\
&&\hspace{2.5cm}\times
\Gamma(D-1-n_{_1}+n_{_3})\Gamma({D\over2}-1-n_{_2}-n_{_4})
\Gamma(2-{D\over2}+n_{_1})
\nonumber\\
&&\hspace{2.5cm}\times
\Gamma(2-{D\over2}+n_{_2})
\Gamma(1-{D\over2}-n_{_3})\Gamma({D\over2}+n_{_4})\Big\}^{-1}\;.
\label{GKZ21e-15-3}
\end{eqnarray}

\item   $I_{_{8}}=\{1,\cdots,4,6\cdots,10,14\}$, i.e. the implement $J_{_{8}}=[1,14]\setminus I_{_{8}}=\{5,11,12,13\}$.
The choice implies the power numbers $\alpha_{_5}=\alpha_{_{11}}=\alpha_{_{12}}=\alpha_{_{13}}=0$, and
\begin{eqnarray}
&&\alpha_{_1}=a_{_5}-a_{_1},\;\alpha_{_2}=a_{_5}-a_{_2},\;\alpha_{_3}=-a_{_3},\;\alpha_{_4}=-a_{_4},
\nonumber\\
&&\alpha_{_{6}}=b_{_1}-1,\;\alpha_{_{7}}=b_{_2}-1,\;\alpha_{_{8}}=b_{_3}-1,
\nonumber\\
&&\alpha_{_9}=b_{_4}-a_{_5}-1,\;\alpha_{_{10}}=b_{_5}-a_{_5}-1,\;\alpha_{_{14}}=-a_{_5}\;.
\label{GKZ21e-16-1}
\end{eqnarray}
The corresponding hypergeometric series is written as
\begin{eqnarray}
&&\Phi_{_{[\tilde{1}\tilde{3}57]}}^{(8)}(\alpha,z)=
y_{_4}^{-1}\sum\limits_{n_{_1}=0}^\infty
\sum\limits_{n_{_2}=0}^\infty\sum\limits_{n_{_3}=0}^\infty\sum\limits_{n_{_4}=0}^\infty
c_{_{[\tilde{1}\tilde{3}57]}}^{(8)}(\alpha,{\bf n})
\nonumber\\
&&\hspace{2.5cm}\times
y_{_1}^{n_{_1}}y_{_3}^{n_{_2}}\Big({1\over y_{_4}}\Big)^{n_{_3}}y_{_2}^{n_{_4}}\;,
\label{GKZ21e-16-2}
\end{eqnarray}
with
\begin{eqnarray}
&&c_{_{[\tilde{1}\tilde{3}57]}}^{(8)}(\alpha,{\bf n})=
(-)^{n_{_3}}\Big\{n_{_1}!n_{_2}!n_{_4}!\Gamma(D-2-n_{_1}-n_{_2}+n_{_3}-n_{_4})
\nonumber\\
&&\hspace{2.5cm}\times
\Gamma({3D\over2}-3-n_{_1}-n_{_2}+n_{_3}-n_{_4})
\Gamma({D\over2}-1-n_{_2}-n_{_4})
\nonumber\\
&&\hspace{2.5cm}\times
\Gamma(D-2-n_{_2}-n_{_4})\Gamma(2-{D\over2}+n_{_1})
\Gamma(2-{D\over2}+n_{_4})
\nonumber\\
&&\hspace{2.5cm}\times
\Gamma(2-{D\over2}+n_{_2})\Gamma(2-{D\over2}+n_{_2}-n_{_3}+n_{_4})
\nonumber\\
&&\hspace{2.5cm}\times
\Gamma(3-D+n_{_2}-n_{_3}+n_{_4})\Big\}^{-1}\;.
\label{GKZ21e-16-3}
\end{eqnarray}

\item   $I_{_{9}}=\{1,2,3,4,8,10,\cdots,14\}$, i.e. the implement $J_{_{9}}=[1,14]\setminus I_{_{9}}=\{5,6,7,9\}$.
The choice implies the power numbers $\alpha_{_5}=\alpha_{_{6}}=\alpha_{_{7}}=\alpha_{_{9}}=0$, and
\begin{eqnarray}
&&\alpha_{_1}=b_{_1}+b_{_4}-a_{_1}-2,\;\alpha_{_2}=b_{_1}+b_{_4}-a_{_2}-2,
\nonumber\\
&&\alpha_{_3}=b_{_4}-a_{_3}-a_{_5}-1,\;\alpha_{_4}=b_{_4}-a_{_4}-a_{_5}-1,
\nonumber\\
&&\alpha_{_{8}}=a_{_5}+b_{_2}+b_{_3}-b_{_4}-1,\;\alpha_{_{10}}=b_{_5}-b_{_4},\;\alpha_{_{11}}=1-b_{_1},
\nonumber\\
&&\alpha_{_{12}}=1-b_{_2},\;\alpha_{_{13}}=a_{_5}+b_{_2}-b_{_4},\;
\alpha_{_{14}}=-a_{_5}\;.
\label{GKZ21e-3-1}
\end{eqnarray}
The corresponding hypergeometric functions are written as
\begin{eqnarray}
&&\Phi_{_{[\tilde{1}\tilde{3}57]}}^{(9),a}(\alpha,z)=
y_{_1}^{{D\over2}-1}y_{_2}^{{D\over2}-1}y_{_4}^{-1}\sum\limits_{n_{_1}=0}^\infty
\sum\limits_{n_{_2}=0}^\infty\sum\limits_{n_{_3}=0}^\infty\sum\limits_{n_{_4}=0}^\infty
c_{_{[\tilde{1}\tilde{3}57]}}^{(9),a}(\alpha,{\bf n})
\nonumber\\
&&\hspace{2.5cm}\times
y_{_1}^{n_{_1}}y_{_4}^{n_{_2}}
\Big({y_{_3}\over y_{_4}}\Big)^{n_{_3}}\Big({y_{_2}\over y_{_4}}\Big)^{n_{_4}}
\;,\nonumber\\
&&\Phi_{_{[\tilde{1}\tilde{3}57]}}^{(9),b}(\alpha,z)=
y_{_1}^{{D\over2}-1}y_{_2}^{{D\over2}}y_{_3}^{-1}y_{_4}^{-1}\sum\limits_{n_{_1}=0}^\infty
\sum\limits_{n_{_2}=0}^\infty\sum\limits_{n_{_3}=0}^\infty\sum\limits_{n_{_4}=0}^\infty
c_{_{[\tilde{1}\tilde{3}57]}}^{(9),b}(\alpha,{\bf n})
\nonumber\\
&&\hspace{2.5cm}\times
y_{_1}^{n_{_1}}y_{_2}^{n_{_2}}
\Big({y_{_2}\over y_{_4}}\Big)^{n_{_3}}\Big({y_{_2}\over y_{_3}}\Big)^{n_{_4}}\;.
\label{GKZ21e-3-2a}
\end{eqnarray}
Where the coefficients are
\begin{eqnarray}
&&c_{_{[\tilde{1}\tilde{3}57]}}^{(9),a}(\alpha,{\bf n})=
(-)^{n_{_1}}\Gamma(1+n_{_1}+n_{_2})\Gamma(1+n_{_3}+n_{_4})
\Big\{n_{_1}!n_{_2}!n_{_3}!n_{_4}!
\nonumber\\
&&\hspace{2.5cm}\times
\Gamma({D\over2}-1-n_{_1}-n_{_2})\Gamma({D\over2}-1-n_{_3}-n_{_4})
\Gamma(2-{D\over2}+n_{_3})
\nonumber\\
&&\hspace{2.5cm}\times
\Gamma(2-{D\over2}+n_{_2})
\Gamma({D\over2}+n_{_1})\Gamma({D\over2}+n_{_4})\Big\}^{-1}
\;,\nonumber\\
&&c_{_{[\tilde{1}\tilde{3}57]}}^{(9),b}(\alpha,{\bf n})=
(-)^{n_{_1}+n_{_4}}\Gamma(1+n_{_1}+n_{_2})\Gamma(1+n_{_2}+n_{_3})\Gamma(1+n_{_4})
\Big\{n_{_1}!n_{_2}!
\nonumber\\
&&\hspace{2.5cm}\times
\Gamma(2+n_{_2}+n_{_3}+n_{_4})\Gamma({D\over2}-1-n_{_1}-n_{_2})
\nonumber\\
&&\hspace{2.5cm}\times
\Gamma({D\over2}-1-n_{_2}-n_{_3})\Gamma(1-{D\over2}-n_{_4})\Gamma(2-{D\over2}+n_{_2})
\nonumber\\
&&\hspace{2.5cm}\times
\Gamma({D\over2}+n_{_1})\Gamma({D\over2}+1+n_{_2}+n_{_3}+n_{_4})\Big\}^{-1}\;.
\label{GKZ21e-3-3}
\end{eqnarray}

\item   $I_{_{10}}=\{1,2,3,4,7,8,10,11,13,14\}$, i.e. the implement $J_{_{10}}=[1,14]\setminus I_{_{10}}=\{5,6,9,12\}$.
The choice implies the power numbers $\alpha_{_5}=\alpha_{_{6}}=\alpha_{_{9}}=\alpha_{_{12}}=0$, and
\begin{eqnarray}
&&\alpha_{_1}=b_{_1}+b_{_4}-a_{_1}-2,\;\alpha_{_2}=b_{_1}+b_{_4}-a_{_2}-2,
\nonumber\\
&&\alpha_{_3}=b_{_4}-a_{_3}-a_{_5}-1,\;\alpha_{_4}=b_{_4}-a_{_4}-a_{_5}-1,
\nonumber\\
&&\alpha_{_{7}}=b_{_2}-1,\;\alpha_{_{8}}=a_{_5}+b_{_3}-b_{_4},\;\alpha_{_{10}}=b_{_5}-b_{_4},
\nonumber\\
&&\alpha_{_{11}}=1-b_{_1},\;\alpha_{_{13}}=a_{_5}-b_{_4}+1,\;
\alpha_{_{14}}=-a_{_5}\;.
\label{GKZ21e-4-1}
\end{eqnarray}
The corresponding hypergeometric series solutions are written as
\begin{eqnarray}
&&\Phi_{_{[\tilde{1}\tilde{3}57]}}^{(10),a}(\alpha,z)=
y_{_1}^{{D\over2}-1}y_{_3}^{{D\over2}-1}y_{_4}^{-1}\sum\limits_{n_{_1}=0}^\infty
\sum\limits_{n_{_2}=0}^\infty\sum\limits_{n_{_3}=0}^\infty\sum\limits_{n_{_4}=0}^\infty
c_{_{[\tilde{1}\tilde{3}57]}}^{(10),a}(\alpha,{\bf n})
\nonumber\\
&&\hspace{2.5cm}\times
y_{_1}^{n_{_1}}y_{_4}^{n_{_2}}
\Big({y_{_3}\over y_{_4}}\Big)^{n_{_3}}\Big({y_{_2}\over y_{_4}}\Big)^{n_{_4}}
\;,\nonumber\\
&&\Phi_{_{[\tilde{1}\tilde{3}57]}}^{(10),b}(\alpha,z)=
y_{_1}^{{D\over2}-1}y_{_2}y_{_3}^{{D\over2}-2}y_{_4}^{-1}\sum\limits_{n_{_1}=0}^\infty
\sum\limits_{n_{_2}=0}^\infty\sum\limits_{n_{_3}=0}^\infty\sum\limits_{n_{_4}=0}^\infty
c_{_{[\tilde{1}\tilde{3}57]}}^{(10),b}(\alpha,{\bf n})
\nonumber\\
&&\hspace{2.5cm}\times
y_{_1}^{n_{_1}}y_{_2}^{n_{_2}}
\Big({y_{_2}\over y_{_4}}\Big)^{n_{_3}}\Big({y_{_2}\over y_{_3}}\Big)^{n_{_4}}\;.
\label{GKZ21e-4-2a}
\end{eqnarray}
Where the coefficients are
\begin{eqnarray}
&&c_{_{[\tilde{1}\tilde{3}57]}}^{(10),a}(\alpha,{\bf n})=(-)^{n_{_1}}
\Gamma(1+n_{_1}+n_{_2})\Gamma(1+n_{_3}+n_{_4})\Big\{n_{_1}!n_{_2}!n_{_3}!n_{_4}!
\nonumber\\
&&\hspace{2.5cm}\times
\Gamma({D\over2}-1-n_{_1}-n_{_2})\Gamma({D\over2}-1-n_{_3}-n_{_4})\Gamma(2-{D\over2}+n_{_4})
\nonumber\\
&&\hspace{2.5cm}\times
\Gamma(2-{D\over2}+n_{_2})\Gamma({D\over2}+n_{_1})\Gamma({D\over2}+n_{_3})\Big\}^{-1}
\;,\nonumber\\
&&c_{_{[\tilde{1}\tilde{3}57]}}^{(10),b}(\alpha,{\bf n})=(-)^{n_{_1}+n_{_4}}
\Gamma(1+n_{_1}+n_{_2})\Gamma(1+n_{_2}+n_{_3})\Gamma(1+n_{_4})\Big\{n_{_1}!n_{_2}!
\nonumber\\
&&\hspace{2.5cm}\times
\Gamma(2+n_{_2}+n_{_3}+n_{_4})\Gamma({D\over2}-1-n_{_1}-n_{_2})
\nonumber\\
&&\hspace{2.5cm}\times
\Gamma({D\over2}-1-n_{_2}-n_{_3})\Gamma(3-{D\over2}+n_{_2}+n_{_3}+n_{_4})
\nonumber\\
&&\hspace{2.5cm}\times
\Gamma(2-{D\over2}+n_{_2})\Gamma({D\over2}+n_{_1})\Gamma({D\over2}-1-n_{_4})\Big\}^{-1}\;.
\label{GKZ21e-4-3}
\end{eqnarray}

\item   $I_{_{11}}=\{1,2,3,4,8,9,11,\cdots,14\}$, i.e. the implement $J_{_{11}}=[1,14]\setminus I_{_{11}}=\{5,6,7,10\}$.
The choice implies the power numbers $\alpha_{_5}=\alpha_{_{6}}=\alpha_{_{7}}=\alpha_{_{10}}=0$, and
\begin{eqnarray}
&&\alpha_{_1}=b_{_1}+b_{_5}-a_{_1}-2,\;\alpha_{_2}=b_{_1}+b_{_5}-a_{_2}-2,
\nonumber\\
&&\alpha_{_3}=b_{_5}-a_{_3}-a_{_5}-1,\;\alpha_{_4}=b_{_5}-a_{_4}-a_{_5}-1,
\nonumber\\
&&\alpha_{_{8}}=a_{_5}+b_{_2}+b_{_3}-b_{_5}-1,\;\alpha_{_{9}}=b_{_4}-b_{_5},\;\alpha_{_{11}}=1-b_{_1},
\nonumber\\
&&\alpha_{_{12}}=1-b_{_2},\;\alpha_{_{13}}=a_{_5}+b_{_2}-b_{_5},\;
\alpha_{_{14}}=-a_{_5}\;.
\label{GKZ21e-5-1}
\end{eqnarray}
The corresponding hypergeometric functions are written as
\begin{eqnarray}
&&\Phi_{_{[\tilde{1}\tilde{3}57]}}^{(11),a}(\alpha,z)=
y_{_1}^{{D\over2}-1}y_{_2}^{{D\over2}-1}y_{_3}^{{D\over2}-1}y_{_4}^{-1}
\sum\limits_{n_{_1}=0}^\infty
\sum\limits_{n_{_2}=0}^\infty\sum\limits_{n_{_3}=0}^\infty\sum\limits_{n_{_4}=0}^\infty
c_{_{[\tilde{1}\tilde{3}57]}}^{(11),a}(\alpha,{\bf n})
\nonumber\\
&&\hspace{2.5cm}\times
y_{_1}^{n_{_1}}y_{_4}^{n_{_2}}
\Big({y_{_3}\over y_{_4}}\Big)^{n_{_3}}\Big({y_{_2}\over y_{_4}}\Big)^{n_{_4}}
\;,\nonumber\\
&&\Phi_{_{[\tilde{1}\tilde{3}57]}}^{(11),b}(\alpha,z)=
y_{_1}^{{D\over2}-1}y_{_2}^{{D\over2}}y_{_3}^{{D\over2}-2}y_{_4}^{-1}
\sum\limits_{n_{_1}=0}^\infty
\sum\limits_{n_{_2}=0}^\infty\sum\limits_{n_{_3}=0}^\infty\sum\limits_{n_{_4}=0}^\infty
c_{_{[\tilde{1}\tilde{3}57]}}^{(11),b}(\alpha,{\bf n})
\nonumber\\
&&\hspace{2.5cm}\times
y_{_1}^{n_{_1}}y_{_2}^{n_{_2}}
\Big({y_{_2}\over y_{_4}}\Big)^{n_{_3}}\Big({y_{_2}\over y_{_3}}\Big)^{n_{_4}}\;.
\label{GKZ21e-5-2a}
\end{eqnarray}
Where the coefficients are
\begin{eqnarray}
&&c_{_{[\tilde{1}\tilde{3}57]}}^{(11),a}(\alpha,{\bf n})=
(-)^{n_{_1}}\Gamma(1+n_{_1}+n_{_2})\Gamma(1+n_{_3}+n_{_4})
\Big\{n_{_1}!n_{_2}!n_{_3}!n_{_4}!
\nonumber\\
&&\hspace{2.5cm}\times
\Gamma(1-{D\over2}-n_{_1}-n_{_2})\Gamma(1-{D\over2}-n_{_3}-n_{_4})
\Gamma({D\over2}+n_{_2})
\nonumber\\
&&\hspace{2.5cm}\times
\Gamma({D\over2}+n_{_1})
\Gamma({D\over2}+n_{_4})\Gamma({D\over2}+n_{_3})\Big\}^{-1}
\;,\nonumber\\
&&c_{_{[\tilde{1}\tilde{3}57]}}^{(11),b}(\alpha,{\bf n})=
(-)^{n_{_1}+n_{_4}}\Gamma(1+n_{_1}+n_{_2})\Gamma(1+n_{_2}+n_{_3})\Gamma(1+n_{_4})
\Big\{n_{_1}!n_{_2}!
\nonumber\\
&&\hspace{2.5cm}\times
\Gamma(2+n_{_2}+n_{_3}+n_{_4})
\Gamma(1-{D\over2}-n_{_1}-n_{_2})
\nonumber\\
&&\hspace{2.5cm}\times
\Gamma(1-{D\over2}-n_{_2}-n_{_3})
\Gamma({D\over2}+n_{_2})\Gamma({D\over2}+n_{_1})
\nonumber\\
&&\hspace{2.5cm}\times
\Gamma({D\over2}+1+n_{_2}+n_{_3}+n_{_4})\Gamma({D\over2}-1-n_{_4})\Big\}^{-1}\;.
\label{GKZ21e-5-3}
\end{eqnarray}

\item   $I_{_{12}}=\{1,2,3,4,7,8,9,11,13,14\}$, i.e. the implement $J_{_{12}}=[1,14]\setminus I_{_{12}}=\{5,6,10,12\}$.
The choice implies the power numbers $\alpha_{_5}=\alpha_{_{6}}=\alpha_{_{10}}=\alpha_{_{12}}=0$, and
\begin{eqnarray}
&&\alpha_{_1}=b_{_1}+b_{_5}-a_{_1}-2,\;\alpha_{_2}=b_{_1}+b_{_5}-a_{_2}-2,
\nonumber\\
&&\alpha_{_3}=b_{_5}-a_{_3}-a_{_5}-1,\;\alpha_{_4}=b_{_5}-a_{_4}-a_{_5}-1,
\nonumber\\
&&\alpha_{_{7}}=b_{_2}-1,\;\alpha_{_{8}}=a_{_5}+b_{_3}-b_{_5},\;\alpha_{_{9}}=b_{_4}-b_{_5},
\nonumber\\
&&\alpha_{_{11}}=1-b_{_1},\;\alpha_{_{13}}=a_{_5}-b_{_5}+1,\;
\alpha_{_{14}}=-a_{_5}\;.
\label{GKZ21e-6-1}
\end{eqnarray}
The corresponding hypergeometric series is
\begin{eqnarray}
&&\Phi_{_{[\tilde{1}\tilde{3}57]}}^{(12)}(\alpha,z)=
y_{_1}^{{D\over2}-1}y_{_3}^{{D}-2}y_{_4}^{-1}\sum\limits_{n_{_1}=0}^\infty
\sum\limits_{n_{_2}=0}^\infty\sum\limits_{n_{_3}=0}^\infty\sum\limits_{n_{_4}=0}^\infty
c_{_{[\tilde{1}\tilde{3}57]}}^{(12)}(\alpha,{\bf n})
\nonumber\\
&&\hspace{2.5cm}\times
y_{_1}^{n_{_1}}y_{_3}^{n_{_2}}\Big({y_{_3}\over y_{_4}}\Big)^{n_{_3}}
\Big({y_{_2}\over y_{_3}}\Big)^{n_{_4}}\;,
\label{GKZ21e-6-2}
\end{eqnarray}
with
\begin{eqnarray}
&&c_{_{[\tilde{1}\tilde{3}57]}}^{(12)}(\alpha,{\bf n})=
(-)^{n_{_1}}\Gamma(1+n_{_1}+n_{_2})\Gamma(1+n_{_2}+n_{_3})
\nonumber\\
&&\hspace{2.5cm}\times
\Big\{n_{_1}!n_{_2}!n_{_4}!\Gamma(1-{D\over2}-n_{_1}-n_{_2})
\Gamma(1-{D\over2}-n_{_2}-n_{_3})
\nonumber\\
&&\hspace{2.5cm}\times
\Gamma(2-{D\over2}+n_{_4})\Gamma({D\over2}+n_{_2}+n_{_3}-n_{_4})
\Gamma({D\over2}+n_{_2})
\nonumber\\
&&\hspace{2.5cm}\times
\Gamma({D\over2}+n_{_1})\Gamma(D-1+n_{_2}+n_{_3}-n_{_4})\Big\}^{-1}\;.
\label{GKZ21e-6-3}
\end{eqnarray}

\item   $I_{_{13}}=\{1,2,3,4,6,8,10,12,13,14\}$, i.e. the implement $J_{_{13}}=[1,14]\setminus I_{_{13}}=\{5,7,9,11\}$.
The choice implies the power numbers $\alpha_{_5}=\alpha_{_{7}}=\alpha_{_{9}}=\alpha_{_{11}}=0$, and
\begin{eqnarray}
&&\alpha_{_1}=b_{_4}-a_{_1}-1,\;\alpha_{_2}=b_{_4}-a_{_2}-1,
\nonumber\\
&&\alpha_{_3}=b_{_4}-a_{_3}-a_{_5}-1,\;
\alpha_{_4}=b_{_4}-a_{_4}-a_{_5}-1,
\nonumber\\
&&\alpha_{_{6}}=b_{_1}-1,\;\alpha_{_{8}}=a_{_5}+b_{_2}+b_{_3}-b_{_4}-1,\;\alpha_{_{10}}=b_{_5}-b_{_4},
\nonumber\\
&&\alpha_{_{12}}=1-b_{_2},\;\alpha_{_{13}}=a_{_5}+b_{_2}-b_{_4},\;\alpha_{_{14}}=-a_{_5}\;.
\label{GKZ21e-11-1}
\end{eqnarray}
The corresponding hypergeometric functions are written as
\begin{eqnarray}
&&\Phi_{_{[\tilde{1}\tilde{3}57]}}^{(13),a}(\alpha,z)=
y_{_2}^{{D\over2}-1}y_{_4}^{-1}\sum\limits_{n_{_1}=0}^\infty
\sum\limits_{n_{_2}=0}^\infty\sum\limits_{n_{_3}=0}^\infty\sum\limits_{n_{_4}=0}^\infty
c_{_{[\tilde{1}\tilde{3}57]}}^{(13),a}(\alpha,{\bf n})
\nonumber\\
&&\hspace{2.5cm}\times
y_{_1}^{n_{_1}}y_{_4}^{n_{_2}}
\Big({y_{_3}\over y_{_4}}\Big)^{n_{_3}}\Big({y_{_2}\over y_{_4}}\Big)^{n_{_4}}
\;,\nonumber\\
&&\Phi_{_{[\tilde{1}\tilde{3}57]}}^{(13),b}(\alpha,z)=
y_{_2}^{{D\over2}}y_{_3}^{-1}y_{_4}^{-1}\sum\limits_{n_{_1}=0}^\infty
\sum\limits_{n_{_2}=0}^\infty\sum\limits_{n_{_3}=0}^\infty\sum\limits_{n_{_4}=0}^\infty
c_{_{[\tilde{1}\tilde{3}57]}}^{(13),b}(\alpha,{\bf n})
\nonumber\\
&&\hspace{2.5cm}\times
y_{_1}^{n_{_1}}y_{_2}^{n_{_2}}
\Big({y_{_2}\over y_{_4}}\Big)^{n_{_3}}\Big({y_{_2}\over y_{_3}}\Big)^{n_{_4}}\;.
\label{GKZ21e-11-2a}
\end{eqnarray}
Where the coefficients are
\begin{eqnarray}
&&c_{_{[\tilde{1}\tilde{3}57]}}^{(13),a}(\alpha,{\bf n})=
(-)^{n_{_2}}\Gamma(1+n_{_3}+n_{_4})\Big\{n_{_1}!n_{_2}!n_{_3}!n_{_4}!
\Gamma({D\over2}-1-n_{_1}-n_{_2})
\nonumber\\
&&\hspace{2.5cm}\times
\Gamma(D-2-n_{_1}-n_{_2})\Gamma({D\over2}-1-n_{_3}-n_{_4})
\Gamma(2-{D\over2}+n_{_1})
\nonumber\\
&&\hspace{2.5cm}\times
\Gamma(2-{D\over2}+n_{_3})\Gamma(2-{D\over2}+n_{_2})
\Gamma({D\over2}+n_{_4})\Big\}^{-1}
\;,\nonumber\\
&&c_{_{[\tilde{1}\tilde{3}57]}}^{(13),b}(\alpha,{\bf n})=
(-)^{n_{_2}+n_{_4}}\Gamma(1+n_{_2}+n_{_3})\Gamma(1+n_{_4})\Big\{n_{_1}!n_{_2}!
\Gamma(2+n_{_2}+n_{_3}+n_{_4})
\nonumber\\
&&\hspace{2.5cm}\times
\Gamma({D\over2}-1-n_{_1}-n_{_2})\Gamma(D-2-n_{_1}-n_{_2})\Gamma({D\over2}-1-n_{_2}-n_{_3})
\nonumber\\
&&\hspace{2.5cm}\times
\Gamma(2-{D\over2}+n_{_1})\Gamma(1-{D\over2}-n_{_4})\Gamma(2-{D\over2}+n_{_2})
\nonumber\\
&&\hspace{2.5cm}\times
\Gamma(1+{D\over2}+n_{_2}+n_{_3}+n_{_4})\Big\}^{-1}\;.
\label{GKZ21e-11-3}
\end{eqnarray}

\item   $I_{_{14}}=\{1,2,3,4,7,8,10,11,13,14\}$, i.e. the implement $J_{_{14}}=[1,14]\setminus I_{_{14}}=\{5,9,11,12\}$.
The choice implies the power numbers $\alpha_{_5}=\alpha_{_{9}}=\alpha_{_{11}}=\alpha_{_{12}}=0$, and
\begin{eqnarray}
&&\alpha_{_1}=b_{_4}-a_{_1}-1,\;\alpha_{_2}=b_{_4}-a_{_2}-1,
\nonumber\\
&&\alpha_{_3}=b_{_4}-a_{_3}-a_{_5}-1,\;\alpha_{_4}=b_{_4}-a_{_4}-a_{_5}-1,
\nonumber\\
&&\alpha_{_{6}}=b_{_1}-1,\;\alpha_{_{7}}=b_{_2}-1,\;\alpha_{_{8}}=a_{_5}+b_{_3}-b_{_4},
\nonumber\\
&&\alpha_{_{10}}=b_{_5}-b_{_4},\;\alpha_{_{13}}=a_{_5}-b_{_4}+1,\;\alpha_{_{14}}=-a_{_5}\;.
\label{GKZ21e-12-1}
\end{eqnarray}
The corresponding hypergeometric functions are
\begin{eqnarray}
&&\Phi_{_{[\tilde{1}\tilde{3}57]}}^{(14),a}(\alpha,z)=
y_{_3}^{{D\over2}-1}y_{_4}^{-1}\sum\limits_{n_{_1}=0}^\infty
\sum\limits_{n_{_2}=0}^\infty\sum\limits_{n_{_3}=0}^\infty\sum\limits_{n_{_4}=0}^\infty
c_{_{[\tilde{1}\tilde{3}57]}}^{(14),a}(\alpha,{\bf n})
\nonumber\\
&&\hspace{2.5cm}\times
y_{_1}^{n_{_1}}y_{_4}^{n_{_2}}
\Big({y_{_3}\over y_{_4}}\Big)^{n_{_3}}\Big({y_{_2}\over y_{_4}}\Big)^{n_{_4}}
\;,\nonumber\\
&&\Phi_{_{[\tilde{1}\tilde{3}57]}}^{(14),b}(\alpha,z)=
y_{_2}y_{_3}^{{D\over2}-2}y_{_4}^{-1}\sum\limits_{n_{_1}=0}^\infty
\sum\limits_{n_{_2}=0}^\infty\sum\limits_{n_{_3}=0}^\infty\sum\limits_{n_{_4}=0}^\infty
c_{_{[\tilde{1}\tilde{3}57]}}^{(14),b}(\alpha,{\bf n})
\nonumber\\
&&\hspace{2.5cm}\times
y_{_1}^{n_{_1}}y_{_2}^{n_{_2}}
\Big({y_{_2}\over y_{_4}}\Big)^{n_{_3}}\Big({y_{_2}\over y_{_3}}\Big)^{n_{_4}}\;.
\label{GKZ21e-12-2a}
\end{eqnarray}
Where the coefficients are
\begin{eqnarray}
&&c_{_{[\tilde{1}\tilde{3}57]}}^{(14),a}(\alpha,{\bf n})=
(-)^{n_{_2}}\Gamma(1+n_{_3}+n_{_4})\Big\{n_{_1}!n_{_2}!n_{_3}!n_{_4}!
\Gamma({D\over2}-1-n_{_1}-n_{_2})
\nonumber\\
&&\hspace{2.5cm}\times
\Gamma(D-2-n_{_1}-n_{_2})\Gamma({D\over2}-1-n_{_3}-n_{_4})
\Gamma(2-{D\over2}+n_{_1})
\nonumber\\
&&\hspace{2.5cm}\times
\Gamma(2-{D\over2}+n_{_4})
\Gamma(2-{D\over2}+n_{_2})\Gamma({D\over2}+n_{_3})\Big\}^{-1}
\;,\nonumber\\
&&c_{_{[\tilde{1}\tilde{3}57]}}^{(14),b}(\alpha,{\bf n})=
(-)^{n_{_2}+n_{_4}}\Gamma(1+n_{_2}+n_{_3})\Gamma(1+n_{_4})
\Big\{n_{_1}!n_{_2}!\Gamma(2+n_{_2}+n_{_3}+n_{_4})
\nonumber\\
&&\hspace{2.5cm}\times
\Gamma({D\over2}-1-n_{_1}-n_{_2})\Gamma(D-2-n_{_1}-n_{_2})
\Gamma({D\over2}-1-n_{_2}-n_{_3})
\nonumber\\
&&\hspace{2.5cm}\times
\Gamma(2-{D\over2}+n_{_1})\Gamma(3-{D\over2}+n_{_2}+n_{_3}+n_{_4})
\Gamma(2-{D\over2}+n_{_2})
\nonumber\\
&&\hspace{2.5cm}\times
\Gamma({D\over2}-1-n_{_4})\Big\}^{-1}\;.
\label{GKZ21e-12-3}
\end{eqnarray}

\item   $I_{_{15}}=\{1,2,3,4,6,8,9,12,13,14\}$, i.e. the implement $J_{_{15}}=[1,14]\setminus I_{_{15}}=\{5,7,10,11\}$.
The choice implies the power numbers $\alpha_{_5}=\alpha_{_{7}}=\alpha_{_{10}}=\alpha_{_{11}}=0$, and
\begin{eqnarray}
&&\alpha_{_1}=b_{_5}-a_{_1}-1,\;\alpha_{_2}=b_{_5}-a_{_2}-1,
\nonumber\\
&&\alpha_{_3}=b_{_5}-a_{_3}-a_{_5}-1,\;\alpha_{_4}=b_{_5}-a_{_4}-a_{_5}-1,
\nonumber\\
&&\alpha_{_{6}}=b_{_1}-1,\;\alpha_{_{8}}=a_{_5}+b_{_2}+b_{_3}-b_{_5}-1,\;\alpha_{_{9}}=b_{_4}-b_{_5},
\nonumber\\
&&\alpha_{_{12}}=1-b_{_2},\;\alpha_{_{13}}=a_{_5}+b_{_2}-b_{_5},\;
\alpha_{_{14}}=-a_{_5}\;.
\label{GKZ21e-13-1}
\end{eqnarray}
The corresponding hypergeometric functions are given as
\begin{eqnarray}
&&\Phi_{_{[\tilde{1}\tilde{3}57]}}^{(15),a}(\alpha,z)=
y_{_2}^{{D\over2}-1}y_{_3}^{{D\over2}-1}y_{_4}^{-1}\sum\limits_{n_{_1}=0}^\infty
\sum\limits_{n_{_2}=0}^\infty\sum\limits_{n_{_3}=0}^\infty\sum\limits_{n_{_4}=0}^\infty
c_{_{[\tilde{1}\tilde{3}57]}}^{(15),a}(\alpha,{\bf n})
\nonumber\\
&&\hspace{2.5cm}\times
y_{_1}^{n_{_1}}y_{_4}^{n_{_2}}
\Big({y_{_3}\over y_{_4}}\Big)^{n_{_3}}\Big({y_{_2}\over y_{_4}}\Big)^{n_{_4}}
\;,\nonumber\\
&&\Phi_{_{[\tilde{1}\tilde{3}57]}}^{(15),b}(\alpha,z)=
y_{_2}^{{D\over2}}y_{_3}^{{D\over2}-2}y_{_4}^{-1}\sum\limits_{n_{_1}=0}^\infty
\sum\limits_{n_{_2}=0}^\infty\sum\limits_{n_{_3}=0}^\infty\sum\limits_{n_{_4}=0}^\infty
c_{_{[\tilde{1}\tilde{3}57]}}^{(15),b}(\alpha,{\bf n})
\nonumber\\
&&\hspace{2.5cm}\times
y_{_1}^{n_{_1}}y_{_2}^{n_{_2}}
\Big({y_{_2}\over y_{_4}}\Big)^{n_{_3}}\Big({y_{_2}\over y_{_3}}\Big)^{n_{_4}}\;.
\label{GKZ21e-13-2a}
\end{eqnarray}
Where the coefficients are
\begin{eqnarray}
&&c_{_{[\tilde{1}\tilde{3}57]}}^{(15),a}(\alpha,{\bf n})=
(-)^{n_{_1}}\Gamma(1+n_{_1}+n_{_2})\Gamma(1+n_{_3}+n_{_4})
\nonumber\\
&&\hspace{2.5cm}\times
\Big\{n_{_1}!n_{_2}!n_{_3}!n_{_4}!
\Gamma({D\over2}-1-n_{_1}-n_{_2})\Gamma(1-{D\over2}-n_{_3}-n_{_4})
\nonumber\\
&&\hspace{2.5cm}\times
\Gamma(2-{D\over2}+n_{_1})\Gamma({D\over2}+n_{_2})
\Gamma({D\over2}+n_{_4})\Gamma({D\over2}+n_{_3})\Big\}^{-1}
\;,\nonumber\\
&&c_{_{[\tilde{1}\tilde{3}57]}}^{(15),b}(\alpha,{\bf n})=
(-)^{n_{_1}+n_{_4}}\Gamma(1+n_{_1}+n_{_2})\Gamma(1+n_{_2}+n_{_3})\Gamma(1+n_{_4})
\Big\{n_{_1}!n_{_2}!
\nonumber\\
&&\hspace{2.5cm}\times
\Gamma(2+n_{_2}+n_{_3}+n_{_4})\Gamma({D\over2}-1-n_{_1}-n_{_2})
\nonumber\\
&&\hspace{2.5cm}\times
\Gamma(1-{D\over2}-n_{_2}-n_{_3})\Gamma(2-{D\over2}+n_{_1})\Gamma({D\over2}+n_{_2})
\nonumber\\
&&\hspace{2.5cm}\times
\Gamma({D\over2}+1+n_{_2}+n_{_3}+n_{_4})\Gamma({D\over2}-1-n_{_4})\Big\}^{-1}\;.
\label{GKZ21e-13-3}
\end{eqnarray}

\item   $I_{_{16}}=\{1,\cdots,4,6,\cdots,9,13,14\}$, i.e. the implement $J_{_{16}}=[1,14]\setminus I_{_{16}}=\{5,10,11,12\}$.
The choice implies the power numbers $\alpha_{_5}=\alpha_{_{10}}=\alpha_{_{11}}=\alpha_{_{12}}=0$, and
\begin{eqnarray}
&&\alpha_{_1}=b_{_5}-a_{_1}-1,\;\alpha_{_2}=b_{_5}-a_{_2}-1,
\nonumber\\
&&\alpha_{_3}=b_{_5}-a_{_3}-a_{_5}-1,\;\alpha_{_4}=b_{_5}-a_{_4}-a_{_5}-1,
\nonumber\\
&&\alpha_{_{6}}=b_{_1}-1,\;\alpha_{_{7}}=b_{_2}-1,\;\alpha_{_{8}}=a_{_5}+b_{_3}-b_{_5},
\nonumber\\
&&\alpha_{_{9}}=b_{_4}-b_{_5},\;\alpha_{_{13}}=a_{_5}-b_{_5}+1,\;\alpha_{_{14}}=-a_{_5}\;.
\label{GKZ21e-14-1}
\end{eqnarray}
The corresponding hypergeometric series is
\begin{eqnarray}
&&\Phi_{_{[\tilde{1}\tilde{3}57]}}^{(16)}(\alpha,z)=
y_{_3}^{{D}-2}y_{_4}^{-1}\sum\limits_{n_{_1}=0}^\infty
\sum\limits_{n_{_2}=0}^\infty\sum\limits_{n_{_3}=0}^\infty\sum\limits_{n_{_4}=0}^\infty
c_{_{[\tilde{1}\tilde{3}57]}}^{(16)}(\alpha,{\bf n})
\nonumber\\
&&\hspace{2.5cm}\times
y_{_1}^{n_{_1}}y_{_3}^{n_{_2}}\Big({y_{_3}\over y_{_4}}\Big)^{n_{_3}}
\Big({y_{_2}\over y_{_3}}\Big)^{n_{_4}}\;,
\label{GKZ21e-14-2}
\end{eqnarray}
with
\begin{eqnarray}
&&c_{_{[\tilde{1}\tilde{3}57]}}^{(16)}(\alpha,{\bf n})=
(-)^{n_{_1}}\Gamma(1+n_{_1}+n_{_2})\Gamma(1+n_{_2}+n_{_3})
\Big\{n_{_1}!n_{_2}!n_{_4}!
\nonumber\\
&&\hspace{2.5cm}\times
\Gamma({D\over2}-1-n_{_1}-n_{_2})\Gamma(1-{D\over2}-n_{_2}-n_{_3})
\Gamma(2-{D\over2}+n_{_1})
\nonumber\\
&&\hspace{2.5cm}\times
\Gamma(2-{D\over2}+n_{_4})
\Gamma({D\over2}+n_{_2}+n_{_3}-n_{_4})\Gamma({D\over2}+n_{_2})
\nonumber\\
&&\hspace{2.5cm}\times
\Gamma(D-1+n_{_2}+n_{_3}-n_{_4})\Big\}^{-1}\;.
\label{GKZ21e-14-3}
\end{eqnarray}
\end{itemize}

\section{The hypergeometric solutions of the integer lattice ${\bf B}_{_{\tilde{1}3\tilde{5}7}}$\label{app7}}
\indent\indent

\begin{itemize}
\item   $I_{_{1}}=\{1,2,4,5,8,10,\cdots,14\}$, i.e. the implement $J_{_{1}}=[1,14]\setminus I_{_{1}}=\{3,6,7,9\}$.
The choice implies the power numbers $\alpha_{_3}=\alpha_{_{6}}=\alpha_{_{7}}=\alpha_{_{9}}=0$, and
\begin{eqnarray}
&&\alpha_{_1}=b_{_1}+b_{_4}-a_{_1}-2,\;\alpha_{_2}=b_{_1}+b_{_4}-a_{_2}-2,
\nonumber\\
&&\alpha_{_4}=a_{_3}-a_{_4},\;\alpha_{_5}=b_{_4}-a_{_3}-a_{_5}-1,
\nonumber\\
&&\alpha_{_8}=b_{_2}+b_{_3}-a_{_3}-2,\;
\alpha_{_{10}}=b_{_5}-b_{_4},\;\alpha_{_{11}}=1-b_{_1},
\nonumber\\
&&\alpha_{_{12}}=1-b_{_2},\;\alpha_{_{13}}=b_{_2}-a_{_3}-1,\;\alpha_{_{14}}=a_{_3}-b_{_4}+1.
\label{GKZ21f-1-1}
\end{eqnarray}
The corresponding hypergeometric series is written as
\begin{eqnarray}
&&\Phi_{_{[\tilde{1}3\tilde{5}7]}}^{(1)}(\alpha,z)=
y_{_1}^{{D\over2}-1}y_{_2}^{{D\over2}-1}y_{_3}^{-1}\sum\limits_{n_{_1}=0}^\infty
\sum\limits_{n_{_2}=0}^\infty\sum\limits_{n_{_3}=0}^\infty\sum\limits_{n_{_4}=0}^\infty
c_{_{[\tilde{1}3\tilde{5}7]}}^{(1)}(\alpha,{\bf n})
\nonumber\\
&&\hspace{2.5cm}\times
y_{_1}^{n_{_1}}\Big({y_{_4}\over y_{_3}}\Big)^{n_{_2}}y_{_4}^{n_{_3}}\Big({y_{_2}\over y_{_3}}\Big)^{n_{_4}}\;,
\label{GKZ21f-1-2}
\end{eqnarray}
with
\begin{eqnarray}
&&c_{_{[\tilde{1}3\tilde{5}7]}}^{(1)}(\alpha,{\bf n})=
(-)^{n_{_1}+n_{_4}}\Gamma(1+n_{_1}+n_{_3})\Gamma(1+n_{_2}+n_{_4})
\Big\{n_{_1}!n_{_2}!n_{_3}!n_{_4}!
\nonumber\\
&&\hspace{2.5cm}\times
\Gamma({D\over2}-1-n_{_1}-n_{_3})\Gamma({D\over2}+n_{_2})\Gamma(1-{D\over2}-n_{_2}-n_{_4})
\nonumber\\
&&\hspace{2.5cm}\times
\Gamma(2-{D\over2}+n_{_3})\Gamma({D\over2}+n_{_1})\Gamma({D\over2}+n_{_4})\Big\}^{-1}\;.
\label{GKZ21f-1-3}
\end{eqnarray}

\item   $I_{_{2}}=\{1,2,4,5,7,8,10,11,13,14\}$, i.e. the implement $J_{_{2}}=[1,14]\setminus I_{_{2}}=\{3,6,9,12\}$.
The choice implies the power numbers $\alpha_{_3}=\alpha_{_{6}}=\alpha_{_{9}}=\alpha_{_{12}}=0$, and
\begin{eqnarray}
&&\alpha_{_1}=b_{_1}+b_{_4}-a_{_1}-2,\;\alpha_{_2}=b_{_1}+b_{_4}-a_{_2}-2,
\nonumber\\
&&\alpha_{_4}=a_{_3}-a_{_4},\;\alpha_{_5}=b_{_4}-a_{_3}-a_{_5}-1,\;
\alpha_{_{7}}=b_{_2}-1,
\nonumber\\
&&\alpha_{_8}=b_{_3}-a_{_3}-1,\;\alpha_{_{10}}=b_{_5}-b_{_4},\;\alpha_{_{11}}=1-b_{_1},
\nonumber\\
&&\alpha_{_{13}}=-a_{_3},\;\alpha_{_{14}}=a_{_3}-b_{_4}+1.
\label{GKZ21f-2-1}
\end{eqnarray}
The corresponding hypergeometric series is
\begin{eqnarray}
&&\Phi_{_{[\tilde{1}3\tilde{5}7]}}^{(2)}(\alpha,z)=
y_{_1}^{{D\over2}-1}y_{_3}^{{D\over2}-2}\sum\limits_{n_{_1}=0}^\infty
\sum\limits_{n_{_2}=0}^\infty\sum\limits_{n_{_3}=0}^\infty\sum\limits_{n_{_4}=0}^\infty
c_{_{[\tilde{1}3\tilde{5}7]}}^{(2)}(\alpha,{\bf n})
\nonumber\\
&&\hspace{2.5cm}\times
y_{_1}^{n_{_1}}\Big({y_{_4}\over y_{_3}}\Big)^{n_{_2}}y_{_4}^{n_{_3}}\Big({y_{_2}\over y_{_3}}\Big)^{n_{_4}}\;,
\label{GKZ21f-2-2}
\end{eqnarray}
with
\begin{eqnarray}
&&c_{_{[\tilde{1}3\tilde{5}7]}}^{(2)}(\alpha,{\bf n})=
(-)^{n_{_1}+n_{_4}}\Gamma(1+n_{_1}+n_{_3})\Gamma(1+n_{_2}+n_{_4})
\Big\{n_{_1}!n_{_2}!n_{_3}!n_{_4}!
\nonumber\\
&&\hspace{2.5cm}\times
\Gamma({D\over2}-1-n_{_1}-n_{_3})\Gamma({D\over2}+n_{_2})
\Gamma(2-{D\over2}+n_{_4})
\nonumber\\
&&\hspace{2.5cm}\times
\Gamma(2-{D\over2}+n_{_3})\Gamma({D\over2}+n_{_1})
\Gamma({D\over2}-1-n_{_2}-n_{_4})\Big\}^{-1}\;.
\label{GKZ21f-2-3}
\end{eqnarray}

\item   $I_{_{3}}=\{1,2,4,5,8,9,11,\cdots,14\}$, i.e. the implement $J_{_{3}}=[1,14]\setminus I_{_{3}}=\{3,6,7,10\}$.
The choice implies the power numbers $\alpha_{_3}=\alpha_{_{6}}=\alpha_{_{7}}=\alpha_{_{10}}=0$, and
\begin{eqnarray}
&&\alpha_{_1}=b_{_1}+b_{_5}-a_{_1}-2,\;\alpha_{_2}=b_{_1}+b_{_5}-a_{_2}-2,
\nonumber\\
&&\alpha_{_4}=a_{_3}-a_{_4},\;\alpha_{_5}=b_{_5}-a_{_3}-a_{_5}-1,
\nonumber\\
&&\alpha_{_8}=b_{_2}+b_{_3}-a_{_3}-2,\;
\alpha_{_{9}}=b_{_4}-b_{_5},\;\alpha_{_{11}}=1-b_{_1},
\nonumber\\
&&\alpha_{_{12}}=1-b_{_2},\;\alpha_{_{13}}=b_{_2}-a_{_3}-1,\;\alpha_{_{14}}=a_{_3}-b_{_5}+1.
\label{GKZ21f-3-1}
\end{eqnarray}
The corresponding hypergeometric series is written as
\begin{eqnarray}
&&\Phi_{_{[\tilde{1}3\tilde{5}7]}}^{(3)}(\alpha,z)=
y_{_1}^{{D\over2}-1}y_{_2}^{{D\over2}-1}y_{_3}^{-1}y_{_4}^{{D\over2}-1}\sum\limits_{n_{_1}=0}^\infty
\sum\limits_{n_{_2}=0}^\infty\sum\limits_{n_{_3}=0}^\infty\sum\limits_{n_{_4}=0}^\infty
c_{_{[\tilde{1}3\tilde{5}7]}}^{(3)}(\alpha,{\bf n})
\nonumber\\
&&\hspace{2.5cm}\times
y_{_1}^{n_{_1}}\Big({y_{_4}\over y_{_3}}\Big)^{n_{_2}}y_{_4}^{n_{_3}}\Big({y_{_2}\over y_{_3}}\Big)^{n_{_4}}\;,
\label{GKZ21f-3-2}
\end{eqnarray}
with
\begin{eqnarray}
&&c_{_{[\tilde{1}3\tilde{5}7]}}^{(3)}(\alpha,{\bf n})=
(-)^{n_{_1}+n_{_4}}\Gamma(1+n_{_1}+n_{_3})\Gamma(1+n_{_2}+n_{_4})
\nonumber\\
&&\hspace{2.5cm}\times
\Big\{n_{_1}!n_{_2}!n_{_3}!n_{_4}!\Gamma(1-{D\over2}-n_{_1}-n_{_3})
\Gamma({D\over2}+n_{_2})
\nonumber\\
&&\hspace{2.5cm}\times
\Gamma(1-{D\over2}-n_{_2}-n_{_4})\Gamma({D\over2}+n_{_3})
\Gamma({D\over2}+n_{_1})\Gamma({D\over2}+n_{_4})\Big\}^{-1}\;.
\label{GKZ21f-3-3}
\end{eqnarray}

\item   $I_{_{4}}=\{1,2,4,5,7,8,9,11,13,14\}$, i.e. the implement $J_{_{4}}=[1,14]\setminus I_{_{4}}=\{3,6,10,12\}$.
The choice implies the power numbers $\alpha_{_3}=\alpha_{_{6}}=\alpha_{_{10}}=\alpha_{_{12}}=0$, and
\begin{eqnarray}
&&\alpha_{_1}=b_{_1}+b_{_5}-a_{_1}-2,\;\alpha_{_2}=b_{_1}+b_{_5}-a_{_2}-2,
\nonumber\\
&&\alpha_{_4}=a_{_3}-a_{_4},\;\alpha_{_5}=b_{_5}-a_{_3}-a_{_5}-1,\;
\alpha_{_{7}}=b_{_2}-1,
\nonumber\\
&&\alpha_{_8}=b_{_3}-a_{_3}-1,\;\alpha_{_{9}}=b_{_4}-b_{_5},\;\alpha_{_{11}}=1-b_{_1},
\nonumber\\
&&\alpha_{_{13}}=-a_{_3},\;\alpha_{_{14}}=a_{_3}-b_{_5}+1.
\label{GKZ21f-4-1}
\end{eqnarray}
The corresponding hypergeometric solution is written as
\begin{eqnarray}
&&\Phi_{_{[\tilde{1}3\tilde{5}7]}}^{(4)}(\alpha,z)=
y_{_1}^{{D\over2}-1}y_{_3}^{{D\over2}-2}y_{_4}^{{D\over2}-1}\sum\limits_{n_{_1}=0}^\infty
\sum\limits_{n_{_2}=0}^\infty\sum\limits_{n_{_3}=0}^\infty\sum\limits_{n_{_4}=0}^\infty
c_{_{[\tilde{1}3\tilde{5}7]}}^{(4)}(\alpha,{\bf n})
\nonumber\\
&&\hspace{2.5cm}\times
y_{_1}^{n_{_1}}\Big({y_{_4}\over y_{_3}}\Big)^{n_{_2}}y_{_4}^{n_{_3}}\Big({y_{_2}\over y_{_3}}\Big)^{n_{_4}}\;,
\label{GKZ21f-4-2}
\end{eqnarray}
with
\begin{eqnarray}
&&c_{_{[\tilde{1}3\tilde{5}7]}}^{(4)}(\alpha,{\bf n})=
(-)^{n_{_1}+n_{_4}}\Gamma(1+n_{_1}+n_{_3})\Gamma(1+n_{_2}+n_{_4})
\Big\{n_{_1}!n_{_2}!n_{_3}!n_{_4}!
\nonumber\\
&&\hspace{2.5cm}\times
\Gamma(1-{D\over2}-n_{_1}-n_{_3})
\Gamma({D\over2}+n_{_2})\Gamma(2-{D\over2}+n_{_4})
\nonumber\\
&&\hspace{2.5cm}\times
\Gamma({D\over2}+n_{_3})\Gamma({D\over2}+n_{_1})\Gamma({D\over2}-1-n_{_2}-n_{_4})\Big\}^{-1}\;.
\label{GKZ21f-4-3}
\end{eqnarray}

\item   $I_{_{5}}=\{1,2,3,5,8,10,\cdots,14\}$, i.e. the implement $J_{_{5}}=[1,14]\setminus I_{_{5}}=\{4,6,7,9\}$.
The choice implies the power numbers $\alpha_{_4}=\alpha_{_{6}}=\alpha_{_{7}}=\alpha_{_{9}}=0$, and
\begin{eqnarray}
&&\alpha_{_1}=b_{_1}+b_{_4}-a_{_1}-2,\;\alpha_{_2}=b_{_1}+b_{_4}-a_{_2}-2,
\nonumber\\
&&\alpha_{_3}=a_{_4}-a_{_3},\;\alpha_{_5}=b_{_4}-a_{_4}-a_{_5}-1,
\nonumber\\
&&\alpha_{_8}=b_{_2}+b_{_3}-a_{_4}-2,\;
\alpha_{_{10}}=b_{_5}-b_{_4},\;\alpha_{_{11}}=1-b_{_1},
\nonumber\\
&&\alpha_{_{12}}=1-b_{_2},\;\alpha_{_{13}}=b_{_2}-a_{_4}-1,\;\alpha_{_{14}}=a_{_4}-b_{_4}+1.
\label{GKZ21f-7-1}
\end{eqnarray}
The corresponding hypergeometric series is written as
\begin{eqnarray}
&&\Phi_{_{[\tilde{1}3\tilde{5}7]}}^{(5)}(\alpha,z)=
y_{_1}^{{D\over2}-1}y_{_2}^{{D\over2}-1}y_{_3}^{{D\over2}-2}y_{_4}^{1-{D\over2}}
\sum\limits_{n_{_1}=0}^\infty
\sum\limits_{n_{_2}=0}^\infty\sum\limits_{n_{_3}=0}^\infty\sum\limits_{n_{_4}=0}^\infty
c_{_{[\tilde{1}3\tilde{5}7]}}^{(5)}(\alpha,{\bf n})
\nonumber\\
&&\hspace{2.5cm}\times
y_{_1}^{n_{_1}}\Big({y_{_4}\over y_{_3}}\Big)^{n_{_2}}y_{_4}^{n_{_3}}\Big({y_{_2}\over y_{_3}}\Big)^{n_{_4}}\;,
\label{GKZ21f-7-2}
\end{eqnarray}
with
\begin{eqnarray}
&&c_{_{[\tilde{1}3\tilde{5}7]}}^{(5)}(\alpha,{\bf n})=
(-)^{1+n_{_1}+n_{_4}}\Gamma(1+n_{_1}+n_{_3})\Gamma(1+n_{_2}+n_{_4})
\Big\{n_{_1}!n_{_2}!n_{_3}!n_{_4}!
\nonumber\\
&&\hspace{2.5cm}\times
\Gamma({D\over2}-1-n_{_1}-n_{_3})
\Gamma(2-{D\over2}+n_{_2})\Gamma(2-{D\over2}+n_{_3})
\nonumber\\
&&\hspace{2.5cm}\times
\Gamma({D\over2}+n_{_1})\Gamma({D\over2}+n_{_4})\Gamma({D\over2}-1-n_{_2}-n_{_4})\Big\}^{-1}\;.
\label{GKZ21f-7-3}
\end{eqnarray}

\item   $I_{_{6}}=\{1,2,3,5,7,8,10,11,13,14\}$, i.e. the implement $J_{_{6}}=[1,14]\setminus I_{_{6}}=\{4,6,9,12\}$.
The choice implies the power numbers $\alpha_{_4}=\alpha_{_{6}}=\alpha_{_{9}}=\alpha_{_{12}}=0$, and
\begin{eqnarray}
&&\alpha_{_1}=b_{_1}+b_{_4}-a_{_1}-2,\;\alpha_{_2}=b_{_1}+b_{_4}-a_{_2}-2,
\nonumber\\
&&\alpha_{_3}=a_{_4}-a_{_3},\;\alpha_{_5}=b_{_4}-a_{_4}-a_{_5}-1,\;
\alpha_{_{7}}=b_{_2}-1,
\nonumber\\
&&\alpha_{_8}=b_{_3}-a_{_4}-1,\;\alpha_{_{10}}=b_{_5}-b_{_4},\;\alpha_{_{11}}=1-b_{_1},
\nonumber\\
&&\alpha_{_{13}}=-a_{_4},\;\alpha_{_{14}}=a_{_4}-b_{_4}+1.
\label{GKZ21f-8-1}
\end{eqnarray}
The corresponding hypergeometric solution is written as
\begin{eqnarray}
&&\Phi_{_{[\tilde{1}3\tilde{5}7]}}^{(6)}(\alpha,z)=
y_{_1}^{{D\over2}-1}y_{_3}^{{D}-3}y_{_4}^{1-{D\over2}}\sum\limits_{n_{_1}=0}^\infty
\sum\limits_{n_{_2}=0}^\infty\sum\limits_{n_{_3}=0}^\infty\sum\limits_{n_{_4}=0}^\infty
c_{_{[\tilde{1}3\tilde{5}7]}}^{(6)}(\alpha,{\bf n})
\nonumber\\
&&\hspace{2.5cm}\times
y_{_1}^{n_{_1}}\Big({y_{_4}\over y_{_3}}\Big)^{n_{_2}}y_{_4}^{n_{_3}}\Big({y_{_2}\over y_{_3}}\Big)^{n_{_4}}\;,
\label{GKZ21f-8-2}
\end{eqnarray}
with
\begin{eqnarray}
&&c_{_{[\tilde{1}3\tilde{5}7]}}^{(6)}(\alpha,{\bf n})=
(-)^{1+n_{_1}+n_{_2}}\Gamma(1+n_{_1}+n_{_3})\Big\{n_{_1}!n_{_2}!n_{_3}!n_{_4}!
\Gamma({D\over2}-1-n_{_1}-n_{_3})
\nonumber\\
&&\hspace{2.5cm}\times
\Gamma(2-{D\over2}+n_{_2})\Gamma(2-{D\over2}+n_{_4})
\Gamma({D\over2}-1-n_{_2}-n_{_4})
\nonumber\\
&&\hspace{2.5cm}\times
\Gamma(2-{D\over2}+n_{_3})\Gamma({D\over2}+n_{_1})\Gamma(D-2-n_{_2}-n_{_4})\Big\}^{-1}\;.
\label{GKZ21f-8-3}
\end{eqnarray}

\item   $I_{_{7}}=\{1,2,3,5,8,9,11,\cdots,14\}$, i.e. the implement $J_{_{7}}=[1,14]\setminus I_{_{7}}=\{4,6,7,10\}$.
The choice implies the power numbers $\alpha_{_4}=\alpha_{_{6}}=\alpha_{_{7}}=\alpha_{_{10}}=0$, and
\begin{eqnarray}
&&\alpha_{_1}=b_{_1}+b_{_5}-a_{_1}-2,\;\alpha_{_2}=b_{_1}+b_{_5}-a_{_2}-2,
\nonumber\\
&&\alpha_{_3}=a_{_4}-a_{_3},\;\alpha_{_5}=b_{_5}-a_{_4}-a_{_5}-1,
\nonumber\\
&&\alpha_{_8}=b_{_2}+b_{_3}-a_{_4}-2,\;
\alpha_{_{9}}=b_{_4}-b_{_5},\;\alpha_{_{11}}=1-b_{_1},
\nonumber\\
&&\alpha_{_{12}}=1-b_{_2},\;\alpha_{_{13}}=b_{_2}-a_{_4}-1,\;\alpha_{_{14}}=a_{_4}-b_{_5}+1.
\label{GKZ21f-9-1}
\end{eqnarray}
The corresponding hypergeometric series is
\begin{eqnarray}
&&\Phi_{_{[\tilde{1}3\tilde{5}7]}}^{(7)}(\alpha,z)=
y_{_1}^{{D\over2}-1}y_{_2}^{{D\over2}-1}y_{_3}^{{D\over2}-2}\sum\limits_{n_{_1}=0}^\infty
\sum\limits_{n_{_2}=0}^\infty\sum\limits_{n_{_3}=0}^\infty\sum\limits_{n_{_4}=0}^\infty
c_{_{[\tilde{1}3\tilde{5}7]}}^{(7)}(\alpha,{\bf n})
\nonumber\\
&&\hspace{2.5cm}\times
y_{_1}^{n_{_1}}\Big({y_{_4}\over y_{_3}}\Big)^{n_{_2}}y_{_4}^{n_{_3}}
\Big({y_{_2}\over y_{_3}}\Big)^{n_{_4}}\;,
\label{GKZ21f-9-2}
\end{eqnarray}
with
\begin{eqnarray}
&&c_{_{[\tilde{1}3\tilde{5}7]}}^{(7)}(\alpha,{\bf n})=
(-)^{n_{_1}+n_{_4}}\Gamma(1+n_{_1}+n_{_3})\Gamma(1+n_{_2}+n_{_4})
\Big\{n_{_1}!n_{_2}!n_{_3}!n_{_4}!
\nonumber\\
&&\hspace{2.5cm}\times
\Gamma(1-{D\over2}-n_{_1}-n_{_3})
\Gamma(2-{D\over2}+n_{_2})\Gamma({D\over2}+n_{_3})
\nonumber\\
&&\hspace{2.5cm}\times
\Gamma({D\over2}+n_{_1})\Gamma({D\over2}+n_{_4})\Gamma({D\over2}-1-n_{_2}-n_{_4})\Big\}\;.
\label{GKZ21f-9-3}
\end{eqnarray}

\item   $I_{_{8}}=\{1,2,3,5,7,8,9,11,13,14\}$, i.e. the implement $J_{_{8}}=[1,14]\setminus I_{_{8}}=\{4,6,10,12\}$.
The choice implies the power numbers $\alpha_{_4}=\alpha_{_{6}}=\alpha_{_{10}}=\alpha_{_{12}}=0$, and
\begin{eqnarray}
&&\alpha_{_1}=b_{_1}+b_{_5}-a_{_1}-2,\;\alpha_{_2}=b_{_1}+b_{_5}-a_{_2}-2,
\nonumber\\
&&\alpha_{_3}=a_{_4}-a_{_3},\;\alpha_{_5}=b_{_5}-a_{_4}-a_{_5}-1,\;
\alpha_{_{7}}=b_{_2}-1,
\nonumber\\
&&\alpha_{_8}=b_{_3}-a_{_4}-1,\;\alpha_{_{9}}=b_{_4}-b_{_5},\;\alpha_{_{11}}=1-b_{_1},
\nonumber\\
&&\alpha_{_{13}}=-a_{_4},\;\alpha_{_{14}}=a_{_4}-b_{_5}+1.
\label{GKZ21f-10-1}
\end{eqnarray}
The corresponding hypergeometric series is written as
\begin{eqnarray}
&&\Phi_{_{[\tilde{1}3\tilde{5}7]}}^{(8)}(\alpha,z)=
y_{_1}^{{D\over2}-1}y_{_3}^{{D}-3}\sum\limits_{n_{_1}=0}^\infty
\sum\limits_{n_{_2}=0}^\infty\sum\limits_{n_{_3}=0}^\infty\sum\limits_{n_{_4}=0}^\infty
c_{_{[\tilde{1}3\tilde{5}7]}}^{(8)}(\alpha,{\bf n})
\nonumber\\
&&\hspace{2.5cm}\times
y_{_1}^{n_{_1}}\Big({y_{_4}\over y_{_3}}\Big)^{n_{_2}}y_{_4}^{n_{_3}}
\Big({y_{_2}\over y_{_3}}\Big)^{n_{_4}}\;,
\label{GKZ21f-10-2}
\end{eqnarray}
with
\begin{eqnarray}
&&c_{_{[\tilde{1}3\tilde{5}7]}}^{(8)}(\alpha,{\bf n})=
(-)^{n_{_1}+n_{_2}}\Gamma(1+n_{_1}+n_{_3})
\Big\{n_{_1}!n_{_2}!n_{_3}!n_{_4}!\Gamma(1-{D\over2}-n_{_1}-n_{_3})
\nonumber\\
&&\hspace{2.5cm}\times
\Gamma(2-{D\over2}+n_{_2})\Gamma(2-{D\over2}+n_{_4})
\Gamma({D\over2}-1-n_{_2}-n_{_4})
\nonumber\\
&&\hspace{2.5cm}\times
\Gamma({D\over2}+n_{_3})\Gamma({D\over2}+n_{_1})\Gamma(D-2-n_{_2}-n_{_4})\Big\}^{-1}\;.
\label{GKZ21f-10-3}
\end{eqnarray}

\item   $I_{_{9}}=\{1,2,4,5,6,8,10,12,13,14\}$, i.e. the implement $J_{_{9}}=[1,14]\setminus I_{_{9}}=\{3,7,9,11\}$.
The choice implies the power numbers $\alpha_{_3}=\alpha_{_{7}}=\alpha_{_{9}}=\alpha_{_{11}}=0$, and
\begin{eqnarray}
&&\alpha_{_1}=b_{_4}-a_{_1}-1,\;\alpha_{_2}=b_{_4}-a_{_2}-1,\;\alpha_{_4}=a_{_3}-a_{_4},
\nonumber\\
&&\alpha_{_5}=b_{_4}-a_{_3}-a_{_5}-1,
\;\alpha_{_{6}}=b_{_1}-1,
\nonumber\\
&&\alpha_{_8}=b_{_2}+b_{_3}-a_{_3}-2,\;
\alpha_{_{10}}=b_{_5}-b_{_4},\;\alpha_{_{12}}=1-b_{_2},
\nonumber\\
&&\alpha_{_{13}}=b_{_2}-a_{_3}-1,\;\alpha_{_{14}}=a_{_3}-b_{_4}+1.
\label{GKZ21f-13-1}
\end{eqnarray}
The corresponding hypergeometric solution is written as
\begin{eqnarray}
&&\Phi_{_{[\tilde{1}3\tilde{5}7]}}^{(9)}(\alpha,z)=
y_{_2}^{{D\over2}-1}y_{_3}^{-1}\sum\limits_{n_{_1}=0}^\infty
\sum\limits_{n_{_2}=0}^\infty\sum\limits_{n_{_3}=0}^\infty\sum\limits_{n_{_4}=0}^\infty
c_{_{[\tilde{1}3\tilde{5}7]}}^{(9)}(\alpha,{\bf n})
\nonumber\\
&&\hspace{2.5cm}\times
y_{_1}^{n_{_1}}\Big({y_{_4}\over y_{_3}}\Big)^{n_{_2}}y_{_4}^{n_{_3}}
\Big({y_{_2}\over y_{_3}}\Big)^{n_{_4}}\;,
\label{GKZ21f-13-2}
\end{eqnarray}
with
\begin{eqnarray}
&&c_{_{[\tilde{1}3\tilde{5}7]}}^{(9)}(\alpha,{\bf n})=
(-)^{n_{_3}+n_{_4}}\Gamma(1+n_{_2}+n_{_4})
\Big\{n_{_1}!n_{_2}!n_{_3}!n_{_4}!\Gamma({D\over2}-1-n_{_1}-n_{_3})
\nonumber\\
&&\hspace{2.5cm}\times
\Gamma(D-2-n_{_1}-n_{_3})\Gamma({D\over2}+n_{_2})\Gamma(2-{D\over2}+n_{_1})
\nonumber\\
&&\hspace{2.5cm}\times
\Gamma(1-{D\over2}-n_{_2}-n_{_4})
\Gamma(2-{D\over2}+n_{_3})\Gamma({D\over2}+n_{_4})\Big\}^{-1}\;.
\label{GKZ21f-13-3}
\end{eqnarray}

\item   $I_{_{10}}=\{1,2,4,\cdots,8,10,13,14\}$, i.e. the implement $J_{_{10}}=[1,14]\setminus I_{_{10}}=\{3,9,11,12\}$.
The choice implies the power numbers $\alpha_{_3}=\alpha_{_{9}}=\alpha_{_{11}}=\alpha_{_{12}}=0$, and
\begin{eqnarray}
&&\alpha_{_1}=b_{_4}-a_{_1}-1,\;\alpha_{_2}=b_{_4}-a_{_2}-1,\;\alpha_{_4}=a_{_3}-a_{_4},
\nonumber\\
&&\alpha_{_5}=b_{_4}-a_{_3}-a_{_5}-1,\;\alpha_{_{6}}=b_{_1}-1,\;
\alpha_{_{7}}=b_{_2}-1,
\nonumber\\
&&\alpha_{_8}=b_{_3}-a_{_3}-1,\;\alpha_{_{10}}=b_{_5}-b_{_4},\;\alpha_{_{13}}=-a_{_3},
\nonumber\\
&&\alpha_{_{14}}=a_{_3}-b_{_4}+1.
\label{GKZ21f-14-1}
\end{eqnarray}
The corresponding hypergeometric series is
\begin{eqnarray}
&&\Phi_{_{[\tilde{1}3\tilde{5}7]}}^{(10)}(\alpha,z)=
y_{_3}^{{D\over2}-2}\sum\limits_{n_{_1}=0}^\infty
\sum\limits_{n_{_2}=0}^\infty\sum\limits_{n_{_3}=0}^\infty\sum\limits_{n_{_4}=0}^\infty
c_{_{[\tilde{1}3\tilde{5}7]}}^{(10)}(\alpha,{\bf n})
\nonumber\\
&&\hspace{2.5cm}\times
y_{_1}^{n_{_1}}\Big({y_{_4}\over y_{_3}}\Big)^{n_{_2}}y_{_4}^{n_{_3}}
\Big({y_{_2}\over y_{_3}}\Big)^{n_{_4}}\;,
\label{GKZ21f-14-2}
\end{eqnarray}
with
\begin{eqnarray}
&&c_{_{[\tilde{1}3\tilde{5}7]}}^{(10)}(\alpha,{\bf n})=
(-)^{n_{_3}+n_{_4}}\Gamma(1+n_{_2}+n_{_4})
\Big\{n_{_1}!n_{_2}!n_{_3}!n_{_4}!\Gamma({D\over2}-1-n_{_1}-n_{_3})
\nonumber\\
&&\hspace{2.5cm}\times
\Gamma(D-2-n_{_1}-n_{_3})\Gamma({D\over2}+n_{_2})
\Gamma(2-{D\over2}+n_{_1})\Gamma(2-{D\over2}+n_{_4})
\nonumber\\
&&\hspace{2.5cm}\times
\Gamma(2-{D\over2}+n_{_3})\Gamma({D\over2}-1-n_{_2}-n_{_4})\Big\}^{-1}\;.
\label{GKZ21f-14-3}
\end{eqnarray}

\item   $I_{_{11}}=\{1,2,4,5,6,8,9,12,13,14\}$, i.e. the implement $J_{_{11}}=[1,14]\setminus I_{_{11}}=\{3,7,10,11\}$.
The choice implies the power numbers $\alpha_{_3}=\alpha_{_{7}}=\alpha_{_{10}}=\alpha_{_{11}}=0$, and
\begin{eqnarray}
&&\alpha_{_1}=b_{_5}-a_{_1}-1,\;\alpha_{_2}=b_{_5}-a_{_2}-1,
\nonumber\\
&&\alpha_{_4}=a_{_3}-a_{_4},\;\alpha_{_5}=b_{_5}-a_{_3}-a_{_5}-1,
\nonumber\\
&&\alpha_{_{6}}=b_{_1}-1,\;\alpha_{_8}=b_{_2}+b_{_3}-a_{_3}-2,\;
\alpha_{_{9}}=b_{_4}-b_{_5},
\nonumber\\
&&\alpha_{_{12}}=1-b_{_2},\;\alpha_{_{13}}=b_{_2}-a_{_3}-1,\;
\alpha_{_{14}}=a_{_3}-b_{_5}+1.
\label{GKZ21f-15-1}
\end{eqnarray}
The corresponding hypergeometric solution is
\begin{eqnarray}
&&\Phi_{_{[\tilde{1}3\tilde{5}7]}}^{(11)}(\alpha,z)=
y_{_2}^{{D\over2}-1}y_{_3}^{-1}y_{_4}^{{D\over2}-1}\sum\limits_{n_{_1}=0}^\infty
\sum\limits_{n_{_2}=0}^\infty\sum\limits_{n_{_3}=0}^\infty\sum\limits_{n_{_4}=0}^\infty
c_{_{[\tilde{1}3\tilde{5}7]}}^{(11)}(\alpha,{\bf n})
\nonumber\\
&&\hspace{2.5cm}\times
y_{_1}^{n_{_1}}\Big({y_{_4}\over y_{_3}}\Big)^{n_{_2}}y_{_4}^{n_{_3}}
\Big({y_{_2}\over y_{_3}}\Big)^{n_{_4}}\;,
\label{GKZ21f-15-2}
\end{eqnarray}
with
\begin{eqnarray}
&&c_{_{[\tilde{1}3\tilde{5}7]}}^{(11)}(\alpha,{\bf n})=
(-)^{n_{_1}+n_{_4}}\Gamma(1+n_{_1}+n_{_3})\Gamma(1+n_{_2}+n_{_4})
\Big\{n_{_1}!n_{_2}!n_{_3}!n_{_4}!
\nonumber\\
&&\hspace{2.5cm}\times
\Gamma({D\over2}-1-n_{_1}-n_{_3})
\Gamma({D\over2}+n_{_2})\Gamma(2-{D\over2}+n_{_1})
\nonumber\\
&&\hspace{2.5cm}\times
\Gamma(1-{D\over2}-n_{_2}-n_{_4})
\Gamma({D\over2}+n_{_3})\Gamma({D\over2}+n_{_4})\Big\}^{-1}\;.
\label{GKZ21f-15-3}
\end{eqnarray}

\item   $I_{_{12}}=\{1,2,4,\cdots,9,13,14\}$, i.e. the implement $J_{_{12}}=[1,14]\setminus I_{_{12}}=\{3,10,11,12\}$.
The choice implies the power numbers $\alpha_{_3}=\alpha_{_{10}}=\alpha_{_{11}}=\alpha_{_{12}}=0$, and
\begin{eqnarray}
&&\alpha_{_1}=b_{_5}-a_{_1}-1,\;\alpha_{_2}=b_{_5}-a_{_2}-1,
\nonumber\\
&&\alpha_{_4}=a_{_3}-a_{_4},\;\alpha_{_5}=b_{_5}-a_{_3}-a_{_5}-1,\;\alpha_{_{6}}=b_{_1}-1,
\nonumber\\
&&\alpha_{_{7}}=b_{_2}-1,\;
\alpha_{_8}=b_{_3}-a_{_3}-1,\;\alpha_{_{9}}=b_{_4}-b_{_5},
\nonumber\\
&&\alpha_{_{13}}=-a_{_3},\;\alpha_{_{14}}=a_{_3}-b_{_5}+1.
\label{GKZ21f-16-1}
\end{eqnarray}
The corresponding hypergeometric solution is written as
\begin{eqnarray}
&&\Phi_{_{[\tilde{1}3\tilde{5}7]}}^{(12)}(\alpha,z)=
y_{_3}^{{D\over2}-2}y_{_4}^{{D\over2}-1}\sum\limits_{n_{_1}=0}^\infty
\sum\limits_{n_{_2}=0}^\infty\sum\limits_{n_{_3}=0}^\infty\sum\limits_{n_{_4}=0}^\infty
c_{_{[\tilde{1}3\tilde{5}7]}}^{(12)}(\alpha,{\bf n})
\nonumber\\
&&\hspace{2.5cm}\times
y_{_1}^{n_{_1}}\Big({y_{_4}\over y_{_3}}\Big)^{n_{_2}}y_{_4}^{n_{_3}}
\Big({y_{_2}\over y_{_3}}\Big)^{n_{_4}}\;,
\label{GKZ21f-16-2}
\end{eqnarray}
with
\begin{eqnarray}
&&c_{_{[\tilde{1}3\tilde{5}7]}}^{(12)}(\alpha,{\bf n})=
(-)^{n_{_1}+n_{_4}}\Gamma(1+n_{_1}+n_{_3})\Gamma(1+n_{_2}+n_{_4})
\Big\{n_{_1}!n_{_2}!n_{_3}!n_{_4}!
\nonumber\\
&&\hspace{2.5cm}\times
\Gamma({D\over2}-1-n_{_1}-n_{_3})\Gamma({D\over2}+n_{_2})\Gamma(2-{D\over2}+n_{_1})
\Gamma(2-{D\over2}+n_{_4})
\nonumber\\
&&\hspace{2.5cm}\times
\Gamma({D\over2}+n_{_3})\Gamma({D\over2}-1-n_{_2}-n_{_4})\Big\}^{-1}\;.
\label{GKZ21f-16-3}
\end{eqnarray}

\item   $I_{_{13}}=\{1,2,3,5,6,8,10,12,13,14\}$, i.e. the implement $J_{_{13}}=[1,14]\setminus I_{_{13}}=\{4,7,9,11\}$.
The choice implies the power numbers $\alpha_{_4}=\alpha_{_{7}}=\alpha_{_{9}}=\alpha_{_{11}}=0$, and
\begin{eqnarray}
&&\alpha_{_1}=b_{_4}-a_{_1}-1,\;\alpha_{_2}=b_{_4}-a_{_2}-1,
\nonumber\\
&&\alpha_{_3}=a_{_4}-a_{_3},\;\alpha_{_5}=b_{_4}-a_{_4}-a_{_5}-1,\;\alpha_{_{6}}=b_{_1}-1,
\nonumber\\
&&\alpha_{_8}=b_{_2}+b_{_3}-a_{_4}-2,\;\alpha_{_{10}}=b_{_5}-b_{_4},
\nonumber\\
&&\alpha_{_{12}}=1-b_{_2},\;\alpha_{_{13}}=b_{_2}-a_{_4}-1,\;\alpha_{_{14}}=a_{_4}-b_{_4}+1.
\label{GKZ21f-19-1}
\end{eqnarray}
The corresponding hypergeometric series is given as
\begin{eqnarray}
&&\Phi_{_{[\tilde{1}3\tilde{5}7]}}^{(13)}(\alpha,z)=
y_{_2}^{{D\over2}-1}y_{_3}^{{D\over2}-2}y_{_4}^{1-{D\over2}}\sum\limits_{n_{_1}=0}^\infty
\sum\limits_{n_{_2}=0}^\infty\sum\limits_{n_{_3}=0}^\infty\sum\limits_{n_{_4}=0}^\infty
c_{_{[\tilde{1}3\tilde{5}7]}}^{(13)}(\alpha,{\bf n})
\nonumber\\
&&\hspace{2.5cm}\times
y_{_1}^{n_{_1}}\Big({y_{_4}\over y_{_3}}\Big)^{n_{_2}}y_{_4}^{n_{_3}}
\Big({y_{_2}\over y_{_3}}\Big)^{n_{_4}}\;,
\label{GKZ21f-19-2}
\end{eqnarray}
with
\begin{eqnarray}
&&c_{_{[\tilde{1}3\tilde{5}7]}}^{(13)}(\alpha,{\bf n})=
(-)^{1+n_{_3}+n_{_4}}\Gamma(1+n_{_2}+n_{_4})\Big\{n_{_1}!n_{_2}!n_{_3}!n_{_4}!
\Gamma({D\over2}-1-n_{_1}-n_{_3})
\nonumber\\
&&\hspace{2.5cm}\times
\Gamma(D-2-n_{_1}-n_{_3})\Gamma(2-{D\over2}+n_{_2})
\Gamma(2-{D\over2}+n_{_1})
\nonumber\\
&&\hspace{2.5cm}\times
\Gamma(2-{D\over2}+n_{_3})\Gamma({D\over2}+n_{_4})
\Gamma({D\over2}-1-n_{_2}-n_{_4})\Big\}^{-1}\;.
\label{GKZ21f-19-3}
\end{eqnarray}

\item   $I_{_{14}}=\{1,2,3,5,6,7,8,10,13,14\}$, i.e. the implement $J_{_{14}}=[1,14]\setminus I_{_{14}}=\{4,9,11,12\}$.
The choice implies the power numbers $\alpha_{_4}=\alpha_{_{9}}=\alpha_{_{11}}=\alpha_{_{12}}=0$, and
\begin{eqnarray}
&&\alpha_{_1}=b_{_4}-a_{_1}-1,\;\alpha_{_2}=b_{_4}-a_{_2}-1,\;\alpha_{_3}=a_{_4}-a_{_3},
\nonumber\\
&&\alpha_{_5}=b_{_4}-a_{_4}-a_{_5}-1,\;\alpha_{_{6}}=b_{_1}-1,\;
\alpha_{_{7}}=b_{_2}-1,
\nonumber\\
&&\alpha_{_8}=b_{_3}-a_{_4}-1,\;\alpha_{_{10}}=b_{_5}-b_{_4},\;\alpha_{_{13}}=-a_{_4},
\nonumber\\
&&\alpha_{_{14}}=a_{_4}-b_{_4}+1.
\label{GKZ21f-20-1}
\end{eqnarray}
The corresponding hypergeometric series is written as
\begin{eqnarray}
&&\Phi_{_{[\tilde{1}3\tilde{5}7]}}^{(14)}(\alpha,z)=
y_{_3}^{{D}-3}y_{_4}^{1-{D\over2}}\sum\limits_{n_{_1}=0}^\infty
\sum\limits_{n_{_2}=0}^\infty\sum\limits_{n_{_3}=0}^\infty\sum\limits_{n_{_4}=0}^\infty
c_{_{[\tilde{1}3\tilde{5}7]}}^{(14)}(\alpha,{\bf n})
\nonumber\\
&&\hspace{2.5cm}\times
y_{_1}^{n_{_1}}\Big({y_{_4}\over y_{_3}}\Big)^{n_{_2}}y_{_4}^{n_{_3}}
\Big({y_{_2}\over y_{_3}}\Big)^{n_{_4}}\;,
\label{GKZ21f-20-2}
\end{eqnarray}
with
\begin{eqnarray}
&&c_{_{[\tilde{1}3\tilde{5}7]}}^{(14)}(\alpha,{\bf n})=
(-)^{1+n_{_2}+n_{_3}}\Big\{n_{_1}!n_{_2}!n_{_3}!n_{_4}!\Gamma({D\over2}-1-n_{_1}-n_{_3})
\Gamma(D-2-n_{_1}-n_{_3})
\nonumber\\
&&\hspace{2.5cm}\times
\Gamma(2-{D\over2}+n_{_2})\Gamma(2-{D\over2}+n_{_1})\Gamma(2-{D\over2}+n_{_4})
\nonumber\\
&&\hspace{2.5cm}\times
\Gamma({D\over2}-1-n_{_2}-n_{_4})\Gamma(2-{D\over2}+n_{_3})
\Gamma(D-2-n_{_2}-n_{_4})\Big\}^{-1}\;.
\label{GKZ21f-20-3}
\end{eqnarray}

\item   $I_{_{15}}=\{1,2,3,5,6,8,9,12,13,14\}$, i.e. the implement $J_{_{15}}=[1,14]\setminus I_{_{15}}=\{4,7,10,11\}$.
The choice implies the power numbers $\alpha_{_4}=\alpha_{_{7}}=\alpha_{_{10}}=\alpha_{_{11}}=0$, and
\begin{eqnarray}
&&\alpha_{_1}=b_{_5}-a_{_1}-1,\;\alpha_{_2}=b_{_5}-a_{_2}-1,
\nonumber\\
&&\alpha_{_3}=a_{_4}-a_{_3},\;\alpha_{_5}=b_{_5}-a_{_4}-a_{_5}-1,\;\alpha_{_{6}}=b_{_1}-1,
\nonumber\\
&&\alpha_{_8}=b_{_2}+b_{_3}-a_{_4}-2,\;\alpha_{_{9}}=b_{_4}-b_{_5},
\nonumber\\
&&\alpha_{_{12}}=1-b_{_2},\;\alpha_{_{13}}=b_{_2}-a_{_4}-1,\;\alpha_{_{14}}=a_{_4}-b_{_5}+1.
\label{GKZ21f-21-1}
\end{eqnarray}
The corresponding hypergeometric solution is written as
\begin{eqnarray}
&&\Phi_{_{[\tilde{1}3\tilde{5}7]}}^{(15)}(\alpha,z)=
y_{_2}^{{D\over2}-1}y_{_3}^{{D\over2}-2}\sum\limits_{n_{_1}=0}^\infty
\sum\limits_{n_{_2}=0}^\infty\sum\limits_{n_{_3}=0}^\infty\sum\limits_{n_{_4}=0}^\infty
c_{_{[\tilde{1}3\tilde{5}7]}}^{(15)}(\alpha,{\bf n})
\nonumber\\
&&\hspace{2.5cm}\times
y_{_1}^{n_{_1}}\Big({y_{_4}\over y_{_3}}\Big)^{n_{_2}}y_{_4}^{n_{_3}}
\Big({y_{_2}\over y_{_3}}\Big)^{n_{_4}}\;,
\label{GKZ21f-21-2}
\end{eqnarray}
with
\begin{eqnarray}
&&c_{_{[\tilde{1}3\tilde{5}7]}}^{(15)}(\alpha,{\bf n})=
(-)^{n_{_1}+n_{_4}}\Gamma(1+n_{_1}+n_{_3})\Gamma(1+n_{_2}+n_{_4})
\Big\{n_{_1}!n_{_2}!n_{_3}!n_{_4}!
\nonumber\\
&&\hspace{2.5cm}\times
\Gamma({D\over2}-1-n_{_1}-n_{_3})\Gamma(2-{D\over2}+n_{_2})\Gamma(2-{D\over2}+n_{_1})
\nonumber\\
&&\hspace{2.5cm}\times
\Gamma({D\over2}+n_{_3})\Gamma({D\over2}+n_{_4})\Gamma({D\over2}-1-n_{_2}-n_{_4})\Big\}^{-1}\;.
\label{GKZ21f-21-3}
\end{eqnarray}

\item   $I_{_{16}}=\{1,2,3,5,\cdots,9,13,14\}$, i.e. the implement $J_{_{16}}=[1,14]\setminus I_{_{16}}=\{4,10,11,12\}$.
The choice implies the power numbers $\alpha_{_4}=\alpha_{_{10}}=\alpha_{_{11}}=\alpha_{_{12}}=0$, and
\begin{eqnarray}
&&\alpha_{_1}=b_{_5}-a_{_1}-1,\;\alpha_{_2}=b_{_5}-a_{_2}-1,\;\alpha_{_3}=a_{_4}-a_{_3},
\nonumber\\
&&\alpha_{_5}=b_{_5}-a_{_4}-a_{_5}-1,\;\alpha_{_{6}}=b_{_1}-1,\;\alpha_{_{7}}=b_{_2}-1,
\nonumber\\
&&\alpha_{_8}=b_{_3}-a_{_4}-1,\;\alpha_{_{9}}=b_{_4}-b_{_5},\;\alpha_{_{13}}=-a_{_4},
\nonumber\\
&&\alpha_{_{14}}=a_{_4}-b_{_5}+1.
\label{GKZ21f-22-1}
\end{eqnarray}
The corresponding hypergeometric series is written as
\begin{eqnarray}
&&\Phi_{_{[\tilde{1}3\tilde{5}7]}}^{(16)}(\alpha,z)=
y_{_3}^{{D}-3}\sum\limits_{n_{_1}=0}^\infty
\sum\limits_{n_{_2}=0}^\infty\sum\limits_{n_{_3}=0}^\infty\sum\limits_{n_{_4}=0}^\infty
c_{_{[\tilde{1}3\tilde{5}7]}}^{(16)}(\alpha,{\bf n})
\nonumber\\
&&\hspace{2.5cm}\times
y_{_1}^{n_{_1}}\Big({y_{_4}\over y_{_3}}\Big)^{n_{_2}}y_{_4}^{n_{_3}}
\Big({y_{_2}\over y_{_3}}\Big)^{n_{_4}}\;,
\label{GKZ21f-22-2}
\end{eqnarray}
with
\begin{eqnarray}
&&c_{_{[\tilde{1}3\tilde{5}7]}}^{(16)}(\alpha,{\bf n})=
(-)^{n_{_1}+n_{_2}}\Gamma(1+n_{_1}+n_{_3})
\Big\{n_{_1}!n_{_2}!n_{_3}!n_{_4}!
\Gamma({D\over2}-1-n_{_1}-n_{_3})
\nonumber\\
&&\hspace{2.5cm}\times
\Gamma(2-{D\over2}+n_{_2})\Gamma(2-{D\over2}+n_{_1})
\Gamma(2-{D\over2}+n_{_4})
\nonumber\\
&&\hspace{2.5cm}\times
\Gamma({D\over2}-1-n_{_2}-n_{_4})
\Gamma({D\over2}+n_{_3})\Gamma(D-2-n_{_2}-n_{_4})\Big\}^{-1}\;.
\label{GKZ21f-22-3}
\end{eqnarray}

\item   $I_{_{17}}=\{1,2,4,5,8,9,11,\cdots,14\}$, i.e. the implement $J_{_{17}}=[1,14]\setminus I_{_{17}}=\{3,6,7,14\}$.
The choice implies the power numbers $\alpha_{_3}=\alpha_{_{6}}=\alpha_{_{7}}=\alpha_{_{14}}=0$, and
\begin{eqnarray}
&&\alpha_{_1}=a_{_3}+b_{_1}-a_{_1}-1,\;\alpha_{_2}=a_{_3}+b_{_1}-a_{_2}-1,
\nonumber\\
&&\alpha_{_4}=a_{_3}-a_{_4},\;\alpha_{_5}=-a_{_5},\;\alpha_{_8}=b_{_2}+b_{_3}-a_{_3}-2,
\nonumber\\
&&\alpha_{_{9}}=b_{_4}-a_{_3}-1,\;\alpha_{_{10}}=b_{_5}-a_{_3}-1,
\nonumber\\
&&\alpha_{_{11}}=1-b_{_1},\;\alpha_{_{12}}=1-b_{_2},\;\alpha_{_{13}}=b_{_2}-a_{_3}-1.
\label{GKZ21f-5-1}
\end{eqnarray}
The corresponding hypergeometric series solutions are written as
\begin{eqnarray}
&&\Phi_{_{[\tilde{1}3\tilde{5}7]}}^{(17),a}(\alpha,z)=
y_{_1}^{{D\over2}-1}y_{_2}^{{D\over2}-1}y_{_3}^{-1}\sum\limits_{n_{_1}=0}^\infty
\sum\limits_{n_{_2}=0}^\infty\sum\limits_{n_{_3}=0}^\infty\sum\limits_{n_{_4}=0}^\infty
c_{_{[\tilde{1}3\tilde{5}7]}}^{(17),a}(\alpha,{\bf n})
\nonumber\\
&&\hspace{2.5cm}\times
y_{_1}^{n_{_1}}\Big({y_{_4}\over y_{_3}}\Big)^{n_{_2}}y_{_4}^{n_{_3}}
\Big({y_{_2}\over y_{_3}}\Big)^{n_{_4}}
\;,\nonumber\\
&&\Phi_{_{[\tilde{1}3\tilde{5}7]}}^{(17),b}(\alpha,z)=
y_{_1}^{{D\over2}-1}y_{_2}^{{D\over2}-1}y_{_3}^{-2}\sum\limits_{n_{_1}=0}^\infty
\sum\limits_{n_{_2}=0}^\infty\sum\limits_{n_{_3}=0}^\infty\sum\limits_{n_{_4}=0}^\infty
c_{_{[\tilde{1}3\tilde{5}7]}}^{(17),b}(\alpha,{\bf n})
\nonumber\\
&&\hspace{2.5cm}\times
\Big({y_{_1}\over y_{_3}}\Big)^{n_{_1}}\Big({1\over y_{_3}}\Big)^{n_{_2}}
\Big({y_{_4}\over y_{_3}}\Big)^{n_{_3}}\Big({y_{_2}\over y_{_3}}\Big)^{n_{_4}}
\;,\nonumber\\
&&\Phi_{_{[\tilde{1}3\tilde{5}7]}}^{(17),c}(\alpha,z)=
y_{_1}^{{D\over2}}y_{_2}^{{D\over2}-1}y_{_3}^{-2}\sum\limits_{n_{_1}=0}^\infty
\sum\limits_{n_{_2}=0}^\infty\sum\limits_{n_{_3}=0}^\infty\sum\limits_{n_{_4}=0}^\infty
c_{_{[\tilde{1}3\tilde{5}7]}}^{(17),c}(\alpha,{\bf n})
\nonumber\\
&&\hspace{2.5cm}\times
y_{_1}^{n_{_1}}\Big({y_{_1}\over y_{_3}}\Big)^{n_{_2}}
\Big({y_{_4}\over y_{_3}}\Big)^{n_{_3}}\Big({y_{_2}\over y_{_3}}\Big)^{n_{_4}}\;.
\label{GKZ21f-5-2a}
\end{eqnarray}
Where the coefficients are
\begin{eqnarray}
&&c_{_{[\tilde{1}3\tilde{5}7]}}^{(17),a}(\alpha,{\bf n})=
(-)^{n_{_1}+n_{_4}}\Gamma(1+n_{_1}+n_{_3})\Gamma(1+n_{_2}+n_{_4})
\Big\{n_{_1}!n_{_2}!n_{_3}!n_{_4}!
\nonumber\\
&&\hspace{2.5cm}\times
\Gamma({D\over2}-1-n_{_1}-n_{_3})
\Gamma({D\over2}+n_{_2})\Gamma(1-{D\over2}-n_{_2}-n_{_4})
\nonumber\\
&&\hspace{2.5cm}\times
\Gamma(2-{D\over2}+n_{_3})
\Gamma({D\over2}+n_{_1})\Gamma({D\over2}+n_{_4})\Big\}^{-1}
\;,\nonumber\\
&&c_{_{[\tilde{1}3\tilde{5}7]}}^{(17),b}(\alpha,{\bf n})=
(-)^{1+n_{_4}}\Gamma(1+n_{_1}+n_{_2})\Gamma(2+n_{_1}+n_{_2}+n_{_3}+n_{_4})
\Big\{n_{_1}!n_{_2}!n_{_4}!
\nonumber\\
&&\hspace{2.5cm}\times
\Gamma(2+n_{_1}+n_{_2}+n_{_3})
\Gamma({D\over2}+n_{_2})\Gamma({D\over2}+1+n_{_1}+n_{_2}+n_{_3})
\nonumber\\
&&\hspace{2.5cm}\times
\Gamma(-{D\over2}-n_{_1}-n_{_2}-n_{_3}-n_{_4})
\Gamma(1-{D\over2}-n_{_1}-n_{_2})
\Gamma({D\over2}+n_{_1})
\nonumber\\
&&\hspace{2.5cm}\times
\Gamma({D\over2}+n_{_4})\Big\}^{-1}
\;,\nonumber\\
&&c_{_{[\tilde{1}3\tilde{5}7]}}^{(17),c}(\alpha,{\bf n})=
(-)^{1+n_{_1}+n_{_4}}\Gamma(1+n_{_1})\Gamma(1+n_{_2})\Gamma(2+n_{_2}+n_{_3}+n_{_4})
\nonumber\\
&&\hspace{2.5cm}\times
\Big\{n_{_4}!\Gamma(2+n_{_1}+n_{_2})\Gamma(2+n_{_2}+n_{_3})
\Gamma({D\over2}-1-n_{_1})
\nonumber\\
&&\hspace{2.5cm}\times
\Gamma({D\over2}+1+n_{_2}+n_{_3})\Gamma(-{D\over2}-n_{_2}-n_{_3}-n_{_4})
\nonumber\\
&&\hspace{2.5cm}\times
\Gamma(1-{D\over2}-n_{_2})\Gamma({D\over2}+1+n_{_1}+n_{_2})
\Gamma({D\over2}+n_{_4})\Big\}^{-1}\;.
\label{GKZ21f-5-3}
\end{eqnarray}

\item   $I_{_{18}}=\{1,2,4,5,7,\cdots,11,13\}$, i.e. the implement $J_{_{18}}=[1,14]\setminus I_{_{18}}=\{3,6,12,14\}$.
The choice implies the power numbers $\alpha_{_3}=\alpha_{_{6}}=\alpha_{_{12}}=\alpha_{_{14}}=0$, and
\begin{eqnarray}
&&\alpha_{_1}=a_{_3}+b_{_1}-a_{_1}-1,\;\alpha_{_2}=a_{_3}+b_{_1}-a_{_2}-1,
\nonumber\\
&&\alpha_{_4}=a_{_3}-a_{_4},\;\alpha_{_5}=-a_{_5},\;\alpha_{_{7}}=b_{_2}-1,
\nonumber\\
&&\alpha_{_8}=b_{_3}-a_{_3}-1,\;\alpha_{_{9}}=b_{_4}-a_{_3}-1,
\nonumber\\
&&\alpha_{_{10}}=b_{_5}-a_{_3}-1,\;\alpha_{_{11}}=1-b_{_1},\;\alpha_{_{13}}=-a_{_3}.
\label{GKZ21f-6-1}
\end{eqnarray}
The corresponding hypergeometric solutions are written as
\begin{eqnarray}
&&\Phi_{_{[\tilde{1}3\tilde{5}7]}}^{(18),a}(\alpha,z)=
y_{_1}^{{D\over2}-1}y_{_3}^{{D\over2}-2}\sum\limits_{n_{_1}=0}^\infty
\sum\limits_{n_{_2}=0}^\infty\sum\limits_{n_{_3}=0}^\infty\sum\limits_{n_{_4}=0}^\infty
c_{_{[\tilde{1}3\tilde{5}7]}}^{(18),a}(\alpha,{\bf n})
\nonumber\\
&&\hspace{2.5cm}\times
y_{_1}^{n_{_1}}\Big({y_{_4}\over y_{_3}}\Big)^{n_{_2}}y_{_4}^{n_{_3}}
\Big({y_{_2}\over y_{_3}}\Big)^{n_{_4}}
\;,\nonumber\\
&&\Phi_{_{[\tilde{1}3\tilde{5}7]}}^{(18),b}(\alpha,z)=
y_{_1}^{{D\over2}-1}y_{_3}^{{D\over2}-3}\sum\limits_{n_{_1}=0}^\infty
\sum\limits_{n_{_2}=0}^\infty\sum\limits_{n_{_3}=0}^\infty\sum\limits_{n_{_4}=0}^\infty
c_{_{[\tilde{1}3\tilde{5}7]}}^{(18),b}(\alpha,{\bf n})
\nonumber\\
&&\hspace{2.5cm}\times
\Big({y_{_1}\over y_{_3}}\Big)^{n_{_1}}\Big({1\over y_{_3}}\Big)^{n_{_2}}
\Big({y_{_4}\over y_{_3}}\Big)^{n_{_3}}\Big({y_{_2}\over y_{_3}}\Big)^{n_{_4}}
\;,\nonumber\\
&&\Phi_{_{[\tilde{1}3\tilde{5}7]}}^{(18),c}(\alpha,z)=
y_{_1}^{{D\over2}}y_{_3}^{{D\over2}-3}\sum\limits_{n_{_1}=0}^\infty
\sum\limits_{n_{_2}=0}^\infty\sum\limits_{n_{_3}=0}^\infty\sum\limits_{n_{_4}=0}^\infty
c_{_{[\tilde{1}3\tilde{5}7]}}^{(18),c}(\alpha,{\bf n})
\nonumber\\
&&\hspace{2.5cm}\times
y_{_1}^{n_{_1}}\Big({y_{_1}\over y_{_3}}\Big)^{n_{_2}}
\Big({y_{_4}\over y_{_3}}\Big)^{n_{_3}}\Big({y_{_2}\over y_{_3}}\Big)^{n_{_4}}\;.
\label{GKZ21f-6-2a}
\end{eqnarray}
Where the coefficients are
\begin{eqnarray}
&&c_{_{[\tilde{1}3\tilde{5}7]}}^{(18),a}(\alpha,{\bf n})=
(-)^{n_{_1}+n_{_4}}\Gamma(1+n_{_1}+n_{_3})\Gamma(1+n_{_2}+n_{_4})
\Big\{n_{_1}!n_{_2}!n_{_3}!n_{_4}!
\nonumber\\
&&\hspace{2.5cm}\times
\Gamma({D\over2}-1-n_{_1}-n_{_3})
\Gamma({D\over2}+n_{_2})\Gamma({D\over2}-1-n_{_2}-n_{_4})
\nonumber\\
&&\hspace{2.5cm}\times
\Gamma(2-{D\over2}+n_{_3})
\Gamma({D\over2}+n_{_1})\Gamma(2-{D\over2}+n_{_4})\Big\}^{-1}
\;,\nonumber\\
&&c_{_{[\tilde{1}3\tilde{5}7]}}^{(18),b}(\alpha,{\bf n})=
(-)^{1+n_{_4}}\Gamma(1+n_{_1}+n_{_2})\Gamma(2+n_{_1}+n_{_2}+n_{_3}+n_{_4})
\Big\{n_{_1}!n_{_2}!n_{_4}!
\nonumber\\
&&\hspace{2.5cm}\times
\Gamma(2+n_{_1}+n_{_2}+n_{_3})
\Gamma({D\over2}+n_{_2})\Gamma({D\over2}+1+n_{_1}+n_{_2}+n_{_3})
\nonumber\\
&&\hspace{2.5cm}\times
\Gamma({D\over2}-n_{_1}-n_{_2}-n_{_3}-n_{_4})
\Gamma(1-{D\over2}-n_{_1}-n_{_2})
\Gamma({D\over2}+n_{_1})
\nonumber\\
&&\hspace{2.5cm}\times
\Gamma(2-{D\over2}+n_{_4})\Big\}^{-1}
\;,\nonumber\\
&&c_{_{[\tilde{1}3\tilde{5}7]}}^{(18),c}(\alpha,{\bf n})=
(-)^{1+n_{_1}+n_{_4}}\Gamma(1+n_{_1})\Gamma(1+n_{_2})\Gamma(2+n_{_2}+n_{_3}+n_{_4})
\nonumber\\
&&\hspace{2.5cm}\times
\Big\{n_{_4}!\Gamma(2+n_{_1}+n_{_2})\Gamma(2+n_{_2}+n_{_3})
\Gamma({D\over2}-1-n_{_1})
\nonumber\\
&&\hspace{2.5cm}\times
\Gamma({D\over2}+1+n_{_2}+n_{_3})\Gamma({D\over2}-n_{_2}-n_{_3}-n_{_4})
\nonumber\\
&&\hspace{2.5cm}\times
\Gamma(1-{D\over2}-n_{_2})\Gamma({D\over2}+1+n_{_1}+n_{_2})
\Gamma(2-{D\over2}+n_{_4})\Big\}^{-1}\;.
\label{GKZ21f-6-3}
\end{eqnarray}

\item   $I_{_{19}}=\{1,2,3,5,8,9,11,\cdots,14\}$, i.e. the implement $J_{_{19}}=[1,14]\setminus I_{_{19}}=\{4,6,7,14\}$.
The choice implies the power numbers $\alpha_{_4}=\alpha_{_{6}}=\alpha_{_{7}}=\alpha_{_{14}}=0$, and
\begin{eqnarray}
&&\alpha_{_1}=a_{_4}+b_{_1}-a_{_1}-1,\;\alpha_{_2}=a_{_4}+b_{_1}-a_{_2}-1,
\nonumber\\
&&\alpha_{_3}=a_{_4}-a_{_3},\;\alpha_{_5}=-a_{_5},\;\alpha_{_8}=b_{_2}+b_{_3}-a_{_4}-2,
\nonumber\\
&&\alpha_{_{9}}=b_{_4}-a_{_4}-1,\;\alpha_{_{10}}=b_{_5}-a_{_4}-1,
\nonumber\\
&&\alpha_{_{11}}=1-b_{_1},\;\alpha_{_{12}}=1-b_{_2},\;\alpha_{_{13}}=b_{_2}-a_{_4}-1.
\label{GKZ21f-11-1}
\end{eqnarray}
The corresponding hypergeometric functions are
\begin{eqnarray}
&&\Phi_{_{[\tilde{1}3\tilde{5}7]}}^{(19),a}(\alpha,z)=
y_{_1}^{{D\over2}-1}y_{_2}^{{D\over2}-1}y_{_3}^{{D\over2}-2}\sum\limits_{n_{_1}=0}^\infty
\sum\limits_{n_{_2}=0}^\infty\sum\limits_{n_{_3}=0}^\infty\sum\limits_{n_{_4}=0}^\infty
c_{_{[\tilde{1}3\tilde{5}7]}}^{(19),a}(\alpha,{\bf n})
\nonumber\\
&&\hspace{2.5cm}\times
y_{_1}^{n_{_1}}\Big({y_{_4}\over y_{_3}}\Big)^{n_{_2}}y_{_4}^{n_{_3}}
\Big({y_{_2}\over y_{_3}}\Big)^{n_{_4}}
\;,\nonumber\\
&&\Phi_{_{[\tilde{1}3\tilde{5}7]}}^{(19),b}(\alpha,z)=
y_{_1}^{{D\over2}-1}y_{_2}^{{D\over2}-1}y_{_3}^{{D\over2}-3}\sum\limits_{n_{_1}=0}^\infty
\sum\limits_{n_{_2}=0}^\infty\sum\limits_{n_{_3}=0}^\infty\sum\limits_{n_{_4}=0}^\infty
c_{_{[\tilde{1}3\tilde{5}7]}}^{(19),b}(\alpha,{\bf n})
\nonumber\\
&&\hspace{2.5cm}\times
\Big({y_{_1}\over y_{_3}}\Big)^{n_{_1}}\Big({1\over y_{_3}}\Big)^{n_{_2}}
\Big({y_{_4}\over y_{_3}}\Big)^{n_{_3}}\Big({y_{_2}\over y_{_3}}\Big)^{n_{_4}}
\;,\nonumber\\
&&\Phi_{_{[\tilde{1}3\tilde{5}7]}}^{(19),c}(\alpha,z)=
y_{_1}^{{D\over2}}y_{_2}^{{D\over2}-1}y_{_3}^{{D\over2}-3}\sum\limits_{n_{_1}=0}^\infty
\sum\limits_{n_{_2}=0}^\infty\sum\limits_{n_{_3}=0}^\infty\sum\limits_{n_{_4}=0}^\infty
c_{_{[\tilde{1}3\tilde{5}7]}}^{(19),c}(\alpha,{\bf n})
\nonumber\\
&&\hspace{2.5cm}\times
y_{_1}^{n_{_1}}\Big({y_{_1}\over y_{_3}}\Big)^{n_{_2}}
\Big({y_{_4}\over y_{_3}}\Big)^{n_{_3}}\Big({y_{_2}\over y_{_3}}\Big)^{n_{_4}}\;.
\label{GKZ21f-11-2a}
\end{eqnarray}
Where the coefficients are
\begin{eqnarray}
&&c_{_{[\tilde{1}3\tilde{5}7]}}^{(19),a}(\alpha,{\bf n})=
(-)^{n_{_1}+n_{_4}}\Gamma(1+n_{_1}+n_{_3})\Gamma(1+n_{_2}+n_{_4})
\Big\{n_{_1}!n_{_2}!n_{_3}!n_{_4}!
\nonumber\\
&&\hspace{2.5cm}\times
\Gamma(1-{D\over2}-n_{_1}-n_{_3})\Gamma(2-{D\over2}+n_{_2})\Gamma({D\over2}+n_{_3})
\nonumber\\
&&\hspace{2.5cm}\times
\Gamma({D\over2}+n_{_1})\Gamma({D\over2}+n_{_4})
\Gamma({D\over2}-1-n_{_2}-n_{_4})\Big\}^{-1}
\;,\nonumber\\
&&c_{_{[\tilde{1}3\tilde{5}7]}}^{(19),b}(\alpha,{\bf n})=
(-)^{1+n_{_4}}\Gamma(1+n_{_1}+n_{_2})\Gamma(2+n_{_1}+n_{_2}+n_{_3}+n_{_4})
\nonumber\\
&&\hspace{2.5cm}\times
\Big\{n_{_1}!n_{_2}!n_{_4}!\Gamma(2+n_{_1}+n_{_2}+n_{_3})
\Gamma(2-{D\over2}+n_{_2})
\nonumber\\
&&\hspace{2.5cm}\times
\Gamma(3-{D\over2}+n_{_1}+n_{_2}+n_{_3})\Gamma({D\over2}-1-n_{_1}-n_{_2})
\nonumber\\
&&\hspace{2.5cm}\times
\Gamma({D\over2}+n_{_1})\Gamma({D\over2}+n_{_4})
\Gamma({D\over2}-2-n_{_1}-n_{_2}-n_{_3}-n_{_4})\Big\}^{-1}
\;,\nonumber\\
&&c_{_{[\tilde{1}3\tilde{5}7]}}^{(19),c}(\alpha,{\bf n})=
(-)^{1+n_{_1}+n_{_4}}\Gamma(1+n_{_1})\Gamma(1+n_{_2})\Gamma(2+n_{_2}+n_{_3}+n_{_4})
\nonumber\\
&&\hspace{2.5cm}\times
\Big\{n_{_4}!\Gamma(2+n_{_1}+n_{_2})\Gamma(2+n_{_2}+n_{_3})
\Gamma(1-{D\over2}-n_{_1})
\nonumber\\
&&\hspace{2.5cm}\times
\Gamma(3-{D\over2}+n_{_2}+n_{_3})\Gamma({D\over2}-1-n_{_2})
\Gamma({D\over2}+1+n_{_1}+n_{_2})
\nonumber\\
&&\hspace{2.5cm}\times
\Gamma({D\over2}+n_{_4})
\Gamma({D\over2}-2-n_{_2}-n_{_3}-n_{_4})\Big\}^{-1}\;.
\label{GKZ21f-11-3}
\end{eqnarray}

\item   $I_{_{20}}=\{1,2,3,5,7,\cdots,11,13\}$, i.e. the implement $J_{_{20}}=[1,14]\setminus I_{_{20}}=\{4,6,12,14\}$.
The choice implies the power numbers $\alpha_{_4}=\alpha_{_{6}}=\alpha_{_{12}}=\alpha_{_{14}}=0$, and
\begin{eqnarray}
&&\alpha_{_1}=a_{_4}+b_{_1}-a_{_1}-1,\;\alpha_{_2}=a_{_4}+b_{_1}-a_{_2}-1,
\nonumber\\
&&\alpha_{_3}=a_{_4}-a_{_3},\;\alpha_{_5}=-a_{_5},\;\alpha_{_{7}}=b_{_2}-1,
\nonumber\\
&&\alpha_{_8}=b_{_3}-a_{_4}-1,\;\alpha_{_{9}}=b_{_4}-a_{_4}-1,
\nonumber\\
&&\alpha_{_{10}}=b_{_5}-a_{_4}-1,\;\alpha_{_{11}}=1-b_{_1},\;\alpha_{_{13}}=-a_{_4}.
\label{GKZ21f-12-1}
\end{eqnarray}
The corresponding hypergeometric functions are written as
\begin{eqnarray}
&&\Phi_{_{[\tilde{1}3\tilde{5}7]}}^{(20),a}(\alpha,z)=
y_{_1}^{{D\over2}-1}y_{_3}^{{D}-3}\sum\limits_{n_{_1}=0}^\infty
\sum\limits_{n_{_2}=0}^\infty\sum\limits_{n_{_3}=0}^\infty\sum\limits_{n_{_4}=0}^\infty
c_{_{[\tilde{1}3\tilde{5}7]}}^{(20),a}(\alpha,{\bf n})
\nonumber\\
&&\hspace{2.5cm}\times
y_{_1}^{n_{_1}}\Big({y_{_4}\over y_{_3}}\Big)^{n_{_2}}y_{_4}^{n_{_3}}
\Big({y_{_2}\over y_{_3}}\Big)^{n_{_4}}
\;,\nonumber\\
&&\Phi_{_{[\tilde{1}3\tilde{5}7]}}^{(20),b}(\alpha,z)=
y_{_1}^{{D\over2}-1}y_{_3}^{{D}-4}\sum\limits_{n_{_1}=0}^\infty
\sum\limits_{n_{_2}=0}^\infty\sum\limits_{n_{_3}=0}^\infty\sum\limits_{n_{_4}=0}^\infty
c_{_{[\tilde{1}3\tilde{5}7]}}^{(20),b}(\alpha,{\bf n})
\nonumber\\
&&\hspace{2.5cm}\times
\Big({y_{_1}\over y_{_3}}\Big)^{n_{_1}}\Big({1\over y_{_3}}\Big)^{n_{_2}}
\Big({y_{_4}\over y_{_3}}\Big)^{n_{_3}}\Big({y_{_2}\over y_{_3}}\Big)^{n_{_4}}
\;,\nonumber\\
&&\Phi_{_{[\tilde{1}3\tilde{5}7]}}^{(20),c}(\alpha,z)=
y_{_1}^{{D\over2}}y_{_3}^{{D}-4}\sum\limits_{n_{_1}=0}^\infty
\sum\limits_{n_{_2}=0}^\infty\sum\limits_{n_{_3}=0}^\infty\sum\limits_{n_{_4}=0}^\infty
c_{_{[\tilde{1}3\tilde{5}7]}}^{(20),c}(\alpha,{\bf n})
\nonumber\\
&&\hspace{2.5cm}\times
y_{_1}^{n_{_1}}\Big({y_{_1}\over y_{_3}}\Big)^{n_{_2}}
\Big({y_{_4}\over y_{_3}}\Big)^{n_{_3}}\Big({y_{_2}\over y_{_3}}\Big)^{n_{_4}}\;.
\label{GKZ21f-12-2a}
\end{eqnarray}
Where the coefficients are
\begin{eqnarray}
&&c_{_{[\tilde{1}3\tilde{5}7]}}^{(20),a}(\alpha,{\bf n})=
(-)^{n_{_1}+n_{_2}}\Gamma(1+n_{_1}+n_{_3})\Big\{n_{_1}!n_{_2}!n_{_3}!n_{_4}!
\Gamma(1-{D\over2}-n_{_1}-n_{_3})
\nonumber\\
&&\hspace{2.5cm}\times
\Gamma(2-{D\over2}+n_{_2})\Gamma(2-{D\over2}+n_{_4})
\Gamma({D\over2}-1-n_{_2}-n_{_4})
\nonumber\\
&&\hspace{2.5cm}\times
\Gamma({D\over2}+n_{_3})\Gamma({D\over2}+n_{_1})\Gamma(D-2-n_{_2}-n_{_4})\Big\}^{-1}
\;,\nonumber\\
&&c_{_{[\tilde{1}3\tilde{5}7]}}^{(20),b}(\alpha,{\bf n})=
(-)^{n_{_1}+n_{_2}+n_{_3}}\Gamma(1+n_{_1}+n_{_2})\Big\{n_{_1}!n_{_2}!n_{_4}!
\Gamma(2+n_{_1}+n_{_2}+n_{_3})
\nonumber\\
&&\hspace{2.5cm}\times
\Gamma(2-{D\over2}+n_{_2})\Gamma(3-{D\over2}+n_{_1}+n_{_2}+n_{_3})\Gamma(2-{D\over2}+n_{_4})
\nonumber\\
&&\hspace{2.5cm}\times
\Gamma({D\over2}-2-n_{_1}-n_{_2}-n_{_3}-n_{_4})\Gamma({D\over2}-1-n_{_1}-n_{_2})
\nonumber\\
&&\hspace{2.5cm}\times
\Gamma({D\over2}+n_{_1})\Gamma(D-3-n_{_1}-n_{_2}-n_{_3}-n_{_4})\Big\}^{-1}
\;,\nonumber\\
&&c_{_{[\tilde{1}3\tilde{5}7]}}^{(20),c}(\alpha,{\bf n})=
(-)^{n_{_1}+n_{_2}+n_{_3}}\Gamma(1+n_{_1})\Gamma(1+n_{_2})\Big\{n_{_4}!
\Gamma(2+n_{_1}+n_{_2})
\nonumber\\
&&\hspace{2.5cm}\times
\Gamma(2+n_{_2}+n_{_3})\Gamma(1-{D\over2}-n_{_1})\Gamma(3-{D\over2}+n_{_2}+n_{_3})
\nonumber\\
&&\hspace{2.5cm}\times
\Gamma(2-{D\over2}+n_{_4})\Gamma({D\over2}-2-n_{_2}-n_{_3}-n_{_4})
\Gamma({D\over2}-1-n_{_2})
\nonumber\\
&&\hspace{2.5cm}\times
\Gamma({D\over2}+1+n_{_1}+n_{_2})\Gamma(D-3-n_{_2}-n_{_3}-n_{_4})\Big\}^{-1}\;.
\label{GKZ21f-12-3}
\end{eqnarray}

\item   $I_{_{21}}=\{1,2,4,5,6,8,9,10,12,13\}$, i.e. the implement $J_{_{21}}=[1,14]\setminus I_{_{21}}=\{3,7,11,14\}$.
The choice implies the power numbers $\alpha_{_3}=\alpha_{_{7}}=\alpha_{_{11}}=\alpha_{_{14}}=0$, and
\begin{eqnarray}
&&\alpha_{_1}=a_{_3}-a_{_1},\;\alpha_{_2}=a_{_3}-a_{_2},\;\alpha_{_4}=a_{_3}-a_{_4},
\nonumber\\
&&\alpha_{_5}=-a_{_5},\;\alpha_{_{6}}=b_{_1}-1,\;\alpha_{_8}=b_{_2}+b_{_3}-a_{_3}-2,
\nonumber\\
&&\alpha_{_{9}}=b_{_4}-a_{_3}-1,\;\alpha_{_{10}}=b_{_5}-a_{_3}-1,
\nonumber\\
&&\alpha_{_{12}}=1-b_{_2},\;\alpha_{_{13}}=b_{_2}-a_{_3}-1.
\label{GKZ21f-17-1}
\end{eqnarray}
The corresponding hypergeometric solutions are
\begin{eqnarray}
&&\Phi_{_{[\tilde{1}3\tilde{5}7]}}^{(21),a}(\alpha,z)=
y_{_2}^{{D\over2}-1}y_{_3}^{-1}\sum\limits_{n_{_1}=0}^\infty
\sum\limits_{n_{_2}=0}^\infty\sum\limits_{n_{_3}=0}^\infty\sum\limits_{n_{_4}=0}^\infty
c_{_{[\tilde{1}3\tilde{5}7]}}^{(21),a}(\alpha,{\bf n})
\nonumber\\
&&\hspace{2.5cm}\times
y_{_1}^{n_{_1}}\Big({y_{_4}\over y_{_3}}\Big)^{n_{_2}}y_{_4}^{n_{_3}}
\Big({y_{_2}\over y_{_3}}\Big)^{n_{_4}}
\;,\nonumber\\
&&\Phi_{_{[\tilde{1}3\tilde{5}7]}}^{(21),b}(\alpha,z)=
y_{_2}^{{D\over2}-1}y_{_3}^{-2}\sum\limits_{n_{_1}=0}^\infty
\sum\limits_{n_{_2}=0}^\infty\sum\limits_{n_{_3}=0}^\infty\sum\limits_{n_{_4}=0}^\infty
c_{_{[\tilde{1}3\tilde{5}7]}}^{(21),b}(\alpha,{\bf n})
\nonumber\\
&&\hspace{2.5cm}\times
y_{_1}^{n_{_1}}\Big({1\over y_{_3}}\Big)^{n_{_2}}
\Big({y_{_4}\over y_{_3}}\Big)^{n_{_3}}\Big({y_{_2}\over y_{_3}}\Big)^{n_{_4}}\;.
\label{GKZ21f-17-2a}
\end{eqnarray}
Where the coefficients are
\begin{eqnarray}
&&c_{_{[\tilde{1}3\tilde{5}7]}}^{(21),a}(\alpha,{\bf n})=
(-)^{n_{_3}+n_{_4}}\Gamma(1+n_{_2}+n_{_4})
\Big\{n_{_1}!n_{_2}!n_{_3}!n_{_4}!\Gamma({D\over2}-1-n_{_1}-n_{_3})
\nonumber\\
&&\hspace{2.5cm}\times
\Gamma(D-2-n_{_1}-n_{_3})\Gamma({D\over2}+n_{_2})\Gamma(2-{D\over2}+n_{_1})
\nonumber\\
&&\hspace{2.5cm}\times
\Gamma(1-{D\over2}-n_{_2}-n_{_4})\Gamma(2-{D\over2}+n_{_3})
\Gamma({D\over2}+n_{_4})\Big\}^{-1}
\;,\nonumber\\
&&c_{_{[\tilde{1}3\tilde{5}7]}}^{(21),b}(\alpha,{\bf n})=
(-)^{1+n_{_4}}\Gamma(1+n_{_2})\Gamma(2+n_{_2}+n_{_3}+n_{_4})
\Big\{n_{_1}!n_{_4}!\Gamma(2+n_{_2}+n_{_3})
\nonumber\\
&&\hspace{2.5cm}\times
\Gamma({D\over2}-n_{_1}+n_{_2})\Gamma(D-1-n_{_1}+n_{_2})
\Gamma({D\over2}+1+n_{_2}+n_{_3})
\nonumber\\
&&\hspace{2.5cm}\times
\Gamma(2-{D\over2}+n_{_1})
\Gamma(-{D\over2}-n_{_2}-n_{_3}-n_{_4})\Gamma(1-{D\over2}-n_{_2})
\nonumber\\
&&\hspace{2.5cm}\times
\Gamma({D\over2}+n_{_4})\Big\}^{-1}\;.
\label{GKZ21f-17-3}
\end{eqnarray}

\item   $I_{_{22}}=\{1,2,4,\cdots,10,13\}$, i.e. the implement $J_{_{22}}=[1,14]\setminus I_{_{22}}=\{3,11,12,14\}$.
The choice implies the power numbers $\alpha_{_3}=\alpha_{_{11}}=\alpha_{_{12}}=\alpha_{_{14}}=0$, and
\begin{eqnarray}
&&\alpha_{_1}=a_{_3}-a_{_1},\;\alpha_{_2}=a_{_3}-a_{_2},
\nonumber\\
&&\alpha_{_4}=a_{_3}-a_{_4},\;\alpha_{_5}=-a_{_5},\;\alpha_{_{6}}=b_{_1}-1,
\nonumber\\
&&\alpha_{_{7}}=b_{_2}-1,\;\alpha_{_8}=b_{_3}-a_{_3}-1,\;\alpha_{_{9}}=b_{_4}-a_{_3}-1,
\nonumber\\
&&\alpha_{_{10}}=b_{_5}-a_{_3}-1,\;\alpha_{_{13}}=-a_{_3}.
\label{GKZ21f-18-1}
\end{eqnarray}
The corresponding hypergeometric solutions are
\begin{eqnarray}
&&\Phi_{_{[\tilde{1}3\tilde{5}7]}}^{(22),a}(\alpha,z)=
y_{_3}^{{D\over2}-2}\sum\limits_{n_{_1}=0}^\infty
\sum\limits_{n_{_2}=0}^\infty\sum\limits_{n_{_3}=0}^\infty\sum\limits_{n_{_4}=0}^\infty
c_{_{[\tilde{1}3\tilde{5}7]}}^{(22),a}(\alpha,{\bf n})
\nonumber\\
&&\hspace{2.5cm}\times
y_{_1}^{n_{_1}}\Big({y_{_4}\over y_{_3}}\Big)^{n_{_2}}y_{_4}^{n_{_3}}
\Big({y_{_2}\over y_{_3}}\Big)^{n_{_4}}
\;,\nonumber\\
&&\Phi_{_{[\tilde{1}3\tilde{5}7]}}^{(22),b}(\alpha,z)=
y_{_3}^{{D\over2}-3}\sum\limits_{n_{_1}=0}^\infty
\sum\limits_{n_{_2}=0}^\infty\sum\limits_{n_{_3}=0}^\infty\sum\limits_{n_{_4}=0}^\infty
c_{_{[\tilde{1}3\tilde{5}7]}}^{(22),b}(\alpha,{\bf n})
\nonumber\\
&&\hspace{2.5cm}\times
y_{_1}^{n_{_1}}\Big({1\over y_{_3}}\Big)^{n_{_2}}
\Big({y_{_4}\over y_{_3}}\Big)^{n_{_3}}\Big({y_{_2}\over y_{_3}}\Big)^{n_{_4}}\;.
\label{GKZ21f-18-2a}
\end{eqnarray}
Where the coefficients are
\begin{eqnarray}
&&c_{_{[\tilde{1}3\tilde{5}7]}}^{(22),a}(\alpha,{\bf n})=
(-)^{n_{_3}+n_{_4}}\Gamma(1+n_{_2}+n_{_4})
\Big\{n_{_1}!n_{_2}!n_{_3}!n_{_4}!\Gamma({D\over2}-1-n_{_1}-n_{_3})
\nonumber\\
&&\hspace{2.5cm}\times
\Gamma(D-2-n_{_1}-n_{_3})\Gamma({D\over2}+n_{_2})\Gamma(2-{D\over2}+n_{_1})
\nonumber\\
&&\hspace{2.5cm}\times
\Gamma({D\over2}-1-n_{_2}-n_{_4})\Gamma(2-{D\over2}+n_{_3})
\Gamma(2-{D\over2}+n_{_4})\Big\}^{-1}
\;,\nonumber\\
&&c_{_{[\tilde{1}3\tilde{5}7]}}^{(22),b}(\alpha,{\bf n})=
(-)^{1+n_{_4}}\Gamma(1+n_{_2})\Gamma(2+n_{_2}+n_{_3}+n_{_4})
\Big\{n_{_1}!n_{_4}!\Gamma(2+n_{_2}+n_{_3})
\nonumber\\
&&\hspace{2.5cm}\times
\Gamma({D\over2}-n_{_1}+n_{_2})\Gamma(D-1-n_{_1}+n_{_2})
\Gamma({D\over2}+1+n_{_2}+n_{_3})
\nonumber\\
&&\hspace{2.5cm}\times
\Gamma(2-{D\over2}+n_{_1})
\Gamma({D\over2}-2-n_{_2}-n_{_3}-n_{_4})\Gamma(1-{D\over2}-n_{_2})
\nonumber\\
&&\hspace{2.5cm}\times
\Gamma(2-{D\over2}+n_{_4})\Big\}^{-1}\;.
\label{GKZ21f-18-3}
\end{eqnarray}

\item   $I_{_{23}}=\{1,2,3,5,6,8,9,10,12,13\}$, i.e. the implement $J_{_{23}}=[1,14]\setminus I_{_{23}}=\{4,7,11,14\}$.
The choice implies the power numbers $\alpha_{_4}=\alpha_{_{7}}=\alpha_{_{11}}=\alpha_{_{14}}=0$, and
\begin{eqnarray}
&&\alpha_{_1}=a_{_4}-a_{_1},\;\alpha_{_2}=a_{_4}-a_{_2},\;\alpha_{_3}=a_{_4}-a_{_3},
\nonumber\\
&&\alpha_{_5}=-a_{_5},\;\alpha_{_{6}}=b_{_1}-1,\;\alpha_{_8}=b_{_2}+b_{_3}-a_{_4}-2,
\nonumber\\
&&\alpha_{_{9}}=b_{_4}-a_{_4}-1,\;\alpha_{_{10}}=b_{_5}-a_{_4}-1,
\nonumber\\
&&\alpha_{_{12}}=1-b_{_2},\;\alpha_{_{13}}=b_{_2}-a_{_4}-1.
\label{GKZ21f-23-1}
\end{eqnarray}
The corresponding hypergeometric functions are presented as
\begin{eqnarray}
&&\Phi_{_{[\tilde{1}3\tilde{5}7]}}^{(23),a}(\alpha,z)=
y_{_2}^{{D\over2}-1}y_{_3}^{{D\over2}-2}\sum\limits_{n_{_1}=0}^\infty
\sum\limits_{n_{_2}=0}^\infty\sum\limits_{n_{_3}=0}^\infty\sum\limits_{n_{_4}=0}^\infty
c_{_{[\tilde{1}3\tilde{5}7]}}^{(23),a}(\alpha,{\bf n})
\nonumber\\
&&\hspace{2.5cm}\times
y_{_1}^{n_{_1}}\Big({y_{_4}\over y_{_3}}\Big)^{n_{_2}}y_{_4}^{n_{_3}}
\Big({y_{_2}\over y_{_3}}\Big)^{n_{_4}}
\;,\nonumber\\
&&\Phi_{_{[\tilde{1}3\tilde{5}7]}}^{(23),b}(\alpha,z)=
y_{_2}^{{D\over2}-1}y_{_3}^{{D\over2}-3}\sum\limits_{n_{_1}=0}^\infty
\sum\limits_{n_{_2}=0}^\infty\sum\limits_{n_{_3}=0}^\infty\sum\limits_{n_{_4}=0}^\infty
c_{_{[\tilde{1}3\tilde{5}7]}}^{(23),b}(\alpha,{\bf n})
\nonumber\\
&&\hspace{2.5cm}\times
\Big({y_{_1}\over y_{_3}}\Big)^{n_{_1}}\Big({1\over y_{_3}}\Big)^{n_{_2}}
\Big({y_{_4}\over y_{_3}}\Big)^{n_{_3}}\Big({y_{_2}\over y_{_3}}\Big)^{n_{_4}}
\;,\nonumber\\
&&\Phi_{_{[\tilde{1}3\tilde{5}7]}}^{(23),c}(\alpha,z)=
y_{_1}y_{_2}^{{D\over2}-1}y_{_3}^{{D\over2}-3}\sum\limits_{n_{_1}=0}^\infty
\sum\limits_{n_{_2}=0}^\infty\sum\limits_{n_{_3}=0}^\infty\sum\limits_{n_{_4}=0}^\infty
c_{_{[\tilde{1}3\tilde{5}7]}}^{(23),c}(\alpha,{\bf n})
\nonumber\\
&&\hspace{2.5cm}\times
y_{_1}^{n_{_1}}\Big({y_{_1}\over y_{_3}}\Big)^{n_{_2}}
\Big({y_{_4}\over y_{_3}}\Big)^{n_{_3}}\Big({y_{_2}\over y_{_3}}\Big)^{n_{_4}}\;.
\label{GKZ21f-23-2a}
\end{eqnarray}
Where the coefficients are
\begin{eqnarray}
&&c_{_{[\tilde{1}3\tilde{5}7]}}^{(23),a}(\alpha,{\bf n})=
(-)^{n_{_1}+n_{_4}}\Gamma(1+n_{_1}+n_{_3})\Gamma(1+n_{_2}+n_{_4})\Big\{n_{_1}!n_{_2}!n_{_3}!n_{_4}!
\nonumber\\
&&\hspace{2.5cm}\times
\Gamma({D\over2}-1-n_{_1}-n_{_3})\Gamma(2-{D\over2}+n_{_2})\Gamma(2-{D\over2}+n_{_1})
\nonumber\\
&&\hspace{2.5cm}\times
\Gamma({D\over2}+n_{_3})
\Gamma({D\over2}+n_{_4})\Gamma({D\over2}-1-n_{_2}-n_{_4})\Big\}^{-1}
\;,\nonumber\\
&&c_{_{[\tilde{1}3\tilde{5}7]}}^{(23),b}(\alpha,{\bf n})=
(-)^{1+n_{_4}}\Gamma(1+n_{_1}+n_{_2})\Gamma(2+n_{_1}+n_{_2}+n_{_3}+n_{_4})
\Big\{n_{_1}!n_{_2}!n_{_4}!
\nonumber\\
&&\hspace{2.5cm}\times
\Gamma(2+n_{_1}+n_{_2}+n_{_3})\Gamma({D\over2}+n_{_2})
\Gamma({D\over2}+1+n_{_1}+n_{_2}+n_{_3})
\nonumber\\
&&\hspace{2.5cm}\times
\Gamma(2-{D\over2}+n_{_1})\Gamma({D\over2}-1-n_{_1}-n_{_2})
\Gamma({D\over2}+n_{_4})
\nonumber\\
&&\hspace{2.5cm}\times
\Gamma({D\over2}-2-n_{_1}-n_{_2}-n_{_3}-n_{_4})\Big\}^{-1}
\;,\nonumber\\
&&c_{_{[\tilde{1}3\tilde{5}7]}}^{(23),c}(\alpha,{\bf n})=
(-)^{1+n_{_1}+n_{_4}}\Gamma(1+n_{_1})\Gamma(1+n_{_2})\Gamma(2+n_{_2}+n_{_3}+n_{_4})
\nonumber\\
&&\hspace{2.5cm}\times
\Big\{n_{_4}!\Gamma(2+n_{_1}+n_{_2})\Gamma(2+n_{_2}+n_{_3})\Gamma({D\over2}-1-n_{_1})
\nonumber\\
&&\hspace{2.5cm}\times
\Gamma({D\over2}+1+n_{_2}+n_{_3})\Gamma(3-{D\over2}+n_{_1}+n_{_2})\Gamma({D\over2}-1-n_{_2})
\nonumber\\
&&\hspace{2.5cm}\times
\Gamma({D\over2}+n_{_4})\Gamma({D\over2}-2-n_{_2}-n_{_3}-n_{_4})\Big\}^{-1}\;.
\label{GKZ21f-23-3}
\end{eqnarray}

\item   $I_{_{24}}=\{1,2,3,5,\cdots,10,13\}$, i.e. the implement $J_{_{24}}=[1,14]\setminus I_{_{24}}=\{4,11,12,14\}$.
The choice implies the power numbers $\alpha_{_4}=\alpha_{_{11}}=\alpha_{_{12}}=\alpha_{_{14}}=0$, and
\begin{eqnarray}
&&\alpha_{_1}=a_{_4}-a_{_1},\;\alpha_{_2}=a_{_4}-a_{_2},\;\alpha_{_3}=a_{_4}-a_{_3},
\nonumber\\
&&\alpha_{_5}=-a_{_5},\;\alpha_{_{6}}=b_{_1}-1,\;\alpha_{_{7}}=b_{_2}-1,
\nonumber\\
&&\alpha_{_8}=b_{_3}-a_{_4}-1,\;\alpha_{_{9}}=b_{_4}-a_{_4}-1,
\nonumber\\
&&\alpha_{_{10}}=b_{_5}-a_{_4}-1,\;\alpha_{_{13}}=-a_{_4}.
\label{GKZ21f-24-1}
\end{eqnarray}
The corresponding hypergeometric solutions are written as
\begin{eqnarray}
&&\Phi_{_{[\tilde{1}3\tilde{5}7]}}^{(24),a}(\alpha,z)=
y_{_3}^{{D}-3}\sum\limits_{n_{_1}=0}^\infty
\sum\limits_{n_{_2}=0}^\infty\sum\limits_{n_{_3}=0}^\infty\sum\limits_{n_{_4}=0}^\infty
c_{_{[\tilde{1}3\tilde{5}7]}}^{(24),a}(\alpha,{\bf n})
\nonumber\\
&&\hspace{2.5cm}\times
y_{_1}^{n_{_1}}\Big({y_{_4}\over y_{_3}}\Big)^{n_{_2}}y_{_4}^{n_{_3}}
\Big({y_{_2}\over y_{_3}}\Big)^{n_{_4}}
\;,\nonumber\\
&&\Phi_{_{[\tilde{1}3\tilde{5}7]}}^{(24),b}(\alpha,z)=
y_{_3}^{{D}-4}\sum\limits_{n_{_1}=0}^\infty
\sum\limits_{n_{_2}=0}^\infty\sum\limits_{n_{_3}=0}^\infty\sum\limits_{n_{_4}=0}^\infty
c_{_{[\tilde{1}3\tilde{5}7]}}^{(24),b}(\alpha,{\bf n})
\nonumber\\
&&\hspace{2.5cm}\times
\Big({y_{_1}\over y_{_3}}\Big)^{n_{_1}}\Big({1\over y_{_3}}\Big)^{n_{_2}}
\Big({y_{_4}\over y_{_3}}\Big)^{n_{_3}}\Big({y_{_2}\over y_{_3}}\Big)^{n_{_4}}
\;,\nonumber\\
&&\Phi_{_{[\tilde{1}3\tilde{5}7]}}^{(24),c}(\alpha,z)=
y_{_1}y_{_3}^{{D}-4}\sum\limits_{n_{_1}=0}^\infty
\sum\limits_{n_{_2}=0}^\infty\sum\limits_{n_{_3}=0}^\infty\sum\limits_{n_{_4}=0}^\infty
c_{_{[\tilde{1}3\tilde{5}7]}}^{(24),c}(\alpha,{\bf n})
\nonumber\\
&&\hspace{2.5cm}\times
y_{_1}^{n_{_1}}\Big({y_{_1}\over y_{_3}}\Big)^{n_{_2}}
\Big({y_{_4}\over y_{_3}}\Big)^{n_{_3}}\Big({y_{_2}\over y_{_3}}\Big)^{n_{_4}}\;.
\label{GKZ21f-24-2a}
\end{eqnarray}
Where the coefficients are
\begin{eqnarray}
&&c_{_{[\tilde{1}3\tilde{5}7]}}^{(24),a}(\alpha,{\bf n})=
(-)^{n_{_1}}\Gamma(1+n_{_1}+n_{_3})\Big\{n_{_1}!n_{_2}!n_{_3}!n_{_4}!
\Gamma({D\over2}-1-n_{_1}-n_{_3})
\nonumber\\
&&\hspace{2.5cm}\times
\Gamma(2-{D\over2}+n_{_2})\Gamma(2-{D\over2}+n_{_1})
\Gamma(2-{D\over2}+n_{_4})\Gamma({D\over2}+n_{_3})
\nonumber\\
&&\hspace{2.5cm}\times
\Gamma({D\over2}-1-n_{_2}-n_{_4})\Gamma(D-2-n_{_2}-n_{_4})\Big\}^{-1}
\;,\nonumber\\
&&c_{_{[\tilde{1}3\tilde{5}7]}}^{(24),b}(\alpha,{\bf n})=
(-)^{n_{_1}+n_{_2}+n_{_3}}\Gamma(1+n_{_1}+n_{_2})\Big\{n_{_1}!n_{_2}!n_{_4}!\Gamma(2+n_{_1}+n_{_2}+n_{_3})
\nonumber\\
&&\hspace{2.5cm}\times
\Gamma({D\over2}+n_{_2})\Gamma(3-{D\over2}+n_{_1}+n_{_2}+n_{_3})
\Gamma(2-{D\over2}+n_{_1})
\nonumber\\
&&\hspace{2.5cm}\times
\Gamma(2-{D\over2}+n_{_4})\Gamma({D\over2}-2-n_{_1}-n_{_2}-n_{_3}-n_{_4})
\nonumber\\
&&\hspace{2.5cm}\times
\Gamma({D\over2}-1-n_{_1}-n_{_2})\Gamma(D-3-n_{_1}-n_{_2}-n_{_3}-n_{_4})\Big\}^{-1}
\;,\nonumber\\
&&c_{_{[\tilde{1}3\tilde{5}7]}}^{(24),c}(\alpha,{\bf n})=
(-)^{n_{_1}+n_{_2}+n_{_3}}\Gamma(1+n_{_1})\Gamma(1+n_{_2})\Big\{n_{_4}!
\Gamma(2+n_{_1}+n_{_2})\Gamma(2+n_{_2}+n_{_3})
\nonumber\\
&&\hspace{2.5cm}\times
\Gamma({D\over2}-1-n_{_1})\Gamma(3-{D\over2}+n_{_2}+n_{_3})
\Gamma(3-{D\over2}+n_{_1}+n_{_2})
\nonumber\\
&&\hspace{2.5cm}\times
\Gamma(2-{D\over2}+n_{_4})\Gamma({D\over2}-2-n_{_2}-n_{_3}-n_{_4})
\Gamma({D\over2}-1-n_{_2})
\nonumber\\
&&\hspace{2.5cm}\times
\Gamma(D-3-n_{_2}-n_{_3}-n_{_4})\Big\}^{-1}\;.
\label{GKZ21f-24-3}
\end{eqnarray}
\end{itemize}

\section{The hypergeometric solutions of the integer lattice ${\bf B}_{_{\tilde{1}35\tilde{7}}}$\label{app8}}
\indent\indent

\begin{itemize}
\item   $I_{_{1}}=\{1,2,4,7,9,\cdots,14\}$, i.e. the implement $J_{_{1}}=[1,14]\setminus I_{_{1}}=\{3,5,6,8\}$.
The choice implies the power numbers $\alpha_{_3}=\alpha_{_{5}}=\alpha_{_{6}}=\alpha_{_{8}}=0$, and
\begin{eqnarray}
&&\alpha_{_1}=a_{_3}+a_{_5}+b_{_1}-a_{_1}-1,\;\alpha_{_2}=a_{_3}+a_{_5}+b_{_1}-a_{_2}-1,
\nonumber\\
&&\alpha_{_4}=a_{_3}-a_{_4},\;\alpha_{_7}=b_{_2}+b_{_3}-a_{_3}-2,
\nonumber\\
&&\alpha_{_9}=b_{_4}-a_{_3}-a_{_5}-1,\;\alpha_{_{10}}=b_{_5}-a_{_3}-a_{_5}-1,
\nonumber\\
&&\alpha_{_{11}}=1-b_{_1},\;\alpha_{_{12}}=b_{_3}-a_{_3}-1,\;\alpha_{_{13}}=1-b_{_3},\;\alpha_{_{14}}=-a_{_5}.
\label{GKZ21g-1-1}
\end{eqnarray}
The corresponding hypergeometric functions are written as
\begin{eqnarray}
&&\Phi_{_{[\tilde{1}35\tilde{7}]}}^{(1),a}(\alpha,z)=
y_{_1}^{{D\over2}-1}y_{_2}^{-1}y_{_3}^{{D\over2}-1}y_{_4}^{-1}\sum\limits_{n_{_1}=0}^\infty
\sum\limits_{n_{_2}=0}^\infty\sum\limits_{n_{_3}=0}^\infty\sum\limits_{n_{_4}=0}^\infty
c_{_{[\tilde{1}35\tilde{7}]}}^{(1),a}(\alpha,{\bf n})
\nonumber\\
&&\hspace{2.5cm}\times
\Big({y_{_1}\over y_{_4}}\Big)^{n_{_1}}\Big({y_{_4}\over y_{_2}}\Big)^{n_{_2}}
\Big({1\over y_{_4}}\Big)^{n_{_3}}\Big({y_{_3}\over y_{_2}}\Big)^{n_{_4}}
\;,\nonumber\\
&&\Phi_{_{[\tilde{1}35\tilde{7}]}}^{(1),b}(\alpha,z)=
y_{_1}^{{D\over2}}y_{_2}^{-1}y_{_3}^{{D\over2}-1}y_{_4}^{-1}\sum\limits_{n_{_1}=0}^\infty
\sum\limits_{n_{_2}=0}^\infty\sum\limits_{n_{_3}=0}^\infty\sum\limits_{n_{_4}=0}^\infty
c_{_{[\tilde{1}35\tilde{7}]}}^{(1),b}(\alpha,{\bf n})
\nonumber\\
&&\hspace{2.5cm}\times
y_{_1}^{n_{_1}}\Big({y_{_1}\over y_{_2}}\Big)^{n_{_2}}
\Big({y_{_1}\over y_{_4}}\Big)^{n_{_3}}\Big({y_{_3}\over y_{_2}}\Big)^{n_{_4}}\;.
\label{GKZ21g-1-2a}
\end{eqnarray}
Where the coefficients are
\begin{eqnarray}
&&c_{_{[\tilde{1}35\tilde{7}]}}^{(1),a}(\alpha,{\bf n})=
(-)^{n_{_4}}\Gamma(1+n_{_1}+n_{_3})\Gamma(1+n_{_2}+n_{_4})
\Big\{n_{_1}!n_{_2}!n_{_3}!n_{_4}!
\nonumber\\
&&\hspace{2.5cm}\times
\Gamma({D\over2}+n_{_3})\Gamma({D\over2}+n_{_2})\Gamma(1-{D\over2}-n_{_2}-n_{_4})
\nonumber\\
&&\hspace{2.5cm}\times
\Gamma(1-{D\over2}-n_{_1}-n_{_3})\Gamma({D\over2}+n_{_1})\Gamma({D\over2}+n_{_4})\Big\}^{-1}
\;,\nonumber\\
&&c_{_{[\tilde{1}35\tilde{7}]}}^{(1),b}(\alpha,{\bf n})=
(-)^{n_{_1}+n_{_4}}\Gamma(1+n_{_1})\Gamma(1+n_{_2}+n_{_3})\Gamma(1+n_{_2}+n_{_4})
\Big\{n_{_2}!n_{_4}!
\nonumber\\
&&\hspace{2.5cm}\times
\Gamma(2+n_{_1}+n_{_2}+n_{_3})\Gamma({D\over2}-1-n_{_1})
\Gamma({D\over2}+n_{_2})
\nonumber\\
&&\hspace{2.5cm}\times
\Gamma(1-{D\over2}-n_{_2}-n_{_4})\Gamma(1-{D\over2}-n_{_2}-n_{_3})
\nonumber\\
&&\hspace{2.5cm}\times
\Gamma({D\over2}+1+n_{_1}+n_{_2}+n_{_3})\Gamma({D\over2}+n_{_4})\Big\}^{-1}\;.
\label{GKZ21g-1-3}
\end{eqnarray}

\item   $I_{_{2}}=\{1,2,4,7,\cdots,12,14\}$, i.e. the implement $J_{_{2}}=[1,14]\setminus I_{_{2}}=\{3,5,6,13\}$.
The choice implies the power numbers $\alpha_{_3}=\alpha_{_{5}}=\alpha_{_{6}}=\alpha_{_{13}}=0$, and
\begin{eqnarray}
&&\alpha_{_1}=a_{_3}+a_{_5}+b_{_1}-a_{_1}-1,\;\alpha_{_2}=a_{_3}+a_{_5}+b_{_1}-a_{_2}-1,
\nonumber\\
&&\alpha_{_4}=a_{_3}-a_{_4},\;\alpha_{_7}=b_{_2}-a_{_3}-1,\;\alpha_{_{8}}=b_{_3}-1,
\nonumber\\
&&\alpha_{_9}=b_{_4}-a_{_3}-a_{_5}-1,\;\alpha_{_{10}}=b_{_5}-a_{_3}-a_{_5}-1,
\nonumber\\
&&\alpha_{_{11}}=1-b_{_1},\;\alpha_{_{12}}=-a_{_3},\;\alpha_{_{14}}=-a_{_5}.
\label{GKZ21g-2-1}
\end{eqnarray}
The corresponding hypergeometric solutions are written as
\begin{eqnarray}
&&\Phi_{_{[\tilde{1}35\tilde{7}]}}^{(2),a}(\alpha,z)=
y_{_1}^{{D\over2}-1}y_{_2}^{{D\over2}-2}y_{_4}^{-1}\sum\limits_{n_{_1}=0}^\infty
\sum\limits_{n_{_2}=0}^\infty\sum\limits_{n_{_3}=0}^\infty\sum\limits_{n_{_4}=0}^\infty
c_{_{[\tilde{1}35\tilde{7}]}}^{(2),a}(\alpha,{\bf n})
\nonumber\\
&&\hspace{2.5cm}\times
\Big({y_{_1}\over y_{_4}}\Big)^{n_{_1}}\Big({y_{_4}\over y_{_2}}\Big)^{n_{_2}}
\Big({1\over y_{_4}}\Big)^{n_{_3}}\Big({y_{_3}\over y_{_2}}\Big)^{n_{_4}}
\;,\nonumber\\
&&\Phi_{_{[\tilde{1}35\tilde{7}]}}^{(2),b}(\alpha,z)=
y_{_1}^{{D\over2}}y_{_2}^{{D\over2}-2}y_{_4}^{-1}\sum\limits_{n_{_1}=0}^\infty
\sum\limits_{n_{_2}=0}^\infty\sum\limits_{n_{_3}=0}^\infty\sum\limits_{n_{_4}=0}^\infty
c_{_{[\tilde{1}35\tilde{7}]}}^{(2),b}(\alpha,{\bf n})
\nonumber\\
&&\hspace{2.5cm}\times
y_{_1}^{n_{_1}}\Big({y_{_1}\over y_{_2}}\Big)^{n_{_2}}
\Big({y_{_1}\over y_{_4}}\Big)^{n_{_3}}\Big({y_{_3}\over y_{_2}}\Big)^{n_{_4}}\;.
\label{GKZ21g-2-2a}
\end{eqnarray}
Where the coefficients are
\begin{eqnarray}
&&c_{_{[\tilde{1}35\tilde{7}]}}^{(2),a}(\alpha,{\bf n})=
(-)^{n_{_4}}\Gamma(1+n_{_1}+n_{_3})\Gamma(1+n_{_2}+n_{_4})
\Big\{n_{_1}!n_{_2}!n_{_3}!n_{_4}!
\nonumber\\
&&\hspace{2.5cm}\times
\Gamma({D\over2}+n_{_3})\Gamma({D\over2}+n_{_2})\Gamma({D\over2}-1-n_{_2}-n_{_4})
\nonumber\\
&&\hspace{2.5cm}\times
\Gamma(1-{D\over2}-n_{_1}-n_{_3})\Gamma({D\over2}+n_{_1})\Gamma(2-{D\over2}+n_{_4})\Big\}^{-1}
\;,\nonumber\\
&&c_{_{[\tilde{1}35\tilde{7}]}}^{(2),b}(\alpha,{\bf n})=
(-)^{n_{_1}+n_{_4}}\Gamma(1+n_{_1})\Gamma(1+n_{_2}+n_{_3})\Gamma(1+n_{_2}+n_{_4})
\Big\{n_{_2}!n_{_4}!
\nonumber\\
&&\hspace{2.5cm}\times
\Gamma(2+n_{_1}+n_{_2}+n_{_3})\Gamma({D\over2}-1-n_{_1})
\Gamma({D\over2}+n_{_2})
\nonumber\\
&&\hspace{2.5cm}\times
\Gamma({D\over2}-1-n_{_2}-n_{_4})\Gamma(1-{D\over2}-n_{_2}-n_{_3})
\nonumber\\
&&\hspace{2.5cm}\times
\Gamma({D\over2}+1+n_{_1}+n_{_2}+n_{_3})\Gamma(2-{D\over2}+n_{_4})\Big\}^{-1}\;.
\label{GKZ21g-2-3}
\end{eqnarray}

\item   $I_{_{3}}=\{1,2,3,7,9,\cdots,14\}$, i.e. the implement $J_{_{3}}=[1,14]\setminus I_{_{3}}=\{4,5,6,8\}$.
The choice implies the power numbers $\alpha_{_4}=\alpha_{_{5}}=\alpha_{_{6}}=\alpha_{_{8}}=0$, and
\begin{eqnarray}
&&\alpha_{_1}=a_{_4}+a_{_5}+b_{_1}-a_{_1}-1,\;\alpha_{_2}=a_{_4}+a_{_5}+b_{_1}-a_{_2}-1,
\nonumber\\
&&\alpha_{_3}=a_{_4}-a_{_3},\;\alpha_{_7}=b_{_2}+b_{_3}-a_{_4}-2,
\nonumber\\
&&\alpha_{_9}=b_{_4}-a_{_4}-a_{_5}-1,\;\alpha_{_{10}}=b_{_5}-a_{_4}-a_{_5}-1,
\nonumber\\
&&\alpha_{_{11}}=1-b_{_1},\;\alpha_{_{12}}=b_{_3}-a_{_4}-1,\;\alpha_{_{13}}=1-b_{_3},\;\alpha_{_{14}}=-a_{_5}.
\label{GKZ21g-3-1}
\end{eqnarray}
The corresponding hypergeometric solutions are
\begin{eqnarray}
&&\Phi_{_{[\tilde{1}35\tilde{7}]}}^{(3),a}(\alpha,z)=
y_{_1}^{{D\over2}-1}y_{_2}^{{D\over2}-2}y_{_3}^{{D\over2}-1}y_{_4}^{-1}
\sum\limits_{n_{_1}=0}^\infty
\sum\limits_{n_{_2}=0}^\infty\sum\limits_{n_{_3}=0}^\infty\sum\limits_{n_{_4}=0}^\infty
c_{_{[\tilde{1}35\tilde{7}]}}^{(3),a}(\alpha,{\bf n})
\nonumber\\
&&\hspace{2.5cm}\times
\Big({y_{_1}\over y_{_4}}\Big)^{n_{_1}}\Big({y_{_4}\over y_{_2}}\Big)^{n_{_2}}
\Big({1\over y_{_4}}\Big)^{n_{_3}}\Big({y_{_3}\over y_{_2}}\Big)^{n_{_4}}
\;,\nonumber\\
&&\Phi_{_{[\tilde{1}35\tilde{7}]}}^{(3),b}(\alpha,z)=
y_{_1}^{{D\over2}}y_{_2}^{{D\over2}-2}y_{_3}^{{D\over2}-1}y_{_4}^{-1}
\sum\limits_{n_{_1}=0}^\infty
\sum\limits_{n_{_2}=0}^\infty\sum\limits_{n_{_3}=0}^\infty\sum\limits_{n_{_4}=0}^\infty
c_{_{[\tilde{1}35\tilde{7}]}}^{(3),b}(\alpha,{\bf n})
\nonumber\\
&&\hspace{2.5cm}\times
y_{_1}^{n_{_1}}\Big({y_{_1}\over y_{_2}}\Big)^{n_{_2}}
\Big({y_{_1}\over y_{_4}}\Big)^{n_{_3}}\Big({y_{_3}\over y_{_2}}\Big)^{n_{_4}}\;.
\label{GKZ21g-3-2a}
\end{eqnarray}
Where the coefficients are
\begin{eqnarray}
&&c_{_{[\tilde{1}35\tilde{7}]}}^{(3),a}(\alpha,{\bf n})=
(-)^{n_{_4}}\Gamma(1+n_{_1}+n_{_3})\Gamma(1+n_{_2}+n_{_4})
\Big\{n_{_1}!n_{_2}!n_{_3}!n_{_4}!
\nonumber\\
&&\hspace{2.5cm}\times
\Gamma(2-{D\over2}+n_{_3})\Gamma(2-{D\over2}+n_{_2})
\Gamma({D\over2}-1-n_{_1}-n_{_3})
\nonumber\\
&&\hspace{2.5cm}\times
\Gamma({D\over2}+n_{_1})
\Gamma({D\over2}-1-n_{_2}-n_{_4})\Gamma({D\over2}+n_{_4})\Big\}^{-1}
\;,\nonumber\\
&&c_{_{[\tilde{1}35\tilde{7}]}}^{(3),b}(\alpha,{\bf n})=
(-)^{n_{_1}+n_{_4}}\Gamma(1+n_{_1})\Gamma(1+n_{_2}+n_{_3})\Gamma(1+n_{_2}+n_{_4})
\Big\{n_{_2}!n_{_4}!
\nonumber\\
&&\hspace{2.5cm}\times
\Gamma(2+n_{_1}+n_{_2}+n_{_3})\Gamma(1-{D\over2}-n_{_1})\Gamma(2-{D\over2}+n_{_2})
\nonumber\\
&&\hspace{2.5cm}\times
\Gamma({D\over2}-1-n_{_2}-n_{_3})\Gamma({D\over2}+1+n_{_1}+n_{_2}+n_{_3})
\nonumber\\
&&\hspace{2.5cm}\times
\Gamma({D\over2}-1-n_{_2}-n_{_4})\Gamma({D\over2}+n_{_4})\Big\}^{-1}\;.
\label{GKZ21g-3-3}
\end{eqnarray}

\item   $I_{_{4}}=\{1,2,3,7,\cdots,12,14\}$, i.e. the implement $J_{_{4}}=[1,14]\setminus I_{_{4}}=\{4,5,6,13\}$.
The choice implies the power numbers $\alpha_{_4}=\alpha_{_{5}}=\alpha_{_{6}}=\alpha_{_{13}}=0$, and
\begin{eqnarray}
&&\alpha_{_1}=a_{_4}+a_{_5}+b_{_1}-a_{_1}-1,\;\alpha_{_2}=a_{_4}+a_{_5}+b_{_1}-a_{_2}-1,
\nonumber\\
&&\alpha_{_3}=a_{_4}-a_{_3},\;\alpha_{_7}=b_{_2}-a_{_4}-1,\;\alpha_{_{8}}=b_{_3}-1,
\nonumber\\
&&\alpha_{_9}=b_{_4}-a_{_4}-a_{_5}-1,\;\alpha_{_{10}}=b_{_5}-a_{_4}-a_{_5}-1,
\nonumber\\
&&\alpha_{_{11}}=1-b_{_1},\;\alpha_{_{12}}=-a_{_4},\;\alpha_{_{14}}=-a_{_5}.
\label{GKZ21g-4-1}
\end{eqnarray}
The corresponding hypergeometric functions are written as
\begin{eqnarray}
&&\Phi_{_{[\tilde{1}35\tilde{7}]}}^{(4),a}(\alpha,z)=
y_{_1}^{{D\over2}-1}y_{_2}^{{D}-3}y_{_4}^{-1}\sum\limits_{n_{_1}=0}^\infty
\sum\limits_{n_{_2}=0}^\infty\sum\limits_{n_{_3}=0}^\infty\sum\limits_{n_{_4}=0}^\infty
c_{_{[\tilde{1}35\tilde{7}]}}^{(4),a}(\alpha,{\bf n})
\nonumber\\
&&\hspace{2.5cm}\times
\Big({y_{_1}\over y_{_4}}\Big)^{n_{_1}}\Big({y_{_4}\over y_{_2}}\Big)^{n_{_2}}
\Big({1\over y_{_4}}\Big)^{n_{_3}}\Big({y_{_3}\over y_{_2}}\Big)^{n_{_4}}
\;,\nonumber\\
&&\Phi_{_{[\tilde{1}35\tilde{7}]}}^{(4),b}(\alpha,z)=
y_{_1}^{{D\over2}}y_{_2}^{{D}-3}y_{_4}^{-1}\sum\limits_{n_{_1}=0}^\infty
\sum\limits_{n_{_2}=0}^\infty\sum\limits_{n_{_3}=0}^\infty\sum\limits_{n_{_4}=0}^\infty
c_{_{[\tilde{1}35\tilde{7}]}}^{(4),b}(\alpha,{\bf n})
\nonumber\\
&&\hspace{2.5cm}\times
y_{_1}^{n_{_1}}\Big({y_{_1}\over y_{_2}}\Big)^{n_{_2}}
\Big({y_{_1}\over y_{_4}}\Big)^{n_{_3}}\Big({y_{_3}\over y_{_2}}\Big)^{n_{_4}}\;.
\label{GKZ21g-4-2a}
\end{eqnarray}
Where the coefficients are
\begin{eqnarray}
&&c_{_{[\tilde{1}35\tilde{7}]}}^{(4),a}(\alpha,{\bf n})=
(-)^{n_{_2}}\Gamma(1+n_{_1}+n_{_3})\Big\{n_{_1}!n_{_2}!n_{_3}!n_{_4}!
\Gamma(2-{D\over2}+n_{_3})
\nonumber\\
&&\hspace{2.5cm}\times
\Gamma(2-{D\over2}+n_{_2})\Gamma({D\over2}-1-n_{_2}-n_{_4})
\Gamma(2-{D\over2}+n_{_4})
\nonumber\\
&&\hspace{2.5cm}\times
\Gamma({D\over2}-1-n_{_1}-n_{_3})
\Gamma({D\over2}+n_{_1})\Gamma(D-2-n_{_2}-n_{_4})\Big\}^{-1}
\;,\nonumber\\
&&c_{_{[\tilde{1}35\tilde{7}]}}^{(4),b}(\alpha,{\bf n})=
(-)^{n_{_1}+n_{_2}}\Gamma(1+n_{_1})\Gamma(1+n_{_2}+n_{_3})
\Big\{n_{_2}!n_{_4}!
\nonumber\\
&&\hspace{2.5cm}\times
\Gamma(2+n_{_1}+n_{_2}+n_{_3})\Gamma(1-{D\over2}-n_{_1})\Gamma(2-{D\over2}+n_{_2})
\nonumber\\
&&\hspace{2.5cm}\times
\Gamma({D\over2}-1-n_{_2}-n_{_4})\Gamma(2-{D\over2}+n_{_4})
\Gamma({D\over2}-1-n_{_2}-n_{_3})
\nonumber\\
&&\hspace{2.5cm}\times
\Gamma({D\over2}+1+n_{_1}+n_{_2}+n_{_3})\Gamma(D-2-n_{_2}-n_{_4})\Big\}^{-1}\;.
\label{GKZ21g-4-3}
\end{eqnarray}

\item   $I_{_{5}}=\{1,2,4,6,7,9,10,12,13,14\}$, i.e. the implement $J_{_{5}}=[1,14]\setminus I_{_{5}}=\{3,5,8,11\}$.
The choice implies the power numbers $\alpha_{_3}=\alpha_{_{5}}=\alpha_{_{8}}=\alpha_{_{11}}=0$, and
\begin{eqnarray}
&&\alpha_{_1}=a_{_3}+a_{_5}-a_{_1},\;\alpha_{_2}=a_{_3}+a_{_5}-a_{_2},
\nonumber\\
&&\alpha_{_4}=a_{_3}-a_{_4},\;\alpha_{_{6}}=b_{_1}-1,\;\alpha_{_7}=b_{_2}+b_{_3}-a_{_3}-2,
\nonumber\\
&&\alpha_{_9}=b_{_4}-a_{_3}-a_{_5}-1,\;\alpha_{_{10}}=b_{_5}-a_{_3}-a_{_5}-1,
\nonumber\\
&&\alpha_{_{12}}=b_{_3}-a_{_3}-1,\;\alpha_{_{13}}=1-b_{_3},\;\alpha_{_{14}}=-a_{_5}.
\label{GKZ21g-5-1}
\end{eqnarray}
The corresponding hypergeometric series is
\begin{eqnarray}
&&\Phi_{_{[\tilde{1}35\tilde{7}]}}^{(5)}(\alpha,z)=
y_{_2}^{-1}y_{_3}^{{D\over2}-1}y_{_4}^{-1}\sum\limits_{n_{_1}=0}^\infty
\sum\limits_{n_{_2}=0}^\infty\sum\limits_{n_{_3}=0}^\infty\sum\limits_{n_{_4}=0}^\infty
c_{_{[\tilde{1}35\tilde{7}]}}^{(5)}(\alpha,{\bf n})
\nonumber\\
&&\hspace{2.5cm}\times
y_{_1}^{n_{_1}}\Big({1\over y_{_2}}\Big)^{n_{_2}}\Big({1\over y_{_4}}\Big)^{n_{_3}}
\Big({y_{_3}\over y_{_2}}\Big)^{n_{_4}}\;,
\label{GKZ21g-5-2}
\end{eqnarray}
with
\begin{eqnarray}
&&c_{_{[\tilde{1}35\tilde{7}]}}^{(5)}(\alpha,{\bf n})=
(-)^{n_{_4}}\Gamma(1+n_{_2}+n_{_3})\Gamma(1+n_{_2}+n_{_4})
\Big\{n_{_1}!n_{_2}!n_{_4}!
\nonumber\\
&&\hspace{2.5cm}\times
\Gamma({D\over2}-n_{_1}+n_{_2}+n_{_3})
\Gamma(D-1-n_{_1}+n_{_2}+n_{_3})\Gamma({D\over2}+n_{_2})
\nonumber\\
&&\hspace{2.5cm}\times
\Gamma(2-{D\over2}+n_{_1})\Gamma(1-{D\over2}-n_{_2}-n_{_4})
\Gamma(1-{D\over2}-n_{_2}-n_{_3})
\nonumber\\
&&\hspace{2.5cm}\times
\Gamma({D\over2}+n_{_4})\Big\}^{-1}\;.
\label{GKZ21g-5-3}
\end{eqnarray}

\item   $I_{_{6}}=\{1,2,4,6,\cdots,10,12,14\}$, i.e. the implement $J_{_{6}}=[1,14]\setminus I_{_{6}}=\{3,5,11,13\}$.
The choice implies the power numbers $\alpha_{_3}=\alpha_{_{5}}=\alpha_{_{11}}=\alpha_{_{13}}=0$, and
\begin{eqnarray}
&&\alpha_{_1}=a_{_3}+a_{_5}-a_{_1},\;\alpha_{_2}=a_{_3}+a_{_5}-a_{_2},\;\alpha_{_4}=a_{_3}-a_{_4},
\nonumber\\
&&\alpha_{_{6}}=b_{_1}-1,\;\alpha_{_7}=b_{_2}-a_{_3}-1,\;\alpha_{_{8}}=b_{_3}-1,
\nonumber\\
&&\alpha_{_9}=b_{_4}-a_{_3}-a_{_5}-1,\;\alpha_{_{10}}=b_{_5}-a_{_3}-a_{_5}-1,
\nonumber\\
&&\alpha_{_{12}}=-a_{_3},\;\alpha_{_{14}}=-a_{_5}.
\label{GKZ21g-6-1}
\end{eqnarray}
The corresponding hypergeometric solution is written as
\begin{eqnarray}
&&\Phi_{_{[\tilde{1}35\tilde{7}]}}^{(6)}(\alpha,z)=
y_{_2}^{{D\over2}-2}y_{_4}^{-1}\sum\limits_{n_{_1}=0}^\infty
\sum\limits_{n_{_2}=0}^\infty\sum\limits_{n_{_3}=0}^\infty\sum\limits_{n_{_4}=0}^\infty
c_{_{[\tilde{1}35\tilde{7}]}}^{(6)}(\alpha,{\bf n})
\nonumber\\
&&\hspace{2.5cm}\times
y_{_1}^{n_{_1}}\Big({1\over y_{_2}}\Big)^{n_{_2}}\Big({1\over y_{_4}}\Big)^{n_{_3}}
\Big({y_{_3}\over y_{_2}}\Big)^{n_{_4}}\;,
\label{GKZ21g-6-2}
\end{eqnarray}
with
\begin{eqnarray}
&&c_{_{[\tilde{1}35\tilde{7}]}}^{(6)}(\alpha,{\bf n})=
(-)^{n_{_4}}\Gamma(1+n_{_2}+n_{_3})\Gamma(1+n_{_2}+n_{_4})
\Big\{n_{_1}!n_{_2}!n_{_4}!
\nonumber\\
&&\hspace{2.5cm}\times
\Gamma({D\over2}-n_{_1}+n_{_2}+n_{_3})
\Gamma(D-1-n_{_1}+n_{_2}+n_{_3})\Gamma({D\over2}+n_{_2})
\nonumber\\
&&\hspace{2.5cm}\times
\Gamma(2-{D\over2}+n_{_1})\Gamma(2-{D\over2}+n_{_4})
\Gamma(1-{D\over2}-n_{_2}-n_{_3})
\nonumber\\
&&\hspace{2.5cm}\times
\Gamma({D\over2}-1-n_{_2}-n_{_4})\Big\}^{-1}\;.
\label{GKZ21g-6-3}
\end{eqnarray}

\item   $I_{_{7}}=\{1,2,3,6,7,9,10,12,13,14\}$, i.e. the implement $J_{_{7}}=[1,14]\setminus I_{_{7}}=\{4,5,8,11\}$.
The choice implies the power numbers $\alpha_{_4}=\alpha_{_{5}}=\alpha_{_{8}}=\alpha_{_{11}}=0$, and
\begin{eqnarray}
&&\alpha_{_1}=a_{_4}+a_{_5}-a_{_1},\;\alpha_{_2}=a_{_4}+a_{_5}-a_{_2},
\nonumber\\
&&\alpha_{_3}=a_{_4}-a_{_3},\;\alpha_{_{6}}=b_{_1}-1,\;\alpha_{_7}=b_{_2}+b_{_3}-a_{_4}-2,
\nonumber\\
&&\alpha_{_9}=b_{_4}-a_{_4}-a_{_5}-1,\;\alpha_{_{10}}=b_{_5}-a_{_4}-a_{_5}-1,
\nonumber\\
&&\alpha_{_{12}}=b_{_3}-a_{_4}-1,\;\alpha_{_{13}}=1-b_{_3},\;\alpha_{_{14}}=-a_{_5}.
\label{GKZ21g-7-1}
\end{eqnarray}
The corresponding hypergeometric functions are given as
\begin{eqnarray}
&&\Phi_{_{[\tilde{1}35\tilde{7}]}}^{(7),a}(\alpha,z)=
y_{_2}^{{D\over2}-2}y_{_3}^{{D\over2}-1}y_{_4}^{-1}\sum\limits_{n_{_1}=0}^\infty
\sum\limits_{n_{_2}=0}^\infty\sum\limits_{n_{_3}=0}^\infty\sum\limits_{n_{_4}=0}^\infty
c_{_{[\tilde{1}35\tilde{7}]}}^{(7),a}(\alpha,{\bf n})
\nonumber\\
&&\hspace{2.5cm}\times
\Big({y_{_1}\over y_{_4}}\Big)^{n_{_1}}\Big({y_{_4}\over y_{_2}}\Big)^{n_{_2}}
\Big({1\over y_{_4}}\Big)^{n_{_3}}\Big({y_{_3}\over y_{_2}}\Big)^{n_{_4}}
\;,\nonumber\\
&&\Phi_{_{[\tilde{1}35\tilde{7}]}}^{(7),b}(\alpha,z)=
y_{_1}y_{_2}^{{D\over2}-2}y_{_3}^{{D\over2}-1}y_{_4}^{-1}\sum\limits_{n_{_1}=0}^\infty
\sum\limits_{n_{_2}=0}^\infty\sum\limits_{n_{_3}=0}^\infty\sum\limits_{n_{_4}=0}^\infty
c_{_{[\tilde{1}35\tilde{7}]}}^{(7),b}(\alpha,{\bf n})
\nonumber\\
&&\hspace{2.5cm}\times
y_{_1}^{n_{_1}}\Big({y_{_1}\over y_{_2}}\Big)^{n_{_2}}
\Big({y_{_1}\over y_{_4}}\Big)^{n_{_3}}\Big({y_{_3}\over y_{_2}}\Big)^{n_{_4}}\;.
\label{GKZ21g-7-2a}
\end{eqnarray}
Where the coefficients are
\begin{eqnarray}
&&c_{_{[\tilde{1}35\tilde{7}]}}^{(7),a}(\alpha,{\bf n})=
(-)^{n_{_4}}\Gamma(1+n_{_1}+n_{_3})\Gamma(1+n_{_2}+n_{_4})
\Big\{n_{_1}!n_{_2}!n_{_3}!n_{_4}!\Gamma({D\over2}+n_{_3})
\nonumber\\
&&\hspace{2.5cm}\times
\Gamma(2-{D\over2}+n_{_2})\Gamma(2-{D\over2}+n_{_1})\Gamma({D\over2}-1-n_{_1}-n_{_3})
\nonumber\\
&&\hspace{2.5cm}\times
\Gamma({D\over2}-1-n_{_2}-n_{_4})\Gamma({D\over2}+n_{_4})\Big\}^{-1}
\;,\nonumber\\
&&c_{_{[\tilde{1}35\tilde{7}]}}^{(7),b}(\alpha,{\bf n})=
(-)^{n_{_1}+n_{_4}}\Gamma(1+n_{_1})\Gamma(1+n_{_2}+n_{_3})\Gamma(1+n_{_2}+n_{_4})
\Big\{n_{_2}!n_{_4}!
\nonumber\\
&&\hspace{2.5cm}\times
\Gamma(2+n_{_1}+n_{_2}+n_{_3})\Gamma({D\over2}-1-n_{_1})
\Gamma(2-{D\over2}+n_{_2})
\nonumber\\
&&\hspace{2.5cm}\times
\Gamma(3-{D\over2}+n_{_1}+n_{_2}+n_{_3})\Gamma({D\over2}-1-n_{_2}-n_{_3})
\nonumber\\
&&\hspace{2.5cm}\times
\Gamma({D\over2}-1-n_{_2}-n_{_4})\Gamma({D\over2}+n_{_4})\Big\}^{-1}\;.
\label{GKZ21g-7-3}
\end{eqnarray}

\item   $I_{_{8}}=\{1,2,3,6,\cdots,10,12,14\}$, i.e. the implement $J_{_{8}}=[1,14]\setminus I_{_{8}}=\{4,5,11,13\}$.
The choice implies the power numbers $\alpha_{_4}=\alpha_{_{5}}=\alpha_{_{11}}=\alpha_{_{13}}=0$, and
\begin{eqnarray}
&&\alpha_{_1}=a_{_4}+a_{_5}-a_{_1},\;\alpha_{_2}=a_{_4}+a_{_5}-a_{_2},\;\alpha_{_3}=a_{_4}-a_{_3},
\nonumber\\
&&\alpha_{_{6}}=b_{_1}-1,\;\alpha_{_7}=b_{_2}-a_{_4}-1,\;\alpha_{_{8}}=b_{_3}-1,
\nonumber\\
&&\alpha_{_9}=b_{_4}-a_{_4}-a_{_5}-1,\;\alpha_{_{10}}=b_{_5}-a_{_4}-a_{_5}-1,
\nonumber\\
&&\alpha_{_{12}}=-a_{_4},\;\alpha_{_{14}}=-a_{_5}.
\label{GKZ21g-8-1}
\end{eqnarray}
The corresponding hypergeometric solutions are given as
\begin{eqnarray}
&&\Phi_{_{[\tilde{1}35\tilde{7}]}}^{(8),a}(\alpha,z)=
y_{_2}^{{D}-3}y_{_4}^{-1}\sum\limits_{n_{_1}=0}^\infty
\sum\limits_{n_{_2}=0}^\infty\sum\limits_{n_{_3}=0}^\infty\sum\limits_{n_{_4}=0}^\infty
c_{_{[\tilde{1}35\tilde{7}]}}^{(8),a}(\alpha,{\bf n})
\nonumber\\
&&\hspace{2.5cm}\times
\Big({y_{_1}\over y_{_4}}\Big)^{n_{_1}}\Big({y_{_4}\over y_{_2}}\Big)^{n_{_2}}
\Big({1\over y_{_4}}\Big)^{n_{_3}}\Big({y_{_3}\over y_{_2}}\Big)^{n_{_4}}
\;,\nonumber\\
&&\Phi_{_{[\tilde{1}35\tilde{7}]}}^{(8),b}(\alpha,z)=
y_{_1}y_{_2}^{{D}-3}y_{_4}^{-1}\sum\limits_{n_{_1}=0}^\infty
\sum\limits_{n_{_2}=0}^\infty\sum\limits_{n_{_3}=0}^\infty\sum\limits_{n_{_4}=0}^\infty
c_{_{[\tilde{1}35\tilde{7}]}}^{(8),b}(\alpha,{\bf n})
\nonumber\\
&&\hspace{2.5cm}\times
y_{_1}^{n_{_1}}\Big({y_{_1}\over y_{_2}}\Big)^{n_{_2}}
\Big({y_{_1}\over y_{_4}}\Big)^{n_{_3}}\Big({y_{_3}\over y_{_2}}\Big)^{n_{_4}}\;.
\label{GKZ21g-8-2a}
\end{eqnarray}
Where the coefficients are
\begin{eqnarray}
&&c_{_{[\tilde{1}35\tilde{7}]}}^{(8),a}(\alpha,{\bf n})=
(-)^{n_{_2}}\Gamma(1+n_{_1}+n_{_3})\Big\{n_{_1}!n_{_2}!n_{_3}!n_{_4}!
\Gamma({D\over2}+n_{_3})
\nonumber\\
&&\hspace{2.5cm}\times
\Gamma(2-{D\over2}+n_{_2})\Gamma(2-{D\over2}+n_{_1})
\Gamma({D\over2}-1-n_{_2}-n_{_4})
\nonumber\\
&&\hspace{2.5cm}\times
\Gamma(2-{D\over2}+n_{_4})\Gamma({D\over2}-1-n_{_1}-n_{_3})
\Gamma(D-2-n_{_2}-n_{_4})\Big\}
\;,\nonumber\\
&&c_{_{[\tilde{1}35\tilde{7}]}}^{(8),b}(\alpha,{\bf n})=
(-)^{n_{_1}+n_{_2}}\Gamma(1+n_{_1})\Gamma(1+n_{_2}+n_{_3})\Big\{n_{_2}!n_{_4}!
\Gamma(2+n_{_1}+n_{_2}+n_{_3})
\nonumber\\
&&\hspace{2.5cm}\times
\Gamma({D\over2}-1-n_{_1})
\Gamma(2-{D\over2}+n_{_2})\Gamma(3-{D\over2}+n_{_1}+n_{_2}+n_{_3})
\nonumber\\
&&\hspace{2.5cm}\times
\Gamma({D\over2}-1-n_{_2}-n_{_4})\Gamma(2-{D\over2}+n_{_4})
\Gamma({D\over2}-1-n_{_2}-n_{_3})
\nonumber\\
&&\hspace{2.5cm}\times
\Gamma(D-2-n_{_2}-n_{_4})\Big\}\;.
\label{GKZ21g-8-3}
\end{eqnarray}
\end{itemize}

\section{The hypergeometric solutions of the integer lattice ${\bf B}_{_{1\widetilde{35}7}}$\label{app9}}
\indent\indent
\begin{itemize}
\item   $I_{_{1}}=\{2,\cdots,6,10,\cdots,14\}$, i.e. the implement $J_{_{1}}=[1,14]\setminus I_{_{1}}=\{1,7,8,9\}$.
The choice implies the power numbers $\alpha_{_1}=\alpha_{_{7}}=\alpha_{_{8}}=\alpha_{_{9}}=0$, and
\begin{eqnarray}
&&\alpha_{_2}=a_{_1}-a_{_2},\;\alpha_{_3}=b_{_2}+b_{_3}-a_{_3}-2,
\nonumber\\
&&\alpha_{_4}=b_{_2}+b_{_3}-a_{_4}-2,\;\alpha_{_5}=b_{_4}-a_{_5}-b_{_2}-b_{_3}+1,
\nonumber\\
&&\alpha_{_6}=b_{_1}+b_{_4}-a_{_1}-2,\;\alpha_{_{10}}=b_{_5}-b_{_4},\;\alpha_{_{11}}=b_{_4}-a_{_1}-1,
\nonumber\\
&&\alpha_{_{12}}=1-b_{_2},\;\alpha_{_{13}}=1-b_{_3},\;\alpha_{_{14}}=b_{_2}+b_{_3}-b_{_4}-1.
\label{GKZ21h-1-1}
\end{eqnarray}
The corresponding hypergeometric series is written as
\begin{eqnarray}
&&\Phi_{_{[1\tilde{3}\tilde{5}7]}}^{(1)}(\alpha,z)=
y_{_1}^{{D\over2}-2}y_{_2}^{{D\over2}-1}y_{_3}^{{D\over2}-1}y_{_4}^{-{D\over2}}
\sum\limits_{n_{_1}=0}^\infty
\sum\limits_{n_{_2}=0}^\infty\sum\limits_{n_{_3}=0}^\infty\sum\limits_{n_{_4}=0}^\infty
c_{_{[1\tilde{3}\tilde{5}7]}}^{(1)}(\alpha,{\bf n})
\nonumber\\
&&\hspace{2.5cm}\times
\Big({1\over y_{_1}}\Big)^{n_{_1}}\Big({y_{_3}\over y_{_4}}\Big)^{n_{_2}}
\Big({y_{_4}\over y_{_1}}\Big)^{n_{_3}}\Big({y_{_2}\over y_{_4}}\Big)^{n_{_4}}\;,
\label{GKZ21h-1-2}
\end{eqnarray}
with
\begin{eqnarray}
&&c_{_{[1\tilde{3}\tilde{5}7]}}^{(1)}(\alpha,{\bf n})=
(-)^{n_{_1}}\Gamma(1+n_{_1}+n_{_3})\Gamma(1+n_{_2}+n_{_4})
\Big\{n_{_1}!n_{_2}!n_{_3}!n_{_4}!
\nonumber\\
&&\hspace{2.5cm}\times
\Gamma({D\over2}+n_{_1})
\Gamma(1-{D\over2}-n_{_2}-n_{_4})\Gamma(2-{D\over2}+n_{_3})
\nonumber\\
&&\hspace{2.5cm}\times
\Gamma({D\over2}-1-n_{_1}-n_{_3})\Gamma({D\over2}+n_{_4})
\Gamma({D\over2}+n_{_2})\Big\}^{-1}\;.
\label{GKZ21h-1-3}
\end{eqnarray}

\item   $I_{_{2}}=\{2,\cdots,7,10,11,13,14\}$, i.e. the implement $J_{_{2}}=[1,14]\setminus I_{_{2}}=\{1,8,9,12\}$.
The choice implies the power numbers $\alpha_{_1}=\alpha_{_{8}}=\alpha_{_{9}}=\alpha_{_{12}}=0$, and
\begin{eqnarray}
&&\alpha_{_2}=a_{_1}-a_{_2},\;\alpha_{_3}=b_{_3}-a_{_3}-1,
\nonumber\\
&&\alpha_{_4}=b_{_3}-a_{_4}-1,\;\alpha_{_5}=b_{_4}-a_{_5}-b_{_3},
\nonumber\\
&&\alpha_{_6}=b_{_1}+b_{_4}-a_{_1}-2,\;\alpha_{_{7}}=b_{_2}-1,\;\alpha_{_{10}}=b_{_5}-b_{_4},
\nonumber\\
&&\alpha_{_{11}}=b_{_4}-a_{_1}-1,\;\alpha_{_{13}}=1-b_{_3},\;\alpha_{_{14}}=b_{_3}-b_{_4}.
\label{GKZ21h-2-1}
\end{eqnarray}
The corresponding hypergeometric solution is written as
\begin{eqnarray}
&&\Phi_{_{[1\tilde{3}\tilde{5}7]}}^{(2)}(\alpha,z)=
y_{_1}^{{D\over2}-2}y_{_3}^{{D\over2}-1}y_{_4}^{-1}\sum\limits_{n_{_1}=0}^\infty
\sum\limits_{n_{_2}=0}^\infty\sum\limits_{n_{_3}=0}^\infty\sum\limits_{n_{_4}=0}^\infty
c_{_{[1\tilde{3}\tilde{5}7]}}^{(2)}(\alpha,{\bf n})
\nonumber\\
&&\hspace{2.5cm}\times
\Big({1\over y_{_1}}\Big)^{n_{_1}}\Big({y_{_3}\over y_{_4}}\Big)^{n_{_2}}
\Big({y_{_4}\over y_{_1}}\Big)^{n_{_3}}\Big({y_{_2}\over y_{_4}}\Big)^{n_{_4}}\;,
\label{GKZ21h-2-2}
\end{eqnarray}
with
\begin{eqnarray}
&&c_{_{[1\tilde{3}\tilde{5}7]}}^{(2)}(\alpha,{\bf n})=
(-)^{n_{_1}}\Gamma(1+n_{_1}+n_{_3})\Gamma(1+n_{_2}+n_{_4})
\Big\{n_{_1}!n_{_2}!n_{_3}!n_{_4}!\Gamma({D\over2}+n_{_1})
\nonumber\\
&&\hspace{2.5cm}\times
\Gamma({D\over2}-1-n_{_2}-n_{_4})
\Gamma(2-{D\over2}+n_{_4})\Gamma(2-{D\over2}+n_{_3})
\nonumber\\
&&\hspace{2.5cm}\times
\Gamma({D\over2}-1-n_{_1}-n_{_3})\Gamma({D\over2}+n_{_2})\Big\}^{-1}\;.
\label{GKZ21h-2-3}
\end{eqnarray}

\item   $I_{_{3}}=\{2,\cdots,6,9,11,12,13,14\}$, i.e. the implement $J_{_{3}}=[1,14]\setminus I_{_{3}}=\{1,7,8,10\}$.
The choice implies the power numbers $\alpha_{_1}=\alpha_{_{7}}=\alpha_{_{8}}=\alpha_{_{10}}=0$, and
\begin{eqnarray}
&&\alpha_{_2}=a_{_1}-a_{_2},\;\alpha_{_3}=b_{_2}+b_{_3}-a_{_3}-2,
\nonumber\\
&&\alpha_{_4}=b_{_2}+b_{_3}-a_{_4}-2,\;\alpha_{_5}=b_{_5}-a_{_5}-b_{_2}-b_{_3}+1,
\nonumber\\
&&\alpha_{_6}=b_{_1}+b_{_5}-a_{_1}-2,\;\alpha_{_{9}}=b_{_4}-b_{_5},\;\alpha_{_{11}}=b_{_5}-a_{_1}-1,
\nonumber\\
&&\alpha_{_{12}}=1-b_{_2},\;\alpha_{_{13}}=1-b_{_3},\;\alpha_{_{14}}=b_{_2}+b_{_3}-b_{_5}-1.
\label{GKZ21h-3-1}
\end{eqnarray}
The corresponding hypergeometric series is written as
\begin{eqnarray}
&&\Phi_{_{[1\tilde{3}\tilde{5}7]}}^{(3)}(\alpha,z)=
y_{_1}^{-1}y_{_2}^{{D\over2}-1}y_{_3}^{{D\over2}-1}y_{_4}^{-1}\sum\limits_{n_{_1}=0}^\infty
\sum\limits_{n_{_2}=0}^\infty\sum\limits_{n_{_3}=0}^\infty\sum\limits_{n_{_4}=0}^\infty
c_{_{[1\tilde{3}\tilde{5}7]}}^{(3)}(\alpha,{\bf n})
\nonumber\\
&&\hspace{2.5cm}\times
\Big({1\over y_{_1}}\Big)^{n_{_1}}\Big({y_{_3}\over y_{_4}}\Big)^{n_{_2}}
\Big({y_{_4}\over y_{_1}}\Big)^{n_{_3}}\Big({y_{_2}\over y_{_4}}\Big)^{n_{_4}}\;,
\label{GKZ21h-3-2}
\end{eqnarray}
with
\begin{eqnarray}
&&c_{_{[1\tilde{3}\tilde{5}7]}}^{(3)}(\alpha,{\bf n})=
(-)^{n_{_1}}\Gamma(1+n_{_1}+n_{_3})\Gamma(1+n_{_2}+n_{_4})
\Big\{n_{_1}!n_{_2}!n_{_3}!n_{_4}!
\nonumber\\
&&\hspace{2.5cm}\times
\Gamma({D\over2}+n_{_1})\Gamma(1-{D\over2}-n_{_2}-n_{_4})\Gamma(1-{D\over2}-n_{_1}-n_{_3})
\nonumber\\
&&\hspace{2.5cm}\times
\Gamma({D\over2}+n_{_3})\Gamma({D\over2}+n_{_4})\Gamma({D\over2}+n_{_2})\Big\}^{-1}\;.
\label{GKZ21h-3-3}
\end{eqnarray}

\item   $I_{_{4}}=\{2,\cdots,7,9,11,13,14\}$, i.e. the implement $J_{_{4}}=[1,14]\setminus I_{_{4}}=\{1,8,10,12\}$.
The choice implies the power numbers $\alpha_{_1}=\alpha_{_{8}}=\alpha_{_{10}}=\alpha_{_{12}}=0$, and
\begin{eqnarray}
&&\alpha_{_2}=a_{_1}-a_{_2},\;\alpha_{_3}=b_{_3}-a_{_3}-1,
\nonumber\\
&&\alpha_{_4}=b_{_3}-a_{_4}-1,\;\alpha_{_5}=b_{_5}-a_{_5}-b_{_3},
\nonumber\\
&&\alpha_{_6}=b_{_1}+b_{_5}-a_{_1}-2,\;\alpha_{_{7}}=b_{_2}-1,\;\alpha_{_{9}}=b_{_4}-b_{_5},
\nonumber\\
&&\alpha_{_{11}}=b_{_5}-a_{_1}-1,\;\alpha_{_{13}}=1-b_{_3},\;\alpha_{_{14}}=b_{_3}-b_{_5}.
\label{GKZ21h-4-1}
\end{eqnarray}
The corresponding hypergeometric solution is
\begin{eqnarray}
&&\Phi_{_{[1\tilde{3}\tilde{5}7]}}^{(4)}(\alpha,z)=
y_{_1}^{-1}y_{_3}^{{D\over2}-1}y_{_4}^{{D\over2}-2}\sum\limits_{n_{_1}=0}^\infty
\sum\limits_{n_{_2}=0}^\infty\sum\limits_{n_{_3}=0}^\infty\sum\limits_{n_{_4}=0}^\infty
c_{_{[1\tilde{3}\tilde{5}7]}}^{(4)}(\alpha,{\bf n})
\nonumber\\
&&\hspace{2.5cm}\times
\Big({1\over y_{_1}}\Big)^{n_{_1}}\Big({y_{_3}\over y_{_4}}\Big)^{n_{_2}}
\Big({y_{_4}\over y_{_1}}\Big)^{n_{_3}}\Big({y_{_2}\over y_{_4}}\Big)^{n_{_4}}\;,
\label{GKZ21h-4-2}
\end{eqnarray}
with
\begin{eqnarray}
&&c_{_{[1\tilde{3}\tilde{5}7]}}^{(4)}(\alpha,{\bf n})=
(-)^{1+n_{_1}}\Gamma(1+n_{_1}+n_{_3})\Gamma(1+n_{_2}+n_{_4})\Big\{n_{_1}!n_{_2}!n_{_3}!n_{_4}!
\nonumber\\
&&\hspace{2.5cm}\times
\Gamma({D\over2}+n_{_1})\Gamma({D\over2}-1-n_{_2}-n_{_4})\Gamma(1-{D\over2}-n_{_1}-n_{_3})
\nonumber\\
&&\hspace{2.5cm}\times
\Gamma(2-{D\over2}+n_{_4})\Gamma({D\over2}+n_{_3})\Gamma({D\over2}+n_{_2})\Big\}^{-1}\;.
\label{GKZ21h-4-3}
\end{eqnarray}

\item   $I_{_{5}}=\{2,\cdots,6,8,10,11,12,14\}$, i.e. the implement $J_{_{5}}=[1,14]\setminus I_{_{5}}=\{1,7,9,13\}$.
The choice implies the power numbers $\alpha_{_1}=\alpha_{_{7}}=\alpha_{_{9}}=\alpha_{_{13}}=0$, and
\begin{eqnarray}
&&\alpha_{_2}=a_{_1}-a_{_2},\;\alpha_{_3}=b_{_2}-a_{_3}-1,\;\alpha_{_4}=b_{_2}-a_{_4}-1,
\nonumber\\
&&\alpha_{_5}=b_{_4}-a_{_5}-b_{_2},\;\alpha_{_6}=b_{_1}+b_{_4}-a_{_1}-2,\;\alpha_{_{8}}=b_{_3}-1,
\nonumber\\
&&\alpha_{_{10}}=b_{_5}-b_{_4},\;\alpha_{_{11}}=b_{_4}-a_{_1}-1,\;
\alpha_{_{12}}=1-b_{_2},\;\alpha_{_{14}}=b_{_2}-b_{_4}.
\label{GKZ21h-11-1}
\end{eqnarray}
The corresponding hypergeometric solution is given as
\begin{eqnarray}
&&\Phi_{_{[1\tilde{3}\tilde{5}7]}}^{(5)}(\alpha,z)=
y_{_1}^{{D\over2}-2}y_{_2}^{{D\over2}-1}y_{_4}^{-1}\sum\limits_{n_{_1}=0}^\infty
\sum\limits_{n_{_2}=0}^\infty\sum\limits_{n_{_3}=0}^\infty\sum\limits_{n_{_4}=0}^\infty
c_{_{[1\tilde{3}\tilde{5}7]}}^{(5)}(\alpha,{\bf n})
\nonumber\\
&&\hspace{2.5cm}\times
\Big({1\over y_{_1}}\Big)^{n_{_1}}\Big({y_{_3}\over y_{_4}}\Big)^{n_{_2}}
\Big({y_{_4}\over y_{_1}}\Big)^{n_{_3}}\Big({y_{_2}\over y_{_4}}\Big)^{n_{_4}}\;,
\label{GKZ21h-11-2}
\end{eqnarray}
with
\begin{eqnarray}
&&c_{_{[1\tilde{3}\tilde{5}7]}}^{(5)}(\alpha,{\bf n})=
(-)^{n_{_1}}\Gamma(1+n_{_1}+n_{_3})\Gamma(1+n_{_2}+n_{_4})
\Big\{n_{_1}!n_{_2}!n_{_3}!n_{_4}!
\nonumber\\
&&\hspace{2.5cm}\times
\Gamma({D\over2}+n_{_1})\Gamma({D\over2}-1-n_{_2}-n_{_4})
\Gamma(2-{D\over2}+n_{_2})
\nonumber\\
&&\hspace{2.5cm}\times
\Gamma(2-{D\over2}+n_{_3})\Gamma({D\over2}-1-n_{_1}-n_{_3})
\Gamma({D\over2}+n_{_4})\Big\}^{-1}\;.
\label{GKZ21h-11-3}
\end{eqnarray}

\item   $I_{_{6}}=\{2,\cdots,8,10,11,14\}$, i.e. the implement $J_{_{6}}=[1,14]\setminus I_{_{6}}=\{1,9,12,13\}$.
The choice implies the power numbers $\alpha_{_1}=\alpha_{_{9}}=\alpha_{_{12}}=\alpha_{_{13}}=0$, and
\begin{eqnarray}
&&\alpha_{_2}=a_{_1}-a_{_2},\;\alpha_{_3}=-a_{_3},\;\alpha_{_4}=-a_{_4},\;\alpha_{_5}=b_{_4}-a_{_5}-1,
\nonumber\\
&&\alpha_{_6}=b_{_1}+b_{_4}-a_{_1}-2,\;\alpha_{_{7}}=b_{_2}-1,\;\alpha_{_{8}}=b_{_3}-1,
\nonumber\\
&&\alpha_{_{10}}=b_{_5}-b_{_4},\;\alpha_{_{11}}=b_{_4}-a_{_1}-1,\;\alpha_{_{14}}=1-b_{_4}.
\label{GKZ21h-12-1}
\end{eqnarray}
The corresponding hypergeometric function is written as
\begin{eqnarray}
&&\Phi_{_{[1\tilde{3}\tilde{5}7]}}^{(6)}(\alpha,z)=
y_{_1}^{{D\over2}-2}y_{_4}^{{D\over2}-2}\sum\limits_{n_{_1}=0}^\infty
\sum\limits_{n_{_2}=0}^\infty\sum\limits_{n_{_3}=0}^\infty\sum\limits_{n_{_4}=0}^\infty
c_{_{[1\tilde{3}\tilde{5}7]}}^{(6)}(\alpha,{\bf n})
\nonumber\\
&&\hspace{2.5cm}\times
\Big({1\over y_{_1}}\Big)^{n_{_1}}\Big({y_{_3}\over y_{_4}}\Big)^{n_{_2}}
\Big({y_{_4}\over y_{_1}}\Big)^{n_{_3}}\Big({y_{_2}\over y_{_4}}\Big)^{n_{_4}}\;,
\label{GKZ21h-12-2}
\end{eqnarray}
with
\begin{eqnarray}
&&c_{_{[1\tilde{3}\tilde{5}7]}}^{(6)}(\alpha,{\bf n})=
(-)^{1+n_{_1}+n_{_2}+n_{_4}}\Gamma(1+n_{_1}+n_{_3})
\Big\{n_{_1}!n_{_2}!n_{_3}!n_{_4}!\Gamma({D\over2}+n_{_1})
\nonumber\\
&&\hspace{2.5cm}\times
\Gamma({D\over2}-1-n_{_2}-n_{_4})\Gamma(D-2-n_{_2}-n_{_4})
\Gamma(2-{D\over2}+n_{_4})
\nonumber\\
&&\hspace{2.5cm}\times
\Gamma(2-{D\over2}+n_{_2})\Gamma(2-{D\over2}+n_{_3})
\Gamma({D\over2}-1-n_{_1}-n_{_3})\Big\}^{-1}\;.
\label{GKZ21h-12-3}
\end{eqnarray}

\item   $I_{_{7}}=\{2,\cdots,6,8,9,11,12,14\}$, i.e. the implement $J_{_{7}}=[1,14]\setminus I_{_{7}}=\{1,7,10,13\}$.
The choice implies the power numbers $\alpha_{_1}=\alpha_{_{7}}=\alpha_{_{10}}=\alpha_{_{13}}=0$, and
\begin{eqnarray}
&&\alpha_{_2}=a_{_1}-a_{_2},\;\alpha_{_3}=b_{_2}-a_{_3}-1,\;\alpha_{_4}=b_{_2}-a_{_4}-1,
\nonumber\\
&&\alpha_{_5}=b_{_5}-a_{_5}-b_{_2},\;\alpha_{_6}=b_{_1}+b_{_5}-a_{_1}-2,\;\alpha_{_{8}}=b_{_3}-1,
\nonumber\\
&&\alpha_{_{9}}=b_{_4}-b_{_5},\;\alpha_{_{11}}=b_{_5}-a_{_1}-1,\;
\alpha_{_{12}}=1-b_{_2},\;\alpha_{_{14}}=b_{_2}-b_{_5}.
\label{GKZ21h-13-1}
\end{eqnarray}
The corresponding hypergeometric solution is written as
\begin{eqnarray}
&&\Phi_{_{[1\tilde{3}\tilde{5}7]}}^{(7)}(\alpha,z)=
y_{_1}^{-1}y_{_2}^{{D\over2}-1}y_{_4}^{{D\over2}-2}\sum\limits_{n_{_1}=0}^\infty
\sum\limits_{n_{_2}=0}^\infty\sum\limits_{n_{_3}=0}^\infty\sum\limits_{n_{_4}=0}^\infty
c_{_{[1\tilde{3}\tilde{5}7]}}^{(7)}(\alpha,{\bf n})
\nonumber\\
&&\hspace{2.5cm}\times
\Big({1\over y_{_1}}\Big)^{n_{_1}}\Big({y_{_3}\over y_{_4}}\Big)^{n_{_2}}
\Big({y_{_4}\over y_{_1}}\Big)^{n_{_3}}\Big({y_{_2}\over y_{_4}}\Big)^{n_{_4}}\;,
\label{GKZ21h-13-2}
\end{eqnarray}
with
\begin{eqnarray}
&&c_{_{[1\tilde{3}\tilde{5}7]}}^{(7)}(\alpha,{\bf n})=
(-)^{1+n_{_1}}\Gamma(1+n_{_1}+n_{_3})\Gamma(1+n_{_2}+n_{_4})
\Big\{n_{_1}!n_{_2}!n_{_3}!n_{_4}!
\nonumber\\
&&\hspace{2.5cm}\times
\Gamma({D\over2}+n_{_1})\Gamma({D\over2}-1-n_{_2}-n_{_4})
\Gamma(1-{D\over2}-n_{_1}-n_{_3})
\nonumber\\
&&\hspace{2.5cm}\times
\Gamma(2-{D\over2}+n_{_2})\Gamma({D\over2}+n_{_3})\Gamma({D\over2}+n_{_4})\Big\}^{-1}\;.
\label{GKZ21h-13-3}
\end{eqnarray}

\item   $I_{_{8}}=\{2,\cdots,7,8,9,11,14\}$, i.e. the implement $J_{_{8}}=[1,14]\setminus I_{_{8}}=\{1,10,12,13\}$.
The choice implies the power numbers $\alpha_{_1}=\alpha_{_{10}}=\alpha_{_{12}}=\alpha_{_{13}}=0$, and
\begin{eqnarray}
&&\alpha_{_2}=a_{_1}-a_{_2},\;\alpha_{_3}=-a_{_3},\;
\alpha_{_4}=-a_{_4},\;\alpha_{_5}=b_{_5}-a_{_5}-1,
\nonumber\\
&&\alpha_{_6}=b_{_1}+b_{_5}-a_{_1}-2,\;\alpha_{_{7}}=b_{_2}-1,\;\alpha_{_{8}}=b_{_3}-1,
\nonumber\\
&&\alpha_{_{9}}=b_{_4}-b_{_5},\;\alpha_{_{11}}=b_{_5}-a_{_1}-1,\;\alpha_{_{14}}=1-b_{_5}.
\label{GKZ21h-14-1}
\end{eqnarray}
The corresponding hypergeometric series is given as
\begin{eqnarray}
&&\Phi_{_{[1\tilde{3}\tilde{5}7]}}^{(8)}(\alpha,z)=
y_{_1}^{-1}y_{_4}^{{D}-3}\sum\limits_{n_{_1}=0}^\infty
\sum\limits_{n_{_2}=0}^\infty\sum\limits_{n_{_3}=0}^\infty\sum\limits_{n_{_4}=0}^\infty
c_{_{[1\tilde{3}\tilde{5}7]}}^{(8)}(\alpha,{\bf n})
\nonumber\\
&&\hspace{2.5cm}\times
\Big({1\over y_{_1}}\Big)^{n_{_1}}\Big({y_{_3}\over y_{_4}}\Big)^{n_{_2}}
\Big({y_{_4}\over y_{_1}}\Big)^{n_{_3}}\Big({y_{_2}\over y_{_4}}\Big)^{n_{_4}}\;,
\label{GKZ21h-14-2}
\end{eqnarray}
with
\begin{eqnarray}
&&c_{_{[1\tilde{3}\tilde{5}7]}}^{(8)}(\alpha,{\bf n})=
(-)^{n_{_1}+n_{_2}+n_{_4}}\Gamma(1+n_{_1}+n_{_3})
\Big\{n_{_1}!n_{_2}!n_{_3}!n_{_4}!\Gamma({D\over2}+n_{_1})
\nonumber\\
&&\hspace{2.5cm}\times
\Gamma({D\over2}-1-n_{_2}-n_{_4})\Gamma(D-2-n_{_2}-n_{_4})
\Gamma(1-{D\over2}-n_{_1}-n_{_3})
\nonumber\\
&&\hspace{2.5cm}\times
\Gamma(2-{D\over2}+n_{_4})
\Gamma(2-{D\over2}+n_{_2})\Gamma({D\over2}+n_{_3})\Big\}^{-1}\;.
\label{GKZ21h-14-3}
\end{eqnarray}

\item   $I_{_{9}}=\{1,3,4,5,6,10,\cdots,14\}$, i.e. the implement $J_{_{9}}=[1,14]\setminus I_{_{9}}=\{2,7,8,9\}$.
The choice implies the power numbers $\alpha_{_2}=\alpha_{_{7}}=\alpha_{_{8}}=\alpha_{_{9}}=0$, and
\begin{eqnarray}
&&\alpha_{_1}=a_{_2}-a_{_1},\;\alpha_{_3}=b_{_2}+b_{_3}-a_{_3}-2,
\nonumber\\
&&\alpha_{_4}=b_{_2}+b_{_3}-a_{_4}-2,\;\alpha_{_5}=b_{_4}-a_{_5}-b_{_2}-b_{_3}+1,
\nonumber\\
&&\alpha_{_6}=b_{_1}+b_{_4}-a_{_2}-2,\;\alpha_{_{10}}=b_{_5}-b_{_4},\;\alpha_{_{11}}=b_{_4}-a_{_2}-1,
\nonumber\\
&&\alpha_{_{12}}=1-b_{_2},\;\alpha_{_{13}}=1-b_{_3},\;\alpha_{_{14}}=b_{_2}+b_{_3}-b_{_4}-1.
\label{GKZ21h-17-1}
\end{eqnarray}
The corresponding hypergeometric solution is written as
\begin{eqnarray}
&&\Phi_{_{[1\tilde{3}\tilde{5}7]}}^{(9)}(\alpha,z)=
y_{_1}^{{D}-3}y_{_2}^{{D\over2}-1}y_{_3}^{{D\over2}-1}y_{_4}^{-{D\over2}}
\sum\limits_{n_{_1}=0}^\infty
\sum\limits_{n_{_2}=0}^\infty\sum\limits_{n_{_3}=0}^\infty\sum\limits_{n_{_4}=0}^\infty
c_{_{[1\tilde{3}\tilde{5}7]}}^{(9)}(\alpha,{\bf n})
\nonumber\\
&&\hspace{2.5cm}\times
\Big({1\over y_{_1}}\Big)^{n_{_1}}\Big({y_{_3}\over y_{_4}}\Big)^{n_{_2}}
\Big({y_{_4}\over y_{_1}}\Big)^{n_{_3}}\Big({y_{_2}\over y_{_4}}\Big)^{n_{_4}}\;,
\label{GKZ21h-17-2}
\end{eqnarray}
with
\begin{eqnarray}
&&c_{_{[1\tilde{3}\tilde{5}7]}}^{(9)}(\alpha,{\bf n})=
(-)^{n_{_3}}\Gamma(1+n_{_2}+n_{_4})\Big\{n_{_1}!n_{_2}!n_{_3}!n_{_4}!
\Gamma(2-{D\over2}+n_{_1})
\nonumber\\
&&\hspace{2.5cm}\times
\Gamma(1-{D\over2}-n_{_2}-n_{_4})
\Gamma({D\over2}-1-n_{_1}-n_{_3})\Gamma(2-{D\over2}+n_{_3})
\nonumber\\
&&\hspace{2.5cm}\times
\Gamma(D-2-n_{_1}-n_{_3})\Gamma({D\over2}+n_{_4})\Gamma({D\over2}+n_{_2})\Big\}^{-1}\;.
\label{GKZ21h-17-3}
\end{eqnarray}

\item   $I_{_{10}}=\{1,3,\cdots,7,10,11,13,14\}$, i.e. the implement $J_{_{10}}=[1,14]\setminus I_{_{10}}=\{2,8,9,12\}$.
The choice implies the power numbers $\alpha_{_2}=\alpha_{_{8}}=\alpha_{_{9}}=\alpha_{_{12}}=0$, and
\begin{eqnarray}
&&\alpha_{_1}=a_{_2}-a_{_1},\;\alpha_{_3}=b_{_3}-a_{_3}-1,
\nonumber\\
&&\alpha_{_4}=b_{_3}-a_{_4}-1,\;\alpha_{_5}=b_{_4}-a_{_5}-b_{_3},
\nonumber\\
&&\alpha_{_6}=b_{_1}+b_{_4}-a_{_2}-2,\;\alpha_{_{7}}=b_{_2}-1,\;\alpha_{_{10}}=b_{_5}-b_{_4},
\nonumber\\
&&\alpha_{_{11}}=b_{_4}-a_{_2}-1,\;\alpha_{_{13}}=1-b_{_3},\;\alpha_{_{14}}=b_{_3}-b_{_4}.
\label{GKZ21h-18-1}
\end{eqnarray}
The corresponding hypergeometric function is written as
\begin{eqnarray}
&&\Phi_{_{[1\tilde{3}\tilde{5}7]}}^{(10)}(\alpha,z)=
y_{_1}^{{D}-3}y_{_3}^{{D\over2}-1}y_{_4}^{-1}\sum\limits_{n_{_1}=0}^\infty
\sum\limits_{n_{_2}=0}^\infty\sum\limits_{n_{_3}=0}^\infty\sum\limits_{n_{_4}=0}^\infty
c_{_{[1\tilde{3}\tilde{5}7]}}^{(10)}(\alpha,{\bf n})
\nonumber\\
&&\hspace{2.5cm}\times
\Big({1\over y_{_1}}\Big)^{n_{_1}}\Big({y_{_3}\over y_{_4}}\Big)^{n_{_2}}
\Big({y_{_4}\over y_{_1}}\Big)^{n_{_3}}\Big({y_{_2}\over y_{_4}}\Big)^{n_{_4}}\;,
\label{GKZ21h-18-2}
\end{eqnarray}
with
\begin{eqnarray}
&&c_{_{[1\tilde{3}\tilde{5}7]}}^{(10)}(\alpha,{\bf n})=
(-)^{n_{_3}}\Gamma(1+n_{_2}+n_{_4})\Big\{n_{_1}!n_{_2}!n_{_3}!n_{_4}!
\Gamma(2-{D\over2}+n_{_1})
\nonumber\\
&&\hspace{2.5cm}\times
\Gamma({D\over2}-1-n_{_2}-n_{_4})\Gamma({D\over2}-1-n_{_1}-n_{_3})\Gamma(2-{D\over2}+n_{_4})
\nonumber\\
&&\hspace{2.5cm}\times
\Gamma(2-{D\over2}+n_{_3})\Gamma(D-2-n_{_1}-n_{_3})\Gamma({D\over2}+n_{_2})\Big\}^{-1}\;.
\label{GKZ21h-18-3}
\end{eqnarray}

\item   $I_{_{11}}=\{1,3,4,5,6,9,11,12,13,14\}$, i.e. the implement $J_{_{11}}=[1,14]\setminus I_{_{11}}=\{2,7,8,10\}$.
The choice implies the power numbers $\alpha_{_2}=\alpha_{_{7}}=\alpha_{_{8}}=\alpha_{_{10}}=0$, and
\begin{eqnarray}
&&\alpha_{_1}=a_{_2}-a_{_1},\;\alpha_{_3}=b_{_2}+b_{_3}-a_{_3}-2,
\nonumber\\
&&\alpha_{_4}=b_{_2}+b_{_3}-a_{_4}-2,\;\alpha_{_5}=b_{_5}-a_{_5}-b_{_2}-b_{_3}+1,
\nonumber\\
&&\alpha_{_6}=b_{_1}+b_{_5}-a_{_2}-2,\;\alpha_{_{9}}=b_{_4}-b_{_5},\;\alpha_{_{11}}=b_{_5}-a_{_2}-1,
\nonumber\\
&&\alpha_{_{12}}=1-b_{_2},\;\alpha_{_{13}}=1-b_{_3},\;\alpha_{_{14}}=b_{_2}+b_{_3}-b_{_5}-1.
\label{GKZ21h-19-1}
\end{eqnarray}
The corresponding hypergeometric series is given as
\begin{eqnarray}
&&\Phi_{_{[1\tilde{3}\tilde{5}7]}}^{(11)}(\alpha,z)=
y_{_1}^{{D\over2}-2}y_{_2}^{{D\over2}-1}y_{_3}^{{D\over2}-1}y_{_4}^{-1}
\sum\limits_{n_{_1}=0}^\infty
\sum\limits_{n_{_2}=0}^\infty\sum\limits_{n_{_3}=0}^\infty\sum\limits_{n_{_4}=0}^\infty
c_{_{[1\tilde{3}\tilde{5}7]}}^{(11)}(\alpha,{\bf n})
\nonumber\\
&&\hspace{2.5cm}\times
\Big({1\over y_{_1}}\Big)^{n_{_1}}\Big({y_{_3}\over y_{_4}}\Big)^{n_{_2}}
\Big({y_{_4}\over y_{_1}}\Big)^{n_{_3}}\Big({y_{_2}\over y_{_4}}\Big)^{n_{_4}}\;,
\label{GKZ21h-19-2}
\end{eqnarray}
with
\begin{eqnarray}
&&c_{_{[1\tilde{3}\tilde{5}7]}}^{(11)}(\alpha,{\bf n})=
(-)^{n_{_1}}\Gamma(1+n_{_1}+n_{_3})\Gamma(1+n_{_2}+n_{_4})
\Big\{n_{_1}!n_{_2}!n_{_3}!n_{_4}!
\nonumber\\
&&\hspace{2.5cm}\times
\Gamma(2-{D\over2}+n_{_1})\Gamma(1-{D\over2}-n_{_2}-n_{_4})\Gamma({D\over2}+n_{_3})
\nonumber\\
&&\hspace{2.5cm}\times
\Gamma({D\over2}-1-n_{_1}-n_{_3})\Gamma({D\over2}+n_{_4})\Gamma({D\over2}+n_{_2})\Big\}^{-1}\;.
\label{GKZ21h-19-3}
\end{eqnarray}

\item   $I_{_{12}}=\{1,3,\cdots,7,9,11,13,14\}$, i.e. the implement $J_{_{12}}=[1,14]\setminus I_{_{12}}=\{2,8,10,12\}$.
The choice implies the power numbers $\alpha_{_2}=\alpha_{_{8}}=\alpha_{_{10}}=\alpha_{_{12}}=0$, and
\begin{eqnarray}
&&\alpha_{_1}=a_{_2}-a_{_1},\;\alpha_{_3}=b_{_3}-a_{_3}-1,
\nonumber\\
&&\alpha_{_4}=b_{_3}-a_{_4}-1,\;\alpha_{_5}=b_{_5}-a_{_5}-b_{_3},
\nonumber\\
&&\alpha_{_6}=b_{_1}+b_{_5}-a_{_2}-2,\;\alpha_{_{7}}=b_{_2}-1,\;\alpha_{_{9}}=b_{_4}-b_{_5},
\nonumber\\
&&\alpha_{_{11}}=b_{_5}-a_{_2}-1,\;\alpha_{_{13}}=1-b_{_3},\;\alpha_{_{14}}=b_{_3}-b_{_5}.
\label{GKZ21h-20-1}
\end{eqnarray}
The corresponding hypergeometric solution is written as
\begin{eqnarray}
&&\Phi_{_{[1\tilde{3}\tilde{5}7]}}^{(12)}(\alpha,z)=
y_{_1}^{{D\over2}-2}y_{_3}^{{D\over2}-1}y_{_4}^{{D\over2}-2}\sum\limits_{n_{_1}=0}^\infty
\sum\limits_{n_{_2}=0}^\infty\sum\limits_{n_{_3}=0}^\infty\sum\limits_{n_{_4}=0}^\infty
c_{_{[1\tilde{3}\tilde{5}7]}}^{(12)}(\alpha,{\bf n})
\nonumber\\
&&\hspace{2.5cm}\times
\Big({1\over y_{_1}}\Big)^{n_{_1}}\Big({y_{_3}\over y_{_4}}\Big)^{n_{_2}}
\Big({y_{_4}\over y_{_1}}\Big)^{n_{_3}}\Big({y_{_2}\over y_{_4}}\Big)^{n_{_4}}\;,
\label{GKZ21h-20-2}
\end{eqnarray}
with
\begin{eqnarray}
&&c_{_{[1\tilde{3}\tilde{5}7]}}^{(12)}(\alpha,{\bf n})=
(-)^{1+n_{_1}}\Gamma(1+n_{_1}+n_{_3})\Gamma(1+n_{_2}+n_{_4})
\Big\{n_{_1}!n_{_2}!n_{_3}!n_{_4}!
\nonumber\\
&&\hspace{2.5cm}\times
\Gamma(2-{D\over2}+n_{_1})\Gamma({D\over2}-1-n_{_2}-n_{_4})
\Gamma(2-{D\over2}+n_{_4})
\nonumber\\
&&\hspace{2.5cm}\times
\Gamma({D\over2}+n_{_3})\Gamma({D\over2}-1-n_{_1}-n_{_3})
\Gamma({D\over2}+n_{_2})\Big\}^{-1}
\nonumber\\\;.
\label{GKZ21h-20-3}
\end{eqnarray}

\item   $I_{_{13}}=\{1,3,\cdots,6,8,10,11,12,14\}$, i.e. the implement $J_{_{13}}=[1,14]\setminus I_{_{13}}=\{2,7,9,13\}$.
The choice implies the power numbers $\alpha_{_2}=\alpha_{_{7}}=\alpha_{_{9}}=\alpha_{_{13}}=0$, and
\begin{eqnarray}
&&\alpha_{_1}=a_{_2}-a_{_1},\;\alpha_{_3}=b_{_2}-a_{_3}-1,\;\alpha_{_4}=b_{_2}-a_{_4}-1,
\nonumber\\
&&\alpha_{_5}=b_{_4}-a_{_5}-b_{_2},\;\alpha_{_6}=b_{_1}+b_{_4}-a_{_2}-2,\;\alpha_{_{8}}=b_{_3}-1,
\nonumber\\
&&\alpha_{_{10}}=b_{_5}-b_{_4},\;\alpha_{_{11}}=b_{_4}-a_{_2}-1,\;
\alpha_{_{12}}=1-b_{_2},\;\alpha_{_{14}}=b_{_2}-b_{_4}.
\label{GKZ21h-27-1}
\end{eqnarray}
The corresponding hypergeometric solution is
\begin{eqnarray}
&&\Phi_{_{[1\tilde{3}\tilde{5}7]}}^{(13)}(\alpha,z)=
y_{_1}^{{D}-3}y_{_2}^{{D\over2}-1}y_{_4}^{-1}\sum\limits_{n_{_1}=0}^\infty
\sum\limits_{n_{_2}=0}^\infty\sum\limits_{n_{_3}=0}^\infty\sum\limits_{n_{_4}=0}^\infty
c_{_{[1\tilde{3}\tilde{5}7]}}^{(13)}(\alpha,{\bf n})
\nonumber\\
&&\hspace{2.5cm}\times
\Big({1\over y_{_1}}\Big)^{n_{_1}}\Big({y_{_3}\over y_{_4}}\Big)^{n_{_2}}
\Big({y_{_4}\over y_{_1}}\Big)^{n_{_3}}\Big({y_{_2}\over y_{_4}}\Big)^{n_{_4}}\;,
\label{GKZ21h-27-2}
\end{eqnarray}
with
\begin{eqnarray}
&&c_{_{[1\tilde{3}\tilde{5}7]}}^{(13)}(\alpha,{\bf n})=
(-)^{n_{_3}}\Gamma(1+n_{_2}+n_{_4})\Big\{n_{_1}!n_{_2}!n_{_3}!n_{_4}!
\Gamma(2-{D\over2}+n_{_1})
\nonumber\\
&&\hspace{2.5cm}\times
\Gamma({D\over2}-1-n_{_2}-n_{_4})
\Gamma({D\over2}-1-n_{_1}-n_{_3})\Gamma(2-{D\over2}+n_{_2})
\nonumber\\
&&\hspace{2.5cm}\times
\Gamma(2-{D\over2}+n_{_3})\Gamma(D-2-n_{_1}-n_{_3})
\Gamma({D\over2}+n_{_4})\Big\}^{-1}\;.
\label{GKZ21h-27-3}
\end{eqnarray}

\item   $I_{_{14}}=\{1,3,\cdots,8,10,11,14\}$, i.e. the implement $J_{_{14}}=[1,14]\setminus I_{_{14}}=\{2,9,12,13\}$.
The choice implies the power numbers $\alpha_{_2}=\alpha_{_{9}}=\alpha_{_{12}}=\alpha_{_{13}}=0$, and
\begin{eqnarray}
&&\alpha_{_1}=a_{_2}-a_{_1},\;\alpha_{_3}=-a_{_3},\;\alpha_{_4}=-a_{_4},\;\alpha_{_5}=b_{_4}-a_{_5}-1,
\nonumber\\
&&\alpha_{_6}=b_{_1}+b_{_4}-a_{_2}-2,\;\alpha_{_{7}}=b_{_2}-1,\;\alpha_{_{8}}=b_{_3}-1,
\nonumber\\
&&\alpha_{_{10}}=b_{_5}-b_{_4},\;\alpha_{_{11}}=b_{_4}-a_{_2}-1,\;\alpha_{_{14}}=1-b_{_4}.
\label{GKZ21h-28-1}
\end{eqnarray}
The corresponding hypergeometric series is
\begin{eqnarray}
&&\Phi_{_{[1\tilde{3}\tilde{5}7]}}^{(14)}(\alpha,z)=
y_{_1}^{{D}-3}y_{_4}^{{D\over2}-2} \sum\limits_{n_{_1}=0}^\infty
\sum\limits_{n_{_2}=0}^\infty\sum\limits_{n_{_3}=0}^\infty\sum\limits_{n_{_4}=0}^\infty
c_{_{[1\tilde{3}\tilde{5}7]}}^{(14)}(\alpha,{\bf n})
\nonumber\\
&&\hspace{2.5cm}\times
\Big({1\over y_{_1}}\Big)^{n_{_1}}\Big({y_{_3}\over y_{_4}}\Big)^{n_{_2}}
\Big({y_{_4}\over y_{_1}}\Big)^{n_{_3}}\Big({y_{_2}\over y_{_4}}\Big)^{n_{_4}}\;,
\label{GKZ21h-28-2}
\end{eqnarray}
with
\begin{eqnarray}
&&c_{_{[1\tilde{3}\tilde{5}7]}}^{(14)}(\alpha,{\bf n})=
(-)^{1+n_{_2}+n_{_3}+n_{_4}}\Big\{n_{_1}!n_{_2}!n_{_3}!n_{_4}!\Gamma(2-{D\over2}+n_{_1})
\Gamma({D\over2}-1-n_{_2}-n_{_4})
\nonumber\\
&&\hspace{2.5cm}\times
\Gamma(D-2-n_{_2}-n_{_4})
\Gamma({D\over2}-1-n_{_1}-n_{_3})\Gamma(2-{D\over2}+n_{_4})
\nonumber\\
&&\hspace{2.5cm}\times
\Gamma(2-{D\over2}+n_{_2})\Gamma(2-{D\over2}+n_{_3})
\Gamma(D-2-n_{_1}-n_{_3})\Big\}^{-1}\;.
\label{GKZ21h-28-3}
\end{eqnarray}

\item   $I_{_{15}}=\{1,3,\cdots,6,8,9,11,12,14\}$, i.e. the implement $J_{_{15}}=[1,14]\setminus I_{_{15}}=\{2,7,10,13\}$.
The choice implies the power numbers $\alpha_{_2}=\alpha_{_{7}}=\alpha_{_{10}}=\alpha_{_{13}}=0$, and
\begin{eqnarray}
&&\alpha_{_1}=a_{_2}-a_{_1},\;\alpha_{_3}=b_{_2}-a_{_3}-1,\;\alpha_{_4}=b_{_2}-a_{_4}-1,
\nonumber\\
&&\alpha_{_5}=b_{_5}-a_{_5}-b_{_2},\;\alpha_{_6}=b_{_1}+b_{_5}-a_{_2}-2,\;\alpha_{_{8}}=b_{_3}-1,
\nonumber\\
&&\alpha_{_{9}}=b_{_4}-b_{_5},\;\alpha_{_{11}}=b_{_5}-a_{_2}-1,\;
\alpha_{_{12}}=1-b_{_2},\;\alpha_{_{14}}=b_{_2}-b_{_5}.
\label{GKZ21h-29-1}
\end{eqnarray}
The corresponding hypergeometric series is written as
\begin{eqnarray}
&&\Phi_{_{[1\tilde{3}\tilde{5}7]}}^{(15)}(\alpha,z)=
y_{_1}^{{D\over2}-2}y_{_2}^{{D\over2}-1}y_{_4}^{{D\over2}-2}\sum\limits_{n_{_1}=0}^\infty
\sum\limits_{n_{_2}=0}^\infty\sum\limits_{n_{_3}=0}^\infty\sum\limits_{n_{_4}=0}^\infty
c_{_{[1\tilde{3}\tilde{5}7]}}^{(15)}(\alpha,{\bf n})
\nonumber\\
&&\hspace{2.5cm}\times
\Big({1\over y_{_1}}\Big)^{n_{_1}}\Big({y_{_3}\over y_{_4}}\Big)^{n_{_2}}
\Big({y_{_4}\over y_{_1}}\Big)^{n_{_3}}\Big({y_{_2}\over y_{_4}}\Big)^{n_{_4}}\;,
\label{GKZ21h-29-2}
\end{eqnarray}
with
\begin{eqnarray}
&&c_{_{[1\tilde{3}\tilde{5}7]}}^{(15)}(\alpha,{\bf n})=
(-)^{1+n_{_1}}\Gamma(1+n_{_1}+n_{_3})\Gamma(1+n_{_2}+n_{_4})
\Big\{n_{_1}!n_{_2}!n_{_3}!n_{_4}!
\nonumber\\
&&\hspace{2.5cm}\times
\Gamma(2-{D\over2}+n_{_1})\Gamma({D\over2}-1-n_{_2}-n_{_4})
\Gamma(2-{D\over2}+n_{_2})
\nonumber\\
&&\hspace{2.5cm}\times
\Gamma({D\over2}+n_{_3})\Gamma({D\over2}-1-n_{_1}-n_{_3})
\Gamma({D\over2}+n_{_4})\Big\}^{-1}\;.
\label{GKZ21h-29-3}
\end{eqnarray}

\item   $I_{_{16}}=\{1,3,\cdots,7,8,9,11,14\}$, i.e. the implement $J_{_{16}}=[1,14]\setminus I_{_{16}}=\{2,10,12,13\}$.
The choice implies the power numbers $\alpha_{_2}=\alpha_{_{10}}=\alpha_{_{12}}=\alpha_{_{13}}=0$, and
\begin{eqnarray}
&&\alpha_{_1}=a_{_2}-a_{_1},\;\alpha_{_3}=-a_{_3},\;
\alpha_{_4}=-a_{_4},\;\alpha_{_5}=b_{_5}-a_{_5}-1,
\nonumber\\
&&\alpha_{_6}=b_{_1}+b_{_5}-a_{_2}-2,\;\alpha_{_{7}}=b_{_2}-1,\;\alpha_{_{8}}=b_{_3}-1,
\nonumber\\
&&\alpha_{_{9}}=b_{_4}-b_{_5},\;\alpha_{_{11}}=b_{_5}-a_{_2}-1,\;\alpha_{_{14}}=1-b_{_5}.
\label{GKZ21h-30-1}
\end{eqnarray}
The corresponding hypergeometric series is written as
\begin{eqnarray}
&&\Phi_{_{[1\tilde{3}\tilde{5}7]}}^{(16)}(\alpha,z)=
y_{_1}^{{D\over2}-2}y_{_4}^{{D}-3}\sum\limits_{n_{_1}=0}^\infty
\sum\limits_{n_{_2}=0}^\infty\sum\limits_{n_{_3}=0}^\infty\sum\limits_{n_{_4}=0}^\infty
c_{_{[1\tilde{3}\tilde{5}7]}}^{(16)}(\alpha,{\bf n})
\nonumber\\
&&\hspace{2.5cm}\times
\Big({1\over y_{_1}}\Big)^{n_{_1}}\Big({y_{_3}\over y_{_4}}\Big)^{n_{_2}}
\Big({y_{_4}\over y_{_1}}\Big)^{n_{_3}}\Big({y_{_2}\over y_{_4}}\Big)^{n_{_4}}\;,
\label{GKZ21h-30-2}
\end{eqnarray}
with
\begin{eqnarray}
&&c_{_{[1\tilde{3}\tilde{5}7]}}^{(16)}(\alpha,{\bf n})=
(-)^{n_{_1}+n_{_2}+n_{_4}}
\Gamma(1+n_{_1}+n_{_3})\Big\{n_{_1}!n_{_2}!n_{_3}!n_{_4}!
\Gamma(2-{D\over2}+n_{_1})
\nonumber\\
&&\hspace{2.5cm}\times
\Gamma({D\over2}-1-n_{_2}-n_{_4})\Gamma(D-2-n_{_2}-n_{_4})
\Gamma(2-{D\over2}+n_{_4})
\nonumber\\
&&\hspace{2.5cm}\times
\Gamma(2-{D\over2}+n_{_2})\Gamma({D\over2}+n_{_3})
\Gamma({D\over2}-1-n_{_1}-n_{_3})\Big\}^{-1}\;.
\label{GKZ21h-30-3}
\end{eqnarray}

\item   $I_{_{17}}=\{2,\cdots,6,9,\cdots,13\}$, i.e. the implement $J_{_{17}}=[1,14]\setminus I_{_{17}}=\{1,7,8,14\}$.
The choice implies the power numbers $\alpha_{_1}=\alpha_{_{7}}=\alpha_{_{8}}=\alpha_{_{14}}=0$, and
\begin{eqnarray}
&&\alpha_{_2}=a_{_1}-a_{_2},\;\alpha_{_3}=b_{_2}+b_{_3}-a_{_3}-2,\;\alpha_{_4}=b_{_2}+b_{_3}-a_{_4}-2,
\nonumber\\
&&\alpha_{_5}=-a_{_5},\;\alpha_{_6}=b_{_1}+b_{_2}+b_{_3}-a_{_1}-3,
\nonumber\\
&&\alpha_{_{9}}=b_{_4}-b_{_2}-b_{_3}+1,\;\alpha_{_{10}}=b_{_5}-b_{_2}-b_{_3}+1,
\nonumber\\
&&\alpha_{_{11}}=b_{_2}+b_{_3}-a_{_1}-2,\;\alpha_{_{12}}=1-b_{_2},\;\alpha_{_{13}}=1-b_{_3}.
\label{GKZ21h-5-1}
\end{eqnarray}
The corresponding hypergeometric solution is
\begin{eqnarray}
&&\Phi_{_{[1\tilde{3}\tilde{5}7]}}^{(17)}(\alpha,z)=
y_{_1}^{-2}y_{_2}^{{D\over2}-1}y_{_3}^{{D\over2}-1}\sum\limits_{n_{_1}=0}^\infty
\sum\limits_{n_{_2}=0}^\infty\sum\limits_{n_{_3}=0}^\infty\sum\limits_{n_{_4}=0}^\infty
c_{_{[1\tilde{3}\tilde{5}7]}}^{(17)}(\alpha,{\bf n})
\nonumber\\
&&\hspace{2.5cm}\times
\Big({1\over y_{_1}}\Big)^{n_{_1}}\Big({y_{_3}\over y_{_1}}\Big)^{n_{_2}}
\Big({y_{_4}\over y_{_1}}\Big)^{n_{_3}}\Big({y_{_2}\over y_{_1}}\Big)^{n_{_4}}\;,
\label{GKZ21h-5-2}
\end{eqnarray}
with
\begin{eqnarray}
&&c_{_{[1\tilde{3}\tilde{5}7]}}^{(17)}(\alpha,{\bf n})=
(-)^{1+n_{_1}}\Gamma(2+n_{_1}+n_{_2}+n_{_3}+n_{_4})\Gamma(1+n_{_2}+n_{_4})
\nonumber\\
&&\hspace{2.5cm}\times
\Big\{n_{_1}!n_{_2}!n_{_4}!\Gamma({D\over2}+n_{_1})
\Gamma(1-{D\over2}-n_{_2}-n_{_4})
\nonumber\\
&&\hspace{2.5cm}\times
\Gamma(-{D\over2}-n_{_1}-n_{_2}-n_{_3}-n_{_4})\Gamma(1+{D\over2}+n_{_2}+n_{_3}+n_{_4})
\nonumber\\
&&\hspace{2.5cm}\times
\Gamma(2+n_{_2}+n_{_3}+n_{_4})\Gamma({D\over2}+n_{_4})\Gamma({D\over2}+n_{_2})\Big\}^{-1}\;.
\label{GKZ21h-5-3}
\end{eqnarray}

\item   $I_{_{18}}=\{2,\cdots,7,9,10,11,13\}$, i.e. the implement $J_{_{18}}=[1,14]\setminus I_{_{18}}=\{1,8,12,14\}$.
The choice implies the power numbers $\alpha_{_1}=\alpha_{_{8}}=\alpha_{_{12}}=\alpha_{_{14}}=0$, and
\begin{eqnarray}
&&\alpha_{_2}=a_{_1}-a_{_2},\;\alpha_{_3}=b_{_3}-a_{_3}-1,\;\alpha_{_4}=b_{_3}-a_{_4}-1,
\nonumber\\
&&\alpha_{_5}=-a_{_5},\;\alpha_{_6}=b_{_1}+b_{_3}-a_{_1}-2,\;\alpha_{_{7}}=b_{_2}-1,\;\alpha_{_{9}}=b_{_4}-b_{_3},
\nonumber\\
&&\alpha_{_{10}}=b_{_5}-b_{_3},\;
\alpha_{_{11}}=b_{_3}-a_{_1}-1,\;\alpha_{_{13}}=1-b_{_3}.
\label{GKZ21h-6-1}
\end{eqnarray}
The corresponding hypergeometric series is written as
\begin{eqnarray}
&&\Phi_{_{[1\tilde{3}\tilde{5}7]}}^{(18)}(\alpha,z)=
y_{_1}^{{D\over2}-3}y_{_3}^{{D\over2}-1}\sum\limits_{n_{_1}=0}^\infty
\sum\limits_{n_{_2}=0}^\infty\sum\limits_{n_{_3}=0}^\infty\sum\limits_{n_{_4}=0}^\infty
c_{_{[1\tilde{3}\tilde{5}7]}}^{(18)}(\alpha,{\bf n})
\nonumber\\
&&\hspace{2.5cm}\times
\Big({1\over y_{_1}}\Big)^{n_{_1}}\Big({y_{_3}\over y_{_1}}\Big)^{n_{_2}}
\Big({y_{_4}\over y_{_1}}\Big)^{n_{_3}}\Big({y_{_2}\over y_{_1}}\Big)^{n_{_4}}\;,
\label{GKZ21h-6-2}
\end{eqnarray}
with
\begin{eqnarray}
&&c_{_{[1\tilde{3}\tilde{5}7]}}^{(18)}(\alpha,{\bf n})=
(-)^{1+n_{_1}}\Gamma(2+n_{_1}+n_{_2}+n_{_3}+n_{_4})\Gamma(1+n_{_2}+n_{_4})
\nonumber\\
&&\hspace{2.5cm}\times
\Big\{n_{_1}!n_{_2}!n_{_4}!\Gamma({D\over2}+n_{_1})
\Gamma({D\over2}-1-n_{_2}-n_{_4})\Gamma(2-{D\over2}+n_{_4})
\nonumber\\
&&\hspace{2.5cm}\times
\Gamma(2+n_{_2}+n_{_3}+n_{_4})\Gamma(3-{D\over2}+n_{_2}+n_{_3}+n_{_4})
\nonumber\\
&&\hspace{2.5cm}\times
\Gamma({D\over2}-2-n_{_1}-n_{_2}-n_{_3}-n_{_4})
\Gamma({D\over2}+n_{_2})\Big\}^{-1}\;.
\label{GKZ21h-6-3}
\end{eqnarray}

\item   $I_{_{19}}=\{2,\cdots,6,8,\cdots,12\}$, i.e. the implement $J_{_{19}}=[1,14]\setminus I_{_{19}}=\{1,7,13,14\}$.
The choice implies the power numbers $\alpha_{_1}=\alpha_{_{7}}=\alpha_{_{13}}=\alpha_{_{14}}=0$, and
\begin{eqnarray}
&&\alpha_{_2}=a_{_1}-a_{_2},\;\alpha_{_3}=b_{_2}-a_{_3}-1,\;\alpha_{_4}=b_{_2}-a_{_4}-1,
\nonumber\\
&&\alpha_{_5}=-a_{_5},\;\alpha_{_6}=b_{_1}+b_{_2}-a_{_1}-2,\;\alpha_{_{8}}=b_{_3}-1,\;\alpha_{_{9}}=b_{_4}-b_{_2},
\nonumber\\
&&\alpha_{_{10}}=b_{_5}-b_{_2},\;\alpha_{_{11}}=b_{_2}-a_{_1}-1,\;\alpha_{_{12}}=1-b_{_2}.
\label{GKZ21h-15-1}
\end{eqnarray}
The corresponding hypergeometric function is written as
\begin{eqnarray}
&&\Phi_{_{[1\tilde{3}\tilde{5}7]}}^{(19)}(\alpha,z)=
y_{_1}^{{D\over2}-3}y_{_2}^{{D\over2}-1}\sum\limits_{n_{_1}=0}^\infty
\sum\limits_{n_{_2}=0}^\infty\sum\limits_{n_{_3}=0}^\infty\sum\limits_{n_{_4}=0}^\infty
c_{_{[1\tilde{3}\tilde{5}7]}}^{(19)}(\alpha,{\bf n})
\nonumber\\
&&\hspace{2.5cm}\times
\Big({1\over y_{_1}}\Big)^{n_{_1}}\Big({y_{_3}\over y_{_1}}\Big)^{n_{_2}}
\Big({y_{_4}\over y_{_1}}\Big)^{n_{_3}}\Big({y_{_2}\over y_{_1}}\Big)^{n_{_4}}\;,
\label{GKZ21h-15-2}
\end{eqnarray}
with
\begin{eqnarray}
&&c_{_{[1\tilde{3}\tilde{5}7]}}^{(19)}(\alpha,{\bf n})=
(-)^{1+n_{_1}}\Gamma(2+n_{_1}+n_{_2}+n_{_3}+n_{_4})\Gamma(1+n_{_2}+n_{_4})
\nonumber\\
&&\hspace{2.5cm}\times
\Big\{n_{_1}!n_{_2}!n_{_4}!\Gamma({D\over2}+n_{_1})\Gamma({D\over2}-1-n_{_2}-n_{_4})
\Gamma(2-{D\over2}+n_{_2})
\nonumber\\
&&\hspace{2.5cm}\times
\Gamma(2+n_{_2}+n_{_3}+n_{_4})\Gamma(3-{D\over2}+n_{_2}+n_{_3}+n_{_4})
\nonumber\\
&&\hspace{2.5cm}\times
\Gamma({D\over2}-2-n_{_1}-n_{_2}-n_{_3}-n_{_4})\Gamma({D\over2}+n_{_4})\Big\}^{-1}\;.
\label{GKZ21h-15-3}
\end{eqnarray}

\item   $I_{_{20}}=\{2,\cdots,11\}$, i.e. the implement $J_{_{20}}=[1,14]\setminus I_{_{20}}=\{1,12,13,14\}$.
The choice implies the power numbers $\alpha_{_1}=\alpha_{_{12}}=\alpha_{_{13}}=\alpha_{_{14}}=0$, and
\begin{eqnarray}
&&\alpha_{_2}=a_{_1}-a_{_2},\;\alpha_{_3}=-a_{_3},\;\alpha_{_4}=-a_{_4},\;
\alpha_{_5}=-a_{_5},
\nonumber\\
&&\alpha_{_6}=b_{_1}-a_{_1}-1,\;\alpha_{_{7}}=b_{_2}-1,\;\alpha_{_{8}}=b_{_3}-1,
\nonumber\\
&&\alpha_{_{9}}=b_{_4}-1,\;\alpha_{_{10}}=b_{_5}-1,\;\alpha_{_{11}}=-a_{_1}.
\label{GKZ21h-16-1}
\end{eqnarray}
The corresponding hypergeometric solution is written as
\begin{eqnarray}
&&\Phi_{_{[1\tilde{3}\tilde{5}7]}}^{(20)}(\alpha,z)=
y_{_1}^{{D}-4}\sum\limits_{n_{_1}=0}^\infty
\sum\limits_{n_{_2}=0}^\infty\sum\limits_{n_{_3}=0}^\infty\sum\limits_{n_{_4}=0}^\infty
c_{_{[1\tilde{3}\tilde{5}7]}}^{(20)}(\alpha,{\bf n})
\nonumber\\
&&\hspace{2.5cm}\times
\Big({1\over y_{_1}}\Big)^{n_{_1}}\Big({y_{_3}\over y_{_1}}\Big)^{n_{_2}}
\Big({y_{_4}\over y_{_1}}\Big)^{n_{_3}}\Big({y_{_2}\over y_{_1}}\Big)^{n_{_4}}\;,
\label{GKZ21h-16-2}
\end{eqnarray}
with
\begin{eqnarray}
&&c_{_{[1\tilde{3}\tilde{5}7]}}^{(20)}(\alpha,{\bf n})=
(-)^{n_{_3}}\Big\{n_{_1}!n_{_2}!n_{_4}!
\Gamma({D\over2}+n_{_1})\Gamma({D\over2}-1-n_{_2}-n_{_4})
\nonumber\\
&&\hspace{2.5cm}\times
\Gamma(D-2-n_{_2}-n_{_4})\Gamma({D\over2}-2-n_{_1}-n_{_2}-n_{_3}-n_{_4})
\nonumber\\
&&\hspace{2.5cm}\times
\Gamma(2-{D\over2}+n_{_4})\Gamma(2-{D\over2}+n_{_2})
\Gamma(3-{D\over2}+n_{_2}+n_{_3}+n_{_4})
\nonumber\\
&&\hspace{2.5cm}\times
\Gamma(4-D+n_{_2}+n_{_3}+n_{_4})\Gamma(D-3-n_{_1}-n_{_2}-n_{_3}-n_{_4})\Big\}^{-1}\;.
\label{GKZ21h-16-3}
\end{eqnarray}

\item   $I_{_{21}}=\{1,3,\cdots,6,9,\cdots,13\}$, i.e. the implement $J_{_{21}}=[1,14]\setminus I_{_{21}}=\{2,7,8,14\}$.
The choice implies the power numbers $\alpha_{_2}=\alpha_{_{7}}=\alpha_{_{8}}=\alpha_{_{14}}=0$, and
\begin{eqnarray}
&&\alpha_{_1}=a_{_2}-a_{_1},\;\alpha_{_3}=b_{_2}+b_{_3}-a_{_3}-2,\;\alpha_{_4}=b_{_2}+b_{_3}-a_{_4}-2,
\nonumber\\
&&\alpha_{_5}=-a_{_5},\;\alpha_{_6}=b_{_1}+b_{_2}+b_{_3}-a_{_2}-3,
\nonumber\\
&&\alpha_{_{9}}=b_{_4}-b_{_2}-b_{_3}+1,\;\alpha_{_{10}}=b_{_5}-b_{_2}-b_{_3}+1,
\nonumber\\
&&\alpha_{_{11}}=b_{_2}+b_{_3}-a_{_2}-2,\;\alpha_{_{12}}=1-b_{_2},\;\alpha_{_{13}}=1-b_{_3}.
\label{GKZ21h-21-1}
\end{eqnarray}
The corresponding hypergeometric series is written as
\begin{eqnarray}
&&\Phi_{_{[1\tilde{3}\tilde{5}7]}}^{(21)}(\alpha,z)=
y_{_1}^{{D\over2}-3}y_{_2}^{{D\over2}-1}y_{_3}^{{D\over2}-1}\sum\limits_{n_{_1}=0}^\infty
\sum\limits_{n_{_2}=0}^\infty\sum\limits_{n_{_3}=0}^\infty\sum\limits_{n_{_4}=0}^\infty
c_{_{[1\tilde{3}\tilde{5}7]}}^{(21)}(\alpha,{\bf n})
\nonumber\\
&&\hspace{2.5cm}\times
\Big({1\over y_{_1}}\Big)^{n_{_1}}\Big({y_{_3}\over y_{_1}}\Big)^{n_{_2}}
\Big({y_{_4}\over y_{_1}}\Big)^{n_{_3}}\Big({y_{_2}\over y_{_1}}\Big)^{n_{_4}}\;,
\label{GKZ21h-21-2}
\end{eqnarray}
with
\begin{eqnarray}
&&c_{_{[1\tilde{3}\tilde{5}7]}}^{(21)}(\alpha,{\bf n})=
(-)^{1+n_{_1}}
\Gamma(2+n_{_1}+n_{_2}+n_{_3}+n_{_4})\Gamma(1+n_{_2}+n_{_4})
\nonumber\\
&&\hspace{2.5cm}\times
\Big\{n_{_1}!n_{_2}!n_{_4}!\Gamma(2-{D\over2}+n_{_1})
\Gamma(1-{D\over2}-n_{_2}-n_{_4})
\nonumber\\
&&\hspace{2.5cm}\times
\Gamma(1+{D\over2}+n_{_2}+n_{_3}+n_{_4})
\Gamma(2+n_{_2}+n_{_3}+n_{_4})
\nonumber\\
&&\hspace{2.5cm}\times
\Gamma({D\over2}-2-n_{_1}-n_{_2}-n_{_3}-n_{_4})
\Gamma({D\over2}+n_{_4})\Gamma({D\over2}+n_{_2})\Big\}^{-1}\;.
\label{GKZ21h-21-3}
\end{eqnarray}

\item   $I_{_{22}}=\{1,3,\cdots,7,9,10,11,13\}$, i.e. the implement $J_{_{22}}=[1,14]\setminus I_{_{22}}=\{2,8,12,14\}$.
The choice implies the power numbers $\alpha_{_2}=\alpha_{_{8}}=\alpha_{_{12}}=\alpha_{_{14}}=0$, and
\begin{eqnarray}
&&\alpha_{_1}=a_{_2}-a_{_1},\;\alpha_{_3}=b_{_3}-a_{_3}-1,\;\alpha_{_4}=b_{_3}-a_{_4}-1,
\nonumber\\
&&\alpha_{_5}=-a_{_5},\;\alpha_{_6}=b_{_1}+b_{_3}-a_{_2}-2,\;\alpha_{_{7}}=b_{_2}-1,\;\alpha_{_{9}}=b_{_4}-b_{_3},
\nonumber\\
&&\alpha_{_{10}}=b_{_5}-b_{_3},\;
\alpha_{_{11}}=b_{_3}-a_{_2}-1,\;\alpha_{_{13}}=1-b_{_3}.
\label{GKZ21h-22-1}
\end{eqnarray}
The corresponding hypergeometric solution is
\begin{eqnarray}
&&\Phi_{_{[1\tilde{3}\tilde{5}7]}}^{(22)}(\alpha,z)=
y_{_1}^{{D}-4}y_{_3}^{{D\over2}-1}\sum\limits_{n_{_1}=0}^\infty
\sum\limits_{n_{_2}=0}^\infty\sum\limits_{n_{_3}=0}^\infty\sum\limits_{n_{_4}=0}^\infty
c_{_{[1\tilde{3}\tilde{5}7]}}^{(22)}(\alpha,{\bf n})
\nonumber\\
&&\hspace{2.5cm}\times
\Big({1\over y_{_1}}\Big)^{n_{_1}}\Big({y_{_3}\over y_{_1}}\Big)^{n_{_2}}
\Big({y_{_4}\over y_{_1}}\Big)^{n_{_3}}\Big({y_{_2}\over y_{_1}}\Big)^{n_{_4}}\;,
\label{GKZ21h-22-2}
\end{eqnarray}
with
\begin{eqnarray}
&&c_{_{[1\tilde{3}\tilde{5}7]}}^{(22)}(\alpha,{\bf n})=
(-)^{n_{_2}+n_{_3}+n_{_4}}\Gamma(1+n_{_2}+n_{_4})\Big\{n_{_1}!n_{_2}!n_{_4}!
\Gamma(2-{D\over2}+n_{_1})
\nonumber\\
&&\hspace{2.5cm}\times
\Gamma({D\over2}-1-n_{_2}-n_{_4})\Gamma({D\over2}-2-n_{_1}-n_{_2}-n_{_3}-n_{_4})
\nonumber\\
&&\hspace{2.5cm}\times
\Gamma(2-{D\over2}+n_{_4})\Gamma(2+n_{_2}+n_{_3}+n_{_4})
\Gamma(3-{D\over2}+n_{_2}+n_{_3}+n_{_4})
\nonumber\\
&&\hspace{2.5cm}\times
\Gamma(D-3-n_{_1}-n_{_2}-n_{_3}-n_{_4})\Gamma({D\over2}+n_{_2})\Big\}^{-1}\;.
\label{GKZ21h-22-3}
\end{eqnarray}

\item   $I_{_{23}}=\{1,3,\cdots,6,8,\cdots,12\}$, i.e. the implement $J_{_{23}}=[1,14]\setminus I_{_{23}}=\{2,7,13,14\}$.
The choice implies the power numbers $\alpha_{_2}=\alpha_{_{7}}=\alpha_{_{13}}=\alpha_{_{14}}=0$, and
\begin{eqnarray}
&&\alpha_{_1}=a_{_2}-a_{_1},\;\alpha_{_3}=b_{_2}-a_{_3}-1,\;\alpha_{_4}=b_{_2}-a_{_4}-1,
\nonumber\\
&&\alpha_{_5}=-a_{_5},\;\alpha_{_6}=b_{_1}+b_{_2}-a_{_2}-2,\;\alpha_{_{8}}=b_{_3}-1,\;\alpha_{_{9}}=b_{_4}-b_{_2},
\nonumber\\
&&\alpha_{_{10}}=b_{_5}-b_{_2},\;\alpha_{_{11}}=b_{_2}-a_{_2}-1,\;\alpha_{_{12}}=1-b_{_2}.
\label{GKZ21h-31-1}
\end{eqnarray}
The corresponding hypergeometric function is written as
\begin{eqnarray}
&&\Phi_{_{[1\tilde{3}\tilde{5}7]}}^{(23)}(\alpha,z)=
y_{_1}^{{D}-4}y_{_2}^{{D\over2}-1}\sum\limits_{n_{_1}=0}^\infty
\sum\limits_{n_{_2}=0}^\infty\sum\limits_{n_{_3}=0}^\infty\sum\limits_{n_{_4}=0}^\infty
c_{_{[1\tilde{3}\tilde{5}7]}}^{(23)}(\alpha,{\bf n})
\nonumber\\
&&\hspace{2.5cm}\times
\Big({1\over y_{_1}}\Big)^{n_{_1}}\Big({y_{_3}\over y_{_1}}\Big)^{n_{_2}}
\Big({y_{_4}\over y_{_1}}\Big)^{n_{_3}}\Big({y_{_2}\over y_{_1}}\Big)^{n_{_4}}\;,
\label{GKZ21h-31-2}
\end{eqnarray}
with
\begin{eqnarray}
&&c_{_{[1\tilde{3}\tilde{5}7]}}^{(23)}(\alpha,{\bf n})=
(-)^{n_{_2}+n_{_3}+n_{_4}}\Gamma(1+n_{_2}+n_{_4})\Big\{n_{_1}!n_{_2}!n_{_4}!
\Gamma(2-{D\over2}+n_{_1})
\nonumber\\
&&\hspace{2.5cm}\times
\Gamma({D\over2}-1-n_{_2}-n_{_4})
\Gamma({D\over2}-2-n_{_1}-n_{_2}-n_{_3}-n_{_4})
\nonumber\\
&&\hspace{2.5cm}\times
\Gamma(2-{D\over2}+n_{_2})\Gamma(2+n_{_2}+n_{_3}+n_{_4})
\Gamma(3-{D\over2}+n_{_2}+n_{_3}+n_{_4})
\nonumber\\
&&\hspace{2.5cm}\times
\Gamma(D-3-n_{_1}-n_{_2}-n_{_3}-n_{_4})\Gamma({D\over2}+n_{_4})\Big\}^{-1}\;.
\label{GKZ21h-31-3}
\end{eqnarray}

\item   $I_{_{24}}=\{1,3,\cdots,11\}$, i.e. the implement $J_{_{24}}=[1,14]\setminus I_{_{24}}=\{2,12,13,14\}$.
The choice implies the power numbers $\alpha_{_2}=\alpha_{_{12}}=\alpha_{_{13}}=\alpha_{_{14}}=0$, and
\begin{eqnarray}
&&\alpha_{_1}=a_{_2}-a_{_1},\;\alpha_{_3}=-a_{_3},\;\alpha_{_4}=-a_{_4},\;
\alpha_{_5}=-a_{_5},
\nonumber\\
&&\alpha_{_6}=b_{_1}-a_{_2}-1,\;\alpha_{_{7}}=b_{_2}-1,\;\alpha_{_{8}}=b_{_3}-1,
\nonumber\\
&&\alpha_{_{9}}=b_{_4}-1,\;\alpha_{_{10}}=b_{_5}-1,\;\alpha_{_{11}}=-a_{_2}.
\label{GKZ21h-32-1}
\end{eqnarray}
The corresponding hypergeometric series is
\begin{eqnarray}
&&\Phi_{_{[1\tilde{3}\tilde{5}7]}}^{(24)}(\alpha,z)=
y_{_1}^{{3D\over2}-5}\sum\limits_{n_{_1}=0}^\infty
\sum\limits_{n_{_2}=0}^\infty\sum\limits_{n_{_3}=0}^\infty\sum\limits_{n_{_4}=0}^\infty
c_{_{[1\tilde{3}\tilde{5}7]}}^{(24)}(\alpha,{\bf n})
\nonumber\\
&&\hspace{2.5cm}\times
\Big({1\over y_{_1}}\Big)^{n_{_1}}\Big({y_{_3}\over y_{_1}}\Big)^{n_{_2}}
\Big({y_{_4}\over y_{_1}}\Big)^{n_{_3}}\Big({y_{_2}\over y_{_1}}\Big)^{n_{_4}}\;,
\label{GKZ21h-32-2}
\end{eqnarray}
with
\begin{eqnarray}
&&c_{_{[1\tilde{3}\tilde{5}7]}}^{(24)}(\alpha,{\bf n})=
(-)^{n_{_3}}\Big\{n_{_1}!n_{_2}!n_{_4}!\Gamma(2-{D\over2}+n_{_1})
\Gamma({D\over2}-1-n_{_2}-n_{_4})
\nonumber\\
&&\hspace{2.5cm}\times
\Gamma(D-2-n_{_2}-n_{_4})\Gamma(D-3-n_{_1}-n_{_2}-n_{_3}-n_{_4})
\nonumber\\
&&\hspace{2.5cm}\times
\Gamma(2-{D\over2}+n_{_4})\Gamma(2-{D\over2}+n_{_2})
\Gamma(3-{D\over2}+n_{_2}+n_{_3}+n_{_4})
\nonumber\\
&&\hspace{2.5cm}\times
\Gamma(4-D+n_{_2}+n_{_3}+n_{_4})
\Gamma({3D\over2}-4-n_{_1}-n_{_2}-n_{_3}-n_{_4})\Big\}^{-1}\;.
\label{GKZ21h-32-3}
\end{eqnarray}

\item   $I_{_{25}}=\{2,\cdots,6,8,10,\cdots,13\}$, i.e. the implement $J_{_{25}}=[1,14]\setminus I_{_{25}}=\{1,7,9,14\}$.
The choice implies the power numbers $\alpha_{_1}=\alpha_{_{7}}=\alpha_{_{9}}=\alpha_{_{14}}=0$, and
\begin{eqnarray}
&&\alpha_{_2}=a_{_1}-a_{_2},\;\alpha_{_3}=b_{_4}-a_{_3}-1,\;\alpha_{_4}=b_{_4}-a_{_4}-1,
\nonumber\\
&&\alpha_{_5}=-a_{_5},\;\alpha_{_6}=b_{_1}+b_{_4}-a_{_1}-2,
\nonumber\\
&&\alpha_{_{8}}=b_{_2}+b_{_3}-b_{_4}-1,\;\alpha_{_{10}}=b_{_5}-b_{_4},
\nonumber\\
&&\alpha_{_{11}}=b_{_4}-a_{_1}-1,\;\alpha_{_{12}}=1-b_{_2},\;\alpha_{_{13}}=b_{_2}-b_{_4}.
\label{GKZ21h-7-1}
\end{eqnarray}
The corresponding hypergeometric functions are given as
\begin{eqnarray}
&&\Phi_{_{[1\tilde{3}\tilde{5}7]}}^{(25),a}(\alpha,z)=
y_{_1}^{{D\over2}-2}y_{_2}^{{D\over2}-1}y_{_3}^{-1}\sum\limits_{n_{_1}=0}^\infty
\sum\limits_{n_{_2}=0}^\infty\sum\limits_{n_{_3}=0}^\infty\sum\limits_{n_{_4}=0}^\infty
c_{_{[1\tilde{3}\tilde{5}7]}}^{(25),a}(\alpha,{\bf n})
\nonumber\\
&&\hspace{2.5cm}\times
\Big({1\over y_{_1}}\Big)^{n_{_1}}\Big({y_{_4}\over y_{_1}}\Big)^{n_{_2}}
\Big({y_{_4}\over y_{_3}}\Big)^{n_{_3}}\Big({y_{_2}\over y_{_3}}\Big)^{n_{_4}}
\;,\nonumber\\
&&\Phi_{_{[1\tilde{3}\tilde{5}7]}}^{(25),b}(\alpha,z)=
y_{_1}^{{D\over2}-3}y_{_2}^{{D\over2}-1}\sum\limits_{n_{_1}=0}^\infty
\sum\limits_{n_{_2}=0}^\infty\sum\limits_{n_{_3}=0}^\infty\sum\limits_{n_{_4}=0}^\infty
c_{_{[1\tilde{3}\tilde{5}7]}}^{(25),b}(\alpha,{\bf n})
\nonumber\\
&&\hspace{2.5cm}\times
\Big({1\over y_{_1}}\Big)^{n_{_1}}\Big({y_{_3}\over y_{_1}}\Big)^{n_{_2}}
\Big({y_{_4}\over y_{_1}}\Big)^{n_{_3}}\Big({y_{_2}\over y_{_1}}\Big)^{n_{_4}}
\;,\nonumber\\
&&\Phi_{_{[1\tilde{3}\tilde{5}7]}}^{(25),c}(\alpha,z)=
y_{_1}^{{D\over2}-3}y_{_2}^{{D\over2}}y_{_3}^{-1}\sum\limits_{n_{_1}=0}^\infty
\sum\limits_{n_{_2}=0}^\infty\sum\limits_{n_{_3}=0}^\infty\sum\limits_{n_{_4}=0}^\infty
c_{_{[1\tilde{3}\tilde{5}7]}}^{(25),c}(\alpha,{\bf n})
\nonumber\\
&&\hspace{2.5cm}\times
\Big({1\over y_{_1}}\Big)^{n_{_1}}\Big({y_{_2}\over y_{_1}}\Big)^{n_{_2}}
\Big({y_{_4}\over y_{_1}}\Big)^{n_{_3}}\Big({y_{_2}\over y_{_3}}\Big)^{n_{_4}}\;.
\label{GKZ21h-7-2a}
\end{eqnarray}
Where the coefficients are
\begin{eqnarray}
&&c_{_{[1\tilde{3}\tilde{5}7]}}^{(25),a}(\alpha,{\bf n})=
(-)^{n_{_1}+n_{_4}}\Gamma(1+n_{_1}+n_{_2})\Gamma(1+n_{_3}+n_{_4})
\Big\{n_{_1}!n_{_2}!n_{_3}!n_{_4}!
\nonumber\\
&&\hspace{2.5cm}\times
\Gamma({D\over2}+n_{_1})\Gamma({D\over2}+n_{_3})
\Gamma({D\over2}-1-n_{_1}-n_{_2})
\nonumber\\
&&\hspace{2.5cm}\times
\Gamma(1-{D\over2}-n_{_3}-n_{_4})
\Gamma(2-{D\over2}+n_{_2})\Gamma({D\over2}+n_{_4})\Big\}^{-1}
\;,\nonumber\\
&&c_{_{[1\tilde{3}\tilde{5}7]}}^{(25),b}(\alpha,{\bf n})=
(-)^{1+n_{_1}}\Gamma(2+n_{_1}+n_{_2}+n_{_3}+n_{_4})\Gamma(1+n_{_2}+n_{_4})
\Big\{n_{_1}!n_{_2}!n_{_4}!
\nonumber\\
&&\hspace{2.5cm}\times
\Gamma(2+n_{_2}+n_{_3}+n_{_4})\Gamma({D\over2}+n_{_1})\Gamma({D\over2}-1-n_{_2}-n_{_4})
\nonumber\\
&&\hspace{2.5cm}\times
\Gamma({D\over2}-2-n_{_1}-n_{_2}-n_{_3}-n_{_4})\Gamma(2-{D\over2}+n_{_2})
\nonumber\\
&&\hspace{2.5cm}\times
\Gamma(3-{D\over2}+n_{_2}+n_{_3}+n_{_4})\Gamma({D\over2}+n_{_4})\Big\}^{-1}
\;,\nonumber\\
&&c_{_{[1\tilde{3}\tilde{5}7]}}^{(25),c}(\alpha,{\bf n})=
(-)^{1+n_{_1}+n_{_4}}\Gamma(2+n_{_1}+n_{_2}+n_{_3})\Gamma(1+n_{_2})\Gamma(1+n_{_4})
\nonumber\\
&&\hspace{2.5cm}\times
\Big\{n_{_1}!\Gamma(2+n_{_2}+n_{_4})\Gamma(2+n_{_2}+n_{_3})
\Gamma({D\over2}+n_{_1})
\nonumber\\
&&\hspace{2.5cm}\times
\Gamma({D\over2}-1-n_{_2})
\Gamma({D\over2}-2-n_{_1}-n_{_2}-n_{_3})\Gamma(1-{D\over2}-n_{_4})
\nonumber\\
&&\hspace{2.5cm}\times
\Gamma(3-{D\over2}+n_{_2}+n_{_3})\Gamma({D\over2}+1+n_{_2}+n_{_4})\Big\}^{-1}\;.
\label{GKZ21h-7-3}
\end{eqnarray}

\item   $I_{_{26}}=\{2,\cdots,8,10,11,13\}$, i.e. the implement $J_{_{26}}=[1,14]\setminus I_{_{26}}=\{1,9,12,14\}$.
The choice implies the power numbers $\alpha_{_1}=\alpha_{_{9}}=\alpha_{_{12}}=\alpha_{_{14}}=0$, and
\begin{eqnarray}
&&\alpha_{_2}=a_{_1}-a_{_2},\;\alpha_{_3}=b_{_4}-a_{_3}-1,\;\alpha_{_4}=b_{_4}-a_{_4}-1,
\nonumber\\
&&\alpha_{_5}=-a_{_5},\;\alpha_{_6}=b_{_1}+b_{_4}-a_{_1}-2,\;\alpha_{_{7}}=b_{_2}-1,\;\alpha_{_{8}}=b_{_3}-b_{_4},
\nonumber\\
&&\alpha_{_{10}}=b_{_5}-b_{_4},\;\alpha_{_{11}}=b_{_4}-a_{_1}-1,\;\alpha_{_{13}}=1-b_{_4}.
\label{GKZ21h-8-1}
\end{eqnarray}
The corresponding hypergeometric solutions are written as
\begin{eqnarray}
&&\Phi_{_{[1\tilde{3}\tilde{5}7]}}^{(26),a}(\alpha,z)=
y_{_1}^{{D\over2}-2}y_{_3}^{{D\over2}-2}\sum\limits_{n_{_1}=0}^\infty
\sum\limits_{n_{_2}=0}^\infty\sum\limits_{n_{_3}=0}^\infty\sum\limits_{n_{_4}=0}^\infty
c_{_{[1\tilde{3}\tilde{5}7]}}^{(26),a}(\alpha,{\bf n})
\nonumber\\
&&\hspace{2.5cm}\times
\Big({1\over y_{_1}}\Big)^{n_{_1}}\Big({y_{_4}\over y_{_1}}\Big)^{n_{_2}}
\Big({y_{_4}\over y_{_3}}\Big)^{n_{_3}}\Big({y_{_2}\over y_{_3}}\Big)^{n_{_4}}
\;,\nonumber\\
&&\Phi_{_{[1\tilde{3}\tilde{5}7]}}^{(26),b}(\alpha,z)=
y_{_1}^{{D\over2}-3}y_{_3}^{{D\over2}-1}\sum\limits_{n_{_1}=0}^\infty
\sum\limits_{n_{_2}=0}^\infty\sum\limits_{n_{_3}=0}^\infty\sum\limits_{n_{_4}=0}^\infty
c_{_{[1\tilde{3}\tilde{5}7]}}^{(26),b}(\alpha,{\bf n})
\nonumber\\
&&\hspace{2.5cm}\times
\Big({1\over y_{_1}}\Big)^{n_{_1}}\Big({y_{_3}\over y_{_1}}\Big)^{n_{_2}}
\Big({y_{_4}\over y_{_1}}\Big)^{n_{_3}}\Big({y_{_2}\over y_{_1}}\Big)^{n_{_4}}
\;,\nonumber\\
&&\Phi_{_{[1\tilde{3}\tilde{5}7]}}^{(26),c}(\alpha,z)=
y_{_1}^{{D\over2}-3}y_{_2}y_{_3}^{{D\over2}-2}\sum\limits_{n_{_1}=0}^\infty
\sum\limits_{n_{_2}=0}^\infty\sum\limits_{n_{_3}=0}^\infty\sum\limits_{n_{_4}=0}^\infty
c_{_{[1\tilde{3}\tilde{5}7]}}^{(26),c}(\alpha,{\bf n})
\nonumber\\
&&\hspace{2.5cm}\times
\Big({1\over y_{_1}}\Big)^{n_{_1}}\Big({y_{_2}\over y_{_1}}\Big)^{n_{_2}}
\Big({y_{_4}\over y_{_1}}\Big)^{n_{_3}}\Big({y_{_2}\over y_{_3}}\Big)^{n_{_4}}\;.
\label{GKZ21h-8-2a}
\end{eqnarray}
Where the coefficients are
\begin{eqnarray}
&&c_{_{[1\tilde{3}\tilde{5}7]}}^{(26),a}(\alpha,{\bf n})=
(-)^{n_{_1}+n_{_4}}\Gamma(1+n_{_1}+n_{_2})\Gamma(1+n_{_3}+n_{_4})
\Big\{n_{_1}!n_{_2}!n_{_3}!n_{_4}!
\nonumber\\
&&\hspace{2.5cm}\times
\Gamma({D\over2}+n_{_1})\Gamma({D\over2}+n_{_3})\Gamma({D\over2}-1-n_{_1}-n_{_2})
\nonumber\\
&&\hspace{2.5cm}\times
\Gamma({D\over2}-1-n_{_3}-n_{_4})\Gamma(2-{D\over2}+n_{_2})\Gamma(2-{D\over2}+n_{_4})\Big\}^{-1}
\;,\nonumber\\
&&c_{_{[1\tilde{3}\tilde{5}7]}}^{(26),b}(\alpha,{\bf n})=
(-)^{1+n_{_1}}\Gamma(2+n_{_1}+n_{_2}+n_{_3}+n_{_4})\Gamma(1+n_{_2}+n_{_4})
\Big\{n_{_1}!n_{_2}!n_{_4}!
\nonumber\\
&&\hspace{2.5cm}\times
\Gamma(2+n_{_2}+n_{_3}+n_{_4})\Gamma({D\over2}+n_{_1})\Gamma({D\over2}-1-n_{_2}-n_{_4})
\nonumber\\
&&\hspace{2.5cm}\times
\Gamma({D\over2}-2-n_{_1}-n_{_2}-n_{_3}-n_{_4})\Gamma({D\over2}+n_{_2})
\nonumber\\
&&\hspace{2.5cm}\times
\Gamma(3-{D\over2}+n_{_2}+n_{_3}+n_{_4})\Gamma(2-{D\over2}+n_{_4})\Big\}^{-1}
\;,\nonumber\\
&&c_{_{[1\tilde{3}\tilde{5}7]}}^{(26),c}(\alpha,{\bf n})=
(-)^{1+n_{_1}+n_{_4}}\Gamma(2+n_{_1}+n_{_2}+n_{_3})\Gamma(1+n_{_2})\Gamma(1+n_{_4})
\nonumber\\
&&\hspace{2.5cm}\times
\Big\{n_{_1}!\Gamma(2+n_{_2}+n_{_4})\Gamma(2+n_{_2}+n_{_3})
\Gamma({D\over2}+n_{_1})
\nonumber\\
&&\hspace{2.5cm}\times
\Gamma({D\over2}-1-n_{_2})
\Gamma({D\over2}-2-n_{_1}-n_{_2}-n_{_3})\Gamma({D\over2}-1-n_{_4})
\nonumber\\
&&\hspace{2.5cm}\times
\Gamma(3-{D\over2}+n_{_2}+n_{_3})\Gamma(3-{D\over2}+n_{_2}+n_{_4})\Big\}^{-1}\;.
\label{GKZ21h-8-3}
\end{eqnarray}

\item   $I_{_{27}}=\{2,\cdots,6,8,9,11,12,13\}$, i.e. the implement $J_{_{27}}=[1,14]\setminus I_{_{27}}=\{1,7,10,14\}$.
The choice implies the power numbers $\alpha_{_1}=\alpha_{_{7}}=\alpha_{_{10}}=\alpha_{_{14}}=0$, and
\begin{eqnarray}
&&\alpha_{_2}=a_{_1}-a_{_2},\;\alpha_{_3}=b_{_5}-a_{_3}-1,\;\alpha_{_4}=b_{_5}-a_{_4}-1,
\nonumber\\
&&\alpha_{_5}=-a_{_5},\;\alpha_{_6}=b_{_1}+b_{_5}-a_{_1}-2,
\nonumber\\
&&\alpha_{_{8}}=b_{_2}+b_{_3}-b_{_5}-1,\;\alpha_{_{9}}=b_{_4}-b_{_5},
\nonumber\\
&&\alpha_{_{11}}=b_{_5}-a_{_1}-1,\;\alpha_{_{12}}=1-b_{_2},\;\alpha_{_{13}}=b_{_2}-b_{_5}.
\label{GKZ21h-9-1}
\end{eqnarray}
The corresponding hypergeometric functions are
\begin{eqnarray}
&&\Phi_{_{[1\tilde{3}\tilde{5}7]}}^{(27),a}(\alpha,z)=
y_{_1}^{-1}y_{_2}^{{D\over2}-1}y_{_3}^{{D\over2}-2}\sum\limits_{n_{_1}=0}^\infty
\sum\limits_{n_{_2}=0}^\infty\sum\limits_{n_{_3}=0}^\infty\sum\limits_{n_{_4}=0}^\infty
c_{_{[1\tilde{3}\tilde{5}7]}}^{(27),a}(\alpha,{\bf n})
\nonumber\\
&&\hspace{2.5cm}\times
\Big({1\over y_{_1}}\Big)^{n_{_1}}\Big({y_{_4}\over y_{_1}}\Big)^{n_{_2}}
\Big({y_{_4}\over y_{_3}}\Big)^{n_{_3}}\Big({y_{_2}\over y_{_3}}\Big)^{n_{_4}}
\;,\nonumber\\
&&\Phi_{_{[1\tilde{3}\tilde{5}7]}}^{(27),b}(\alpha,z)=
y_{_1}^{-2}y_{_2}^{{D\over2}-1}y_{_3}^{{D\over2}-1}\sum\limits_{n_{_1}=0}^\infty
\sum\limits_{n_{_2}=0}^\infty\sum\limits_{n_{_3}=0}^\infty\sum\limits_{n_{_4}=0}^\infty
c_{_{[1\tilde{3}\tilde{5}7]}}^{(27),b}(\alpha,{\bf n})
\nonumber\\
&&\hspace{2.5cm}\times
\Big({1\over y_{_1}}\Big)^{n_{_1}}\Big({y_{_3}\over y_{_1}}\Big)^{n_{_2}}
\Big({y_{_4}\over y_{_1}}\Big)^{n_{_3}}\Big({y_{_2}\over y_{_1}}\Big)^{n_{_4}}
\;,\nonumber\\
&&\Phi_{_{[1\tilde{3}\tilde{5}7]}}^{(27),c}(\alpha,z)=
y_{_1}^{-2}y_{_2}^{{D\over2}}y_{_3}^{{D\over2}-2}\sum\limits_{n_{_1}=0}^\infty
\sum\limits_{n_{_2}=0}^\infty\sum\limits_{n_{_3}=0}^\infty\sum\limits_{n_{_4}=0}^\infty
c_{_{[1\tilde{3}\tilde{5}7]}}^{(27),c}(\alpha,{\bf n})
\nonumber\\
&&\hspace{2.5cm}\times
\Big({1\over y_{_1}}\Big)^{n_{_1}}\Big({y_{_2}\over y_{_1}}\Big)^{n_{_2}}
\Big({y_{_4}\over y_{_1}}\Big)^{n_{_3}}\Big({y_{_2}\over y_{_3}}\Big)^{n_{_4}}\;.
\label{GKZ21h-9-2a}
\end{eqnarray}
Where the coefficients are
\begin{eqnarray}
&&c_{_{[1\tilde{3}\tilde{5}7]}}^{(27),a}(\alpha,{\bf n})=
(-)^{n_{_1}+n_{_4}}\Gamma(1+n_{_1}+n_{_2})\Gamma(1+n_{_3}+n_{_4})
\Big\{n_{_1}!n_{_2}!n_{_3}!n_{_4}!
\Gamma({D\over2}+n_{_1})
\nonumber\\
&&\hspace{2.5cm}\times
\Gamma(2-{D\over2}+n_{_3})\Gamma(1-{D\over2}-n_{_1}-n_{_2})
\Gamma({D\over2}+n_{_2})\Gamma({D\over2}+n_{_4})
\nonumber\\
&&\hspace{2.5cm}\times
\Gamma({D\over2}-1-n_{_3}-n_{_4})\Big\}^{-1}
\;,\nonumber\\
&&c_{_{[1\tilde{3}\tilde{5}7]}}^{(27),b}(\alpha,{\bf n})=
(-)^{1+n_{_1}}\Gamma(2+n_{_1}+n_{_2}+n_{_3}+n_{_4})\Gamma(1+n_{_2}+n_{_4})
\Big\{n_{_1}!n_{_2}!n_{_4}!
\nonumber\\
&&\hspace{2.5cm}\times
\Gamma(2+n_{_2}+n_{_3}+n_{_4})\Gamma({D\over2}+n_{_1})
\Gamma(1-{D\over2}-n_{_2}-n_{_4})
\nonumber\\
&&\hspace{2.5cm}\times
\Gamma(-{D\over2}-n_{_1}-n_{_2}-n_{_3}-n_{_4})\Gamma({D\over2}+1+n_{_2}+n_{_3}+n_{_4})
\nonumber\\
&&\hspace{2.5cm}\times
\Gamma({D\over2}+n_{_4})\Gamma({D\over2}+n_{_2})\Big\}^{-1}
\;,\nonumber\\
&&c_{_{[1\tilde{3}\tilde{5}7]}}^{(27),c}(\alpha,{\bf n})=
(-)^{1+n_{_1}+n_{_4}}\Gamma(2+n_{_1}+n_{_2}+n_{_3})\Gamma(1+n_{_2})\Gamma(1+n_{_4})
\nonumber\\
&&\hspace{2.5cm}\times
\Big\{n_{_1}!\Gamma(2+n_{_2}+n_{_4})\Gamma(2+n_{_2}+n_{_3})\Gamma({D\over2}+n_{_1})
\nonumber\\
&&\hspace{2.5cm}\times
\Gamma(1-{D\over2}-n_{_2})
\Gamma(-{D\over2}-n_{_1}-n_{_2}-n_{_3})\Gamma({D\over2}+1+n_{_2}+n_{_3})
\nonumber\\
&&\hspace{2.5cm}\times
\Gamma({D\over2}+1+n_{_2}+n_{_4})\Gamma({D\over2}-1-n_{_4})\Big\}^{-1}\;.
\label{GKZ21h-9-3}
\end{eqnarray}

\item   $I_{_{28}}=\{2,\cdots,9,11,13\}$, i.e. the implement $J_{_{28}}=[1,14]\setminus I_{_{28}}=\{1,10,12,14\}$.
The choice implies the power numbers $\alpha_{_1}=\alpha_{_{10}}=\alpha_{_{12}}=\alpha_{_{14}}=0$, and
\begin{eqnarray}
&&\alpha_{_2}=a_{_1}-a_{_2},\;\alpha_{_3}=b_{_5}-a_{_3}-1,\;\alpha_{_4}=b_{_5}-a_{_4}-1,
\nonumber\\
&&\alpha_{_5}=-a_{_5},\;\alpha_{_6}=b_{_1}+b_{_5}-a_{_1}-2,\;\alpha_{_{7}}=b_{_2}-1,\;\alpha_{_{8}}=b_{_3}-b_{_5},
\nonumber\\
&&\alpha_{_{9}}=b_{_4}-b_{_5},\;\alpha_{_{11}}=b_{_5}-a_{_1}-1,\;\alpha_{_{13}}=1-b_{_5}.
\label{GKZ21h-10-1}
\end{eqnarray}
The corresponding hypergeometric solutions are written as
\begin{eqnarray}
&&\Phi_{_{[1\tilde{3}\tilde{5}7]}}^{(28),a}(\alpha,z)=
y_{_1}^{-1}y_{_3}^{{D}-3}\sum\limits_{n_{_1}=0}^\infty
\sum\limits_{n_{_2}=0}^\infty\sum\limits_{n_{_3}=0}^\infty\sum\limits_{n_{_4}=0}^\infty
c_{_{[1\tilde{3}\tilde{5}7]}}^{(28),a}(\alpha,{\bf n})
\nonumber\\
&&\hspace{2.5cm}\times
\Big({1\over y_{_1}}\Big)^{n_{_1}}\Big({y_{_4}\over y_{_1}}\Big)^{n_{_2}}
\Big({y_{_4}\over y_{_3}}\Big)^{n_{_3}}\Big({y_{_2}\over y_{_3}}\Big)^{n_{_4}}
\;,\nonumber\\
&&\Phi_{_{[1\tilde{3}\tilde{5}7]}}^{(28),b}(\alpha,z)=
y_{_1}^{-2}y_{_3}^{{D}-2}\sum\limits_{n_{_1}=0}^\infty
\sum\limits_{n_{_2}=0}^\infty\sum\limits_{n_{_3}=0}^\infty\sum\limits_{n_{_4}=0}^\infty
c_{_{[1\tilde{3}\tilde{5}7]}}^{(28),b}(\alpha,{\bf n})
\nonumber\\
&&\hspace{2.5cm}\times
\Big({1\over y_{_1}}\Big)^{n_{_1}}\Big({y_{_3}\over y_{_1}}\Big)^{n_{_2}}
\Big({y_{_4}\over y_{_1}}\Big)^{n_{_3}}\Big({y_{_2}\over y_{_3}}\Big)^{n_{_4}}\;.
\label{GKZ21h-10-2a}
\end{eqnarray}
Where the coefficients are
\begin{eqnarray}
&&c_{_{[1\tilde{3}\tilde{5}7]}}^{(28),a}(\alpha,{\bf n})=
(-)^{n_{_1}+n_{_3}}\Gamma(1+n_{_1}+n_{_2})\Big\{n_{_1}!n_{_2}!n_{_3}!n_{_4}!
\Gamma({D\over2}+n_{_1})\Gamma(2-{D\over2}+n_{_3})
\nonumber\\
&&\hspace{2.5cm}\times
\Gamma(1-{D\over2}-n_{_1}-n_{_2})\Gamma(2-{D\over2}+n_{_4})
\Gamma({D\over2}-1-n_{_3}-n_{_4})
\nonumber\\
&&\hspace{2.5cm}\times
\Gamma({D\over2}+n_{_2})\Gamma(D-3-n_{_3}-n_{_4})\Big\}^{-1}
\;,\nonumber\\
&&c_{_{[1\tilde{3}\tilde{5}7]}}^{(28),b}(\alpha,{\bf n})=
(-)^{1+n_{_1}}\Gamma(2+n_{_1}+n_{_2}+n_{_3})\Gamma(1+n_{_2})\Big\{n_{_1}!n_{_4}!
\Gamma(2+n_{_2}+n_{_3})
\nonumber\\
&&\hspace{2.5cm}\times
\Gamma({D\over2}+n_{_1})\Gamma(1-{D\over2}-n_{_2})
\Gamma(-{D\over2}-n_{_1}-n_{_2}-n_{_3})
\nonumber\\
&&\hspace{2.5cm}\times
\Gamma(2-{D\over2}+n_{_4})\Gamma({D\over2}+n_{_2}-n_{_4})\Gamma({D\over2}+1+n_{_2}+n_{_3})
\nonumber\\
&&\hspace{2.5cm}\times
\Gamma(D-1+n_{_2}-n_{_4})\Big\}^{-1}\;.
\label{GKZ21h-10-3}
\end{eqnarray}

\item   $I_{_{29}}=\{1,3,\cdots,6,8,10,\cdots,13\}$, i.e. the implement $J_{_{29}}=[1,14]\setminus I_{_{29}}=\{2,7,9,14\}$.
The choice implies the power numbers $\alpha_{_2}=\alpha_{_{7}}=\alpha_{_{9}}=\alpha_{_{14}}=0$, and
\begin{eqnarray}
&&\alpha_{_1}=a_{_2}-a_{_1},\;\alpha_{_3}=b_{_4}-a_{_3}-1,\;\alpha_{_4}=b_{_4}-a_{_4}-1,
\nonumber\\
&&\alpha_{_5}=-a_{_5},\;\alpha_{_6}=b_{_1}+b_{_4}-a_{_2}-2,
\nonumber\\
&&\alpha_{_{8}}=b_{_2}+b_{_3}-b_{_4}-1,\;\alpha_{_{10}}=b_{_5}-b_{_4},
\nonumber\\
&&\alpha_{_{11}}=b_{_4}-a_{_2}-1,\;\alpha_{_{12}}=1-b_{_2},\;\alpha_{_{13}}=b_{_2}-b_{_4}.
\label{GKZ21h-23-1}
\end{eqnarray}
The corresponding hypergeometric solutions are written as
\begin{eqnarray}
&&\Phi_{_{[1\tilde{3}\tilde{5}7]}}^{(29),a}(\alpha,z)=
y_{_1}^{{D}-3}y_{_2}^{{D\over2}-1}y_{_3}^{-1}\sum\limits_{n_{_1}=0}^\infty
\sum\limits_{n_{_2}=0}^\infty\sum\limits_{n_{_3}=0}^\infty\sum\limits_{n_{_4}=0}^\infty
c_{_{[1\tilde{3}\tilde{5}7]}}^{(29),a}(\alpha,{\bf n})
\nonumber\\
&&\hspace{2.5cm}\times
\Big({1\over y_{_1}}\Big)^{n_{_1}}\Big({y_{_4}\over y_{_1}}\Big)^{n_{_2}}
\Big({y_{_4}\over y_{_3}}\Big)^{n_{_3}}\Big({y_{_2}\over y_{_3}}\Big)^{n_{_4}}
\;,\nonumber\\
&&\Phi_{_{[1\tilde{3}\tilde{5}7]}}^{(29),b}(\alpha,z)=
y_{_1}^{{D}-4}y_{_2}^{{D\over2}-1}\sum\limits_{n_{_1}=0}^\infty
\sum\limits_{n_{_2}=0}^\infty\sum\limits_{n_{_3}=0}^\infty\sum\limits_{n_{_4}=0}^\infty
c_{_{[1\tilde{3}\tilde{5}7]}}^{(29),b}(\alpha,{\bf n})
\nonumber\\
&&\hspace{2.5cm}\times
\Big({1\over y_{_1}}\Big)^{n_{_1}}\Big({y_{_3}\over y_{_1}}\Big)^{n_{_2}}
\Big({y_{_4}\over y_{_1}}\Big)^{n_{_3}}\Big({y_{_2}\over y_{_1}}\Big)^{n_{_4}}
\;,\nonumber\\
&&\Phi_{_{[1\tilde{3}\tilde{5}7]}}^{(29),c}(\alpha,z)=
y_{_1}^{{D}-4}y_{_2}^{{D\over2}}y_{_3}^{-1}\sum\limits_{n_{_1}=0}^\infty
\sum\limits_{n_{_2}=0}^\infty\sum\limits_{n_{_3}=0}^\infty\sum\limits_{n_{_4}=0}^\infty
c_{_{[1\tilde{3}\tilde{5}7]}}^{(29),c}(\alpha,{\bf n})
\nonumber\\
&&\hspace{2.5cm}\times
\Big({1\over y_{_1}}\Big)^{n_{_1}}\Big({y_{_2}\over y_{_1}}\Big)^{n_{_2}}
\Big({y_{_4}\over y_{_1}}\Big)^{n_{_3}}\Big({y_{_2}\over y_{_3}}\Big)^{n_{_4}}\;.
\label{GKZ21h-23-2a}
\end{eqnarray}
Where the coefficients are
\begin{eqnarray}
&&c_{_{[1\tilde{3}\tilde{5}7]}}^{(29),a}(\alpha,{\bf n})=
(-)^{n_{_2}+n_{_4}}\Gamma(1+n_{_3}+n_{_4})
\Big\{n_{_1}!n_{_2}!n_{_3}!n_{_4}!\Gamma(2-{D\over2}+n_{_1})
\nonumber\\
&&\hspace{2.5cm}\times
\Gamma({D\over2}+n_{_3})\Gamma({D\over2}-1-n_{_1}-n_{_2})\Gamma(1-{D\over2}-n_{_3}-n_{_4})
\nonumber\\
&&\hspace{2.5cm}\times
\Gamma(2-{D\over2}+n_{_2})\Gamma(D-2-n_{_1}-n_{_2})\Gamma({D\over2}+n_{_4})\Big\}^{-1}
\;,\nonumber\\
&&c_{_{[1\tilde{3}\tilde{5}7]}}^{(29),b}(\alpha,{\bf n})=
(-)^{n_{_2}+n_{_3}+n_{_4}}\Gamma(1+n_{_2}+n_{_4})\Big\{n_{_1}!n_{_2}!n_{_4}!
\nonumber\\
&&\hspace{2.5cm}\times
\Gamma(2+n_{_2}+n_{_3}+n_{_4})\Gamma(2-{D\over2}+n_{_1})
\nonumber\\
&&\hspace{2.5cm}\times
\Gamma({D\over2}-1-n_{_2}-n_{_4})\Gamma({D\over2}-2-n_{_1}-n_{_2}-n_{_3}-n_{_4})
\nonumber\\
&&\hspace{2.5cm}\times
\Gamma(2-{D\over2}+n_{_2})\Gamma(3-{D\over2}+n_{_2}+n_{_3}+n_{_4})
\nonumber\\
&&\hspace{2.5cm}\times
\Gamma(D-3-n_{_1}-n_{_2}-n_{_3}-n_{_4})\Gamma({D\over2}+n_{_4})\Big\}^{-1}
\;,\nonumber\\
&&c_{_{[1\tilde{3}\tilde{5}7]}}^{(29),c}(\alpha,{\bf n})=
(-)^{n_{_2}+n_{_3}+n_{_4}}\Gamma(1+n_{_2})\Gamma(1+n_{_4})
\Big\{n_{_1}!\Gamma(2+n_{_2}+n_{_4})
\nonumber\\
&&\hspace{2.5cm}\times
\Gamma(2+n_{_2}+n_{_3})\Gamma(2-{D\over2}+n_{_1})\Gamma({D\over2}-1-n_{_2})
\nonumber\\
&&\hspace{2.5cm}\times
\Gamma({D\over2}-2-n_{_1}-n_{_2}-n_{_3})\Gamma(1-{D\over2}-n_{_4})
\nonumber\\
&&\hspace{2.5cm}\times
\Gamma(3-{D\over2}+n_{_2}+n_{_3})\Gamma(D-3-n_{_1}-n_{_2}-n_{_3})
\nonumber\\
&&\hspace{2.5cm}\times
\Gamma({D\over2}+1+n_{_2}+n_{_4})\Big\}^{-1}\;.
\label{GKZ21h-23-3}
\end{eqnarray}

\item   $I_{_{30}}=\{1,3,\cdots,8,10,11,13\}$, i.e. the implement $J_{_{30}}=[1,14]\setminus I_{_{30}}=\{2,9,12,14\}$.
The choice implies the power numbers $\alpha_{_2}=\alpha_{_{9}}=\alpha_{_{12}}=\alpha_{_{14}}=0$, and
\begin{eqnarray}
&&\alpha_{_1}=a_{_2}-a_{_1},\;\alpha_{_3}=b_{_4}-a_{_3}-1,\;\alpha_{_4}=b_{_4}-a_{_4}-1,
\nonumber\\
&&\alpha_{_5}=-a_{_5},\;\alpha_{_6}=b_{_1}+b_{_4}-a_{_2}-2,\;\alpha_{_{7}}=b_{_2}-1,\;\alpha_{_{8}}=b_{_3}-b_{_4},
\nonumber\\
&&\alpha_{_{10}}=b_{_5}-b_{_4},\;\alpha_{_{11}}=b_{_4}-a_{_2}-1,\;\alpha_{_{13}}=1-b_{_4}.
\label{GKZ21h-24-1}
\end{eqnarray}
The corresponding hypergeometric functions are
\begin{eqnarray}
&&\Phi_{_{[1\tilde{3}\tilde{5}7]}}^{(30),a}(\alpha,z)=
y_{_1}^{{D}-3}y_{_3}^{{D\over2}-2}\sum\limits_{n_{_1}=0}^\infty
\sum\limits_{n_{_2}=0}^\infty\sum\limits_{n_{_3}=0}^\infty\sum\limits_{n_{_4}=0}^\infty
c_{_{[1\tilde{3}\tilde{5}7]}}^{(30),a}(\alpha,{\bf n})
\nonumber\\
&&\hspace{2.5cm}\times
\Big({1\over y_{_1}}\Big)^{n_{_1}}\Big({y_{_4}\over y_{_1}}\Big)^{n_{_2}}
\Big({y_{_4}\over y_{_3}}\Big)^{n_{_3}}\Big({y_{_2}\over y_{_3}}\Big)^{n_{_4}}
\;,\nonumber\\
&&\Phi_{_{[1\tilde{3}\tilde{5}7]}}^{(30),b}(\alpha,z)=
y_{_1}^{{D}-4}y_{_3}^{{D\over2}-1}\sum\limits_{n_{_1}=0}^\infty
\sum\limits_{n_{_2}=0}^\infty\sum\limits_{n_{_3}=0}^\infty\sum\limits_{n_{_4}=0}^\infty
c_{_{[1\tilde{3}\tilde{5}7]}}^{(30),b}(\alpha,{\bf n})
\nonumber\\
&&\hspace{2.5cm}\times
\Big({1\over y_{_1}}\Big)^{n_{_1}}\Big({y_{_3}\over y_{_1}}\Big)^{n_{_2}}
\Big({y_{_4}\over y_{_1}}\Big)^{n_{_3}}\Big({y_{_2}\over y_{_1}}\Big)^{n_{_4}}
\;,\nonumber\\
&&\Phi_{_{[1\tilde{3}\tilde{5}7]}}^{(30),c}(\alpha,z)=
y_{_1}^{{D}-4}y_{_2}y_{_3}^{{D\over2}-2}\sum\limits_{n_{_1}=0}^\infty
\sum\limits_{n_{_2}=0}^\infty\sum\limits_{n_{_3}=0}^\infty\sum\limits_{n_{_4}=0}^\infty
c_{_{[1\tilde{3}\tilde{5}7]}}^{(30),c}(\alpha,{\bf n})
\nonumber\\
&&\hspace{2.5cm}\times
\Big({1\over y_{_1}}\Big)^{n_{_1}}\Big({y_{_2}\over y_{_1}}\Big)^{n_{_2}}
\Big({y_{_4}\over y_{_1}}\Big)^{n_{_3}}\Big({y_{_2}\over y_{_3}}\Big)^{n_{_4}}\;.
\label{GKZ21h-24-2a}
\end{eqnarray}
Where the coefficients are
\begin{eqnarray}
&&c_{_{[1\tilde{3}\tilde{5}7]}}^{(30),a}(\alpha,{\bf n})=
(-)^{n_{_2}+n_{_4}}\Gamma(1+n_{_3}+n_{_4})\Big\{n_{_1}!n_{_2}!n_{_3}!n_{_4}!\Gamma(2-{D\over2}+n_{_1})
\nonumber\\
&&\hspace{2.5cm}\times
\Gamma({D\over2}+n_{_3})
\Gamma({D\over2}-1-n_{_1}-n_{_2})\Gamma(2-{D\over2}+n_{_4})
\nonumber\\
&&\hspace{2.5cm}\times
\Gamma(2-{D\over2}+n_{_2})
\Gamma(D-2-n_{_1}-n_{_2})\Gamma({D\over2}-1-n_{_3}-n_{_4})\Big\}^{-1}
\;,\nonumber\\
&&c_{_{[1\tilde{3}\tilde{5}7]}}^{(30),b}(\alpha,{\bf n})=
(-)^{n_{_2}+n_{_3}+n_{_4}}\Gamma(1+n_{_2}+n_{_4})\Big\{n_{_1}!n_{_2}!n_{_4}!
\nonumber\\
&&\hspace{2.5cm}\times
\Gamma(2+n_{_2}+n_{_3}+n_{_4})\Gamma(2-{D\over2}+n_{_1})
\nonumber\\
&&\hspace{2.5cm}\times
\Gamma({D\over2}-1-n_{_2}-n_{_4})
\Gamma({D\over2}-2-n_{_1}-n_{_2}-n_{_3}-n_{_4})
\nonumber\\
&&\hspace{2.5cm}\times
\Gamma(2-{D\over2}+n_{_4})\Gamma(3-{D\over2}+n_{_2}+n_{_3}+n_{_4})
\nonumber\\
&&\hspace{2.5cm}\times
\Gamma(D-3-n_{_1}-n_{_2}-n_{_3}-n_{_4})\Gamma({D\over2}+n_{_2})\Big\}^{-1}
\;,\nonumber\\
&&c_{_{[1\tilde{3}\tilde{5}7]}}^{(30),c}(\alpha,{\bf n})=
(-)^{n_{_2}+n_{_3}+n_{_4}}\Gamma(1+n_{_2})\Gamma(1+n_{_4})\Big\{n_{_1}!\Gamma(+n_{_2}+n_{_4})
\nonumber\\
&&\hspace{2.5cm}\times
\Gamma(2+n_{_2}+n_{_3})\Gamma(2-{D\over2}+n_{_1})\Gamma({D\over2}-1-n_{_2})
\nonumber\\
&&\hspace{2.5cm}\times
\Gamma({D\over2}-2-n_{_1}-n_{_2}-n_{_3})\Gamma(3-{D\over2}+n_{_2}+n_{_4})
\nonumber\\
&&\hspace{2.5cm}\times
\Gamma(3-{D\over2}+n_{_2}+n_{_3})\Gamma(D-3-n_{_1}-n_{_2}-n_{_3})
\nonumber\\
&&\hspace{2.5cm}\times
\Gamma({D\over2}-1-n_{_4})\Big\}^{-1}\;.
\label{GKZ21h-24-3}
\end{eqnarray}

\item   $I_{_{31}}=\{1,3,\cdots,6,8,9,11,12,13\}$, i.e. the implement $J_{_{31}}=[1,14]\setminus I_{_{31}}=\{2,7,10,14\}$.
The choice implies the power numbers $\alpha_{_2}=\alpha_{_{7}}=\alpha_{_{10}}=\alpha_{_{14}}=0$, and
\begin{eqnarray}
&&\alpha_{_1}=a_{_2}-a_{_1},\;\alpha_{_3}=b_{_5}-a_{_3}-1,\;\alpha_{_4}=b_{_5}-a_{_4}-1,
\nonumber\\
&&\alpha_{_5}=-a_{_5},\;\alpha_{_6}=b_{_1}+b_{_5}-a_{_2}-2,
\nonumber\\
&&\alpha_{_{8}}=b_{_2}+b_{_3}-b_{_5}-1,\;\alpha_{_{9}}=b_{_4}-b_{_5},
\nonumber\\
&&\alpha_{_{11}}=b_{_5}-a_{_2}-1,\;\alpha_{_{12}}=1-b_{_2},\;\alpha_{_{13}}=b_{_2}-b_{_5}.
\label{GKZ21h-25-1}
\end{eqnarray}
The corresponding hypergeometric solutions are given as
\begin{eqnarray}
&&\Phi_{_{[1\tilde{3}\tilde{5}7]}}^{(31),a}(\alpha,z)=
y_{_1}^{{D\over2}-2}y_{_2}^{{D\over2}-1}y_{_3}^{{D\over2}-2}\sum\limits_{n_{_1}=0}^\infty
\sum\limits_{n_{_2}=0}^\infty\sum\limits_{n_{_3}=0}^\infty\sum\limits_{n_{_4}=0}^\infty
c_{_{[1\tilde{3}\tilde{5}7]}}^{(31),a}(\alpha,{\bf n})
\nonumber\\
&&\hspace{2.5cm}\times
\Big({1\over y_{_1}}\Big)^{n_{_1}}\Big({y_{_4}\over y_{_1}}\Big)^{n_{_2}}
\Big({y_{_4}\over y_{_3}}\Big)^{n_{_3}}\Big({y_{_2}\over y_{_3}}\Big)^{n_{_4}}
\;,\nonumber\\
&&\Phi_{_{[1\tilde{3}\tilde{5}7]}}^{(31),b}(\alpha,z)=
y_{_1}^{{D\over2}-3}y_{_2}^{{D\over2}-1}y_{_3}^{{D\over2}-1}\sum\limits_{n_{_1}=0}^\infty
\sum\limits_{n_{_2}=0}^\infty\sum\limits_{n_{_3}=0}^\infty\sum\limits_{n_{_4}=0}^\infty
c_{_{[1\tilde{3}\tilde{5}7]}}^{(31),b}(\alpha,{\bf n})
\nonumber\\
&&\hspace{2.5cm}\times
\Big({1\over y_{_1}}\Big)^{n_{_1}}\Big({y_{_3}\over y_{_1}}\Big)^{n_{_2}}
\Big({y_{_4}\over y_{_1}}\Big)^{n_{_3}}\Big({y_{_2}\over y_{_1}}\Big)^{n_{_4}}
\;,\nonumber\\
&&\Phi_{_{[1\tilde{3}\tilde{5}7]}}^{(31),c}(\alpha,z)=
y_{_1}^{{D\over2}-3}y_{_2}^{{D\over2}}y_{_3}^{{D\over2}-2}\sum\limits_{n_{_1}=0}^\infty
\sum\limits_{n_{_2}=0}^\infty\sum\limits_{n_{_3}=0}^\infty\sum\limits_{n_{_4}=0}^\infty
c_{_{[1\tilde{3}\tilde{5}7]}}^{(31),c}(\alpha,{\bf n})
\nonumber\\
&&\hspace{2.5cm}\times
\Big({1\over y_{_1}}\Big)^{n_{_1}}\Big({y_{_2}\over y_{_1}}\Big)^{n_{_2}}
\Big({y_{_4}\over y_{_1}}\Big)^{n_{_3}}\Big({y_{_2}\over y_{_3}}\Big)^{n_{_4}}\;.
\label{GKZ21h-25-2a}
\end{eqnarray}
Where the coefficients are
\begin{eqnarray}
&&c_{_{[1\tilde{3}\tilde{5}7]}}^{(31),a}(\alpha,{\bf n})=
(-)^{n_{_1}+n_{_4}}\Gamma(1+n_{_1}+n_{_2})\Gamma(1+n_{_3}+n_{_4})
\nonumber\\
&&\hspace{2.5cm}\times
\Big\{n_{_1}!n_{_2}!n_{_3}!n_{_4}!
\Gamma(2-{D\over2}+n_{_1})\Gamma(2-{D\over2}+n_{_3})
\nonumber\\
&&\hspace{2.5cm}\times
\Gamma({D\over2}+n_{_2})\Gamma({D\over2}-1-n_{_1}-n_{_2})
\nonumber\\
&&\hspace{2.5cm}\times
\Gamma({D\over2}+n_{_4})\Gamma({D\over2}-1-n_{_3}-n_{_4})\Big\}^{-1}
\;,\nonumber\\
&&c_{_{[1\tilde{3}\tilde{5}7]}}^{(31),b}(\alpha,{\bf n})=
(-)^{1+n_{_1}}\Gamma(2+n_{_1}+n_{_2}+n_{_3}+n_{_4})\Gamma(1+n_{_2}+n_{_4})
\nonumber\\
&&\hspace{2.5cm}\times
\Big\{n_{_1}!n_{_2}!n_{_4}!\Gamma(2+n_{_2}+n_{_3}+n_{_4})\Gamma(2-{D\over2}+n_{_1})
\nonumber\\
&&\hspace{2.5cm}\times
\Gamma(1-{D\over2}-n_{_2}-n_{_4})\Gamma({D\over2}+1+n_{_2}+n_{_3}+n_{_4})
\nonumber\\
&&\hspace{2.5cm}\times
\Gamma({D\over2}-2-n_{_1}-n_{_2}-n_{_3}-n_{_4})
\Gamma({D\over2}+n_{_4})\Gamma({D\over2}+n_{_2})\Big\}^{-1}
\;,\nonumber\\
&&c_{_{[1\tilde{3}\tilde{5}7]}}^{(31),c}(\alpha,{\bf n})=
(-)^{1+n_{_1}+n_{_4}}\Gamma(2+n_{_1}+n_{_2}+n_{_3})\Gamma(1+n_{_2})\Gamma(1+n_{_4})
\nonumber\\
&&\hspace{2.5cm}\times
\Big\{n_{_1}!\Gamma(2+n_{_2}+n_{_4})\Gamma(2+n_{_2}+n_{_3})\Gamma(2-D+n_{_1})
\nonumber\\
&&\hspace{2.5cm}\times
\Gamma(1-{D\over2}-n_{_2})\Gamma({D\over2}+1+n_{_2}+n_{_3})
\nonumber\\
&&\hspace{2.5cm}\times
\Gamma({D\over2}-2-n_{_1}-n_{_2}-n_{_3})
\Gamma({D\over2}+n_{_2}+n_{_4})\Gamma({D\over2}-1-n_{_4})\Big\}^{-1}\;.
\label{GKZ21h-25-3}
\end{eqnarray}

\item   $I_{_{32}}=\{1,3,\cdots,9,11,13\}$, i.e. the implement $J_{_{32}}=[1,14]\setminus I_{_{32}}=\{2,10,12,14\}$.
The choice implies the power numbers $\alpha_{_2}=\alpha_{_{10}}=\alpha_{_{12}}=\alpha_{_{14}}=0$, and
\begin{eqnarray}
&&\alpha_{_1}=a_{_2}-a_{_1},\;\alpha_{_3}=b_{_5}-a_{_3}-1,\;\alpha_{_4}=b_{_5}-a_{_4}-1,
\nonumber\\
&&\alpha_{_5}=-a_{_5},\;\alpha_{_6}=b_{_1}+b_{_5}-a_{_2}-2,\;\alpha_{_{7}}=b_{_2}-1,\;\alpha_{_{8}}=b_{_3}-b_{_5},
\nonumber\\
&&\alpha_{_{9}}=b_{_4}-b_{_5},\;\alpha_{_{11}}=b_{_5}-a_{_2}-1,\;\alpha_{_{13}}=1-b_{_5}.
\label{GKZ21h-26-1}
\end{eqnarray}
The corresponding hypergeometric solutions are written as
\begin{eqnarray}
&&\Phi_{_{[1\tilde{3}\tilde{5}7]}}^{(32),a}(\alpha,z)=
y_{_1}^{{D\over2}-2}y_{_3}^{{D}-3}\sum\limits_{n_{_1}=0}^\infty
\sum\limits_{n_{_2}=0}^\infty\sum\limits_{n_{_3}=0}^\infty\sum\limits_{n_{_4}=0}^\infty
c_{_{[1\tilde{3}\tilde{5}7]}}^{(32),a}(\alpha,{\bf n})
\nonumber\\
&&\hspace{2.5cm}\times
\Big({1\over y_{_1}}\Big)^{n_{_1}}\Big({y_{_4}\over y_{_1}}\Big)^{n_{_2}}
\Big({y_{_4}\over y_{_3}}\Big)^{n_{_3}}\Big({y_{_2}\over y_{_3}}\Big)^{n_{_4}}
\;,\nonumber\\
&&\Phi_{_{[1\tilde{3}\tilde{5}7]}}^{(32),b}(\alpha,z)=
y_{_1}^{{D\over2}-3}y_{_3}^{{D}-2}\sum\limits_{n_{_1}=0}^\infty
\sum\limits_{n_{_2}=0}^\infty\sum\limits_{n_{_3}=0}^\infty\sum\limits_{n_{_4}=0}^\infty
c_{_{[1\tilde{3}\tilde{5}7]}}^{(32),b}(\alpha,{\bf n})
\nonumber\\
&&\hspace{2.5cm}\times
\Big({1\over y_{_1}}\Big)^{n_{_1}}\Big({y_{_3}\over y_{_1}}\Big)^{n_{_2}}
\Big({y_{_4}\over y_{_1}}\Big)^{n_{_3}}\Big({y_{_2}\over y_{_1}}\Big)^{n_{_4}}\;.
\label{GKZ21h-26-2a}
\end{eqnarray}
Where the coefficients are
\begin{eqnarray}
&&c_{_{[1\tilde{3}\tilde{5}7]}}^{(32),a}(\alpha,{\bf n})=
(-)^{n_{_1}+n_{_3}}\Gamma(1+n_{_1}+n_{_2})\Big\{n_{_1}!n_{_2}!n_{_3}!n_{_4}!
\nonumber\\
&&\hspace{2.5cm}\times
\Gamma(2-{D\over2}+n_{_1})\Gamma(2-{D\over2}+n_{_3})
\nonumber\\
&&\hspace{2.5cm}\times
\Gamma(2-{D\over2}+n_{_4})
\Gamma({D\over2}-1-n_{_3}-n_{_4})\Gamma({D\over2}+n_{_2})
\nonumber\\
&&\hspace{2.5cm}\times
\Gamma({D\over2}-1-n_{_1}-n_{_2})\Gamma(D-2-n_{_3}-n_{_4})\Big\}^{-1}
\;,\nonumber\\
&&c_{_{[1\tilde{3}\tilde{5}7]}}^{(32),b}(\alpha,{\bf n})=
(-)^{1+n_{_1}}\Gamma(2+n_{_1}+n_{_2}+n_{_3})\Gamma(1+n_{_2})\Big\{n_{_1}!n_{_4}!
\nonumber\\
&&\hspace{2.5cm}\times
\Gamma(2+n_{_2}+n_{_3})\Gamma(2-D+n_{_1})\Gamma(1-{D\over2}-n_{_2})
\nonumber\\
&&\hspace{2.5cm}\times
\Gamma(2-{D\over2}+n_{_4})
\Gamma({D\over2}+n_{_2}-n_{_4})\Gamma({D\over2}+1+n_{_2}+n_{_3})
\nonumber\\
&&\hspace{2.5cm}\times
\Gamma({D\over2}-2-n_{_1}-n_{_2}-n_{_3})\Gamma(D-1+n_{_2}-n_{_4})\Big\}^{-1}\;.
\label{GKZ21h-26-3}
\end{eqnarray}
\end{itemize}

\section{The hypergeometric solutions of the integer lattice ${\bf B}_{_{1\tilde{3}5\tilde{7}}}$\label{app10}}
\indent\indent
\begin{itemize}
\item   $I_{_{1}}=\{2,3,4,6,7,10,\cdots,14\}$, i.e. the implement $J_{_{1}}=[1,14]\setminus I_{_{1}}=\{1,5,8,9\}$.
The choice implies the power numbers $\alpha_{_1}=\alpha_{_{5}}=\alpha_{_{8}}=\alpha_{_{9}}=0$, and
\begin{eqnarray}
&&\alpha_{_2}=a_{_1}-a_{_2},\;\alpha_{_3}=b_{_4}-a_{_3}-a_{_5}-1,\;\alpha_{_4}=b_{_4}-a_{_4}-a_{_5}-1,
\nonumber\\
&&\alpha_{_6}=b_{_1}+b_{_4}-a_{_1}-2,\;\alpha_{_7}=a_{_5}+b_{_2}+b_{_3}-b_{_4}-1,\;\alpha_{_{10}}=b_{_5}-b_{_4},
\nonumber\\
&&\alpha_{_{11}}=b_{_4}-a_{_1}-1,\;\alpha_{_{12}}=a_{_5}+b_{_3}-b_{_4},
\;\alpha_{_{13}}=1-b_{_3},\;\alpha_{_{14}}=-a_{_5}.
\label{GKZ21i-1-1}
\end{eqnarray}
The corresponding hypergeometric solutions are given as
\begin{eqnarray}
&&\Phi_{_{[1\tilde{3}5\tilde{7}]}}^{(1),a}(\alpha,z)=
y_{_1}^{{D\over2}-2}y_{_3}^{{D\over2}-1}y_{_4}^{-1}\sum\limits_{n_{_1}=0}^\infty
\sum\limits_{n_{_2}=0}^\infty\sum\limits_{n_{_3}=0}^\infty\sum\limits_{n_{_4}=0}^\infty
c_{_{[1\tilde{3}5\tilde{7}]}}^{(1),a}(\alpha,{\bf n})
\nonumber\\
&&\hspace{2.5cm}\times
\Big({1\over y_{_1}}\Big)^{n_{_1}}\Big({y_{_4}\over y_{_1}}\Big)^{n_{_2}}
\Big({y_{_2}\over y_{_4}}\Big)^{n_{_3}}\Big({y_{_3}\over y_{_4}}\Big)^{n_{_4}}
\;,\nonumber\\
&&\Phi_{_{[1\tilde{3}5\tilde{7}]}}^{(1),b}(\alpha,z)=
y_{_1}^{{D\over2}-2}y_{_2}^{-1}y_{_3}^{{D\over2}}y_{_4}^{-1}\sum\limits_{n_{_1}=0}^\infty
\sum\limits_{n_{_2}=0}^\infty\sum\limits_{n_{_3}=0}^\infty\sum\limits_{n_{_4}=0}^\infty
c_{_{[1\tilde{3}5\tilde{7}]}}^{(1),b}(\alpha,{\bf n})
\nonumber\\
&&\hspace{2.5cm}\times
\Big({1\over y_{_1}}\Big)^{n_{_1}}\Big({y_{_3}\over y_{_1}}\Big)^{n_{_2}}
\Big({y_{_3}\over y_{_4}}\Big)^{n_{_3}}\Big({y_{_3}\over y_{_2}}\Big)^{n_{_4}}\;.
\label{GKZ21i-1-2a}
\end{eqnarray}
Where the coefficients are
\begin{eqnarray}
&&c_{_{[1\tilde{3}5\tilde{7}]}}^{(1),a}(\alpha,{\bf n})=
(-)^{n_{_1}}\Gamma(1+n_{_1}+n_{_2})\Gamma(1+n_{_3}+n_{_4})
\Big\{n_{_1}!n_{_2}!n_{_3}!n_{_4}!
\nonumber\\
&&\hspace{2.5cm}\times
\Gamma({D\over2}+n_{_1})
\Gamma({D\over2}-1-n_{_3}-n_{_4})\Gamma({D\over2}-1-n_{_1}-n_{_2})
\nonumber\\
&&\hspace{2.5cm}\times
\Gamma(2-{D\over2}+n_{_3})\Gamma(2-{D\over2}+n_{_2})\Gamma({D\over2}+n_{_4})\Big\}^{-1}
\;,\nonumber\\
&&c_{_{[1\tilde{3}5\tilde{7}]}}^{(1),b}(\alpha,{\bf n})=
(-)^{n_{_1}+n_{_4}}\Gamma(1+n_{_1}+n_{_2})\Gamma(1+n_{_2}+n_{_3})\Gamma(1+n_{_4})
\nonumber\\
&&\hspace{2.5cm}\times
\Big\{n_{_1}!n_{_2}!\Gamma(2+n_{_2}+n_{_3}+n_{_4})\Gamma({D\over2}+n_{_1})
\nonumber\\
&&\hspace{2.5cm}\times
\Gamma({D\over2}-1-n_{_2}-n_{_3})\Gamma({D\over2}-1-n_{_1}-n_{_2})
\Gamma(1-{D\over2}-n_{_4})
\nonumber\\
&&\hspace{2.5cm}\times
\Gamma(2-{D\over2}+n_{_2})
\Gamma({D\over2}+1+n_{_2}+n_{_3}+n_{_4})\Big\}^{-1}\;.
\label{GKZ21i-1-3}
\end{eqnarray}

\item   $I_{_{2}}=\{2,3,4,6,7,8,10,11,12,14\}$, i.e. the implement $J_{_{2}}=[1,14]\setminus I_{_{2}}=\{1,5,9,13\}$.
The choice implies the power numbers $\alpha_{_1}=\alpha_{_{5}}=\alpha_{_{9}}=\alpha_{_{13}}=0$, and
\begin{eqnarray}
&&\alpha_{_2}=a_{_1}-a_{_2},\;\alpha_{_3}=b_{_4}-a_{_3}-a_{_5}-1,\;\alpha_{_4}=b_{_4}-a_{_4}-a_{_5}-1,
\nonumber\\
&&\alpha_{_6}=b_{_1}+b_{_4}-a_{_1}-2,\;\alpha_{_7}=a_{_5}+b_{_2}-b_{_4},\;\alpha_{_{8}}=b_{_3}-1,
\nonumber\\
&&\alpha_{_{10}}=b_{_5}-b_{_4},\;\alpha_{_{11}}=b_{_4}-a_{_1}-1,\;\alpha_{_{12}}=a_{_5}-b_{_4}+1,\;
\alpha_{_{14}}=-a_{_5}.
\label{GKZ21i-2-1}
\end{eqnarray}
The corresponding hypergeometric solutions are given as
\begin{eqnarray}
&&\Phi_{_{[1\tilde{3}5\tilde{7}]}}^{(2),a}(\alpha,z)=
y_{_1}^{{D\over2}-2}y_{_2}^{{D\over2}-1}y_{_4}^{-1}\sum\limits_{n_{_1}=0}^\infty
\sum\limits_{n_{_2}=0}^\infty\sum\limits_{n_{_3}=0}^\infty\sum\limits_{n_{_4}=0}^\infty
c_{_{[1\tilde{3}5\tilde{7}]}}^{(2),a}(\alpha,{\bf n})
\nonumber\\
&&\hspace{2.5cm}\times
\Big({1\over y_{_1}}\Big)^{n_{_1}}\Big({y_{_4}\over y_{_1}}\Big)^{n_{_2}}
\Big({y_{_2}\over y_{_4}}\Big)^{n_{_3}}\Big({y_{_3}\over y_{_4}}\Big)^{n_{_4}}
\;,\nonumber\\
&&\Phi_{_{[1\tilde{3}5\tilde{7}]}}^{(2),b}(\alpha,z)=
y_{_1}^{{D\over2}-2}y_{_2}^{{D\over2}-2}y_{_3}y_{_4}^{-1}\sum\limits_{n_{_1}=0}^\infty
\sum\limits_{n_{_2}=0}^\infty\sum\limits_{n_{_3}=0}^\infty\sum\limits_{n_{_4}=0}^\infty
c_{_{[1\tilde{3}5\tilde{7}]}}^{(2),b}(\alpha,{\bf n})
\nonumber\\
&&\hspace{2.5cm}\times
\Big({1\over y_{_1}}\Big)^{n_{_1}}\Big({y_{_3}\over y_{_1}}\Big)^{n_{_2}}
\Big({y_{_3}\over y_{_4}}\Big)^{n_{_3}}\Big({y_{_3}\over y_{_2}}\Big)^{n_{_4}}\;.
\label{GKZ21i-2-2a}
\end{eqnarray}
Where the coefficients are
\begin{eqnarray}
&&c_{_{[1\tilde{3}5\tilde{7}]}}^{(2),a}(\alpha,{\bf n})=
(-)^{n_{_1}}\Gamma(1+n_{_1}+n_{_2})\Gamma(1+n_{_3}+n_{_4})
\Big\{n_{_1}!n_{_2}!n_{_3}!n_{_4}!
\nonumber\\
&&\hspace{2.5cm}\times
\Gamma({D\over}+n_{_1})
\Gamma({D\over2}-1-n_{_3}-n_{_4})\Gamma({D\over2}-1-n_{_1}-n_{_2})
\nonumber\\
&&\hspace{2.5cm}\times
\Gamma(2-{D\over2}+n_{_4})\Gamma(2-{D\over2}+n_{_2})\Gamma({D\over2}+n_{_3})\Big\}^{-1}
\;,\nonumber\\
&&c_{_{[1\tilde{3}5\tilde{7}]}}^{(2),b}(\alpha,{\bf n})=
(-)^{n_{_1}+n_{_4}}\Gamma(1+n_{_1}+n_{_2})\Gamma(1+n_{_2}+n_{_3})\Gamma(1+n_{_4})
\nonumber\\
&&\hspace{2.5cm}\times
\Big\{n_{_1}!n_{_2}!\Gamma(2+n_{_2}+n_{_3}+n_{_4})\Gamma({D\over2}+n_{_1})
\nonumber\\
&&\hspace{2.5cm}\times
\Gamma({D\over2}-1-n_{_2}-n_{_3})\Gamma({D\over2}-1-n_{_1}-n_{_2})
\Gamma({D\over2}-1-n_{_4})
\nonumber\\
&&\hspace{2.5cm}\times
\Gamma(2-{D\over2}+n_{_2})
\Gamma(3-{D\over2}+n_{_2}+n_{_3}+n_{_4})\Big\}^{-1}\;.
\label{GKZ21i-2-3}
\end{eqnarray}

\item   $I_{_{3}}=\{2,3,4,6,7,9,11,\cdots,14\}$, i.e. the implement $J_{_{3}}=[1,14]\setminus I_{_{3}}=\{1,5,8,10\}$.
The choice implies the power numbers $\alpha_{_1}=\alpha_{_{5}}=\alpha_{_{8}}=\alpha_{_{10}}=0$, and
\begin{eqnarray}
&&\alpha_{_2}=a_{_1}-a_{_2},\;\alpha_{_3}=b_{_5}-a_{_3}-a_{_5}-1,\;\alpha_{_4}=b_{_5}-a_{_4}-a_{_5}-1,
\nonumber\\
&&\alpha_{_6}=b_{_1}+b_{_5}-a_{_1}-2,\;\alpha_{_7}=a_{_5}+b_{_2}+b_{_3}-b_{_5}-1,\;\alpha_{_{9}}=b_{_4}-b_{_5},
\nonumber\\
&&\alpha_{_{11}}=b_{_5}-a_{_1}-1,\;\alpha_{_{12}}=a_{_5}+b_{_3}-b_{_5},
\;\alpha_{_{13}}=1-b_{_3},\;\alpha_{_{14}}=-a_{_5}.
\label{GKZ21i-3-1}
\end{eqnarray}
The corresponding hypergeometric functions are presented as
\begin{eqnarray}
&&\Phi_{_{[1\tilde{3}5\tilde{7}]}}^{(3),a}(\alpha,z)=
y_{_1}^{-1}y_{_2}^{{D\over2}-1}y_{_3}^{{D\over2}-1}y_{_4}^{-1}\sum\limits_{n_{_1}=0}^\infty
\sum\limits_{n_{_2}=0}^\infty\sum\limits_{n_{_3}=0}^\infty\sum\limits_{n_{_4}=0}^\infty
c_{_{[1\tilde{3}5\tilde{7}]}}^{(3),a}(\alpha,{\bf n})
\nonumber\\
&&\hspace{2.5cm}\times
\Big({1\over y_{_1}}\Big)^{n_{_1}}\Big({y_{_4}\over y_{_1}}\Big)^{n_{_2}}
\Big({y_{_2}\over y_{_4}}\Big)^{n_{_3}}\Big({y_{_3}\over y_{_4}}\Big)^{n_{_4}}
\;,\nonumber\\
&&\Phi_{_{[1\tilde{3}5\tilde{7}]}}^{(3),b}(\alpha,z)=
y_{_1}^{-1}y_{_2}^{{D\over2}-2}y_{_3}^{{D\over2}}y_{_4}^{-1}\sum\limits_{n_{_1}=0}^\infty
\sum\limits_{n_{_2}=0}^\infty\sum\limits_{n_{_3}=0}^\infty\sum\limits_{n_{_4}=0}^\infty
c_{_{[1\tilde{3}5\tilde{7}]}}^{(3),b}(\alpha,{\bf n})
\nonumber\\
&&\hspace{2.5cm}\times
\Big({1\over y_{_1}}\Big)^{n_{_1}}\Big({y_{_3}\over y_{_1}}\Big)^{n_{_2}}
\Big({y_{_3}\over y_{_4}}\Big)^{n_{_3}}\Big({y_{_3}\over y_{_2}}\Big)^{n_{_4}}\;.
\label{GKZ21i-3-2a}
\end{eqnarray}
Where the coefficients are
\begin{eqnarray}
&&c_{_{[1\tilde{3}5\tilde{7}]}}^{(3),a}(\alpha,{\bf n})=
(-)^{n_{_1}}\Gamma(1+n_{_1}+n_{_2})\Gamma(1+n_{_3}+n_{_4})\Big\{n_{_1}!n_{_2}!n_{_3}!n_{_4}!
\Gamma({D\over2}+n_{_1})
\nonumber\\
&&\hspace{2.5cm}\times
\Gamma(1-{D\over2}-n_{_3}-n_{_4})
\Gamma(1-{D\over2}-n_{_1}-n_{_2})\Gamma({D\over2}+n_{_2})\Gamma({D\over2}+n_{_3})
\nonumber\\
&&\hspace{2.5cm}\times
\Gamma({D\over2}+n_{_4})\Big\}^{-1}
\;,\nonumber\\
&&c_{_{[1\tilde{3}5\tilde{7}]}}^{(3),b}(\alpha,{\bf n})=
(-)^{n_{_1}+n_{_4}}\Gamma(1+n_{_1}+n_{_2})\Gamma(1+n_{_2}+n_{_3})\Gamma(1+n_{_4})\Big\{n_{_1}!n_{_2}!
\nonumber\\
&&\hspace{2.5cm}\times
\Gamma(2+n_{_2}+n_{_3}+n_{_4})\Gamma({D\over2}+n_{_1})\Gamma(1-{D\over2}-n_{_2}-n_{_3})
\nonumber\\
&&\hspace{2.5cm}\times
\Gamma(1-{D\over2}-n_{_1}-n_{_2})\Gamma({D\over2}+n_{_2})
\Gamma({D\over2}-1-n_{_4})
\nonumber\\
&&\hspace{2.5cm}\times
\Gamma({D\over2}+1+n_{_2}+n_{_3}+n_{_4})\Big\}^{-1}\;.
\label{GKZ21i-3-3}
\end{eqnarray}

\item   $I_{_{4}}=\{2,3,4,6,7,8,9,11,12,14\}$, i.e. the implement $J_{_{4}}=[1,14]\setminus I_{_{4}}=\{1,5,10,13\}$.
The choice implies the power numbers $\alpha_{_1}=\alpha_{_{5}}=\alpha_{_{10}}=\alpha_{_{13}}=0$, and
\begin{eqnarray}
&&\alpha_{_2}=a_{_1}-a_{_2},\;\alpha_{_3}=b_{_5}-a_{_3}-a_{_5}-1,\;\alpha_{_4}=b_{_5}-a_{_4}-a_{_5}-1,
\nonumber\\
&&\alpha_{_6}=b_{_1}+b_{_5}-a_{_1}-2,\;\alpha_{_7}=a_{_5}+b_{_2}-b_{_5},\;\alpha_{_{8}}=b_{_3}-1,
\nonumber\\
&&\alpha_{_{9}}=b_{_4}-b_{_5},\;\alpha_{_{11}}=b_{_5}-a_{_1}-1,\;\alpha_{_{12}}=a_{_5}-b_{_5}+1,\;
\alpha_{_{14}}=-a_{_5}.
\label{GKZ21i-4-1}
\end{eqnarray}
The corresponding hypergeometric series is written as
\begin{eqnarray}
&&\Phi_{_{[1\tilde{3}5\tilde{7}]}}^{(4)}(\alpha,z)=
y_{_1}^{-1}y_{_2}^{{D}-2}y_{_4}^{-1}\sum\limits_{n_{_1}=0}^\infty
\sum\limits_{n_{_2}=0}^\infty\sum\limits_{n_{_3}=0}^\infty\sum\limits_{n_{_4}=0}^\infty
c_{_{[1\tilde{3}5\tilde{7}]}}^{(4)}(\alpha,{\bf n})
\nonumber\\
&&\hspace{2.5cm}\times
\Big({1\over y_{_1}}\Big)^{n_{_1}}\Big({y_{_2}\over y_{_1}}\Big)^{n_{_2}}
\Big({y_{_2}\over y_{_4}}\Big)^{n_{_3}}\Big({y_{_3}\over y_{_2}}\Big)^{n_{_4}}\;,
\label{GKZ21i-4-2}
\end{eqnarray}
with
\begin{eqnarray}
&&c_{_{[1\tilde{3}5\tilde{7}]}}^{(4)}(\alpha,{\bf n})=
(-)^{n_{_1}}\Gamma(1+n_{_1}+n_{_2})\Gamma(1+n_{_2}+n_{_3})
\Big\{n_{_1}!n_{_2}!n_{_4}!\Gamma({D\over2}+n_{_1})
\nonumber\\
&&\hspace{2.5cm}\times
\Gamma(1-{D\over2}-n_{_2}-n_{_3})\Gamma(1-{D\over2}-n_{_1}-n_{_2})
\Gamma({D\over2}+n_{_2}+n_{_3}-n_{_4})
\nonumber\\
&&\hspace{2.5cm}\times
\Gamma(2-{D\over2}+n_{_4})\Gamma({D\over2}+n_{_2})
\Gamma(D-1+n_{_2}+n_{_3}-n_{_4})\Big\}^{-1}\;.
\label{GKZ21i-4-3}
\end{eqnarray}

\item   $I_{_{5}}=\{1,3,4,6,7,10,\cdots,14\}$, i.e. the implement $J_{_{5}}=[1,14]\setminus I_{_{5}}=\{2,5,8,9\}$.
The choice implies the power numbers $\alpha_{_2}=\alpha_{_{5}}=\alpha_{_{8}}=\alpha_{_{9}}=0$, and
\begin{eqnarray}
&&\alpha_{_1}=a_{_2}-a_{_1},\;\alpha_{_3}=b_{_4}-a_{_3}-a_{_5}-1,\;\alpha_{_4}=b_{_4}-a_{_4}-a_{_5}-1,
\nonumber\\
&&\alpha_{_6}=b_{_1}+b_{_4}-a_{_2}-2,\;\alpha_{_7}=a_{_5}+b_{_2}+b_{_3}-b_{_4}-1,\;\alpha_{_{10}}=b_{_5}-b_{_4},
\nonumber\\
&&\alpha_{_{11}}=b_{_4}-a_{_2}-1,\;\alpha_{_{12}}=a_{_5}+b_{_3}-b_{_4},
\;\alpha_{_{13}}=1-b_{_3},\;\alpha_{_{14}}=-a_{_5}.
\label{GKZ21i-5-1}
\end{eqnarray}
The corresponding hypergeometric solutions are given as
\begin{eqnarray}
&&\Phi_{_{[1\tilde{3}5\tilde{7}]}}^{(5),a}(\alpha,z)=
y_{_1}^{{D}-3}y_{_3}^{{D\over2}-1}y_{_4}^{-1}\sum\limits_{n_{_1}=0}^\infty
\sum\limits_{n_{_2}=0}^\infty\sum\limits_{n_{_3}=0}^\infty\sum\limits_{n_{_4}=0}^\infty
c_{_{[1\tilde{3}5\tilde{7}]}}^{(5),a}(\alpha,{\bf n})
\nonumber\\
&&\hspace{2.5cm}\times
\Big({1\over y_{_1}}\Big)^{n_{_1}}\Big({y_{_4}\over y_{_1}}\Big)^{n_{_2}}
\Big({y_{_2}\over y_{_4}}\Big)^{n_{_3}}\Big({y_{_3}\over y_{_4}}\Big)^{n_{_4}}
\;,\nonumber\\
&&\Phi_{_{[1\tilde{3}5\tilde{7}]}}^{(5),b}(\alpha,z)=
y_{_1}^{{D}-3}y_{_2}^{-1}y_{_3}^{{D\over2}}y_{_4}^{-1}\sum\limits_{n_{_1}=0}^\infty
\sum\limits_{n_{_2}=0}^\infty\sum\limits_{n_{_3}=0}^\infty\sum\limits_{n_{_4}=0}^\infty
c_{_{[1\tilde{3}5\tilde{7}]}}^{(5),b}(\alpha,{\bf n})
\nonumber\\
&&\hspace{2.5cm}\times
\Big({1\over y_{_1}}\Big)^{n_{_1}}\Big({y_{_3}\over y_{_1}}\Big)^{n_{_2}}
\Big({y_{_3}\over y_{_4}}\Big)^{n_{_3}}\Big({y_{_3}\over y_{_2}}\Big)^{n_{_4}}\;.
\label{GKZ21i-5-2a}
\end{eqnarray}
Where the coefficients are
\begin{eqnarray}
&&c_{_{[1\tilde{3}5\tilde{7}]}}^{(5),a}(\alpha,{\bf n})=
(-)^{n_{_2}}\Gamma(1+n_{_3}+n_{_4})\Big\{n_{_1}!n_{_2}!n_{_3}!n_{_4}!
\nonumber\\
&&\hspace{2.5cm}\times
\Gamma(2-{D\over2}+n_{_1})\Gamma({D\over2}-1-n_{_3}-n_{_4})
\nonumber\\
&&\hspace{2.5cm}\times
\Gamma({D\over2}-1-n_{_1}-n_{_2})\Gamma(2-{D\over2}+n_{_3})
\nonumber\\
&&\hspace{2.5cm}\times
\Gamma(2-{D\over2}+n_{_2})\Gamma(D-2-n_{_1}-n_{_2})\Gamma({D\over2}+n_{_4})\Big\}^{-1}
\;,\nonumber\\
&&c_{_{[1\tilde{3}5\tilde{7}]}}^{(5),b}(\alpha,{\bf n})=
(-)^{n_{_2}+n_{_4}}\Gamma(1+n_{_2}+n_{_3})\Gamma(1+n_{_4})\Big\{n_{_1}!n_{_2}!
\nonumber\\
&&\hspace{2.5cm}\times
\Gamma(2+n_{_2}+n_{_3}+n_{_4})\Gamma(2-{D\over2}+n_{_1})\Gamma({D\over2}-1-n_{_2}-n_{_3})
\nonumber\\
&&\hspace{2.5cm}\times
\Gamma({D\over2}-1-n_{_1}-n_{_2})\Gamma(1-{D\over2}-n_{_4})
\Gamma(2-{D\over2}+n_{_2})
\nonumber\\
&&\hspace{2.5cm}\times
\Gamma(D-2-n_{_1}-n_{_2})\Gamma({D\over2}+1+n_{_2}+n_{_3}+n_{_4})\Big\}^{-1}\;.
\label{GKZ21i-5-3}
\end{eqnarray}

\item   $I_{_{6}}=\{1,3,4,6,7,8,10,11,12,14\}$, i.e. the implement $J_{_{6}}=[1,14]\setminus I_{_{6}}=\{2,5,9,13\}$.
The choice implies the power numbers $\alpha_{_2}=\alpha_{_{5}}=\alpha_{_{9}}=\alpha_{_{13}}=0$, and
\begin{eqnarray}
&&\alpha_{_1}=a_{_2}-a_{_1},\;\alpha_{_3}=b_{_4}-a_{_3}-a_{_5}-1,\;\alpha_{_4}=b_{_4}-a_{_4}-a_{_5}-1,
\nonumber\\
&&\alpha_{_6}=b_{_1}+b_{_4}-a_{_2}-2,\;\alpha_{_7}=a_{_5}+b_{_2}-b_{_4},\;\alpha_{_{8}}=b_{_3}-1,
\nonumber\\
&&\alpha_{_{10}}=b_{_5}-b_{_4},\;\alpha_{_{11}}=b_{_4}-a_{_2}-1,\;\alpha_{_{12}}=a_{_5}-b_{_4}+1,\;
\alpha_{_{14}}=-a_{_5}.
\label{GKZ21i-6-1}
\end{eqnarray}
The corresponding hypergeometric solutions are written as
\begin{eqnarray}
&&\Phi_{_{[1\tilde{3}5\tilde{7}]}}^{(6),a}(\alpha,z)=
y_{_1}^{{D}-3}y_{_2}^{{D\over2}-1}y_{_4}^{-1}\sum\limits_{n_{_1}=0}^\infty
\sum\limits_{n_{_2}=0}^\infty\sum\limits_{n_{_3}=0}^\infty\sum\limits_{n_{_4}=0}^\infty
c_{_{[1\tilde{3}5\tilde{7}]}}^{(6),a}(\alpha,{\bf n})
\nonumber\\
&&\hspace{2.5cm}\times
\Big({1\over y_{_1}}\Big)^{n_{_1}}\Big({y_{_4}\over y_{_1}}\Big)^{n_{_2}}
\Big({y_{_2}\over y_{_4}}\Big)^{n_{_3}}\Big({y_{_3}\over y_{_4}}\Big)^{n_{_4}}
\;,\nonumber\\
&&\Phi_{_{[1\tilde{3}5\tilde{7}]}}^{(6),b}(\alpha,z)=
y_{_1}^{{D}-3}y_{_2}^{{D\over2}-2}y_{_3}y_{_4}^{-1}\sum\limits_{n_{_1}=0}^\infty
\sum\limits_{n_{_2}=0}^\infty\sum\limits_{n_{_3}=0}^\infty\sum\limits_{n_{_4}=0}^\infty
c_{_{[1\tilde{3}5\tilde{7}]}}^{(6),b}(\alpha,{\bf n})
\nonumber\\
&&\hspace{2.5cm}\times
\Big({1\over y_{_1}}\Big)^{n_{_1}}\Big({y_{_3}\over y_{_1}}\Big)^{n_{_2}}
\Big({y_{_3}\over y_{_4}}\Big)^{n_{_3}}\Big({y_{_3}\over y_{_2}}\Big)^{n_{_4}}\;.
\label{GKZ21i-6-2a}
\end{eqnarray}
Where the coefficients are
\begin{eqnarray}
&&c_{_{[1\tilde{3}5\tilde{7}]}}^{(6),a}(\alpha,{\bf n})=
(-)^{n_{_2}}\Gamma(1+n_{_3}+n_{_4})\Big\{n_{_1}!n_{_2}!n_{_3}!n_{_4}!
\Gamma(2-{D\over2}+n_{_1})
\nonumber\\
&&\hspace{2.5cm}\times
\Gamma({D\over2}-1-n_{_3}-n_{_4})
\Gamma({D\over2}-1-n_{_1}-n_{_2})\Gamma(2-{D\over2}+n_{_4})
\nonumber\\
&&\hspace{2.5cm}\times
\Gamma(2-{D\over2}+n_{_2})\Gamma(D-2-n_{_1}-n_{_2})\Gamma({D\over2}+n_{_3})\Big\}^{-1}
\;,\nonumber\\
&&c_{_{[1\tilde{3}5\tilde{7}]}}^{(6),b}(\alpha,{\bf n})=
(-)^{n_{_2}+n_{_4}}\Gamma(1+n_{_2}+n_{_3})\Gamma(1+n_{_4})\Big\{n_{_1}!n_{_2}!
\nonumber\\
&&\hspace{2.5cm}\times
\Gamma(2+n_{_2}+n_{_3}+n_{_4})\Gamma(2-{D\over2}+n_{_1})\Gamma({D\over2}-1-n_{_2}-n_{_3})
\nonumber\\
&&\hspace{2.5cm}\times
\Gamma({D\over2}-1-n_{_1}-n_{_2})\Gamma({D\over2}-1-n_{_4})
\Gamma(2-{D\over2}+n_{_2})
\nonumber\\
&&\hspace{2.5cm}\times
\Gamma(D-2-n_{_1}-n_{_2})\Gamma(3-{D\over2}+n_{_2}+n_{_3}+n_{_4})\Big\}^{-1}\;.
\label{GKZ21i-6-3}
\end{eqnarray}

\item   $I_{_{7}}=\{1,3,4,6,7,9,11,\cdots,14\}$, i.e. the implement $J_{_{7}}=[1,14]\setminus I_{_{7}}=\{2,5,8,10\}$.
The choice implies the power numbers $\alpha_{_2}=\alpha_{_{5}}=\alpha_{_{8}}=\alpha_{_{10}}=0$, and
\begin{eqnarray}
&&\alpha_{_1}=a_{_2}-a_{_1},\;\alpha_{_3}=b_{_5}-a_{_3}-a_{_5}-1,\;\alpha_{_4}=b_{_5}-a_{_4}-a_{_5}-1,
\nonumber\\
&&\alpha_{_6}=b_{_1}+b_{_5}-a_{_2}-2,\;\alpha_{_7}=a_{_5}+b_{_2}+b_{_3}-b_{_5}-1,\;\alpha_{_{9}}=b_{_4}-b_{_5},
\nonumber\\
&&\alpha_{_{11}}=b_{_5}-a_{_2}-1,\;\alpha_{_{12}}=a_{_5}+b_{_3}-b_{_5},
\;\alpha_{_{13}}=1-b_{_3},\;\alpha_{_{14}}=-a_{_5}.
\label{GKZ21i-7-1}
\end{eqnarray}
The corresponding hypergeometric solutions are
\begin{eqnarray}
&&\Phi_{_{[1\tilde{3}5\tilde{7}]}}^{(7),a}(\alpha,z)=
y_{_1}^{{D\over2}-2}y_{_2}^{{D\over2}-1}y_{_3}^{{D\over2}-1}y_{_4}^{-1}
\sum\limits_{n_{_1}=0}^\infty
\sum\limits_{n_{_2}=0}^\infty\sum\limits_{n_{_3}=0}^\infty\sum\limits_{n_{_4}=0}^\infty
c_{_{[1\tilde{3}5\tilde{7}]}}^{(7),a}(\alpha,{\bf n})
\nonumber\\
&&\hspace{2.5cm}\times
\Big({1\over y_{_1}}\Big)^{n_{_1}}\Big({y_{_4}\over y_{_1}}\Big)^{n_{_2}}
\Big({y_{_2}\over y_{_4}}\Big)^{n_{_3}}\Big({y_{_3}\over y_{_4}}\Big)^{n_{_4}}
\;,\nonumber\\
&&\Phi_{_{[1\tilde{3}5\tilde{7}]}}^{(7),b}(\alpha,z)=
y_{_1}^{{D\over2}-2}y_{_2}^{{D\over2}-2}y_{_3}^{{D\over2}}y_{_4}^{-1}
\sum\limits_{n_{_1}=0}^\infty
\sum\limits_{n_{_2}=0}^\infty\sum\limits_{n_{_3}=0}^\infty\sum\limits_{n_{_4}=0}^\infty
c_{_{[1\tilde{3}5\tilde{7}]}}^{(7),b}(\alpha,{\bf n})
\nonumber\\
&&\hspace{2.5cm}\times
\Big({1\over y_{_1}}\Big)^{n_{_1}}\Big({y_{_3}\over y_{_1}}\Big)^{n_{_2}}
\Big({y_{_3}\over y_{_4}}\Big)^{n_{_3}}\Big({y_{_3}\over y_{_2}}\Big)^{n_{_4}}\;.
\label{GKZ21i-7-2a}
\end{eqnarray}
Where the coefficients are
\begin{eqnarray}
&&c_{_{[1\tilde{3}5\tilde{7}]}}^{(7),a}(\alpha,{\bf n})=
(-)^{n_{_1}}\Gamma(1+n_{_1}+n_{_2})\Gamma(1+n_{_3}+n_{_4})
\Big\{n_{_1}!n_{_2}!n_{_3}!n_{_4}!
\nonumber\\
&&\hspace{2.5cm}\times
\Gamma(2-{D\over2}+n_{_1})\Gamma(1-{D\over2}-n_{_3}-n_{_4})
\Gamma({D\over2}+n_{_2})
\nonumber\\
&&\hspace{2.5cm}\times
\Gamma({D\over2}-1-n_{_1}-n_{_2})
\Gamma({D\over2}+n_{_3})\Gamma({D\over2}+n_{_4})\Big\}^{-1}
\;,\nonumber\\
&&c_{_{[1\tilde{3}5\tilde{7}]}}^{(7),b}(\alpha,{\bf n})=
(-)^{n_{_1}+n_{_4}}\Gamma(1+n_{_1}+n_{_2})\Gamma(1+n_{_2}+n_{_3})\Gamma(1+n_{_4})
\nonumber\\
&&\hspace{2.5cm}\times
\Big\{n_{_1}!n_{_2}!\Gamma(2+n_{_2}+n_{_3}+n_{_4})\Gamma(2-{D\over2}+n_{_1})
\nonumber\\
&&\hspace{2.5cm}\times
\Gamma(1-{D\over2}-n_{_2}-n_{_3})
\Gamma({D\over2}+n_{_2})\Gamma({D\over2}-1-n_{_1}-n_{_2})
\nonumber\\
&&\hspace{2.5cm}\times
\Gamma({D\over2}+1+n_{_2}+n_{_3}+n_{_4})\Gamma({D\over2}-1-n_{_4})\Big\}^{-1}\;.
\label{GKZ21i-7-3}
\end{eqnarray}

\item   $I_{_{8}}=\{1,3,4,6,7,8,9,11,12,14\}$, i.e. the implement $J_{_{8}}=[1,14]\setminus I_{_{8}}=\{2,5,10,13\}$.
The choice implies the power numbers $\alpha_{_2}=\alpha_{_{5}}=\alpha_{_{10}}=\alpha_{_{13}}=0$, and
\begin{eqnarray}
&&\alpha_{_1}=a_{_2}-a_{_1},\;\alpha_{_3}=b_{_5}-a_{_3}-a_{_5}-1,\;\alpha_{_4}=b_{_5}-a_{_4}-a_{_5}-1,
\nonumber\\
&&\alpha_{_6}=b_{_1}+b_{_5}-a_{_2}-2,\;\alpha_{_7}=a_{_5}+b_{_2}-b_{_5},\;\alpha_{_{8}}=b_{_3}-1,
\nonumber\\
&&\alpha_{_{9}}=b_{_4}-b_{_5},\;\alpha_{_{11}}=b_{_5}-a_{_2}-1,\;\alpha_{_{12}}=a_{_5}-b_{_5}+1,\;
\alpha_{_{14}}=-a_{_5}.
\label{GKZ21i-8-1}
\end{eqnarray}
The corresponding hypergeometric solution is
\begin{eqnarray}
&&\Phi_{_{[1\tilde{3}5\tilde{7}]}}^{(8)}(\alpha,z)=
y_{_1}^{{D\over2}-2}y_{_2}^{{D}-2}y_{_4}^{-1}\sum\limits_{n_{_1}=0}^\infty
\sum\limits_{n_{_2}=0}^\infty\sum\limits_{n_{_3}=0}^\infty\sum\limits_{n_{_4}=0}^\infty
c_{_{[1\tilde{3}5\tilde{7}]}}^{(8)}(\alpha,{\bf n})
\nonumber\\
&&\hspace{2.5cm}\times
\Big({1\over y_{_1}}\Big)^{n_{_1}}\Big({y_{_2}\over y_{_1}}\Big)^{n_{_2}}
\Big({y_{_2}\over y_{_4}}\Big)^{n_{_3}}\Big({y_{_3}\over y_{_2}}\Big)^{n_{_4}}\;,
\label{GKZ21i-8-2}
\end{eqnarray}
with
\begin{eqnarray}
&&c_{_{[1\tilde{3}5\tilde{7}]}}^{(8)}(\alpha,{\bf n})=
(-)^{n_{_1}}\Gamma(1+n_{_1}+n_{_2})\Gamma(1+n_{_2}+n_{_3})
\nonumber\\
&&\hspace{2.5cm}\times
\Big\{n_{_1}!n_{_2}!n_{_4}!\Gamma(2-{D\over2}+n_{_1})
\Gamma(1-{D\over2}-n_{_2}-n_{_3})
\nonumber\\
&&\hspace{2.5cm}\times
\Gamma({D\over2}+n_{_2}+n_{_3}-n_{_4})\Gamma(2-{D\over2}+n_{_4})
\Gamma({D\over2}+n_{_2})
\nonumber\\
&&\hspace{2.5cm}\times
\Gamma({D\over2}-1-n_{_1}-n_{_2})\Gamma(D-1+n_{_2}+n_{_3}-n_{_4})\Big\}^{-1}\;.
\label{GKZ21i-8-3}
\end{eqnarray}
\end{itemize}

\section{The hypergeometric solutions of the integer lattice ${\bf B}_{_{13\widetilde{57}}}$\label{app11}}
\indent\indent
\begin{itemize}
\item   $I_{_{1}}=\{2,4,5,6,7,10,\cdots,14\}$, i.e. the implement $J_{_{1}}=[1,14]\setminus I_{_{1}}=\{1,3,8,9\}$.
The choice implies the power numbers $\alpha_{_1}=\alpha_{_{3}}=\alpha_{_{8}}=\alpha_{_{9}}=0$, and
\begin{eqnarray}
&&\alpha_{_2}=a_{_1}-a_{_2},\;\alpha_{_4}=a_{_3}-a_{_4},\;\alpha_{_5}=b_{_4}-a_{_3}-a_{_5}-1,
\nonumber\\
&&\alpha_{_6}=b_{_1}+b_{_4}-a_{_1}-2,\;\alpha_{_7}=b_{_2}+b_{_3}-a_{_3}-2,\;\alpha_{_{10}}=b_{_5}-b_{_4},
\nonumber\\
&&\alpha_{_{11}}=b_{_4}-a_{_1}-1,\;\alpha_{_{12}}=b_{_3}-a_{_3}-1,\;\alpha_{_{13}}=1-b_{_3},
\nonumber\\
&&\alpha_{_{14}}=a_{_3}-b_{_4}+1.
\label{GKZ21j-1-1}
\end{eqnarray}
The corresponding hypergeometric series is written as
\begin{eqnarray}
&&\Phi_{_{[13\tilde{5}\tilde{7}]}}^{(1)}(\alpha,z)=
y_{_1}^{{D\over2}-2}y_{_2}^{-1}y_{_3}^{{D\over2}-1}\sum\limits_{n_{_1}=0}^\infty
\sum\limits_{n_{_2}=0}^\infty\sum\limits_{n_{_3}=0}^\infty\sum\limits_{n_{_4}=0}^\infty
c_{_{[13\tilde{5}\tilde{7}]}}^{(1)}(\alpha,{\bf n})
\nonumber\\
&&\hspace{2.5cm}\times
\Big({1\over y_{_1}}\Big)^{n_{_1}}\Big({y_{_4}\over y_{_2}}\Big)^{n_{_2}}
\Big({y_{_4}\over y_{_1}}\Big)^{n_{_3}}\Big({y_{_3}\over y_{_2}}\Big)^{n_{_4}},
\label{GKZ21j-1-2}
\end{eqnarray}
with
\begin{eqnarray}
&&c_{_{[13\tilde{5}\tilde{7}]}}^{(1)}(\alpha,{\bf n})=
(-)^{n_{_1}+n_{_4}}\Gamma(1+n_{_1}+n_{_3})\Gamma(1+n_{_2}+n_{_4}))
\Big\{n_{_1}!n_{_2}!n_{_3}!n_{_4}!
\nonumber\\
&&\hspace{2.5cm}\times
\Gamma({D\over2}+n_{_1})\Gamma({D\over2}+n_{_2})\Gamma(1-{D\over2}-n_{_2}-n_{_4})
\nonumber\\
&&\hspace{2.5cm}\times
\Gamma(2-{D\over2}+n_{_3})\Gamma({D\over2}-1-n_{_1}-n_{_3})\Gamma({D\over2}+n_{_4})\Big\}^{-1}\;.
\label{GKZ21j-1-3}
\end{eqnarray}

\item   $I_{_{2}}=\{2,4,5,6,7,8,10,11,12,14\}$, i.e. the implement $J_{_{2}}=[1,14]\setminus I_{_{2}}=\{1,3,9,13\}$.
The choice implies the power numbers $\alpha_{_1}=\alpha_{_{3}}=\alpha_{_{9}}=\alpha_{_{13}}=0$, and
\begin{eqnarray}
&&\alpha_{_2}=a_{_1}-a_{_2},\;\alpha_{_4}=a_{_3}-a_{_4},\;\alpha_{_5}=b_{_4}-a_{_3}-a_{_5}-1,
\nonumber\\
&&\alpha_{_6}=b_{_1}+b_{_4}-a_{_1}-2,\;\alpha_{_7}=b_{_2}-a_{_3}-1,\;\alpha_{_{8}}=b_{_3}-1,
\nonumber\\
&&\alpha_{_{10}}=b_{_5}-b_{_4},\;\alpha_{_{11}}=b_{_4}-a_{_1}-1,\;\alpha_{_{12}}=-a_{_3},\;
\alpha_{_{14}}=a_{_3}-b_{_4}+1.
\label{GKZ21j-2-1}
\end{eqnarray}
The corresponding hypergeometric series is written as
\begin{eqnarray}
&&\Phi_{_{[13\tilde{5}\tilde{7}]}}^{(2)}(\alpha,z)=
y_{_1}^{{D\over2}-2}y_{_2}^{{D\over2}-2}\sum\limits_{n_{_1}=0}^\infty
\sum\limits_{n_{_2}=0}^\infty\sum\limits_{n_{_3}=0}^\infty\sum\limits_{n_{_4}=0}^\infty
c_{_{[13\tilde{5}\tilde{7}]}}^{(2)}(\alpha,{\bf n})
\nonumber\\
&&\hspace{2.5cm}\times
\Big({1\over y_{_1}}\Big)^{n_{_1}}\Big({y_{_4}\over y_{_2}}\Big)^{n_{_2}}
\Big({y_{_4}\over y_{_1}}\Big)^{n_{_3}}\Big({y_{_3}\over y_{_2}}\Big)^{n_{_4}}\;.
\label{GKZ21j-2-2}
\end{eqnarray}
with
\begin{eqnarray}
&&c_{_{[13\tilde{5}\tilde{7}]}}^{(2)}(\alpha,{\bf n})=
(-)^{n_{_1}+n_{_4}}\Gamma(1+n_{_1}+n_{_3})\Gamma(1+n_{_2}+n_{_4}))
\Big\{n_{_1}!n_{_2}!n_{_3}!n_{_4}!
\nonumber\\
&&\hspace{2.5cm}\times
\Gamma({D\over2}+n_{_1})\Gamma({D\over2}+n_{_2})
\Gamma(2-{D\over2}+n_{_4})\Gamma(2-{D\over2}+n_{_3})
\nonumber\\
&&\hspace{2.5cm}\times
\Gamma({D\over2}-1-n_{_1}-n_{_3})\Gamma({D\over2}-1-n_{_2}-n_{_4})\Big\}^{-1}\;.
\label{GKZ21j-2-3}
\end{eqnarray}

\item   $I_{_{3}}=\{2,4,5,6,7,9,11,\cdots,14\}$, i.e. the implement $J_{_{3}}=[1,14]\setminus I_{_{3}}=\{1,3,8,10\}$.
The choice implies the power numbers $\alpha_{_1}=\alpha_{_{3}}=\alpha_{_{8}}=\alpha_{_{10}}=0$, and
\begin{eqnarray}
&&\alpha_{_2}=a_{_1}-a_{_2},\;\alpha_{_4}=a_{_3}-a_{_4},\;\alpha_{_5}=b_{_5}-a_{_3}-a_{_5}-1,
\nonumber\\
&&\alpha_{_6}=b_{_1}+b_{_5}-a_{_1}-2,\;\alpha_{_7}=b_{_2}+b_{_3}-a_{_3}-2,\;\alpha_{_{9}}=b_{_4}-b_{_5},
\nonumber\\
&&\alpha_{_{11}}=b_{_5}-a_{_1}-1,\;\alpha_{_{12}}=b_{_3}-a_{_3}-1,\;\alpha_{_{13}}=1-b_{_3},
\nonumber\\
&&\alpha_{_{14}}=a_{_3}-b_{_5}+1.
\label{GKZ21j-3-1}
\end{eqnarray}
The corresponding hypergeometric solution is
\begin{eqnarray}
&&\Phi_{_{[13\tilde{5}\tilde{7}]}}^{(3)}(\alpha,z)=
y_{_1}^{-1}y_{_2}^{-1}y_{_3}^{{D\over2}-1}y_{_4}^{{D\over2}-1}
\sum\limits_{n_{_1}=0}^\infty
\sum\limits_{n_{_2}=0}^\infty\sum\limits_{n_{_3}=0}^\infty\sum\limits_{n_{_4}=0}^\infty
c_{_{[13\tilde{5}\tilde{7}]}}^{(3)}(\alpha,{\bf n})
\nonumber\\
&&\hspace{2.5cm}\times
\Big({1\over y_{_1}}\Big)^{n_{_1}}\Big({y_{_4}\over y_{_2}}\Big)^{n_{_2}}
\Big({y_{_4}\over y_{_1}}\Big)^{n_{_3}}\Big({y_{_3}\over y_{_2}}\Big)^{n_{_4}}\;,
\label{GKZ21j-3-2}
\end{eqnarray}
with
\begin{eqnarray}
&&c_{_{[13\tilde{5}\tilde{7}]}}^{(3)}(\alpha,{\bf n})=
(-)^{n_{_1}+n_{_4}}\Gamma(1+n_{_1}+n_{_3})\Gamma(1+n_{_2}+n_{_4})
\Big\{n_{_1}!n_{_2}!n_{_3}!n_{_4}!
\nonumber\\
&&\hspace{2.5cm}\times
\Gamma({D\over2}+n_{_1})\Gamma({D\over2}+n_{_2})
\Gamma(1-{D\over2}-n_{_1}-n_{_3})
\nonumber\\
&&\hspace{2.5cm}\times
\Gamma(1-{D\over2}-n_{_2}-n_{_4})\Gamma({D\over2}+n_{_3})
\Gamma({D\over2}+n_{_4})\Big\}^{-1}\;.
\label{GKZ21j-3-3}
\end{eqnarray}

\item   $I_{_{4}}=\{2,4,\cdots,9,11,12,14\}$, i.e. the implement $J_{_{4}}=[1,14]\setminus I_{_{4}}=\{1,3,10,13\}$.
The choice implies the power numbers $\alpha_{_1}=\alpha_{_{3}}=\alpha_{_{10}}=\alpha_{_{13}}=0$, and
\begin{eqnarray}
&&\alpha_{_2}=a_{_1}-a_{_2},\;\alpha_{_4}=a_{_3}-a_{_4},\;\alpha_{_5}=b_{_5}-a_{_3}-a_{_5}-1,
\nonumber\\
&&\alpha_{_6}=b_{_1}+b_{_5}-a_{_1}-2,\;\alpha_{_7}=b_{_2}-a_{_3}-1,\;\alpha_{_{8}}=b_{_3}-1,
\nonumber\\
&&\alpha_{_{9}}=b_{_4}-b_{_5},\;\alpha_{_{11}}=b_{_5}-a_{_1}-1,\;\alpha_{_{12}}=-a_{_3},\;
\alpha_{_{14}}=a_{_3}-b_{_5}+1.
\label{GKZ21j-4-1}
\end{eqnarray}
The corresponding hypergeometric solution is written as
\begin{eqnarray}
&&\Phi_{_{[13\tilde{5}\tilde{7}]}}^{(4)}(\alpha,z)=
y_{_1}^{-1}y_{_2}^{{D\over2}-2}y_{_4}^{{D\over2}-1}\sum\limits_{n_{_1}=0}^\infty
\sum\limits_{n_{_2}=0}^\infty\sum\limits_{n_{_3}=0}^\infty\sum\limits_{n_{_4}=0}^\infty
c_{_{[13\tilde{5}\tilde{7}]}}^{(4)}(\alpha,{\bf n})
\nonumber\\
&&\hspace{2.5cm}\times
\Big({1\over y_{_1}}\Big)^{n_{_1}}\Big({y_{_4}\over y_{_2}}\Big)^{n_{_2}}
\Big({y_{_4}\over y_{_1}}\Big)^{n_{_3}}\Big({y_{_3}\over y_{_2}}\Big)^{n_{_4}}\;,
\label{GKZ21j-4-2}
\end{eqnarray}
with
\begin{eqnarray}
&&c_{_{[13\tilde{5}\tilde{7}]}}^{(4)}(\alpha,{\bf n})=
(-)^{n_{_1}+n_{_4}}\Gamma(1+n_{_1}+n_{_3})\Gamma(1+n_{_2}+n_{_4})
\Big\{n_{_1}!n_{_2}!n_{_3}!n_{_4}!
\nonumber\\
&&\hspace{2.5cm}\times
\Gamma({D\over2}+n_{_1})\Gamma({D\over2}+n_{_2})
\Gamma(1-{D\over2}-n_{_1}-n_{_3})
\nonumber\\
&&\hspace{2.5cm}\times
\Gamma(2-{D\over2}+n_{_4})\Gamma({D\over2}+n_{_3})
\Gamma({D\over2}-1-n_{_2}-n_{_4})\Big\}^{-1}\;.
\label{GKZ21j-4-3}
\end{eqnarray}

\item   $I_{_{5}}=\{2,3,5,6,7,10,\cdots,14\}$, i.e. the implement $J_{_{5}}=[1,14]\setminus I_{_{5}}=\{1,4,8,9\}$.
The choice implies the power numbers $\alpha_{_1}=\alpha_{_{4}}=\alpha_{_{8}}=\alpha_{_{9}}=0$, and
\begin{eqnarray}
&&\alpha_{_2}=a_{_1}-a_{_2},\;\alpha_{_3}=a_{_4}-a_{_3},\;\alpha_{_5}=b_{_4}-a_{_4}-a_{_5}-1,
\nonumber\\
&&\alpha_{_6}=b_{_1}+b_{_4}-a_{_1}-2,\;\alpha_{_7}=b_{_2}+b_{_3}-a_{_4}-2,\;\alpha_{_{10}}=b_{_5}-b_{_4},
\nonumber\\
&&\alpha_{_{11}}=b_{_4}-a_{_1}-1,\;\alpha_{_{12}}=b_{_3}-a_{_4}-1,\;\alpha_{_{13}}=1-b_{_3},
\nonumber\\
&&\alpha_{_{14}}=a_{_4}-b_{_4}+1.
\label{GKZ21j-7-1}
\end{eqnarray}
The corresponding hypergeometric solution is written as
\begin{eqnarray}
&&\Phi_{_{[13\tilde{5}\tilde{7}]}}^{(5)}(\alpha,z)=
y_{_1}^{{D\over2}-2}y_{_2}^{{D\over2}-2}y_{_3}^{{D\over2}-1}y_{_4}^{1-{D\over2}}
\sum\limits_{n_{_1}=0}^\infty
\sum\limits_{n_{_2}=0}^\infty\sum\limits_{n_{_3}=0}^\infty\sum\limits_{n_{_4}=0}^\infty
c_{_{[13\tilde{5}\tilde{7}]}}^{(5)}(\alpha,{\bf n})
\nonumber\\
&&\hspace{2.5cm}\times
\Big({1\over y_{_1}}\Big)^{n_{_1}}\Big({y_{_4}\over y_{_2}}\Big)^{n_{_2}}
\Big({y_{_4}\over y_{_1}}\Big)^{n_{_3}}\Big({y_{_3}\over y_{_2}}\Big)^{n_{_4}}\;,
\label{GKZ21j-7-2}
\end{eqnarray}
with
\begin{eqnarray}
&&c_{_{[13\tilde{5}\tilde{7}]}}^{(5)}(\alpha,{\bf n})=
(-)^{1+n_{_1}+n_{_4}}\Gamma(1+n_{_1}+n_{_3})\Gamma(1+n_{_2}+n_{_4})
\nonumber\\
&&\hspace{2.5cm}\times
\Big\{n_{_1}!n_{_2}!n_{_3}!n_{_4}!
\Gamma({D\over2}+n_{_1})\Gamma(2-{D\over2}+n_{_2})
\Gamma(2-{D\over2}+n_{_3})
\nonumber\\
&&\hspace{2.5cm}\times
\Gamma({D\over2}-1-n_{_1}-n_{_3})\Gamma({D\over2}-1-n_{_2}-n_{_4})
\Gamma({D\over2}+n_{_4})\Big\}^{-1}\;.
\label{GKZ21j-7-3}
\end{eqnarray}

\item   $I_{_{6}}=\{2,3,5,6,7,8,10,11,12,14\}$, i.e. the implement $J_{_{6}}=[1,14]\setminus I_{_{6}}=\{1,4,9,13\}$.
The choice implies the power numbers $\alpha_{_1}=\alpha_{_{4}}=\alpha_{_{9}}=\alpha_{_{13}}=0$, and
\begin{eqnarray}
&&\alpha_{_2}=a_{_1}-a_{_2},\;\alpha_{_3}=a_{_4}-a_{_3},\;\alpha_{_5}=b_{_4}-a_{_4}-a_{_5}-1,
\nonumber\\
&&\alpha_{_6}=b_{_1}+b_{_4}-a_{_1}-2,\;\alpha_{_7}=b_{_2}-a_{_4}-1,\;\alpha_{_{8}}=b_{_3}-1,
\nonumber\\
&&\alpha_{_{10}}=b_{_5}-b_{_4},\;\alpha_{_{11}}=b_{_4}-a_{_1}-1,\;\alpha_{_{12}}=-a_{_4},\;
\alpha_{_{14}}=a_{_4}-b_{_4}+1.
\label{GKZ21j-8-1}
\end{eqnarray}
The corresponding hypergeometric solution is written as
\begin{eqnarray}
&&\Phi_{_{[13\tilde{5}\tilde{7}]}}^{(6)}(\alpha,z)=
y_{_1}^{{D\over2}-2}y_{_2}^{{D}-3}y_{_4}^{1-{D\over2}}\sum\limits_{n_{_1}=0}^\infty
\sum\limits_{n_{_2}=0}^\infty\sum\limits_{n_{_3}=0}^\infty\sum\limits_{n_{_4}=0}^\infty
c_{_{[13\tilde{5}\tilde{7}]}}^{(6)}(\alpha,{\bf n})
\nonumber\\
&&\hspace{2.5cm}\times
\Big({1\over y_{_1}}\Big)^{n_{_1}}\Big({y_{_4}\over y_{_2}}\Big)^{n_{_2}}
\Big({y_{_4}\over y_{_1}}\Big)^{n_{_3}}\Big({y_{_3}\over y_{_2}}\Big)^{n_{_4}}\;,
\label{GKZ21j-8-2}
\end{eqnarray}
with
\begin{eqnarray}
&&c_{_{[13\tilde{5}\tilde{7}]}}^{(6)}(\alpha,{\bf n})=
(-)^{1+n_{_1}+n_{_2}}\Gamma(1+n_{_1}+n_{_3})
\Big\{n_{_1}!n_{_2}!n_{_3}!n_{_4}!\Gamma({D\over2}+n_{_1})
\nonumber\\
&&\hspace{2.5cm}\times
\Gamma(2-{D\over2}+n_{_2})
\Gamma({D\over2}-1-n_{_2}-n_{_4})\Gamma(2-{D\over2}+n_{_4})
\nonumber\\
&&\hspace{2.5cm}\times
\Gamma(2-{D\over2}+n_{_3})\Gamma({D\over2}-1-n_{_1}-n_{_3})
\Gamma(D-2-n_{_2}-n_{_4})\Big\}^{-1}\;.
\label{GKZ21j-8-3}
\end{eqnarray}

\item   $I_{_{7}}=\{2,3,5,6,7,9,11,\cdots,14\}$, i.e. the implement $J_{_{7}}=[1,14]\setminus I_{_{7}}=\{1,4,8,10\}$.
The choice implies the power numbers $\alpha_{_1}=\alpha_{_{4}}=\alpha_{_{8}}=\alpha_{_{10}}=0$, and
\begin{eqnarray}
&&\alpha_{_2}=a_{_1}-a_{_2},\;\alpha_{_3}=a_{_4}-a_{_3},\;\alpha_{_5}=b_{_5}-a_{_4}-a_{_5}-1,
\nonumber\\
&&\alpha_{_6}=b_{_1}+b_{_5}-a_{_1}-2,\;\alpha_{_7}=b_{_2}+b_{_3}-a_{_4}-2,\;\alpha_{_{9}}=b_{_4}-b_{_5},
\nonumber\\
&&\alpha_{_{11}}=b_{_5}-a_{_1}-1,\;\alpha_{_{12}}=b_{_3}-a_{_4}-1,\;\alpha_{_{13}}=1-b_{_3},
\nonumber\\
&&\alpha_{_{14}}=a_{_4}-b_{_5}+1.
\label{GKZ21j-9-1}
\end{eqnarray}
The corresponding hypergeometric solution is given as
\begin{eqnarray}
&&\Phi_{_{[13\tilde{5}\tilde{7}]}}^{(7)}(\alpha,z)=
y_{_1}^{-1}y_{_2}^{{D\over2}-2}y_{_3}^{{D\over2}-1}\sum\limits_{n_{_1}=0}^\infty
\sum\limits_{n_{_2}=0}^\infty\sum\limits_{n_{_3}=0}^\infty\sum\limits_{n_{_4}=0}^\infty
c_{_{[13\tilde{5}\tilde{7}]}}^{(7)}(\alpha,{\bf n})
\nonumber\\
&&\hspace{2.5cm}\times
\Big({1\over y_{_1}}\Big)^{n_{_1}}\Big({y_{_4}\over y_{_2}}\Big)^{n_{_2}}
\Big({y_{_4}\over y_{_1}}\Big)^{n_{_3}}\Big({y_{_3}\over y_{_2}}\Big)^{n_{_4}}\;,
\label{GKZ21j-9-2}
\end{eqnarray}
with
\begin{eqnarray}
&&c_{_{[13\tilde{5}\tilde{7}]}}^{(7)}(\alpha,{\bf n})=
(-)^{n_{_1}+n_{_4}}\Gamma(1+n_{_1}+n_{_3})\Gamma(1+n_{_2}+n_{_4})
\Big\{n_{_1}!n_{_2}!n_{_3}!n_{_4}!
\nonumber\\
&&\hspace{2.5cm}\times
\Gamma({D\over2}+n_{_1})\Gamma(2-{D\over2}+n_{_2})
\Gamma(1-{D\over2}-n_{_1}-n_{_3})
\nonumber\\
&&\hspace{2.5cm}\times
\Gamma({D\over2}+n_{_3})\Gamma({D\over2}-1-n_{_2}-n_{_4})
\Gamma({D\over2}+n_{_4})\Big\}^{-1}\;.
\label{GKZ21j-9-3}
\end{eqnarray}

\item   $I_{_{8}}=\{2,3,5,\cdots,9,11,12,14\}$, i.e. the implement $J_{_{8}}=[1,14]\setminus I_{_{8}}=\{1,4,10,13\}$.
The choice implies the power numbers $\alpha_{_1}=\alpha_{_{4}}=\alpha_{_{10}}=\alpha_{_{13}}=0$, and
\begin{eqnarray}
&&\alpha_{_2}=a_{_1}-a_{_2},\;\alpha_{_3}=a_{_4}-a_{_3},\;\alpha_{_5}=b_{_5}-a_{_4}-a_{_5}-1,
\nonumber\\
&&\alpha_{_6}=b_{_1}+b_{_5}-a_{_1}-2,\;\alpha_{_7}=b_{_2}-a_{_4}-1,\;\alpha_{_{8}}=b_{_3}-1,
\nonumber\\
&&\alpha_{_{9}}=b_{_4}-b_{_5},\;\alpha_{_{11}}=b_{_5}-a_{_1}-1,\;\alpha_{_{12}}=-a_{_4},\;
\alpha_{_{14}}=a_{_4}-b_{_5}+1.
\label{GKZ21j-10-1}
\end{eqnarray}
The corresponding hypergeometric series is
\begin{eqnarray}
&&\Phi_{_{[13\tilde{5}\tilde{7}]}}^{(8)}(\alpha,z)=
y_{_1}^{-1}y_{_2}^{{D}-3}\sum\limits_{n_{_1}=0}^\infty
\sum\limits_{n_{_2}=0}^\infty\sum\limits_{n_{_3}=0}^\infty\sum\limits_{n_{_4}=0}^\infty
c_{_{[13\tilde{5}\tilde{7}]}}^{(8)}(\alpha,{\bf n})
\nonumber\\
&&\hspace{2.5cm}\times
\Big({1\over y_{_1}}\Big)^{n_{_1}}\Big({y_{_4}\over y_{_2}}\Big)^{n_{_2}}
\Big({y_{_4}\over y_{_1}}\Big)^{n_{_3}}\Big({y_{_3}\over y_{_2}}\Big)^{n_{_4}}\;,
\label{GKZ21j-10-2}
\end{eqnarray}
with
\begin{eqnarray}
&&c_{_{[13\tilde{5}\tilde{7}]}}^{(8)}(\alpha,{\bf n})=
(-)^{n_{_1}+n_{_2}}\Gamma(1+n_{_1}+n_{_3})
\Big\{n_{_1}!n_{_2}!n_{_3}!n_{_4}!\Gamma({D\over2}+n_{_1})
\nonumber\\
&&\hspace{2.5cm}\times
\Gamma(2-{D\over2}+n_{_2})\Gamma(1-{D\over2}-n_{_1}-n_{_3})
\Gamma({D\over2}-1-n_{_2}-n_{_4})
\nonumber\\
&&\hspace{2.5cm}\times
\Gamma(2-{D\over2}+n_{_4})
\Gamma({D\over2}+n_{_3})\Gamma(D-2-n_{_2}-n_{_4})\Big\}^{-1}\;.
\label{GKZ21j-10-3}
\end{eqnarray}

\item   $I_{_{9}}=\{1,4,5,6,7,10,\cdots,14\}$, i.e. the implement $J_{_{9}}=[1,14]\setminus I_{_{9}}=\{2,3,8,9\}$.
The choice implies the power numbers $\alpha_{_2}=\alpha_{_{3}}=\alpha_{_{8}}=\alpha_{_{9}}=0$, and
\begin{eqnarray}
&&\alpha_{_1}=a_{_2}-a_{_1},\;\alpha_{_4}=a_{_3}-a_{_4},\;\alpha_{_5}=b_{_4}-a_{_3}-a_{_5}-1,
\nonumber\\
&&\alpha_{_6}=b_{_1}+b_{_4}-a_{_2}-2,\;\alpha_{_7}=b_{_2}+b_{_3}-a_{_3}-2,\;\alpha_{_{10}}=b_{_5}-b_{_4},
\nonumber\\
&&\alpha_{_{11}}=b_{_4}-a_{_2}-1,\;\alpha_{_{12}}=b_{_3}-a_{_3}-1,\;\alpha_{_{13}}=1-b_{_3},
\nonumber\\
&&\alpha_{_{14}}=a_{_3}-b_{_4}+1.
\label{GKZ21j-17-1}
\end{eqnarray}
The corresponding hypergeometric solution is presented as
\begin{eqnarray}
&&\Phi_{_{[13\tilde{5}\tilde{7}]}}^{(9)}(\alpha,z)=
y_{_1}^{{D}-3}y_{_2}^{-1}y_{_3}^{{D\over2}-1}\sum\limits_{n_{_1}=0}^\infty
\sum\limits_{n_{_2}=0}^\infty\sum\limits_{n_{_3}=0}^\infty\sum\limits_{n_{_4}=0}^\infty
c_{_{[13\tilde{5}\tilde{7}]}}^{(9)}(\alpha,{\bf n})
\nonumber\\
&&\hspace{2.5cm}\times
\Big({1\over y_{_1}}\Big)^{n_{_1}}\Big({y_{_4}\over y_{_2}}\Big)^{n_{_2}}
\Big({y_{_4}\over y_{_1}}\Big)^{n_{_3}}\Big({y_{_3}\over y_{_2}}\Big)^{n_{_4}}\;,
\label{GKZ21j-17-2}
\end{eqnarray}
with
\begin{eqnarray}
&&c_{_{[13\tilde{5}\tilde{7}]}}^{(9)}(\alpha,{\bf n})=
(-)^{n_{_3}+n_{_4}}\Gamma(1+n_{_2}+n_{_4})
\Big\{n_{_1}!n_{_2}!n_{_3}!n_{_4}!\Gamma(2-{D\over2}+n_{_1})
\nonumber\\
&&\hspace{2.5cm}\times
\Gamma({D\over2}+n_{_2})\Gamma({D\over2}-1-n_{_1}-n_{_3})
\Gamma(1-{D\over2}-n_{_2}-n_{_4})
\nonumber\\
&&\hspace{2.5cm}\times
\Gamma(2-{D\over2}+n_{_3})\Gamma(D-2-n_{_1}-n_{_3})
\Gamma({D\over2}+n_{_4})\Big\}^{-1}\;.
\label{GKZ21j-17-3}
\end{eqnarray}

\item   $I_{_{10}}=\{1,4,5,6,7,8,10,11,12,14\}$, i.e. the implement $J_{_{10}}=[1,14]\setminus I_{_{10}}=\{2,3,9,13\}$.
The choice implies the power numbers $\alpha_{_2}=\alpha_{_{3}}=\alpha_{_{9}}=\alpha_{_{13}}=0$, and
\begin{eqnarray}
&&\alpha_{_1}=a_{_2}-a_{_1},\;\alpha_{_4}=a_{_3}-a_{_4},\;\alpha_{_5}=b_{_4}-a_{_3}-a_{_5}-1,
\nonumber\\
&&\alpha_{_6}=b_{_1}+b_{_4}-a_{_2}-2,\;\alpha_{_7}=b_{_2}-a_{_3}-1,\;\alpha_{_{8}}=b_{_3}-1,
\nonumber\\
&&\alpha_{_{10}}=b_{_5}-b_{_4},\;\alpha_{_{11}}=b_{_4}-a_{_2}-1,\;\alpha_{_{12}}=-a_{_3},\;
\alpha_{_{14}}=a_{_3}-b_{_4}+1.
\label{GKZ21j-18-1}
\end{eqnarray}
The corresponding hypergeometric series is written as
\begin{eqnarray}
&&\Phi_{_{[13\tilde{5}\tilde{7}]}}^{(10)}(\alpha,z)=
y_{_1}^{{D}-3}y_{_2}^{{D\over2}-2}\sum\limits_{n_{_1}=0}^\infty
\sum\limits_{n_{_2}=0}^\infty\sum\limits_{n_{_3}=0}^\infty\sum\limits_{n_{_4}=0}^\infty
c_{_{[13\tilde{5}\tilde{7}]}}^{(10)}(\alpha,{\bf n})
\nonumber\\
&&\hspace{2.5cm}\times
\Big({1\over y_{_1}}\Big)^{n_{_1}}\Big({y_{_4}\over y_{_2}}\Big)^{n_{_2}}
\Big({y_{_4}\over y_{_1}}\Big)^{n_{_3}}\Big({y_{_3}\over y_{_2}}\Big)^{n_{_4}}\;,
\label{GKZ21j-18-2}
\end{eqnarray}
with
\begin{eqnarray}
&&c_{_{[13\tilde{5}\tilde{7}]}}^{(10)}(\alpha,{\bf n})=
(-)^{n_{_3}+n_{_4}}\Gamma(1+n_{_2}+n_{_4})\Big\{n_{_1}!n_{_2}!n_{_3}!n_{_4}!
\Gamma(2-{D\over2}+n_{_1})
\nonumber\\
&&\hspace{2.5cm}\times
\Gamma({D\over2}+n_{_2})\Gamma({D\over2}-1-n_{_1}-n_{_3})
\Gamma(2-{D\over2}+n_{_4})
\nonumber\\
&&\hspace{2.5cm}\times
\Gamma(2-{D\over2}+n_{_3})\Gamma(D-2-n_{_1}-n_{_3})
\Gamma({D\over2}-1-n_{_2}-n_{_4})\Big\}^{-1}\;.
\label{GKZ21j-18-3}
\end{eqnarray}

\item   $I_{_{11}}=\{1,4,5,6,7,9,11,\cdots,14\}$, i.e. the implement $J_{_{11}}=[1,14]\setminus I_{_{11}}=\{2,3,8,10\}$.
The choice implies the power numbers $\alpha_{_2}=\alpha_{_{3}}=\alpha_{_{8}}=\alpha_{_{10}}=0$, and
\begin{eqnarray}
&&\alpha_{_1}=a_{_2}-a_{_1},\;\alpha_{_4}=a_{_3}-a_{_4},\;\alpha_{_5}=b_{_5}-a_{_3}-a_{_5}-1,
\nonumber\\
&&\alpha_{_6}=b_{_1}+b_{_5}-a_{_2}-2,\;\alpha_{_7}=b_{_2}+b_{_3}-a_{_3}-2,\;\alpha_{_{9}}=b_{_4}-b_{_5},
\nonumber\\
&&\alpha_{_{11}}=b_{_5}-a_{_2}-1,\;\alpha_{_{12}}=b_{_3}-a_{_3}-1,\;\alpha_{_{13}}=1-b_{_3},
\nonumber\\
&&\alpha_{_{14}}=a_{_3}-b_{_5}+1.
\label{GKZ21j-19-1}
\end{eqnarray}
The corresponding hypergeometric solution is given as
\begin{eqnarray}
&&\Phi_{_{[13\tilde{5}\tilde{7}]}}^{(11)}(\alpha,z)=
y_{_1}^{{D\over2}-2}y_{_2}^{-1}y_{_3}^{{D\over2}-1}y_{_4}^{{D\over2}-1}
\sum\limits_{n_{_1}=0}^\infty
\sum\limits_{n_{_2}=0}^\infty\sum\limits_{n_{_3}=0}^\infty\sum\limits_{n_{_4}=0}^\infty
c_{_{[13\tilde{5}\tilde{7}]}}^{(11)}(\alpha,{\bf n})
\nonumber\\
&&\hspace{2.5cm}\times
\Big({1\over y_{_1}}\Big)^{n_{_1}}\Big({y_{_4}\over y_{_2}}\Big)^{n_{_2}}
\Big({y_{_4}\over y_{_1}}\Big)^{n_{_3}}\Big({y_{_3}\over y_{_2}}\Big)^{n_{_4}}\;,
\label{GKZ21j-19-2}
\end{eqnarray}
with
\begin{eqnarray}
&&c_{_{[13\tilde{5}\tilde{7}]}}^{(11)}(\alpha,{\bf n})=
(-)^{n_{_1}+n_{_4}}\Gamma(1+n_{_1}+n_{_3})\Gamma(1+n_{_2}+n_{_4})
\Big\{n_{_1}!n_{_2}!n_{_3}!n_{_4}!
\nonumber\\
&&\hspace{2.5cm}\times
\Gamma(2-{D\over2}+n_{_1})\Gamma({D\over2}+n_{_2})
\Gamma(1-{D\over2}-n_{_2}-n_{_4})\Gamma({D\over2}+n_{_3})
\nonumber\\
&&\hspace{2.5cm}\times
\Gamma({D\over2}-1-n_{_1}-n_{_3})\Gamma({D\over2}+n_{_4})\Big\}^{-1}\;.
\label{GKZ21j-19-3}
\end{eqnarray}

\item   $I_{_{12}}=\{1,4,\cdots,9,11,12,14\}$, i.e. the implement $J_{_{12}}=[1,14]\setminus I_{_{12}}=\{2,3,10,13\}$.
The choice implies the power numbers $\alpha_{_2}=\alpha_{_{3}}=\alpha_{_{10}}=\alpha_{_{13}}=0$, and
\begin{eqnarray}
&&\alpha_{_1}=a_{_2}-a_{_1},\;\alpha_{_4}=a_{_3}-a_{_4},\;\alpha_{_5}=b_{_5}-a_{_3}-a_{_5}-1,
\nonumber\\
&&\alpha_{_6}=b_{_1}+b_{_5}-a_{_2}-2,\;\alpha_{_7}=b_{_2}-a_{_3}-1,\;\alpha_{_{8}}=b_{_3}-1,
\nonumber\\
&&\alpha_{_{9}}=b_{_4}-b_{_5},\;\alpha_{_{11}}=b_{_5}-a_{_2}-1,\;\alpha_{_{12}}=-a_{_3},\;
\alpha_{_{14}}=a_{_3}-b_{_5}+1.
\label{GKZ21j-20-1}
\end{eqnarray}
The corresponding hypergeometric series is written as
\begin{eqnarray}
&&\Phi_{_{[13\tilde{5}\tilde{7}]}}^{(12)}(\alpha,z)=
y_{_1}^{{D\over2}-2}y_{_2}^{{D\over2}-2}y_{_4}^{{D\over2}-1}\sum\limits_{n_{_1}=0}^\infty
\sum\limits_{n_{_2}=0}^\infty\sum\limits_{n_{_3}=0}^\infty\sum\limits_{n_{_4}=0}^\infty
c_{_{[13\tilde{5}\tilde{7}]}}^{(12)}(\alpha,{\bf n})
\nonumber\\
&&\hspace{2.5cm}\times
\Big({1\over y_{_1}}\Big)^{n_{_1}}\Big({y_{_4}\over y_{_2}}\Big)^{n_{_2}}
\Big({y_{_4}\over y_{_1}}\Big)^{n_{_3}}\Big({y_{_3}\over y_{_2}}\Big)^{n_{_4}}\;,
\label{GKZ21j-20-2}
\end{eqnarray}
with
\begin{eqnarray}
&&c_{_{[13\tilde{5}\tilde{7}]}}^{(12)}(\alpha,{\bf n})=
(-)^{n_{_1}+n_{_4}}\Gamma(1+n_{_1}+n_{_3})\Gamma(1+n_{_2}+n_{_4})
\nonumber\\
&&\hspace{2.5cm}\times
\Big\{n_{_1}!n_{_2}!n_{_3}!n_{_4}!
\Gamma(2-{D\over2}+n_{_1})\Gamma({D\over2}+n_{_2})\Gamma(1+n_{_4})
\nonumber\\
&&\hspace{2.5cm}\times
\Gamma({D\over2}+n_{_3})\Gamma({D\over2}-1-n_{_1}-n_{_3})
\Gamma({D\over2}-1-n_{_2}-n_{_4})\Big\}^{-1}\;.
\label{GKZ21j-20-3}
\end{eqnarray}

\item   $I_{_{13}}=\{1,3,5,6,7,10,\cdots,14\}$, i.e. the implement $J_{_{13}}=[1,14]\setminus I_{_{13}}=\{2,4,8,9\}$.
The choice implies the power numbers $\alpha_{_2}=\alpha_{_{4}}=\alpha_{_{8}}=\alpha_{_{9}}=0$, and
\begin{eqnarray}
&&\alpha_{_1}=a_{_2}-a_{_1},\;\alpha_{_3}=a_{_4}-a_{_3},\;\alpha_{_5}=b_{_4}-a_{_4}-a_{_5}-1,
\nonumber\\
&&\alpha_{_6}=b_{_1}+b_{_4}-a_{_2}-2,\;\alpha_{_7}=b_{_2}+b_{_3}-a_{_4}-2,\;\alpha_{_{10}}=b_{_5}-b_{_4},
\nonumber\\
&&\alpha_{_{11}}=b_{_4}-a_{_2}-1,\;\alpha_{_{12}}=b_{_3}-a_{_4}-1,\;\alpha_{_{13}}=1-b_{_3},
\nonumber\\
&&\alpha_{_{14}}=a_{_4}-b_{_4}+1.
\label{GKZ21j-23-1}
\end{eqnarray}
The corresponding hypergeometric series is written as
\begin{eqnarray}
&&\Phi_{_{[13\tilde{5}\tilde{7}]}}^{(13)}(\alpha,z)=
y_{_1}^{{D}-3}y_{_2}^{{D\over2}-2}y_{_3}^{{D\over2}-1}y_{_4}^{1-{D\over2}}
\sum\limits_{n_{_1}=0}^\infty
\sum\limits_{n_{_2}=0}^\infty\sum\limits_{n_{_3}=0}^\infty\sum\limits_{n_{_4}=0}^\infty
c_{_{[13\tilde{5}\tilde{7}]}}^{(13)}(\alpha,{\bf n})
\nonumber\\
&&\hspace{2.5cm}\times
\Big({1\over y_{_1}}\Big)^{n_{_1}}\Big({y_{_4}\over y_{_2}}\Big)^{n_{_2}}
\Big({y_{_4}\over y_{_1}}\Big)^{n_{_3}}\Big({y_{_3}\over y_{_2}}\Big)^{n_{_4}}\;,
\label{GKZ21j-23-2}
\end{eqnarray}
with
\begin{eqnarray}
&&c_{_{[13\tilde{5}\tilde{7}]}}^{(13)}(\alpha,{\bf n})=
(-)^{1+n_{_3}+n_{_4}}\Gamma(1+n_{_2}+n_{_4})
\Big\{n_{_1}!n_{_2}!n_{_3}!n_{_4}!\Gamma(2-{D\over2}+n_{_1})
\nonumber\\
&&\hspace{2.5cm}\times
\Gamma(2-{D\over2}+n_{_2})\Gamma({D\over2}-1-n_{_1}-n_{_3})
\Gamma(2-{D\over2}+n_{_3})
\nonumber\\
&&\hspace{2.5cm}\times
\Gamma(D-2-n_{_1}-n_{_3})\Gamma({D\over2}-1-n_{_2}-n_{_4})
\Gamma({D\over2}+n_{_4})\Big\}^{-1}\;.
\label{GKZ21j-23-3}
\end{eqnarray}

\item   $I_{_{14}}=\{1,3,5,6,7,8,10,11,12,14\}$, i.e. the implement $J_{_{14}}=[1,14]\setminus I_{_{14}}=\{2,4,9,13\}$.
The choice implies the power numbers $\alpha_{_2}=\alpha_{_{4}}=\alpha_{_{9}}=\alpha_{_{13}}=0$, and
\begin{eqnarray}
&&\alpha_{_1}=a_{_2}-a_{_1},\;\alpha_{_3}=a_{_4}-a_{_3},\;\alpha_{_5}=b_{_4}-a_{_4}-a_{_5}-1,
\nonumber\\
&&\alpha_{_6}=b_{_1}+b_{_4}-a_{_2}-2,\;\alpha_{_7}=b_{_2}-a_{_4}-1,\;\alpha_{_{8}}=b_{_3}-1,
\nonumber\\
&&\alpha_{_{10}}=b_{_5}-b_{_4},\;\alpha_{_{11}}=b_{_4}-a_{_2}-1,\;\alpha_{_{12}}=-a_{_4},\;
\alpha_{_{14}}=a_{_4}-b_{_4}+1.
\label{GKZ21j-24-1}
\end{eqnarray}
The corresponding hypergeometric solution is given as
\begin{eqnarray}
&&\Phi_{_{[13\tilde{5}\tilde{7}]}}^{(14)}(\alpha,z)=
y_{_1}^{{D}-3}y_{_2}^{{D}-3}y_{_4}^{1-{D\over2}}\sum\limits_{n_{_1}=0}^\infty
\sum\limits_{n_{_2}=0}^\infty\sum\limits_{n_{_3}=0}^\infty\sum\limits_{n_{_4}=0}^\infty
c_{_{[13\tilde{5}\tilde{7}]}}^{(14)}(\alpha,{\bf n})
\nonumber\\
&&\hspace{2.5cm}\times
\Big({1\over y_{_1}}\Big)^{n_{_1}}\Big({y_{_4}\over y_{_2}}\Big)^{n_{_2}}
\Big({y_{_4}\over y_{_1}}\Big)^{n_{_3}}\Big({y_{_3}\over y_{_2}}\Big)^{n_{_4}}\;,
\label{GKZ21j-24-2}
\end{eqnarray}
with
\begin{eqnarray}
&&c_{_{[13\tilde{5}\tilde{7}]}}^{(14)}(\alpha,{\bf n})=
(-)^{1+n_{_2}+n_{_3}}
\Big\{n_{_1}!n_{_2}!n_{_3}!n_{_4}!\Gamma(2-{D\over2}+n_{_1})\Gamma(2-{D\over2}+n_{_2})
\nonumber\\
&&\hspace{2.5cm}\times
\Gamma({D\over2}-1-n_{_1}-n_{_3})
\Gamma({D\over2}-1-n_{_2}-n_{_4})\Gamma(2-{D\over2}+n_{_4})
\nonumber\\
&&\hspace{2.5cm}\times
\Gamma(2-{D\over2}+n_{_3})\Gamma(D-2-n_{_1}-n_{_3})
\Gamma(D-2-n_{_2}-n_{_4})\Big\}^{-1}\;.
\label{GKZ21j-24-3}
\end{eqnarray}

\item   $I_{_{15}}=\{1,3,5,6,7,9,11,\cdots,14\}$, i.e. the implement $J_{_{15}}=[1,14]\setminus I_{_{15}}=\{2,4,8,10\}$.
The choice implies the power numbers $\alpha_{_2}=\alpha_{_{4}}=\alpha_{_{8}}=\alpha_{_{10}}=0$, and
\begin{eqnarray}
&&\alpha_{_1}=a_{_2}-a_{_1},\;\alpha_{_3}=a_{_4}-a_{_3},\;\alpha_{_5}=b_{_5}-a_{_4}-a_{_5}-1,
\nonumber\\
&&\alpha_{_6}=b_{_1}+b_{_5}-a_{_2}-2,\;\alpha_{_7}=b_{_2}+b_{_3}-a_{_4}-2,\;\alpha_{_{9}}=b_{_4}-b_{_5},
\nonumber\\
&&\alpha_{_{11}}=b_{_5}-a_{_2}-1,\;\alpha_{_{12}}=b_{_3}-a_{_4}-1,\;\alpha_{_{13}}=1-b_{_3},
\nonumber\\
&&\alpha_{_{14}}=a_{_4}-b_{_5}+1.
\label{GKZ21j-25-1}
\end{eqnarray}
The corresponding hypergeometric series is written as
\begin{eqnarray}
&&\Phi_{_{[13\tilde{5}\tilde{7}]}}^{(15)}(\alpha,z)=
y_{_1}^{{D\over2}-2}y_{_2}^{{D\over2}-2}y_{_3}^{{D\over2}-1}\sum\limits_{n_{_1}=0}^\infty
\sum\limits_{n_{_2}=0}^\infty\sum\limits_{n_{_3}=0}^\infty\sum\limits_{n_{_4}=0}^\infty
c_{_{[13\tilde{5}\tilde{7}]}}^{(15)}(\alpha,{\bf n})
\nonumber\\
&&\hspace{2.5cm}\times
\Big({1\over y_{_1}}\Big)^{n_{_1}}\Big({y_{_4}\over y_{_2}}\Big)^{n_{_2}}
\Big({y_{_4}\over y_{_1}}\Big)^{n_{_3}}\Big({y_{_3}\over y_{_2}}\Big)^{n_{_4}}\;,
\label{GKZ21j-25-2}
\end{eqnarray}
with
\begin{eqnarray}
&&c_{_{[13\tilde{5}\tilde{7}]}}^{(15)}(\alpha,{\bf n})=
(-)^{n_{_1}+n_{_4}}\Gamma(1+n_{_1}+n_{_3})\Gamma(1+n_{_2}+n_{_4})
\Big\{n_{_1}!n_{_2}!n_{_3}!n_{_4}!
\nonumber\\
&&\hspace{2.5cm}\times
\Gamma(2-{D\over2}+n_{_1})\Gamma(2-{D\over2}+n_{_2})\Gamma({D\over2}+n_{_3})
\nonumber\\
&&\hspace{2.5cm}\times
\Gamma({D\over2}-1-n_{_1}-n_{_3})\Gamma({D\over2}-1-n_{_2}-n_{_4})
\Gamma({D\over2}+n_{_4})\Big\}^{-1}\;.
\label{GKZ21j-25-3}
\end{eqnarray}

\item   $I_{_{16}}=\{1,3,5,\cdots,9,11,12,14\}$, i.e. the implement $J_{_{16}}=[1,14]\setminus I_{_{16}}=\{2,4,10,13\}$.
The choice implies the power numbers $\alpha_{_2}=\alpha_{_{4}}=\alpha_{_{10}}=\alpha_{_{13}}=0$, and
\begin{eqnarray}
&&\alpha_{_1}=a_{_2}-a_{_1},\;\alpha_{_3}=a_{_4}-a_{_3},\;\alpha_{_5}=b_{_5}-a_{_4}-a_{_5}-1,
\nonumber\\
&&\alpha_{_6}=b_{_1}+b_{_5}-a_{_2}-2,\;\alpha_{_7}=b_{_2}-a_{_4}-1,\;\alpha_{_{8}}=b_{_3}-1,
\nonumber\\
&&\alpha_{_{9}}=b_{_4}-b_{_5},\;\alpha_{_{11}}=b_{_5}-a_{_2}-1,\;\alpha_{_{12}}=-a_{_4},\;
\alpha_{_{14}}=a_{_4}-b_{_5}+1.
\label{GKZ21j-26-1}
\end{eqnarray}
The corresponding hypergeometric series is written as
\begin{eqnarray}
&&\Phi_{_{[13\tilde{5}\tilde{7}]}}^{(16)}(\alpha,z)=
y_{_1}^{{D\over2}-2}y_{_2}^{{D}-3}\sum\limits_{n_{_1}=0}^\infty
\sum\limits_{n_{_2}=0}^\infty\sum\limits_{n_{_3}=0}^\infty\sum\limits_{n_{_4}=0}^\infty
c_{_{[13\tilde{5}\tilde{7}]}}^{(16)}(\alpha,{\bf n})
\nonumber\\
&&\hspace{2.5cm}\times
\Big({1\over y_{_1}}\Big)^{n_{_1}}\Big({y_{_4}\over y_{_2}}\Big)^{n_{_2}}
\Big({y_{_4}\over y_{_1}}\Big)^{n_{_3}}\Big({y_{_3}\over y_{_2}}\Big)^{n_{_4}}\;,
\label{GKZ21j-26-2}
\end{eqnarray}
with
\begin{eqnarray}
&&c_{_{[13\tilde{5}\tilde{7}]}}^{(16)}(\alpha,{\bf n})=
(-)^{n_{_1}+n_{_2}}\Gamma(1+n_{_1}+n_{_3})
\Big\{n_{_1}!n_{_2}!n_{_3}!n_{_4}!\Gamma(2-{D\over2}+n_{_1})
\nonumber\\
&&\hspace{2.5cm}\times
\Gamma(2-{D\over2}+n_{_2})
\Gamma({D\over2}-1-n_{_2}-n_{_4})\Gamma(2-{D\over2}+n_{_4})
\nonumber\\
&&\hspace{2.5cm}\times
\Gamma({D\over2}+n_{_3})\Gamma({D\over2}-1-n_{_1}-n_{_3})
\Gamma(D-2-n_{_2}-n_{_4})\Big\}^{-1}\;.
\label{GKZ21j-26-3}
\end{eqnarray}

\item   $I_{_{17}}=\{2,4,5,6,7,9,\cdots,13\}$, i.e. the implement $J_{_{17}}=[1,14]\setminus I_{_{17}}=\{1,3,8,14\}$.
The choice implies the power numbers $\alpha_{_1}=\alpha_{_{3}}=\alpha_{_{8}}=\alpha_{_{14}}=0$, and
\begin{eqnarray}
&&\alpha_{_2}=a_{_1}-a_{_2},\;\alpha_{_4}=a_{_3}-a_{_4},\;\alpha_{_5}=-a_{_5},
\nonumber\\
&&\alpha_{_6}=a_{_3}+b_{_1}-a_{_1}-1,\;\alpha_{_7}=b_{_2}+b_{_3}-a_{_3}-2,
\nonumber\\
&&\alpha_{_{9}}=b_{_4}-a_{_3}-1,\;\alpha_{_{10}}=b_{_5}-a_{_3}-1,\;\alpha_{_{11}}=a_{_3}-a_{_1},
\nonumber\\
&&\alpha_{_{12}}=b_{_3}-a_{_3}-1,\;\alpha_{_{13}}=1-b_{_3}.
\label{GKZ21j-5-1}
\end{eqnarray}
The corresponding hypergeometric solutions are written as
\begin{eqnarray}
&&\Phi_{_{[13\tilde{5}\tilde{7}]}}^{(17),a}(\alpha,z)=
y_{_1}^{{D\over2}-2}y_{_2}^{-1}y_{_3}^{{D\over2}-1}\sum\limits_{n_{_1}=0}^\infty
\sum\limits_{n_{_2}=0}^\infty\sum\limits_{n_{_3}=0}^\infty\sum\limits_{n_{_4}=0}^\infty
c_{_{[13\tilde{5}\tilde{7}]}}^{(17),a}(\alpha,{\bf n})
\nonumber\\
&&\hspace{2.5cm}\times
\Big({1\over y_{_1}}\Big)^{n_{_1}}\Big({y_{_4}\over y_{_2}}\Big)^{n_{_2}}
\Big({y_{_4}\over y_{_1}}\Big)^{n_{_3}}\Big({y_{_3}\over y_{_2}}\Big)^{n_{_4}}
\;,\nonumber\\
&&\Phi_{_{[13\tilde{5}\tilde{7}]}}^{(17),b}(\alpha,z)=
y_{_1}^{{D\over2}-1}y_{_2}^{-2}y_{_3}^{{D\over2}-1}\sum\limits_{n_{_1}=0}^\infty
\sum\limits_{n_{_2}=0}^\infty\sum\limits_{n_{_3}=0}^\infty\sum\limits_{n_{_4}=0}^\infty
c_{_{[13\tilde{5}\tilde{7}]}}^{(17),b}(\alpha,{\bf n})
\nonumber\\
&&\hspace{2.5cm}\times
\Big({1\over y_{_2}}\Big)^{n_{_1}}\Big({y_{_1}\over y_{_2}}\Big)^{n_{_2}}
\Big({y_{_4}\over y_{_2}}\Big)^{n_{_3}}\Big({y_{_3}\over y_{_2}}\Big)^{n_{_4}}
\;,\nonumber\\
&&\Phi_{_{[13\tilde{5}\tilde{7}]}}^{(17),c}(\alpha,z)=
y_{_1}^{{D\over2}-2}y_{_2}^{-2}y_{_3}^{{D\over2}-1}\sum\limits_{n_{_1}=0}^\infty
\sum\limits_{n_{_2}=0}^\infty\sum\limits_{n_{_3}=0}^\infty\sum\limits_{n_{_4}=0}^\infty
c_{_{[13\tilde{5}\tilde{7}]}}^{(17),c}(\alpha,{\bf n})
\nonumber\\
&&\hspace{2.5cm}\times
\Big({1\over y_{_1}}\Big)^{n_{_1}}\Big({1\over y_{_2}}\Big)^{n_{_2}}
\Big({y_{_4}\over y_{_2}}\Big)^{n_{_3}}\Big({y_{_3}\over y_{_2}}\Big)^{n_{_4}}\;.
\label{GKZ21j-5-2a}
\end{eqnarray}
Where the coefficients are
\begin{eqnarray}
&&c_{_{[13\tilde{5}\tilde{7}]}}^{(17),a}(\alpha,{\bf n})=
(-)^{n_{_1}+n_{_4}}\Gamma(1+n_{_1}+n_{_3})\Gamma(1+n_{_2}+n_{_4})
\Big\{n_{_1}!n_{_2}!n_{_3}!n_{_4}!
\nonumber\\
&&\hspace{2.5cm}\times
\Gamma({D\over2}+n_{_1})\Gamma({D\over2}+n_{_2})\Gamma({D\over2}-1-n_{_1}-n_{_3})
\nonumber\\
&&\hspace{2.5cm}\times
\Gamma(1-{D\over2}-n_{_2}-n_{_4})
\Gamma(2-{D\over2}+n_{_3})\Gamma({D\over2}+n_{_4})\Big\}^{-1}
\;,\nonumber\\
&&c_{_{[13\tilde{5}\tilde{7}]}}^{(17),b}(\alpha,{\bf n})=
(-)^{1+n_{_4}}\Gamma(1+n_{_1}+n_{_2})\Gamma(2+n_{_1}+n_{_2}+n_{_3}+n_{_4})
\nonumber\\
&&\hspace{2.5cm}\times
\Big\{n_{_1}!n_{_2}!n_{_4}!\Gamma(2+n_{_1}+n_{_2}+n_{_3})\Gamma({D\over2}+n_{_1})
\nonumber\\
&&\hspace{2.5cm}\times
\Gamma({D\over2}+1+n_{_1}+n_{_2}+n_{_3})\Gamma({D\over2}+n_{_2})\Gamma({D\over2}+n_{_4})
\nonumber\\
&&\hspace{2.5cm}\times
\Gamma(-{D\over2}-n_{_1}-n_{_2}-n_{_3}-n_{_4})
\Gamma(1-{D\over2}-n_{_1}-n_{_2})\Big\}^{-1}
\;,\nonumber\\
&&c_{_{[13\tilde{5}\tilde{7}]}}^{(17),c}(\alpha,{\bf n})=
(-)^{1+n_{_1}+n_{_4}}\Gamma(1+n_{_1})\Gamma(1+n_{_2})\Gamma(2+n_{_2}+n_{_3}+n_{_4})
\nonumber\\
&&\hspace{2.5cm}\times
\Big\{n_{_4}!\Gamma(2+n_{_1}+n_{_2})\Gamma(2+n_{_2}+n_{_3})\Gamma({D\over2}+1+n_{_1}+n_{_2})
\nonumber\\
&&\hspace{2.5cm}\times
\Gamma({D\over2}+1+n_{_2}+n_{_3})\Gamma({D\over2}-1-n_{_1})
\nonumber\\
&&\hspace{2.5cm}\times
\Gamma(-{D\over2}-n_{_2}-n_{_3}-n_{_4})
\Gamma(1-{D\over2}-n_{_2})\Gamma({D\over2}+n_{_4})\Big\}^{-1}\;.
\label{GKZ21j-5-3}
\end{eqnarray}

\item   $I_{_{18}}=\{2,4,\cdots,12\}$, i.e. the implement $J_{_{18}}=[1,14]\setminus I_{_{18}}=\{1,3,13,14\}$.
The choice implies the power numbers $\alpha_{_1}=\alpha_{_{3}}=\alpha_{_{13}}=\alpha_{_{14}}=0$, and
\begin{eqnarray}
&&\alpha_{_2}=a_{_1}-a_{_2},\;\alpha_{_4}=a_{_3}-a_{_4},\;\alpha_{_5}=-a_{_5},
\nonumber\\
&&\alpha_{_6}=a_{_3}+b_{_1}-a_{_1}-1,\;\alpha_{_7}=b_{_2}-a_{_3}-1,\;\alpha_{_{8}}=b_{_3}-1,
\nonumber\\
&&\alpha_{_{9}}=b_{_4}-a_{_3}-1,\;\alpha_{_{10}}=b_{_5}-a_{_3}-1,\;\alpha_{_{11}}=a_{_3}-a_{_1},\;
\alpha_{_{12}}=-a_{_3}.
\label{GKZ21j-6-1}
\end{eqnarray}
The corresponding hypergeometric functions are
\begin{eqnarray}
&&\Phi_{_{[13\tilde{5}\tilde{7}]}}^{(18),a}(\alpha,z)=
y_{_1}^{{D\over2}-2}y_{_2}^{{D\over2}-2}\sum\limits_{n_{_1}=0}^\infty
\sum\limits_{n_{_2}=0}^\infty\sum\limits_{n_{_3}=0}^\infty\sum\limits_{n_{_4}=0}^\infty
c_{_{[13\tilde{5}\tilde{7}]}}^{(18),a}(\alpha,{\bf n})
\nonumber\\
&&\hspace{2.5cm}\times
\Big({1\over y_{_1}}\Big)^{n_{_1}}\Big({y_{_4}\over y_{_2}}\Big)^{n_{_2}}
\Big({y_{_4}\over y_{_1}}\Big)^{n_{_3}}\Big({y_{_3}\over y_{_2}}\Big)^{n_{_4}}
\;,\nonumber\\
&&\Phi_{_{[13\tilde{5}\tilde{7}]}}^{(18),b}(\alpha,z)=
y_{_1}^{{D\over2}-1}y_{_2}^{{D\over2}-3}\sum\limits_{n_{_1}=0}^\infty
\sum\limits_{n_{_2}=0}^\infty\sum\limits_{n_{_3}=0}^\infty\sum\limits_{n_{_4}=0}^\infty
c_{_{[13\tilde{5}\tilde{7}]}}^{(18),b}(\alpha,{\bf n})
\nonumber\\
&&\hspace{2.5cm}\times
\Big({1\over y_{_2}}\Big)^{n_{_1}}\Big({y_{_1}\over y_{_2}}\Big)^{n_{_2}}
\Big({y_{_4}\over y_{_2}}\Big)^{n_{_3}}\Big({y_{_3}\over y_{_2}}\Big)^{n_{_4}}
\;,\nonumber\\
&&\Phi_{_{[13\tilde{5}\tilde{7}]}}^{(18),c}(\alpha,z)=
y_{_1}^{{D\over2}-2}y_{_2}^{{D\over2}-3}\sum\limits_{n_{_1}=0}^\infty
\sum\limits_{n_{_2}=0}^\infty\sum\limits_{n_{_3}=0}^\infty\sum\limits_{n_{_4}=0}^\infty
c_{_{[13\tilde{5}\tilde{7}]}}^{(18),c}(\alpha,{\bf n})
\nonumber\\
&&\hspace{2.5cm}\times
\Big({1\over y_{_1}}\Big)^{n_{_1}}\Big({1\over y_{_2}}\Big)^{n_{_2}}
\Big({y_{_4}\over y_{_2}}\Big)^{n_{_3}}\Big({y_{_3}\over y_{_2}}\Big)^{n_{_4}}\;.
\label{GKZ21j-6-2a}
\end{eqnarray}
Where the coefficients are
\begin{eqnarray}
&&c_{_{[13\tilde{5}\tilde{7}]}}^{(18),a}(\alpha,{\bf n})=
(-)^{n_{_1}+n_{_4}}\Gamma(1+n_{_1}+n_{_3})\Gamma(1+n_{_2}+n_{_4})
\Big\{n_{_1}!n_{_2}!n_{_3}!n_{_4}!
\nonumber\\
&&\hspace{2.5cm}\times
\Gamma({D\over2}+n_{_1})\Gamma({D\over2}+n_{_2})\Gamma({D\over2}-1-n_{_1}-n_{_3})
\nonumber\\
&&\hspace{2.5cm}\times
\Gamma({D\over2}-1-n_{_2}-n_{_4})
\Gamma(2-{D\over2}+n_{_3})\Gamma(2-{D\over2}+n_{_4})\Big\}^{-1}
\;,\nonumber\\
&&c_{_{[13\tilde{5}\tilde{7}]}}^{(18),b}(\alpha,{\bf n})=
(-)^{1+n_{_4}}\Gamma(1+n_{_1}+n_{_2})\Gamma(2+n_{_1}+n_{_2}+n_{_3}+n_{_4})
\nonumber\\
&&\hspace{2.5cm}\times
\Big\{n_{_1}!n_{_2}!n_{_4}!\Gamma(2+n_{_1}+n_{_2}+n_{_3})\Gamma({D\over2}+n_{_1})
\nonumber\\
&&\hspace{2.5cm}\times
\Gamma({D\over2}+1+n_{_1}+n_{_2}+n_{_3})\Gamma({D\over2}+n_{_2})\Gamma(2-{D\over2}+n_{_4})
\nonumber\\
&&\hspace{2.5cm}\times
\Gamma({D\over2}-n_{_1}-n_{_2}-n_{_3}-n_{_4})
\Gamma(1-{D\over2}-n_{_1}-n_{_2})\Big\}^{-1}
\;,\nonumber\\
&&c_{_{[13\tilde{5}\tilde{7}]}}^{(18),c}(\alpha,{\bf n})=
(-)^{1+n_{_1}+n_{_4}}\Gamma(1+n_{_1})\Gamma(1+n_{_2})\Gamma(2+n_{_2}+n_{_3}+n_{_4})
\nonumber\\
&&\hspace{2.5cm}\times
\Big\{n_{_4}!\Gamma(2+n_{_1}+n_{_2})\Gamma(2+n_{_2}+n_{_3})\Gamma({D\over2}+1+n_{_1}+n_{_2})
\nonumber\\
&&\hspace{2.5cm}\times
\Gamma({D\over2}+1+n_{_2}+n_{_3})\Gamma({D\over2}-1-n_{_1})
\nonumber\\
&&\hspace{2.5cm}\times
\Gamma({D\over2}-2-n_{_2}-n_{_3}-n_{_4})
\Gamma(1-{D\over2}-n_{_2})\Gamma(2-{D\over2}+n_{_4})\Big\}^{-1}\;.
\label{GKZ21j-6-3}
\end{eqnarray}

\item   $I_{_{19}}=\{2,3,5,6,7,9,\cdots,13\}$, i.e. the implement $J_{_{19}}=[1,14]\setminus I_{_{19}}=\{1,4,8,14\}$.
The choice implies the power numbers $\alpha_{_1}=\alpha_{_{4}}=\alpha_{_{8}}=\alpha_{_{14}}=0$, and
\begin{eqnarray}
&&\alpha_{_2}=a_{_1}-a_{_2},\;\alpha_{_3}=a_{_4}-a_{_3},\;\alpha_{_5}=-a_{_5},
\nonumber\\
&&\alpha_{_6}=a_{_4}+b_{_1}-a_{_1}-1,\;\alpha_{_7}=b_{_2}+b_{_3}-a_{_4}-2,
\nonumber\\
&&\alpha_{_{9}}=b_{_4}-a_{_4}-1,\;\alpha_{_{10}}=b_{_5}-a_{_4}-1,\;\alpha_{_{11}}=a_{_4}-a_{_1},
\nonumber\\
&&\alpha_{_{12}}=b_{_3}-a_{_4}-1,\;\alpha_{_{13}}=1-b_{_3}.
\label{GKZ21j-11-1}
\end{eqnarray}
The corresponding hypergeometric functions are written as
\begin{eqnarray}
&&\Phi_{_{[13\tilde{5}\tilde{7}]}}^{(19),a}(\alpha,z)=
y_{_1}^{-1}y_{_2}^{{D\over2}-2}y_{_3}^{{D\over2}-1}\sum\limits_{n_{_1}=0}^\infty
\sum\limits_{n_{_2}=0}^\infty\sum\limits_{n_{_3}=0}^\infty\sum\limits_{n_{_4}=0}^\infty
c_{_{[13\tilde{5}\tilde{7}]}}^{(19),a}(\alpha,{\bf n})
\nonumber\\
&&\hspace{2.5cm}\times
\Big({1\over y_{_1}}\Big)^{n_{_1}}\Big({y_{_4}\over y_{_2}}\Big)^{n_{_2}}
\Big({y_{_4}\over y_{_1}}\Big)^{n_{_3}}\Big({y_{_3}\over y_{_2}}\Big)^{n_{_4}}
\;,\nonumber\\
&&\Phi_{_{[13\tilde{5}\tilde{7}]}}^{(19),b}(\alpha,z)=
y_{_2}^{{D\over2}-3}y_{_3}^{{D\over2}-1}\sum\limits_{n_{_1}=0}^\infty
\sum\limits_{n_{_2}=0}^\infty\sum\limits_{n_{_3}=0}^\infty\sum\limits_{n_{_4}=0}^\infty
c_{_{[13\tilde{5}\tilde{7}]}}^{(19),b}(\alpha,{\bf n})
\nonumber\\
&&\hspace{2.5cm}\times
\Big({1\over y_{_2}}\Big)^{n_{_1}}\Big({y_{_1}\over y_{_2}}\Big)^{n_{_2}}
\Big({y_{_4}\over y_{_2}}\Big)^{n_{_3}}\Big({y_{_3}\over y_{_2}}\Big)^{n_{_4}}
\;,\nonumber\\
&&\Phi_{_{[13\tilde{5}\tilde{7}]}}^{(19),c}(\alpha,z)=
y_{_1}^{-1}y_{_2}^{{D\over2}-3}y_{_3}^{{D\over2}-1}\sum\limits_{n_{_1}=0}^\infty
\sum\limits_{n_{_2}=0}^\infty\sum\limits_{n_{_3}=0}^\infty\sum\limits_{n_{_4}=0}^\infty
c_{_{[13\tilde{5}\tilde{7}]}}^{(19),c}(\alpha,{\bf n})
\nonumber\\
&&\hspace{2.5cm}\times
\Big({1\over y_{_1}}\Big)^{n_{_1}}\Big({1\over y_{_2}}\Big)^{n_{_2}}
\Big({y_{_4}\over y_{_2}}\Big)^{n_{_3}}\Big({y_{_3}\over y_{_2}}\Big)^{n_{_4}}\;.
\label{GKZ21j-11-2a}
\end{eqnarray}
Where the coefficients are
\begin{eqnarray}
&&c_{_{[13\tilde{5}\tilde{7}]}}^{(19),a}(\alpha,{\bf n})=
(-)^{n_{_1}+n_{_4}}\Gamma(1+n_{_1}+n_{_3})\Gamma(1+n_{_2}+n_{_4})
\Big\{n_{_1}!n_{_2}!n_{_3}!n_{_4}!
\nonumber\\
&&\hspace{2.5cm}\times
\Gamma({D\over2}+n_{_1})
\Gamma(2-{D\over2}+n_{_2})\Gamma(1-{D\over2}-n_{_1}-n_{_3})
\nonumber\\
&&\hspace{2.5cm}\times
\Gamma({D\over2}+n_{_3})
\Gamma({D\over2}-1-n_{_2}-n_{_4})\Gamma({D\over2}+n_{_4})\Big\}^{-1}
\;,\nonumber\\
&&c_{_{[13\tilde{5}\tilde{7}]}}^{(19),b}(\alpha,{\bf n})=
(-)^{1+n_{_4}}\Gamma(1+n_{_1}+n_{_2})\Gamma(2+n_{_1}+n_{_2}+n_{_3}+n_{_4})
\Big\{n_{_1}!n_{_2}!n_{_4}!
\nonumber\\
&&\hspace{2.5cm}\times
\Gamma(2+n_{_1}+n_{_2}+n_{_3})
\Gamma({D\over2}+n_{_1})\Gamma(3-{D\over2}+n_{_1}+n_{_2}+n_{_3})
\nonumber\\
&&\hspace{2.5cm}\times
\Gamma(2-{D\over2}+n_{_2})
\Gamma({D\over2}-1-n_{_1}-n_{_2})\Gamma({D\over2}+n_{_4})
\nonumber\\
&&\hspace{2.5cm}\times
\Gamma({D\over2}-2-n_{_1}-n_{_2}-n_{_3}-n_{_4})\Big\}^{-1}
\;,\nonumber\\
&&c_{_{[13\tilde{5}\tilde{7}]}}^{(19),c}(\alpha,{\bf n})=
(-)^{1+n_{_1}+n_{_4}}\Gamma(1+n_{_1})\Gamma(1+n_{_2})\Gamma(2+n_{_2}+n_{_3}+n_{_4})
\Big\{n_{_4}!
\nonumber\\
&&\hspace{2.5cm}\times
\Gamma(2+n_{_1}+n_{_2})\Gamma(2+n_{_2}+n_{_3})
\Gamma({D\over2}+1+n_{_1}+n_{_2})
\nonumber\\
&&\hspace{2.5cm}\times
\Gamma(3-{D\over2}+n_{_2}+n_{_3})
\Gamma(1-{D\over2}-n_{_1})\Gamma({D\over2}-1-n_{_2})
\nonumber\\
&&\hspace{2.5cm}\times
\Gamma({D\over2}+n_{_4})\Gamma({D\over2}-2-n_{_2}-n_{_3}-n_{_4})\Big\}^{-1}\;.
\label{GKZ21j-11-3}
\end{eqnarray}

\item   $I_{_{20}}=\{2,3,5,\cdots,12\}$, i.e. the implement $J_{_{20}}=[1,14]\setminus I_{_{20}}=\{1,4,13,14\}$.
The choice implies the power numbers $\alpha_{_1}=\alpha_{_{4}}=\alpha_{_{13}}=\alpha_{_{14}}=0$, and
\begin{eqnarray}
&&\alpha_{_2}=a_{_1}-a_{_2},\;\alpha_{_3}=a_{_4}-a_{_3},\;\alpha_{_5}=-a_{_5},
\nonumber\\
&&\alpha_{_6}=a_{_4}+b_{_1}-a_{_1}-1,\;\alpha_{_7}=b_{_2}-a_{_4}-1,\;\alpha_{_{8}}=b_{_3}-1,
\nonumber\\
&&\alpha_{_{9}}=b_{_4}-a_{_4}-1,\;\alpha_{_{10}}=b_{_5}-a_{_4}-1,\;\alpha_{_{11}}=a_{_4}-a_{_1},\;
\alpha_{_{12}}=-a_{_4}.
\label{GKZ21j-12-1}
\end{eqnarray}
The corresponding hypergeometric solutions are
\begin{eqnarray}
&&\Phi_{_{[13\tilde{5}\tilde{7}]}}^{(20),a}(\alpha,z)=
y_{_1}^{-1}y_{_2}^{{D}-3}\sum\limits_{n_{_1}=0}^\infty
\sum\limits_{n_{_2}=0}^\infty\sum\limits_{n_{_3}=0}^\infty\sum\limits_{n_{_4}=0}^\infty
c_{_{[13\tilde{5}\tilde{7}]}}^{(20),a}(\alpha,{\bf n})
\nonumber\\
&&\hspace{2.5cm}\times
\Big({1\over y_{_1}}\Big)^{n_{_1}}\Big({y_{_4}\over y_{_2}}\Big)^{n_{_2}}
\Big({y_{_4}\over y_{_1}}\Big)^{n_{_3}}\Big({y_{_3}\over y_{_2}}\Big)^{n_{_4}}
\;,\nonumber\\
&&\Phi_{_{[13\tilde{5}\tilde{7}]}}^{(20),b}(\alpha,z)=
y_{_2}^{{D}-4}\sum\limits_{n_{_1}=0}^\infty
\sum\limits_{n_{_2}=0}^\infty\sum\limits_{n_{_3}=0}^\infty\sum\limits_{n_{_4}=0}^\infty
c_{_{[13\tilde{5}\tilde{7}]}}^{(20),b}(\alpha,{\bf n})
\nonumber\\
&&\hspace{2.5cm}\times
\Big({1\over y_{_2}}\Big)^{n_{_1}}\Big({y_{_1}\over y_{_2}}\Big)^{n_{_2}}
\Big({y_{_4}\over y_{_2}}\Big)^{n_{_3}}\Big({y_{_3}\over y_{_2}}\Big)^{n_{_4}}
\;,\nonumber\\
&&\Phi_{_{[13\tilde{5}\tilde{7}]}}^{(20),c}(\alpha,z)=
y_{_1}^{-1}y_{_2}^{{D}-4}\sum\limits_{n_{_1}=0}^\infty
\sum\limits_{n_{_2}=0}^\infty\sum\limits_{n_{_3}=0}^\infty\sum\limits_{n_{_4}=0}^\infty
c_{_{[13\tilde{5}\tilde{7}]}}^{(20),c}(\alpha,{\bf n})\nonumber\\
&&\hspace{2.5cm}\times
\Big({1\over y_{_1}}\Big)^{n_{_1}}\Big({1\over y_{_2}}\Big)^{n_{_2}}
\Big({y_{_4}\over y_{_2}}\Big)^{n_{_3}}\Big({y_{_3}\over y_{_2}}\Big)^{n_{_4}}\;.
\label{GKZ21j-12-2a}
\end{eqnarray}
Where the coefficients are
\begin{eqnarray}
&&c_{_{[13\tilde{5}\tilde{7}]}}^{(20),a}(\alpha,{\bf n})=
(-)^{n_{_1}}\Gamma(1+n_{_1}+n_{_3})\Big\{n_{_1}!n_{_2}!n_{_3}!n_{_4}!
\Gamma({D\over2}+n_{_1})
\nonumber\\
&&\hspace{2.5cm}\times
\Gamma(2-{D\over2}+n_{_2})
\Gamma(1-{D\over2}-n_{_1}-n_{_3})\Gamma({D\over2}-1-n_{_2}-n_{_4})
\nonumber\\
&&\hspace{2.5cm}\times
\Gamma(2-{D\over2}+n_{_4})
\Gamma({D\over2}+n_{_3})\Gamma(D-2-n_{_2}-n_{_4})\Big\}^{-1}
\;,\nonumber\\
&&c_{_{[13\tilde{5}\tilde{7}]}}^{(20),b}(\alpha,{\bf n})=
(-)^{n_{_1}+n_{_2}+n_{_3}}\Gamma(1+n_{_1}+n_{_2})\Big\{n_{_1}!n_{_2}!n_{_4}!
\Gamma(2+n_{_1}+n_{_2}+n_{_3})
\nonumber\\
&&\hspace{2.5cm}\times
\Gamma({D\over2}+n_{_1})
\Gamma(3-{D\over2}+n_{_1}+n_{_2}+n_{_3})\Gamma(2-{D\over2}+n_{_2})
\nonumber\\
&&\hspace{2.5cm}\times
\Gamma({D\over2}-2-n_{_1}-n_{_2}-n_{_3}-n_{_4})
\Gamma(2-{D\over2}+n_{_4})
\nonumber\\
&&\hspace{2.5cm}\times
\Gamma({D\over2}-1-n_{_1}-n_{_2})
\Gamma(D-3-n_{_1}-n_{_2}-n_{_3}-n_{_4})\Big\}^{-1}
\;,\nonumber\\
&&c_{_{[13\tilde{5}\tilde{7}]}}^{(20),c}(\alpha,{\bf n})=
(-)^{n_{_1}+n_{_2}+n_{_3}}\Gamma(1+n_{_1})\Gamma(1+n_{_2})\Big\{n_{_4}!
\Gamma(2+n_{_1}+n_{_2})
\nonumber\\
&&\hspace{2.5cm}\times
\Gamma(2+n_{_2}+n_{_3})\Gamma({D\over2}+1+n_{_1}+n_{_2})
\Gamma(3-{D\over2}+n_{_2}+n_{_3})
\nonumber\\
&&\hspace{2.5cm}\times
\Gamma(1-{D\over2}-n_{_1})\Gamma({D\over2}-2-n_{_2}-n_{_3}-n_{_4})
\Gamma(2-{D\over2}+n_{_4})
\nonumber\\
&&\hspace{2.5cm}\times
\Gamma({D\over2}-1-n_{_2})\Gamma(D-3-n_{_2}-n_{_3}-n_{_4})\Big\}^{-1}\;.
\label{GKZ21j-12-3}
\end{eqnarray}

\item   $I_{_{21}}=\{1,4,5,6,7,9,\cdots,13\}$, i.e. the implement $J_{_{21}}=[1,14]\setminus I_{_{21}}=\{2,3,8,14\}$.
The choice implies the power numbers $\alpha_{_2}=\alpha_{_{3}}=\alpha_{_{8}}=\alpha_{_{14}}=0$, and
\begin{eqnarray}
&&\alpha_{_1}=a_{_2}-a_{_1},\;\alpha_{_4}=a_{_3}-a_{_4},\;\alpha_{_5}=-a_{_5},
\nonumber\\
&&\alpha_{_6}=a_{_3}+b_{_1}-a_{_2}-1,\;\alpha_{_7}=b_{_2}+b_{_3}-a_{_3}-2,
\nonumber\\
&&\alpha_{_{9}}=b_{_4}-a_{_3}-1,\;\alpha_{_{10}}=b_{_5}-a_{_3}-1,\;\alpha_{_{11}}=a_{_3}-a_{_2},
\nonumber\\
&&\alpha_{_{12}}=b_{_3}-a_{_3}-1,\;\alpha_{_{13}}=1-b_{_3}.
\label{GKZ21j-21-1}
\end{eqnarray}
The corresponding hypergeometric solutions are written as
\begin{eqnarray}
&&\Phi_{_{[13\tilde{5}\tilde{7}]}}^{(21),a}(\alpha,z)=
y_{_1}^{{D}-3}y_{_2}^{-1}y_{_3}^{{D\over2}-1}\sum\limits_{n_{_1}=0}^\infty
\sum\limits_{n_{_2}=0}^\infty\sum\limits_{n_{_3}=0}^\infty\sum\limits_{n_{_4}=0}^\infty
c_{_{[13\tilde{5}\tilde{7}]}}^{(21),a}(\alpha,{\bf n})
\nonumber\\
&&\hspace{2.5cm}\times
\Big({1\over y_{_1}}\Big)^{n_{_1}}\Big({y_{_4}\over y_{_2}}\Big)^{n_{_2}}
\Big({y_{_4}\over y_{_1}}\Big)^{n_{_3}}\Big({y_{_3}\over y_{_2}}\Big)^{n_{_4}}
\;,\nonumber\\
&&\Phi_{_{[13\tilde{5}\tilde{7}]}}^{(21),b}(\alpha,z)=
y_{_1}^{{D}-2}y_{_2}^{-2}y_{_3}^{{D\over2}-1}\sum\limits_{n_{_1}=0}^\infty
\sum\limits_{n_{_2}=0}^\infty\sum\limits_{n_{_3}=0}^\infty\sum\limits_{n_{_4}=0}^\infty
c_{_{[13\tilde{5}\tilde{7}]}}^{(21),b}(\alpha,{\bf n})
\nonumber\\
&&\hspace{2.5cm}\times
\Big({1\over y_{_1}}\Big)^{n_{_1}}\Big({y_{_1}\over y_{_2}}\Big)^{n_{_2}}
\Big({y_{_4}\over y_{_2}}\Big)^{n_{_3}}\Big({y_{_3}\over y_{_2}}\Big)^{n_{_4}}\;.
\label{GKZ21j-21-2a}
\end{eqnarray}
Where the coefficients are
\begin{eqnarray}
&&c_{_{[13\tilde{5}\tilde{7}]}}^{(21),a}(\alpha,{\bf n})=
(-)^{n_{_3}+n_{_4}}\Gamma(1+n_{_2}+n_{_4})\Big\{n_{_1}!n_{_2}!n_{_3}!n_{_4}!
\Gamma(2-{D\over2}+n_{_1})
\nonumber\\
&&\hspace{2.5cm}\times
\Gamma({D\over2}+n_{_2})\Gamma({D\over2}-1-n_{_1}-n_{_3})
\Gamma(1-{D\over2}-n_{_2}-n_{_4})
\nonumber\\
&&\hspace{2.5cm}\times
\Gamma(2-{D\over2}+n_{_3})
\Gamma(D-2-n_{_1}-n_{_3})\Gamma({D\over2}+n_{_4})\Big\}^{-1}
\;,\nonumber\\
&&c_{_{[13\tilde{5}\tilde{7}]}}^{(21),b}(\alpha,{\bf n})=
(-)^{1+n_{_4}}\Gamma(2+n_{_2}+n_{_3}+n_{_4})\Gamma(1+n_{_2})\Big\{n_{_1}!n_{_4}!
\nonumber\\
&&\hspace{2.5cm}\times
\Gamma(2+n_{_2}+n_{_3})\Gamma(2-{D\over2}+n_{_1})\Gamma({D\over2}+1+n_{_2}+n_{_3})
\nonumber\\
&&\hspace{2.5cm}\times
\Gamma({D\over2}-n_{_1}+n_{_2})
\Gamma(-{D\over2}-n_{_2}-n_{_3}-n_{_4})\Gamma(1-{D\over2}-n_{_2})
\nonumber\\
&&\hspace{2.5cm}\times
\Gamma(D-1-n_{_1}+n_{_2})\Gamma({D\over2}+n_{_4})\Big\}^{-1}\;.
\label{GKZ21j-21-3}
\end{eqnarray}

\item   $I_{_{22}}=\{1,4,\cdots,12\}$, i.e. the implement $J_{_{22}}=[1,14]\setminus I_{_{22}}=\{2,3,13,14\}$.
The choice implies the power numbers $\alpha_{_2}=\alpha_{_{3}}=\alpha_{_{13}}=\alpha_{_{14}}=0$, and
\begin{eqnarray}
&&\alpha_{_1}=a_{_2}-a_{_1},\;\alpha_{_4}=a_{_3}-a_{_4},\;\alpha_{_5}=-a_{_5},
\nonumber\\
&&\alpha_{_6}=a_{_3}+b_{_1}-a_{_2}-1,\;\alpha_{_7}=b_{_2}-a_{_3}-1,\;\alpha_{_{8}}=b_{_3}-1,
\nonumber\\
&&\alpha_{_{9}}=b_{_4}-a_{_3}-1,\;\alpha_{_{10}}=b_{_5}-a_{_3}-1,\;\alpha_{_{11}}=a_{_3}-a_{_2},\;
\alpha_{_{12}}=-a_{_3}.
\label{GKZ21j-22-1}
\end{eqnarray}
The corresponding hypergeometric solutions are given as
\begin{eqnarray}
&&\Phi_{_{[13\tilde{5}\tilde{7}]}}^{(22),a}(\alpha,z)=
y_{_1}^{{D}-3}y_{_2}^{{D\over2}-2}\sum\limits_{n_{_1}=0}^\infty
\sum\limits_{n_{_2}=0}^\infty\sum\limits_{n_{_3}=0}^\infty\sum\limits_{n_{_4}=0}^\infty
c_{_{[13\tilde{5}\tilde{7}]}}^{(22),a}(\alpha,{\bf n})
\nonumber\\
&&\hspace{2.5cm}\times
\Big({1\over y_{_1}}\Big)^{n_{_1}}\Big({y_{_4}\over y_{_2}}\Big)^{n_{_2}}
\Big({y_{_4}\over y_{_1}}\Big)^{n_{_3}}\Big({y_{_3}\over y_{_2}}\Big)^{n_{_4}}
\;,\nonumber\\
&&\Phi_{_{[13\tilde{5}\tilde{7}]}}^{(22),b}(\alpha,z)=
y_{_1}^{{D}-2}y_{_2}^{{D\over2}-3}\sum\limits_{n_{_1}=0}^\infty
\sum\limits_{n_{_2}=0}^\infty\sum\limits_{n_{_3}=0}^\infty\sum\limits_{n_{_4}=0}^\infty
c_{_{[13\tilde{5}\tilde{7}]}}^{(22),b}(\alpha,{\bf n})
\nonumber\\
&&\hspace{2.5cm}\times
\Big({1\over y_{_1}}\Big)^{n_{_1}}\Big({y_{_1}\over y_{_2}}\Big)^{n_{_2}}
\Big({y_{_4}\over y_{_2}}\Big)^{n_{_3}}\Big({y_{_3}\over y_{_2}}\Big)^{n_{_4}}\;.
\label{GKZ21j-22-2a}
\end{eqnarray}
Where the coefficients are
\begin{eqnarray}
&&c_{_{[13\tilde{5}\tilde{7}]}}^{(22),a}(\alpha,{\bf n})=
(-)^{n_{_3}+n_{_4}}\Gamma(1+n_{_2}+n_{_4})\Big\{n_{_1}!n_{_2}!n_{_3}!n_{_4}!
\Gamma(2-{D\over2}+n_{_1})
\nonumber\\
&&\hspace{2.5cm}\times
\Gamma({D\over2}+n_{_2})\Gamma({D\over2}-1-n_{_1}-n_{_3})
\Gamma({D\over2}-1-n_{_2}-n_{_4})
\nonumber\\
&&\hspace{2.5cm}\times
\Gamma(2-{D\over2}+n_{_3})
\Gamma(D-2-n_{_1}-n_{_3})\Gamma(2-{D\over2}+n_{_4})\Big\}^{-1}
\;,\nonumber\\
&&c_{_{[13\tilde{5}\tilde{7}]}}^{(22),b}(\alpha,{\bf n})=
(-)^{1+n_{_4}}\Gamma(2+n_{_2}+n_{_3}+n_{_4})\Gamma(1+n_{_2})\Big\{n_{_1}!n_{_4}!
\nonumber\\
&&\hspace{2.5cm}\times
\Gamma(2+n_{_2}+n_{_3})\Gamma(2-{D\over2}+n_{_1})\Gamma({D\over2}+1+n_{_2}+n_{_3})
\nonumber\\
&&\hspace{2.5cm}\times
\Gamma({D\over2}-n_{_1}+n_{_2})
\Gamma({D\over2}-2-n_{_2}-n_{_3}-n_{_4})\Gamma(1-{D\over2}-n_{_2})
\nonumber\\
&&\hspace{2.5cm}\times
\Gamma(D-1-n_{_1}+n_{_2})\Gamma(2-{D\over2}+n_{_4})\Big\}^{-1}\;.
\label{GKZ21j-22-3}
\end{eqnarray}

\item   $I_{_{23}}=\{1,3,5,6,7,9,\cdots,13\}$, i.e. the implement $J_{_{23}}=[1,14]\setminus I_{_{23}}=\{2,4,8,14\}$.
The choice implies the power numbers $\alpha_{_2}=\alpha_{_{4}}=\alpha_{_{8}}=\alpha_{_{14}}=0$, and
\begin{eqnarray}
&&\alpha_{_1}=a_{_2}-a_{_1},\;\alpha_{_3}=a_{_4}-a_{_3},\;\alpha_{_5}=-a_{_5},
\nonumber\\
&&\alpha_{_6}=a_{_4}+b_{_1}-a_{_2}-1,\;\alpha_{_7}=b_{_2}+b_{_3}-a_{_4}-2,
\nonumber\\
&&\alpha_{_{9}}=b_{_4}-a_{_4}-1,\;\alpha_{_{10}}=b_{_5}-a_{_4}-1,\;\alpha_{_{11}}=a_{_4}-a_{_2},
\nonumber\\
&&\alpha_{_{12}}=b_{_3}-a_{_4}-1,\;\alpha_{_{13}}=1-b_{_3}.
\label{GKZ21j-27-1}
\end{eqnarray}
The corresponding hypergeometric functions are written as
\begin{eqnarray}
&&\Phi_{_{[13\tilde{5}\tilde{7}]}}^{(23),a}(\alpha,z)=
y_{_1}^{{D\over2}-2}y_{_2}^{{D\over2}-2}y_{_3}^{{D\over2}-1}\sum\limits_{n_{_1}=0}^\infty
\sum\limits_{n_{_2}=0}^\infty\sum\limits_{n_{_3}=0}^\infty\sum\limits_{n_{_4}=0}^\infty
c_{_{[13\tilde{5}\tilde{7}]}}^{(23),a}(\alpha,{\bf n})
\nonumber\\
&&\hspace{2.5cm}\times
\Big({1\over y_{_1}}\Big)^{n_{_1}}\Big({y_{_4}\over y_{_2}}\Big)^{n_{_2}}
\Big({y_{_4}\over y_{_1}}\Big)^{n_{_3}}\Big({y_{_3}\over y_{_2}}\Big)^{n_{_4}}
\;,\nonumber\\
&&\Phi_{_{[13\tilde{5}\tilde{7}]}}^{(23),b}(\alpha,z)=
y_{_1}^{{D\over2}-1}y_{_2}^{{D\over2}-3}y_{_3}^{{D\over2}-1}\sum\limits_{n_{_1}=0}^\infty
\sum\limits_{n_{_2}=0}^\infty\sum\limits_{n_{_3}=0}^\infty\sum\limits_{n_{_4}=0}^\infty
c_{_{[13\tilde{5}\tilde{7}]}}^{(23),b}(\alpha,{\bf n})
\nonumber\\
&&\hspace{2.5cm}\times
\Big({1\over y_{_2}}\Big)^{n_{_1}}\Big({y_{_1}\over y_{_2}}\Big)^{n_{_2}}
\Big({y_{_4}\over y_{_2}}\Big)^{n_{_3}}\Big({y_{_3}\over y_{_2}}\Big)^{n_{_4}}
\;,\nonumber\\
&&\Phi_{_{[13\tilde{5}\tilde{7}]}}^{(23),c}(\alpha,z)=
y_{_1}^{{D\over2}-2}y_{_2}^{{D\over2}-3}y_{_3}^{{D\over2}-1}\sum\limits_{n_{_1}=0}^\infty
\sum\limits_{n_{_2}=0}^\infty\sum\limits_{n_{_3}=0}^\infty\sum\limits_{n_{_4}=0}^\infty
c_{_{[13\tilde{5}\tilde{7}]}}^{(23),c}(\alpha,{\bf n})
\nonumber\\
&&\hspace{2.5cm}\times
\Big({1\over y_{_1}}\Big)^{n_{_1}}\Big({1\over y_{_2}}\Big)^{n_{_2}}
\Big({y_{_4}\over y_{_2}}\Big)^{n_{_3}}\Big({y_{_3}\over y_{_2}}\Big)^{n_{_4}}
\label{GKZ21j-27-2a}
\end{eqnarray}
Where the coefficients are
\begin{eqnarray}
&&c_{_{[13\tilde{5}\tilde{7}]}}^{(23),a}(\alpha,{\bf n})=
(-)^{n_{_1}+n_{_4}}\Gamma(1+n_{_1}+n_{_3})\Gamma(1+n_{_2}+n_{_4})\Big\{n_{_1}!n_{_2}!n_{_3}!n_{_4}!
\nonumber\\
&&\hspace{2.5cm}\times
\Gamma(2-{D\over2}+n_{_1})\Gamma(2-{D\over2}+n_{_2})\Gamma({D\over2}+n_{_3})
\nonumber\\
&&\hspace{2.5cm}\times
\Gamma({D\over2}-1-n_{_1}-n_{_3})
\Gamma({D\over2}-1-n_{_2}-n_{_4})\Gamma({D\over2}+n_{_4})\Big\}^{-1}
\;,\nonumber\\
&&c_{_{[13\tilde{5}\tilde{7}]}}^{(23),b}(\alpha,{\bf n})=
(-)^{1+n_{_4}}\Gamma(1+n_{_1}+n_{_2})\Gamma(2+n_{_1}+n_{_2}+n_{_3}+n_{_4})
\nonumber\\
&&\hspace{2.5cm}\times
\Big\{n_{_1}!n_{_2}!n_{_4}!\Gamma(2+n_{_1}+n_{_2}+n_{_3})\Gamma(2-{D\over2}+n_{_1})
\nonumber\\
&&\hspace{2.5cm}\times
\Gamma(3-{D\over2}+n_{_1}+n_{_2}+n_{_3})\Gamma({D\over2}-1-n_{_1}-n_{_2})
\nonumber\\
&&\hspace{2.5cm}\times
\Gamma({D\over2}+n_{_2})
\Gamma({D\over2}-2-n_{_1}-n_{_2}-n_{_3}-n_{_4})\Gamma({D\over2}+n_{_4})\Big\}^{-1}
\;,\nonumber\\
&&c_{_{[13\tilde{5}\tilde{7}]}}^{(23),c}(\alpha,{\bf n})=
(-)^{1+n_{_1}+n_{_4}}\Gamma(1+n_{_1})\Gamma(1+n_{_2})\Gamma(2+n_{_2}+n_{_3}+n_{_4})
\nonumber\\
&&\hspace{2.5cm}\times
\Big\{n_{_4}!\Gamma(2+n_{_1}+n_{_2})\Gamma(2+n_{_2}+n_{_3})\Gamma(3-{D\over2}+n_{_1}+n_{_2})
\nonumber\\
&&\hspace{2.5cm}\times
\Gamma(3-{D\over2}+n_{_2}+n_{_3})\Gamma({D\over2}-1-n_{_2})\Gamma({D\over2}-1-n_{_1})
\nonumber\\
&&\hspace{2.5cm}\times
\Gamma({D\over2}-2-n_{_2}-n_{_3}-n_{_4})\Gamma({D\over2}+n_{_4})\Big\}^{-1}\;.
\label{GKZ21j-27-3}
\end{eqnarray}

\item   $I_{_{24}}=\{1,3,5,\cdots,12\}$, i.e. the implement $J_{_{24}}=[1,14]\setminus I_{_{24}}=\{2,4,13,14\}$.
The choice implies the power numbers $\alpha_{_2}=\alpha_{_{4}}=\alpha_{_{13}}=\alpha_{_{14}}=0$, and
\begin{eqnarray}
&&\alpha_{_1}=a_{_2}-a_{_1},\;\alpha_{_3}=a_{_4}-a_{_3},\;\alpha_{_5}=-a_{_5},
\nonumber\\
&&\alpha_{_6}=a_{_4}+b_{_1}-a_{_2}-1,\;\alpha_{_7}=b_{_2}-a_{_4}-1,\;\alpha_{_{8}}=b_{_3}-1,
\nonumber\\
&&\alpha_{_{9}}=b_{_4}-a_{_4}-1,\;\alpha_{_{10}}=b_{_5}-a_{_4}-1,\;\alpha_{_{11}}=a_{_4}-a_{_2},\;
\alpha_{_{12}}=-a_{_4}.
\label{GKZ21j-28-1}
\end{eqnarray}
The corresponding hypergeometric solutions are
\begin{eqnarray}
&&\Phi_{_{[13\tilde{5}\tilde{7}]}}^{(24),a}(\alpha,z)=
y_{_1}^{{D\over2}-2}y_{_2}^{{D}-3}\sum\limits_{n_{_1}=0}^\infty
\sum\limits_{n_{_2}=0}^\infty\sum\limits_{n_{_3}=0}^\infty\sum\limits_{n_{_4}=0}^\infty
c_{_{[13\tilde{5}\tilde{7}]}}^{(24),a}(\alpha,{\bf n})
\nonumber\\
&&\hspace{2.5cm}\times
\Big({1\over y_{_1}}\Big)^{n_{_1}}\Big({y_{_4}\over y_{_2}}\Big)^{n_{_2}}
\Big({y_{_4}\over y_{_1}}\Big)^{n_{_3}}\Big({y_{_3}\over y_{_2}}\Big)^{n_{_4}}
\;,\nonumber\\
&&\Phi_{_{[13\tilde{5}\tilde{7}]}}^{(24),b}(\alpha,z)=
y_{_1}^{{D\over2}-1}y_{_2}^{{D}-4}\sum\limits_{n_{_1}=0}^\infty
\sum\limits_{n_{_2}=0}^\infty\sum\limits_{n_{_3}=0}^\infty\sum\limits_{n_{_4}=0}^\infty
c_{_{[13\tilde{5}\tilde{7}]}}^{(24),b}(\alpha,{\bf n})
\nonumber\\
&&\hspace{2.5cm}\times
\Big({1\over y_{_2}}\Big)^{n_{_1}}\Big({y_{_1}\over y_{_2}}\Big)^{n_{_2}}
\Big({y_{_4}\over y_{_2}}\Big)^{n_{_3}}\Big({y_{_3}\over y_{_2}}\Big)^{n_{_4}}
\;,\nonumber\\
&&\Phi_{_{[13\tilde{5}\tilde{7}]}}^{(24),c}(\alpha,z)=
y_{_1}^{{D\over2}-2}y_{_2}^{{D}-4}\sum\limits_{n_{_1}=0}^\infty
\sum\limits_{n_{_2}=0}^\infty\sum\limits_{n_{_3}=0}^\infty\sum\limits_{n_{_4}=0}^\infty
c_{_{[13\tilde{5}\tilde{7}]}}^{(24),c}(\alpha,{\bf n})
\nonumber\\
&&\hspace{2.5cm}\times
\Big({1\over y_{_1}}\Big)^{n_{_1}}\Big({1\over y_{_2}}\Big)^{n_{_2}}
\Big({y_{_4}\over y_{_2}}\Big)^{n_{_3}}\Big({y_{_3}\over y_{_2}}\Big)^{n_{_4}}
\label{GKZ21j-28-2a}
\end{eqnarray}
Where the coefficients are
\begin{eqnarray}
&&c_{_{[13\tilde{5}\tilde{7}]}}^{(24),a}(\alpha,{\bf n})=
(-)^{n_{_1}+n_{_2}}\Gamma(1+n_{_1}+n_{_3})\Big\{n_{_1}!n_{_2}!n_{_3}!n_{_4}!
\Gamma(2-{D\over2}+n_{_1})
\nonumber\\
&&\hspace{2.5cm}\times
\Gamma(2-{D\over2}+n_{_2})\Gamma({D\over2}+n_{_3})\Gamma(D-2-n_{_2}-n_{_4})
\nonumber\\
&&\hspace{2.5cm}\times
\Gamma({D\over2}-1-n_{_1}-n_{_3})
\Gamma({D\over2}-1-n_{_2}-n_{_4})\Gamma(2-{D\over2}+n_{_4})\Big\}^{-1}
\;,\nonumber\\
&&c_{_{[13\tilde{5}\tilde{7}]}}^{(24),b}(\alpha,{\bf n})=
(-)^{n_{_1}+n_{_2}+n_{_3}}\Gamma(1+n_{_1}+n_{_2})\Big\{n_{_1}!n_{_2}!n_{_4}!
\Gamma(2-{D\over2}+n_{_1})
\nonumber\\
&&\hspace{2.5cm}\times
\Gamma(2+n_{_1}+n_{_2}+n_{_3})\Gamma(3-{D\over2}+n_{_1}+n_{_2}+n_{_3})
\Gamma({D\over2}+n_{_2})
\nonumber\\
&&\hspace{2.5cm}\times
\Gamma({D\over2}-1-n_{_1}-n_{_2})
\Gamma({D\over2}-2-n_{_1}-n_{_2}-n_{_3}-n_{_4})
\nonumber\\
&&\hspace{2.5cm}\times
\Gamma(2-{D\over2}+n_{_4})\Gamma(D-3-n_{_1}-n_{_2}-n_{_3}-n_{_4})\Big\}^{-1}
\;,\nonumber\\
&&c_{_{[13\tilde{5}\tilde{7}]}}^{(24),c}(\alpha,{\bf n})=
(-)^{n_{_1}+n_{_2}+n_{_3}}\Gamma(1+n_{_1})\Gamma(1+n_{_2})
\Big\{n_{_4}!\Gamma(2+n_{_1}+n_{_2})
\nonumber\\
&&\hspace{2.5cm}\times
\Gamma(2+n_{_2}+n_{_3})\Gamma(3-{D\over2}+n_{_1}+n_{_2})\Gamma(3-{D\over2}+n_{_2}+n_{_3})
\nonumber\\
&&\hspace{2.5cm}\times
\Gamma({D\over2}-1-n_{_2})\Gamma({D\over2}-1-n_{_1})\Gamma({D\over2}-2-n_{_2}-n_{_3}-n_{_4})
\nonumber\\
&&\hspace{2.5cm}\times
\Gamma(2-{D\over2}+n_{_4})\Gamma(D-3-n_{_2}-n_{_3}-n_{_4})\Big\}^{-1}\;.
\label{GKZ21j-28-3}
\end{eqnarray}

\item   $I_{_{25}}=\{2,3,4,5,7,9,\cdots,13\}$, i.e. the implement $J_{_{25}}=[1,14]\setminus I_{_{25}}=\{1,6,8,14\}$.
The choice implies the power numbers $\alpha_{_1}=\alpha_{_{6}}=\alpha_{_{8}}=\alpha_{_{14}}=0$, and
\begin{eqnarray}
&&\alpha_{_2}=a_{_1}-a_{_2},\;\alpha_{_3}=a_{_1}-a_{_3}-b_{_1}+1,\;\alpha_{_4}=a_{_1}-a_{_4}-b_{_1}+1,
\nonumber\\
&&\alpha_{_5}=-a_{_5},\;\alpha_{_7}=b_{_1}+b_{_2}+b_{_3}-a_{_1}-3,\;\alpha_{_{9}}=b_{_1}+b_{_4}-a_{_1}-2,
\nonumber\\
&&\alpha_{_{10}}=b_{_1}+b_{_5}-a_{_1}-2,\;\alpha_{_{11}}=1-b_{_1},\;
\alpha_{_{12}}=b_{_1}+b_{_3}-a_{_1}-2,
\nonumber\\
&&\alpha_{_{13}}=1-b_{_3}.
\label{GKZ21j-13-1}
\end{eqnarray}
The corresponding hypergeometric solution is written as
\begin{eqnarray}
&&\Phi_{_{[13\tilde{5}\tilde{7}]}}^{(25)}(\alpha,z)=
y_{_1}^{{D\over2}-1}y_{_2}^{-2}y_{_3}^{{D\over2}-1}\sum\limits_{n_{_1}=0}^\infty
\sum\limits_{n_{_2}=0}^\infty\sum\limits_{n_{_3}=0}^\infty\sum\limits_{n_{_4}=0}^\infty
c_{_{[13\tilde{5}\tilde{7}]}}^{(25)}(\alpha,{\bf n})
\nonumber\\
&&\hspace{2.5cm}\times
\Big({1\over y_{_2}}\Big)^{n_{_1}}\Big({y_{_1}\over y_{_2}}\Big)^{n_{_2}}
\Big({y_{_4}\over y_{_2}}\Big)^{n_{_3}}\Big({y_{_3}\over y_{_2}}\Big)^{n_{_4}}\;,
\label{GKZ21j-13-2}
\end{eqnarray}
with
\begin{eqnarray}
&&c_{_{[13\tilde{5}\tilde{7}]}}^{(25)}(\alpha,{\bf n})=
(-)^{1+n_{_4}}\Gamma(1+n_{_1}+n_{_2})\Gamma(2+n_{_1}+n_{_2}+n_{_3}+n_{_4})
\Big\{n_{_1}!n_{_2}!n_{_4}!
\nonumber\\
&&\hspace{2.5cm}\times
\Gamma({D\over2}+n_{_1})\Gamma(2+n_{_1}+n_{_2}+n_{_3})
\Gamma(1+{D\over2}+n_{_1}+n_{_2}+n_{_3})
\nonumber\\
&&\hspace{2.5cm}\times
\Gamma(-{D\over2}-n_{_1}-n_{_2}-n_{_3}-n_{_4})
\Gamma(1-{D\over2}-n_{_1}-n_{_2})\Gamma({D\over2}+n_{_2})
\nonumber\\
&&\hspace{2.5cm}\times
\Gamma({D\over2}+n_{_4})\Big\}^{-1}\;.
\label{GKZ21j-13-3}
\end{eqnarray}

\item   $I_{_{26}}=\{2,3,4,5,7,8,9,\cdots,12\}$, i.e. the implement $J_{_{26}}=[1,14]\setminus I_{_{26}}=\{1,6,13,14\}$.
The choice implies the power numbers $\alpha_{_1}=\alpha_{_{6}}=\alpha_{_{13}}=\alpha_{_{14}}=0$, and
\begin{eqnarray}
&&\alpha_{_2}=a_{_1}-a_{_2},\;\alpha_{_3}=a_{_1}-a_{_3}-b_{_1}+1,\;\alpha_{_4}=a_{_1}-a_{_4}-b_{_1}+1,
\nonumber\\
&&\alpha_{_5}=-a_{_5},\;\alpha_{_7}=b_{_1}+b_{_2}-a_{_1}-2,\;\alpha_{_{8}}=b_{_3}-1,
\nonumber\\
&&\alpha_{_{9}}=b_{_1}+b_{_4}-a_{_1}-2,\;\alpha_{_{10}}=b_{_1}+b_{_5}-a_{_1}-2,\;\alpha_{_{11}}=1-b_{_1},
\nonumber\\
&&\alpha_{_{12}}=b_{_1}-a_{_1}-1.
\label{GKZ21j-14-1}
\end{eqnarray}
The corresponding hypergeometric series is written as
\begin{eqnarray}
&&\Phi_{_{[13\tilde{5}\tilde{7}]}}^{(26)}(\alpha,z)=
y_{_1}^{{D\over2}-1}y_{_2}^{{D\over2}-3}\sum\limits_{n_{_1}=0}^\infty
\sum\limits_{n_{_2}=0}^\infty\sum\limits_{n_{_3}=0}^\infty\sum\limits_{n_{_4}=0}^\infty
c_{_{[13\tilde{5}\tilde{7}]}}^{(26)}(\alpha,{\bf n})
\nonumber\\
&&\hspace{2.5cm}\times
\Big({1\over y_{_2}}\Big)^{n_{_1}}\Big({y_{_1}\over y_{_2}}\Big)^{n_{_2}}
\Big({y_{_4}\over y_{_2}}\Big)^{n_{_3}}\Big({y_{_3}\over y_{_2}}\Big)^{n_{_4}}\;,
\label{GKZ21j-14-2}
\end{eqnarray}
with
\begin{eqnarray}
&&c_{_{[13\tilde{5}\tilde{7}]}}^{(26)}(\alpha,{\bf n})=
(-)^{1+n_{_4}}\Gamma(1+n_{_1}+n_{_2})\Gamma(2+n_{_1}+n_{_2}+n_{_3}+n_{_4})
\Big\{n_{_1}!n_{_2}!n_{_4}!
\nonumber\\
&&\hspace{2.5cm}\times
\Gamma({D\over2}+n_{_1})\Gamma(2+n_{_1}+n_{_2}+n_{_3})
\Gamma(1+{D\over2}+n_{_1}+n_{_2}+n_{_3})
\nonumber\\
&&\hspace{2.5cm}\times
\Gamma(2-{D\over2}+n_{_4})\Gamma(1-{D\over2}-n_{_1}-n_{_2})
\Gamma({D\over2}+n_{_2})
\nonumber\\
&&\hspace{2.5cm}\times
\Gamma({D\over2}-2-n_{_1}-n_{_2}-n_{_3}-n_{_4})\Big\}^{-1}\;.
\label{GKZ21j-14-3}
\end{eqnarray}

\item   $I_{_{27}}=\{2,\cdots,7,9,10,12,13\}$, i.e. the implement $J_{_{27}}=[1,14]\setminus I_{_{27}}=\{1,8,11,14\}$.
The choice implies the power numbers $\alpha_{_1}=\alpha_{_{8}}=\alpha_{_{11}}=\alpha_{_{14}}=0$, and
\begin{eqnarray}
&&\alpha_{_2}=a_{_1}-a_{_2},\;\alpha_{_3}=a_{_1}-a_{_3},\;\alpha_{_4}=a_{_1}-a_{_4},\;\alpha_{_5}=-a_{_5},
\nonumber\\
&&\alpha_{_{6}}=b_{_1}-1,\;\alpha_{_7}=b_{_2}+b_{_3}-a_{_1}-2,\;\alpha_{_{9}}=b_{_4}-a_{_1}-1,
\nonumber\\
&&\alpha_{_{10}}=b_{_5}-a_{_1}-1,\;\alpha_{_{12}}=b_{_3}-a_{_1}-1,\;\alpha_{_{13}}=1-b_{_3}.
\label{GKZ21j-15-1}
\end{eqnarray}
The corresponding hypergeometric series is written as
\begin{eqnarray}
&&\Phi_{_{[13\tilde{5}\tilde{7}]}}^{(27)}(\alpha,z)=
y_{_2}^{{D\over2}-3}y_{_3}^{{D\over2}-1}\sum\limits_{n_{_1}=0}^\infty
\sum\limits_{n_{_2}=0}^\infty\sum\limits_{n_{_3}=0}^\infty\sum\limits_{n_{_4}=0}^\infty
c_{_{[13\tilde{5}\tilde{7}]}}^{(27)}(\alpha,{\bf n})
\nonumber\\
&&\hspace{2.5cm}\times
\Big({1\over y_{_2}}\Big)^{n_{_1}}\Big({y_{_1}\over y_{_2}}\Big)^{n_{_2}}
\Big({y_{_4}\over y_{_2}}\Big)^{n_{_3}}\Big({y_{_3}\over y_{_2}}\Big)^{n_{_4}}\;,
\label{GKZ21j-15-2}
\end{eqnarray}
with
\begin{eqnarray}
&&c_{_{[13\tilde{5}\tilde{7}]}}^{(27)}(\alpha,{\bf n})=
(-)^{1+n_{_4}}\Gamma(1+n_{_1}+n_{_2})\Gamma(2+n_{_1}+n_{_2}+n_{_3}+n_{_4})
\nonumber\\
&&\hspace{2.5cm}\times
\Big\{n_{_1}!n_{_2}!n_{_4}!\Gamma({D\over2}+n_{_1})
\Gamma(3-{D\over2}+n_{_1}+n_{_2}+n_{_3})
\nonumber\\
&&\hspace{2.5cm}\times
\Gamma(2+n_{_1}+n_{_2}+n_{_3})\Gamma(2-{D\over2}+n_{_2})
\Gamma({D\over2}-1-n_{_1}-n_{_2})
\nonumber\\
&&\hspace{2.5cm}\times
\Gamma({D\over2}-2-n_{_1}-n_{_2}-n_{_3}-n_{_4})\Gamma({D\over2}+n_{_4})\Big\}^{-1}\;.
\label{GKZ21j-15-3}
\end{eqnarray}

\item   $I_{_{28}}=\{2,\cdots,10,12\}$, i.e. the implement $J_{_{28}}=[1,14]\setminus I_{_{28}}=\{1,11,13,14\}$.
The choice implies the power numbers $\alpha_{_1}=\alpha_{_{11}}=\alpha_{_{13}}=\alpha_{_{14}}=0$, and
\begin{eqnarray}
&&\alpha_{_2}=a_{_1}-a_{_2},\;\alpha_{_3}=a_{_1}-a_{_3},\;\alpha_{_4}=a_{_1}-a_{_4},
\nonumber\\
&&\alpha_{_5}=-a_{_5},\;\alpha_{_{6}}=b_{_1}-1,\;\alpha_{_7}=b_{_2}-a_{_1}-1,\;\alpha_{_{8}}=b_{_3}-1,
\nonumber\\
&&\alpha_{_{9}}=b_{_4}-a_{_1}-1,\;\alpha_{_{10}}=b_{_5}-a_{_1}-1,\;\alpha_{_{12}}=-a_{_1}.
\label{GKZ21j-16-1}
\end{eqnarray}
The corresponding hypergeometric solution is written as
\begin{eqnarray}
&&\Phi_{_{[13\tilde{5}\tilde{7}]}}^{(28)}(\alpha,z)=
y_{_2}^{{D}-4}\sum\limits_{n_{_1}=0}^\infty
\sum\limits_{n_{_2}=0}^\infty\sum\limits_{n_{_3}=0}^\infty\sum\limits_{n_{_4}=0}^\infty
c_{_{[13\tilde{5}\tilde{7}]}}^{(28)}(\alpha,{\bf n})
\nonumber\\
&&\hspace{2.5cm}\times
\Big({1\over y_{_2}}\Big)^{n_{_1}}\Big({y_{_1}\over y_{_2}}\Big)^{n_{_2}}
\Big({y_{_4}\over y_{_2}}\Big)^{n_{_3}}\Big({y_{_3}\over y_{_2}}\Big)^{n_{_4}}\;,
\label{GKZ21j-16-2}
\end{eqnarray}
with
\begin{eqnarray}
&&c_{_{[13\tilde{5}\tilde{7}]}}^{(28)}(\alpha,{\bf n})=
(-)^{n_{_1}+n_{_2}+n_{_3}}\Gamma(1+n_{_1}+n_{_2})
\Big\{n_{_1}!n_{_2}!n_{_4}!\Gamma({D\over2}+n_{_1})
\nonumber\\
&&\hspace{2.5cm}\times
\Gamma(3-{D\over2}+n_{_1}+n_{_2}+n_{_3})\Gamma(2+n_{_1}+n_{_2}+n_{_3})
\nonumber\\
&&\hspace{2.5cm}\times
\Gamma(2-{D\over2}+n_{_2})\Gamma({D\over2}-2-n_{_1}-n_{_2}-n_{_3}-n_{_4})
\nonumber\\
&&\hspace{2.5cm}\times
\Gamma(2-{D\over2}+n_{_4})\Gamma({D\over2}-1-n_{_1}-n_{_2})
\nonumber\\
&&\hspace{2.5cm}\times
\Gamma(D-3-n_{_1}-n_{_2}-n_{_3}-n_{_4})\Big\}^{-1}\;.
\label{GKZ21j-16-3}
\end{eqnarray}

\item   $I_{_{29}}=\{1,3,4,5,7,9,\cdots,13\}$, i.e. the implement $J_{_{29}}=[1,14]\setminus I_{_{29}}=\{2,6,8,14\}$.
The choice implies the power numbers $\alpha_{_2}=\alpha_{_{6}}=\alpha_{_{8}}=\alpha_{_{14}}=0$, and
\begin{eqnarray}
&&\alpha_{_1}=a_{_2}-a_{_1},\;\alpha_{_3}=a_{_2}-a_{_3}-b_{_1}+1,\;\alpha_{_4}=a_{_2}-a_{_4}-b_{_1}+1,
\nonumber\\
&&\alpha_{_5}=-a_{_5},\;\alpha_{_7}=b_{_1}+b_{_2}+b_{_3}-a_{_2}-3,\;\alpha_{_{9}}=b_{_1}+b_{_4}-a_{_2}-2,
\nonumber\\
&&\alpha_{_{10}}=b_{_1}+b_{_5}-a_{_2}-2,\;\alpha_{_{11}}=1-b_{_1},\;
\alpha_{_{12}}=b_{_1}+b_{_3}-a_{_2}-2,
\nonumber\\
&&\alpha_{_{13}}=1-b_{_3}.
\label{GKZ21j-29-1}
\end{eqnarray}
The corresponding hypergeometric series is written as
\begin{eqnarray}
&&\Phi_{_{[13\tilde{5}\tilde{7}]}}^{(29)}(\alpha,z)=
y_{_1}^{{D\over2}-1}y_{_2}^{{D\over2}-3}y_{_3}^{{D\over2}-1}\sum\limits_{n_{_1}=0}^\infty
\sum\limits_{n_{_2}=0}^\infty\sum\limits_{n_{_3}=0}^\infty\sum\limits_{n_{_4}=0}^\infty
c_{_{[13\tilde{5}\tilde{7}]}}^{(29)}(\alpha,{\bf n})
\nonumber\\
&&\hspace{2.5cm}\times
\Big({1\over y_{_2}}\Big)^{n_{_1}}\Big({y_{_1}\over y_{_2}}\Big)^{n_{_2}}
\Big({y_{_4}\over y_{_2}}\Big)^{n_{_3}}\Big({y_{_3}\over y_{_2}}\Big)^{n_{_4}}\;,
\label{GKZ21j-29-2}
\end{eqnarray}
with
\begin{eqnarray}
&&c_{_{[13\tilde{5}\tilde{7}]}}^{(29)}(\alpha,{\bf n})=
(-)^{1+n_{_4}}\Gamma(1+n_{_1}+n_{_2})\Gamma(2+n_{_1}+n_{_2}+n_{_3}+n_{_4})
\nonumber\\
&&\hspace{2.5cm}\times
\Big\{n_{_1}!n_{_2}!n_{_4}!\Gamma(2-{D\over2}+n_{_1})
\Gamma(3-{D\over2}+n_{_1}+n_{_2}+n_{_3})
\nonumber\\
&&\hspace{2.5cm}\times
\Gamma(2+n_{_1}+n_{_2}+n_{_3})\Gamma({D\over2}-1-n_{_1}-n_{_2})
\Gamma({D\over2}+n_{_2})
\nonumber\\
&&\hspace{2.5cm}\times
\Gamma({D\over2}-2-n_{_1}-n_{_2}-n_{_3}-n_{_4})\Gamma({D\over2}+n_{_4})\Big\}^{-1}\;.
\label{GKZ21j-29-3}
\end{eqnarray}

\item   $I_{_{30}}=\{1,3,4,5,7,8,9,\cdots,12\}$, i.e. the implement $J_{_{30}}=[1,14]\setminus I_{_{30}}=\{2,6,13,14\}$.
The choice implies the power numbers $\alpha_{_2}=\alpha_{_{6}}=\alpha_{_{13}}=\alpha_{_{14}}=0$, and
\begin{eqnarray}
&&\alpha_{_1}=a_{_2}-a_{_1},\;\alpha_{_3}=a_{_2}-a_{_3}-b_{_1}+1,\;\alpha_{_4}=a_{_2}-a_{_4}-b_{_1}+1,
\nonumber\\
&&\alpha_{_5}=-a_{_5},\;\alpha_{_7}=b_{_1}+b_{_2}-a_{_2}-2,\;\alpha_{_{8}}=b_{_3}-1,
\nonumber\\
&&\alpha_{_{9}}=b_{_1}+b_{_4}-a_{_2}-2,\;\alpha_{_{10}}=b_{_1}+b_{_5}-a_{_2}-2,\;\alpha_{_{11}}=1-b_{_1},
\nonumber\\
&&\alpha_{_{12}}=b_{_1}-a_{_2}-1.
\label{GKZ21j-30-1}
\end{eqnarray}
The corresponding hypergeometric solution is written as
\begin{eqnarray}
&&\Phi_{_{[13\tilde{5}\tilde{7}]}}^{(30)}(\alpha,z)=
y_{_1}^{{D\over2}-1}y_{_2}^{{D}-4}\sum\limits_{n_{_1}=0}^\infty
\sum\limits_{n_{_2}=0}^\infty\sum\limits_{n_{_3}=0}^\infty\sum\limits_{n_{_4}=0}^\infty
c_{_{[13\tilde{5}\tilde{7}]}}^{(30)}(\alpha,{\bf n})
\nonumber\\
&&\hspace{2.5cm}\times
\Big({1\over y_{_2}}\Big)^{n_{_1}}\Big({y_{_1}\over y_{_2}}\Big)^{n_{_2}}
\Big({y_{_4}\over y_{_2}}\Big)^{n_{_3}}\Big({y_{_3}\over y_{_2}}\Big)^{n_{_4}}\;,
\label{GKZ21j-30-2}
\end{eqnarray}
with
\begin{eqnarray}
&&\hspace{-1cm}c_{_{[13\tilde{5}\tilde{7}]}}^{(30)}(\alpha,{\bf n})=
(-)^{n_{_1}+n_{_2}+n_{_3}}\Gamma(1+n_{_1}+n_{_2})\Big\{n_{_1}!n_{_2}!n_{_4}!
\Gamma(2-{D\over2}+n_{_1})
\nonumber\\
&&\hspace{1.5cm}\times
\Gamma(3-{D\over2}+n_{_1}+n_{_2}+n_{_3})\Gamma(2+n_{_1}+n_{_2}+n_{_3})
\nonumber\\
&&\hspace{1.5cm}\times
\Gamma({D\over2}-2-n_{_1}-n_{_2}-n_{_3}-n_{_4})\Gamma(2-{D\over2}+n_{_4})
\nonumber\\
&&\hspace{1.5cm}\times
\Gamma({D\over2}-1-n_{_1}-n_{_2})
\Gamma({D\over2}+n_{_2})\Gamma(D-3-n_{_1}-n_{_2}-n_{_3}-n_{_4})\Big\}^{-1}\;.
\label{GKZ21j-30-3}
\end{eqnarray}

\item   $I_{_{31}}=\{1,3,\cdots,7,9,10,12,13\}$, i.e. the implement $J_{_{31}}=[1,14]\setminus I_{_{31}}=\{2,8,11,14\}$.
The choice implies the power numbers $\alpha_{_2}=\alpha_{_{8}}=\alpha_{_{11}}=\alpha_{_{14}}=0$, and
\begin{eqnarray}
&&\alpha_{_1}=a_{_2}-a_{_1},\;\alpha_{_3}=a_{_2}-a_{_3},\;\alpha_{_4}=a_{_2}-a_{_4},\;\alpha_{_5}=-a_{_5},
\nonumber\\
&&\alpha_{_{6}}=b_{_1}-1,\;\alpha_{_7}=b_{_2}+b_{_3}-a_{_2}-2,\;\alpha_{_{9}}=b_{_4}-a_{_2}-1,
\nonumber\\
&&\alpha_{_{10}}=b_{_5}-a_{_2}-1,\;\alpha_{_{12}}=b_{_3}-a_{_2}-1,\;\alpha_{_{13}}=1-b_{_3}.
\label{GKZ21j-31-1}
\end{eqnarray}
The corresponding hypergeometric solution is
\begin{eqnarray}
&&\Phi_{_{[13\tilde{5}\tilde{7}]}}^{(31)}(\alpha,z)=
y_{_2}^{{D}-4}y_{_3}^{{D\over2}-1}\sum\limits_{n_{_1}=0}^\infty
\sum\limits_{n_{_2}=0}^\infty\sum\limits_{n_{_3}=0}^\infty\sum\limits_{n_{_4}=0}^\infty
c_{_{[13\tilde{5}\tilde{7}]}}^{(31)}(\alpha,{\bf n})
\nonumber\\
&&\hspace{2.5cm}\times
\Big({1\over y_{_2}}\Big)^{n_{_1}}\Big({y_{_1}\over y_{_2}}\Big)^{n_{_2}}
\Big({y_{_4}\over y_{_2}}\Big)^{n_{_3}}\Big({y_{_3}\over y_{_2}}\Big)^{n_{_4}}\;,
\label{GKZ21j-31-2}
\end{eqnarray}
with
\begin{eqnarray}
&&\hspace{-1cm}c_{_{[13\tilde{5}\tilde{7}]}}^{(31)}(\alpha,{\bf n})=
(-)^{n_{_3}}\Big\{n_{_1}!n_{_2}!n_{_4}!\Gamma(2-{D\over2}+n_{_1})
\Gamma(4-D+n_{_1}+n_{_2}+n_{_3})
\nonumber\\
&&\hspace{1.5cm}\times
\Gamma(3-{D\over2}+n_{_1}+n_{_2}+n_{_3})\Gamma(2-{D\over2}+n_{_2})
\nonumber\\
&&\hspace{1.5cm}\times
\Gamma({D\over2}-2-n_{_1}-n_{_2}-n_{_3}-n_{_4})\Gamma(D-2-n_{_1}-n_{_2})
\nonumber\\
&&\hspace{1.5cm}\times
\Gamma({D\over2}-1-n_{_1}-n_{_2})
\Gamma(D-3-n_{_1}-n_{_2}-n_{_3}-n_{_4})\Gamma({D\over2}+n_{_4})\Big\}^{-1}\;.
\label{GKZ21j-31-3}
\end{eqnarray}

\item   $I_{_{32}}=\{1,3,\cdots,10,12\}$, i.e. the implement $J_{_{32}}=[1,14]\setminus I_{_{32}}=\{2,11,13,14\}$.
The choice implies the power numbers $\alpha_{_2}=\alpha_{_{11}}=\alpha_{_{13}}=\alpha_{_{14}}=0$, and
\begin{eqnarray}
&&\alpha_{_1}=a_{_2}-a_{_1},\;\alpha_{_3}=a_{_2}-a_{_3},\;\alpha_{_4}=a_{_2}-a_{_4},
\nonumber\\
&&\alpha_{_5}=-a_{_5},\;\alpha_{_{6}}=b_{_1}-1,\;\alpha_{_7}=b_{_2}-a_{_2}-1,\;\alpha_{_{8}}=b_{_3}-1,
\nonumber\\
&&\alpha_{_{9}}=b_{_4}-a_{_2}-1,\;\alpha_{_{10}}=b_{_5}-a_{_2}-1,\;\alpha_{_{12}}=-a_{_2}.
\label{GKZ21j-32-1}
\end{eqnarray}
The corresponding hypergeometric series is written as
\begin{eqnarray}
&&\Phi_{_{[13\tilde{5}\tilde{7}]}}^{(32)}(\alpha,z)=
y_{_2}^{{3D\over2}-5}\sum\limits_{n_{_1}=0}^\infty
\sum\limits_{n_{_2}=0}^\infty\sum\limits_{n_{_3}=0}^\infty\sum\limits_{n_{_4}=0}^\infty
c_{_{[13\tilde{5}\tilde{7}]}}^{(32)}(\alpha,{\bf n})
\nonumber\\
&&\hspace{2.5cm}\times
\Big({1\over y_{_2}}\Big)^{n_{_1}}\Big({y_{_1}\over y_{_2}}\Big)^{n_{_2}}
\Big({y_{_4}\over y_{_2}}\Big)^{n_{_3}}\Big({y_{_3}\over y_{_2}}\Big)^{n_{_4}}\;,
\label{GKZ21j-32-2}
\end{eqnarray}
with
\begin{eqnarray}
&&c_{_{[13\tilde{5}\tilde{7}]}}^{(32)}(\alpha,{\bf n})=
(-)^{n_{_3}}\Big\{n_{_1}!n_{_2}!n_{_4}!\Gamma(2-{D\over2}+n_{_1})
\Gamma(4-D+n_{_1}+n_{_2}+n_{_3})
\nonumber\\
&&\hspace{2.5cm}\times
\Gamma(3-{D\over2}+n_{_1}+n_{_2}+n_{_3})\Gamma(D-3-n_{_1}-n_{_2}-n_{_3}-n_{_4})
\nonumber\\
&&\hspace{2.5cm}\times
\Gamma(2-{D\over2}+n_{_2})\Gamma(2-{D\over2}+n_{_4})\Gamma(D-2-n_{_1}-n_{_2})
\nonumber\\
&&\hspace{2.5cm}\times
\Gamma({D\over2}-1-n_{_1}-n_{_2})\Gamma({3D\over2}-4-n_{_1}-n_{_2}-n_{_3}-n_{_4})\Big\}^{-1}\;.
\label{GKZ21j-32-3}
\end{eqnarray}
\end{itemize}

\section{The hypergeometric solutions of the integer lattice ${\bf B}_{_{\widetilde{135}7}}$\label{app12}}
\indent\indent

\begin{itemize}
\item   $I_{_{1}}=\{1,\cdots,5,10,\cdots,14\}$, i.e. the implement $J_{_{1}}=[1,14]\setminus I_{_{1}}=\{6,7,8,9\}$.
The choice implies the power numbers $\alpha_{_6}=\alpha_{_{7}}=\alpha_{_{8}}=\alpha_{_{9}}=0$, and
\begin{eqnarray}
&&\alpha_{_1}=b_{_1}+b_{_4}-a_{_1}-2,\;\alpha_{_2}=b_{_1}+b_{_4}-a_{_2}-2,
\nonumber\\
&&\alpha_{_3}=b_{_2}+b_{_3}-a_{_3}-2,\;\alpha_{_4}=b_{_2}+b_{_3}-a_{_4}-2,
\nonumber\\
&&\alpha_{_5}=b_{_4}-a_{_5}-b_{_2}-b_{_3}+1,\;\alpha_{_{10}}=b_{_5}-b_{_4},
\nonumber\\
&&\alpha_{_{11}}=1-b_{_1},\;\alpha_{_{12}}=1-b_{_2},\;\alpha_{_{13}}=1-b_{_3},
\nonumber\\
&&\alpha_{_{14}}=b_{_2}+b_{_3}-b_{_4}-1.
\label{GKZ21k-1-1}
\end{eqnarray}
The corresponding hypergeometric series is given as
\begin{eqnarray}
&&\Phi_{_{[\tilde{1}\tilde{3}\tilde{5}7]}}^{(1)}(\alpha,z)=
y_{_1}^{{D\over2}-1}y_{_2}^{{D\over2}-1}y_{_3}^{{D\over2}-1}y_{_4}^{-{D\over2}}
\sum\limits_{n_{_1}=0}^\infty
\sum\limits_{n_{_2}=0}^\infty\sum\limits_{n_{_3}=0}^\infty\sum\limits_{n_{_4}=0}^\infty
c_{_{[\tilde{1}\tilde{3}\tilde{5}7]}}^{(1)}(\alpha,{\bf n})
\nonumber\\
&&\hspace{2.5cm}\times
y_{_1}^{n_{_1}}\Big({y_{_3}\over y_{_4}}\Big)^{n_{_2}}
y_{_4}^{n_{_3}}\Big({y_{_2}\over y_{_4}}\Big)^{n_{_4}}\;,
\label{GKZ21k-1-2}
\end{eqnarray}
with
\begin{eqnarray}
&&c_{_{[\tilde{1}\tilde{3}\tilde{5}7]}}^{(1)}(\alpha,{\bf n})=
(-)^{n_{_1}}\Gamma(1+n_{_1}+n_{_3})\Gamma(1+n_{_2}+n_{_4})
\Big\{n_{_1}!n_{_2}!n_{_3}!n_{_4}!
\nonumber\\
&&\hspace{2.5cm}\times
\Gamma({D\over2}-1-n_{_1}-n_{_3})\Gamma(1-{D\over2}-n_{_2}-n_{_4})
\Gamma(2-{D\over2}+n_{_3})
\nonumber\\
&&\hspace{2.5cm}\times
\Gamma({D\over2}+n_{_1})\Gamma({D\over2}+n_{_4})\Gamma({D\over2}+n_{_2})\Big\}^{-1}\;.
\label{GKZ21k-1-3}
\end{eqnarray}

\item   $I_{_{2}}=\{1,\cdots,5,7,10,11,13,14\}$, i.e. the implement $J_{_{2}}=[1,14]\setminus I_{_{2}}=\{6,8,9,12\}$.
The choice implies the power numbers $\alpha_{_6}=\alpha_{_{8}}=\alpha_{_{9}}=\alpha_{_{12}}=0$, and
\begin{eqnarray}
&&\alpha_{_1}=b_{_1}+b_{_4}-a_{_1}-2,\;\alpha_{_2}=b_{_1}+b_{_4}-a_{_2}-2,
\nonumber\\
&&\alpha_{_3}=b_{_3}-a_{_3}-1,\;\alpha_{_4}=b_{_3}-a_{_4}-1,\;\alpha_{_5}=b_{_4}-a_{_5}-b_{_3},
\nonumber\\
&&\alpha_{_{7}}=b_{_2}-1,\;\alpha_{_{10}}=b_{_5}-b_{_4},\;
\alpha_{_{11}}=1-b_{_1},\;\alpha_{_{13}}=1-b_{_3},
\nonumber\\
&&\alpha_{_{14}}=b_{_3}-b_{_4}.
\label{GKZ21k-2-1}
\end{eqnarray}
The corresponding hypergeometric solution is written as
\begin{eqnarray}
&&\Phi_{_{[\tilde{1}\tilde{3}\tilde{5}7]}}^{(2)}(\alpha,z)=
y_{_1}^{{D\over2}-1}y_{_3}^{{D\over2}-1}y_{_4}^{-1}\sum\limits_{n_{_1}=0}^\infty
\sum\limits_{n_{_2}=0}^\infty\sum\limits_{n_{_3}=0}^\infty\sum\limits_{n_{_4}=0}^\infty
c_{_{[\tilde{1}\tilde{3}\tilde{5}7]}}^{(2)}(\alpha,{\bf n})
\nonumber\\
&&\hspace{2.5cm}\times
y_{_1}^{n_{_1}}\Big({y_{_3}\over y_{_4}}\Big)^{n_{_2}}
y_{_4}^{n_{_3}}\Big({y_{_2}\over y_{_4}}\Big)^{n_{_4}}\;,
\label{GKZ21k-2-2}
\end{eqnarray}
with
\begin{eqnarray}
&&c_{_{[\tilde{1}\tilde{3}\tilde{5}7]}}^{(2)}(\alpha,{\bf n})=
(-)^{n_{_1}}\Gamma(1+n_{_1}+n_{_3})\Gamma(1+n_{_2}+n_{_4})
\Big\{n_{_1}!n_{_2}!n_{_3}!n_{_4}!
\nonumber\\
&&\hspace{2.5cm}\times
\Gamma({D\over2}-1-n_{_1}-n_{_3})\Gamma({D\over2}-1-n_{_2}-n_{_4})
\Gamma(2-{D\over2}+n_{_4})
\nonumber\\
&&\hspace{2.5cm}\times
\Gamma(2-{D\over2}+n_{_3})\Gamma({D\over2}+n_{_1})\Gamma({D\over2}+n_{_2})\Big\}^{-1}\;.
\label{GKZ21k-2-3}
\end{eqnarray}

\item   $I_{_{3}}=\{1,\cdots,5,9,11,\cdots,14\}$, i.e. the implement $J_{_{3}}=[1,14]\setminus I_{_{3}}=\{6,7,8,10\}$.
The choice implies the power numbers $\alpha_{_6}=\alpha_{_{7}}=\alpha_{_{8}}=\alpha_{_{10}}=0$, and
\begin{eqnarray}
&&\alpha_{_1}=b_{_1}+b_{_5}-a_{_1}-2,\;\alpha_{_2}=b_{_1}+b_{_5}-a_{_2}-2,
\nonumber\\
&&\alpha_{_3}=b_{_2}+b_{_3}-a_{_3}-2,\;\alpha_{_4}=b_{_2}+b_{_3}-a_{_4}-2,
\nonumber\\
&&\alpha_{_5}=b_{_5}-a_{_5}-b_{_2}-b_{_3}+1,\;\alpha_{_{9}}=b_{_4}-b_{_5},
\nonumber\\
&&\alpha_{_{11}}=1-b_{_1},\;\alpha_{_{12}}=1-b_{_2},\;\alpha_{_{13}}=1-b_{_3},
\nonumber\\
&&\alpha_{_{14}}=b_{_2}+b_{_3}-b_{_5}-1.
\label{GKZ21k-3-1}
\end{eqnarray}
The corresponding hypergeometric solution is
\begin{eqnarray}
&&\Phi_{_{[\tilde{1}\tilde{3}\tilde{5}7]}}^{(3)}(\alpha,z)=
y_{_1}^{{D\over2}-1}y_{_2}^{{D\over2}-1}y_{_3}^{{D\over2}-1}y_{_4}^{-1}
\sum\limits_{n_{_1}=0}^\infty
\sum\limits_{n_{_2}=0}^\infty\sum\limits_{n_{_3}=0}^\infty\sum\limits_{n_{_4}=0}^\infty
c_{_{[\tilde{1}\tilde{3}\tilde{5}7]}}^{(3)}(\alpha,{\bf n})
\nonumber\\
&&\hspace{2.5cm}\times
y_{_1}^{n_{_1}}\Big({y_{_3}\over y_{_4}}\Big)^{n_{_2}}
y_{_4}^{n_{_3}}\Big({y_{_2}\over y_{_4}}\Big)^{n_{_4}}\;,
\label{GKZ21k-3-2}
\end{eqnarray}
with
\begin{eqnarray}
&&c_{_{[\tilde{1}\tilde{3}\tilde{5}7]}}^{(3)}(\alpha,{\bf n})=
(-)^{n_{_1}}\Gamma(1+n_{_1}+n_{_3})\Gamma(1+n_{_2}+n_{_4})
\Big\{n_{_1}!n_{_2}!n_{_3}!n_{_4}!
\nonumber\\
&&\hspace{2.5cm}\times
\Gamma(1-{D\over2}-n_{_1}-n_{_3})\Gamma(1-{D\over2}-n_{_2}-n_{_4})
\nonumber\\
&&\hspace{2.5cm}\times
\Gamma({D\over2}+n_{_3})\Gamma({D\over2}+n_{_1})
\Gamma({D\over2}+n_{_4})\Gamma({D\over2}+n_{_2})\Big\}^{-1}\;.
\label{GKZ21k-3-3}
\end{eqnarray}

\item   $I_{_{4}}=\{1,\cdots,5,7,9,11,13,14\}$, i.e. the implement $J_{_{4}}=[1,14]\setminus I_{_{4}}=\{6,8,10,12\}$.
The choice implies the power numbers $\alpha_{_6}=\alpha_{_{8}}=\alpha_{_{10}}=\alpha_{_{12}}=0$, and
\begin{eqnarray}
&&\alpha_{_1}=b_{_1}+b_{_5}-a_{_1}-2,\;\alpha_{_2}=b_{_1}+b_{_5}-a_{_2}-2,
\nonumber\\
&&\alpha_{_3}=b_{_3}-a_{_3}-1,\;\alpha_{_4}=b_{_3}-a_{_4}-1,\;\alpha_{_5}=b_{_5}-a_{_5}-b_{_3},
\nonumber\\
&&\alpha_{_{7}}=b_{_2}-1,\;\alpha_{_{9}}=b_{_4}-b_{_5},\;
\alpha_{_{11}}=1-b_{_1},\;\alpha_{_{13}}=1-b_{_3},
\nonumber\\
&&\alpha_{_{14}}=b_{_3}-b_{_5}.
\label{GKZ21k-4-1}
\end{eqnarray}
The corresponding hypergeometric series is written as
\begin{eqnarray}
&&\Phi_{_{[\tilde{1}\tilde{3}\tilde{5}7]}}^{(4)}(\alpha,z)=
y_{_1}^{{D\over2}-1}y_{_3}^{{D\over2}-1}y_{_4}^{{D\over2}-2}\sum\limits_{n_{_1}=0}^\infty
\sum\limits_{n_{_2}=0}^\infty\sum\limits_{n_{_3}=0}^\infty\sum\limits_{n_{_4}=0}^\infty
c_{_{[\tilde{1}\tilde{3}\tilde{5}7]}}^{(4)}(\alpha,{\bf n})
\nonumber\\
&&\hspace{2.5cm}\times
y_{_1}^{n_{_1}}\Big({y_{_3}\over y_{_4}}\Big)^{n_{_2}}
y_{_4}^{n_{_3}}\Big({y_{_2}\over y_{_4}}\Big)^{n_{_4}}\;,
\label{GKZ21k-4-2}
\end{eqnarray}
with
\begin{eqnarray}
&&c_{_{[\tilde{1}\tilde{3}\tilde{5}7]}}^{(4)}(\alpha,{\bf n})=
(-)^{1+n_{_1}}\Gamma(1+n_{_1}+n_{_3})\Gamma(1+n_{_2}+n_{_4})
\Big\{n_{_1}!n_{_2}!n_{_3}!n_{_4}!
\nonumber\\
&&\hspace{2.5cm}\times
\Gamma(1-{D\over2}-n_{_1}-n_{_3})\Gamma({D\over2}-1-n_{_2}-n_{_4})
\nonumber\\
&&\hspace{2.5cm}\times
\Gamma(2-{D\over2}+n_{_4})\Gamma({D\over2}+n_{_3})
\Gamma({D\over2}+n_{_1})
\Gamma({D\over2}+n_{_2})\Big\}^{-1}\;.
\label{GKZ21k-4-3}
\end{eqnarray}

\item   $I_{_{5}}=\{1,\cdots,5,8,10,11,12,14\}$, i.e. the implement $J_{_{5}}=[1,14]\setminus I_{_{5}}=\{6,7,9,13\}$.
The choice implies the power numbers $\alpha_{_6}=\alpha_{_{7}}=\alpha_{_{9}}=\alpha_{_{13}}=0$, and
\begin{eqnarray}
&&\alpha_{_1}=b_{_1}+b_{_4}-a_{_1}-2,\;\alpha_{_2}=b_{_1}+b_{_4}-a_{_2}-2,
\nonumber\\
&&\alpha_{_3}=b_{_2}-a_{_3}-1,\;\alpha_{_4}=b_{_2}-a_{_4}-1,
\nonumber\\
&&\alpha_{_5}=b_{_4}-a_{_5}-b_{_2},\;\alpha_{_{8}}=b_{_3}-1,\;\alpha_{_{10}}=b_{_5}-b_{_4},
\nonumber\\
&&\alpha_{_{11}}=1-b_{_1},\;\alpha_{_{12}}=1-b_{_2},\;
\alpha_{_{14}}=b_{_2}-b_{_4}.
\label{GKZ21k-11-1}
\end{eqnarray}
The corresponding hypergeometric solution is
\begin{eqnarray}
&&\Phi_{_{[\tilde{1}\tilde{3}\tilde{5}7]}}^{(5)}(\alpha,z)=
y_{_1}^{{D\over2}-1}y_{_2}^{{D\over2}-1}y_{_4}^{-1}\sum\limits_{n_{_1}=0}^\infty
\sum\limits_{n_{_2}=0}^\infty\sum\limits_{n_{_3}=0}^\infty\sum\limits_{n_{_4}=0}^\infty
c_{_{[\tilde{1}\tilde{3}\tilde{5}7]}}^{(5)}(\alpha,{\bf n})
\nonumber\\
&&\hspace{2.5cm}\times
y_{_1}^{n_{_1}}\Big({y_{_3}\over y_{_4}}\Big)^{n_{_2}}
y_{_4}^{n_{_3}}\Big({y_{_2}\over y_{_4}}\Big)^{n_{_4}}\;,
\label{GKZ21k-11-2}
\end{eqnarray}
with
\begin{eqnarray}
&&c_{_{[\tilde{1}\tilde{3}\tilde{5}7]}}^{(5)}(\alpha,{\bf n})=
(-)^{n_{_1}}\Gamma(1+n_{_1}+n_{_3})\Gamma(1+n_{_2}+n_{_4})
\Big\{n_{_1}!n_{_2}!n_{_3}!n_{_4}!
\nonumber\\
&&\hspace{2.5cm}\times
\Gamma({D\over2}-1-n_{_1}-n_{_3})\Gamma({D\over2}-1-n_{_2}-n_{_4})
\nonumber\\
&&\hspace{2.5cm}\times
\Gamma(2-{D\over2}+n_{_2})\Gamma(2-{D\over2}+n_{_3})\Gamma({D\over2}+n_{_1})
\Gamma({D\over2}+n_{_4})\Big\}^{-1}\;.
\label{GKZ21k-11-3}
\end{eqnarray}

\item   $I_{_{6}}=\{1,\cdots,5,7,8,10,11,14\}$, i.e. the implement $J_{_{6}}=[1,14]\setminus I_{_{6}}=\{6,9,12,13\}$.
The choice implies the power numbers $\alpha_{_6}=\alpha_{_{9}}=\alpha_{_{12}}=\alpha_{_{13}}=0$, and
\begin{eqnarray}
&&\alpha_{_1}=b_{_1}+b_{_4}-a_{_1}-2,\;\alpha_{_2}=b_{_1}+b_{_4}-a_{_2}-2,
\nonumber\\
&&\alpha_{_3}=-a_{_3},\;\alpha_{_4}=-a_{_4},\;\alpha_{_5}=b_{_4}-a_{_5}-1,\;\alpha_{_{7}}=b_{_2}-1,
\nonumber\\
&&\alpha_{_{8}}=b_{_3}-1,\;\alpha_{_{10}}=b_{_5}-b_{_4},\;\alpha_{_{11}}=1-b_{_1},
\;\alpha_{_{14}}=1-b_{_4}.
\label{GKZ21k-12-1}
\end{eqnarray}
The corresponding hypergeometric solution is written as
\begin{eqnarray}
&&\Phi_{_{[\tilde{1}\tilde{3}\tilde{5}7]}}^{(6)}(\alpha,z)=
y_{_1}^{{D\over2}-1}y_{_4}^{{D\over2}-2}\sum\limits_{n_{_1}=0}^\infty
\sum\limits_{n_{_2}=0}^\infty\sum\limits_{n_{_3}=0}^\infty\sum\limits_{n_{_4}=0}^\infty
c_{_{[\tilde{1}\tilde{3}\tilde{5}7]}}^{(6)}(\alpha,{\bf n})
\nonumber\\
&&\hspace{2.5cm}\times
y_{_1}^{n_{_1}}\Big({y_{_3}\over y_{_4}}\Big)^{n_{_2}}
y_{_4}^{n_{_3}}\Big({y_{_2}\over y_{_4}}\Big)^{n_{_4}}\;,
\label{GKZ21k-12-2}
\end{eqnarray}
with
\begin{eqnarray}
&&c_{_{[\tilde{1}\tilde{3}\tilde{5}7]}}^{(6)}(\alpha,{\bf n})=
(-)^{1+n_{_1}+n_{_2}+n_{_3}}\Gamma(1+n_{_1}+n_{_3})\Big\{n_{_1}!n_{_2}!n_{_3}!n_{_4}!
\Gamma({D\over2}-1-n_{_1}-n_{_3})
\nonumber\\
&&\hspace{2.5cm}\times
\Gamma({D\over2}-1-n_{_2}-n_{_4})\Gamma(D-2-n_{_2}-n_{_4})
\Gamma(2-{D\over2}+n_{_4})
\nonumber\\
&&\hspace{2.5cm}\times
\Gamma(2-{D\over2}+n_{_2})\Gamma(2-{D\over2}+n_{_3})
\Gamma({D\over2}+n_{_1})\Big\}^{-1}\;.
\label{GKZ21k-12-3}
\end{eqnarray}

\item   $I_{_{7}}=\{1,\cdots,5,8,9,11,12,14\}$, i.e. the implement $J_{_{7}}=[1,14]\setminus I_{_{7}}=\{6,7,10,13\}$.
The choice implies the power numbers $\alpha_{_6}=\alpha_{_{7}}=\alpha_{_{10}}=\alpha_{_{13}}=0$, and
\begin{eqnarray}
&&\alpha_{_1}=b_{_1}+b_{_5}-a_{_1}-2,\;\alpha_{_2}=b_{_1}+b_{_5}-a_{_2}-2,
\nonumber\\
&&\alpha_{_3}=b_{_2}-a_{_3}-1,\;\alpha_{_4}=b_{_2}-a_{_4}-1,
\nonumber\\
&&\alpha_{_5}=b_{_5}-a_{_5}-b_{_2},\;\alpha_{_{8}}=b_{_3}-1,\;\alpha_{_{9}}=b_{_4}-b_{_5},
\nonumber\\
&&\alpha_{_{11}}=1-b_{_1},\;\alpha_{_{12}}=1-b_{_2},\;\alpha_{_{14}}=b_{_2}-b_{_5}.
\label{GKZ21k-13-1}
\end{eqnarray}
The corresponding hypergeometric series is written as
\begin{eqnarray}
&&\Phi_{_{[\tilde{1}\tilde{3}\tilde{5}7]}}^{(7)}(\alpha,z)=
y_{_1}^{{D\over2}-1}y_{_2}^{{D\over2}-1}y_{_4}^{{D\over2}-2}\sum\limits_{n_{_1}=0}^\infty
\sum\limits_{n_{_2}=0}^\infty\sum\limits_{n_{_3}=0}^\infty\sum\limits_{n_{_4}=0}^\infty
c_{_{[\tilde{1}\tilde{3}\tilde{5}7]}}^{(7)}(\alpha,{\bf n})
\nonumber\\
&&\hspace{2.5cm}\times
y_{_1}^{n_{_1}}\Big({y_{_3}\over y_{_4}}\Big)^{n_{_2}}
y_{_4}^{n_{_3}}\Big({y_{_2}\over y_{_4}}\Big)^{n_{_4}}\;,
\label{GKZ21k-13-2}
\end{eqnarray}
with
\begin{eqnarray}
&&c_{_{[\tilde{1}\tilde{3}\tilde{5}7]}}^{(7)}(\alpha,{\bf n})=
(-)^{1+n_{_1}}\Gamma(1+n_{_1}+n_{_3})\Gamma(1+n_{_2}+n_{_4})
\Big\{n_{_1}!n_{_2}!n_{_3}!n_{_4}!
\nonumber\\
&&\hspace{2.5cm}\times
\Gamma(1-{D\over2}-n_{_1}-n_{_3})\Gamma({D\over2}-1-n_{_2}-n_{_4})
\nonumber\\
&&\hspace{2.5cm}\times
\Gamma(2-{D\over2}+n_{_2})\Gamma({D\over2}+n_{_3})
\Gamma({D\over2}+n_{_1})\Gamma({D\over2}+n_{_4})\Big\}^{-1}\;.
\label{GKZ21k-13-3}
\end{eqnarray}

\item   $I_{_{8}}=\{1,\cdots,5,7,8,9,11,14\}$, i.e. the implement $J_{_{8}}=[1,14]\setminus I_{_{8}}=\{6,10,12,13\}$.
The choice implies the power numbers $\alpha_{_6}=\alpha_{_{10}}=\alpha_{_{12}}=\alpha_{_{13}}=0$, and
\begin{eqnarray}
&&\alpha_{_1}=b_{_1}+b_{_5}-a_{_1}-2,\;\alpha_{_2}=b_{_1}+b_{_5}-a_{_2}-2,
\nonumber\\
&&\alpha_{_3}=-a_{_3},\;\alpha_{_4}=-a_{_4},\;\alpha_{_5}=b_{_5}-a_{_5}-1,\;\alpha_{_{7}}=b_{_2}-1,
\nonumber\\
&&\alpha_{_{8}}=b_{_3}-1,\;\alpha_{_{9}}=b_{_4}-b_{_5},\;\alpha_{_{11}}=1-b_{_1},
\;\alpha_{_{14}}=1-b_{_5}.
\label{GKZ21k-14-1}
\end{eqnarray}
The corresponding hypergeometric solution is
\begin{eqnarray}
&&\Phi_{_{[\tilde{1}\tilde{3}\tilde{5}7]}}^{(8)}(\alpha,z)=
y_{_1}^{{D\over2}-1}y_{_4}^{{D}-3}\sum\limits_{n_{_1}=0}^\infty
\sum\limits_{n_{_2}=0}^\infty\sum\limits_{n_{_3}=0}^\infty\sum\limits_{n_{_4}=0}^\infty
c_{_{[\tilde{1}\tilde{3}\tilde{5}7]}}^{(8)}(\alpha,{\bf n})
\nonumber\\
&&\hspace{2.5cm}\times
y_{_1}^{n_{_1}}\Big({y_{_3}\over y_{_4}}\Big)^{n_{_2}}
y_{_4}^{n_{_3}}\Big({y_{_2}\over y_{_4}}\Big)^{n_{_4}}\;,
\label{GKZ21k-14-2}
\end{eqnarray}
with
\begin{eqnarray}
&&c_{_{[\tilde{1}\tilde{3}\tilde{5}7]}}^{(8)}(\alpha,{\bf n})=
(-)^{n_{_1}+n_{_2}+n_{_4}}\Gamma(1+n_{_1}+n_{_3})\Big\{n_{_1}!n_{_2}!n_{_3}!n_{_4}!
\Gamma(1-{D\over2}-n_{_1}-n_{_3})
\nonumber\\
&&\hspace{2.5cm}\times
\Gamma({D\over2}-1-n_{_2}-n_{_4})\Gamma(D-2-n_{_2}-n_{_4})
\Gamma(2-{D\over2}+n_{_4})
\nonumber\\
&&\hspace{2.5cm}\times
\Gamma(2-{D\over2}+n_{_2})\Gamma({D\over2}+n_{_3})
\Gamma({D\over2}+n_{_1})\Big\}^{-1}\;.
\label{GKZ21k-14-3}
\end{eqnarray}

\item   $I_{_{9}}=\{1,\cdots,6,10,12,13,14\}$, i.e. the implement $J_{_{9}}=[1,14]\setminus I_{_{9}}=\{7,8,9,11\}$.
The choice implies the power numbers $\alpha_{_7}=\alpha_{_{8}}=\alpha_{_{9}}=\alpha_{_{11}}=0$, and
\begin{eqnarray}
&&\alpha_{_1}=b_{_4}-a_{_1}-1,\;\alpha_{_2}=b_{_4}-a_{_2}-1,
\nonumber\\
&&\alpha_{_3}=b_{_2}+b_{_3}-a_{_3}-2,\;\alpha_{_4}=b_{_2}+b_{_3}-a_{_4}-2,
\nonumber\\
&&\alpha_{_5}=b_{_4}-a_{_5}-b_{_2}-b_{_3}+1,\;\alpha_{_{6}}=b_{_1}-1,\;\alpha_{_{10}}=b_{_5}-b_{_4},
\nonumber\\
&&\alpha_{_{12}}=1-b_{_2},\;\alpha_{_{13}}=1-b_{_3},\;\alpha_{_{14}}=b_{_2}+b_{_3}-b_{_4}-1.
\label{GKZ21k-17-1}
\end{eqnarray}
The corresponding hypergeometric series is given as
\begin{eqnarray}
&&\Phi_{_{[\tilde{1}\tilde{3}\tilde{5}7]}}^{(9)}(\alpha,z)=
y_{_2}^{{D\over2}-1}y_{_3}^{{D\over2}-1}y_{_4}^{-{D\over2}}\sum\limits_{n_{_1}=0}^\infty
\sum\limits_{n_{_2}=0}^\infty\sum\limits_{n_{_3}=0}^\infty\sum\limits_{n_{_4}=0}^\infty
c_{_{[\tilde{1}\tilde{3}\tilde{5}7]}}^{(9)}(\alpha,{\bf n})
\nonumber\\
&&\hspace{2.5cm}\times
y_{_1}^{n_{_1}}\Big({y_{_3}\over y_{_4}}\Big)^{n_{_2}}
y_{_4}^{n_{_3}}\Big({y_{_2}\over y_{_4}}\Big)^{n_{_4}}\;,
\label{GKZ21k-17-2}
\end{eqnarray}
with
\begin{eqnarray}
&&c_{_{[\tilde{1}\tilde{3}\tilde{5}7]}}^{(9)}(\alpha,{\bf n})=
(-)^{n_{_3}}\Gamma(1+n_{_2}+n_{_4})\Big\{n_{_1}!n_{_2}!n_{_3}!n_{_4}!
\Gamma({D\over2}-1-n_{_1}-n_{_3})
\nonumber\\
&&\hspace{2.5cm}\times
\Gamma(D-2-n_{_1}-n_{_3})\Gamma(1-{D\over2}-n_{_2}-n_{_4})
\Gamma(2-{D\over2}+n_{_1})
\nonumber\\
&&\hspace{2.5cm}\times
\Gamma(2-{D\over2}+n_{_3})\Gamma({D\over2}+n_{_4})
\Gamma({D\over2}+n_{_2})\Big\}^{-1}\;.
\label{GKZ21k-17-3}
\end{eqnarray}

\item   $I_{_{10}}=\{1,\cdots,7,10,13,14\}$, i.e. the implement $J_{_{10}}=[1,14]\setminus I_{_{10}}=\{8,9,11,12\}$.
The choice implies the power numbers $\alpha_{_8}=\alpha_{_{9}}=\alpha_{_{11}}=\alpha_{_{12}}=0$, and
\begin{eqnarray}
&&\alpha_{_1}=b_{_4}-a_{_1}-1,\;\alpha_{_2}=b_{_4}-a_{_2}-1,\;\alpha_{_3}=b_{_3}-a_{_3}-1,
\nonumber\\
&&\alpha_{_4}=b_{_3}-a_{_4}-1,\;\alpha_{_5}=b_{_4}-a_{_5}-b_{_3},\;\alpha_{_{6}}=b_{_1}-1,
\nonumber\\
&&\alpha_{_{7}}=b_{_2}-1,\;\alpha_{_{10}}=b_{_5}-b_{_4},\;
\alpha_{_{13}}=1-b_{_3},\;\alpha_{_{14}}=b_{_3}-b_{_4}.
\label{GKZ21k-18-1}
\end{eqnarray}
The corresponding hypergeometric series is
\begin{eqnarray}
&&\Phi_{_{[\tilde{1}\tilde{3}\tilde{5}7]}}^{(10)}(\alpha,z)=
y_{_3}^{{D\over2}-1}y_{_4}^{-1}\sum\limits_{n_{_1}=0}^\infty
\sum\limits_{n_{_2}=0}^\infty\sum\limits_{n_{_3}=0}^\infty\sum\limits_{n_{_4}=0}^\infty
c_{_{[\tilde{1}\tilde{3}\tilde{5}7]}}^{(10)}(\alpha,{\bf n})
\nonumber\\
&&\hspace{2.5cm}\times
y_{_1}^{n_{_1}}\Big({y_{_3}\over y_{_4}}\Big)^{n_{_2}}
y_{_4}^{n_{_3}}\Big({y_{_2}\over y_{_4}}\Big)^{n_{_4}}\;,
\label{GKZ21k-18-2}
\end{eqnarray}
with
\begin{eqnarray}
&&c_{_{[\tilde{1}\tilde{3}\tilde{5}7]}}^{(10)}(\alpha,{\bf n})=
(-)^{n_{_3}}\Gamma(1+n_{_2}+n_{_4})\Big\{n_{_1}!n_{_2}!n_{_3}!n_{_4}!
\Gamma({D\over2}-1-n_{_1}-n_{_3})
\nonumber\\
&&\hspace{2.5cm}\times
\Gamma(D-2-n_{_1}-n_{_3})\Gamma({D\over2}-1-n_{_2}-n_{_4})\Gamma(2-{D\over2}+n_{_1})
\nonumber\\
&&\hspace{2.5cm}\times
\Gamma(2-{D\over2}+n_{_4})\Gamma(2-{D\over2}+n_{_3})\Gamma({D\over2}+n_{_2})\Big\}^{-1}\;.
\label{GKZ21k-18-3}
\end{eqnarray}

\item   $I_{_{11}}=\{1,\cdots,6,9,12,13,14\}$, i.e. the implement $J_{_{11}}=[1,14]\setminus I_{_{11}}=\{7,8,10,11\}$.
The choice implies the power numbers $\alpha_{_7}=\alpha_{_{8}}=\alpha_{_{10}}=\alpha_{_{11}}=0$, and
\begin{eqnarray}
&&\alpha_{_1}=b_{_5}-a_{_1}-1,\;\alpha_{_2}=b_{_5}-a_{_2}-1,
\nonumber\\
&&\alpha_{_3}=b_{_2}+b_{_3}-a_{_3}-2,\;\alpha_{_4}=b_{_2}+b_{_3}-a_{_4}-2,
\nonumber\\
&&\alpha_{_5}=b_{_5}-a_{_5}-b_{_2}-b_{_3}+1,\;\alpha_{_{6}}=b_{_1}-1,\;\alpha_{_{9}}=b_{_4}-b_{_5},
\nonumber\\
&&\alpha_{_{12}}=1-b_{_2},\;\alpha_{_{13}}=1-b_{_3},\;\alpha_{_{14}}=b_{_2}+b_{_3}-b_{_5}-1.
\label{GKZ21k-19-1}
\end{eqnarray}
The corresponding hypergeometric solution is written as
\begin{eqnarray}
&&\Phi_{_{[\tilde{1}\tilde{3}\tilde{5}7]}}^{(11)}(\alpha,z)=
y_{_2}^{{D\over2}-1}y_{_3}^{{D\over2}-1}y_{_4}^{-1}\sum\limits_{n_{_1}=0}^\infty
\sum\limits_{n_{_2}=0}^\infty\sum\limits_{n_{_3}=0}^\infty\sum\limits_{n_{_4}=0}^\infty
c_{_{[\tilde{1}\tilde{3}\tilde{5}7]}}^{(11)}(\alpha,{\bf n})
\nonumber\\
&&\hspace{2.5cm}\times
y_{_1}^{n_{_1}}\Big({y_{_3}\over y_{_4}}\Big)^{n_{_2}}
y_{_4}^{n_{_3}}\Big({y_{_2}\over y_{_4}}\Big)^{n_{_4}}\;,
\label{GKZ21k-19-2}
\end{eqnarray}
with
\begin{eqnarray}
&&c_{_{[\tilde{1}\tilde{3}\tilde{5}7]}}^{(11)}(\alpha,{\bf n})=
(-)^{n_{_1}}\Gamma(1+n_{_1}+n_{_3})\Gamma(1+n_{_2}+n_{_4})
\Big\{n_{_1}!n_{_2}!n_{_3}!n_{_4}!
\nonumber\\
&&\hspace{2.5cm}\times
\Gamma({D\over2}-1-n_{_1}-n_{_3})\Gamma(1-{D\over2}-n_{_2}-n_{_4})
\Gamma(2-{D\over2}+n_{_1})
\nonumber\\
&&\hspace{2.5cm}\times
\Gamma({D\over2}+n_{_3})\Gamma({D\over2}+n_{_4})\Gamma({D\over2}+n_{_2})\Big\}^{-1}\;.
\label{GKZ21k-19-3}
\end{eqnarray}

\item   $I_{_{12}}=\{1,\cdots,7,9,13,14\}$, i.e. the implement $J_{_{12}}=[1,14]\setminus I_{_{12}}=\{8,10,11,12\}$.
The choice implies the power numbers $\alpha_{_8}=\alpha_{_{10}}=\alpha_{_{11}}=\alpha_{_{12}}=0$, and
\begin{eqnarray}
&&\alpha_{_1}=b_{_5}-a_{_1}-1,\;\alpha_{_2}=b_{_5}-a_{_2}-1,\;\alpha_{_3}=b_{_3}-a_{_3}-1,
\nonumber\\
&&\alpha_{_4}=b_{_3}-a_{_4}-1,\;\alpha_{_5}=b_{_5}-a_{_5}-b_{_3},\;\alpha_{_{6}}=b_{_1}-1,
\nonumber\\
&&\alpha_{_{7}}=b_{_2}-1,\;\alpha_{_{9}}=b_{_4}-b_{_5},\;\alpha_{_{13}}=1-b_{_3},\;\alpha_{_{14}}=b_{_3}-b_{_5}.
\label{GKZ21k-20-1}
\end{eqnarray}
The corresponding hypergeometric solution is written as
\begin{eqnarray}
&&\Phi_{_{[\tilde{1}\tilde{3}\tilde{5}7]}}^{(12)}(\alpha,z)=
y_{_3}^{{D\over2}-1}y_{_4}^{{D\over2}-2}\sum\limits_{n_{_1}=0}^\infty
\sum\limits_{n_{_2}=0}^\infty\sum\limits_{n_{_3}=0}^\infty\sum\limits_{n_{_4}=0}^\infty
c_{_{[\tilde{1}\tilde{3}\tilde{5}7]}}^{(12)}(\alpha,{\bf n})
\nonumber\\
&&\hspace{2.5cm}\times
y_{_1}^{n_{_1}}\Big({y_{_3}\over y_{_4}}\Big)^{n_{_2}}
y_{_4}^{n_{_3}}\Big({y_{_2}\over y_{_4}}\Big)^{n_{_4}}\;,
\label{GKZ21k-20-2}
\end{eqnarray}
with
\begin{eqnarray}
&&c_{_{[\tilde{1}\tilde{3}\tilde{5}7]}}^{(12)}(\alpha,{\bf n})=
(-)^{1+n_{_1}}\Gamma(1+n_{_1}+n_{_3})\Gamma(1+n_{_2}+n_{_4})
\Big\{n_{_1}!n_{_2}!n_{_3}!n_{_4}!
\nonumber\\
&&\hspace{2.5cm}\times
\Gamma({D\over2}-1-n_{_1}-n_{_3})\Gamma({D\over2}-1-n_{_2}-n_{_4})
\nonumber\\
&&\hspace{2.5cm}\times
\Gamma(2-{D\over2}+n_{_1})\Gamma(2-{D\over2}+n_{_4})
\Gamma({D\over2}+n_{_3})\Gamma({D\over2}+n_{_2})\Big\}^{-1}\;.
\label{GKZ21k-20-3}
\end{eqnarray}

\item   $I_{_{13}}=\{1,\cdots,6,8,10,12,14\}$, i.e. the implement $J_{_{13}}=[1,14]\setminus I_{_{13}}=\{7,9,11,13\}$.
The choice implies the power numbers $\alpha_{_7}=\alpha_{_{9}}=\alpha_{_{11}}=\alpha_{_{13}}=0$, and
\begin{eqnarray}
&&\alpha_{_1}=b_{_4}-a_{_1}-1,\;\alpha_{_2}=b_{_4}-a_{_2}-1,\;\alpha_{_3}=b_{_2}-a_{_3}-1,
\nonumber\\
&&\alpha_{_4}=b_{_2}-a_{_4}-1,\;\alpha_{_5}=b_{_4}-a_{_5}-b_{_2},\;\alpha_{_{6}}=b_{_1}-1,
\nonumber\\
&&\alpha_{_{8}}=b_{_3}-1,\;\alpha_{_{10}}=b_{_5}-b_{_4},\;\alpha_{_{12}}=1-b_{_2},\;\alpha_{_{14}}=b_{_2}-b_{_4}.
\label{GKZ21k-27-1}
\end{eqnarray}
The corresponding hypergeometric series is
\begin{eqnarray}
&&\Phi_{_{[\tilde{1}\tilde{3}\tilde{5}7]}}^{(13)}(\alpha,z)=
y_{_2}^{{D\over2}-1}y_{_4}^{-1}\sum\limits_{n_{_1}=0}^\infty
\sum\limits_{n_{_2}=0}^\infty\sum\limits_{n_{_3}=0}^\infty\sum\limits_{n_{_4}=0}^\infty
c_{_{[\tilde{1}\tilde{3}\tilde{5}7]}}^{(13)}(\alpha,{\bf n})
\nonumber\\
&&\hspace{2.5cm}\times
y_{_1}^{n_{_1}}\Big({y_{_3}\over y_{_4}}\Big)^{n_{_2}}
y_{_4}^{n_{_3}}\Big({y_{_2}\over y_{_4}}\Big)^{n_{_4}}\;,
\label{GKZ21k-27-2}
\end{eqnarray}
with
\begin{eqnarray}
&&c_{_{[\tilde{1}\tilde{3}\tilde{5}7]}}^{(13)}(\alpha,{\bf n})=
(-)^{n_{_3}}\Gamma(1+n_{_2}+n_{_4})
\Big\{n_{_1}!n_{_2}!n_{_3}!n_{_4}!\Gamma({D\over2}-1-n_{_1}-n_{_3})
\nonumber\\
&&\hspace{2.5cm}\times
\Gamma(D-2-n_{_1}-n_{_3})\Gamma({D\over2}-1-n_{_2}-n_{_4})
\Gamma(2-{D\over2}+n_{_1})
\nonumber\\
&&\hspace{2.5cm}\times
\Gamma(2-{D\over2}+n_{_2})\Gamma(2-{D\over2}+n_{_3})
\Gamma({D\over2}+n_{_4})\Big\}^{-1}\;.
\label{GKZ21k-27-3}
\end{eqnarray}

\item   $I_{_{14}}=\{1,\cdots,8,10,14\}$, i.e. the implement $J_{_{14}}=[1,14]\setminus I_{_{14}}=\{9,11,12,13\}$.
The choice implies the power numbers $\alpha_{_9}=\alpha_{_{11}}=\alpha_{_{12}}=\alpha_{_{13}}=0$, and
\begin{eqnarray}
&&\alpha_{_1}=b_{_4}-a_{_1}-1,\;\alpha_{_2}=b_{_4}-a_{_2}-1,\;\alpha_{_3}=-a_{_3},
\nonumber\\
&&\alpha_{_4}=-a_{_4},\;\alpha_{_5}=b_{_4}-a_{_5}-1,\;\alpha_{_{6}}=b_{_1}-1,\;\alpha_{_{7}}=b_{_2}-1,
\nonumber\\
&&\alpha_{_{8}}=b_{_3}-1,\;\alpha_{_{10}}=b_{_5}-b_{_4},\;\alpha_{_{14}}=1-b_{_4}.
\label{GKZ21k-28-1}
\end{eqnarray}
The corresponding hypergeometric solution is given as
\begin{eqnarray}
&&\Phi_{_{[\tilde{1}\tilde{3}\tilde{5}7]}}^{(14)}(\alpha,z)=
y_{_4}^{{D\over2}-2}\sum\limits_{n_{_1}=0}^\infty
\sum\limits_{n_{_2}=0}^\infty\sum\limits_{n_{_3}=0}^\infty\sum\limits_{n_{_4}=0}^\infty
c_{_{[\tilde{1}\tilde{3}\tilde{5}7]}}^{(14)}(\alpha,{\bf n})
\nonumber\\
&&\hspace{2.5cm}\times
y_{_1}^{n_{_1}}\Big({y_{_3}\over y_{_4}}\Big)^{n_{_2}}
y_{_4}^{n_{_3}}\Big({y_{_2}\over y_{_4}}\Big)^{n_{_4}}\;,
\label{GKZ21k-28-2}
\end{eqnarray}
with
\begin{eqnarray}
&&c_{_{[\tilde{1}\tilde{3}\tilde{5}7]}}^{(14)}(\alpha,{\bf n})=
(-)^{1+n_{_2}+n_{_3}+n_{_4}}\Big\{n_{_1}!n_{_2}!n_{_3}!n_{_4}!\Gamma({D\over2}-1-n_{_1}-n_{_3})
\Gamma(D-2-n_{_1}-n_{_3})
\nonumber\\
&&\hspace{2.5cm}\times
\Gamma({D\over2}-1-n_{_2}-n_{_4})\Gamma(D-2-n_{_2}-n_{_4})\Gamma(2-{D\over2}+n_{_1})
\nonumber\\
&&\hspace{2.5cm}\times
\Gamma(2-{D\over2}+n_{_4})\Gamma(2-{D\over2}+n_{_2})
\Gamma(2-{D\over2}+n_{_3})\Big\}^{-1}\;.
\label{GKZ21k-28-3}
\end{eqnarray}

\item   $I_{_{15}}=\{1,\cdots,5,8,9,11,12,14\}$, i.e. the implement $J_{_{15}}=[1,14]\setminus I_{_{15}}=\{7,10,11,13\}$.
The choice implies the power numbers $\alpha_{_7}=\alpha_{_{10}}=\alpha_{_{11}}=\alpha_{_{13}}=0$, and
\begin{eqnarray}
&&\alpha_{_1}=b_{_5}-a_{_1}-1,\;\alpha_{_2}=b_{_5}-a_{_2}-1,\;\alpha_{_3}=b_{_2}-a_{_3}-1,
\nonumber\\
&&\alpha_{_4}=b_{_2}-a_{_4}-1,\;\alpha_{_5}=b_{_5}-a_{_5}-b_{_2},\;\alpha_{_{6}}=b_{_1}-1,\;
\nonumber\\
&&\alpha_{_{8}}=b_{_3}-1,\;\alpha_{_{9}}=b_{_4}-b_{_5},\;\alpha_{_{12}}=1-b_{_2},\;\alpha_{_{14}}=b_{_2}-b_{_5}.
\label{GKZ21k-29-1}
\end{eqnarray}
The corresponding hypergeometric series is written as
\begin{eqnarray}
&&\Phi_{_{[\tilde{1}\tilde{3}\tilde{5}7]}}^{(15)}(\alpha,z)=
y_{_2}^{{D\over2}-1}y_{_4}^{{D\over2}-2}\sum\limits_{n_{_1}=0}^\infty
\sum\limits_{n_{_2}=0}^\infty\sum\limits_{n_{_3}=0}^\infty\sum\limits_{n_{_4}=0}^\infty
c_{_{[\tilde{1}\tilde{3}\tilde{5}7]}}^{(15)}(\alpha,{\bf n})
\nonumber\\
&&\hspace{2.5cm}\times
y_{_1}^{n_{_1}}\Big({y_{_3}\over y_{_4}}\Big)^{n_{_2}}
y_{_4}^{n_{_3}}\Big({y_{_2}\over y_{_4}}\Big)^{n_{_4}}\;,
\label{GKZ21k-29-2}
\end{eqnarray}
with
\begin{eqnarray}
&&c_{_{[\tilde{1}\tilde{3}\tilde{5}7]}}^{(15)}(\alpha,{\bf n})=
(-)^{1+n_{_1}}\Gamma(1+n_{_1}+n_{_3})\Gamma(1+n_{_2}+n_{_4})
\Big\{n_{_1}!n_{_2}!n_{_3}!n_{_4}!
\nonumber\\
&&\hspace{2.5cm}\times
\Gamma({D\over2}-1-n_{_1}-n_{_3})\Gamma({D\over2}-1-n_{_2}-n_{_4})
\Gamma(2-{D\over2}+n_{_1})
\nonumber\\
&&\hspace{2.5cm}\times
\Gamma(2-{D\over2}+n_{_2})\Gamma({D\over2}+n_{_3})\Gamma({D\over2}+n_{_4})\Big\}^{-1}\;.
\label{GKZ21k-29-3}
\end{eqnarray}

\item   $I_{_{16}}=\{1,\cdots,9,14\}$, i.e. the implement $J_{_{16}}=[1,14]\setminus I_{_{16}}=\{10,11,12,13\}$.
The choice implies the power numbers $\alpha_{_{10}}=\alpha_{_{11}}=\alpha_{_{12}}=\alpha_{_{13}}=0$, and
\begin{eqnarray}
&&\alpha_{_1}=b_{_5}-a_{_1}-1,\;\alpha_{_2}=b_{_5}-a_{_2}-1,\;\alpha_{_3}=-a_{_3},
\nonumber\\
&&\alpha_{_4}=-a_{_4},\;\alpha_{_5}=b_{_5}-a_{_5}-1,\;\alpha_{_{6}}=b_{_1}-1,
\nonumber\\
&&\alpha_{_{7}}=b_{_2}-1,\;\alpha_{_{8}}=b_{_3}-1,\;\alpha_{_{9}}=b_{_4}-b_{_5},\;\alpha_{_{14}}=1-b_{_5}.
\label{GKZ21k-30-1}
\end{eqnarray}
The corresponding hypergeometric solution is
\begin{eqnarray}
&&\Phi_{_{[\tilde{1}\tilde{3}\tilde{5}7]}}^{(16)}(\alpha,z)=
y_{_4}^{{D}-3}\sum\limits_{n_{_1}=0}^\infty
\sum\limits_{n_{_2}=0}^\infty\sum\limits_{n_{_3}=0}^\infty\sum\limits_{n_{_4}=0}^\infty
c_{_{[\tilde{1}\tilde{3}\tilde{5}7]}}^{(16)}(\alpha,{\bf n})
\nonumber\\
&&\hspace{2.5cm}\times
y_{_1}^{n_{_1}}\Big({y_{_3}\over y_{_4}}\Big)^{n_{_2}}
y_{_4}^{n_{_3}}\Big({y_{_2}\over y_{_4}}\Big)^{n_{_4}}\;,
\label{GKZ21k-30-2}
\end{eqnarray}
with
\begin{eqnarray}
&&c_{_{[\tilde{1}\tilde{3}\tilde{5}7]}}^{(16)}(\alpha,{\bf n})=
(-)^{n_{_1}+n_{_2}+n_{_4}}\Gamma(1+n_{_1}+n_{_3})\Big\{n_{_1}!n_{_2}!n_{_3}!n_{_4}!
\Gamma({D\over2}-1-n_{_1}-n_{_3})
\nonumber\\
&&\hspace{2.5cm}\times
\Gamma({D\over2}-1-n_{_2}-n_{_4})\Gamma(D-2-n_{_2}-n_{_4})
\Gamma(2-{D\over2}+n_{_1})
\nonumber\\
&&\hspace{2.5cm}\times
\Gamma(2-{D\over2}+n_{_4})\Gamma(2-{D\over2}+n_{_2})
\Gamma({D\over2}+n_{_3})\Big\}^{-1}\;.
\label{GKZ21k-30-3}
\end{eqnarray}

\item   $I_{_{17}}=\{1,\cdots,5,9,\cdots,13\}$, i.e. the implement $J_{_{17}}=[1,14]\setminus I_{_{17}}=\{6,7,8,14\}$.
The choice implies the power numbers $\alpha_{_6}=\alpha_{_{7}}=\alpha_{_{8}}=\alpha_{_{14}}=0$, and
\begin{eqnarray}
&&\alpha_{_1}=b_{_1}+b_{_2}+b_{_3}-a_{_1}-3,\;\alpha_{_2}=b_{_1}+b_{_2}+b_{_3}-a_{_2}-3,
\nonumber\\
&&\alpha_{_3}=b_{_2}+b_{_3}-a_{_3}-2,\;\alpha_{_4}=b_{_2}+b_{_3}-a_{_4}-2,\;\alpha_{_5}=-a_{_5},
\nonumber\\
&&\alpha_{_{9}}=b_{_4}-b_{_2}-b_{_3}+1,\;\alpha_{_{10}}=b_{_5}-b_{_2}-b_{_3}+1,\;\alpha_{_{11}}=1-b_{_1},
\nonumber\\
&&\alpha_{_{12}}=1-b_{_2},\;\alpha_{_{13}}=1-b_{_3}.
\label{GKZ21k-5-1}
\end{eqnarray}
The corresponding hypergeometric series is written as
\begin{eqnarray}
&&\Phi_{_{[\tilde{1}\tilde{3}\tilde{5}7]}}^{(17)}(\alpha,z)=
y_{_1}^{{D\over2}-1}y_{_2}^{{D\over2}-1}y_{_3}^{{D\over2}-1}\sum\limits_{n_{_1}=0}^\infty
\sum\limits_{n_{_2}=0}^\infty\sum\limits_{n_{_3}=0}^\infty\sum\limits_{n_{_4}=0}^\infty
c_{_{[\tilde{1}\tilde{3}\tilde{5}7]}}^{(17)}(\alpha,{\bf n})
\nonumber\\
&&\hspace{2.5cm}\times
y_{_1}^{n_{_1}}y_{_3}^{n_{_2}}y_{_4}^{n_{_3}}y_{_2}^{n_{_4}}\;,
\label{GKZ21k-5-2}
\end{eqnarray}
with
\begin{eqnarray}
&&c_{_{[\tilde{1}\tilde{3}\tilde{5}7]}}^{(17)}(\alpha,{\bf n})=
(-)^{1+n_{_1}}\Gamma(2+n_{_1}+n_{_2}+n_{_3}+n_{_4})\Gamma(1+n_{_2}+n_{_4})
\nonumber\\
&&\hspace{2.5cm}\times
\Big\{n_{_1}!n_{_2}!n_{_4}!\Gamma(-{D\over2}-n_{_1}-n_{_2}-n_{_3}-n_{_4})
\Gamma(1-{D\over2}-n_{_2}-n_{_4})
\nonumber\\
&&\hspace{2.5cm}\times
\Gamma({D\over2}+1+n_{_2}+n_{_3}+n_{_4})\Gamma(2-{D\over2}+n_{_2}+n_{_3}+n_{_4})
\nonumber\\
&&\hspace{2.5cm}\times
\Gamma({D\over2}+n_{_1})\Gamma({D\over2}+n_{_4})\Gamma({D\over2}+n_{_2})\Big\}^{-1}\;.
\label{GKZ21k-5-3}
\end{eqnarray}

\item   $I_{_{18}}=\{1,\cdots,5,7,9,10,11,13\}$, i.e. the implement $J_{_{18}}=[1,14]\setminus I_{_{18}}=\{6,8,12,14\}$.
The choice implies the power numbers $\alpha_{_6}=\alpha_{_{8}}=\alpha_{_{12}}=\alpha_{_{14}}=0$, and
\begin{eqnarray}
&&\alpha_{_1}=b_{_1}+b_{_3}-a_{_1}-2,\;\alpha_{_2}=b_{_1}+b_{_3}-a_{_2}-2,
\nonumber\\
&&\alpha_{_3}=b_{_3}-a_{_3}-1,\;\alpha_{_4}=b_{_3}-a_{_4}-1,\;\alpha_{_5}=-a_{_5},
\nonumber\\
&&\alpha_{_{7}}=b_{_2}-1,\;\alpha_{_{9}}=b_{_4}-b_{_3},\;\alpha_{_{10}}=b_{_5}-b_{_3},\;\alpha_{_{11}}=1-b_{_1},
\nonumber\\
&&\alpha_{_{13}}=1-b_{_3}.
\label{GKZ21k-6-1}
\end{eqnarray}
The corresponding hypergeometric series is written as
\begin{eqnarray}
&&\Phi_{_{[\tilde{1}\tilde{3}\tilde{5}7]}}^{(18)}(\alpha,z)=
y_{_1}^{{D\over2}-1}y_{_3}^{{D\over2}-1}\sum\limits_{n_{_1}=0}^\infty
\sum\limits_{n_{_2}=0}^\infty\sum\limits_{n_{_3}=0}^\infty\sum\limits_{n_{_4}=0}^\infty
c_{_{[\tilde{1}\tilde{3}\tilde{5}7]}}^{(18)}(\alpha,{\bf n})
\nonumber\\
&&\hspace{2.5cm}\times
y_{_1}^{n_{_1}}y_{_3}^{n_{_2}}y_{_4}^{n_{_3}}y_{_2}^{n_{_4}}\;,
\label{GKZ21k-6-2}
\end{eqnarray}
with
\begin{eqnarray}
&&c_{_{[\tilde{1}\tilde{3}\tilde{5}7]}}^{(18)}(\alpha,{\bf n})=
(-)^{1+n_{_1}}\Gamma(2+n_{_1}+n_{_2}+n_{_3}+n_{_4})\Gamma(1+n_{_2}+n_{_4})
\Big\{n_{_1}!n_{_2}!n_{_4}!
\nonumber\\
&&\hspace{2.5cm}\times
\Gamma({D\over2}-2-n_{_1}-n_{_2}-n_{_3}-n_{_4})\Gamma({D\over2}-1-n_{_2}-n_{_4})
\nonumber\\
&&\hspace{2.5cm}\times
\Gamma(2-{D\over2}+n_{_4})\Gamma(2+n_{_2}+n_{_3}+n_{_4})
\nonumber\\
&&\hspace{2.5cm}\times
\Gamma(3-{D\over2}+n_{_2}+n_{_3}+n_{_4})
\Gamma({D\over2}+n_{_1})\Gamma({D\over2}+n_{_2})\Big\}^{-1}\;.
\label{GKZ21k-6-3}
\end{eqnarray}

\item   $I_{_{19}}=\{1,\cdots,5,8,\cdots,12\}$, i.e. the implement $J_{_{19}}=[1,14]\setminus I_{_{19}}=\{6,7,13,14\}$.
The choice implies the power numbers $\alpha_{_6}=\alpha_{_{7}}=\alpha_{_{13}}=\alpha_{_{14}}=0$, and
\begin{eqnarray}
&&\alpha_{_1}=b_{_1}+b_{_2}-a_{_1}-2,\;\alpha_{_2}=b_{_1}+b_{_2}-a_{_2}-2,
\nonumber\\
&&\alpha_{_3}=b_{_2}-a_{_3}-1,\;\alpha_{_4}=b_{_2}-a_{_4}-1,\;\alpha_{_5}=-a_{_5},\;\alpha_{_{8}}=b_{_3}-1,
\nonumber\\
&&\alpha_{_{9}}=b_{_4}-b_{_2},\;\alpha_{_{10}}=b_{_5}-b_{_2},\;\alpha_{_{11}}=1-b_{_1},
\;\alpha_{_{12}}=1-b_{_2}.
\label{GKZ21k-15-1}
\end{eqnarray}
The corresponding hypergeometric solution is written as
\begin{eqnarray}
&&\Phi_{_{[\tilde{1}\tilde{3}\tilde{5}7]}}^{(19)}(\alpha,z)=
y_{_1}^{{D\over2}-1}y_{_2}^{{D\over2}-1}\sum\limits_{n_{_1}=0}^\infty
\sum\limits_{n_{_2}=0}^\infty\sum\limits_{n_{_3}=0}^\infty\sum\limits_{n_{_4}=0}^\infty
c_{_{[\tilde{1}\tilde{3}\tilde{5}7]}}^{(19)}(\alpha,{\bf n})
\nonumber\\
&&\hspace{2.5cm}\times
y_{_1}^{n_{_1}}y_{_3}^{n_{_2}}y_{_4}^{n_{_3}}y_{_2}^{n_{_4}}\;,
\label{GKZ21k-15-2}
\end{eqnarray}
with
\begin{eqnarray}
&&c_{_{[\tilde{1}\tilde{3}\tilde{5}7]}}^{(19)}(\alpha,{\bf n})=
(-)^{1+n_{_1}}\Gamma(2+n_{_1}+n_{_2}+n_{_3}+n_{_4})\Gamma(1+n_{_2}+n_{_4})
\Big\{n_{_1}!n_{_2}!n_{_4}!
\nonumber\\
&&\hspace{2.5cm}\times
\Gamma({D\over2}-2-n_{_1}-n_{_2}-n_{_3}-n_{_4})\Gamma({D\over2}-1-n_{_2}-n_{_4})
\Gamma(2-{D\over2}+n_{_2})
\nonumber\\
&&\hspace{2.5cm}\times
\Gamma(2+n_{_2}+n_{_3}+n_{_4})\Gamma(3-{D\over2}+n_{_2}+n_{_3}+n_{_4})
\Gamma({D\over2}+n_{_1})
\nonumber\\
&&\hspace{2.5cm}\times
\Gamma({D\over2}+n_{_4})\Big\}^{-1}\;.
\label{GKZ21k-15-3}
\end{eqnarray}

\item   $I_{_{20}}=\{1,\cdots,5,7,\cdots,11\}$, i.e. the implement $J_{_{20}}=[1,14]\setminus I_{_{20}}=\{6,12,13,14\}$.
The choice implies the power numbers $\alpha_{_6}=\alpha_{_{12}}=\alpha_{_{13}}=\alpha_{_{14}}=0$, and
\begin{eqnarray}
&&\alpha_{_1}=b_{_1}-a_{_1}-1,\;\alpha_{_2}=b_{_1}-a_{_2}-1,\;\alpha_{_3}=-a_{_3},\;\alpha_{_4}=-a_{_4},
\nonumber\\
&&\alpha_{_5}=-a_{_5},\;\alpha_{_{7}}=b_{_2}-1,\;\alpha_{_{8}}=b_{_3}-1,\;\alpha_{_{9}}=b_{_4}-1,
\nonumber\\
&&\alpha_{_{10}}=b_{_5}-1,\;\alpha_{_{11}}=1-b_{_1}.
\label{GKZ21k-16-1}
\end{eqnarray}
The corresponding hypergeometric series is
\begin{eqnarray}
&&\Phi_{_{[\tilde{1}\tilde{3}\tilde{5}7]}}^{(20)}(\alpha,z)=
y_{_1}^{{D\over2}-1}\sum\limits_{n_{_1}=0}^\infty
\sum\limits_{n_{_2}=0}^\infty\sum\limits_{n_{_3}=0}^\infty\sum\limits_{n_{_4}=0}^\infty
c_{_{[\tilde{1}\tilde{3}\tilde{5}7]}}^{(20)}(\alpha,{\bf n})
\nonumber\\
&&\hspace{2.5cm}\times
y_{_1}^{n_{_1}}y_{_3}^{n_{_2}}y_{_4}^{n_{_3}}y_{_2}^{n_{_4}}\;,
\label{GKZ21k-16-2}
\end{eqnarray}
with
\begin{eqnarray}
&&c_{_{[\tilde{1}\tilde{3}\tilde{5}7]}}^{(20)}(\alpha,{\bf n})=
(-)^{n_{_3}}\Big\{n_{_1}!n_{_2}!n_{_4}!\Gamma({D\over2}-2-n_{_1}-n_{_2}-n_{_3}-n_{_4})
\nonumber\\
&&\hspace{2.5cm}\times
\Gamma(D-3-n_{_1}-n_{_2}-n_{_3}-n_{_4})\Gamma({D\over2}-1-n_{_2}-n_{_4})
\nonumber\\
&&\hspace{2.5cm}\times
\Gamma(D-2-n_{_2}-n_{_4})\Gamma(2-{D\over2}+n_{_4})
\Gamma(2-{D\over2}+n_{_2})
\nonumber\\
&&\hspace{2.5cm}\times
\Gamma(3-{D\over2}+n_{_2}+n_{_3}+n_{_4})\Gamma(4-D+n_{_2}+n_{_3}+n_{_4})
\nonumber\\
&&\hspace{2.5cm}\times
\Gamma({D\over2}+n_{_1})\Big\}^{-1}\;.
\label{GKZ21k-16-3}
\end{eqnarray}

\item   $I_{_{21}}=\{1,\cdots,6,9,10,12,13\}$, i.e. the implement $J_{_{21}}=[1,14]\setminus I_{_{21}}=\{7,8,11,14\}$.
The choice implies the power numbers $\alpha_{_7}=\alpha_{_{8}}=\alpha_{_{11}}=\alpha_{_{14}}=0$, and
\begin{eqnarray}
&&\alpha_{_1}=b_{_2}+b_{_3}-a_{_1}-2,\;\alpha_{_2}=b_{_2}+b_{_3}-a_{_2}-2,
\nonumber\\
&&\alpha_{_3}=b_{_2}+b_{_3}-a_{_3}-2,\;\alpha_{_4}=b_{_2}+b_{_3}-a_{_4}-2,\;\alpha_{_5}=-a_{_5},
\nonumber\\
&&\alpha_{_{6}}=b_{_1}-1,\;\alpha_{_{9}}=b_{_4}-b_{_2}-b_{_3}+1,\;\alpha_{_{10}}=b_{_5}-b_{_2}-b_{_3}+1,
\nonumber\\
&&\alpha_{_{12}}=1-b_{_2},\;\alpha_{_{13}}=1-b_{_3}.
\label{GKZ21k-21-1}
\end{eqnarray}
The corresponding hypergeometric solution is written as
\begin{eqnarray}
&&\Phi_{_{[\tilde{1}\tilde{3}\tilde{5}7]}}^{(21)}(\alpha,z)=
y_{_2}^{{D\over2}-1}y_{_3}^{{D\over2}-1}\sum\limits_{n_{_1}=0}^\infty
\sum\limits_{n_{_2}=0}^\infty\sum\limits_{n_{_3}=0}^\infty\sum\limits_{n_{_4}=0}^\infty
c_{_{[\tilde{1}\tilde{3}\tilde{5}7]}}^{(21)}(\alpha,{\bf n})
\nonumber\\
&&\hspace{2.5cm}\times
y_{_1}^{n_{_1}}y_{_3}^{n_{_2}}y_{_4}^{n_{_3}}y_{_2}^{n_{_4}}\;,
\label{GKZ21k-21-2}
\end{eqnarray}
with
\begin{eqnarray}
&&c_{_{[\tilde{1}\tilde{3}\tilde{5}7]}}^{(21)}(\alpha,{\bf n})=
(-)^{1+n_{_1}}\Gamma(2+n_{_1}+n_{_2}+n_{_3}+n_{_4})\Gamma(1+n_{_2}+n_{_4})
\Big\{n_{_1}!n_{_2}!n_{_4}!
\nonumber\\
&&\hspace{2.5cm}\times
\Gamma({D\over2}-2-n_{_1}-n_{_2}-n_{_3}-n_{_4})\Gamma(1-{D\over2}-n_{_2}-n_{_4})
\nonumber\\
&&\hspace{2.5cm}\times
\Gamma(2-{D\over2}+n_{_1})\Gamma(1+{D\over2}+n_{_2}+n_{_3}+n_{_4})
\nonumber\\
&&\hspace{2.5cm}\times
\Gamma(2+n_{_2}+n_{_3}+n_{_4})\Gamma({D\over2}+n_{_4})\Gamma({D\over2}+n_{_2})\Big\}^{-1}\;.
\label{GKZ21k-21-3}
\end{eqnarray}

\item   $I_{_{22}}=\{1,\cdots,7,9,10,13\}$, i.e. the implement $J_{_{22}}=[1,14]\setminus I_{_{22}}=\{8,11,12,14\}$.
The choice implies the power numbers $\alpha_{_8}=\alpha_{_{11}}=\alpha_{_{12}}=\alpha_{_{14}}=0$, and
\begin{eqnarray}
&&\alpha_{_1}=b_{_3}-a_{_1}-1,\;\alpha_{_2}=b_{_3}-a_{_2}-1,\;\alpha_{_3}=b_{_3}-a_{_3}-1,
\nonumber\\
&&\alpha_{_4}=b_{_3}-a_{_4}-1,\;\alpha_{_5}=-a_{_5},\;\alpha_{_{6}}=b_{_1}-1,\;\alpha_{_{7}}=b_{_2}-1,
\nonumber\\
&&\alpha_{_{9}}=b_{_4}-b_{_3},\;\alpha_{_{10}}=b_{_5}-b_{_3},\;\alpha_{_{13}}=1-b_{_3}.
\label{GKZ21k-22-1}
\end{eqnarray}
The corresponding hypergeometric function is written as
\begin{eqnarray}
&&\Phi_{_{[\tilde{1}\tilde{3}\tilde{5}7]}}^{(22)}(\alpha,z)=
y_{_3}^{{D\over2}-1}\sum\limits_{n_{_1}=0}^\infty
\sum\limits_{n_{_2}=0}^\infty\sum\limits_{n_{_3}=0}^\infty\sum\limits_{n_{_4}=0}^\infty
c_{_{[\tilde{1}\tilde{3}\tilde{5}7]}}^{(22)}(\alpha,{\bf n})
\nonumber\\
&&\hspace{2.5cm}\times
y_{_1}^{n_{_1}}y_{_3}^{n_{_2}}y_{_4}^{n_{_3}}y_{_2}^{n_{_4}}\;,
\label{GKZ21k-22-2}
\end{eqnarray}
with
\begin{eqnarray}
&&c_{_{[\tilde{1}\tilde{3}\tilde{5}7]}}^{(22)}(\alpha,{\bf n})=
(-)^{n_{_2}+n_{_3}+n_{_4}}\Gamma(1+n_{_2}+n_{_4})
\Big\{n_{_1}!n_{_2}!n_{_4}!
\nonumber\\
&&\hspace{2.5cm}\times
\Gamma({D\over2}-2-n_{_1}-n_{_2}-n_{_3}-n_{_4})
\Gamma(D-3-n_{_1}-n_{_2}-n_{_3}-n_{_4})
\nonumber\\
&&\hspace{2.5cm}\times
\Gamma({D\over2}-1-n_{_2}-n_{_4})\Gamma(2-{D\over2}+n_{_1})
\Gamma(2-{D\over2}+n_{_4})
\nonumber\\
&&\hspace{2.5cm}\times
\Gamma(2+n_{_2}+n_{_3}+n_{_4})
\Gamma(3-{D\over2}+n_{_2}+n_{_3}+n_{_4})\Gamma({D\over2}+n_{_2})\Big\}^{-1}\;.
\label{GKZ21k-22-3}
\end{eqnarray}

\item   $I_{_{23}}=\{1,\cdots,6,8,9,10,12\}$, i.e. the implement $J_{_{23}}=[1,14]\setminus I_{_{23}}=\{7,11,13,14\}$.
The choice implies the power numbers $\alpha_{_7}=\alpha_{_{11}}=\alpha_{_{13}}=\alpha_{_{14}}=0$, and
\begin{eqnarray}
&&\alpha_{_1}=b_{_2}-a_{_1}-1,\;\alpha_{_2}=b_{_2}-a_{_2}-1,\;\alpha_{_3}=b_{_2}-a_{_3}-1,
\nonumber\\
&&\alpha_{_4}=b_{_2}-a_{_4}-1,\;\alpha_{_5}=-a_{_5},\;\alpha_{_{6}}=b_{_1}-1,\;\alpha_{_{8}}=b_{_3}-1,
\nonumber\\
&&\alpha_{_{9}}=b_{_4}-b_{_2},\;\alpha_{_{10}}=b_{_5}-b_{_2},\;\alpha_{_{12}}=1-b_{_2}.
\label{GKZ21k-31-1}
\end{eqnarray}
The corresponding hypergeometric series is
\begin{eqnarray}
&&\Phi_{_{[\tilde{1}\tilde{3}\tilde{5}7]}}^{(23)}(\alpha,z)=
y_{_2}^{{D\over2}-1}\sum\limits_{n_{_1}=0}^\infty
\sum\limits_{n_{_2}=0}^\infty\sum\limits_{n_{_3}=0}^\infty\sum\limits_{n_{_4}=0}^\infty
c_{_{[\tilde{1}\tilde{3}\tilde{5}7]}}^{(23)}(\alpha,{\bf n})
\nonumber\\
&&\hspace{2.5cm}\times
y_{_1}^{n_{_1}}y_{_3}^{n_{_2}}y_{_4}^{n_{_3}}y_{_2}^{n_{_4}}\;,
\label{GKZ21k-31-2}
\end{eqnarray}
with
\begin{eqnarray}
&&c_{_{[\tilde{1}\tilde{3}\tilde{5}7]}}^{(23)}(\alpha,{\bf n})=
(-)^{n_{_2}+n_{_3}+n_{_4}}\Gamma(1+n_{_2}+n_{_4})
\Big\{n_{_1}!n_{_2}!n_{_4}!
\nonumber\\
&&\hspace{2.5cm}\times
\Gamma({D\over2}-2-n_{_1}-n_{_2}-n_{_3}-n_{_4})
\Gamma(D-3-n_{_1}-n_{_2}-n_{_3}-n_{_4})
\nonumber\\
&&\hspace{2.5cm}\times
\Gamma({D\over2}-1-n_{_2}-n_{_4})\Gamma(2-{D\over2}+n_{_1})
\Gamma(2-{D\over2}+n_{_2})
\nonumber\\
&&\hspace{2.5cm}\times
\Gamma(2+n_{_2}+n_{_3}+n_{_4})
\Gamma(3-{D\over2}+n_{_2}+n_{_3}+n_{_4})\Gamma({D\over2}+n_{_4})\Big\}^{-1}\;.
\label{GKZ21k-31-3}
\end{eqnarray}

\item   $I_{_{24}}=\{1,\cdots,10\}$, i.e. the implement $J_{_{24}}=[1,14]\setminus I_{_{24}}=\{11,12,13,14\}$.
The choice implies the power numbers $\alpha_{_{11}}=\alpha_{_{12}}=\alpha_{_{13}}=\alpha_{_{14}}=0$, and
\begin{eqnarray}
&&\alpha_{_1}=-a_{_1},\;\alpha_{_2}=-a_{_2},\;\alpha_{_3}=-a_{_3},\;\alpha_{_4}=-a_{_4},\;\alpha_{_5}=-a_{_5},
\nonumber\\
&&\alpha_{_{6}}=b_{_1}-1,\;\alpha_{_{7}}=b_{_2}-1,\;\alpha_{_{8}}=b_{_3}-1,\;\alpha_{_{9}}=b_{_4}-1,\;
\alpha_{_{10}}=b_{_5}-1.
\label{GKZ21k-32-1}
\end{eqnarray}
The corresponding hypergeometric solution is written as
\begin{eqnarray}
&&\Phi_{_{[\tilde{1}\tilde{3}\tilde{5}7]}}^{(24)}(\alpha,z)=
\sum\limits_{n_{_1}=0}^\infty
\sum\limits_{n_{_2}=0}^\infty\sum\limits_{n_{_3}=0}^\infty\sum\limits_{n_{_4}=0}^\infty
c_{_{[\tilde{1}\tilde{3}\tilde{5}7]}}^{(24)}(\alpha,{\bf n})
\nonumber\\
&&\hspace{2.5cm}\times
y_{_1}^{n_{_1}}y_{_3}^{n_{_2}}y_{_4}^{n_{_3}}y_{_2}^{n_{_4}}\;,
\label{GKZ21k-32-2}
\end{eqnarray}
with
\begin{eqnarray}
&&c_{_{[\tilde{1}\tilde{3}\tilde{5}7]}}^{(24)}(\alpha,{\bf n})=
(-)^{n_{_3}}\Big\{n_{_1}!n_{_2}!n_{_4}!\Gamma(D-3-n_{_1}-n_{_2}-n_{_3}-n_{_4})
\nonumber\\
&&\hspace{2.5cm}\times
\Gamma({3D\over2}-4-n_{_1}-n_{_2}-n_{_3}-n_{_4})
\Gamma({D\over2}-1-n_{_2}-n_{_4})
\nonumber\\
&&\hspace{2.5cm}\times
\Gamma(D-2-n_{_2}-n_{_4})\Gamma(2-{D\over2}+n_{_1})
\Gamma(2-{D\over2}+n_{_4})\Gamma(2-{D\over2}+n_{_2})
\nonumber\\
&&\hspace{2.5cm}\times
\Gamma(3-{D\over2}+n_{_2}+n_{_3}+n_{_4})
\Gamma(4-D+n_{_2}+n_{_3}+n_{_4})\Big\}^{-1}\;.
\label{GKZ21k-32-3}
\end{eqnarray}

\item   $I_{_{25}}=\{1,\cdots,5,8,10,\cdots,13\}$, i.e. the implement $J_{_{25}}=[1,14]\setminus I_{_{25}}=\{6,7,9,14\}$.
The choice implies the power numbers $\alpha_{_6}=\alpha_{_{7}}=\alpha_{_{9}}=\alpha_{_{14}}=0$, and
\begin{eqnarray}
&&\alpha_{_1}=b_{_1}+b_{_4}-a_{_1}-2,\;\alpha_{_2}=b_{_1}+b_{_4}-a_{_2}-2,
\nonumber\\
&&\alpha_{_3}=b_{_4}-a_{_3}-1,\;\alpha_{_4}=b_{_4}-a_{_4}-1,\;\alpha_{_5}=-a_{_5},
\nonumber\\
&&\alpha_{_{8}}=b_{_2}+b_{_3}-b_{_4}-1,\;\alpha_{_{10}}=b_{_5}-b_{_4},\;\alpha_{_{11}}=1-b_{_1},
\nonumber\\
&&\alpha_{_{12}}=1-b_{_2},\;\alpha_{_{13}}=b_{_2}-b_{_4}.
\label{GKZ21k-7-1}
\end{eqnarray}
The corresponding hypergeometric series solutions are written as
\begin{eqnarray}
&&\Phi_{_{[\tilde{1}\tilde{3}\tilde{5}7]}}^{(25),a}(\alpha,z)=
y_{_1}^{{D\over2}-1}y_{_2}^{{D\over2}-1}y_{_3}^{-1}\sum\limits_{n_{_1}=0}^\infty
\sum\limits_{n_{_2}=0}^\infty\sum\limits_{n_{_3}=0}^\infty\sum\limits_{n_{_4}=0}^\infty
c_{_{[\tilde{1}\tilde{3}\tilde{5}7]}}^{(25),a}(\alpha,{\bf n})
\nonumber\\
&&\hspace{2.5cm}\times
y_{_1}^{n_{_1}}y_{_4}^{n_{_2}}\Big({y_{_4}\over y_{_3}}\Big)^{n_{_3}}
\Big({y_{_2}\over y_{_3}}\Big)^{n_{_4}}
\;,\nonumber\\
&&\Phi_{_{[\tilde{1}\tilde{3}\tilde{5}7]}}^{(25),b}(\alpha,z)=
y_{_1}^{{D\over2}-1}y_{_2}^{{D\over2}-1}\sum\limits_{n_{_1}=0}^\infty
\sum\limits_{n_{_2}=0}^\infty\sum\limits_{n_{_3}=0}^\infty\sum\limits_{n_{_4}=0}^\infty
c_{_{[\tilde{1}\tilde{3}\tilde{5}7]}}^{(25),b}(\alpha,{\bf n})
\nonumber\\
&&\hspace{2.5cm}\times
y_{_1}^{n_{_1}}y_{_3}^{n_{_2}}y_{_4}^{n_{_3}}y_{_2}^{n_{_4}}
\;,\nonumber\\
&&\Phi_{_{[\tilde{1}\tilde{3}\tilde{5}7]}}^{(25),c}(\alpha,z)=
y_{_1}^{{D\over2}-1}y_{_2}^{{D\over2}}y_{_3}^{-1}\sum\limits_{n_{_1}=0}^\infty
\sum\limits_{n_{_2}=0}^\infty\sum\limits_{n_{_3}=0}^\infty\sum\limits_{n_{_4}=0}^\infty
c_{_{[\tilde{1}\tilde{3}\tilde{5}7]}}^{(25),c}(\alpha,{\bf n})
\nonumber\\
&&\hspace{2.5cm}\times
y_{_1}^{n_{_1}}y_{_2}^{n_{_2}}y_{_4}^{n_{_3}}\Big({y_{_2}\over y_{_3}}\Big)^{n_{_4}}\;.
\label{GKZ21k-7-2a}
\end{eqnarray}
Where the coefficients are
\begin{eqnarray}
&&c_{_{[\tilde{1}\tilde{3}\tilde{5}7]}}^{(25),a}(\alpha,{\bf n})=
(-)^{n_{_1}+n_{_4}}\Gamma(1+n_{_1}+n_{_2})\Gamma(1+n_{_3}+n_{_4})\Big\{n_{_1}!n_{_2}!n_{_3}!n_{_4}!
\nonumber\\
&&\hspace{2.5cm}\times
\Gamma({D\over2}-1-n_{_1}-n_{_2})\Gamma({D\over2}+n_{_3})\Gamma(1-{D\over2}-n_{_3}-n_{_4})
\nonumber\\
&&\hspace{2.5cm}\times
\Gamma(2-{D\over2}+n_{_2})\Gamma({D\over2}+n_{_1})\Gamma({D\over2}+n_{_4})\Big\}^{-1}
\;,\nonumber\\
&&c_{_{[\tilde{1}\tilde{3}\tilde{5}7]}}^{(25),b}(\alpha,{\bf n})=
(-)^{1+n_{_1}}\Gamma(2+n_{_1}+n_{_2}+n_{_3}+n_{_4})\Gamma(1+n_{_2}+n_{_4})
\nonumber\\
&&\hspace{2.5cm}\times
\Big\{n_{_1}!n_{_2}!n_{_4}!\Gamma(2+n_{_2}+n_{_3}+n_{_4})
\Gamma({D\over2}-1-n_{_2}-n_{_4})
\nonumber\\
&&\hspace{2.5cm}\times
\Gamma({D\over2}-2-n_{_1}-n_{_2}-n_{_3}-n_{_4})\Gamma(2-{D\over2}+n_{_2})
\nonumber\\
&&\hspace{2.5cm}\times
\Gamma(3-{D\over2}+n_{_2}+n_{_3}+n_{_4})\Gamma({D\over2}+n_{_1})
\Gamma({D\over2}+n_{_4})\Big\}^{-1}
\;,\nonumber\\
&&c_{_{[\tilde{1}\tilde{3}\tilde{5}7]}}^{(25),c}(\alpha,{\bf n})=
(-)^{1+n_{_1}+n_{_4}}\Gamma(2+n_{_1}+n_{_2}+n_{_3})\Gamma(1+n_{_2})\Gamma(1+n_{_4})
\nonumber\\
&&\hspace{2.5cm}\times
\Big\{n_{_1}!\Gamma(2+n_{_2}+n_{_3})\Gamma(2+n_{_2}+n_{_4})
\nonumber\\
&&\hspace{2.5cm}\times
\Gamma({D\over2}-2-n_{_1}-n_{_2}-n_{_3})\Gamma({D\over2}-1-n_{_2})
\Gamma(1-{D\over2}-n_{_4})
\nonumber\\
&&\hspace{2.5cm}\times
\Gamma(3-{D\over2}+n_{_2}+n_{_3})\Gamma({D\over2}+n_{_1})
\Gamma({D\over2}+1+n_{_2}+n_{_4})\Big\}^{-1}\;.
\label{GKZ21k-7-3}
\end{eqnarray}

\item   $I_{_{26}}=\{1,\cdots,5,7,8,10,11,13\}$, i.e. the implement $J_{_{26}}=[1,14]\setminus I_{_{26}}=\{6,9,12,14\}$.
The choice implies the power numbers $\alpha_{_6}=\alpha_{_{9}}=\alpha_{_{12}}=\alpha_{_{14}}=0$, and
\begin{eqnarray}
&&\alpha_{_1}=b_{_1}+b_{_4}-a_{_1}-2,\;\alpha_{_2}=b_{_1}+b_{_4}-a_{_2}-2,
\nonumber\\
&&\alpha_{_3}=b_{_4}-a_{_3}-1,\;\alpha_{_4}=b_{_4}-a_{_4}-1,\;\alpha_{_5}=-a_{_5},
\nonumber\\
&&\alpha_{_{7}}=b_{_2}-1,\;\alpha_{_{8}}=b_{_3}-b_{_4},\;\alpha_{_{10}}=b_{_5}-b_{_4},\;\alpha_{_{11}}=1-b_{_1},
\nonumber\\
&&\alpha_{_{13}}=1-b_{_4}.
\label{GKZ21k-8-1}
\end{eqnarray}
The corresponding hypergeometric solutions are
\begin{eqnarray}
&&\Phi_{_{[\tilde{1}\tilde{3}\tilde{5}7]}}^{(26),a}(\alpha,z)=
y_{_1}^{{D\over2}-1}y_{_3}^{{D\over2}-2}\sum\limits_{n_{_1}=0}^\infty
\sum\limits_{n_{_2}=0}^\infty\sum\limits_{n_{_3}=0}^\infty\sum\limits_{n_{_4}=0}^\infty
c_{_{[\tilde{1}\tilde{3}\tilde{5}7]}}^{(26),a}(\alpha,{\bf n})
\nonumber\\
&&\hspace{2.5cm}\times
y_{_1}^{n_{_1}}y_{_4}^{n_{_2}}\Big({y_{_4}\over y_{_3}}\Big)^{n_{_3}}
\Big({y_{_2}\over y_{_3}}\Big)^{n_{_4}}
\;,\nonumber\\
&&\Phi_{_{[\tilde{1}\tilde{3}\tilde{5}7]}}^{(26),b}(\alpha,z)=
y_{_1}^{{D\over2}-1}y_{_3}^{{D\over2}-1}\sum\limits_{n_{_1}=0}^\infty
\sum\limits_{n_{_2}=0}^\infty\sum\limits_{n_{_3}=0}^\infty\sum\limits_{n_{_4}=0}^\infty
c_{_{[\tilde{1}\tilde{3}\tilde{5}7]}}^{(26),b}(\alpha,{\bf n})
\nonumber\\
&&\hspace{2.5cm}\times
y_{_1}^{n_{_1}}y_{_3}^{n_{_2}}y_{_4}^{n_{_3}}y_{_2}^{n_{_4}}
\;,\nonumber\\
&&\Phi_{_{[\tilde{1}\tilde{3}\tilde{5}7]}}^{(26),c}(\alpha,z)=
y_{_1}^{{D\over2}-1}y_{_2}y_{_3}^{{D\over2}-2}\sum\limits_{n_{_1}=0}^\infty
\sum\limits_{n_{_2}=0}^\infty\sum\limits_{n_{_3}=0}^\infty\sum\limits_{n_{_4}=0}^\infty
c_{_{[\tilde{1}\tilde{3}\tilde{5}7]}}^{(26),c}(\alpha,{\bf n})
\nonumber\\
&&\hspace{2.5cm}\times
y_{_1}^{n_{_1}}y_{_2}^{n_{_2}}y_{_4}^{n_{_3}}\Big({y_{_2}\over y_{_3}}\Big)^{n_{_4}}\;.
\label{GKZ21k-8-2a}
\end{eqnarray}
Where the coefficients are
\begin{eqnarray}
&&c_{_{[\tilde{1}\tilde{3}\tilde{5}7]}}^{(26),a}(\alpha,{\bf n})=
(-)^{n_{_1}+n_{_4}}\Gamma(1+n_{_1}+n_{_2})\Gamma(1+n_{_3}+n_{_4})\Big\{n_{_1}!n_{_2}!n_{_3}!n_{_4}!
\nonumber\\
&&\hspace{2.5cm}\times
\Gamma({D\over2}-1-n_{_1}-n_{_2})\Gamma({D\over2}+n_{_3})\Gamma({D\over2}-1-n_{_3}-n_{_4})
\nonumber\\
&&\hspace{2.5cm}\times
\Gamma(2-{D\over2}+n_{_2})\Gamma({D\over2}+n_{_1})\Gamma(2-{D\over2}+n_{_4})\Big\}^{-1}
\;,\nonumber\\
&&c_{_{[\tilde{1}\tilde{3}\tilde{5}7]}}^{(26),b}(\alpha,{\bf n})=
(-)^{1+n_{_1}}\Gamma(2+n_{_1}+n_{_2}+n_{_3}+n_{_4})\Gamma(1+n_{_2}+n_{_4})
\nonumber\\
&&\hspace{2.5cm}\times
\Big\{n_{_1}!n_{_2}!n_{_4}!\Gamma(2+n_{_2}+n_{_3}+n_{_4})
\Gamma({D\over2}-1-n_{_2}-n_{_4})
\nonumber\\
&&\hspace{2.5cm}\times
\Gamma({D\over2}-2-n_{_1}-n_{_2}-n_{_3}-n_{_4})\Gamma(2-{D\over2}+n_{_4})
\nonumber\\
&&\hspace{2.5cm}\times
\Gamma(3-{D\over2}+n_{_2}+n_{_3}+n_{_4})\Gamma({D\over2}+n_{_1})
\Gamma({D\over2}+n_{_2})\Big\}^{-1}
\;,\nonumber\\
&&c_{_{[\tilde{1}\tilde{3}\tilde{5}7]}}^{(26),c}(\alpha,{\bf n})=
(-)^{1+n_{_1}+n_{_4}}\Gamma(2+n_{_1}+n_{_2}+n_{_3})\Gamma(1+n_{_2})\Gamma(1+n_{_4})
\nonumber\\
&&\hspace{2.5cm}\times
\Big\{n_{_1}!\Gamma(2+n_{_2}+n_{_3})\Gamma(2+n_{_2}+n_{_4})
\nonumber\\
&&\hspace{2.5cm}\times
\Gamma({D\over2}-2-n_{_1}-n_{_2}-n_{_3})\Gamma({D\over2}-1-n_{_2})
\Gamma({D\over2}-1-n_{_4})
\nonumber\\
&&\hspace{2.5cm}\times
\Gamma(3-{D\over2}+n_{_2}+n_{_3})\Gamma({D\over2}+n_{_1})
\Gamma(3-{D\over2}+n_{_2}+n_{_4})\Big\}^{-1}\;.
\label{GKZ21k-8-3}
\end{eqnarray}

\item   $I_{_{27}}=\{1,\cdots,5,8,9,11,12,13\}$, i.e. the implement $J_{_{27}}=[1,14]\setminus I_{_{27}}=\{6,7,10,14\}$.
The choice implies the power numbers $\alpha_{_6}=\alpha_{_{7}}=\alpha_{_{10}}=\alpha_{_{14}}=0$, and
\begin{eqnarray}
&&\alpha_{_1}=b_{_1}+b_{_5}-a_{_1}-2,\;\alpha_{_2}=b_{_1}+b_{_5}-a_{_2}-2,
\nonumber\\
&&\alpha_{_3}=b_{_5}-a_{_3}-1,\;\alpha_{_4}=b_{_5}-a_{_4}-1,\;\alpha_{_5}=-a_{_5},
\nonumber\\
&&\alpha_{_{8}}=b_{_2}+b_{_3}-b_{_5}-1,\;\alpha_{_{9}}=b_{_4}-b_{_5},\;\alpha_{_{11}}=1-b_{_1},
\nonumber\\
&&\alpha_{_{12}}=1-b_{_2},\;\alpha_{_{13}}=b_{_2}-b_{_5}.
\label{GKZ21k-9-1}
\end{eqnarray}
The corresponding hypergeometric functions are
\begin{eqnarray}
&&\Phi_{_{[\tilde{1}\tilde{3}\tilde{5}7]}}^{(27),a}(\alpha,z)=
y_{_1}^{{D\over2}-1}y_{_2}^{{D\over2}-1}y_{_3}^{{D\over2}-2}\sum\limits_{n_{_1}=0}^\infty
\sum\limits_{n_{_2}=0}^\infty\sum\limits_{n_{_3}=0}^\infty\sum\limits_{n_{_4}=0}^\infty
c_{_{[\tilde{1}\tilde{3}\tilde{5}7]}}^{(27),a}(\alpha,{\bf n})
\nonumber\\
&&\hspace{2.5cm}\times
y_{_1}^{n_{_1}}y_{_4}^{n_{_2}}\Big({y_{_4}\over y_{_3}}\Big)^{n_{_3}}
\Big({y_{_2}\over y_{_3}}\Big)^{n_{_4}}
\;,\nonumber\\
&&\Phi_{_{[\tilde{1}\tilde{3}\tilde{5}7]}}^{(27),b}(\alpha,z)=
y_{_1}^{{D\over2}-1}y_{_2}^{{D\over2}-1}y_{_3}^{{D\over2}-1}\sum\limits_{n_{_1}=0}^\infty
\sum\limits_{n_{_2}=0}^\infty\sum\limits_{n_{_3}=0}^\infty\sum\limits_{n_{_4}=0}^\infty
c_{_{[\tilde{1}\tilde{3}\tilde{5}7]}}^{(27),b}(\alpha,{\bf n})
\nonumber\\
&&\hspace{2.5cm}\times
y_{_1}^{n_{_1}}y_{_3}^{n_{_2}}y_{_4}^{n_{_3}}y_{_2}^{n_{_4}}
\;,\nonumber\\
&&\Phi_{_{[\tilde{1}\tilde{3}\tilde{5}7]}}^{(27),c}(\alpha,z)=
y_{_1}^{{D\over2}-1}y_{_2}^{{D\over2}}y_{_3}^{{D\over2}-2}\sum\limits_{n_{_1}=0}^\infty
\sum\limits_{n_{_2}=0}^\infty\sum\limits_{n_{_3}=0}^\infty\sum\limits_{n_{_4}=0}^\infty
c_{_{[\tilde{1}\tilde{3}\tilde{5}7]}}^{(27),c}(\alpha,{\bf n})
\nonumber\\
&&\hspace{2.5cm}\times
y_{_1}^{n_{_1}}y_{_2}^{n_{_2}}y_{_4}^{n_{_3}}\Big({y_{_2}\over y_{_3}}\Big)^{n_{_4}}\;.
\label{GKZ21k-9-2a}
\end{eqnarray}
Where the coefficients are
\begin{eqnarray}
&&c_{_{[\tilde{1}\tilde{3}\tilde{5}7]}}^{(27),a}(\alpha,{\bf n})=
(-)^{n_{_1}+n_{_4}}\Gamma(1+n_{_1}+n_{_2})\Gamma(1+n_{_3}+n_{_4})
\nonumber\\
&&\hspace{2.5cm}\times
\Big\{n_{_1}!n_{_2}!n_{_3}!n_{_4}!\Gamma(1-{D\over2}-n_{_1}-n_{_2})\Gamma(2-{D\over2}+n_{_3})
\nonumber\\
&&\hspace{2.5cm}\times
\Gamma({D\over2}+n_{_2})\Gamma({D\over2}+n_{_1})
\Gamma({D\over2}+n_{_4})\Gamma({D\over2}-1-n_{_3}-n_{_4})\Big\}^{-1}
\;,\nonumber\\
&&c_{_{[\tilde{1}\tilde{3}\tilde{5}7]}}^{(27),b}(\alpha,{\bf n})=
(-)^{1+n_{_1}}\Gamma(2+n_{_1}+n_{_2}+n_{_3}+n_{_4})\Gamma(1+n_{_2}+n_{_4})
\nonumber\\
&&\hspace{2.5cm}\times
\Big\{n_{_1}!n_{_2}!n_{_4}!\Gamma(2+n_{_2}+n_{_3}+n_{_4})
\Gamma(-{D\over2}-n_{_1}-n_{_2}-n_{_3}-n_{_4})
\nonumber\\
&&\hspace{2.5cm}\times
\Gamma(1-{D\over2}-n_{_2}-n_{_4})\Gamma({D\over2}+1+n_{_2}+n_{_3}+n_{_4})
\nonumber\\
&&\hspace{2.5cm}\times
\Gamma({D\over2}+n_{_1})\Gamma({D\over2}+n_{_4})\Gamma({D\over2}+n_{_2})\Big\}^{-1}
\;,\nonumber\\
&&c_{_{[\tilde{1}\tilde{3}\tilde{5}7]}}^{(27),c}(\alpha,{\bf n})=
(-)^{1+n_{_1}+n_{_4}}\Gamma(2+n_{_1}+n_{_2}+n_{_3})\Gamma(1+n_{_2})\Gamma(1+n_{_4})
\nonumber\\
&&\hspace{2.5cm}\times
\Big\{n_{_1}!\Gamma(2+n_{_2}+n_{_3})\Gamma(2+n_{_2}+n_{_4})\Gamma(1-{D\over2}-n_{_2})
\nonumber\\
&&\hspace{2.5cm}\times
\Gamma(-{D\over2}-n_{_1}-n_{_2}-n_{_3})\Gamma({D\over2}+1+n_{_2}+n_{_3})
\nonumber\\
&&\hspace{2.5cm}\times
\Gamma({D\over2}+n_{_1})
\Gamma({D\over2}+1+n_{_2}+n_{_4})\Gamma({D\over2}-1-n_{_4})\Big\}^{-1}\;.
\label{GKZ21k-9-3}
\end{eqnarray}

\item   $I_{_{28}}=\{1,\cdots,5,7,8,9,11,13\}$, i.e. the implement $J_{_{28}}=[1,14]\setminus I_{_{28}}=\{6,10,12,14\}$.
The choice implies the power numbers $\alpha_{_6}=\alpha_{_{10}}=\alpha_{_{12}}=\alpha_{_{14}}=0$, and
\begin{eqnarray}
&&\alpha_{_1}=b_{_1}+b_{_5}-a_{_1}-2,\;\alpha_{_2}=b_{_1}+b_{_5}-a_{_2}-2,
\nonumber\\
&&\alpha_{_3}=b_{_5}-a_{_3}-1,\;\alpha_{_4}=b_{_5}-a_{_4}-1,\;\alpha_{_5}=-a_{_5},
\nonumber\\
&&\alpha_{_{7}}=b_{_2}-1,\;\alpha_{_{8}}=b_{_3}-b_{_5},\;\alpha_{_{9}}=b_{_4}-b_{_5},\;\alpha_{_{11}}=1-b_{_1},
\nonumber\\
&&\alpha_{_{13}}=1-b_{_5}.
\label{GKZ21k-10-1}
\end{eqnarray}
The corresponding hypergeometric series solutions are written as
\begin{eqnarray}
&&\Phi_{_{[\tilde{1}\tilde{3}\tilde{5}7]}}^{(28),a}(\alpha,z)=
y_{_1}^{{D\over2}-1}y_{_3}^{{D}-3}\sum\limits_{n_{_1}=0}^\infty
\sum\limits_{n_{_2}=0}^\infty\sum\limits_{n_{_3}=0}^\infty\sum\limits_{n_{_4}=0}^\infty
c_{_{[\tilde{1}\tilde{3}\tilde{5}7]}}^{(28),a}(\alpha,{\bf n})
\nonumber\\
&&\hspace{2.5cm}\times
y_{_1}^{n_{_1}}y_{_4}^{n_{_2}}\Big({y_{_4}\over y_{_3}}\Big)^{n_{_3}}
\Big({y_{_2}\over y_{_3}}\Big)^{n_{_4}}
\;,\nonumber\\
&&\Phi_{_{[\tilde{1}\tilde{3}\tilde{5}7]}}^{(28),b}(\alpha,z)=
y_{_1}^{{D\over2}-1}y_{_3}^{{D}-2}\sum\limits_{n_{_1}=0}^\infty
\sum\limits_{n_{_2}=0}^\infty\sum\limits_{n_{_3}=0}^\infty\sum\limits_{n_{_4}=0}^\infty
c_{_{[\tilde{1}\tilde{3}\tilde{5}7]}}^{(28),b}(\alpha,{\bf n})
\nonumber\\
&&\hspace{2.5cm}\times
y_{_1}^{n_{_1}}y_{_3}^{n_{_2}}y_{_4}^{n_{_3}}\Big({y_{_2}\over y_{_3}}\Big)^{n_{_4}}\;.
\label{GKZ21k-10-2a}
\end{eqnarray}
Where the coefficients are
\begin{eqnarray}
&&c_{_{[\tilde{1}\tilde{3}\tilde{5}7]}}^{(28),a}(\alpha,{\bf n})=
(-)^{n_{_1}+n_{_3}}\Gamma(1+n_{_1}+n_{_2})\Big\{n_{_1}!n_{_2}!n_{_3}!n_{_4}!
\Gamma(1-{D\over2}-n_{_1}-n_{_2})
\nonumber\\
&&\hspace{2.5cm}\times
\Gamma(2-{D\over2}+n_{_3})\Gamma(2-{D\over2}+n_{_4})\Gamma({D\over2}-1-n_{_3}-n_{_4})
\nonumber\\
&&\hspace{2.5cm}\times
\Gamma({D\over2}+n_{_2})\Gamma({D\over2}+n_{_1})\Gamma(D-2-n_{_3}-n_{_4})\Big\}^{-1}
\;,\nonumber\\
&&c_{_{[\tilde{1}\tilde{3}\tilde{5}7]}}^{(28),b}(\alpha,{\bf n})=
(-)^{1+n_{_1}}\Gamma(2+n_{_1}+n_{_2}+n_{_3})\Gamma(1+n_{_2})\Big\{n_{_1}!n_{_4}!
\nonumber\\
&&\hspace{2.5cm}\times
\Gamma(2+n_{_2}+n_{_3})\Gamma(-{D\over2}-n_{_1}-n_{_2}-n_{_3})
\Gamma(1-{D\over2}-n_{_2})
\nonumber\\
&&\hspace{2.5cm}\times
\Gamma(2-{D\over2}+n_{_4})
\Gamma({D\over2}+n_{_2}-n_{_4})\Gamma({D\over2}+1+n_{_2}+n_{_3})
\nonumber\\
&&\hspace{2.5cm}\times
\Gamma({D\over2}+n_{_1})\Gamma(D-1+n_{_2}-n_{_4})\Big\}^{-1}\;.
\label{GKZ21k-10-3}
\end{eqnarray}

\item   $I_{_{29}}=\{1,\cdots,6,8,10,12,13\}$, i.e. the implement $J_{_{29}}=[1,14]\setminus I_{_{29}}=\{7,9,11,14\}$.
The choice implies the power numbers $\alpha_{_7}=\alpha_{_{9}}=\alpha_{_{11}}=\alpha_{_{14}}=0$, and
\begin{eqnarray}
&&\alpha_{_1}=b_{_4}-a_{_1}-1,\;\alpha_{_2}=b_{_4}-a_{_2}-1,\;\alpha_{_3}=b_{_4}-a_{_3}-1,
\nonumber\\
&&\alpha_{_4}=b_{_4}-a_{_4}-1,\;\alpha_{_5}=-a_{_5},\;\alpha_{_{6}}=b_{_1}-1,\;\alpha_{_{8}}=b_{_2}+b_{_3}-b_{_4}-1,
\nonumber\\
&&\alpha_{_{10}}=b_{_5}-b_{_4},\;\alpha_{_{12}}=1-b_{_2},\;\alpha_{_{13}}=b_{_2}-b_{_4}.
\label{GKZ21k-23-1}
\end{eqnarray}
The corresponding hypergeometric solutions are written as
\begin{eqnarray}
&&\Phi_{_{[\tilde{1}\tilde{3}\tilde{5}7]}}^{(29),a}(\alpha,z)=
y_{_2}^{{D\over2}-1}y_{_3}^{-1}\sum\limits_{n_{_1}=0}^\infty
\sum\limits_{n_{_2}=0}^\infty\sum\limits_{n_{_3}=0}^\infty\sum\limits_{n_{_4}=0}^\infty
c_{_{[\tilde{1}\tilde{3}\tilde{5}7]}}^{(29),a}(\alpha,{\bf n})
\nonumber\\
&&\hspace{2.5cm}\times
y_{_1}^{n_{_1}}y_{_4}^{n_{_2}}\Big({y_{_4}\over y_{_3}}\Big)^{n_{_3}}
\Big({y_{_2}\over y_{_3}}\Big)^{n_{_4}}
\;,\nonumber\\
&&\Phi_{_{[\tilde{1}\tilde{3}\tilde{5}7]}}^{(29),b}(\alpha,z)=
y_{_2}^{{D\over2}-1}\sum\limits_{n_{_1}=0}^\infty
\sum\limits_{n_{_2}=0}^\infty\sum\limits_{n_{_3}=0}^\infty\sum\limits_{n_{_4}=0}^\infty
c_{_{[\tilde{1}\tilde{3}\tilde{5}7]}}^{(29),b}(\alpha,{\bf n})
\nonumber\\
&&\hspace{2.5cm}\times
y_{_1}^{n_{_1}}y_{_3}^{n_{_2}}y_{_4}^{n_{_3}}y_{_2}^{n_{_4}}
\;,\nonumber\\
&&\Phi_{_{[\tilde{1}\tilde{3}\tilde{5}7]}}^{(29),c}(\alpha,z)=
y_{_2}^{{D\over2}}y_{_3}^{-1}\sum\limits_{n_{_1}=0}^\infty
\sum\limits_{n_{_2}=0}^\infty\sum\limits_{n_{_3}=0}^\infty\sum\limits_{n_{_4}=0}^\infty
c_{_{[\tilde{1}\tilde{3}\tilde{5}7]}}^{(29),c}(\alpha,{\bf n})
\nonumber\\
&&\hspace{2.5cm}\times
y_{_1}^{n_{_1}}y_{_2}^{n_{_2}}y_{_4}^{n_{_3}}\Big({y_{_2}\over y_{_3}}\Big)^{n_{_4}}\;.
\label{GKZ21k-23-2a}
\end{eqnarray}
Where the coefficients are
\begin{eqnarray}
&&c_{_{[\tilde{1}\tilde{3}\tilde{5}7]}}^{(29),a}(\alpha,{\bf n})=
(-)^{n_{_2}+n_{_4}}\Gamma(1+n_{_3}+n_{_4})
\Big\{n_{_1}!n_{_2}!n_{_3}!n_{_4}!\Gamma({D\over2}-1-n_{_1}-n_{_2})
\nonumber\\
&&\hspace{2.5cm}\times
\Gamma(D-2-n_{_1}-n_{_2})\Gamma({D\over2}+n_{_3})
\Gamma(2-{D\over2}+n_{_1})
\nonumber\\
&&\hspace{2.5cm}\times
\Gamma(1-{D\over2}-n_{_3}-n_{_4})\Gamma(2-{D\over2}+n_{_2})
\Gamma({D\over2}+n_{_4})\Big\}^{-1}
\;,\nonumber\\
&&c_{_{[\tilde{1}\tilde{3}\tilde{5}7]}}^{(29),b}(\alpha,{\bf n})=
(-)^{1+n_{_2}+n_{_3}+n_{_4}}\Gamma(1+n_{_2}+n_{_4})
\Big\{n_{_1}!n_{_2}!n_{_4}!\Gamma(2+n_{_2}+n_{_3}+n_{_4})
\nonumber\\
&&\hspace{2.5cm}\times
\Gamma({D\over2}-2-n_{_1}-n_{_2}-n_{_3}-n_{_4})\Gamma(D-3-n_{_1}-n_{_2}-n_{_3}-n_{_4})
\nonumber\\
&&\hspace{2.5cm}\times
\Gamma({D\over2}-1-n_{_2}-n_{_4})\Gamma(2-{D\over2}+n_{_1})
\Gamma(3-{D\over2}+n_{_2}+n_{_3}+n_{_4})
\nonumber\\
&&\hspace{2.5cm}\times
\Gamma(2-{D\over2}+n_{_2})\Gamma({D\over2}+n_{_4})\Big\}^{-1}
\;,\nonumber\\
&&c_{_{[\tilde{1}\tilde{3}\tilde{5}7]}}^{(29),c}(\alpha,{\bf n})=
(-)^{1+n_{_2}+n_{_3}+n_{_4}}\Gamma(1+n_{_2})\Gamma(1+n_{_4})
\Big\{n_{_1}!\Gamma(2+n_{_2}+n_{_4})
\nonumber\\
&&\hspace{2.5cm}\times
\Gamma(2+n_{_2}+n_{_3})\Gamma({D\over2}-2-n_{_1}-n_{_2}-n_{_3})
\nonumber\\
&&\hspace{2.5cm}\times
\Gamma(D-3-n_{_1}-n_{_2}-n_{_3})
\Gamma({D\over2}-1-n_{_2})\Gamma(2-{D\over2}+n_{_1})
\nonumber\\
&&\hspace{2.5cm}\times
\Gamma(1-{D\over2}-n_{_4})
\Gamma(3-{D\over2}+n_{_2}+n_{_3})\Gamma({D\over2}+1+n_{_2}+n_{_4})\Big\}^{-1}\;.
\label{GKZ21k-23-3}
\end{eqnarray}

\item   $I_{_{30}}=\{1,\cdots,8,10,13\}$, i.e. the implement $J_{_{30}}=[1,14]\setminus I_{_{30}}=\{9,11,12,14\}$.
The choice implies the power numbers $\alpha_{_9}=\alpha_{_{11}}=\alpha_{_{12}}=\alpha_{_{14}}=0$, and
\begin{eqnarray}
&&\alpha_{_1}=b_{_4}-a_{_1}-1,\;\alpha_{_2}=b_{_4}-a_{_2}-1,\;\alpha_{_3}=b_{_4}-a_{_3}-1,
\nonumber\\
&&\alpha_{_4}=b_{_4}-a_{_4}-1,\;\alpha_{_5}=-a_{_5},\;\alpha_{_{6}}=b_{_1}-1,\;\alpha_{_{7}}=b_{_2}-1,
\nonumber\\
&&\alpha_{_{8}}=b_{_3}-b_{_4},\;\alpha_{_{10}}=b_{_5}-b_{_4},\;\alpha_{_{13}}=1-b_{_4}.
\label{GKZ21k-24-1}
\end{eqnarray}
The corresponding hypergeometric solutions are written as
\begin{eqnarray}
&&\Phi_{_{[\tilde{1}\tilde{3}\tilde{5}7]}}^{(30),a}(\alpha,z)=
y_{_3}^{{D\over2}-2}\sum\limits_{n_{_1}=0}^\infty
\sum\limits_{n_{_2}=0}^\infty\sum\limits_{n_{_3}=0}^\infty\sum\limits_{n_{_4}=0}^\infty
c_{_{[\tilde{1}\tilde{3}\tilde{5}7]}}^{(30),a}(\alpha,{\bf n})
\nonumber\\
&&\hspace{2.5cm}\times
y_{_1}^{n_{_1}}y_{_4}^{n_{_2}}\Big({y_{_4}\over y_{_3}}\Big)^{n_{_3}}
\Big({y_{_2}\over y_{_3}}\Big)^{n_{_4}}
\;,\nonumber\\
&&\Phi_{_{[\tilde{1}\tilde{3}\tilde{5}7]}}^{(30),b}(\alpha,z)=
y_{_3}^{{D\over2}-1}\sum\limits_{n_{_1}=0}^\infty
\sum\limits_{n_{_2}=0}^\infty\sum\limits_{n_{_3}=0}^\infty\sum\limits_{n_{_4}=0}^\infty
c_{_{[\tilde{1}\tilde{3}\tilde{5}7]}}^{(30),b}(\alpha,{\bf n})
\nonumber\\
&&\hspace{2.5cm}\times
y_{_1}^{n_{_1}}y_{_3}^{n_{_2}}y_{_4}^{n_{_3}}y_{_2}^{n_{_4}}
\;,\nonumber\\
&&\Phi_{_{[\tilde{1}\tilde{3}\tilde{5}7]}}^{(30),c}(\alpha,z)=y_{_2}
y_{_3}^{{D\over2}-2}\sum\limits_{n_{_1}=0}^\infty
\sum\limits_{n_{_2}=0}^\infty\sum\limits_{n_{_3}=0}^\infty\sum\limits_{n_{_4}=0}^\infty
c_{_{[\tilde{1}\tilde{3}\tilde{5}7]}}^{(30),c}(\alpha,{\bf n})
\nonumber\\
&&\hspace{2.5cm}\times
y_{_1}^{n_{_1}}y_{_2}^{n_{_2}}y_{_4}^{n_{_3}}\Big({y_{_2}\over y_{_3}}\Big)^{n_{_4}}\;.
\label{GKZ21k-24-2a}
\end{eqnarray}
Where the coefficients are
\begin{eqnarray}
&&c_{_{[\tilde{1}\tilde{3}\tilde{5}7]}}^{(30),a}(\alpha,{\bf n})=
(-)^{n_{_2}+n_{_4}}\Gamma(1+n_{_3}+n_{_4})
\Big\{n_{_1}!n_{_2}!n_{_3}!n_{_4}!\Gamma({D\over2}-1-n_{_1}-n_{_2})
\nonumber\\
&&\hspace{2.5cm}\times
\Gamma(D-2-n_{_1}-n_{_2})\Gamma({D\over2}+n_{_3})
\Gamma(2-{D\over2}+n_{_1})
\nonumber\\
&&\hspace{2.5cm}\times
\Gamma({D\over2}-1-n_{_3}-n_{_4})\Gamma(2-{D\over2}+n_{_2})
\Gamma(2-{D\over2}+n_{_4})\Big\}^{-1}
\;,\nonumber\\
&&c_{_{[\tilde{1}\tilde{3}\tilde{5}7]}}^{(30),b}(\alpha,{\bf n})=
(-)^{1+n_{_2}+n_{_3}+n_{_4}}\Gamma(1+n_{_2}+n_{_4})
\Big\{n_{_1}!n_{_2}!n_{_4}!\Gamma(2+n_{_2}+n_{_3}+n_{_4})
\nonumber\\
&&\hspace{2.5cm}\times
\Gamma({D\over2}-2-n_{_1}-n_{_2}-n_{_3}-n_{_4})\Gamma(D-3-n_{_1}-n_{_2}-n_{_3}-n_{_4})
\nonumber\\
&&\hspace{2.5cm}\times
\Gamma({D\over2}-1-n_{_2}-n_{_4})\Gamma(2-{D\over2}+n_{_1})
\Gamma(3-{D\over2}+n_{_2}+n_{_3}+n_{_4})
\nonumber\\
&&\hspace{2.5cm}\times
\Gamma({D\over2}+n_{_2})\Gamma(2-{D\over2}+n_{_4})\Big\}^{-1}
\;,\nonumber\\
&&c_{_{[\tilde{1}\tilde{3}\tilde{5}7]}}^{(30),c}(\alpha,{\bf n})=
(-)^{1+n_{_2}+n_{_3}+n_{_4}}\Gamma(1+n_{_2})\Gamma(1+n_{_4})
\Big\{n_{_1}!\Gamma(2+n_{_2}+n_{_4})
\nonumber\\
&&\hspace{2.5cm}\times
\Gamma(2+n_{_2}+n_{_3})\Gamma({D\over2}-2-n_{_1}-n_{_2}-n_{_3})
\nonumber\\
&&\hspace{2.5cm}\times
\Gamma(D-3-n_{_1}-n_{_2}-n_{_3})\Gamma({D\over2}-1-n_{_2})\Gamma(2-{D\over2}+n_{_1})
\nonumber\\
&&\hspace{2.5cm}\times
\Gamma({D\over2}-1-n_{_4})
\Gamma(3-{D\over2}+n_{_2}+n_{_3})\Gamma(3-{D\over2}+n_{_2}+n_{_4})\Big\}^{-1}\;.
\label{GKZ21k-24-3}
\end{eqnarray}

\item   $I_{_{31}}=\{1,\cdots,6,8,9,12,13\}$, i.e. the implement $J_{_{31}}=[1,14]\setminus I_{_{31}}=\{7,10,11,14\}$.
The choice implies the power numbers $\alpha_{_7}=\alpha_{_{10}}=\alpha_{_{11}}=\alpha_{_{14}}=0$, and
\begin{eqnarray}
&&\alpha_{_1}=b_{_5}-a_{_1}-1,\;\alpha_{_2}=b_{_5}-a_{_2}-1,\;\alpha_{_3}=b_{_5}-a_{_3}-1,
\nonumber\\
&&\alpha_{_4}=b_{_5}-a_{_4}-1,\;\alpha_{_5}=-a_{_5},\;\alpha_{_{6}}=b_{_1}-1,
\nonumber\\
&&\alpha_{_{8}}=b_{_2}+b_{_3}-b_{_5}-1,\;\alpha_{_{9}}=b_{_4}-b_{_5},\;
\alpha_{_{12}}=1-b_{_2},\;\alpha_{_{13}}=b_{_2}-b_{_5}.
\label{GKZ21k-25-1}
\end{eqnarray}
The corresponding hypergeometric solutions are
\begin{eqnarray}
&&\Phi_{_{[\tilde{1}\tilde{3}\tilde{5}7]}}^{(31),a}(\alpha,z)=
y_{_2}^{{D\over2}-1}y_{_3}^{{D\over2}-2}\sum\limits_{n_{_1}=0}^\infty
\sum\limits_{n_{_2}=0}^\infty\sum\limits_{n_{_3}=0}^\infty\sum\limits_{n_{_4}=0}^\infty
c_{_{[\tilde{1}\tilde{3}\tilde{5}7]}}^{(31),a}(\alpha,{\bf n})
\nonumber\\
&&\hspace{2.5cm}\times
y_{_1}^{n_{_1}}y_{_4}^{n_{_2}}\Big({y_{_4}\over y_{_3}}\Big)^{n_{_3}}
\Big({y_{_2}\over y_{_3}}\Big)^{n_{_4}}
\;,\nonumber\\
&&\Phi_{_{[\tilde{1}\tilde{3}\tilde{5}7]}}^{(31),b}(\alpha,z)=
y_{_2}^{{D\over2}-1}y_{_3}^{{D\over2}-1}\sum\limits_{n_{_1}=0}^\infty
\sum\limits_{n_{_2}=0}^\infty\sum\limits_{n_{_3}=0}^\infty\sum\limits_{n_{_4}=0}^\infty
c_{_{[\tilde{1}\tilde{3}\tilde{5}7]}}^{(31),b}(\alpha,{\bf n})
\nonumber\\
&&\hspace{2.5cm}\times
y_{_1}^{n_{_1}}y_{_3}^{n_{_2}}y_{_4}^{n_{_3}}y_{_2}^{n_{_4}}
\;,\nonumber\\
&&\Phi_{_{[\tilde{1}\tilde{3}\tilde{5}7]}}^{(31),c}(\alpha,z)=
y_{_2}^{{D\over2}}y_{_3}^{{D\over2}-2}\sum\limits_{n_{_1}=0}^\infty
\sum\limits_{n_{_2}=0}^\infty\sum\limits_{n_{_3}=0}^\infty\sum\limits_{n_{_4}=0}^\infty
c_{_{[\tilde{1}\tilde{3}\tilde{5}7]}}^{(31),c}(\alpha,{\bf n})
\nonumber\\
&&\hspace{2.5cm}\times
y_{_1}^{n_{_1}}y_{_2}^{n_{_2}}y_{_4}^{n_{_3}}\Big({y_{_2}\over y_{_3}}\Big)^{n_{_4}}\;.
\label{GKZ21k-25-2a}
\end{eqnarray}
Where the coefficients are
\begin{eqnarray}
&&c_{_{[\tilde{1}\tilde{3}\tilde{5}7]}}^{(31),a}(\alpha,{\bf n})=
(-)^{n_{_1}+n_{_4}}\Gamma(1+n_{_1}+n_{_2})\Gamma(1+n_{_3}+n_{_4})\Big\{n_{_1}!n_{_2}!n_{_3}!n_{_4}!
\nonumber\\
&&\hspace{2.5cm}\times
\Gamma({D\over2}-1-n_{_1}-n_{_2})\Gamma(2-{D\over2}+n_{_3})\Gamma(2-{D\over2}+n_{_1})
\nonumber\\
&&\hspace{2.5cm}\times
\Gamma({D\over2}+n_{_2})\Gamma({D\over2}+n_{_4})\Gamma({D\over2}-1-n_{_3}-n_{_4})\Big\}^{-1}
\;,\nonumber\\
&&c_{_{[\tilde{1}\tilde{3}\tilde{5}7]}}^{(31),b}(\alpha,{\bf n})=
(-)^{1+n_{_1}}\Gamma(2+n_{_1}+n_{_2}+n_{_3}+n_{_4})\Gamma(1+n_{_2}+n_{_4})
\nonumber\\
&&\hspace{2.5cm}\times
\Big\{n_{_1}!n_{_2}!n_{_4}!\Gamma(2+n_{_2}+n_{_3}+n_{_4})
\Gamma(1-{D\over2}-n_{_2}-n_{_4})
\nonumber\\
&&\hspace{2.5cm}\times
\Gamma({D\over2}-2-n_{_1}-n_{_2}-n_{_3}-n_{_4})\Gamma(2-{D\over2}+n_{_1})
\nonumber\\
&&\hspace{2.5cm}\times
\Gamma({D\over2}+1+n_{_2}+n_{_3}+n_{_4})\Gamma({D\over2}+n_{_4})\Gamma({D\over2}+n_{_2})\Big\}^{-1}
\;,\nonumber\\
&&c_{_{[\tilde{1}\tilde{3}\tilde{5}7]}}^{(31),c}(\alpha,{\bf n})=
(-)^{1+n_{_1}+n_{_4}}\Gamma(2+n_{_1}+n_{_2}+n_{_3})\Gamma(1+n_{_2})\Gamma(1+n_{_4})
\nonumber\\
&&\hspace{2.5cm}\times
\Big\{n_{_1}!\Gamma(2+n_{_2}+n_{_3})\Gamma(2+n_{_2}+n_{_4})
\nonumber\\
&&\hspace{2.5cm}\times
\Gamma({D\over2}-2-n_{_1}-n_{_2}-n_{_3})\Gamma(1-{D\over2}-n_{_2})
\Gamma(2-{D\over2}+n_{_1})
\nonumber\\
&&\hspace{2.5cm}\times
\Gamma({D\over2}+1+n_{_2}+n_{_3})\Gamma({D\over2}+1+n_{_2}+n_{_4})\Gamma({D\over2}-1-n_{_4})\Big\}^{-1}\;.
\label{GKZ21k-25-3}
\end{eqnarray}

\item   $I_{_{32}}=\{1,\cdots,9,13\}$, i.e. the implement $J_{_{32}}=[1,14]\setminus I_{_{32}}=\{10,11,12,14\}$.
The choice implies the power numbers $\alpha_{_{10}}=\alpha_{_{11}}=\alpha_{_{12}}=\alpha_{_{14}}=0$, and
\begin{eqnarray}
&&\alpha_{_1}=b_{_5}-a_{_1}-1,\;\alpha_{_2}=b_{_5}-a_{_2}-1,\;\alpha_{_3}=b_{_5}-a_{_3}-1,
\nonumber\\
&&\alpha_{_4}=b_{_5}-a_{_4}-1,\;\alpha_{_5}=-a_{_5},\;\alpha_{_{6}}=b_{_1}-1,\;\alpha_{_{7}}=b_{_2}-1,
\nonumber\\
&&\alpha_{_{8}}=b_{_3}-b_{_5},\;\alpha_{_{9}}=b_{_4}-b_{_5},\;
\alpha_{_{13}}=1-b_{_5}.
\label{GKZ21k-26-1}
\end{eqnarray}
The corresponding hypergeometric functions are written as
\begin{eqnarray}
&&\Phi_{_{[\tilde{1}\tilde{3}\tilde{5}7]}}^{(32),a}(\alpha,z)=
y_{_3}^{{D}-3}\sum\limits_{n_{_1}=0}^\infty
\sum\limits_{n_{_2}=0}^\infty\sum\limits_{n_{_3}=0}^\infty\sum\limits_{n_{_4}=0}^\infty
c_{_{[\tilde{1}\tilde{3}\tilde{5}7]}}^{(32),a}(\alpha,{\bf n})
\nonumber\\
&&\hspace{2.5cm}\times
y_{_1}^{n_{_1}}y_{_4}^{n_{_2}}\Big({y_{_4}\over y_{_3}}\Big)^{n_{_3}}
\Big({y_{_2}\over y_{_3}}\Big)^{n_{_4}}
\;,\nonumber\\
&&\Phi_{_{[\tilde{1}\tilde{3}\tilde{5}7]}}^{(32),b}(\alpha,z)=
y_{_3}^{{D}-2}\sum\limits_{n_{_1}=0}^\infty
\sum\limits_{n_{_2}=0}^\infty\sum\limits_{n_{_3}=0}^\infty\sum\limits_{n_{_4}=0}^\infty
c_{_{[\tilde{1}\tilde{3}\tilde{5}7]}}^{(32),b}(\alpha,{\bf n})
\nonumber\\
&&\hspace{2.5cm}\times
y_{_1}^{n_{_1}}y_{_3}^{n_{_2}}y_{_4}^{n_{_3}}\Big({y_{_2}\over y_{_3}}\Big)^{n_{_4}}\;.
\label{GKZ21k-26-2a}
\end{eqnarray}
Where the coefficients are
\begin{eqnarray}
&&c_{_{[\tilde{1}\tilde{3}\tilde{5}7]}}^{(32),a}(\alpha,{\bf n})=
(-)^{n_{_1}+n_{_3}}\Gamma(1+n_{_1}+n_{_2})\Big\{n_{_1}!n_{_2}!n_{_3}n_{_4}!
\Gamma(2-{D\over2}+n_{_1})
\nonumber\\
&&\hspace{2.5cm}\times
\Gamma({D\over2}-1-n_{_1}-n_{_2})\Gamma(2-{D\over2}+n_{_3})
\Gamma(2-{D\over2}+n_{_4})
\nonumber\\
&&\hspace{2.5cm}\times
\Gamma({D\over2}-1-n_{_3}-n_{_4})
\Gamma({D\over2}+n_{_2})\Gamma(D-2-n_{_3}-n_{_4})\Big\}^{-1}
\;,\nonumber\\
&&c_{_{[\tilde{1}\tilde{3}\tilde{5}7]}}^{(32),b}(\alpha,{\bf n})=
(-)^{1+n_{_1}}\Gamma(2+n_{_1}+n_{_2}+n_{_3})\Gamma(1+n_{_2})
\Big\{n_{_1}!n_{_4}!\Gamma(2+n_{_2}+n_{_3})
\nonumber\\
&&\hspace{2.5cm}\times
\Gamma(2-{D\over2}+n_{_1})
\Gamma({D\over2}-1-n_{_1}-n_{_2}-n_{_3})\Gamma(1-{D\over2}-n_{_2})
\nonumber\\
&&\hspace{2.5cm}\times
\Gamma(2-{D\over2}+n_{_4})\Gamma({D\over2}+n_{_2}-n_{_4})
\Gamma({D\over2}+1+n_{_2}+n_{_3})
\nonumber\\
&&\hspace{2.5cm}\times
\Gamma(D-1+n_{_2}-n_{_4})\Big\}^{-1}\;.
\label{GKZ21k-26-3}
\end{eqnarray}
\end{itemize}

\section{The hypergeometric solutions of the integer lattice ${\bf B}_{_{\widetilde{13}5\tilde{7}}}$\label{app13}}
\indent\indent

\begin{itemize}
\item   $I_{_{1}}=\{1,2,3,4,7,10,\cdots,14\}$, i.e. the implement $J_{_{1}}=[1,14]\setminus I_{_{1}}=\{5,6,8,9\}$.
The choice implies the power numbers $\alpha_{_5}=\alpha_{_{6}}=\alpha_{_{8}}=\alpha_{_{9}}=0$, and
\begin{eqnarray}
&&\alpha_{_1}=b_{_1}+b_{_4}-a_{_1}-2,\;\alpha_{_2}=b_{_1}+b_{_4}-a_{_2}-2,
\nonumber\\
&&\alpha_{_3}=b_{_4}-a_{_3}-a_{_5}-1,\;\alpha_{_4}=b_{_4}-a_{_4}-a_{_5}-1,
\nonumber\\
&&\alpha_{_7}=a_{_5}+b_{_2}+b_{_3}-b_{_4}-1,\;\alpha_{_{10}}=b_{_5}-b_{_4},\;\alpha_{_{11}}=1-b_{_1},
\nonumber\\
&&\alpha_{_{12}}=a_{_5}+b_{_3}-b_{_4},\;\alpha_{_{13}}=1-b_{_3},\;
\alpha_{_{14}}=-a_{_5}.
\label{GKZ21l-1-1}
\end{eqnarray}
The corresponding hypergeometric solutions are written as
\begin{eqnarray}
&&\Phi_{_{[\tilde{1}\tilde{3}5\tilde{7}]}}^{(1),a}(\alpha,z)=
y_{_1}^{{D\over2}-1}y_{_3}^{{D\over2}-1}y_{_4}^{-1}\sum\limits_{n_{_1}=0}^\infty
\sum\limits_{n_{_2}=0}^\infty\sum\limits_{n_{_3}=0}^\infty\sum\limits_{n_{_4}=0}^\infty
c_{_{[\tilde{1}\tilde{3}5\tilde{7}]}}^{(1),a}(\alpha,{\bf n})
\nonumber\\
&&\hspace{2.5cm}\times
y_{_1}^{n_{_1}}y_{_4}^{n_{_2}}\Big({y_{_2}\over y_{_4}}\Big)^{n_{_3}}
\Big({y_{_3}\over y_{_4}}\Big)^{n_{_4}}
\;,\nonumber\\
&&\Phi_{_{[\tilde{1}\tilde{3}5\tilde{7}]}}^{(1),b}(\alpha,z)=
y_{_1}^{{D\over2}-1}y_{_2}^{-1}y_{_3}^{{D\over2}}y_{_4}^{-1}\sum\limits_{n_{_1}=0}^\infty
\sum\limits_{n_{_2}=0}^\infty\sum\limits_{n_{_3}=0}^\infty\sum\limits_{n_{_4}=0}^\infty
c_{_{[\tilde{1}\tilde{3}5\tilde{7}]}}^{(1),b}(\alpha,{\bf n})
\nonumber\\
&&\hspace{2.5cm}\times
y_{_1}^{n_{_1}}y_{_3}^{n_{_2}}\Big({y_{_3}\over y_{_4}}\Big)^{n_{_3}}
\Big({y_{_3}\over y_{_2}}\Big)^{n_{_4}}\;.
\label{GKZ21l-1-2a}
\end{eqnarray}
Where the coefficients are
\begin{eqnarray}
&&c_{_{[\tilde{1}\tilde{3}5\tilde{7}]}}^{(1),a}(\alpha,{\bf n})=
(-)^{n_{_1}}\Gamma(1+n_{_1}+n_{_2})\Gamma(1+n_{_3}+n_{_4})
\Big\{n_{_1}!n_{_2}!n_{_3}!n_{_4}!
\nonumber\\
&&\hspace{2.5cm}\times
\Gamma({D\over2}-1-n_{_1}-n_{_2})\Gamma({D\over2}-1-n_{_3}-n_{_4})
\nonumber\\
&&\hspace{2.5cm}\times
\Gamma(2-{D\over2}+n_{_3})\Gamma(2-{D\over2}+n_{_2})
\Gamma({D\over2}+n_{_1})\Gamma({D\over2}+n_{_4})\Big\}^{-1}
\;,\nonumber\\
&&c_{_{[\tilde{1}\tilde{3}5\tilde{7}]}}^{(1),b}(\alpha,{\bf n})=
(-)^{n_{_1}+n_{_4}}\Gamma(1+n_{_1}+n_{_2})\Gamma(1+n_{_2}+n_{_3})\Gamma(1+n_{_4})
\nonumber\\
&&\hspace{2.5cm}\times
\Big\{n_{_1}!n_{_2}!\Gamma(2+n_{_2}+n_{_3}+n_{_4})
\Gamma({D\over2}-1-n_{_1}-n_{_2})
\nonumber\\
&&\hspace{2.5cm}\times
\Gamma({D\over2}-1-n_{_2}-n_{_3})
\Gamma(1-{D\over2}-n_{_4})\Gamma(2-{D\over2}+n_{_2})
\nonumber\\
&&\hspace{2.5cm}\times
\Gamma({D\over2}+n_{_1})\Gamma({D\over2}+1+n_{_2}+n_{_3}+n_{_4})\Big\}^{-1}\;.
\label{GKZ21l-1-3}
\end{eqnarray}

\item   $I_{_{2}}=\{1,2,3,4,7,8,10,11,12,14\}$, i.e. the implement $J_{_{2}}=[1,14]\setminus I_{_{2}}=\{5,6,9,13\}$.
The choice implies the power numbers $\alpha_{_5}=\alpha_{_{6}}=\alpha_{_{9}}=\alpha_{_{13}}=0$, and
\begin{eqnarray}
&&\alpha_{_1}=b_{_1}+b_{_4}-a_{_1}-2,\;\alpha_{_2}=b_{_1}+b_{_4}-a_{_2}-2,
\nonumber\\
&&\alpha_{_3}=b_{_4}-a_{_3}-a_{_5}-1,\;\alpha_{_4}=b_{_4}-a_{_4}-a_{_5}-1,
\nonumber\\
&&\alpha_{_7}=a_{_5}+b_{_2}-b_{_4},\;\alpha_{_{8}}=b_{_3}-1,\;\alpha_{_{10}}=b_{_5}-b_{_4},
\nonumber\\
&&\alpha_{_{11}}=1-b_{_1},\;\alpha_{_{12}}=a_{_5}-b_{_4}+1,\;\alpha_{_{14}}=-a_{_5}.
\label{GKZ21l-2-1}
\end{eqnarray}
The corresponding hypergeometric solutions are
\begin{eqnarray}
&&\Phi_{_{[\tilde{1}\tilde{3}5\tilde{7}]}}^{(2),a}(\alpha,z)=
y_{_1}^{{D\over2}-1}y_{_2}^{{D\over2}-1}y_{_4}^{-1}\sum\limits_{n_{_1}=0}^\infty
\sum\limits_{n_{_2}=0}^\infty\sum\limits_{n_{_3}=0}^\infty\sum\limits_{n_{_4}=0}^\infty
c_{_{[\tilde{1}\tilde{3}5\tilde{7}]}}^{(2),a}(\alpha,{\bf n})
\nonumber\\
&&\hspace{2.5cm}\times
y_{_1}^{n_{_1}}y_{_4}^{n_{_2}}\Big({y_{_2}\over y_{_4}}\Big)^{n_{_3}}
\Big({y_{_3}\over y_{_4}}\Big)^{n_{_4}}
\;,\nonumber\\
&&\Phi_{_{[\tilde{1}\tilde{3}5\tilde{7}]}}^{(2),b}(\alpha,z)=
y_{_1}^{{D\over2}-1}y_{_2}^{{D\over2}-2}y_{_3}y_{_4}^{-1}\sum\limits_{n_{_1}=0}^\infty
\sum\limits_{n_{_2}=0}^\infty\sum\limits_{n_{_3}=0}^\infty\sum\limits_{n_{_4}=0}^\infty
c_{_{[\tilde{1}\tilde{3}5\tilde{7}]}}^{(2),b}(\alpha,{\bf n})
\nonumber\\
&&\hspace{2.5cm}\times
y_{_1}^{n_{_1}}y_{_3}^{n_{_2}}\Big({y_{_3}\over y_{_4}}\Big)^{n_{_3}}
\Big({y_{_3}\over y_{_2}}\Big)^{n_{_4}}\;.
\label{GKZ21l-2-2a}
\end{eqnarray}
Where the coefficients are
\begin{eqnarray}
&&c_{_{[\tilde{1}\tilde{3}5\tilde{7}]}}^{(2),a}(\alpha,{\bf n})=
(-)^{n_{_1}}\Gamma(1+n_{_1}+n_{_2})\Gamma(1+n_{_3}+n_{_4})
\Big\{n_{_1}!n_{_2}!n_{_3}!n_{_4}!
\nonumber\\
&&\hspace{2.5cm}\times
\Gamma({D\over2}-1-n_{_1}-n_{_2})\Gamma({D\over2}-1-n_{_3}-n_{_4})
\nonumber\\
&&\hspace{2.5cm}\times
\Gamma(2-{D\over2}+n_{_4})\Gamma(2-{D\over2}+n_{_2})
\Gamma({D\over2}+n_{_1})\Gamma({D\over2}+n_{_3})\Big\}^{-1}
\;,\nonumber\\
&&c_{_{[\tilde{1}\tilde{3}5\tilde{7}]}}^{(2),b}(\alpha,{\bf n})=
(-)^{n_{_1}+n_{_4}}\Gamma(1+n_{_1}+n_{_2})\Gamma(1+n_{_2}+n_{_3})\Gamma(1+n_{_4})
\Big\{n_{_1}!n_{_2}!
\nonumber\\
&&\hspace{2.5cm}\times
\Gamma(2+n_{_2}+n_{_3}+n_{_4})
\Gamma({D\over2}-1-n_{_1}-n_{_2})\Gamma({D\over2}-1-n_{_2}-n_{_3})
\nonumber\\
&&\hspace{2.5cm}\times
\Gamma({D\over2}-1-n_{_4})\Gamma(2-{D\over2}+n_{_2})
\Gamma({D\over2}+n_{_1})
\nonumber\\
&&\hspace{2.5cm}\times
\Gamma(3-{D\over2}+n_{_2}+n_{_3}+n_{_4})\Big\}^{-1}\;.
\label{GKZ21l-2-3}
\end{eqnarray}

\item   $I_{_{3}}=\{1,2,3,4,7,9,11,\cdots,14\}$, i.e. the implement $J_{_{3}}=[1,14]\setminus I_{_{3}}=\{5,6,8,10\}$.
The choice implies the power numbers $\alpha_{_5}=\alpha_{_{6}}=\alpha_{_{8}}=\alpha_{_{10}}=0$, and
\begin{eqnarray}
&&\alpha_{_1}=b_{_1}+b_{_5}-a_{_1}-2,\;\alpha_{_2}=b_{_1}+b_{_5}-a_{_2}-2,
\nonumber\\
&&\alpha_{_3}=b_{_5}-a_{_3}-a_{_5}-1,\;\alpha_{_4}=b_{_5}-a_{_4}-a_{_5}-1,
\nonumber\\
&&\alpha_{_7}=a_{_5}+b_{_2}+b_{_3}-b_{_5}-1,\;\alpha_{_{9}}=b_{_4}-b_{_5},\;\alpha_{_{11}}=1-b_{_1},
\nonumber\\
&&\alpha_{_{12}}=a_{_5}+b_{_3}-b_{_5},\;\alpha_{_{13}}=1-b_{_3},\;
\alpha_{_{14}}=-a_{_5}.
\label{GKZ21l-3-1}
\end{eqnarray}
The corresponding hypergeometric functions are given as
\begin{eqnarray}
&&\Phi_{_{[\tilde{1}\tilde{3}5\tilde{7}]}}^{(3),a}(\alpha,z)=
y_{_1}^{{D\over2}-1}y_{_2}^{{D\over2}-1}y_{_3}^{{D\over2}-1}y_{_4}^{-1}
\sum\limits_{n_{_1}=0}^\infty
\sum\limits_{n_{_2}=0}^\infty\sum\limits_{n_{_3}=0}^\infty\sum\limits_{n_{_4}=0}^\infty
c_{_{[\tilde{1}\tilde{3}5\tilde{7}]}}^{(3),a}(\alpha,{\bf n})
\nonumber\\
&&\hspace{2.5cm}\times
y_{_1}^{n_{_1}}y_{_4}^{n_{_2}}\Big({y_{_2}\over y_{_4}}\Big)^{n_{_3}}
\Big({y_{_3}\over y_{_4}}\Big)^{n_{_4}}
\;,\nonumber\\
&&\Phi_{_{[\tilde{1}\tilde{3}5\tilde{7}]}}^{(3),b}(\alpha,z)=
y_{_1}^{{D\over2}-1}y_{_2}^{{D\over2}-2}y_{_3}^{{D\over2}}y_{_4}^{-1}\sum\limits_{n_{_1}=0}^\infty
\sum\limits_{n_{_2}=0}^\infty\sum\limits_{n_{_3}=0}^\infty\sum\limits_{n_{_4}=0}^\infty
c_{_{[\tilde{1}\tilde{3}5\tilde{7}]}}^{(3),b}(\alpha,{\bf n})
\nonumber\\
&&\hspace{2.5cm}\times
y_{_1}^{n_{_1}}y_{_3}^{n_{_2}}\Big({y_{_3}\over y_{_4}}\Big)^{n_{_3}}
\Big({y_{_3}\over y_{_2}}\Big)^{n_{_4}}\;.
\label{GKZ21l-3-2a}
\end{eqnarray}
Where the coefficients are
\begin{eqnarray}
&&c_{_{[\tilde{1}\tilde{3}5\tilde{7}]}}^{(3),a}(\alpha,{\bf n})=
(-)^{n_{_1}}\Gamma(1+n_{_1}+n_{_2})\Gamma(1+n_{_3}+n_{_4})
\Big\{n_{_1}!n_{_2}!n_{_3}!n_{_4}!
\nonumber\\
&&\hspace{2.5cm}\times
\Gamma(1-{D\over2}-n_{_1}-n_{_2})\Gamma(1-{D\over2}-n_{_3}-n_{_4})
\nonumber\\
&&\hspace{2.5cm}\times
\Gamma({D\over2}+n_{_2})\Gamma({D\over2}+n_{_1})
\Gamma({D\over2}+n_{_3})\Gamma({D\over2}+n_{_4})\Big\}^{-1}
\;,\nonumber\\
&&c_{_{[\tilde{1}\tilde{3}5\tilde{7}]}}^{(3),b}(\alpha,{\bf n})=
(-)^{n_{_1}+n_{_4}}\Gamma(1+n_{_1}+n_{_2})\Gamma(1+n_{_2}+n_{_3})
\Gamma(1+n_{_4})
\nonumber\\
&&\hspace{2.5cm}\times
\Big\{n_{_1}!n_{_2}!\Gamma(2+n_{_2}+n_{_3}+n_{_4})\Gamma(1-{D\over2}-n_{_1}-n_{_2})
\nonumber\\
&&\hspace{2.5cm}\times
\Gamma(1-{D\over2}-n_{_2}-n_{_3})\Gamma({D\over2}+n_{_2})\Gamma({D\over2}+n_{_1})
\nonumber\\
&&\hspace{2.5cm}\times
\Gamma({D\over2}-1-n_{_4})
\Gamma({D\over2}+1+n_{_2}+n_{_3}+n_{_4})\Big\}^{-1}\;.
\label{GKZ21l-3-3}
\end{eqnarray}

\item   $I_{_{4}}=\{1,2,3,4,7,8,9,11,12,14\}$, i.e. the implement $J_{_{4}}=[1,14]\setminus I_{_{4}}=\{5,6,10,13\}$.
The choice implies the power numbers $\alpha_{_5}=\alpha_{_{6}}=\alpha_{_{10}}=\alpha_{_{13}}=0$, and
\begin{eqnarray}
&&\alpha_{_1}=b_{_1}+b_{_5}-a_{_1}-2,\;\alpha_{_2}=b_{_1}+b_{_5}-a_{_2}-2,
\nonumber\\
&&\alpha_{_3}=b_{_5}-a_{_3}-a_{_5}-1,\;\alpha_{_4}=b_{_5}-a_{_4}-a_{_5}-1,
\nonumber\\
&&\alpha_{_7}=a_{_5}+b_{_2}-b_{_5},\;\alpha_{_{8}}=b_{_3}-1,\;\alpha_{_{9}}=b_{_4}-b_{_5},
\nonumber\\
&&\alpha_{_{11}}=1-b_{_1},\;\alpha_{_{12}}=a_{_5}-b_{_5}+1,\;\alpha_{_{14}}=-a_{_5}.
\label{GKZ21l-4-1}
\end{eqnarray}
The corresponding hypergeometric series is written as
\begin{eqnarray}
&&\Phi_{_{[\tilde{1}\tilde{3}5\tilde{7}]}}^{(4)}(\alpha,z)=
y_{_1}^{{D\over2}-1}y_{_2}^{{D}-2}y_{_4}^{-1}
\sum\limits_{n_{_1}=0}^\infty
\sum\limits_{n_{_2}=0}^\infty\sum\limits_{n_{_3}=0}^\infty\sum\limits_{n_{_4}=0}^\infty
c_{_{[\tilde{1}\tilde{3}5\tilde{7}]}}^{(4)}(\alpha,{\bf n})
\nonumber\\
&&\hspace{2.5cm}\times
y_{_1}^{n_{_1}}y_{_2}^{n_{_2}}\Big({y_{_2}\over y_{_4}}\Big)^{n_{_3}}\Big({y_{_3}\over y_{_2}}\Big)^{n_{_4}}\;,
\label{GKZ21l-4-2}
\end{eqnarray}
with
\begin{eqnarray}
&&c_{_{[\tilde{1}\tilde{3}5\tilde{7}]}}^{(4)}(\alpha,{\bf n})=
(-)^{n_{_1}}\Gamma(1+n_{_1}+n_{_2})\Gamma(1+n_{_2}+n_{_3})
\nonumber\\
&&\hspace{2.5cm}\times
\Big\{n_{_1}!n_{_2}!n_{_4}!\Gamma(1-{D\over2}-n_{_1}-n_{_2})
\Gamma(1-{D\over2}-n_{_2}-n_{_3})
\nonumber\\
&&\hspace{2.5cm}\times
\Gamma({D\over2}+n_{_2}+n_{_3}-n_{_4})\Gamma(2-{D\over2}+n_{_4})
\Gamma({D\over2}+n_{_2})
\nonumber\\
&&\hspace{2.5cm}\times
\Gamma({D\over2}+n_{_1})\Gamma(D-1+n_{_2}+n_{_3}-n_{_4})\Big\}^{-1}\;.
\label{GKZ21l-4-3}
\end{eqnarray}

\item   $I_{_{5}}=\{1,2,3,4,6,7,10,12,13,14\}$, i.e. the implement $J_{_{5}}=[1,14]\setminus I_{_{5}}=\{5,8,9,11\}$.
The choice implies the power numbers $\alpha_{_5}=\alpha_{_{8}}=\alpha_{_{9}}=\alpha_{_{11}}=0$, and
\begin{eqnarray}
&&\alpha_{_1}=b_{_4}-a_{_1}-1,\;\alpha_{_2}=b_{_4}-a_{_2}-1,
\nonumber\\
&&\alpha_{_3}=b_{_4}-a_{_3}-a_{_5}-1,\;\alpha_{_4}=b_{_4}-a_{_4}-a_{_5}-1,
\nonumber\\
&&\alpha_{_{6}}=b_{_1}-1,\;\alpha_{_7}=a_{_5}+b_{_2}+b_{_3}-b_{_4}-1,\;\alpha_{_{10}}=b_{_5}-b_{_4},\;
\nonumber\\
&&\alpha_{_{12}}=a_{_5}+b_{_3}-b_{_4},\;\alpha_{_{13}}=1-b_{_3},\;\alpha_{_{14}}=-a_{_5}.
\label{GKZ21l-5-1}
\end{eqnarray}
The corresponding hypergeometric functions are written as
\begin{eqnarray}
&&\Phi_{_{[\tilde{1}\tilde{3}5\tilde{7}]}}^{(5),a}(\alpha,z)=
y_{_3}^{{D\over2}-1}y_{_4}^{-1}\sum\limits_{n_{_1}=0}^\infty
\sum\limits_{n_{_2}=0}^\infty\sum\limits_{n_{_3}=0}^\infty\sum\limits_{n_{_4}=0}^\infty
c_{_{[\tilde{1}\tilde{3}5\tilde{7}]}}^{(5),a}(\alpha,{\bf n})
\nonumber\\
&&\hspace{2.5cm}\times
y_{_1}^{n_{_1}}y_{_4}^{n_{_2}}\Big({y_{_2}\over y_{_4}}\Big)^{n_{_3}}
\Big({y_{_3}\over y_{_4}}\Big)^{n_{_4}}
\;,\nonumber\\
&&\Phi_{_{[\tilde{1}\tilde{3}5\tilde{7}]}}^{(5),b}(\alpha,z)=
y_{_2}^{-1}y_{_3}^{{D\over2}}y_{_4}^{-1}\sum\limits_{n_{_1}=0}^\infty
\sum\limits_{n_{_2}=0}^\infty\sum\limits_{n_{_3}=0}^\infty\sum\limits_{n_{_4}=0}^\infty
c_{_{[\tilde{1}\tilde{3}5\tilde{7}]}}^{(5),b}(\alpha,{\bf n})
\nonumber\\
&&\hspace{2.5cm}\times
y_{_1}^{n_{_1}}y_{_3}^{n_{_2}}\Big({y_{_3}\over y_{_4}}\Big)^{n_{_3}}
\Big({y_{_3}\over y_{_2}}\Big)^{n_{_4}}\;.
\label{GKZ21l-5-2a}
\end{eqnarray}
Where the coefficients are
\begin{eqnarray}
&&c_{_{[\tilde{1}\tilde{3}5\tilde{7}]}}^{(5),a}(\alpha,{\bf n})=
(-)^{n_{_2}}\Gamma(1+n_{_3}+n_{_4})\Big\{n_{_1}!n_{_2}!n_{_3}!n_{_4}!
\Gamma({D\over2}-1-n_{_1}-n_{_2})
\nonumber\\
&&\hspace{2.5cm}\times
\Gamma(D-2-n_{_1}-n_{_2})\Gamma({D\over2}-1-n_{_3}-n_{_4})
\Gamma(2-{D\over2}+n_{_1})
\nonumber\\
&&\hspace{2.5cm}\times
\Gamma(2-{D\over2}+n_{_3})
\Gamma(2-{D\over2}+n_{_2})\Gamma({D\over2}+n_{_4})\Big\}^{-1}
\;,\nonumber\\
&&c_{_{[\tilde{1}\tilde{3}5\tilde{7}]}}^{(5),b}(\alpha,{\bf n})=
(-)^{n_{_2}+n_{_4}}\Gamma(1+n_{_2}+n_{_3})\Gamma(1+n_{_4})
\Big\{n_{_1}!n_{_2}!\Gamma(2+n_{_2}+n_{_3}+n_{_4})
\nonumber\\
&&\hspace{2.5cm}\times
\Gamma({D\over2}-1-n_{_1}-n_{_2})\Gamma(D-2-n_{_1}-n_{_2})
\nonumber\\
&&\hspace{2.5cm}\times
\Gamma({D\over2}-1-n_{_2}-n_{_3})
\Gamma(2-{D\over2}+n_{_1})\Gamma(1-{D\over2}-n_{_4})
\nonumber\\
&&\hspace{2.5cm}\times
\Gamma(2-{D\over2}+n_{_2})\Gamma({D\over2}+1+n_{_2}+n_{_3}+n_{_4})\Big\}^{-1}\;.
\label{GKZ21l-5-3}
\end{eqnarray}

\item   $I_{_{6}}=\{1,2,3,4,6,7,8,10,12,14\}$, i.e. the implement $J_{_{6}}=[1,14]\setminus I_{_{6}}=\{5,9,11,13\}$.
The choice implies the power numbers $\alpha_{_5}=\alpha_{_{9}}=\alpha_{_{11}}=\alpha_{_{13}}=0$, and
\begin{eqnarray}
&&\alpha_{_1}=b_{_4}-a_{_1}-1,\;\alpha_{_2}=b_{_4}-a_{_2}-1,
\nonumber\\
&&\alpha_{_3}=b_{_4}-a_{_3}-a_{_5}-1,\;\alpha_{_4}=b_{_4}-a_{_4}-a_{_5}-1,
\nonumber\\
&&\alpha_{_{6}}=b_{_1}-1,\;\alpha_{_7}=a_{_5}+b_{_2}-b_{_4},\;\alpha_{_{8}}=b_{_3}-1,
\nonumber\\
&&\alpha_{_{10}}=b_{_5}-b_{_4},\;\alpha_{_{12}}=a_{_5}-b_{_4}+1,\;\alpha_{_{14}}=-a_{_5}.
\label{GKZ21l-6-1}
\end{eqnarray}
The corresponding hypergeometric solutions are written as
\begin{eqnarray}
&&\Phi_{_{[\tilde{1}\tilde{3}5\tilde{7}]}}^{(6),a}(\alpha,z)=
y_{_2}^{{D\over2}-1}y_{_4}^{-1}\sum\limits_{n_{_1}=0}^\infty
\sum\limits_{n_{_2}=0}^\infty\sum\limits_{n_{_3}=0}^\infty\sum\limits_{n_{_4}=0}^\infty
c_{_{[\tilde{1}\tilde{3}5\tilde{7}]}}^{(6),a}(\alpha,{\bf n})
\nonumber\\
&&\hspace{2.5cm}\times
y_{_1}^{n_{_1}}y_{_4}^{n_{_2}}\Big({y_{_2}\over y_{_4}}\Big)^{n_{_3}}
\Big({y_{_3}\over y_{_4}}\Big)^{n_{_4}}
\;,\nonumber\\
&&\Phi_{_{[\tilde{1}\tilde{3}5\tilde{7}]}}^{(6),b}(\alpha,z)=
y_{_2}^{{D\over2}-2}y_{_3}y_{_4}^{-1}\sum\limits_{n_{_1}=0}^\infty
\sum\limits_{n_{_2}=0}^\infty\sum\limits_{n_{_3}=0}^\infty\sum\limits_{n_{_4}=0}^\infty
c_{_{[\tilde{1}\tilde{3}5\tilde{7}]}}^{(6),b}(\alpha,{\bf n})
\nonumber\\
&&\hspace{2.5cm}\times
y_{_1}^{n_{_1}}y_{_3}^{n_{_2}}\Big({y_{_3}\over y_{_4}}\Big)^{n_{_3}}
\Big({y_{_3}\over y_{_2}}\Big)^{n_{_4}}\;.
\label{GKZ21l-6-2a}
\end{eqnarray}
Where the coefficients are
\begin{eqnarray}
&&c_{_{[\tilde{1}\tilde{3}5\tilde{7}]}}^{(6),a}(\alpha,{\bf n})=
(-)^{n_{_2}}\Gamma(1+n_{_3}+n_{_4})
\Big\{n_{_1}!n_{_2}!n_{_3}!n_{_4}!\Gamma({D\over2}-1-n_{_1}-n_{_2})
\nonumber\\
&&\hspace{2.5cm}\times
\Gamma(D-2-n_{_1}-n_{_2})\Gamma({D\over2}-1-n_{_3}-n_{_4})
\Gamma(2-{D\over2}+n_{_1})
\nonumber\\
&&\hspace{2.5cm}\times
\Gamma(2-{D\over2}+n_{_4})
\Gamma(2-{D\over2}+n_{_2})\Gamma({D\over2}+n_{_3})\Big\}^{-1}
\;,\nonumber\\
&&c_{_{[\tilde{1}\tilde{3}5\tilde{7}]}}^{(6),b}(\alpha,{\bf n})=
(-)^{n_{_2}+n_{_4}}\Gamma(1+n_{_2}+n_{_3})\Gamma(1+n_{_4})
\Big\{n_{_1}!n_{_2}!\Gamma(2+n_{_2}+n_{_3}+n_{_4})
\nonumber\\
&&\hspace{2.5cm}\times
\Gamma({D\over2}-1-n_{_1}-n_{_2})\Gamma(D-2-n_{_1}-n_{_2})
\nonumber\\
&&\hspace{2.5cm}\times
\Gamma({D\over2}-1-n_{_2}-n_{_3})
\Gamma(2-{D\over2}+n_{_1})\Gamma({D\over2}-1-n_{_4})
\nonumber\\
&&\hspace{2.5cm}\times
\Gamma(2-{D\over2}+n_{_2})\Gamma(3-{D\over2}+n_{_2}+n_{_3}+n_{_4})\Big\}^{-1}\;.
\label{GKZ21l-6-3}
\end{eqnarray}

\item   $I_{_{7}}=\{1,2,3,4,6,7,9,12,13,14\}$, i.e. the implement $J_{_{7}}=[1,14]\setminus I_{_{7}}=\{5,8,10,11\}$.
The choice implies the power numbers $\alpha_{_5}=\alpha_{_{8}}=\alpha_{_{10}}=\alpha_{_{11}}=0$, and
\begin{eqnarray}
&&\alpha_{_1}=b_{_5}-a_{_1}-1,\;\alpha_{_2}=b_{_5}-a_{_2}-1,
\nonumber\\
&&\alpha_{_3}=b_{_5}-a_{_3}-a_{_5}-1,\;\alpha_{_4}=b_{_5}-a_{_4}-a_{_5}-1,
\nonumber\\
&&\alpha_{_{6}}=b_{_1}-1,\;\alpha_{_7}=a_{_5}+b_{_2}+b_{_3}-b_{_5}-1,\;\alpha_{_{9}}=b_{_4}-b_{_5},\;
\nonumber\\
&&\alpha_{_{12}}=a_{_5}+b_{_3}-b_{_5},\;\alpha_{_{13}}=1-b_{_3},\;
\alpha_{_{14}}=-a_{_5}.
\label{GKZ21l-7-1}
\end{eqnarray}
The corresponding hypergeometric solutions are
\begin{eqnarray}
&&\Phi_{_{[\tilde{1}\tilde{3}5\tilde{7}]}}^{(7),a}(\alpha,z)=
y_{_2}^{{D\over2}-1}y_{_3}^{{D\over2}-1}y_{_4}^{-1}
\sum\limits_{n_{_1}=0}^\infty
\sum\limits_{n_{_2}=0}^\infty\sum\limits_{n_{_3}=0}^\infty\sum\limits_{n_{_4}=0}^\infty
c_{_{[\tilde{1}\tilde{3}5\tilde{7}]}}^{(7),a}(\alpha,{\bf n})
\nonumber\\
&&\hspace{2.5cm}\times
y_{_1}^{n_{_1}}y_{_4}^{n_{_2}}\Big({y_{_2}\over y_{_4}}\Big)^{n_{_3}}
\Big({y_{_3}\over y_{_4}}\Big)^{n_{_4}}
\;,\nonumber\\
&&\Phi_{_{[\tilde{1}\tilde{3}5\tilde{7}]}}^{(7),b}(\alpha,z)=
y_{_2}^{{D\over2}-2}y_{_3}^{{D\over2}}y_{_4}^{-1}\sum\limits_{n_{_1}=0}^\infty
\sum\limits_{n_{_2}=0}^\infty\sum\limits_{n_{_3}=0}^\infty\sum\limits_{n_{_4}=0}^\infty
c_{_{[\tilde{1}\tilde{3}5\tilde{7}]}}^{(7),b}(\alpha,{\bf n})
\nonumber\\
&&\hspace{2.5cm}\times
y_{_1}^{n_{_1}}y_{_3}^{n_{_2}}\Big({y_{_3}\over y_{_4}}\Big)^{n_{_3}}
\Big({y_{_3}\over y_{_2}}\Big)^{n_{_4}}\;.
\label{GKZ21l-7-2a}
\end{eqnarray}
Where the coefficients are
\begin{eqnarray}
&&c_{_{[\tilde{1}\tilde{3}5\tilde{7}]}}^{(7),a}(\alpha,{\bf n})=
(-)^{n_{_1}}\Gamma(1+n_{_1}+n_{_2})
\Gamma(1+n_{_3}+n_{_4})\Big\{n_{_1}!n_{_2}!n_{_3}!n_{_4}!
\nonumber\\
&&\hspace{2.5cm}\times
\Gamma({D\over2}-1-n_{_1}-n_{_2})
\Gamma(1-{D\over2}-n_{_3}-n_{_4})\Gamma(2-{D\over2}+n_{_1})
\nonumber\\
&&\hspace{2.5cm}\times
\Gamma({D\over2}+n_{_2})\Gamma({D\over2}+n_{_3})\Gamma({D\over2}+n_{_4})\Big\}^{-1}
\;,\nonumber\\
&&c_{_{[\tilde{1}\tilde{3}5\tilde{7}]}}^{(7),b}(\alpha,{\bf n})=
(-)^{n_{_1}+n_{_4}}\Gamma(1+n_{_1}+n_{_2})\Gamma(1+n_{_2}+n_{_3})\Gamma(1+n_{_4})\Big\{n_{_1}!n_{_2}!
\nonumber\\
&&\hspace{2.5cm}\times
\Gamma(2+n_{_2}+n_{_3}+n_{_4})\Gamma({D\over2}-1-n_{_1}-n_{_2})
\Gamma(1-{D\over2}-n_{_2}-n_{_3})
\nonumber\\
&&\hspace{2.5cm}\times
\Gamma(2-{D\over2}+n_{_1})\Gamma({D\over2}+n_{_2})
\Gamma({D\over2}-n_{_4})\Gamma({D\over2}+1+n_{_2}+n_{_3}+n_{_4})\Big\}^{-1}\;.
\label{GKZ21l-7-3}
\end{eqnarray}

\item   $I_{_{8}}=\{1,2,3,4,6,7,8,9,12,14\}$, i.e. the implement $J_{_{8}}=[1,14]\setminus I_{_{8}}=\{5,10,11,13\}$.
The choice implies the power numbers $\alpha_{_5}=\alpha_{_{10}}=\alpha_{_{11}}=\alpha_{_{13}}=0$, and
\begin{eqnarray}
&&\alpha_{_1}=b_{_5}-a_{_1}-1,\;\alpha_{_2}=b_{_5}-a_{_2}-1,
\nonumber\\
&&\alpha_{_3}=b_{_5}-a_{_3}-a_{_5}-1,\;\alpha_{_4}=b_{_5}-a_{_4}-a_{_5}-1,
\nonumber\\
&&\alpha_{_{6}}=b_{_1}-1,\;\alpha_{_7}=a_{_5}+b_{_2}-b_{_5},\;\alpha_{_{8}}=b_{_3}-1,
\nonumber\\
&&\alpha_{_{9}}=b_{_4}-b_{_5},\;\alpha_{_{12}}=a_{_5}-b_{_5}+1,\;\alpha_{_{14}}=-a_{_5}.
\label{GKZ21l-8-1}
\end{eqnarray}
The corresponding hypergeometric series is written as
\begin{eqnarray}
&&\Phi_{_{[\tilde{1}\tilde{3}5\tilde{7}]}}^{(8)}(\alpha,z)=
y_{_2}^{{D}-2}y_{_4}^{-1}\sum\limits_{n_{_1}=0}^\infty
\sum\limits_{n_{_2}=0}^\infty\sum\limits_{n_{_3}=0}^\infty\sum\limits_{n_{_4}=0}^\infty
c_{_{[\tilde{1}\tilde{3}5\tilde{7}]}}^{(8)}(\alpha,{\bf n})
\nonumber\\
&&\hspace{2.5cm}\times
y_{_1}^{n_{_1}}y_{_2}^{n_{_2}}\Big({y_{_2}\over y_{_4}}\Big)^{n_{_3}}
\Big({y_{_3}\over y_{_2}}\Big)^{n_{_4}}\;,
\label{GKZ21l-8-2}
\end{eqnarray}
with
\begin{eqnarray}
&&c_{_{[\tilde{1}\tilde{3}5\tilde{7}]}}^{(8)}(\alpha,{\bf n})=
(-)^{n_{_2}}\Gamma(1+n_{_1}+n_{_2})
\Gamma(1+n_{_2}+n_{_3})\Big\{n_{_1}!n_{_2}!n_{_4}!
\nonumber\\
&&\hspace{2.5cm}\times
\Gamma({D\over2}-1-n_{_1}-n_{_2})\Gamma(1-{D\over2}-n_{_2}-n_{_3})
\Gamma(2-{D\over2}+n_{_1})
\nonumber\\
&&\hspace{2.5cm}\times
\Gamma({D\over2}+n_{_2}+n_{_3}-n_{_4})\Gamma(2-{D\over2}+n_{_4})\Gamma({D\over2}+n_{_2})
\nonumber\\
&&\hspace{2.5cm}\times
\Gamma(D-1+n_{_2}+n_{_3}-n_{_4})\Big\}^{-1}\;.
\label{GKZ21l-8-3}
\end{eqnarray}
\end{itemize}

\section{The hypergeometric solutions of the integer lattice ${\bf B}_{_{\tilde{1}3\widetilde{57}}}$\label{app14}}
\indent\indent

\begin{itemize}
\item   $I_{_{1}}=\{1,2,4,5,7,10,\cdots,14\}$, i.e. the implement $J_{_{1}}=[1,14]\setminus I_{_{1}}=\{3,6,8,9\}$.
The choice implies the power numbers $\alpha_{_3}=\alpha_{_{6}}=\alpha_{_{8}}=\alpha_{_{9}}=0$, and
\begin{eqnarray}
&&\alpha_{_1}=b_{_1}+b_{_4}-a_{_1}-2,\;\alpha_{_2}=b_{_1}+b_{_4}-a_{_2}-2,
\nonumber\\
&&\alpha_{_4}=a_{_3}-a_{_4},\;\alpha_{_5}=b_{_4}-a_{_3}-a_{_5}-1,
\nonumber\\
&&\alpha_{_7}=b_{_2}+b_{_3}-a_{_3}-2,\;\alpha_{_{10}}=b_{_5}-b_{_4},\;\alpha_{_{11}}=1-b_{_1},
\nonumber\\
&&\alpha_{_{12}}=b_{_3}-a_{_3}-1,\;\alpha_{_{13}}=1-b_{_3},\;
\alpha_{_{14}}=a_{_3}-b_{_4}+1.
\label{GKZ21m-1-1}
\end{eqnarray}
The corresponding hypergeometric series is written as
\begin{eqnarray}
&&\Phi_{_{[\tilde{1}3\tilde{5}\tilde{7}]}}^{(1)}(\alpha,z)=
y_{_1}^{{D\over2}-1}y_{_2}^{-1}y_{_3}^{{D\over2}-1}\sum\limits_{n_{_1}=0}^\infty
\sum\limits_{n_{_2}=0}^\infty\sum\limits_{n_{_3}=0}^\infty\sum\limits_{n_{_4}=0}^\infty
c_{_{[\tilde{1}3\tilde{5}\tilde{7}]}}^{(1)}(\alpha,{\bf n})
\nonumber\\
&&\hspace{2.5cm}\times
y_{_1}^{n_{_1}}\Big({y_{_4}\over y_{_2}}\Big)^{n_{_2}}
y_{_4}^{n_{_3}}\Big({y_{_3}\over y_{_2}}\Big)^{n_{_4}}\;,
\label{GKZ21m-1-2}
\end{eqnarray}
with
\begin{eqnarray}
&&c_{_{[\tilde{1}3\tilde{5}\tilde{7}]}}^{(1)}(\alpha,{\bf n})=
(-)^{n_{_1}+n_{_4}}\Gamma(1+n_{_1}+n_{_3})\Gamma(1+n_{_2}+n_{_4})\Big\{n_{_1}!n_{_2}!n_{_3}!n_{_4}!
\nonumber\\
&&\hspace{2.5cm}\times
\Gamma({D\over2}-1-n_{_1}-n_{_3})\Gamma({D\over2}+n_{_2})
\Gamma(1-{D\over2}-n_{_2}-n_{_4})
\nonumber\\
&&\hspace{2.5cm}\times
\Gamma(2-{D\over2}+n_{_3})\Gamma({D\over2}+n_{_1})\Gamma({D\over2}+n_{_4})\Big\}^{-1}\;.
\label{GKZ21m-1-3}
\end{eqnarray}

\item   $I_{_{2}}=\{1,2,4,5,7,8,10,11,12,14\}$, i.e. the implement $J_{_{2}}=[1,14]\setminus I_{_{2}}=\{3,6,9,13\}$.
The choice implies the power numbers $\alpha_{_3}=\alpha_{_{6}}=\alpha_{_{9}}=\alpha_{_{13}}=0$, and
\begin{eqnarray}
&&\alpha_{_1}=b_{_1}+b_{_4}-a_{_1}-2,\;\alpha_{_2}=b_{_1}+b_{_4}-a_{_2}-2,\;\alpha_{_4}=a_{_3}-a_{_4},
\nonumber\\
&&\alpha_{_5}=b_{_4}-a_{_3}-a_{_5}-1,\;\alpha_{_7}=b_{_2}-a_{_3}-1,\;\alpha_{_{8}}=b_{_3}-1,
\nonumber\\
&&\alpha_{_{10}}=b_{_5}-b_{_4},\;\alpha_{_{11}}=1-b_{_1},\;
\alpha_{_{12}}=-a_{_3},\;\alpha_{_{14}}=a_{_3}-b_{_4}+1.
\label{GKZ21m-2-1}
\end{eqnarray}
The corresponding hypergeometric solution is written as
\begin{eqnarray}
&&\Phi_{_{[\tilde{1}3\tilde{5}\tilde{7}]}}^{(2)}(\alpha,z)=
y_{_1}^{{D\over2}-1}y_{_2}^{{D\over2}-2}\sum\limits_{n_{_1}=0}^\infty
\sum\limits_{n_{_2}=0}^\infty\sum\limits_{n_{_3}=0}^\infty\sum\limits_{n_{_4}=0}^\infty
c_{_{[\tilde{1}3\tilde{5}\tilde{7}]}}^{(2)}(\alpha,{\bf n})
\nonumber\\
&&\hspace{2.5cm}\times
y_{_1}^{n_{_1}}\Big({y_{_4}\over y_{_2}}\Big)^{n_{_2}}
y_{_4}^{n_{_3}}\Big({y_{_3}\over y_{_2}}\Big)^{n_{_4}}\;,
\label{GKZ21m-2-2}
\end{eqnarray}
with
\begin{eqnarray}
&&c_{_{[\tilde{1}3\tilde{5}\tilde{7}]}}^{(2)}(\alpha,{\bf n})=
(-)^{n_{_1}+n_{_4}}\Gamma(1+n_{_1}+n_{_3})\Gamma(1+n_{_2}+n_{_4})
\Big\{n_{_1}!n_{_2}!n_{_3}!n_{_4}!
\nonumber\\
&&\hspace{2.5cm}\times
\Gamma({D\over2}-1-n_{_1}-n_{_3})\Gamma({D\over2}+n_{_2})
\Gamma(2-{D\over2}+n_{_4})
\nonumber\\
&&\hspace{2.5cm}\times
\Gamma(2-{D\over2}+n_{_3})\Gamma({D\over2}+n_{_1})
\Gamma({D\over2}-1-n_{_2}-n_{_4})\Big\}^{-1}\;.
\label{GKZ21m-2-3}
\end{eqnarray}

\item   $I_{_{3}}=\{1,2,4,5,7,9,11,12,13,14\}$, i.e. the implement $J_{_{3}}=[1,14]\setminus I_{_{3}}=\{3,6,8,10\}$.
The choice implies the power numbers $\alpha_{_3}=\alpha_{_{6}}=\alpha_{_{8}}=\alpha_{_{10}}=0$, and
\begin{eqnarray}
&&\alpha_{_1}=b_{_1}+b_{_5}-a_{_1}-2,\;\alpha_{_2}=b_{_1}+b_{_5}-a_{_2}-2,
\nonumber\\
&&\alpha_{_4}=a_{_3}-a_{_4},\;\alpha_{_5}=b_{_5}-a_{_3}-a_{_5}-1,
\nonumber\\
&&\alpha_{_7}=b_{_2}+b_{_3}-a_{_3}-2,\;\alpha_{_{9}}=b_{_4}-b_{_5},\;\alpha_{_{11}}=1-b_{_1},
\nonumber\\
&&\alpha_{_{12}}=b_{_3}-a_{_3}-1,\;\alpha_{_{13}}=1-b_{_3},\;
\alpha_{_{14}}=a_{_3}-b_{_5}+1.
\label{GKZ21m-3-1}
\end{eqnarray}
The corresponding hypergeometric function is written as
\begin{eqnarray}
&&\Phi_{_{[\tilde{1}3\tilde{5}\tilde{7}]}}^{(3)}(\alpha,z)=
y_{_1}^{{D\over2}-1}y_{_2}^{-1}y_{_3}^{{D\over2}-1}y_{_4}^{{D\over2}-1}
\sum\limits_{n_{_1}=0}^\infty
\sum\limits_{n_{_2}=0}^\infty\sum\limits_{n_{_3}=0}^\infty\sum\limits_{n_{_4}=0}^\infty
c_{_{[\tilde{1}3\tilde{5}\tilde{7}]}}^{(3)}(\alpha,{\bf n})
\nonumber\\
&&\hspace{2.5cm}\times
y_{_1}^{n_{_1}}\Big({y_{_4}\over y_{_2}}\Big)^{n_{_2}}
y_{_4}^{n_{_3}}\Big({y_{_3}\over y_{_2}}\Big)^{n_{_4}}\;,
\label{GKZ21m-3-2}
\end{eqnarray}
with
\begin{eqnarray}
&&c_{_{[\tilde{1}3\tilde{5}\tilde{7}]}}^{(3)}(\alpha,{\bf n})=
\Gamma(1+n_{_1}+n_{_3})\Gamma(1+n_{_2}+n_{_4})
\nonumber\\
&&\hspace{2.5cm}\times
\Big\{n_{_1}!n_{_2}!n_{_3}!n_{_4}!\Gamma(1-{D\over2}-n_{_1}-n_{_3})\Gamma({D\over2}+n_{_2})
\nonumber\\
&&\hspace{2.5cm}\times
\Gamma(1-{D\over2}-n_{_2}-n_{_4})\Gamma({D\over2}+n_{_3})\Gamma({D\over2}+n_{_1})
\Gamma({D\over2}+n_{_4})\Big\}^{-1}\;.
\label{GKZ21m-3-3}
\end{eqnarray}

\item   $I_{_{4}}=\{1,2,4,5,7,8,9,11,12,14\}$, i.e. the implement $J_{_{4}}=[1,14]\setminus I_{_{4}}=\{3,6,10,13\}$.
The choice implies the power numbers $\alpha_{_3}=\alpha_{_{6}}=\alpha_{_{10}}=\alpha_{_{13}}=0$, and
\begin{eqnarray}
&&\alpha_{_1}=b_{_1}+b_{_5}-a_{_1}-2,\;\alpha_{_2}=b_{_1}+b_{_5}-a_{_2}-2,\;\alpha_{_4}=a_{_3}-a_{_4},
\nonumber\\
&&\alpha_{_5}=b_{_5}-a_{_3}-a_{_5}-1,\;\alpha_{_7}=b_{_2}-a_{_3}-1,\;\alpha_{_{8}}=b_{_3}-1,
\nonumber\\
&&\alpha_{_{9}}=b_{_4}-b_{_5},\;\alpha_{_{11}}=1-b_{_1},\;
\alpha_{_{12}}=-a_{_3},\;\alpha_{_{14}}=a_{_3}-b_{_5}+1.
\label{GKZ21m-4-1}
\end{eqnarray}
The corresponding hypergeometric series is
\begin{eqnarray}
&&\Phi_{_{[\tilde{1}3\tilde{5}\tilde{7}]}}^{(4)}(\alpha,z)=
y_{_1}^{{D\over2}-1}y_{_2}^{{D\over2}-2}y_{_4}^{{D\over2}-1}\sum\limits_{n_{_1}=0}^\infty
\sum\limits_{n_{_2}=0}^\infty\sum\limits_{n_{_3}=0}^\infty\sum\limits_{n_{_4}=0}^\infty
c_{_{[\tilde{1}3\tilde{5}\tilde{7}]}}^{(4)}(\alpha,{\bf n})
\nonumber\\
&&\hspace{2.5cm}\times
y_{_1}^{n_{_1}}\Big({y_{_4}\over y_{_2}}\Big)^{n_{_2}}
y_{_4}^{n_{_3}}\Big({y_{_3}\over y_{_2}}\Big)^{n_{_4}}\;,
\label{GKZ21m-4-2}
\end{eqnarray}
with
\begin{eqnarray}
&&c_{_{[\tilde{1}3\tilde{5}\tilde{7}]}}^{(4)}(\alpha,{\bf n})=
\Gamma(1+n_{_1}+n_{_3})\Gamma(1+n_{_2}+n_{_4})
\Big\{n_{_1}!n_{_2}!n_{_3}!n_{_4}!
\nonumber\\
&&\hspace{2.5cm}\times
\Gamma(1-{D\over2}-n_{_1}-n_{_3})
\Gamma({D\over2}+n_{_2})\Gamma(2-{D\over2}+n_{_4})
\nonumber\\
&&\hspace{2.5cm}\times
\Gamma({D\over2}+n_{_3})\Gamma({D\over2}+n_{_1})\Gamma({D\over2}-1-n_{_2}-n_{_4})\Big\}^{-1}\;.
\label{GKZ21m-4-3}
\end{eqnarray}

\item   $I_{_{5}}=\{1,2,3,5,7,10,\cdots,14\}$, i.e. the implement $J_{_{5}}=[1,14]\setminus I_{_{5}}=\{4,6,8,9\}$.
The choice implies the power numbers $\alpha_{_4}=\alpha_{_{6}}=\alpha_{_{8}}=\alpha_{_{9}}=0$, and
\begin{eqnarray}
&&\alpha_{_1}=b_{_1}+b_{_4}-a_{_1}-2,\;\alpha_{_2}=b_{_1}+b_{_4}-a_{_2}-2,
\nonumber\\
&&\alpha_{_3}=a_{_4}-a_{_3},\;\alpha_{_5}=b_{_4}-a_{_4}-a_{_5}-1,
\nonumber\\
&&\alpha_{_7}=b_{_2}+b_{_3}-a_{_4}-2,\;\alpha_{_{10}}=b_{_5}-b_{_4},\;\alpha_{_{11}}=1-b_{_1},
\nonumber\\
&&\alpha_{_{12}}=b_{_3}-a_{_4}-1,\;\alpha_{_{13}}=1-b_{_3},\;
\alpha_{_{14}}=a_{_4}-b_{_4}+1.
\label{GKZ21m-7-1}
\end{eqnarray}
The corresponding hypergeometric series is written as
\begin{eqnarray}
&&\Phi_{_{[\tilde{1}3\tilde{5}\tilde{7}]}}^{(5)}(\alpha,z)=
y_{_1}^{{D\over2}-1}y_{_2}^{{D\over2}-2}y_{_3}^{{D\over2}-1}y_{_4}^{1-{D\over2}}
\sum\limits_{n_{_1}=0}^\infty
\sum\limits_{n_{_2}=0}^\infty\sum\limits_{n_{_3}=0}^\infty\sum\limits_{n_{_4}=0}^\infty
c_{_{[\tilde{1}3\tilde{5}\tilde{7}]}}^{(5)}(\alpha,{\bf n})
\nonumber\\
&&\hspace{2.5cm}\times
y_{_1}^{n_{_1}}\Big({y_{_4}\over y_{_2}}\Big)^{n_{_2}}
y_{_4}^{n_{_3}}\Big({y_{_3}\over y_{_2}}\Big)^{n_{_4}}\;,
\label{GKZ21m-7-2}
\end{eqnarray}
with
\begin{eqnarray}
&&c_{_{[\tilde{1}3\tilde{5}\tilde{7}]}}^{(5)}(\alpha,{\bf n})=
(-)^{1+n_{_1}+n_{_4}}\Gamma(1+n_{_1}+n_{_3})\Gamma(1+n_{_2}+n_{_4})
\Big\{n_{_1}!n_{_2}!n_{_3}!n_{_4}!
\nonumber\\
&&\hspace{2.5cm}\times
\Gamma({D\over2}-1-n_{_1}-n_{_3})\Gamma(2-{D\over2}+n_{_2})\Gamma(2-{D\over2}+n_{_3})
\nonumber\\
&&\hspace{2.5cm}\times
\Gamma({D\over2}+n_{_1})\Gamma({D\over2}-1-n_{_2}-n_{_4})\Gamma({D\over2}+n_{_4})\Big\}^{-1}\;.
\label{GKZ21m-7-3}
\end{eqnarray}

\item   $I_{_{6}}=\{1,2,3,5,7,8,10,11,12,14\}$, i.e. the implement $J_{_{6}}=[1,14]\setminus I_{_{6}}=\{4,6,9,13\}$.
The choice implies the power numbers $\alpha_{_4}=\alpha_{_{6}}=\alpha_{_{9}}=\alpha_{_{13}}=0$, and
\begin{eqnarray}
&&\alpha_{_1}=b_{_1}+b_{_4}-a_{_1}-2,\;\alpha_{_2}=b_{_1}+b_{_4}-a_{_2}-2,\;\alpha_{_3}=a_{_4}-a_{_3},
\nonumber\\
&&\alpha_{_5}=b_{_4}-a_{_4}-a_{_5}-1,\;\alpha_{_7}=b_{_2}-a_{_4}-1,\;\alpha_{_{8}}=b_{_3}-1,
\nonumber\\
&&\alpha_{_{10}}=b_{_5}-b_{_4},\;\alpha_{_{11}}=1-b_{_1},\;
\alpha_{_{12}}=-a_{_4},\;\alpha_{_{14}}=a_{_4}-b_{_4}+1.
\label{GKZ21m-8-1}
\end{eqnarray}
The corresponding hypergeometric solution is
\begin{eqnarray}
&&\Phi_{_{[\tilde{1}3\tilde{5}\tilde{7}]}}^{(6)}(\alpha,z)=
y_{_1}^{{D\over2}-1}y_{_2}^{{D}-3}y_{_4}^{1-{D\over2}}\sum\limits_{n_{_1}=0}^\infty
\sum\limits_{n_{_2}=0}^\infty\sum\limits_{n_{_3}=0}^\infty\sum\limits_{n_{_4}=0}^\infty
c_{_{[\tilde{1}3\tilde{5}\tilde{7}]}}^{(6)}(\alpha,{\bf n})
\nonumber\\
&&\hspace{2.5cm}\times
y_{_1}^{n_{_1}}\Big({y_{_4}\over y_{_2}}\Big)^{n_{_2}}
y_{_4}^{n_{_3}}\Big({y_{_3}\over y_{_2}}\Big)^{n_{_4}}\;,
\label{GKZ21m-8-2}
\end{eqnarray}
with
\begin{eqnarray}
&&c_{_{[\tilde{1}3\tilde{5}\tilde{7}]}}^{(6)}(\alpha,{\bf n})=
(-)^{1+n_{_1}+n_{_2}}\Gamma(1+n_{_1}+n_{_3})\Big\{n_{_1}!n_{_2}!n_{_3}!n_{_4}!
\Gamma({D\over2}-1-n_{_1}-n_{_3})
\nonumber\\
&&\hspace{2.5cm}\times
\Gamma(2-{D\over2}+n_{_2})
\Gamma({D\over2}-1-n_{_2}-n_{_4})\Gamma(2-{D\over2}+n_{_4})
\nonumber\\
&&\hspace{2.5cm}\times
\Gamma(2-{D\over2}+n_{_3})\Gamma({D\over2}+n_{_1})
\Gamma(D-2-n_{_2}-n_{_4})\Big\}^{-1}\;.
\label{GKZ21m-8-3}
\end{eqnarray}

\item   $I_{_{7}}=\{1,2,3,5,7,9,11,12,13,14\}$, i.e. the implement $J_{_{7}}=[1,14]\setminus I_{_{7}}=\{4,6,8,10\}$.
The choice implies the power numbers $\alpha_{_4}=\alpha_{_{6}}=\alpha_{_{8}}=\alpha_{_{10}}=0$, and
\begin{eqnarray}
&&\alpha_{_1}=b_{_1}+b_{_5}-a_{_1}-2,\;\alpha_{_2}=b_{_1}+b_{_5}-a_{_2}-2,
\nonumber\\
&&\alpha_{_3}=a_{_4}-a_{_3},\;\alpha_{_5}=b_{_5}-a_{_4}-a_{_5}-1,
\nonumber\\
&&\alpha_{_7}=b_{_2}+b_{_3}-a_{_4}-2,\;\alpha_{_{9}}=b_{_4}-b_{_5},\;\alpha_{_{11}}=1-b_{_1},
\nonumber\\
&&\alpha_{_{12}}=b_{_3}-a_{_4}-1,\;\alpha_{_{13}}=1-b_{_3},\;
\alpha_{_{14}}=a_{_4}-b_{_5}+1.
\label{GKZ21m-9-1}
\end{eqnarray}
The corresponding hypergeometric solution is written as
\begin{eqnarray}
&&\Phi_{_{[\tilde{1}3\tilde{5}\tilde{7}]}}^{(7)}(\alpha,z)=
y_{_1}^{{D\over2}-1}y_{_2}^{{D\over2}-2}y_{_3}^{{D\over2}-1}\sum\limits_{n_{_1}=0}^\infty
\sum\limits_{n_{_2}=0}^\infty\sum\limits_{n_{_3}=0}^\infty\sum\limits_{n_{_4}=0}^\infty
c_{_{[\tilde{1}3\tilde{5}\tilde{7}]}}^{(7)}(\alpha,{\bf n})
\nonumber\\
&&\hspace{2.5cm}\times
y_{_1}^{n_{_1}}\Big({y_{_4}\over y_{_2}}\Big)^{n_{_2}}
y_{_4}^{n_{_3}}\Big({y_{_3}\over y_{_2}}\Big)^{n_{_4}}\;,
\label{GKZ21m-9-2}
\end{eqnarray}
with
\begin{eqnarray}
&&c_{_{[\tilde{1}3\tilde{5}\tilde{7}]}}^{(7)}(\alpha,{\bf n})=
(-)^{n_{_1}+n_{_4}}\Gamma(1+n_{_1}+n_{_3})\Gamma(1+n_{_2}+n_{_4})
\Big\{n_{_1}!n_{_2}!n_{_3}!n_{_4}!
\nonumber\\
&&\hspace{2.5cm}\times
\Gamma(1-{D\over2}-n_{_1}-n_{_3})\Gamma(2-{D\over2}+n_{_2})
\Gamma({D\over2}+n_{_3})
\nonumber\\
&&\hspace{2.5cm}\times
\Gamma({D\over2}+n_{_1})\Gamma({D\over2}-1-n_{_2}-n_{_4})
\Gamma({D\over2}+n_{_4})\Big\}^{-1}\;.
\label{GKZ21m-9-3}
\end{eqnarray}

\item   $I_{_{8}}=\{1,2,3,5,7,8,9,11,12,14\}$, i.e. the implement $J_{_{8}}=[1,14]\setminus I_{_{8}}=\{4,6,10,13\}$.
The choice implies the power numbers $\alpha_{_4}=\alpha_{_{6}}=\alpha_{_{10}}=\alpha_{_{13}}=0$, and
\begin{eqnarray}
&&\alpha_{_1}=b_{_1}+b_{_5}-a_{_1}-2,\;\alpha_{_2}=b_{_1}+b_{_5}-a_{_2}-2,\;\alpha_{_3}=a_{_4}-a_{_3},
\nonumber\\
&&\alpha_{_5}=b_{_5}-a_{_4}-a_{_5}-1,\;\alpha_{_7}=b_{_2}-a_{_4}-1,\;\alpha_{_{8}}=b_{_3}-1,
\nonumber\\
&&\alpha_{_{9}}=b_{_4}-b_{_5},\;\alpha_{_{11}}=1-b_{_1},\;
\alpha_{_{12}}=-a_{_4},\;\alpha_{_{14}}=a_{_4}-b_{_5}+1.
\label{GKZ21m-10-1}
\end{eqnarray}
The corresponding hypergeometric function is written as
\begin{eqnarray}
&&\Phi_{_{[\tilde{1}3\tilde{5}\tilde{7}]}}^{(8)}(\alpha,z)=
y_{_1}^{{D\over2}-1}y_{_2}^{{D}-3}\sum\limits_{n_{_1}=0}^\infty
\sum\limits_{n_{_2}=0}^\infty\sum\limits_{n_{_3}=0}^\infty\sum\limits_{n_{_4}=0}^\infty
c_{_{[\tilde{1}3\tilde{5}\tilde{7}]}}^{(8)}(\alpha,{\bf n})
\nonumber\\
&&\hspace{2.5cm}\times
y_{_1}^{n_{_1}}\Big({y_{_4}\over y_{_2}}\Big)^{n_{_2}}
y_{_4}^{n_{_3}}\Big({y_{_3}\over y_{_2}}\Big)^{n_{_4}}\;,
\label{GKZ21m-10-2}
\end{eqnarray}
with
\begin{eqnarray}
&&c_{_{[\tilde{1}3\tilde{5}\tilde{7}]}}^{(8)}(\alpha,{\bf n})=
(-)^{n_{_1}+n_{_2}}\Gamma(1+n_{_1}+n_{_3})\Big\{n_{_1}!n_{_2}!n_{_3}!n_{_4}!
\Gamma(1-{D\over2}-n_{_1}-n_{_3})
\nonumber\\
&&\hspace{2.5cm}\times
\Gamma(2-{D\over2}+n_{_2})\Gamma({D\over2}-1-n_{_2}-n_{_4})\Gamma(2-{D\over2}+n_{_4})
\nonumber\\
&&\hspace{2.5cm}\times
\Gamma({D\over2}+n_{_3})\Gamma({D\over2}+n_{_1})\Gamma(D-2-n_{_2}-n_{_4})\Big\}^{-1}\;.
\label{GKZ21m-10-3}
\end{eqnarray}

\item   $I_{_{9}}=\{1,2,4,5,6,7,10,12,13,14\}$, i.e. the implement $J_{_{9}}=[1,14]\setminus I_{_{9}}=\{3,8,9,11\}$.
The choice implies the power numbers $\alpha_{_3}=\alpha_{_{8}}=\alpha_{_{9}}=\alpha_{_{11}}=0$, and
\begin{eqnarray}
&&\alpha_{_1}=b_{_4}-a_{_1}-1,\;\alpha_{_2}=b_{_4}-a_{_2}-1,\;\alpha_{_4}=a_{_3}-a_{_4},
\nonumber\\
&&\alpha_{_5}=b_{_4}-a_{_3}-a_{_5}-1,\;\alpha_{_{6}}=b_{_1}-1,
\nonumber\\
&&\alpha_{_7}=b_{_2}+b_{_3}-a_{_3}-2,\;\alpha_{_{10}}=b_{_5}-b_{_4},
\nonumber\\
&&\alpha_{_{12}}=b_{_3}-a_{_3}-1,\;\alpha_{_{13}}=1-b_{_3},\;
\alpha_{_{14}}=a_{_3}-b_{_4}+1.
\label{GKZ21m-13-1}
\end{eqnarray}
The corresponding hypergeometric series is written as
\begin{eqnarray}
&&\Phi_{_{[\tilde{1}3\tilde{5}\tilde{7}]}}^{(9)}(\alpha,z)=
y_{_2}^{-1}y_{_3}^{{D\over2}-1}\sum\limits_{n_{_1}=0}^\infty
\sum\limits_{n_{_2}=0}^\infty\sum\limits_{n_{_3}=0}^\infty\sum\limits_{n_{_4}=0}^\infty
c_{_{[\tilde{1}3\tilde{5}\tilde{7}]}}^{(9)}(\alpha,{\bf n})
\nonumber\\
&&\hspace{2.5cm}\times
y_{_1}^{n_{_1}}\Big({y_{_4}\over y_{_2}}\Big)^{n_{_2}}
y_{_4}^{n_{_3}}\Big({y_{_3}\over y_{_2}}\Big)^{n_{_4}}\;,
\label{GKZ21m-13-2}
\end{eqnarray}
with
\begin{eqnarray}
&&c_{_{[\tilde{1}3\tilde{5}\tilde{7}]}}^{(9)}(\alpha,{\bf n})=
(-)^{n_{_3}+n_{_4}}\Gamma(1+n_{_2}+n_{_4})
\Big\{n_{_1}!n_{_2}!n_{_3}!n_{_4}!\Gamma({D\over2}-1-n_{_1}-n_{_3})
\nonumber\\
&&\hspace{2.5cm}\times
\Gamma(D-2-n_{_1}-n_{_3})\Gamma({D\over2}+n_{_2})\Gamma(2-{D\over2}+n_{_1})
\nonumber\\
&&\hspace{2.5cm}\times
\Gamma(1-{D\over2}-n_{_2}-n_{_4})
\Gamma(2-{D\over2}+n_{_3})\Gamma({D\over2}+n_{_4})\Big\}^{-1}\;.
\label{GKZ21m-13-3}
\end{eqnarray}

\item   $I_{_{10}}=\{1,2,4,\cdots,8,10,12,14\}$, i.e. the implement $J_{_{10}}=[1,14]\setminus I_{_{10}}=\{3,9,11,13\}$.
The choice implies the power numbers $\alpha_{_3}=\alpha_{_{9}}=\alpha_{_{11}}=\alpha_{_{13}}=0$, and
\begin{eqnarray}
&&\alpha_{_1}=b_{_4}-a_{_1}-1,\;\alpha_{_2}=b_{_4}-a_{_2}-1,\;\alpha_{_4}=a_{_3}-a_{_4},
\nonumber\\
&&\alpha_{_5}=b_{_4}-a_{_3}-a_{_5}-1,\;\alpha_{_{6}}=b_{_1}-1,\;\alpha_{_7}=b_{_2}-a_{_3}-1,
\nonumber\\
&&\alpha_{_{8}}=b_{_3}-1,\;\alpha_{_{10}}=b_{_5}-b_{_4},\;\alpha_{_{12}}=-a_{_3},\;\alpha_{_{14}}=a_{_3}-b_{_4}+1.
\label{GKZ21m-14-1}
\end{eqnarray}
The corresponding hypergeometric series is
\begin{eqnarray}
&&\Phi_{_{[\tilde{1}3\tilde{5}\tilde{7}]}}^{(10)}(\alpha,z)=
y_{_2}^{{D\over2}-2}\sum\limits_{n_{_1}=0}^\infty
\sum\limits_{n_{_2}=0}^\infty\sum\limits_{n_{_3}=0}^\infty\sum\limits_{n_{_4}=0}^\infty
c_{_{[\tilde{1}3\tilde{5}\tilde{7}]}}^{(10)}(\alpha,{\bf n})
\nonumber\\
&&\hspace{2.5cm}\times
y_{_1}^{n_{_1}}\Big({y_{_4}\over y_{_2}}\Big)^{n_{_2}}
y_{_4}^{n_{_3}}\Big({y_{_3}\over y_{_2}}\Big)^{n_{_4}}\;,
\label{GKZ21m-14-2}
\end{eqnarray}
with
\begin{eqnarray}
&&c_{_{[\tilde{1}3\tilde{5}\tilde{7}]}}^{(10)}(\alpha,{\bf n})=
(-)^{n_{_3}+n_{_4}}\Gamma(1+n_{_2}+n_{_4})
\Big\{n_{_1}!n_{_2}!n_{_3}!n_{_4}!\Gamma({D\over2}-1-n_{_1}-n_{_3})
\nonumber\\
&&\hspace{2.5cm}\times
\Gamma(D-2-n_{_1}-n_{_3})\Gamma({D\over2}+n_{_2})\Gamma(2-{D\over2}+n_{_1})
\nonumber\\
&&\hspace{2.5cm}\times
\Gamma(2-{D\over2}+n_{_4})\Gamma(2-{D\over2}+n_{_3})
\Gamma({D\over2}-1-n_{_2}-n_{_4})\Big\}^{-1}\;.
\label{GKZ21m-14-3}
\end{eqnarray}

\item   $I_{_{11}}=\{1,2,4,5,6,7,9,12,13,14\}$, i.e. the implement $J_{_{11}}=[1,14]\setminus I_{_{11}}=\{3,8,10,11\}$.
The choice implies the power numbers $\alpha_{_3}=\alpha_{_{8}}=\alpha_{_{10}}=\alpha_{_{11}}=0$, and
\begin{eqnarray}
&&\alpha_{_1}=b_{_5}-a_{_1}-1,\;\alpha_{_2}=b_{_5}-a_{_2}-1,
\nonumber\\
&&\alpha_{_4}=a_{_3}-a_{_4},\;\alpha_{_5}=b_{_5}-a_{_3}-a_{_5}-1,\;\alpha_{_{6}}=b_{_1}-1,
\nonumber\\
&&\alpha_{_7}=b_{_2}+b_{_3}-a_{_3}-2,\;\alpha_{_{9}}=b_{_4}-b_{_5},
\nonumber\\
&&\alpha_{_{12}}=b_{_3}-a_{_3}-1,\;\alpha_{_{13}}=1-b_{_3},\;
\alpha_{_{14}}=a_{_3}-b_{_5}+1.
\label{GKZ21m-15-1}
\end{eqnarray}
The corresponding hypergeometric series is written as
\begin{eqnarray}
&&\Phi_{_{[\tilde{1}3\tilde{5}\tilde{7}]}}^{(11)}(\alpha,z)=
y_{_2}^{-1}y_{_3}^{{D\over2}-1}y_{_4}^{{D\over2}-1}\sum\limits_{n_{_1}=0}^\infty
\sum\limits_{n_{_2}=0}^\infty\sum\limits_{n_{_3}=0}^\infty\sum\limits_{n_{_4}=0}^\infty
c_{_{[\tilde{1}3\tilde{5}\tilde{7}]}}^{(11)}(\alpha,{\bf n})
\nonumber\\
&&\hspace{2.5cm}\times
y_{_1}^{n_{_1}}\Big({y_{_4}\over y_{_2}}\Big)^{n_{_2}}
y_{_4}^{n_{_3}}\Big({y_{_3}\over y_{_2}}\Big)^{n_{_4}}\;,
\label{GKZ21m-15-2}
\end{eqnarray}
with
\begin{eqnarray}
&&c_{_{[\tilde{1}3\tilde{5}\tilde{7}]}}^{(11)}(\alpha,{\bf n})=
(-)^{n_{_1}+n_{_4}}\Gamma(1+n_{_1}+n_{_3})\Gamma(1+n_{_2}+n_{_4})
\Big\{n_{_1}!n_{_2}!n_{_3}!n_{_4}!
\nonumber\\
&&\hspace{2.5cm}\times
\Gamma({D\over2}-1-n_{_1}-n_{_3})\Gamma({D\over2}+n_{_2})\Gamma(2-{D\over2}+n_{_1})
\nonumber\\
&&\hspace{2.5cm}\times
\Gamma(1-{D\over2}-n_{_2}-n_{_4})\Gamma({D\over2}+n_{_3})\Gamma({D\over2}+n_{_4})\Big\}^{-1}\;.
\label{GKZ21m-15-3}
\end{eqnarray}

\item   $I_{_{12}}=\{1,2,4,\cdots,9,12,14\}$, i.e. the implement $J_{_{12}}=[1,14]\setminus I_{_{12}}=\{3,10,11,13\}$.
The choice implies the power numbers $\alpha_{_3}=\alpha_{_{10}}=\alpha_{_{11}}=\alpha_{_{13}}=0$, and
\begin{eqnarray}
&&\alpha_{_1}=b_{_5}-a_{_1}-1,\;\alpha_{_2}=b_{_5}-a_{_2}-1,\;\alpha_{_4}=a_{_3}-a_{_4},
\nonumber\\
&&\alpha_{_5}=b_{_5}-a_{_3}-a_{_5}-1,\;\alpha_{_{6}}=b_{_1}-1,\;\alpha_{_7}=b_{_2}-a_{_3}-1,
\nonumber\\
&&\alpha_{_{8}}=b_{_3}-1,\;\alpha_{_{9}}=b_{_4}-b_{_5},\;
\alpha_{_{12}}=-a_{_3},\;\alpha_{_{14}}=a_{_3}-b_{_5}+1.
\label{GKZ21m-16-1}
\end{eqnarray}
The corresponding hypergeometric solution is
\begin{eqnarray}
&&\Phi_{_{[\tilde{1}3\tilde{5}\tilde{7}]}}^{(12)}(\alpha,z)=
y_{_2}^{{D\over2}-2}y_{_4}^{{D\over2}-1}\sum\limits_{n_{_1}=0}^\infty
\sum\limits_{n_{_2}=0}^\infty\sum\limits_{n_{_3}=0}^\infty\sum\limits_{n_{_4}=0}^\infty
c_{_{[\tilde{1}3\tilde{5}\tilde{7}]}}^{(12)}(\alpha,{\bf n})
\nonumber\\
&&\hspace{2.5cm}\times
y_{_1}^{n_{_1}}\Big({y_{_4}\over y_{_2}}\Big)^{n_{_2}}
y_{_4}^{n_{_3}}\Big({y_{_3}\over y_{_2}}\Big)^{n_{_4}}\;,
\label{GKZ21m-16-2}
\end{eqnarray}
with
\begin{eqnarray}
&&c_{_{[\tilde{1}3\tilde{5}\tilde{7}]}}^{(12)}(\alpha,{\bf n})=
(-)^{n_{_1}+n_{_4}}\Gamma(1+n_{_1}+n_{_3})\Gamma(1+n_{_2}+n_{_4})
\Big\{n_{_1}!n_{_2}!n_{_3}!n_{_4}!
\nonumber\\
&&\hspace{2.5cm}\times
\Gamma({D\over2}-1-n_{_1}-n_{_3})\Gamma({D\over2}+n_{_2})\Gamma(2-{D\over2}+n_{_1})
\nonumber\\
&&\hspace{2.5cm}\times
\Gamma(2-{D\over2}+n_{_4})\Gamma({D\over2}+n_{_3})
\Gamma({D\over2}-1-n_{_2}-n_{_4})\Big\}^{-1}\;.
\label{GKZ21m-16-3}
\end{eqnarray}

\item   $I_{_{13}}=\{1,2,3,5,6,7,10,12,13,14\}$, i.e. the implement $J_{_{13}}=[1,14]\setminus I_{_{13}}=\{4,8,9,11\}$.
The choice implies the power numbers $\alpha_{_4}=\alpha_{_{8}}=\alpha_{_{9}}=\alpha_{_{11}}=0$, and
\begin{eqnarray}
&&\alpha_{_1}=b_{_4}-a_{_1}-1,\;\alpha_{_2}=b_{_4}-a_{_2}-1,\;\alpha_{_3}=a_{_4}-a_{_3},
\nonumber\\
&&\alpha_{_5}=b_{_4}-a_{_4}-a_{_5}-1,\;\alpha_{_{6}}=b_{_1}-1,
\nonumber\\
&&\alpha_{_7}=b_{_2}+b_{_3}-a_{_4}-2,\;\alpha_{_{10}}=b_{_5}-b_{_4},
\nonumber\\
&&\alpha_{_{12}}=b_{_3}-a_{_4}-1,\;\alpha_{_{13}}=1-b_{_3},\;
\alpha_{_{14}}=a_{_4}-b_{_4}+1.
\label{GKZ21m-19-1}
\end{eqnarray}
The corresponding hypergeometric solution is given as
\begin{eqnarray}
&&\Phi_{_{[\tilde{1}3\tilde{5}\tilde{7}]}}^{(13)}(\alpha,z)=
y_{_2}^{{D\over2}-2}y_{_3}^{{D\over2}-1}y_{_4}^{1-{D\over2}}\sum\limits_{n_{_1}=0}^\infty
\sum\limits_{n_{_2}=0}^\infty\sum\limits_{n_{_3}=0}^\infty\sum\limits_{n_{_4}=0}^\infty
c_{_{[\tilde{1}3\tilde{5}\tilde{7}]}}^{(13)}(\alpha,{\bf n})
\nonumber\\
&&\hspace{2.5cm}\times
y_{_1}^{n_{_1}}\Big({y_{_4}\over y_{_2}}\Big)^{n_{_2}}
y_{_4}^{n_{_3}}\Big({y_{_3}\over y_{_2}}\Big)^{n_{_4}}\;,
\label{GKZ21m-19-2}
\end{eqnarray}
with
\begin{eqnarray}
&&c_{_{[\tilde{1}3\tilde{5}\tilde{7}]}}^{(13)}(\alpha,{\bf n})=
(-)^{1+n_{_3}+n_{_4}}\Gamma(1+n_{_2}+n_{_4})\Big\{n_{_1}!n_{_2}!n_{_3}!n_{_4}!
\Gamma({D\over2}-1-n_{_1}-n_{_3})
\nonumber\\
&&\hspace{2.5cm}\times
\Gamma(D-2-n_{_1}-n_{_3})\Gamma(2-{D\over2}+n_{_2})
\Gamma(2-{D\over2}+n_{_1})
\nonumber\\
&&\hspace{2.5cm}\times
\Gamma(2-{D\over2}+n_{_3})\Gamma({D\over2}-1-n_{_2}-n_{_4})
\Gamma({D\over2}+n_{_4})\Big\}^{-1}\;.
\label{GKZ21m-19-3}
\end{eqnarray}

\item   $I_{_{14}}=\{1,2,3,5,6,7,8,10,12,14\}$, i.e. the implement $J_{_{14}}=[1,14]\setminus I_{_{14}}=\{4,9,11,13\}$.
The choice implies the power numbers $\alpha_{_4}=\alpha_{_{9}}=\alpha_{_{11}}=\alpha_{_{13}}=0$, and
\begin{eqnarray}
&&\alpha_{_1}=b_{_4}-a_{_1}-1,\;\alpha_{_2}=b_{_4}-a_{_2}-1,\;\alpha_{_3}=a_{_4}-a_{_3},
\nonumber\\
&&\alpha_{_5}=b_{_4}-a_{_4}-a_{_5}-1,\;\alpha_{_{6}}=b_{_1}-1,\;\alpha_{_7}=b_{_2}-a_{_4}-1,
\nonumber\\
&&\alpha_{_{8}}=b_{_3}-1,\;\alpha_{_{10}}=b_{_5}-b_{_4},\;
\alpha_{_{12}}=-a_{_4},\;\alpha_{_{14}}=a_{_4}-b_{_4}+1.
\label{GKZ21m-20-1}
\end{eqnarray}
The corresponding hypergeometric solution is
\begin{eqnarray}
&&\Phi_{_{[\tilde{1}3\tilde{5}\tilde{7}]}}^{(14)}(\alpha,z)=
y_{_2}^{{D}-3}y_{_4}^{1-{D\over2}}\sum\limits_{n_{_1}=0}^\infty
\sum\limits_{n_{_2}=0}^\infty\sum\limits_{n_{_3}=0}^\infty\sum\limits_{n_{_4}=0}^\infty
c_{_{[\tilde{1}3\tilde{5}\tilde{7}]}}^{(14)}(\alpha,{\bf n})
\nonumber\\
&&\hspace{2.5cm}\times
y_{_1}^{n_{_1}}\Big({y_{_4}\over y_{_2}}\Big)^{n_{_2}}
y_{_4}^{n_{_3}}\Big({y_{_3}\over y_{_2}}\Big)^{n_{_4}}\;,
\label{GKZ21m-20-2}
\end{eqnarray}
with
\begin{eqnarray}
&&c_{_{[\tilde{1}3\tilde{5}\tilde{7}]}}^{(14)}(\alpha,{\bf n})=
(-)^{1+n_{_2}+n_{_3}}
\Big\{n_{_1}!n_{_2}!n_{_3}!n_{_4}!\Gamma({D\over2}-1-n_{_1}-n_{_3})
\Gamma(D-2-n_{_1}-n_{_3})
\nonumber\\
&&\hspace{2.5cm}\times
\Gamma(2-{D\over2}+n_{_2})\Gamma(2-{D\over2}+n_{_1})
\Gamma({D\over2}-1-n_{_2}-n_{_4})
\nonumber\\
&&\hspace{2.5cm}\times
\Gamma(2-{D\over2}+n_{_4})\Gamma(2-{D\over2}+n_{_3})
\Gamma(D-2-n_{_2}-n_{_4})\Big\}^{-1}\;.
\label{GKZ21m-20-3}
\end{eqnarray}

\item   $I_{_{15}}=\{1,2,3,5,6,7,9,12,13,14\}$, i.e. the implement $J_{_{15}}=[1,14]\setminus I_{_{15}}=\{4,8,10,11\}$.
The choice implies the power numbers $\alpha_{_4}=\alpha_{_{8}}=\alpha_{_{10}}=\alpha_{_{11}}=0$, and
\begin{eqnarray}
&&\alpha_{_1}=b_{_5}-a_{_1}-1,\;\alpha_{_2}=b_{_5}-a_{_2}-1,
\nonumber\\
&&\alpha_{_3}=a_{_4}-a_{_3},\;\alpha_{_5}=b_{_5}-a_{_4}-a_{_5}-1,\;\alpha_{_{6}}=b_{_1}-1,
\nonumber\\
&&\alpha_{_7}=b_{_2}+b_{_3}-a_{_4}-2,\;\alpha_{_{9}}=b_{_4}-b_{_5},
\nonumber\\
&&\alpha_{_{12}}=b_{_3}-a_{_4}-1,\;\alpha_{_{13}}=1-b_{_3},\;
\alpha_{_{14}}=a_{_4}-b_{_5}+1.
\label{GKZ21m-21-1}
\end{eqnarray}
The corresponding hypergeometric series is written as
\begin{eqnarray}
&&\Phi_{_{[\tilde{1}3\tilde{5}\tilde{7}]}}^{(15)}(\alpha,z)=
y_{_2}^{{D\over2}-2}y_{_3}^{{D\over2}-1}\sum\limits_{n_{_1}=0}^\infty
\sum\limits_{n_{_2}=0}^\infty\sum\limits_{n_{_3}=0}^\infty\sum\limits_{n_{_4}=0}^\infty
c_{_{[\tilde{1}3\tilde{5}\tilde{7}]}}^{(15)}(\alpha,{\bf n})
\nonumber\\
&&\hspace{2.5cm}\times
y_{_1}^{n_{_1}}\Big({y_{_4}\over y_{_2}}\Big)^{n_{_2}}
y_{_4}^{n_{_3}}\Big({y_{_3}\over y_{_2}}\Big)^{n_{_4}}\;,
\label{GKZ21m-21-2}
\end{eqnarray}
with
\begin{eqnarray}
&&c_{_{[\tilde{1}3\tilde{5}\tilde{7}]}}^{(15)}(\alpha,{\bf n})=
(-)^{n_{_1}+n_{_4}}\Gamma(1+n_{_1}+n_{_3})\Gamma(1+n_{_2}+n_{_4})
\Big\{n_{_1}!n_{_2}!n_{_3}!n_{_4}!
\nonumber\\
&&\hspace{2.5cm}\times
\Gamma({D\over2}-1-n_{_1}-n_{_3})\Gamma(2-{D\over2}+n_{_2})
\Gamma(2-{D\over2}+n_{_1})
\nonumber\\
&&\hspace{2.5cm}\times
\Gamma({D\over2}+n_{_3})\Gamma({D\over2}-1-n_{_2}-n_{_4})
\Gamma({D\over2}+n_{_4})\Big\}^{-1}\;.
\label{GKZ21m-21-3}
\end{eqnarray}

\item   $I_{_{16}}=\{1,2,3,5,6,7,8,9,12,14\}$, i.e. the implement $J_{_{16}}=[1,14]\setminus I_{_{16}}=\{4,10,11,13\}$.
The choice implies the power numbers $\alpha_{_4}=\alpha_{_{10}}=\alpha_{_{11}}=\alpha_{_{13}}=0$, and
\begin{eqnarray}
&&\alpha_{_1}=b_{_5}-a_{_1}-1,\;\alpha_{_2}=b_{_5}-a_{_2}-1,\;\alpha_{_3}=a_{_4}-a_{_3},
\nonumber\\
&&\alpha_{_5}=b_{_5}-a_{_4}-a_{_5}-1,\;\alpha_{_{6}}=b_{_1}-1,\;\alpha_{_7}=b_{_2}-a_{_4}-1,
\nonumber\\
&&\alpha_{_{8}}=b_{_3}-1,\;\alpha_{_{9}}=b_{_4}-b_{_5},\;
\alpha_{_{12}}=-a_{_4},\;\alpha_{_{14}}=a_{_4}-b_{_5}+1.
\label{GKZ21m-22-1}
\end{eqnarray}
The corresponding hypergeometric series is given as
\begin{eqnarray}
&&\Phi_{_{[\tilde{1}3\tilde{5}\tilde{7}]}}^{(16)}(\alpha,z)=
y_{_2}^{{D}-3}\sum\limits_{n_{_1}=0}^\infty
\sum\limits_{n_{_2}=0}^\infty\sum\limits_{n_{_3}=0}^\infty\sum\limits_{n_{_4}=0}^\infty
c_{_{[\tilde{1}3\tilde{5}\tilde{7}]}}^{(16)}(\alpha,{\bf n})
\nonumber\\
&&\hspace{2.5cm}\times
y_{_1}^{n_{_1}}\Big({y_{_4}\over y_{_2}}\Big)^{n_{_2}}
y_{_4}^{n_{_3}}\Big({y_{_3}\over y_{_2}}\Big)^{n_{_4}}\;,
\label{GKZ21m-22-2}
\end{eqnarray}
with
\begin{eqnarray}
&&c_{_{[\tilde{1}3\tilde{5}\tilde{7}]}}^{(16)}(\alpha,{\bf n})=
(-)^{n_{_1}+n_{_2}}\Gamma(1+n_{_1}+n_{_3})\Big\{n_{_1}!n_{_2}!n_{_3}!n_{_4}!
\Gamma({D\over2}-1-n_{_1}-n_{_3})
\nonumber\\
&&\hspace{2.5cm}\times
\Gamma(2-{D\over2}+n_{_2})\Gamma(2-{D\over2}+n_{_1})
\Gamma({D\over2}-1-n_{_2}-n_{_4})
\nonumber\\
&&\hspace{2.5cm}\times
\Gamma(2-{D\over2}+n_{_4})\Gamma({D\over2}+n_{_3})
\Gamma(D-2-n_{_2}-n_{_4})\Big\}^{-1}\;.
\label{GKZ21m-22-3}
\end{eqnarray}

\item   $I_{_{17}}=\{1,2,4,5,7,9,\cdots,13\}$, i.e. the implement $J_{_{17}}=[1,14]\setminus I_{_{17}}=\{3,6,8,14\}$.
The choice implies the power numbers $\alpha_{_3}=\alpha_{_{6}}=\alpha_{_{8}}=\alpha_{_{14}}=0$, and
\begin{eqnarray}
&&\alpha_{_1}=a_{_3}+b_{_3}-a_{_1}-1,\;\alpha_{_2}=a_{_3}+b_{_3}-a_{_2}-1,\;\alpha_{_4}=a_{_3}-a_{_4},
\nonumber\\
&&\alpha_{_5}=-a_{_5},\;\alpha_{_7}=b_{_1}+b_{_2}-a_{_3}-2,\;\alpha_{_{9}}=b_{_1}+b_{_4}-a_{_3}-b_{_3}-1,
\nonumber\\
&&\alpha_{_{10}}=b_{_1}+b_{_5}-a_{_3}-b_{_3}-1,\;\alpha_{_{11}}=1-b_{_1},\;\alpha_{_{12}}=b_{_1}-a_{_3}-1,\;
\nonumber\\
&&\alpha_{_{13}}=1-b_{_3}.
\label{GKZ21m-5-1}
\end{eqnarray}
The corresponding hypergeometric solutions are written as
\begin{eqnarray}
&&\Phi_{_{[\tilde{1}3\tilde{5}\tilde{7}]}}^{(17),a}(\alpha,z)=
y_{_1}^{{D\over2}-1}y_{_2}^{-1}y_{_3}^{{D\over2}-1}\sum\limits_{n_{_1}=0}^\infty
\sum\limits_{n_{_2}=0}^\infty\sum\limits_{n_{_3}=0}^\infty\sum\limits_{n_{_4}=0}^\infty
c_{_{[\tilde{1}3\tilde{5}\tilde{7}]}}^{(17),a}(\alpha,{\bf n})
\nonumber\\
&&\hspace{2.5cm}\times
y_{_1}^{n_{_1}}\Big({y_{_4}\over y_{_2}}\Big)^{n_{_2}}
y_{_4}^{n_{_3}}\Big({y_{_3}\over y_{_2}}\Big)^{n_{_4}}
\;,\nonumber\\
&&\Phi_{_{[\tilde{1}3\tilde{5}\tilde{7}]}}^{(17),b}(\alpha,z)=
y_{_1}^{{D\over2}-1}y_{_2}^{-2}y_{_3}^{{D\over2}-1}\sum\limits_{n_{_1}=0}^\infty
\sum\limits_{n_{_2}=0}^\infty\sum\limits_{n_{_3}=0}^\infty\sum\limits_{n_{_4}=0}^\infty
c_{_{[\tilde{1}3\tilde{5}\tilde{7}]}}^{(17),b}(\alpha,{\bf n})
\nonumber\\
&&\hspace{2.5cm}\times
\Big({y_{_1}\over y_{_2}}\Big)^{n_{_1}}\Big({1\over y_{_2}}\Big)^{n_{_2}}
\Big({y_{_4}\over y_{_2}}\Big)^{n_{_3}}\Big({y_{_3}\over y_{_2}}\Big)^{n_{_4}}
\;,\nonumber\\
&&\Phi_{_{[\tilde{1}3\tilde{5}\tilde{7}]}}^{(17),c}(\alpha,z)=
y_{_1}^{{D\over2}}y_{_2}^{-2}y_{_3}^{{D\over2}-1}\sum\limits_{n_{_1}=0}^\infty
\sum\limits_{n_{_2}=0}^\infty\sum\limits_{n_{_3}=0}^\infty\sum\limits_{n_{_4}=0}^\infty
c_{_{[\tilde{1}3\tilde{5}\tilde{7}]}}^{(17),c}(\alpha,{\bf n})
\nonumber\\
&&\hspace{2.5cm}\times
y_{_1}^{n_{_1}}\Big({y_{_1}\over y_{_2}}\Big)^{n_{_2}}
\Big({y_{_4}\over y_{_2}}\Big)^{n_{_3}}\Big({y_{_3}\over y_{_2}}\Big)^{n_{_4}}\;.
\label{GKZ21m-5-2a}
\end{eqnarray}
Where the coefficients are
\begin{eqnarray}
&&c_{_{[\tilde{1}3\tilde{5}\tilde{7}]}}^{(17),a}(\alpha,{\bf n})=
(-)^{n_{_1}+n_{_4}}\Gamma(1+n_{_1}+n_{_3})\Gamma(1+n_{_2}+n_{_4})\Big\{n_{_1}!n_{_2}!n_{_3}!n_{_4}!
\nonumber\\
&&\hspace{2.5cm}\times
\Gamma({D\over2}-1-n_{_1}-n_{_3})
\Gamma({D\over2}+n_{_2})\Gamma(1-{D\over2}-n_{_2}-n_{_4})
\nonumber\\
&&\hspace{2.5cm}\times
\Gamma(2-{D\over2}+n_{_3})\Gamma({D\over2}+n_{_1})
\Gamma({D\over2}+n_{_4})\Big\}^{-1}
\;,\nonumber\\
&&c_{_{[\tilde{1}3\tilde{5}\tilde{7}]}}^{(17),b}(\alpha,{\bf n})=
(-)^{n_{_4}}\Gamma(1+n_{_1}+n_{_2})\Gamma(2+n_{_1}+n_{_2}+n_{_3}+n_{_4})\Big\{n_{_1}!n_{_2}!n_{_4}!
\nonumber\\
&&\hspace{2.5cm}\times
\Gamma(2+n_{_1}+n_{_2}+n_{_3})\Gamma({D\over2}+n_{_2})
\Gamma({D\over2}+1+n_{_1}+n_{_2}+n_{_3})
\nonumber\\
&&\hspace{2.5cm}\times
\Gamma(-{D\over2}-n_{_1}-n_{_2}-n_{_3}-n_{_4})
\Gamma(1-{D\over2}-n_{_1}-n_{_2})\Gamma({D\over2}+n_{_1})
\nonumber\\
&&\hspace{2.5cm}\times
\Gamma({D\over2}+n_{_4})\Big\}^{-1}
\;,\nonumber\\
&&c_{_{[\tilde{1}3\tilde{5}\tilde{7}]}}^{(17),c}(\alpha,{\bf n})=
(-)^{1+n_{_1}+n_{_4}}\Gamma(1+n_{_1})\Gamma(1+n_{_2})\Gamma(2+n_{_2}+n_{_3}+n_{_4})
\nonumber\\
&&\hspace{2.5cm}\times
\Big\{n_{_4}!\Gamma(2+n_{_1}+n_{_2})
\Gamma(2+n_{_2}+n_{_3})\Gamma({D\over2}-1-n_{_1})
\nonumber\\
&&\hspace{2.5cm}\times
\Gamma({D\over2}+1+n_{_2}+n_{_3})\Gamma(-{D\over2}-n_{_2}-n_{_3}-n_{_4})
\nonumber\\
&&\hspace{2.5cm}\times
\Gamma(1-{D\over2}-n_{_2})\Gamma({D\over2}+1+n_{_1}+n_{_2})\Gamma({D\over2}+n_{_4})\Big\}^{-1}\;.
\label{GKZ21m-5-3}
\end{eqnarray}

\item   $I_{_{18}}=\{1,2,4,5,7,\cdots,12\}$, i.e. the implement $J_{_{18}}=[1,14]\setminus I_{_{18}}=\{3,6,13,14\}$.
The choice implies the power numbers $\alpha_{_3}=\alpha_{_{6}}=\alpha_{_{13}}=\alpha_{_{14}}=0$, and
\begin{eqnarray}
&&\alpha_{_1}=a_{_3}+b_{_1}-a_{_1}-1,\;\alpha_{_2}=a_{_3}+b_{_1}-a_{_2}-1,\;\alpha_{_4}=a_{_3}-a_{_4},
\nonumber\\
&&\alpha_{_5}=-a_{_5},\;\alpha_{_7}=b_{_2}-a_{_3}-1,\;\alpha_{_{8}}=b_{_3}-1,\;\alpha_{_{9}}=b_{_4}-a_{_3}-1,
\nonumber\\
&&\alpha_{_{10}}=b_{_5}-a_{_3}-1,\;\alpha_{_{11}}=1-b_{_1},\;\alpha_{_{12}}=-a_{_3}.
\label{GKZ21m-6-1}
\end{eqnarray}
The corresponding hypergeometric functions are
\begin{eqnarray}
&&\Phi_{_{[\tilde{1}3\tilde{5}\tilde{7}]}}^{(18),a}(\alpha,z)=
y_{_1}^{{D\over2}-1}y_{_2}^{{D\over2}-2}\sum\limits_{n_{_1}=0}^\infty
\sum\limits_{n_{_2}=0}^\infty\sum\limits_{n_{_3}=0}^\infty\sum\limits_{n_{_4}=0}^\infty
c_{_{[\tilde{1}3\tilde{5}\tilde{7}]}}^{(18),a}(\alpha,{\bf n})
\nonumber\\
&&\hspace{2.5cm}\times
y_{_1}^{n_{_1}}\Big({y_{_4}\over y_{_2}}\Big)^{n_{_2}}
y_{_4}^{n_{_3}}\Big({y_{_3}\over y_{_2}}\Big)^{n_{_4}}
\;,\nonumber\\
&&\Phi_{_{[\tilde{1}3\tilde{5}\tilde{7}]}}^{(18),b}(\alpha,z)=
y_{_1}^{{D\over2}-1}y_{_2}^{{D\over2}-3}\sum\limits_{n_{_1}=0}^\infty
\sum\limits_{n_{_2}=0}^\infty\sum\limits_{n_{_3}=0}^\infty\sum\limits_{n_{_4}=0}^\infty
c_{_{[\tilde{1}3\tilde{5}\tilde{7}]}}^{(18),b}(\alpha,{\bf n})
\nonumber\\
&&\hspace{2.5cm}\times
\Big({y_{_1}\over y_{_2}}\Big)^{n_{_1}}\Big({1\over y_{_2}}\Big)^{n_{_2}}
\Big({y_{_4}\over y_{_2}}\Big)^{n_{_3}}\Big({y_{_3}\over y_{_2}}\Big)^{n_{_4}}
\;,\nonumber\\
&&\Phi_{_{[\tilde{1}3\tilde{5}\tilde{7}]}}^{(18),c}(\alpha,z)=
y_{_1}^{{D\over2}}y_{_2}^{{D\over2}-3}\sum\limits_{n_{_1}=0}^\infty
\sum\limits_{n_{_2}=0}^\infty\sum\limits_{n_{_3}=0}^\infty\sum\limits_{n_{_4}=0}^\infty
c_{_{[\tilde{1}3\tilde{5}\tilde{7}]}}^{(18),c}(\alpha,{\bf n})
\nonumber\\
&&\hspace{2.5cm}\times
y_{_1}^{n_{_1}}\Big({y_{_1}\over y_{_2}}\Big)^{n_{_2}}
\Big({y_{_4}\over y_{_2}}\Big)^{n_{_3}}\Big({y_{_3}\over y_{_2}}\Big)^{n_{_4}}\;.
\label{GKZ21m-6-2a}
\end{eqnarray}
Where the coefficients are
\begin{eqnarray}
&&c_{_{[\tilde{1}3\tilde{5}\tilde{7}]}}^{(18),a}(\alpha,{\bf n})=
(-)^{n_{_1}+n_{_4}}\Gamma(1+n_{_1}+n_{_3})\Gamma(1+n_{_2}+n_{_4})\Big\{n_{_1}!n_{_2}!n_{_3}!n_{_4}!
\nonumber\\
&&\hspace{2.5cm}\times
\Gamma({D\over2}-1-n_{_1}-n_{_3})
\Gamma({D\over2}+n_{_2})\Gamma({D\over2}-1-n_{_2}-n_{_4})
\nonumber\\
&&\hspace{2.5cm}\times
\Gamma(2-{D\over2}+n_{_3})\Gamma({D\over2}+n_{_1})
\Gamma(2-{D\over2}+n_{_4})\Big\}^{-1}
\;,\nonumber\\
&&c_{_{[\tilde{1}3\tilde{5}\tilde{7}]}}^{(18),b}(\alpha,{\bf n})=
(-)^{n_{_4}}\Gamma(1+n_{_1}+n_{_2})\Gamma(2+n_{_1}+n_{_2}+n_{_3}+n_{_4})\Big\{n_{_1}!n_{_2}!n_{_4}!
\nonumber\\
&&\hspace{2.5cm}\times
\Gamma(2+n_{_1}+n_{_2}+n_{_3})\Gamma({D\over2}+n_{_2})
\Gamma({D\over2}+1+n_{_1}+n_{_2}+n_{_3})
\nonumber\\
&&\hspace{2.5cm}\times
\Gamma({D\over2}-2-n_{_1}-n_{_2}-n_{_3}-n_{_4})
\Gamma(1-{D\over2}-n_{_1}-n_{_2})\Gamma({D\over2}+n_{_1})
\nonumber\\
&&\hspace{2.5cm}\times
\Gamma(2-{D\over2}+n_{_4})\Big\}^{-1}
\;,\nonumber\\
&&c_{_{[\tilde{1}3\tilde{5}\tilde{7}]}}^{(18),c}(\alpha,{\bf n})=
(-)^{1+n_{_1}+n_{_4}}\Gamma(1+n_{_1})\Gamma(1+n_{_2})\Gamma(2+n_{_2}+n_{_3}+n_{_4})
\nonumber\\
&&\hspace{2.5cm}\times
\Big\{n_{_4}!\Gamma(2+n_{_1}+n_{_2})
\Gamma(2+n_{_2}+n_{_3})\Gamma({D\over2}-1-n_{_1})
\nonumber\\
&&\hspace{2.5cm}\times
\Gamma({D\over2}+1+n_{_2}+n_{_3})\Gamma({D\over2}-2-n_{_2}-n_{_3}-n_{_4})
\nonumber\\
&&\hspace{2.5cm}\times
\Gamma(1-{D\over2}-n_{_2})\Gamma({D\over2}+1+n_{_1}+n_{_2})\Gamma(2-{D\over2}+n_{_4})\Big\}^{-1}\;.
\label{GKZ21m-6-3}
\end{eqnarray}

\item   $I_{_{19}}=\{1,2,3,5,7,9,\cdots,13\}$, i.e. the implement $J_{_{19}}=[1,14]\setminus I_{_{19}}=\{4,6,8,14\}$.
The choice implies the power numbers $\alpha_{_4}=\alpha_{_{6}}=\alpha_{_{8}}=\alpha_{_{14}}=0$, and
\begin{eqnarray}
&&\alpha_{_1}=a_{_4}+b_{_3}-a_{_1}-1,\;\alpha_{_2}=a_{_4}+b_{_3}-a_{_2}-1,\;\alpha_{_3}=a_{_4}-a_{_3},
\nonumber\\
&&\alpha_{_5}=-a_{_5},\;\alpha_{_7}=b_{_1}+b_{_2}-a_{_4}-2,\;\alpha_{_{9}}=b_{_1}+b_{_4}-a_{_4}-b_{_3}-1,
\nonumber\\
&&\alpha_{_{10}}=b_{_1}+b_{_5}-a_{_4}-b_{_3}-1,\;\alpha_{_{11}}=1-b_{_1},\;\alpha_{_{12}}=b_{_1}-a_{_4}-1,\;
\nonumber\\
&&\alpha_{_{13}}=1-b_{_3}.
\label{GKZ21m-11-1}
\end{eqnarray}
The corresponding hypergeometric functions are presented as
\begin{eqnarray}
&&\Phi_{_{[\tilde{1}3\tilde{5}\tilde{7}]}}^{(19),a}(\alpha,z)=
y_{_1}^{{D\over2}-1}y_{_2}^{{D\over2}-2}y_{_3}^{{D\over2}-1}\sum\limits_{n_{_1}=0}^\infty
\sum\limits_{n_{_2}=0}^\infty\sum\limits_{n_{_3}=0}^\infty\sum\limits_{n_{_4}=0}^\infty
c_{_{[\tilde{1}3\tilde{5}\tilde{7}]}}^{(19),a}(\alpha,{\bf n})
\nonumber\\
&&\hspace{2.5cm}\times
y_{_1}^{n_{_1}}\Big({y_{_4}\over y_{_2}}\Big)^{n_{_2}}
y_{_4}^{n_{_3}}\Big({y_{_3}\over y_{_2}}\Big)^{n_{_4}}
\;,\nonumber\\
&&\Phi_{_{[\tilde{1}3\tilde{5}\tilde{7}]}}^{(19),b}(\alpha,z)=
y_{_1}^{{D\over2}-1}y_{_2}^{{D\over2}-3}y_{_3}^{{D\over2}-1}\sum\limits_{n_{_1}=0}^\infty
\sum\limits_{n_{_2}=0}^\infty\sum\limits_{n_{_3}=0}^\infty\sum\limits_{n_{_4}=0}^\infty
c_{_{[\tilde{1}3\tilde{5}\tilde{7}]}}^{(19),b}(\alpha,{\bf n})
\nonumber\\
&&\hspace{2.5cm}\times
\Big({y_{_1}\over y_{_2}}\Big)^{n_{_1}}\Big({1\over y_{_2}}\Big)^{n_{_2}}
\Big({y_{_4}\over y_{_2}}\Big)^{n_{_3}}\Big({y_{_3}\over y_{_2}}\Big)^{n_{_4}}
\;,\nonumber\\
&&\Phi_{_{[\tilde{1}3\tilde{5}\tilde{7}]}}^{(19),c}(\alpha,z)=
y_{_1}^{{D\over2}}y_{_2}^{{D\over2}-3}y_{_3}^{{D\over2}-1}\sum\limits_{n_{_1}=0}^\infty
\sum\limits_{n_{_2}=0}^\infty\sum\limits_{n_{_3}=0}^\infty\sum\limits_{n_{_4}=0}^\infty
c_{_{[\tilde{1}3\tilde{5}\tilde{7}]}}^{(19),c}(\alpha,{\bf n})
\nonumber\\
&&\hspace{2.5cm}\times
y_{_1}^{n_{_1}}\Big({y_{_1}\over y_{_2}}\Big)^{n_{_2}}
\Big({y_{_4}\over y_{_2}}\Big)^{n_{_3}}\Big({y_{_3}\over y_{_2}}\Big)^{n_{_4}}\;.
\label{GKZ21m-11-2a}
\end{eqnarray}
Where the coefficients are
\begin{eqnarray}
&&c_{_{[\tilde{1}3\tilde{5}\tilde{7}]}}^{(19),a}(\alpha,{\bf n})=
(-)^{n_{_1}+n_{_4}}\Gamma(1+n_{_1}+n_{_3})\Gamma(1+n_{_2}+n_{_4})
\Big\{n_{_1}!n_{_2}!n_{_3}!n_{_4}!
\nonumber\\
&&\hspace{2.5cm}\times
\Gamma(1-{D\over2}-n_{_1}-n_{_3})\Gamma(2-{D\over2}+n_{_2})\Gamma({D\over2}+n_{_3})
\nonumber\\
&&\hspace{2.5cm}\times
\Gamma({D\over2}+n_{_1})
\Gamma({D\over2}-1-n_{_2}-n_{_4})\Gamma({D\over2}+n_{_4})\Big\}^{-1}
\;,\nonumber\\
&&c_{_{[\tilde{1}3\tilde{5}\tilde{7}]}}^{(19),b}(\alpha,{\bf n})=
(-)^{1+n_{_4}}\Gamma(2+n_{_1}+n_{_2}+n_{_3}+n_{_4})\Gamma(1+n_{_1}+n_{_2})
\nonumber\\
&&\hspace{2.5cm}\times
\Big\{n_{_1}!n_{_2}!n_{_4}!\Gamma(2+n_{_1}+n_{_2}+n_{_3})\Gamma(2-{D\over2}+n_{_2})
\nonumber\\
&&\hspace{2.5cm}\times
\Gamma(3-{D\over2}+n_{_1}+n_{_2}+n_{_3})\Gamma({D\over2}-1-n_{_1}-n_{_2})
\nonumber\\
&&\hspace{2.5cm}\times
\Gamma({D\over2}+n_{_1})
\Gamma({D\over2}-2-n_{_1}-n_{_2}-n_{_3}-n_{_4})\Gamma({D\over2}+n_{_4})\Big\}^{-1}
\;,\nonumber\\
&&c_{_{[\tilde{1}3\tilde{5}\tilde{7}]}}^{(19),c}(\alpha,{\bf n})=
(-)^{1+n_{_1}+n_{_4}}\Gamma(1+n_{_1})\Gamma(1+n_{_2})\Gamma(2+n_{_2}+n_{_3}+n_{_4})
\nonumber\\
&&\hspace{2.5cm}\times
\Big\{n_{_4}!\Gamma(2+n_{_1}+n_{_2})\Gamma(2+n_{_2}+n_{_3})
\Gamma(1-{D\over2}-n_{_1})
\nonumber\\
&&\hspace{2.5cm}\times
\Gamma(3-{D\over2}+n_{_2}+n_{_3})\Gamma({D\over2}-1-n_{_2})
\nonumber\\
&&\hspace{2.5cm}\times
\Gamma({D\over2}+1+n_{_1}+n_{_2})
\Gamma({D\over2}-2-n_{_2}-n_{_3}-n_{_4})\Gamma({D\over2}+n_{_4})\Big\}^{-1}\;.
\label{GKZ21m-11-3}
\end{eqnarray}

\item   $I_{_{20}}=\{1,2,3,5,7,\cdots,12\}$, i.e. the implement $J_{_{20}}=[1,14]\setminus I_{_{20}}=\{4,6,13,14\}$.
The choice implies the power numbers $\alpha_{_4}=\alpha_{_{6}}=\alpha_{_{13}}=\alpha_{_{14}}=0$, and
\begin{eqnarray}
&&\alpha_{_1}=a_{_4}+b_{_1}-a_{_1}-1,\;\alpha_{_2}=a_{_4}+b_{_1}-a_{_2}-1,\;\alpha_{_3}=a_{_4}-a_{_3},
\nonumber\\
&&\alpha_{_5}=-a_{_5},\;\alpha_{_7}=b_{_2}-a_{_4}-1,\;\alpha_{_{8}}=b_{_3}-1,\;\alpha_{_{9}}=b_{_4}-a_{_4}-1,
\nonumber\\
&&\alpha_{_{10}}=b_{_5}-a_{_4}-1,\;\alpha_{_{11}}=1-b_{_1},\;\alpha_{_{12}}=-a_{_4}.
\label{GKZ21m-12-1}
\end{eqnarray}
The corresponding hypergeometric solutions are
\begin{eqnarray}
&&\Phi_{_{[\tilde{1}3\tilde{5}\tilde{7}]}}^{(20),a}(\alpha,z)=
y_{_1}^{{D\over2}-1}y_{_2}^{{D}-3}\sum\limits_{n_{_1}=0}^\infty
\sum\limits_{n_{_2}=0}^\infty\sum\limits_{n_{_3}=0}^\infty\sum\limits_{n_{_4}=0}^\infty
c_{_{[\tilde{1}3\tilde{5}\tilde{7}]}}^{(20),a}(\alpha,{\bf n})
\nonumber\\
&&\hspace{2.5cm}\times
y_{_1}^{n_{_1}}\Big({y_{_4}\over y_{_2}}\Big)^{n_{_2}}
y_{_4}^{n_{_3}}\Big({y_{_3}\over y_{_2}}\Big)^{n_{_4}}
\;,\nonumber\\
&&\Phi_{_{[\tilde{1}3\tilde{5}\tilde{7}]}}^{(20),b}(\alpha,z)=
y_{_1}^{{D\over2}-1}y_{_2}^{{D}-4}\sum\limits_{n_{_1}=0}^\infty
\sum\limits_{n_{_2}=0}^\infty\sum\limits_{n_{_3}=0}^\infty\sum\limits_{n_{_4}=0}^\infty
c_{_{[\tilde{1}3\tilde{5}\tilde{7}]}}^{(20),b}(\alpha,{\bf n})
\nonumber\\
&&\hspace{2.5cm}\times
\Big({y_{_1}\over y_{_2}}\Big)^{n_{_1}}\Big({1\over y_{_2}}\Big)^{n_{_2}}
\Big({y_{_4}\over y_{_2}}\Big)^{n_{_3}}\Big({y_{_3}\over y_{_2}}\Big)^{n_{_4}}
\;,\nonumber\\
&&\Phi_{_{[\tilde{1}3\tilde{5}\tilde{7}]}}^{(20),c}(\alpha,z)=
y_{_1}^{{D\over2}}y_{_2}^{{D}-4}\sum\limits_{n_{_1}=0}^\infty
\sum\limits_{n_{_2}=0}^\infty\sum\limits_{n_{_3}=0}^\infty\sum\limits_{n_{_4}=0}^\infty
c_{_{[\tilde{1}3\tilde{5}\tilde{7}]}}^{(20),c}(\alpha,{\bf n})
\nonumber\\
&&\hspace{2.5cm}\times
y_{_1}^{n_{_1}}\Big({y_{_1}\over y_{_2}}\Big)^{n_{_2}}
\Big({y_{_4}\over y_{_2}}\Big)^{n_{_3}}\Big({y_{_3}\over y_{_2}}\Big)^{n_{_4}}\;.
\label{GKZ21m-12-2a}
\end{eqnarray}
Where the coefficients are
\begin{eqnarray}
&&c_{_{[\tilde{1}3\tilde{5}\tilde{7}]}}^{(20),a}(\alpha,{\bf n})=
(-)^{n_{_1}+n_{_2}}\Gamma(1+n_{_1}+n_{_3})\Big\{n_{_1}!n_{_2}!n_{_3}!n_{_4}!
\Gamma(1-{D\over2}-n_{_1}-n_{_3})
\nonumber\\
&&\hspace{2.5cm}\times
\Gamma(2-{D\over2}+n_{_2})\Gamma({D\over2}+n_{_3})\Gamma({D\over2}+n_{_1})
\nonumber\\
&&\hspace{2.5cm}\times
\Gamma({D\over2}-1-n_{_2}-n_{_4})\Gamma(2-{D\over2}+n_{_4})\Gamma(D-2-n_{_2}-n_{_4})\Big\}^{-1}
\;,\nonumber\\
&&c_{_{[\tilde{1}3\tilde{5}\tilde{7}]}}^{(20),b}(\alpha,{\bf n})=
(-)^{n_{_1}+n_{_2}+n_{_3}}\Gamma(1+n_{_1}+n_{_2})\Big\{n_{_1}!n_{_2}!n_{_4}!
\Gamma(2+n_{_1}+n_{_2}+n_{_3})
\nonumber\\
&&\hspace{2.5cm}\times
\Gamma(2-{D\over2}+n_{_2})\Gamma(3-{D\over2}+n_{_1}+n_{_2}+n_{_3})\Gamma({D\over2}+n_{_1})
\nonumber\\
&&\hspace{2.5cm}\times
\Gamma({D\over2}-1-n_{_1}-n_{_2})\Gamma({D\over2}-2-n_{_1}-n_{_2}-n_{_3}-n_{_4})
\nonumber\\
&&\hspace{2.5cm}\times
\Gamma(2-{D\over2}+n_{_4})\Gamma(D-3-n_{_1}-n_{_2}-n_{_3}-n_{_4})\Big\}^{-1}
\;,\nonumber\\
&&c_{_{[\tilde{1}3\tilde{5}\tilde{7}]}}^{(20),c}(\alpha,{\bf n})=
(-)^{n_{_1}+n_{_2}+n_{_3}}\Gamma(1+n_{_1})\Gamma(1+n_{_2})
\Big\{n_{_4}!\Gamma(2+n_{_1}+n_{_2})
\nonumber\\
&&\hspace{2.5cm}\times
\Gamma(2+n_{_2}+n_{_3})\Gamma(1-{D\over2}-n_{_1})\Gamma(3-{D\over2}+n_{_2}+n_{_3})
\nonumber\\
&&\hspace{2.5cm}\times
\Gamma({D\over2}-1-n_{_2})\Gamma({D\over2}+1+n_{_1}+n_{_2})
\nonumber\\
&&\hspace{2.5cm}\times
\Gamma({D\over2}-2-n_{_2}-n_{_3}-n_{_4})\Gamma(2-{D\over2}+n_{_4})
\nonumber\\
&&\hspace{2.5cm}\times
\Gamma(D-3-n_{_2}-n_{_3}-n_{_4})\Big\}^{-1}\;.
\label{GKZ21m-12-3}
\end{eqnarray}

\item   $I_{_{21}}=\{1,2,4,5,6,7,9,10,12,13\}$, i.e. the implement $J_{_{21}}=[1,14]\setminus I_{_{21}}=\{3,8,11,14\}$.
The choice implies the power numbers $\alpha_{_3}=\alpha_{_{8}}=\alpha_{_{11}}=\alpha_{_{14}}=0$, and
\begin{eqnarray}
&&\alpha_{_1}=a_{_3}-a_{_1},\;\alpha_{_2}=a_{_3}-a_{_2},\;\alpha_{_4}=a_{_3}-a_{_4},\;\alpha_{_5}=-a_{_5},
\nonumber\\
&&\alpha_{_{6}}=b_{_1}-1,\;\alpha_{_7}=b_{_2}+b_{_3}-a_{_3}-2,\;\alpha_{_{9}}=b_{_4}-a_{_3}-1,
\nonumber\\
&&\alpha_{_{10}}=b_{_5}-a_{_3}-1,\;\alpha_{_{12}}=b_{_3}-a_{_3}-1,\;\alpha_{_{13}}=1-b_{_3}.
\label{GKZ21m-17-1}
\end{eqnarray}
The corresponding hypergeometric solutions are given as
\begin{eqnarray}
&&\Phi_{_{[\tilde{1}3\tilde{5}\tilde{7}]}}^{(21),a}(\alpha,z)=
y_{_2}^{-1}y_{_3}^{{D\over2}-1}\sum\limits_{n_{_1}=0}^\infty
\sum\limits_{n_{_2}=0}^\infty\sum\limits_{n_{_3}=0}^\infty\sum\limits_{n_{_4}=0}^\infty
c_{_{[\tilde{1}3\tilde{5}\tilde{7}]}}^{(21),a}(\alpha,{\bf n})
\nonumber\\
&&\hspace{2.5cm}\times
y_{_1}^{n_{_1}}\Big({y_{_4}\over y_{_2}}\Big)^{n_{_2}}
y_{_4}^{n_{_3}}\Big({y_{_3}\over y_{_2}}\Big)^{n_{_4}}
\;,\nonumber\\
&&\Phi_{_{[\tilde{1}3\tilde{5}\tilde{7}]}}^{(21),b}(\alpha,z)=
y_{_2}^{-2}y_{_3}^{{D\over2}-1}\sum\limits_{n_{_1}=0}^\infty
\sum\limits_{n_{_2}=0}^\infty\sum\limits_{n_{_3}=0}^\infty\sum\limits_{n_{_4}=0}^\infty
c_{_{[\tilde{1}3\tilde{5}\tilde{7}]}}^{(21),b}(\alpha,{\bf n})
\nonumber\\
&&\hspace{2.5cm}\times
y_{_1}^{n_{_1}}\Big({1\over y_{_2}}\Big)^{n_{_2}}
\Big({y_{_4}\over y_{_2}}\Big)^{n_{_3}}\Big({y_{_3}\over y_{_2}}\Big)^{n_{_4}}\;.
\label{GKZ21m-17-2a}
\end{eqnarray}
Where the coefficients are
\begin{eqnarray}
&&c_{_{[\tilde{1}3\tilde{5}\tilde{7}]}}^{(21),a}(\alpha,{\bf n})=
(-)^{n_{_3}+n_{_4}}\Gamma(1+n_{_2}+n_{_4})
\Big\{n_{_1}!n_{_2}!n_{_3}!n_{_4}!\Gamma({D\over2}-1-n_{_1}-n_{_3})
\nonumber\\
&&\hspace{2.5cm}\times
\Gamma(D-2-n_{_1}-n_{_3})\Gamma({D\over2}+n_{_2})
\Gamma(2-{D\over2}+n_{_1})
\nonumber\\
&&\hspace{2.5cm}\times
\Gamma(1-{D\over2}-n_{_2}-n_{_4})
\Gamma(2-{D\over2}+n_{_3})\Gamma({D\over2}+n_{_4})\Big\}^{-1}
\;,\nonumber\\
&&c_{_{[\tilde{1}3\tilde{5}\tilde{7}]}}^{(21),b}(\alpha,{\bf n})=
(-)^{1+n_{_4}}\Gamma(1+n_{_2})\Gamma(2+n_{_2}+n_{_3}+n_{_4})
\Big\{n_{_1}!n_{_4}!\Gamma(2+n_{_2}+n_{_3})
\nonumber\\
&&\hspace{2.5cm}\times
\Gamma({D\over2}-n_{_1}+n_{_2})\Gamma(D-1-n_{_1}+n_{_2})
\Gamma({D\over2}+1+n_{_2}+n_{_3})
\nonumber\\
&&\hspace{2.5cm}\times
\Gamma(2-{D\over2}+n_{_1})\Gamma(-{D\over2}-n_{_2}-n_{_3}-n_{_4})
\Gamma(1-{D\over2}-n_{_2})
\nonumber\\
&&\hspace{2.5cm}\times
\Gamma({D\over2}+n_{_4})\Big\}^{-1}\;.
\label{GKZ21m-17-3}
\end{eqnarray}

\item   $I_{_{22}}=\{1,2,4,\cdots,10,12\}$, i.e. the implement $J_{_{22}}=[1,14]\setminus I_{_{22}}=\{3,11,13,14\}$.
The choice implies the power numbers $\alpha_{_3}=\alpha_{_{11}}=\alpha_{_{13}}=\alpha_{_{14}}=0$, and
\begin{eqnarray}
&&\alpha_{_1}=a_{_3}-a_{_1},\;\alpha_{_2}=a_{_3}-a_{_2},\;\alpha_{_4}=a_{_3}-a_{_4},
\nonumber\\
&&\alpha_{_5}=-a_{_5},\;\alpha_{_{6}}=b_{_1}-1,\;\alpha_{_7}=b_{_2}-a_{_3}-1,\;\alpha_{_{8}}=b_{_3}-1,
\nonumber\\
&&\alpha_{_{9}}=b_{_4}-a_{_3}-1,\;\alpha_{_{10}}=b_{_5}-a_{_3}-1,\;\alpha_{_{12}}=-a_{_3}.
\label{GKZ21m-18-1}
\end{eqnarray}
The corresponding hypergeometric solutions are written as
\begin{eqnarray}
&&\Phi_{_{[\tilde{1}3\tilde{5}\tilde{7}]}}^{(22),a}(\alpha,z)=
y_{_2}^{{D\over2}-2}\sum\limits_{n_{_1}=0}^\infty
\sum\limits_{n_{_2}=0}^\infty\sum\limits_{n_{_3}=0}^\infty\sum\limits_{n_{_4}=0}^\infty
c_{_{[\tilde{1}3\tilde{5}\tilde{7}]}}^{(22),a}(\alpha,{\bf n})
\nonumber\\
&&\hspace{2.5cm}\times
y_{_1}^{n_{_1}}\Big({y_{_4}\over y_{_2}}\Big)^{n_{_2}}
y_{_4}^{n_{_3}}\Big({y_{_3}\over y_{_2}}\Big)^{n_{_4}}
\;,\nonumber\\
&&\Phi_{_{[\tilde{1}3\tilde{5}\tilde{7}]}}^{(22),b}(\alpha,z)=
y_{_2}^{{D\over2}-3}\sum\limits_{n_{_1}=0}^\infty
\sum\limits_{n_{_2}=0}^\infty\sum\limits_{n_{_3}=0}^\infty\sum\limits_{n_{_4}=0}^\infty
c_{_{[\tilde{1}3\tilde{5}\tilde{7}]}}^{(22),b}(\alpha,{\bf n})
\nonumber\\
&&\hspace{2.5cm}\times
y_{_1}^{n_{_1}}\Big({1\over y_{_2}}\Big)^{n_{_2}}
\Big({y_{_4}\over y_{_2}}\Big)^{n_{_3}}\Big({y_{_3}\over y_{_2}}\Big)^{n_{_4}}\;.
\label{GKZ21m-18-2a}
\end{eqnarray}
Where the coefficients are
\begin{eqnarray}
&&c_{_{[\tilde{1}3\tilde{5}\tilde{7}]}}^{(22),a}(\alpha,{\bf n})=
(-)^{n_{_3}+n_{_4}}\Gamma(1+n_{_2}+n_{_4})
\Big\{n_{_1}!n_{_2}!n_{_3}!n_{_4}!\Gamma({D\over2}-1-n_{_1}-n_{_3})
\nonumber\\
&&\hspace{2.5cm}\times
\Gamma(D-2-n_{_1}-n_{_3})\Gamma({D\over2}+n_{_2})
\Gamma(2-{D\over2}+n_{_1})
\nonumber\\
&&\hspace{2.5cm}\times
\Gamma({D\over2}-1-n_{_2}-n_{_4})
\Gamma(2-{D\over2}+n_{_3})\Gamma(2-{D\over2}+n_{_4})\Big\}^{-1}
\;,\nonumber\\
&&c_{_{[\tilde{1}3\tilde{5}\tilde{7}]}}^{(22),b}(\alpha,{\bf n})=
(-)^{1+n_{_4}}\Gamma(1+n_{_2})\Gamma(2+n_{_2}+n_{_3}+n_{_4})
\Big\{n_{_1}!n_{_4}!\Gamma(2+n_{_2}+n_{_3})
\nonumber\\
&&\hspace{2.5cm}\times
\Gamma({D\over2}-n_{_1}+n_{_2})\Gamma(D-1-n_{_1}+n_{_2})
\Gamma({D\over2}+1+n_{_2}+n_{_3})
\nonumber\\
&&\hspace{2.5cm}\times
\Gamma(2-{D\over2}+n_{_1})\Gamma({D\over2}-2-n_{_2}-n_{_3}-n_{_4})
\Gamma(1-{D\over2}-n_{_2})
\nonumber\\
&&\hspace{2.5cm}\times
\Gamma(2-{D\over2}+n_{_4})\Big\}^{-1}\;.
\label{GKZ21m-18-3}
\end{eqnarray}

\item   $I_{_{23}}=\{1,2,3,5,6,7,9,10,12,13\}$, i.e. the implement $J_{_{23}}=[1,14]\setminus I_{_{23}}=\{4,8,11,14\}$.
The choice implies the power numbers $\alpha_{_4}=\alpha_{_{8}}=\alpha_{_{11}}=\alpha_{_{14}}=0$, and
\begin{eqnarray}
&&\alpha_{_1}=a_{_4}-a_{_1},\;\alpha_{_2}=a_{_4}-a_{_2},\;\alpha_{_3}=a_{_4}-a_{_3},\;\alpha_{_5}=-a_{_5},
\nonumber\\
&&\alpha_{_{6}}=b_{_1}-1,\;\alpha_{_7}=b_{_2}+b_{_3}-a_{_4}-2,\;\alpha_{_{9}}=b_{_4}-a_{_4}-1,
\nonumber\\
&&\alpha_{_{10}}=b_{_5}-a_{_4}-1,\;\alpha_{_{12}}=b_{_3}-a_{_4}-1,\;
\alpha_{_{13}}=1-b_{_3}.
\label{GKZ21m-23-1}
\end{eqnarray}
The corresponding hypergeometric functions are given as
\begin{eqnarray}
&&\Phi_{_{[\tilde{1}3\tilde{5}\tilde{7}]}}^{(23),a}(\alpha,z)=
y_{_2}^{{D\over2}-2}y_{_3}^{{D\over2}-1}\sum\limits_{n_{_1}=0}^\infty
\sum\limits_{n_{_2}=0}^\infty\sum\limits_{n_{_3}=0}^\infty\sum\limits_{n_{_4}=0}^\infty
c_{_{[\tilde{1}3\tilde{5}\tilde{7}]}}^{(23),a}(\alpha,{\bf n})
\nonumber\\
&&\hspace{2.5cm}\times
y_{_1}^{n_{_1}}\Big({y_{_4}\over y_{_2}}\Big)^{n_{_2}}
y_{_4}^{n_{_3}}\Big({y_{_3}\over y_{_2}}\Big)^{n_{_4}}
\;,\nonumber\\
&&\Phi_{_{[\tilde{1}3\tilde{5}\tilde{7}]}}^{(23),b}(\alpha,z)=
y_{_2}^{{D\over2}-3}y_{_3}^{{D\over2}-1}\sum\limits_{n_{_1}=0}^\infty
\sum\limits_{n_{_2}=0}^\infty\sum\limits_{n_{_3}=0}^\infty\sum\limits_{n_{_4}=0}^\infty
c_{_{[\tilde{1}3\tilde{5}\tilde{7}]}}^{(23),b}(\alpha,{\bf n})
\nonumber\\
&&\hspace{2.5cm}\times
\Big({y_{_1}\over y_{_2}}\Big)^{n_{_1}}\Big({1\over y_{_2}}\Big)^{n_{_2}}
\Big({y_{_4}\over y_{_2}}\Big)^{n_{_3}}\Big({y_{_3}\over y_{_2}}\Big)^{n_{_4}}
\;,\nonumber\\
&&\Phi_{_{[\tilde{1}3\tilde{5}\tilde{7}]}}^{(23),c}(\alpha,z)=
y_{_1}y_{_2}^{{D\over2}-3}y_{_3}^{{D\over2}-1}\sum\limits_{n_{_1}=0}^\infty
\sum\limits_{n_{_2}=0}^\infty\sum\limits_{n_{_3}=0}^\infty\sum\limits_{n_{_4}=0}^\infty
c_{_{[\tilde{1}3\tilde{5}\tilde{7}]}}^{(23),c}(\alpha,{\bf n})
\nonumber\\
&&\hspace{2.5cm}\times
y_{_1}^{n_{_1}}\Big({y_{_1}\over y_{_2}}\Big)^{n_{_2}}
\Big({y_{_4}\over y_{_2}}\Big)^{n_{_3}}\Big({y_{_3}\over y_{_2}}\Big)^{n_{_4}}\;.
\label{GKZ21m-23-2a}
\end{eqnarray}
Where the coefficients are
\begin{eqnarray}
&&c_{_{[\tilde{1}3\tilde{5}\tilde{7}]}}^{(23),a}(\alpha,{\bf n})=
(-)^{n_{_1}+n_{_4}}\Gamma(1+n_{_1}+n_{_3})
\Gamma(1+n_{_2}+n_{_4})\Big\{n_{_1}!n_{_2}!n_{_3}!n_{_4}!
\nonumber\\
&&\hspace{2.5cm}\times
\Gamma({D\over2}-1-n_{_1}-n_{_3})
\Gamma(2-{D\over2}+n_{_2})\Gamma(2-{D\over2}+n_{_1})
\nonumber\\
&&\hspace{2.5cm}\times
\Gamma({D\over2}+n_{_3})
\Gamma({D\over2}-1-n_{_2}-n_{_4})\Gamma({D\over2}+n_{_4})\Big\}^{-1}
\;,\nonumber\\
&&c_{_{[\tilde{1}3\tilde{5}\tilde{7}]}}^{(23),b}(\alpha,{\bf n})=
(-)^{1+n_{_4}}\Gamma(1+n_{_1}+n_{_2})\Gamma(2+n_{_1}+n_{_2}+n_{_3}+n_{_4})\Big\{n_{_1}!n_{_2}!n_{_4}!
\nonumber\\
&&\hspace{2.5cm}\times
\Gamma(2+n_{_1}+n_{_2}+n_{_3})\Gamma({D\over2}+n_{_2})
\Gamma(3-{D\over2}+n_{_1}+n_{_2}+n_{_3})
\nonumber\\
&&\hspace{2.5cm}\times
\Gamma(2-{D\over2}+n_{_1})\Gamma({D\over2}-1-n_{_1}-n_{_2})
\nonumber\\
&&\hspace{2.5cm}\times
\Gamma({D\over2}-2-n_{_1}-n_{_2}-n_{_3}-n_{_4})\Gamma({D\over2}+n_{_4})\Big\}^{-1}
\;,\nonumber\\
&&c_{_{[\tilde{1}3\tilde{5}\tilde{7}]}}^{(23),c}(\alpha,{\bf n})=
(-)^{1+n_{_1}+n_{_4}}\Gamma(1+n_{_1})\Gamma(1+n_{_2})\Gamma(2+n_{_2}+n_{_3}+n_{_4})\Big\{n_{_4}!
\nonumber\\
&&\hspace{2.5cm}\times
\Gamma(2+n_{_1}+n_{_2})\Gamma(2+n_{_2}+n_{_3})\Gamma({D\over2}-1-n_{_1})
\nonumber\\
&&\hspace{2.5cm}\times
\Gamma(3-{D\over2}+n_{_2}+n_{_3})\Gamma(3-{D\over2}+n_{_1}+n_{_2})\Gamma({D\over2}-1-n_{_2})
\nonumber\\
&&\hspace{2.5cm}\times
\Gamma({D\over2}-2-n_{_2}-n_{_3}-n_{_4})\Gamma({D\over2}+n_{_4})\Big\}^{-1}\;.
\label{GKZ21m-23-3}
\end{eqnarray}

\item   $I_{_{24}}=\{1,2,3,5,\cdots,10,12\}$, i.e. the implement $J_{_{24}}=[1,14]\setminus I_{_{24}}=\{4,11,13,14\}$.
The choice implies the power numbers $\alpha_{_4}=\alpha_{_{11}}=\alpha_{_{13}}=\alpha_{_{14}}=0$, and
\begin{eqnarray}
&&\alpha_{_1}=a_{_4}-a_{_1},\;\alpha_{_2}=a_{_4}-a_{_2},\;\alpha_{_3}=a_{_4}-a_{_3},\;\alpha_{_5}=-a_{_5},
\nonumber\\
&&\alpha_{_{6}}=b_{_1}-1,\;\alpha_{_7}=b_{_2}-a_{_4}-1,\;\alpha_{_{8}}=b_{_3}-1,
\nonumber\\
&&\alpha_{_{9}}=b_{_4}-a_{_4}-1,\;\alpha_{_{10}}=b_{_5}-a_{_4}-1,\;\alpha_{_{12}}=-a_{_4}.
\label{GKZ21m-24-1}
\end{eqnarray}
The corresponding hypergeometric solutions are written as
\begin{eqnarray}
&&\Phi_{_{[\tilde{1}3\tilde{5}\tilde{7}]}}^{(24),a}(\alpha,z)=
y_{_2}^{{D}-3}\sum\limits_{n_{_1}=0}^\infty
\sum\limits_{n_{_2}=0}^\infty\sum\limits_{n_{_3}=0}^\infty\sum\limits_{n_{_4}=0}^\infty
c_{_{[\tilde{1}3\tilde{5}\tilde{7}]}}^{(24),a}(\alpha,{\bf n})
\nonumber\\
&&\hspace{2.5cm}\times
y_{_1}^{n_{_1}}\Big({y_{_4}\over y_{_2}}\Big)^{n_{_2}}
y_{_4}^{n_{_3}}\Big({y_{_3}\over y_{_2}}\Big)^{n_{_4}}
\;,\nonumber\\
&&\Phi_{_{[\tilde{1}3\tilde{5}\tilde{7}]}}^{(24),b}(\alpha,z)=
y_{_2}^{{D}-4}\sum\limits_{n_{_1}=0}^\infty
\sum\limits_{n_{_2}=0}^\infty\sum\limits_{n_{_3}=0}^\infty\sum\limits_{n_{_4}=0}^\infty
c_{_{[\tilde{1}3\tilde{5}\tilde{7}]}}^{(24),b}(\alpha,{\bf n})
\nonumber\\
&&\hspace{2.5cm}\times
\Big({y_{_1}\over y_{_2}}\Big)^{n_{_1}}\Big({1\over y_{_2}}\Big)^{n_{_2}}
\Big({y_{_4}\over y_{_2}}\Big)^{n_{_3}}\Big({y_{_3}\over y_{_2}}\Big)^{n_{_4}}
\;,\nonumber\\
&&\Phi_{_{[\tilde{1}3\tilde{5}\tilde{7}]}}^{(24),c}(\alpha,z)=
y_{_1}y_{_2}^{{D}-4}\sum\limits_{n_{_1}=0}^\infty
\sum\limits_{n_{_2}=0}^\infty\sum\limits_{n_{_3}=0}^\infty\sum\limits_{n_{_4}=0}^\infty
c_{_{[\tilde{1}3\tilde{5}\tilde{7}]}}^{(24),c}(\alpha,{\bf n})
\nonumber\\
&&\hspace{2.5cm}\times
y_{_1}^{n_{_1}}\Big({y_{_1}\over y_{_2}}\Big)^{n_{_2}}
\Big({y_{_4}\over y_{_2}}\Big)^{n_{_3}}\Big({y_{_3}\over y_{_2}}\Big)^{n_{_4}}\;.
\label{GKZ21m-24-2a}
\end{eqnarray}
Where the coefficients are
\begin{eqnarray}
&&c_{_{[\tilde{1}3\tilde{5}\tilde{7}]}}^{(24),a}(\alpha,{\bf n})=
(-)^{n_{_1}+n_{_2}}\Gamma(1+n_{_1}+n_{_3})\Big\{n_{_1}!n_{_2}!n_{_3}!n_{_4}!
\Gamma(2-{D\over2}+n_{_1})
\nonumber\\
&&\hspace{2.5cm}\times
\Gamma({D\over2}-1-n_{_1}-n_{_3})
\Gamma(2-{D\over2}+n_{_2})\Gamma({D\over2}+n_{_3})
\nonumber\\
&&\hspace{2.5cm}\times
\Gamma({D\over2}-1-n_{_2}-n_{_4})\Gamma(2-{D\over2}+n_{_4})
\Gamma(D-2-n_{_2}-n_{_4})\Big\}^{-1}
\;,\nonumber\\
&&c_{_{[\tilde{1}3\tilde{5}\tilde{7}]}}^{(24),b}(\alpha,{\bf n})=
(-)^{n_{_1}+n_{_2}+n_{_3}}\Gamma(1+n_{_1}+n_{_2})\Big\{n_{_1}!n_{_2}!n_{_4}!
\Gamma(2+n_{_1}+n_{_2}+n_{_3})
\nonumber\\
&&\hspace{2.5cm}\times
\Gamma({D\over2}+n_{_2})
\Gamma(3-{D\over2}+n_{_1}+n_{_2}+n_{_3})\Gamma(2-{D\over2}+n_{_1})
\nonumber\\
&&\hspace{2.5cm}\times
\Gamma({D\over2}-1-n_{_1}-n_{_2})\Gamma({D\over2}-2-n_{_1}-n_{_2}-n_{_3}-n_{_4})
\nonumber\\
&&\hspace{2.5cm}\times
\Gamma(2-{D\over2}+n_{_4})\Gamma(D-3-n_{_1}-n_{_2}-n_{_3}-n_{_4})\Big\}^{-1}
\;,\nonumber\\
&&c_{_{[\tilde{1}3\tilde{5}\tilde{7}]}}^{(24),c}(\alpha,{\bf n})=
(-)^{n_{_1}+n_{_2}+n_{_3}}\Gamma(1+n_{_1})\Gamma(1+n_{_2})\Big\{n_{_4}!
\Gamma(2+n_{_1}+n_{_2})\Gamma(2+n_{_2}+n_{_3})
\nonumber\\
&&\hspace{2.5cm}\times
\Gamma({D\over2}-n_{_1})
\Gamma(3-{D\over2}+n_{_2}+n_{_3})\Gamma(3-{D\over2}+n_{_1}+n_{_2})
\nonumber\\
&&\hspace{2.5cm}\times
\Gamma({D\over2}-1-n_{_2})\Gamma({D\over2}-2-n_{_2}-n_{_3}-n_{_4})
\nonumber\\
&&\hspace{2.5cm}\times
\Gamma(2-{D\over2}+n_{_4})\Gamma(D-3-n_{_2}-n_{_3}-n_{_4})\Big\}^{-1}\;.
\label{GKZ21m-24-3}
\end{eqnarray}
\end{itemize}

\section{The hypergeometric solutions of the integer lattice ${\bf B}_{_{1\widetilde{357}}}$\label{app15}}
\indent\indent

\begin{itemize}
\item   $I_{_{1}}=\{2,\cdots,7,10,\cdots,13\}$, i.e. the implement $J_{_{1}}=[1,14]\setminus I_{_{1}}=\{1,8,9,14\}$.
The choice implies the power numbers $\alpha_{_1}=\alpha_{_{8}}=\alpha_{_{9}}=\alpha_{_{14}}=0$, and
\begin{eqnarray}
&&\alpha_{_2}=a_{_1}-a_{_2},\;\alpha_{_3}=b_{_4}-a_{_3}-1,\;\alpha_{_4}=b_{_4}-a_{_4}-1,
\nonumber\\
&&\alpha_{_5}=-a_{_5},\;\alpha_{_6}=b_{_1}+b_{_4}-a_{_1}-2,\;\alpha_{_7}=b_{_2}+b_{_3}-b_{_4}-1,
\nonumber\\
&&\alpha_{_{10}}=b_{_5}-b_{_4},\;\alpha_{_{11}}=b_{_4}-a_{_1}-1,\;\alpha_{_{12}}=b_{_3}-b_{_4},\;\alpha_{_{13}}=1-b_{_3}.
\label{GKZ21n-1-1}
\end{eqnarray}
The corresponding hypergeometric solutions are
\begin{eqnarray}
&&\Phi_{_{[1\tilde{3}\tilde{5}\tilde{7}]}}^{(1),a}(\alpha,z)=
y_{_1}^{{D\over2}-2}y_{_2}^{-1}y_{_3}^{{D\over2}-1}\sum\limits_{n_{_1}=0}^\infty
\sum\limits_{n_{_2}=0}^\infty\sum\limits_{n_{_3}=0}^\infty\sum\limits_{n_{_4}=0}^\infty
c_{_{[1\tilde{3}\tilde{5}\tilde{7}]}}^{(1),a}(\alpha,{\bf n})
\nonumber\\
&&\hspace{2.5cm}\times
\Big({1\over y_{_1}}\Big)^{n_{_1}}\Big({y_{_4}\over y_{_1}}\Big)^{n_{_2}}
\Big({y_{_4}\over y_{_2}}\Big)^{n_{_3}}\Big({y_{_3}\over y_{_2}}\Big)^{n_{_4}}
\;,\nonumber\\
&&\Phi_{_{[1\tilde{3}\tilde{5}\tilde{7}]}}^{(1),b}(\alpha,z)=
y_{_1}^{{D\over2}-3}y_{_3}^{{D\over2}-1}\sum\limits_{n_{_1}=0}^\infty
\sum\limits_{n_{_2}=0}^\infty\sum\limits_{n_{_3}=0}^\infty\sum\limits_{n_{_4}=0}^\infty
c_{_{[1\tilde{3}\tilde{5}\tilde{7}]}}^{(1),b}(\alpha,{\bf n})
\nonumber\\
&&\hspace{2.5cm}\times
\Big({1\over y_{_1}}\Big)^{n_{_1}}\Big({y_{_2}\over y_{_1}}\Big)^{n_{_2}}
\Big({y_{_4}\over y_{_1}}\Big)^{n_{_3}}\Big({y_{_3}\over y_{_1}}\Big)^{n_{_4}}
\;,\nonumber\\
&&\Phi_{_{[1\tilde{3}\tilde{5}\tilde{7}]}}^{(1),c}(\alpha,z)=
y_{_1}^{{D\over2}-3}y_{_2}^{-1}y_{_3}^{{D\over2}}\sum\limits_{n_{_1}=0}^\infty
\sum\limits_{n_{_2}=0}^\infty\sum\limits_{n_{_3}=0}^\infty\sum\limits_{n_{_4}=0}^\infty
c_{_{[1\tilde{3}\tilde{5}\tilde{7}]}}^{(1),c}(\alpha,{\bf n})
\nonumber\\
&&\hspace{2.5cm}\times
\Big({1\over y_{_1}}\Big)^{n_{_1}}\Big({y_{_3}\over y_{_1}}\Big)^{n_{_2}}
\Big({y_{_4}\over y_{_1}}\Big)^{n_{_3}}\Big({y_{_3}\over y_{_2}}\Big)^{n_{_4}}
\label{GKZ21n-1-2a}
\end{eqnarray}
Where the coefficients are
\begin{eqnarray}
&&c_{_{[1\tilde{3}\tilde{5}\tilde{7}]}}^{(1),a}(\alpha,{\bf n})=
(-)^{n_{_1}+n_{_4}}\Gamma(1+n_{_1}+n_{_2})\Gamma(1+n_{_3}+n_{_4})\Big\{n_{_1}!n_{_2}!n_{_3}!n_{_4}!
\nonumber\\
&&\hspace{2.5cm}\times
\Gamma({D\over2}+n_{_1})\Gamma({D\over2}+n_{_3})\Gamma({D\over2}-1-n_{_1}-n_{_2})
\nonumber\\
&&\hspace{2.5cm}\times
\Gamma(1-{D\over2}-n_{_3}-n_{_4})\Gamma(2-{D\over2}+n_{_2})\Gamma({D\over2}+n_{_4})\Big\}^{-1}
\;,\nonumber\\
&&c_{_{[1\tilde{3}\tilde{5}\tilde{7}]}}^{(1),b}(\alpha,{\bf n})=
(-)^{1+n_{_1}}\Gamma(2+n_{_1}+n_{_2}+n_{_3}+n_{_4})\Gamma(1+n_{_2}+n_{_4})\Big\{n_{_1}!n_{_2}!n_{_4}!
\nonumber\\
&&\hspace{2.5cm}\times
\Gamma(2+n_{_2}+n_{_3}+n_{_4})\Gamma({D\over2}+n_{_1})\Gamma({D\over2}-1-n_{_2}-n_{_4})
\nonumber\\
&&\hspace{2.5cm}\times
\Gamma({D\over2}-2-n_{_1}-n_{_2}-n_{_3}-n_{_4})\Gamma(2-{D\over2}+n_{_2})
\nonumber\\
&&\hspace{2.5cm}\times
\Gamma(3-{D\over2}+n_{_2}+n_{_3}+n_{_4})\Gamma({D\over2}+n_{_4})\Big\}^{-1}
\;,\nonumber\\
&&c_{_{[1\tilde{3}\tilde{5}\tilde{7}]}}^{(1),c}(\alpha,{\bf n})=
(-)^{1+n_{_1}+n_{_4}}\Gamma(1+n_{_2})\Gamma(1+n_{_4})\Gamma(2+n_{_1}+n_{_2}+n_{_3})
\nonumber\\
&&\hspace{2.5cm}\times
\Big\{n_{_1}!\Gamma(2+n_{_2}+n_{_4})\Gamma(2+n_{_2}+n_{_3})\Gamma({D\over2}+n_{_1})
\nonumber\\
&&\hspace{2.5cm}\times
\Gamma({D\over2}-1-n_{_2})\Gamma({D\over2}-2-n_{_1}-n_{_2}-n_{_3})
\nonumber\\
&&\hspace{2.5cm}\times
\Gamma(1-{D\over2}-n_{_4})\Gamma(3-{D\over2}+n_{_2}+n_{_3})\Gamma({D\over2}+1+n_{_2}+n_{_4})\Big\}^{-1}\;.
\label{GKZ21n-1-3}
\end{eqnarray}

\item   $I_{_{2}}=\{2,\cdots,8,10,11,12\}$, i.e. the implement $J_{_{2}}=[1,14]\setminus I_{_{2}}=\{1,9,13,14\}$.
The choice implies the power numbers $\alpha_{_1}=\alpha_{_{9}}=\alpha_{_{13}}=\alpha_{_{14}}=0$, and
\begin{eqnarray}
&&\alpha_{_2}=a_{_1}-a_{_2},\;\alpha_{_3}=b_{_4}-a_{_3}-1,\;\alpha_{_4}=b_{_4}-a_{_4}-1,\;\alpha_{_5}=-a_{_5},
\nonumber\\
&&\alpha_{_6}=b_{_1}+b_{_4}-a_{_1}-2,\;\alpha_{_7}=b_{_2}-b_{_4},\;\alpha_{_{8}}=b_{_3}-1,
\nonumber\\
&&\alpha_{_{10}}=b_{_5}-b_{_4},\;\alpha_{_{11}}=b_{_4}-a_{_1}-1,\;\alpha_{_{12}}=1-b_{_4}.
\label{GKZ21n-2-1}
\end{eqnarray}
The corresponding hypergeometric solutions are written as
\begin{eqnarray}
&&\Phi_{_{[1\tilde{3}\tilde{5}\tilde{7}]}}^{(2),a}(\alpha,z)=
y_{_1}^{{D\over2}-2}y_{_2}^{{D\over2}-2}\sum\limits_{n_{_1}=0}^\infty
\sum\limits_{n_{_2}=0}^\infty\sum\limits_{n_{_3}=0}^\infty\sum\limits_{n_{_4}=0}^\infty
c_{_{[1\tilde{3}\tilde{5}\tilde{7}]}}^{(2),a}(\alpha,{\bf n})
\nonumber\\
&&\hspace{2.5cm}\times
\Big({1\over y_{_1}}\Big)^{n_{_1}}\Big({y_{_4}\over y_{_1}}\Big)^{n_{_2}}
\Big({y_{_4}\over y_{_2}}\Big)^{n_{_3}}\Big({y_{_3}\over y_{_2}}\Big)^{n_{_4}}
\;,\nonumber\\
&&\Phi_{_{[1\tilde{3}\tilde{5}\tilde{7}]}}^{(2),b}(\alpha,z)=
y_{_1}^{{D\over2}-3}y_{_2}^{{D\over2}-1}\sum\limits_{n_{_1}=0}^\infty
\sum\limits_{n_{_2}=0}^\infty\sum\limits_{n_{_3}=0}^\infty\sum\limits_{n_{_4}=0}^\infty
c_{_{[1\tilde{3}\tilde{5}\tilde{7}]}}^{(2),b}(\alpha,{\bf n})
\nonumber\\
&&\hspace{2.5cm}\times
\Big({1\over y_{_1}}\Big)^{n_{_1}}\Big({y_{_2}\over y_{_1}}\Big)^{n_{_2}}
\Big({y_{_4}\over y_{_1}}\Big)^{n_{_3}}\Big({y_{_3}\over y_{_1}}\Big)^{n_{_4}}
\;,\nonumber\\
&&\Phi_{_{[1\tilde{3}\tilde{5}\tilde{7}]}}^{(2),c}(\alpha,z)=
y_{_1}^{{D\over2}-3}y_{_2}^{{D\over2}-2}y_{_3}\sum\limits_{n_{_1}=0}^\infty
\sum\limits_{n_{_2}=0}^\infty\sum\limits_{n_{_3}=0}^\infty\sum\limits_{n_{_4}=0}^\infty
c_{_{[1\tilde{3}\tilde{5}\tilde{7}]}}^{(2),c}(\alpha,{\bf n})
\nonumber\\
&&\hspace{2.5cm}\times
\Big({1\over y_{_1}}\Big)^{n_{_1}}\Big({y_{_3}\over y_{_1}}\Big)^{n_{_2}}
\Big({y_{_4}\over y_{_1}}\Big)^{n_{_3}}\Big({y_{_3}\over y_{_2}}\Big)^{n_{_4}}
\label{GKZ21n-2-2a}
\end{eqnarray}
Where the coefficients are
\begin{eqnarray}
&&c_{_{[1\tilde{3}\tilde{5}\tilde{7}]}}^{(2),a}(\alpha,{\bf n})=
(-)^{n_{_1}+n_{_4}}\Gamma(1+n_{_1}+n_{_2})\Gamma(1+n_{_3}+n_{_4})\Big\{n_{_1}!n_{_2}!n_{_3}!n_{_4}!
\nonumber\\
&&\hspace{2.5cm}\times
\Gamma({D\over2}+n_{_1})\Gamma({D\over2}+n_{_3})\Gamma({D\over2}-1-n_{_1}-n_{_2})
\nonumber\\
&&\hspace{2.5cm}\times
\Gamma({D\over2}-1-n_{_3}-n_{_4})\Gamma(2-{D\over2}+n_{_2})\Gamma(2-{D\over2}+n_{_4})\Big\}^{-1}
\;,\nonumber\\
&&c_{_{[1\tilde{3}\tilde{5}\tilde{7}]}}^{(2),b}(\alpha,{\bf n})=
(-)^{1+n_{_1}}\Gamma(2+n_{_1}+n_{_2}+n_{_3}+n_{_4})\Gamma(1+n_{_2}+n_{_4})\Big\{n_{_1}!n_{_2}!n_{_4}!
\nonumber\\
&&\hspace{2.5cm}\times
\Gamma(2+n_{_2}+n_{_3}+n_{_4})\Gamma({D\over2}+n_{_1})\Gamma({D\over2}-1-n_{_2}-n_{_4})
\nonumber\\
&&\hspace{2.5cm}\times
\Gamma({D\over2}-2-n_{_1}-n_{_2}-n_{_3}-n_{_4})\Gamma({D\over2}+n_{_2})
\nonumber\\
&&\hspace{2.5cm}\times
\Gamma(3-{D\over2}+n_{_2}+n_{_3}+n_{_4})\Gamma(2-{D\over2}+n_{_4})\Big\}^{-1}
\;,\nonumber\\
&&c_{_{[1\tilde{3}\tilde{5}\tilde{7}]}}^{(2),c}(\alpha,{\bf n})=
(-)^{1+n_{_1}+n_{_4}}\Gamma(1+n_{_2})\Gamma(1+n_{_4})\Gamma(2+n_{_1}+n_{_2}+n_{_3})
\nonumber\\
&&\hspace{2.5cm}\times
\Big\{n_{_1}!\Gamma(2+n_{_2}+n_{_4})\Gamma(2+n_{_2}+n_{_3})\Gamma({D\over2}+n_{_1})
\nonumber\\
&&\hspace{2.5cm}\times
\Gamma({D\over2}-1-n_{_2})\Gamma({D\over2}-2-n_{_1}-n_{_2}-n_{_3})
\nonumber\\
&&\hspace{2.5cm}\times
\Gamma({D\over2}-1-n_{_4})\Gamma(3-{D\over2}+n_{_2}+n_{_3})\Gamma(3-{D\over2}+n_{_2}+n_{_4})\Big\}^{-1}\;.
\label{GKZ21n-2-3}
\end{eqnarray}

\item   $I_{_{3}}=\{2,\cdots,7,9,11,12,13\}$, i.e. the implement $J_{_{3}}=[1,14]\setminus I_{_{3}}=\{1,8,10,14\}$.
The choice implies the power numbers $\alpha_{_1}=\alpha_{_{8}}=\alpha_{_{10}}=\alpha_{_{14}}=0$, and
\begin{eqnarray}
&&\alpha_{_2}=a_{_1}-a_{_2},\;\alpha_{_3}=b_{_5}-a_{_3}-1,\;\alpha_{_4}=b_{_5}-a_{_4}-1,
\nonumber\\
&&\alpha_{_5}=-a_{_5},\;\alpha_{_6}=b_{_1}+b_{_5}-a_{_1}-2,\;\alpha_{_7}=b_{_2}+b_{_3}-b_{_5}-1,
\nonumber\\
&&\alpha_{_{9}}=b_{_4}-b_{_5},\;\alpha_{_{11}}=b_{_5}-a_{_1}-1,\;\alpha_{_{12}}=b_{_3}-b_{_5},\;\alpha_{_{13}}=1-b_{_3}.
\label{GKZ21n-3-1}
\end{eqnarray}
The corresponding hypergeometric functions are
\begin{eqnarray}
&&\Phi_{_{[1\tilde{3}\tilde{5}\tilde{7}]}}^{(3),a}(\alpha,z)=
y_{_1}^{-1}y_{_2}^{{D\over2}-2}y_{_3}^{{D\over2}-1}\sum\limits_{n_{_1}=0}^\infty
\sum\limits_{n_{_2}=0}^\infty\sum\limits_{n_{_3}=0}^\infty\sum\limits_{n_{_4}=0}^\infty
c_{_{[1\tilde{3}\tilde{5}\tilde{7}]}}^{(3),a}(\alpha,{\bf n})
\nonumber\\
&&\hspace{2.5cm}\times
\Big({1\over y_{_1}}\Big)^{n_{_1}}\Big({y_{_4}\over y_{_1}}\Big)^{n_{_2}}
\Big({y_{_4}\over y_{_2}}\Big)^{n_{_3}}\Big({y_{_3}\over y_{_2}}\Big)^{n_{_4}}
\;,\nonumber\\
&&\Phi_{_{[1\tilde{3}\tilde{5}\tilde{7}]}}^{(3),b}(\alpha,z)=
y_{_1}^{-2}y_{_2}^{{D\over2}-1}y_{_3}^{{D\over2}-1}\sum\limits_{n_{_1}=0}^\infty
\sum\limits_{n_{_2}=0}^\infty\sum\limits_{n_{_3}=0}^\infty\sum\limits_{n_{_4}=0}^\infty
c_{_{[1\tilde{3}\tilde{5}\tilde{7}]}}^{(3),b}(\alpha,{\bf n})
\nonumber\\
&&\hspace{2.5cm}\times
\Big({1\over y_{_1}}\Big)^{n_{_1}}\Big({y_{_2}\over y_{_1}}\Big)^{n_{_2}}
\Big({y_{_4}\over y_{_1}}\Big)^{n_{_3}}\Big({y_{_3}\over y_{_1}}\Big)^{n_{_4}}
\;,\nonumber\\
&&\Phi_{_{[1\tilde{3}\tilde{5}\tilde{7}]}}^{(3),c}(\alpha,z)=
y_{_1}^{-2}y_{_2}^{{D\over2}-2}y_{_3}^{{D\over2}}\sum\limits_{n_{_1}=0}^\infty
\sum\limits_{n_{_2}=0}^\infty\sum\limits_{n_{_3}=0}^\infty\sum\limits_{n_{_4}=0}^\infty
c_{_{[1\tilde{3}\tilde{5}\tilde{7}]}}^{(3),c}(\alpha,{\bf n})
\nonumber\\
&&\hspace{2.5cm}\times
\Big({1\over y_{_1}}\Big)^{n_{_1}}\Big({y_{_3}\over y_{_1}}\Big)^{n_{_2}}
\Big({y_{_4}\over y_{_1}}\Big)^{n_{_3}}\Big({y_{_3}\over y_{_2}}\Big)^{n_{_4}}\;.
\label{GKZ21n-3-2a}
\end{eqnarray}
Where the coefficients are
\begin{eqnarray}
&&c_{_{[1\tilde{3}\tilde{5}\tilde{7}]}}^{(3),a}(\alpha,{\bf n})=
(-)^{n_{_1}+n_{_4}}\Gamma(1+n_{_1}+n_{_2})
\Gamma(1+n_{_3}+n_{_4})\Big\{n_{_1}!n_{_2}!n_{_3}!n_{_4}!
\nonumber\\
&&\hspace{2.5cm}\times
\Gamma({D\over2}+n_{_1})
\Gamma(2-{D\over2}+n_{_3})\Gamma(1-{D\over2}-n_{_1}-n_{_2})\Gamma({D\over2}+n_{_2})
\nonumber\\
&&\hspace{2.5cm}\times
\Gamma({D\over2}-1-n_{_3}-n_{_4})\Gamma({D\over2}+n_{_4})\Big\}^{-1}
\;,\nonumber\\
&&c_{_{[1\tilde{3}\tilde{5}\tilde{7}]}}^{(3),b}(\alpha,{\bf n})=
(-)^{1+n_{_1}}\Gamma(2+n_{_1}+n_{_2}+n_{_3}+n_{_4})
\Gamma(1+n_{_2}+n_{_4})\Big\{n_{_1}!n_{_2}!n_{_4}!
\nonumber\\
&&\hspace{2.5cm}\times
\Gamma(2+n_{_2}+n_{_3}+n_{_4})\Gamma(1-{D\over2}-n_{_2}-n_{_4})
\nonumber\\
&&\hspace{2.5cm}\times
\Gamma(-{D\over2}-n_{_1}-n_{_2}-n_{_3}-n_{_4})\Gamma({D\over2}+n_{_1})
\nonumber\\
&&\hspace{2.5cm}\times
\Gamma({D\over2}+1+n_{_2}+n_{_3}+n_{_4})\Gamma({D\over2}+n_{_2})
\Gamma({D\over2}+n_{_4})\Big\}^{-1}
\;,\nonumber\\
&&c_{_{[1\tilde{3}\tilde{5}\tilde{7}]}}^{(3),c}(\alpha,{\bf n})=
(-)^{1+n_{_1}+n_{_4}}\Gamma(2+n_{_1}+n_{_2}+n_{_3})
\Gamma(1+n_{_2})\Gamma(1+n_{_4})\Big\{n_{_1}!
\nonumber\\
&&\hspace{2.5cm}\times
\Gamma(2+n_{_2}+n_{_4})
\Gamma(2+n_{_2}+n_{_3})\Gamma({D\over2}+n_{_1})\Gamma(1-{D\over2}-n_{_2})
\nonumber\\
&&\hspace{2.5cm}\times
\Gamma(-{D\over2}-n_{_1}-n_{_2}-n_{_3})\Gamma({D\over2}+1+n_{_2}+n_{_3})
\nonumber\\
&&\hspace{2.5cm}\times
\Gamma({D\over2}-1-n_{_4})\Gamma({D\over2}+n_{_2}+n_{_4})\Big\}^{-1}\;.
\label{GKZ21n-3-3}
\end{eqnarray}

\item   $I_{_{4}}=\{2,\cdots,9,11,12\}$, i.e. the implement $J_{_{4}}=[1,14]\setminus I_{_{4}}=\{1,10,13,14\}$.
The choice implies the power numbers $\alpha_{_1}=\alpha_{_{10}}=\alpha_{_{13}}=\alpha_{_{14}}=0$, and
\begin{eqnarray}
&&\alpha_{_2}=a_{_1}-a_{_2},\;\alpha_{_3}=b_{_5}-a_{_3}-1,\;\alpha_{_4}=b_{_5}-a_{_4}-1,\;\alpha_{_5}=-a_{_5},
\nonumber\\
&&\alpha_{_6}=b_{_1}+b_{_5}-a_{_1}-2,\;\alpha_{_7}=b_{_2}-b_{_5},\;\alpha_{_{8}}=b_{_3}-1,
\nonumber\\
&&\alpha_{_{9}}=b_{_4}-b_{_5},\;\alpha_{_{11}}=b_{_5}-a_{_1}-1,\;\alpha_{_{12}}=1-b_{_5}.
\label{GKZ21n-4-1}
\end{eqnarray}
The corresponding hypergeometric solutions are written as
\begin{eqnarray}
&&\Phi_{_{[1\tilde{3}\tilde{5}\tilde{7}]}}^{(4),a}(\alpha,z)=
y_{_1}^{-1}y_{_2}^{{D}-3}\sum\limits_{n_{_1}=0}^\infty
\sum\limits_{n_{_2}=0}^\infty\sum\limits_{n_{_3}=0}^\infty\sum\limits_{n_{_4}=0}^\infty
c_{_{[1\tilde{3}\tilde{5}\tilde{7}]}}^{(4),a}(\alpha,{\bf n})
\nonumber\\
&&\hspace{2.5cm}\times
\Big({1\over y_{_1}}\Big)^{n_{_1}}\Big({y_{_4}\over y_{_1}}\Big)^{n_{_2}}
\Big({y_{_4}\over y_{_2}}\Big)^{n_{_3}}\Big({y_{_3}\over y_{_2}}\Big)^{n_{_4}}
\;,\nonumber\\
&&\Phi_{_{[1\tilde{3}\tilde{5}\tilde{7}]}}^{(4),b}(\alpha,z)=
y_{_1}^{-2}y_{_2}^{{D}-2}\sum\limits_{n_{_1}=0}^\infty
\sum\limits_{n_{_2}=0}^\infty\sum\limits_{n_{_3}=0}^\infty\sum\limits_{n_{_4}=0}^\infty
c_{_{[1\tilde{3}\tilde{5}\tilde{7}]}}^{(4),b}(\alpha,{\bf n})
\nonumber\\
&&\hspace{2.5cm}\times
\Big({1\over y_{_1}}\Big)^{n_{_1}}\Big({y_{_2}\over y_{_1}}\Big)^{n_{_2}}
\Big({y_{_4}\over y_{_1}}\Big)^{n_{_3}}\Big({y_{_3}\over y_{_2}}\Big)^{n_{_4}}\;.
\label{GKZ21n-4-2a}
\end{eqnarray}
Where the coefficients are
\begin{eqnarray}
&&c_{_{[1\tilde{3}\tilde{5}\tilde{7}]}}^{(4),a}(\alpha,{\bf n})=
(-)^{n_{_1}+n_{_3}}\Gamma(1+n_{_1}+n_{_2})\Big\{n_{_1}!n_{_2}!n_{_3}!n_{_4}!
\Gamma({D\over2}+n_{_1})\Gamma(2-{D\over2}+n_{_3})
\nonumber\\
&&\hspace{2.5cm}\times
\Gamma(1-{D\over2}-n_{_1}-n_{_2})
\Gamma({D\over2}-1-n_{_3}-n_{_4})\Gamma(2-{D\over2}+n_{_4})
\nonumber\\
&&\hspace{2.5cm}\times
\Gamma({D\over2}+n_{_2})
\Gamma(D-2-n_{_3}-n_{_4})\Big\}^{-1}
\;,\nonumber\\
&&c_{_{[1\tilde{3}\tilde{5}\tilde{7}]}}^{(4),b}(\alpha,{\bf n})=
(-)^{1+n_{_1}}\Gamma(2+n_{_1}+n_{_2}+n_{_3})\Gamma(1+n_{_2})\Big\{n_{_1}!n_{_4}!
\Gamma(2+n_{_2}+n_{_3})
\nonumber\\
&&\hspace{2.5cm}\times
\Gamma({D\over2}+n_{_1})
\Gamma(1-{D\over2}-n_{_2})\Gamma(-{D\over2}-n_{_1}-n_{_2}-n_{_3})
\nonumber\\
&&\hspace{2.5cm}\times
\Gamma({D\over2}+n_{_2}-n_{_4})\Gamma(2-{D\over2}+n_{_4})\Gamma({D\over2}+1+n_{_2}+n_{_3})
\nonumber\\
&&\hspace{2.5cm}\times
\Gamma(D-1+n_{_2}-n_{_4})\Big\}^{-1}\;.
\label{GKZ21n-4-3}
\end{eqnarray}

\item   $I_{_{5}}=\{1,3,\cdots,7,10,\cdots,13\}$, i.e. the implement $J_{_{5}}=[1,14]\setminus I_{_{5}}=\{2,8,9,14\}$.
The choice implies the power numbers $\alpha_{_2}=\alpha_{_{8}}=\alpha_{_{9}}=\alpha_{_{14}}=0$, and
\begin{eqnarray}
&&\alpha_{_1}=a_{_2}-a_{_1},\;\alpha_{_3}=b_{_4}-a_{_3}-1,\;\alpha_{_4}=b_{_4}-a_{_4}-1,
\nonumber\\
&&\alpha_{_5}=-a_{_5},\;\alpha_{_6}=b_{_1}+b_{_4}-a_{_2}-2,\;\alpha_{_7}=b_{_2}+b_{_3}-b_{_4}-1,
\nonumber\\
&&\alpha_{_{10}}=b_{_5}-b_{_4},\;\alpha_{_{11}}=b_{_4}-a_{_2}-1,\;\alpha_{_{12}}=b_{_3}-b_{_4},\;\alpha_{_{13}}=1-b_{_3}.
\label{GKZ21n-5-1}
\end{eqnarray}
The corresponding hypergeometric solutions are
\begin{eqnarray}
&&\Phi_{_{[1\tilde{3}\tilde{5}\tilde{7}]}}^{(5),a}(\alpha,z)=
y_{_1}^{{D}-3}y_{_2}^{-1}y_{_3}^{{D\over2}-1}\sum\limits_{n_{_1}=0}^\infty
\sum\limits_{n_{_2}=0}^\infty\sum\limits_{n_{_3}=0}^\infty\sum\limits_{n_{_4}=0}^\infty
c_{_{[1\tilde{3}\tilde{5}\tilde{7}]}}^{(5),a}(\alpha,{\bf n})
\nonumber\\
&&\hspace{2.5cm}\times
\Big({1\over y_{_1}}\Big)^{n_{_1}}\Big({y_{_4}\over y_{_1}}\Big)^{n_{_2}}
\Big({y_{_4}\over y_{_2}}\Big)^{n_{_3}}\Big({y_{_3}\over y_{_2}}\Big)^{n_{_4}}
\;,\nonumber\\
&&\Phi_{_{[1\tilde{3}\tilde{5}\tilde{7}]}}^{(5),b}(\alpha,z)=
y_{_1}^{{D}-4}y_{_3}^{{D\over2}-1}\sum\limits_{n_{_1}=0}^\infty
\sum\limits_{n_{_2}=0}^\infty\sum\limits_{n_{_3}=0}^\infty\sum\limits_{n_{_4}=0}^\infty
c_{_{[1\tilde{3}\tilde{5}\tilde{7}]}}^{(5),b}(\alpha,{\bf n})
\nonumber\\
&&\hspace{2.5cm}\times
\Big({1\over y_{_1}}\Big)^{n_{_1}}\Big({y_{_2}\over y_{_1}}\Big)^{n_{_2}}
\Big({y_{_4}\over y_{_1}}\Big)^{n_{_3}}\Big({y_{_3}\over y_{_1}}\Big)^{n_{_4}}
\;,\nonumber\\
&&\Phi_{_{[1\tilde{3}\tilde{5}\tilde{7}]}}^{(5),c}(\alpha,z)=
y_{_1}^{{D}-4}y_{_2}^{-1}y_{_3}^{{D\over2}}\sum\limits_{n_{_1}=0}^\infty
\sum\limits_{n_{_2}=0}^\infty\sum\limits_{n_{_3}=0}^\infty\sum\limits_{n_{_4}=0}^\infty
c_{_{[1\tilde{3}\tilde{5}\tilde{7}]}}^{(5),c}(\alpha,{\bf n})
\nonumber\\
&&\hspace{2.5cm}\times
\Big({1\over y_{_1}}\Big)^{n_{_1}}\Big({y_{_3}\over y_{_1}}\Big)^{n_{_2}}
\Big({y_{_4}\over y_{_1}}\Big)^{n_{_3}}\Big({y_{_3}\over y_{_2}}\Big)^{n_{_4}}\;.
\label{GKZ21n-5-2a}
\end{eqnarray}
Where the coefficients are
\begin{eqnarray}
&&c_{_{[1\tilde{3}\tilde{5}\tilde{7}]}}^{(5),a}(\alpha,{\bf n})=
(-)^{n_{_2}+n_{_4}}\Gamma(1+n_{_3}+n_{_4})\Big\{n_{_1}!n_{_2}!n_{_3}!n_{_4}!
\Gamma(2-{D\over2}+n_{_1})
\nonumber\\
&&\hspace{2.5cm}\times
\Gamma({D\over2}-1-n_{_1}-n_{_2})\Gamma(1-{D\over2}-n_{_3}-n_{_4})
\Gamma(2-{D\over2}+n_{_2})
\nonumber\\
&&\hspace{2.5cm}\times
\Gamma(D-2-n_{_1}-n_{_2})\Gamma({D\over2}+n_{_3})\Gamma({D\over2}+n_{_4})\Big\}^{-1}
\;,\nonumber\\
&&c_{_{[1\tilde{3}\tilde{5}\tilde{7}]}}^{(5),b}(\alpha,{\bf n})=
(-)^{n_{_2}+n_{_3}+n_{_4}}\Gamma(1+n_{_2}+n_{_4})\Big\{n_{_1}!n_{_2}!n_{_4}!
\Gamma(2+n_{_2}+n_{_3}+n_{_4})
\nonumber\\
&&\hspace{2.5cm}\times
\Gamma(2-{D\over2}+n_{_1})\Gamma({D\over2}-2-n_{_1}-n_{_2}-n_{_3}-n_{_4})
\nonumber\\
&&\hspace{2.5cm}\times
\Gamma(3-{D\over2}+n_{_2}+n_{_3}+n_{_4})\Gamma(D-3-n_{_1}-n_{_2}-n_{_3}-n_{_4})
\nonumber\\
&&\hspace{2.5cm}\times
\Gamma(2-{D\over2}+n_{_2})\Gamma({D\over2}-1-n_{_2}-n_{_4})\Gamma({D\over2}+n_{_4})\Big\}^{-1}
\;,\nonumber\\
&&c_{_{[1\tilde{3}\tilde{5}\tilde{7}]}}^{(5),c}(\alpha,{\bf n})=
(-)^{n_{_2}+n_{_3}+n_{_4}}\Gamma(1+n_{_2})\Gamma(1+n_{_4})\Big\{n_{_1}!
\Gamma(2+n_{_2}+n_{_4})
\nonumber\\
&&\hspace{2.5cm}\times
\Gamma(2+n_{_2}+n_{_3})
\Gamma(2-{D\over2}+n_{_1})\Gamma({D\over2}-2-n_{_1}-n_{_2}-n_{_3})
\nonumber\\
&&\hspace{2.5cm}\times
\Gamma(1-{D\over2}-n_{_4})\Gamma(D-3-n_{_1}-n_{_2}-n_{_3})
\nonumber\\
&&\hspace{2.5cm}\times
\Gamma(3-{D\over2}+n_{_2}+n_{_3})\Gamma({D\over2}-1-n_{_2})\Gamma({D\over2}+1+n_{_2}+n_{_4})\Big\}^{-1}\;.
\label{GKZ21n-5-3}
\end{eqnarray}

\item   $I_{_{6}}=\{1,3,\cdots,8,10,11,12\}$, i.e. the implement $J_{_{6}}=[1,14]\setminus I_{_{6}}=\{2,9,13,14\}$.
The choice implies the power numbers $\alpha_{_2}=\alpha_{_{9}}=\alpha_{_{13}}=\alpha_{_{14}}=0$, and
\begin{eqnarray}
&&\alpha_{_1}=a_{_2}-a_{_1},\;\alpha_{_3}=b_{_4}-a_{_3}-1,\;\alpha_{_4}=b_{_4}-a_{_4}-1,\;\alpha_{_5}=-a_{_5},
\nonumber\\
&&\alpha_{_6}=b_{_1}+b_{_4}-a_{_2}-2,\;\alpha_{_7}=b_{_2}-b_{_4},\;\alpha_{_{8}}=b_{_3}-1,
\nonumber\\
&&\alpha_{_{10}}=b_{_5}-b_{_4},\;\alpha_{_{11}}=b_{_4}-a_{_2}-1,\;\alpha_{_{12}}=1-b_{_4}.
\label{GKZ21n-6-1}
\end{eqnarray}
The corresponding hypergeometric solutions are written as
\begin{eqnarray}
&&\Phi_{_{[1\tilde{3}\tilde{5}\tilde{7}]}}^{(6),a}(\alpha,z)=
y_{_1}^{{D}-3}y_{_2}^{{D\over2}-2}\sum\limits_{n_{_1}=0}^\infty
\sum\limits_{n_{_2}=0}^\infty\sum\limits_{n_{_3}=0}^\infty\sum\limits_{n_{_4}=0}^\infty
c_{_{[1\tilde{3}\tilde{5}\tilde{7}]}}^{(6),a}(\alpha,{\bf n})
\nonumber\\
&&\hspace{2.5cm}\times
\Big({1\over y_{_1}}\Big)^{n_{_1}}\Big({y_{_4}\over y_{_1}}\Big)^{n_{_2}}
\Big({y_{_4}\over y_{_2}}\Big)^{n_{_3}}\Big({y_{_3}\over y_{_2}}\Big)^{n_{_4}}
\;,\nonumber\\
&&\Phi_{_{[1\tilde{3}\tilde{5}\tilde{7}]}}^{(6),b}(\alpha,z)=
y_{_1}^{{D}-4}y_{_2}^{{D\over2}-1}\sum\limits_{n_{_1}=0}^\infty
\sum\limits_{n_{_2}=0}^\infty\sum\limits_{n_{_3}=0}^\infty\sum\limits_{n_{_4}=0}^\infty
c_{_{[1\tilde{3}\tilde{5}\tilde{7}]}}^{(6),b}(\alpha,{\bf n})
\nonumber\\
&&\hspace{2.5cm}\times
\Big({1\over y_{_1}}\Big)^{n_{_1}}\Big({y_{_2}\over y_{_1}}\Big)^{n_{_2}}
\Big({y_{_4}\over y_{_1}}\Big)^{n_{_3}}\Big({y_{_3}\over y_{_1}}\Big)^{n_{_4}}
\;,\nonumber\\
&&\Phi_{_{[1\tilde{3}\tilde{5}\tilde{7}]}}^{(6),c}(\alpha,z)=
y_{_1}^{{D}-4}y_{_2}^{{D\over2}-2}y_{_3}\sum\limits_{n_{_1}=0}^\infty
\sum\limits_{n_{_2}=0}^\infty\sum\limits_{n_{_3}=0}^\infty\sum\limits_{n_{_4}=0}^\infty
c_{_{[1\tilde{3}\tilde{5}\tilde{7}]}}^{(6),c}(\alpha,{\bf n})
\nonumber\\
&&\hspace{2.5cm}\times
\Big({1\over y_{_1}}\Big)^{n_{_1}}\Big({y_{_3}\over y_{_1}}\Big)^{n_{_2}}
\Big({y_{_4}\over y_{_1}}\Big)^{n_{_3}}\Big({y_{_3}\over y_{_2}}\Big)^{n_{_4}}\;.
\label{GKZ21n-6-2a}
\end{eqnarray}
Where the coefficients are
\begin{eqnarray}
&&c_{_{[1\tilde{3}\tilde{5}\tilde{7}]}}^{(6),a}(\alpha,{\bf n})=
(-)^{n_{_2}+n_{_4}}\Gamma(1+n_{_3}+n_{_4})\Big\{n_{_1}!n_{_2}!n_{_3}!n_{_4}!
\Gamma(2-{D\over2}+n_{_1})
\nonumber\\
&&\hspace{2.5cm}\times
\Gamma({D\over2}-1-n_{_1}-n_{_2})\Gamma({D\over2}-1-n_{_3}-n_{_4})
\Gamma(2-{D\over2}+n_{_2})
\nonumber\\
&&\hspace{2.5cm}\times
\Gamma(D-2-n_{_1}-n_{_2})\Gamma({D\over2}+n_{_3})\Gamma(2-{D\over2}+n_{_4})\Big\}^{-1}
\;,\nonumber\\
&&c_{_{[1\tilde{3}\tilde{5}\tilde{7}]}}^{(6),b}(\alpha,{\bf n})=
(-)^{n_{_2}+n_{_3}+n_{_4}}\Gamma(1+n_{_2}+n_{_4})\Big\{n_{_1}!n_{_2}!n_{_4}!
\Gamma(2+n_{_2}+n_{_3}+n_{_4})
\nonumber\\
&&\hspace{2.5cm}\times
\Gamma(2-{D\over2}+n_{_1})\Gamma({D\over2}-2-n_{_1}-n_{_2}-n_{_3}-n_{_4})
\nonumber\\
&&\hspace{2.5cm}\times
\Gamma(3-{D\over2}+n_{_2}+n_{_3}+n_{_4})\Gamma(D-3-n_{_1}-n_{_2}-n_{_3}-n_{_4})
\nonumber\\
&&\hspace{2.5cm}\times
\Gamma({D\over2}+n_{_2})\Gamma({D\over2}-1-n_{_2}-n_{_4})\Gamma(2-{D\over2}+n_{_4})\Big\}^{-1}
\;,\nonumber\\
&&c_{_{[1\tilde{3}\tilde{5}\tilde{7}]}}^{(6),c}(\alpha,{\bf n})=
(-)^{n_{_2}+n_{_3}+n_{_4}}\Gamma(1+n_{_2})\Gamma(1+n_{_4})\Big\{n_{_1}!
\Gamma(2+n_{_2}+n_{_4})
\nonumber\\
&&\hspace{2.5cm}\times
\Gamma(2+n_{_2}+n_{_3})
\Gamma(2-{D\over2}+n_{_1})\Gamma({D\over2}-2-n_{_1}-n_{_2}-n_{_3})
\nonumber\\
&&\hspace{2.5cm}\times
\Gamma({D\over2}-1-n_{_4})\Gamma(D-3-n_{_1}-n_{_2}-n_{_3})
\nonumber\\
&&\hspace{2.5cm}\times
\Gamma(3-{D\over2}+n_{_2}+n_{_3})\Gamma({D\over2}-1-n_{_2})\Gamma(3-{D\over2}+n_{_2}+n_{_4})\Big\}^{-1}\;.
\label{GKZ21n-6-3}
\end{eqnarray}

\item   $I_{_{7}}=\{1,3,\cdots,7,9,11,12,13\}$, i.e. the implement $J_{_{7}}=[1,14]\setminus I_{_{7}}=\{2,8,10,14\}$.
The choice implies the power numbers $\alpha_{_1}=\alpha_{_{8}}=\alpha_{_{10}}=\alpha_{_{14}}=0$, and
\begin{eqnarray}
&&\alpha_{_1}=a_{_2}-a_{_1},\;\alpha_{_3}=b_{_5}-a_{_3}-1,\;\alpha_{_4}=b_{_5}-a_{_4}-1,
\nonumber\\
&&\alpha_{_5}=-a_{_5},\;\alpha_{_6}=b_{_1}+b_{_5}-a_{_2}-2,\;\alpha_{_7}=b_{_2}+b_{_3}-b_{_5}-1,
\nonumber\\
&&\alpha_{_{9}}=b_{_4}-b_{_5},\;\alpha_{_{11}}=b_{_5}-a_{_2}-1,\;\alpha_{_{12}}=b_{_3}-b_{_5},\;\alpha_{_{13}}=1-b_{_3}.
\label{GKZ21n-7-1}
\end{eqnarray}
The corresponding hypergeometric functions are given as
\begin{eqnarray}
&&\Phi_{_{[1\tilde{3}\tilde{5}\tilde{7}]}}^{(7),a}(\alpha,z)=
y_{_1}^{{D\over2}-2}y_{_2}^{{D\over2}-2}y_{_3}^{{D\over2}-1}\sum\limits_{n_{_1}=0}^\infty
\sum\limits_{n_{_2}=0}^\infty\sum\limits_{n_{_3}=0}^\infty\sum\limits_{n_{_4}=0}^\infty
c_{_{[1\tilde{3}\tilde{5}\tilde{7}]}}^{(7),a}(\alpha,{\bf n})
\nonumber\\
&&\hspace{2.5cm}\times
\Big({1\over y_{_1}}\Big)^{n_{_1}}\Big({y_{_4}\over y_{_1}}\Big)^{n_{_2}}
\Big({y_{_4}\over y_{_2}}\Big)^{n_{_3}}\Big({y_{_3}\over y_{_2}}\Big)^{n_{_4}}
\;,\nonumber\\
&&\Phi_{_{[1\tilde{3}\tilde{5}\tilde{7}]}}^{(7),b}(\alpha,z)=
y_{_1}^{{D\over2}-3}y_{_2}^{{D\over2}-1}y_{_3}^{{D\over2}-1}\sum\limits_{n_{_1}=0}^\infty
\sum\limits_{n_{_2}=0}^\infty\sum\limits_{n_{_3}=0}^\infty\sum\limits_{n_{_4}=0}^\infty
c_{_{[1\tilde{3}\tilde{5}\tilde{7}]}}^{(7),b}(\alpha,{\bf n})
\nonumber\\
&&\hspace{2.5cm}\times
\Big({1\over y_{_1}}\Big)^{n_{_1}}\Big({y_{_2}\over y_{_1}}\Big)^{n_{_2}}
\Big({y_{_4}\over y_{_1}}\Big)^{n_{_3}}\Big({y_{_3}\over y_{_1}}\Big)^{n_{_4}}
\;,\nonumber\\
&&\Phi_{_{[1\tilde{3}\tilde{5}\tilde{7}]}}^{(7),c}(\alpha,z)=
y_{_1}^{{D\over2}-3}y_{_2}^{{D\over2}-2}y_{_3}^{{D\over2}}\sum\limits_{n_{_1}=0}^\infty
\sum\limits_{n_{_2}=0}^\infty\sum\limits_{n_{_3}=0}^\infty\sum\limits_{n_{_4}=0}^\infty
c_{_{[1\tilde{3}\tilde{5}\tilde{7}]}}^{(7),c}(\alpha,{\bf n})
\nonumber\\
&&\hspace{2.5cm}\times
\Big({1\over y_{_1}}\Big)^{n_{_1}}\Big({y_{_3}\over y_{_1}}\Big)^{n_{_2}}
\Big({y_{_4}\over y_{_1}}\Big)^{n_{_3}}\Big({y_{_3}\over y_{_2}}\Big)^{n_{_4}}\;.
\label{GKZ21n-7-2a}
\end{eqnarray}
Where the coefficients are
\begin{eqnarray}
&&c_{_{[1\tilde{3}\tilde{5}\tilde{7}]}}^{(7),a}(\alpha,{\bf n})=
(-)^{n_{_1}+n_{_4}}\Gamma(1+n_{_1}+n_{_2})\Gamma(1+n_{_3}+n_{_4})
\nonumber\\
&&\hspace{2.5cm}\times
\Big\{n_{_1}!n_{_2}!n_{_3}!n_{_4}!\Gamma(2-{D\over2}+n_{_1})\Gamma(2-{D\over2}+n_{_3})
\Gamma({D\over2}+n_{_2})
\nonumber\\
&&\hspace{2.5cm}\times
\Gamma({D\over2}-1-n_{_1}-n_{_2})
\Gamma({D\over2}-1-n_{_3}-n_{_4})\Gamma({D\over2}+n_{_4})\Big\}^{-1}
\;,\nonumber\\
&&c_{_{[1\tilde{3}\tilde{5}\tilde{7}]}}^{(7),b}(\alpha,{\bf n})=
(-)^{1+n_{_1}}\Gamma(2+n_{_1}+n_{_2}+n_{_3}+n_{_4})\Gamma(1+n_{_2}+n_{_4})\Big\{n_{_1}!n_{_2}!n_{_4}!
\nonumber\\
&&\hspace{2.5cm}\times
\Gamma(2+n_{_2}+n_{_3}+n_{_4})\Gamma(2-{D\over2}+n_{_1})\Gamma(1-{D\over2}-n_{_2}-n_{_4})
\nonumber\\
&&\hspace{2.5cm}\times
\Gamma({D\over2}+1+n_{_2}+n_{_3}+n_{_4})\Gamma({D\over2}-2-n_{_1}-n_{_2}-n_{_3}-n_{_4})
\nonumber\\
&&\hspace{2.5cm}\times
\Gamma({D\over2}+n_{_2})\Gamma({D\over2}+n_{_4})\Big\}^{-1}
\;,\nonumber\\
&&c_{_{[1\tilde{3}\tilde{5}\tilde{7}]}}^{(7),c}(\alpha,{\bf n})=
(-)^{1+n_{_1}+n_{_4}}\Gamma(2+n_{_1}+n_{_2}+n_{_3})\Gamma(1+n_{_2})\Gamma(1+n_{_4})
\Big\{n_{_1}!
\nonumber\\
&&\hspace{2.5cm}\times
\Gamma(2+n_{_2}+n_{_4})\Gamma(2+n_{_2}+n_{_3})\Gamma(2-{D\over2}+n_{_1})
\Gamma(1-{D\over2}-n_{_2})
\nonumber\\
&&\hspace{2.5cm}\times
\Gamma({D\over2}+1+n_{_2}+n_{_3})\Gamma({D\over2}-2-n_{_1}-n_{_2}-n_{_3})
\nonumber\\
&&\hspace{2.5cm}\times
\Gamma({D\over2}-1-n_{_4})\Gamma({D\over2}+1+n_{_2}+n_{_4})\Big\}^{-1}\;.
\label{GKZ21n-7-3}
\end{eqnarray}

\item   $I_{_{8}}=\{1,3,\cdots,9,11,12\}$, i.e. the implement $J_{_{8}}=[1,14]\setminus I_{_{8}}=\{2,10,13,14\}$.
The choice implies the power numbers $\alpha_{_1}=\alpha_{_{10}}=\alpha_{_{13}}=\alpha_{_{14}}=0$, and
\begin{eqnarray}
&&\alpha_{_1}=a_{_2}-a_{_1},\;\alpha_{_3}=b_{_5}-a_{_3}-1,\;\alpha_{_4}=b_{_5}-a_{_4}-1,\;\alpha_{_5}=-a_{_5},
\nonumber\\
&&\alpha_{_6}=b_{_1}+b_{_5}-a_{_2}-2,\;\alpha_{_7}=b_{_2}-b_{_5},\;\alpha_{_{8}}=b_{_3}-1,
\nonumber\\
&&\alpha_{_{9}}=b_{_4}-b_{_5},\;\alpha_{_{11}}=b_{_5}-a_{_2}-1,\;\alpha_{_{12}}=1-b_{_5}.
\label{GKZ21n-8-1}
\end{eqnarray}
The corresponding hypergeometric solutions are written as
\begin{eqnarray}
&&\Phi_{_{[1\tilde{3}\tilde{5}\tilde{7}]}}^{(8),a}(\alpha,z)=
y_{_1}^{{D\over2}-2}y_{_2}^{{D}-3}\sum\limits_{n_{_1}=0}^\infty
\sum\limits_{n_{_2}=0}^\infty\sum\limits_{n_{_3}=0}^\infty\sum\limits_{n_{_4}=0}^\infty
c_{_{[1\tilde{3}\tilde{5}\tilde{7}]}}^{(8),a}(\alpha,{\bf n})
\nonumber\\
&&\hspace{2.5cm}\times
\Big({1\over y_{_1}}\Big)^{n_{_1}}\Big({y_{_4}\over y_{_1}}\Big)^{n_{_2}}
\Big({y_{_4}\over y_{_2}}\Big)^{n_{_3}}\Big({y_{_3}\over y_{_2}}\Big)^{n_{_4}}
\;,\nonumber\\
&&\Phi_{_{[1\tilde{3}\tilde{5}\tilde{7}]}}^{(8),b}(\alpha,z)=
y_{_1}^{{D\over2}-1}y_{_2}^{{D}-2}\sum\limits_{n_{_1}=0}^\infty
\sum\limits_{n_{_2}=0}^\infty\sum\limits_{n_{_3}=0}^\infty\sum\limits_{n_{_4}=0}^\infty
c_{_{[1\tilde{3}\tilde{5}\tilde{7}]}}^{(8),b}(\alpha,{\bf n})
\nonumber\\
&&\hspace{2.5cm}\times
\Big({1\over y_{_1}}\Big)^{n_{_1}}\Big({y_{_2}\over y_{_1}}\Big)^{n_{_2}}
\Big({y_{_4}\over y_{_1}}\Big)^{n_{_3}}\Big({y_{_3}\over y_{_2}}\Big)^{n_{_4}}\;.
\label{GKZ21n-8-2a}
\end{eqnarray}
Where the coefficients are
\begin{eqnarray}
&&c_{_{[1\tilde{3}\tilde{5}\tilde{7}]}}^{(8),a}(\alpha,{\bf n})=
(-)^{n_{_1}+n_{_3}}\Gamma(1+n_{_1}+n_{_2})\Big\{n_{_1}!n_{_2}!n_{_3}!n_{_4}!
\Gamma(2-{D\over2}+n_{_1})
\nonumber\\
&&\hspace{2.5cm}\times
\Gamma(2-{D\over2}+n_{_3})\Gamma({D\over2}-1-n_{_3}-n_{_4})\Gamma(2-{D\over2}+n_{_4})
\nonumber\\
&&\hspace{2.5cm}\times
\Gamma({D\over2}+n_{_2})\Gamma({D\over2}-1-n_{_1}-n_{_2})\Gamma(D-2-n_{_3}-n_{_4})\Big\}^{-1}
\;,\nonumber\\
&&c_{_{[1\tilde{3}\tilde{5}\tilde{7}]}}^{(8),b}(\alpha,{\bf n})=
(-)^{1+n_{_1}}\Gamma(2+n_{_1}+n_{_2}+n_{_3})\Gamma(1+n_{_2})\Big\{n_{_1}!n_{_4}!
\nonumber\\
&&\hspace{2.5cm}\times
\Gamma(2+n_{_2}+n_{_3})\Gamma(2-{D\over2}+n_{_1})\Gamma(1-{D\over2}-n_{_2})
\nonumber\\
&&\hspace{2.5cm}\times
\Gamma({D\over2}+n_{_2}-n_{_4})\Gamma(2-{D\over2}+n_{_4})\Gamma({D\over2}+1+n_{_2}+n_{_3})
\nonumber\\
&&\hspace{2.5cm}\times
\Gamma({D\over2}-1-n_{_1}-n_{_2}-n_{_3})\Gamma(D-1+n_{_2}-n_{_4})\Big\}^{-1}\;.
\label{GKZ21n-8-3}
\end{eqnarray}
\end{itemize}

\section{The hypergeometric solutions of the integer lattice ${\bf B}_{_{\widetilde{1357}}}$\label{app16}}
\indent\indent

\begin{itemize}
\item   $I_{_{1}}=\{1,\cdots,5,7,10,\cdots,13\}$, i.e. the implement $J_{_{1}}=[1,14]\setminus I_{_{1}}=\{6,8,9,14\}$.
The choice implies the power numbers $\alpha_{_6}=\alpha_{_{8}}=\alpha_{_{9}}=\alpha_{_{14}}=0$, and
\begin{eqnarray}
&&\alpha_{_1}=b_{_1}+b_{_4}-a_{_1}-2,\;\alpha_{_2}=b_{_1}+b_{_4}-a_{_2}-2,\;\alpha_{_3}=b_{_4}-a_{_3}-1,
\nonumber\\
&&\alpha_{_4}=b_{_4}-a_{_4}-1,\;\alpha_{_5}=-a_{_5},\;\alpha_{_7}=b_{_2}+b_{_3}-b_{_4}-1,
\nonumber\\
&&\alpha_{_{10}}=b_{_5}-b_{_4},\;\alpha_{_{11}}=1-b_{_1},\;\alpha_{_{12}}=b_{_3}-b_{_4},\;\alpha_{_{13}}=1-b_{_3}.
\label{GKZ21o-1-1}
\end{eqnarray}
The corresponding hypergeometric solutions are
\begin{eqnarray}
&&\Phi_{_{[\tilde{1}\tilde{3}\tilde{5}\tilde{7}]}}^{(1),a}(\alpha,z)=
y_{_1}^{{D\over2}-1}y_{_2}^{-1}y_{_3}^{{D\over2}-1}\sum\limits_{n_{_1}=0}^\infty
\sum\limits_{n_{_2}=0}^\infty\sum\limits_{n_{_3}=0}^\infty\sum\limits_{n_{_4}=0}^\infty
c_{_{[\tilde{1}\tilde{3}\tilde{5}\tilde{7}]}}^{(1),a}(\alpha,{\bf n})
\nonumber\\
&&\hspace{2.5cm}\times
y_{_1}^{n_{_1}}y_{_4}^{n_{_2}}\Big({y_{_4}\over y_{_2}}\Big)^{n_{_3}}
\Big({y_{_3}\over y_{_2}}\Big)^{n_{_4}}
\;,\nonumber\\
&&\Phi_{_{[\tilde{1}\tilde{3}\tilde{5}\tilde{7}]}}^{(1),b}(\alpha,z)=
y_{_1}^{{D\over2}-1}y_{_3}^{{D\over2}-1}\sum\limits_{n_{_1}=0}^\infty
\sum\limits_{n_{_2}=0}^\infty\sum\limits_{n_{_3}=0}^\infty\sum\limits_{n_{_4}=0}^\infty
c_{_{[\tilde{1}\tilde{3}\tilde{5}\tilde{7}]}}^{(1),b}(\alpha,{\bf n})
\nonumber\\
&&\hspace{2.5cm}\times
y_{_1}^{n_{_1}}y_{_2}^{n_{_2}}y_{_4}^{n_{_3}}y_{_3}^{n_{_4}}
\;,\nonumber\\
&&\Phi_{_{[\tilde{1}\tilde{3}\tilde{5}\tilde{7}]}}^{(1),c}(\alpha,z)=
y_{_1}^{{D\over2}-1}y_{_2}^{-1}y_{_3}^{{D\over2}}\sum\limits_{n_{_1}=0}^\infty
\sum\limits_{n_{_2}=0}^\infty\sum\limits_{n_{_3}=0}^\infty\sum\limits_{n_{_4}=0}^\infty
c_{_{[\tilde{1}\tilde{3}\tilde{5}\tilde{7}]}}^{(1),c}(\alpha,{\bf n})
\nonumber\\
&&\hspace{2.5cm}\times
y_{_1}^{n_{_1}}y_{_3}^{n_{_2}}y_{_4}^{n_{_3}}\Big({y_{_3}\over y_{_2}}\Big)^{n_{_4}}\;.
\label{GKZ21o-1-2a}
\end{eqnarray}
Where the coefficients are
\begin{eqnarray}
&&c_{_{[\tilde{1}\tilde{3}\tilde{5}\tilde{7}]}}^{(1),a}(\alpha,{\bf n})=
(-)^{n_{_1}+n_{_4}}\Gamma(1+n_{_1}+n_{_2})
\Gamma(1+n_{_3}+n_{_4})\Big\{n_{_1}!n_{_2}!n_{_3}!n_{_4}!
\nonumber\\
&&\hspace{2.5cm}\times
\Gamma({D\over2}-1-n_{_1}-n_{_2})\Gamma({D\over2}+n_{_3})
\Gamma(1-{D\over2}-n_{_3}-n_{_4})
\nonumber\\
&&\hspace{2.5cm}\times
\Gamma(2-{D\over2}+n_{_2})\Gamma({D\over2}+n_{_1})\Gamma({D\over2}+n_{_4})\Big\}^{-1}
\;,\nonumber\\
&&c_{_{[\tilde{1}\tilde{3}\tilde{5}\tilde{7}]}}^{(1),b}(\alpha,{\bf n})=
(-)^{1+n_{_1}}\Gamma(2+n_{_1}+n_{_2}+n_{_3}+n_{_4})\Gamma(1+n_{_2}+n_{_4})\Big\{n_{_1}!n_{_2}!n_{_4}!
\nonumber\\
&&\hspace{2.5cm}\times
\Gamma(2+n_{_2}+n_{_3}+n_{_4})\Gamma({D\over2}+n_{_1})
\Gamma({D\over2}-2-n_{_1}-n_{_2}-n_{_3}-n_{_4})
\nonumber\\
&&\hspace{2.5cm}\times
\Gamma({D\over2}+n_{_4})\Gamma(2-{D\over2}+n_{_2})
\Gamma({D\over2}-1-n_{_2}-n_{_4})
\nonumber\\
&&\hspace{2.5cm}\times
\Gamma(3-{D\over2}+n_{_2}+n_{_3}+n_{_4})\Big\}^{-1}
\;,\nonumber\\
&&c_{_{[\tilde{1}\tilde{3}\tilde{5}\tilde{7}]}}^{(1),c}(\alpha,{\bf n})=
(-)^{1+n_{_1}+n_{_4}}\Gamma(2+n_{_1}+n_{_2}+n_{_3})\Gamma(1+n_{_2})\Gamma(1+n_{_4})
\Big\{n_{_1}!
\nonumber\\
&&\hspace{2.5cm}\times
\Gamma(2+n_{_2}+n_{_4})\Gamma(2+n_{_2}+n_{_3})
\Gamma({D\over2}-2-n_{_1}-n_{_2}-n_{_3})
\nonumber\\
&&\hspace{2.5cm}\times
\Gamma({D\over2}-1-n_{_2})
\Gamma(1-{D\over2}-n_{_4})\Gamma(3-{D\over2}+n_{_2}+n_{_3})
\nonumber\\
&&\hspace{2.5cm}\times
\Gamma({D\over2}+n_{_1})\Gamma({D\over2}+1+n_{_2}+n_{_4})\Big\}^{-1}\;.
\label{GKZ21o-1-3}
\end{eqnarray}

\item   $I_{_{2}}=\{1,\cdots,5,7,8,10,11,12\}$, i.e. the implement $J_{_{2}}=[1,14]\setminus I_{_{2}}=\{6,9,13,14\}$.
The choice implies the power numbers $\alpha_{_6}=\alpha_{_{9}}=\alpha_{_{13}}=\alpha_{_{14}}=0$, and
\begin{eqnarray}
&&\alpha_{_1}=b_{_1}+b_{_4}-a_{_1}-2,\;\alpha_{_2}=b_{_1}+b_{_4}-a_{_2}-2,
\nonumber\\
&&\alpha_{_3}=b_{_4}-a_{_3}-1,\;\alpha_{_4}=b_{_4}-a_{_4}-1,\;\alpha_{_5}=-a_{_5},\;\alpha_{_7}=b_{_2}-b_{_4},
\nonumber\\
&&\alpha_{_{8}}=b_{_3}-1,\;\alpha_{_{10}}=b_{_5}-b_{_4},\;\alpha_{_{11}}=1-b_{_1},\;\alpha_{_{12}}=1-b_{_4}.
\label{GKZ21o-2-1}
\end{eqnarray}
The corresponding hypergeometric functions are written as
\begin{eqnarray}
&&\Phi_{_{[\tilde{1}\tilde{3}\tilde{5}\tilde{7}]}}^{(2),a}(\alpha,z)=
y_{_1}^{{D\over2}-1}y_{_2}^{{D\over2}-2}\sum\limits_{n_{_1}=0}^\infty
\sum\limits_{n_{_2}=0}^\infty\sum\limits_{n_{_3}=0}^\infty\sum\limits_{n_{_4}=0}^\infty
c_{_{[\tilde{1}\tilde{3}\tilde{5}\tilde{7}]}}^{(2),a}(\alpha,{\bf n})
\nonumber\\
&&\hspace{2.5cm}\times
y_{_1}^{n_{_1}}y_{_4}^{n_{_2}}\Big({y_{_4}\over y_{_2}}\Big)^{n_{_3}}
\Big({y_{_3}\over y_{_2}}\Big)^{n_{_4}}
\;,\nonumber\\
&&\Phi_{_{[\tilde{1}\tilde{3}\tilde{5}\tilde{7}]}}^{(2),b}(\alpha,z)=
y_{_1}^{{D\over2}-1}y_{_2}^{{D\over2}-1}\sum\limits_{n_{_1}=0}^\infty
\sum\limits_{n_{_2}=0}^\infty\sum\limits_{n_{_3}=0}^\infty\sum\limits_{n_{_4}=0}^\infty
c_{_{[\tilde{1}\tilde{3}\tilde{5}\tilde{7}]}}^{(2),b}(\alpha,{\bf n})
\nonumber\\
&&\hspace{2.5cm}\times
y_{_1}^{n_{_1}}y_{_2}^{n_{_2}}y_{_4}^{n_{_3}}y_{_3}^{n_{_4}}
\;,\nonumber\\
&&\Phi_{_{[\tilde{1}\tilde{3}\tilde{5}\tilde{7}]}}^{(2),c}(\alpha,z)=
y_{_1}^{{D\over2}-1}y_{_2}^{{D\over2}-2}y_{_3}\sum\limits_{n_{_1}=0}^\infty
\sum\limits_{n_{_2}=0}^\infty\sum\limits_{n_{_3}=0}^\infty\sum\limits_{n_{_4}=0}^\infty
c_{_{[\tilde{1}\tilde{3}\tilde{5}\tilde{7}]}}^{(2),c}(\alpha,{\bf n})
\nonumber\\
&&\hspace{2.5cm}\times
y_{_1}^{n_{_1}}y_{_3}^{n_{_2}}y_{_4}^{n_{_3}}\Big({y_{_3}\over y_{_2}}\Big)^{n_{_4}}\;.
\label{GKZ21o-2-2a}
\end{eqnarray}
Where the coefficients are
\begin{eqnarray}
&&c_{_{[\tilde{1}\tilde{3}\tilde{5}\tilde{7}]}}^{(2),a}(\alpha,{\bf n})=
(-)^{n_{_1}+n_{_4}}\Gamma(1+n_{_1}+n_{_2})
\Gamma(1+n_{_3}+n_{_4})\Big\{n_{_1}!n_{_2}!n_{_3}!n_{_4}!
\nonumber\\
&&\hspace{2.5cm}\times
\Gamma({D\over2}-1-n_{_1}-n_{_2})\Gamma({D\over2}+n_{_3})
\Gamma({D\over2}-1-n_{_3}-n_{_4})
\nonumber\\
&&\hspace{2.5cm}\times
\Gamma(2-{D\over2}+n_{_2})\Gamma({D\over2}+n_{_1})\Gamma(2-{D\over2}+n_{_4})\Big\}^{-1}
\;,\nonumber\\
&&c_{_{[\tilde{1}\tilde{3}\tilde{5}\tilde{7}]}}^{(2),b}(\alpha,{\bf n})=
(-)^{1+n_{_1}}\Gamma(2+n_{_1}+n_{_2}+n_{_3}+n_{_4})\Gamma(1+n_{_2}+n_{_4})\Big\{n_{_1}!n_{_2}!n_{_4}!
\nonumber\\
&&\hspace{2.5cm}\times
\Gamma(2+n_{_2}+n_{_3}+n_{_4})\Gamma({D\over2}+n_{_1})
\Gamma({D\over2}-2-n_{_1}-n_{_2}-n_{_3}-n_{_4})
\nonumber\\
&&\hspace{2.5cm}\times
\Gamma(2-{D\over2}+n_{_4})\Gamma({D\over2}+n_{_2})
\Gamma({D\over2}-1-n_{_2}-n_{_4})
\nonumber\\
&&\hspace{2.5cm}\times
\Gamma(3-{D\over2}+n_{_2}+n_{_3}+n_{_4})\Big\}^{-1}
\;,\nonumber\\
&&c_{_{[\tilde{1}\tilde{3}\tilde{5}\tilde{7}]}}^{(2),c}(\alpha,{\bf n})=
(-)^{1+n_{_1}+n_{_4}}\Gamma(2+n_{_1}+n_{_2}+n_{_3})\Gamma(1+n_{_2})\Gamma(1+n_{_4})\Big\{n_{_1}!
\nonumber\\
&&\hspace{2.5cm}\times
\Gamma(2+n_{_2}+n_{_4})\Gamma(2+n_{_2}+n_{_3})
\Gamma({D\over2}-2-n_{_1}-n_{_2}-n_{_3})
\nonumber\\
&&\hspace{2.5cm}\times
\Gamma({D\over2}-1-n_{_2})\Gamma({D\over2}-1-n_{_4})\Gamma(3-{D\over2}+n_{_2}+n_{_3})
\nonumber\\
&&\hspace{2.5cm}\times
\Gamma({D\over2}+n_{_1})\Gamma(3-{D\over2}+n_{_2}+n_{_4})\Big\}^{-1}\;.
\label{GKZ21o-2-3}
\end{eqnarray}

\item   $I_{_{3}}=\{1,\cdots,5,7,9,11,12,13\}$, i.e. the implement $J_{_{3}}=[1,14]\setminus I_{_{3}}=\{6,8,10,14\}$.
The choice implies the power numbers $\alpha_{_6}=\alpha_{_{8}}=\alpha_{_{10}}=\alpha_{_{14}}=0$, and
\begin{eqnarray}
&&\alpha_{_1}=b_{_1}+b_{_5}-a_{_1}-2,\;\alpha_{_2}=b_{_1}+b_{_5}-a_{_2}-2,\;\alpha_{_3}=b_{_5}-a_{_3}-1,
\nonumber\\
&&\alpha_{_4}=b_{_5}-a_{_4}-1,\;\alpha_{_5}=-a_{_5},\;\alpha_{_7}=b_{_2}+b_{_3}-b_{_5}-1,
\nonumber\\
&&\alpha_{_{9}}=b_{_4}-b_{_5},\;\alpha_{_{11}}=1-b_{_1},\;\alpha_{_{12}}=b_{_3}-b_{_5},\;\alpha_{_{13}}=1-b_{_3}.
\label{GKZ21o-3-1}
\end{eqnarray}
The corresponding hypergeometric solutions are
\begin{eqnarray}
&&\Phi_{_{[\tilde{1}\tilde{3}\tilde{5}\tilde{7}]}}^{(3),a}(\alpha,z)=
y_{_1}^{{D\over2}-1}y_{_2}^{{D\over2}-2}y_{_3}^{{D\over2}-1}\sum\limits_{n_{_1}=0}^\infty
\sum\limits_{n_{_2}=0}^\infty\sum\limits_{n_{_3}=0}^\infty\sum\limits_{n_{_4}=0}^\infty
c_{_{[\tilde{1}\tilde{3}\tilde{5}\tilde{7}]}}^{(3),a}(\alpha,{\bf n})
\nonumber\\
&&\hspace{2.5cm}\times
y_{_1}^{n_{_1}}y_{_4}^{n_{_2}}\Big({y_{_4}\over y_{_2}}\Big)^{n_{_3}}
\Big({y_{_3}\over y_{_2}}\Big)^{n_{_4}}
\;,\nonumber\\
&&\Phi_{_{[\tilde{1}\tilde{3}\tilde{5}\tilde{7}]}}^{(3),b}(\alpha,z)=
y_{_1}^{{D\over2}-1}y_{_2}^{{D\over2}-1}y_{_3}^{{D\over2}-1}\sum\limits_{n_{_1}=0}^\infty
\sum\limits_{n_{_2}=0}^\infty\sum\limits_{n_{_3}=0}^\infty\sum\limits_{n_{_4}=0}^\infty
c_{_{[\tilde{1}\tilde{3}\tilde{5}\tilde{7}]}}^{(3),b}(\alpha,{\bf n})
\nonumber\\
&&\hspace{2.5cm}\times
y_{_1}^{n_{_1}}y_{_2}^{n_{_2}}y_{_4}^{n_{_3}}y_{_3}^{n_{_4}}
\;,\nonumber\\
&&\Phi_{_{[\tilde{1}\tilde{3}\tilde{5}\tilde{7}]}}^{(3),c}(\alpha,z)=
y_{_1}^{{D\over2}-1}y_{_2}^{{D\over2}-2}y_{_3}^{{D\over2}}\sum\limits_{n_{_1}=0}^\infty
\sum\limits_{n_{_2}=0}^\infty\sum\limits_{n_{_3}=0}^\infty\sum\limits_{n_{_4}=0}^\infty
c_{_{[\tilde{1}\tilde{3}\tilde{5}\tilde{7}]}}^{(3),c}(\alpha,{\bf n})
\nonumber\\
&&\hspace{2.5cm}\times
y_{_1}^{n_{_1}}y_{_3}^{n_{_2}}y_{_4}^{n_{_3}}\Big({y_{_3}\over y_{_2}}\Big)^{n_{_4}}\;.
\label{GKZ21o-3-2a}
\end{eqnarray}
Where the coefficients are
\begin{eqnarray}
&&c_{_{[\tilde{1}\tilde{3}\tilde{5}\tilde{7}]}}^{(3),a}(\alpha,{\bf n})=
(-)^{n_{_1}+n_{_4}}\Gamma(1+n_{_1}+n_{_2})\Gamma(1+n_{_3}+n_{_4})
\Big\{n_{_1}!n_{_2}!n_{_3}!n_{_4}!
\nonumber\\
&&\hspace{2.5cm}\times
\Gamma(1-{D\over2}-n_{_1}-n_{_2})\Gamma(2-{D\over2}+n_{_3})\Gamma({D\over2}+n_{_2})
\nonumber\\
&&\hspace{2.5cm}\times
\Gamma({D\over2}+n_{_1})\Gamma({D\over2}-1-n_{_3}-n_{_4})
\Gamma({D\over2}+n_{_4})\Big\}^{-1}
\;,\nonumber\\
&&c_{_{[\tilde{1}\tilde{3}\tilde{5}\tilde{7}]}}^{(3),b}(\alpha,{\bf n})=
(-)^{1+n_{_1}}\Gamma(2+n_{_1}+n_{_2}+n_{_3}+n_{_4})\Gamma(1+n_{_2}+n_{_4})\Big\{n_{_1}!n_{_2}!n_{_4}!
\nonumber\\
&&\hspace{2.5cm}\times
\Gamma(2+n_{_2}+n_{_3}+n_{_4})\Gamma(-{D\over2}-n_{_1}-n_{_2}-n_{_3}-n_{_4})
\nonumber\\
&&\hspace{2.5cm}\times
\Gamma(1-{D\over2}-n_{_2}-n_{_4})\Gamma({D\over2}+1+n_{_2}+n_{_3}+n_{_4})
\nonumber\\
&&\hspace{2.5cm}\times
\Gamma({D\over2}+n_{_1})\Gamma({D\over2}+n_{_2})\Gamma({D\over2}+n_{_4})\Big\}^{-1}
\;,\nonumber\\
&&c_{_{[\tilde{1}\tilde{3}\tilde{5}\tilde{7}]}}^{(3),c}(\alpha,{\bf n})=
(-)^{1+n_{_1}+n_{_4}}\Gamma(2+n_{_1}+n_{_2}+n_{_3})\Gamma(1+n_{_2})\Gamma(1+n_{_4})
\nonumber\\
&&\hspace{2.5cm}\times
\Big\{n_{_1}!\Gamma(2+n_{_2}+n_{_3})\Gamma(-{D\over2}-n_{_1}-n_{_2}-n_{_3})
\nonumber\\
&&\hspace{2.5cm}\times
\Gamma(2+n_{_2}+n_{_4})\Gamma(1-{D\over2}-n_{_2})\Gamma({D\over2}+1+n_{_2}+n_{_3})
\nonumber\\
&&\hspace{2.5cm}\times
\Gamma({D\over2}+n_{_1})\Gamma({D\over2}-1-n_{_4})\Gamma({D\over2}+1+n_{_2}+n_{_4})\Big\}^{-1}\;.
\label{GKZ21o-3-3}
\end{eqnarray}

\item   $I_{_{4}}=\{1,\cdots,5,7,8,9,11,12\}$, i.e. the implement $J_{_{4}}=[1,14]\setminus I_{_{4}}=\{6,10,13,14\}$.
The choice implies the power numbers $\alpha_{_6}=\alpha_{_{10}}=\alpha_{_{13}}=\alpha_{_{14}}=0$, and
\begin{eqnarray}
&&\alpha_{_1}=b_{_1}+b_{_5}-a_{_1}-2,\;\alpha_{_2}=b_{_1}+b_{_5}-a_{_2}-2,
\nonumber\\
&&\alpha_{_3}=b_{_5}-a_{_3}-1,\;\alpha_{_4}=b_{_5}-a_{_4}-1,\;\alpha_{_5}=-a_{_5},\;\alpha_{_7}=b_{_2}-b_{_5},
\nonumber\\
&&\alpha_{_{8}}=b_{_3}-1,\;\alpha_{_{9}}=b_{_4}-b_{_5},\;\alpha_{_{11}}=1-b_{_1},\;\alpha_{_{12}}=1-b_{_5}.
\label{GKZ21o-4-1}
\end{eqnarray}
The corresponding hypergeometric solutions are
\begin{eqnarray}
&&\Phi_{_{[\tilde{1}\tilde{3}\tilde{5}\tilde{7}]}}^{(4),a}(\alpha,z)=
y_{_1}^{{D\over2}-1}y_{_2}^{{D}-3}\sum\limits_{n_{_1}=0}^\infty
\sum\limits_{n_{_2}=0}^\infty\sum\limits_{n_{_3}=0}^\infty\sum\limits_{n_{_4}=0}^\infty
c_{_{[\tilde{1}\tilde{3}\tilde{5}\tilde{7}]}}^{(4),a}(\alpha,{\bf n})
\nonumber\\
&&\hspace{2.5cm}\times
y_{_1}^{n_{_1}}y_{_4}^{n_{_2}}\Big({y_{_4}\over y_{_2}}\Big)^{n_{_3}}
\Big({y_{_3}\over y_{_2}}\Big)^{n_{_4}}
\;,\nonumber\\
&&\Phi_{_{[\tilde{1}\tilde{3}\tilde{5}\tilde{7}]}}^{(4),b}(\alpha,z)=
y_{_1}^{{D\over2}-1}y_{_2}^{{D}-2}\sum\limits_{n_{_1}=0}^\infty
\sum\limits_{n_{_2}=0}^\infty\sum\limits_{n_{_3}=0}^\infty\sum\limits_{n_{_4}=0}^\infty
c_{_{[\tilde{1}\tilde{3}\tilde{5}\tilde{7}]}}^{(4),b}(\alpha,{\bf n})
\nonumber\\
&&\hspace{2.5cm}\times
y_{_1}^{n_{_1}}y_{_2}^{n_{_2}}y_{_4}^{n_{_3}}\Big({y_{_3}\over y_{_2}}\Big)^{n_{_4}}\;.
\label{GKZ21o-4-2a}
\end{eqnarray}
Where the coefficients are
\begin{eqnarray}
&&c_{_{[\tilde{1}\tilde{3}\tilde{5}\tilde{7}]}}^{(4),a}(\alpha,{\bf n})=
(-)^{n_{_1}+n_{_3}}\Gamma(1+n_{_1}+n_{_2})\Big\{n_{_1}!n_{_2}!n_{_3}!n_{_4}!
\nonumber\\
&&\hspace{2.5cm}\times
\Gamma(1-{D\over2}-n_{_1}-n_{_2})\Gamma(2-{D\over2}+n_{_3})
\Gamma({D\over2}-1-n_{_3}-n_{_4})
\nonumber\\
&&\hspace{2.5cm}\times
\Gamma(2-{D\over2}+n_{_4})\Gamma({D\over2}+n_{_2})\Gamma({D\over2}+n_{_1})
\Gamma(D-2-n_{_3}-n_{_4})\Big\}^{-1}
\;,\nonumber\\
&&c_{_{[\tilde{1}\tilde{3}\tilde{5}\tilde{7}]}}^{(4),b}(\alpha,{\bf n})=
(-)^{1+n_{_1}}\Gamma(2+n_{_1}+n_{_2}+n_{_3})\Gamma(1+n_{_2})\Big\{n_{_1}!n_{_4}!
\nonumber\\
&&\hspace{2.5cm}\times
\Gamma(2+n_{_2}+n_{_3})\Gamma(-{D\over2}-n_{_1}-n_{_2}-n_{_3})\Gamma(1-{D\over2}-n_{_2})
\nonumber\\
&&\hspace{2.5cm}\times
\Gamma({D\over2}+n_{_2}-n_{_4})\Gamma(2-{D\over2}+n_{_4})\Gamma({D\over2}+1+n_{_2}+n_{_3})
\nonumber\\
&&\hspace{2.5cm}\times
\Gamma({D\over2}+n_{_1})\Gamma(D-1+n_{_2}-n_{_4})\Big\}^{-1}\;.
\label{GKZ21o-4-3}
\end{eqnarray}

\item   $I_{_{5}}=\{1,\cdots,7,10,12,13\}$, i.e. the implement $J_{_{5}}=[1,14]\setminus I_{_{5}}=\{8,9,11,14\}$.
The choice implies the power numbers $\alpha_{_8}=\alpha_{_{9}}=\alpha_{_{11}}=\alpha_{_{14}}=0$, and
\begin{eqnarray}
&&\alpha_{_1}=b_{_4}-a_{_1}-1,\;\alpha_{_2}=b_{_4}-a_{_2}-1,\;\alpha_{_3}=b_{_4}-a_{_3}-1,
\nonumber\\
&&\alpha_{_4}=b_{_4}-a_{_4}-1,\;\alpha_{_5}=-a_{_5},\;\alpha_{_{6}}=b_{_1}-1,\;\alpha_{_7}=b_{_2}+b_{_3}-b_{_4}-1,
\nonumber\\
&&\alpha_{_{10}}=b_{_5}-b_{_4},\;\alpha_{_{12}}=b_{_3}-b_{_4},\;\alpha_{_{13}}=1-b_{_3}.
\label{GKZ21o-5-1}
\end{eqnarray}
The corresponding hypergeometric functions are given as
\begin{eqnarray}
&&\Phi_{_{[\tilde{1}\tilde{3}\tilde{5}\tilde{7}]}}^{(5),a}(\alpha,z)=
y_{_2}^{-1}y_{_3}^{{D\over2}-1}\sum\limits_{n_{_1}=0}^\infty
\sum\limits_{n_{_2}=0}^\infty\sum\limits_{n_{_3}=0}^\infty\sum\limits_{n_{_4}=0}^\infty
c_{_{[\tilde{1}\tilde{3}\tilde{5}\tilde{7}]}}^{(5),a}(\alpha,{\bf n})
\nonumber\\
&&\hspace{2.5cm}\times
y_{_1}^{n_{_1}}y_{_4}^{n_{_2}}\Big({y_{_4}\over y_{_2}}\Big)^{n_{_3}}
\Big({y_{_3}\over y_{_2}}\Big)^{n_{_4}}
\;,\nonumber\\
&&\Phi_{_{[\tilde{1}\tilde{3}\tilde{5}\tilde{7}]}}^{(5),b}(\alpha,z)=
y_{_3}^{{D\over2}-1}\sum\limits_{n_{_1}=0}^\infty
\sum\limits_{n_{_2}=0}^\infty\sum\limits_{n_{_3}=0}^\infty\sum\limits_{n_{_4}=0}^\infty
c_{_{[\tilde{1}\tilde{3}\tilde{5}\tilde{7}]}}^{(5),b}(\alpha,{\bf n})
\nonumber\\
&&\hspace{2.5cm}\times
y_{_1}^{n_{_1}}y_{_2}^{n_{_2}}y_{_4}^{n_{_3}}y_{_3}^{n_{_4}}
\;,\nonumber\\
&&\Phi_{_{[\tilde{1}\tilde{3}\tilde{5}\tilde{7}]}}^{(5),c}(\alpha,z)=
y_{_2}^{-1}y_{_3}^{{D\over2}}\sum\limits_{n_{_1}=0}^\infty
\sum\limits_{n_{_2}=0}^\infty\sum\limits_{n_{_3}=0}^\infty\sum\limits_{n_{_4}=0}^\infty
c_{_{[\tilde{1}\tilde{3}\tilde{5}\tilde{7}]}}^{(5),c}(\alpha,{\bf n})
\nonumber\\
&&\hspace{2.5cm}\times
y_{_1}^{n_{_1}}y_{_3}^{n_{_2}}y_{_4}^{n_{_3}}\Big({y_{_3}\over y_{_2}}\Big)^{n_{_4}}\;.
\label{GKZ21o-5-2a}
\end{eqnarray}
Where the coefficients are
\begin{eqnarray}
&&c_{_{[\tilde{1}\tilde{3}\tilde{5}\tilde{7}]}}^{(5),a}(\alpha,{\bf n})=
(-)^{n_{_2}+n_{_4}}\Gamma(1+n_{_3}+n_{_4})
\Big\{n_{_1}!n_{_2}!n_{_3}!n_{_4}!\Gamma({D\over2}-1-n_{_1}-n_{_2})
\nonumber\\
&&\hspace{2.5cm}\times
\Gamma(D-2-n_{_1}-n_{_2})\Gamma({D\over2}+n_{_3})\Gamma(2-{D\over2}+n_{_1})
\nonumber\\
&&\hspace{2.5cm}\times
\Gamma(1-{D\over2}-n_{_3}-n_{_4})\Gamma(2-{D\over2}+n_{_2})
\Gamma({D\over2}+n_{_4})\Big\}^{-1}
\;,\nonumber\\
&&c_{_{[\tilde{1}\tilde{3}\tilde{5}\tilde{7}]}}^{(5),b}(\alpha,{\bf n})=
(-)^{n_{_2}+n_{_3}+n_{_4}}\Gamma(1+n_{_2}+n_{_4})\Big\{n_{_1}!n_{_2}!n_{_4}!
\Gamma(2+n_{_2}+n_{_3}+n_{_4})
\nonumber\\
&&\hspace{2.5cm}\times
\Gamma({D\over2}-2-n_{_1}-n_{_2}-n_{_3}-n_{_4})\Gamma(D-3-n_{_1}-n_{_2}-n_{_3}-n_{_4})
\nonumber\\
&&\hspace{2.5cm}\times
\Gamma({D\over2}-1-n_{_2}-n_{_4})\Gamma(2-{D\over2}+n_{_1})
\Gamma(2-{D\over2}+n_{_2})
\nonumber\\
&&\hspace{2.5cm}\times
\Gamma(3-{D\over2}+n_{_2}+n_{_3}+n_{_4})\Gamma({D\over2}+n_{_4})\Big\}^{-1}
\;,\nonumber\\
&&c_{_{[\tilde{1}\tilde{3}\tilde{5}\tilde{7}]}}^{(5),c}(\alpha,{\bf n})=
(-)^{n_{_2}+n_{_3}+n_{_4}}\Gamma(1+n_{_2})\Gamma(1+n_{_4})
\Big\{n_{_1}!\Gamma(2+n_{_2}+n_{_3})\Gamma(2+n_{_2}+n_{_4})
\nonumber\\
&&\hspace{2.5cm}\times
\Gamma(D-3-n_{_1}-n_{_2}-n_{_3})\Gamma({D\over2}-2-n_{_1}-n_{_2}-n_{_3})
\nonumber\\
&&\hspace{2.5cm}\times
\Gamma({D\over2}-1-n_{_2})\Gamma(2-{D\over2}+n_{_1})\Gamma(1-{D\over2}-n_{_4})
\nonumber\\
&&\hspace{2.5cm}\times
\Gamma(3-{D\over2}+n_{_2}+n_{_3})\Gamma({D\over2}+1+n_{_2}+n_{_4})\Big\}^{-1}\;.
\label{GKZ21o-5-3}
\end{eqnarray}

\item   $I_{_{6}}=\{1,\cdots,8,10,12\}$, i.e. the implement $J_{_{6}}=[1,14]\setminus I_{_{6}}=\{9,11,13,14\}$.
The choice implies the power numbers $\alpha_{_9}=\alpha_{_{11}}=\alpha_{_{13}}=\alpha_{_{14}}=0$, and
\begin{eqnarray}
&&\alpha_{_1}=b_{_4}-a_{_1}-1,\;\alpha_{_2}=b_{_4}-a_{_2}-1,\;\alpha_{_3}=b_{_4}-a_{_3}-1,
\nonumber\\
&&\alpha_{_4}=b_{_4}-a_{_4}-1,\;\alpha_{_5}=-a_{_5},\;\alpha_{_{6}}=b_{_1}-1,\;\alpha_{_7}=b_{_2}-b_{_4},
\nonumber\\
&&\alpha_{_{8}}=b_{_3}-1,\;\alpha_{_{10}}=b_{_5}-b_{_4},\;\alpha_{_{12}}=1-b_{_4}.
\label{GKZ21o-6-1}
\end{eqnarray}
The corresponding hypergeometric solutions are
\begin{eqnarray}
&&\Phi_{_{[\tilde{1}\tilde{3}\tilde{5}\tilde{7}]}}^{(6),a}(\alpha,z)=
y_{_2}^{{D\over2}-2}\sum\limits_{n_{_1}=0}^\infty
\sum\limits_{n_{_2}=0}^\infty\sum\limits_{n_{_3}=0}^\infty\sum\limits_{n_{_4}=0}^\infty
c_{_{[\tilde{1}\tilde{3}\tilde{5}\tilde{7}]}}^{(6),a}(\alpha,{\bf n})
\nonumber\\
&&\hspace{2.5cm}\times
y_{_1}^{n_{_1}}y_{_4}^{n_{_2}}\Big({y_{_4}\over y_{_2}}\Big)^{n_{_3}}
\Big({y_{_3}\over y_{_2}}\Big)^{n_{_4}}
\;,\nonumber\\
&&\Phi_{_{[\tilde{1}\tilde{3}\tilde{5}\tilde{7}]}}^{(6),b}(\alpha,z)=
y_{_2}^{{D\over2}-1}\sum\limits_{n_{_1}=0}^\infty
\sum\limits_{n_{_2}=0}^\infty\sum\limits_{n_{_3}=0}^\infty\sum\limits_{n_{_4}=0}^\infty
c_{_{[\tilde{1}\tilde{3}\tilde{5}\tilde{7}]}}^{(6),b}(\alpha,{\bf n})
\nonumber\\
&&\hspace{2.5cm}\times
y_{_1}^{n_{_1}}y_{_2}^{n_{_2}}y_{_4}^{n_{_3}}y_{_3}^{n_{_4}}
\;,\nonumber\\
&&\Phi_{_{[\tilde{1}\tilde{3}\tilde{5}\tilde{7}]}}^{(6),c}(\alpha,z)=
y_{_2}^{{D\over2}-2}y_{_3}\sum\limits_{n_{_1}=0}^\infty
\sum\limits_{n_{_2}=0}^\infty\sum\limits_{n_{_3}=0}^\infty\sum\limits_{n_{_4}=0}^\infty
c_{_{[\tilde{1}\tilde{3}\tilde{5}\tilde{7}]}}^{(6),c}(\alpha,{\bf n})
\nonumber\\
&&\hspace{2.5cm}\times
y_{_1}^{n_{_1}}y_{_3}^{n_{_2}}y_{_4}^{n_{_3}}\Big({y_{_3}\over y_{_2}}\Big)^{n_{_4}}\;.
\label{GKZ21o-6-2a}
\end{eqnarray}
Where the coefficients are
\begin{eqnarray}
&&c_{_{[\tilde{1}\tilde{3}\tilde{5}\tilde{7}]}}^{(6),a}(\alpha,{\bf n})=
(-)^{n_{_2}+n_{_4}}\Gamma(1+n_{_3}+n_{_4})
\Big\{n_{_1}!n_{_2}!n_{_3}!n_{_4}!\Gamma({D\over2}-1-n_{_1}-n_{_2})
\nonumber\\
&&\hspace{2.5cm}\times
\Gamma(D-2-n_{_1}-n_{_2})\Gamma({D\over2}+n_{_3})\Gamma(2-{D\over2}+n_{_1})
\nonumber\\
&&\hspace{2.5cm}\times
\Gamma({D\over2}-1-n_{_3}-n_{_4})\Gamma(2-{D\over2}+n_{_2})
\Gamma(2-{D\over2}+n_{_4})\Big\}^{-1}
\;,\nonumber\\
&&c_{_{[\tilde{1}\tilde{3}\tilde{5}\tilde{7}]}}^{(6),b}(\alpha,{\bf n})=
(-)^{n_{_2}+n_{_3}+n_{_4}}\Gamma(1+n_{_2}+n_{_4})\Big\{n_{_1}!n_{_2}!n_{_4}!
\Gamma(2+n_{_2}+n_{_3}+n_{_4})
\nonumber\\
&&\hspace{2.5cm}\times
\Gamma({D\over2}-2-n_{_1}-n_{_2}-n_{_3}-n_{_4})\Gamma(D-3-n_{_1}-n_{_2}-n_{_3}-n_{_4})
\nonumber\\
&&\hspace{2.5cm}\times
\Gamma({D\over2}-1-n_{_2}-n_{_4})\Gamma(2-{D\over2}+n_{_1})
\Gamma(2-{D\over2}+n_{_4})
\nonumber\\
&&\hspace{2.5cm}\times
\Gamma(3-{D\over2}+n_{_2}+n_{_3}+n_{_4})\Gamma({D\over2}+n_{_2})\Big\}^{-1}
\;,\nonumber\\
&&c_{_{[\tilde{1}\tilde{3}\tilde{5}\tilde{7}]}}^{(6),c}(\alpha,{\bf n})=
(-)^{n_{_2}+n_{_3}+n_{_4}}\Gamma(1+n_{_2})\Gamma(1+n_{_4})
\Big\{n_{_1}!\Gamma(2+n_{_2}+n_{_3})\Gamma(2+n_{_2}+n_{_4})
\nonumber\\
&&\hspace{2.5cm}\times
\Gamma({D\over2}-2-n_{_1}-n_{_2}-n_{_3})\Gamma(D-3-n_{_1}-n_{_2}-n_{_3})
\nonumber\\
&&\hspace{2.5cm}\times
\Gamma({D\over2}-1-n_{_2})\Gamma(2-{D\over2}+n_{_1})\Gamma({D\over2}-1-n_{_4})
\nonumber\\
&&\hspace{2.5cm}\times
\Gamma(3-{D\over2}+n_{_2}+n_{_3})\Gamma(3-{D\over2}+n_{_2}+n_{_4})\Big\}^{-1}\;.
\label{GKZ21o-6-3}
\end{eqnarray}

\item   $I_{_{7}}=\{1,\cdots,7,9,12,13\}$, i.e. the implement $J_{_{7}}=[1,14]\setminus I_{_{7}}=\{8,10,11,14\}$.
The choice implies the power numbers $\alpha_{_8}=\alpha_{_{10}}=\alpha_{_{11}}=\alpha_{_{14}}=0$, and
\begin{eqnarray}
&&\alpha_{_1}=b_{_5}-a_{_1}-1,\;\alpha_{_2}=b_{_5}-a_{_2}-1,\;\alpha_{_3}=b_{_5}-a_{_3}-1,
\nonumber\\
&&\alpha_{_4}=b_{_5}-a_{_4}-1,\;\alpha_{_5}=-a_{_5},\;\alpha_{_{6}}=b_{_1}-1,\;\alpha_{_7}=b_{_2}+b_{_3}-b_{_5}-1,
\nonumber\\
&&\alpha_{_{9}}=b_{_4}-b_{_5},\;\alpha_{_{12}}=b_{_3}-b_{_5},\;\alpha_{_{13}}=1-b_{_3}.
\label{GKZ21o-7-1}
\end{eqnarray}
The corresponding hypergeometric solutions are written as
\begin{eqnarray}
&&\Phi_{_{[\tilde{1}\tilde{3}\tilde{5}\tilde{7}]}}^{(7),a}(\alpha,z)=
y_{_2}^{{D\over2}-2}y_{_3}^{{D\over2}-1}\sum\limits_{n_{_1}=0}^\infty
\sum\limits_{n_{_2}=0}^\infty\sum\limits_{n_{_3}=0}^\infty\sum\limits_{n_{_4}=0}^\infty
c_{_{[\tilde{1}\tilde{3}\tilde{5}\tilde{7}]}}^{(7),a}(\alpha,{\bf n})
\nonumber\\
&&\hspace{2.5cm}\times
y_{_1}^{n_{_1}}y_{_4}^{n_{_2}}\Big({y_{_4}\over y_{_2}}\Big)^{n_{_3}}
\Big({y_{_3}\over y_{_2}}\Big)^{n_{_4}}
\;,\nonumber\\
&&\Phi_{_{[\tilde{1}\tilde{3}\tilde{5}\tilde{7}]}}^{(7),b}(\alpha,z)=
y_{_2}^{{D\over2}-1}y_{_3}^{{D\over2}-1}\sum\limits_{n_{_1}=0}^\infty
\sum\limits_{n_{_2}=0}^\infty\sum\limits_{n_{_3}=0}^\infty\sum\limits_{n_{_4}=0}^\infty
c_{_{[\tilde{1}\tilde{3}\tilde{5}\tilde{7}]}}^{(7),b}(\alpha,{\bf n})
\nonumber\\
&&\hspace{2.5cm}\times
y_{_1}^{n_{_1}}y_{_2}^{n_{_2}}y_{_4}^{n_{_3}}y_{_3}^{n_{_4}}
\;,\nonumber\\
&&\Phi_{_{[\tilde{1}\tilde{3}\tilde{5}\tilde{7}]}}^{(7),c}(\alpha,z)=
y_{_2}^{{D\over2}-2}y_{_3}^{{D\over2}}\sum\limits_{n_{_1}=0}^\infty
\sum\limits_{n_{_2}=0}^\infty\sum\limits_{n_{_3}=0}^\infty\sum\limits_{n_{_4}=0}^\infty
c_{_{[\tilde{1}\tilde{3}\tilde{5}\tilde{7}]}}^{(7),c}(\alpha,{\bf n})
\nonumber\\
&&\hspace{2.5cm}\times
y_{_1}^{n_{_1}}y_{_3}^{n_{_2}}y_{_4}^{n_{_3}}\Big({y_{_3}\over y_{_2}}\Big)^{n_{_4}}\;.
\label{GKZ21o-7-2a}
\end{eqnarray}
Where the coefficients are
\begin{eqnarray}
&&c_{_{[\tilde{1}\tilde{3}\tilde{5}\tilde{7}]}}^{(7),a}(\alpha,{\bf n})=
(-)^{n_{_1}+n_{_2}+n_{_4}}\Gamma(1+n_{_1}+n_{_2})
\Gamma(1+n_{_3}+n_{_4})\Big\{n_{_1}!n_{_2}!n_{_3}!n_{_4}!
\nonumber\\
&&\hspace{2.5cm}\times
\Gamma({D\over2}-1-n_{_1}-n_{_2})\Gamma(2-{D\over2}+n_{_3})
\Gamma(2-{D\over2}+n_{_1})
\nonumber\\
&&\hspace{2.5cm}\times
\Gamma({D\over2}+n_{_2})
\Gamma({D\over2}-1-n_{_3}-n_{_4})\Gamma({D\over2}+n_{_4})\Big\}^{-1}
\;,\nonumber\\
&&c_{_{[\tilde{1}\tilde{3}\tilde{5}\tilde{7}]}}^{(7),b}(\alpha,{\bf n})=
(-)^{1+n_{_1}}\Gamma(2+n_{_1}+n_{_2}+n_{_3}+n_{_4})\Gamma(1+n_{_2}+n_{_4})
\nonumber\\
&&\hspace{2.5cm}\times
\Big\{n_{_1}!n_{_2}!n_{_4}!\Gamma(2+n_{_2}+n_{_3}+n_{_4})\Gamma(1-{D\over2}-n_{_2}-n_{_4})
\nonumber\\
&&\hspace{2.5cm}\times
\Gamma({D\over2}-2-n_{_1}-n_{_2}-n_{_3}-n_{_4})\Gamma(2-{D\over2}+n_{_1})
\nonumber\\
&&\hspace{2.5cm}\times
\Gamma({D\over2}+1+n_{_2}+n_{_3}+n_{_4})
\Gamma({D\over2}+n_{_2})\Gamma({D\over2}+n_{_4})\Big\}^{-1}
\;,\nonumber\\
&&c_{_{[\tilde{1}\tilde{3}\tilde{5}\tilde{7}]}}^{(7),c}(\alpha,{\bf n})=
(-)^{1+n_{_1}+n_{_4}}\Gamma(2+n_{_1}+n_{_2}+n_{_3})\Gamma(1+n_{_2})\Gamma(1+n_{_4})
\nonumber\\
&&\hspace{2.5cm}\times
\Big\{n_{_1}!\Gamma(2+n_{_2}+n_{_4})\Gamma(2+n_{_2}+n_{_3})
\nonumber\\
&&\hspace{2.5cm}\times
\Gamma({D\over2}-2-n_{_1}-n_{_2}-n_{_3})\Gamma(1-{D\over2}-n_{_2})\Gamma(2-{D\over2}+n_{_1})
\nonumber\\
&&\hspace{2.5cm}\times
\Gamma({D\over2}+1+n_{_2}+n_{_3})
\Gamma({D\over2}+1+n_{_2}+n_{_4})\Gamma({D\over2}-1-n_{_4})\Big\}^{-1}\;.
\label{GKZ21o-7-3}
\end{eqnarray}

\item   $I_{_{8}}=\{1,\cdots,9,12\}$, i.e. the implement $J_{_{8}}=[1,14]\setminus I_{_{8}}=\{10,11,13,14\}$.
The choice implies the power numbers $\alpha_{_{10}}=\alpha_{_{11}}=\alpha_{_{13}}=\alpha_{_{14}}=0$, and
\begin{eqnarray}
&&\alpha_{_1}=b_{_5}-a_{_1}-1,\;\alpha_{_2}=b_{_5}-a_{_2}-1,\;\alpha_{_3}=b_{_5}-a_{_3}-1,
\nonumber\\
&&\alpha_{_4}=b_{_5}-a_{_4}-1,\;\alpha_{_5}=-a_{_5},\;\alpha_{_{6}}=b_{_1}-1,\;\alpha_{_7}=b_{_2}-b_{_5},
\nonumber\\
&&\alpha_{_{8}}=b_{_3}-1,\;\alpha_{_{9}}=b_{_4}-b_{_5},\;\alpha_{_{12}}=1-b_{_5}.
\label{GKZ21o-8-1}
\end{eqnarray}
The corresponding hypergeometric solutions are written as
\begin{eqnarray}
&&\Phi_{_{[\tilde{1}\tilde{3}\tilde{5}\tilde{7}]}}^{(8),a}(\alpha,z)=
y_{_2}^{{D}-3}\sum\limits_{n_{_1}=0}^\infty
\sum\limits_{n_{_2}=0}^\infty\sum\limits_{n_{_3}=0}^\infty\sum\limits_{n_{_4}=0}^\infty
c_{_{[\tilde{1}\tilde{3}\tilde{5}\tilde{7}]}}^{(8),a}(\alpha,{\bf n})
\nonumber\\
&&\hspace{2.5cm}\times
y_{_1}^{n_{_1}}y_{_4}^{n_{_2}}\Big({y_{_4}\over y_{_2}}\Big)^{n_{_3}}
\Big({y_{_3}\over y_{_2}}\Big)^{n_{_4}}
\;,\nonumber\\
&&\Phi_{_{[\tilde{1}\tilde{3}\tilde{5}\tilde{7}]}}^{(8),b}(\alpha,z)=
y_{_2}^{{D}-2}\sum\limits_{n_{_1}=0}^\infty
\sum\limits_{n_{_2}=0}^\infty\sum\limits_{n_{_3}=0}^\infty\sum\limits_{n_{_4}=0}^\infty
c_{_{[\tilde{1}\tilde{3}\tilde{5}\tilde{7}]}}^{(8),b}(\alpha,{\bf n})
\nonumber\\
&&\hspace{2.5cm}\times
y_{_1}^{n_{_1}}y_{_2}^{n_{_2}}y_{_4}^{n_{_3}}\Big({y_{_3}\over y_{_2}}\Big)^{n_{_4}}\;.
\label{GKZ21o-8-2a}
\end{eqnarray}
Where the coefficients are
\begin{eqnarray}
&&c_{_{[\tilde{1}\tilde{3}\tilde{5}\tilde{7}]}}^{(8),a}(\alpha,{\bf n})=
(-)^{n_{_1}+n_{_2}+n_{_3}}\Gamma(1+n_{_1}+n_{_2})\Big\{n_{_1}!n_{_2}!n_{_3}!n_{_4}!
\Gamma({D\over2}-1-n_{_1}-n_{_2})
\nonumber\\
&&\hspace{2.5cm}\times
\Gamma(2-{D\over2}+n_{_3})\Gamma(2-{D\over2}+n_{_1})\Gamma({D\over2}-1-n_{_3}-n_{_4})
\nonumber\\
&&\hspace{2.5cm}\times
\Gamma(2-{D\over2}+n_{_4})\Gamma({D\over2}+n_{_2})\Gamma(D-2-n_{_3}-n_{_4})\Big\}^{-1}
\;,\nonumber\\
&&c_{_{[\tilde{1}\tilde{3}\tilde{5}\tilde{7}]}}^{(8),b}(\alpha,{\bf n})=
(-)^{1+n_{_1}}\Gamma(2+n_{_1}+n_{_2}+n_{_3})\Gamma(1+n_{_2})\Big\{n_{_1}!n_{_4}!\Gamma(2+n_{_2}+n_{_3})
\nonumber\\
&&\hspace{2.5cm}\times
\Gamma({D\over2}-2-n_{_1}-n_{_2}-n_{_3})
\Gamma(1-{D\over2}-n_{_2})\Gamma(2-{D\over2}+n_{_1})
\nonumber\\
&&\hspace{2.5cm}\times
\Gamma({D\over2}+n_{_2}-n_{_4})
\Gamma(2-{D\over2}+n_{_4})\Gamma({D\over2}+1+n_{_2}+n_{_3})
\nonumber\\
&&\hspace{2.5cm}\times
\Gamma(D-1+n_{_2}-n_{_4})\Big\}^{-1}\;.
\label{GKZ21o-8-3}
\end{eqnarray}
\end{itemize}

\section{The convergent regions of the hypergeometric functions\label{app-con}}
\indent\indent

\begin{itemize}
\item The 28 hypergeometric functions $\Phi_{_{[1357]}}^{(i),a}$,
$\Phi_{_{[1357]}}^{(j)}$, $i=1,\cdots,4,7,\cdots,13,15$, $j=17,\cdots,32$ are convergent in the nonempty proper subset
of the whole parameter space
\begin{eqnarray}
&&\Xi_{_{[1357]}}^{1}=\{(y_{_1},\;y_{_2},\;y_{_3},\;y_{_4})
\Big|1<|y_{_4}|,\;|y_{_4}|<|y_{_3}|,\;|y_{_1}|<|y_{_4}|,\;|y_{_2}|<|y_{_3}|\}\;.
\label{GKZ22-1-1}
\end{eqnarray}
\item The 6 hypergeometric functions $\Phi_{_{[1357]}}^{(i),b}$, $i=1,\cdots,4,7,8$
are convergent in the nonempty proper subset of the whole parameter space
\begin{eqnarray}
&&\Xi_{_{[1357]}}^{2}=\{(y_{_1},\;y_{_2},\;y_{_3},\;y_{_4})
\Big|1<|y_{_1}|,\;1<|y_{_3}|,\;1<|y_{_4}|,\;|y_{_2}|<|y_{_3}|\}\;.
\label{GKZ22-1-2}
\end{eqnarray}
\item The 2 hypergeometric functions $\Phi_{_{[1357]}}^{(i)}$, $i=5,6$
are convergent in the nonempty proper subset of the whole parameter space
\begin{eqnarray}
&&\Xi_{_{[1357]}}^{3}=\{(y_{_1},\;y_{_2},\;y_{_3},\;y_{_4})
\Big|1<|y_{_1}|,\;|y_{_1}|<|y_{_3}|,\;|y_{_1}|<|y_{_4}|,\;|y_{_2}|<|y_{_3}|\}\;.
\label{GKZ22-1-3}
\end{eqnarray}
\item The 5 hypergeometric functions $\Phi_{_{[1357]}}^{(i),b}$, $i=9,\cdots,13$
are convergent in the nonempty proper subset of the whole parameter space
\begin{eqnarray}
&&\Xi_{_{[1357]}}^{4}=\{(y_{_1},\;y_{_2},\;y_{_3},\;y_{_4})
\Big|1<|y_{_4}|,\;|y_{_1}|<|y_{_4}|,\;|y_{_3}|<|y_{_4}|,\;|y_{_2}|<|y_{_4}|\}\;.
\label{GKZ22-1-4}
\end{eqnarray}
\item The 4 hypergeometric functions $\Phi_{_{[1357]}}^{(i),c}$, $i=9,10,11,13$
are convergent in the nonempty proper subset of the whole parameter space
\begin{eqnarray}
&&\Xi_{_{[1357]}}^{5}=\{(y_{_1},\;y_{_2},\;y_{_3},\;y_{_4})
\Big|1<|y_{_4}|,\;|y_{_1}|<|y_{_4}|,\;|y_{_2}|<|y_{_4}|,\;|y_{_2}|<|y_{_3}|\}\;.
\label{GKZ22-1-5}
\end{eqnarray}
\item The 2 hypergeometric functions $\Phi_{_{[1357]}}^{(i)}$, $i=14,16$
are convergent in the nonempty proper subset of the whole parameter space
\begin{eqnarray}
&&\Xi_{_{[1357]}}^{6}=\{(y_{_1},\;y_{_2},\;y_{_3},\;y_{_4})
\Big|1<|y_{_3}|,\;|y_{_1}|<|y_{_3}|,\;|y_{_3}|<|y_{_4}|,\;|y_{_2}|<|y_{_3}|\}\;.
\label{GKZ22-1-6}
\end{eqnarray}
\item The hypergeometric function $\Phi_{_{[1357]}}^{(15),b}$
is convergent in the nonempty proper subset of the whole parameter space
\begin{eqnarray}
&&\Xi_{_{[1357]}}^{7}=\{(y_{_1},\;y_{_2},\;y_{_3},\;y_{_4})
\Big|1<|y_{_4}|,\;|y_{_1}|<|y_{_4}|,\;|y_{_3}|<|y_{_4}|,\;|y_{_2}|<|y_{_3}|\}\;.
\label{GKZ22-1-7}
\end{eqnarray}

\item The 6 hypergeometric functions $\Phi_{_{[\tilde{1}357]}}^{(i),a}$,
$i=1,\cdots,4,7,8$ are convergent in the nonempty proper subset of the whole parameter space
\begin{eqnarray}
&&\Xi_{_{[\tilde{1}357]}}^{1}=\{(y_{_1},\;y_{_2},\;y_{_3},\;y_{_4})
\Big||y_{_1}|<|y_{_4}|,\;|y_{_4}|<|y_{_3}|,\;1<|y_{_4}|,\;|y_{_2}|<|y_{_3}|\}\;.
\label{GKZ22-2-1}
\end{eqnarray}
\item The 6 hypergeometric functions $\Phi_{_{[\tilde{1}357]}}^{(i),b}$,
$i=1,\cdots,4,7,8$
are convergent in the nonempty proper subset  of the whole parameter space
\begin{eqnarray}
&&\Xi_{_{[\tilde{1}357]}}^{2}=\{(y_{_1},\;y_{_2},\;y_{_3},\;y_{_4})
\Big||y_{_1}|<1,\;|y_{_1}|<|y_{_3}|,\;|y_{_1}|<|y_{_4}|,\;|y_{_2}|<|y_{_3}|\}\;.
\label{GKZ22-2-2}
\end{eqnarray}
\item The 2 hypergeometric functions $\Phi_{_{[\tilde{1}357]}}^{(5,6)}$
are convergent in the nonempty proper subset of the whole parameter space
\begin{eqnarray}
&&\Xi_{_{[\tilde{1}357]}}^{3}=\{(y_{_1},\;y_{_2},\;y_{_3},\;y_{_4})
\Big||y_{_1}|<1,\;1<|y_{_3}|,\;1<|y_{_4}|,\;|y_{_2}|<|y_{_3}|\}\;.
\label{GKZ22-2-3}
\end{eqnarray}

\item The 12 hypergeometric functions $\Phi_{_{[1\tilde{3}57]}}^{(i),a}$,
$i=1,2,3,5,6,7,25,26,27,29,30,31$
are convergent in the nonempty proper subset of the whole parameter space
\begin{eqnarray}
&&\Xi_{_{[1\tilde{3}57]}}^{1}=\{(y_{_1},\;y_{_2},\;y_{_3},\;y_{_4})
\Big|1<|y_{_1}|,\;|y_{_3}|<|y_{_4}|,\;|y_{_4}|<|y_{_1}|,\;|y_{_2}|<|y_{_4}|\}\;.
\label{GKZ22-3-1}
\end{eqnarray}
\item The 19 hypergeometric functions $\Phi_{_{[1\tilde{3}57]}}^{(i),b}$,
$\Phi_{_{[1\tilde{3}57]}}^{(j)}$, $i=1,2,3$, $j=9,\cdots,24$
are convergent in the nonempty proper subset of the whole parameter space
\begin{eqnarray}
&&\Xi_{_{[1\tilde{3}57]}}^{2}=\{(y_{_1},\;y_{_2},\;y_{_3},\;y_{_4})
\Big|1<|y_{_4}|,\;|y_{_3}|<|y_{_4}|,\;|y_{_1}|<|y_{_4}|,\;|y_{_2}|<|y_{_4}|\}\;.
\label{GKZ22-3-2}
\end{eqnarray}
\item The 4 hypergeometric functions $\Phi_{_{[1\tilde{3}57]}}^{(i),c}$,
$i=1,2,3,5$
are convergent in the nonempty proper subset of the whole parameter space
\begin{eqnarray}
&&\Xi_{_{[1\tilde{3}57]}}^{3}=\{(y_{_1},\;y_{_2},\;y_{_3},\;y_{_4})
\Big|1<|y_{_1}|,\;|y_{_3}|<|y_{_4}|,\;1<|y_{_4}|,\;|y_{_2}|<|y_{_4}|\}\;.
\label{GKZ22-3-3}
\end{eqnarray}
\item The 2 hypergeometric functions $\Phi_{_{[1\tilde{3}57]}}^{(4,8)}$
are convergent in the nonempty proper subset of the whole parameter space
\begin{eqnarray}
&&\Xi_{_{[1\tilde{3}57]}}^{4}=\{(y_{_1},\;y_{_2},\;y_{_3},\;y_{_4})
\Big|1<|y_{_1}|,\;|y_{_3}|<|y_{_1}|,\;|y_{_1}|<|y_{_4}|,\;|y_{_2}|<|y_{_1}|\}\;.
\label{GKZ22-3-4}
\end{eqnarray}
\item The 3 hypergeometric functions $\Phi_{_{[1\tilde{3}57]}}^{(i),b}$,
$i=5,6,7$ are convergent in the nonempty proper subset of the whole parameter space
\begin{eqnarray}
&&\Xi_{_{[1\tilde{3}57]}}^{5}=\{(y_{_1},\;y_{_2},\;y_{_3},\;y_{_4})
\Big|1<|y_{_1}|,\;|y_{_3}|<|y_{_4}|,\;|y_{_1}|<|y_{_4}|,\;|y_{_2}|<|y_{_4}|\}\;.
\label{GKZ22-3-5}
\end{eqnarray}
\item The 6 hypergeometric functions $\Phi_{_{[1\tilde{3}57]}}^{(i),b}$,
$i=25,26,27,29,30,31$
are convergent in the nonempty proper subset of the whole parameter space
\begin{eqnarray}
&&\Xi_{_{[1\tilde{3}57]}}^{6}=\{(y_{_1},\;y_{_2},\;y_{_3},\;y_{_4})
\Big|1<|y_{_1}|,\;|y_{_2}|<|y_{_1}|,\;|y_{_2}|<|y_{_4}|,\;|y_{_2}|<|y_{_3}|\}\;.
\label{GKZ22-3-6}
\end{eqnarray}
\item The 2 hypergeometric functions $\Phi_{_{[1\tilde{3}57]}}^{(28,32)}$
are convergent in the nonempty proper subset of the whole parameter space
\begin{eqnarray}
&&\Xi_{_{[1\tilde{3}57]}}^{7}=\{(y_{_1},\;y_{_2},\;y_{_3},\;y_{_4})
\Big|1<|y_{_1}|,\;|y_{_3}|<|y_{_1}|,\;|y_{_3}|<|y_{_4}|,\;|y_{_2}|<|y_{_3}|\}\;.
\label{GKZ22-3-7}
\end{eqnarray}

\item The 24 hypergeometric functions $\Phi_{_{[13\tilde{5}7]}}^{(i)}$,
$\Phi_{_{[13\tilde{5}7]}}^{(j),a}$, $i=1,\cdots,16$, $j=17,\cdots,24$
are convergent in the nonempty proper subset of the whole parameter space
\begin{eqnarray}
&&\Xi_{_{[13\tilde{5}7]}}^{1}=\{(y_{_1},\;y_{_2},\;y_{_3},\;y_{_4})
\Big|1<|y_{_1}|,\;|y_{_4}|<|y_{_3}|,\;|y_{_4}|<|y_{_1}|,\;|y_{_2}|<|y_{_3}|\}\;.
\label{GKZ22-4-1}
\end{eqnarray}
\item The 16 hypergeometric functions $\Phi_{_{[13\tilde{5}7]}}^{(i),b}$,
$\Phi_{_{[13\tilde{5}7]}}^{(j)}$, $i=17,\cdots,24$, $j=25,\cdots,32$
are convergent in the nonempty proper subset of the whole parameter space
\begin{eqnarray}
&&\Xi_{_{[13\tilde{5}7]}}^{2}=\{(y_{_1},\;y_{_2},\;y_{_3},\;y_{_4})
\Big|1<|y_{_3}|,\;|y_{_1}|<|y_{_3}|,\;|y_{_4}|<|y_{_3}|,\;|y_{_2}|<|y_{_3}|\}\;.
\label{GKZ22-4-2}
\end{eqnarray}
\item The 6 hypergeometric functions $\Phi_{_{[13\tilde{5}7]}}^{(i),c}$,
$i=17,\cdots,20,23,24$
are convergent in the nonempty proper subset of the whole parameter space
\begin{eqnarray}
&&\Xi_{_{[13\tilde{5}7]}}^{3}=\{(y_{_1},\;y_{_2},\;y_{_3},\;y_{_4})
\Big|1<|y_{_1}|,\;1<|y_{_3}|,\;|y_{_4}|<|y_{_3}|,\;|y_{_2}|<|y_{_3}|\}\;.
\label{GKZ22-4-3}
\end{eqnarray}

\item The 28 hypergeometric functions $\Phi_{_{[135\tilde{7}]}}^{(i),a}$,
$i=1,2,3,4,7,\cdots,30$
are convergent in the nonempty proper subset of the whole parameter space
\begin{eqnarray}
&&\Xi_{_{[135\tilde{7}]}}^{1}=\{(y_{_1},\;y_{_2},\;y_{_3},\;y_{_4})
\Big|1<|y_{_4}|,\;|y_{_4}|<|y_{_2}|,\;|y_{_1}|<|y_{_4}|,\;|y_{_3}|<|y_{_2}|\}\;.
\label{GKZ22-5-1}
\end{eqnarray}
\item The 6 hypergeometric functions $\Phi_{_{[135\tilde{7}]}}^{(i),b}$,
$i=1,\cdots,4,7,8$
are convergent in the nonempty proper subset of the whole parameter space
\begin{eqnarray}
&&\Xi_{_{[135\tilde{7}]}}^{2}=\{(y_{_1},\;y_{_2},\;y_{_3},\;y_{_4})
\Big|1<|y_{_1}|,\;1<|y_{_2}|,\;1<|y_{_4}|,\;|y_{_3}|<|y_{_2}|\}\;.
\label{GKZ22-5-2}
\end{eqnarray}
\item The 2 hypergeometric functions $\Phi_{_{[135\tilde{7}]}}^{(5,6)}$
are convergent in the nonempty proper subset of the whole parameter space
\begin{eqnarray}
&&\Xi_{_{[135\tilde{7}]}}^{3}=\{(y_{_1},\;y_{_2},\;y_{_3},\;y_{_4})
\Big|1<|y_{_1}|,\;|y_{_1}|<|y_{_2}|,\;|y_{_1}|<|y_{_4}|,\;|y_{_3}|<|y_{_2}|\}\;.
\label{GKZ22-5-3}
\end{eqnarray}
\item The 2 hypergeometric functions $\Phi_{_{[135\tilde{7}]}}^{(i),b}$
$i=28,30$ are convergent in the nonempty proper subset of the whole parameter space
\begin{eqnarray}
&&\Xi_{_{[135\tilde{7}]}}^{4}=\{(y_{_1},\;y_{_2},\;y_{_3},\;y_{_4})
\Big|1<|y_{_4}|,\;|y_{_1}|<|y_{_4}|,\;|y_{_2}|<|y_{_4}|,\;|y_{_3}|<|y_{_2}|\}\;.
\label{GKZ22-5-4}
\end{eqnarray}
\item The 4 hypergeometric functions $\Phi_{_{[135\tilde{7}]}}^{(i),c}$
$i=25,26,27,29$
are convergent in the nonempty proper subset of the whole parameter space
\begin{eqnarray}
&&\Xi_{_{[135\tilde{7}]}}^{5}=\{(y_{_1},\;y_{_2},\;y_{_3},\;y_{_4})
\Big|1<|y_{_4}|,\;|y_{_1}|<|y_{_4}|,\;|y_{_3}|<|y_{_4}|,\;|y_{_3}|<|y_{_2}|\}\;.
\label{GKZ22-5-5}
\end{eqnarray}
\item The 4 hypergeometric functions $\Phi_{_{[135\tilde{7}]}}^{(i),b}$,
$i=25,26,27,29$
are convergent in the nonempty proper subset of the whole parameter space
\begin{eqnarray}
&&\Xi_{_{[135\tilde{7}]}}^{6}=\{(y_{_1},\;y_{_2},\;y_{_3},\;y_{_4})
\Big|1<|y_{_4}|,\;|y_{_1}|<|y_{_4}|,\;|y_{_2}|<|y_{_4}|,\;|y_{_3}|<|y_{_4}|\}\;.
\label{GKZ22-5-6}
\end{eqnarray}
\item The 2 hypergeometric functions $\Phi_{_{[135\tilde{7}]}}^{(31,32)}$
are convergent in the nonempty proper subset of the whole parameter space
\begin{eqnarray}
&&\Xi_{_{[135\tilde{7}]}}^{7}=\{(y_{_1},\;y_{_2},\;y_{_3},\;y_{_4})
\Big|1<|y_{_2}|,\;|y_{_1}|<|y_{_2}|,\;|y_{_2}|<|y_{_4}|,\;|y_{_3}|<|y_{_2}|\}\;.
\label{GKZ22-5-7}
\end{eqnarray}

\item The 12 hypergeometric functions $\Phi_{_{[\tilde{1}\tilde{3}57]}}^{(i),a}$,
$i=1,2,3,5,6,7,9,10,11,13,14,15$
are convergent in the nonempty proper subset of the whole parameter space
\begin{eqnarray}
&&\Xi_{_{[\tilde{1}\tilde{3}57]}}^{1}=\{(y_{_1},\;y_{_2},\;y_{_3},\;y_{_4})
\Big||y_{_1}|<1,\;|y_{_3}|<|y_{_4}|,\;|y_{_4}|<1,\;|y_{_2}|<|y_{_4}|\}\;.
\label{GKZ22-6-1}
\end{eqnarray}
\item The 4 hypergeometric functions $\Phi_{_{[\tilde{1}\tilde{3}57]}}^{(i),b}$,
$i=1,2,3,5$
are convergent in the nonempty proper subset of the whole parameter space
\begin{eqnarray}
&&\Xi_{_{[\tilde{1}\tilde{3}57]}}^{2}=\{(y_{_1},\;y_{_2},\;y_{_3},\;y_{_4})
\Big||y_{_1}|<|y_{_4}|,\;|y_{_3}|<|y_{_4}|,\;1<|y_{_4}|,\;|y_{_2}|<|y_{_4}|\}\;.
\label{GKZ22-6-2}
\end{eqnarray}
\item The 4 hypergeometric functions $\Phi_{_{[\tilde{1}\tilde{3}57]}}^{(i),c}$,
$i=1,2,3,5$
are convergent in the nonempty proper subset of the whole parameter space
\begin{eqnarray}
&&\Xi_{_{[\tilde{1}\tilde{3}57]}}^{3}=\{(y_{_1},\;y_{_2},\;y_{_3},\;y_{_4})
\Big||y_{_1}|<1,\;|y_{_3}|<|y_{_4}|,\;|y_{_1}|<|y_{_4}|,\;|y_{_2}|<|y_{_4}|\}\;.
\label{GKZ22-6-3}
\end{eqnarray}
\item The 2 hypergeometric functions $\Phi_{_{[\tilde{1}\tilde{3}57]}}^{(4,8)}$
are convergent in the nonempty proper subset of the whole parameter space
\begin{eqnarray}
&&\Xi_{_{[\tilde{1}\tilde{3}57]}}^{4}=\{(y_{_1},\;y_{_2},\;y_{_3},\;y_{_4})
\Big||y_{_1}|<1,\;|y_{_3}|<1,\;1<|y_{_4}|,\;|y_{_2}|<1\}\;.
\label{GKZ22-6-4}
\end{eqnarray}
\item The 2 hypergeometric functions $\Phi_{_{[\tilde{1}\tilde{3}57]}}^{(i),b}$,
$i=6,7$ are convergent in the nonempty proper subset of the whole parameter space
\begin{eqnarray}
&&\Xi_{_{[\tilde{1}\tilde{3}57]}}^{5}=\{(y_{_1},\;y_{_2},\;y_{_3},\;y_{_4})
\Big||y_{_1}|<1,\;|y_{_3}|<|y_{_4}|,\;1<|y_{_4}|,\;|y_{_2}|<|y_{_4}|\}\;.
\label{GKZ22-6-5}
\end{eqnarray}
\item The 6 hypergeometric functions $\Phi_{_{[\tilde{1}\tilde{3}57]}}^{(i),b}$,
$i=9,10,11,13,14,15$
are convergent in the nonempty proper subset of the whole parameter space
\begin{eqnarray}
&&\Xi_{_{[\tilde{1}\tilde{3}57]}}^{6}=\{(y_{_1},\;y_{_2},\;y_{_3},\;y_{_4})
\Big||y_{_1}|<1,\;|y_{_2}|<1,\;|y_{_2}|<|y_{_4}|,\;|y_{_2}|<|y_{_3}|\}\;.
\label{GKZ22-6-6}
\end{eqnarray}
\item The 2 hypergeometric functions $\Phi_{_{[\tilde{1}\tilde{3}57]}}^{(12,16)}$
are convergent in the nonempty proper subset of the whole parameter space
\begin{eqnarray}
&&\Xi_{_{[\tilde{1}\tilde{3}57]}}^{7}=\{(y_{_1},\;y_{_2},\;y_{_3},\;y_{_4})
\Big||y_{_1}|<1,\;|y_{_3}|<1,\;|y_{_3}|<|y_{_4}|,\;|y_{_2}|<|y_{_3}|\}\;.
\label{GKZ22-6-7}
\end{eqnarray}

\item The 24 hypergeometric functions $\Phi_{_{[\tilde{1}3\tilde{5}7]}}^{(i)}$,
$\Phi_{_{[\tilde{1}3\tilde{5}7]}}^{(j),a}$, $i=1,\cdots,16$, $j=17,\cdots,24$
are convergent in the nonempty proper subset of the whole parameter space
\begin{eqnarray}
&&\Xi_{_{[\tilde{1}3\tilde{5}7]}}^{1}=\{(y_{_1},\;y_{_2},\;y_{_3},\;y_{_4})
\Big||y_{_1}|<1,\;|y_{_4}|<|y_{_3}|,\;|y_{_4}|<1,\;|y_{_2}|<|y_{_3}|\}\;.
\label{GKZ22-7-1}
\end{eqnarray}
\item The 6 hypergeometric functions $\Phi_{_{[\tilde{1}3\tilde{5}7]}}^{(i),b}$
$i=17,\cdots,20,23,24$
are convergent in the nonempty proper subset of the whole parameter space
\begin{eqnarray}
&&\Xi_{_{[\tilde{1}3\tilde{5}7]}}^{2}=\{(y_{_1},\;y_{_2},\;y_{_3},\;y_{_4})
\Big||y_{_1}|<|y_{_3}|,\;1<|y_{_3}|,\;|y_{_4}|<|y_{_3}|,\;|y_{_2}|<|y_{_3}|\}\;.
\label{GKZ22-7-2}
\end{eqnarray}
\item The 6 hypergeometric functions $\Phi_{_{[\tilde{1}3\tilde{5}7]}}^{(i),c}$,
$i=17,\cdots,20,23,24$
are convergent in the nonempty proper subset of the whole parameter space
\begin{eqnarray}
&&\Xi_{_{[\tilde{1}3\tilde{5}7]}}^{3}=\{(y_{_1},\;y_{_2},\;y_{_3},\;y_{_4})
\Big||y_{_1}|<1,\;|y_{_1}|<|y_{_3}|,\;|y_{_4}|<|y_{_3}|,\;|y_{_2}|<|y_{_3}|\}\;.
\label{GKZ22-7-3}
\end{eqnarray}
\item The 2 hypergeometric functions $\Phi_{_{[\tilde{1}3\tilde{5}7]}}^{(i),b}$,
$i=21,22$
are convergent in the nonempty proper subset of the whole parameter space
\begin{eqnarray}
&&\Xi_{_{[\tilde{1}3\tilde{5}7]}}^{4}=\{(y_{_1},\;y_{_2},\;y_{_3},\;y_{_4})
\Big||y_{_1}|<1,\;1<|y_{_3}|,\;|y_{_4}|<|y_{_3}|,\;|y_{_2}|<|y_{_3}|\}\;.
\label{GKZ22-7-4}
\end{eqnarray}

\item The 6 hypergeometric functions $\Phi_{_{[\tilde{1}35\tilde{7}]}}^{(i),a}$
$i=1,\cdots,4,7,8$
are convergent in the nonempty proper subset of the whole parameter space
\begin{eqnarray}
&&\Xi_{_{[\tilde{1}35\tilde{7}]}}^{1}=\{(y_{_1},\;y_{_2},\;y_{_3},\;y_{_4})
\Big||y_{_1}|<|y_{_4}|,\;|y_{_4}|<|y_{_2}|,\;1<|y_{_4}|,\;|y_{_3}|<|y_{_2}|\}\;.
\label{GKZ22-8-1}
\end{eqnarray}
\item The 6 hypergeometric functions $\Phi_{_{[\tilde{1}35\tilde{7}]}}^{(i),b}$,
$i=1,\cdots,4,7,8$
are convergent in the nonempty proper subset of the whole parameter space
\begin{eqnarray}
&&\Xi_{_{[\tilde{1}35\tilde{7}]}}^{2}=\{(y_{_1},\;y_{_2},\;y_{_3},\;y_{_4})
\Big||y_{_1}|<1,\;|y_{_1}|<|y_{_2}|,\;|y_{_1}|<|y_{_4}|,\;|y_{_3}|<|y_{_2}|\}\;.
\label{GKZ22-8-2}
\end{eqnarray}
\item The 2 hypergeometric functions $\Phi_{_{[\tilde{1}35\tilde{7}]}}^{(5,6)}$
are convergent in the nonempty proper subset of the whole parameter space
\begin{eqnarray}
&&\Xi_{_{[\tilde{1}35\tilde{7}]}}^{3}=\{(y_{_1},\;y_{_2},\;y_{_3},\;y_{_4})
\Big||y_{_1}|<1,\;1<|y_{_2}|,\;1<|y_{_4}|,\;|y_{_3}|<|y_{_2}|\}\;.
\label{GKZ22-8-3}
\end{eqnarray}

\item The 16 hypergeometric functions $\Phi_{_{[1\tilde{3}\tilde{5}7]}}^{(i)}$,
$i=1,\cdots,16$
are convergent in the nonempty proper subset of the whole parameter space
\begin{eqnarray}
&&\Xi_{_{[1\tilde{3}\tilde{5}7]}}^{1}=\{(y_{_1},\;y_{_2},\;y_{_3},\;y_{_4})
\Big|1<|y_{_1}|,\;|y_{_3}|<|y_{_4}|,\;|y_{_4}|<|y_{_1}|,\;|y_{_2}|<|y_{_4}|\}\;.
\label{GKZ22-9-1}
\end{eqnarray}
\item The 15 hypergeometric functions $\Phi_{_{[1\tilde{3}\tilde{5}7]}}^{(i)}$,
$\Phi_{_{[1\tilde{3}\tilde{5}7]}}^{(j),b}$, $i=17,\cdots,24$, $j=25,26,27,29,\cdots,32$
are convergent in the nonempty proper subset of the whole parameter space
\begin{eqnarray}
&&\Xi_{_{[1\tilde{3}\tilde{5}7]}}^{2}=\{(y_{_1},\;y_{_2},\;y_{_3},\;y_{_4})
\Big|1<|y_{_1}|,\;|y_{_3}|<|y_{_1}|,\;|y_{_4}|<|y_{_1}|,\;|y_{_2}|<|y_{_1}|\}\;.
\label{GKZ22-9-2}
\end{eqnarray}
\item The 8 hypergeometric functions $\Phi_{_{[1\tilde{3}\tilde{5}7]}}^{(i),a}$,
$i=25,\cdots,32$ are convergent in the nonempty proper subset of the whole parameter space
\begin{eqnarray}
&&\Xi_{_{[1\tilde{3}\tilde{5}7]}}^{3}=\{(y_{_1},\;y_{_2},\;y_{_3},\;y_{_4})
\Big|1<|y_{_1}|,\;|y_{_4}|<|y_{_1}|,\;|y_{_4}|<|y_{_3}|,\;|y_{_2}|<|y_{_3}|\}\;.
\label{GKZ22-9-3}
\end{eqnarray}
\item The 3 hypergeometric functions $\Phi_{_{[1\tilde{3}\tilde{5}7]}}^{(i),c}$,
$i=25,26,27$ are convergent in the nonempty proper subset of the whole parameter space
\begin{eqnarray}
&&\Xi_{_{[1\tilde{3}\tilde{5}7]}}^{4}=\{(y_{_1},\;y_{_2},\;y_{_3},\;y_{_4})
\Big|1<|y_{_1}|,\;|y_{_2}|<|y_{_1}|,\;|y_{_4}|<1,\;|y_{_2}|<|y_{_3}|\}\;.
\label{GKZ22-9-4}
\end{eqnarray}
\item The hypergeometric function $\Phi_{_{[1\tilde{3}\tilde{5}7]}}^{(28),b}$
is convergent in the nonempty proper subset of the whole parameter space
\begin{eqnarray}
&&\Xi_{_{[1\tilde{3}\tilde{5}7]}}^{5}=\{(y_{_1},\;y_{_2},\;y_{_3},\;y_{_4})
\Big|1<|y_{_1}|,\;|y_{_3}|<|y_{_1}|,\;|y_{_4}|<|y_{_1}|,\;|y_{_2}|<|y_{_3}|\}\;.
\label{GKZ22-9-5}
\end{eqnarray}
\item The 3 hypergeometric functions $\Phi_{_{[1\tilde{3}\tilde{5}7]}}^{(i),c}$,
$i=29,30,31$
are convergent in the nonempty proper subset of the whole parameter space
\begin{eqnarray}
&&\Xi_{_{[1\tilde{3}\tilde{5}7]}}^{6}=\{(y_{_1},\;y_{_2},\;y_{_3},\;y_{_4})
\Big|1<|y_{_1}|,\;|y_{_2}|<|y_{_1}|,\;|y_{_4}|<|y_{_1}|,\;|y_{_2}|<|y_{_3}|\}\;.
\label{GKZ22-9-6}
\end{eqnarray}

\item The 6 hypergeometric functions $\Phi_{_{[1\tilde{3}5\tilde{7}]}}^{(i),a}$,
$i=1,2,3,5,6,7$
are convergent in the nonempty proper subset of the whole parameter space
\begin{eqnarray}
&&\Xi_{_{[1\tilde{3}5\tilde{7}]}}^{1}=\{(y_{_1},\;y_{_2},\;y_{_3},\;y_{_4})
\Big|1<|y_{_1}|,\;|y_{_4}|<|y_{_1}|,\;|y_{_2}|<|y_{_4}|,\;|y_{_3}|<|y_{_4}|\}\;.
\label{GKZ22-10-1}
\end{eqnarray}
\item The 6 hypergeometric functions $\Phi_{_{[1\tilde{3}5\tilde{7}]}}^{(i),b}$,
$i=1,2,3,5,6,7$
are convergent in the nonempty proper subset of the whole parameter space
\begin{eqnarray}
&&\Xi_{_{[1\tilde{3}5\tilde{7}]}}^{2}=\{(y_{_1},\;y_{_2},\;y_{_3},\;y_{_4})
\Big|1<|y_{_1}|,\;|y_{_3}|<|y_{_1}|,\;|y_{_3}|<|y_{_4}|,\;|y_{_3}|<|y_{_2}|\}\;.
\label{GKZ22-10-2}
\end{eqnarray}
\item The 2 hypergeometric functions $\Phi_{_{[1\tilde{3}5\tilde{7}]}}^{(4,8)}$
are convergent in the nonempty proper subset of the whole parameter space
\begin{eqnarray}
&&\Xi_{_{[1\tilde{3}5\tilde{7}]}}^{3}=\{(y_{_1},\;y_{_2},\;y_{_3},\;y_{_4})
\Big|1<|y_{_1}|,\;|y_{_2}|<|y_{_1}|,\;|y_{_2}|<|y_{_4}|,\;|y_{_3}|<|y_{_2}|\}\;.
\label{GKZ22-10-3}
\end{eqnarray}

\item The 24 hypergeometric functions $\Phi_{_{[13\tilde{5}\tilde{7}]}}^{(i)}$,
$\Phi_{_{[13\tilde{5}\tilde{7}]}}^{(j),a}$, $i=1,\cdots,16$, $j=17,\cdots,24$
are convergent in the nonempty proper subset of the whole parameter space
\begin{eqnarray}
&&\Xi_{_{[13\tilde{5}\tilde{7}]}}^{1}=\{(y_{_1},\;y_{_2},\;y_{_3},\;y_{_4})
\Big|1<|y_{_1}|,\;|y_{_4}|<|y_{_2}|,\;|y_{_4}|<|y_{_1}|,\;|y_{_3}|<|y_{_2}|\}\;.
\label{GKZ22-11-1}
\end{eqnarray}
\item The 8 hypergeometric functions $\Phi_{_{[13\tilde{5}\tilde{7}]}}^{(i),b}$,
$i=17,\cdots,24$
are convergent in the nonempty proper subset of the whole parameter space
\begin{eqnarray}
&&\Xi_{_{[13\tilde{5}\tilde{7}]}}^{2}=\{(y_{_1},\;y_{_2},\;y_{_3},\;y_{_4})
\Big|1<|y_{_1}|,\;|y_{_1}|<|y_{_2}|,\;|y_{_4}|<|y_{_2}|,\;|y_{_3}|<|y_{_2}|\}\;.
\label{GKZ22-11-2}
\end{eqnarray}
\item The 6 hypergeometric functions $\Phi_{_{[13\tilde{5}\tilde{7}]}}^{(i),c}$,
$i=17,\cdots,20,23,24$
are convergent in the nonempty proper subset of the whole parameter space
\begin{eqnarray}
&&\Xi_{_{[13\tilde{5}\tilde{7}]}}^{3}=\{(y_{_1},\;y_{_2},\;y_{_3},\;y_{_4})
\Big|1<|y_{_1}|,\;1<|y_{_2}|,\;|y_{_4}|<|y_{_2}|,\;|y_{_3}|<|y_{_2}|\}\;.
\label{GKZ22-11-3}
\end{eqnarray}
\item The 8 hypergeometric functions $\Phi_{_{[13\tilde{5}\tilde{7}]}}^{(i)}$,
$i=25,\cdots,32$
are convergent in the nonempty proper subset of the whole parameter space
\begin{eqnarray}
&&\Xi_{_{[13\tilde{5}\tilde{7}]}}^{4}=\{(y_{_1},\;y_{_2},\;y_{_3},\;y_{_4})
\Big|1<|y_{_2}|,\;|y_{_1}|<|y_{_2}|,\;|y_{_4}|<|y_{_2}|,\;|y_{_3}|<|y_{_2}|\}\;.
\label{GKZ22-11-4}
\end{eqnarray}

\item The 16 hypergeometric functions $\Phi_{_{[\tilde{1}\tilde{3}\tilde{5}7]}}^{(i)}$,
$i=1,\cdots,16$
are convergent in the nonempty proper subset of the whole parameter space
\begin{eqnarray}
&&\Xi_{_{[\tilde{1}\tilde{3}\tilde{5}7]}}^{1}=\{(y_{_1},\;y_{_2},\;y_{_3},\;y_{_4})
\Big||y_{_1}|<1,\;|y_{_3}|<|y_{_4}|,\;|y_{_4}|<1,\;|y_{_2}|<|y_{_4}|\}\;.
\label{GKZ22-12-1}
\end{eqnarray}
\item The 14 hypergeometric functions $\Phi_{_{[\tilde{1}\tilde{3}\tilde{5}7]}}^{(i)}$,
$\Phi_{_{[\tilde{1}\tilde{3}\tilde{5}7]}}^{(j),b}$, $i=17,\cdots,24$, $j=25,26,27,29,30,31$
are convergent in the nonempty proper subset of the whole parameter space
\begin{eqnarray}
&&\Xi_{_{[\tilde{1}\tilde{3}\tilde{5}7]}}^{2}=\{(y_{_1},\;y_{_2},\;y_{_3},\;y_{_4})
\Big||y_{_1}|<1,\;|y_{_2}|<1,\;|y_{_3}|<1,\;|y_{_4}|<1\}\;.
\label{GKZ22-12-2}
\end{eqnarray}
\item The 8 hypergeometric functions $\Phi_{_{[\tilde{1}\tilde{3}\tilde{5}7]}}^{(i),a}$,
$i=25,\cdots,32$
are convergent in the nonempty proper subset of the whole parameter space
\begin{eqnarray}
&&\Xi_{_{[\tilde{1}\tilde{3}\tilde{5}7]}}^{3}=\{(y_{_1},\;y_{_2},\;y_{_3},\;y_{_4})
\Big||y_{_1}|<1,\;|y_{_4}|<1,\;|y_{_4}|<|y_{_3}|,\;|y_{_2}|<|y_{_3}|\}\;.
\label{GKZ22-12-3}
\end{eqnarray}
\item The 8 hypergeometric functions $\Phi_{_{[\tilde{1}\tilde{3}\tilde{5}7]}}^{(i),c}$,
$\Phi_{_{[\tilde{1}\tilde{3}\tilde{5}7]}}^{(j),b}$, $i=25,26,27,29,30,31$, $j=28,32$
are convergent in the nonempty proper subset of the whole parameter space
\begin{eqnarray}
&&\Xi_{_{[\tilde{1}\tilde{3}\tilde{5}7]}}^{4}=\{(y_{_1},\;y_{_2},\;y_{_3},\;y_{_4})
\Big||y_{_1}|<1,\;|y_{_2}|<1,\;|y_{_4}|<1,\;|y_{_2}|<|y_{_3}|\}\;.
\label{GKZ22-12-4}
\end{eqnarray}

\item The 6 hypergeometric functions $\Phi_{_{[\tilde{1}\tilde{3}5\tilde{7}]}}^{(i),a}$,
$i=1,2,3,5,6,7$
are convergent in the nonempty proper subset of the whole parameter space
\begin{eqnarray}
&&\Xi_{_{[\tilde{1}\tilde{3}5\tilde{7}]}}^{1}=\{(y_{_1},\;y_{_2},\;y_{_3},\;y_{_4})
\Big||y_{_1}|<1,\;|y_{_4}|<1,\;|y_{_2}|<|y_{_4}|,\;|y_{_3}|<|y_{_4}|\}\;.
\label{GKZ22-13-1}
\end{eqnarray}
\item The 6 hypergeometric functions $\Phi_{_{[\tilde{1}\tilde{3}5\tilde{7}]}}^{(i),b}$,
$i=1,2,3,5,6,7$
are convergent in the nonempty proper subset of the whole parameter space
\begin{eqnarray}
&&\Xi_{_{[\tilde{1}\tilde{3}5\tilde{7}]}}^{2}=\{(y_{_1},\;y_{_2},\;y_{_3},\;y_{_4})
\Big||y_{_1}|<1,\;|y_{_3}|<1,\;|y_{_3}|<|y_{_4}|,\;|y_{_3}|<|y_{_2}|\}\;.
\label{GKZ22-13-2}
\end{eqnarray}
\item The 2 hypergeometric functions $\Phi_{_{[\tilde{1}\tilde{3}5\tilde{7}]}}^{(4,8)}$
are convergent in the nonempty proper subset of the whole parameter space
\begin{eqnarray}
&&\Xi_{_{[\tilde{1}\tilde{3}5\tilde{7}]}}^{3}=\{(y_{_1},\;y_{_2},\;y_{_3},\;y_{_4})
\Big||y_{_1}|<1,\;|y_{_2}|<1,\;|y_{_2}|<|y_{_4}|,\;|y_{_3}|<|y_{_2}|\}\;.
\label{GKZ22-13-3}
\end{eqnarray}

\item The 24 hypergeometric functions $\Phi_{_{[\tilde{1}3\tilde{5}\tilde{7}]}}^{(i)}$,
$\Phi_{_{[\tilde{1}3\tilde{5}\tilde{7}]}}^{(j),a}$, $i=1,\cdots,16$, $j=17,\cdots,24$
are convergent in the nonempty proper subset of the whole parameter space
\begin{eqnarray}
&&\Xi_{_{[\tilde{1}3\tilde{5}\tilde{7}]}}^{1}=\{(y_{_1},\;y_{_2},\;y_{_3},\;y_{_4})
\Big||y_{_1}|<1,\;|y_{_4}|<|y_{_2}|,\;|y_{_4}|<1,\;|y_{_3}|<|y_{_2}|\}\;.
\label{GKZ22-14-1}
\end{eqnarray}
\item The 6 hypergeometric functions $\Phi_{_{[\tilde{1}3\tilde{5}\tilde{7}]}}^{(i),b}$,
$i=17,\cdots,20,23,24$
are convergent in the nonempty proper subset of the whole parameter space
\begin{eqnarray}
&&\Xi_{_{[\tilde{1}3\tilde{5}\tilde{7}]}}^{2}=\{(y_{_1},\;y_{_2},\;y_{_3},\;y_{_4})
\Big||y_{_1}|<|y_{_2}|,\;1<|y_{_2}|,\;|y_{_4}|<|y_{_2}|,\;|y_{_3}|<|y_{_2}|\}\;.
\label{GKZ22-14-2}
\end{eqnarray}
\item The 6 hypergeometric functions $\Phi_{_{[\tilde{1}3\tilde{5}\tilde{7}]}}^{(i),c}$,
$i=17,\cdots,20,23,24$
are convergent in the nonempty proper subset of the whole parameter space
\begin{eqnarray}
&&\Xi_{_{[\tilde{1}3\tilde{5}\tilde{7}]}}^{3}=\{(y_{_1},\;y_{_2},\;y_{_3},\;y_{_4})
\Big||y_{_1}|<1,\;|y_{_1}|<|y_{_2}|,\;|y_{_4}|<|y_{_2}|,\;|y_{_3}|<|y_{_2}|\}\;.
\label{GKZ22-14-3}
\end{eqnarray}
\item The 2 hypergeometric functions $\Phi_{_{[\tilde{1}3\tilde{5}\tilde{7}]}}^{(i),b}$,
$i=21,22$
are convergent in the nonempty proper subset of the whole parameter space
\begin{eqnarray}
&&\Xi_{_{[\tilde{1}3\tilde{5}\tilde{7}]}}^{4}=\{(y_{_1},\;y_{_2},\;y_{_3},\;y_{_4})
\Big||y_{_1}|<1,\;1<|y_{_2}|,\;|y_{_4}|<|y_{_2}|,\;|y_{_3}|<|y_{_2}|\}\;.
\label{GKZ22-14-4}
\end{eqnarray}

\item The 8 hypergeometric functions $\Phi_{_{[1\tilde{3}\tilde{5}\tilde{7}]}}^{(i),a}$,
$i=1,\cdots,8$
are convergent in the nonempty proper subset of the whole parameter space
\begin{eqnarray}
&&\Xi_{_{[1\tilde{3}\tilde{5}\tilde{7}]}}^{1}=\{(y_{_1},\;y_{_2},\;y_{_3},\;y_{_4})
\Big|1<|y_{_1}|,\;|y_{_4}|<|y_{_1}|,\;|y_{_4}|<|y_{_2}|,\;|y_{_3}|<|y_{_2}|\}\;.
\label{GKZ22-15-1}
\end{eqnarray}
\item The 6 hypergeometric functions $\Phi_{_{[1\tilde{3}\tilde{5}\tilde{7}]}}^{(i),b}$,
$i=1,2,3,5,6,7$
are convergent in the nonempty proper subset of the whole parameter space
\begin{eqnarray}
&&\Xi_{_{[1\tilde{3}\tilde{5}\tilde{7}]}}^{2}=\{(y_{_1},\;y_{_2},\;y_{_3},\;y_{_4})
\Big|1<|y_{_1}|,\;|y_{_2}|<|y_{_1}|,\;|y_{_4}|<|y_{_1}|,\;|y_{_3}|<|y_{_1}|\}\;.
\label{GKZ22-15-2}
\end{eqnarray}
\item The 6 hypergeometric functions $\Phi_{_{[1\tilde{3}\tilde{5}\tilde{7}]}}^{(i),c}$,
$i=1,2,3,5,6,7$
are convergent in the nonempty proper subset of the whole parameter space
\begin{eqnarray}
&&\Xi_{_{[1\tilde{3}\tilde{5}\tilde{7}]}}^{3}=\{(y_{_1},\;y_{_2},\;y_{_3},\;y_{_4})
\Big|1<|y_{_1}|,\;|y_{_3}|<|y_{_1}|,\;|y_{_4}|<|y_{_1}|,\;|y_{_3}|<|y_{_2}|\}\;.
\label{GKZ22-15-3}
\end{eqnarray}
\item The 2 hypergeometric functions $\Phi_{_{[1\tilde{3}\tilde{5}\tilde{7}]}}^{(i),b}$,
$i=4,8$ are convergent in the nonempty proper subset of the whole parameter space
\begin{eqnarray}
&&\Xi_{_{[1\tilde{3}\tilde{5}\tilde{7}]}}^{4}=\{(y_{_1},\;y_{_2},\;y_{_3},\;y_{_4})
\Big|1<|y_{_1}|,\;|y_{_2}|<|y_{_1}|,\;|y_{_4}|<|y_{_1}|,\;|y_{_3}|<|y_{_2}|\}\;.
\label{GKZ22-15-4}
\end{eqnarray}

\item The 8 hypergeometric functions $\Phi_{_{[\tilde{1}\tilde{3}\tilde{5}\tilde{7}]}}^{(i),a}$,
$i=1,\cdots,8$ are convergent in the nonempty proper subset of the whole parameter space
\begin{eqnarray}
&&\Xi_{_{[\tilde{1}\tilde{3}\tilde{5}\tilde{7}]}}^{1}=\{(y_{_1},\;y_{_2},\;y_{_3},\;y_{_4})
\Big||y_{_1}|<1,\;|y_{_4}|<1,\;|y_{_4}|<|y_{_2}|,\;|y_{_3}|<|y_{_2}|\}\;.
\label{GKZ22-16-1}
\end{eqnarray}
\item The 6 hypergeometric functions $\Phi_{_{[\tilde{1}\tilde{3}\tilde{5}\tilde{7}]}}^{(i),b}$,
$i=1,2,3,5,6,7$ are convergent in the nonempty proper subset of the whole parameter space
\begin{eqnarray}
&&\Xi_{_{[\tilde{1}\tilde{3}\tilde{5}\tilde{7}]}}^{2}=\{(y_{_1},\;y_{_2},\;y_{_3},\;y_{_4})
\Big||y_{_1}|<1,\;|y_{_2}|<1,\;|y_{_3}|<1,\;|y_{_4}|<1\}\;.
\label{GKZ22-16-2}
\end{eqnarray}
\item The 6 hypergeometric functions $\Phi_{_{[\tilde{1}\tilde{3}\tilde{5}\tilde{7}]}}^{(i),c}$,
$i=1,2,3,5,6,7$ are convergent in the nonempty proper subset of the whole parameter space
\begin{eqnarray}
&&\Xi_{_{[\tilde{1}\tilde{3}\tilde{5}\tilde{7}]}}^{3}=\{(y_{_1},\;y_{_2},\;y_{_3},\;y_{_4})
\Big||y_{_1}|<1,\;|y_{_3}|<1,\;|y_{_4}|<1,\;|y_{_3}|<|y_{_2}|\}\;.
\label{GKZ22-16-3}
\end{eqnarray}
\item The 2 hypergeometric functions $\Phi_{_{[\tilde{1}\tilde{3}\tilde{5}\tilde{7}]}}^{(i),b}$,
$i=4,8$ are convergent in the nonempty proper subset of the whole parameter space
\begin{eqnarray}
&&\Xi_{_{[\tilde{1}\tilde{3}\tilde{5}\tilde{7}]}}^{4}=\{(y_{_1},\;y_{_2},\;y_{_3},\;y_{_4})
\Big||y_{_1}|<1,\;|y_{_2}|<1,\;|y_{_4}|<1,\;|y_{_3}|<|y_{_2}|\}\;.
\label{GKZ22-16-4}
\end{eqnarray}
\end{itemize}


\begin{thebibliography}{99}
\bibitem{CEPC-SPPC}CEPC-SPPC study group,
{\it CEPC-SPPC preliminary conceptual design report. 1. Physics and detector}, IHEP-CEPC-DR-2015-01, 2015.
\bibitem{ILC}T.~Behnke et al., {\it The International Linear Collider Technical Design Report - Volume I: Executive Summary},
arXiv:1306.6327 [physics.acc-ph].
\bibitem{HI-LHC}G.~Apollinari, et al., {\it High-Luminosity Large Hadron Collider (HL-LHC):
Preliminary Design Report}, Technical Report CERN-2015-005 (2015).

\bibitem{CMS2012}{\rm CMS}~Collaboration, S.~Chatrchyan et al.,
{\it Observation of a New Boson at a Mass of 125 GeV with the CMS Experiment at the LHC},
published in Phys.~Lett.~B{\bf 716}(2012)30.
\bibitem{ATLAS2012}{\rm  ATLAS}~Collaboration, G.~Aad et al., {\it Observation of a new
particle in the search for the Standard Model Higgs boson with the ATLAS detector at the LHC}
published in Phys.~Lett.~B{\bf716}(2012)1.


\bibitem{Heinrich2021}G.~Heinrich, {\it Collider Physics at the Precision Frontier},
published in Phys.~Rept.~{\bf922}(2021)1-69.

\bibitem{tHooft1979}G.~t'Hooft, M.~Veltman, {\it Scalar One Loop Integrals},
published in Nucl.~Phys.~B{\bf153}(1979)365.

\bibitem{Regge1967}T.~Regge, {\it Algebraic Topology Methods in the Theory of Feynman Relativistic
Amplitudes}, In: {\it Battelle Rencontres - 1967 Lectures in Math. and Phys.,} edited by C.~M.~DeWitt
and J.~A.~Wheeler, 1967, pp.~433-458.
\bibitem{Nasrollahpoursamami2016}E.~Nasrollahpoursamami,
{\it Periods of Feynman Diagrams and GKZ D-modules},
arXiv: 1605.04970 [math-ph].

\bibitem{Gelfand1987}I.~M.~Gel'fand, {\it General theory of hypergeometric functions},
published in Soviet~Math.~Dokl.~{\bf 33}(1986)573.
\bibitem{Gelfand1988}I.~M.~Gel'fand, M.~I.~Graev, and A.~V.~Zelevinsky,
{\it Holonomic systems of equations and series of hypergeometric type},
published in Soviet~Math.~Dokl.~{\bf 36}(1988)5.
\bibitem{Gelfand1988a}I.~M.~Gel'fand, A.~V.~Zelevinsky, and M.~M.~Kapranov,
{\it Hypergeometric functions and toral manifold}, published in Soviet~Math.~Dokl.~{\bf 37}(1988)678.
\bibitem{Gelfand1990}I.~M.~Gel'fand, M.~M.~Kapranov, and A.~V.~Zelevinsky,
{\it Generalized Euler integrals and A-hypergeometric systems},
published in Adv.~in~Math.~{\bf 84}(1990)255.
\bibitem{Gelfand1989}I.~M.~Gelfand, A.~V.~Zelevinskii, and M.~M.~kapranov,
{\it Hypergeometric functions and toric varieties},
published in Functional.~Anal.~Appl.~{\bf23}(1989)94-106.
\bibitem{Kashiwara1976}M.~Kashiwara, T.~Kawai, {\it Holonomic Systems of Linear Differential Equations
and Feynman Integrals}, Publications of the Research Institute for Math. Sci. {\bf 12}(1976)131.
\bibitem{Davydychev1}E.~E.~Boos, and A.~I.~Davydychev,
{\it A method for calculating vertex-type Feynman integrals},
published in Vestn.~Mosk.~Univ.~{\bf28}(1987)8.
\bibitem{Davydychev2000}A.~I.~Davydychev, {\it Explicit results for all orders of
the epsilon expansion of certain massive and massless diagrams}, published in Phys.~Rev.~D{\bf61}(2000)087701.
\bibitem{Davydychev1993NPB}A.~I.~Davydychev, and J.~B.~Tausk,
{\it Two loop selfenergy diagrams with different masses and the momentum expansion},
published in Nucl.~Phys.~B{\bf397}(1993)123.
\bibitem{Davydychev3}E.~E.~Boos, and A.~I.~Davydychev,
{\it A Method of evaluating massive Feynman integrals}, published in Theor.~Math.~Phys.~{\bf89}(1991)1052.
\bibitem{Davydychev1992JPA}A.~I.~Davydychev, {\it Recursive algorithm of evaluating
vertex type Feynman integrals}, published in J.~Phys.~A{\bf25}(1992)5587.
\bibitem{Davydychev2006}A.~I.~Davydychev, {\it Geometrical methods in loop calculations
and the three-point function}, published in Nucl.~Instrum.~Meth.~A{\bf559}(2006)293.
\bibitem{Davydychev1992JMP}A.~I.~Davydychev, {\it General results for massive N point
Feynman diagrams with different masses}, published in J.~Math.~Phys.{\bf33}(1992)358.
\bibitem{Davydychev1991JMP}A.~I.~Davydychev, {\it Some exact results for N point
massive Feynman integrals}, published in J.~Math.~Phys.{\bf32}(1991)1052.
\bibitem{Davydychev1993}N.~I.~Ussyukina, and A.~I.~Davydychev,
{\it An Approach to the evaluation of three and four point ladder diagrams},
published in Phys.~Lett.~B{\bf298}(1993)363.
\bibitem{Tarasov2000}O.~V.~Tarasov, {\it Application and explicit solution of recurrence
relations with respect to space-time dimension}, published in Nucl.~Phys.~B~(Proc.~Suppl){\bf 89}(2000)237.
\bibitem{Tarasov2003}J.~Fleischer, F.~Jegerlehner, and O.~V.~Tarasov,
{\it A New hypergeometric representation of one loop scalar integrals in d dimensions},
published in Nucl.~Phys.~B{\bf 672}(2003)303.
\bibitem{Kalmykov2009}M.~Yu.~Kalmykov, B.~A.~Kniehl,
{\it Towards all-order Laurent expansion of generalized hypergeometric functions
around rational values of parameters}, published in Nucl.~Phys.~B{\bf 809} (2009)365.
\bibitem{Bytev2010}V.~V.~Bytev, M.~Yu.~Kalmykov, B.~A.~Kniehl,
{\it Differential reduction of generalized hypergeometric functions from
Feynman diagrams: One-variable case}, published in Nucl.~Phys.~B{\bf 836} (2010)129.
\bibitem{Kalmykov2011}M.~Yu.~Kalmykov, B.~A.~Kniehl, {\it Counting master integrals:
Integration by parts versus differential reduction}, published in Phys.~Lett.~B{\bf 702}(2011)268.
\bibitem{Bytev2013}V.~V.~Bytev, M.~Yu.~Kalmykov, B.~A.~Kniehl,
{\it HYPERDIRE, HYPERgeometric functions DIfferential REduction:
MATHEMATICA-based packages for differential reduction of generalized
hypergeometric functions $_pF_{p-1}$, $F_1$, $F_2$, $F_3$, $F_4$},
published in Comput.~Phys.~Commun{\bf 184}(2013)2332.
\bibitem{Bytev2015}V.~V.~Bytev, M.~Yu.~Kalmykov,
{\it HYPERDIRE, HYPERgeometric functions DIfferential REduction:
Mathematica-based packages for the differential reduction of generalized
hypergeometric functions: Horn-type hypergeometric functions of two variables},
published in Comput.~Phys.~Commun{\bf 189}(2015)128.
\bibitem{Bytev2016}V.~V.~Bytev, M.~Yu.~Kalmykov, {\it HYPERDIRE,
HYPERgeometric functions DIfferential REduction: Mathematica-based
packages for the differential reduction of generalized hypergeometric
functions: Lauricella function Fc of three variables},
published in Comput.~Phys.~Commun{\bf 206}(2016)78.
\bibitem{Kalmykov2012}M.~Y.~Kalmykov, and B.~A.~Kniehl, {\it Mellin-Barnes representations
of Feynman diagrams, linear systems of differential equations, and polynomial solutions},
published in Phys.~Lett.~B{\bf 714}(2012)103.
\bibitem{Kalmykov2017}M.~Y.~Kalmykov, and B.~A.~Kniehl, {\it Counting the number of
master integrals for sunrise diagrams via the Mellin-Barnes representation},
published in JHEP{\bf 1707}(2017)031.

\bibitem{Davydychev2000a}A.~I.~Davydychev, and M.~Yu.~Kalmykov,
{\it Some remarks on the $\varepsilon$-expandsion of dimensionally regulated Feynman diagrams},
published in Nucl.~Phys.~B~(Proc.~Suppl){\bf 89}(2000)283.
\bibitem{Davydychev2001a}A.~I.~Davydychev, and M.~Yu.~Kalmykov,
{\it New results for the $\varepsilon$-expandsion of certain one-, two-, and
three-loop Feynman diagrams}, published in Nucl.~Phys.~B{\bf 605}(2001)266.
\bibitem{Suzuki2002}A.~T.~Suzuki, E.~S.~Santos, and A.~G.~M.~Schmidt,
{\it Massless and massive one-loop three point functions in negative dimensional approach},
published in  Eur.~Phys.~J.~C{\bf26}(2002)125.
\bibitem{Suzuki2003a}A.~T.~Suzuki, E.~S.~Santos, and A.~G.~M.~Schmidt,
{\it General massive one-loop off-shell three-point functions},
published in  J.~Phys.~A: Math.~Gen.~{\bf36}(2003)4465.
\bibitem{Suzuki2003b}A.~T.~Suzuki, E.~S.~Santos, and A.~G.~M.~Schmidt,
{\it One-loop n-point equivalence among negative-dimensional Mellin-Barnes and
Feynman parametrization approaches to Feynman integrals},
published in  J.~Phys.~A: Math.~Gen.~{\bf36}(2003)11859.
\bibitem{Davydychev2004}A.~I.~Davydychev, and M.~Yu.~Kalmykov,
{\it Massive Feynman diagrams and inverse binomial sums},
published in Nucl.~Phys.~B{\bf 699}(2004)3.
\bibitem{Jegerlehner2004}F.~Jegerlehner, and M.~Yu.~Kalmykov,
{\it The $\cal O(\alpha\alpha_{_s})$ correction to the pole mass of the t-quark
within the standard model}, published in Nucl.~Phys.~B{\bf 676}(2004)365.

\bibitem{Cruz2019}L.~Cruz, {\it Feynman integrals as A-hypergeometric functions},
published in JHEP{\bf1912}(2019)123 (arXiv:1907.00507~[math-ph]).
\bibitem{Klausen2019}R.~Klausen, {\it Hypergeometric Series Representations of Feynman
integrals by GKZ Hypergeometric Systems}, published in JHEP{\bf2004}(2020)121 (arXiv:1910.08651~[hep-th]).
\bibitem{Lee2013}R.~N.~Lee, A.~A.~Pomeransky, {\it Critical points and number
of master integrals}, published in JHEP{\bf 1311}(2013)165.

\bibitem{Oaku1997}T.~Oaku, Adv.~in~Appl.~Math.~{\bf19}(1997)61.
\bibitem{Walther1999}U.~Walther, J.~of~Pure~and~Applied~Algebra.~{\bf139}(1999)303.
\bibitem{Oaku2001}T.~Oaku, N.~Takayama, J.~of~Pure~and~Applied~Algebra.~{\bf156}(2001)267.
\bibitem{Feng2020}T.-F.~Feng, C.-H.~Chang, J.-B.~Chen, and H.-B.~Zhang,
{\it GKZ-hypergeometric systems for Feynman integrals}, published in
Nucl.~Phys.~B{\bf953}(2020)114952.

\bibitem{Feng2018}T.-F.~Feng, C.-H.~Chang, J.-B.~Chen, Z.-H.~Gu, and H.-B.~Zhang,
{\it Evaluating Feynman integrals by the hypergeometry}, published in
Nucl.~Phys.~B{\bf927}(2018)516.
\bibitem{Feng2019}T.-F.~Feng, C.-H.~Chang, J.-B.~Chen, and H.-B.~Zhang,
{\it The system of partial differential equations for the $C_{_0}$ function},
published in Nucl.~Phys.~B{\bf940}(2019)130.

\bibitem{Miller68}W.~Miller~Jr., J.~Math.~Mech.~{\bf17}(1968)1143.
\bibitem{Miller72}W.~Miller~Jr., SIAM.~J.~Math.~Anal.~{\bf3}(1972)31.

\bibitem{Loebbert2020}F.~Loebbert, D.~M\"{u}ller, and H.~M\"{u}nkler,
{\it Yangian Bootstrap for Conformal Feynman Integrals},  published
in Phys.~Rev~.D{\bf101}(2020)066006.
\bibitem{Klemm2020}A.~Klemm, C.~Nega, and R.~Safari,
{\it The l-loop Banana Amplitude from GKZ Systems
and relative Calabi-Yau Periods}, published in JHEP{\bf2004}(2020)088.
\bibitem{Bonisch2021}K.~B\"{o}nisch, F.~Fischbach, A.~Klemm, C.~Nega, and R.~Safari,
{\it Analytic Structure of all Loop Banana Amplitudes}, published in
JHEP{\bf2105}(2021)066.
\bibitem{Reichelt2020}T.~Reichelt, M.~Schulze, C.~Sevenheck, and U.~Walther,
{\it Algebraic aspects of hypergeometric differential equations},
e-Print: 2004.07262[math.AG].
\bibitem{Hidding2021}M.~Hidding, {\it DiffExp, a Mathematica package for computing
Feynman integrals in terms of one-dimensional series expansions}, published in
Comput.~Phys.~Commun.~{\bf269}(2021)108125.
\bibitem{Borinsky2020}M.~Borinsky, {\it Tropical Monte Carlo quadrature for Feynman integrals},
arXiv:~2008.12310[math-ph].
\bibitem{Tellander2021}F.~Tellander, M.~Helmer, {\it Cohen-Macaulay
Property of Feynman Integrals}, arXiv: 2108.01410 [hep-th].
\bibitem{Mizera2021}S.~Mizera, and S.~Telen, {\it Landau Discriminants},
arXiv:~2109.08036[math-ph].
\bibitem{Arkani-Hamed2022}N.~Arkani-Hamed, A.~Hillman, and S.~Mizera,
{\it Feynman Polytopes and the tropical geometry of UV and IR divergence},
published in Phys.~Rev.~D{\bf 105}(2022)125013.
\bibitem{Chestnov2022}V.~Chestnov, F.~Gasparotto, M.~K.~Mandal, P.~Mastrolia,
S.~J.~Matsubara-Heo, H.~J.~Munch, and N.~Takayamac,
{\it Macaulay Matrix for Feynman Integrals: Linear Relations and Intersection Numbers},
arXiv:~2204.12983[hep-th].
\bibitem{Ananthanarayan2022}B.~Ananthanarayan, S.~Bera, S.~Friot, T.~Pathak,
{\it Olsson.wl: a MathematicaMathematica package for the computation of linear
transformations of multivariable hypergeometric functions},
arXiv:~2201.01189[cs.MS].
\bibitem{Berends1994}F.~A.~Berends, M.~B\"{o}hm, M.~Buza, and R.~Scharf, Z.~Phys.~C{\bf63}, 227(1994).
\bibitem{Ananthanarayan2019}B.~Ananthanarayan, S.~Friot, and S.~Ghosh, {\it New series representations
for the two-loop massive sunset diagram}, published in  Eur.~Phys.~J.~C{\bf80}, 606(2020).
\bibitem{Kalmykov2021}M.~Kalmykov, V.~Bytev, B.~Kniehl, S.-O.~Moch, B.~Ward,
and S.~Yost, {\it Hypergeometric Functions and Feynman Diagrams},
arXiv:2012.14492~[hep-th].
\bibitem{Munch2022}H.~Munch, {\it Feynman Integral relations from GKZ-hypergeometric systems},
arXiv: 2207.09780 [hep-th].
\bibitem{Gu2020}Z.-H.~Gu, H.-B.~Zhang, and T.-F.~Feng,
{\it Hypergeometric expression for a three-loop vacuum integral},
published in Int.~J.~Mod.~Phys.~A{\bf35}, 2050089(2020).
\bibitem{Bera2022}S.~Bera, {\it $\epsilon$-Expansion of Multivariables Hypergeometric Functions
Appearing in Feynman Integral Calculas}, arXiv: 2208.01000 [math-ph].
\bibitem{Lairez2022}P.~Lairez, and P. Vanhove, {\it Algorithms for minimal Picard-Fuchs
operators of Feynman integrals}, arXiv: 2209.10962 [hep-th].






\bibitem{Eden1966}R.~J.~Eden, P.~V.~Landshoff, D.~I.~Olive, and J.C.~Polkinghorne,
{\it The analytic S-matrix}, Cambridge Univ. Press, Cambridge 1966.
\bibitem{Feng2022}T.-F.~Feng, H.-B.~Zhang, C.-H.~Chang, {\it Feynman integrals of Grassmannians},
Phys.~Rev.~D{\bf 106}, 116025(2022).



\bibitem{V.A.Smirnov2012}V.~A.~Smirnov, {\it Analytic Tools for Feynman Integrals},
(Springer, Heidelberg 2012), and references therein.


\bibitem{Berends1995}F.~A.~Berends, A.~I.~Davydychev, V.~A.~Smirnov, J.~B.~Tausk,
{\it Zero-threshold expansion of two-loop self-energy diagrams},
Nucl.~Phys.~B{\bf439},536(1995).
\bibitem{Bauberger1995}S.~Bauberger, F.~A.~Berends, M.~B\"ohm, and M.~Buza,
{\it Analytical and numerical methods for massive two-loop self-energy diagrams},
Nucl.~Phys.~B{\bf434}, 383(1995).

\end{thebibliography}
\end{document}